\documentclass[12pt,a4paper,twoside]{book}
\usepackage[T1]{fontenc} % Better font encoding for output (important for accented characters)
\usepackage[utf8]{inputenc} % Allows input of UTF-8 characters
\usepackage{amssymb} % Additional mathematical symbols
\usepackage{mathrsfs} % Script fonts for math mode
\usepackage{amsmath} % Enhanced math environments and features
\usepackage{amsthm} % Theorem and proof environments
\usepackage{dsfont} % Double stroke font (e.g., for indicator function)
\usepackage{enumitem} % Customization of lists (itemize, enumerate)
\usepackage{float} % Improved control over floating objects (figures, tables)
\usepackage{physics} % Physics notation and operators
\usepackage{cancel} % Cancel lines in math expressions
\usepackage[pdftex]{graphicx} % Include graphics/images
\usepackage[top=2.5cm, bottom=2.5cm, left=2.5cm, right=2.5cm]{geometry} % Page layout and margins
\usepackage{mdframed} % Framed boxes (e.g., for theorems, remarks)
\usepackage{fancyhdr} % Custom headers and footers
\usepackage{caption} % Customization of captions
\usepackage{mathtools} % Extensions to amsmath (e.g., \coloneqq)
\usepackage{xcolor} % Colors for text and elements
\usepackage{tikzit} % TikZ-based diagrams (with Tikzit)
\tikzstyle{red dot}=[fill=red, draw=none, shape=circle]
\tikzstyle{green dot}=[fill={rgb,255: red,70; green,176; blue,20}, draw=none, shape=circle]
\tikzstyle{Resize}=[font={\scriptsize}]
\tikzstyle{Big}=[font={\huge}]
\tikzstyle{Photon}=[fill={rgb,255: red,255; green,230; blue,103}, draw=none, shape=circle, minimum size=4pt]
\tikzstyle{Big node}=[fill={rgb,255: red,191; green,191; blue,191}, draw=black, shape=circle]
\tikzstyle{small node}=[fill={rgb,255: red,191; green,191; blue,191}, draw=black, shape=circle, inner sep=0, minimum size=300pt]
\tikzstyle{Blue dot}=[fill=blue, draw=black, shape=circle]
\tikzstyle{Black dot}=[fill=black, draw=none, shape=circle, inner sep=0pt, minimum size=4pt]

\tikzstyle{Fill red}=[-, fill={rgb,255: red,255; green,184; blue,184}]
\tikzstyle{dashes}=[-, dashed, dash pattern=on 1mm off 1mm]
\tikzstyle{Fill grey}=[-, fill={rgb,255: red,166; green,166; blue,166}]
\tikzstyle{Thick}=[-, thick]
\tikzstyle{Arrow}=[->]
\tikzstyle{red line}=[-, fill=none, draw=red]
\tikzstyle{Blue line}=[-, fill=none, draw=blue]
\tikzstyle{Gray line}=[-, fill=none, draw=gray]
\tikzstyle{Thick arrow}=[thick, ->]
\tikzstyle{Fill pink}=[-, fill={rgb,255: red,255; green,140; blue,140}]
\tikzstyle{dashed arrow}=[->, dashed, dash pattern=on 1mm off 0.5mm]
\tikzstyle{arrow}=[->, very thick, draw=red]
\tikzstyle{dashes red}=[-, dashed, dash pattern=on 1mm off 1mm, fill=none, draw=red]
\tikzstyle{dashes blue}=[-, dashed, dash pattern=on 1mm off 1mm, fill=none, draw=blue]
\tikzstyle{Thick dashed arrow}=[->, thick, dashed, dash pattern=on 1mm off 0.5mm]
\tikzstyle{Fill green}=[-, fill={rgb,255: red,70; green,176; blue,20}, draw=none]
\tikzstyle{Fill real red}=[-, fill=red, draw=none]
\tikzstyle{fill grey borderless}=[-, draw=none, fill={rgb,255: red,166; green,166; blue,166}]
\usepackage{quantikz} % Quantum circuit diagrams
\usepackage{etoc} %For smaller custom tables of contents
\usepackage{accents} %For dealing with double superscripts
\usepackage{cite} %Better bibliography management
\usepackage{etoolbox} %For patching commands (used for the footrule)
\usepackage[nottoc]{tocbibind} %Correctly print the bibliography in the TOC
\usepackage{hyperref} % Hyperlinks in PDF
\usepackage{fontawesome5} % For icons 

\newcounter{ShowFront}
\newcounter{ShowChapone}
\newcounter{ShowChaptwo}
\newcounter{ShowChapthree}
\newcounter{ShowChapfour}
\newcounter{ShowChapfive}
\newcounter{Paperversion}
\ifnum \thePaperversion=0
    \hypersetup{colorlinks=true,linkcolor=blue,citecolor=red,linktoc=page} % Hyperref setup: colored links
\else
    \hypersetup{hidelinks}  % Hyperref setup: colored links
\fi

\newcounter{shortversion}
\newcounter{Symbols}
\newcommand{\IconIfSymbolsZero}[1]{%
  \ifnum\theSymbols=0 #1\fi
}

\DeclareRobustCommand{\lit}{%
    \IconIfSymbolsZero{%
    \hyperlink{icon-lit}{\textcolor{black}{\faBook}}~%
  }%
}

\DeclareRobustCommand{\publi}{%
    \IconIfSymbolsZero{%
        \hyperlink{icon-publi}{\textcolor{black}{\faNewspaper[regular]}}~
    }%
}

\DeclareRobustCommand{\newc}{%
    \IconIfSymbolsZero{%  
        \hyperlink{icon-new}{\textcolor{black}{\faLightbulb[regular]}}~
    }%
}

\newcommand{\currentanchor}{none}
\newcommand{\setcurrentanchor}[1]{%
  \renewcommand{\currentanchor}{#1}%
  \vspace{-6cm}
  \hypertarget{#1}{}%
  \vspace{6cm}
}

\newcommand{\navfooter}{%
  \small
  \hyperlink{maintoc}{\textsc{Contents}}
  \,\textbar\,
  \hyperlink{\currentanchor}{\textsc{Chapter}}
}

\fancypagestyle{mainmatterstyle}{
    \fancyhf{}

    \fancyhead[LE]{\nouppercase\leftmark}   % Chapter on Even pages (Left side)
    \fancyhead[RO]{\nouppercase\rightmark}  % Section on Odd pages (Right side)

    \ifnum \thePaperversion = 0
        \fancyfoot[C]{\navfooter} % centered navigation
    \fi
    
    \fancyfoot[LE]{\hspace{5em}\thepage} % even pages
    \fancyfoot[RO]{\thepage\hspace{5em}} % odd pages

}

\fancypagestyle{plain}{ 
    \fancyhf{} % clear header and footer 

    \fancyfoot[LE]{\hspace{5em}\thepage} % even pages 
    \fancyfoot[RO]{\thepage\hspace{5em}} % odd pages 
}

\makeatletter
\renewcommand{\footrule}{%
  \if@twoside
    \ifodd\c@page
      \hbox to\headwidth{%
        \hfill
        \rule[0\baselineskip]{\dimexpr\headwidth-32em\relax}{0.4pt}%
      }
    \else
      \hbox to\headwidth{%
        \rule[0\baselineskip]{\dimexpr\headwidth-32em\relax}{0.4pt}%
        \hfill
      }
    \fi
  \else
    \rule{\headwidth}{0.4pt}
  \fi
}
\makeatother

\fancypagestyle{frontmatterstyle}{
  \fancyhf{}  % clear everything
  
  \fancyfoot[LE]{\hspace{5em}\thepage}
  \fancyfoot[RO]{\thepage\hspace{5em}}

  \ifnum \thePaperversion = 0
        \fancyfoot[C]{\small  \hyperlink{maintoc}{\textsc{Contents}}} % centered navigation
    \fi

}

\fancypagestyle{backmatterstyle}{
    \fancyhf{}

    \fancyhead[LE]{\nouppercase\leftmark}   % Chapter on Even pages (Left side)
    \fancyhead[RO]{\nouppercase\rightmark}  % Section on Odd pages (Right side)
    \ifnum \thePaperversion = 0
        \fancyfoot[C]{\small  \hyperlink{maintoc}{\textsc{Contents}}} % centered navigation
    \fi
    \fancyfoot[LE]{\hspace{5em}\thepage} % even pages
    \fancyfoot[RO]{\thepage\hspace{5em}} % odd pages

}

\title{PhD Manuscript}
\author{Éloi Descamps}
\date{\today}

\newcommand{\N}{\mathbb{N}}
\newcommand{\R}{\mathbb{R}}
\newcommand{\C}{\mathbb{C}}
\newcommand{\Z}{\mathbb{Z}}
\newcommand{\1}{\mathds{1}}
\renewcommand{\P}{\mathbb{P}}

\newcommand{\Cov}{\operatorname{Cov}}
\newcommand{\ie}{\textit{i.e.}}
\newcommand{\vac}{\ket{\text{vac}}}

\newenvironment{derivation}{
    \begin{mdframed}[topline=false,bottomline=false,rightline=false]
        \begin{proof}
    }{
        \end{proof}
    \end{mdframed}
}

\newtheorem{result}{Result}

\newcommand{\resultentry}[2]{%
  \par
  \textbf{Result}~\ref{#1}~--~#2%
  \dotfill
  \hyperref[#1]{\pageref{#1}}%
  \par
}

\begin{document}

\frontmatter
\pagestyle{frontmatterstyle}

%----------------------------------------------------------------------------------------
%	TITLE PAGE
%----------------------------------------------------------------------------------------
\begin{titlepage}
    \begin{center}
        \begin{tabular}{c@{\hskip 7cm}c@{\hskip 1cm}}
            \includegraphics[height=2.5cm]{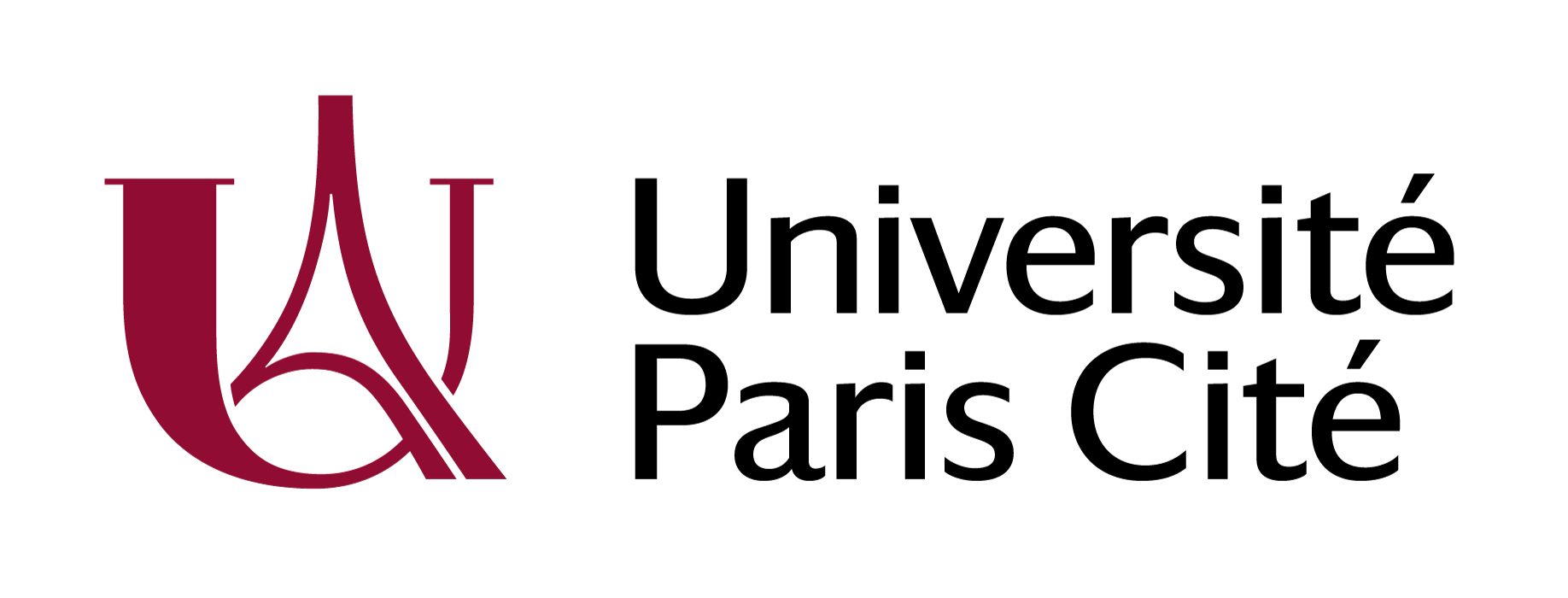} &
        \end{tabular}
    \end{center}

    \begin{center}
    
    \vspace*{.03\textheight}
    \textsc{\LARGE Université Paris Cité}\\[0.2cm]
    \large École doctorale ED 564 -- Physique en Île-de-France\\
    Laboratoire Matériaux et Phénomènes Quantiques\\

        \vfill

        \rule{\textwidth}{0.8pt} \\ 
        \vspace{10pt}
            { \LARGE \bfseries Modes, States, and Symmetries in Quantum Optics for Quantum Information and Metrology}
            \vspace{10pt}
            \rule{\textwidth}{0.8pt} \\ 
    \end{center}
    
    \vfill
    
    \begin{center}
        Par \textsc{\Large Éloi DESCAMPS}\\[1cm] 
        Thèse de doctorat de \textsc{\large Physique}\\[1.2cm]
        Dirigée par \textsc{\large Pérola MILMAN}\\[0.2cm]
        Et par \textsc{\large Arne KELLER}\\[0.2cm] 
        Présentée et soutenue publiquement le 04/06/2026
    \end{center}
    
    \vfill
    Devant un jury composé de : 
    \vspace{-0.9em}
    \begin{center}
        \rule{\textwidth}{0.4pt}		
        \begin{tabular}{l@{\hskip 0.7cm}l@{\hskip 0.7cm}l}
            Lorenzo Maccone, Ass. Prof. & Univ. Pavia & Rapporteur \\
            Tommaso Roscilde, MCF HDR  & ENS Lyon & Rapporteur \\
            Vincenzo D'Ambrosio, Ass. Prof.  & Univ. Napoli Federico II & Examinateur \\
            Giulia Ferrini, Prof.  & Univ. Chalmers & Examinatrice \\
            Sara Ducci, Prof. HDR  & Univ. Paris Cité  & Membre invitée \\
            Pérola Milman, DR HDR & Univ. Paris Cité & Directrice de thèse \\
            Arne Keller, Prof. HDR  & Univ. Paris-Saclay & Codirecteur de thèse \\
        \end{tabular}
        \rule{\textwidth}{0.4pt}		
    \end{center}

    %\vspace{-0.5cm}
    \begin{center}
        \includegraphics[width=3cm]{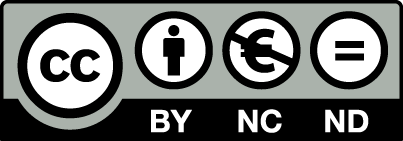}
    \end{center}
    \newpage
    \thispagestyle{empty}
\end{titlepage}

\ifnum \theShowFront = 1
\cleardoublepage
\addcontentsline{toc}{chapter}{Preliminary content}

\vspace*{1em}

%--------------------------------------------------------
% Short abstracts
%--------------------------------------------------------
\section*{Résumé}
\addcontentsline{toc}{section}{Résumé court}
    \noindent\textbf{Titre de la thèse :}
    Modes, états et symétries en optique quantique pour l'information quantique et la métrologie

    \vspace{0.5cm}

    Cette thèse explore le rôle des modes, états et symétries en optique quantique, dans le contexte de l'information quantique et de la métrologie. Elle propose un cadre unifié pour analyser comment la structure modale des champs photoniques, la nature statistique des états et leurs propriétés de symétrie déterminent les ressources physiques exploitables pour le traitement de l'information et l'estimation quantiques. On développe une description des degrés de liberté temps-fréquence en tant que variables quantiques continues, mettant en évidence leur richesse pour l'encodage et la manipulation de l'information. Un deuxième axe étudie l'intrication et les variables collectives, en clarifiant le lien entre ressources physiques et gains métrologiques, notamment dans l'atteinte des limites ultimes de précision. On analyse ensuite des situations interférométriques de type Hong-Ou-Mandel, développant un formalisme général, axé sur la notion de symétrie, permettant d'analyser une large gamme de situations et menant à diverses généralisations. Enfin, on étudie les symétries imposées par les règles de super-sélection optiques et leurs conséquences sur la structure des états et leurs performances opérationnelles afin de dégager une meilleure compréhension des fondements de l'optique quantique.

    \vspace{0.5cm}
    \noindent\textbf{Mots-clés :}
    Optique quantique, Métrologie, Interférométrie, Information quantique, Règles de super-sélection, Variables temps-fréquence

\clearpage
\section*{Abstract}
\addcontentsline{toc}{section}{Short abstract}
    \noindent\textbf{Thesis title:}
    Modes, States, and Symmetries in Quantum Optics for Quantum Information and Metrology

    \vspace{0.5cm}

    This thesis explores the role of modes, states, and symmetries in quantum optics, within the context of quantum information and quantum metrology. It proposes a unified framework to analyze how the modal structure of photonic fields, the statistical nature of states, and their symmetry properties determine the physical resources that can be exploited for quantum information processing and quantum parameter estimation. A first line of investigation develops a description of time-frequency degrees of freedom as continuous quantum variables, highlighting their richness for encoding and manipulating information. A second axis studies entanglement and collective variables, clarifying the link between physical resources and metrological gains, particularly in reaching ultimate precision limits. Interferometric scenarios of the Hong-Ou-Mandel type are then analyzed, and a general formalism centered on the notion of symmetry is developed. This framework enables the analysis of a broad range of situations and leads to several generalizations. Finally, the thesis examines the symmetries imposed by optical superselection rules and their consequences for the structure of quantum states and their operational performance, with the aim of providing a deeper understanding of the foundations of quantum optics.

    \vspace{0.5cm}
    \noindent\textbf{Keywords:}
    Quantum optics, Metrology, Interferometry, Quantum information, Superselection rules, Time-frequency variables

\ifnum \theshortversion=0    

\clearpage

\section*{How to use this thesis}
\addcontentsline{toc}{section}{How to use this thesis}
This thesis is structured to accommodate both readers seeking a gentle introduction and those primarily interested in the original research contributions.

\medbreak

Chapter \ref{chap: Formalism and framework} serves mainly as a recap of standard concepts in quantum information, quantum optics, and quantum metrology. It establishes notation and conventions used throughout the manuscript. Readers already familiar with these topics may choose to skim or skip this chapter, referring back to it only when clarification is needed.

\medbreak

Chapter \ref{chap: Time-Frequency Systems} introduces the general framework for time-frequency quantum optical systems. While the formalism itself is not fundamentally new, a clear understanding of it is essential, as it underpins much of the analysis developed in the remainder of the thesis.

\medbreak

Chapters \ref{chap: collective entanglement}, \ref{chap: HOM interferometry and metrology}, and \ref{chap: SSR} form the core of the thesis. These chapters present and unify the main research contributions, including the results from the articles incorporated into this manuscript. In addition, several smaller, previously unpublished results are sprinkled throughout these chapters, providing complementary insights and extensions.

\medbreak

The appendices collect technical material, including detailed derivations, proofs, and supplementary results. They are intended to support the main text by providing rigorous justification of the computations and statements, while allowing the core chapters to remain focused on conceptual development and key results.

\medbreak

Depending on their background and interests, readers may therefore choose to focus directly on Chapters 3--5, using Chapters 1--2 and the appendices as reference material when needed.

\ifnum \theSymbols =0
\medbreak

Much care has been taken to write this thesis in a pedagogical manner. As such, it combines the presentation of established facts, results published during the course of the PhD, and small original contributions. To help the reader distinguish between these different types of content, each paragraph is preceded by an icon with the following meaning:
\begin{itemize}
    \item \hypertarget{icon-lit}{\faBook} -- The content of the paragraph is well established and can be found in standard textbooks or review articles. This also includes such standard material reformulated in unusual contexts.
    
    \item \hypertarget{icon-publi}{\faNewspaper[regular]} -- The content of the paragraph corresponds to results published during the PhD thesis~\cite{descamps_quantum_2023, descamps_time-frequency_2023, descamps_gottesman-kitaev-preskill_2024, meskine_approaching_2024, descamps_superselection_2024, descamps_measuring_2025, descamps_role_2026, saharyan_resources_2026, descamps_unified_2026}.
    
    \item \hypertarget{icon-new}{\faLightbulb[regular]} -- The content of the paragraph corresponds to small original contributions that have not been published in the articles included in this thesis, but which complement them.
\end{itemize}
\fi
\fi

\clearpage
\section*{Résumé substantiel}
\addcontentsline{toc}{section}{Résumé substantiel}

\paragraph{Introduction}
La mécanique quantique, initialement développée pour décrire le comportement de la matière à l'échelle microscopique, a progressivement dépassé le cadre de la physique fondamentale pour devenir le socle de nombreuses avancées technologiques. Les premières applications issues de cette théorie (souvent regroupées sous le terme de première révolution quantique) ont conduit à des dispositifs aujourd'hui omniprésents, tels que les transistors, les lasers, l'imagerie par résonance magnétique ou encore les horloges atomiques. Bien que reposant sur des principes quantiques, et parfois sur des phénomènes de cohérence et d'interférence liés à la superposition quantique, ces technologies ne nécessitent généralement pas le contrôle individuel et la manipulation précise d'états quantiques complexes, notamment intriqués.

La situation a profondément évolué avec l'émergence de la seconde révolution quantique, qui vise cette fois à préparer, manipuler et mesurer de manière contrôlée des états quantiques individuels ou fortement corrélés, en exploitant explicitement des ressources telles que la cohérence quantique et l'intrication. Dans ce contexte, les systèmes quantiques sont envisagés comme des ressources permettant d'améliorer ou de renouveler des tâches fondamentales comme le calcul, la simulation, la communication ou encore la mesure de précision.

Parmi les différentes plateformes physiques envisageables, la lumière occupe une place privilégiée. Dans le régime quantique, elle est décrite dans le cadre de l'optique quantique, où les états du champ électromagnétique présentent des propriétés remarquables, tant du point de vue fondamental que pour les applications. La lumière constitue en particulier un support naturel pour la transmission d'information, du fait de sa propagation rapide et de la relative facilité avec laquelle ses degrés de liberté peuvent être contrôlés expérimentalement. Les systèmes photoniques présentent en outre des avantages pratiques importants, tels qu'un faible couplage à l'environnement, une compatibilité avec des techniques optiques bien établies, et une grande flexibilité dans la manipulation des degrés de liberté spatiaux, temporels ou fréquentiels.

La description quantique de la lumière peut cependant être abordée selon différents cadres conceptuels, correspondant à des régimes physiques et des formalismes mathématiques distincts. D'une part, la nature corpusculaire de la lumière conduit à une description en termes de photons, dans laquelle l'information quantique est encodée dans des systèmes de dimension finie, comme le nombre de photons ou leur polarisation. D'autre part, sa nature ondulatoire motive une approche complémentaire, où le champ électromagnétique est caractérisé par des variables continues, telles que les quadratures du champ. Ces deux points de vue, souvent respectivement désignés comme les paradigmes à variables discrètes et à variables continues, possèdent chacun leurs avantages et leurs limitations. Ils sont généralement mobilisés dans des contextes différents, mais peuvent aussi conduire à des descriptions apparemment disjointes d'un même système physique.

Dans ce travail, une attention particulière est portée à la mise en relation de ces différentes descriptions. Plutôt que de les considérer comme fondamentalement distinctes, l'objectif est de dégager des principes communs permettant de comprendre de manière unifiée les propriétés de la lumière quantique et leur rôle dans les protocoles d'information quantique.

Cette thèse s'articule autour de trois notions centrales : les modes, les états et les symétries. Ceux-ci offrent des perspectives complémentaires sur la structure et le comportement de la lumière quantique. Les modes décrivent la manière dont le champ électromagnétique est décomposé selon ses différents degrés de liberté, qu'ils soient spatiaux, temporels ou spectraux. Ils constituent le cadre naturel dans lequel sont identifiés les supports physiques de l'information et jouent un rôle déterminant dans la description des systèmes multimodes, où des corrélations peuvent apparaître entre différentes composantes du champ.

Les états quantiques, quant à eux, encapsulent l'ensemble des propriétés statistiques et de cohérence des photons distribués dans ces modes. Ils déterminent la manière dont les excitations du champ sont réparties et corrélées, et peuvent être adaptés en fonction des tâches envisagées. Dans ce contexte, une question centrale consiste à identifier quelles propriétés des états quantiques constituent des ressources exploitables, en particulier lorsque plusieurs modes sont impliqués et que les contributions modales et statistiques s'entremêlent.

Enfin, les symétries jouent un rôle fondamental en imposant des contraintes générales sur les systèmes physiques. Dans le cas de la lumière, la nature bosonique des photons implique une symétrie intrinsèque sous échange de particules, qui a des conséquences directes sur les phénomènes d'interférence et sur la structure des états accessibles. Ces considérations permettent souvent de dégager des résultats généraux indépendamment des détails dynamiques des systèmes étudiés.

L'un des fils directeurs de cette thèse est d'analyser l'interaction entre ces trois aspects afin de mieux comprendre les propriétés de la lumière quantique et les ressources qu'elle offre. Cette approche conduit à revisiter différents phénomènes et cadres théoriques sous un angle unificateur, en mettant en évidence les mécanismes fondamentaux qui sous-tendent leur comportement.

Une attention particulière est accordée aux applications en métrologie quantique, c'est-à-dire à l'étude des stratégies de mesure dans le régime quantique. L'objectif de ce domaine est de déterminer dans quelle mesure les propriétés quantiques des systèmes physiques peuvent être exploitées pour améliorer la précision des mesures au-delà des limites classiques. La lumière constitue un outil privilégié dans ce contexte, en raison de son rôle central dans de nombreuses techniques de mesure et de la grande contrôlabilité de ses états quantiques.

Dans ce cadre, la notion de ressource joue un rôle essentiel. Selon les situations, différentes quantités physiques peuvent être considérées comme des ressources, telles que le nombre de photons, le nombre de modes, ou encore l'énergie totale du champ. Ces différentes définitions, souvent associées respectivement aux approches à variables discrètes et continues, reflètent en réalité la diversité des mécanismes permettant d'améliorer la précision des mesures. Elles mettent en évidence le fait que les propriétés modales et les propriétés statistiques de la lumière peuvent toutes deux contribuer aux performances des protocoles de métrologie.

L'analyse développée dans cette thèse vise ainsi à clarifier ces différentes contributions et à proposer un cadre conceptuel permettant de les comparer et de les exploiter de manière cohérente. En mettant en lumière le rôle conjoint des modes, des états et des symétries, ce travail cherche à fournir une compréhension plus profonde des ressources offertes par la lumière quantique, tant du point de vue fondamental que pour leurs applications en information et en métrologie quantiques.

\bigskip

\paragraph{Chapitre 1 --- Cadre formel et outils conceptuels}

Le premier chapitre a pour objectif d'introduire les bases théoriques nécessaires à l'ensemble des développements présentés dans cette thèse. Il ne constitue pas un apport original en soi, mais établit le cadre conceptuel commun aux différents chapitres, en précisant les notations, les conventions et les outils qui seront utilisés par la suite. Il s'inscrit à l'interface de trois domaines étroitement liés : la théorie de l'information quantique, la métrologie quantique et l'optique quantique, qui forment les trois piliers du travail présenté.

Dans un premier temps, les fondements de la mécanique quantique sont rappelés à travers le formalisme standard des états et des observables. Cette introduction permet de fixer les idées et de définir les objets mathématiques essentiels à la description des systèmes quantiques. Sur cette base, les concepts centraux de la théorie de l'information quantique sont introduits. Une attention particulière est portée aux systèmes paradigmatiques que sont les qubits, correspondant à des systèmes de dimension finie, ainsi qu'aux systèmes à variables continues, qui jouent un rôle central en optique quantique. Les notions élémentaires de traitement de l'information quantique sont également abordées, notamment les principes du calcul quantique et les idées générales de correction d'erreurs, afin de situer les développements ultérieurs dans un cadre plus large.

Le chapitre se poursuit par une introduction détaillée à la métrologie, puis à la métrologie quantique. L'objectif est de formaliser le problème général de l'estimation de paramètres physiques à partir de données expérimentales. Les notions d'estimateur, de biais et de variance sont introduites, conduisant à la borne de Cramér-Rao, qui fixe une limite fondamentale à la précision des estimations. Ce cadre est ensuite étendu au régime quantique, où les propriétés des états et des mesures doivent être prises en compte explicitement. Les concepts d'information de Fisher et d'information de Fisher quantique sont présentés comme des outils centraux pour quantifier la précision atteignable. Cette partie met ainsi en évidence le lien profond entre les propriétés statistiques des systèmes quantiques et leurs performances en métrologie.

Enfin, les éléments fondamentaux de l'optique quantique sont introduits. Partant de la description classique du champ électromagnétique, le chapitre présente la décomposition en modes, qui constitue une étape essentielle pour comprendre la structure des systèmes optiques. Cette description conduit naturellement à modéliser chaque mode comme un oscillateur harmonique indépendant, fournissant ainsi la base de la quantification du champ. Le formalisme de l'espace de Fock est ensuite introduit pour décrire les états à nombre de photons bien défini, ainsi que les transformations de modes, qui jouent un rôle central dans de nombreux dispositifs optiques. Cette présentation permet de relier de manière explicite les degrés de liberté physiques du champ aux structures mathématiques utilisées dans le reste de la thèse.

Le chapitre se conclut par une discussion de certaines questions conceptuelles, en particulier celles liées à la définition de la phase optique globale et aux règles de super-sélection associées. Ces considérations, bien que parfois techniques, annoncent des problématiques plus profondes qui seront abordées dans les chapitres ultérieurs, notamment en ce qui concerne la définition des ressources en information quantique bosonique.

Dans son ensemble, ce premier chapitre fournit ainsi un socle unifié pour les développements qui suivent, en mettant en place les outils nécessaires à l'analyse des systèmes quantiques de lumière du point de vue de l'information et de la métrologie.

\bigskip

\paragraph{Chapitre 2 --- Optique quantique temps-fréquence}

Afin d'explorer plus concrètement l'articulation entre propriétés modales et propriétés statistiques de la lumière quantique, le deuxième chapitre introduit un cadre d'étude central pour l'ensemble de cette thèse : celui de l'optique quantique dans le domaine temps-fréquence. Ce formalisme fournit une description complète des degrés de liberté temporels et spectraux du champ électromagnétique, ainsi que des photons qui l'occupent. Il constitue le terrain principal sur lequel seront développés et illustrés de nombreux résultats dans les chapitres suivants.

Les variables de temps et de fréquence apparaissent naturellement comme des degrés de liberté fondamentaux des champs optiques. Si, dans de nombreuses situations, une description monochromatique de la lumière est suffisante, il devient indispensable, dans de nombreux contextes, de prendre en compte la structure fine des états optiques dans ces variables. En particulier, dans des domaines tels que la communication quantique, le calcul quantique optique ou encore l'interférométrie, l'information peut être encodée directement dans des modes temporels ou spectraux, et manipulée via les corrélations qui existent entre eux. De plus, les performances métrologiques associées à l'estimation de paramètres temporels sont intrinsèquement liées à la distribution temps-fréquence de l'état utilisé comme sonde, ce qui rend ce cadre particulièrement pertinent pour l'étude des protocoles de mesure quantique.

D'un point de vue théorique, les variables de temps et de fréquence présentent une structure analogue à celle de variables conjuguées en mécanique quantique, à l'image de la position et de l'impulsion. Cette analogie devient particulièrement féconde dans le régime à un photon, qui constitue le cas principalement étudié dans ce chapitre. Dans ce contexte, les degrés de liberté temps-fréquence d'un photon peuvent être traités de manière proche des variables de quadrature en optique à variables continues. Cette correspondance permet d'introduire des outils issus de l'analyse en espace de phase, tels que les représentations de Wigner, qui offrent une description intuitive et puissante des états quantiques, de leurs transformations et de leurs propriétés.

Le chapitre commence par poser les définitions fondamentales des modes temporels et fréquentiels en optique quantique. Les opérateurs et les états associés sont introduits, ainsi que leur lien avec des situations expérimentales concrètes, notamment en ce qui concerne la génération et la manipulation des états. Cette première étape permet d'établir un langage commun pour décrire les systèmes temps-fréquence.

Une attention particulière est ensuite portée à la diversité des états quantiques pouvant être définis dans ce cadre. Différentes classes d'états à un photon sont présentées et comparées, en mettant en évidence leurs propriétés caractéristiques. Cette « zoologie » d'états permet de dégager des analogies fortes avec les états gaussiens et non gaussiens de l'optique à variables continues, tout en soulignant les différences conceptuelles qui subsistent entre ces deux cadres.

Le chapitre aborde également les transformations et évolutions utiles dans l'espace temps-fréquence. Parmi celles-ci figurent les translations, les effets de dispersion (pouvant être interprétés comme des opérations de cisaillement) ainsi que des transformations analogues à des rotations dans l'espace de phase. Ces opérations jouent un rôle essentiel dans la manipulation des états et dans la compréhension des dispositifs expérimentaux, en particulier dans les protocoles interférométriques.

Enfin, le lien entre les systèmes temps-fréquence et la métrologie quantique est discuté. Le formalisme introduit dans ce chapitre permet en effet d'analyser de manière fine les propriétés des états utilisés comme sondes, ainsi que leur sensibilité à différents paramètres. Il fournit également des outils conceptuels et calculatoires, notamment via les représentations en espace de phase, qui seront mobilisés dans les chapitres suivants pour étudier les performances métrologiques de différents protocoles.

Dans son ensemble, ce chapitre établit ainsi un cadre formel riche et flexible, qui permet de relier de manière explicite les degrés de liberté physiques de la lumière à des outils issus de l'information quantique et de la métrologie, et qui servira de base aux analyses développées dans la suite de la thèse.

\bigskip

\paragraph{Chapitre 3 --- Variables collectives et corrélations multimodes}

Le troisième chapitre est consacré à l'étude des variables collectives et des corrélations multimodes dans les systèmes optiques quantiques. Lorsqu'un système est composé de plusieurs sous-systèmes, comme c'est naturellement le cas pour différents modes du champ électromagnétique, les quantités physiques pertinentes ne sont plus seulement locales, mais peuvent dépendre de manière conjointe de l'ensemble des composantes. Ces observables collectives jouent un rôle central en physique, et en particulier en optique quantique, où elles permettent de mettre en évidence des propriétés globales du système.

Dans ce contexte, des propriétés dites non locales peuvent émerger, c'est-à-dire des propriétés qui ne peuvent être attribuées à aucun sous-système pris isolément, mais qui résultent des corrélations entre eux. Parmi celles-ci, les corrélations quantiques, et en particulier l'intrication, constituent des ressources essentielles pour de nombreuses applications en information et en métrologie quantiques. Si le rôle de l'intrication multipartite est bien établi dans des systèmes discrets, comme les ensembles d'atomes, son rôle dans les systèmes optiques multimodes reste plus subtil, en particulier lorsqu'il s'agit de degrés de liberté continus tels que le temps et la fréquence.

Ce chapitre s'attache précisément à clarifier ce point, en étudiant le rôle des corrélations temps-fréquence dans des protocoles de métrologie quantique. Alors que ces corrélations sont souvent considérées comme des propriétés techniques ou classiques du champ optique, l'analyse développée ici montre qu'elles peuvent, dans certaines situations, constituer de véritables ressources quantiques permettant d'améliorer la précision des mesures. En particulier, il est mis en évidence que l'intrication fréquentielle peut conduire à des gains collectifs dans des problèmes d'estimation de paramètres temporels.

À partir de ces observations, un cadre théorique général est introduit pour décrire l'intrication associée à des variables collectives. Cette approche permet de dépasser la distinction traditionnelle entre systèmes à variables discrètes et continues, en proposant une description unifiée des corrélations multimodes. Dans ce cadre, des inégalités générales portant sur les observables collectives sont établies, permettant de caractériser les limites fondamentales des protocoles de mesure. Ces résultats peuvent être interprétés comme des généralisations des bornes usuelles de la métrologie quantique, et mettent en évidence le rôle structurant des corrélations entre modes.

Le chapitre introduit également des outils pour quantifier l'intrication dans ce contexte, en proposant des mesures adaptées aux variables collectives. Une notion d'intrication partielle, appelée $k$-intrication, est notamment discutée afin de caractériser plus finement la structure des corrélations présentes dans les systèmes multimodes. Ces développements permettent de relier de manière explicite la structure des corrélations à la précision atteignable dans des protocoles métrologiques.

Enfin, les applications de ce formalisme sont étendues au-delà de la métrologie, en particulier dans le domaine de l'encodage de l'information quantique. Il est montré que les variables collectives peuvent servir de support à des schémas d'encodage robustes, notamment à travers des généralisations des codes de Gottesman-Kitaev-Preskill (GKP) adaptées aux degrés de liberté temps-fréquence des photons uniques. Cette approche ouvre la voie à des stratégies de correction d'erreurs exploitant directement la structure multimode des systèmes optiques.

Dans son ensemble, ce chapitre propose ainsi une nouvelle perspective sur les corrélations multimodes en optique quantique, en mettant en évidence leur rôle en tant que ressources physiques, et en fournissant un cadre unifié pour leur analyse et leur exploitation.

\bigskip

\paragraph{Chapitre 4 --- Interférométrie de Hong-Ou-Mandel et métrologie}

Le quatrième chapitre est consacré à l'étude de l'interférométrie linéaire en optique quantique et à ses applications en métrologie. Les interféromètres constituent des outils fondamentaux pour manipuler et sonder les états de la lumière, en exploitant la nature ondulatoire du champ électromagnétique. En redistribuant les photons entre différents modes de sortie par interférence, ils permettent de mettre en évidence des effets extrêmement sensibles à de faibles variations de paramètres physiques, ce qui en fait des dispositifs privilégiés pour les mesures de haute précision.

Parmi les phénomènes d'interférence quantique, l'effet de Hong-Ou-Mandel occupe une place centrale. Il met en jeu l'interférence de deux photons arrivant par des ports d'entrée distincts sur un séparateur de faisceau équilibré et se manifeste par une suppression caractéristique des coïncidences en sortie lorsque les photons sont indiscernables. Ce phénomène est généralement interprété comme une conséquence directe de l'indiscernabilité modale des photons, combinée à leur nature bosonique. Il constitue ainsi un outil essentiel pour sonder les propriétés des états photoniques et intervient comme brique élémentaire dans de nombreux protocoles d'information quantique.

Dans ce chapitre, une analyse approfondie de l'interféromètre de Hong-Ou-Mandel est développée, en mettant en évidence ses propriétés fondamentales ainsi que ses applications en métrologie. Au-delà de l'interprétation standard en termes d'indiscernabilité, une perspective plus générale est introduite, fondée sur les propriétés de symétrie des états d'entrée par rapport aux modes de l'interféromètre. Cette approche permet de reformuler l'effet de Hong-Ou-Mandel comme une manifestation de symétries du système multimode, établissant un lien direct entre la structure des états et les motifs d'interférence observés.

Cette reformulation conceptuelle conduit à une meilleure compréhension des capacités métrologiques des protocoles basés sur l'interférence de Hong-Ou-Mandel. En particulier, elle permet d'analyser de manière systématique la précision atteignable dans des tâches d'estimation de paramètres, en reliant explicitement la visibilité des interférences, la structure modale des états et les limites fondamentales de précision. L'étude est menée en détail dans le cadre des états temps-fréquence, ce qui permet de dépasser les hypothèses usuelles d'états indépendants et d'explorer des situations plus générales où des corrélations sont présentes.

Le chapitre examine également l'impact de conditions expérimentales réalistes, notamment la présence d'une visibilité imparfaite, sur les performances métrologiques. Cette analyse permet de quantifier de manière précise la dégradation des performances et d'identifier les conditions dans lesquelles des gains quantiques peuvent néanmoins être préservés.

Au-delà du cas à deux photons, les résultats sont généralisés à des régimes multiphotoniques. Dans ce contexte, le rôle des symétries devient encore plus structurant, en particulier à travers des liens entre symétrie spatiale et propriétés de parité. Ces considérations permettent de dériver des expressions générales décrivant les résultats interférométriques ainsi que les précisions associées. Des configurations plus générales, telles que l'interféromètre de Mach-Zehnder, sont également discutées dans ce cadre unifié.

Enfin, le formalisme est étendu à des interféromètres impliquant un nombre arbitraire de modes. Cette généralisation conduit à une famille d'interféromètres multimodes qui peuvent être vus comme des extensions naturelles de l'interféromètre de Hong-Ou-Mandel. Là encore, la notion de symétrie joue un rôle central, permettant de dériver des bornes générales sur la précision des estimations et d'identifier les structures d'états optimales. L'extension à des états mixtes et l'intégration explicite des degrés de liberté temporels complètent ce cadre général.

Dans son ensemble, ce chapitre met en lumière le rôle fondamental des symétries et de la structure modale dans les phénomènes d'interférence quantique. Il montre comment ces éléments peuvent être exploités de manière systématique pour concevoir et analyser des protocoles interférométriques performants, en particulier dans une perspective de métrologie quantique.

\bigskip

\paragraph{Chapitre 5 --- Règles de super-sélection et systèmes optiques}

Le cinquième et dernier chapitre aborde la question des règles de super-sélection (SSR pour {\it superselection rules}) et de leur rôle dans la compréhension des systèmes bosoniques, en particulier en optique quantique. Alors que les chapitres précédents ont clarifié l'importance des propriétés modales et statistiques de la lumière, certaines questions fondamentales restaient ouvertes, notamment le statut de l'intrication modale en tant que ressource et la relation entre descriptions à variables discrètes et à variables continues.

Pour traiter ces problématiques, un cadre unifiant fondé sur les règles de super-sélection est introduit. Les SSR découlent de symétries fondamentales des systèmes quantiques et imposent des contraintes sur l'ensemble des états et opérations accessibles. Dans les systèmes optiques, la symétrie pertinente est celle des transformations de phase globale du champ électromagnétique. En l'absence d'une référence de phase partagée, cette symétrie se traduit par une règle de super-sélection excluant les cohérences observables entre secteurs de nombres totaux de photons différents, tout en autorisant les mélanges statistiques de ces secteurs. L'adoption explicite de ce cadre permet non seulement de rendre opérationnel le rôle des références de phase dans les expériences, mais aussi de relier de manière naturelle les descriptions discrètes et continues de l'optique quantique.

Le chapitre commence par présenter le formalisme mathématique des systèmes compatibles avec les SSR, en s'appuyant sur des situations bimodales paradigmiques et la représentation de Schwinger. Cette approche permet de transposer des outils issus de la physique des solides, tels que les états cohérents de spin, les transformations de Möbius, les polynômes de Majorana et les fonctions de Wigner sphériques, à l'étude des systèmes optiques. Les notions de transformation de modes, de rotations et de compression de spin y sont également analysées, fournissant une base solide pour les développements ultérieurs.

Une discussion est ensuite consacrée aux systèmes à variables continues considérés comme des limites de systèmes soumis aux SSR. Cette construction fournit une justification formelle de la pratique courante consistant à traiter les systèmes à variables continues comme des approximations idéalisées de systèmes à dimension finie et renforce l'importance du cadre SSR pour comprendre les propriétés physiques des systèmes optiques.

Le chapitre examine ensuite les implications des SSR pour l'informatique quantique bosonique. En adoptant le formalisme SSR, il est possible de clarifier comment les contraintes physiques sur les états et opérations bosoniques définissent les ressources computationnelles, et comment des notions telles que la gaussianité, la non-classicalité et l'universalité dépendent de la représentation choisie. Ce cadre permet de comprendre de manière unifiée les encodages à variables discrètes et continues, et montre que les systèmes continus peuvent être vus comme des limites particulières de systèmes à nombre de particules fixe. Ces résultats offrent une caractérisation physiquement fondée des ressources nécessaires à la réalisation d'un calcul quantique universel avec des plateformes bosoniques.

Enfin, le chapitre explore les applications métrologiques de ce cadre. L'adoption du point de vue SSR clarifie la notion de ressource en métrologie quantique, en identifiant explicitement le rôle de l'intrication modale et de l'intrication entre particules. Elle permet également de décrire de manière unifiée des régimes bosoniques apparemment incompatibles, fournissant un langage cohérent pour l'analyse de la précision quantique et des performances des protocoles de mesure. Ainsi, le cadre des SSR offre non seulement une compréhension conceptuelle plus profonde des systèmes optiques, mais fournit également des outils pratiques pour l'étude des applications computationnelles et métrologiques des systèmes bosoniques.

Dans son ensemble, ce chapitre fournit une perspective unifiée sur l'information quantique bosonique, reliant symétrie, intrication, ressource et métrologie, et complète la thèse en offrant un cadre conceptuel qui relie et généralise l'ensemble des développements présentés dans les chapitres précédents.

\clearpage

\section*{Remerciements}
\addcontentsline{toc}{section}{Remerciements}

Avant toute chose, je voudrais très chaleureusement remercier mes directeurs de thèse qui m'ont accueilli dans leur équipe de recherche et m'ont permis de découvrir les subtilités de l'optique quantique. Votre encadrement présent, attentif et d'une grande gentillesse ainsi que votre participation active et motrice dans le processus de recherche sont d'une grande richesse. Tout d'abord, merci à Pérola Milman, ma directrice, qui m'a tant transmis. Même si les discussions n'ont pas toujours été de tout repos, on a réussi à grandement progresser, ce qui a mené à cette prolifique production scientifique. Il ne faut jamais oublier que même si ce que tu dis n'est pas toujours très clair, on finit toujours par comprendre que tu avais raison depuis le début. Merci chaudement à Arne Keller qui a codirigé ma thèse. Ton caractère posé et ta rigueur complètent parfaitement Pérola, et j'ai beaucoup appris à tes côtés.

\vspace{0.1em}

I also want to deeply thank all the jury members, Vincenzo D'Ambrosio, Giulia Ferrini and especially the referees Lorenzo Maccone and Tommaso Roscilde who had the patience to read this thick manuscript, provided interesting feedback and participated in the defense.

\vspace{0.1em}

Merci aux autres permanents de l'équipe QITe, surtout Sara Ducci et Florent Baboux avec qui on a eu de très nombreuses et passionnantes discussions entre théorie et expériences. Une telle proximité avec la recherche expérimentale est extrêmement enrichissante, et j'espère qu'elle continuera à être cultivée au sein de l'équipe. Merci pour le temps et la confiance que vous m'avez accordés en prenant soin d'écouter nos délires théoriques. Merci également aux autres permanents de l'équipe, Maria Amanti, Luca Guidoni, Valentin Cambier, Jean-Pierre Likforman, Marco Ravaro et Albane Douillet qui participent à créer une ambiance de travail agréable et stimulante.

\vspace{0.1em}

Merci aux autres chercheurs qui m'ont accompagné dans le monde de la recherche. Tout d'abord Nicolas Fabre, mon prédécesseur, qui m'a laissé un terreau extrêmement fertile pour débuter mon travail de thèse. Je suis toujours heureux de discuter avec toi, et surtout d'entendre tes critiques éclairantes sur la communauté. Thanks to Borivoje Daki\'c and Damian Markham who welcomed me in their research teams for internships and were the first to shape my understanding of research and quantum information. Et enfin merci à Jean-Pierre Delville, Thomas Salez et Raphaël Saiseau qui ont été les premiers à m'accueillir dans une équipe de recherche.

\vspace{0.1em}

Merci également à tous les jeunes chercheurs de l'équipe QITe. Tout d'abord un grand merci à Astghik qui nous a accompagnés dans tout ce travail théorique pendant une grande partie de ma thèse. Je te souhaite un excellent post-doc à Liège. Merci aux stagiaires accueillis par Pérola et Arne, notamment Nicolas et Adrien, qui ont temporairement partagé mes sujets de recherche. Merci également aux expérimentateurs avec qui j'ai eu le plaisir de converser et de réfléchir. Othmane d'abord, dont la collaboration a mené à un des articles présentés dans cette thèse. Lorenzo, ensuite, avec qui j'ai eu plein de discussions éclairantes sur les subtilités de la manipulation et de l'interférence des photons. Italo qui a eu la patience d'écouter nos explications théoriques. Et enfin Till, dont je me demande parfois si Pérola n'a pas finalement réussi son coup en le convertissant à la théorie. Merci, aussi, aux anciens doctorants qui m'ont accueilli à mon arrivée dans l'équipe : Théo, Jérémie, Arnault, Iannis et Derwell. Et merci enfin à tous les autres doctorants de l'équipe qui ont commencé en même temps ou après moi ainsi qu'aux post-docs (l'équipe a bien grossi depuis mon arrivée), votre présence et votre bonne humeur participent à la bonne ambiance : Jean, Alessandro, Célia, Valentin, Maxime, Sacha, Lilay, Gautier, Michael, Alyssa, Pierre ainsi que tous les nouveaux stagiaires.

\vspace{0.1em}

Un très grand merci à l'équipe administrative du laboratoire sans qui il se serait effondré depuis bien longtemps. Merci à Nathalie Merlet, à Sandrine Di Concetto et à Jocelyne Moreau qui font toujours preuve de disponibilité, de gentillesse et d'efficacité pour répondre à toutes nos requêtes et problèmes.

\vspace{0.1em}

Merci à la musique qui m'accompagne depuis toujours et dont la pratique régulière me sort un peu de la physique. Merci aux orchestres, à tous les super projets auxquels j'ai eu la chance de participer et à tous les gens chouettes que j'ai pu y croiser : OSIUP, Furiante, Fuga Furiosa, Studi'Orchestra, Elektra, Ut Cinquieme et RSO (oui il y en a beaucoup ; un jour j'arriverai à être raisonnable). La musique m'a aussi accompagné dans la rédaction de ce pavé alors merci à Mahler (3 symphonies sur 11 ; je triche dans le décompte et sur la temporalité), Stravinsky (à quand le Sacre ?), Tchaikovsky (un jour un ballet complet ?), Brahms (big up à l'interprétation du quatuor Agate), Rachmaninov (un jour les concertos pour piano ?)\dots

\vspace{0.1em}

Merci à mes amis qui m'accompagnent depuis longtemps maintenant, surtout Téo et Lilian, qui prennent du plaisir à écouter mes digressions physico-mathématiques. Je ne suis pas très bavard d'habitude, mais quand on me lance sur mes sujets, je peux souvent être inarrêtable. Merci aussi à tous mes amis de l'ENS Ulm.

\vspace{0.1em}

Merci à mes divers enseignants de maths et de physique qui ont participé à la construction de ma curiosité scientifique et m'ont introduit à la pensée rigoureuse : Flavie Loyzance Lapouge, Bernard Froget, Nicolas Pazat, Valérie Monturet, Guillaume Roussel, Laurence Petitjean et Denis Favennec.

\vspace{0.1em}

Une avant-dernière pensée pour ma famille. À mes parents avant tout, qui nous ont toujours accompagnés, mes frères et moi, et ont tout fait pour respecter nos projets et nous accompagner dans leur réalisation. À mes frères évidemment, Alexandre, Thomas et Pierre, avec qui on continue de partager une grande complicité ; à leurs femmes, compagne et familles qui s'agrandissent. À mes généreux grands-parents surtout, qui m'ont donné l'amour du Bassin et de la voile ainsi qu'à mes oncles, tantes, cousins et cousines.

\vspace{0.1em}

Et enfin, pour terminer, un énorme merci à ma compagne Marie-Caroline (ma star !) qui arrive à être encore plus hypée que moi pour ma thèse. L'intérêt que tu portes à ce monde qui t'est étranger est ma plus grande source de motivation.
\begin{flushright}
    \it Hasta el fin del mundo, cari\~no.
\end{flushright}

\clearpage
\section*{List of publications}
\addcontentsline{toc}{section}{Publications}
\subsection*{Presented in the thesis}
\begin{itemize}
    \item [\cite{descamps_quantum_2023}] \hypertarget{Article: prl metro}{\'E. Descamps, N. Fabre, A. Keller, and P. Milman. Quantum Metrology Using Time-Frequency as Quantum Continuous Variables: Resources, Sub-Shot-Noise Precision and Phase Space Representation. {\it Physical Review Letters}, 131(3):030801, 2023.}
    \item [\cite{descamps_time-frequency_2023}] \hypertarget{Article: pra HOM}{\'E. Descamps, A. Keller, and P. Milman. Time-frequency metrology with two single-photon states: Phase-space picture and the Hong-Ou-Mandel interferometer. {\it Physical Review A}, 108(1):013707, 2023.}
    \item [\cite{descamps_gottesman-kitaev-preskill_2024}] \hypertarget{Article: prl GKP}{\'E. Descamps, A. Keller, and P. Milman. Gottesman-Kitaev-Preskill encoding in continuous modal variables of single photons. {\it Physical Review Letters}, 132(17):170601, 2024.}
    \item [\cite{meskine_approaching_2024}] \hypertarget{Article: prl visibility}{O. Meskine, \'E. Descamps, A. Keller, A. Lemaître, F. Baboux, S. Ducci, and P. Milman. Approaching Maximal Precision of Hong-Ou-Mandel Interferometry with Non-perfect Visibility. {\it Physical Review Letters}, 132(19):193603, 2024.}
    \item[\cite{descamps_superselection_2024}] \hypertarget{Article: prl SSRC}{\'E. Descamps, N. Fabre, A. Saharyan, A. Keller, and P. Milman. Superselection Rules and Bosonic Quantum Computational Resources. {\it Physical Review Letters}, 133(26):260605, 2024.}
    \item [\cite{descamps_measuring_2025}] \hypertarget{Article: pra entanglement}{\'E. Descamps, A. Keller, and P. Milman. Measuring entanglement along collective operators. {\it Physical Review A}, 111(5):052428, 2025.}
    \item [\cite{descamps_role_2026}] \hypertarget{Article: prl extended HOM}{\'E. Descamps, A. Keller, and P. Milman. Role of Symmetry in Generalized Hong-Ou-Mandel Interference and Quantum Metrology. {\it Physical Review Letters}, 136(6):060807, 2026.}
    \item[\cite{saharyan_resources_2026}] \hypertarget{Article: PRA SSRC metro}{ A. Saharyan, \'E. Descamps, A. Keller, and P. Milman. Resources for bosonic metrology: Quantum-enhanced precision from a superselection rule perspective. {\it Physical Review A}, 113(2):022428, 2026.}
    \item[\cite{descamps_unified_2026}] \hypertarget{Article: optica SSRC}{\'E. Descamps, A. Saharyan, A. Chivet, A. Keller, and P. Milman. Unified framework for bosonic quantum information encoding, resources and universality from superselection rules. {\it Optica Quantum}, 4(2):148--161, 2026.}
\end{itemize}

\subsection*{Not included in the thesis}
\begin{itemize}
    \item[\cite{descamps_heisenberg-weyl_2025}] \hypertarget{Article: SSRC phase space}{A. Saharyan, \'E. Descamps, A. Keller, and P. Milman. Heisenberg-Weyl bosonic phase spaces: Emergence, constraints and quantum informational resources, {\it arXiv}, arXiv:2512.05603, 2025.}
    \item[\cite{moulonguet_non-gaussianity_2026}] \hypertarget{Article: SSR and non gaussianity}{N. Moulonguet, \'E. Descamps, J. Lorgeré, A. Saharyan, A. Keller, and P. Milman. Non-Gaussianity from superselection rules, {\it arXiv}, arXiv:2603.20810, 2026.}
\end{itemize}
\subsection*{Outside of the PhD}
\begin{itemize}
    \item [\cite{descamps_stabilizer_2024}] \hypertarget{Article: JPhysA Bori}{\'E. Descamps and B. Daki\'c. On the stabilizer formalism and its generalization. {\it Journal of Physics A: Mathematical and Theoretical}, 57(45):455301, 2024.}
    \item [\cite{descamps_noisy_2025}] \hypertarget{Article: pra Damian}{É. Descamps and D. Markham. Noisy certification of continuous-variable graph states. {\it Physical Review A}, 111(2):022413, 2025.}
\end{itemize}

\fi

\ifnum \theshortversion=0

\clearpage
\section*{List of notations}
\addcontentsline{toc}{section}{Notations}

\begin{itemize}
    \item $\mathcal{H}$: Hilbert space
    \item $\R$: set of real numbers
    \item $\C$: set of complex numbers
    \item $i$: imaginary unit
    \item $\delta_{j,k}$: Kronecker delta, equal to $1$ when $j=k$ and $0$ otherwise
    \item $\delta(x)$: Dirac delta distribution
    \item $\abs{z}$: absolute value (modulus) of the complex number $z$
    \item $z^*$: complex conjugate of the complex number $z$
    \item $\hat A$: hat notation denotes a (linear) operator
    \item $\Tr(\hat A)$: trace of the operator $\hat A$
    \item $[\hat A,\hat B]$: commutator of the operators $\hat A$ and $\hat B$, defined as
    \begin{equation}
        [\hat A,\hat B] = \hat A\hat B - \hat B\hat A
    \end{equation}
    \item $\expval*{\hat H}_{\ket{\psi}}$: expectation value of the operator $\hat H$ in the normalized state $\ket{\psi}$, defined as
    \begin{equation}
        \expval*{\hat H}_{\ket{\psi}} = \bra{\psi}\hat H\ket{\psi}
    \end{equation}
    \item $\expval*{\hat H}$: expectation value of $\hat H$ when the state $\ket{\psi}$ is clear from the context
    \item $\Delta^2_{\ket{\psi}}\hat H$: variance of the Hermitian observable $\hat H$ in the normalized state $\ket{\psi}$, defined as
    \begin{equation}
        \Delta^2_{\ket{\psi}}\hat H = \expval*{\hat H^2}_{\ket{\psi}} - \expval*{\hat H}_{\ket{\psi}}^2
    \end{equation}
    \item $\Delta^2\hat H$: variance of $\hat H$ when the state $\ket{\psi}$ is clear from the context
    \item $\1$: identity operator
    \item $\vec n$: vector with scalar components; a vector of operators is written $\vec{\hat n}$
    \item $\vec n \cdot \vec m$: standard dot product between $\vec n$ and $\vec m$
    \item $\vec n \times \vec m$: standard cross product between $\vec n$ and $\vec m$
    \item $\norm{\vec n}$: norm of the vector $\vec n$
    \item $a \equiv b \,[c]$: congruence relation modulo $c$
\end{itemize}

\clearpage
\section*{Acronyms}
\addcontentsline{toc}{section}{List of acronyms}
\begin{itemize}
    \item BQC: Bosonic quantum computer
    \item BS: Beam Splitter
    \item CNOT: Controlled-NOT
    \item CRB: Cramér-Rao Bound
    \item CV: Continuous Variables
    \item DFT: Discrete Fourier Transform
    \item DOF: Degree of Freedom
    \item DV: Discrete Variables
    \item FI: Fisher Information
    \item GHZ: Greenberger-Horne-Zeilinger
    \item GKP: Gottesman-Kitaev-Preskill
    \item HOM: Hong-Ou-Mandel
    \item JSA: Joint Spectral Amplitude
    \item JSI: Joint Spectral Intensity
    \item JTA: Joint Temporal Amplitude
    \item JTI: Joint Temporal Intensity
    \item MBQC: Measurement-Based Quantum Computing
    \item MLE: Maximum Likelihood Estimator
    \item MZI: Mach-Zehnder interferometer
    \item OAT: One-Axis Twisting
    \item POVM: Positive Operator-Valued Measurement
    \item QCRB: Quantum Cramér-Rao Bound
    \item QEC: Quantum Error Correction
    \item QFI: Quantum Fisher Information
    \item QG: Quadrature-Gaussian
    \item QNG: Quadrature non-Gaussian
    \item SG: Superselection rules compliant Gaussian
    \item SLD: Symmetric Logarithmic Derivative
    \item SNG: Superselection rules compliant non-Gaussian
    \item SPDC: Spontaneous Parametric Down-Conversion
    \item SQL: Standard Quantum Limit
    \item SSR: Superselection Rules
    \item SSRC: Superselection Rules Compliant
    \item TACT: Two-Axis Countertwisting
\end{itemize}

\clearpage
\section*{List of figures}
\addcontentsline{toc}{section}{List of figures}
\makeatletter
\@starttoc{lof}
\makeatother

\clearpage
\section*{List of tables}
\addcontentsline{toc}{section}{List of tables}
\makeatletter
\@starttoc{lot}
\makeatother

\fi

\cleardoublepage
\hypertarget{maintoc}{}
\tableofcontents

%------------ Style of the small table of contents-------------
\etocsettocstyle
  {%
    \vspace*{1em}
    \par\noindent
    \textbf{\Large Contents}\par
    \vspace{-0.8em}
    \noindent\rule{\textwidth}{0.8pt}\par
  }
  {%
    \vspace{-0.2em}
    \par\noindent\rule{\textwidth}{0.8pt}\par
  }

%--------------------------------------------------------------

\cleardoublepage

\mainmatter
\pagestyle{mainmatterstyle}

\ifnum \theShowFront = 1
\chapter*{Introduction}
\setcurrentanchor{intro}
\addcontentsline{toc}{chapter}{Introduction}
\markboth{Introduction}{Introduction}

\section*{Quantum optics and quantum information}

Although quantum mechanics was initially developed to describe the behavior of matter at the microscopic scale~\cite{dirac_principles_1981, neumann_mathematical_1955, sakurai_modern_2017, landau_quantum_1991, schrodinger_undulatory_1926, heisenberg_uber_1927}, it has long since moved beyond the realm of fundamental physics and enabled numerous technological developments. During the so-called \emph{first quantum revolution}, the understanding of quantum mechanics led to the creation of technologies that are now ubiquitous in everyday life, such as transistors, lasers, magnetic resonance imaging, and atomic clocks. Although some of these technologies already rely on genuinely quantum effects, including coherence and interference phenomena arising from quantum superposition, their operation does not require the precise preparation, manipulation, and readout of complex quantum states at the individual quantum-system level. The \emph{second quantum revolution} refers to the current era of technological development that aims precisely at achieving such control, in particular by exploiting coherence and entanglement as operational resources. Among other applications, it explores the use of quantum systems to improve the ways we perform computation~\cite{nielsen_quantum_2010}, simulation~\cite{feynman_simulating_1982}, communication~\cite{bennett_teleporting_1993}, and sensing~\cite{giovannetti_advances_2011, caves_quantum-mechanical_1981}.

Among the many physical systems that can be considered for technological applications, light occupies a particularly prominent position. In the quantum regime, light is described within the framework of \emph{quantum optics}~\cite{glauber_quantum_1963, walls_quantum_2008, gerry_introductory_2005, gardiner_quantum_2004}. When properly prepared and manipulated, quantum states of light exhibit remarkable properties that make them indispensable for a wide range of applications. Because it propagates at the speed of light, optical radiation is the natural carrier for tasks involving communication or the transmission of information between distant parties. Moreover, photonic platforms benefit from several practical advantages, including operation at room temperature, the reuse of well-established linear and nonlinear optical techniques, low levels of decoherence, precise control of spatial and temporal modes, and their natural suitability for interferometry and sensing. For these reasons, the study of quantum states of light is of crucial importance, both from a fundamental perspective and for practical implementations.

Importantly, quantum light can be described within different theoretical frameworks that correspond to distinct physical regimes and mathematical formalisms. Because light can be viewed as being composed of discrete quanta called \emph{photons}, one widely used framework is the \emph{discrete-variable} (DV) paradigm. In this approach, light is described in terms of discrete excitations of the electromagnetic field, and quantum information is encoded in finite-dimensional systems such as photon-number states or polarization modes. Conversely, motivated by the wave nature of light, another major framework is the \emph{continuous-variable} (CV) paradigm~\cite{weedbrook_gaussian_2012}. In this description, the electromagnetic field is characterized by observables that take values in continuous spectra, such as field quadratures. Quantum information is then encoded in continuous degrees of freedom of the optical field. Both paradigms possess their own advantages and limitations, and they are often employed in different contexts depending on the requirements of a given application. Nevertheless, they are sometimes presented as fundamentally distinct approaches to the description of quantum light. As a consequence, many results formulated in one framework appear to have no direct counterpart in the other, occasionally leading to apparent inconsistencies or even contradictions.

One of the main goals of this thesis is to clarify the relationships between these different perspectives. In particular, we aim to develop a clearer understanding of the key ingredients of quantum optics that govern the general properties of quantum light and underlie its applications in quantum information science.

\section*{Modes, states, and symmetries}

This thesis focuses on three central concepts that are essential for understanding quantum optics and its applications: \emph{modes}, \emph{states}, and \emph{symmetries}. These notions provide complementary perspectives on quantum light and are deeply intertwined in determining both its physical properties and the ways it can be harnessed for quantum applications.

\emph{Modes} are a concept that originates from the classical description of the electromagnetic field. They describe how light is distributed among different physical degrees of freedom, such as spatial, temporal, or frequency components. In quantum optics, modes provide the natural framework for identifying and labeling the degrees of freedom of the electromagnetic field~\cite{fabre_modes_2020,walls_quantum_2008,gerry_introductory_2005}. They therefore play a central role in determining how quantum information can be encoded, manipulated, and measured.

Understanding the modal structure of light is particularly important when dealing with multimode systems, where correlations between different modes may arise. Such correlations constitute valuable resources for many applications, including quantum communication, quantum computing, and quantum metrology~\cite{fabre_modes_2020}. Consequently, identifying and exploiting modal correlations is a key aspect of modern quantum optical research.

Beyond the modal description, the quantum nature of light must also be taken into account. In quantum optics, light is understood as being composed of discrete quanta known as \emph{photons}. Different photon-number distributions can therefore be considered across the available modes. The \emph{quantum state} of light contains the complete information about the statistical (\ie, the distribution of photon numbers across modes) and coherence properties of these photons~\cite{glauber_quantum_1963,walls_quantum_2008,gerry_introductory_2005}. It describes how photons are distributed among the different modes and how they are correlated.

Quantum states can be engineered and tailored to specific tasks, and identifying which properties of these states constitute useful resources is a central question in quantum information science. In practice, modal properties and statistical properties are often deeply intertwined. Disentangling the respective roles played by these different contributions is therefore essential for gaining a deeper understanding of the properties and capabilities of quantum light.

Finally, \emph{symmetries} play a fundamental role in physics in general, and quantum optics is no exception. A symmetry corresponds to a transformation that leaves certain properties of a system invariant. Symmetry principles often provide powerful constraints and insights, allowing one to derive general results without relying on the detailed dynamics of a system.

In quantum optics, symmetries are particularly significant because they underlie the very structure of the theoretical description. Photons are bosonic particles and are therefore intrinsically symmetric under particle exchange~\cite{dirac_principles_1981,sakurai_modern_2017}. This fundamental symmetry is not merely a mathematical property; it has direct physical consequences and lies at the heart of many quantum optical phenomena. A well-known example is the Hong-Ou-Mandel interference effect~\cite{hong_measurement_1987}, which will be discussed in detail in Chap.~\ref{chap: HOM interferometry and metrology}.

Throughout this thesis, we investigate the interplay between these three fundamental concepts by analyzing how modes, states, and symmetries jointly determine the behavior of quantum light. In particular, we explore three main directions. First, we study collective variables and their relation to multimode correlations (Chap.~\ref{chap: collective entanglement}). Second, we investigate interferometric phenomena, focusing on the role of symmetry and modal indistinguishability in multiphoton interference (Chap.~\ref{chap: HOM interferometry and metrology}). Finally, we analyze the role of superselection rules and their implications for bosonic quantum information~\cite{wick_intrinsic_1952, bartlett_reference_2007} (Chap.~\ref{chap: SSR}).

\section*{Quantum metrology}

Among the many possible applications of quantum optics, this thesis focuses in particular on \emph{quantum metrology} with optical systems. While classical metrology is the science of measurement, concerned with understanding and improving the precision of measurement procedures, quantum metrology studies these questions in the quantum regime. More specifically, it aims at determining how quantum properties of physical systems can be exploited to enhance measurement precision beyond classical limits~\cite{giovannetti_quantum_2006,giovannetti_advances_2011}.

Quantum optics provides a natural platform for quantum metrology. Light is one of the most widely used physical probes in precision measurements, appearing in applications ranging from spectroscopy and imaging to interferometry~\cite{dorfman_hong-ou-mandel_2021} and gravitational-wave detection~\cite{ligo_scientific_collaboration_and_virgo_collaboration_observation_2016,ligo_scientific_collaboration_and_virgo_collaboration_quantum-enhanced_2019}. The ability to precisely engineer quantum states of light and control their modal structure makes optical systems particularly well suited for exploring quantum-enhanced measurement strategies~\cite{caves_quantum-mechanical_1981,walls_quantum_2008}.

A central question in quantum metrology concerns the notion of \emph{resources}: which physical resources can be exploited to improve the precision of a measurement, and how can they be used most efficiently? The notion of resource depends on the scenario considered and is generally associated with the physical cost of implementing a measurement protocol. Typical examples include the number of repetitions of the experiment, the number of probes interacting with the system, or the total energy used during the measurement procedure.

Within this framework, two characteristic scaling regimes are often distinguished. In the \emph{shot-noise} regime, the estimation uncertainty (standard deviation) scales as $1/\sqrt{N}$, where $N$ denotes the amount of resources used. This behavior corresponds to the classical situation in which independent probes are employed, and the precision is ultimately limited by the statistical fluctuations of individual measurement outcomes. In contrast, quantum correlations between probes can in principle enable the \emph{Heisenberg} scaling, where this uncertainty decreases as $1/N$, representing the ultimate quantum limit for many estimation problems~\cite{giovannetti_quantum_2006,giovannetti_advances_2011,paris_quantum_2009}. In practice, approaching this regime typically requires the preparation of nonclassical states of light, such as entangled or squeezed states~\cite{caves_quantum-mechanical_1981}.

In quantum optics, and particularly in light of the discrete-variable and continuous-variable paradigms introduced above, the definition of resources is not always uniform. In many DV protocols, the relevant resource is the number of photons or, when each occupied mode contains exactly one photon, the number of occupied modes. In contrast, in CV approaches the relevant quantity is often the average photon number or the total optical energy of the field. Rather than constituting a simple inconsistency, this difference highlights the fact that both \emph{modal} and \emph{statistical} properties of quantum light can serve as resources for metrological protocols.

Understanding how these different contributions arise and how they can be exploited constitutes one of the central themes of this thesis. In particular, the perspective introduced in the previous section, based on the interplay between modes, states, and symmetries, provides a natural framework for analyzing these questions.

In Chap.~\ref{chap: collective entanglement}, we study the role of multimode correlations in metrology by introducing the notion of collective variables and the associated forms of entanglement. In Chap.~\ref{chap: HOM interferometry and metrology}, we investigate interferometric protocols and analyze how symmetry and modal indistinguishability give rise to multiphoton interference effects that can be exploited for precision measurements. Finally, in Chap.~\ref{chap: SSR}, we introduce a unifying framework for bosonic quantum information based on superselection rules~\cite{wick_intrinsic_1952, bartlett_reference_2007}, which clarifies the notion of resources in bosonic systems and provides further insight into the contributions that can be harnessed for quantum metrology.

\section*{Time-frequency quantum optics}

To further investigate the interplay between modal and statistical properties of quantum light, we introduce a concrete framework that will serve as a central playground for many of the concepts and results developed in this thesis, particularly in Chap.~\ref{chap: collective entanglement} and Chap.~\ref{chap: HOM interferometry and metrology}. This framework is that of \emph{time-frequency quantum optics}. Its purpose is to provide a complete description of the temporal and spectral degrees of freedom of the electromagnetic field and of the photons that populate it.

The time and frequency degrees of freedom constitute natural and fundamental variables of optical fields. In many situations a monochromatic description of light is sufficient. However, in a wide range of applications the detailed time-frequency structure of optical states plays a crucial role. For instance, in quantum communication, quantum computation, and interferometric protocols, information can be encoded directly in temporal or spectral modes and manipulated through their correlations~\cite{brecht_photon_2015,weiner_ultrafast_2011}. More importantly, the precision achievable in time-parameter estimation is fundamentally linked to the time-frequency distribution of the probe state. As a consequence, a rigorous description of the time-frequency structure of quantum optical states is essential for understanding their metrological capabilities~\cite{giovannetti_advances_2011,paris_quantum_2009}. We explore these connections in detail in Chap.~\ref{chap: HOM interferometry and metrology}.

From a theoretical perspective, time and frequency variables exhibit a structure analogous to conjugate variables in quantum mechanics, in much the same way as position and momentum. This analogy becomes particularly clear in the single-photon regime, where different spatial modes are each populated by a single photon. In this setting, the time-frequency variables of a photon can be treated in close analogy with quadrature variables in continuous-variable quantum optics. This correspondence enables the application of phase-space tools to the analysis of time-frequency quantum states. In particular, techniques such as Wigner representations provide powerful insights into the structure of single-photon states, their transformations under optical operations, and their usefulness as resources for quantum information and quantum metrology~\cite{wigner_quantum_1932, austin_measuring_2010}. The formal framework underlying these descriptions is introduced and developed in detail in Chap.~\ref{chap: Time-Frequency Systems}.

\section*{Collective variables and multimode correlations}

When several distinct systems are considered, for example spatially separated subsystems or different external optical modes, the relevant physical quantities are often \emph{collective observables}, that is, observables that depend on the properties of each individual subsystem. Such collective quantities naturally arise in many areas of physics and play a particularly important role in quantum optics, where different modes of the electromagnetic field define the elementary subsystems of interest. In these intrinsically multimode systems, \emph{nonlocal} properties can emerge. These are properties that cannot be attributed to any single subsystem but instead arise from correlations between them. Such correlations, and in particular quantum correlations such as entanglement, can be exploited as valuable resources for quantum information processing and quantum metrology~\cite{horodecki_quantum_2009,giovannetti_advances_2011}.

In the context of discrete systems, such as ensembles of atomic probes, multipartite entanglement has long been identified as a key resource for quantum-enhanced metrology. In particular, entangled collective states can enable the Heisenberg scaling of precision, surpassing the classical shot-noise limit~\cite{giovannetti_quantum-enhanced_2004,giovannetti_advances_2011}. However, the role played by multimode correlations in quantum optical systems remains less well understood. This question becomes especially relevant when considering the time-frequency structure of optical states introduced in the previous section. Time-frequency correlations are ubiquitous in quantum optical systems, for example in photon-pair sources, but they are often treated primarily as technical or classical properties of the optical field. As a consequence, their potential role as genuine quantum resources for metrology has received comparatively limited attention, and a clear framework for analyzing the contribution of time-frequency entanglement to ultimate precision limits is still lacking.

Chap.~\ref{chap: collective entanglement} addresses these questions by first clarifying the role of time-frequency distributions and entanglement in collective time-parameter estimation. Building on these results, we then develop a more general framework that unifies observations originating from discrete systems with those arising in continuous time-frequency descriptions. This approach provides a unified perspective for studying collective variables and their associated entanglement. Within this framework, we derive general inequalities for collective observables, identify classes of optimal states for metrological protocols based on collective measurements, and establish explicit connections between the structure of multimode correlations and the achievable measurement precision. Finally, we discuss how the concept of collective variables can be extended beyond metrology, highlighting its relevance for quantum information processing and for encoding information in collective-variable quantum codes.

\section*{Linear interferometry and metrology}

\emph{Interferometry} is one of the central tools of quantum optics and plays a crucial role in many applications, ranging from quantum metrology to quantum information processing. Interferometers exploit the wave nature of light to manipulate the distribution of photons across different output modes through controlled interference. As a result, they constitute extremely sensitive probes for small physical parameters. Interferometric techniques are therefore widely used in precision measurements, for instance in gravitational-wave detection, and also play an important role in optical quantum information protocols such as quantum teleportation and linear-optical quantum computation~\cite{kok_linear_2007,nielsen_quantum_2010,giovannetti_advances_2011}.

One of the most emblematic examples of multiphoton interference is the \emph{Hong-Ou-Mandel} (HOM) effect~\cite{hong_measurement_1987}. This phenomenon arises from the modal indistinguishability of two photons entering distinct input ports of a balanced beam splitter. Due to the bosonic exchange symmetry of photons, the two-particle amplitudes interfere destructively for the case where the photons leave through different output ports, leading to the characteristic HOM dip in the coincidence probability. The HOM effect has become a cornerstone of quantum optics, providing a powerful probe of modal indistinguishability and a key building block for many optical quantum information protocols. Beyond its conceptual importance, the HOM effect can also serve as a sensitive interferometric tool. In particular, HOM interference can be used for high-resolution temporal measurements~\cite{hong_measurement_1987,chen_hong-ou-mandel_2019,jing_hongoumandel_2025}, with experimental demonstrations reaching attosecond-scale precision using modern photonic technologies~\cite{lyons_attosecond-resolution_2018}. Despite its widespread use, however, several aspects of the HOM effect remain only partially understood. In particular, the roles played by modal distributions and entanglement, as well as the precise relationship between interference visibility and estimation precision are still subjects of ongoing investigation.

Traditionally, the HOM effect is interpreted as a direct consequence of the indistinguishability of the incoming photon states. In Chap.~\ref{chap: HOM interferometry and metrology}, we show that this interpretation can be generalized by identifying a broader notion of symmetry with respect to the input modes of the interferometer. From this perspective, the HOM effect emerges as a manifestation of symmetry properties of the multimode quantum state rather than solely from modal indistinguishability. This viewpoint reveals a direct connection between the modal structure of the input state and the resulting interference pattern. Beyond providing a conceptual clarification of the HOM effect, the introduction of symmetry considerations offers powerful tools for analyzing the metrological capabilities of HOM-based estimation protocols. In particular, it enables the study of estimation tasks beyond the usual phase or time-delay measurements. Building on this symmetry-based framework, we further demonstrate that HOM-type interference and its metrological applications can be extended to more general settings. In particular, the formalism naturally generalizes to situations involving more than two interfering modes and to regimes beyond one photon per input mode, thereby opening new perspectives for exploiting multiphoton interference phenomena in quantum metrology.

\section*{Superselection rules for optical systems}

While the previous lines of investigation provide a clearer understanding of the roles played by modal and statistical properties in quantum optical systems, they do not yet offer a fully comprehensive picture. In particular, the status of modal entanglement as a resource raises several fundamental questions. For instance, passive linear optical elements, typically regarded as operations that do not generate resources, can nevertheless create modal entanglement. Understanding how such correlations arise and how they should be interpreted within a resource-theoretic framework remains a subtle issue.

Furthermore, the relationship between discrete-variable (DV) and continuous-variable (CV) descriptions of quantum light is still not fully clarified. Although both frameworks describe the same underlying physical system, many results obtained in one formalism do not have an immediate counterpart in the other. Developing a unified perspective capable of bridging these descriptions is therefore an important step toward a more complete understanding of bosonic quantum information.

To address these questions, in Chap.~\ref{chap: SSR} we introduce a unifying framework for bosonic quantum information based on the notion of \emph{superselection rules} (SSR)~\cite{wick_intrinsic_1952, bartlett_reference_2007, bartlett_dialogue_2006,sanders_photon-number_2003,molmer_optical_1997}. Superselection rules arise from fundamental symmetries in quantum systems and impose constraints on the set of physically accessible states and operations. In optical systems, the relevant symmetry is associated with global phase transformations of the electromagnetic field. In the absence of a shared phase reference, this symmetry leads to an effective superselection rule excluding observable coherence between different total-photon-number sectors, while allowing statistical mixtures of these sectors.

At first sight, such a constraint might appear to limit the range of accessible quantum states. However, the SSR framework has the advantage of explicitly incorporating the notion of a phase reference into the formalism. In practice, the availability of a phase reference is an experimental requirement that is often implicitly assumed in theoretical descriptions of quantum optical systems. By treating it explicitly, the SSR framework provides a more operationally meaningful description of optical experiments. Moreover, it naturally establishes a bridge between the discrete-variable and continuous-variable descriptions of quantum optics. Importantly, this perspective highlights the central role played by the bosonic symmetry of photons. The exchange symmetry of identical particles, which is sometimes treated as a purely formal property, becomes a key ingredient for understanding the structure of optical quantum states and the transformations they can undergo.

Beyond clarifying the properties of optical systems, the analogy between quantum optics constrained by superselection rules and other quantum systems with symmetries, such as spin systems, allows us to import powerful tools and concepts from these fields. This connection provides new insights into the structure of bosonic quantum information and clarifies the notion of resources for both quantum information processing and quantum metrology. More broadly, the superselection rule framework offers a unifying perspective that applies to bosonic systems in general. It sheds new light on the relationship between discrete-variable and continuous-variable descriptions of quantum optics and clarifies the respective roles played by modal and statistical properties in determining the capabilities of optical quantum systems.

In this context, superselection rules should not be viewed merely as an alternative description that is equivalent to the usual formalism for most experimental situations. Rather, they provide a complementary viewpoint that reveals new structural features of optical quantum systems. In Chap.~\ref{chap: SSR}, we demonstrate how this framework leads to new insights into several problems in quantum optics and quantum information. In particular, we revisit computational questions in bosonic quantum information and clarify the roles played by resources such as Gaussianity and particle entanglement, thereby providing new perspectives on the classical-quantum boundary for bosonic quantum computation. Finally, we show how the SSR framework can also be used to better understand the resources relevant for metrological applications in optical systems.

\section*{Thesis overview}

The thesis is organized as follows:

\begin{itemize}

    \item In Chap.~\ref{chap: Formalism and framework}, we introduce the basic formalism used throughout the thesis. This chapter reviews the essential concepts of quantum information theory, quantum metrology, and quantum optics, and establishes the notation and tools employed in the subsequent chapters. Readers familiar with these concepts can safely skip this chapter, although it may be useful for clarifying the specific definitions and conventions used in this thesis.

    \item In Chap.~\ref{chap: Time-Frequency Systems}, we develop the framework for describing time-frequency quantum optical systems. We introduce the formalism for single-photon time-frequency states, study their transformations under relevant operations, and analyze their metrological properties. This chapter provides the concrete framework used in the following investigations.

    \item In Chap.~\ref{chap: collective entanglement}, we investigate the role of collective variables and multimode correlations in quantum metrology. Starting from time-frequency single-photon systems, we introduce a general framework for collective observables and their associated entanglement, derive inequalities for collective variables, and establish connections between multimode correlations and achievable metrological precision. We also discuss applications to quantum information encoding using collective-variable codes.

    \item In Chap.~\ref{chap: HOM interferometry and metrology}, we study the role of symmetry and indistinguishability in linear interferometry, focusing on the Hong-Ou-Mandel effect. We show that this interference phenomenon can be understood in terms of a broader symmetry of the input state with respect to the interferometer modes, and we analyze the metrological capabilities of HOM-based estimation protocols. The framework is further generalized to multiphoton regimes and to interferometers involving multiple modes.

    \item In Chap.~\ref{chap: SSR}, we introduce a unifying framework for bosonic quantum information based on superselection rules. We show how this approach clarifies the notion of resources in optical systems, establishes connections between discrete-variable and continuous-variable descriptions, and provides new insights into computational and metrological tasks in bosonic quantum systems.

\end{itemize}

\fi

\ifnum \theShowChapone = 1
\chapter{Formalism and framework}
\setcurrentanchor{chap1}
\label{chap: Formalism and framework}
\emph{This chapter introduces the three fundamental building blocks of this thesis: quantum information theory, quantum metrology, and quantum optics. The aim is primarily to introduce the main conceptual tools and notations used throughout the manuscript. }

\localtableofcontents

\section{Introduction to quantum information}
\label{sec: Introduction to quantum Information}
\emph{In this section, we introduce the fundamental building blocks of quantum mechanics. We then present essential concepts from quantum information theory that form the foundation for the developments in the subsequent chapters. In particular, we define paradigmatic quantum systems, such as qubits and continuous-variable (CV) systems, and introduce the basic notions of quantum computation and quantum error correction.}
\subsection{Hilbert spaces and quantum states}
\label{subsec: Hilbert spaces and quantum states}
\paragraph{\lit Pure states}
At a formal level, the field of quantum mechanics is built upon the notion of \emph{Hilbert spaces}, which provide the mathematical framework necessary for modeling the physical properties of quantum systems. A Hilbert space is a linear vector space $\mathcal H$ endowed with a complex inner product $\langle\, \cdot\, , \,\cdot \,\rangle : \mathcal H \times \mathcal H \to \mathbb{C}$ whose naturally associated norm induces a metric that makes $\mathcal H$ a complete metric space. This completeness property is a technical condition required for important structural results, such as the diagonalisability of Hermitian operators~\cite{reed_methods_1980}, but will not be considered explicitly in the following. An element $\psi \in \mathcal H$ is called a \emph{quantum state} and provides a mathematical representation of all the physical information of a specific configuration of the quantum system. Only vectors of unit norm, \ie, those satisfying $\langle \psi,\psi \rangle=1$, are physically allowed. Furthermore, states that differ only by a global phase factor are considered physically equivalent, so in the following, any global phase will be ignored.\footnote{As such, the correct mathematical construct for describing quantum states is given by the projective Hilbert space, which is the set of equivalence classes of vectors in $\mathcal H$ under the equivalence relation $\psi \sim z\psi$ for $z \in \mathbb{C}^*$. However, since this distinction does not play a significant role in the developments of this thesis, we will simply treat quantum states as unit vectors in $\mathcal H$, with the understanding that global phases are physically irrelevant.}

In quantum mechanics, it is standard to use \emph{Dirac notation}~\cite{dirac_principles_1981} for convenience. Formally, this can be introduced via the notion of duality. For $\psi \in \mathcal H$, we define the \emph{ket} vector as
\begin{align}
    \ket{\psi} : \,& \mathbb{C} \to \mathcal H, \notag \\
    & z \mapsto z \psi.
\end{align}
It can be verified that every element of $\mathcal L(\mathbb{C},\mathcal H)$, the set of all linear maps from $\mathbb{C}$ to $\mathcal H$, takes this form for some $\psi \in \mathcal H$. Consequently, there is a natural isomorphism $\mathcal H \simeq \mathcal L(\mathbb{C},\mathcal H)$, allowing us to treat $\psi$ and $\ket{\psi}$ interchangeably. Henceforth, all elements of $\mathcal H$ will be denoted as $\ket{\psi}$.

The \emph{bra} vector $\bra{\psi}$ is defined as an element of the topological dual $\mathcal H^\dagger$, \ie, the set of all bounded linear functionals from $\mathcal H$ to $\mathbb{C}$
\begin{align}
    \bra{\psi} : \,& \mathcal H \to \mathbb{C}, \notag \\
    & \ket{\phi} \mapsto \langle \psi,\phi \rangle.
\end{align}
This formal definition allows mathematically rigorous manipulations: the composition of maps corresponds to familiar operations on quantum states. For instance, for two states $\ket{\psi}$ and $\ket{\phi}$, the composition $\bra{\psi} \circ \ket{\phi}$ defines the map $z \mapsto \langle \psi,\phi \rangle z$, which is identified with the complex number $\langle \psi,\phi \rangle$. As such, $\bra{\psi} \circ \ket{\phi}$, shortened to $\braket{\psi}{\phi}$, corresponds to the scalar product between the two vectors. As such, in the following, the scalar product will be denoted $\braket{\psi}{\phi}$. On the other hand, $ \ket{\phi}\circ \bra{\psi}$ corresponds to the outer product $\ketbra{\phi}{\psi}$. Hilbert spaces can usually be endowed with a (possibly infinite) orthonormal basis $\{\ket{\psi_j}\}_{j \in J}$, meaning any state $\ket{\psi}$ can be uniquely expressed as
\begin{equation}
    \ket{\psi} = \sum_{j \in J} c_j \ket{\psi_j}, \quad \text{with } \braket{\psi_j}{\psi_k} = \delta_{j,k},
\end{equation}
where for $j\in J$, $c_j$ is a complex coefficient. In the following, all bases are assumed to be orthonormal. The normalization condition $\braket{\psi}{\psi} = 1$ implies $\sum_{j \in J} \abs{c_j}^2 = 1,$ and the inner product between two states $\ket{\psi} = \sum_j c_j \ket{\psi_j}$ and $\ket{\phi} = \sum_j d_j \ket{\psi_j}$ is
\begin{equation}
    \braket{\psi}{\phi} = \sum_{j \in J} c_j^* d_j.
\end{equation}

\paragraph{\lit Mixed states}
However, elements of $\mathcal H$ alone cannot describe all quantum scenarios. For instance, given two states $\ket{\psi}$ and $\ket{\phi}$, the present formalism is unable to represent a situation where a source probabilistically emits either $\ket{\psi}$ or $\ket{\phi}$, {\it e.g.}, according to a coin toss. To describe such statistical mixtures, we introduce \emph{density matrices}.  

A \emph{density matrix} $\hat{\rho}$ is a linear operator on $\mathcal H$ that is positive semi-definite ($\bra{\psi} \hat{\rho} \ket{\psi} \ge 0$ for all $\ket{\psi} \in \mathcal H$) and has unit trace ($\Tr(\hat{\rho}) = 1$). By the spectral theorem~\cite{reed_methods_1980}, any density operator can be diagonalized in some orthonormal basis $\{\ket{\psi_j}\}_{j \in J}$ as
\begin{equation}
    \hat{\rho} = \sum_{j \in J} p_j \ketbra{\psi_j}{\psi_j},
\end{equation}
where the $p_j$ are non-negative real numbers summing to one. Physically, this decomposition corresponds to a probabilistic preparation of the state $\ket{\psi_j}$ with probability $p_j$. For instance, the above hypothetical coin-toss scenario can be represented as
\begin{equation}
    \hat{\rho} = \frac{1}{2} \left( \ketbra{\psi}{\psi} + \ketbra{\phi}{\phi} \right).
\end{equation}

Density matrices of the form $\ketbra{\psi}{\psi}$ are called \emph{pure states} and correspond exactly to the state $\ket{\psi}$. In this case, the ket notation $\ket{\psi}$ is usually preferred. Notice that any global phase in $e^{i\varphi}\ket{\psi}$ cancels out when constructing the corresponding density matrix. In contrast, density matrices that cannot be written in this form correspond to \emph{mixed states}, which typically encode statistical uncertainty or imperfections in the preparation of the quantum state. In the following, we will primarily use the pure-state formalism, introducing mixed states only when necessary.

\subsection{Evolution and measurement}
\label{subsec: Evolution and measurement}
One can say that a general physical theory, regardless of its domain of application, is described by three main ingredients. First, one must provide the mathematical objects that encode the properties of the physical system under consideration. In the context of this work, these objects are the pure or mixed quantum states introduced above. Second, one must specify how this description evolves over time under given interactions. Finally, one must establish how the mathematical representation relates to experimentally observable quantities. In quantum mechanics, evolution is encoded via the Schrödinger equation and unitary transformations, while measurement is described by observables.

\paragraph{\lit Evolution} 
A linear operator $\hat H$ acting on a Hilbert space $\mathcal H$ is called \emph{Hermitian} if $\hat H=\hat H^\dagger$, where $\hat A^\dagger$ denotes the Hermitian conjugate of an operator $\hat A$, uniquely defined by the relation
\begin{equation}
    \bra{\psi}\hat A\ket{\phi} = \bra{\phi}\hat A^\dagger \ket{\psi}^*, \quad \forall \ket{\psi}, \ket{\phi}\in\mathcal H.
\end{equation}
Such an operator $\hat H$ is usually referred to as a \emph{Hamiltonian} and governs the evolution of a pure state via the Schrödinger equation~\cite{schrodinger_undulatory_1926,landau_quantum_1991,sakurai_modern_2017}
\begin{equation}
    i\hbar\frac{\dd \ket{\psi}}{\dd t} = \hat H \ket{\psi}.
\end{equation}
The Hamiltonian $\hat H$ typically encodes the energy of the system and is specified by the physical context. This equation can be formally solved using the exponential of operators, yielding
\begin{equation}
    \ket{\psi(t)} = e^{-i(t-t_0)\hat H/\hbar} \ket{\psi(t_0)}.
\end{equation}
For mixed states, the corresponding evolution is described by the von Neumann equation~\cite{neumann_mathematical_1955}
\begin{equation}
    i\hbar \frac{\dd \hat \rho}{\dd t} = [\hat H, \hat \rho],
\end{equation}
which admits the solution
\begin{equation}
    \hat \rho(t) = e^{-i(t-t_0)\hat H/\hbar} \hat \rho(t_0) e^{i(t-t_0)\hat H/\hbar}.
\end{equation}
In the following, we adopt the convention $\hbar = 1$, which simplifies the above expressions. The operator $\hat U = e^{-i(t-t_0)\hat H}$ is \emph{unitary}, satisfying $\hat U \hat U^\dagger=\hat U^\dagger \hat U = \1$. Consequently, when precise time labels are not needed, quantum states can be evolved using an arbitrary unitary operator $\hat U$ via
\begin{align}
    \ket{\psi} \mapsto \hat U \ket{\psi}, && \hat \rho \mapsto \hat U \hat \rho \hat U^\dagger.
\end{align}
For pure states, unitary evolution encompasses all physically allowed transformations. For mixed states, however, more general evolutions are possible, {\it e.g.}, via master equations or quantum channels~\cite{nielsen_quantum_2010}. These generalizations will not be needed in the remainder of this manuscript.

\paragraph{\lit Measurement} 
Unlike classical physics, where the link between mathematical representation and measurable quantities is often direct, quantum mechanics presents a subtler connection. The quantum state $\ket{\psi}$ (or $\hat \rho$) is not directly observable, limiting the accessible information. 

The simplest notion of measurement involves \emph{observables}. An operator $\hat A$ is a quantum observable if it is Hermitian. Quantum mechanics postulates that $\hat A$ corresponds to a measurable quantity, with possible outcomes given by its eigenvalues $\lambda$. Measurements are intrinsically probabilistic: the probability $p_\lambda$ of obtaining outcome $\lambda$ on a state $\ket{\psi}$ is
\begin{equation}
    p_\lambda = \bra{\psi} \hat \Pi_\lambda \ket{\psi},
\end{equation}
where $\hat \Pi_\lambda$ is the projector onto the eigenspace of $\hat A$ associated with $\lambda$. The \emph{expectation value} of $\hat A$ is 
\begin{equation}
    \ev*{\hat A}_{\ket{\psi}} = \bra{\psi} \hat A \ket{\psi},
\end{equation}
and the \emph{variance} is
\begin{equation}
    \Delta^2_{\ket{\psi}} \hat A = \ev{(\hat A - \ev*{\hat A})^2} = \ev{\hat A^2} - \ev{\hat A}^2,
\end{equation}
where the index will be omitted in the following, when the state considered is clear from the context. For mixed states, these quantities generalize naturally
\begin{align}
    p_\lambda = \Tr(\hat \Pi_\lambda \hat \rho), && \ev*{\hat A}_{\hat \rho} = \Tr(\hat A \hat \rho), && \Delta^2_{\hat \rho} \hat A = \Tr(\hat A^2 \hat \rho) - \Tr(\hat A \hat \rho)^2.
\end{align}
Measurement generally changes the quantum state by projecting it onto the corresponding eigenspace
\begin{align}
    \ket{\psi} \mapsto \frac{\hat \Pi_\lambda \ket{\psi}}{\sqrt{\bra{\psi}\hat \Pi_\lambda \ket{\psi}}}, &&
    \hat \rho \mapsto \frac{\hat \Pi_\lambda \hat \rho \hat \Pi_\lambda}{\Tr(\hat \Pi_\lambda \hat \rho)}.
\end{align}

The notion of observables is not sufficient to describe all physically implementable measurements. This motivates the concept of \emph{Positive Operator-Valued Measures (POVMs)}. A set of Hermitian operators $\{\hat E_\lambda\}_{\lambda \in \Lambda}$ is a POVM if each $\hat E_\lambda$ is positive semi-definite and
\begin{equation}
    \sum_{\lambda \in \Lambda} \hat E_\lambda = \1.
\end{equation}
Here, $\Lambda$ labels possible measurement outcomes, with associated probabilities
\begin{equation}
    p_\lambda = \bra{\psi} \hat E_\lambda \ket{\psi}.
\end{equation}
Projective measurements correspond to the special case $\hat E_\lambda = \hat \Pi_\lambda$, where $\Lambda$ is the spectrum of $\hat A$.

To fully determine the post-measurement state, one can introduce \emph{Kraus operators} $\{\hat K_\lambda\}_{\lambda \in \Lambda}$ satisfying
\begin{equation}
    \sum_{\lambda \in \Lambda} \hat K_\lambda^\dagger \hat K_\lambda = \1.
\end{equation}
This implies that $\{\hat K_\lambda^\dagger \hat K_\lambda\}$ forms a POVM. The post-measurement state is then given by Born's rule~\cite{nielsen_quantum_2010}
\begin{align}
    \ket{\psi} \mapsto \frac{\hat K_\lambda \ket{\psi}}{\sqrt{\bra{\psi} \hat K_\lambda^\dagger \hat K_\lambda \ket{\psi}}}, &&
    \hat \rho \mapsto \frac{\hat K_\lambda \hat \rho \hat K_\lambda^\dagger}{\Tr(\hat K_\lambda^\dagger \hat K_\lambda \hat \rho)}.
\end{align}

\subsection{Composite Quantum Systems and Entanglement}
\label{subsec: Composite Quantum Systems and Entanglement}
\paragraph{\lit Tensor product} When considering two interacting quantum systems, each described by its own Hilbert space $\mathcal H_A$ and $\mathcal H_B$, it is often useful to describe them as a single composite system. The appropriate mathematical structure for this purpose is the tensor product of the individual Hilbert spaces
\begin{equation}
    \mathcal H_\text{total} = \mathcal H_A \otimes \mathcal H_B.
\end{equation}
If both spaces are equipped with orthonormal bases $\{\ket*{\psi_j^{(A)}}\}_{j\in J_A}$ and $\{\ket*{\psi_k^{(B)}}\}_{k\in J_B}$, then the set
\begin{equation}
    \left\{ \ket{\psi_j^{(A)}} \otimes \ket{\psi_k^{(B)}} \,\middle|\, j\in J_A,\, k\in J_B \right\}
\end{equation}
forms a basis for the tensor product space $\mathcal H_\text{total}$. Consequently, for finite-dimensional systems, the total Hilbert space is also finite-dimensional and the corresponding dimensions are related by $n_\text{total} = n_A \times n_B$. Any state in the composite system can then be expressed as
\begin{equation}
    \ket{\psi} = \sum_{j\in J_A,\, k\in J_B} c_{jk} \, \ket{\psi_j^{(A)}} \otimes \ket{\psi_k^{(B)}}.
\end{equation}

\paragraph{\lit Product and entangled states} A particularly important class of pure states consists of \emph{product states} (or \emph{separable states}), which describe situations where the quantum states of each subsystem are independent of the other
\begin{equation}
    \ket{\psi} = \ket{\psi_A} \otimes \ket{\psi_B}.
\end{equation}
If such a decomposition is not possible, the state is called \emph{entangled}. The extension of this notion to mixed states is slightly more involved. A mixed state $\hat \rho$ is said to be separable if it can be written as a convex sum of product states:
\begin{equation}
    \hat \rho = \sum_{j\in J} p_j \, \hat \rho_j^{(A)} \otimes \hat \rho_j^{(B)},
\end{equation}
where $\hat \rho_j^{(A)}$ and $\hat \rho_j^{(B)}$ are density matrices defined on the individual subsystems, and $p_j \ge 0$ with $\sum_j p_j = 1$~\cite{werner_quantum_1989, horodecki_quantum_2009}. The presence of the sum allows correlation between the two systems but only of a classical nature. If no such decomposition exists, the mixed state is entangled.

Entanglement is a fundamental concept in quantum information theory, both conceptually and practically. It not only implies that the subsystems cannot be described independently, but also that their properties exhibit correlations beyond what is possible classically~\cite{horodecki_quantum_2009}. Entanglement serves as a key resource in applications such as quantum metrology~\cite{giovannetti_advances_2011}, quantum communication~\cite{bennett_teleporting_1993}, quantum computation~\cite{nielsen_quantum_2010}, and quantum simulation~\cite{feynman_simulating_1982,georgescu_quantum_2014}.

\paragraph{\lit Reduced Density Matrices}
Given the state of a composite system $\hat \rho_\text{total}$, one can focus on a subsystem by computing its \emph{reduced density matrix}, obtained via the partial trace over the complementary subsystem:
\begin{equation}
    \hat \rho_A = \Tr_B(\hat \rho_\text{total}).
\end{equation}
The reduced density matrix contains all information relevant to local measurements and dynamics. Interestingly, if the global state $\hat \rho_\text{total}$ is a pure entangled state, the reduced state $\hat \rho_A$ is generally mixed. Therefore, the mixedness of a subsystem can indicate entanglement with its environment, often arising due to imperfect isolation~\cite{nielsen_quantum_2010}.

\paragraph{\lit Multipartite Systems}

More than two subsystems can be combined in an analogous way. For $n$ subsystems, the total Hilbert space is given by
\begin{equation}
    \mathcal H_\text{total} = \bigotimes_{j=1}^n \mathcal H_j.
\end{equation}
In this multipartite setting, the notions of entanglement and separability generalize naturally. However, with more than two parties, entanglement becomes richer and more complex, giving rise to various definitions and quantification measures~\cite{horodecki_quantum_2009, guhne_entanglement_2009}. We discuss some of these aspects in Sec.~\ref{sec: Collective entanglement}.

\subsection{Qubit systems}
\label{subsec: qubits}
\paragraph{\lit Definitions}
A fundamental quantum system, relevant both for the description of physically important systems, such as spin-$1/2$ particles~\cite{sakurai_modern_2017} and light polarization~\cite{born_principles_1999}, and for abstract computational models~\cite{nielsen_quantum_2010}, is the \emph{qubit}. A qubit is associated with the simplest non-trivial Hilbert space $\mathcal H_{\text{qubit}} = \mathbb{C}^2$, which is two-dimensional. We introduce the vectors
\begin{align}
    \ket{0} = (1,0), && \ket{1} = (0,1),
\end{align}
which form an orthonormal basis of $\mathcal H_{\text{qubit}}$, commonly referred to as the \emph{computational basis}. Any pure qubit state can therefore be written as
\begin{equation}
    \ket{\psi} = \alpha \ket{0} + \beta \ket{1},
\end{equation}
where $\alpha,\beta \in \mathbb{C}$ and normalization requires $\abs{\alpha}^2 + \abs{\beta}^2 = 1$.

Important operators acting on qubits are the \emph{Pauli matrices}, whose matrix representations in the computational basis are
\begin{align}
    \hat X=\hat \sigma_x &= 
    \begin{pmatrix}
        0 & 1 \\
        1 & 0
    \end{pmatrix}, &
    \hat Y=\hat\sigma_y &= 
    \begin{pmatrix}
        0 & -i \\
        i & 0
    \end{pmatrix}, &
    \hat Z=\hat \sigma_z &=
    \begin{pmatrix}
        1 & 0 \\
        0 & -1
    \end{pmatrix}.
\end{align}
It is straightforward to verify that these operators are both Hermitian and unitary. It is often convenient to group them into a vector of operators,
\begin{equation}
    \vec{\hat \sigma} = (\hat X, \hat Y, \hat Z).
\end{equation}
While $\hat Z$ is diagonal in the computational basis, the operators $\hat X$ and $\hat Y$ are diagonal in the bases $\{\ket{\pm}\}$ and $\{\ket{\pm_i}\}$, respectively, defined by
\begin{align}
    \ket{\pm} &= \frac{1}{\sqrt{2}} \left( \ket{0} \pm \ket{1} \right), &
    \ket{\pm_i} &= \frac{1}{\sqrt{2}} \left( \ket{0} \pm i \ket{1} \right).
\end{align}

\paragraph{\lit The Bloch sphere} Qubits admit a particularly intuitive geometrical representation via the \emph{Bloch sphere}~\cite{bloch_nuclear_1946,nielsen_quantum_2010}. Since quantum states are normalized and defined only up to a global phase, any pure qubit state can be parametrized as
\begin{equation}
    \ket{\psi} = \cos\!\left(\frac{\theta}{2}\right)\ket{0}
    + e^{i\varphi}\sin\!\left(\frac{\theta}{2}\right)\ket{1},
\end{equation}
where $\theta \in [0,\pi]$ and $\varphi \in [0,2\pi)$. The pair $(\theta,\varphi)$ can be interpreted as the spherical coordinates of a point on the unit sphere in $\mathbb{R}^3$, establishing a one-to-one correspondence between pure qubit states and points on the two-dimensional sphere. This provides a powerful geometric visualization of qubit states. In this picture, $\ket{0}$ and $\ket{1}$ correspond to the north and south poles of the sphere, while the states $\ket{\pm}$ and $\ket{\pm_i}$ lie along the $X$ and $Y$ axes, respectively (see Fig.~\ref{fig: bloch sphere}).

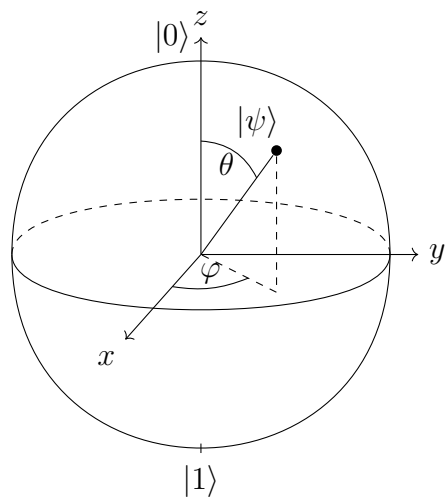
\begin{figure}[ht]
    \centering
    \scalebox{1}{\begin{tikzpicture}
	\begin{pgfonlayer}{nodelayer}
		\node [style=none] (0) at (-5, 0) {};
		\node [style=none] (1) at (5, 0) {};
		\node [style=none] (2) at (-0.75, 5.75) {$\ket{0}$};
		\node [style=none] (3) at (0, -6) {$\ket{1}$};
		\node [style=none] (4) at (0, 0) {};
		\node [style=none] (5) at (0, 5.75) {};
		\node [style=none] (6) at (5.75, 0) {};
		\node [style=none] (7) at (-2, -2.25) {};
		\node [style=none] (8) at (0, 6.25) {$z$};
		\node [style=none] (9) at (6.25, 0) {$y$};
		\node [style=none] (10) at (-2.5, -2.75) {$x$};
		\node [style=none] (11) at (0, -5) {};
		\node [style=none] (12) at (0, -5.25) {};
		\node [style=Black dot] (13) at (2, 2.75) {};
		\node [style=none] (14) at (2, -1) {};
		\node [style=none] (15) at (1.5, 3.5) {$\ket{\psi}$};
		\node [style=none] (22) at (0, 3) {};
		\node [style=none] (23) at (1.475, 2.025) {};
		\node [style=none] (24) at (-0.75, -0.85) {};
		\node [style=none] (25) at (1.25, -0.625) {};
		\node [style=none] (26) at (0.675, 2.4) {$\theta$};
		\node [style=none] (27) at (0.25, -0.5) {$\varphi$};
	\end{pgfonlayer}
	\begin{pgfonlayer}{edgelayer}
		\draw [bend left=90, looseness=1.75] (0.center) to (1.center);
		\draw [bend right=90, looseness=1.75] (0.center) to (1.center);
		\draw [bend right=90, looseness=0.50] (0.center) to (1.center);
		\draw [style=dashes, bend left=90, looseness=0.50] (0.center) to (1.center);
		\draw [style=Arrow] (4.center) to (5.center);
		\draw [style=Arrow] (4.center) to (6.center);
		\draw [style=Arrow] (4.center) to (7.center);
		\draw (11.center) to (12.center);
		\draw (4.center) to (13);
		\draw [style=dashes] (13)
			 to (14.center)
			 to (4.center);
		\draw [bend left] (22.center) to (23.center);
		\draw [bend right=15] (24.center) to (25.center);
	\end{pgfonlayer}
\end{tikzpicture}}
    \caption[Geometric representation of a pure qubit state on the Bloch sphere]{Geometric representation of a pure qubit state on the Bloch sphere.}
    \label{fig: bloch sphere}
\end{figure}

The Bloch sphere representation can be naturally extended to mixed states, leading to the so-called \emph{Bloch ball}. Indeed, any qubit density matrix $\hat \rho$ can be uniquely written as
\begin{equation}
    \hat \rho = \frac{1}{2}\left( \1 + \vec n \cdot \vec{\hat \sigma} \right),
\end{equation}
where $\vec n \in \mathbb{R}^3$ is the \emph{Bloch vector}, satisfying $\norm{\vec n} \leq 1$. Each point inside the unit ball corresponds to a unique mixed qubit state, while pure states are precisely those for which $\norm{\vec n} = 1$, lying on the surface of the sphere.

\paragraph{\lit Extensions} The use of a single qubit is limiting for most applications. More general quantum systems can be constructed by considering multiple qubits, described by the tensor-product Hilbert space $(\mathbb{C}^2)^{\otimes n}$ which has dimension $2^n$. In this setting, computational basis states are conveniently written in the compact form
\begin{equation}
    \ket{i_1} \otimes \cdots \otimes \ket{i_n} =\ket{i_1, \dots, i_n},
\end{equation}
where $i_k \in \{0,1\}$ for all $k=1,\dots,n$.

Qubit systems, which are described by two-dimensional Hilbert spaces, can be naturally generalized to \emph{qudits}. In this more general setting, each subsystem is associated with a $d$-dimensional Hilbert space $\mathbb{C}^d$, with $d>2$. A convenient choice of basis is the \emph{computational basis}, defined as the orthonormal set
\begin{equation}
    \{\ket{0}, \ket{1}, \dots, \ket{d-1}\}.
\end{equation}
Qudit systems provide a natural framework for describing physical platforms with more than two accessible energy levels and allow for a richer structure of quantum states and operations compared to qubits.

The use of higher-dimensional quantum systems can offer significant advantages in a variety of quantum information processing tasks. These include increased information density, enhanced robustness against certain types of noise, and improved security and efficiency in quantum communication protocols such as quantum key distribution. Moreover, qudits play an important role in foundational studies and in the development of generalized quantum error-correcting codes and entanglement theory~\cite{nielsen_quantum_2010,bechmann-pasquinucci_quantum_2000,durt_security_2003}. Despite these potential benefits, qubits remain the most widely studied and experimentally implemented quantum systems, largely due to their conceptual simplicity and the maturity of existing qubit-based technologies.

\subsection{Continuous variables}
\label{subsec: continuous variables}
\paragraph{\lit Definitions}
Another fundamental Hilbert space encountered in quantum mechanics is the space of square-integrable functions on the real line \footnote{More precisely, the correct mathematical framework is more complex, since operators such as $\hat x$ and $\hat p$ are unbounded and do not leave $L^{2}(\mathbb{R})$ invariant. This technical subtlety has no direct consequence for the physical discussion presented here and will therefore not be developed further.}
\begin{equation}
    \mathcal H = L^{2}(\mathbb{R}) = \left\{ f:\mathbb{R}\to\mathbb{C} \,\middle|\, \int_{\mathbb{R}} \abs{f(x)}^{2}\,\dd x < \infty \right\},
\end{equation}
equipped with the canonical inner product
\begin{equation}
    \braket{f}{g} = \int_{\mathbb{R}} f^{*}(x) g(x)\,\dd x.
\end{equation}
This Hilbert space is typically used to model the quantum state of a particle moving along a single spatial dimension, where $\abs{f(x)}^{2}$ represents the probability density for finding the particle at position $x$ \cite{galindo_quantum_2012,garrison_quantum_2014,landau_quantum_1991,merzbacher_quantum_1998}. In this manuscript, we will mostly work in natural units where $\hbar=1$.

Two fundamental operators acting on $\mathcal H$ are the (dimensionless) position and momentum operators, defined formally by
\begin{align}
    \begin{array}{rl}
        \hat x:\, \mathcal H &\to \mathcal H, \\
        f &\mapsto \bigl(x \mapsto x\,f(x)\bigr),
    \end{array}
    &&
    \begin{array}{rl}
        \hat p:\, \mathcal H &\to \mathcal H, \\
        f &\mapsto -i\,\dfrac{\dd f}{\dd x}.
    \end{array}
\end{align}
With appropriate domains, both $\hat x$ and $\hat p$ are Hermitian operators. They play a dual role in quantum theory. On the one hand, as Hermitian operators, they correspond to observables associated with position and momentum measurements. On the other hand, as generators of unitary transformations, they implement translations in momentum and position space, respectively.

Neither $\hat x$ nor $\hat p$ admits normalizable eigenvectors within $\mathcal H$. However, by extending the formalism to include distributions, one can introduce generalized eigenstates of position and momentum, denoted by $\{\ket{x}\}_{x\in\mathbb{R}}$ and $\{\ket{p}\}_{p\in\mathbb{R}}$, defined by
\begin{align}
    \ket{x}: y &\mapsto  \delta(y-x), & \ket{p}: x &\mapsto \frac{1}{\sqrt{2\pi}}e^{i p x},
\end{align}
which satisfy the eigenvalue equations
\begin{align}
    \hat x \ket{x} = x \ket{x}, && \hat p \ket{p} = p \ket{p}.
\end{align}
These generalized states do not belong to $\mathcal H$ and are not normalizable. Nevertheless, they obey the resolution of the identity
\begin{align}
    \int_{\mathbb{R}} \ketbra{x}{x}\,\dd x = \1, &&
    \int_{\mathbb{R}} \ketbra{p}{p}\,\dd p = \1,
\end{align}
which allows any state in $\mathcal H$ to be expanded in either the position or momentum representation. The position and momentum eigenstates are related by Fourier transformation,
\begin{align}
    \ket{p} &= \frac{1}{\sqrt{2\pi}} \int_{\mathbb{R}} e^{i p x}\ket{x}\,\dd x, &
    \ket{x} &= \frac{1}{\sqrt{2\pi}} \int_{\mathbb{R}} e^{-i p x}\ket{p}\,\dd p.
\end{align}

A central property of the operators $\hat x$ and $\hat p$ is the canonical commutation relation
\begin{equation}
    [\hat x,\hat p] = i\,\1,
\end{equation}
which lies at the heart of many distinctly quantum phenomena. In particular, it implies that position and momentum cannot be simultaneously measured with arbitrary precision, as quantified by the Heisenberg uncertainty principle~\cite{heisenberg_uber_1927,robertson_uncertainty_1929}
\begin{equation}
    \Delta^{2}\hat x \, \Delta^{2}\hat p \geq \frac{1}{4}.
\end{equation}

Finally, the interpretation of $\hat x$ and $\hat p$ as generators of translations is made explicit by the relations
\begin{align}\label{eq: momentum and postion translation}
    e^{-i p_{0}\hat x}\ket{p} &= \ket{p - p_{0}}, &
    e^{-i x_{0}\hat p}\ket{x} &= \ket{x + x_{0}},
\end{align}
showing that $\hat x$ generates momentum translations, while $\hat p$ generates position translations.

\paragraph{\lit Harmonic oscillator}
A central Hamiltonian in quantum physics is the quadratic operator
\begin{equation}
    \hat H=\frac{1}{2}\left(\hat p^2+\hat x^2\right),
\end{equation}
which describes the energy of the quantum harmonic oscillator. In these units the mass and angular frequency of the oscillator are both set to unity, so that no free parameters appear in the Hamiltonian. Solving this model amounts to understanding the behavior of a quantum particle in a one-dimensional parabolic potential.

The harmonic oscillator plays a fundamental role across many areas of quantum physics. In particular, as will be discussed in Sec.~\ref{subsec: Field Quantization}, quantized electromagnetic fields can be modeled as collections of independent quantum harmonic oscillators: each mode of the field corresponds to a single oscillator. This makes continuous-variable systems and the harmonic-oscillator Hamiltonian essential tools in quantum optics and quantum information processing~\cite{gerry_introductory_2005,weedbrook_gaussian_2012,fabre_modes_2020}.

The Hamiltonian can be diagonalized by introducing the ladder (or bosonic) operators
\begin{align}
    \hat a=\frac{1}{\sqrt{2}}(\hat x+i\hat p), &&  
    \hat a^\dagger=\frac{1}{\sqrt{2}}(\hat x-i\hat p),
\end{align}
which satisfy the canonical commutation relation $[\hat a,\hat a^\dagger]=\1$. In terms of these operators, the Hamiltonian takes the simple form
\begin{equation}
    \hat H=\hat a^\dagger \hat a+\frac{1}{2}\1.
\end{equation}
Defining the number operator $\hat n=\hat a^\dagger \hat a$, one can show using the commutation relations that its spectrum is given by the non-negative integers. Its eigenstates $\{\ket{n}\}_{n\in\mathbb{N}}$, called Fock states or number states, satisfy
\begin{align}
    \hat n\ket{n}=n\ket{n}, && 
    \ket{n}=\frac{1}{\sqrt{n!}}(\hat a^\dagger)^n\ket{0},
\end{align}
where $\ket{0}$ denotes the ground state. The operators $\hat a$ and $\hat a^\dagger$ act as lowering and raising operators on the Fock basis,
\begin{align}
    \hat a\ket{n}=\sqrt{n}\ket{n-1}, &&
    \hat a^\dagger\ket{n}=\sqrt{n+1}\ket{n+1},
\end{align}
and are thus referred to as the annihilation and creation operators. The ground state, also called the vacuum state, is uniquely defined by $\hat a\ket{0}=0$. In the position representation it is given by
\begin{equation}
    \braket{x}{0}=\pi^{-\frac{1}{4}}e^{-\frac{x^2}{2}}.
\end{equation}
The higher Fock states can be obtained by repeated application of the creation operator. Their explicit wavefunction expressions involve Hermite polynomials~\cite{sakurai_modern_2017,merzbacher_quantum_1998}.

\paragraph{\lit Quantum states}
Beyond the Fock states introduced above, there exists a large variety of physically relevant continuous-variable quantum states. Among the most important are the Glauber coherent states~\cite{glauber_quantum_1963}. They form a family of states $\ket{\alpha}$ parametrized by a complex number $\alpha\in\mathbb{C}$ and describe states whose dynamics most closely resemble those of a classical harmonic oscillator, with oscillation amplitude $\abs{\alpha}$ and phase $\arg(\alpha)$.

Coherent states admit several equivalent definitions. They are eigenstates of the annihilation operator,
\begin{equation}
    \hat a\ket{\alpha}=\alpha\ket{\alpha},
\end{equation}
and can also be generated by applying the displacement operator
\begin{equation}
    \hat D(\alpha)=e^{\alpha \hat a^\dagger-\alpha^*\hat a}
\end{equation}
to the vacuum state, $\ket{\alpha}=\hat D(\alpha)\ket{0}$. Their expansion in the Fock basis reads
\begin{equation}
    \ket{\alpha}=e^{-\frac{\abs{\alpha}^2}{2}}\sum_{n=0}^\infty \frac{\alpha^n}{\sqrt{n!}}\ket{n}.
\end{equation}
The overlap between two coherent states is given by
\begin{align}
    \braket{\alpha}{\beta} = e^{-\frac{1}{2}(\abs{\alpha}^2 + \abs{\beta}^2 - 2 \alpha^* \beta)}, && \abs{\braket{\alpha}{\beta}}^2=e^{-\abs{\alpha - \beta}^2}.
\end{align}

Although coherent states are often regarded as the most ``classical'' pure states of the harmonic oscillator, they can be used as building blocks for genuinely nonclassical states. A paradigmatic example is given by Schrödinger cat states~\cite{yurke_generating_1986}, defined as coherent superpositions of two coherent states with opposite phases,
\begin{equation}
    \ket{\text{cat}_\alpha^{\pm}}=\mathcal N_\pm\left(\ket{\alpha}\pm\ket{-\alpha}\right),
\end{equation}
where $\mathcal N_\pm$ is a normalization constant. These states represent superpositions of two quasi-classical states and provide a valuable platform to study the quantum-to-classical transition and decoherence~\cite{zurek_decoherence_2003}. Moreover, cat states play an important role in continuous-variable and bosonic quantum computing architectures~\cite{mirrahimi_dynamically_2014,leghtas_confining_2015}. Generalizations involving superpositions of more than two coherent states lead to so-called compass states.

Another important class of pure states is formed by squeezed states. Originally introduced in the context of precision measurements and quantum metrology~\cite{caves_quantum-mechanical_1981}, squeezed states have since found widespread applications in quantum optics~\cite{walls_quantum_2008} and quantum information theory~\cite{braunstein_squeezing_2005}. They are defined by applying the squeezing operator
\begin{equation}
    \hat S(z)=e^{\frac{1}{2}(z^* \hat a^2-z \hat a^{\dagger 2})},
\end{equation}
with $z=re^{i\theta}\in\mathbb{C}$, to the vacuum state: $\ket{z}=\hat S(z)\ket{0}$. Physically, these states exhibit reduced quantum uncertainty in one quadrature (position- or momentum-like observable) below the vacuum level, at the expense of increased uncertainty in the conjugate quadrature. Their Fock-basis expansion is given by
\begin{equation}
    \ket{z}=\frac{1}{\sqrt{\cosh r}}\sum_{n=0}^\infty 
    \frac{\sqrt{(2n)!}}{2^n n!}\left(-e^{i\theta}\tanh r\right)^n\ket{2n},
\end{equation}
where $r$ quantifies the amount of squeezing and $\theta$ denotes the squeezing angle.

All states discussed so far are pure states. In practice, however, mixed states are of central importance. A particularly relevant family is that of thermal states, which describe a quantum harmonic oscillator in thermal equilibrium with a bath at temperature $T$. Their density operator is given by
\begin{equation}
    \hat \rho_\text{th}=\frac{1}{\overline n+1}\sum_{n=0}^\infty 
    \left(\frac{\overline n}{\overline n+1}\right)^n \ketbra{n},
\end{equation}
where
\begin{equation}
    \overline n=\frac{1}{e^{1/T}-1}
\end{equation}
is the mean excitation number at temperature $T$ (in units where $k_B=1$). Thermal states provide a simple and physically motivated model for noise, losses, and finite-temperature effects in continuous-variable quantum systems~\cite{gardiner_quantum_2004}.

\paragraph{\lit Wigner function}
A final and central concept in the theory of continuous-variable quantum systems is the \emph{Wigner function}. We will revisit and extend this construction in more detail in Sec.~\ref{subsec: TF Wigner function}, where we introduce the notion of a chrono-cyclic Wigner function. Here, we briefly present its standard definition and main properties.

The Wigner function was originally introduced by E.~Wigner as a means to provide an alternative, phase-space representation of quantum states~\cite{wigner_quantum_1932}. Expansions of a quantum state in either the position or momentum basis are intrinsically biased toward one of the two conjugate variables. In contrast, the Wigner function defines a quasi-probability distribution over the phase space $(x,p)\in\mathbb{R}^2$, treating position and momentum on an equal footing.

For a quantum state described by a density operator $\hat\rho$, the Wigner function is defined as
\begin{equation}
    W(x,p)=\frac{1}{\pi}\int_{-\infty}^{+\infty} 
    e^{2 i p y}\,\bra{x-y}\hat \rho\ket{x+y}\,\dd y.
\end{equation}
The function $W(x,p)$ is real-valued, bounded within the interval $[-1/\pi,\,1/\pi]$ and has total mass 1. However, it is not a genuine probability distribution, since it may take negative values for certain quantum states. This non-positivity reflects the intrinsically quantum nature of the state and is closely related to the impossibility of simultaneously defining sharp values for both position and momentum. Despite this, the Wigner function reproduces the correct marginal probability distributions:
\begin{align}
    \int W(x,p)\,\dd p &= \bra{x}\hat \rho\ket{x}, &
    \int W(x,p)\,\dd x &= \bra{p}\hat \rho\ket{p}.
\end{align}

The Wigner function possesses many useful properties, of which we mention only a few. First, for pure states, it allows one to compute the overlap between two quantum states directly in phase space:
\begin{equation}
    \abs{\braket{\psi}{\phi}}^2
    =2\pi \int W_\psi(x,p)\,W_\phi(x,p)\,\dd x\,\dd p.
\end{equation}
Most importantly, the Wigner function provides a complete characterization of the quantum state: knowing $W(x,p)$ is equivalent to knowing the density operator $\hat\rho$. Consequently, all physical observables can be computed from it. In particular, the expectation value of an operator $\hat A$ can be expressed as
\begin{equation}
    \expval{\hat A} =2\pi \int W_A(x,p)\,W(x,p)\,\dd x\,\dd p,
\end{equation}
where $W_A(x,p)$ is the Wigner function associated with the operator $\hat A$, defined analogously as
\begin{equation}
    W_A(x,p)=\frac{1}{\pi}\int_{-\infty}^{+\infty}
    e^{2 i p y}\,\bra{x-y}\hat A\ket{x+y}\,\dd y.
\end{equation}

Beyond its formal properties, the Wigner function provides an intuitive and powerful way to visualize quantum states and their evolution in phase space. The appearance of negative regions in the Wigner function is often interpreted as a signature of non-classicality~\cite{hudson_when_1974,kenfack_negativity_2004}. Phase-space methods based on the Wigner function play a crucial role in a wide range of applications, including quantum metrology~\cite{braunstein_statistical_1994}, quantum computing and complexity~\cite{mari_positive_2012}, and quantum optics~\cite{schleich_quantum_2011,gerry_introductory_2005}.

The Wigner functions of several important continuous-variable states introduced previously can be computed explicitly. Illustrative examples are shown in Fig.~\ref{fig: example wigner}.
\begin{itemize}
    \item The Wigner function of a Fock state has a comparatively involved analytical expression. It exhibits circular symmetry in phase space, with radially decaying oscillations. As a consequence, all Fock states with $n\geq1$ display regions of negativity.
    \item The Wigner function of a coherent state $\ket{\alpha}$ is a circularly symmetric Gaussian centered at the phase-space point $(x,p)=(\sqrt{2}\Re(\alpha),\sqrt{2}\Im(\alpha))$. It is strictly positive and saturates the Heisenberg uncertainty bound.
    \item The Wigner function of a cat state $\mathcal N_\pm(\ket{\alpha}\pm\ket{-\alpha})$ consists of two Gaussian peaks centered at $\pm(\sqrt{2}\Re(\alpha),\sqrt{2}\Im(\alpha))$, corresponding to the two coherent components. An interference pattern of alternating positive and negative fringes appears between the peaks, reflecting the coherent superposition.
    \item The Wigner function of a squeezed state $\ket{z}$ is a Gaussian centered at the origin, but with an elliptical shape. The orientation and eccentricity of the ellipse are determined by the squeezing angle $\theta$ and strength $r$. One quadrature exhibits reduced fluctuations below the vacuum level, compensated by increased fluctuations in the conjugate quadrature.
    \item The Wigner function of a thermal state is a circularly symmetric Gaussian centered at the origin. Its width increases with temperature, reflecting the growing uncertainty and mixedness of the state, while remaining positive for all temperatures.
\end{itemize}

\begin{figure}[ht]
    \centering
    \begin{tabular}{ccc}
        \includegraphics[width=0.3\linewidth]{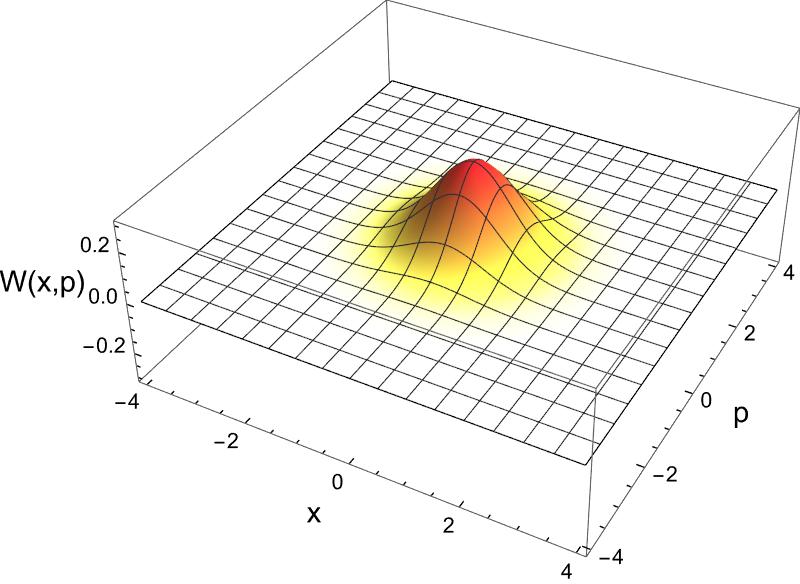}
        &
        \includegraphics[width=0.3\linewidth]{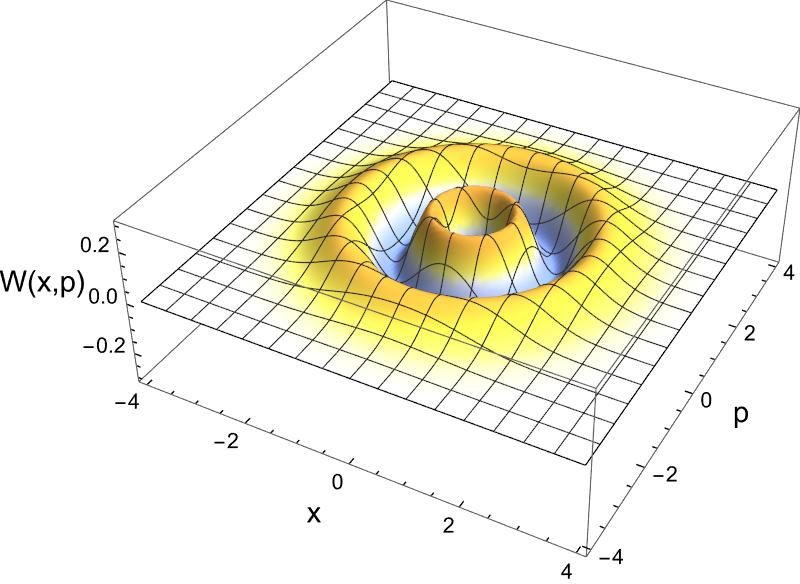}
        &
        \includegraphics[width=0.3\linewidth]{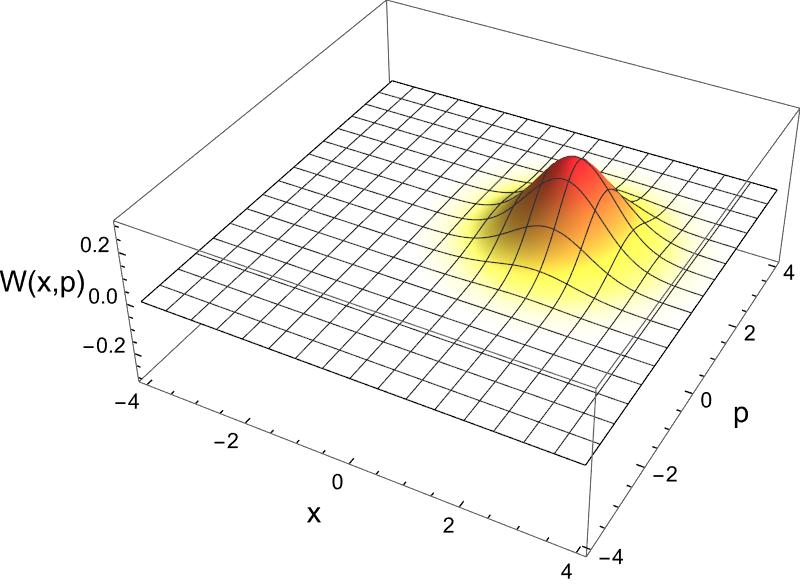}\\
        Vacuum state $\ket{0}$ & Fock state $\ket{3}$ & Coherent state $\ket{1+0.6 i}$\\[1ex]
        \includegraphics[width=0.3\linewidth]{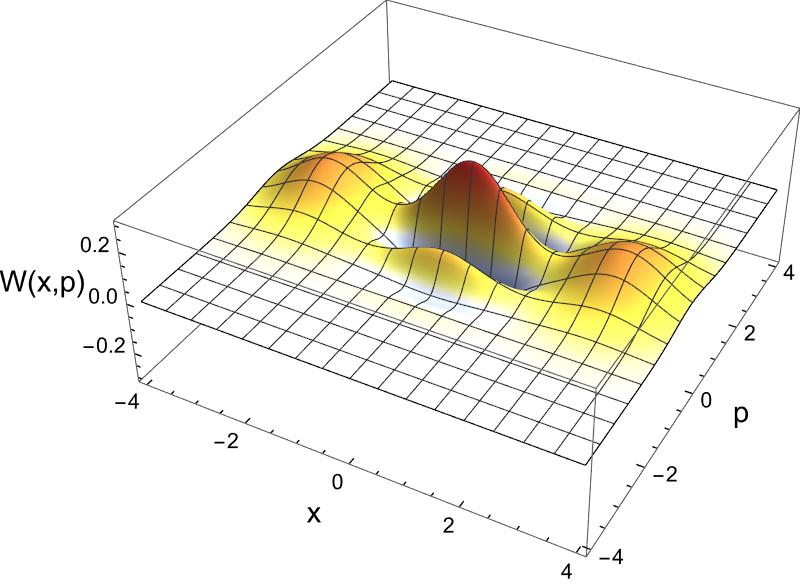}
        &
        \includegraphics[width=0.3\linewidth]{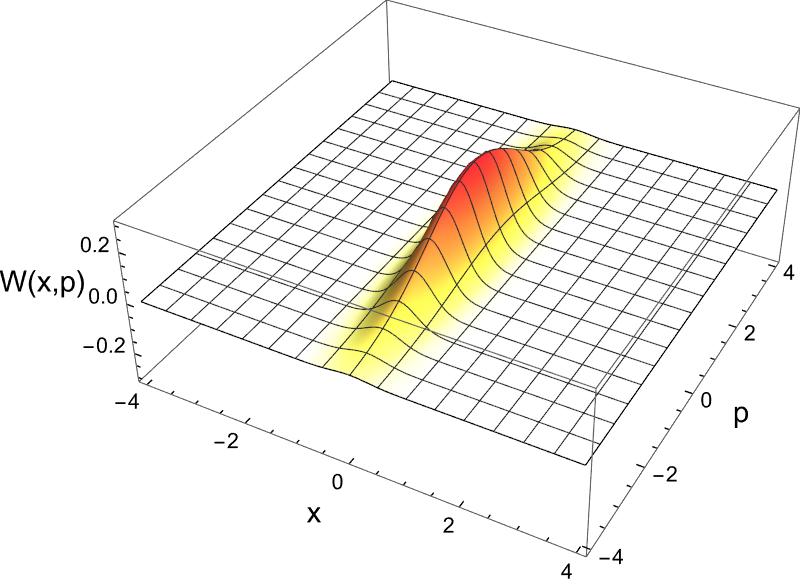}
        &
        \includegraphics[width=0.3\linewidth]{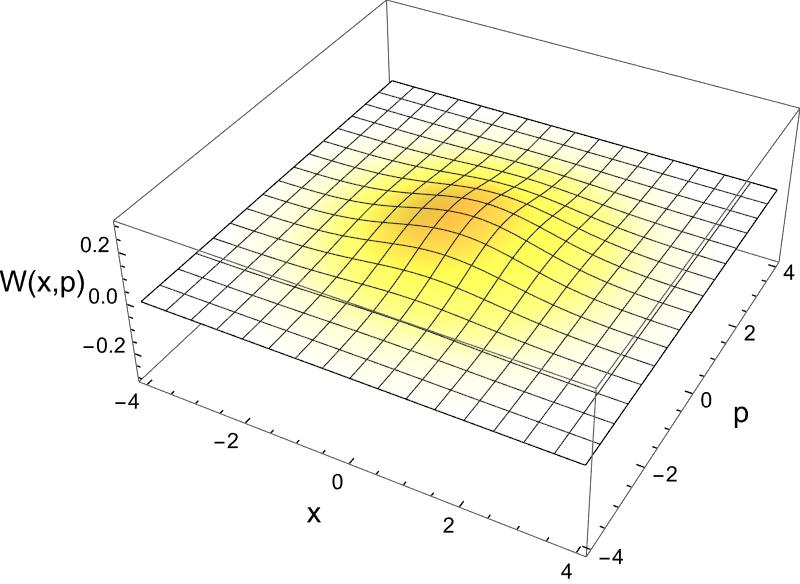}\\
        Cat state with $\alpha=4$ & Squeezed state $\ket{z=0.8}$ & Thermal state with $\overline n=1$
    \end{tabular}
    \caption[Wigner functions of typical continuous-variable quantum states]{Wigner functions of typical continuous-variable quantum states.}
    \label{fig: example wigner}
\end{figure}

\subsection{Quantum computation}
\label{subsec: quantum computation}
\paragraph{\lit Encoding and manipulating information}
Classical information processing, present in every modern computer, relies on the manipulation of digital information stored in bits taking two distinct values, usually denoted $0$ and $1$. Such a simple abstraction allows for reliable storage, transmission, and processing of information, and has enabled the development of modern classical computing technologies. While classical computing is universal in the sense that any computable function can, in principle, be evaluated on a classical computer, the use of quantum computing devices promises significant advantages in terms of speed and efficiency for certain computational tasks~\cite{nielsen_quantum_2010, preskill_quantum_2018}. 

Quantum computing is based on the same overarching idea of information processing as its classical counterpart, but the specific rules of quantum mechanics render quantum information processing fundamentally different. Most notably, the impossibility of measuring a quantum system without disturbing it, together with the fact that one cannot generally extract the full information contained in a quantum state from a single measurement, imposes new constraints on how information can be stored, processed, and retrieved.

At a fundamental level, qubits provide the natural model for encoding quantum information. Any collection of $n$ qubits can encode a classical bit string $s$ of length $n$ by associating the classical bit values $0$ and $1$ with the computational basis states $\ket{0}$ and $\ket{1}$. The bit string $s=s_1 s_2 \cdots s_n$ is then represented by the quantum state
\begin{equation}
    \ket{s} = \ket{s_1} \otimes \ket{s_2} \otimes \cdots \otimes \ket{s_n}
    = \ket{s_1, s_2, \ldots, s_n}.
\end{equation}
Classical logical operations can be implemented using sequences of simple gates such as the NOT gate and the controlled-NOT (CNOT) gate. These operations have direct analogues in quantum computing, given respectively by the Pauli-$X$ gate and the CNOT (or controlled $X$) gate
\begin{align}
    \hat X &= \begin{pmatrix}
        0 & 1 \\
        1 & 0
    \end{pmatrix}, &
    \hat C_X &= \begin{pmatrix}
        1 & 0 & 0 & 0 \\
        0 & 1 & 0 & 0 \\
        0 & 0 & 0 & 1 \\
        0 & 0 & 1 & 0
    \end{pmatrix}.
\end{align}

However, qubits can also be prepared in states beyond the computational basis. The phase gate
\begin{equation}
    \hat P_\theta=\begin{pmatrix}
    1 & 0 \\
    0 & e^{i\theta}
    \end{pmatrix}
\end{equation}
modifies the relative phase between the computational basis states while leaving their populations unchanged. The Hadamard gate
\begin{equation}
    \hat H=\frac{1}{\sqrt{2}}
    \begin{pmatrix}
    1 & 1 \\
    1 & -1
    \end{pmatrix}
\end{equation}
creates coherent superpositions of computational basis states and, conversely, maps such superpositions back to the computational basis. Additionally, the CNOT gate can be generalized to a controlled-unitary gate
\begin{equation}
    \hat C_U=
    \begin{pmatrix}
    \1 & 0 \\
    0 & \hat U
    \end{pmatrix},
\end{equation}
where the unitary $\hat U$ is conditionally applied to one qubit depending on the state of another. Combining single-qubit gates with an entangling two-qubit gate such as the CNOT is sufficient to perform arbitrary quantum computations, in the sense that any unitary transformation on a finite-dimensional Hilbert space can be approximated to arbitrary accuracy by a suitable sequence of such gates~\cite{nielsen_quantum_2010}.

\paragraph{\lit Computational models}
There exist several models to describe quantum computation, each with its own advantages and drawbacks depending on the problem considered and the physical system used to implement the computation. The circuit model provides a particularly convenient and intuitive framework~\cite{nielsen_quantum_2010}. In this model, a quantum computation is represented as a sequence of quantum gates acting on an initial state, typically chosen to be a computational basis state. The computation proceeds through the application of unitary operations (quantum gates), followed by measurements that extract classical information from the final quantum state. This procedure is commonly represented using circuit diagrams, where qubits are depicted as horizontal lines and gates as symbols acting on these lines (see Fig.~\ref{fig: circuit model ex}).

\begin{figure}[ht]
    \centering
    \begin{tabular}{ccccc}
        \begin{quantikz}
            &\gate{\hat H}&
        \end{quantikz} &
        \begin{quantikz}
            &\gate{\hat U}&
        \end{quantikz} &
        \begin{quantikz}
            &\ctrl{1}&\\
            &\gate{\hat U}& 
        \end{quantikz} &
        \begin{quantikz}
            &\meter{}
        \end{quantikz} &
        \begin{quantikz}
            \lstick{$\ket{\psi}$}& &
        \end{quantikz}\\
        Hadamard gate & Single-qubit gate & Controlled gate & Measurement & Input state
    \end{tabular}
    \begin{quantikz}
        &\gate{\hat H} & \ctrl{1}& & \permute{2,1} & & \ctrl{2} & &\permute{2,3,1}&\\
        & & \gate{\hat P_{\pi/2}} & \gate{\hat H}& & \ctrl{1} & & & & \\
        & & & & & \gate{\hat P_{\pi/4}} & \gate{\hat P_{\pi/2}} & \gate{\hat H} & &
    \end{quantikz}
    \caption[Quantum gates and circuit representations]{Top: basic quantum gates and their circuit representations. Bottom: example of a circuit implementing the quantum Fourier transform on three qubits. Intersecting lines represent swaps of qubits.}
    \label{fig: circuit model ex}
\end{figure}
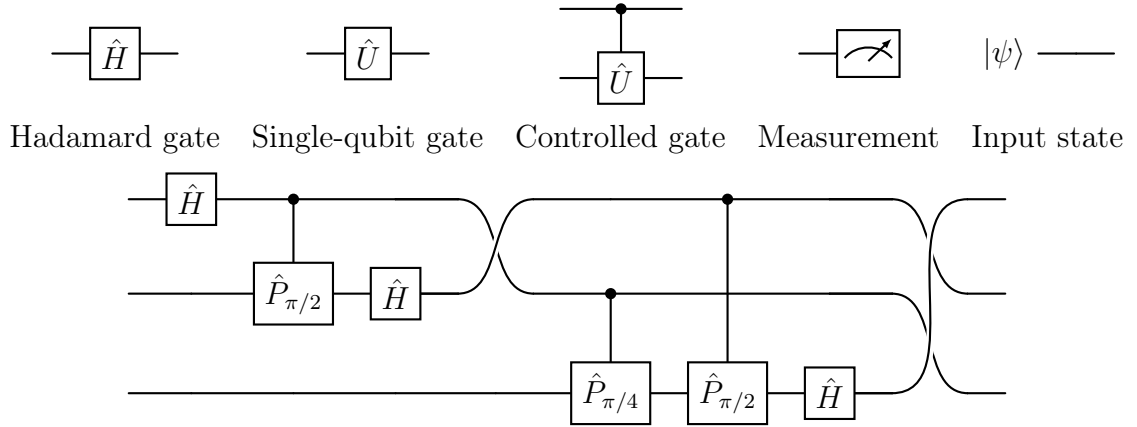

Another important computational paradigm is measurement-based quantum computation (MBQC). In the circuit model, one may assume that the initial state is a simple, fixed, unentangled state and that measurements are performed in the computational basis; the computational power is then entirely contained in the choice of the gate sequence. In contrast, MBQC relies on the preparation of a highly entangled resource state, followed by a sequence of adaptive local measurements~\cite{raussendorf_measurement-based_2003, briegel_persistent_2001}. The most prominent examples of such resource states are cluster states, which enable the implementation of arbitrary quantum computations using measurements alone. This model offers advantages in terms of conceptual clarity, fault tolerance, and scalability, making it a promising approach for practical quantum computing architectures. The basic building block of MBQC is the graph state.

A mathematical graph $G=(V,E)$ consists of a set of vertices $V$ connected by edges $E$. A graph state is constructed by associating each vertex with a qubit initialized in the state $\ket{+}$ and by applying a controlled-$Z$ gate between every pair of qubits connected by an edge. Explicitly, the graph state reads
\begin{equation}
    \ket{G}=\prod_{(i,j)\in E}\hat C_{Z}^{(i,j)} \ket{+}^{\otimes \abs{V}},
\end{equation}
where the superscript $(i,j)$ specifies the pair of qubits on which the controlled-$Z$ gate acts. The central idea of MBQC is that local measurements on individual qubits of the graph state can effectively implement quantum gates on the remaining, unmeasured qubits. This mechanism is closely related to the concept of gate teleportation~\cite{gottesman_demonstrating_1999}. The choice of measurement bases, together with the order in which measurements are performed, determines the implemented quantum operations. Suitable choices of the graph geometry and size are required to ensure universality of the computational scheme~\cite{raussendorf_one-way_2001}. Adaptive measurements, in which later measurement bases depend on earlier outcomes, enable the realization of arbitrary quantum algorithms. It can be shown that the circuit model and the measurement-based model are computationally equivalent~\cite{nielsen_quantum_2010}.

Measurement-based quantum computation can be generalized to CV systems by replacing qubits with bosonic modes. A corresponding notion of graph states exists in this setting: the initial $\ket{+}$ states are replaced by momentum-squeezed states, and the controlled-$Z$ gates used to generate entanglement are replaced by the CV entangling gate $\hat C_Z = e^{i \hat x_i \hat x_j}$~\cite{menicucci_universal_2006, menicucci_graphical_2011}. Such CV graph states can serve as resource states for CV measurement-based quantum computation~\cite{gu_quantum_2009}. Since large-scale CV cluster states can be generated efficiently in optical platforms~\cite{yokoyama_ultra-large-scale_2013, asavanant_generation_2019}, this approach represents a promising route toward scalable quantum information processing.

The use of CV cluster states exploits the full infinite-dimensional Hilbert space of continuous variables for quantum computation. However, an alternative paradigm seeks to recover the mathematical and conceptual simplicity of qubit-based quantum computing by encoding qubits into CV systems via suitable logical states. In this approach, two orthogonal states $\ket{\overline 0}$ and $\ket{\overline 1}$ of the CV system are chosen to represent the logical qubit states, and quantum computation is performed by manipulating these encoded states using appropriate gates and measurements~\cite{nielsen_quantum_2010}. Such encodings can leverage the physical advantages of CV systems while retaining the familiar qubit-based computational framework. Several encoding schemes have been proposed, each with different trade-offs in terms of error correction, resource overhead, and experimental feasibility. Prominent examples include the Gottesman-Kitaev-Preskill (GKP) encoding~\cite{gottesman_encoding_2001} (see Sec.~\ref{subsec: error correction}), the cat code~\cite{mirrahimi_dynamically_2014}, and the binomial code~\cite{michael_new_2016}.

\paragraph{\lit Universality} 
A central concept in quantum computation is that of \emph{universality}~\cite{nielsen_quantum_2010}. Loosely speaking, it reflects the fact that it is neither practical nor necessary to physically implement every possible unitary transformation directly. Instead, one seeks a small set of elementary gates that can be composed to approximate any desired unitary operation to arbitrary accuracy.

Formally, a set of gates $\mathcal U$ is said to be \emph{universal} if, for any unitary operator $\hat U$ acting on the Hilbert space of interest and for any $\epsilon>0$, there exist $\hat U_1,\dots,\hat U_n \in \mathcal U$ such that
\begin{equation}
    \norm{ \hat U - \hat U_1 \hat U_2 \cdots \hat U_n } \leq \epsilon ,
\end{equation}
where $\norm{\cdot}$ denotes some operator norm.\footnote{In finite-dimensional Hilbert spaces all norms are equivalent, so the specific choice does not affect the definition of universality. In infinite-dimensional settings, however, the choice of norm is more delicate and must be physically meaningful in order to obtain a useful notion of universality.} 
If such an approximation exists, the Solovay-Kitaev theorem~\cite{kitaev_quantum_1997,dawson_solovay-kitaev_2005} guarantees that it can be found efficiently, with the number of gates $n$ scaling only polylogarithmically with $1/\epsilon$. This fundamental result implies that a universal gate set allows one to efficiently approximate any desired unitary operation.

For qubit-based systems, many universal gate sets are known. A prominent example consists of the Hadamard gate $\hat H$, the $\pi/4$ phase gate $\hat P_{\pi/4}\equiv \hat T$, and the controlled-NOT gate $\hat C_X$. Any unitary operation on a system of $n$ qubits can be approximated to arbitrary accuracy by a sequence of these gates acting on one or two qubits at a time~\cite{nielsen_quantum_2010,barenco_universal_1995}. 

Universal gate sets can also be constructed for continuous-variable (CV) systems. Braunstein and Lloyd~\cite{lloyd_quantum_1999} showed that the set composed of all Gaussian unitaries together with any single non-Gaussian unitary is universal for CV quantum computation. Gaussian unitaries are those generated by Hamiltonians that are at most quadratic in the mode operators $\hat a$ and $\hat a^\dagger$. They include displacements, phase-space rotations, squeezing operations, and beam splitters (BS). A paradigmatic example of a non-Gaussian unitary is the cubic phase gate $\hat V(\gamma)=e^{i \gamma \hat x^3}$, generated by the Hamiltonian $\hat H=\gamma \hat x^3$. Consequently, for instance, the set
\begin{equation}
    \{e^{is\hat x_j},\, e^{is\hat x_j^2},\, e^{is\hat x_j \hat x_k},\, e^{i \gamma \hat x_j^3},\, e^{i\frac{\pi}{4}(\hat x_j^2+\hat p_j^2)} \mid s\in\mathbb{R},\ j,k=1,\dots,N \}
\end{equation}
is universal for CV quantum computation on $N$ modes.

\paragraph{\lit Efficiency and quantum advantage}
A central question in quantum computation is whether a given quantum algorithm provides a \emph{genuine} advantage over the best possible classical algorithms for the same computational task. While several celebrated examples of quantum speedup are known, such as Shor's polynomial-time algorithm for integer factorization~\cite{shor_polynomial-time_1997} and Grover's quadratic speedup for unstructured search~\cite{grover_fast_1996}, a general and rigorous characterization of when and why quantum computers outperform classical ones remains elusive. In practice, a quantum computation fails to demonstrate an advantage whenever its outcome can be efficiently reproduced by a classical simulation.

A simple but instructive observation is that quantum computations involving only product states, local operations, and local measurements can always be simulated efficiently on a classical computer: one merely needs to track the state of each subsystem independently. This immediately implies that \emph{entanglement} is a necessary resource for quantum advantage. However, it is not sufficient. A striking illustration of this fact is provided by the Gottesman-Knill theorem~\cite{gottesman_stabilizer_1997}. According to this result, any quantum computation that starts from computational-basis states, uses only gates from the set
\begin{equation}
        \{\hat H,\ \hat P_{\pi/2},\ \hat C_X\},
    \end{equation}
known collectively as \emph{Clifford gates}, and involves measurements in the computational basis, can be efficiently simulated on a classical computer, even though such circuits may generate highly entangled many-qubit states. The key to this efficient simulation is the stabilizer formalism~\cite{gottesman_stabilizer_1997,aaronson_improved_2004}. While a brute-force classical simulation would require keeping track of exponentially many complex amplitudes, the stabilizer description encodes the quantum state using only a number of parameters that scales polynomially with the system size. This demonstrates unambiguously that entanglement alone does not guarantee quantum computational advantage. Remarkably, the situation changes dramatically if the Clifford gate set is augmented by a single non-Clifford operation. For example, replacing the $\pi/2$ phase gate by the $\pi/4$ phase gate $\hat T$ promotes the gate set to a universal one, enabling quantum computations that are not known to admit efficient classical simulation. This sharp transition highlights the subtle boundary between classically tractable and genuinely quantum computational models.

Closely related questions arise in continuous-variable quantum systems. Here, an analogue of the Gottesman-Knill theorem exists: quantum computations that involve only Gaussian states, Gaussian unitary operations, and Gaussian measurements can be efficiently simulated classically~\cite{bartlett_efficient_2002}. In this correspondence, Gaussian states play a role analogous to stabilizer states, while Gaussian operations correspond to Clifford operations. The underlying reason for efficient simulability is that Gaussian states are fully characterized by their first and second moments of the quadrature operators, allowing the entire computation to be tracked via covariance matrices and mean-value vectors whose size grows only polynomially with the number of modes. As in the qubit case, Gaussian operations can generate highly entangled states, yet this entanglement alone is insufficient to yield a quantum advantage.

Further insight into classical simulability is provided by phase-space methods. In particular, it has been shown that negativity of the Wigner function constitutes a necessary resource for quantum computational advantage~\cite{veitch_negative_2012,mari_positive_2012}. If the initial states, the operations, and the measurements all admit a positive Wigner-function representation, then the entire quantum computation can be efficiently simulated by classical probabilistic sampling from the corresponding phase-space distributions. This result applies both to qudit systems of odd prime dimension~\cite{veitch_negative_2012} and to continuous-variable systems~\cite{mari_positive_2012}. Wigner-function negativity therefore acquires a clear operational meaning: it is a signature of nonclassicality that is indispensable for achieving a genuine quantum computational advantage. Phase-space approaches can be further generalized by considering alternative quasi-probability representations beyond the Wigner function~\cite{ferrie_quasi-probability_2011}, or by combining multiple phase-space representations to analyze classical simulability and quantum advantage in a unified framework~\cite{pashayan_estimating_2015,frigerio_resourcefulness_2025}.

\subsection{Error correction}
\label{subsec: error correction}
While quantum computation and quantum devices offer promising advantages over their classical counterparts, they are inherently fragile and susceptible to errors arising from decoherence, imperfect control, and unwanted interactions with the environment. Quantum error correction (QEC) provides a framework to protect quantum information from such errors, enabling reliable quantum computation even in the presence of noise~\cite{nielsen_quantum_2010}. 

At a conceptual level, quantum error correction follows ideas analogous to those of classical error correction: redundancy is used to encode information in a larger system. Quantum systems are therefore described on two distinct levels. On the \emph{physical} level, information is stored in physical qubits, which are directly affected by noise. On the \emph{logical} level, information is encoded into logical qubits using several physical qubits. The goal of QEC is to detect and correct errors acting on the physical level without disturbing the logical information. While the richness of quantum mechanics enables powerful computational advantages, it also introduces unique challenges for error correction, most notably the impossibility of directly copying or completely measuring unknown quantum states.

\paragraph{\lit Repetition code}
A naive approach to error correction consists in repeating the information multiple times to create redundancy. In the classical case, the repetition code encodes a single bit $b \in \{0,1\}$ into three bits by mapping $b \mapsto bbb$. A majority vote then allows one to correct a single bit-flip error. Adapting this idea to quantum information, one can define the following three-qubit code
\begin{align}
    \ket{\overline 0} = \ket{0,0,0}, && \ket{\overline 1} = \ket{1,1,1},
\end{align}
where the bar notation distinguishes logical states from physical ones. If any one of the local bit-flip operators $\hat X_1$, $\hat X_2$, or $\hat X_3$ acts on the state, the original logical information can be recovered using a majority rule. However, the correction procedure must not measure the explicit value of the qubits, as this would collapse an unknown superposition state
\begin{equation}
    \ket{\psi} = \alpha \ket{\overline 0} + \beta \ket{\overline 1}.
\end{equation}
This illustrates a first central challenge of QEC: errors must be corrected using measurements that do not reveal or destroy the encoded quantum information. In this example, this is achieved by measuring the commuting operators $\hat Z_1 \hat Z_2$ and $\hat Z_2 \hat Z_3$. The corresponding measurement outcomes, called the \emph{error syndrome}, reveal which error occurred without distinguishing between $\ket{\overline 0}$ and $\ket{\overline 1}$. Based on the syndrome, applying the appropriate $\hat X$ operator allows one to correct the error (see Tab.~\ref{tab: repetition code}). 

This repetition code, however, only protects against bit-flip errors. Phase-flip errors, corresponding to the local application of the $\hat Z$ operator, cannot be corrected. This highlights a second major challenge of QEC: a useful quantum error-correcting code must be able to protect against both bit-flip and phase-flip errors. Remarkably, correcting these two types of errors is sufficient to protect against arbitrary single-qubit errors, since any single-qubit error can be expanded as a linear combination of Pauli operators~\cite{nielsen_quantum_2010}.

\begin{table}
    \setlength{\tabcolsep}{10pt}
    \renewcommand{\arraystretch}{1.7}
    \centering
    \begin{tabular}{|c|c|c|c|c|}
        \hline
        Error & Quantum state & $\hat Z_1\hat Z_2$ & $\hat Z_2\hat Z_3$ & Correction \\
        \hline\hline
        None & $\alpha\ket{0,0,0}+\beta\ket{1,1,1}$ & $+1$ & $+1$ & None \\
        \hline
        $\hat X_1$ & $\alpha\ket{1,0,0}+\beta\ket{0,1,1}$ & $-1$ & $+1$ & $\hat X_1$ \\
        \hline
        $\hat X_2$ & $\alpha\ket{0,1,0}+\beta\ket{1,0,1}$ & $-1$ & $-1$ & $\hat X_2$ \\
        \hline
        $\hat X_3$ & $\alpha\ket{0,0,1}+\beta\ket{1,1,0}$ & $+1$ & $-1$ & $\hat X_3$ \\
        \hline
    \end{tabular}
    \caption[Syndrome table for the three-qubit repetition code]{Syndrome table for the three-qubit repetition code correcting a single bit-flip error. Each syndrome uniquely identifies the error without revealing the logical information, allowing unambiguous correction.}
    \label{tab: repetition code}
\end{table}

\paragraph{\lit Shor 9-qubit code}
The first quantum error-correcting code capable of correcting both bit-flip and phase-flip errors was introduced by Shor~\cite{shor_scheme_1995}. The key idea is that phase-flip errors can be corrected by a repetition code defined in the $\hat X$ basis. In particular, a three-qubit phase-flip code is given by
\begin{align}
    \ket{\overline 0} = \ket{+,+,+}, && \ket{\overline 1} = \ket{-,-,-}.
\end{align}
By reasoning analogous to the bit-flip case, this code can correct a single phase-flip error. Shor's crucial insight was to concatenate the bit-flip and phase-flip repetition codes, thereby protecting against both types of errors. This leads to the nine-qubit Shor code, defined as
\begin{align}
    \ket{\overline 0} &= \frac{1}{2\sqrt{2}}(\ket{000}+\ket{111})^{\otimes 3}, &  \ket{\overline 1} &= \frac{1}{2\sqrt{2}}(\ket{000}-\ket{111})^{\otimes 3}.
\end{align}
This code can correct an arbitrary error acting on any single physical qubit. However, it requires nine physical qubits to encode a single logical qubit, making it highly resource-intensive. More efficient codes were later discovered, most notably the five-qubit code, which is the smallest possible code capable of correcting an arbitrary single-qubit error~\cite{laflamme_perfect_1996}.

\paragraph{\lit Stabilizer codes}
To provide a unified and systematic framework for constructing and analyzing quantum error-correcting codes, Gottesman introduced the stabilizer formalism~\cite{gottesman_stabilizer_1997}. On $n$ qubits, the Pauli group is defined as
\begin{equation}
    \mathcal P_n = \{\pm 1, \pm i\} \times \{\1, \hat X, \hat Y, \hat Z\}^{\otimes n}.
\end{equation}
A stabilizer code is specified by an abelian subgroup $\mathcal S \subset \mathcal P_n$ that does not contain the element $-\1$. The code space $\mathcal C$ is defined as the subspace of states stabilized by all elements of $\mathcal S$,
\begin{equation}
    \ket{\psi} \in \mathcal C \iff \hat S \ket{\psi} = \ket{\psi}, \quad \forall \hat S \in \mathcal S.
\end{equation}
If the stabilizer group is generated by $n-k$ independent generators $\hat S_1, \dots, \hat S_{n-k}$, then the code encodes $k$ logical qubits into $n$ physical qubits.

Errors are detected by measuring the stabilizer generators. The resulting outcomes constitute the error syndrome, which indicates whether an error has occurred and identifies it up to equivalence within the code. Error correction is achieved by applying an appropriate recovery operation conditioned on the measured syndrome. Logical operations on the encoded qubits correspond to physical operators that commute with all stabilizer generators but act non-trivially within the code space.

The stabilizer formalism encompasses many important quantum codes, including the Shor code and the Steane code~\cite{steane_multiple-particle_1996}. Furthermore, it provides the foundation for more advanced classes of codes, such as topological codes~\cite{kitaev_fault-tolerant_2003} and subsystem codes~\cite{poulin_stabilizer_2005}, which play a central role in modern approaches to fault-tolerant quantum computation.

\paragraph{\lit Gottesman-Kitaev-Preskill code}

A prominent example of a quantum error-correcting code for continuous-variable systems is the Gottesman-Kitaev-Preskill (GKP) code~\cite{gottesman_encoding_2001}. The GKP code encodes a logical qubit into the infinite-dimensional Hilbert space of a single bosonic mode using highly non-classical states, known as GKP states, such that quantum information is protected against sufficiently small shifts in position and momentum. Owing to the central role played by GKP codes in Sec.~\ref{sec: collective GKP codes}, we provide here an intuitive introduction in which the structure of the GKP logical states arises naturally from the stabilizer formalism.

The goal is to protect against displacement errors of the form $e^{-i \delta \hat p}$ (position shifts) and $e^{-i \delta \hat x}$ (momentum shifts). It is therefore natural to define a stabilizer code whose stabilizers are displacement operators. We introduce the two stabilizers
\begin{align}
    \hat S_1 = e^{-2 i \alpha \hat p}, &&
    \hat S_2 = e^{-2 i \beta \hat x}.
\end{align}
Since stabilizers must commute, the parameters $\alpha$ and $\beta$ must be chosen appropriately. A particular case of the Baker-Campbell-Hausdorff formula (see Appendix~\ref{app: formalism and framework}, Result~\ref{res: BCH}) states that if operators $\hat A$ and $\hat B$ commute with their commutator $[\hat A, \hat B]$, then
\begin{equation}
    e^{\hat A} e^{\hat B} = e^{\hat B} e^{\hat A} e^{[\hat A, \hat B]}.
\end{equation}
Applying this identity yields
\begin{equation}
    \hat S_1 \hat S_2
    = \hat S_2 \hat S_1 e^{[2 i \alpha \hat p, 2 i \beta \hat x]}
    = \hat S_2 \hat S_1 e^{4 \alpha \beta [\hat x, \hat p]}
    = \hat S_2 \hat S_1 e^{4 i \alpha \beta},
\end{equation}
where we used $[\hat x, \hat p] = i\1$. The stabilizers thus commute if and only if
\begin{equation}
    4 \alpha \beta = 2 \pi m
\end{equation}
for some integer $m$, implying $\beta = \frac{m \pi}{2 \alpha}$. The choice $m=1$ would lead to a stabilized subspace of dimension one and is therefore insufficient to encode a qubit. Larger values of $m$ would allow encoding higher-dimensional logical systems. In the following, we choose $m=2$, which yields
\begin{equation}
    \beta = \frac{\pi}{\alpha},
\end{equation}
and the stabilizers take the form
\begin{align}
    \hat S_1 = e^{-2 i \alpha \hat p}, &&  \hat S_2 = e^{-2 i \pi \hat x / \alpha}.
\end{align}

We now derive the form of the states stabilized by $\hat S_1$ and $\hat S_2$. Let $\ket{\psi}$ satisfy $\hat S_1 \ket{\psi} = \hat S_2 \ket{\psi} = \ket{\psi}$ and write it in the position basis as $\ket{\psi} = \int \dd x\, f(x) \ket{x}$. Acting with $\hat S_2$ gives
\begin{equation}
    \hat S_2 \ket{\psi}
    = \int \dd x\, e^{-2 i \pi x / \alpha} f(x) \ket{x}
    = \ket{\psi}.
\end{equation}
This implies that $e^{-2 i \pi x / \alpha} f(x) = f(x)$ for all $x$, which, for nonzero $f(x)$, is only possible when $x = k \alpha$ with $k \in \mathbb{Z}$. Therefore,
\begin{equation}
    \ket{\psi} = \sum_{k \in \mathbb{Z}} a_k \ket{k \alpha},
\end{equation}
for some coefficients $a_k \in \mathbb{C}$. Next, we impose stabilization under $\hat S_1$. Using the fact that $e^{-i x_0 \hat p}$ translates position eigenstates by $x_0$, we find
\begin{align}
    \hat S_1 \ket{\psi}
    &= \sum_{k \in \mathbb{Z}} a_k \ket{k \alpha + 2 \alpha}
    = \sum_{k \in \mathbb{Z}} a_{k-2} \ket{k \alpha}.
\end{align}
Requiring $\hat S_1 \ket{\psi} = \ket{\psi}$ implies $a_k = a_{k-2}$ for all $k$, so the sequence $\{a_k\}$ is $2$-periodic. The stabilized subspace is therefore two-dimensional, as expected for a logical qubit. Grouping even and odd indices, we obtain the logical basis states
\begin{align}
    \ket{\overline 0} = \sum_{k \in \mathbb{Z}} \ket{2 k \alpha}, && \ket{\overline 1} = \sum_{k \in \mathbb{Z}} \ket{(2 k + 1) \alpha}.
\end{align}

These are the ideal GKP logical states. They are not normalizable and hence unphysical. Physical approximations are obtained by replacing position eigenstates with sharply peaked wave packets, such as finitely squeezed states~\cite{gottesman_encoding_2001, menicucci_fault-tolerant_2014}. To fully specify a logical qubit, we define logical Pauli operators. Searching for translations that reproduce the action of the Pauli matrices, we introduce
\begin{align}
    \hat X = e^{-i \alpha \hat p}, && \hat Z = e^{-i \pi \hat x / \alpha}.
\end{align}
These satisfy $\hat X^2 = \hat S_1$ and $\hat Z^2 = \hat S_2$. Their commutation relation is $\hat Z \hat X = - \hat X \hat Z$ since $[-i \pi \hat x / \alpha, -i \alpha \hat p] = - i \pi$. One readily verifies their action on the logical states:
\begin{subequations}
\begin{align}
    \hat X \ket{\overline 0} &= \ket{\overline 1}, &
    \hat X \ket{\overline 1} &= \ket{\overline 0}, \\
    \hat Z \ket{\overline 0} &= \ket{\overline 0}, &
    \hat Z \ket{\overline 1} &= -\ket{\overline 1}.
\end{align}
\end{subequations}

The GKP code is designed to correct small displacement errors in position and momentum. Consider a momentum displacement error $\hat D(\delta) = e^{-i \delta \hat x}$. Acting on a logical state $\ket{\overline k}$ and measuring the stabilizer $\hat S_1$ yields
\begin{subequations}
\begin{align}
    \hat S_1 \hat D(\delta) \ket{\overline k} &= e^{-2 i \alpha \hat p} e^{-i \delta \hat x} \ket{\overline k}, \\
    &= e^{2 i \alpha \delta} \hat D(\delta) \hat S_1 \ket{\overline k}, \\
    &= e^{2 i \alpha \delta} \hat D(\delta) \ket{\overline k}.
\end{align}
\end{subequations}
Thus $\hat D(\delta) \ket{\overline k}$ is an eigenstate of $\hat S_1$ with eigenvalue $e^{2 i \alpha \delta}$. Measuring $\hat S_1$ reveals information about $\delta$ without disturbing the logical information, since the outcome is independent of $k$. Because the map $\delta \mapsto e^{2 i \alpha \delta}$ is periodic, $\delta$ can be uniquely inferred only if $2 \alpha \delta \in (-\pi, \pi)$, \ie,
\begin{equation}
    \abs{\delta} < \frac{\pi}{2 \alpha}.
\end{equation}
In this regime, the error can be corrected by applying $\hat D(-\delta)$. An analogous argument shows that position displacement errors $e^{i \delta \hat p}$ are detected by measuring $\hat S_2$ and are correctable provided $\abs{\delta} < \alpha / 2$.

These correctability thresholds can also be understood geometrically from the wavefunction representations of the logical states, shown in Fig.~\ref{fig: 1 mode gkp error correction}. In the position representation, the spacing between peaks is $\alpha$, so errors smaller than $\alpha/2$ are correctable. By symmetry, the momentum representation of the logical $\ket{\overline \pm}$ states exhibits the same structure with $\alpha$ and $\pi/\alpha$ exchanged~\cite{terhal_quantum_2015, weedbrook_gaussian_2012}. The parameter $\alpha$ can therefore be tuned to trade off protection against position versus momentum displacement errors.

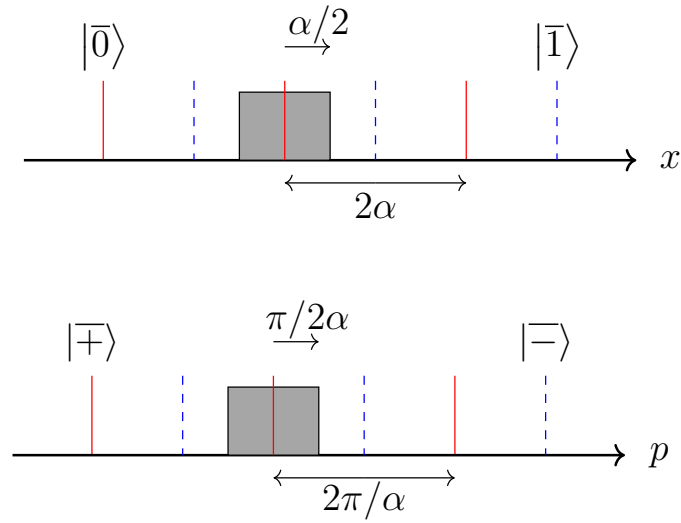
\begin{figure}[ht]
    \centering
    \scalebox{1.2}{\begin{tikzpicture}
	\begin{pgfonlayer}{nodelayer}
		\node [style=none] (0) at (6.75, 0) {};
		\node [style=none] (1) at (-6.75, 0) {};
		\node [style=none] (4) at (-5, 1.75) {};
		\node [style=none] (5) at (-3, 1.75) {};
		\node [style=none] (6) at (-1, 1.75) {};
		\node [style=none] (7) at (1, 1.75) {};
		\node [style=none] (8) at (3, 1.75) {};
		\node [style=none] (9) at (5, 1.75) {};
		\node [style=none] (13) at (-5, 0) {};
		\node [style=none] (14) at (-3, 0) {};
		\node [style=none] (15) at (-1, 0) {};
		\node [style=none] (16) at (1, 0) {};
		\node [style=none] (17) at (3, 0) {};
		\node [style=none] (18) at (5, 0) {};
		\node [style=none] (20) at (-2, 1.5) {};
		\node [style=none] (21) at (0, 1.5) {};
		\node [style=none] (22) at (0, 0) {};
		\node [style=none] (23) at (-2, 0) {};
		\node [style=none] (24) at (1, -0.5) {};
		\node [style=none] (25) at (-1, -0.5) {};
		\node [style=none] (26) at (3, -0.5) {};
		\node [style=none] (27) at (3, -0.5) {};
		\node [style=none] (28) at (1, -1) {$2\alpha$};
		\node [style=none] (29) at (-1, 2.5) {};
		\node [style=none] (30) at (0, 2.5) {};
		\node [style=none] (31) at (-0.25, 3) {$\alpha/2$};
		\node [style=none] (32) at (7.5, 0) {$x$};
		\node [style=none] (33) at (6.5, -6.5) {};
		\node [style=none] (34) at (-7, -6.5) {};
		\node [style=none] (35) at (-5.25, -4.75) {};
		\node [style=none] (36) at (-3.25, -4.75) {};
		\node [style=none] (37) at (-1.25, -4.75) {};
		\node [style=none] (38) at (0.75, -4.75) {};
		\node [style=none] (39) at (2.75, -4.75) {};
		\node [style=none] (40) at (4.75, -4.75) {};
		\node [style=none] (41) at (-5.25, -6.5) {};
		\node [style=none] (42) at (-3.25, -6.5) {};
		\node [style=none] (43) at (-1.25, -6.5) {};
		\node [style=none] (44) at (0.75, -6.5) {};
		\node [style=none] (45) at (2.75, -6.5) {};
		\node [style=none] (46) at (4.75, -6.5) {};
		\node [style=none] (47) at (-2.25, -5) {};
		\node [style=none] (48) at (-0.25, -5) {};
		\node [style=none] (49) at (-0.25, -6.5) {};
		\node [style=none] (50) at (-2.25, -6.5) {};
		\node [style=none] (51) at (0.75, -7) {};
		\node [style=none] (52) at (-1.25, -7) {};
		\node [style=none] (53) at (2.75, -7) {};
		\node [style=none] (54) at (2.75, -7) {};
		\node [style=none] (55) at (0.75, -7.5) {$2\pi/\alpha$};
		\node [style=none] (56) at (-1.25, -4) {};
		\node [style=none] (57) at (-0.25, -4) {};
		\node [style=none] (58) at (-0.5, -3.5) {$\pi/2\alpha$};
		\node [style=none] (59) at (7.25, -6.5) {$p$};
		\node [style=none] (60) at (-5, 2.5) {$\ket{\overline 0}$};
		\node [style=none] (62) at (5, 2.5) {$\ket{\overline 1}$};
		\node [style=none] (63) at (-5.25, -4) {$\ket{\overline +}$};
		\node [style=none] (64) at (4.75, -4) {$\ket{\overline -}$};
	\end{pgfonlayer}
	\begin{pgfonlayer}{edgelayer}
		\draw [style=red line] (4.center) to (13.center);
		\draw [style=red line] (8.center) to (17.center);
		\draw [style=Thick arrow] (1.center) to (0.center);
		\draw [style=dashes blue] (5.center) to (14.center);
		\draw [style=dashes blue] (7.center) to (16.center);
		\draw [style=dashes blue] (9.center) to (18.center);
		\draw [style=Fill grey] (20.center)
			 to [in=180, out=0] (21.center)
			 to (22.center)
			 to (23.center)
			 to cycle;
		\draw [style=red line] (6.center) to (15.center);
		\draw [style=Arrow] (24.center) to (27.center);
		\draw [style=Arrow] (24.center) to (25.center);
		\draw [style=Arrow] (29.center) to (30.center);
		\draw [style=red line] (35.center) to (41.center);
		\draw [style=red line] (39.center) to (45.center);
		\draw [style=Thick arrow] (34.center) to (33.center);
		\draw [style=dashes blue] (36.center) to (42.center);
		\draw [style=dashes blue] (38.center) to (44.center);
		\draw [style=dashes blue] (40.center) to (46.center);
		\draw [style=Fill grey] (47.center)
			 to [in=180, out=0] (48.center)
			 to (49.center)
			 to (50.center)
			 to cycle;
		\draw [style=red line] (37.center) to (43.center);
		\draw [style=Arrow] (51.center) to (54.center);
		\draw [style=Arrow] (51.center) to (52.center);
		\draw [style=Arrow] (56.center) to (57.center);
	\end{pgfonlayer}
\end{tikzpicture}}
    \caption[Position and momentum representation of GKP logical states]{Top: position-space wavefunctions of the logical states $\ket{\overline 0}$ and $\ket{\overline 1}$. Bottom: momentum-space wavefunctions of $\ket{\overline \pm}$. The shaded regions indicate the correctable displacement errors.}
    \label{fig: 1 mode gkp error correction}
\end{figure}

\clearpage
\section{Metrology and quantum metrology}
\label{sec: metrology}
\emph{This section provides a detailed introduction to metrology and quantum metrology. It introduces the fundamental concepts of parameter estimation, estimators, the Cramér-Rao bound, Fisher information, and their quantum generalizations. This presentation is largely inspired by two hybrid lecture-tutorial sessions I delivered for first-year master's students in quantum information at Sorbonne Université (Paris) in 2025.}

\subsection{Classical metrology: formalism and estimators}
\label{subsec: classical metrology}
\paragraph{\lit Context and definition}

\emph{Metrology} is the science of measurement and, more precisely, of estimating unknown parameters characterizing physical systems or processes. A central goal of metrology is to determine such parameters with the highest possible precision and to identify the fundamental limits imposed by statistics and physical laws~\cite{helstrom_quantum_1976, kay_fundamentals_1998}. At the most fundamental level, metrology addresses the following problem: given access to a physical process depending on an unknown parameter $\theta$, how precisely can $\theta$ be estimated?\footnote{The ideas presented in this section can be generalized to a multi-parameter setting, where multiple parameters $\theta_1,\dots,\theta_n$ have to be simultaneously determined. However, throughout this thesis only the single-parameter case will be introduced and studied.} In practice, two broad scenarios can be distinguished:
\begin{itemize}
    \item The parameter can be measured directly. For instance, estimating a length using a ruler, where the achievable precision is directly limited by the device resolution (\ie, the spacing between markings).
    
    \item The parameter cannot be accessed directly but influences other measurable quantities. In this case, estimating the uncertainty becomes nontrivial. Only when the parameter depends on a single measured quantity can one straightforwardly apply standard error-propagation formulas.
\end{itemize}

To formalize this setting, we assume that we wish to estimate a real parameter $\theta \in \mathbb{R}$. A measurement produces an outcome $x$ belonging to a set $X$ of possible outcomes. The outcome is assumed to be random, with a probability distribution that depends on the value of $\theta$. We denote by $P_\theta(x)$ the probability of obtaining outcome $x$ when the true value of the parameter is $\theta$.

If the measurement is repeated several times, one obtains a sequence of outcomes $x_1, x_2, \dots$. The task is then to infer $\theta$ from these data. This is achieved through the definition of an \emph{estimator}. An estimator $\overline{\theta}$ is a function
\begin{equation}
    \begin{aligned}
        \overline{\theta}: X &\to \mathbb{R},\\
        x &\mapsto \overline{\theta}(x),
    \end{aligned}
\end{equation}
which assigns an estimated value of the parameter to each possible measurement outcome. For an estimator to be useful, it should satisfy certain desirable properties
\begin{itemize}
    \item \textbf{Unbiased (or accurate) estimator:} On average, the estimator should return the true value of the parameter:
    \begin{equation}
        \mathbb{E}(\overline{\theta}) = \sum_{x \in X} \overline{\theta}(x) P_\theta(x) = \theta.
    \end{equation}
    
    \item \textbf{Precise estimator:} The estimator should exhibit minimal statistical fluctuations around its mean value. This is quantified by its variance
    \begin{equation}
        \mathbb{V}(\overline{\theta}) = \mathbb{E}(\overline{\theta}^2) - \mathbb{E}(\overline{\theta})^2.
    \end{equation}
\end{itemize}

\paragraph{\lit Coin-flip example}

As a simple illustrative example, consider a coin with an unknown probability $\theta$ of landing heads. The set of outcomes is $X=\{H,T\}$, with probabilities
\begin{align}
    P_\theta(H) = \theta, && P_\theta(T) = 1-\theta.
\end{align}
A natural estimator is defined by $\overline{\theta}(H)=1$ and $\overline{\theta}(T)=0$. Its expectation value is
\begin{equation}
    \mathbb{E}(\overline{\theta}) = 1 \times \theta + 0 \times (1-\theta) = \theta,
\end{equation}
so the estimator is unbiased. Its variance is
\begin{align}
    \mathbb{V}(\overline{\theta}) 
    &= \mathbb{E}(\overline{\theta}^2) - \mathbb{E}(\overline{\theta})^2, \notag \\
    &= \theta - \theta^2 = \theta(1-\theta),
\end{align}
which corresponds to the variance of a Bernoulli random variable.

\paragraph{\lit Independent repetitions}

As shown in Appendix~\ref{app: formalism and framework}, Result~\ref{res: unbaised estimator may not exist}, no unbiased estimator exists for estimating $\theta$ in a coin-flipping scenario where $P_\theta(H)=\theta^2$. This highlights a subtle point: is estimating $\theta$ fundamentally different from estimating $\theta^2$? Intuitively, one might estimate $\theta^2$ first and then infer $\theta$ via error propagation. This apparent paradox arises because the estimator defined above is a \emph{single-shot estimator}, depending on only one outcome. In practice, however, parameters are estimated from many repetitions of the same experiment. Importantly, repeated measurements can always be reinterpreted as a single measurement with a larger outcome space. Formally, suppose the experiment is repeated $n$ times independently. The joint outcome space is $\tilde{X} = X^n$, and the joint probability distribution is
\begin{equation}
    \tilde{P}_\theta(x_1,\dots,x_n) = \prod_{i=1}^n P_\theta(x_i).
\end{equation}
Given an estimator $\overline{\theta}$ defined on $X$, a natural estimator on $\tilde{X}$ is
\begin{equation}
    \overline{\tilde{\theta}}(x_1,\dots,x_n) = \frac{1}{n}\sum_{i=1}^n \overline{\theta}(x_i).
\end{equation}
If $\overline{\theta}$ is unbiased, then so is $\overline{\tilde{\theta}}$, and their variances are related by
\begin{equation}
    \mathbb{V}(\overline{\tilde{\theta}}) = \frac{\mathbb{V}(\overline{\theta})}{n},
\end{equation}
which we verify in Appendix~\ref{app: formalism and framework}, Result~\ref{res: bias and variance of a repeated measurement}. This $1/\sqrt{n}$ scaling of the standard deviation is a hallmark of classical statistics and will reappear later when discussing quantum-enhanced strategies.

\subsection{The Cramér-Rao bound}
\label{subsec: cramer-rao bound}
\paragraph{\lit Statement}
A central question in metrology and statistics is to determine the fundamental limits on the precision with which an unknown parameter can be estimated from measurement data. As mentioned above, precision is typically quantified by the variance of an estimator. The Cramér-Rao bound (CRB) provides a universal lower bound on this variance, thereby characterizing the best achievable precision for a given statistical model. The bound is expressed in terms of the \emph{Fisher information}, an information-theoretic quantity that measures how sensitively the probability distribution depends on the parameter of interest~\cite{harald_cramer_mathematical_1946, kay_fundamentals_1998}. Using the notation introduced above, consider the estimation of a parameter $\theta$ using an arbitrary estimator $\overline{\theta}$. Its variance is bounded from below as
\begin{equation}
    \mathbb V(\overline{\theta}) \geq 
    \frac{\left[\frac{\partial}{\partial\theta}\mathbb E(\overline{\theta})\right]^2}{\mathcal F_\theta},
\end{equation}
where $\mathcal F_\theta$ denotes the Fisher information, defined for a discrete\footnote{For continuous measurement data, the sum is simply replaced by an integral:
\begin{equation}
    \mathcal F_\theta = \int \dd x\, \frac{1}{P_\theta(x)}
    \left(\frac{\partial P_\theta(x)}{\partial \theta}\right)^2.
\end{equation}
} outcome space $X$ as
\begin{equation}\label{eq: Fisher information expression}
    \mathcal F_\theta = \sum_{x\in X} \frac{1}{P_\theta(x)}
    \left(\frac{\partial P_\theta(x)}{\partial \theta}\right)^2 .
\end{equation}
Informally, the Fisher information quantifies the amount of information about the parameter $\theta$ contained in the probability distribution $P_\theta(x)$. Intuitively, it measures how strongly the distribution changes when $\theta$ is varied. If the estimator $\overline{\theta}$ is \emph{unbiased}, \ie, $\mathbb E(\overline{\theta})=\theta$, then
\begin{equation}
    \frac{\partial}{\partial\theta}\mathbb E(\overline{\theta}) = 1,
\end{equation}
and the Cramér-Rao bound simplifies to
\begin{equation}
    \mathbb V(\overline{\theta}) \geq \frac{1}{\mathcal F_\theta}.
\end{equation}
This form of the CRB is particularly important in metrology: it shows that the precision of any unbiased estimator is fundamentally limited by the inverse of the Fisher information. A larger Fisher information implies a tighter bound and hence potentially higher precision.

As a simple illustration, consider again the coin-tossing example where the probability of obtaining heads is $\theta$. As shown in Appendix~\ref{app: formalism and framework}, Result~\ref{res: Fi coin flip}, the Fisher information associated with the estimator introduced there is
\begin{equation}
    \mathcal F_\theta = \frac{1}{\theta(1-\theta)} = \frac{1}{\mathbb V(\overline{\theta})},
\end{equation}
which shows that the Cramér-Rao bound is saturated in this case.

\paragraph{\lit Properties of the Fisher information}
An important property of the Fisher information is its \emph{additivity} for independent experiments. Let $X = X_1 \times X_2$ and assume that the joint probability distribution factorizes as
\begin{equation}
P_\theta(x_1,x_2) = P_\theta^{(1)}(x_1)\,P_\theta^{(2)}(x_2),
\end{equation}
where $P_\theta^{(1)}$ and $P_\theta^{(2)}$ correspond to two independent experiments. Denoting by $\mathcal F_\theta$, $\mathcal F_\theta^{(1)}$, and $\mathcal F_\theta^{(2)}$ the associated Fisher information values, one finds (see Appendix~\ref{app: formalism and framework}, Result~\ref{res: additivity of Fi} for a proof)
\begin{equation}
    \mathcal F_\theta = \mathcal F_\theta^{(1)} + \mathcal F_\theta^{(2)}.
\end{equation}
Intuitively, this additivity reflects the fact that independent data sets contribute independent pieces of information about the parameter $\theta$.

Now consider $n$ identical and independent repetitions of the same measurement. If $\mathcal F_\theta$ denotes the Fisher information associated with a single measurement and $\mathcal F_\theta^{(n)}$ that associated with all $n$ outcomes jointly, repeated application of the additivity property yields
\begin{equation}
    \mathcal F_\theta^{(n)} = n\, \mathcal F_\theta .
\end{equation}
Suppose that the measurements yield unbiased estimators $\overline{\theta}_1,\dots,\overline{\theta}_n$, each satisfying
\begin{equation}
    \mathbb V(\overline{\theta}_i) \geq \frac{1}{\mathcal F_\theta}.
\end{equation}
Consider now the averaged estimator
\begin{equation}
    \overline{\theta} = \frac{1}{n}\sum_{i=1}^n \overline{\theta}_i .
\end{equation}
By independence, its variance is
\begin{equation}
    \mathbb V(\overline{\theta})
    = \frac{\mathbb V(\overline{\theta}_1)+\cdots+\mathbb V(\overline{\theta}_n)}{n^2}
    = \frac{\mathbb V(\overline{\theta}_1)}{n}.
\end{equation}
Using the CRB for each $\overline{\theta}_i$ then gives
\begin{equation}
    \mathbb V(\overline{\theta}) \geq \frac{1}{n \mathcal F_\theta}
    = \frac{1}{\mathcal F_\theta^{(n)}}.
\end{equation}
This coincides with the Cramér-Rao bound obtained by treating $\overline{\theta}$ as a single unbiased estimator based on all $n$ measurement outcomes. The conclusion is that, when unbiased estimators exist, repeating the experiment and simply averaging the individual estimates already achieves the optimal scaling of precision. No more elaborate estimator can do better.

Another key property of the Fisher information is its \emph{convexity}. Mixing probability distributions can only decrease the Fisher information. More precisely, consider a distribution of the form
\begin{equation}
    P_\theta(x) = \sum_j p_j P_\theta^{(j)}(x),
\end{equation}
where $\{p_i\}$ is a probability distribution and each $P_\theta^{(j)}(x)$ depends on $\theta$. Then one has (see Appendix~\ref{app: formalism and framework}, Result~\ref{res: convexity of Fi} for a proof)
\begin{equation}
    \mathcal F_\theta \leq \sum_j p_j \mathcal F_\theta^{(j)},
\end{equation}
where $\mathcal F_\theta$ and $\mathcal F_\theta^{(j)}$ are the Fisher information values associated with $P_\theta$ and $P_\theta^{(j)}$, respectively. This property has important implications in metrology: introducing classical randomness or uncertainty into the measurement process cannot improve the achievable precision.

\paragraph{\lit Proof}
The proof of the Cramér-Rao bound relies on the \emph{score function}, which quantifies the sensitivity of the log-likelihood to changes in the parameter $\theta$~\cite{kay_fundamentals_1998}. The main steps of the proof are presented in Appendix~\ref{app: formalism and framework}, Result~\ref{res: CRB}. The score function $S$ is defined as
\begin{equation}
    \begin{aligned}
        S: X &\to \mathbb R,\\
        x &\mapsto S(x) = \frac{\partial}{\partial \theta}\ln P_\theta(x)
        = \frac{1}{P_\theta(x)}\frac{\partial P_\theta(x)}{\partial \theta}.
    \end{aligned}
\end{equation}
One can verify that the variance of the score equals the Fisher information,
\begin{equation}
\mathbb V(S) = \mathcal F_\theta,
\end{equation}
and that its covariance with any estimator $\overline{\theta}$ satisfies
\begin{equation}
    \Cov (S,\overline{\theta})
    = \frac{\partial}{\partial \theta}\mathbb E(\overline{\theta}).
\end{equation}
Applying the Cauchy-Schwarz inequality to the covariance yields
\begin{equation}
    \abs{\Cov (S,\overline{\theta})}^2
    \leq \mathbb V(S)\,\mathbb V(\overline{\theta}),
\end{equation}
which, upon substitution, becomes
\begin{equation}
    \left(\frac{\partial}{\partial\theta}\mathbb E(\overline{\theta})\right)^2
    \leq \mathcal F_\theta\, \mathbb V(\overline{\theta}).
\end{equation}
Rearranging the terms directly leads to the Cramér-Rao bound.

\subsection{Maximum likelihood estimator}
\label{subsec: maximum likelihood estimator}
\paragraph{\lit Definition}
In practical situations, it is often unclear how to construct an estimator that efficiently estimates the parameter $\theta$. Moreover, as seen with the example above, unbiased estimators do not always exist. The problem becomes even more intricate when multiple measurement rounds are performed, since the estimator may, in principle, depend on all outcomes jointly. A powerful and widely used approach is provided by the \emph{maximum likelihood estimator} (MLE)~\cite{fisher_mathematical_1922, casella_statistical_2002}. The MLE offers a systematic method for constructing estimators and enjoys strong optimality properties: under suitable regularity conditions and for independent and identical repetitions of the experiment, the MLE is asymptotically unbiased and attains the Cramér-Rao bound.

The basic idea of the MLE is simple. Upon observing an outcome $x\in X$, one chooses as estimate the value of $\theta$ that makes this outcome most likely. Formally, the estimator is defined as
\begin{equation}
    \overline{\theta}(x) = \operatorname*{argmax}_{\theta} P_\theta(x).
\end{equation}
The function $P_\theta(x)$ is also called the \emph{likelihood}, and maximizing it explains the name of the method. If the map $\theta\mapsto P_\theta(x)$ is sufficiently regular, the optimization problem can be reduced to solving
\begin{equation}
\left.\frac{\partial}{\partial\theta}P_\theta(x)\right|_{\overline{\theta}} = 0.
\end{equation}
Since the logarithm is an increasing function, this is equivalent to solving
\begin{equation}
\left.\frac{\partial}{\partial\theta}\ln P_\theta(x)\right|_{\overline{\theta}} = 0,
\end{equation}
which is often more convenient in practice.

\paragraph{\lit Examples}
For the coin-tossing example with $P_\theta(H)=\theta$, the estimator considered previously coincides with the MLE. Indeed, restricting to $\theta\in[0,1]$, one finds
\begin{equation}
    \overline{\theta}(H)
    = \operatorname*{argmax}_{\theta\in[0,1]} \theta
    = 1,
\end{equation}
and similarly
\begin{equation}
\overline{\theta}(T) = \operatorname*{argmax}_{\theta\in[0,1]} (1-\theta) = 0.
\end{equation}

For $n$ identical and independent repetitions of a measurement, the joint probability of obtaining outcomes $x_1,\dots,x_n$ factorizes as
\begin{equation}
    \tilde{P}_\theta(x_1,\dots,x_n)
    = P_\theta(x_1)\cdots P_\theta(x_n).
\end{equation}
Taking the logarithm, the MLE condition becomes
\begin{equation}
    \frac{P'_{\overline{\theta}}(x_1)}{P_{\overline{\theta}}(x_1)}
    + \cdots +
    \frac{P'_{\overline{\theta}}(x_n)}{P_{\overline{\theta}}(x_n)}
    = 0.
\end{equation}
Consider again the coin-tossing example, but now assume $P_\theta(H)=f(\theta)$ for some smooth function $f:\mathbb R\to(0,1)$. If $k$ denotes the number of heads observed among the $n$ trials, the above condition reduces to
\begin{equation*}
    k\frac{f'(\overline{\theta})}{f(\overline{\theta})}
    - (n-k)\frac{f'(\overline{\theta})}{1-f(\overline{\theta})}
    = 0,
\end{equation*}
which, provided $f'$ does not vanish, is solved by
\begin{equation}
f(\overline{\theta}) = \frac{k}{n},
\end{equation}
or, if $f$ is invertible,
\begin{equation}
\overline{\theta} = f^{-1}\!\left(\frac{k}{n}\right).
\end{equation}
The quantity $k/n$ is simply the observed frequency of heads, which for large $n$ provides a good estimate of the true probability $f(\theta)$. The MLE thus implements the intuitive strategy of first estimating the probability of heads and then inferring the parameter $\theta$.

\paragraph{\lit Convergence}
In most cases, the expectation value and variance of the MLE cannot be computed exactly. Nevertheless, under suitable regularity assumptions on $P_\theta(x)$, one can show that the MLE based on $n$ independent repetitions is asymptotically unbiased and achieves the optimal variance~\cite{harald_cramer_mathematical_1946, jeganathan_extension_1980, cam_locally_1960}. More precisely,
\begin{equation}
    \sqrt{n}\,(\overline{\theta}-\theta)
    \overset{d}{\longrightarrow}
    \mathcal N\!\left(0,\frac{1}{\mathcal F_\theta}\right),
\end{equation}
meaning that $\sqrt{n}\,(\overline{\theta}-\theta)$ converges in distribution to a centered normal random variable with variance equal to the inverse Fisher information, thereby saturating the Cramér-Rao bound in the asymptotic limit.

\subsection{Metrology in a quantum setting}
\label{subsec: quantum metrology}
Quantum metrology studies the ultimate precision limits achievable in the estimation of physical parameters when quantum resources are employed. In its most general formulation, one considers a family of quantum states
\begin{equation}
    \theta \longmapsto \hat\rho_\theta,
\end{equation}
where $\theta \in \mathbb{R}$ denotes an unknown parameter to be estimated. It is assumed that the functional dependence of the state $\hat\rho_\theta$ on $\theta$ is known \emph{a priori}, while the actual value of $\theta$ is not. The goal of quantum metrology is to infer $\theta$ with the highest possible accuracy by performing suitable measurements on $\hat\rho_\theta$.

\paragraph{\lit Measurement model and outcome statistics}

A quantum measurement is described by a POVM $\{\hat E_x\}_{x \in X}$, where each $\hat E_x \geq 0$ and $\sum_{x \in X} \hat E_x = \1$. Upon measuring the state $\hat\rho_\theta$, an outcome $x$ is obtained with probability given by Born's rule,
\begin{equation}
    P_\theta(x) = \Tr(\hat E_x \hat\rho_\theta).
    \label{eq:born-rule}
\end{equation}
The probability distribution $P_\theta(x)$ encodes the information about $\theta$ that can be extracted from the measurement outcomes. The map $\theta \mapsto \hat\rho_\theta$ is, in principle, arbitrary, subject to mild regularity assumptions such as injectivity (to ensure identifiability of the parameter) and sufficient smoothness (to allow differentiation with respect to $\theta$). In practice, however, physically motivated models are often considered. Two particularly important cases are discussed below.

\paragraph{\lit Pure states and unitary parameter encoding}
A first simplification arises when the probe state is pure, namely $\hat\rho_\theta = \ketbra{\psi_\theta}$. In this case, the measurement statistics reduce to
\begin{equation}
    P_\theta(x) = \bra{\psi_\theta} \hat E_x \ket{\psi_\theta}.
\end{equation}
A second, and very common, assumption is that the dependence on the parameter $\theta$ is generated by a unitary evolution,
\begin{equation}
    \hat\rho_\theta = e^{-i \theta \hat H} \hat \rho_0 e^{i \theta \hat H},
    \label{eq:unitary-encoding}
\end{equation}
where $H$ is a Hermitian operator. For a mixed initial state $\rho_0$, the measurement probabilities are given by
\begin{equation}
    P_\theta(x) = \Tr(\hat E_x e^{-i \theta \hat H} \rho_0 e^{i \theta \hat H}).
\end{equation}
If, in addition, the initial state is pure, $\rho_0 = \ketbra{\psi_0}$, one obtains
\begin{equation}
    P_\theta(x) = \bra{\psi_0} e^{i \theta \hat H} \hat E_x e^{-i \theta \hat H} \ket{\psi_0}.
\end{equation}
While the treatment can be generalized to more complex evolution, in the following we will restrict ourselves to unitary evolution, studying first the case of pure states and then generalizing to mixed states.

\paragraph{\lit Quantum metrological protocol}
A typical quantum metrology protocol can be summarized as follows. A probe state $\ket{\psi_0}$ is first prepared and subsequently interacts with the system or field encoding the parameter $\theta$, resulting in a parameter-dependent state $\ket{\psi_\theta}$. A measurement described by a suitably chosen POVM is then performed, producing a set of classical outcomes. Finally, these outcomes are post-processed to construct an estimator $\overline{\theta}$ for the unknown parameter. This procedure is schematically illustrated in Fig.~\ref{fig: quantum-metrology-scheme}.

\begin{figure}[ht]
    \centering
    \footnotesize
    \scalebox{1.2}{\input{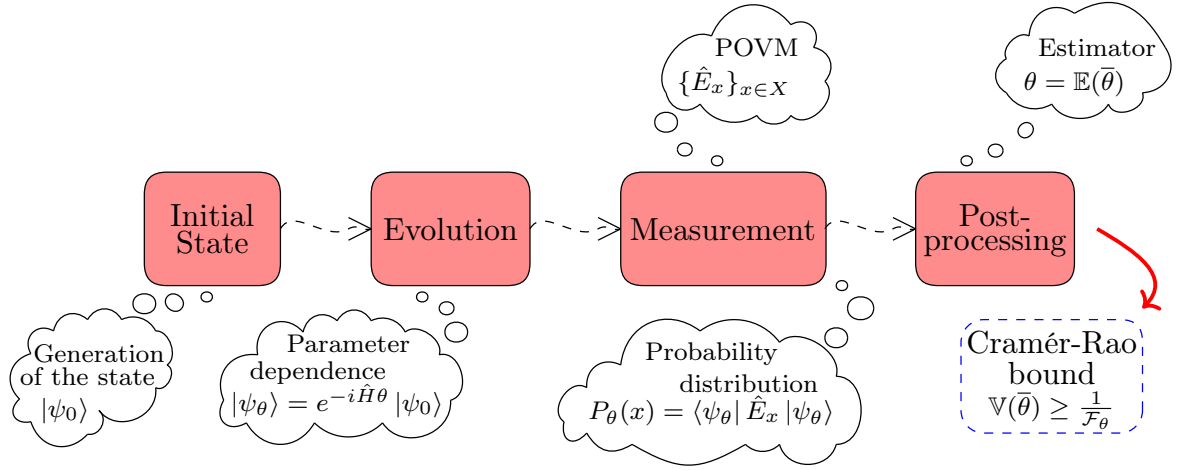}}
    \caption[Schematic representation of a quantum metrology protocol]{Schematic representation of a quantum metrology protocol: probe preparation, parameter encoding, measurement, and classical data processing.}
    \label{fig: quantum-metrology-scheme}
\end{figure}

A central question in quantum metrology is how the choice of measurement affects the attainable precision, and whether there exist fundamental limits that cannot be surpassed, regardless of the measurement strategy.

\paragraph{\lit Fundamental precision limits and the quantum Cramér-Rao bound}
For a given POVM and estimator, as explained previously, the Cramér-Rao inequality provides a lower bound on this variance in terms of the Fisher information $\mathcal F_\theta$ associated with the probability distribution $P_\theta(x)$. In the quantum setting, the Fisher information depends on the chosen POVM. Remarkably, quantum mechanics imposes an ultimate, measurement-independent upper bound on $\mathcal F_\theta$, known as the quantum Fisher information (QFI) $Q(\hat\rho_\theta)$~\cite{helstrom_quantum_1976, holevo_probabilistic_2011}. Specifically,
\begin{equation}
    \mathcal F_\theta \leq Q(\hat\rho_\theta),
\end{equation}
for any POVM. Combining this inequality with the Cramér-Rao bound yields the quantum Cramér-Rao bound (QCRB): for any unbiased estimator $\overline{\theta}$ constructed from the outcomes of a quantum measurement performed on the state $\hat\rho_\theta$, the variance satisfies
\begin{equation}
    \mathbb{V}(\overline{\theta}) \geq \frac{1}{Q(\hat\rho_\theta)}.
\end{equation}
The quantum Cramér-Rao bound represents a fundamental limit to precision imposed by quantum mechanics itself. In the following we will rederive the QCRB for the cases of pure and mixed states parametrized unitarily, discuss the optimality condition, provide explicit expressions for the QFI as well as analyze some of its properties.

\subsection{Pure-state case}
\label{subsec: quantum metrology pure states}
We now specialize to a paradigmatic and analytically tractable scenario in quantum metrology, namely the estimation of a parameter encoded through a unitary evolution acting on a pure probe state. This setting captures a wide range of physically relevant situations and already illustrates many of the essential features of quantum-enhanced metrology.

\paragraph{\lit Model and statement of the result}

Let the probe state depend on the unknown parameter $\theta$ according to
\begin{equation}
    \ket{\psi_\theta} = \hat U(\theta)\ket{\psi_0} = e^{-i \theta \hat H}\ket{\psi_0},
    \label{eq: pure-unitary-encoding}
\end{equation}
where $H$ is a Hermitian operator generating the unitary encoding. Measurements are described by a POVM $\{\hat E_x\}_{x\in X}$, leading to outcome probabilities
\begin{equation}
    P_\theta(x) = \bra{\psi_0} e^{i\theta \hat H} \hat E_x e^{-i\theta \hat H} \ket{\psi_0}.
\end{equation}

In this context, the quantum Cramér-Rao bound admits a particularly simple and physically transparent form: The associated Fisher information satisfies
\begin{equation}
    \mathcal F_\theta \leq 4\,\Delta^2 \hat H.
\end{equation}
Consequently, the quantum Fisher information is given by
\begin{equation}
    Q(\hat\rho_\theta) = Q(\hat H,\ket{\psi_0}) = 4\,\Delta^2 \hat H.
\end{equation}
Moreover, for any fixed value $\theta_0$, there exists a POVM such that
\begin{equation}
    \mathcal F_{\theta_0} = 4\,\Delta^2 \hat H,
\end{equation}
meaning that the bound is locally saturated. The quantity $\Delta^2 \hat H$ depends only on the initial state $\ket{\psi_0}$ and the generator $\hat H$. Furthermore, owing to the unitary encoding, it is independent of the parameter $\theta$.

\paragraph{\lit Bounding the Fisher information}
The proof of the quantum Cramér-Rao bound for pure states proceeds in two main steps. First, one derives a bound on the derivative $\partial_\theta P_\theta(x)$ in terms of expectation values of the generator $\hat H$. Second, this bound is inserted into the definition of the classical Fisher information,
\begin{equation}
    \mathcal F_\theta = \sum_{x\in X} \frac{1}{P_\theta(x)} \left( \partial_\theta P_\theta(x) \right)^2,
\end{equation}
and optimized over a suitable real parameter. The detailed derivation relies on elementary operator identities and the Cauchy-Schwarz inequality, and is presented in Appendix~\ref{app: formalism and framework}, Result~\ref{res: QCRB pure states}.

The resulting inequality shows that, regardless of the chosen measurement, the sensitivity of parameter estimation is fundamentally limited by the fluctuations of the generator $\hat H$ in the probe state. This provides a clear operational interpretation of the quantum Fisher information: it quantifies the susceptibility of the state to infinitesimal changes of the parameter.

\paragraph{\lit Optimal measurements and saturation of the bound}

Beyond establishing an upper bound, it is crucial to determine whether the quantum Cramér-Rao bound can actually be achieved. For pure states under unitary encoding, this is indeed the case, at least locally in the parameter.

For a fixed value $\theta_0$, one can construct a two-outcome POVM tailored to the state $\ket{\psi_{\theta_0}}$, consisting of a projection onto the probe state and its orthogonal complement. Evaluating the Fisher information associated with this measurement and taking the limit $\theta \to \theta_0$, one finds that
\begin{equation}
    \lim_{\theta \to \theta_0} \mathcal F_\theta = 4\,\Delta^2 \hat H.
\end{equation}
The explicit construction of this POVM and the corresponding asymptotic expansion of the Fisher information are given in Appendix~\ref{app: formalism and framework}, Result~\ref{res: saturation QCRB pure states}. While the dependence of the optimal measurement on the unknown parameter may appear problematic, this issue can be resolved in practice by adaptive or locally optimized measurement strategies. 

\paragraph{\lit Illustrative Example: Qubit Probes}
We begin with the simplest nontrivial system, a single qubit. Let the initial probe state be
\begin{equation}
    \ket{\psi_0} = \ket{+} = \frac{1}{\sqrt{2}}\left(\ket{0} + \ket{1}\right),
\end{equation}
and consider a unitary parameter encoding generated by the Pauli operator $\hat Z$, $\ket{\psi_\theta} = e^{-i \hat Z \theta}\ket{\psi_0}$. In this case, it is straightforward to obtain $\Delta^2 \hat Z = 1$ and as a consequence, the quantum Fisher information reads
\begin{equation}
    Q(\hat Z,\ket{\psi_0}) = 4\,\Delta^2 \hat Z = 4.
\end{equation}
This example already shows that even a single qubit can provide nontrivial sensitivity to the parameter $\theta$. We now consider a system of $n$ qubits and compare two qualitatively different probe states. The parameter encoding is generated by the collective Hamiltonian
\begin{equation}
    \hat H = \hat Z_1 + \hat Z_2 + \cdots + \hat Z_n,
\end{equation}
where $\hat Z_j$ denotes the Pauli-$Z$ operator acting on the $j$-th qubit. We compare the following probe states
\begin{align}
    \ket{\phi_1} = \frac{1}{(\sqrt{2})^n}\left(\ket{0}+\ket{1}\right)^{\otimes n}, &&
    \ket{\phi_2} = \frac{1}{\sqrt{2}}\left(\ket{0\cdots 0}+\ket{1\cdots 1}\right). 
\end{align}
The state $\ket{\phi_1}$ is a fully separable product state, while $\ket{\phi_2}$ is a maximally entangled Greenberger-Horne-Zeilinger (GHZ) state. For product states, the variance of a sum of local observables is additive. As such, one finds $\Delta^2 \hat H = n$. The corresponding quantum Fisher information is therefore
\begin{equation}
    Q(\hat H,\ket{\phi_1}) = 4n.
\end{equation}
This linear scaling with the number of probes $n$ is known as the \emph{shot-noise scaling} (or standard quantum limit). It is characteristic of strategies employing independent probes and classical resources only. We now turn to the entangled state $\ket{\phi_2}$. One finds $\Delta^2 \hat H = n^2$, leading to the quantum Fisher information
\begin{equation}
    Q(\hat H,\ket{\phi_2}) = 4n^2.
\end{equation}
All the corresponding derivations are presented in detail in Appendix~\ref{app: formalism and framework}, Result~\ref{res: qubit variance computations}. This quadratic scaling with $n$ is referred to as \emph{Heisenberg scaling}. It represents the ultimate precision limit allowed by quantum mechanics for unitary parameter estimation and can only be achieved by exploiting genuine multipartite entanglement.

The comparison between $\ket{\phi_1}$ and $\ket{\phi_2}$ clearly illustrates one of the central messages of quantum metrology: entanglement can be used as a resource to enhance parameter estimation beyond the shot-noise limit. While separable states lead to a precision scaling as $1/\sqrt{n}$, entangled states such as GHZ states enable a scaling as $1/n$, corresponding to the Heisenberg limit. In realistic scenarios, decoherence and experimental imperfections may limit the practical advantage of highly entangled states. Nevertheless, these simple qubit examples provide a clean and intuitive benchmark for understanding the role of quantum correlations in metrology. Further discussion of the role of entanglement, collective variables, and precision in quantum metrology is provided in Sec.~\ref{sec: TF entanglement and metrology}.

\subsection{Mixed-state case}
\label{subsec: quantum metrology mixed states}
Now, the probe state is assumed to be initially prepared in a mixed state $\hat \rho_0$, and the parameter-dependent state takes the form
\begin{equation}
    \hat \rho_\theta = e^{-i \hat H \theta}\, \hat \rho_0 \, e^{i \hat H \theta}.
\end{equation}

\paragraph{\lit Quantum Cramér-Rao bound for mixed states}
The central object governing precision in the mixed-state case is the \emph{symmetric logarithmic derivative} (SLD), denoted $\hat L_\theta$. It is defined implicitly by the operator equation
\begin{equation}
    \partial_\theta \hat \rho_\theta
    = \frac{1}{2}\bigl( \hat L_\theta \hat \rho_\theta + \hat \rho_\theta \hat L_\theta \bigr).
    \label{eq:SLD_def}
\end{equation}
The SLD is Hermitian but not necessarily unique when $\hat \rho_\theta$ has a nontrivial kernel. For any POVM $\{\hat E_x\}$ performed on the state $\hat \rho_\theta$, the associated classical Fisher information satisfies
\begin{equation}
    \mathcal F_\theta \le Q(\hat \rho_\theta),
\end{equation}
where the quantum Fisher information (QFI) is given by
\begin{equation}
    Q(\hat \rho_\theta) = \Tr(\hat \rho_\theta \hat L_\theta^2).
\end{equation}
The proof proceeds by relating $\partial_\theta P_\theta(x)$ to the SLD and applying the Cauchy-Schwarz inequality at the operator level. The full derivation is deferred to Appendix~\ref{app: formalism and framework}, Result~\ref{res: QCRB mixed states}. Conceptually, the result shows that $Q(\hat \rho_\theta)$ plays the role of an intrinsic, measurement-independent measure of statistical distinguishability between nearby quantum states~\cite{helstrom_quantum_1976,holevo_probabilistic_2011}.

\paragraph{\lit Properties of the SLD and expression of the quantum Fisher information}

A direct consequence of the definition of the SLD is that $\Tr(\hat \rho_\theta \hat L_\theta) = 0$, which implies that the QFI can be interpreted as a variance,
\begin{equation}
    Q(\hat \rho_\theta) = \Delta^2 \hat L_\theta .
\end{equation}
For unitary parameter encoding, the SLD and the QFI admit explicit expressions in terms of the spectral decomposition of the initial state $\hat \rho_0$. Writing
\begin{equation}
    \hat \rho_0 = \sum_j p_j \ketbra{\psi_j}{\psi_j},
\end{equation}
one finds that the SLD at $\theta=0$ can be expressed in this eigenbasis, and the corresponding QFI takes the closed form
\begin{equation}
    Q(\hat \rho_0)
    = 2 \sum_{j,k \,:\, p_j+p_k>0}
    \frac{(p_j-p_k)^2}{p_j+p_k}
    \abs{\bra{\psi_j}\hat H\ket{\psi_k}}^2 .
    \label{eq:QFI_mixed_unitary}
\end{equation}
The derivation is presented in Appendix~\ref{app: formalism and framework}, Result~\ref{res: SLD and QFI mixed states}. This expression, originally derived in Refs.~\cite{helstrom_quantum_1976,braunstein_statistical_1994}, reduces to $Q=4\Delta^2\hat H$ in the special case of pure states, in agreement with the result obtained in the previous section on pure states. Furthermore, for unitary families of the form considered here, the SLD transforms covariantly,
\begin{equation}
    \hat L_\theta = e^{-i\hat H\theta}\, \hat L_0 \, e^{i\hat H\theta},
\end{equation}
implying that the quantum Fisher information is independent of $\theta$.

\paragraph{\lit Saturation of the bound}
The quantum Cramér-Rao bound is locally tight. For a given value $\theta_0$, an optimal measurement can be constructed by diagonalizing the SLD,
\begin{equation}
    \hat L_{\theta_0} = \sum_j \lambda_j \hat\Pi_j,
\end{equation}
and choosing the POVM elements to be the corresponding orthogonal projectors $\{\hat\Pi_j\}$. The associated classical Fisher information then coincides with the QFI,
\begin{equation}
    \mathcal F_{\theta_0} = Q(\hat \rho_{\theta_0}),
\end{equation}
showing that the QCRB is achievable in principle. The explicit verification of this saturation is presented in Appendix~\ref{app: formalism and framework}, Result~\ref{res: saturation QCRB mixed states}.

\subsection{Properties of the quantum Fisher information}
\label{subsec: qfi properties}
In this last section, we establish some fundamental properties of the quantum Fisher information: additivity on product states, convexity under classical mixing, the relation to quantum variance, and the convex roof property.

\paragraph{\lit Additivity on product states.}
Consider a bipartite system with state $\hat\rho = \hat\rho_1 \otimes \hat\rho_2$ and a generator of the form $\hat H = \hat H_1 \otimes \1 + \1 \otimes \hat H_2$.
Let $\hat L$, $\hat L_1$, and $\hat L_2$ denote the SLDs associated with $\hat\rho$, $\hat\rho_1$, and $\hat\rho_2$, respectively. We have the relation
\begin{equation}
    \mathcal Q(\hat H,\hat \rho_1\otimes \hat\rho_2)=\mathcal Q(\hat H_1,\hat \rho_1)+\mathcal Q(\hat H_2,\hat \rho_2).
\end{equation}
Intuitively, this property reflects the fact that independent subsystems contribute additively to the overall sensitivity to parameter changes. In particular, it means that for such scenarios, local measurement strategies are already optimal, and no advantage may be obtained by using an entangled POVM. This property is the direct analog of the additivity of the classical Fisher information. The detailed proof is presented in Appendix~\ref{app: formalism and framework}, Result~\ref{res: additivity QFI}.

\paragraph{\lit Convexity} 
The quantum Fisher information associated with a generator $\hat H$ is a convex function of the quantum state. Explicitly, if a state admits a convex decomposition $\hat\rho = \sum_j p_j \hat\rho_j$ with $p_j\ge 0$, then we have
\begin{equation}
    \mathcal Q(\hat H,\hat\rho) \leq \sum_j p_j \mathcal Q(\hat H,\hat\rho_j).
\end{equation}
This implies that for estimation purposes, classically mixing states will not provide an increase in measurement precision. This can be intuitively understood as follows. The right-hand side corresponds to the situation where we have access to the label $j$ and know with which state $\hat\rho_j$ the measurement is performed. In this case, we can optimize the measurement for each $\hat\rho_j$ separately and obtain the sum of the QFIs. If we do not have access to this label, we can only perform a measurement on the mixed state $\hat\rho$ and obtain a QFI that is upper bounded by the previous case, which corresponds to the left-hand side. As such, it is natural to expect that the second scenario will provide a less precise estimation, corresponding to the inequality. The detailed and formal proof is presented in Appendix~\ref{app: formalism and framework}, Result~\ref{res: convexity QFI}.

\paragraph{\lit Inequality with the quantum variance}
Defining the quantum variance of any observable $\hat H$ on a mixed state $\hat\rho$ by
\begin{equation}
    \Delta^2 \hat H = \Tr(\hat \rho\hat H^2)-\Tr(\hat \rho\hat H)^2,
\end{equation}
we have the following inequality
\begin{equation}
    \mathcal Q(\hat H,\hat\rho) \leq 4\,\Delta^2 \hat H.
\end{equation}
This inequality can be interpreted in the following way. The quantum variance quantifies the total fluctuations of the observable $\hat H$ on the state $\hat\rho$. However, these fluctuations can arise from two distinct sources: classical uncertainty due to the mixed nature of the state, and quantum uncertainty inherent to the non-commutativity of observables in quantum mechanics. The quantum Fisher information captures only the quantum part of these fluctuations that are relevant for parameter estimation. Therefore, it is natural that the QFI is upper bounded by four times the total variance, reflecting that only a portion of the total fluctuations contribute to the sensitivity to parameter changes. The detailed proof is presented in Appendix~\ref{app: formalism and framework}, Result~\ref{res: QFI variance inequality}. This property is mainly a consequence of the convexity of the QFI as well as the concavity of the quantum variance.

\paragraph{\lit Convex roof property}
We have discussed and shown above that the quantum Fisher information is a convex function of the state, and that it coincides with the usual variance for pure states. It turns out that other quantities share these two properties, such as for example the Wigner-Yanase skew information~\cite{wigner_information_1963}, defined for an observable $\hat H$ and a state $\hat \rho$ as
\begin{equation}
    I(\hat H,\hat\rho) = -\frac{1}{2}\Tr\!\left( \left[\sqrt{\hat\rho},\hat H\right]^2 \right).
\end{equation}

Beyond the role played by the quantum Fisher information in quantum metrology, a natural question arises: does a fundamental property of the quantum Fisher information distinguish it from other convex functions that reduce to the variance on pure states? The answer is affirmative. This distinguishing feature is known as the convex roof property, which states that the quantum Fisher information is the maximal convex function that reduces to the variance on pure states. More precisely, one has
\begin{equation}
    \mathcal Q(\hat H,\hat \rho)=4\inf_{\{p_j,\ket{\psi_j}\}} \sum_j p_j \, \Delta^2_{\ket{\psi_j}} \hat H,
\end{equation}
where the infimum is taken over all, not necessarily orthogonal, convex decompositions of $\hat\rho$ into pure states, $\hat\rho = \sum_j p_j \ketbra{\psi_j}{\psi_j}$, and where $\Delta^2 \hat H$ denotes the variance of $\hat H$ evaluated in the pure state $\ket{\psi_j}$.

This property admits a clear interpretation. Intuitively, the quantum Fisher information quantifies the amount of purely quantum fluctuations of the observable $\hat H$ in the state $\hat\rho$. The convex roof construction formalizes this idea by considering all possible pure-state decompositions of $\hat\rho$ and selecting the one that minimizes the average variance. In this way, any additional fluctuations that can be attributed to classical mixing are removed, and only the intrinsic quantum contribution is retained. Such decompositions therefore highlight the fundamental nature of the quantum Fisher information as a measure of quantum uncertainty that is not artificially increased by classical probabilistic mixtures.

In addition to its conceptual significance, this characterization also provides a practical tool. It allows one to compute or bound the QFI in situations where a direct evaluation is difficult, by exploiting known expressions for pure states and optimizing over decompositions. This result was initially conjectured in~\cite{toth_extremal_2013} and later proven in~\cite{yu_quantum_2013}. In Appendix~\ref{app: formalism and framework}, Result~\ref{res: convex roof QFI}, we present a reformulation of this proof, based on the same fundamental ingredient, but with an emphasis on a more progressive and motivated construction.

\clearpage
\section{Quantum optics}
\label{sec: quantum optics}
\emph{This section presents the fundamental elements required to describe quantum optical systems. Following the standard approach reviewed in~\cite{fabre_modes_2020,cohen-tannoudji_photons_1989,loudon_quantum_2000}, we first introduce the classical mode decomposition of the electromagnetic field. This naturally leads to a description in terms of independent harmonic oscillators, which forms the basis for field quantization. We provide the Fock space construction and a brief description of mode transformations. We conclude with remarks anticipating the conceptual issues associated with the definition of the global optical phase and related superselection rules~\cite{mandel_optical_1995,walls_quantum_2008}, which are the main concerns of Chap.~\ref{chap: SSR}.}

\subsection{Classical Electromagnetic Field and Mode Decomposition}
\label{subsec: Classical Electromagnetism}
In vacuum and in the absence of charges and currents, the classical electromagnetic field is described by the electric and magnetic fields
$\vec{E}(\vec{r},t)$ and $\vec{B}(\vec{r},t)$, which satisfy Maxwell's equations
\begin{align}
    \nabla \cdot \vec{E} &= 0, &
    \nabla \cdot \vec{B} &= 0, \\
    \nabla \times \vec{E} &= -\frac{\partial \vec{B}}{\partial t}, &
    \nabla \times \vec{B} &= \frac{1}{c^{2}}\,\frac{\partial \vec{E}}{\partial t},
\end{align}
where $c = 1/\sqrt{\varepsilon_0\mu_0}$ is the speed of light in vacuum, and $\varepsilon_0$ and $\mu_0$ denote the vacuum permittivity and permeability, respectively.

\paragraph{\lit Potentials and Gauge Choice}
It is convenient to express the electromagnetic field in terms of the scalar and vector potentials, $\phi(\vec{r},t)$ and $\vec{A}(\vec{r},t)$, defined through
\begin{align}
    \vec{E} = -\nabla \phi - \frac{\partial \vec{A}}{\partial t},
    &&
    \vec{B} = \nabla \times \vec{A}.
\end{align}
These potentials are not unique, reflecting the gauge freedom of electrodynamics. In free space, a particularly convenient choice is the Coulomb gauge,
\begin{align}
    \nabla \cdot \vec{A} = 0, && \phi = 0,
\end{align}
in which the electromagnetic field is entirely described by the vector potential $\vec{A}(\vec{r},t)$. In this gauge, Maxwell's equations reduce to the wave equation
\begin{equation}
    \left( \nabla^{2} - \frac{1}{c^{2}}\frac{\partial^{2}}{\partial t^{2}} \right)
    \vec{A}(\vec{r},t) = 0.
\end{equation}

\paragraph{\lit Normal Modes of the Electromagnetic Field}
To define the notion of normal modes in a precise manner, it is convenient to consider the field confined to a finite volume $V$ (for instance an optical cavity),
together with boundary conditions on $\partial V$ that ensure the electromagnetic energy remains finite~\cite{loudon_quantum_2000,cohen-tannoudji_photons_1989}. We first seek separable solutions of the form
\begin{equation}
    \vec{A}(\vec{r},t) = q(t)\vec{u}(\vec{r}),
\end{equation}
where $\vec{u}(\vec{r})$ is a spatial mode function. Standard separation of variables techniques then show that there exists a constant $\Lambda$ such that
\begin{align}
    \ddot{q}(t) = \Lambda q(t), && \nabla^{2}\vec{u}(\vec{r}) = \frac{\Lambda}{c^{2}}\vec{u}(\vec{r}).
\end{align}
The total electromagnetic energy is given by
\begin{equation}
    \mathcal{E} = \frac{1}{2}\int_{V} \dd^{3}r\, \left[ \varepsilon_0 \abs{\vec{E}(\vec{r},t)}^{2} + \frac{1}{\mu_0} \abs{\vec{B}(\vec{r},t)}^{2} \right].
\end{equation}
Requiring this energy to remain finite excludes exponentially growing solutions and imposes $\Lambda = -\omega^{2}$ with $\omega > 0$. The temporal equation therefore describes harmonic motion,
\begin{equation}
    \ddot{q}(t) + \omega^{2} q(t) = 0,
\end{equation}
while the spatial equation becomes the vector Helmholtz equation
\begin{align}
    \nabla^{2}\vec{u}(\vec{r}) + \frac{\omega^{2}}{c^{2}}\vec{u}(\vec{r}) = 0, && \nabla \cdot \vec{u}(\vec{r}) = 0.
\end{align}
Nontrivial solutions exist only for a discrete set of eigenfrequencies $\omega = \omega_\lambda$, determined by the geometry of the volume $V$ and the boundary conditions. Each solution $\vec{u}_{\lambda}(\vec{r})$ defines a \emph{normal mode} of the electromagnetic field. Each mode is characterized by:
\begin{itemize}
    \item a spatial mode function $\vec{u}_{\lambda}(\vec{r})$,
    \item a polarization encoded in its vector structure,
    \item a real and positive eigenfrequency $\omega_{\lambda}$.
\end{itemize}
The collective index $\lambda$ labels all these properties ({\it e.g.}, wave vector, polarization, and mode number). The modes can be chosen orthonormal with respect to the inner product
\begin{equation}
    \int_{V} \dd^{3}r\,
    \vec{u}_{\lambda}^{\,*}(\vec{r}) \cdot \vec{u}_{\lambda'}(\vec{r})
    = \delta_{\lambda,\lambda'}.
\end{equation}
Completeness of the modes allows the vector potential to be expanded as
\begin{equation}
    \vec{A}(\vec{r},t)
    = \sum_{\lambda} q_{\lambda}(t)\,\vec{u}_{\lambda}(\vec{r}),
\end{equation}
where the real-valued functions $q_{\lambda}(t)$ describe the time-dependent amplitudes. Substitution into the wave equation and projection onto each mode yield independent equations of motion,
\begin{equation}
    \ddot{q}_{\lambda}(t) + \omega_{\lambda}^{2} q_{\lambda}(t) = 0.
\end{equation}
Each normal mode thus behaves dynamically as a classical harmonic oscillator of frequency $\omega_{\lambda}$.

\paragraph{\lit Hilbert Space of Modes}
The set of mode functions $\{\vec{u}_{\lambda}\}$ forms a Hilbert space with inner product
\begin{equation}
    \langle \vec{u}_{\lambda}, \vec{u}_{\lambda'} \rangle
    = \int_V \dd^{3}r\,
    \vec{u}_{\lambda}^{\,*}(\vec{r}) \cdot \vec{u}_{\lambda'}(\vec{r}),
\end{equation}
with respect to which a mode basis may be chosen orthonormal. Different choices of mode basis functions correspond to different orthonormal bases of this Hilbert space. A mode transformation corresponds to a unitary change of basis,
\begin{equation}
    \vec{v}_{\mu}(\vec{r})
    = \sum_{\lambda} U_{\mu\lambda}\,\vec{u}_{\lambda}(\vec{r}),
\end{equation}
which leaves the physical electromagnetic field invariant while redistributing its description among the mode amplitudes~\cite{mandel_optical_1995}.

\paragraph{\lit Energy of the Electromagnetic Field}
Expressed in terms of the vector potential, the total electromagnetic energy is
\begin{equation}
    \mathcal{E}
    = \frac{1}{2}\int_V \dd^{3}r\,
    \left(
    \varepsilon_0 \abs{\frac{\partial \vec{A}}{\partial t} }^{2}
    + \frac{1}{\mu_0} \abs{ \nabla \times \vec{A} }^{2}
    \right).
\end{equation}
Using the mode expansion and orthonormality, this energy separates into independent contributions,
\begin{equation}
    \mathcal{E}
    = \sum_{\lambda}
    \left[
    \frac{1}{2}\varepsilon_0 \dot{q}_{\lambda}^{2}(t)
    + \frac{1}{2}\varepsilon_0 \omega_{\lambda}^{2} q_{\lambda}^{2}(t)
    \right].
\end{equation}
Each mode therefore contributes as an independent classical harmonic oscillator.

\subsection{Canonical Quantization of the Electromagnetic Field}
\label{subsec: Field Quantization}
The mode decomposition introduced above shows that the classical electromagnetic field in vacuum is dynamically equivalent to an infinite set of independent harmonic oscillators. This observation provides a natural route to quantization via canonical quantization~\cite{cohen-tannoudji_photons_1989,loudon_quantum_2000,walls_quantum_2008}.

\paragraph{\lit Canonical Variables}
Starting from the classical energy of the field,
\begin{equation}
    \mathcal{E}
    = \sum_{\lambda}
    \left[
    \frac{1}{2}\varepsilon_0 \dot{q}_{\lambda}^{2}(t)
    + \frac{1}{2}\varepsilon_0 \omega_{\lambda}^{2} q_{\lambda}^{2}(t)
    \right],
\end{equation}
we identify $q_{\lambda}(t)$ as the generalized coordinate associated with mode $\lambda$. The corresponding canonical momentum is defined as
\begin{equation}
    p_{\lambda}(t)
    \equiv
    \frac{\partial \mathcal{L}}{\partial \dot{q}_{\lambda}}
    = \varepsilon_0 \dot{q}_{\lambda}(t),
\end{equation}
where $\mathcal{L}$ is the Lagrangian of the electromagnetic field in the Coulomb gauge. In terms of the canonical variables $(q_{\lambda},p_{\lambda})$, the Hamiltonian becomes
\begin{equation}
    H
    = \sum_{\lambda}
    \left[
    \frac{p_{\lambda}^{2}}{2\varepsilon_0}
    + \frac{1}{2}\varepsilon_0 \omega_{\lambda}^{2} q_{\lambda}^{2}
    \right].
\end{equation}

Canonical quantization proceeds by promoting the classical canonical variables to operators acting on a Hilbert space and imposing the usual commutation relations
\begin{align}
    [\hat{q}_{\lambda},\hat{p}_{\lambda'}]  = i\hbar\,\delta_{\lambda,\lambda'}\1,  &&  [\hat{q}_{\lambda},\hat{q}_{\lambda'}] = [\hat{p}_{\lambda},\hat{p}_{\lambda'}] = 0.
\end{align}
Each mode is therefore quantized independently as a quantum harmonic oscillator.

\paragraph{\lit Creation and Annihilation Operators}
It is convenient to introduce dimensionless ladder operators for each mode,
\begin{align}
    \hat{a}_{\lambda}
    =
    \frac{1}{\sqrt{2\hbar\varepsilon_0\omega_{\lambda}}}
    \left(
    \varepsilon_0 \omega_{\lambda} \hat{q}_{\lambda}
    + i \hat{p}_{\lambda}
    \right), &&
    \hat{a}_{\lambda}^{\dagger}
    =
    \frac{1}{\sqrt{2\hbar\varepsilon_0\omega_{\lambda}}}
    \left(
    \varepsilon_0 \omega_{\lambda} \hat{q}_{\lambda}
    - i \hat{p}_{\lambda}
    \right),
\end{align}
which satisfy the bosonic commutation relations
\begin{align}
    [\hat{a}_{\lambda},\hat{a}_{\lambda'}^{\dagger}]= \delta_{\lambda,\lambda'}, &&[\hat{a}_{\lambda},\hat{a}_{\lambda'}]= [\hat{a}_{\lambda}^{\dagger},\hat{a}_{\lambda'}^{\dagger}] = 0.
\end{align}
Inverting these relations yields
\begin{align}
    \hat{q}_{\lambda} = \sqrt{\frac{\hbar}{2\varepsilon_0\omega_{\lambda}}} \left( \hat{a}_{\lambda} + \hat{a}_{\lambda}^{\dagger} \right), && \hat{p}_{\lambda} = -i\sqrt{\frac{\hbar\varepsilon_0\omega_{\lambda}}{2}} \left( \hat{a}_{\lambda} - \hat{a}_{\lambda}^{\dagger} \right).
\end{align}

\paragraph{\lit Quantized Hamiltonian}

Expressed in terms of the ladder operators, the Hamiltonian becomes
\begin{equation}
    \hat{H} = \sum_{\lambda} \hbar\omega_{\lambda} \left( \hat{a}_{\lambda}^{\dagger}\hat{a}_{\lambda} + \frac{1}{2} \right).
\end{equation}
The operator
\begin{equation}
    \hat{N}_{\lambda} = \hat{a}_{\lambda}^{\dagger}\hat{a}_{\lambda}
\end{equation}
is the photon number operator associated with mode $\lambda$.
Its eigenvalues $n_{\lambda}=0,1,2,\dots$ correspond to the number of photons occupying that mode.

\paragraph{\lit Quantized Field Operators}

The vector potential becomes an operator-valued field,
\begin{equation}
    \hat{\vec{A}}(\vec{r})
    = \sum_{\lambda}
    \sqrt{\frac{\hbar}{2\varepsilon_0\omega_{\lambda}}}
    \left[
    \hat{a}_{\lambda}\,\vec{u}_{\lambda}(\vec{r})
    + \hat{a}_{\lambda}^{\dagger}\,\vec{u}_{\lambda}^{*}(\vec{r})
    \right].
\end{equation}
Using the classical expressions for the electric and magnetic fields
\begin{equation}
    \vec E(\vec r)=-\frac{\partial \vec A}{\partial t}=-\sum_\lambda \dot{q}_\lambda(t) \vec u_\lambda(\vec r),
\end{equation}
\begin{equation}
    \vec B(\vec r)=\nabla\times \vec A(\vec r)=\sum_\lambda q_\lambda(t) \nabla\times \vec u_\lambda(\vec r),
\end{equation}
and quantizing the classical conjugate variables $q_\lambda$ and $\dot{q}_\lambda=p_\lambda /\epsilon_0$, we obtain the electric and magnetic field operators as
\begin{align}
    \hat{\vec{E}}(\vec{r}) &= i\sum_{\lambda} \sqrt{\frac{\hbar\omega_{\lambda}}{2\varepsilon_0}} \left[ \hat{a}_{\lambda}\,\vec{u}_{\lambda}(\vec{r}) - \hat{a}_{\lambda}^{\dagger}\,\vec{u}_{\lambda}^{*}(\vec{r}) \right], \\
    \hat{\vec{B}}(\vec{r}) &= \nabla \times \hat{\vec{A}}(\vec{r}) = \sum_{\lambda} \sqrt{\frac{\hbar}{2\varepsilon_0\omega_{\lambda}}} \left[ \hat{a}_{\lambda}\,\nabla \times \vec{u}_{\lambda}(\vec{r}) + \hat{a}_{\lambda}^{\dagger}\,\nabla \times \vec{u}_{\lambda}^{*}(\vec{r}) \right].
\end{align}
Splitting each field operator into two parts based on the above formulas, we define
\begin{align}
    \vec{\hat E}^{(+)}(\vec r)= i\sum_{\lambda} \sqrt{\frac{\hbar\omega_{\lambda}}{2\varepsilon_0}} \hat{a}_{\lambda}\,\vec{u}_{\lambda}(\vec{r}), && \vec{\hat E}^{(-)}(\vec r)= -i\sum_{\lambda} \sqrt{\frac{\hbar\omega_{\lambda}}{2\varepsilon_0}} \hat{a}_{\lambda}^{\dagger}\,\vec{u}_{\lambda}^{*}(\vec{r}),
\end{align}
such that $\vec{\hat E}=\vec{\hat E}^{(+)}+\vec{\hat E}^{(-)}$. The operators $\vec{\hat E}^{(+)}$ and $\vec{\hat E}^{(-)}$ are called the positive- and negative-frequency parts of the electric field operator, respectively.

\paragraph{\lit Relation to continuous variables}
Quantum optical systems provide the ideal way to implement CV physics. This is made explicit by renormalizing the quadrature operators
\begin{align}
    \hat q_\lambda\mapsto \hat x_\lambda = \sqrt{\frac{\omega_\lambda \varepsilon_0}{\hbar}} \hat q_\lambda, && \hat p_\lambda \mapsto \frac{1}{\sqrt{\hbar \epsilon_0\omega_\lambda}}\hat p_\lambda.
\end{align}
These dimensionless operators then follow the canonical commutation relation $[\hat x_\lambda, \hat p_{\lambda'}]=i\delta_{\lambda,\lambda'}\1$ and are directly related to the ladder operators by
\begin{align}
    \hat a_\lambda = \frac{1}{\sqrt{2}}(\hat x_\lambda + i \hat p_\lambda), && \hat a_\lambda^\dagger = \frac{1}{\sqrt{2}}(\hat x_\lambda - i \hat p_\lambda).
\end{align}
as introduced in section \ref{subsec: continuous variables}.

\subsection{Fock Space Construction}
\label{subsec: FockSpace}

In Sec.~\ref{subsec: continuous variables} we described how CV can be mathematically formalized, while in Sec.~\ref{subsec: Field Quantization} we showed that such a framework naturally arises when quantizing the electromagnetic field. Although a single CV system (or mode) already exhibits rich physics, many applications require the description of multiple DOF or of systems with a variable number of excitations. 

If one considers a finite collection of modes $\lambda = 1,\dots,n$, the total Hilbert space is simply given by the tensor product of the individual mode Hilbert spaces. However, when the number of modes becomes infinite, this naive tensor product construction is no longer mathematically well-defined. To overcome this difficulty, one introduces the Fock space construction, which provides a rigorous framework to describe quantum systems with a variable number of indistinguishable particles~\cite{reed_methods_1980, bratteli_operator_1987}.

\paragraph{\lit Single-particle Hilbert space}
The Fock space construction starts from a Hilbert space $\mathcal H$, called the \emph{single-particle Hilbert space}. This space encodes the internal or external DOF of a single particle. Typical examples include frequency, polarization, spatial mode, or spin. The dimension of $\mathcal H$ may be finite or infinite.

As a simple illustration, the trivial case $\mathcal H = \C$ corresponds to a particle without any degree of freedom.\footnote{Or equivalently a degree of freedom, with only one possible state.} The associated Fock space then reduces to the one-mode space $\operatorname{Span}\{\ket{n}\}_{n\in\mathbb N}$, where the only degree of freedom is the particle (or photon) number. Another instructive example is $\mathcal H = \mathbb C^2$, which may represent the Hilbert space of modes of an interferometer with two spatial modes. In this case, a single photon can propagate in either arm or in a coherent superposition of both, reflecting the Hilbert space structure of $\mathcal H$.

In quantum optics, the single-particle Hilbert space is typically taken as the space spanned by the normal modes of the electromagnetic field,
\begin{equation}
    \mathcal H = \operatorname{Span}\{\vec u_\lambda\}_{\lambda \in \Lambda},
\end{equation}
where $\Lambda$ is an index set that may be discrete or continuous, for instance when describing frequency or spatial DOF~\cite{loudon_quantum_2000, mandel_optical_1995}.

\paragraph{\lit Definition of the Fock space}
The (bosonic) Fock space associated with $\mathcal H$ is defined as
\begin{equation}
    \mathcal F(\mathcal H) = \bigoplus_{n=0}^\infty \operatorname{Sym}^n(\mathcal H),
\end{equation}
where $\operatorname{Sym}^n(\mathcal H)$ denotes the symmetric subspace of the $n$-fold tensor product $\mathcal H^{\otimes n}$. This subspace corresponds to states of exactly $n$ indistinguishable bosons, each described by $\mathcal H$. The direct sum structure allows for a superposition of states with different particle numbers.

Starting from the inner product $\langle \,\cdot\, ,\, \cdot \,\rangle$ on $\mathcal H$, the Fock space $\mathcal F(\mathcal H)$ is naturally endowed with an inner product defined by
\begin{equation}
    \langle \psi, \phi \rangle_{\mathcal F}
    =
    \sum_{n=0}^\infty
    \langle \psi_n, \phi_n \rangle_n,
\end{equation}
where $\psi_n$ and $\phi_n$ denote the components of $\psi$ and $\phi$ in $\operatorname{Sym}^n(\mathcal H)$, and $\langle \cdot, \cdot \rangle_n$ is the inner product inherited from $\mathcal H^{\otimes n}$. Explicitly, for symmetrized tensors,
\begin{equation}
    \Big\langle 
    \mathcal S(\psi_1 \otimes \cdots \otimes \psi_n),
    \mathcal S(\phi_1 \otimes \cdots \otimes \phi_n)
    \Big\rangle
    =
    \frac{1}{n!}
    \sum_{\sigma \in S_n}
    \prod_{i=1}^n
    \langle \psi_i, \phi_{\sigma(i)} \rangle,
\end{equation}
where $S_n$ is the symmetric group on $n$ elements and $\mathcal S$ denotes the symmetrization operator defined as
\begin{equation}
    \mathcal S(\psi_1 \otimes \cdots \otimes \psi_n)=\frac{1}{n!}\sum_{\sigma \in S_n}\psi_{\sigma(1)} \otimes \cdots \otimes \psi_{\sigma(n)}.
\end{equation}
With this definition, $\mathcal F(\mathcal H)$ becomes a well-defined Hilbert space~\cite{reed_methods_1980}.

\paragraph{\lit Creation and annihilation operators}
Given a vector $\psi \in \mathcal H$, one defines the corresponding creation operator $\hat a_\psi^\dagger$ by its action on symmetrized tensors,
\begin{equation}
    \hat a_\psi^\dagger \,\mathcal S(\psi_1 \otimes \cdots \otimes \psi_n)=\sqrt{n+1}\,\mathcal S(\psi \otimes \psi_1 \otimes \cdots \otimes \psi_n).
\end{equation}
By linearity and continuity, this uniquely defines $\hat a_\psi^\dagger$ on the whole Fock space. The annihilation operator $\hat a_\psi$ is defined as the adjoint of $\hat a_\psi^\dagger$. The action of $\hat a_\psi$ is given by\footnote{Notice the normalisation factor which differs from the single-mode case $\hat a_\psi \ket{n} = \sqrt{n}\ket{n-1}$. This is due to the fact that we are summing over each tensor factor, which, when the space $\mathcal H$ has dimension 1, yields $n$ identical terms.}
\begin{equation}
    \hat a_\psi \,\mathcal S(\psi_1 \otimes \cdots \otimes \psi_n)=\frac{1}{\sqrt{n}}\sum_{j=1}^n\langle \psi, \psi_j \rangle\mathcal S(\psi_1 \otimes \cdots \otimes \widehat{\psi_j} \otimes \cdots \otimes \psi_n),
\end{equation}
where the notation $\widehat{\psi_j}$ indicates that the term $\psi_j$ is omitted from the symmetrized tensor. These operators satisfy the canonical commutation relations,
\begin{align}
    [\hat a_\psi, \hat a_\phi^\dagger] = \langle \psi, \phi \rangle, && [\hat a_\psi, \hat a_\phi] = [\hat a_\psi^\dagger, \hat a_\phi^\dagger] = 0,
\end{align}
which characterize bosonic statistics~\cite{bratteli_operator_1987}. For a discrete orthonormal basis $\{\ket{\psi_j}\}_{j\in J}$ of $\mathcal H$, one introduces mode operators $\hat a_j = \hat a_{\psi_j}$, satisfying
\begin{equation}
    [\hat a_j, \hat a_k^\dagger] = \delta_{j,k}.
\end{equation}
Similarly, for a continuous basis $\{\ket{\psi_x}\}_{x\in X}$, one defines operator-valued distributions $\hat a(x)$ and $\hat a^\dagger(x)$ obeying
\begin{equation}
    [\hat a(x), \hat a^\dagger(x')] = \delta(x-x').
\end{equation}
Discrete and continuous DOF can be combined, leading to operators $\hat a_j(x)$ satisfying
\begin{equation}
    [\hat a_j(x), \hat a_{j'}^\dagger(x')] = \delta_{j,j'}\,\delta(x-x').
\end{equation}
Throughout this manuscript, indices typically denote discrete DOF, while variables written in parentheses generally refer to continuous ones.

\paragraph{\lit Interpretation}
Beyond providing a rigorous mathematical definition of the Hilbert space describing multimode optical systems, the Fock-space construction also offers an alternative physical perspective on optical fields. Rather than starting from the quantization of the classical electromagnetic field and viewing photons as excitations of its modes, as done in Sec.~\ref{subsec: Field Quantization}, one may instead adopt a particle-based description in which photons are taken as the elementary constituents of the theory. In this picture, photons are treated as addressable particles, in much the same way as massive particles, which can be created or annihilated and occupy the various available optical modes.

The quantum state of the optical field is then specified by the distribution of photons among these modes. The symmetrization operator $\mathcal S$ ensures that all photons play the same role in the description, reflecting their bosonic nature. From this viewpoint, photons are fundamentally indistinguishable particles. Further discussion of the notion of photon distinguishability is provided in Chap.~\ref{chap: HOM interferometry and metrology}, in particular in Sec.~\ref{subsec: distinguishability clarification}.

This particle-based perspective also provides a different interpretation of the quantum properties of optical fields. In particular, it naturally leads to the notion of particle entanglement, which can be studied independently of the mode description. In Chap.~\ref{chap: SSR}, we adopt this viewpoint and show that particle entanglement constitutes a fundamental, yet often overlooked, quantity that sheds new light on several informational and metrological results.

\paragraph{\lit Explicit description of Fock states}
Let $\vac$ denote the vacuum state, \ie, the unique (up to normalization) element of $\operatorname{Sym}^0(\mathcal H)$. For a set of occupation numbers $\{n_j\}_{j\in J}$ with $\sum_j n_j = n < \infty$, the corresponding Fock state is defined as
\begin{equation}
    \ket{\{n_j\}}
    =
    \prod_{j\in J}
    \frac{(\hat a_j^\dagger)^{n_j}}{\sqrt{n_j!}}
    \vac.
\end{equation}
A general normalized state in $\mathcal F(\mathcal H)$ can then be written as
\begin{align}
    \ket{\psi} = \sum_{\{n_j\}:\,\sum_j n_j < \infty} c_{\{n_j\}}\,\ket{\{n_j\}}, && \sum_{\{n_j\}} \abs{c_{\{n_j\}}}^2 = 1.
\end{align}
This decomposition can be generalized to continuous DOF, leading to functional-valued amplitudes and integrals instead of discrete sums~\cite{mandel_optical_1995, walls_quantum_2008}.

\paragraph{\lit Fock space and direct sums}
A key structural property of the Fock space construction concerns its behavior under direct sums. Given two Hilbert spaces $\mathcal H_1$ and $\mathcal H_2$, one has the natural isomorphism
\begin{equation}\label{eq: Fock space tensorization}
    \mathcal F(\mathcal H_1 \oplus \mathcal H_2)
    \cong
    \mathcal F(\mathcal H_1) \otimes \mathcal F(\mathcal H_2).
\end{equation}
This result justifies the common identification of the Fock space of a particle with $n$ independent DOF either as the tensor product of $n$ one-dimensional Fock spaces, or equivalently as the Fock space constructed from the single-particle Hilbert space $\mathbb C^n$~\cite{bratteli_operator_1987,nielsen_quantum_2010}.

\paragraph{\lit Fermionic systems}
A closely related construction exists for fermionic particles. In that case, the symmetric tensor powers $\operatorname{Sym}^n(\mathcal H)$ are replaced by the anti-symmetric subspaces $\Lambda^n \mathcal H$, leading to the fermionic Fock space and to creation and annihilation operators satisfying the canonical \emph{anticommutation} relations~\cite{bratteli_operator_1987, peskin_introduction_1995}. Although fermionic systems will not be considered further in this thesis, which is focused exclusively on optical (bosonic) systems, many of the structural results presented here admit direct fermionic analogues, often differing only by sign factors. Such constructions are central to the quantum description of electrons, for which interference phenomena, analogous to those discussed later in this manuscript, can also be observed and experimentally probed~\cite{bocquillon_electron_2014,roussel_processing_2021,ferraro_wigner_2013}.

\subsection{Mode transformations}
\label{subsec: mode transformation}

\paragraph{\lit Definitions and properties}
A particularly important class of transformations in quantum optics is given by \emph{mode transformations}. These correspond to changes of basis in the single-particle Hilbert space $\mathcal H$ associated with the electromagnetic field. As discussed in Sec.~\ref{subsec: Classical Electromagnetism}, such transformations leave the physical electromagnetic field invariant while redistributing its description among the mode amplitudes~\cite{mandel_optical_1995,loudon_quantum_2000}. In other words, different choices of mode bases provide equivalent descriptions of the same physical field configuration.

From a quantum-mechanical point of view, mode transformations are implemented as linear transformations of the creation and annihilation operators. For simplicity, we restrict our discussion to a finite collection of $n$ orthogonal modes, described by bosonic creation operators $\hat a_j^\dagger$ ($j=0,\dots,n-1$). A general mode transformation is defined by
\begin{align}
    \hat M \hat a_j^\dagger \hat M^\dagger 
    &= \hat b_j^\dagger 
    = \sum_{k=0}^{n-1} M_{kj}\,\hat a_k^\dagger,
    &
    \hat M \hat a_j \hat M^\dagger 
    &= \hat b_j 
    = \sum_{k=0}^{n-1} M_{kj}^*\,\hat a_k ,
\end{align}
where $M$ is an $n\times n$ unitary matrix. The choice of index ordering in $M_{kj}$ is a matter of convention; alternative choices are equally valid but may lead to slightly less compact expressions in subsequent formulas.

We assume that the transformation leaves the vacuum invariant,
\begin{equation}
    \hat M \vac = \vac,
\end{equation}
which ensures that $\hat M$ defines a well-defined transformation on the entire Fock space. The unitarity of $M$ is required to preserve the canonical commutation relations. Indeed, if the original modes satisfy
\begin{equation}
    [\hat a_j, \hat a_k^\dagger] = \delta_{j,k},
\end{equation}
then the transformed modes obey
\begin{equation}
    [\hat b_j, \hat b_k^\dagger] = \delta_{j,k},
\end{equation}
if and only if $M$ is unitary. Physically, this corresponds to the preservation of mode orthogonality and photon number statistics, as expected for passive linear-optical transformations such as beam splitters (BS) and phase shifters~\cite{reck_experimental_1994}.

From now on, we denote by $\hat M$ the unitary operator acting on the Fock space that implements the mode transformation associated with the matrix $M$. Since $\hat M$ is a proper mode transformation, it is automatically unitary and therefore preserves inner products and norms of states.

Let $M$ and $N$ be two unitary matrices describing two-mode transformations. The action of the composed operator $\hat M \hat N$ on the creation operators is given by
\begin{align}
    \hat M \hat N \hat a_j^\dagger \hat N^\dagger \hat M^\dagger
    &= \sum_{k=0}^{n-1} N_{kj} \hat M \hat a_k^\dagger \hat M^\dagger \notag,\\
    &= \sum_{k,l=0}^{n-1} N_{kj} M_{lk} \hat a_l^\dagger \notag,\\
    &= \sum_{l=0}^{n-1} (MN)_{lj} \hat a_l^\dagger .
\end{align}
Thus, the composition of the operators $\hat M$ and $\hat N$ corresponds exactly to the matrix product $MN$. Using a compact notation, one may write
\begin{equation}
    \hat M \hat N = \widehat{MN}.
\end{equation}

In the language of group theory, this construction defines a group homomorphism from the unitary group $U(n)$ to the group of unitary operators acting on the $n$-mode bosonic Fock space. This homomorphism underlies the mathematical description of linear optics and plays a central role in interferometry, quantum state engineering, and photonic quantum information processing~\cite{kok_linear_2007}.

\paragraph{\lit Mode transformation in first quantization}
Above, we defined mode transformations through their action on creation operators. Equivalently, they can be defined directly through their action on the Fock space introduced in Sec.~\ref{subsec: FockSpace}. Let ${\ket{j}}_{j=0,\dots,n-1}$ be a basis of the single-particle Hilbert space associated with the mode-operator basis ${\hat a_j^\dagger}_{j=0,\dots,n-1}$. A mode transformation $\hat M$ can then be described by its action on each individual photon
\begin{equation}
    \ket{l:j}\mapsto \hat M \ket{l:j} = \sum_{k=0}^{n-1} M_{kj}\ket{l:k},
\end{equation}
where $\ket{l:j}$ denotes the state of the $l$-th photon occupying mode $j$. By linearity, this definition extends to the entire single-particle Hilbert space, and subsequently to the full Fock space through symmetrization.

This perspective is particularly useful because it makes explicit that mode transformations act locally on individual photons. Consequently, such transformations cannot generate particle entanglement between photons; they can only redistribute entanglement already present in the state. A derivation of the equivalence between the first- and second-quantized descriptions is provided in Appendix~\ref{app: formalism and framework}, Result~\ref{res: equivalence of mode transformation definitions}.

\paragraph{\lit Passive linear transformations}

Splitting the Hilbert space of modes into \emph{internal} ({\it e.g.}, frequency, polarization\dots) and \emph{external} modes ({\it e.g.}, spatial modes)
\begin{equation}
    \mathcal H=\mathcal H_{\text{int}} \otimes \mathcal H_{\text{ext}},
\end{equation}
one can further define a subclass of mode transformations which acts on the external degree of freedom only. We call these operations \emph{passive linear transformations}, as they correspond to passive linear-optical transformations such as beam splitters and phase shifters. Parametrizing creation operators as
\begin{equation}
    \hat a_j^\dagger(\lambda),
\end{equation}
where $j$ labels the external mode and $\lambda$ denotes the internal DOF, a passive linear transformation is defined analogously by the relation
\begin{equation}
    \hat M \hat a_j^\dagger(\lambda) \hat M^\dagger 
    = \sum_{k=0}^{n-1} M_{kj}\,\hat a_k^\dagger(\lambda),
\end{equation}
and by assuming that $\hat M\vac=\vac$. Such transformations will be extensively used in Chap.~\ref{chap: HOM interferometry and metrology}, and they follow the same compatibility rule for composition as general mode transformations, \emph{i.e.}, $\hat M \hat N = \widehat{MN}$, where $M$ and $N$ are the unitary matrices associated with $\hat M$ and $\hat N$ respectively.

\subsection{Phase Reference and Superselection Rule}
\label{subsec: Phase Reference and SSR}
A recurrent conceptual issue in quantum optics concerns the definition and physical relevance of the \emph{global phase} of a quantum optical state. While global phases are unobservable in isolated systems, the issue becomes operationally meaningful when quantum states are communicated, compared, or jointly processed by spatially separated parties. In such scenarios, the physical interpretation of phase depends crucially on the availability of a shared phase reference~\cite{bartlett_dialogue_2006}.

\paragraph{\lit Historical background.}
The notion of a superselection rule was introduced by Wick, Wightman, and Wigner in their seminal work on the foundations of quantum mechanics~\cite{wick_intrinsic_1952}. They argued that certain observables, such as electric charge, define distinct sectors of Hilbert space between which coherent superpositions are physically inaccessible. An intuitive example is provided by the distinction between bosonic and fermionic states. Let $\ket{B}$ and $\ket{F}$ denote, respectively, a bosonic and a fermionic state. Under a rotation by $2\pi$, the bosonic state remains invariant whereas the fermionic state acquires a minus sign. Consequently, a superposition
\begin{equation}
    \ket{\psi}=\ket{B}+\ket{F}
\end{equation}
would transform as
\begin{equation}
    R(2\pi)\ket{\psi}=\ket{B}-\ket{F}.
\end{equation}
The relative phase between the bosonic and fermionic sectors would therefore appear to be observable, despite the fact that a full $2\pi$ rotation corresponds to the identity operation on the physical configuration of the system. This observation motivates the so-called univalence superselection rule, according to which coherent superpositions of bosonic and fermionic states are forbidden.

More generally, states carrying different values of a conserved quantity, such as electric charge, were long assumed not to interfere, leading to the notion of charge superselection sectors. Similar arguments were later extended to massive particles, for which coherent superpositions of different particle-number sectors were often regarded as unphysical. By contrast, photons, being massless and electrically neutral, were generally considered exempt from such restrictions. Furthermore, the success of coherent states in quantum optics, together with the practical availability of intense laser fields acting as phase references, suggested that superpositions of different photon-number states were physically meaningful.

This viewpoint was challenged by Mølmer's influential analysis of optical coherence~\cite{molmer_optical_1997}, which reignited a broader debate regarding the physical status of coherence and the role of phase references in quantum optics~\cite{rudolph_requirement_2001,wiseman_defending_2004,bartlett_dialogue_2006}. The modern understanding that emerged from these discussions is that many apparent superselection rules are not fundamental constraints, but rather consequences of the absence of an appropriate reference frame. In particular, coherence between different photon-number sectors can only be operationally accessed when a shared phase reference is available. This insight led to the development of the quantum reference frame approach, in which superselection rules are understood as restrictions arising from a lack of shared symmetry-breaking resources rather than as intrinsic limitations of quantum theory~\cite{aharonov_charge_1967,bartlett_reference_2007}. As we explain in detail below, the presence or absence of a photon-number superselection rule is therefore fundamentally linked to the availability of a phase reference.

\paragraph{\lit Motivating example: transmission of a coherent state.}
Consider a sender who prepares a single-mode coherent state~\cite{glauber_quantum_1963,sudarshan_equivalence_1963}
\begin{equation}
    \ket{\alpha} = e^{-\abs{\alpha}^2/2} \sum_{n=0}^{\infty} \frac{\alpha^n}{\sqrt{n!}} \ket{n},
\end{equation}
with $\alpha = \abs{\alpha} e^{i\theta}$, and transmits it to a distant receiver. The phase $\theta$ is defined with respect to a phase reference, typically implemented by an implicit local oscillator. If the receiver does not share this phase reference with the sender, the value of $\theta$ has no operational meaning from the receiver's perspective.

Operationally, the receiver will manipulate a state that must be obtained by averaging over all possible phases, leading to the effective state
\begin{equation}
    \rho = \int_{0}^{2\pi} \frac{\dd\theta}{2\pi} \,
    \ketbra{\alpha e^{i\theta}}{\alpha e^{i\theta}}
    = \sum_{n=0}^{\infty} e^{-\abs{\alpha}^2} \frac{\abs{\alpha}^{2n}}{n!} \ketbra{n}.
\end{equation}
The coherent superpositions between different photon-number sectors are lost, and the state becomes diagonal in the Fock basis. This illustrates that the phase of a coherent state is not an intrinsic property of the state alone, but a \emph{relational} property defined with respect to a reference system.

\paragraph{\lit Generalization to arbitrary single-mode pure states.}
The same reasoning applies to any single-mode pure state of the form
\begin{equation}
    \ket{\psi} = \sum_{n=0}^{\infty} c_n \ket{n}.
\end{equation}
In the absence of a shared phase reference, the physically accessible state is obtained by averaging over all global phase shifts,
\begin{equation}
    \rho = \int_{0}^{2\pi} \frac{\dd\theta}{2\pi} \,
    e^{-i\theta \hat{n}} \ketbra{\psi} e^{i\theta \hat{n}}
    = \sum_{n=0}^{\infty} \abs{c_n}^2 \ketbra{n},
\end{equation}
where $\hat{n}$ is the photon-number operator. All coherence between different photon-number sectors becomes unobservable. Consequently, an initial pure state exhibiting superpositions of different photon numbers is operationally equivalent to a mixed state diagonal in the photon-number basis. In this context, there is no physically accessible coherent superposition of photon numbers, but only a statistical mixture.

\paragraph{\lit Operator viewpoint and photon-number superselection.}
This restriction can equivalently be formulated at the level of observables. In the absence of a phase reference, all physically measurable operators $\hat{O}$ must commute with the total photon-number operator,
\begin{equation}
    [\hat{O}, \hat{n}] = 0.
\end{equation}
Operators capable of detecting relative phases between different photon-number sectors are therefore excluded. This constraint defines a \emph{superselection rule} (SSR) for photon number~\cite{wick_intrinsic_1952}: coherent superpositions of different eigenvalues of the total photon-number operator cannot be operationally accessed. From this perspective, the loss of phase information is not a dynamical decoherence process, but rather a restriction on the set of allowed measurements imposed by the absence of a reference frame.

\paragraph{\lit Superselection rules for multiple systems.}
The photon-number superselection rule extends naturally to multimode systems. Consider $k$ spatially separated modes described by annihilation operators $\hat{a}_1,\dots,\hat{a}_k$. The total photon-number operator is
\begin{equation}
    \hat{N}_{\text{tot}} = \sum_{j=1}^{k} \hat{a}_j^{\dagger} \hat{a}_j.
\end{equation}
In the absence of a shared phase reference, all physically measurable operators $\hat{O}$ must satisfy
\begin{equation}
    [\hat{O}, \hat{N}_{\text{tot}}] = 0.
\end{equation}
As a result, only pure states with a well-defined total photon number $N$,
\begin{equation}
    \ket{\psi} = \sum_{n_1+\cdots+n_k = N}
    c_{n_1,\dots,n_k} \ket{n_1,\dots,n_k},
\end{equation}
can be prepared and distinguished as pure states. More generally, only statistical mixtures of such fixed-total-photon-number states are physically meaningful.

\paragraph{\lit Phase reference as an external quantum system.}
The notion of phase can be restored by explicitly introducing a phase reference, modeled as an additional quantum system~\cite{bartlett_dialogue_2006,wiseman_quantum_2009}. In this enlarged description, the total system is invariant under global phase shifts, while relative phases between subsystems become observable. States that appear mixed when the phase reference is traced out may in fact be pure in the extended Hilbert space.

One may therefore distinguish two equivalent representations:
\begin{itemize}
    \item A description in which the phase reference is explicitly represented in the quantum state, for example
    \begin{equation}
        \ket{\psi} = \sum_{k=0}^{N} c_k \ket{k, N-k},
    \end{equation}
    where the first mode corresponds to the quantum system of interest and the second mode acts as a phase reference. Such states will be referred to as \emph{SSR-compliant} (SSRC).
    
    \item A description in which the phase reference exists physically but is not explicitly represented in the formalism. This allows one to manipulate coherent superpositions of photon numbers, such as
    \begin{equation}
        \ket{\psi} = \sum_{k=0}^{\infty} c_k \ket{k}.
    \end{equation}
    This representation is more concise, but obscures the fact that a phase reference is implicitly required. In this manuscript, we will refer to this description as the continuous-variable (CV) representation introduced in Sec.~\ref{subsec: continuous variables}.
\end{itemize}

\paragraph{\publi Equivalence and practical implications.}
In the limit where the phase reference contains a large number of photons, these two viewpoints become operationally equivalent, yielding identical predictions for all physically allowed measurements. Nevertheless, the superselection rule formulation provides a conceptually transparent and operationally economical framework when no shared phase reference is available.

This perspective highlights which resources are genuinely required to access phase-sensitive quantum features and clarifies the role of reference frames as physical and potentially consumable resources. It is particularly useful in quantum communication and quantum metrology, where an explicit accounting of phase references allows one to identify operational advantages, resource costs, and fundamental limitations arising from symmetry and superselection constraints~\cite{bartlett_dialogue_2006}. These aspects will be developed in greater detail in Chap.~\ref{chap: SSR}.

\subsection{Entanglement in quantum optics}
\label{subsec: entanglement_in_quantum_optics}

We have introduced the basic definitions of entanglement in Sec.~\ref{subsec: Composite Quantum Systems and Entanglement}. In this short section we discuss subtleties of entanglement in quantum optics, focusing on the two associated notions of mode and particle entanglement, and on their connection to phase references and superselection rules.

\paragraph{\lit Two types of entanglement}
In quantum optics, it is often natural to define subsystems in terms of spatial or spectral field modes rather than individual particles. In this description, each mode is associated with a Fock space, and the global Hilbert space factorizes as a tensor product over modes, as shown in Eq.~\eqref{eq: Fock space tensorization}. Entanglement can then be defined with respect to this mode decomposition in the standard way.

A prototypical example is the single-photon state distributed over two spatial modes,
\begin{equation}\label{eq: non local single photon}
    \frac{1}{\sqrt{2}}\left(\ket{1}_a \ket{0}_b + \ket{0}_a \ket{1}_b \right),
\end{equation}
which is formally equivalent to a Bell state in the single-excitation subspace. From this perspective, the state is entangled across modes, despite involving only a single physical excitation.

As discussed in Sec.~\ref{subsec: FockSpace}, one can alternatively describe optical systems in terms of individual indistinguishable photons. One can thus consider entanglement between these particles as a fundamental resource. In this framework, entanglement is defined with respect to tensor products of particle Hilbert spaces rather than field modes. The properties of particle entanglement, and how it can be harnessed for applications, will be a major thread of Chap.~\ref{chap: SSR}.

\paragraph{\lit The single-particle example and superselection rules}
The single-photon state distributed over two modes of Eq.~\eqref{eq: non local single photon} provides a paradigmatic example illustrating the difference between these two points of view, as well as the subtleties involved in harnessing such nonlocal correlations in practice. In the mode description, the state is entangled as written above. However, from a particle-based perspective, the system contains only a single excitation, suggesting that no entanglement between particles is present.

Assuming that the photon is distributed among two spatially separated modes, this distinction raises the question of whether such a single-photon state can be used to experimentally observe nonlocal effects, such as a violation of a Bell inequality. This question has sparked a substantial debate in the literature. By considering implementations based on excitations of two-level atoms, van Enk~\cite{van_enk_single-particle_2005} argued that single-particle states exhibit genuine entanglement when defined with respect to modes, and that this entanglement can, in principle, be used as a resource. Peres~\cite{peres_nonlocal_1995} proposed an alternative perspective, suggesting that direct measurement schemes can already lead to nonlocal correlations. Björk et al.~\cite{bjork_single-particle_2001} further analyzed nonlocal correlations arising from single-photon interference, showing how a coherent phase reference can play a crucial role in revealing nonclassical features.

Taken together, these proposals and their subsequent discussion indicate that the ability to extract nonlocal correlations from a single-photon state is intimately connected to the presence of superselection rules (SSR), in particular those associated with particle-number conservation. Indeed, each of the above approaches relies, either explicitly or implicitly, on the availability of a phase reference (for instance a coherent beam, or an operational mechanism that effectively breaks particle-number conservation) to circumvent the constraints imposed by SSR. In the absence of such a reference, the entanglement present in the mode description may remain operationally inaccessible, highlighting the central role of phase references in quantum optical experiments.

Entanglement and its usability as a resource will be a central theme of this thesis. In particular, Chap.~\ref{chap: collective entanglement} discusses correlations between individual photons, while Chap.~\ref{chap: SSR} analyzes these aspects in the context of photon-number superselection rules.

\fi

\ifnum \theShowChaptwo=1
\chapter{Time-frequency quantum optics}
\setcurrentanchor{chap2}
\label{chap: Time-Frequency Systems}
\emph{This chapter introduces one of the paradigmatic quantum optical systems studied in this thesis: the time and frequency degrees of freedom of photons. We describe the formalism used as well as many of its properties. While this formalism applies to general photon-number distributions, the following discussion will focus mainly on the single-photon case.}

\localtableofcontents

\section{Definitions}
\label{sec: TF definitions}
\emph{In this section we provide the basic definitions of time or frequency modes for quantum optical systems. We introduce the associated states and operators. We also consider the link with continuous variables and define the chrono-cyclic Wigner function.}

\subsection{Motivations}
\label{subsec: TF motivations}
Quantum optical systems are frequently described under the assumption of monochromatic or quasi-monochromatic fields. While this approximation is convenient and often adequate for narrowband sources, it is never strictly valid. Any physically realizable optical field necessarily exhibits a finite spectral bandwidth, as dictated by the Fourier time-bandwidth relation and by practical constraints such as source dynamics, dispersion, and decoherence. In many modern quantum optical platforms, including ultrafast laser systems, spontaneous parametric down-conversion sources, and broadband solid-state emitters, the spectral width is not a small perturbation, but a defining characteristic of the system~\cite{boyd_nonlinear_2008, mandel_optical_1995}.

The limitations of the monochromatic approximation become evident for sources with intrinsically broad spectra. In such cases, treating frequency as a fixed parameter obscures essential physical phenomena, including temporal correlations, dispersion-induced distortions, and spectral entanglement. A description that explicitly incorporates both time and frequency degrees of freedom is therefore required for an accurate and complete characterization of the quantum state~\cite{walmsley_characterization_2009}.

Time- and frequency-bin encodings constitute an important step beyond a monochromatic regime, where spectral and temporal parameters are disregarded, by explicitly exploiting the temporal or spectral structure of optical fields. By discretizing time into localized wave packets or frequency into well-defined spectral modes, these approaches access enlarged Hilbert spaces and enable a variety of powerful applications in quantum communication and information processing, including high-dimensional encoding schemes and increased robustness to certain noise mechanisms~\cite{brendel_pulsed_1999, tagliavacche_frequency-bin_2025}. Nevertheless, such encodings remain intrinsically model-dependent: they impose strict constraints on the admissible temporal or spectral profiles by restricting the description to a predefined set of modes or bins. While this discretization is often well suited to specific experimental implementations, it does not capture the full generality of broadband quantum optical fields. A fully general time-frequency framework should therefore avoid \emph{a priori} restrictions on the spectral or temporal structure of the state, and instead treat arbitrary waveforms on equal footing, with time- and frequency-bin encodings emerging as particular limiting cases or convenient approximations within a more comprehensive description.

Beyond the choice of representation, it is crucial to adopt an explicit operator-based point of view when describing time-frequency degrees of freedom. In practice, many approaches, particularly in experimental contexts, already account for the full spectral amplitude of a photon or optical field. However, working solely at the level of spectral or temporal waveforms is inherently limiting, as it focuses on state descriptions while leaving the underlying physical transformations implicit. This can obscure the structure of dynamical processes and, in some cases, conceal nontrivial effects arising from mode mixing, dispersion, or correlations between time and frequency. By contrast, defining operators that act directly on the time-frequency modes provides a clear and concise framework for describing evolution, modulation, and measurement processes. Such operators make explicit how physical transformations act on the field, allow for systematic composition of operations, and enable the use of all quantum mechanical results written in terms of operators. An operator-based formalism thus clarifies the physical content of time-frequency dynamics and provides a natural language for analyzing and designing quantum optical protocols involving arbitrary broadband states.

From an operational standpoint, time and frequency DOFs are particularly attractive due to their accessibility and controllability. Temporal delays, phase shifts, and spectral filtering can be implemented with high precision using well-established optical components. Recent advances in electro-optic modulation and pulse shaping further enable coherent control across both domains, allowing for sophisticated quantum state preparation and manipulation~\cite{weiner_ultrafast_2011}. Correspondingly, measurement techniques such as time-resolved detection, spectral analysis, and interferometric delay measurements directly probe these variables, making them especially suitable for experimental characterization.

Time-frequency variables are also intrinsically linked to optical metrology. Tasks such as delay estimation, dispersion measurement, and clock synchronization fundamentally rely on the interplay between temporal and spectral properties of light. Quantum-enhanced metrology in these contexts often exploits broadband states and temporal correlations that cannot be captured within narrowband or single-mode approximations~\cite{giovannetti_advances_2011, humphreys_quantum_2013}. Treating time and frequency on equal footing is therefore essential for describing and optimizing such protocols.

These considerations motivate the need for a formal theoretical framework that describes time and frequency as continuous variables equipped with suitably defined operators. A consistent quantum description of time-frequency DOFs must account for their conjugate nature, the structure of the associated mode spaces, and their role in quantum measurement and information processing. Such a framework provides the foundation for accurately modeling broadband quantum optical systems, systematically analyzing time-frequency encodings, and exploring new regimes of quantum control and metrology beyond the scope of monochromatic or single-mode descriptions.

This formalism was first introduced and systematically studied in~\cite{fabre_quantum_2020,fabre_time_2022}.  The purpose of the present chapter is to present the main concepts and technical tools developed within this framework.  These elements will serve as a common foundation for the remainder of the thesis and will be repeatedly used and extended in subsequent chapters. Furthermore, the following chapters (especially Sections~\ref{sec: TF entanglement and metrology}, \ref{sec: collective GKP codes} and \ref{sec: TF HOM}) will reveal how such a framework yields a better understanding of the quantum properties of time-frequency systems.

\subsection{Creation and annihilation operators}
\label{subsec: TF crea anni operators}

As introduced in the previous chapter (see Sections~\ref{subsec: Field Quantization} and~\ref{subsec: FockSpace}), the quantized electromagnetic field can be described as a collection of independent bosonic modes, each associated with annihilation and creation operators $\hat a_\lambda$ and $\hat a_\lambda^\dagger$. The index $\lambda$ labels the distribution of the relevant degrees of freedom of the field, such as frequency, polarization, or spatial mode. In the time-frequency description adopted here, these modes are naturally labeled by their angular frequency distribution, or equivalently by their temporal profile.

\paragraph{\lit Single spatial mode}
We first consider the simplest situation, namely a single propagation mode, where frequency is the only relevant degree of freedom. In this case, the specific frequency distribution completely characterizes the modal properties. The Hilbert space of modes is then given by the set of functions
\begin{equation}
    F: \omega\mapsto F(\omega).
\end{equation}
The associated annihilation and creation operators $\hat a_F$ and $\hat a_F^\dagger$ then follow by definition the corresponding commutation relation
\begin{equation}
    [\hat a_F, \hat a_G^\dagger] = \int \dd\omega\, F^*(\omega) G(\omega),
\end{equation}
where $\int \dd\omega\, F^*(\omega) G(\omega)$ corresponds to the natural scalar product between the frequency distributions $F$ and $G$. In the following, we typically consider the specific distribution 
\begin{equation}
    F(\omega)=\delta(\omega-\omega_0),
\end{equation}
which permits introducing the monochromatic annihilation operator $\hat a(\omega_0)$, which satisfies the canonical bosonic commutation relations
\begin{align}
    [\hat a(\omega), \hat a^\dagger(\omega')] &= \delta(\omega-\omega'), & [\hat a(\omega), \hat a(\omega')] &= [\hat a^\dagger(\omega), \hat a^\dagger (\omega')] = 0.
\end{align}
In this formalism, we allow for negative frequencies $\omega < 0$. These should not be interpreted as unphysical excitations, but rather as a convenient mathematical representation corresponding to propagation at frequency $\abs{\omega}$ in the opposite direction along the chosen propagation axis~\cite{garrison_quantum_2014}. Since time and frequency are related by Fourier transformation, it is natural to introduce the associated time-domain annihilation operator
\begin{equation}
    \hat b(t) = \frac{1}{\sqrt{2\pi}} \int \dd\omega\, e^{-i\omega t} \hat a(\omega).
\end{equation}
The parameter $t$ can be interpreted operationally as the time of arrival of a photon, for instance as measured by an ideal time-resolving detector placed at the output of the propagation mode~\cite{scully_quantum_1997}. Inverting the Fourier transform yields
\begin{equation}
    \hat a(\omega) = \frac{1}{\sqrt{2\pi}} \int\dd t\, e^{i\omega t} \hat b(t).
\end{equation}
As shown explicitly in Appendix~\ref{app: TF basic computations}, Result~\ref{res: Time creation op commutation}, the time-domain operators satisfy the canonical commutation relation
\begin{equation}
    [\hat b(t), \hat b^\dagger(t')] = \delta(t-t').
\end{equation}
This confirms that $\hat b(t)$ and $\hat b^\dagger(t)$ can be interpreted as {\it bona fide} annihilation and creation operators.

\paragraph{\lit Multiple spatial modes}
The restriction to a single propagation mode is often insufficient in realistic scenarios. In particular, when dealing with multiple photons, it is frequently necessary to introduce several auxiliary modes such as position or polarization in order to avoid unwanted photon-photon interactions or to describe multi-port optical networks~\cite{reck_experimental_1994, kok_linear_2007}. We therefore consider the more general case where several such modes are present. Following the monochromatic notation introduced before, a relevant basis for the annihilation and creation operators is now given by
\begin{align}
    \hat a_\alpha(\omega), && \hat a_\alpha^\dagger(\omega),
\end{align}
and these operators satisfy the multimode bosonic commutation relations
\begin{align}
    [\hat a_\alpha(\omega), \hat a_{\alpha'}^\dagger(\omega')] &= \delta_{\alpha,\alpha'} \delta(\omega-\omega'), & [\hat a_\alpha(\omega), \hat a_{\alpha'}(\omega')] &= [\hat a_\alpha^\dagger(\omega), \hat a_{\alpha'}^\dagger(\omega')] = 0.
\end{align}
Here, the discrete index $\alpha = 1,\dots,n$ labels an additional degree of freedom, which may correspond to distinct spatial modes, waveguides, polarization channels, or paths in an interferometer. For simplicity, and unless stated otherwise, we will refer to $\alpha$ as a \emph{spatial mode}. The operator $\hat a_\alpha^\dagger(\omega)$ thus creates a photon of frequency $\omega$ in spatial mode $\alpha$. The corresponding time-domain operators are defined as
\begin{equation}
    \hat b_\alpha(t) = \frac{1}{\sqrt{2\pi}} \int \dd\omega\, e^{-i\omega t} \hat a_\alpha(\omega),
\end{equation}
with inverse relation
\begin{equation}
    \hat a_\alpha(\omega) = \frac{1}{\sqrt{2\pi}} \int \dd t\, e^{i\omega t} \hat b_\alpha(t).
\end{equation}
As shown in Appendix~\ref{app: TF basic computations}, Result~\ref{res: Time creation op commutation}, these operators obey the canonical commutation relations
\begin{equation}
    [\hat b_\alpha(t), \hat b_{\alpha'}^\dagger(t')] = \delta_{\alpha,\alpha'} \delta(t-t'),
\end{equation}
which ensures the independence of different spatial modes and the consistency of the multimode time-frequency description.

\paragraph{\lit Physical interpretation}
In this thesis, photons are described through their spectral annihilation operators $a(\omega)$ and the corresponding temporal operators $b(t)$, related by a Fourier transform. Within the paraxial approximation, the optical field is assumed to propagate predominantly along a fixed longitudinal axis $z$, so that the transverse degrees of freedom can be neglected. The photon frequency $\omega$ is then directly associated with the longitudinal wavevector component $k_z$ through the dispersion relation
\begin{equation}
    k_z \simeq \frac{\omega}{c},
\end{equation}
for propagation in free space (or more generally $k_z = n(\omega)\omega/c$ in a medium). As a consequence, the pair $(\omega,t)$ forms a Fourier-conjugate pair that is formally equivalent to the longitudinal momentum--position pair $(k_z,z)$, since the field amplitudes are related through
\begin{equation}
    e^{-i\omega t}\longleftrightarrow e^{ik_z z}.
\end{equation}
Therefore, manipulations performed in the frequency and arrival-time domains can be interpreted as operations on the longitudinal momentum and position degrees of freedom of the photon, with the correspondence
\begin{equation}
    (\omega,t)\longleftrightarrow(k_z,z).
\end{equation}
This identification provides a convenient physical picture in which spectral shaping corresponds to engineering the longitudinal momentum distribution, while temporal shaping corresponds to controlling the longitudinal spatial wavepacket.

\subsection{Time-frequency states}
\label{subsec: TF states}
In Section~\ref{subsec: FockSpace} we described how general quantum states of light can be constructed in Fock space by repeated application of creation operators. For completeness and for later use, we now reformulate these constructions explicitly in the time-frequency representation, which is particularly convenient for the description of photonic wave packets and correlations in quantum optics~\cite{mandel_optical_1995,loudon_quantum_2000}.

\paragraph{\lit Single-photon states}
A particularly important class of states for the following discussion is given by single-photon states. Physically, these correspond to situations in which exactly one photon occupies each spatial mode $\alpha$, for a total of $n$ photons. Such states can be written as
\begin{equation}
    \ket{\psi} = \int \dd\omega_1 \cdots \dd\omega_n\, F(\omega_1,\dots,\omega_n)\, \hat a_1^\dagger(\omega_1)\cdots \hat a_n^\dagger(\omega_n)\vac,
\end{equation}
where the complex-valued function $F$ is known as the \emph{joint spectral amplitude} (JSA)\footnote{The term joint spectral amplitude is typically reserved for the case $n=2$. For ease of language we will sometimes use it for general $n$, even $n=1$.} of the state~\cite{grice_spectral_1997,law_continuous_2000}. The normalization condition $\braket{\psi}{\psi}=1$ implies
\begin{equation}
    \int \dd\omega_1 \cdots \dd\omega_n\, \abs{F(\omega_1,\dots,\omega_n)}^2 = 1.
\end{equation}
The squared magnitude of the JSA is referred to as the joint spectral intensity (JSI). When manipulating single-photon states, it is convenient to introduce the shorthand notation
\begin{equation}
    \hat a_1^\dagger(\omega_1)\cdots \hat a_n^\dagger(\omega_n)\vac
    \equiv
    \ket{\omega_1,\dots,\omega_n},
\end{equation}
which represents perfectly monochromatic photons with frequencies $\omega_1,\dots,\omega_n$ in spatial modes $1,\dots,n$, respectively. Using this notation, the state $\ket{\psi}$ can be written more compactly as
\begin{equation}
    \ket{\psi} = \int \dd\omega_1 \cdots \dd\omega_n\, F(\omega_1,\dots,\omega_n)\ket{\omega_1,\dots,\omega_n}.
\end{equation}

Perfectly monochromatic states satisfy the scalar product
\begin{equation}
    \braket{\omega_1',\dots,\omega_n'}{\omega_1,\dots,\omega_n} = \prod_{j=1}^n \delta(\omega_j-\omega_j'),
\end{equation}
which shows that such states are not normalizable. This is consistent with the fact that perfectly monochromatic light fields are unphysical; only wave packets with finite spectral width can be realized experimentally~\cite{mandel_optical_1995,loudon_quantum_2000}. Using Fourier-transform relations between frequency and time representations, one can equivalently express the single-photon state in terms of temporal variables,
\begin{equation}
    \ket{\psi} = \int \dd t_1 \cdots \dd t_n\, \tilde F(t_1,\dots,t_n)\, \hat b_1^\dagger(t_1)\cdots \hat b_n^\dagger(t_n)\vac,
\end{equation}
where
\begin{equation}
    \tilde F(t_1,\dots,t_n) = \frac{1}{(2\pi)^{n/2}} \int \dd\omega_1 \cdots \dd\omega_n\, e^{-i(\omega_1 t_1+\cdots+\omega_n t_n)} F(\omega_1,\dots,\omega_n)
\end{equation}
is the $n$-dimensional Fourier transform of the JSA. The function $\tilde F$ is referred to as the \emph{joint temporal amplitude} (JTA)~\cite{kolobov_quantum_2007}. The normalization condition in the temporal domain reads
\begin{equation}
    \int \dd t_1 \cdots \dd t_n\, \abs{\tilde F(t_1,\dots,t_n)}^2 = 1.
\end{equation}
The squared absolute value of $\tilde F$ is named the joint temporal intensity (JTI). By taking the JTA to be a product of Dirac distributions, one is led to the definition of single-photon states perfectly localized in time,
\begin{equation}
    \hat b_1^\dagger(t_1)\cdots \hat b_n^\dagger(t_n)\vac = \ket{t_1,\dots,t_n},
\end{equation}
which, similarly to monochromatic states, are non-normalizable and therefore unphysical idealizations. Both the JSA and the JTA can be recovered by taking the scalar product of the state $\ket{\psi}$ with the corresponding frequency or time eigenstates,
\begin{align}
    F(\omega_1,\dots,\omega_n) &= \braket{\omega_1,\dots,\omega_n}{\psi}, & \tilde F(t_1,\dots,t_n) &= \braket{t_1,\dots,t_n}{\psi}.
\end{align}

\paragraph{\lit General states}
Beyond the single-photon regime, states involving arbitrary superpositions of photon numbers can be considered, as already discussed in Section~\ref{subsec: FockSpace}. For clarity, we first focus on a single spatial mode. A general time-frequency state on a single spatial mode can be written as
\begin{equation}
    \label{eq: general TF state single mode}
    \ket{\psi} = \sum_{k=0}^\infty c_k \ket{\psi_k} = \sum_{k=0}^\infty c_k \int \dd\omega_1\cdots \dd\omega_k\, F_k(\omega_1,\dots,\omega_k) \hat a^\dagger(\omega_1)\cdots\hat a^\dagger(\omega_k)\vac,
\end{equation}
where $\ket{\psi_k}$ denotes the component of the state containing exactly $k$ photons. Each $k$-photon component is characterized by a function $F_k$, which generalizes the notion of a JSA to the $k$-photon sector. The complex coefficients $c_k$ satisfy the normalization condition
\begin{equation}
    \sum_{k=0}^\infty \abs{c_k}^2 = 1.
\end{equation}
Since photons are bosons, the functions $F_k$ can be taken to be symmetric under permutations of their arguments. Under this assumption, their normalization reduces to
\begin{equation}
    k! \int \dd\omega_1\cdots \dd\omega_k\, \abs{F_k(\omega_1,\dots,\omega_k)}^2 = 1.
\end{equation}

We now extend this construction to $n$ spatial modes. A general multimode state can be written as
\begin{equation}
    \label{eq: general TF state multimode}
    \ket{\psi} = \sum_{k_1,\dots,k_n=0}^\infty C_{k_1,\dots,k_n} \ket{\psi_{k_1,\dots,k_n}},
\end{equation}
where $\ket{\psi_{k_1,\dots,k_n}}$ denotes a state with $k_\alpha$ photons in spatial mode $\alpha$. The coefficients $C_{k_1,\dots,k_n}$ satisfy the normalization condition
\begin{equation}
    \sum_{k_1,\dots,k_n=0}^\infty
    \abs{C_{k_1,\dots,k_n}}^2 = 1.
\end{equation}
Each fixed-photon-number component can be expanded as
\begin{equation}
    \ket{\psi_{k_1,\dots,k_n}} = \int \dd\{\omega_\alpha^j\}\, F(\{\omega_\alpha^j\}) \prod_{\alpha=1}^n \prod_{j=1}^{k_\alpha} \hat a_\alpha^\dagger(\omega_\alpha^j)\vac,
\end{equation}
where the integration is performed over the set of $k_1+\cdots+k_n$ frequency variables $\omega_\alpha^j$ with $\alpha=1,\dots,n$ and $j=1,\dots,k_\alpha$. The function $F$ generalizes the JSA to arbitrary photon numbers and modes; it encodes both the spectral distributions of individual photons and their frequency correlations across modes~\cite{kolobov_quantum_2007}.

Without loss of generality, $F$ can be assumed to be symmetric under permutations of the variables $\omega_\alpha^j$ for fixed $\alpha$. Under this assumption, its normalization condition reads
\begin{equation}
    k_1!\cdots k_n! \int \dd\{\omega_\alpha^j\}\, \abs{F(\{\omega_\alpha^j\})}^2 = 1.
\end{equation}
If this symmetry is not imposed, the factorials must be replaced by explicit sums over all permutations of the indices $\omega_\alpha^j$ for each fixed $\alpha$. Although such explicit expressions provide a complete description of general multimode states, they will rarely be required in the remainder of this manuscript. Instead, more compact operator-based or mode-decomposed descriptions will be preferred whenever possible.

\subsection{Time and frequency operators}
\label{subsec: TF operators}
As motivated in the introduction of this section, we aim at providing a general operator-based formalism for the description of time-frequency degrees of freedom in quantum optics. To this end, we introduce the frequency operator $\hat \omega_\alpha$ and the time operator $\hat t_\alpha$, where the index $\alpha$ labels the spatial mode on which each operator acts. For clarity, we first focus on the single-spatial-mode case and omit the index $\alpha$; the extension to the multimode setting is straightforward, although notationally slightly more involved.

\paragraph{\lit Definition}
We define the frequency and time operators respectively as~\cite{fabre_time_2022}
\begin{align}
    \hat\omega=\int\dd\omega\, \omega\hat a^\dagger(\omega)\hat a(\omega), && \hat t=\int \dd t\, t \hat b^\dagger(t)\hat b(t).
\end{align}
These definitions should be understood as follows. For each frequency $\omega$, the operator $\hat a^\dagger(\omega)\hat a(\omega)$ is the number operator counting the number of photons with frequency $\omega$. The integral over $\omega$ averages this number distribution with the weight $\omega$; loosely speaking, $\hat\omega$ corresponds to the measurement of the average frequency. An analogous interpretation holds for the operator $\hat t$.

\paragraph{\lit Warning on the time operator}
The introduction of a time operator has been the subject of long-standing and intense debates in quantum mechanics~\cite{pauli_general_2012,busch_time-energy_2002,galapon_paulis_2002}. The central question concerns whether a self-adjoint time operator can exist that is canonically conjugate to the Hamiltonian and associated with the time evolution generated by the Schrödinger equation. This issue probes deeply into the quantum nature of time itself and is not the purpose of the time operator $\hat t$ introduced here.

In our setting, time should not be understood as the external parameter appearing in the Schrödinger equation and governing the dynamical evolution of the quantum state. Instead, it corresponds to a measurable physical quantity associated with the time of detection. Indeed, taking into account the finite speed of propagation, the time degree of freedom can equivalently be mapped onto the longitudinal position along the propagation direction: a photon located farther along the propagation axis will reach the detector earlier. In this operational sense, the time operator $\hat t$ is well defined and widely used in quantum optics~\cite{mandel_optical_1995,breuer_theory_2002}.

\paragraph{\lit Measurement operators}
By inspection of their Hermitian conjugates, both operators $\hat \omega$ and $\hat t$ are Hermitian and therefore represent legitimate quantum observables. We have argued above that they correspond to frequency and time measurements, respectively. This interpretation can be made precise in the single-photon regime. \footnote{For states more general than single-photon states, it is difficult to assign a simple meaning to $\hat \omega$ and $\hat t$. When more than one photon is present, the question ``what is the frequency of \emph{the} photon?'' becomes ill defined, since individual photons cannot be uniquely addressed. In such cases, the operators describe collective or correlated time-frequency properties of the multiphoton state.}

To proceed, we note the following identities, valid for any single-photon state $\ket{\psi}$ with JSA $F$ and JTA $\tilde F$:
\begin{align}\label{eq: tf op actions}
    \hat \omega\ket{\psi} &=\int \dd\omega \,\omega F(\omega)\ket{\omega}, & \hat t\ket{\psi} &=\int \dd t\, t \tilde F(t)\ket{ t},\\ 
    \hat t \ket{\psi} &=-i\int \dd\omega\, F'(\omega)\ket{\omega}, & \hat \omega\ket{\psi} &=i\int \dd t\,\tilde F'(t)\ket{t},
\end{align}
where $F'$ denotes the derivative of $F$ with respect to its argument. These expressions show that the operators act simply on the JSA and JTA as
\begin{align}
    \hat \omega&: F\mapsto \omega F(\omega), &  \hat t&: \tilde F\mapsto t\,\tilde F(t),\\
    \hat t &: F\mapsto -i F', & \hat \omega&:\tilde F\mapsto i\tilde F'.
\end{align}
From these relations, we immediately see that the fixed-frequency states $\ket{\omega}$ are eigenstates of $\hat \omega$,
\begin{equation}
    \hat \omega\ket{\omega}=\omega\ket{\omega}.
\end{equation}
This result implies that $\omega$ is precisely the measurement outcome of the frequency observable for a single-photon state $\ket{\psi}$. Consequently, when restricted to the single-photon subspace, the operator $\hat \omega$ can be written as
\begin{equation}
    \hat \omega=\int \dd \omega\, \omega \ketbra{\omega}{\omega}.
\end{equation}
Entirely analogous reasoning confirms that $\hat t$ measures the time of arrival of a single photon.

\paragraph{\lit Commutation relation}
Using the action of the operators on the JSA and JTA, we can compute their commutator in the single-photon regime~\cite{fabre_time_2022}
\begin{equation}
    [\hat \omega, \hat t]=i\1.
\end{equation}
This relation is reminiscent of the canonical commutation relation between position and momentum operators in quantum mechanics. It reflects the conjugate nature of time and frequency variables, in close analogy with the position-momentum duality. As a direct consequence, applying the Robertson uncertainty relation~\cite{robertson_uncertainty_1929} (see also Appendix~\ref{app: SSRC}, Result~\ref{res: heisenberg robertson}) yields
\begin{equation}
    \Delta^2 \hat \omega \Delta^2 \hat t \geq \frac{1}{4},
\end{equation}
where $\Delta^2 \hat \omega$ and $\Delta^2 \hat t$ denote the variances of the frequency and time observables, respectively. This inequality formalizes the intuitive notion that a photon cannot be simultaneously well localized in both time and frequency domains, reflecting the fundamental trade-off imposed by their conjugate relationship.

Beyond the single-photon subspace, the commutation relation generalizes to
\begin{equation}
    [\hat \omega, \hat t]=i \hat N,
\end{equation}
where
\begin{equation}
    \hat N=\int\dd \omega\, \hat a^\dagger(\omega)\hat a(\omega)
\end{equation}
is the total photon number operator, which reduces to unity in the single-photon regime. Detailed derivations of these commutation relations are provided in Appendix~\ref{app: TF basic computations}, Results~\ref{res: Single photon TF commutation} and \ref{res: General TF commutator}.

\paragraph{\lit Multimode extension}
In the multimode case, one introduces frequency and time operators for each spatial mode $\alpha$. The above definitions extend naturally to
\begin{align}
    \hat\omega_\alpha=\int\dd\omega\, \omega\hat a_\alpha^\dagger(\omega)\hat a_\alpha(\omega),
    && \hat t_\alpha=\int \dd t\, t \hat b_\alpha^\dagger(t)\hat b_\alpha(t).
\end{align}
The action of these operators on the multimode JSA and JTA generalizes the single-spatial-mode case:
\begin{align}
    \hat \omega_\alpha&: F\mapsto \omega_\alpha F(\omega_1,\dots,\omega_n),
    &
    \hat t_\alpha&: \tilde F\mapsto t_\alpha \tilde F(t_1,\dots,t_n),\\
    \hat t_\alpha &: F\mapsto -i\partial_\alpha F,
    &
    \hat \omega_\alpha&:\tilde F\mapsto i\partial_\alpha \tilde F,
\end{align}
where $\partial_\alpha$ denotes the derivative with respect to the variable associated with mode $\alpha$. Following the same reasoning as in the single-mode case, the operators $\hat \omega_\alpha$ and $\hat t_\alpha$ measure the frequency and time of arrival of the photon in mode $\alpha$. The commutation relations naturally generalize to
\begin{equation}
    [\hat \omega_\alpha, \hat t_{\alpha'}]=i \delta_{\alpha,\alpha'} \hat N_\alpha,
\end{equation}
with $\hat N_\alpha$ the photon number operator associated with mode $\alpha$.

\subsection{Analogy with continuous variables}
\label{subsec: TF analogy CV}

\paragraph{\lit Formal analogy} By comparing the results obtained in the single-photon regime with those of continuous-variable systems, one observes a direct and useful analogy. In particular, the operators $\hat \omega$ and $\hat t$ satisfy the same canonical commutation relation as the position and momentum operators $\hat x$ and $\hat p$, respectively. This correspondence can be made explicit by mapping the frequency degree of freedom onto position and the time degree of freedom onto momentum. More precisely, one can introduce the following mapping
\begin{align}
    \hat \omega \leftrightarrow \hat x, && 
    \hat t \leftrightarrow \hat p, && 
    \ket{\omega} \leftrightarrow \ket{x}, && 
    \ket{t} \leftrightarrow \ket{p},
\end{align}
which is naturally extended to the multimode setting. From a fundamental point of view, this analogy is not surprising. Time and frequency are conjugate variables related by a Fourier transform, in the same way as position and momentum are. As a consequence, they obey analogous commutation relations and uncertainty relations, giving rise to mathematically equivalent structures~\cite{busch_operational_1997, galindo_quantum_2012}. In Appendix~\ref{app: transferring op to Fock}, we present an underlying mathematical framework that makes this connection explicit. 

From a practical perspective, this correspondence allows one to directly exploit many of the results developed in continuous-variable quantum information, simply by replacing position and momentum with frequency and time, respectively. In particular, tools such as Gaussian state analysis, phase-space representations, and uncertainty-based criteria can be readily translated to the time-frequency domain~\cite{braunstein_quantum_2005, weedbrook_gaussian_2012,walschaers_non-gaussian_2021}. This analogy will be used extensively throughout the following chapters. 

\paragraph{\lit Limitations} From an experimental and implementation-oriented viewpoint, this correspondence reveals that time-frequency single photons provide an alternative platform for continuous-variable quantum information, based on a physical mechanism that is fundamentally different from the usual quadrature-based implementations relying on the electromagnetic field amplitude. This difference leads to distinct advantages and drawbacks. While time-frequency encodings are particularly sensitive to losses, since the loss of a photon completely erases the encoded information, they are extremely robust against noise in time and frequency, as small perturbations in these variables do not significantly affect the encoded state. Moreover, the availability of advanced temporal and spectral shaping techniques enables the generation of arbitrary single-photon states in this basis~\cite{kolobov_quantum_2007, brecht_photon_2015}, making time-frequency single photons a highly versatile platform for continuous-variable quantum information processing.

\subsection{Chrono-cyclic Wigner function}
\label{subsec: TF Wigner function}
In continuous variables, the Wigner function quasi-probability distribution, which we introduced in Sec.~\ref{subsec: continuous variables}, offers a remarkably powerful representation of a quantum state. The analogy between time-frequency single-photon states and continuous-variable systems suggests the possibility of defining phase-space representations for time-frequency degrees of freedom. In this section, we introduce the \emph{chrono-cyclic Wigner function}, which provides a convenient quasi-probability distribution in the joint time-frequency phase space~\cite{fabre_time_2022}. The definition we provide is limited to the single-photon case and will be used extensively in the remainder of this manuscript.

\paragraph{\lit Single spatial mode}
Given a single time-frequency photon state $\hat\rho$ we define the chrono-cyclic Wigner function or just Wigner function $W(\varphi,\tau)$ for short as
\begin{equation}
    \label{eq: Wigner function single mode}
    W(\varphi,\tau) = \frac{1}{\pi} \int \dd \omega\, e^{2i\omega \tau} \bra{\varphi - \omega} \hat\rho \ket{\varphi + \omega},
\end{equation}
which in the case of a pure state $\ket{\psi}=\int \dd \omega\, F(\omega)\ket{\omega}$, reduces to
\begin{equation}
    W(\varphi,\tau) = \frac{1}{\pi} \int \dd \omega\, e^{2i\omega \tau} F(\varphi - \omega) F^*(\varphi + \omega).
\end{equation}
Interestingly, using the closure relation $\1=\int\dd t\,\ketbra{t}$, one can derive an alternate expression for the Wigner function (see Appendix~\ref{app: TF basic computations}, Result~\ref{res: alternate wigner})
\begin{equation}
    W(\varphi,\tau) = \frac{1}{\pi} \int \dd t\, e^{-2i\varphi t}\bra{\tau-t}\hat \rho\ket{\tau+t}.
\end{equation}

As a consequence of the direct analogy with CV systems, the chrono-cyclic Wigner function automatically exhibits similar interesting properties. The most important one for us is the fact that, as already mentioned, the overlap of two pure quantum states can be obtained via the overlap of their respective Wigner functions
\begin{equation}
    \abs{\braket{\psi}{\phi}}^2 = 2\pi \int \dd \varphi \dd \tau\, W_\psi(\varphi,\tau) W_\phi(\varphi,\tau).
\end{equation}
where $W_\psi$ and $W_\phi$ denote the Wigner functions associated with the states $\ket{\psi}$ and $\ket{\phi}$ respectively.

It is important to notice that since a single-photon state is in fact a single-mode state, in this single-spatial-mode picture, the Wigner function only characterizes the classical properties of the field and thus the chrono-cyclic Wigner function is an inherently classical object. As a matter of fact, such a distribution could be introduced to describe the properties of classical time-frequency fields.

\paragraph{\lit Multimode case}
The Wigner function admits a straightforward generalization to the multimode case. For any time-frequency single-photon state $\hat\rho$  on $n$ spatial modes, we define the Wigner distribution as a function of the $2n$ variables $\varphi_1,\dots,\varphi_n$ and $\tau_1,\dots,\tau_n$
\begin{align}
    W(\varphi_1,\dots,\varphi_n,\tau_1,\dots,\tau_n)& = \frac{1}{\pi^n} \int \dd\omega_1\cdots\dd\omega_n\, e^{2i\sum_{\alpha=1}^n \omega_\alpha \tau_\alpha}\notag\\
    &\qquad\times \bra{\varphi_1-\omega_1,\dots,\varphi_n-\omega_n} \hat\rho \ket{\varphi_1+\omega_1,\dots,\varphi_n+\omega_n}.
\end{align}

Contrary to the single-spatial-mode case, this multimode Wigner function characterizes a quantum state which has no classical wavelike description. In Section~\ref{sec: TF entanglement and metrology}, via the example of metrology, we reveal the fundamental importance of multiple spatial modes and the associated Wigner function to describe and quantify the quantum properties of time-frequency states.

\paragraph{\lit Chrono-cyclic Wigner function and the quadrature one}
Consider a single-mode time-frequency optical state $\ket{\psi}$, with some spectrum $F$
\begin{equation}
    \ket{\psi} = \sum_{n=0}^\infty c_n \frac{(\hat a_F^\dagger)^n}{\sqrt{n!}}\vac,
\end{equation}
where $\hat a_F^\dagger = \int \dd \omega\, F(\omega) \hat a^\dagger(\omega)$, is the creation operator associated with the mode defined by $F$. In this case, one can define two different Wigner functions. The chrono-cyclic one, associated with the spectral properties
\begin{equation}
    W_\text{s}(\varphi,\tau) = \frac{1}{\pi} \int \dd \omega\, e^{2i\omega \tau} F(\varphi - \omega) F^*(\varphi + \omega),
\end{equation}
and the quadrature one associated with the photon number statistics $\{c_n\}$
\begin{equation}
    W_\text{q}(x,p) = \frac{1}{\pi} \int \dd y\, e^{2i p y} \psi(x-y) \psi^*(x+y),
\end{equation}
where $\psi(x) = \sum_{n=0}^\infty c_n \phi_n(x)$ is the wavefunction in the `$x$' basis, with $\phi_n(x)$ the $n$-th Fock state wavefunction. These two Wigner functions provide two different phase spaces that describe different aspects of the same quantum state. The chrono-cyclic Wigner function captures the time-frequency spectral properties of the state, while the quadrature Wigner function encapsulates the photon number statistics and quadrature properties. Both representations are valuable for understanding different facets of the quantum state, and their interplay can provide deeper insights into the state's overall behavior. Furthermore, as we develop in Section~\ref{sec: TF evolutions}, evolutions have different effects on these phase spaces. For example, the free evolution generated by the operator $\hat \omega$ induces, respectively, a translation in the time-frequency phase space, and a rotation at angular velocity $\overline\omega=\int \dd\omega\, \omega\abs{F(\omega)}^2$ in the quadrature one.

More generally, for arbitrary states, for which the statistical and spectral properties cannot be separated, defining both Wigner functions is more involved, but one should always keep in mind that these two phase spaces coexist and describe different aspects of the same quantum state.

\clearpage
\section{Zoology of states}
\label{sec: TF_zoology_of_states}

\emph{In this section we present several classes of time-frequency quantum states that will play an important role throughout this manuscript. We focus in particular on different examples of single-photon states in the time-frequency degree of freedom. We emphasize their analogy with continuous-variable (CV) states, while carefully discussing the fundamental conceptual differences.}

\subsection{Single-spatial-mode single-photon states}
\label{subsec: ex_TF_single_photon_states}
In this section we consider single-spatial-mode single-photon time-frequency states
\begin{equation}
    \ket{\psi} = \int \dd \omega \, F(\omega) \ket{\omega},
\end{equation} 
for various spectra $F$. Alternatively, one can think of $\ket{\psi}$ as describing a single photon in the mode associated with $\hat a_F^\dagger$, where $\hat a_F^\dagger$ is the creation operator associated with the spectral distribution $F$
\begin{equation}
    \ket{\psi}=\hat a_F^\dagger \vac.
\end{equation} 
Throughout this section, we draw parallels with the formalism of continuous-variable quantum optics, while emphasizing the specific features of time-frequency encodings.

\paragraph{\lit Time-frequency Gaussian states}
A particularly important class of states is obtained by choosing a Gaussian spectral amplitude,
\begin{equation}
    \ket{\psi_G} = \frac{1}{(2\pi\sigma^2)^{1/4}} \int \dd \omega \, \exp[-\frac{(\omega-\omega_0)^2}{4\sigma^2}] \ket{\omega},
\end{equation}
where $\omega_0$ denotes the central frequency and $\sigma$ the spectral width of the photon. Gaussian wave packets naturally arise in many experimental settings and play a role analogous to Gaussian states in the CV framework. Indeed, in the CV setting, pure Gaussian states are precisely those whose wavefunction in phase space is Gaussian; they include coherent and squeezed states of a harmonic oscillator. The state $\ket{\psi_G}$ can therefore be regarded as the time-frequency analogue of a single-mode pure Gaussian CV state. However, two important subtleties must be emphasized
\begin{itemize}
    \item In CV quantum optics, the vacuum state is the ground state of the harmonic oscillator and is itself a Gaussian state centered at the origin of phase space. In contrast, in the present context the vacuum corresponds to the absence of a photon. A forward-propagating single-photon state necessarily occupies positive frequencies and is typically centered around an optical carrier frequency $\omega_0 \geq 0$. As a consequence, a strict analogue of the CV vacuum state does not exist in the time-frequency single-photon Hilbert space. In practice, one often redefines the effective phase space by working in a rotating frame centered at $\omega_0$.
    \item In CV systems, coherent states are defined as Gaussian states with a fixed width set by the ground state of the harmonic oscillator. No such natural scale exists for time and frequency: the spectral width $\sigma$ can be chosen arbitrarily, depending on the physical process used to generate the photon. As a result, notions such as squeezing are inherently relative and must be defined with respect to a chosen reference scale.
\end{itemize}

\paragraph{\lit Time-frequency cat-like states}
In further analogy with CV systems, one can construct time-frequency cat-like states as coherent superpositions of two well-separated time-frequency Gaussian single-photon states. A general definition is
\begin{equation}
    \ket{\psi_{\text{cat}}} = \frac{1}{\mathcal N} \left(\ket{\psi_G(\omega_1,\sigma)}\pm\ket{\psi_G(\omega_2,\sigma)} \right),
\end{equation}
where $\ket{\psi_G(\omega_i,\sigma)}$ denotes a Gaussian time-frequency single-photon state centered at frequency $\omega_i$, $\sigma$ is their common spectral width, and $\mathcal N$ is a normalization factor. These states are the time-frequency counterparts of CV Schrödinger cat states, which are superpositions of coherent states with macroscopically distinct amplitudes. The same conceptual subtleties discussed above apply here as well. In particular, while CV cat states are often defined as superpositions of coherent states with opposite complex amplitudes, time-frequency cat-like states must necessarily be centered around positive frequencies, reflecting the physical constraint imposed by photons.

\paragraph{\lit Frequency combs}
Another important class of single-photon time-frequency states is provided by frequency comb states. These are characterized by a spectral amplitude consisting of a series of equally spaced frequency components, reminiscent of the teeth of a comb. An idealized frequency comb state can be written as
\begin{equation}
    \ket{\psi_{\text{comb}}} = \sum_{k=-\infty}^{\infty} \ket{\omega_0 + k\Delta},
\end{equation}
where $\omega_0$ is a reference frequency and $\Delta$ is the spacing between adjacent spectral lines. Such states are the time-frequency analogue of GKP states in the CV formalism, which we introduced in Sec.~\ref{subsec: error correction}. As in the CV case, ideal frequency comb states are nonphysical: they are not normalizable and would require infinite energy. A physically meaningful approximation can be obtained by assigning a finite width to each spectral peak and introducing an overall envelope,
\begin{equation}
    \ket{\psi_{\text{comb}}^{\text{phys}}} = \frac{1}{\mathcal N} \sum_{k=-\infty}^{\infty} \exp[-\frac{(\omega_0+k\Delta)^2}{2\sigma^2}] \int \dd \omega \, \exp[-\frac{(\omega-\omega_0-k\Delta)^2}{2\delta^2}] \ket{\omega},
\end{equation}
where $\sigma$ sets the width of the global envelope, $\delta$ controls the width of each individual peak, and $\mathcal N$ is a normalization factor. The parameters $\sigma$ and $\delta$ determine how closely the state approximates the ideal comb and thus quantify the quality of the encoding.

\subsection{Spontaneous parametric down-conversion}
\label{subsec: SPDC}

A widely used method to generate entangled photon pairs with correlated time-frequency properties is spontaneous parametric down-conversion (SPDC)~\cite{burnham_observation_1970,rubin_theory_1994,kwiat_new_1995}. While a detailed study of this process is not the aim of this thesis, we provide a concise introduction in order to physically motivate the assumptions that will be made in the following regarding the possible shape of a single-photon joint spectral amplitude (JSA). More precisely, we show that the JSA of a photon pair generated via SPDC can, under very general conditions, be written in the factorized form
\begin{equation}
    F(\omega_s,\omega_i)=F_+(\omega_+)F_-(\omega_-),
\end{equation}
with $\omega_\pm = (\omega_s \pm \omega_i)/\sqrt{2}$, and $F_\pm$ two complex-valued functions. In this expression, $\omega_s$ and $\omega_i$ denote the frequencies of the signal and idler photons, respectively. This decomposition plays a central role in the engineering of spectral correlations and entanglement properties of SPDC photon pairs. More detailed and comprehensive overviews of the SPDC process can be found in~\cite{boyd_nonlinear_2008,orieux_semiconductor_2017,maltese_generation_2019}.

\paragraph{\lit Principles}
SPDC is a particular instance of a nonlinear frequency conversion process known as three-wave mixing~\cite{boyd_nonlinear_2008}, which occurs in optical media for which the induced polarization is a nonlinear function of the applied electric field. In the time domain, the polarization can be expanded as
\begin{equation}
    P(t)=\epsilon_0(\chi^{(1)}E(t)+\chi^{(2)}E^2(t)+\chi^{(3)}E^3(t)+\dots),
\end{equation}
where $\chi^{(1)}$ is the linear susceptibility, $\chi^{(2)}$ the second-order nonlinear susceptibility, and higher-order terms describe progressively weaker nonlinear responses. The presence of the nonlinear terms allows different frequency components of the electric field to interact, thereby enabling frequency conversion processes.

In SPDC, a strong coherent pump beam interacts with a medium possessing a non-vanishing $\chi^{(2)}$ susceptibility. Through this interaction, pump photons at frequency $\omega_p$ are spontaneously converted into pairs of lower-frequency photons, conventionally referred to as signal and idler photons, with frequencies $\omega_s$ and $\omega_i$, respectively. Energy conservation imposes the constraint
\begin{equation}
    \omega_p=\omega_s+\omega_i.
\end{equation}
In addition to energy conservation, an additional requirement known as phase matching must be satisfied. This condition, which is fundamentally related to momentum conservation in the nonlinear medium, imposes further constraints on the allowed frequency and propagation directions of the generated photons. The phase-matching condition will be discussed in more detail below, as it directly determines the form of the phase-matching function $\phi_\text{PM}$ appearing in the JSA.

Depending on the polarization states of the interacting fields, SPDC processes are commonly classified into three distinct types:
\begin{itemize}
    \item Type 0: pump, signal, and idler fields all share the same polarization
    \item Type I: signal and idler photons have the same polarization, which is orthogonal to that of the pump
    \item Type II: signal and idler photons have orthogonal polarizations
\end{itemize}
The specific type of SPDC that can occur in a given setup is determined by the specific properties of the nonlinear medium.

\begin{figure}[ht]
    \centering
    \footnotesize
    \scalebox{1.3}{\begin{tikzpicture}
	\begin{pgfonlayer}{nodelayer}
		\node [style=none] (0) at (-2.25, 1) {};
		\node [style=none] (1) at (2.25, 1) {};
		\node [style=none] (2) at (-2.25, -1) {};
		\node [style=none] (3) at (2.25, -1) {};
		\node [style=none] (4) at (0, 0) {$\chi^{(2)}$};
		\node [style=none] (5) at (-4.25, 0) {};
		\node [style=none] (6) at (-2.5, 0) {};
		\node [style=none] (7) at (2.75, 0.375) {};
		\node [style=none] (8) at (5, 0.375) {};
		\node [style=none] (11) at (2.75, -0.375) {};
		\node [style=none] (12) at (5, -0.375) {};
		\node [style=none] (13) at (-5.25, 0) {\scalebox{0.9}{Pump}};
		\node [style=none] (14) at (4.8, 0.925) {\scalebox{0.9}{Signal}};
		\node [style=none] (15) at (4.75, -0.875) {\scalebox{0.9}{Idler}};
		\node [style=none] (16) at (3.5, 0) {};
		\node [style=none] (17) at (3.1, -0.75) {};
		\node [style=none] (18) at (3.9, -0.75) {};
		\node [style=none] (19) at (3.1, 0.75) {};
		\node [style=none] (20) at (3.9, 0.75) {};
		\node [style=none] (21) at (-2, 1.25) {};
		\node [style=none] (22) at (2.425, 1.25) {};
		\node [style=none] (23) at (2.425, -0.75) {};
	\end{pgfonlayer}
	\begin{pgfonlayer}{edgelayer}
		\draw [style=Fill grey] (22.center)
			 to (1.center)
			 to (0.center)
			 to (21.center)
			 to cycle
			 to (23.center)
			 to (3.center)
			 to (1.center);
		\draw (2.center) to (0.center);
		\draw [style=dashed arrow] (11.center) to (12.center);
		\draw [style=Thick arrow] (5.center) to (6.center);
		\draw [style=dashed arrow] (7.center) to (8.center);
		\draw [style=red line, bend left=90, looseness=1.50] (19.center) to (20.center);
		\draw [style=red line, in=45, out=-90, looseness=0.75] (20.center) to (16.center);
		\draw [style=red line, in=135, out=-90, looseness=0.75] (19.center) to (16.center);
		\draw [style=red line, in=90, out=-135] (16.center) to (17.center);
		\draw [style=red line, bend right=90, looseness=1.50] (17.center) to (18.center);
		\draw [style=red line, in=315, out=90] (18.center) to (16.center);
		\draw (3.center) to (2.center);
	\end{pgfonlayer}
\end{tikzpicture}}
    \caption[Schematics of the nonlinear conversion process in a $\chi^{(2)}$ medium leading to SPDC]{Schematics of the nonlinear conversion process in a $\chi^{(2)}$ medium leading to SPDC. The pump beam $p$ is converted into a signal beam $s$ and an idler beam $i$.}
    \label{fig: SPDC schematics}
\end{figure}
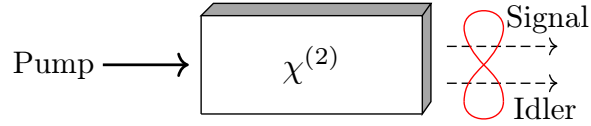

\paragraph{\lit JSA of SPDC photons}

Considering only the nonlinear interaction term, the Hamiltonian governing SPDC can be written as
\begin{equation}
    \hat H_I(t)=\epsilon_0 \chi^{(2)} \int_V \dd^3 \vec r \, \hat E_p^{(+)}(\vec r,t) \hat E_s^{(-)}(\vec r,t) \hat E_i^{(-)}(\vec r,t) + h.c.,
\end{equation}
where $V$ is the interaction volume, $\hat E_p^{(+)}$ is the positive-frequency part of the pump electric field operator, while $\hat E_s^{(-)}$ and $\hat E_i^{(-)}$ are the negative-frequency parts of the signal and idler electric field operators, respectively, which we introduced in section \ref{subsec: Field Quantization}. Notice that we consider explicit time dependence by placing ourselves in the Heisenberg picture. For simplicity, we assume that the polarizations of the three fields are fixed and that a scalar description of the electric fields is sufficient. A more general treatment including vectorial fields and tensorial properties of $\chi^{(2)}$ can be found in standard references on nonlinear optics~\cite{boyd_nonlinear_2008}.

In the SPDC process it is commonly assumed that the pump field can be treated classically and that the conversion efficiency is sufficiently low such that pump depletion can be neglected. Under these assumptions, the interaction Hamiltonian can be approximated as
\begin{equation}
    \hat H_I(t)=\epsilon_0 \chi^{(2)} \int_V \dd^3 \vec r \, E_p^{(+)}(\vec r,t) \hat E_s^{(-)}(\vec r,t) \hat E_i^{(-)}(\vec r,t) + h.c.,
\end{equation}
where $E_p^{(+)}$ now denotes the positive-frequency part of the classical pump electric field. The state generated by the nonlinear interaction is obtained by applying the time-ordered evolution operator $\mathcal T\{e^{-i\int \dd t\, \hat H_I(t)/\hbar}\}$ to the vacuum state $\vac$. Assuming an inefficient conversion process, it is sufficient to retain only the first-order term
\begin{equation}
    \left(1 - \frac{i}{\hbar}\int \dd t \, \hat H_I(t)\right)\vac=\vac + \beta\ket{\psi},
\end{equation}
where $\beta \ll 1$ quantifies the overall conversion efficiency and $\ket{\psi}$ denotes the two-photon component of the generated state. As a result, SPDC sources predominantly emit the vacuum state, with photon pairs being generated only probabilistically. The two-photon component can thus be written as
\begin{equation}
    \ket{\psi}=\frac{1}{i\beta\hbar}\int \dd t \, \hat H_I(t)\vac.
\end{equation}

To obtain an explicit expression for $\ket{\psi}$, we now introduce the field expansions. The pump beam is treated classically and expressed as a superposition of plane waves as
\begin{equation}
    E_p^{(+)}(\vec r,t)=\frac{1}{\sqrt{2\pi}}\int \dd\omega_p \, \mathcal E_p(\vec r,\omega_p)e^{i[k_p(\omega_p)y-\omega_p t]},
\end{equation}
where $y$ is chosen as the propagation direction and $k_p(\omega)$ is the corresponding longitudinal wave vector, which incorporates the material dispersion and ultimately governs phase matching and frequency correlations. The signal and idler electric field operators are expanded as
\begin{align}
    \hat E_s^{(-)}(\vec r,t)&=\frac{1}{\sqrt{2\pi}}\int \dd\omega_s \, \mathcal E_s(\vec r,\omega_s) \hat a_s^\dagger(\omega_s) e^{-i[k_s(\omega_s)y-\omega_s t]},\\
    \hat E_i^{(-)}(\vec r,t)&=\frac{1}{\sqrt{2\pi}}\int \dd\omega_i \, \mathcal E_i(\vec r,\omega_i) \hat a_i^\dagger(\omega_i) e^{-i[k_i(\omega_i)y-\omega_i t]},
\end{align}
where $\hat a_j^\dagger(\omega_j)$ creates a photon of frequency $\omega_j$ in mode $j=s,i$. Substituting these expressions into the definition of $\ket{\psi}$ yields
\begin{align}
    \ket{\psi}&=\frac{\epsilon_0 \chi^{(2)}}{i\beta\hbar (2\pi)^{3/2}} \int_V \dd^3 \vec r \int \dd t \int \dd\omega_p \dd\omega_s \dd\omega_i \, \mathcal E_p(\vec r,\omega_p) \mathcal E_s(\vec r,\omega_s) \mathcal E_i(\vec r,\omega_i)\notag\\
    &\qquad\times e^{i\Delta k y} e^{-i\Delta \omega t} \hat a_s^\dagger(\omega_s) \hat a_i^\dagger(\omega_i)\vac,
\end{align}
where $\Delta k=k_p(\omega_p)-k_s(\omega_s)-k_i(\omega_i)$ and $\Delta \omega=\omega_p-\omega_s-\omega_i$. Performing the time integral yields a Dirac delta function enforcing energy conservation, $\omega_p=\omega_s+\omega_i$. The two-photon state can therefore be written in the general form
\begin{equation}
    \ket{\psi}=\int \dd\omega_s \dd\omega_i \, F(\omega_s,\omega_i) \hat a_s^\dagger(\omega_s) \hat a_i^\dagger(\omega_i)\vac,
\end{equation}
where $F(\omega_s,\omega_i)$ is the JSA, given here by
\begin{equation}
    F(\omega_s,\omega_i)=\frac{\epsilon_0 \chi^{(2)}}{i\beta\hbar (2\pi)^{1/2}} \int_V \dd^3 \vec r \, \mathcal E_p(\vec r,\omega_s+\omega_i) \mathcal E_s(\vec r,\omega_s) \mathcal E_i(\vec r,\omega_i) e^{i\Delta k y}.
\end{equation}

Further insight can be obtained by making additional assumptions on the spatial and spectral structure of the fields. Assuming propagation in a waveguide, the fields can be taken as independent of the longitudinal coordinate apart from the wave vector phase factor. Moreover, if the intrinsic frequency dependence of the output mode is negligible compared to dispersion effects, the field amplitudes can be factorized as
\begin{align}
    \mathcal E_p(\vec r,\omega)&\simeq\mathcal E_p(x,z)\alpha(\omega), &
    \mathcal E_s(\vec r,\omega_s)&\simeq\mathcal E_s(x,z), &
    \mathcal E_i(\vec r,\omega_i)&\simeq\mathcal E_i(x,z).
\end{align}
where $\alpha(\omega)$ is the spectrum of the field. Under these assumptions, the JSA simplifies to
\begin{align}
    F(\omega_s,\omega_i)&=\frac{\epsilon_0 \chi^{(2)}}{i\beta\hbar (2\pi)^{1/2}}\alpha(\omega_s+\omega_i) \int \dd y \, e^{i\Delta k y}\int \dd x \dd z \, \mathcal E_p(x,z) \mathcal E_s(x,z) \mathcal E_i(x,z),\\
    &=\alpha(\omega_s+\omega_i)\phi_{\text{PM}}(\omega_s,\omega_i)\Gamma,
\end{align}
where $\Gamma$ is a constant quantifying the spatial mode overlap and
\begin{equation}
    \phi_{\text{PM}}(\omega_s,\omega_i)=\int \dd y \, e^{i\Delta k y}
\end{equation}
is the phase-matching function, which enforces momentum conservation. The pump spectrum $\alpha(\omega)$ is typically assumed to be Gaussian or Lorentzian, centered at the pump frequency $\omega_p^0$ with bandwidth $\sigma_p$. The phase-matching function depends on the dispersion properties of the nonlinear medium and on the interaction geometry. For a uniform medium of length $L$, one finds
\begin{equation}
    \phi_{\text{PM}}(\omega_s,\omega_i)=\int_0^L \dd y \, e^{i\Delta k y}=L \text{sinc}\Big(\frac{\Delta k L}{2}\Big)e^{i\frac{\Delta k L}{2}}.
\end{equation}

To further analyze the structure of the phase-matching function, it is convenient to introduce the diagonal frequency variables
\begin{align}
    \omega_+=\frac{1}{\sqrt{2}}(\omega_s+\omega_i), && \omega_-=\frac{1}{\sqrt{2}}(\omega_s-\omega_i),
\end{align}
where the normalization by $\sqrt{2}$ ensures a unit Jacobian for the transformation. Expanding $\Delta k$ to first order around the central frequencies $\omega_s^0$ and $\omega_i^0$, defined such that $\omega_p^0=\omega_s^0+\omega_i^0$, yields
\begin{align}
    \Delta k(\omega_s,\omega_i)&\simeq (k_p' - k_s')(\omega_s-\omega_s^0) + (k_p' - k_i')(\omega_i-\omega_i^0)\notag\\
    &= \frac{1}{\sqrt{2}}[(k_p' - k_s') + (k_p' - k_i')](\omega_+ - \omega_+^0)\notag\\
    &\quad+ \frac{1}{\sqrt{2}}[(k_p' - k_s') - (k_p' - k_i')](\omega_- - \omega_-^0),
\end{align}
where $k_j'=\frac{\dd k_j}{\dd \omega}\big|_{\omega_j^0}$ is the inverse group velocity of the $j$-th field. If the pump spectral width is much smaller than the characteristic variation scale of the phase-matching function, $\omega_+$ can be treated as approximately constant, leading to an approximate factorization of the JSA,
\begin{equation}
    F(\omega_s,\omega_i)\simeq F_+(\omega_+)F_-(\omega_-),
\end{equation}
where $F_+$ is primarily determined by the pump spectrum and $F_-$ by the phase-matching function. This factorization provides a powerful framework for analyzing and engineering the time-frequency correlations of SPDC photons, as it allows independent control along the diagonal and anti-diagonal directions in the $(\omega_s,\omega_i)$ plane. Furthermore, as analyzed in detail in Chap.~\ref{chap: HOM interferometry and metrology}, some interferometric effects such as the Hong-Ou-Mandel effect~\cite{hong_measurement_1987} are only sensitive to $F_+$ or $F_-$. Such a factorization then allows a simpler understanding of these effects.

It is worth noting that many physical effects, such as higher-order dispersion, birefringence, spatial walk-off, and cavity enhancement, have been neglected in this simplified treatment. More refined models show that substantial additional control over the shape of the JSA can be achieved by tailoring these effects through material choice, waveguide design, and resonant structures~\cite{orieux_semiconductor_2017}.

\paragraph{\lit Examples}
As argued above, in the following we will rely on the assumption that the JSA can be factorized in terms of the diagonal frequency variables $\omega_\pm=(\omega_s\pm\omega_i)/\sqrt{2}$,
\begin{equation}
    F(\omega_s,\omega_i)=F_+(\omega_+)F_-(\omega_-).
\end{equation}

To conclude this introductory discussion of SPDC, we now present typical forms for $F_\pm$. The function $F_+$ is primarily determined by the spectral profile of the pump field. Common choices include Gaussian and Lorentzian envelopes,
\begin{align}
    F_+^{(G)}(\omega_+)&=\frac{1}{(2\pi\sigma_+^2)^{1/4}}\exp\Big[-\frac{(\omega_+-\omega_+^0)^2}{4\sigma_+^2}\Big], &
    F_+^{(L)}(\omega_+)&=\frac{\sigma_+}{\pi}\frac{1}{(\omega_+-\omega_+^0)^2+\sigma_+^2},
\end{align}
where $\sigma_+$ sets the spectral bandwidth of the pump and $\omega_+^0$ is the central frequency of the pump. In the idealized monochromatic limit $\sigma_+\to 0$, the function $F_+$ reduces to a Dirac delta distribution
\begin{equation}
    F_+^{(\delta)}(\omega_+)=\delta(\omega_+-\omega_+^0).
\end{equation}

As introduced before, the function $F_-$ is determined by the phase-matching conditions inside the nonlinear medium and takes the form of a sinc function originating from the finite crystal length. A commonly used expression is
\begin{align}
    F_-^{(\text{sinc})}(\omega_-)&=\text{sinc}\Big(\frac{\omega_--\omega_-^0}{\sigma_-}\Big)
    \exp\Big[i\frac{\omega_--\omega_-^0}{\sigma_-}\Big],
\end{align}
where $\sigma_-$ characterizes the spectral width associated with phase matching and $\omega_-^0$ is the central difference frequency, which can usually be set to $\omega_0=0$. For analytical convenience, the sinc function is often approximated by a Gaussian of identical width,
\begin{align}
    F_-^{(G)}(\omega_-)&=\frac{1}{(2\pi\sigma_-^2)^{1/4}}\exp\Big[-\frac{(\omega_--\omega_-^0)^2}{4\sigma_-^2}\Big].
\end{align}

Spectral filtering can be employed to engineer the shape of the function $F_-$, at the cost of reducing the overall pair generation efficiency. For instance, filtering that transmits two well-separated frequency modes leads to the generation of a Schrödinger cat-like spectral state~\cite{ou_observation_1988,fabre_producing_2020},
\begin{equation}
    F_-(\omega_-)=\frac{1}{\mathcal N}\Big[F_-^{(G)}(\omega_--\Delta/2)+F_-^{(G)}(\omega_-+\Delta/2)\Big],
\end{equation}
where $\mathcal N$ is a normalization constant and $\Delta$ denotes the frequency separation between the two spectral peaks. As we will see in Sec.~\ref{subsec: visibility impact on precision}, such states provide a powerful advantage for interferometric and metrological applications.

In the special case where both functions $F_\pm$ are Gaussian with identical widths, the resulting JSA is itself Gaussian and becomes separable in the original signal and idler frequencies $\omega_s$ and $\omega_i$,
\begin{equation}
    F^{(G)}(\omega_s,\omega_i)=\frac{1}{\sqrt{2\pi\sigma^2}}
    \exp\Big[-\frac{(\omega_s-\omega_s^0)^2}{4\sigma^2}\Big]
    \exp\Big[-\frac{(\omega_i-\omega_i^0)^2}{4\sigma^2}\Big].
\end{equation}
This situation corresponds to a spectrally unentangled biphoton state, which is particularly desirable for heralded single-photon sources. In most other cases, however, the JSA cannot be factorized in terms of $\omega_s$ and $\omega_i$, resulting in spectral entanglement between the two photons. Representative examples of such JSAs are shown in Fig.~\ref{fig: example JSA}.

\begin{figure}[ht]
    \centering
    \begin{tabular}{cc}
        \includegraphics[width=0.3\linewidth]{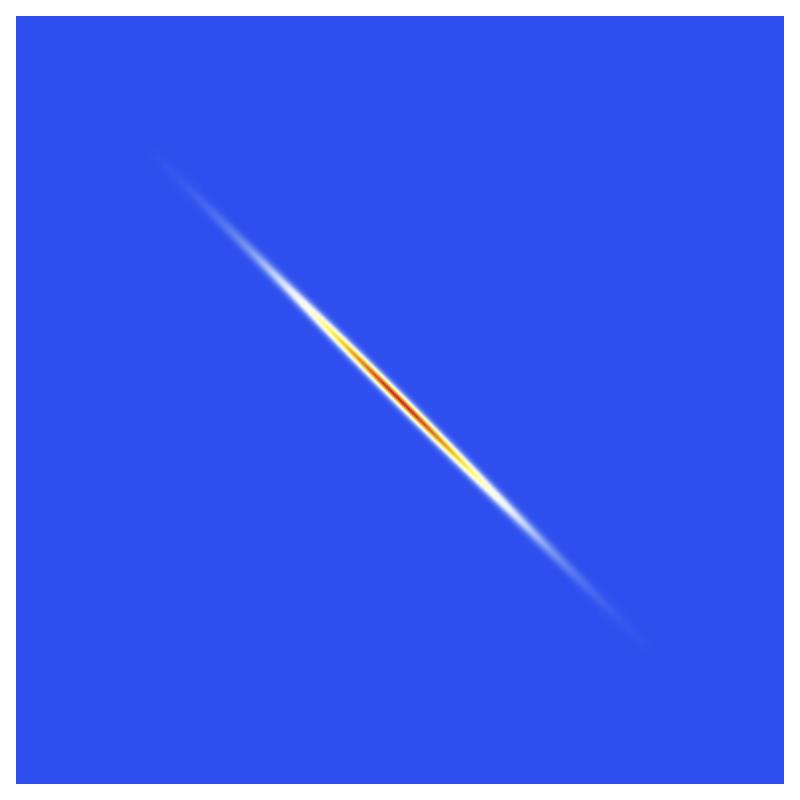}
        &
        \includegraphics[width=0.3\linewidth]{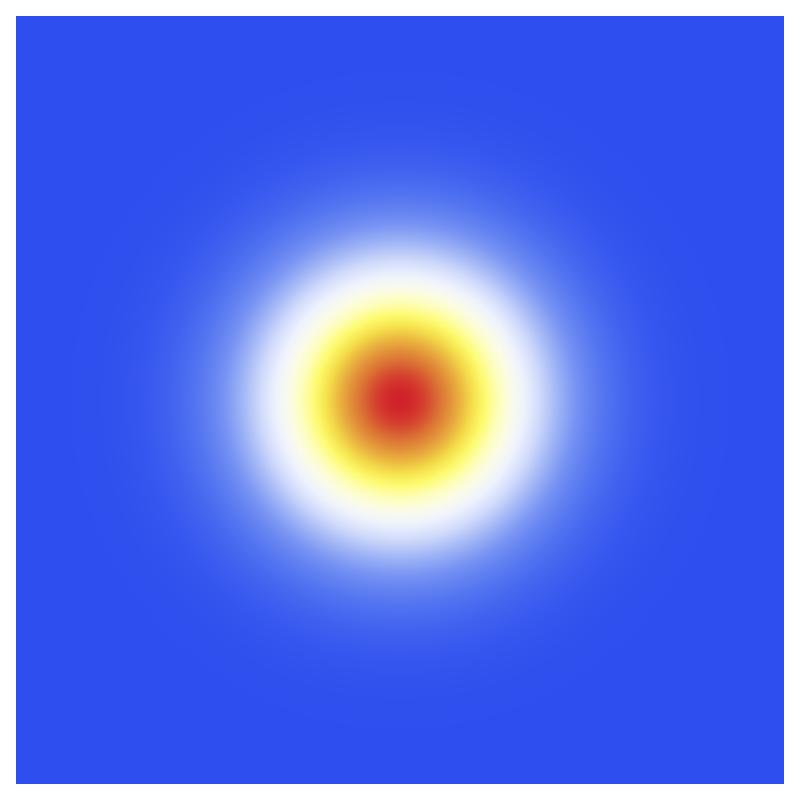}\\
        Gaussian JSA & Circular JSA\\[1ex]
        \includegraphics[width=0.3\linewidth]{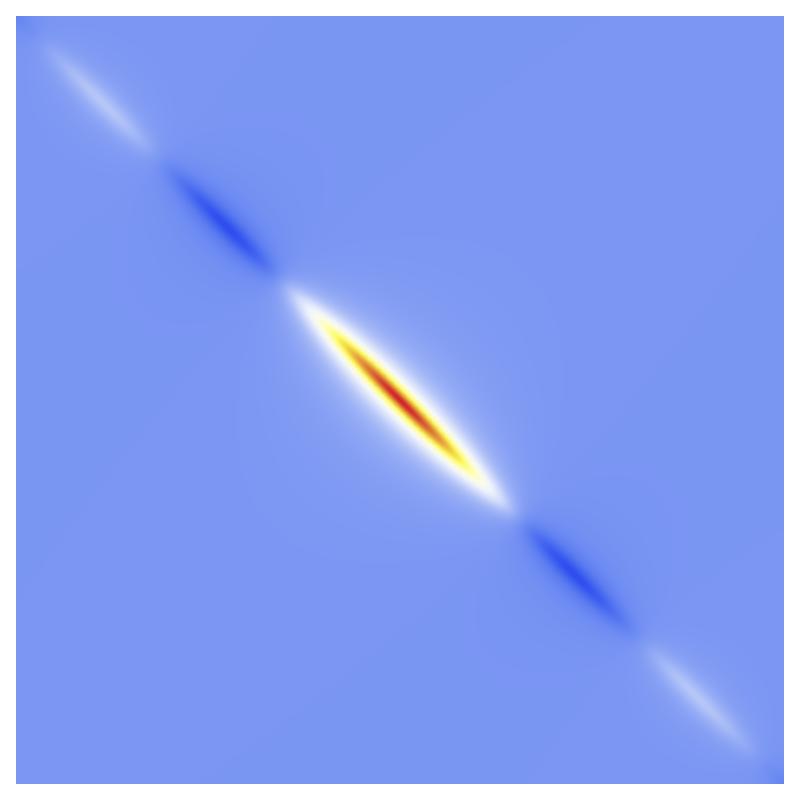}
        &
        \includegraphics[width=0.3\linewidth]{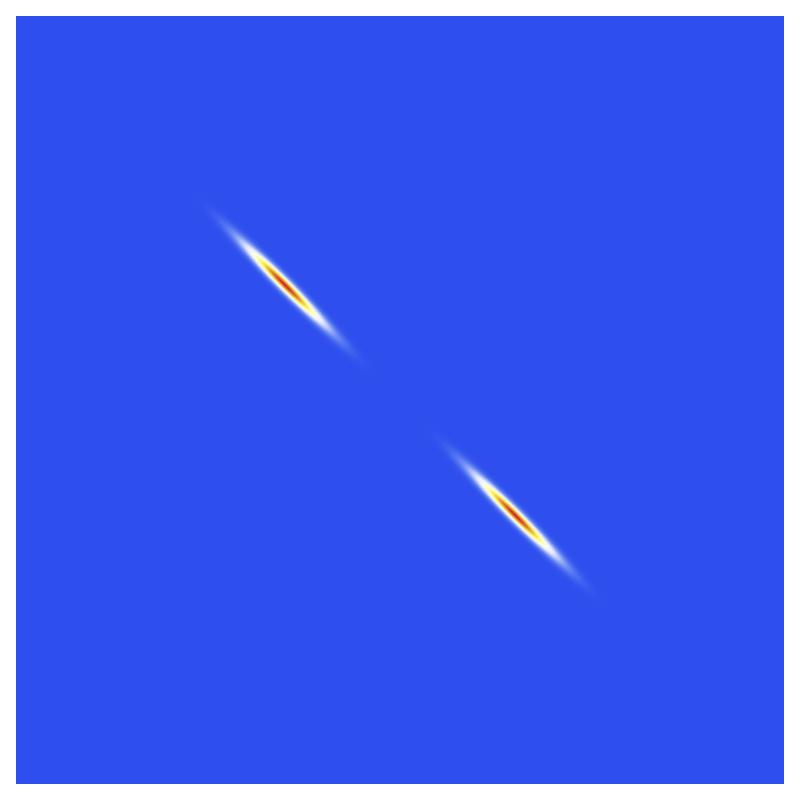}\\
        Lorentzian and sinc JSA & Cat-like JSA
    \end{tabular}
    \caption[Density plots of the JSA for several typical biphoton states encountered in SPDC]{Density plots of the JSA for several typical biphoton states encountered in SPDC.}
    \label{fig: example JSA}
\end{figure}

\subsection{Wigner function for typical states}
\label{subsec: Wigner function for typical TF states}
Now that we have introduced in more detail the structure of single-photon time-frequency states, we turn to the explicit form of their associated chrono-cyclic Wigner functions.

\paragraph{\lit Single-mode example}
The time-frequency single-photon states introduced in Sec.~\ref{subsec: TF states} are, from a formal perspective, direct analogs of continuous-variable single-mode states. While we previously discussed the qualitative features of various Wigner functions in the continuous-variable setting, we now provide explicit expressions in the time-frequency framework. In particular, we compute the chrono-cyclic Wigner functions of the single-spatial-mode single-photon states introduced in Sec.~\ref{subsec: ex_TF_single_photon_states}.

For a single-photon state with Gaussian spectral amplitude,
\begin{equation}
    \ket{\psi_G}=\frac{1}{(2\pi\sigma^2)^{1/4}}\int \dd \omega \, \exp\Big[-\frac{(\omega-\omega_0)^2}{4\sigma^2}\Big]\ket{\omega},
\end{equation}
the corresponding Wigner function reads
\begin{equation}\label{eq: wigner function gaussian}
    W_G(\varphi,\tau)=\frac{1}{\pi}\exp\Big[-\frac{(\varphi-\omega_0)^2}{2\sigma^2}-2\sigma^2\tau^2\Big].
\end{equation}
Geometrically, this Wigner function is a Gaussian distribution in the $(\varphi,\tau)$ phase space, centered at $(\omega_0,0)$, with variances $\sigma^2$ along the frequency axis and $1/(4\sigma^2)$ along the time axis, illustrating the reciprocal localization imposed by the Fourier relation between time and frequency.

For a cat-like state defined as a coherent superposition of two spectrally separated Gaussian wavepackets,
\begin{equation}
    \ket{\psi_\text{cat}}=\frac{1}{\mathcal N}\big(\ket{\psi_G(\omega_0+\Delta/2)}\pm\ket{\psi_G(\omega_0-\Delta/2)}\big),
\end{equation}
where $\mathcal N$ is a normalization constant, the Wigner function takes the form
\begin{align}
    W_\text{cat}(\varphi,\tau) &=\frac{1}{\pi\mathcal N^2}e^{-2\sigma^2\tau^2} \Big( e^{-(\varphi-\omega_0-\Delta/2)^2/2\sigma^2} +e^{-(\varphi-\omega_0+\Delta/2)^2/2\sigma^2} \notag\\
    &\qquad\pm 2\cos(\tau\Delta) e^{-(\varphi-\omega_0)^2/2\sigma^2} \Big).
\end{align}
This expression explicitly reveals that the Wigner function is composed of three distinct contributions:
\begin{itemize}
    \item Two Gaussian peaks centered at $(\omega_0-\Delta/2,0)$ and $(\omega_0+\Delta/2,0)$, each with the same widths as the original Gaussian state. These peaks correspond to the incoherent contributions of the two spectral components.
    \item An interference term oscillating along the time axis with angular frequency $\Delta$, enclosed within a Gaussian envelope centered at the midpoint $(\omega_0,0)$. This interference pattern is the hallmark of quantum coherence between the two components and is directly analogous to Schrödinger cat states in continuous-variable phase space.
\end{itemize}

As a final single-mode example, we consider the GKP or frequency-comb state,
\begin{equation}
    \ket{\psi_\text{GKP}}=\sum_{k=-\infty}^{\infty}\ket{\omega_0+k\Delta},
\end{equation}
where $\omega_0$ is a reference frequency and $\Delta$ denotes the spacing between adjacent spectral lines. The associated Wigner function reads\footnote{Since GKP states are built with Dirac delta functions, they are not normalized. As such, their Wigner function is not bounded, unlike that of a normalized state which is always bounded within the interval $[-1/\pi,1/\pi]$.}
\begin{align}
    W_\text{GKP}(\varphi,\tau) &=\frac{1}{2\Delta}\sum_{u,v\in\Z} \delta(\varphi-\omega_0-v\Delta) \delta\Big(\tau-\frac{\pi}{\Delta}u\Big) \notag\\
    &\qquad +\frac{1}{2\Delta}\sum_{u,v\in\Z} (-1)^u \delta\Big(\varphi-\omega_0-\frac{2v+1}{2}\Delta\Big) \delta\Big(\tau-\frac{\pi}{\Delta}u\Big).
\end{align}
A rigorous derivation is provided in Appendix~\ref{app: TF basic computations}, Result~\ref{res: Wigner GKP}. The Wigner function thus consists of a lattice of Dirac peaks arranged on a grid centered at $(\omega_0,0)$, with spacings $\Delta/2$ along the frequency axis and $\pi/\Delta$ along the time axis. One quarter of these peaks carry a negative sign, reflecting the intrinsically non-classical nature of the state. A schematic representation is shown in Fig.~\ref{fig: wigner function GKP}.

\begin{figure}[ht]
    \centering
    \scalebox{1}{\input{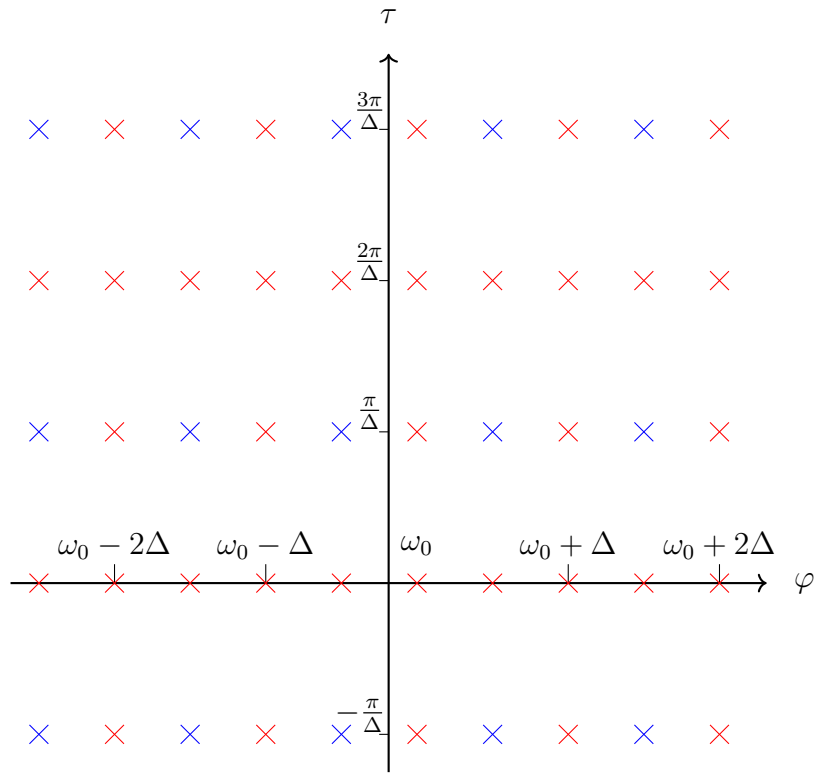}}
    \caption[Wigner function of a GKP state]{Representation of the Wigner function of a GKP state. Each cross corresponds to a Dirac peak. Peaks shown in red have positive weight, while those in blue carry a negative sign.}
    \label{fig: wigner function GKP}
\end{figure}

\paragraph{\lit Two-mode case}
We now extend the discussion to the two-mode case. While the general four-dimensional Wigner function is difficult to visualize, significant simplifications occur under physically relevant assumptions on the JSA $F(\omega_s,\omega_i)$. If the JSA is separable,
\begin{equation}
    F(\omega_s,\omega_i)=F_s(\omega_s)F_i(\omega_i),
\end{equation}
then the Wigner function of the biphoton state factorizes as
\begin{equation}
    W(\varphi_s,\varphi_i,\tau_s,\tau_i)
    =W_s(\varphi_s,\tau_s)W_i(\varphi_i,\tau_i),
\end{equation}
where $W_j$ denotes the single-mode Wigner function associated with the spectral amplitude $F_j$. This situation corresponds to an unentangled time-frequency state. Similarly, if the JSA factorizes in terms of the diagonal variables $\omega_\pm=(\omega_s\pm\omega_i)/\sqrt{2}$, such that
\begin{equation}
    F(\omega_s,\omega_i)=F_+(\omega_+)F_-(\omega_-),
\end{equation}
the Wigner function can be written as
\begin{equation}
    W(\varphi_s,\varphi_i,\tau_s,\tau_i)
    =
    W_+\Big(\frac{\varphi_s+\varphi_i}{\sqrt{2}},\frac{\tau_s+\tau_i}{\sqrt{2}}\Big)
    W_-\Big(\frac{\varphi_s-\varphi_i}{\sqrt{2}},\frac{\tau_s-\tau_i}{\sqrt{2}}\Big).
\end{equation}
Such factorization rules considerably simplify the geometric interpretation of two-mode states. Although the full Wigner function lives in a four-dimensional phase space, each factor is a two-dimensional function that can be independently visualized and analyzed. Moreover, as will be discussed in Sec.~\ref{sec: TF evolutions}, many physically relevant transformations act only on one of these factors, allowing for an intuitive understanding of state evolution and correlations. Detailed proofs of these factorizations are provided in Appendix~\ref{app: TF basic computations}, Results~\ref{res: Wigner local separable} and \ref{res: Wigner diagonal separable}.

\subsection{Time-frequency coherent states}
\label{subsec: TF coherent states}
\paragraph{\lit Definition}
We have introduced the notion of coherent states in CV systems, which provide the quantum-mechanical representation of classical field states. In the time-frequency setting, these states admit a natural generalization to broadband and multimode fields~\cite{loudon_quantum_2000,combescure_coherent_2012,van_enk_quantum_2001}. As in the standard CV case, coherent states can be defined as solutions of an eigenvalue problem. For any complex function $\alpha:\R\to\C$, the associated time-frequency coherent state $\ket{\alpha(\omega)}$ is defined as the solution of
\begin{equation}
    \hat a(\omega)\ket{\alpha(\omega)}=\alpha(\omega)\ket{\alpha(\omega)}.
\end{equation}
This definition generalizes the single-mode coherent state to a continuum of frequency modes. An alternative and often more intuitive view relies on the intrinsic single-mode structure of coherent states. Defining a normalized spectrum
\begin{equation}
    F(\omega)=\frac{\alpha(\omega)}{\sqrt{I}},
\end{equation}
with total intensity
\begin{equation}
    I=\int \dd \omega\, \abs{\alpha(\omega)}^2,
\end{equation}
we introduce the single-mode creation operator $\hat a^\dagger$ associated with the spectral distribution $F$
\begin{equation}
    \hat a_F^\dagger=\int \dd \omega \, F(\omega)\hat a^\dagger(\omega),
\end{equation}
which satisfies $[\hat a_F,\hat a_F^\dagger]=1$. One can then show (see Appendix~\ref{app: TF basic computations}, Result~\ref{res: TF coherent state}) that the coherent state $\ket{\alpha(\omega)}$ admits the explicit expression
\begin{equation}
    \ket{\alpha(\omega)}=e^{-I/2}e^{\sqrt{I}\hat a_F^\dagger}\vac.
\end{equation}
This representation makes explicit that the coherent state $\ket{\alpha}$ is a single-mode coherent state of amplitude $\sqrt{I}$ in the mode given by the spectral distribution $F(\omega)$. In this sense, a time-frequency coherent state is a single-mode coherent state embedded in a continuous frequency space.

The scalar product between two coherent states $\ket{\alpha(\omega)}$ and $\ket{\beta(\omega)}$ can be computed as
\begin{equation}
    \braket{\alpha(\omega)}{\beta(\omega)}=
    \exp\Big(-\frac{1}{2}\int \dd \omega\, \big[\abs{\alpha(\omega)}^2+\abs{\beta(\omega)}^2-2\alpha^\ast(\omega)\beta(\omega)\big]\Big).
\end{equation}
Its squared modulus can be written as
\begin{equation}
    \abs{\braket{\alpha(\omega)}{\beta(\omega)}}^2=
    \exp\Big(-\int \dd \omega\, \abs{\alpha(\omega)-\beta(\omega)}^2\Big).
\end{equation}
A proof is provided in Appendix~\ref{app: TF basic computations}, Result~\ref{res: TF coherent states scalar product}. It is instructive to compare this result with the scalar product of two single-photon states $\ket{\psi_1}$ and $\ket{\psi_2}$,
\begin{equation}
    \braket{\psi_1}{\psi_2}=\int \dd \omega \, F_1^\ast(\omega)F_2(\omega),
\end{equation}
where $F_1$ and $F_2$ are the normalized spectral amplitudes of the two states. Contrary to coherent states, the overlap of single-photon states depends linearly on the mode overlap and not on an intensity-weighted distance.

\paragraph{\lit Time representation}
Time-frequency coherent states can equivalently be characterized in the time domain. Using the Fourier-transform relation between frequency- and time-domain field operators, one finds that the time-domain annihilation operator $\hat b(t)$ acts on $\ket{\alpha(\omega)}$ as
\begin{equation}
    \hat b(t)\ket{\alpha(\omega)}=\tilde \alpha(t)\ket{\alpha(\omega)},
\end{equation}
where $\tilde \alpha(t)$ is the Fourier transform of $\alpha(\omega)$. This shows that coherent states are simultaneous eigenstates of all time-domain annihilation operators, with eigenvalues given by the temporal field envelope.

By normalizing $\tilde \alpha(t)$, one can again separate intensity and temporal structure: the quantity $\frac{1}{I}\abs{\tilde \alpha(t)}^2$ represents the temporal photon distribution of the coherent state, while $I$ is the total mean photon number.

\paragraph{\lit Coherent states and time-frequency operators}
Contrary to the case of single-photon states, there is no simple closed expression for the states $\hat \omega\ket{\alpha(\omega)}$ or $\hat t\ket{\alpha(\omega)}$. Nevertheless, expectation values of the time and frequency operators can be readily computed,
\begin{align}
    \bra{\alpha(\omega)}\hat \omega\ket{\alpha(\omega)}
    &=\int \dd \omega \, \omega \abs{\alpha(\omega)}^2, &
    \bra{\alpha(\omega)}\hat t\ket{\alpha(\omega)}
    &=\int \dd t \, t \abs{\tilde \alpha(t)}^2.
\end{align}
As in the single-photon case, the average frequency and time correspond to intensity-weighted averages of the spectral and temporal profiles, respectively. This reflects the fact that the operators $\hat \omega$ and $\hat t$ simultaneously probe the field energy distribution and the photon number. Only in the single-photon regime can these two aspects be fully disentangled.

\paragraph{\lit Multimode extensions}
Time-frequency coherent states admit a straightforward extension to the setting of multiple spatial modes. For a set of complex functions $\{\alpha_k(\omega)\}$, one for each spatial mode $k=1,\dots,n$, the associated multimode coherent state is defined as the solution of
\begin{equation}
    \hat a_k(\omega)\ket{\{\alpha_k(\omega)\}}=\alpha_k(\omega)\ket{\{\alpha_k(\omega)\}},
\end{equation}
and admits the explicit tensor-product form
\begin{equation}
    \ket{\{\alpha_k(\omega)\}}=\bigotimes_{k=1}^n \ket{\alpha_k(\omega)}.
\end{equation}

\clearpage
\section{Evolutions}
\label{sec: TF evolutions}
\emph{In this section we discuss various important evolutions and transformations of time-frequency systems. While these evolutions apply to general time-frequency states, we will mostly restrict ourselves to the single-photon case. We discuss translation, dispersion (or shear) and rotation.}

\subsection{Translation}
\label{subsec: TF translation}

\paragraph{\lit State transformation}
Since $\hat \omega$ and $\hat t$ are Hermitian operators, they generate unitary transformations, which, as we show below, correspond to translations in time and frequency, respectively. For real parameters $\tau$ and $\varphi$, we consider the unitary operators $e^{-i\hat \omega \tau}$ and $e^{-i\varphi\hat t}$. Their action on the creation operators reads
\begin{align}
    e^{-i\hat \omega \tau}\hat a^\dagger(\omega)e^{i\hat \omega \tau}
    &=e^{-i\omega \tau}\hat a^\dagger(\omega), &
    e^{-i\hat \omega \tau}\hat b^\dagger(t)e^{i\hat \omega \tau}
    &=\hat b^\dagger(t-\tau),\\
    e^{-i\varphi \hat t}\hat a^\dagger(\omega)e^{i\varphi \hat t}
    &=\hat a^\dagger(\omega+\varphi), &
    e^{-i\varphi \hat t}\hat b^\dagger(t)e^{i\varphi\hat t}
    &=e^{-i\varphi t}\hat b^\dagger(t).
\end{align}
The detailed derivations are given in Appendix~\ref{app: TF basic computations}, Result~\ref{res: Translation creation operators}. These relations show that the operator $e^{-i\hat \omega \tau}$ induces a time delay of $\tau$ on the annihilation operator $\hat a(t)$, while $e^{-i\varphi \hat t}$ induces a frequency shift of $\varphi$ on $\hat a(\omega)$. 

Since both unitary operators leave the vacuum invariant, their action on arbitrary states follows directly. For a general single-mode time-frequency state $\ket{\psi}$, defined in Eq.~\eqref{eq: general TF state single mode}, one obtains
\begin{align}
    e^{-i\hat \omega \tau}\ket{\psi}
    &=\sum_k c_k \int \dd t_1\cdots \dd t_k \,
    \tilde F_k(t_1,\dots,t_k)
    \hat b^\dagger(t_1-\tau)\cdots \hat b^\dagger(t_k-\tau)\vac,\\
    e^{-i\varphi \hat t}\ket{\psi}
    &=\sum_k c_k \int \dd\omega_1\cdots \dd\omega_k \,
    F_k(\omega_1,\dots,\omega_k)
    \hat a^\dagger(\omega_1+\varphi)\cdots \hat a^\dagger(\omega_k+\varphi)\vac.
\end{align}
Hence, $e^{-i\hat \omega \tau}$ acts as a rigid translation in time, while $e^{-i\varphi \hat t}$ acts as a rigid translation in frequency on the multiphoton wavefunctions. Such transformations are the optical analogue of phase-space displacements generated by the quadrature operators $\hat x$ and $\hat p$.

The extension to the multimode case is straightforward. The operator $e^{-i\hat \omega_k \tau}$ induces a time delay $\tau$ on mode $k$, while $e^{-i\varphi_k \hat t_k}$ induces a frequency shift $\varphi_k$ on the same mode. In the single-photon time or frequency basis, the corresponding transformation rules read
\begin{align}\label{eq: translation TF basis states}
    e^{-i(\hat\omega_1\tau_1+\cdots+\hat \omega_n\tau_n)} \ket{\omega_1,\dots,\omega_n} &=e^{-i(\omega_1\tau_1+\cdots+\omega_n\tau_n)} \ket{\omega_1,\dots,\omega_n},\\
    e^{-i(\hat\omega_1\tau_1+\cdots+\hat \omega_n\tau_n)} \ket{t_1,\dots,t_n} &=\ket{t_1-\tau_1,\dots,t_n-\tau_n},\\
    e^{-i(\varphi_1 \hat t_1+\cdots+\varphi_n \hat t_n)} \ket{\omega_1,\dots,\omega_n} &=\ket{\omega_1+\varphi_1,\dots,\omega_n+\varphi_n},\\
    e^{-i(\varphi_1 \hat t_1+\cdots+\varphi_n \hat t_n)} \ket{t_1,\dots,t_n} &=e^{-i(\varphi_1 t_1+\cdots+\varphi_n t_n)} \ket{t_1,\dots,t_n}.
\end{align}

\paragraph{\lit Action on the Wigner function}
The Wigner function provides a natural geometric representation in the so-called \emph{chrono-cyclic phase space}, where time and frequency play the role of conjugate variables. Using the definition of the single-mode Wigner function given in Eq.~\eqref{eq: Wigner function single mode}, one finds (see Appendix~\ref{app: TF basic computations}, Result~\ref{res: Wigner translation}) that time and frequency translations act as
\begin{align}
    W(\varphi,\tau) \xrightarrow{e^{-i\hat \omega \tau_0}} W(\varphi,\tau+\tau_0), && W(\varphi,\tau) \xrightarrow{e^{-i\varphi_0 \hat t}} W(\varphi-\varphi_0,\tau).
\end{align}
Thus, $e^{-i\hat \omega \tau_0}$ and $e^{-i\varphi_0 \hat t}$ correspond to rigid translations along the time and frequency axes of the chrono-cyclic phase space, respectively. This interpretation provides a clear geometric picture of these unitary operations and directly generalizes to the multimode setting, where each mode undergoes an independent translation of its associated phase-space coordinates. For illustrative purposes, we plot in Figure~\ref{fig: Wigner translation} the effect of time and frequency translations on the Wigner function of a single-photon cat-like state.

\begin{figure}[ht]
    \centering
    \begin{tabular}{c}
        \includegraphics[width=0.9\linewidth]{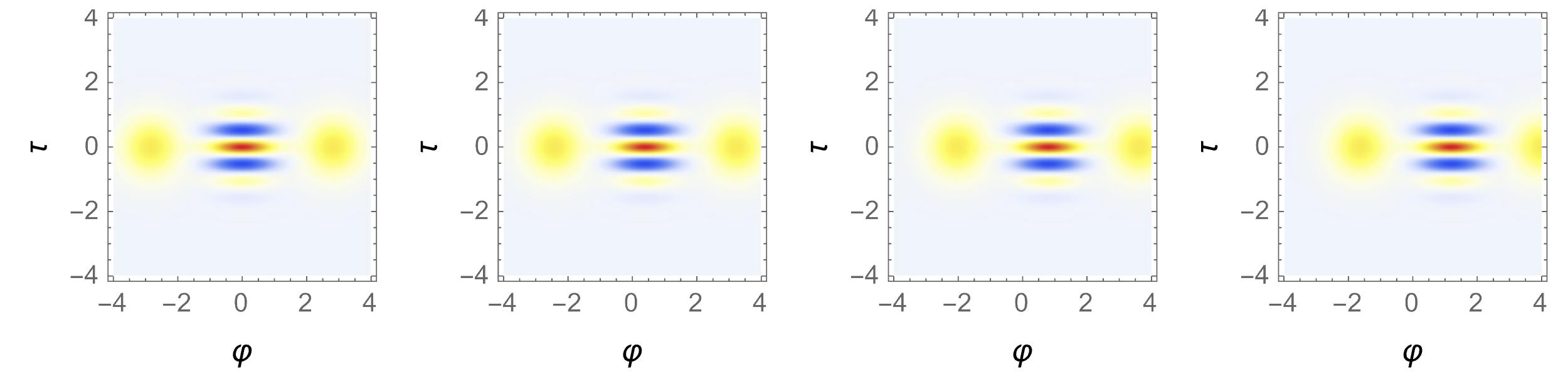}
        \\
        Translation in $\varphi$\\[1ex]
        \includegraphics[width=0.9\linewidth]{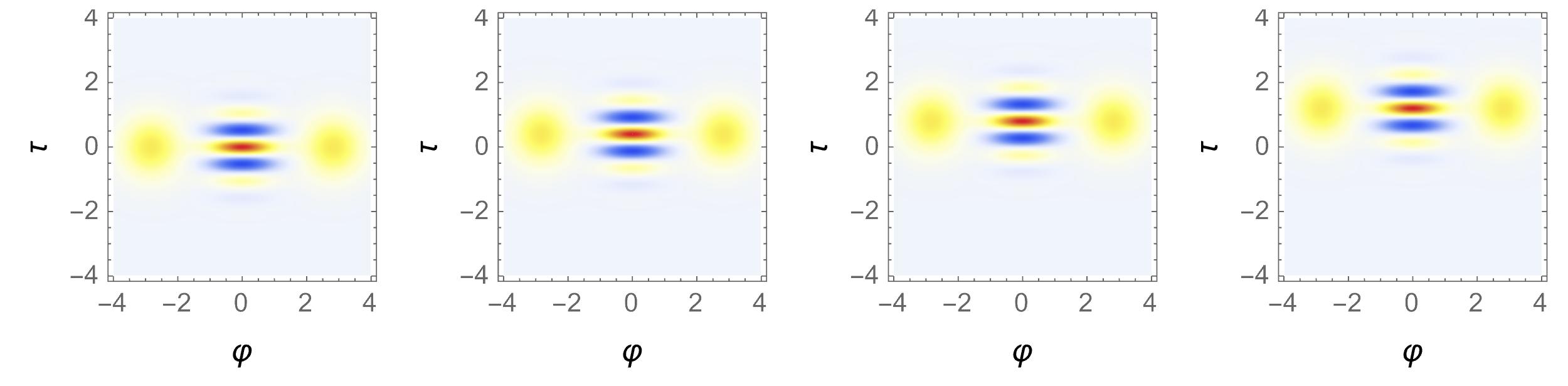}\\
        Translation in $\tau$ 
    \end{tabular}
    \caption[Time and frequency translations of the Wigner function of a single-photon cat-like state]{Effect of time and frequency translations on the Wigner function of a single-photon cat-like state. The top panel shows the effect of a frequency translation, while the bottom panel shows the effect of a time translation.}
    \label{fig: Wigner translation}
\end{figure}

\subsection{Dispersion, time lensing and shearing}
\label{subsec: TF dispersion and time lens}
In this section we study the transformations induced by the quadratic generators $\hat \omega^2$ and $\hat t^2$, which play a central role in dispersive propagation and temporal imaging.

\paragraph{\lit Quadratic frequency gate}
The unitary operator $e^{-i\alpha\hat \omega^2}$ acts diagonally in the frequency basis. Indeed, for a single-photon frequency eigenstate one has
\begin{equation}
    e^{-i\alpha \hat \omega^2}\ket{\omega}=e^{-i\alpha \omega^2}\ket{\omega},
\end{equation}
as verified in Appendix~\ref{app: TF basic computations}, Result~\ref{res: Quadratic gate creation op}. When acting on an arbitrary single-photon wavepacket, this transformation therefore imprints a quadratic spectral phase.

In contrast to free propagation, which induces a phase linear in frequency and corresponds to a uniform time delay, a quadratic spectral phase implements a frequency-dependent delay. This is the hallmark of dispersive propagation, where different frequency components travel with different group velocities. Such an evolution accurately models group velocity dispersion in optical fibers or other dispersive media~\cite{boyd_nonlinear_2008}.

The action of the quadratic frequency gate on creation operators can be written as
\begin{equation}
    e^{-i\alpha \hat \omega^2}\hat a^\dagger(\omega)e^{i\alpha \hat \omega^2}
    =e^{-i\alpha \omega^2}\hat a^\dagger(\omega)e^{-2i\alpha\hat \omega\omega},
\end{equation}
as shown in Appendix~\ref{app: TF basic computations}, Result~\ref{res: Quadratic gate creation op}. This expression is essential to understand the effect of quadratic spectral phases on multiphoton states. One immediately notices the presence of the operator-valued linear term $e^{-2i\alpha\hat \omega\omega}$ on the right-hand side. When cascading the action of $\hat \omega^2$ on several photons, this term leads to the emergence of correlated phases between photons. A more transparent approach to computing the action of the quadratic frequency gate on multiphoton states relies on the identity
\begin{equation}
    \hat \omega \hat a^\dagger(\omega_1)\cdots\hat a^\dagger(\omega_n)\vac =(\omega_1+\cdots+\omega_n)\hat a^\dagger(\omega_1)\cdots\hat a^\dagger(\omega_n)\vac,
\end{equation}
which follows from the commutation relations discussed in Appendix~\ref{app: TF basic computations}, Result~\ref{res: General TF commutator}. Exponentiating this relation yields
\begin{equation}
    e^{-i\alpha\hat\omega^2}\hat a^\dagger(\omega_1)\cdots\hat a^\dagger(\omega_n)\vac =e^{-i\alpha(\omega_1+\cdots+\omega_n)^2} \hat a^\dagger(\omega_1)\cdots\hat a^\dagger(\omega_n)\vac.
\end{equation}
Hence, the quadratic frequency gate imprints a correlated phase that depends on the square of the sum of all photon frequencies.

The effect of this quadratic evolution admits a simple geometric interpretation in terms of the Wigner function. For an initial single-photon state $\hat\rho$ with Wigner function $W(\varphi,\tau)$, the transformation induced by $e^{-i\alpha\hat \omega^2}$ reads
\begin{equation}
    W(\varphi,\tau)\xrightarrow{e^{-i\hat \omega^2 \alpha}}W(\varphi,\tau+2\alpha \varphi).
\end{equation}
This result, derived in Appendix~\ref{app: TF basic computations}, Result~\ref{res: Wigner quadratic transformation}, shows that the phase-space distribution is translated along the time axis by an amount proportional to the frequency coordinate $\varphi$. Geometrically, this corresponds to a shear of phase space, making explicit the interpretation of dispersion as a frequency-dependent time delay.

\paragraph{\lit Time lens}
The quadratic evolution generated by $\hat t^2$ can be treated in a fully analogous manner. In the time basis one finds
\begin{equation}
    e^{-i\beta \hat t^2}\ket{t}=e^{-i\beta t^2}\ket{t},
\end{equation}
as verified in Appendix~\ref{app: TF basic computations}, Result~\ref{res: Quadratic gate creation op}. Acting on a single-photon temporal wavepacket, this operation imprints a quadratic temporal phase.

Such a transformation corresponds physically to a time lens, the temporal analogue of a spatial lens, and is a fundamental building block of temporal imaging systems~\cite{kolner_temporal_1989,kolner_space-time_1994,patera_space-time_2018}. Experimentally, time lenses can be implemented using electro-optic phase modulators or nonlinear optical interactions such as sum-frequency generation.

As in the frequency quadratic case, the action on creation operators is more involved:
\begin{equation}
    e^{-i\beta \hat t^2}\hat b^\dagger(t)e^{i\beta \hat t^2}
    =e^{-i\beta t^2}\hat b^\dagger(t)e^{-2i\beta\hat t t}.
\end{equation}
Because of the operator-valued linear term, interpreting the effect of this gate on multiphoton states is less intuitive, but it can be analyzed using the same strategy as for the quadratic frequency gate.

The Wigner function again provides a clear geometric picture. For a single-photon state $\hat\rho$ with Wigner function $W(\varphi,\tau)$, the action of $e^{-i\beta \hat t^2}$ yields
\begin{equation}
    W(\varphi,\tau)\xrightarrow{e^{-i\beta \hat t^2}}W(\varphi-2\beta \tau,\tau).
\end{equation}
This transformation corresponds to a shear of phase space along the frequency axis, with an amplitude proportional to the time coordinate $\tau$. A detailed proof of this rule is given in Appendix~\ref{app: TF basic computations}, Result~\ref{res: Wigner quadratic transformation}. For illustrative purposes, we plot in Figure~\ref{fig: Wigner shearing} the effect of time and frequency translations on the Wigner function of a single-photon cat-like state.

\begin{figure}[ht]
    \centering
    \begin{tabular}{c}
        \includegraphics[width=0.9\linewidth]{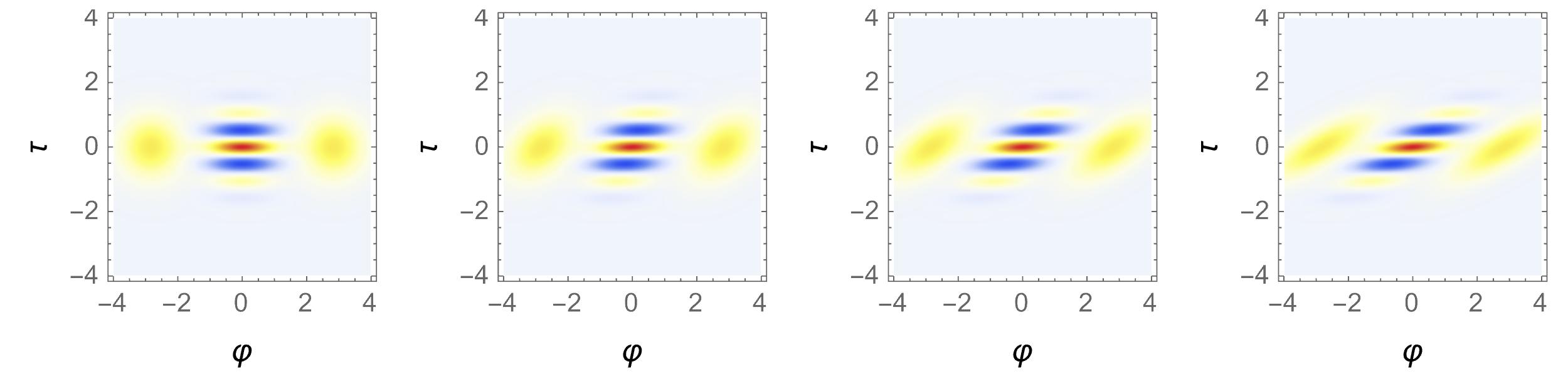}
        \\
        Shear in $\varphi$\\[1ex]
        \includegraphics[width=0.9\linewidth]{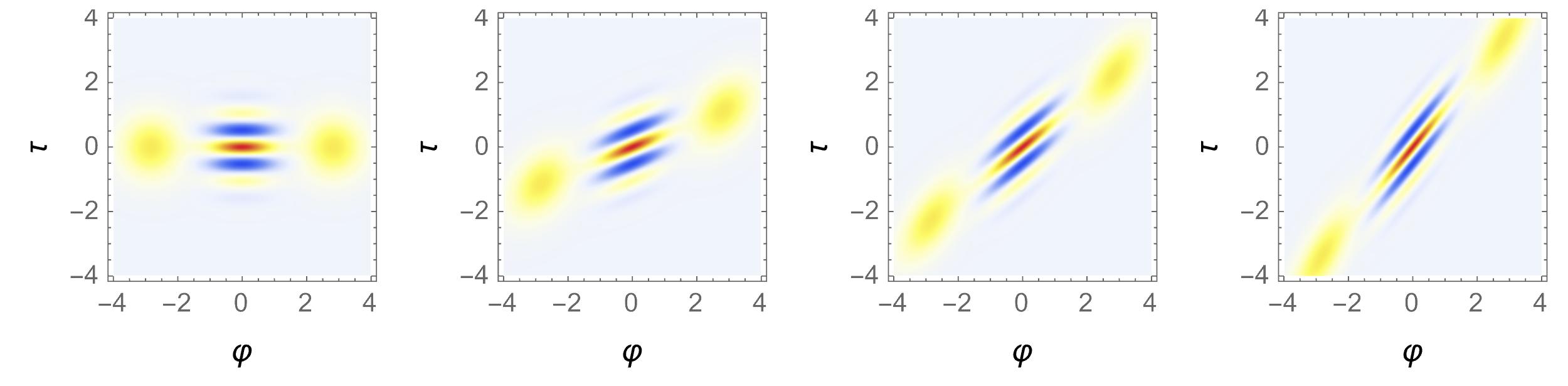}\\
        Shear in $\tau$ 
    \end{tabular}
    \caption[Time and frequency shears of the Wigner function of a single-photon cat-like state]{Effect of time and frequency shears on the Wigner function of a single-photon cat-like state. The top panel shows the effect of a frequency shear, while the bottom panel shows the effect of a time shear.}
    \label{fig: Wigner shearing}
\end{figure}

\subsection{Rotation and partial Fourier transform}
\label{susec: TF rotation}

\paragraph{\lit Definition}
Recall that the quadratic Hamiltonian of continuous-variable systems in a harmonic potential reads
\begin{equation}
    \hat H=\frac{1}{2}(\hat x^2+\hat p^2),
\end{equation}
which, from a phase-space perspective, generates rotations. Following the established analogy between continuous variables and single-photon time-frequency systems, we introduce the operator
\begin{equation}
    \hat R=\frac{1}{2}(\hat t^2+\hat \omega^2).
\end{equation}
By analogy with the CV case, the associated unitary $e^{-i\theta\hat R}$ is expected to implement rotations in the chrono-cyclic phase space. In the following, we rederive the corresponding properties explicitly in the time-frequency setting.

\paragraph{\lit Remark on dimension}
In all previous sections, we have manipulated time and frequency variables, without specifying their dimensions or units. This was possible as long as only products were considered. However, the definition of $\hat R$ involves sums of squared time and frequency operators, which requires both terms to have the same dimension. This can be achieved by introducing a characteristic time scale $T$, and defining dimensionless time and frequency operators as
\begin{align}
    \hat t'=\frac{\hat t}{T}, && \hat \omega' = T \hat \omega,
\end{align}
which we implicitly do in the following. Notice that the choice of $T$ fixes a natural scale in the time-frequency phase space and does not alter the fundamental properties of the operators. For example, the dimensionless operators satisfy the commutation relation $[\hat \omega',\hat t']=i\1$.

\paragraph{\lit Rotation of frequency and time operators}
A first justification for interpreting $\hat R$ as a rotation generator is obtained by examining the transformation of the frequency and time operators $\hat\omega$ and $\hat t$. One finds
\begin{align}
    e^{-i\theta \hat R}\hat \omega e^{i\theta \hat R}&=\cos\theta\,\hat \omega-\sin\theta\,\hat t, &
    e^{-i\theta \hat R}\hat t e^{i\theta \hat R}&=\sin\theta\,\hat \omega+\cos\theta\,\hat t,
\end{align}
which can be written in vector form as
\begin{equation}
    \begin{pmatrix}
        \hat \omega\\
        \hat t
    \end{pmatrix}
    \xrightarrow{e^{-i\theta \hat R}}
    \begin{pmatrix}
        \cos\theta & -\sin\theta\\
        \sin\theta & \cos\theta
    \end{pmatrix}
    \begin{pmatrix}
        \hat \omega\\
        \hat t
    \end{pmatrix}.
\end{equation}
Here $R_\theta=\begin{pmatrix}\cos\theta & -\sin\theta\\ \sin\theta & \cos\theta\end{pmatrix}$ denotes the standard two-dimensional rotation matrix. A detailed derivation is provided in Appendix~\ref{app: TF basic computations}, Result~\ref{res: Rotation of TF operators}. This transformation law makes explicit that $\hat R$ generates rotations in the $(\hat\omega,\hat t)$ operator space, thereby confirming the initial intuition.

\paragraph{\lit Rotation and Wigner function}
A particularly transparent connection between $\hat R$ and geometrical rotations is obtained using the Wigner function and its associated phase-space picture. For a single-photon time-frequency state $\hat\rho$ with Wigner function $W_{\hat\rho}$, we show in Appendix~\ref{app: TF basic computations}, Result~\ref{res: rotation wigner}, that
\begin{equation}
    W_{\hat\rho}(\varphi,\tau)\xrightarrow{e^{-i\theta \hat R}}
    W_{\hat\rho}(\varphi\cos\theta-\tau\sin\theta,\tau\cos\theta+\varphi\sin\theta).
\end{equation}
The phase-space coordinates are therefore rotated by an angle $\theta$ in the chrono-cyclic plane, which once again confirms the interpretation of $\hat R$ as a rotation generator. For illustrative purposes, we plot in Figure~\ref{fig: Wigner rotation} the effect of a rotation on the Wigner function of a single-photon cat-like state.

\begin{figure}[ht]
    \centering
    \includegraphics[width=0.9\linewidth]{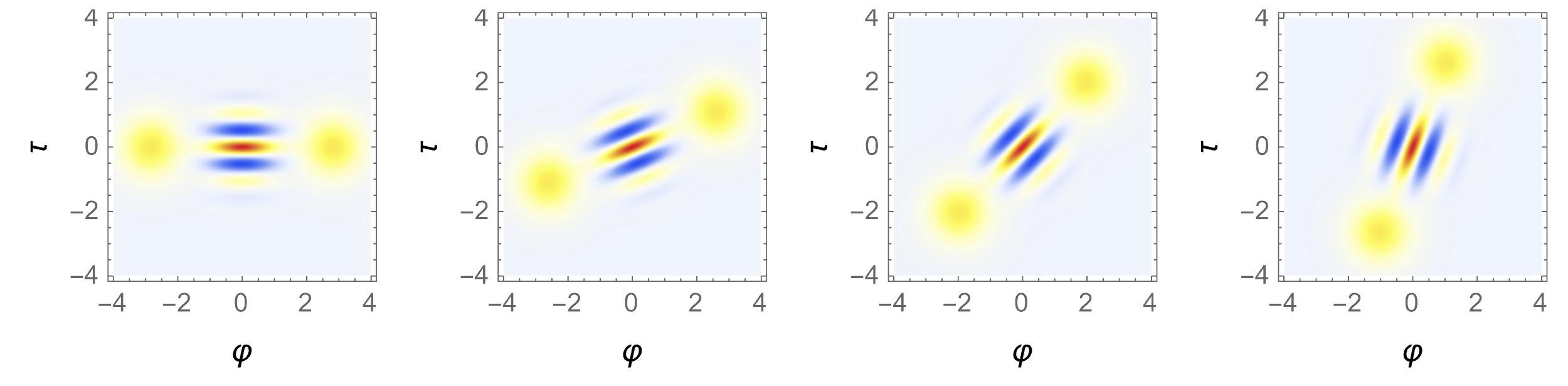}
    \caption[Rotation of the Wigner function of a single-photon cat-like state]{Effect of a rotation on the Wigner function of a single-photon cat-like state.}
    \label{fig: Wigner rotation}
\end{figure}

\paragraph{\lit Shears and Fourier transform}
As will become apparent in the following, the rotation operator is intimately related to the Fourier and partial Fourier transforms. This connection is particularly relevant from an experimental perspective, as it raises the question of how to implement rotations using physically accessible operations.

Consider a single-photon state $\ket{\psi_\text{in}}$ with time-of-arrival wavefunction $\psi_\text{in}$. We seek a sequence of operations $\hat U$ such that
\begin{equation}
    \ket{\psi_\text{out}}=\hat U\ket{\psi_\text{in}}=\int\dd t\,\psi_\text{out}(t)\ket{t},
\end{equation}
where the output wavefunction is the Fourier transform of the input,
\begin{equation}
    \psi_\text{out}(t)=\frac{1}{\sqrt{2\pi}}\int\dd\tau\,e^{-it\tau}\psi_\text{in}(\tau).
\end{equation}
This problem is closely related to the propagation of optical fields in dispersive media and its interpretation in terms of temporal Fourier optics~\cite{lu_arbitrary_2018,goda_theory_2009}.

As discussed in Sec.~\ref{subsec: TF dispersion and time lens}, dispersive evolution is naturally expressed in the spectral domain through a quadratic phase. Introducing Fourier transforms to switch between spectral and temporal representations, one obtains
\begin{equation}
    \psi_\text{out}(t)=\frac{1}{2\pi}\int\dd\omega\dd\tau\,\psi_\text{in}(\tau)
    e^{-i\alpha\omega^2}e^{i\omega(\tau-t)}.
\end{equation}
Performing the integral over $\omega$ yields
\begin{equation}
    \psi_\text{out}(t)\propto\int\dd\tau\,\psi_\text{in}(\tau)e^{i(t-\tau)^2/4\alpha}
    =e^{it^2/4\alpha}\int\dd\tau\,\psi_\text{in}(\tau)
    e^{i\tau^2/4\alpha}e^{-it\tau/2\alpha},
\end{equation}
where time-independent global phase factors have been omitted for simplicity. The term $e^{-it\tau/2\alpha}$ is precisely the kernel of a Fourier transform. If one could implement the replacements
\begin{subequations}
\begin{align}
    \psi_\text{in}(\tau)&\mapsto\psi_\text{in}(\tau)e^{-i\tau^2/4\alpha},& 
    \psi_\text{out}(t)&\mapsto\psi_\text{out}(t)e^{-it^2/4\alpha},
\end{align}
\end{subequations}
one could eliminate the remaining quadratic phases and the resulting transformation would be a pure Fourier transform. Physically, this requires applying a time lens before and after the dispersive evolution~\cite{kolner_temporal_1989}. In operator form, the required sequence reads
\begin{equation}
    e^{-i\hat t^2/4\alpha}e^{-i\alpha\hat \omega^2}e^{-i\hat t^2/4\alpha}.
\end{equation}
Setting $\alpha=1/2$, we expect this composition to implement a rotation, leading to the operator identity
\begin{equation}
    e^{-i\hat t^2/2}e^{-i\hat \omega^2/2}e^{-i\hat t^2/2}
    =e^{-i\lambda\hat R},
\end{equation}
for some constant $\lambda$, which we cannot determine explicitly from this reasoning alone.

\paragraph{\lit Phase space picture for composing shears}
The above result can be understood geometrically using the Wigner function. The action of the quadratic operators $\hat\omega^2$ and $\hat t^2$ on the Wigner function is given by (see Sec.~\ref{subsec: TF dispersion and time lens})
\begin{subequations}
\begin{align}
    W(\varphi,\tau)&\xrightarrow{e^{-i\alpha\hat \omega^2}}W(\varphi,\tau+2\alpha\varphi),\\
    W(\varphi,\tau)&\xrightarrow{e^{-i\beta\hat t^2}}W(\varphi-2\beta\tau,\tau).
\end{align}
\end{subequations}
Applying successively the three operations yields
\begin{subequations}
\begin{align}
    W(\varphi,\tau)&\xrightarrow{e^{-i\hat t^2/2}}W(\varphi-\tau,\tau),\\
    &\xrightarrow{e^{-i\hat\omega^2/2}}W(-\tau,\tau+\varphi),\\
    &\xrightarrow{e^{-i\hat t^2/2}}W(-\tau,\varphi).
\end{align}
\end{subequations}
On the other hand, the rotation generated by $\hat R=(\hat\omega^2+\hat t^2)/2$ acts as
\begin{equation}
    W(\varphi,\tau)\xrightarrow{e^{-i\theta\hat R}}
    W(\cos\theta\,\varphi-\sin\theta\,\tau,\cos\theta\,\tau+\sin\theta\,\varphi),
\end{equation}
which for $\theta=\pi/2$ reduces to $W(-\tau,\varphi)$. This phase-space argument therefore guarantees the relation
\begin{equation}
    e^{-i\hat t^2/2}e^{-i\hat \omega^2/2}e^{-i\hat t^2/2}
    =e^{-i\frac{\pi}{2}\hat R}.
\end{equation}
Notice that in this case the phase-space axes are rotated by $90^\circ$, which is exactly what we expect with a Fourier transform that switches the role of the spectral and temporal variables. For illustrative purposes, we plot in Figure~\ref{fig: Wigner 3 shear} the effect of this sequence of three shears on the Wigner function of a single-photon cat-like state.

\begin{figure}[ht]
    \centering
    \includegraphics[width=0.9\linewidth]{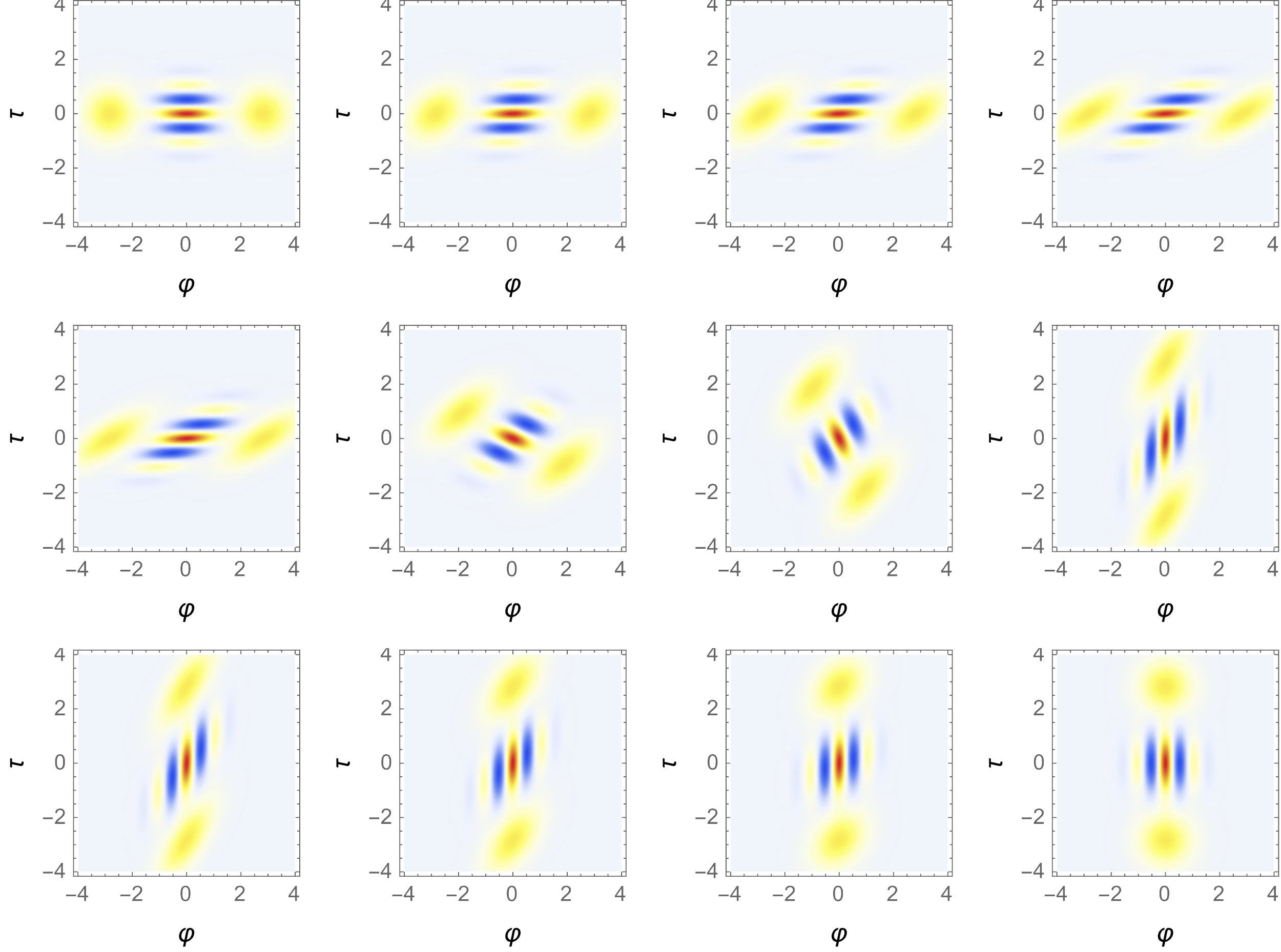}
    \caption[Composition of three shears implementing a rotation of the Wigner function]{Effect of three shears implementing a rotation on the Wigner function of a single-photon cat-like state. The first line implements the frequency shear, the second line the time shear and the third line the final frequency shear. The resulting transformation is equivalent to a rotation by $90^\circ$.}
    \label{fig: Wigner 3 shear}
\end{figure}

\paragraph{\lit Partial Fourier transform}
The previous result shows that the Fourier transform corresponds to a $\pi/2$ rotation in the chrono-cyclic phase space and that it can be implemented as a sequence of three shears. More generally, partial or fractional Fourier transforms correspond to rotations by an arbitrary angle $\theta$~\cite{ozaktas_fractional_2001}. One may therefore expect such operations to be realizable through appropriately tuned shears.

As shown in Appendix~\ref{app: TF basic computations}, Result~\ref{res: Partial Fourier}, the general identity reads
\begin{equation}
    e^{-ia\frac{\hat t^2}{2}}e^{-ib\frac{\hat \omega^2}{2}}e^{-ia\frac{\hat t^2}{2}}
    =e^{-i\theta\hat R},
\end{equation}
provided that
\begin{align}
    a=\frac{1-\cos(\theta)}{\sin(\theta)}, && b=\sin(\theta).
\end{align}
Remarkably, both time lenses must be implemented with the same parameter. By similar reasoning, it is also possible to realize a fractional Fourier transform using two dispersive evolutions and a single time lens. In Figure \ref{fig: Wigner partial Fourier} we illustrate the effect of a partial Fourier transform on the Wigner function of a single-photon cat-like state.

\begin{figure}[ht]
    \centering
    \includegraphics[width=0.9\linewidth]{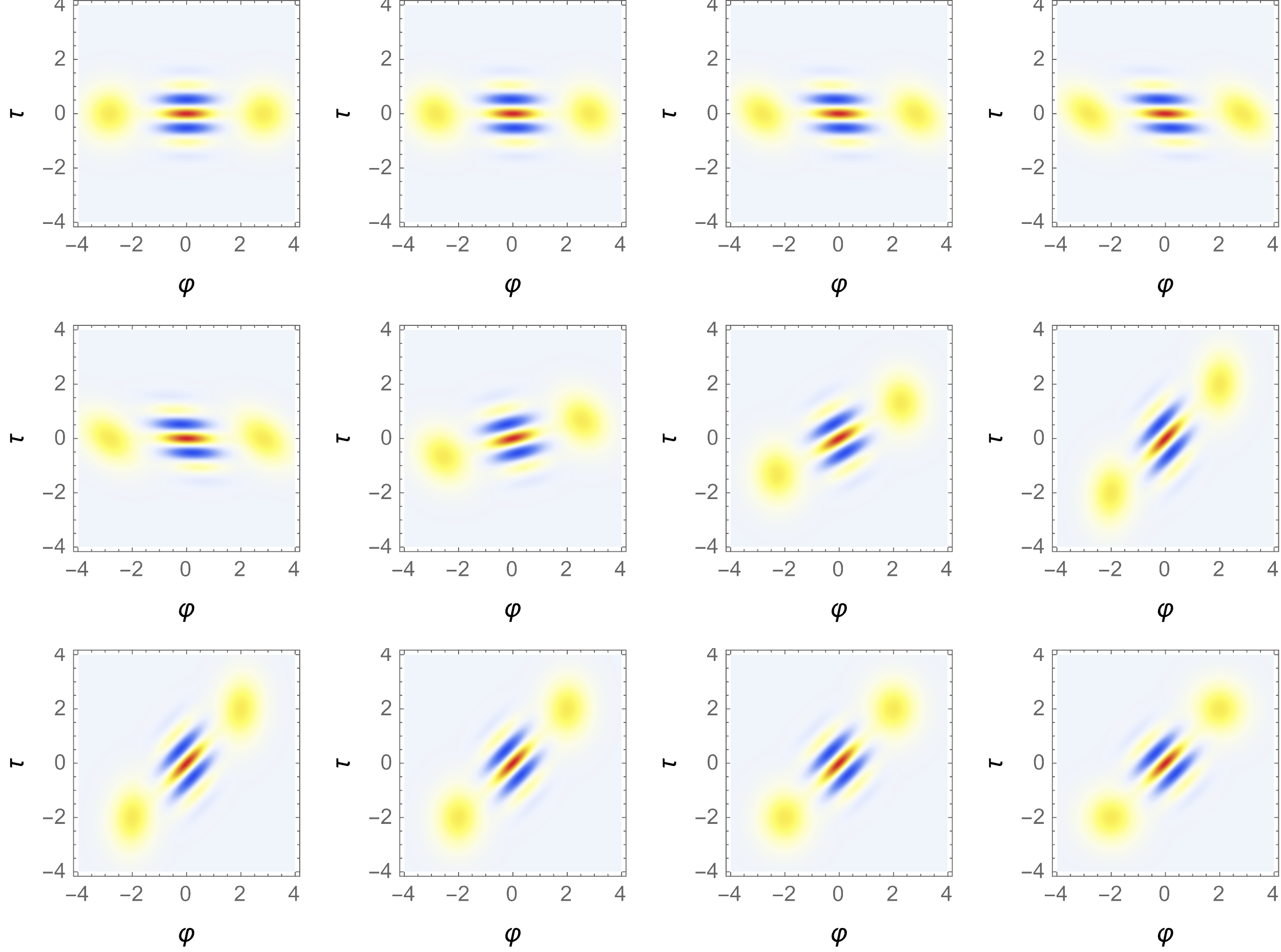}
    \caption[Three successively applied shears implementing a $45^\circ$ rotation of the Wigner function]{Three successively applied shears implementing a $45^\circ$ rotation on the Wigner function of a single-photon cat-like state. The first line implements the frequency shear ($a=1-\sqrt{2}$), the second line the time shear ($b=-1/\sqrt{2}$) and the third line the final frequency shear ($a=1-\sqrt{2}$).}
    \label{fig: Wigner partial Fourier}
\end{figure}

\subsection{Multimode operators}
\label{subsec: Multimode TF operators}

\paragraph{\lit Collective time and frequency operators}
So far, we have introduced only single-mode operators. Naturally, they can be extended to the multimode case by specifying on which mode they act. For example, $\hat{\omega}_1$ induces time translations on mode $1$ only, while $\hat{t}_3^2$ induces time lensing on mode $3$ only, etc. However, one can also consider operators that act non-locally on several modes simultaneously. To illustrate this idea, we restrict ourselves to the case of two modes ($n=2$). Further developments on collective operators are presented in Sec.~\ref{sec: Collective entanglement}.

When working with two spatial modes, the phase space is four-dimensional, which leaves considerable freedom in the choice of coordinates. One may introduce different parametrizations, among which rotated axes provide a natural and physically transparent example. This motivates the introduction of the collective operators
\begin{subequations}
    \begin{align}
        \hat{\omega}_\theta&=\cos(\theta)\hat{\omega}_1+\sin(\theta)\hat{\omega}_2, &
        \hat{t}_\theta&=\cos(\theta)\hat{t}_1+\sin(\theta)\hat{t}_2,\\
        \hat{\omega}_\theta'&=-\sin(\theta)\hat{\omega}_1+\cos(\theta)\hat{\omega}_2, &
        \hat{t}_\theta'&=-\sin(\theta)\hat{t}_1+\cos(\theta)\hat{t}_2.
    \end{align}
\end{subequations}
For these operators to be meaningful, the primed and non-primed sets must correspond to orthogonal (independent) collective directions, \ie, they should commute with each other, and within each mode we should recover the usual canonical commutation relations between $\hat{\omega}$ and $\hat{t}$. We therefore compute
\begin{subequations}
    \begin{align}
        [\hat{\omega}_\theta,\hat{t}_\theta]
        &=[\cos(\theta)\hat{\omega}_1+\sin(\theta)\hat{\omega}_2,\cos(\theta)\hat{t}_1+\sin(\theta)\hat{t}_2],\\
        &=\cos^2(\theta)[\hat{\omega}_1,\hat{t}_1]+\sin^2(\theta)[\hat{\omega}_2,\hat{t}_2],\\
        &=\left(\cos^2(\theta)+\sin^2(\theta)\right)i\1,\\
        &=i\1.
    \end{align}
\end{subequations}
As desired, the canonical commutation relation is preserved. The same result holds for the primed operators, which can be obtained by the substitution $\theta\mapsto \theta+\pi/2$. We also verify the orthogonality of the two collective modes
\begin{subequations}
    \begin{align}
         [\hat{\omega}_\theta,\hat{t}_\theta']
         &=[\cos(\theta)\hat{\omega}_1+\sin(\theta)\hat{\omega}_2,-\sin(\theta)\hat{t}_1+\cos(\theta)\hat{t}_2],\\
        &=-\cos(\theta)\sin(\theta)[\hat{\omega}_1,\hat{t}_1]+\cos(\theta)\sin(\theta)[\hat{\omega}_2,\hat{t}_2],\\
        &=0.
    \end{align}
\end{subequations}
An analogous calculation shows that $[\hat{\omega}_\theta',\hat{t}_\theta]=0$. Hence, the two collective modes are indeed independent.

While any value of $\theta$ is valid, a particularly common and convenient choice is $\theta=\pi/4$, which yields the symmetric and anti-symmetric combinations
\begin{align}
    \hat{\omega}_+=\frac{\hat{\omega}_1+\hat{\omega}_2}{\sqrt{2}}, &&
    \hat{t}_+=\frac{\hat{t}_1+\hat{t}_2}{\sqrt{2}}, &&
    \hat{\omega}_-=\frac{\hat{\omega}_1-\hat{\omega}_2}{\sqrt{2}}, &&
    \hat{t}_-=\frac{\hat{t}_1-\hat{t}_2}{\sqrt{2}}.
\end{align}
These operators correspond to correlated ($+$) and anti-correlated ($-$) evolutions of the two spatial modes. The normalization factor $1/\sqrt{2}$ arises naturally from the orthogonal rotation and is required to preserve the canonical commutation relations.

\paragraph{\newc General collective operators}
The operators introduced above, as well as the underlying construction, naturally lead to a broader class of non-local transformations. In particular, $\hat{\omega}_\theta$ and $\hat{t}_\theta$ can be used as building blocks for more general evolutions, including shears and rotations in collective phase space. One may therefore consider operators such as $\hat{\omega}_\theta^2$, $\hat{t}_\theta^2$, $(\hat{\omega}_\theta')^2$, or $(\hat{t}_\theta')^2$, and define, for instance,
\begin{align}
    \hat{R}_\theta=\frac{1}{2}(\hat{t}_\theta^2+\hat{\omega}_\theta^2), &&
    \hat{R}_\theta'=\frac{1}{2}(\hat{t}_\theta'^2+\hat{\omega}_\theta'^2),
\end{align}
which play a role analogous to harmonic-oscillator rotation generators in the collective variables. Once again, the case $\theta=\pi/4$ is particularly interesting, leading to
\begin{align}
    \hat{R}_+
    &=\frac{1}{2}(\hat{t}_+^2+\hat{\omega}_+^2)
    =\frac{1}{4}(\hat{\omega}_1+\hat{\omega}_2)^2+\frac{1}{4}(\hat{t}_1+\hat{t}_2)^2,\\
    \hat{R}_-
    &=\frac{1}{2}(\hat{t}_-^2+\hat{\omega}_-^2)
    =\frac{1}{4}(\hat{\omega}_1-\hat{\omega}_2)^2+\frac{1}{4}(\hat{t}_1-\hat{t}_2)^2,
\end{align}
and similarly for the associated shear operators. Although it may be difficult to assign a direct physical interpretation to these non-local transformations, they are conceptually valuable. In particular, when the joint spectral amplitude of a bipartite quantum state is separable in the diagonal variables $\omega_\pm$, the corresponding evolution generated by these operators is also separable, which can lead to significant analytical simplifications.

A useful identity follows from the quadratic definitions above:
\begin{align}
    \hat{\omega}_\theta^2+(\hat{\omega}_\theta')^2=\hat{\omega}_1^2+\hat{\omega}_2^2, &&
    \hat{t}_\theta^2+(\hat{t}_\theta')^2=\hat{t}_1^2+\hat{t}_2^2, &&
    \hat{R}_\theta+\hat{R}_\theta'=\hat{R}_1+\hat{R}_2.
\end{align}
These relations can simplify practical calculations and can be interpreted as expressing the invariance of simultaneous correlated actions in orthogonal collective variables with respect to the choice of rotation axis in phase space.

\clearpage
\section{Time-frequency systems and metrology}
\label{sec: TF metrology}
\emph{In this short section, we discuss the relationship between time-frequency quantum systems and metrology, emphasizing the importance and applications of the formalism introduced in this chapter. We also address computational considerations and the connection to the Wigner function.}

\subsection{Relation to experiment}
\label{subsec: relation to experiment}

\paragraph{\lit Metrological protocols}
As introduced in Sec.~\ref{subsec: quantum metrology}, and schematized in Fig.~\ref{fig: quantum-metrology-scheme}, a quantum metrology protocol can generally be decomposed into four steps: probe preparation, parameter-dependent evolution, measurement, and post-processing. In quantum optics, this structure is naturally implemented and is summarized in Fig.~\ref{fig: general metrology interferometer}.
\begin{itemize}
    \item The probe is typically a specific state of light. In the context of time-frequency quantum optics, such states are described using the formalism introduced in this chapter. Examples include single-photon states with tailored spectral or temporal distributions, entangled photon pairs characterized by a joint spectral amplitude, or more generally multimode time-frequency states with arbitrary photon-number distributions, as in \eqref{eq: general TF state multimode}~\cite{humphreys_quantum_2013, brecht_photon_2015}.
    \item Due to their constant propagation speed, photons naturally relate spatial and temporal degrees of freedom, making them ideal probes for distance and time measurements. The time-frequency formalism provides a natural framework for time-delay estimation, modeled through time-translation operators $\hat \omega$ implementing free-space propagation. Other evolutions, such as propagation in dispersive media, can be described using quadratic frequency operators, while more complex transformations including time lenses or rotations in chrono-cyclic phase space are captured by appropriate unitary generators.
    .
    \item Quantum measurements often rely on sophisticated architectures that may involve nonlocal operations. In quantum optics, this is commonly achieved using linear optical interferometers followed by single-mode (local) detection. The interferometer mixes the optical modes such that local measurements suffice to extract information about the parameter of interest. Detection schemes typically fall into two categories: single-photon detection, which registers the presence of photons with or without photon-number resolution and is implemented using avalanche photodiodes or superconducting nanowire single-photon detectors~\cite{hadfield_single-photon_2009}; and homodyne detection, which mixes the signal with a strong local oscillator to measure field quadratures using balanced beam splitters and photodiodes~\cite{bachor_guide_2019}.
    \item The post-processing stage is not specific to time-frequency quantum optics and follows standard approaches used in quantum metrology, such as maximum-likelihood estimation, discussed in Sec.~\ref{subsec: maximum likelihood estimator}. We do not elaborate further on this aspect here.
\end{itemize}

\begin{figure}[ht]
    \centering
    \footnotesize
    \scalebox{1.3}{\input{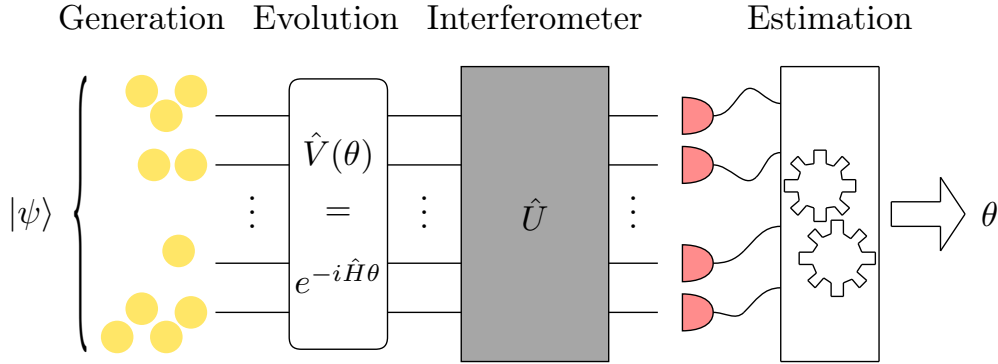}}
    \caption[Schematics of a general optical interferometric protocol]{Schematics of a general optical interferometric protocol. An initial probe state $\ket{\psi}$ undergoes a parameter-dependent evolution $\hat V(\theta)=e^{-i\hat H\theta}$. The evolved state is then processed by an interferometer represented by the unitary $\hat U$ and measured locally on each mode. The parameter $\theta$ is finally estimated from the measurement outcomes.}
    \label{fig: general metrology interferometer}
\end{figure}

\paragraph{\lit Precision}
Within this framework, we consider an initial state $\ket{\psi}$ evolving under $\hat V(\theta)=e^{-i\hat H\theta}$. As reviewed in Sec.~\ref{subsec: quantum metrology pure states}, the quantum Fisher information $\mathcal Q$ is proportional to the variance of the generator $\hat H$,
\begin{equation}
    \mathcal Q=4\Delta^2 \hat H=4\left(\bra{\psi}\hat H^2\ket{\psi}-\bra{\psi}\hat H\ket{\psi}^2\right).
\end{equation}
For a given interferometer $\hat U$ and measurement scheme, the classical Fisher information $\mathcal F$ can be computed from the measurement statistics as introduced in Sec.~\ref{subsec: quantum metrology}. While $\mathcal F$ quantifies the precision of a specific implementation, $\mathcal Q$ provides an upper bound optimized over all possible measurements.

In the simplest case of time-delay estimation, the maximal achievable precision is determined by the variance of the time-translation operator $\hat \omega_j$ acting on mode $j$,
\begin{equation}
    \mathcal Q=4\Delta^2\hat \omega_j.
\end{equation}
More general scenarios involve collective evolutions, where the unknown delay $\theta$ acts on several modes simultaneously. If the delay acts identically on all $n$ modes, the relevant generator is the collective operator $\hat \Omega=\sum_{j=1}^n \hat \omega_j$, yielding
\begin{equation}
    \mathcal Q=4\Delta^2 \hat \Omega.
\end{equation}
The role of such collective operators is discussed in more detail in Sec.~\ref{sec: TF entanglement and metrology}.

More generally, any generator introduced in Sec.~\ref{sec: TF evolutions} can be considered. The corresponding achievable precision is obtained by computing the variance of the relevant operator, for instance
\begin{align}
    \mathcal Q&=4\Delta^2 \hat R_1, &
    \mathcal Q&=4\Delta^2 \hat t_3^2, &
    \mathcal Q&=4\Delta^2 (\hat\omega_2^2+\hat\omega_3^2),
\end{align}
which directly links the structure of the time-frequency evolution to the attainable metrological performance.

\subsection{Variance computation}
\label{subsec: Variance computation}
As argued in the previous section, for pure probe states under unitary evolution the ultimate precision is expressed as variances. In this section we discuss different cases to develop intuition, focusing mostly on time-delay measurements.

\paragraph{\publi Single-mode time-delay measurement}
While interferometric measurements need to involve at least two modes, for mathematical simplicity and since assumptions about the factorizability of the JSA (see Sec.~\ref{subsec: SPDC}) usually allow us to reduce the problem to a single mode, we first consider the simplest case of a single-mode time-delay measurement. The relevant operator is simply $\hat \omega$ and the quantum Fisher information reads $4\Delta^2\hat\omega$. Assuming the probe state has a spectrum $F$, one gets
\begin{equation}
    \Delta^2 \hat \omega=\bra{\psi}\hat \omega^2\ket{\psi}-\bra{\psi}\hat \omega\ket{\psi}^2=\int \dd \omega\, \abs{F(\omega)}^2 \omega^2 -\left(\int \dd \omega\, \abs{F(\omega)}^2 \omega\right)^2,
\end{equation}
which we can interpret as the variance of the frequency distribution $\abs{F(\omega)}^2$. As such, in order to optimize the precision, one needs to maximize the frequency variance of the probe state. Considering typical states, we can explicitly compute the variance.

For a single-photon state with a Gaussian spectrum
\begin{equation}
    F(\omega)=\frac{1}{(2\pi \sigma^2)^{1/4}}e^{-\frac{(\omega-\omega_0)^2}{4\sigma^2}},
\end{equation}
where $\omega_0$ is the central frequency and $\sigma$ the spectral width, one gets
\begin{equation}
    \Delta^2\hat\omega=\sigma^2.
\end{equation}
As expected, the precision increases with the spectral width of the photon, while the value of the central frequency $\omega_0$ does not play any role. Considering a single-photon Schrödinger cat state with spectrum
\begin{equation}
    F(\omega)=\frac{1}{\mathcal N}\left(\frac{1}{(2\pi \sigma^2)^{1/4}}e^{-\frac{(\omega-\omega_0-\Delta/2)^2}{4\sigma^2}}+\frac{1}{(2\pi \sigma^2)^{1/4}}e^{-\frac{(\omega-\omega_0+\Delta/2)^2}{4\sigma^2}}\right),
\end{equation}
where $\mathcal N$ is a normalization constant, $\omega_0$ the central frequency, $\sigma$ the spectral width of each peak and $\Delta$ the distance between the two peaks. Assuming that the two peaks are well separated ($\Delta\gg \sigma$), in which case $\mathcal N\simeq \sqrt{2}$, one gets
\begin{equation}
    \Delta^2\hat\omega=\sigma^2+\frac{\Delta^2}{4}.
\end{equation}
One sees that the variance is increased by the distance between the two peaks, which can be understood as the spectrum being more spread. This shows that using more complex states can help improve the precision of time-delay measurement.

If one uses a coherent state with spectrum $F$ and intensity $I$, one obtains the variance (see Appendix~\ref{app: TF basic computations}, Result~\ref{res: variance coherent states})
\begin{equation}
    \Delta^2\hat\omega=I\int \dd \omega\, \abs{F(\omega)}^2 \omega^2.
\end{equation}
Contrary to the single-photon case, the variance does not involve the square of the mean frequency. Beyond the boost in precision given by the intensity $I$, the mean frequency further enhances the precision. We thus recover the result from classical optical metrology where the precision is mainly set by the mean frequency of the light source.

\paragraph{\publi Single-mode rotation}
To metrologically study rotations, we need to compute the variance of $\hat{R}$. For a general single-photon state $\ket{\psi}=\int \dd\omega\, F(\omega)\ket{\omega}$, we can check that we have
\begin{align}\label{eq: variance of R single mode}
    \Delta^2(\hat{R})=\frac{1}{4}\int \dd\omega\,\abs{\omega^2 F-F^{(2)}}^2-\frac{1}{4}\left(\int \dd\omega \,[\omega^2\abs{F}^2-F^*F^{(2)}]\right)^2.
\end{align}
where the notation $F^{(j)}$ denotes the $j$-th derivative of the function $F$. First applying this to a time-frequency Gaussian state at central frequency $\omega_0$ and spectral width $\sigma$ yields
\begin{equation}\label{eq: Variance rotation gaussian state}
    \Delta^2(\hat{R})=\frac{1}{8}\left[\frac{1}{4\sigma^4}+4\sigma^4-2\right]+\sigma^2\omega_0^2.
\end{equation}
For a cat-like state whose peaks are at the frequencies $\omega_0-\Delta/2$ and $\omega_0+\Delta/2$, with spectral width $\sigma$, and still assuming that $\sigma\ll\Delta$ such that the two peaks are well separated, we obtain
\begin{equation}
    \operatorname{Var}_{\ket{\psi_C}}(\hat{R})=\frac{1}{8}\left[\frac{1}{4\sigma^4}+4\sigma^4-2\right]+\frac{1}{4}\Delta^2(\sigma^2+\omega_0^2)+\sigma^2\omega_0^2.
\end{equation}

\paragraph{\publi Computation methods}
Beyond temporal delay evolution for which variances are relatively easy to compute, other evolutions such as dispersion, time lensing and rotation can be more challenging. Expanding the expression of the variance, one needs to compute many expectation values in the form of products of powers of time and/or frequency operators
\begin{equation}
    \expval{\hat t^k \hat \omega^l \hat t^m \hat \omega^n}.
\end{equation}
These expectation values can be computed using different methods. A first method consists in working directly in the frequency or time domain, expanding the operators as derivatives and multiplications. While this method is straightforward, it can quickly become cumbersome when dealing with high-order operators. A second method consists in using the commutation relation between time and frequency operators to rearrange the operators in a convenient order, such that for example operators acting as derivatives are applied first. This reduces the number of expectation values that have to be computed at the cost of having to manage all the commutation relations. Using symbolic software such as Mathematica to automate this process can be very helpful.

\paragraph{\publi Multimode computation}
From a technical perspective, multimode situations are not more complicated than single-mode ones and similar methods allow the expression of the variance. However, the multimode nature of the states can lead to more interesting phenomena. While a more detailed study of multimode metrology is presented in Sec.~\ref{sec: TF entanglement and metrology} and \ref{sec: multimode generalization}, we briefly illustrate here the interest of multimode states for time-delay measurement.

First, in the case of independent photons, where the JSA factorizes as
\begin{equation}
    \ket{\psi}=\int \dd \omega_1 \dd \omega_2\, F_1(\omega_1)F_2(\omega_2)\ket{\omega_1,\omega_2},
\end{equation}
variances of local operators can simply be computed on the associated local state such as
\begin{equation}
    \Delta^2_{\ket{\psi}} \hat \omega_1=\Delta^2_{\ket{\psi_1}} \hat \omega_1,
\end{equation}
which simplifies the computations. Consider now the case of a two-mode single-photon state with a JSA that can be factored in the diagonal variables $\omega_\pm$
\begin{equation}
    \ket{\psi}=\int \dd \omega_+ \dd \omega_-\, F_+(\omega_+)F_-(\omega_-)\ket{\omega_1=\tfrac{\omega_++\omega_-}{\sqrt{2}},\omega_2=\tfrac{\omega_+-\omega_-}{\sqrt{2}}}.
\end{equation}
Assuming that the evolution is governed by a single delay in the first mode, with associated generator $\hat\omega_1$, by expressing
\begin{equation}
    \hat\omega_1=\frac{\hat\omega_++\hat\omega_-}{\sqrt{2}},
\end{equation}
and using the independence of the variables one gets
\begin{equation}
    \Delta^2_{\ket{\psi}} \hat \omega_1=\frac{1}{2}\left(\Delta^2_{\ket{\psi_+}} \hat \omega_+ +\Delta^2_{\ket{\psi_-}} \hat \omega_-\right),
\end{equation}
where we define the states $\ket{\psi_\pm}=\int \dd \omega_\pm\, F_\pm(\omega_\pm)\ket{\omega_\pm}$. We thus see that spectral profiles $F_+$ and $F_-$ both contribute to the variance, which can help improve the precision of the time-delay measurement. This simple example illustrates how multimode states can be used to enhance the precision of metrological protocols.

\subsection{Metrology and Wigner functions}
\label{subsec: Metrology and Wigner functions}
To conclude this chapter, we discuss the relation between Wigner functions and quantum metrology. Since Wigner functions provide a clear geometrical picture of time-frequency states and their evolution, it is natural to ask whether this phase-space representation can help in understanding and interpreting metrological protocols.

\paragraph{\lit Principle}
As already introduced, quantum metrology aims at estimating a parameter $\theta$ encoded in a quantum state
\begin{equation}
    \ket{\psi(\theta)} = e^{-i\hat H\theta}\ket{\psi}.
\end{equation}
In order to extract information about $\theta$, one must be able to distinguish the states $\ket{\psi(\theta)}$ and $\ket{\psi(\theta')}$ for two close values $\theta$ and $\theta'$. For optimal precision, these two states should be as distinguishable as possible, \ie, their overlap
\begin{equation}
    \abs{\braket{\psi(\theta)}{\psi(\theta')}}^2,
\end{equation}
should decrease rapidly as $\abs{\theta-\theta'}$ increases from zero. This means that the achievable precision is directly tied to how fast this overlap goes to zero. From a geometric point of view, this connects quantum metrology to the geometry of the state space: the quantum Fisher information is directly related to the Bures metric, which for pure states is fully determined by the overlap between neighboring states~\cite{braunstein_statistical_1994}.

Since the overlap between two states can be computed using their Wigner functions (see Sec.~\ref{subsec: TF Wigner function}), this representation naturally provides a geometric picture of quantum metrology. Moreover, as discussed previously, the action of the operators $\hat{\omega}$, $\hat{t}$, $\hat{\omega}^2$, $\hat{t}^2$, or $\hat{R}$ corresponds respectively to translations, shears, or rotations in phase space. As a consequence, the measurement precision is directly related to how much the Wigner function must be transformed in order to become almost orthogonal to its initial, non-transformed version.

As a first example, consider a time-frequency Gaussian single-photon state evolving under the generator $\hat{\omega}$, which corresponds to a translation along the $\tau$ direction in phase space. In order to obtain two quasi-orthogonal Wigner distributions, the state must be displaced by a distance of order $1/2\sigma$. Indeed, the thickness of the Wigner function of a time-frequency Gaussian state along the $\tau$ direction is $1/2\sigma$ (see Eq.~\eqref{eq: wigner function gaussian}), meaning that a displacement of this order is required to significantly reduce the overlap. Since the estimation precision is inversely proportional to the squared distance between the two states, we expect the QFI to scale as
\begin{equation}
    \mathcal{Q}_{\mathrm{est}} \sim 4\sigma^2.
\end{equation}
This heuristic estimate perfectly matches the exact result, since in this case $\mathcal Q=4\Delta^2\hat{\omega} = 4\sigma^2$.

Let us now consider the same evolution, but starting from a single-photon cat-like state. In this case, a displacement of length $\pi/\Delta$ is already sufficient to obtain quasi-orthogonal states. This distance corresponds to half the period of the oscillations in the interference pattern of the Wigner function. When computing the overlap between the two Wigner functions, the interference fringes contribute strongly and negatively, leading to a small overall scalar product. This immediately suggests that metrology performed with a cat-like state can achieve better accuracy than with a time-frequency Gaussian state, and that this accuracy should improve as the separation $\omega_1-\omega_2$ increases. From this geometrical argument, we thus expect the QFI to scale as
\begin{equation}
    \mathcal{Q}_{\mathrm{est}} \sim \frac{\Delta^2}{\pi^2}.
\end{equation}
This estimate should be compared with the exact result
\begin{equation}
    \mathcal Q = 4\sigma^2 + \Delta^2,
\end{equation}
which, although differing by a numerical factor, correctly captures the order of magnitude in the regime $\sigma \ll \Delta$.

All these considerations show that simple geometrical reasoning in phase space allows one to estimate the scaling of the quantum Fisher information~\cite{toscano_sub-planck_2006, zurek_sub-planck_2001}. For the quadrature Wigner function, fine structures whose size is smaller than the Planck constant $\hbar$ are known to play a crucial role in quantum metrology~\cite{praxmeyer_time-frequency_2007, austin_measuring_2010}. These ``sub-Planck'' structures admit analogues in the time-frequency case, which also play a crucial role in enhancing metrological precision.

\paragraph{\publi Example for rotation}
In a previous section, we computed the quantum Fisher information $4\Delta^2\hat R$ for both time-frequency Gaussian and cat-like single-photon states. Using the Wigner function picture, we can now give a clear geometrical interpretation of the different contributions. For a time-frequency Gaussian state, we found
\begin{equation}
    \Delta^2\hat{R} = \frac{1}{8}\left[\frac{1}{4\sigma^4} + 4\sigma^4 - 2\right] + \sigma^2\omega_0^2.
\end{equation}
Two distinct contributions can be identified:
\begin{itemize}
    \item The term $\sigma^2\omega_0^2$ corresponds to the distance from the center of rotation multiplied by the typical size of the state in the direction of rotation. Geometrically, this term is very intuitive. In the Wigner picture, the state must be rotated by an angle $\theta \sim 1/2\sigma\omega_0$ for the overlap between the two Wigner functions to become small. This situation is illustrated in Fig.~\ref{fig: rotation states} (a).
    \item The term $\frac{1}{4\sigma^4} + 4\sigma^4 - 2$ reflects the intrinsic anisotropy of the state. This expression reaches its minimum value $0$ for $\sigma = 1/\sqrt{2}$, corresponding to a perfectly rotationally symmetric Wigner function. Setting $\omega_0 = 0$, the rotation is performed around the center of the state. If the state is fully symmetric, such a rotation has no effect, and the variance vanishes. A non-zero contribution therefore requires an asymmetric state. This interpretation is illustrated in Fig.~\ref{fig: rotation states} (b).
\end{itemize}

For a cat-like state with well-separated peaks, the variance reads
\begin{equation}
    \Delta^2\hat R = \frac{1}{8}\left[\frac{1}{4\sigma^4} + 4\sigma^4 - 2\right] + \frac{1}{4}\Delta^2(\sigma^2 + \omega_0^2) + \sigma^2\omega_0^2.
\end{equation}
Setting $\omega_0 = 0$ yields the simpler expression
\begin{equation}
    \Delta^2\hat R = \frac{1}{8}\left[\frac{1}{4\sigma^4} + 4\sigma^4 - 2\right] + \frac{1}{4}\Delta^2\sigma^2.
\end{equation}
In this case, there is no metrological gain compared to the time-frequency Gaussian state, since $\Delta/2$ plays a role analogous to $\omega_0$. This can again be understood geometrically. As shown in Fig.~\ref{fig: rotation states} (c), the interference pattern is rotated around its center, and although it contains fine sub-Planck structures, these are displaced only by a small amount, resulting in no significant enhancement.

For non-zero $\omega_0$, two additional terms appear: $\sigma^2\omega_0^2$, already present in the time-frequency Gaussian case, and the new term $\Delta^2\omega_0^2/4$, which has no equivalent in the previous situations and leads to a strong metrological enhancement. This contribution can be understood using Fig.~\ref{fig: rotation states} (d). In this configuration, the interference fringes are displaced by an amount proportional to $\omega_0\theta$. Since the fringe spacing is of order $\Delta$, distinguishability requires $\theta \sim 1/\omega_0\Delta$, which directly corresponds to the term $\Delta^2\omega_0^2/4$ in the variance.

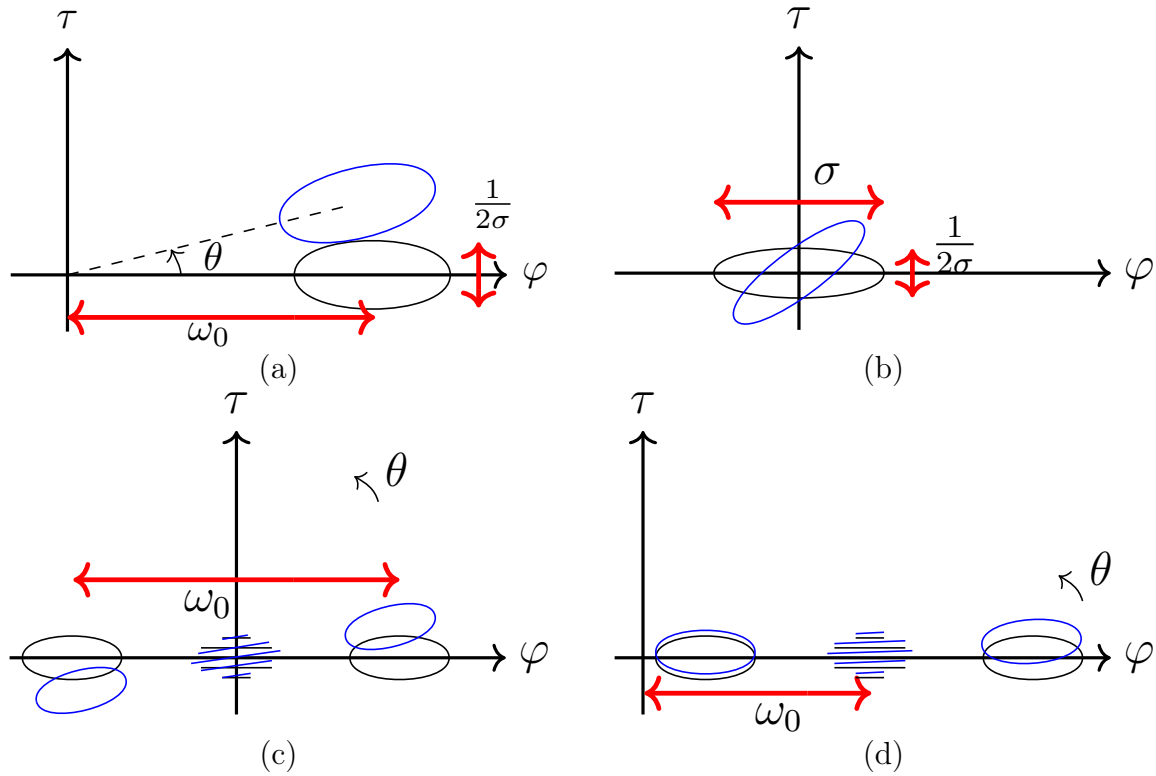
\begin{figure}[ht]
    \centering
    \begin{tabular}{cc}
        \footnotesize
        \scalebox{1.5}{\begin{tikzpicture}
	\begin{pgfonlayer}{nodelayer}
		\node [style=none] (0) at (0, -1) {};
		\node [style=none] (1) at (-1, 0) {};
		\node [style=none] (2) at (0, 4) {};
		\node [style=none] (3) at (7.75, 0) {};
		\node [style=none] (4) at (4, 0) {};
		\node [style=none] (5) at (6.75, 0) {};
		\node [style=none] (6) at (3.75, 0.925) {};
		\node [style=none] (7) at (6.475, 1.6) {};
		\node [style=none] (8) at (8.25, 0) {$\varphi$};
		\node [style=none] (9) at (0, 4.5) {$\tau$};
		\node [style=none] (10) at (4.975, 1.225) {};
		\node [style=none] (11) at (0, 0) {};
		\node [style=none] (12) at (2, 0) {};
		\node [style=none] (13) at (1.825, 0.45) {};
		\node [style=none] (14) at (2.575, 0.325) {$\theta$};
		\node [style=none] (15) at (0, -0.75) {};
		\node [style=none] (16) at (4, -0.75) {};
		\node [style=none] (17) at (5.425, -0.75) {};
		\node [style=none] (18) at (7.25, 0.575) {};
		\node [style=none] (19) at (7.25, -0.25) {};
		\node [style=none] (20) at (7.25, -0.575) {};
		\node [style=none] (21) at (2.5, -1.05) {$\omega_0$};
		\node [style=none] (22) at (7.5, 1.25) {$\frac{1}{2\sigma}$};
	\end{pgfonlayer}
	\begin{pgfonlayer}{edgelayer}
		\draw [style=Thick arrow] (1.center) to (3.center);
		\draw [style=Thick arrow] (0.center) to (2.center);
		\draw [bend right=90, looseness=0.75] (4.center) to (5.center);
		\draw [bend left=90, looseness=0.75] (4.center) to (5.center);
		\draw [style=Blue line] (7.center)
			 to [bend left=90, looseness=0.75] (6.center)
			 to [bend left=90, looseness=0.75] cycle;
		\draw [style=dashes] (11.center) to (10.center);
		\draw [style=Arrow, in=-60, out=90, looseness=0.75] (12.center) to (13.center);
		\draw [style=arrow] (16.center) to (15.center);
		\draw [style=arrow] (16.center) to (17.center);
		\draw [style=arrow] (19.center) to (18.center);
		\draw [style=arrow] (19.center) to (20.center);
	\end{pgfonlayer}
\end{tikzpicture}} & \scalebox{1.5}{\begin{tikzpicture}
	\begin{pgfonlayer}{nodelayer}
		\node [style=none] (0) at (0, -1) {};
		\node [style=none] (1) at (-3.25, 0) {};
		\node [style=none] (2) at (0, 4) {};
		\node [style=none] (3) at (5.5, 0) {};
		\node [style=none] (4) at (-1.5, 0) {};
		\node [style=none] (5) at (1.5, 0) {};
		\node [style=none] (6) at (-1.125, -0.825) {};
		\node [style=none] (7) at (1.125, 0.85) {};
		\node [style=none] (8) at (6, 0) {$\varphi$};
		\node [style=none] (9) at (0, 4.5) {$\tau$};
		\node [style=none] (18) at (2, 0.425) {};
		\node [style=none] (19) at (2, 0) {};
		\node [style=none] (20) at (2, -0.425) {};
		\node [style=none] (22) at (2.75, 0.5) {$\frac{1}{2\sigma}$};
		\node [style=none] (23) at (-1.5, 1.25) {};
		\node [style=none] (24) at (0, 1.25) {};
		\node [style=none] (25) at (1.5, 1.25) {};
		\node [style=none] (26) at (0.5, 1.75) {$\sigma$};
	\end{pgfonlayer}
	\begin{pgfonlayer}{edgelayer}
		\draw [style=Thick arrow] (1.center) to (3.center);
		\draw [style=Thick arrow] (0.center) to (2.center);
		\draw [bend right=90, looseness=0.50] (4.center) to (5.center);
		\draw [bend left=90, looseness=0.50] (4.center) to (5.center);
		\draw [style=Blue line] (7.center)
			 to [bend left=90, looseness=0.50] (6.center)
			 to [bend left=90, looseness=0.50] cycle;
		\draw [style=arrow] (19.center) to (18.center);
		\draw [style=arrow] (19.center) to (20.center);
		\draw [style=arrow] (24.center) to (25.center);
		\draw [style=arrow] (24.center) to (23.center);
	\end{pgfonlayer}
\end{tikzpicture}}\\
        (a) & (b)\\
        \scalebox{1.5}{\begin{tikzpicture}
	\begin{pgfonlayer}{nodelayer}
		\node [style=none] (0) at (0, -1) {};
		\node [style=none] (1) at (-4, 0) {};
		\node [style=none] (2) at (0, 4) {};
		\node [style=none] (3) at (4.75, 0) {};
		\node [style=none] (4) at (2, 0) {};
		\node [style=none] (5) at (3.75, 0) {};
		\node [style=none] (6) at (1.925, 0.35) {};
		\node [style=none] (7) at (3.5, 0.75) {};
		\node [style=none] (8) at (5.25, 0) {$\varphi$};
		\node [style=none] (9) at (0, 4.5) {$\tau$};
		\node [style=none] (12) at (2.5, 2.75) {};
		\node [style=none] (13) at (2.075, 3.2) {};
		\node [style=none] (14) at (2.825, 3.325) {$\theta$};
		\node [style=none] (15) at (-2.875, 1.375) {};
		\node [style=none] (16) at (1, 1.375) {};
		\node [style=none] (17) at (2.875, 1.375) {};
		\node [style=none] (21) at (-0.5, 0.95) {$\omega_0$};
		\node [style=none] (22) at (-3.775, 0) {};
		\node [style=none] (23) at (-2.025, 0) {};
		\node [style=none] (24) at (-1.95, -0.375) {};
		\node [style=none] (25) at (-3.525, -0.775) {};
		\node [style=none] (27) at (0.75, 0) {};
		\node [style=none] (28) at (-0.75, 0) {};
		\node [style=none] (29) at (-0.625, -0.175) {};
		\node [style=none] (30) at (0.625, -0.175) {};
		\node [style=none] (31) at (-0.625, 0.175) {};
		\node [style=none] (32) at (0.625, 0.175) {};
		\node [style=none] (33) at (-0.25, 0.35) {};
		\node [style=none] (34) at (0.25, 0.35) {};
		\node [style=none] (35) at (-0.25, -0.35) {};
		\node [style=none] (36) at (0.25, -0.35) {};
		\node [style=none] (37) at (0.775, 0.125) {};
		\node [style=none] (38) at (-0.8, -0.1) {};
		\node [style=none] (39) at (-0.625, -0.225) {};
		\node [style=none] (40) at (0.625, -0.05) {};
		\node [style=none] (41) at (-0.675, 0.075) {};
		\node [style=none] (42) at (0.575, 0.275) {};
		\node [style=none] (43) at (-0.25, 0.325) {};
		\node [style=none] (44) at (0.2, 0.4) {};
		\node [style=none] (45) at (-0.25, -0.35) {};
		\node [style=none] (46) at (0.25, -0.275) {};
	\end{pgfonlayer}
	\begin{pgfonlayer}{edgelayer}
		\draw [style=Thick arrow] (1.center) to (3.center);
		\draw [style=Thick arrow] (0.center) to (2.center);
		\draw [bend right=90, looseness=0.75] (4.center) to (5.center);
		\draw [bend left=90, looseness=0.75] (4.center) to (5.center);
		\draw [style=Blue line] (7.center)
			 to [bend left=90, looseness=0.75] (6.center)
			 to [bend left=90, looseness=0.75] cycle;
		\draw [style=Arrow, in=-30, out=105] (12.center) to (13.center);
		\draw [style=arrow] (16.center) to (15.center);
		\draw [style=arrow] (16.center) to (17.center);
		\draw [bend right=90, looseness=0.75] (22.center) to (23.center);
		\draw [bend left=90, looseness=0.75] (22.center) to (23.center);
		\draw [style=Blue line, bend left=90, looseness=0.75] (25.center) to (24.center);
		\draw [style=Blue line, bend left=90, looseness=0.75] (24.center) to (25.center);
		\draw (28.center) to (27.center);
		\draw (29.center) to (30.center);
		\draw (33.center) to (34.center);
		\draw (31.center) to (32.center);
		\draw (35.center) to (36.center);
		\draw [style=Blue line] (38.center) to (37.center);
		\draw [style=Blue line] (39.center) to (40.center);
		\draw [style=Blue line] (43.center) to (44.center);
		\draw [style=Blue line] (41.center) to (42.center);
		\draw [style=Blue line] (45.center) to (46.center);
	\end{pgfonlayer}
\end{tikzpicture}} & \scalebox{1.5}{\begin{tikzpicture}
	\begin{pgfonlayer}{nodelayer}
		\node [style=none] (0) at (-3.5, -1) {};
		\node [style=none] (1) at (-4, 0) {};
		\node [style=none] (2) at (-3.5, 4) {};
		\node [style=none] (3) at (4.75, 0) {};
		\node [style=none] (4) at (2.5, 0) {};
		\node [style=none] (5) at (4.25, 0) {};
		\node [style=none] (6) at (2.475, 0.225) {};
		\node [style=none] (7) at (4.225, 0.35) {};
		\node [style=none] (8) at (5.25, 0) {$\varphi$};
		\node [style=none] (9) at (-3.5, 4.5) {$\tau$};
		\node [style=none] (10) at (4.25, 1) {};
		\node [style=none] (11) at (3.825, 1.45) {};
		\node [style=none] (12) at (4.575, 1.575) {$\theta$};
		\node [style=none] (13) at (-3.475, -0.625) {};
		\node [style=none] (14) at (-0.6, -0.625) {};
		\node [style=none] (15) at (0.525, -0.625) {};
		\node [style=none] (16) at (-1.1, -1.05) {$\omega_0$};
		\node [style=none] (17) at (-3.275, 0) {};
		\node [style=none] (18) at (-1.525, 0) {};
		\node [style=none] (19) at (-1.525, 0.1) {};
		\node [style=none] (20) at (-3.275, 0.1) {};
		\node [style=none] (21) at (1.25, 0) {};
		\node [style=none] (22) at (-0.25, 0) {};
		\node [style=none] (23) at (-0.125, -0.175) {};
		\node [style=none] (24) at (1.125, -0.175) {};
		\node [style=none] (25) at (-0.125, 0.175) {};
		\node [style=none] (26) at (1.125, 0.175) {};
		\node [style=none] (27) at (0.25, 0.35) {};
		\node [style=none] (28) at (0.75, 0.35) {};
		\node [style=none] (29) at (0.25, -0.35) {};
		\node [style=none] (30) at (0.75, -0.35) {};
		\node [style=none] (31) at (1.25, 0.125) {};
		\node [style=none] (32) at (-0.25, 0.075) {};
		\node [style=none] (33) at (-0.125, -0.1) {};
		\node [style=none] (34) at (1.125, -0.05) {};
		\node [style=none] (35) at (-0.125, 0.25) {};
		\node [style=none] (36) at (1.125, 0.3) {};
		\node [style=none] (37) at (0.25, 0.425) {};
		\node [style=none] (38) at (0.75, 0.45) {};
		\node [style=none] (39) at (0.25, -0.275) {};
		\node [style=none] (40) at (0.75, -0.25) {};
	\end{pgfonlayer}
	\begin{pgfonlayer}{edgelayer}
		\draw [style=Thick arrow] (1.center) to (3.center);
		\draw [style=Thick arrow] (0.center) to (2.center);
		\draw [bend right=90, looseness=0.75] (4.center) to (5.center);
		\draw [bend left=90, looseness=0.75] (4.center) to (5.center);
		\draw [style=Blue line, bend left=90, looseness=0.75] (7.center) to (6.center);
		\draw [style=Blue line, bend left=90, looseness=0.75] (6.center) to (7.center);
		\draw [style=Arrow, in=-30, out=105] (10.center) to (11.center);
		\draw [style=arrow] (14.center) to (13.center);
		\draw [style=arrow] (14.center) to (15.center);
		\draw [bend right=90, looseness=0.75] (17.center) to (18.center);
		\draw [bend left=90, looseness=0.75] (17.center) to (18.center);
		\draw [style=Blue line, bend left=90, looseness=0.75] (20.center) to (19.center);
		\draw [style=Blue line, bend left=90, looseness=0.75] (19.center) to (20.center);
		\draw (22.center) to (21.center);
		\draw (23.center) to (24.center);
		\draw (27.center) to (28.center);
		\draw (25.center) to (26.center);
		\draw (29.center) to (30.center);
		\draw [style=Blue line] (32.center) to (31.center);
		\draw [style=Blue line] (33.center) to (34.center);
		\draw [style=Blue line] (37.center) to (38.center);
		\draw [style=Blue line] (35.center) to (36.center);
		\draw [style=Blue line] (39.center) to (40.center);
	\end{pgfonlayer}
\end{tikzpicture}}\\
        (c) & (d)
    \end{tabular}
    \caption[Representation of the Wigner functions of various states under rotation]{Schematic representation of the Wigner functions of various states under rotation. The ellipses represent the typical widths of time-frequency Gaussian states. The black lines indicate the initial state and the blue lines indicate the rotated states. (a) Time-frequency Gaussian state centered at $\omega_0$. Distinguishability requires $\theta\omega_0 \sim 1/2\sigma$. (b) Time-frequency Gaussian state centered at the origin. The rotated state becomes distinguishable only if it is not rotationally symmetric. (c) Time-frequency cat state centered at the origin. The interference fringes are weakly displaced and provide little metrological gain. (d) Time-frequency cat state centered at $\omega_0$. The fringes play a crucial role, as distinguishability requires $\theta\omega_0 \sim 1/\Delta$. Adapted from Ref.~\cite{descamps_time-frequency_2023}. \copyright 2023 American Physical Society. }
    \label{fig: rotation states}
\end{figure}

\fi

\ifnum \theShowChapthree=1
\chapter{Collective-variables entanglement}
\setcurrentanchor{chap3}
\label{chap: collective entanglement}
\emph{In this chapter we explore the concept of collective-variables entanglement. Starting from observations on collective metrological properties of single-photon systems, which reveal the role of frequency entanglement for precision enhancement, we then develop a general framework describing single-mode entanglement. Finally, we show how such entanglement can be used to provide more robust error-correcting codes.}

\localtableofcontents

\section{Time-frequency entanglement and metrology}
\label{sec: TF entanglement and metrology}
\emph{This section explores the role of frequency as a resource for quantum metrology. We demonstrate that frequency can play a role beyond a mere classical degree of freedom, by enabling collective enhancements in metrological protocols. This section is mainly based on \hyperlink{Article: prl metro}{Quantum Metrology Using Time-Frequency as Quantum Continuous Variables: Resources, Sub-Shot-Noise Precision and Phase Space Representation}~\cite{descamps_quantum_2023} which was published during the PhD thesis.}

\subsection{Motivations}
\label{subsec: motivations collective metrology}

As discussed in previous chapters, the wave nature of radiation makes it a natural system for high-precision time measurements. Using interferometric techniques, the achievable time resolution is set by the inverse of the optical frequency for monochromatic fields. Leaving aside systematic errors, classical power noise generally limits the precision. This noise can be reduced down to the standard quantum limit (SQL)~\cite{ligo_scientific_collaboration_and_virgo_collaboration_observation_2016}, also known as shot noise, which scales as $1/\sqrt{\langle \hat n \rangle}$, where $\langle \hat n \rangle$ is the average photon number, or equivalently the field intensity. In this regime, photons behave as independent probes, a feature that follows from the Poissonian photon-number statistics of coherent (quasi-classical) states.

By fully exploiting quantum resources, the SQL can be quadratically improved, leading to Heisenberg-limited scaling~\cite{giovannetti_quantum_2006}. In quantum optics, sub-shot-noise precision can be achieved using non-classical field statistics, such as those associated with squeezed states~\cite{pinel_ultimate_2012, pezze_mach-zehnder_2008, zhuang_distributed_2020, ligo_scientific_collaboration_and_virgo_collaboration_quantum-enhanced_2019}, NOON states~\cite{dambrosio_photonic_2013, barbieri_optical_2022, durkin_local_2007}, and Schrödinger cat-like states~\cite{penasa_measurement_2016, zurek_sub-planck_2001, toscano_sub-planck_2006, dalvit_quantum_2006}. Quantum enhancement can also arise from non-local evolutions in multimode states~\cite{benatti_sub-shot_2011}.

In the context of time-precision measurements, two main factors limiting the achievable precision are often distinguished: photon-number statistics, related to the particle nature of light, and modal properties, associated with the wave character of the field~\cite{dittel_wave-particle_2021}. The latter are frequently treated as a classical resource. However, these two aspects of radiation are not generally independent, especially for intrinsically multimode and non-Gaussian states. For instance, it has been shown that frequency-entangled and frequency-squeezed states can lead to quantum-enhanced clock synchronization and position measurements~\cite{fabre_modes_2020, giovannetti_quantum-enhanced_2001}. Despite this progress, providing a clear and unified picture of the interplay between particle and modal aspects of radiation in quantum metrology and quantum optics remains an open problem, despite its fundamental and practical relevance.

This section aims to clarify this interplay by analyzing the ultimate precision limits of time-delay estimation using time-frequency entangled states. Adopting a phase-space perspective, we also unveil the underlying time-frequency phase-space structure associated with quantum precision bounds and introduce a definition of classical resource that applies equally to modal and particle degrees of freedom of the quantum field. This geometrical framework provides a unified description of how precision scaling at fixed resources depends on modal and particle entanglement, or more generally on collective photonic behavior.

To further develop our analysis, we mainly study the single-photon time-frequency systems introduced in the previous Chap.~\ref{chap: Time-Frequency Systems} and demonstrate how quantum-metrological enhancement can arise in such systems. This section will primarily focus on the value of the quantum Fisher information, which for pure states is proportional to the variance of the evolution generator. As such, the main figure of merit will simply be the variance of the generator.

\subsection{Single-mode precision}
\label{subsec: single mode precision}

\paragraph{\lit Monochromatic case}
To provide a fully self-contained discussion, we first consider the simple case of a single monochromatic optical mode of frequency $\omega_0$ propagating in free space. In this setting, quantum states are described by a single pair of creation and annihilation operators $\hat a^\dagger$ and $\hat a$, and a generic pure state can be expanded as
\begin{equation}
    \ket{\psi}=\sum_{n=0}^\infty c_n \ket{n},
\end{equation}
where $\ket{n}=(\hat a^\dagger)^n/\sqrt{n!}\ket{0}$ denotes the $n$-photon Fock state. The free evolution $e^{-i\hat H t}$ is generated by the Hamiltonian $\hat H=\omega_0 \hat a^\dagger \hat a$ (setting $\hbar=1$). The precision for time-delay estimation is then governed by
\begin{equation}
    \Delta^2 \hat H = \omega_0^2 \Delta^2 \hat n,
\end{equation}
where $\hat n=\hat a^\dagger \hat a$ is the photon number operator. A first important observation is that the precision scales linearly with the frequency $\omega_0$, which therefore plays the role of a classical scaling parameter fixing the units of the measurement.

Maximizing the precision thus amounts to maximizing the photon number variance $\Delta^2 \hat n$. Without further assumptions on the quantum state, this variance can be made arbitrarily large, leading to an unbounded QFI. In practice, however, experimental constraints impose limitations on the average photon number, or equivalently on the mean energy $\langle \hat n \rangle$. This naturally raises the question of whether $\Delta^2 \hat n$ can be bounded as a function of $\langle \hat n \rangle$.

A simple pathological counterexample shows that this is not possible in full generality. Consider the family of states
\begin{equation}
    \ket{\psi_N}=\sqrt{1-\frac{\overline n}{N}}\ket{0}+\sqrt{\frac{\overline n}{N}}\ket{N},
\end{equation}
which has a fixed average photon number $\expval{\hat n}=\overline n$, while the variance reads $\Delta^2 \hat n = \overline n (N-\overline n)$. Since $N$ can be taken arbitrarily large, the variance can diverge while keeping $\overline n$ fixed. As a consequence, no general bound of the form
\begin{equation}
    \Delta^2 \hat n \leq f(\langle \hat n \rangle)
\end{equation}
can be established without additional assumptions.

The state $\ket{\psi_N}$ is clearly unphysical, as it involves coherent superpositions of the vacuum and arbitrarily large photon-number states. More generally, in continuous-variable quantum metrology, states with highly fluctuating photon-number distributions can formally lead to sensitivities that appear to surpass the usual Heisenberg scaling~\cite{zwierz_general_2010}.

One convenient way to exclude such pathological situations is to consider families of states whose photon-number moments remain sufficiently well behaved. A family of states ${\ket{\psi_\lambda}}_{\lambda\geq 0}$ is said to have no thick tails if
\begin{equation}
    \expval{\hat n^2}=O(\expval{\hat n}^2),
\end{equation}
as $\lambda\to\infty$. Under this condition,
\begin{equation}
    \Delta^2\hat n=O(\expval{\hat n}^2),
\end{equation}
so that the QFI scales at most quadratically with the average photon number, yielding the conventional Heisenberg scaling. Similar conclusions can also be obtained by imposing, for instance, a hard cutoff on the total photon number.

It is worth noting that states exhibiting apparent sub-Heisenberg scalings do not generally constitute a violation of the Heisenberg limit. As discussed in Ref.~\cite{giovannetti_sub-heisenberg_2012}, such enhancements are typically associated with increasingly restrictive assumptions on the prior knowledge of the parameter to be estimated. This highlights the fact that the identification of meaningful resources and precision scalings in continuous-variable quantum metrology requires some care. These issues will be discussed further in Sec.~\ref{sec: SSRC and metrology}.

\paragraph{\publi Single-mode time-frequency case}
We now turn to the time-frequency description of a single optical mode. In this case, the precision of time-delay estimation is governed by the variance of the frequency operator $\hat \omega$, and the corresponding QFI reads
\begin{equation}
    \mathcal Q = 4 \Delta^2 \hat \omega.
\end{equation}

We consider a quantum state confined to a single time-frequency mode with spectral amplitude $F(\omega)$. Such a mode is described by the creation operator
\begin{equation}
    \hat a_F^\dagger = \int \dd \omega\, F(\omega)\, \hat a^\dagger(\omega),
\end{equation}
and a general pure state can be written as
\begin{equation}
    \ket{\psi}=\sum_{m=0}^\infty c_m \frac{(\hat a_F^\dagger)^m}{\sqrt{m!}}\vac.
\end{equation}

In this setting, the variance of the frequency operator takes the form
\begin{equation}
    \label{eq: single mode variance}
    \Delta^2 \hat \omega = \overline n\, \Delta^2 \omega + \overline \omega^2 \Delta^2 n,
\end{equation}
where $\overline n$ and $\Delta^2 n$ denote respectively the mean photon number and its variance,
\begin{align}
    \overline n &= \sum_{n=0}^\infty n \abs{c_n}^2, &
    \Delta^2 n &= \sum_{n=0}^\infty (n-\overline n)^2 \abs{c_n}^2,
\end{align}
while $\overline \omega$ and $\Delta^2 \omega$ are the mean and variance of the spectral distribution $\abs{F(\omega)}^2$,
\begin{align}
    \overline \omega &= \int \dd \omega\, \abs{F(\omega)}^2 \omega, &
    \Delta^2 \omega &= \int \dd \omega\, \abs{F(\omega)}^2 (\omega-\overline\omega)^2.
\end{align}
A proof of Eq.~\eqref{eq: single mode variance} is given in Appendix~\ref{app: collective var derivation}, Result~\ref{res: single mode variance}. Contrary to the original derivation in~\cite{descamps_quantum_2023}, the present one does not rely on a narrow-band approximation.

Equation~\eqref{eq: single mode variance} clearly separates two distinct contributions to the QFI. The first term originates from spectral fluctuations and scales only linearly with the average photon number. It is therefore often interpreted as stemming from a classical resource. The second term arises from photon number fluctuations and is associated with genuinely quantum metrological advantages, extensively studied in the literature~\cite{giovannetti_quantum-enhanced_2004}, and closely related to the phase-space structure of the field.  

This expression can be readily applied to several relevant limiting cases. In the monochromatic limit $\Delta^2 \omega=0$, one recovers the result discussed previously: only photon number fluctuations contribute, and the carrier frequency acts merely as a scaling factor. At the opposite extreme, single-photon states satisfy $\Delta^2 n=0$, so that the precision is entirely determined by the spectral width, as already discussed in Section~\ref{subsec: relation to experiment}.

Finally, consider a coherent state of intensity $I$ and spectral amplitude $F(\omega)$ in a single spatial mode, as introduced in Section~\ref{subsec: TF coherent states}. In this case, as coherent states have Poissonian statistics, $\overline{n}=\Delta^2 n=I$, Eq.~\eqref{eq: single mode variance} reduces to
\begin{equation}
    \Delta^2 \hat \omega = I \int \omega^2\, \abs{F(\omega)}^2 \dd \omega = I \overline{\omega^2}.
\end{equation}
This result admits several complementary interpretations. First, the variance scales linearly with the intensity $I$, corresponding to the expected shot-noise limit. Second, it does not depend on the total mean energy $I \overline \omega$, often regarded as the relevant classical resource, but rather on the second spectral moment $\overline{\omega^2}$. Consequently, for a fixed total energy, one can freely engineer the spectral distribution to define different time-resolution scales through $\Delta^2 \omega$, while preserving shot-noise scaling, as demonstrated for instance in~\cite{praxmeyer_time-frequency_2007,austin_measuring_2010,fabre_parameter_2021}. This observation suggests that the classical resource relevant for time estimation should not be identified solely with the field energy, but must also include its spectral properties.

\subsection{Multimode case}
\label{subsec: multimode case}

\paragraph{\publi Setting}
To further develop the understanding of the interplay between modal and particle properties of light, we now consider the multimode case. We assume that the same temporal delay is applied to each spatial mode. The corresponding free-evolution Hamiltonian then reads\footnote{Alternate expressions with signs, such as $\hat \Omega = \sum_{j=1}^n \alpha_j \hat \omega_j$, can also be considered, corresponding to different physical situations. The following discussion remains valid in these cases. Further discussion on the choice of the coefficients $\alpha_j$ can be found in Sec.~\ref{subsec: remark def coll op}.}
\begin{equation}
    \hat \Omega = \sum_{j=1}^n \hat \omega_j.
\end{equation}

The analysis of the variance of $\hat \Omega$ in full generality is too involved for our present purposes. However, the single-photon regime already reveals key features concerning the role of frequency entanglement in quantum metrology. As introduced in the previous chapter, we consider $n$ photons, each occupying a different ancillary mode (for instance a spatial mode), and described by a joint frequency wavefunction. In this situation, frequency and intensity properties are no longer independent, and the frequency variance can be exploited to modify the scaling of the QFI even when the photon number variance vanishes.

\paragraph{\publi Separable states}
If the single photons are prepared in a separable state
\begin{equation}
    \ket{\psi}=\int\dd \omega_1\cdots\dd\omega_n\, F_1(\omega_1)\cdots F_n(\omega_n)\ket{\omega_1,\dots,\omega_n},
\end{equation}
where $F_j(\omega)$ denotes the spectrum of the photon in the $j$-th spatial mode, one finds
\begin{equation}
    \Delta^2 \hat \Omega = \sum_{j=1}^n \Delta^2 \omega_j,
\end{equation}
where
\begin{equation}
    \Delta^2 \omega_j
    = \int \omega^2 \abs{F_j(\omega)}^2\, \dd \omega
    - \left(\int \omega \abs{F_j(\omega)}^2\, \dd \omega\right)^2.
\end{equation}
For simplicity, we assume that all photons have the same frequency variance $\Delta^2 \omega$, and that the state is pure.\footnote{A discussion of mixed states is provided in Sec.~\ref{subsec: extension to mixed states}.} One then obtains
\begin{equation}
    \Delta^2 \hat \Omega = n \Delta^2 \omega,
\end{equation}
which corresponds to the shot-noise scaling, as expected for $n$ independent probes. By comparison with the coherent-state case, one can identify $n \Delta^2 \omega = I \overline{\omega^2}$. Both expressions scale linearly with the photon number. This reflects the fact that a coherent state represents the same metrological resource as $n$ independent photons~\cite{giovannetti_quantum-enhanced_2001, jacobson_photonic_1995}. It is nevertheless worth emphasizing that, while for coherent states this scaling originates from the photon number fluctuations $\Delta^2 \hat n = I$, one has $\Delta^2 \hat n = 0$ for single-photon states. In Fig. \ref{fig: JSI eseparable} (a) and (b) we show the Joint Spectral Intensity (JSI) of separable states of two independent (separable) photons ($n=2$).

\begin{figure}[ht]
    \centering
    \begin{tabular}{cc}
        \includegraphics[width=0.4\linewidth]{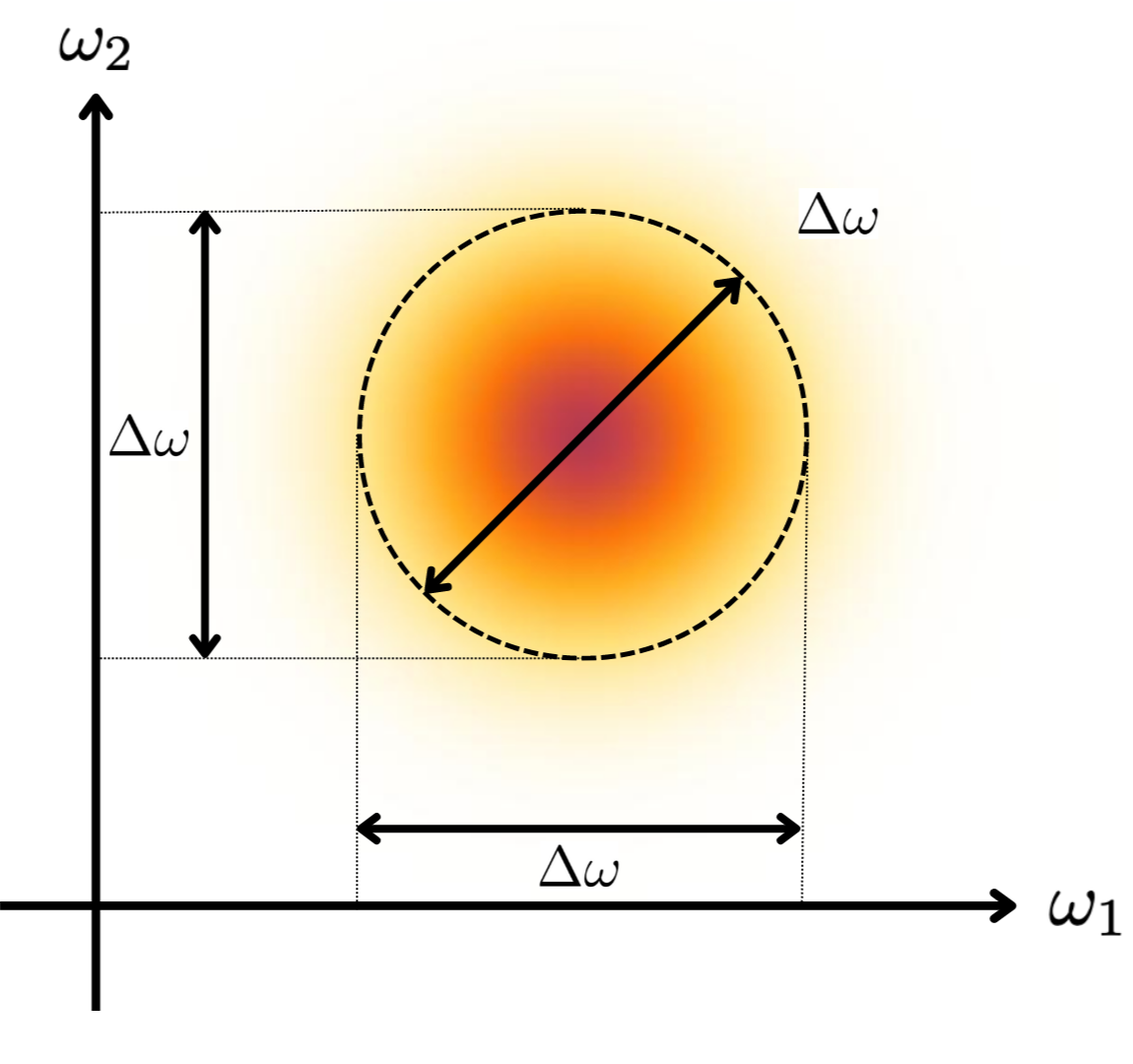} & \includegraphics[width=0.4\linewidth]{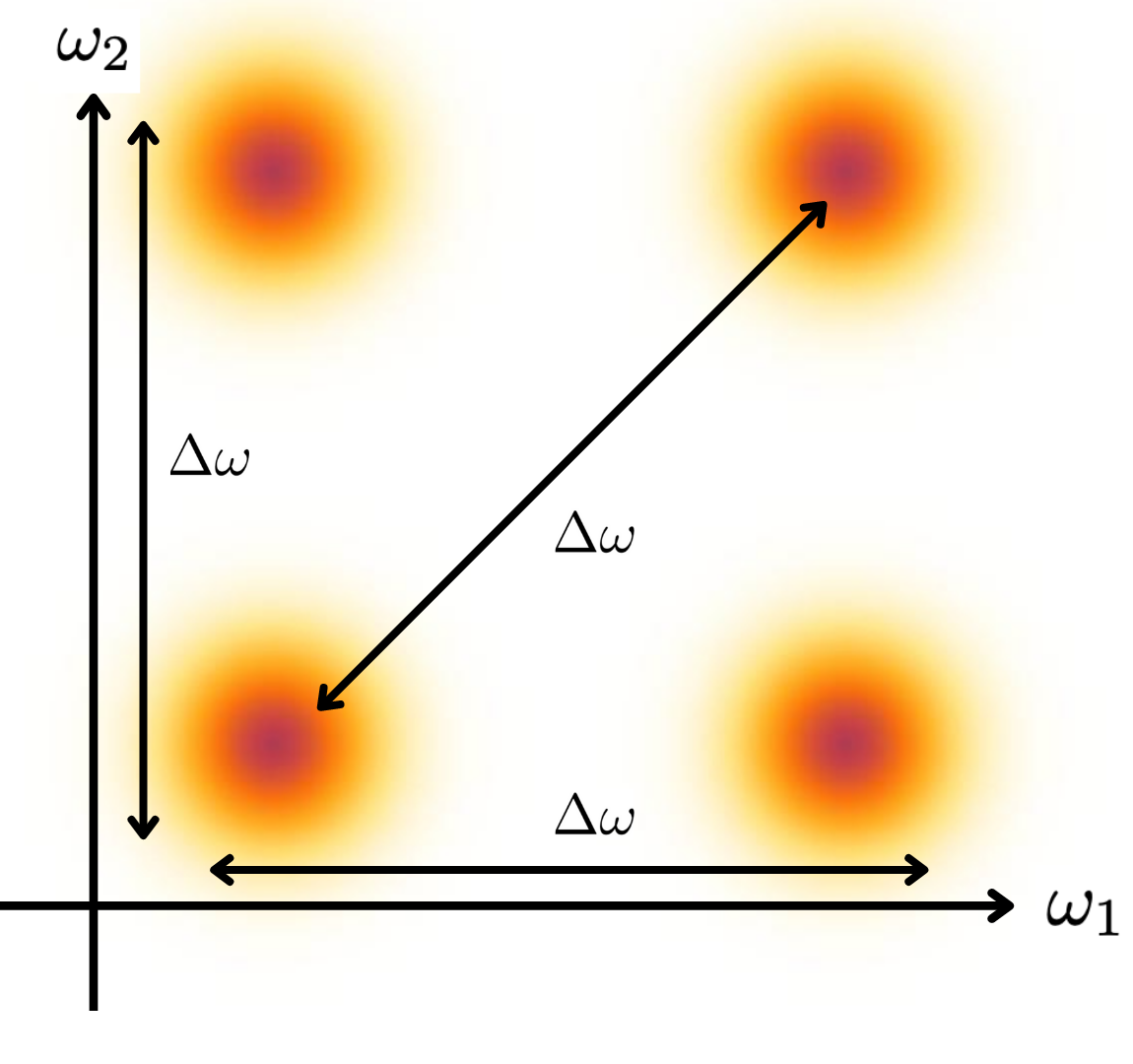}\\
    (a) Circular Shot-noise scaling & (b) Compass shot-noise scaling
    \end{tabular}
    \caption[Joint Spectral Intensity of different separable quantum states]{Joint Spectral Intensity (JSI) of different separable quantum states ($n=2$). In both cases the precision scales as the shot noise, which is reflected in the fact that the width is the same locally and diagonally. The arrows depict the size of the state in various directions as measured by the standard deviation. Adapted from Ref.~\cite{descamps_quantum_2023}. \copyright 2023 American Physical Society.}
    \label{fig: JSI eseparable}
\end{figure}

\paragraph{\publi General single-photon states}
We now consider a general pure, non-separable single-photon state. The variance of the operator $\hat \Omega$ then takes the form
\begin{equation}
    \label{eq: variance Omega general}
    \Delta^2 \hat \Omega
    = \sum_{j=1}^n \Delta^2 \omega_j
    + \sum_{j,k=1,\, j \neq k}^n \Cov(\omega_j,\omega_k),
\end{equation}
where $\Delta^2 \omega_j$ is the local frequency variance and
\begin{equation}
    \Cov(\omega_j,\omega_k)
    = \expval{\hat \omega_j \hat \omega_k}
    - \expval{\hat \omega_j}\expval{\hat \omega_k}
\end{equation}
is the covariance between the frequencies of photons in modes $j$ and $k$. In the case where all single-photon variances are equal, the Cauchy-Schwarz inequality applied to the covariance bilinear form shows that
\begin{equation}
    \Delta^2 \hat \Omega \leq n^2 \Delta^2 \omega,
\end{equation}
which corresponds precisely to the Heisenberg limit. It is instructive to compare this result with the usual analysis of precision limits in optical phase estimation, where mode variance is typically disregarded as a quantum resource and the metrological advantage arises solely from photon number fluctuations. For instance, for NOON states~\cite{kok_quantum_2004, reisner_quantum-limited_2022} or Schrödinger cat states~\cite{gilchrist_schrodinger_2004}, which saturate the Heisenberg limit, one has $\Delta^2 \hat n \propto \expval{\hat n}^2$, where $\expval{\hat n}$ denotes the average photon number. By contrast, for single-photon states the photon number variance is identically zero, and the variance of the global evolution generator $\hat \Omega$ depends exclusively on modal properties.

As a consequence, in the single-photon regime the Heisenberg limit can only be reached by exploiting frequency entanglement and the associated mode-particle correlations of an intrinsically multimode pure state. Interestingly, states that saturate this limit exhibit perfectly correlated frequency measurement outcomes across the different photons, in the sense that a joint frequency measurement yields identical outcomes in each experimental run. These correlations refer to the classical statistics of the measurement results, i.e., the joint probability distribution of detected frequencies, which can be represented as that of maximally correlated random variables.

This classical-looking structure of the measurement statistics does not imply the absence of quantum entanglement in the underlying state. On the contrary, the state remains genuinely entangled in the frequency degree of freedom, and this entanglement is essential for the scaling of the quantum Fisher information. The key point is that, for pure single-photon multimode states, the relevant quantum correlations are fully reflected in the statistics of frequency measurements, which appear classically perfectly correlated, while still encoding non-classical phase coherence across modes.

In this sense, attaining Heisenberg-limited precision requires quantum states whose frequency-resolved detection statistics mimic those of maximally correlated classical random variables, even though the physical origin of these correlations is quantum mechanical and encoded in the multimode entanglement structure of the state.

\subsection{Optimal states}
\label{subsec: TF optimal states}
\paragraph{\publi Expression of the optimal state}
In the previous section, we have shown that, assuming for simplicity that all local variances $\Delta^2\hat\omega_j$ are equal, any pure $n$-mode time-frequency single-photon state satisfies the inequality
\begin{equation}\label{eq: TF local vs global variance ineq}
    \Delta^2\hat\Omega \leq n^2 \Delta^2\omega,
\end{equation}
where $\Delta^2\omega$ denotes the local frequency variance of each photon. We now address the problem of characterizing the states that saturate this inequality. Since this bound is obtained by applying the Cauchy-Schwarz inequality to the covariance bilinear form, the equality condition implies that states saturating the bound must satisfy
\begin{equation}
    \Cov(\omega_j,\omega_k) = \Delta^2\omega,
\end{equation}
for all pairs $(j,k)$. The equality condition of the Cauchy-Schwarz inequality further implies that, for each pair of indices, the random variables $\omega_j$ and $\omega_k$ must differ only by a constant. As a consequence, the joint spectral distribution must contain Dirac delta constraints of the form $\delta(\omega_j - \omega_{j+1} + C_j)$. These constraints reduce the number of independent variables to a single degree of freedom, yielding the general expression of a state saturating the Heisenberg bound,
\begin{equation}
    \ket{\psi} = \int \dd\omega_1 \cdots \dd\omega_n \,
    f(\omega_1 + \cdots + \omega_n)\,
    \delta(\omega_1 - \omega_2 + C_1) \cdots
    \delta(\omega_{n-1} - \omega_n + C_{n-1})
    \ket{\omega_1,\dots,\omega_n},
\end{equation}
where $f$ is an arbitrary complex function of a single variable. After performing the integrations, the state can be expressed in terms of the collective variable $\Omega = \omega_1 + \cdots + \omega_n$ as
\begin{equation}
    \label{eq: optimal diagonal states}
    \ket{\psi} = \int \dd \Omega \, f(\Omega) \,
    \ket{\Omega + \omega_1^0,\dots,\Omega + \omega_n^0},
\end{equation}
where the constants $\omega_j^0$ depend on the coefficients $C_j$ and correspond to fixed frequency offsets. The spectral function $f$ thus depends on a single collective variable, and the state $\ket{\psi}$ possesses a non-physical spectrum that is infinitely localized in all collective variables except for $\Omega$, which is the one associated with the operator $\hat\Omega$. This implies that all photons behave collectively and are associated with an effective de Broglie wavelength $\lambda = 2\pi c / \Omega$~\cite{jacobson_photonic_1995}. As illustrated by the joint spectral intensity (JSI) shown in Fig.~\ref{fig: JSI entangled}(a) for $n=2$, such states correspond to diagonal distributions, where the variance of each individual mode arises as the projection of this diagonal structure onto the corresponding frequency axis. This geometrical picture highlights the crucial role of correlations in the scaling behavior. States of this type, with different spectral functions, are currently produced experimentally for $n=2$ (see, for instance,~\cite{chen_hong-ou-mandel_2019,ramelow_discrete_2009,maltese_generation_2020,olislager_frequency-bin_2010,fabre_generation_2020}). Moreover, Fig.~\ref{fig: JSI entangled}(b) shows that although entanglement is necessary to achieve sub-shot-noise scaling, it is not sufficient. The symmetry of the spectral variance plays a central role in determining the metrological precision. For $n$ photons distributed over $n$ modes, the same geometrical interpretation applies, and the states saturating the Heisenberg limit correspond to diagonals of an $n$-dimensional hypercube.

\begin{figure}[ht]
    \centering
    \begin{tabular}{cc}
        \includegraphics[width=0.4\linewidth]{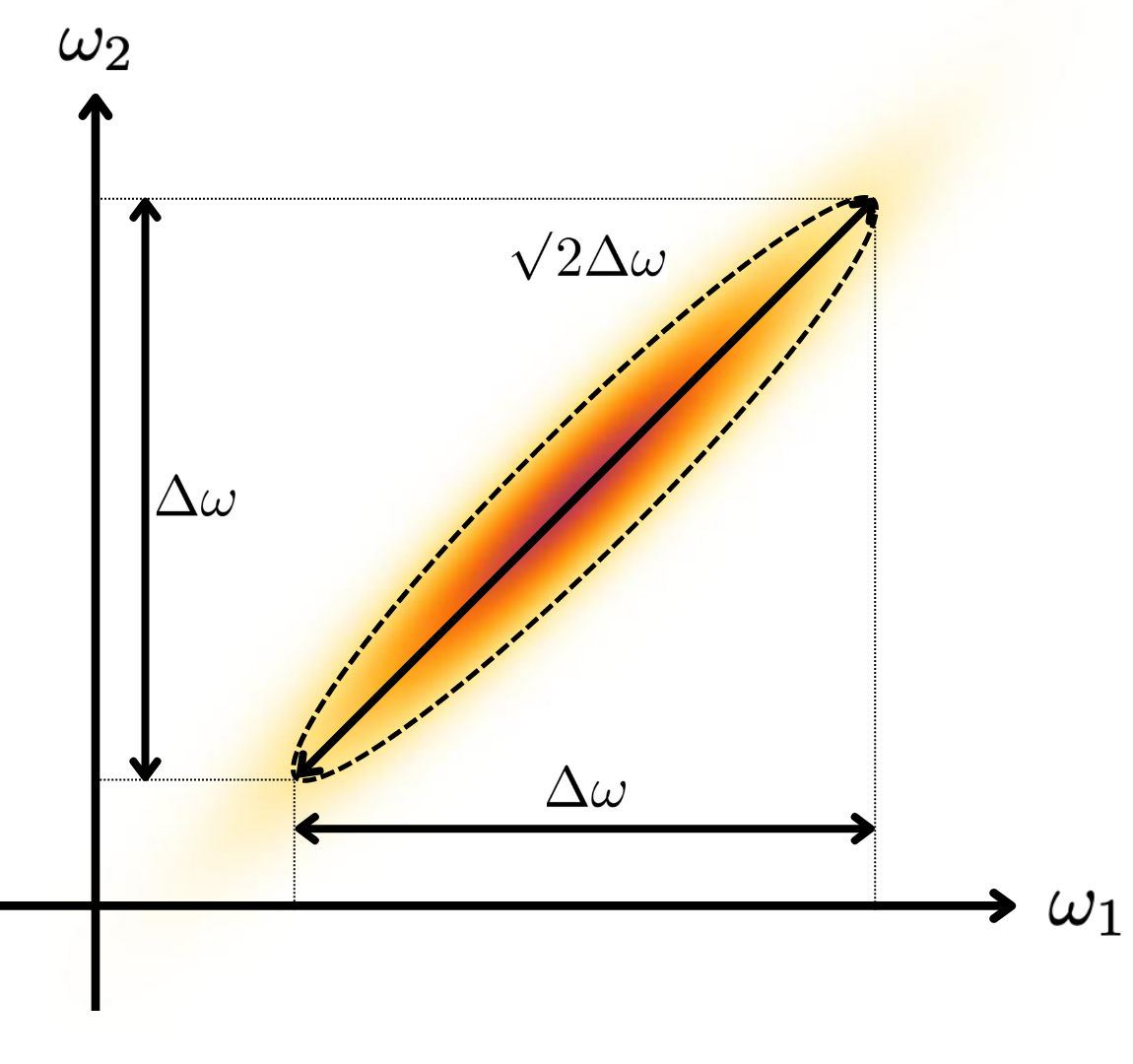} &
        \includegraphics[width=0.4\linewidth]{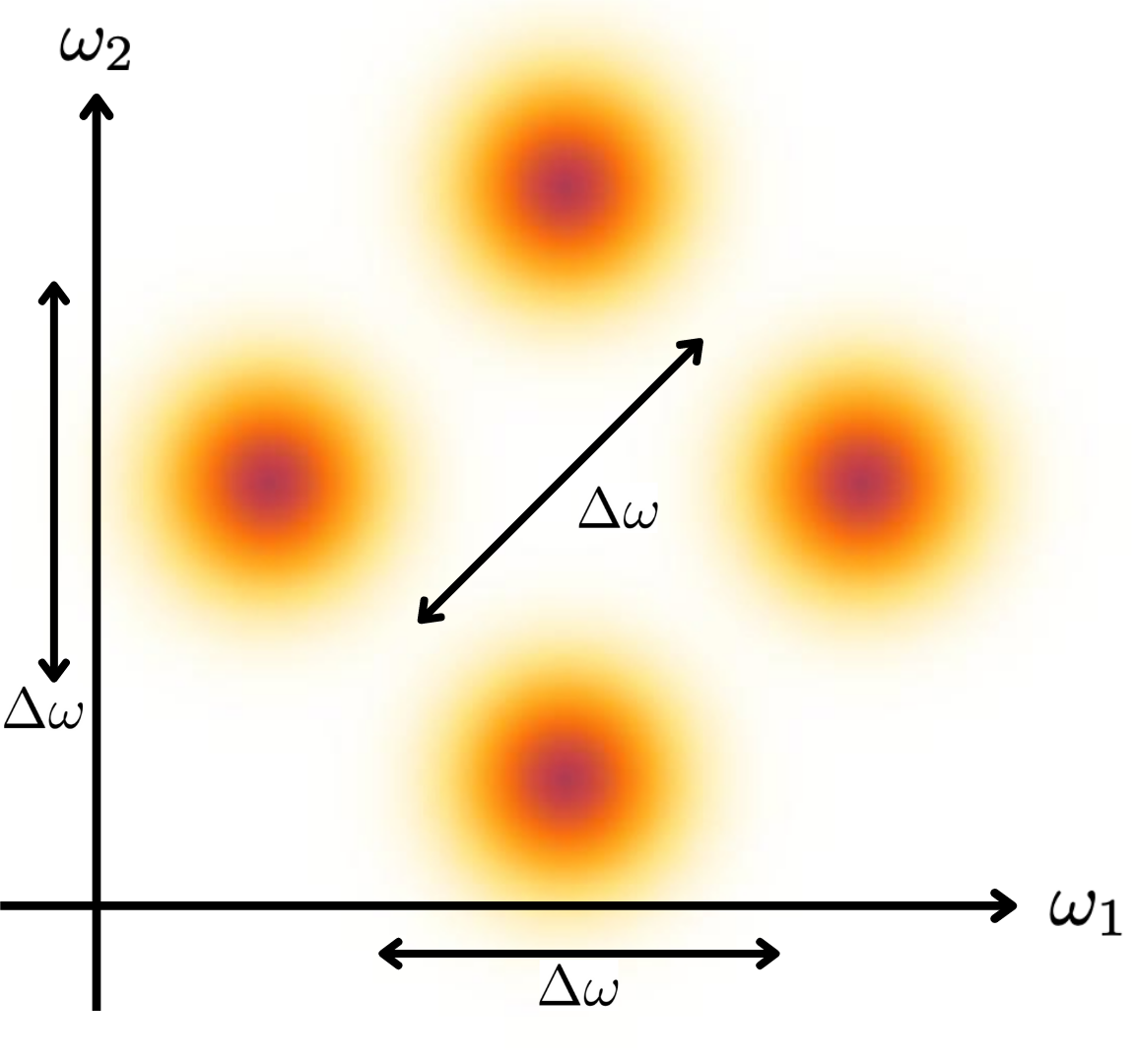} \\
        (a) Diagonal Heisenberg-like scaling. & (b) Compass shot-noise scaling.
    \end{tabular}
    \caption[Joint spectral intensity of different entangled quantum states]{Joint spectral intensity (JSI) of different entangled quantum states ($n=2$). In (a), an approximation of a diagonal state exhibits an effective scaling factor $\simeq \sqrt{2}$ in the diagonal width. In (b), the state is entangled but displays shot-noise scaling. Adapted from Ref.~\cite{descamps_quantum_2023}. \copyright 2023 American Physical Society.}
    \label{fig: JSI entangled}
\end{figure}

\paragraph{\publi Physical approximation}
The states reaching the Heisenberg limit are non-physical, as they rely on perfect correlations represented by Dirac delta distributions. Physical states necessarily possess a finite spectral width. To assess the impact of this finite width, we consider for simplicity that the spectra of the $n-1$ collective variables orthogonal to the direction associated with $\hat\Omega$ are characterized by identical variances $\sigma^2$. We assume that this variance is related to the width along the $\Omega$ direction by
\begin{equation}
    \sigma^2 = (1 - \eta)\Delta^2\hat\Omega.
\end{equation}
The parameter $\eta \in [0,1]$ quantifies the quality of the approximation: $\eta = 1$ corresponds to the ideal non-physical case, while $\eta = 0$ corresponds to separable states. In Appendix~\ref{app: collective var derivation}, Result~\ref{res: physical approx}, we show that under these assumptions
\begin{equation}
    \Delta^2 \hat\Omega = \frac{n^2 \Delta^2 \omega}{n(1 - \eta) + \eta}.
\end{equation}
For such states, the Heisenberg limit is no longer reached when $\sigma^2 \lesssim \Delta^2\omega$. This expression reveals a transition from a quadratic to a linear scaling with $n$. More precisely, the transition occurs around $n \approx \eta/(1 - \eta)$: the variance exhibits predominantly quadratic behavior for $n \ll \eta/(1 - \eta)$ and linear behavior for $n \gg \eta/(1 - \eta)$. To illustrate this effect, let us consider $\eta = 0.99$, a value that can be achieved without difficulty for $n=2$ in many experimental platforms~\cite{maltese_generation_2020,chen_hong-ou-mandel_2019,orieux_semiconductor_2017,le_jeannic_experimental_2021}. In this case, a predominantly quadratic scaling is maintained for $n \lesssim 99$, demonstrating that sub-shot-noise scaling remains robust when more physical states are considered. Figure~\ref{fig: finite width scaling} shows the variance as a function of the number of photons, highlighting the different scaling regimes and the transition point.

\begin{figure}[ht]
    \centering
    \includegraphics[width=9cm]{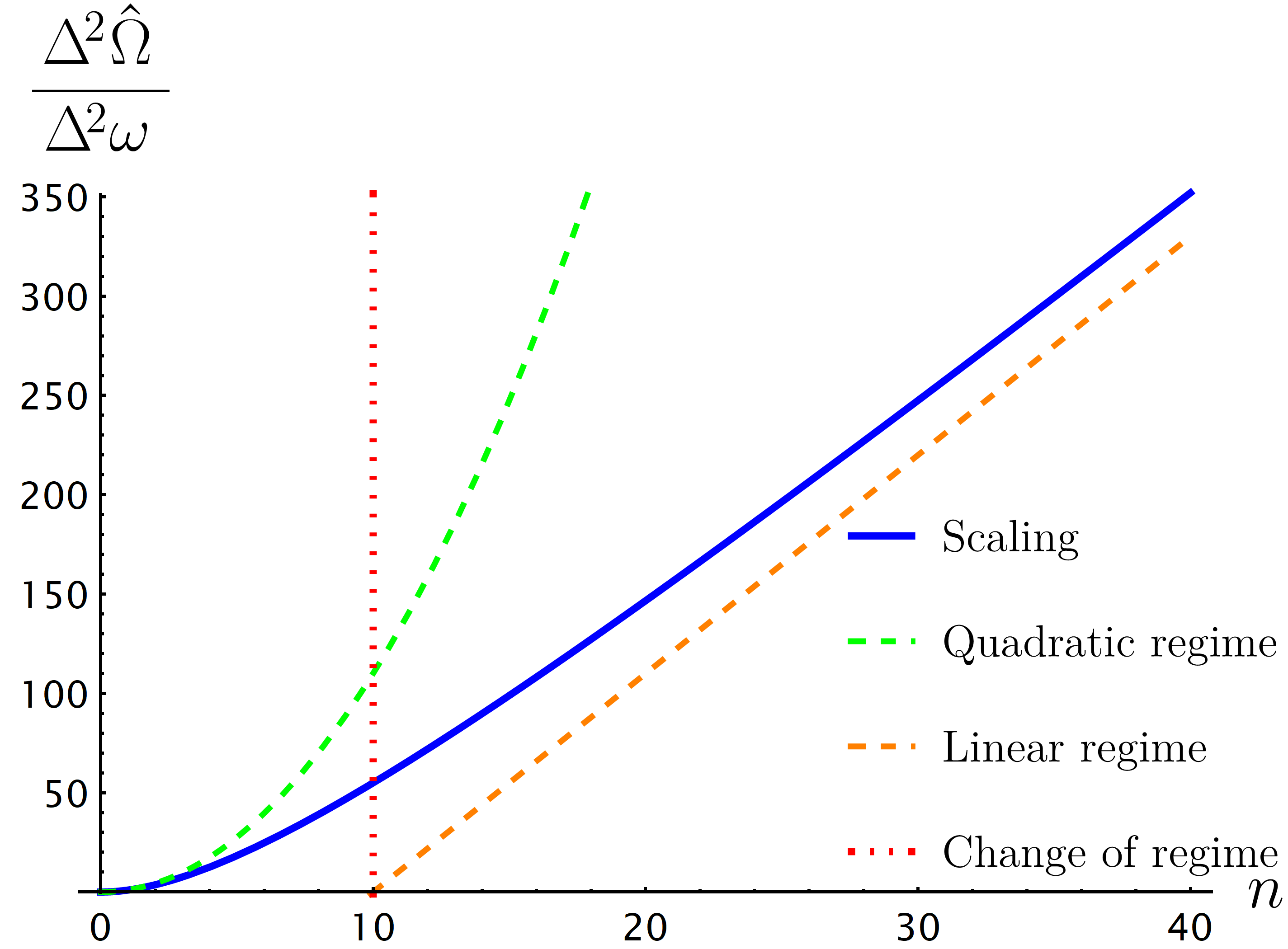}
    \caption[$\Delta^2 \hat\Omega / \Delta^2 \omega$ as a function of the number of photons]{$\Delta^2 \hat\Omega / \Delta^2 \omega$ as a function of the number of photons. The different scaling behaviors (linear and quadratic) and the transition point are displayed. Adapted from Ref.~\cite{descamps_quantum_2023}. \copyright 2023 American Physical Society.}
    \label{fig: finite width scaling}
\end{figure}

The existence of a transition from Heisenberg to shot-noise scaling is reminiscent of the results obtained in~\cite{escher_general_2011} in a different context, where photon loss was modeled by a parameter $\eta$. In that work, $\eta = 0$ corresponded to maximal loss, while $\eta = 1$ represented the absence of loss. In the present case, we consider pure states only, but the finite width of continuous-variable spectra can be interpreted as a continuous superposition of frequency-displaced states. In continuous-variable quantum information and quantum computing models, such displacements and finite spectral widths are regarded as deviations from ideal states~\cite{gottesman_encoding_2001,menicucci_fault-tolerant_2014,fabre_generation_2020}. Although the states remain pure, these effects mimic the impact of noise. This observation suggests a deep connection between physical continuous-variable states, noise models in continuous-variable quantum information, and ultimate precision limits in noisy quantum metrology.

\paragraph{\publi State generation}
As discussed above, the states of interest exhibit perfect frequency correlations and are designed to probe evolutions generated by the collective operator $\hat\Omega = \hat\omega_1 + \cdots + \hat\omega_n$. However, other collective operators involving relative minus signs can also be considered. This extension is particularly relevant for $n=2$, where one may consider the operator $\hat\omega_- = \hat\omega_1 - \hat\omega_2$. In this case, the optimal state follows the same general structure but exhibits perfect frequency anti-correlation. Different experimental platforms, characterized by distinct physical mechanisms controlling the spectral width, can produce either correlated or anti-correlated states, and the variance $\Delta^2\hat \Omega$ must then be interpreted accordingly. This distinction is essential when evaluating the parameter $\eta$ defined through $\sigma^2 = (1 - \eta)\Delta^2\hat \Omega$, with $\Delta^2\hat \Omega \gg \sigma^2$. For simplicity, we focus on examples satisfying the condition $\Cov(\omega_i,\omega_j) = \Delta^2\omega$, which can be achieved experimentally.

\begin{enumerate}
    \item In Ref.~\cite{maltese_generation_2020}, the authors employ an integrated AlGaAs nonlinear optical waveguide operating at room temperature to generate telecom-band frequency-entangled photon pairs via SPDC. The width of the joint spectral amplitude along the $\omega_{+}$ axis is determined by the pump bandwidth and is $\sigma = 2\pi \times 100$ kHz. The width along the $\omega_{-}$ axis is $\Delta = 2\pi \times 10.9$ THz. This yields $\eta \sim 1$, ensuring Heisenberg scaling, since in this case $n \ll \eta/(1 - \eta)$.
    \item In Ref.~\cite{chen_hong-ou-mandel_2019}, photon pairs are generated using a bulk ppKTP nonlinear crystal, where the phase-matching bandwidth can be tuned by adjusting the crystal temperature. The ratio between the spectral width of the two peaks in the joint spectral amplitude along the $\omega_{-}$ axis and the width along the $\omega_{+}$ axis reaches 68, corresponding to $\eta = 0.9998$. In this case, Heisenberg scaling for temporal parameter estimation is confirmed as $n \ll 4999$.
\end{enumerate}

We now briefly discuss experimental strategies for generating entangled states of more than two single photons. One approach relies on cascading nonlinear crystals, although current technological limitations restrict its scalability. An alternative strategy uses multiple independent single-photon sources~\cite{wang_boson_2019}, which are subsequently entangled. This can be achieved by mediating interactions between two photons via an auxiliary system. A promising example has been proposed and experimentally demonstrated in~\cite{le_jeannic_experimental_2021,le_jeannic_dynamical_2022}, where two initially separable photons propagating in the same pulse interact with a quantum dot embedded in a waveguide, resulting in an entangled photon pair. The resulting time-of-arrival distribution exhibits an elliptical shape oriented at 45 degrees, indicating temporal correlations or, equivalently, spectral anti-correlations. Notably, the spectral entanglement produced by this device reproduces that of SPDC-generated photon pairs. The measured value $\eta = 0.99$, corresponding to $\sigma/\Delta \sim 0.1$, confirms the suitability of this platform for achieving Heisenberg scaling in temporal estimation. Once two photons are entangled, larger states can be built by sequentially entangling additional photons. Related techniques, such as feed-forward and multiplexing, have enabled the generation of multiphoton polarization-entangled states~\cite{meyer-scott_scalable_2022} and represent promising avenues for extending frequency-entangled state generation. Recent theoretical developments have further clarified the physical mechanisms underlying single-photon frequency entanglement~\cite{alushi_waveguide_2023}.

\subsection{Phase-space interpretation}
\label{subsec: phase space interpretation}
These scaling effects can also be observed using the chrono-cyclic Wigner function and the associated phase space introduced in Section~\ref{subsec: TF Wigner function}. Recalling (see Section~\ref{subsec: TF translation}) that the operators $\hat\omega_j$ act as translations in the temporal $j$-th direction, the collective operator $\hat \Omega$ therefore implements a translation along the $\tau_1+\cdots+\tau_n$ direction. For separable states, each phase space (one for each spatial mode) and the associated translations can be analyzed independently.

\paragraph{\publi Collective Wigner function and evolution} For states saturating the Heisenberg limit, as given in \eqref{eq: optimal diagonal states}, the description can be reduced to a single wavefunction of the collective variable $\Omega$. In this case, the corresponding Wigner function takes the form (see Appendix~\ref{app: collective var derivation}, Result~\ref{res: ideal correlated wigner})
\begin{align}
    W(\varphi_1+\omega_1^0,&\dots,\varphi_n+\omega_n^0,\tau_1,\dots,\tau_n)\notag\\
    &=\frac{1}{(2\pi)^{n-1}}W_1(\varphi_1,\tau_1+\tau_2+\cdots+\tau_n)\delta(\varphi_2-\varphi_1)\cdots\delta(\varphi_n-\varphi_1),
\end{align}
where
\begin{equation}
W_1(\varphi,t)=\frac{1}{\pi}\int \dd\Omega\, e^{2i\Omega t}f(\varphi-\Omega)f^\ast(\varphi+\Omega),
\end{equation}
is the Wigner function associated with the single-mode state $\int \dd\omega\, f(\omega)\ket{\omega}$. As a consequence, the full Wigner function $W$ is entirely characterized by a two-dimensional phase space.

Since we are interested in an evolution generated by a collective variable, we set $\tau=\tau_1=\tau_2=\cdots=\tau_n$, so that $t=n\tau$. The relevant Wigner function then reads
\begin{equation}
    W_d(\phi_1,\tau)=W_1(\phi_1,n\tau)=\frac{1}{\pi}\int \dd\Omega\, e^{2in\Omega\tau}f(\phi_1-\Omega)f^\ast(\phi_1+\Omega).
\end{equation}
This situation differs from the case in which the evolution is generated by a single local operator, which would correspond to the term $W_1(\phi_1,\tau)$ of the full Wigner function. The relation $W_d(\phi,\tau)=W_1(\phi,n\tau)$ shows that the Wigner function associated with the collective variable is rescaled along the $\tau$ direction by a factor $n$. As a result, frequency measurements become more efficient. This rescaling constitutes the phase-space signature of the Heisenberg scaling of the QFI. A displacement of $\delta t/n$ in the phase space associated with the collective variable $\Omega$ leads to the same change in the Wigner function as a displacement $\delta t$ for independent variables. Equivalently, the Wigner function evolves $n$ times faster under collective translations. This provides a phase-space picture of the metrological quantum advantage, which is a multi-dimensional analogue of the phenomena discussed in~\cite{toscano_sub-planck_2006,zurek_sub-planck_2001}, although it originates from a fundamentally different physical mechanism.

Figure~\ref{fig: Wigner cat state different scales}(a) shows the Wigner function associated with the local variable $\omega_1$ (with all other variables set to zero) for a local cat-like state (see Section~\ref{subsec: TF states}). As discussed in Section~\ref{subsec: Metrology and Wigner functions}, the achievable precision is directly related to the fringe inter-spacing. When considering instead a cat-like state defined in the collective variable $\Omega$, whose Wigner function is shown in Fig.~\ref{fig: Wigner cat state different scales}(b), the fringe inter-spacing scales with $n$, and temporal displacements in the corresponding phase space can be resolved with higher precision. Finally, Fig.~\ref{fig: Wigner cat state different scales}(c) shows the Wigner function of a cat-like maximally correlated state with $n=10$ photons, defined in an (arbitrary) collective variable.

\begin{figure}[ht]
    \begin{tabular}{ccc}
        \includegraphics[width=4.5cm]{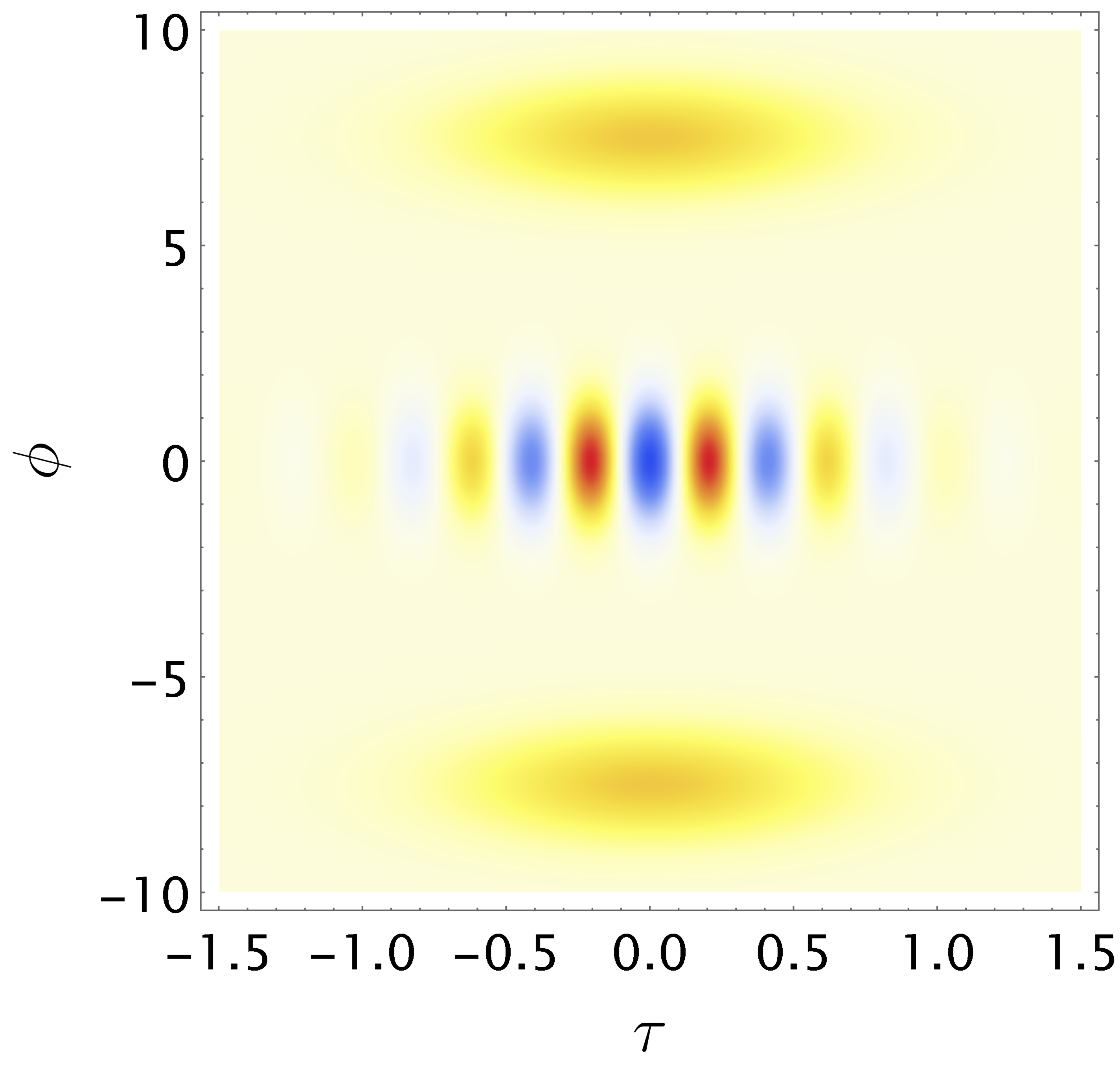} &
        \includegraphics[width=4.5cm]{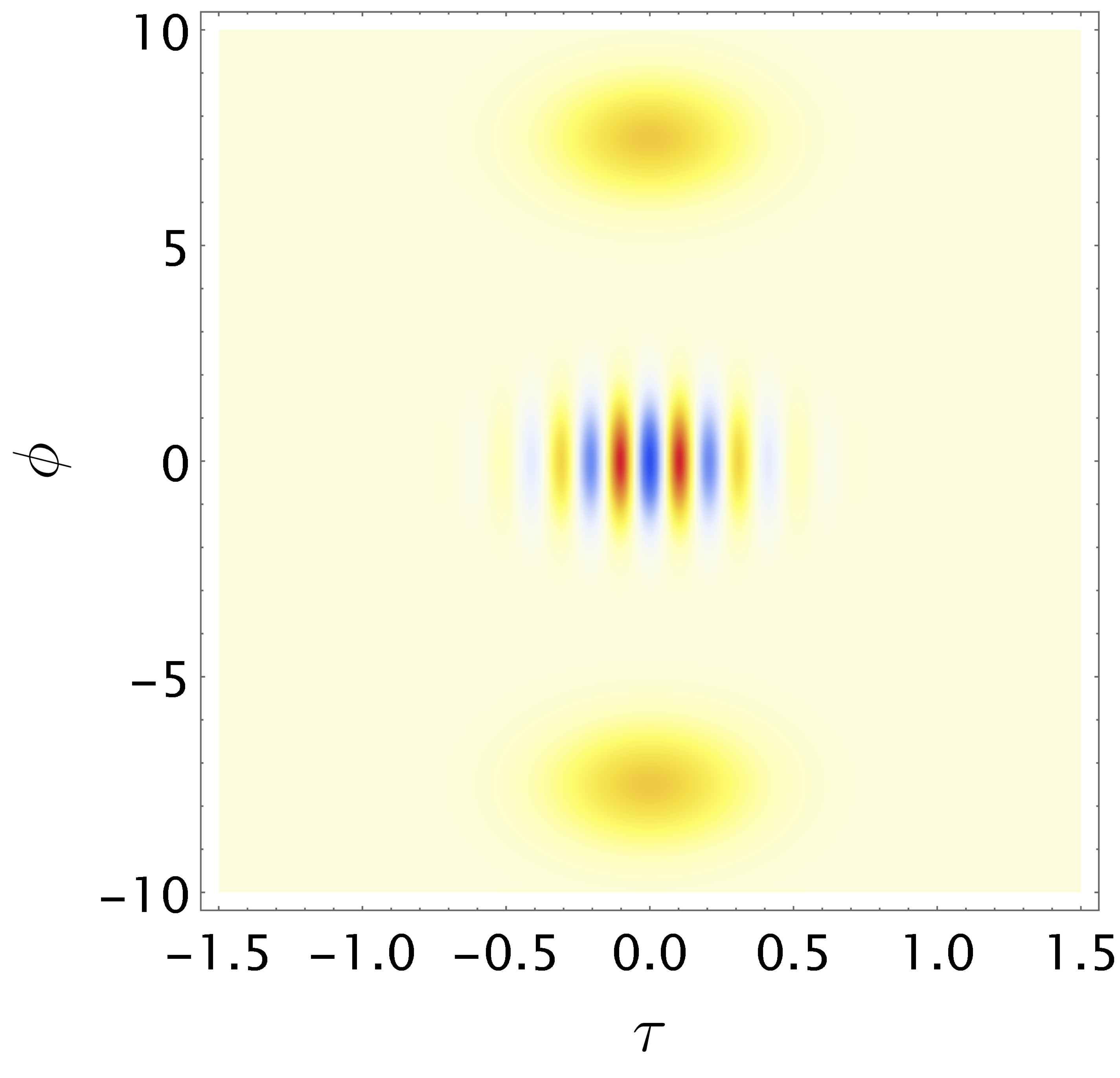} &
        \includegraphics[width=4.5cm]{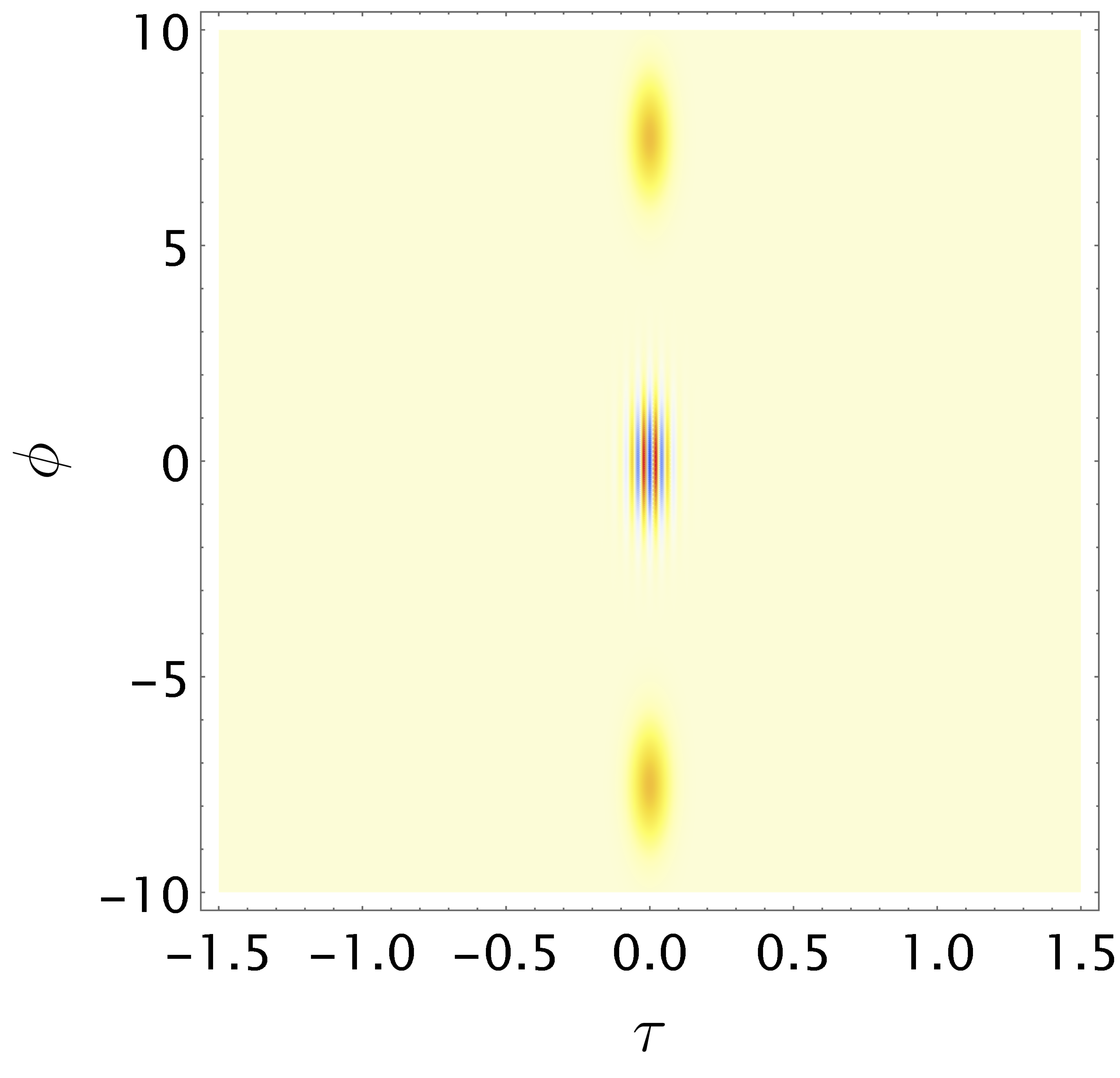} \\
        (a) & (b) & (c)
    \end{tabular}
    \caption[Wigner function of a cat-like state distributed in local or collective variables]{Wigner function of a cat-like state distributed in (a) the local variable $\omega_1$, (b) the collective variable $\Omega$ for $n=2$, and (c) the collective variable $\Omega$ for $n=10$. Adapted from Ref.~\cite{descamps_quantum_2023}. \copyright 2023 American Physical Society.}
    \label{fig: Wigner cat state different scales}
\end{figure}

It is important to recall that, while in the single-photon case a two-peaked spectrum admits a classical interpretation, for two or more entangled photons it reflects the structure of the state entangled in the collective variables and therefore directly contributes to the genuinely quantum properties of the $n$-photon state. The phase-space representation of quantum metrological properties of the field emerges only for multi-mode states and becomes manifest when the spectral properties are described using collective variables. Consequently, sub-Planck-like structures in phase space~\cite{praxmeyer_time-frequency_2007,austin_measuring_2010} cannot, by themselves, be attributed to genuinely quantum effects. Nevertheless, they remain useful for optimizing the variance and improving metrological performance in classical and single-mode fields for a fixed spectral bandwidth.

Finally, it is worth noting that collective variables and a two-dimensional phase space can also be used to describe the shot-noise scaling of independent photons using the same formalism. In that case, however, translations generated by the collective operator $\hat \Omega$ displace the Wigner function by an amount $\sqrt{n}\delta t$ only.

\clearpage
\section{General theory of entanglement along collective variables}
\label{sec: Collective entanglement}
\emph{Following the ideas presented in the previous section, we further explore the notion of single-mode entanglement. We first introduce a general theoretical framework, then derive a bound that can be interpreted as a generalization of the Heisenberg scaling limit. On this basis, we introduce a specific entanglement quantifier. Finally, the notion of $k$-entanglement is discussed. This section is mainly based on \hyperlink{Article: pra entanglement}{Measuring entanglement along collective operators}~\cite{descamps_measuring_2025}, which was published during the PhD thesis.}

\subsection{Motivations}
\label{subsec: collective variable motivation}
Entanglement stands as a cornerstone of quantum physics, providing a decisive advantage over classical systems in a wide range of applications, including sensing~\cite{hyllus_fisher_2012,toth_quantum_2014,pezze_quantum_2014}, computing~\cite{nielsen_quantum_2010,menicucci_universal_2006,shor_polynomial-time_1997}, and communication~\cite{bennett_quantum_2014,gisin_quantum_2002}. Despite extensive investigation, the characterization and quantification of entanglement remain active fields of research, particularly in the multipartite regime. While bipartite entanglement is now well understood~\cite{horodecki_separability_1996,sperling_necessary_2009}, there is no unique or universally accepted framework for multipartite entanglement, even for pure states. Existing approaches differ depending on whether one focuses on specific partitions of the system~\cite{szalay_multipartite_2015} or adopts a more global perspective~\cite{hyllus_fisher_2012,toth_multipartite_2012}. We refer to~\cite{horodecki_quantum_2009,amico_entanglement_2008} for comprehensive reviews. In this context, the development of entanglement quantifiers~\cite{plenio_introduction_2006,ma_multipartite_2024} and entanglement witnesses~\cite{li_entanglement_2013,laurell_witnessing_2024,akbari-kourbolagh_entanglement_2019} plays a central role in revealing the diverse facets of entanglement.

In Section~\ref{sec: TF entanglement and metrology}, we investigated the metrological advantages provided by frequency-entangled single-photon states for the estimation of temporal delays. In particular, our analysis relied on the collective operator $\hat \Omega=\hat\omega_1+\cdots+\hat \omega_n$. We showed that the Heisenberg limit can be attained by exploiting perfect correlations along the associated collective variable. Beyond elucidating the role of frequency and frequency entanglement as resources for quantum metrology, this result suggests that collective variables and metrological considerations provide a natural and powerful framework for the study of the corresponding single-mode entanglement. Motivated by this observation, we develop in this section a general theory of single-mode entanglement.

The idea that the variance of collective observables encodes information about multipartite entanglement was first put forward in the context of spin systems, where the variance of collective spin operators was shown to serve as a meaningful indicator of entanglement~\cite{guhne_multipartite_2005}. This insight was later extended to broader classes of systems~\cite{pezze_quantum_2014,chen_wigner-yanase_2005,hong_detection_2021}, although these generalizations were often restricted to finite-dimensional settings or relied on technically involved constructions~\cite{hong_detecting_2015}. In this work, we aim to unify this line of research with the time-frequency perspective developed earlier. To this end, we introduce a novel entanglement quantifier inspired by these prior works and investigate its extension to mixed quantum states.

A particularly important aspect of multipartite entanglement is $k$-entanglement~\cite{guhne_entanglement_2009}, for which the proposed quantifier proves especially relevant. Building upon the arguments and derivations presented in~\cite{hyllus_fisher_2012,chen_wigner-yanase_2005}, we show that our quantifier can be used to derive bounds whose violation signals the presence of $k$-entanglement, in close analogy with known criteria. We further establish a more general theorem and discuss variants of the resulting inequalities, including situations in which partial information about the single-mode entanglement structure is available.

\subsection{General setting}
\label{subsec: General setting collective ent}
We consider a Hilbert space $\mathcal H$ (finite or infinite-dimensional) and fix a Hamiltonian $\hat H$ acting on $\mathcal H$. For simplicity, we assume throughout this section that the eigenvalues of $\hat H$ are non-degenerate. For a given integer $n$, we consider the $n$-partite Hilbert space $\mathcal H^{\otimes n}$. We denote by $\hat H_j$ the operator obtained from $\hat H$ acting non-trivially on the $j$-th subsystem only, namely
\begin{equation}
    \hat H_j=\1^{\otimes (j-1)}\otimes \hat H \otimes \1 ^{\otimes (n-j)}.
\end{equation}
In close analogy with the definition of $\hat \Omega$, we introduce the collective Hamiltonian
\begin{equation}
    \hat H_\text{coll}=\sum_{j=1}^n \hat H_j,
\end{equation}
which acts on the full $n$-partite system. While this framework is completely general, we will focus on two representative examples that provide valuable intuition for the general case.

\paragraph{\publi Finite-dimensional setting}
We first consider the case where $\mathcal H$ is finite-dimensional. In this situation, the operator $\hat H$ has a finite spectrum, which is therefore bounded. In the following, $\lambda_{\max}$ and $\lambda_{\min}$ denote respectively the maximal and minimal eigenvalues of $\hat H$. A paradigmatic example of this setting is given by $\mathcal{H} = \mathbb{C}^2$ and $\hat{H} = \hat{Z}$, where $\hat{Z}$ is the Pauli-$Z$ operator introduced in Sec.~\ref{subsec: qubits}. This corresponds to the spin-$\frac{1}{2}$ case. The associated collective Hamiltonian
\begin{equation}
    \hat{H}_\text{coll} = \sum_{j=1}^n \hat{Z}_j
\end{equation}
represents the total spin operator along the $z$-axis. This model describes a system of $n$ spin-$\frac{1}{2}$ particles and constitutes the original framework in which many of the concepts discussed here were first developed~\cite{guhne_multipartite_2005}.

\paragraph{\publi Time-frequency case}
\label{subsec: time-frequency case}
In contrast to finite-dimensional systems, infinite-dimensional systems provide a markedly different physical setting. The time-frequency systems introduced earlier constitute a relevant example of such systems. In this context, $\mathcal H$ denotes the Hilbert space of single-photon time-frequency states in a single spatial mode, and the Hamiltonian is given by the frequency operator $\hat H=\hat \omega$. The corresponding collective Hamiltonian reads
\begin{equation}
    \hat H_\text{coll}=\hat\Omega=\sum_{j=1}^n \hat \omega_j.
\end{equation}
It is worth emphasizing that although we use time-frequency variables as a representative example of continuous-variable systems, the analogy discussed in Sec.~\ref{subsec: TF analogy CV} between single-photon time-frequency systems and standard continuous-variable systems implies that an analogous construction also applies to the usual quadrature operators $(\hat x,\hat p)$, via the correspondence $\hat \omega\leftrightarrow \hat x$ and $\hat t\leftrightarrow \hat p$.

\subsection{Spectral space}
\label{subsec: spectral space}
\paragraph{\publi Definition} To provide intuitive graphical representations, we introduce what we call the spectral space. We first give a general definition and then explain how it applies to two important examples: finite-dimensional systems and time-frequency states. Let $\Lambda\subset\R$ denote the spectrum of the operator $\hat H$. Since we assume the spectrum to be non-degenerate, each $\lambda\in\Lambda$ uniquely defines an eigenstate $\ket{\psi_\lambda}$. The family of product states $\ket{\psi_{\lambda_1}}\otimes\cdots\otimes\ket{\psi_{\lambda_n}}$, with $\lambda_i\in\Lambda$, forms a basis of the tensor product space $\mathcal H^{\otimes n}$. As a consequence, any state $\ket{\psi}\in\mathcal H^{\otimes n}$ admits the expansion
\begin{equation}
    \ket{\psi} = \int\limits_{\lambda_1,\dots,\lambda_n\in\Lambda} F(\lambda_1,\dots,\lambda_n) \ket{\psi_{\lambda_1}}\otimes\cdots\otimes\ket{\psi_{\lambda_n}} \, \dd^n \va{\lambda}.
\end{equation}
For notational simplicity, we write $\va{\lambda}=(\lambda_1,\dots,\lambda_n)\in\Lambda^n$ and $\ket{\psi_{\va{\lambda}}}=\ket{\psi_{\lambda_1}}\otimes\cdots\otimes\ket{\psi_{\lambda_n}}$. In the finite-dimensional case, the above integral should be understood as a discrete sum.

\paragraph{\publi Geometric interpretation} The function $F:\Lambda^n\to\C$ is square-normalized and will be referred to as the spectral amplitude. We call $\Lambda^n$ the spectral space associated with the collective operator $\hat H_\text{coll}$. For small values of $n$, the function $F$ can be represented graphically. As will become clear below, our analysis does not depend on the precise values taken by $F$, but only on its support, that is, the set of points $\va{\lambda}\in\Lambda^n$ for which $F(\va{\lambda})\neq 0$. In the following, we will illustrate this representation in the bipartite case $n=2$.

The spectral intensity $\abs*{F(\va{\lambda})}^2$ can naturally be interpreted as a probability distribution over the spectral space $\Lambda^n$.
Within this framework, expectation values of quantum observables that are diagonal in the basis $\left\{\ket{\psi_{\va{\lambda}}}\right\}$ coincide with expectation values of classical random variables defined on $\Lambda^n$. More precisely, to each local operator $\hat H_i$ we associate the random variable $H_i:\Lambda^n\to\R$ defined by $H_i(\va{\lambda})=\lambda_i$. The same correspondence applies to collective operators defined as linear combinations. For instance, the random variable associated with $\hat H_\text{coll}$ is $H_\text{coll}=\sum_{i=1}^n H_i$. A direct computation shows that for a state $\ket{\psi}$ with spectral amplitude $F$,
\begin{equation}
    \bra{\psi}\hat A\ket{\psi}=\expval{A},
\end{equation}
where the left-hand side denotes the quantum expectation value of an operator $\hat A$ that is diagonal in the basis $\{\ket{\psi_{\va{\lambda}}}\}$, and the right-hand side is the classical expectation value of the associated random variable $A$, given by $\expval{A}=\int_{\va{\lambda}\in\Lambda^n} A(\va{\lambda})\abs*{F(\va{\lambda})}^2\,\dd^n \va{\lambda}$. As an important example, for any normalized vector $\va{v}=(v_1,\dots,v_n)$ we define the operator $\hat V=\sum_{i=1}^n v_i \hat H_i$ and the corresponding random variable
$V=\sum_{i=1}^n v_i H_i$. Their expectation values coincide, that is, $\expval*{\hat V}=\expval{V}$. Defining as usual the quantum variance of an operator $\hat A$ on the pure state $\ket{\psi}$ as
\begin{equation}
    \Delta^2_{\ket{\psi}}(\hat A) = \bra{\psi}\hat A^2\ket{\psi} - \bra{\psi}\hat A\ket{\psi}^2,
\end{equation}
and since variances are built from expectation values, the same correspondence holds, and in particular we have $\Delta^2 \hat V=\Delta^2 V$. The latter quantity is simply the variance of the probability distribution $\abs*{F}^2$ along the direction specified by $\va{v}$ in spectral space. This provides a clear geometric interpretation of $\Delta \hat V$ as a measure of the ``thickness'' of the spectral amplitude in that direction. This intuition will be especially useful for local directions associated with $\hat H_i$ and for the collective or diagonal direction associated with $\hat H_\text{coll}$.

In the finite-dimensional case, the spectral space $\Lambda^n$ is a discrete subset of $\R^n$, corresponding to a finite and generally irregular rectangular grid. The support of the spectral amplitude of a state $\ket{\psi}$ is then represented by a finite set of points, as illustrated in Fig.~\ref{fig: example spectral space} (left) for $n=2$. At the opposite extreme lie time-frequency states. In that case, the spectrum of the frequency operator $\hat\omega$ is $\Lambda=\R$, and the eigenstate associated with $\lambda$ is the single-photon state $\ket{\lambda}$ at frequency $\lambda$. The spectral amplitude $F$ is therefore defined on $\R^n$, and the support of a time-frequency state can exhibit arbitrary continuous shapes, as shown in Fig.~\ref{fig: example spectral space} (right) for $n=2$.

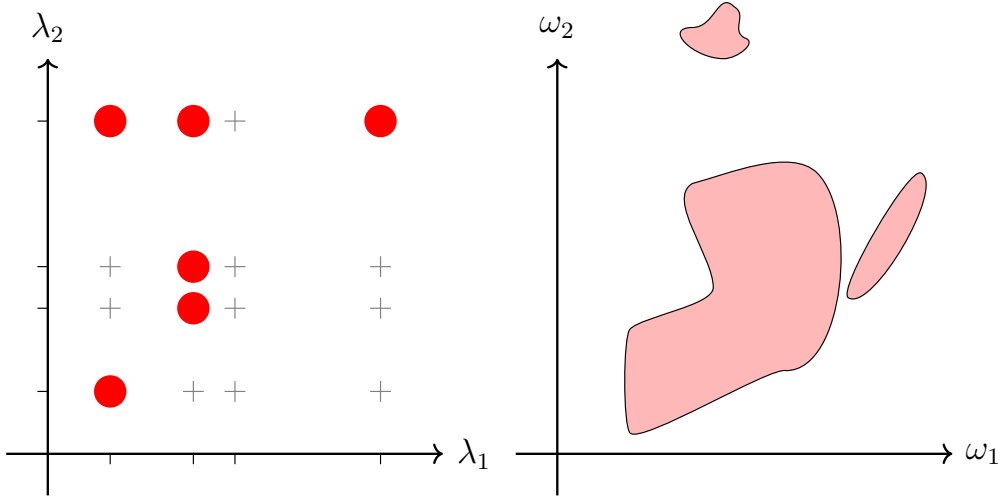
\begin{figure}[ht]
    \centering
    \scalebox{1.1}{\begin{tikzpicture}
	\begin{pgfonlayer}{nodelayer}
		\node [style=none] (0) at (-1, 0) {};
		\node [style=none] (1) at (0, -1) {};
		\node [style=none] (2) at (0, 9.5) {};
		\node [style=none] (3) at (9.5, 0) {};
		\node [style=none] (4) at (1.5, -0.25) {};
		\node [style=none] (5) at (1.5, 0) {};
		\node [style=none] (6) at (3.5, -0.25) {};
		\node [style=none] (7) at (3.5, 0) {};
		\node [style=none] (8) at (4.5, -0.25) {};
		\node [style=none] (9) at (4.5, 0) {};
		\node [style=none] (10) at (8, -0.25) {};
		\node [style=none] (11) at (8, 0) {};
		\node [style=none] (12) at (1.5, -0.25) {};
		\node [style=none] (13) at (0, 1.5) {};
		\node [style=none] (14) at (-0.25, 1.5) {};
		\node [style=none] (15) at (0, 3.5) {};
		\node [style=none] (16) at (-0.25, 3.5) {};
		\node [style=none] (17) at (0, 4.5) {};
		\node [style=none] (18) at (-0.25, 4.5) {};
		\node [style=none] (19) at (0, 8) {};
		\node [style=none] (20) at (-0.25, 8) {};
		\node [style=none] (25) at (1.25, 4.5) {};
		\node [style=none] (26) at (1.75, 4.5) {};
		\node [style=none] (27) at (1.5, 4.75) {};
		\node [style=none] (28) at (1.5, 4.25) {};
		\node [style=none] (29) at (1.25, 3.5) {};
		\node [style=none] (30) at (1.75, 3.5) {};
		\node [style=none] (31) at (1.5, 3.75) {};
		\node [style=none] (32) at (1.5, 3.25) {};
		\node [style=none] (49) at (3.25, 1.5) {};
		\node [style=none] (50) at (3.75, 1.5) {};
		\node [style=none] (51) at (3.5, 1.75) {};
		\node [style=none] (52) at (3.5, 1.25) {};
		\node [style=none] (53) at (4.25, 8) {};
		\node [style=none] (54) at (4.75, 8) {};
		\node [style=none] (55) at (4.5, 8.25) {};
		\node [style=none] (56) at (4.5, 7.75) {};
		\node [style=none] (57) at (4.25, 4.5) {};
		\node [style=none] (58) at (4.75, 4.5) {};
		\node [style=none] (59) at (4.5, 4.75) {};
		\node [style=none] (60) at (4.5, 4.25) {};
		\node [style=none] (61) at (4.25, 3.5) {};
		\node [style=none] (62) at (4.75, 3.5) {};
		\node [style=none] (63) at (4.5, 3.75) {};
		\node [style=none] (64) at (4.5, 3.25) {};
		\node [style=none] (65) at (4.25, 1.5) {};
		\node [style=none] (66) at (4.75, 1.5) {};
		\node [style=none] (67) at (4.5, 1.75) {};
		\node [style=none] (68) at (4.5, 1.25) {};
		\node [style=none] (73) at (7.75, 4.5) {};
		\node [style=none] (74) at (8.25, 4.5) {};
		\node [style=none] (75) at (8, 4.75) {};
		\node [style=none] (76) at (8, 4.25) {};
		\node [style=none] (77) at (7.75, 3.5) {};
		\node [style=none] (78) at (8.25, 3.5) {};
		\node [style=none] (79) at (8, 3.75) {};
		\node [style=none] (80) at (8, 3.25) {};
		\node [style=none] (81) at (7.75, 1.5) {};
		\node [style=none] (82) at (8.25, 1.5) {};
		\node [style=none] (83) at (8, 1.75) {};
		\node [style=none] (84) at (8, 1.25) {};
		\node [style=none] (85) at (0, 10.25) {$\lambda_2$};
		\node [style=none] (86) at (10.25, 0) {$\lambda_1$};
		\node [style=red dot] (87) at (1.5, 1.5) {};
		\node [style=red dot] (88) at (3.5, 3.5) {};
		\node [style=red dot] (89) at (3.5, 4.5) {};
		\node [style=red dot] (90) at (8, 8) {};
		\node [style=none] (102) at (11.25, 0) {};
		\node [style=none] (103) at (12.25, -1) {};
		\node [style=none] (104) at (12.25, 9.5) {};
		\node [style=none] (105) at (21.75, 0) {};
		\node [style=none] (171) at (12.25, 10.25) {$\omega_2$};
		\node [style=none] (172) at (22.5, 0) {$\omega_1$};
		\node [style=none] (173) at (14, 0.5) {};
		\node [style=none] (174) at (16, 4) {};
		\node [style=none] (175) at (14, 3) {};
		\node [style=none] (176) at (17.75, 2) {};
		\node [style=red dot] (185) at (1.5, 8) {};
		\node [style=red dot] (186) at (3.5, 8) {};
		\node [style=none] (187) at (15.5, 6.5) {};
		\node [style=none] (188) at (18.5, 6.75) {};
		\node [style=none] (189) at (19.25, 3.75) {};
		\node [style=none] (190) at (21, 6.75) {};
		\node [style=none] (191) at (15.5, 10.25) {};
		\node [style=none] (192) at (16.25, 9.5) {};
		\node [style=none] (193) at (16.75, 10) {};
		\node [style=none] (194) at (16.5, 10.75) {};
	\end{pgfonlayer}
	\begin{pgfonlayer}{edgelayer}
		\draw [style=Thick arrow] (0.center) to (3.center);
		\draw [style=Thick arrow] (1.center) to (2.center);
		\draw (5.center) to (12.center);
		\draw (7.center) to (6.center);
		\draw (9.center) to (8.center);
		\draw (11.center) to (10.center);
		\draw (20.center) to (19.center);
		\draw (18.center) to (17.center);
		\draw (16.center) to (15.center);
		\draw (14.center) to (13.center);
		\draw [style=Gray line] (27.center) to (28.center);
		\draw [style=Gray line] (25.center) to (26.center);
		\draw [style=Gray line] (31.center) to (32.center);
		\draw [style=Gray line] (29.center) to (30.center);
		\draw [style=Gray line] (51.center) to (52.center);
		\draw [style=Gray line] (49.center) to (50.center);
		\draw [style=Gray line] (55.center) to (56.center);
		\draw [style=Gray line] (53.center) to (54.center);
		\draw [style=Gray line] (59.center) to (60.center);
		\draw [style=Gray line] (57.center) to (58.center);
		\draw [style=Gray line] (63.center) to (64.center);
		\draw [style=Gray line] (61.center) to (62.center);
		\draw [style=Gray line] (67.center) to (68.center);
		\draw [style=Gray line] (65.center) to (66.center);
		\draw [style=Gray line] (75.center) to (76.center);
		\draw [style=Gray line] (73.center) to (74.center);
		\draw [style=Gray line] (79.center) to (80.center);
		\draw [style=Gray line] (77.center) to (78.center);
		\draw [style=Gray line] (83.center) to (84.center);
		\draw [style=Gray line] (81.center) to (82.center);
		\draw [style=Thick arrow] (102.center) to (105.center);
		\draw [style=Thick arrow] (103.center) to (104.center);
		\draw [style=Fill red] (174.center)
			 to [in=-150, out=90, looseness=0.75] (187.center)
			 to [in=135, out=15, looseness=0.75] (188.center)
			 to [in=0, out=-45, looseness=0.75] (176.center)
			 to [in=-45, out=165, looseness=0.25] (173.center)
			 to [in=-135, out=135, looseness=0.25] (175.center)
			 to [in=-90, out=45, looseness=0.50] cycle;
		\draw [style=Fill red] (191.center)
			 to [in=135, out=0, looseness=1.25] (194.center)
			 to [in=150, out=-30, looseness=1.25] (193.center)
			 to [in=0, out=-15, looseness=1.25] (192.center)
			 to [in=-180, out=180, looseness=1.75] cycle;
		\draw [style=Fill red] (190.center)
			 to [bend left=90, looseness=0.50] (189.center)
			 to [bend left=90, looseness=0.25] cycle;
	\end{pgfonlayer}
\end{tikzpicture}}
    \caption[Examples of the representation of the support of the spectrum in the bipartite case]{Examples of the representation of the support of the spectrum in the bipartite $n=2$ case. On the left: the case of finite dimension. On the right: the case of time-frequency variables. In the left panel, crosses represent the spectral space (all the possible pairs $(\lambda_1,\lambda_2)$ of eigenvalues of $\hat H$), and the red filled circles represent the support of a particular state. Adapted from Ref.~\cite{descamps_measuring_2025}. \copyright 2025 American Physical Society.}
    \label{fig: example spectral space}
\end{figure}

\subsection{Entanglement quantifier}
\label{subsec: entanglement measure}

\paragraph{\publi Inequality}
For any pure state $\ket{\psi}\in\mathcal H^{\otimes n}$, one can verify that
\begin{equation}\label{eq: general bound collective global}
    \Delta^2\hat H_\text{coll}\leq n^2 \max_j \Delta^2\hat H_j.
\end{equation}
This inequality establishes a link between the variance of a quantum state along collective and local directions. It constitutes a direct generalization of Eq.~\eqref{eq: TF local vs global variance ineq}, obtained previously for the time-frequency single-photon case. In particular, it shows that the collective variance is bounded by $n^2$ times the largest local variance.

We now provide a proof of this inequality. Although a concise proof can be obtained using the Cauchy-Schwarz inequality,\footnote{$\Delta^2\hat H_\text{coll}=\sum_{j,k=1}^n\Cov (\hat H_j,\hat H_k)\leq \sum_{j,k=1}^n\sqrt{\Delta^2\hat H_j\Delta^2\hat H_k}\leq n^2\max_j\Delta^2\hat H_j$.} applied to the bilinear form of the quantum covariance, we present an alternative derivation. This approach relies on a formalism that will be useful in later sections.
\begin{derivation}
    Let us introduce an orthogonal matrix $A=(\alpha_{j,k})\in\mathcal O_n(\R)$ such that $\alpha_{1,k}=\frac{1}{\sqrt{n}}$ for all $k$. Since $A$ is orthogonal, it satisfies $AA^T=A^TA=I_n$, which implies
    \begin{align}
        \sum_{j=1}^n \alpha_{j,k}\alpha_{j,l}=\delta_{k,l}, 
        && 
        \sum_{j=1}^n \alpha_{k,j}\alpha_{l,j}=\delta_{k,l}.
    \end{align}
    Using the matrix $A$, we define a new set of collective operators
    \begin{equation}\label{eq: definition P operators}
        \hat P_j=\sum_{k=1}^n \alpha_{j,k}\hat H_k.
    \end{equation}
    By construction, the first operator reads
    \begin{equation}
        \hat P_1=\sum_{j=1}^n \alpha_{1,j}\hat H_j=\frac{1}{\sqrt{n}}\hat H_\text{coll}.
    \end{equation}
    Moreover, using the inversion property of $A$, the local operators $\hat H_k$ can be expressed in terms of the $\hat P_j$ as
    \begin{equation}
        \sum_{j=1}^n \alpha_{j,k}\hat P_j
        =
        \sum_{j,l=1}^n \alpha_{j,k}\alpha_{j,l}\hat H_l
        =
        \sum_{l=1}^n \delta_{k,l}\hat H_l
        =
        \hat H_k.
    \end{equation}
    We can now derive the inequality. Starting from the observation that the maximum is greater than or equal to the mean, we have
    \begin{subequations}
        \begin{align}
            \max_j\Delta^2\hat H_j
            &\geq \frac{1}{n}\sum_{j=1}^n \Delta^2\hat H_j,\\
            &=\frac{1}{n}\sum_{j,k,l=1}^n\alpha_{k,j}\alpha_{l,j}\Cov (\hat P_k,\hat P_l),\\
            &=\frac{1}{n}\sum_{k=1}^n\Cov (\hat P_k,\hat P_k),\\
            &=\frac{1}{n}\sum_{k=1}^n\Delta^2\hat P_k,\\
            &\geq \frac{1}{n}\Delta^2\hat P_1
            =\frac{1}{n^2}\Delta^2\hat H_\text{coll},\label{eq: last step proof inequ}
        \end{align}
    \end{subequations}
    where $\Cov _{\ket{\psi}}(\hat A,\hat B)=\bra{\psi}\hat A\hat B\ket{\psi}-\bra{\psi}\hat A\ket{\psi}\bra{\psi}\hat B\ket{\psi}$ denotes the quantum covariance for commuting operators. The last inequality follows from the non-negativity of the variance $\Delta^2\hat P_j$ for $j>1$. Multiplying both sides by $n^2$ yields Eq.~\eqref{eq: general bound collective global}.    
\end{derivation}

In the case of finite-dimensional Hilbert spaces, inequality~\eqref{eq: general bound collective global} can be further simplified by bounding the right-hand side in a state-independent manner. Denoting by $\lambda_{\max}$ (resp. $\lambda_{\min}$) the largest (resp. smallest) eigenvalue of $\hat H$, one finds that for any state $\ket{\psi}\in\mathcal H^{\otimes n}$~\cite{pezze_quantum_2014} (see App.~\ref{app: collective var derivation}, Result~\ref{res: popoviciu's}, for a proof),
\begin{equation}
    \Delta^2_{\ket{\psi}}\hat H_i\leq \frac{(\lambda_{\max}-\lambda_{\min})^2}{4}.
\end{equation}
Substituting this bound into Eq.~\eqref{eq: general bound collective global} yields
\begin{equation}\label{eq: general bound in finite dim}
    \Delta^2\hat H_\text{coll}\leq \frac{n^2(\lambda_{\max}-\lambda_{\min})^2}{4},
\end{equation}
which reproduces the bound already obtained in Ref.~\cite{pezze_quantum_2014}. In particular, for spin observables satisfying $\lambda_{\max}=-\lambda_{\min}=1$, one recovers the standard inequality $\Delta^2(\hat Z_1+\cdots+\hat Z_n)\leq n^2$.

\paragraph{\publi Entanglement quantifier}
Motivated by inequality~\eqref{eq: general bound collective global}, we define the dimensionless quantity
\begin{equation}\label{eq: def quantifier}
    I_{\ket{\psi}}=\frac{\Delta^2\hat H_\text{coll}}{\max_j \Delta^2\hat H_j},
\end{equation}
provided that the denominator is non-zero. If $\max_j\Delta^2\hat H_j=0$, then Eq.~\eqref{eq: general bound collective global} implies that $\Delta^2\hat H_\text{coll}=0$ as well, leading to an indeterminate form. By convention, we define $I=0$ for such states.

Using the additivity of the quantum variance for pure product states~\cite{pezze_quantum_2014},\footnote{This property also follows directly from the additivity of the quantum Fisher information, proven in App.~\ref{app: formalism and framework}, Result~\ref{res: additivity QFI}, since the QFI reduces to the variance for pure states.} one verifies that $I$ is sub-additive on product states. For $\ket{\psi}\in\mathcal H^{\otimes l}$ and $\ket{\phi}\in\mathcal H^{\otimes m}$, one finds
\begin{subequations}\label{eq: subadditivity}
    \begin{align}
        I_{\ket{\psi}\otimes\ket{\phi}}
        &=
        \frac{\Delta^2_{\ket{\psi}\otimes\ket{\phi}}\hat H_\text{coll}}{\max_j\Delta^2_{\ket{\psi}\otimes\ket{\phi}}\hat H_j},\\
        &=
        \frac{\Delta^2_{\ket{\psi}}(\hat H_1+\cdots+\hat H_l)+\Delta^2_{\ket{\phi}}(\hat H_1+\cdots+\hat H_m)}{\max\{\max_j \Delta^2_{\ket{\psi}}\hat H_j,\max_k \Delta^2_{\ket{\phi}}\hat H_k\}},\\
        &\leq
        \frac{\Delta^2_{\ket{\psi}}(\hat H_1+\cdots+\hat H_l)}{\max_j \Delta^2_{\ket{\psi}}\hat H_j}
        +
        \frac{\Delta^2_{\ket{\phi}}(\hat H_1+\cdots+\hat H_m)}{\max_k \Delta^2_{\ket{\phi}}\hat H_k},\\
        &= I_{\ket{\psi}}+I_{\ket{\phi}}.
    \end{align}
\end{subequations}

By recursively applying this result to a fully factorized state $\ket{\psi}=\ket{\psi_1}\otimes\cdots\otimes\ket{\psi_n}$ and noting that $I_{\ket{\psi_j}}=1$ for any single-system state $\ket{\psi_j}\in\mathcal H$, it follows that $I_{\ket{\psi}}\leq n$ for all product states. Consequently, any state satisfying $I_{\ket{\psi}}>n$ must be entangled.

Entanglement witnesses are typically defined as functions $\mathcal{I}(\hat{\rho})$ such that $\mathcal{I}(\hat{\rho})\geq 0$ for all separable states and $\mathcal{I}(\hat{\rho})<0$ for some entangled states. In the present case, the choice $\mathcal{I}(\hat{\rho})=n-I$ fulfills this criterion. While the simplest witnesses are linear and take the form $\mathcal{I}(\hat{\rho})=\Tr(\hat W\hat\rho)$ for a suitable observable $\hat W$, the witness introduced here is non-linear. Such non-linear witnesses constitute a broader and more powerful class, which has been extensively studied, for example in Refs.~\cite{hyllus_optimal_2006,guhne_nonlinear_2006}. Importantly, the present witness admits a clear physical interpretation.

\paragraph{\publi Metrological interpretation}
The inequality~\eqref{eq: general bound collective global} and the definition of $I$ are deeply inspired by the metrological considerations discussed in Sec.~\ref{sec: TF entanglement and metrology}. From this perspective, the bounds $I\leq n$ for product states and $I\leq n^2$ in general naturally generalize the notions of shot-noise scaling and Heisenberg scaling, respectively, to a broad class of systems.

The use of the quantum Fisher information as a tool to detect and quantify entanglement is based on the following observation: if entanglement is required to achieve large values of the quantum Fisher information, then, conversely, observing a large quantum Fisher information certifies the presence of entanglement. This idea has been explored in several works, including Refs.~\cite{hyllus_fisher_2012,chen_wigner-yanase_2005,pezze_quantum_2014}. In the present work, we further develop this viewpoint.

For a given initial state $\ket{\psi}$, the global variance $\Delta^2_{\ket{\psi}}\hat H_\text{coll}$ can generally be increased in two distinct ways. First, one may increase the local variances $\Delta^2\hat H_j$ without changing the amount of entanglement, effectively ``stretching'' the state. Second, one may increase the degree of entanglement while keeping the local variances fixed. Both mechanisms are illustrated in Fig.~\ref{fig: increasing_precision}. Although such transformations are not always possible, depending on the structure of the initial state and the Hilbert space $\mathcal H$, they demonstrate that the global variance alone is not a faithful quantifier of entanglement, since it can be enhanced without introducing additional quantum correlations.

This observation motivates the definition of $I$ as the ratio between the global variance $\Delta^2\hat H_\text{coll}$ and the maximal local variance $\max_j\Delta^2\hat H_j$. By construction, this ratio is insensitive to state stretching that merely rescales local fluctuations and does not affect entanglement. This is fully consistent with the discussion in Sec.~\ref{sec: TF entanglement and metrology}, where we identified the local variance $\Delta^2\hat\omega_j$ as setting a classical scale for estimation precision rather than quantifying genuinely quantum features of single-photon states. The usefulness of comparing local variances with the quantum Fisher information of non-local operators was already emphasized in Ref.~\cite{gessner_efficient_2016}, where a general framework for constructing entanglement criteria based on local observables was introduced, and further developed in Ref.~\cite{lopetegui_detection_2024} to design experimentally accessible witnesses capable of detecting strong forms of entanglement in non-Gaussian continuous-variable states.

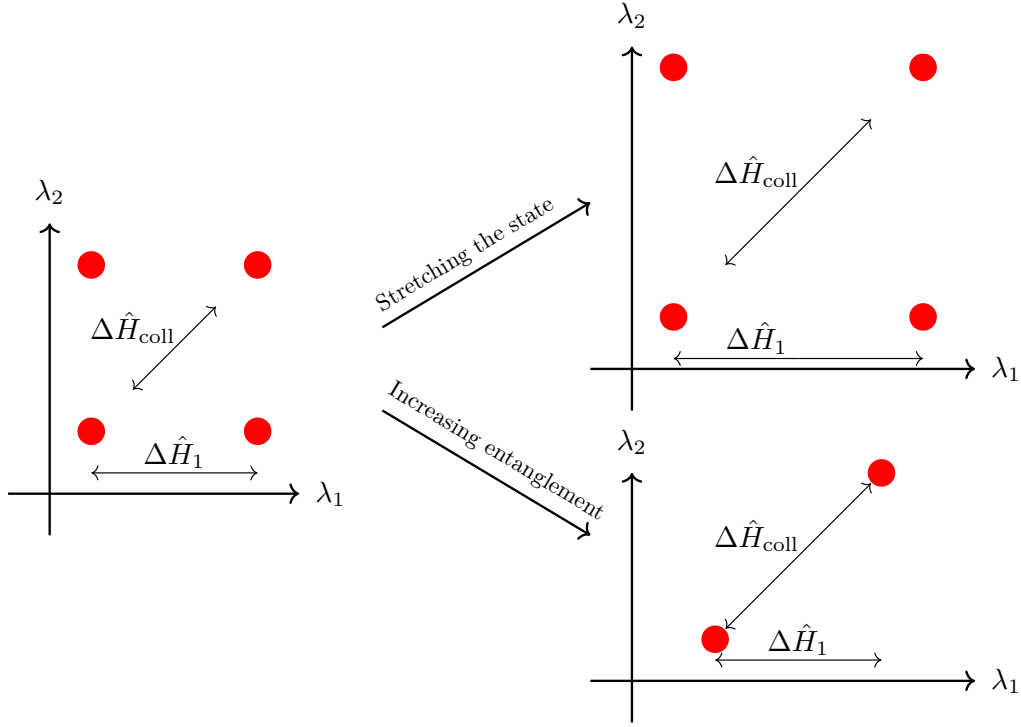
\begin{figure}[ht]
    \centering
    \footnotesize
    \scalebox{1.1}{\begin{tikzpicture}
	\begin{pgfonlayer}{nodelayer}
		\node [style=none] (83) at (-15, -3) {};
		\node [style=none] (84) at (-14, -4) {};
		\node [style=none] (85) at (-14, 3.5) {};
		\node [style=none] (86) at (-8, -3) {};
		\node [style=none] (87) at (-14, 4.25) {$\lambda_2$};
		\node [style=none] (88) at (-7.25, -3) {$\lambda_1$};
		\node [style=none] (106) at (-1, -7.5) {};
		\node [style=none] (107) at (0, -8.5) {};
		\node [style=none] (108) at (0, -2.5) {};
		\node [style=none] (109) at (8.25, -7.5) {};
		\node [style=none] (112) at (-1, 0) {};
		\node [style=none] (113) at (0, -1) {};
		\node [style=none] (114) at (0, 7.75) {};
		\node [style=none] (115) at (8.25, 0) {};
		\node [style=none] (126) at (-13, -2.5) {};
		\node [style=none] (127) at (-10, -2.5) {};
		\node [style=none] (128) at (-9, -2.5) {};
		\node [style=red dot] (129) at (-13, 2.5) {};
		\node [style=red dot] (130) at (-9, 2.5) {};
		\node [style=red dot] (131) at (-13, -1.5) {};
		\node [style=red dot] (132) at (-9, -1.5) {};
		\node [style=none] (133) at (-12, -0.5) {};
		\node [style=none] (134) at (-11, 0.5) {};
		\node [style=none] (135) at (-10, 1.5) {};
		\node [style=none] (136) at (-11, -2) {$\Delta\hat H_1$};
		\node [style=none] (137) at (-12, 1) {$\Delta\hat  H_\text{coll}$};
		\node [style=none] (138) at (1, 0.25) {};
		\node [style=none] (139) at (4, 0.25) {};
		\node [style=none] (140) at (7, 0.25) {};
		\node [style=red dot] (141) at (1, 7.25) {};
		\node [style=red dot] (142) at (7, 7.25) {};
		\node [style=red dot] (143) at (1, 1.25) {};
		\node [style=red dot] (144) at (7, 1.25) {};
		\node [style=none] (148) at (3, 0.75) {$\Delta\hat H_1$};
		\node [style=none] (150) at (2, -7) {};
		\node [style=none] (151) at (5, -7) {};
		\node [style=none] (152) at (6, -7) {};
		\node [style=red dot] (154) at (6, -2.5) {};
		\node [style=red dot] (155) at (2, -6.5) {};
		\node [style=none] (157) at (2.25, -6.25) {};
		\node [style=none] (158) at (4, -4.5) {};
		\node [style=none] (159) at (5.75, -2.75) {};
		\node [style=none] (160) at (4, -6.5) {$\Delta\hat H_1$};
		\node [style=none] (161) at (3, -4) {$\Delta\hat H_\text{coll}$};
		\node [style=none] (162) at (2.25, 2.5) {};
		\node [style=none] (163) at (4, 4.25) {};
		\node [style=none] (164) at (5.75, 6) {};
		\node [style=none] (165) at (3, 4.75) {$\Delta\hat  H_\text{coll}$};
		\node [style=none] (166) at (0, 8.5) {$\lambda_2$};
		\node [style=none] (167) at (0, -1.75) {$\lambda_2$};
		\node [style=none] (168) at (9, -7.5) {$\lambda_1$};
		\node [style=none] (169) at (9, 0) {$\lambda_1$};
		\node [style=none] (170) at (-6, 1) {};
		\node [style=none] (171) at (-1, 4) {};
		\node [style=none] (172) at (-6, -1) {};
		\node [style=none] (173) at (-1, -4) {};
		\node [style=none] (174) at (-4, 2.75) {\rotatebox{30.9638}{\scalebox{0.8}{Stretching the state}}};
		\node [style=none] (175) at (-3.25, -2) {\rotatebox{-30.9638}{\scalebox{0.8}{Increasing entanglement}}};
	\end{pgfonlayer}
	\begin{pgfonlayer}{edgelayer}
		\draw [style=Thick arrow] (83.center) to (86.center);
		\draw [style=Thick arrow] (84.center) to (85.center);
		\draw [style=Thick arrow] (106.center) to (109.center);
		\draw [style=Thick arrow] (107.center) to (108.center);
		\draw [style=Thick arrow] (112.center) to (115.center);
		\draw [style=Thick arrow] (113.center) to (114.center);
		\draw [style=Arrow] (127.center) to (128.center);
		\draw [style=Arrow] (127.center) to (126.center);
		\draw [style=Arrow] (134.center) to (135.center);
		\draw [style=Arrow] (134.center) to (133.center);
		\draw [style=Arrow] (139.center) to (140.center);
		\draw [style=Arrow] (139.center) to (138.center);
		\draw [style=Arrow] (151.center) to (152.center);
		\draw [style=Arrow] (151.center) to (150.center);
		\draw [style=Arrow] (158.center) to (159.center);
		\draw [style=Arrow] (158.center) to (157.center);
		\draw [style=Arrow] (163.center) to (164.center);
		\draw [style=Arrow] (163.center) to (162.center);
		\draw [style=Thick arrow] (170.center) to (171.center);
		\draw [style=Thick arrow] (172.center) to (173.center);
	\end{pgfonlayer}
\end{tikzpicture}}
    \caption[Two ways to increase the collective variance $\Delta^2\hat H_\text{coll}$ of a pure state]{Two ways to increase the collective variance $\Delta^2\hat H_\text{coll}$ of a pure state. The left plot shows the spectral representation of an initial product state $\ket{\psi}$. The upper-right plot depicts the same state after it has been stretched. Although the state remains a product state, it now exhibits a larger $\Delta^2\hat H_\text{coll}$, at the cost of similarly increasing the local variance $\Delta^2\hat H_j$. The lower-right plot illustrates a state of similar size to the original, but with certain points removed, resulting in an entangled state with an unchanged local variance $\Delta^2\hat H_j$. In both cases, the collective variance $\Delta^2\hat H_\text{coll}$ is increased to the same extent. Adapted from Ref.~\cite{descamps_measuring_2025}. \copyright 2025 American Physical Society.}
    \label{fig: increasing_precision}
\end{figure}

\subsection{Remark on the definition of collective operators}
\label{subsec: remark def coll op}
\paragraph{\publi Different definitions} 
Previously, we introduced the collective operators $\hat H_\text{coll}$ and $\hat P_j$. In this subsection, we comment on the specific choices underlying these definitions and discuss possible generalizations. In particular, $\hat H_\text{coll}$ was defined as the sum of the local operators $\hat H_j$,
\begin{equation}
    \hat H_\text{coll} = \hat H_1 + \cdots + \hat H_n,
\end{equation}
that is, with unit prefactors. More generally, one may consider a collective operator of the form
\begin{equation}
    \hat H_\text{coll} = \sum_{j=1}^n c_j \hat H_j,
\end{equation}
where the coefficients satisfy $c_j = \pm 1$. All results presented so far, as well as those developed later in this thesis, can be straightforwardly extended to this more general definition. Nevertheless, in order to keep the presentation clear and uncluttered, we restrict ourselves throughout the manuscript, unless stated otherwise, to the case $c_j = 1$ for all $j$.

The most appropriate definition of $\hat H_\text{coll}$ depends on the physical system and the application under consideration. For example, the original choice $c_j = 1$ for all $j$ is particularly natural in quantum metrology, as discussed in Ref.~\cite{pezze_quantum_2014}. The results of Ref.~\cite{pezze_quantum_2014}, together with inequality~\eqref{eq: general bound collective global}, provide important insights into the ultimate precision bounds and the structure of optimal probe states. A detailed discussion of the states saturating this bound is given in Sec.~\ref{subsec: equality case}.

Collective operators involving negative prefactors can also play a central role in specific scenarios. A notable example is provided by time-frequency states of two single photons ($n=2$). As discussed in Sec.~\ref{subsec: SPDC}, such states can be generated via the SPDC process~\cite{boucher_toolbox_2015} and, under suitable conditions, exhibit strong frequency anti-correlations. In this case, the quantity $I$ takes a large value when evaluated for the collective operator $\hat H_\text{coll} = \hat{\omega}_- = \hat{\omega}_1 - \hat{\omega}_2$, while it remains close to zero for $\hat{\omega}_+ = \hat{\omega}_1 + \hat{\omega}_2$. Moreover, as discussed in Chap.~\ref{chap: HOM interferometry and metrology}, the Hong-Ou-Mandel interferometer is particularly well suited to probing two-photon states and is sensitive primarily to the operator $\hat{\omega}_-$ rather than $\hat{\omega}_+$. From this perspective, analyzing $I$ for the collective operator $\hat{\omega}_-$ provides a meaningful characterization of how well a given two-photon state is tailored to Hong-Ou-Mandel measurements.

\paragraph{\publi Normalization choice} In the proof of Eq.~\eqref{eq: general bound collective global}, we introduce a set of operators $\hat P_j$, defined in Eq.~\eqref{eq: definition P operators}, which are constructed from collective variables orthogonal to the one defining $\hat H_\text{coll}$. Although the operators $\hat P_j$ are themselves collective, in the sense that they act simultaneously on several subsystems, their definition differs from that of $\hat H_\text{coll}$ in two important aspects. First, the normalization conventions are different. For $\hat H_\text{coll}$, the coefficients are chosen such that
\begin{equation}
    \sum_{j=1}^n c_j^2 = n,
\end{equation}
whereas the coefficients $\alpha_{j,k}$ defining $\hat P_j$ satisfy
\begin{equation}
    \sum_{k=1}^n \alpha_{j,k}^2 = 1.
\end{equation}
To place all operators on equal footing, we therefore introduce the normalized operator $\hat P_1 = \hat H_\text{coll} / \sqrt{n}$. Second, while the coefficients $c_j$ entering $\hat H_\text{coll}$ are restricted to the discrete values $\pm 1$, the coefficients $\alpha_{j,k}$ defining $\hat P_j$ are allowed to take arbitrary real values subject to the normalization and orthogonality constraints. The restriction $c_j = \pm 1$ in the definition of $\hat H_\text{coll}$ serves two purposes. On the one hand, it enforces symmetry among all modes, since $\abs{c_j}$ is the same for all $j$. On the other hand, as shown in Appendix~\ref{app: collective var derivation}, Result~\ref{res: cauchy-schwarz general}, this choice is the only one for which states saturating inequality~\eqref{eq: general bound collective global} can exist.

\subsection{Saturation of the inequality \eqref{eq: general bound collective global}}
\label{subsec: equality case}

In the previous section, we argued that $I_{\ket{\psi}}$ quantifies a specific form of entanglement that is relevant in a metrological context. A natural next step to deepen this understanding is to identify the states $\ket{\psi}$ for which $I_{\ket{\psi}}$ is maximal, \ie, those that saturate the inequality~\eqref{eq: general bound collective global}. We first establish the general conditions such states must satisfy. We then discuss the resulting geometrical constraints in spectral space, before illustrating these ideas in two representative examples.

\paragraph{\publi General argument}
Inspecting the proof of Eq.~\eqref{eq: general bound collective global}, one can directly identify the conditions under which the bound is saturated. The inequality is obtained through a sequence of equalities and inequalities, among which only two are inequalities. As a result, Eq.~\eqref{eq: general bound collective global} is saturated if and only if
\begin{equation}
    \left\{\begin{aligned}
        &\forall j,k,\,\Delta^2 \hat H_j = \Delta^2 \hat H_k,\\
        &\forall j \geq 2,\,\Delta^2 \hat P_j = 0.
    \end{aligned}\right.
\end{equation}
The second condition in fact implies the first one. Indeed, by the Cauchy-Schwarz inequality applied to covariances, one has
\begin{equation}
    \abs{\Cov(\hat P_j,\hat P_k)}^2 \leq \Delta^2 \hat P_j \, \Delta^2 \hat P_k = 0,
\end{equation}
whenever $j \geq 2$ or $k \geq 2$. Therefore, when expanding $\Delta^2 \hat H_j$ in terms of the covariances of the operators $\hat P_k$, only a single term contributes. One then directly obtains
\begin{equation}
    \Delta^2 \hat H_j = \frac{1}{n^2} \Delta^2 \hat H_\text{coll}.
\end{equation}

A first, trivial class of states saturating the bound consists of zero-variance states, for which both sides of Eq.~\eqref{eq: general bound collective global} vanish. These are product states built from eigenvectors of $\hat H$,
\begin{equation}
    \ket{\psi} = \ket{\psi_{\lambda_1}} \otimes \cdots \otimes \ket{\psi_{\lambda_n}}.
\end{equation}
By convention, the quantifier $I$, defined in Eq.~\eqref{eq: def quantifier}, is set to zero for such product states. We therefore exclude them from the following discussion. A nontrivial class of states saturating Eq.~\eqref{eq: general bound collective global} is given by
\begin{equation}\label{eq: first diagonal states}
    \ket{\psi} = \int\limits_{\lambda \in \Lambda} F(\lambda) \ket{\psi_\lambda}^{\otimes n}\, \dd\lambda,
\end{equation}
where $F:\Lambda \to \C$ is an arbitrary function. In finite dimension, this integral reduces to a sum of the form $\sum_{\lambda \in \Lambda} C_\lambda \ket{\psi_\lambda}^{\otimes n}$, with arbitrary normalized complex coefficients $C_\lambda$. Depending on the specific geometry of the spectrum $\Lambda$, additional classes of saturating states may also exist.

\paragraph{\publi Geometrical consequence}
Let $\ket{\psi} \in \mathcal H^{\otimes n}$ and denote by $F:\Lambda^n \to \C$ its spectral amplitude. As established above, $\ket{\psi}$ saturates Eq.~\eqref{eq: general bound collective global} if and only if $\Delta^2 \hat P_j = 0$ for all $j \geq 2$. Within the geometric framework introduced in Sec.~\ref{subsec: spectral space}, this condition implies that the distribution $F$ has zero thickness along the directions associated with the variables $P_2,\dots,P_n$.

The remaining orthogonal direction is associated with the variable $P_1$, defined by the vector $\va{u} = (1,\dots,1)$. Consequently, the support of $F$ must lie on a line, \ie, a one-dimensional affine subspace of $\R^n$, directed by $\va{u}$. This line does not necessarily pass through the origin. Depending on the structure of the spectrum $\Lambda$, there may exist more or fewer such lines contained within the spectral space $\Lambda^n$, each corresponding to a different class of states saturating Eq.~\eqref{eq: general bound collective global}.

This geometric picture provides direct insight into the nature of the entanglement quantified by $I$. Saturating states are characterized by spectral representations elongated along the main diagonal, corresponding to positive correlations among all random variables $H_j$. The quantifier $I$ thus captures the extent of this collective spectral stretching along the direction $P_1$.

\paragraph{\publi Finite-dimensional case}
In finite dimension, one may analyze the saturation of either the original inequality~\eqref{eq: general bound collective global} or its simplified version~\eqref{eq: general bound in finite dim}.

$\blacktriangleright$ Since $\Lambda$ is finite, its elements can be enumerated and ordered as $\Lambda = \{\lambda_1 < \cdots < \lambda_d\}$. The spectral space $\Lambda^n$ is then a discrete, generally uneven, rectangular grid of $d^n$ points. As discussed above, the support of $F$ must lie on a line directed by $\va{u} = (1,\dots,1)$. In Fig.~\ref{fig:evenly spaced grid}, we depict the two-dimensional spectral space associated with a Hamiltonian $\hat H$ with evenly spaced eigenvalues. In this case, any diagonal with direction $\va{u}$ passing through at least two grid points corresponds to a nontrivial state saturating the bound.

In contrast, when the eigenvalues of $\hat H$ are not evenly spaced, the number of admissible diagonals is reduced, as illustrated in Fig.~\ref{fig:not evenly spaced grid}. To obtain more possibilities than those associated with the states in Eq.~\eqref{eq: first diagonal states}, at least three eigenvalues of $\hat H$ must be equally spaced, \ie, they must satisfy $\abs{\lambda_a - \lambda_b} = \abs{\lambda_c - \lambda_b}$.

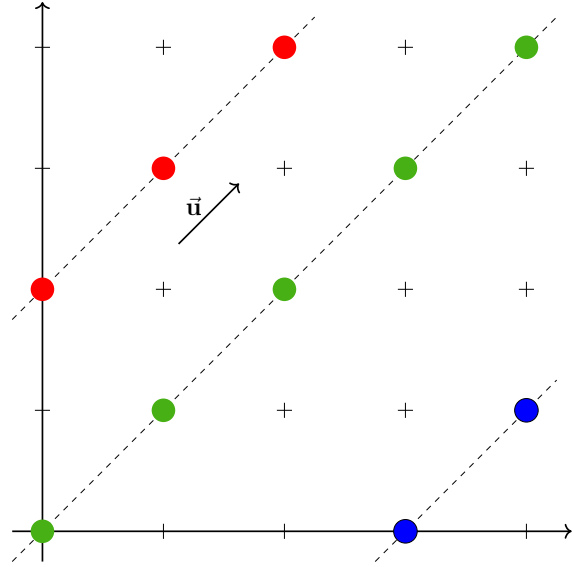
\begin{figure}[ht]
    \centering
    \scalebox{0.8}{\begin{tikzpicture}
	\begin{pgfonlayer}{nodelayer}
		\node [style=green dot] (0) at (0, 0) {};
		\node [style=Blue dot] (3) at (12, 0) {};
		\node [style=red dot] (7) at (0, 8) {};
		\node [style=green dot] (11) at (4, 4) {};
		\node [style=red dot] (13) at (4, 12) {};
		\node [style=green dot] (17) at (8, 8) {};
		\node [style=red dot] (19) at (8, 16) {};
		\node [style=green dot] (23) at (12, 12) {};
		\node [style=Blue dot] (26) at (16, 4) {};
		\node [style=green dot] (28) at (16, 16) {};
		\node [style=none] (34) at (0, 17.5) {};
		\node [style=none] (35) at (0, -1) {};
		\node [style=none] (36) at (-1, 0) {};
		\node [style=none] (37) at (17.5, 0) {};
		\node [style=none] (86) at (-1, -1) {};
		\node [style=none] (87) at (17, 17) {};
		\node [style=none] (88) at (-1, 7) {};
		\node [style=none] (89) at (9, 17) {};
		\node [style=none] (90) at (11, -1) {};
		\node [style=none] (91) at (17, 5) {};
		\node [style=none] (92) at (-0.25, 16) {};
		\node [style=none] (93) at (0.25, 16) {};
		\node [style=none] (94) at (4, 15.75) {};
		\node [style=none] (95) at (4, 16.25) {};
		\node [style=none] (96) at (3.75, 16) {};
		\node [style=none] (97) at (4.25, 16) {};
		\node [style=none] (98) at (12, 15.75) {};
		\node [style=none] (99) at (12, 16.25) {};
		\node [style=none] (100) at (11.75, 16) {};
		\node [style=none] (101) at (12.25, 16) {};
		\node [style=none] (102) at (8, 11.75) {};
		\node [style=none] (103) at (8, 12.25) {};
		\node [style=none] (104) at (7.75, 12) {};
		\node [style=none] (105) at (8.25, 12) {};
		\node [style=none] (106) at (4, 7.75) {};
		\node [style=none] (107) at (4, 8.25) {};
		\node [style=none] (108) at (3.75, 8) {};
		\node [style=none] (109) at (4.25, 8) {};
		\node [style=none] (112) at (-0.25, 12) {};
		\node [style=none] (113) at (0.25, 12) {};
		\node [style=none] (114) at (16, 11.75) {};
		\node [style=none] (115) at (16, 12.25) {};
		\node [style=none] (116) at (15.75, 12) {};
		\node [style=none] (117) at (16.25, 12) {};
		\node [style=none] (118) at (16, 7.75) {};
		\node [style=none] (119) at (16, 8.25) {};
		\node [style=none] (120) at (15.75, 8) {};
		\node [style=none] (121) at (16.25, 8) {};
		\node [style=none] (122) at (12, 7.75) {};
		\node [style=none] (123) at (12, 8.25) {};
		\node [style=none] (124) at (11.75, 8) {};
		\node [style=none] (125) at (12.25, 8) {};
		\node [style=none] (126) at (12, 3.75) {};
		\node [style=none] (127) at (12, 4.25) {};
		\node [style=none] (128) at (11.75, 4) {};
		\node [style=none] (129) at (12.25, 4) {};
		\node [style=none] (130) at (8, 3.75) {};
		\node [style=none] (131) at (8, 4.25) {};
		\node [style=none] (132) at (7.75, 4) {};
		\node [style=none] (133) at (8.25, 4) {};
		\node [style=none] (134) at (-0.25, 4) {};
		\node [style=none] (135) at (0.25, 4) {};
		\node [style=none] (136) at (8, -0.25) {};
		\node [style=none] (137) at (8, 0.25) {};
		\node [style=none] (138) at (4, -0.25) {};
		\node [style=none] (139) at (4, 0.25) {};
		\node [style=none] (140) at (16, -0.25) {};
		\node [style=none] (141) at (16, 0.25) {};
		\node [style=none] (142) at (4.5, 9.5) {};
		\node [style=none] (143) at (6.5, 11.5) {};
		\node [style=none] (144) at (5, 10.75) {$\va{u}$};
	\end{pgfonlayer}
	\begin{pgfonlayer}{edgelayer}
		\draw [style=Thick arrow] (35.center) to (34.center);
		\draw [style=Thick arrow] (36.center) to (37.center);
		\draw [style=dashes] (86.center) to (87.center);
		\draw [style=dashes] (88.center) to (89.center);
		\draw [style=dashes] (90.center) to (91.center);
		\draw (92.center) to (93.center);
		\draw (95.center) to (94.center);
		\draw (96.center) to (97.center);
		\draw (99.center) to (98.center);
		\draw (100.center) to (101.center);
		\draw (103.center) to (102.center);
		\draw (104.center) to (105.center);
		\draw (107.center) to (106.center);
		\draw (108.center) to (109.center);
		\draw (112.center) to (113.center);
		\draw (115.center) to (114.center);
		\draw (116.center) to (117.center);
		\draw (119.center) to (118.center);
		\draw (120.center) to (121.center);
		\draw (123.center) to (122.center);
		\draw (124.center) to (125.center);
		\draw (127.center) to (126.center);
		\draw (128.center) to (129.center);
		\draw (131.center) to (130.center);
		\draw (132.center) to (133.center);
		\draw (134.center) to (135.center);
		\draw (137.center) to (136.center);
		\draw (139.center) to (138.center);
		\draw (141.center) to (140.center);
		\draw [style=Thick arrow] (142.center) to (143.center);
	\end{pgfonlayer}
\end{tikzpicture}}
    \caption[Spectral space of $\mathcal H^{\otimes 2}$ for evenly spaced eigenvalues]{Spectral space of $\mathcal H^{\otimes 2}$ for evenly spaced eigenvalues. Crosses represent all possible pairs $(\lambda_1,\lambda_2)$ of eigenvalues of $\hat H$. The colored lines illustrate examples of supports of states saturating the inequality~\eqref{eq: general bound collective global}. Adapted from Ref.~\cite{descamps_measuring_2025}. \copyright 2025 American Physical Society.}
    \label{fig:evenly spaced grid}
\end{figure}

\begin{figure}[ht]
    \centering
    \scalebox{0.8}{\begin{tikzpicture}
	\begin{pgfonlayer}{nodelayer}
		\node [style=green dot] (0) at (0, 0) {};
		\node [style=green dot] (1) at (6, 0) {};
		\node [style=green dot] (2) at (9, 0) {};
		\node [style=green dot] (3) at (12, 0) {};
		\node [style=green dot] (4) at (17, 0) {};
		\node [style=green dot] (6) at (0, 6) {};
		\node [style=green dot] (7) at (0, 9) {};
		\node [style=green dot] (8) at (0, 12) {};
		\node [style=green dot] (9) at (0, 17) {};
		\node [style=green dot] (11) at (6, 6) {};
		\node [style=green dot] (12) at (6, 9) {};
		\node [style=green dot] (13) at (6, 12) {};
		\node [style=green dot] (14) at (6, 17) {};
		\node [style=green dot] (16) at (9, 6) {};
		\node [style=green dot] (17) at (9, 9) {};
		\node [style=green dot] (18) at (9, 12) {};
		\node [style=green dot] (19) at (9, 17) {};
		\node [style=green dot] (21) at (12, 6) {};
		\node [style=green dot] (22) at (12, 9) {};
		\node [style=green dot] (23) at (12, 12) {};
		\node [style=green dot] (24) at (12, 17) {};
		\node [style=green dot] (26) at (17, 6) {};
		\node [style=green dot] (27) at (17, 12) {};
		\node [style=green dot] (28) at (17, 17) {};
		\node [style=none] (34) at (0, 18.5) {};
		\node [style=none] (35) at (0, -1) {};
		\node [style=none] (36) at (-1, 0) {};
		\node [style=none] (37) at (18.5, 0) {};
		\node [style=none] (40) at (16, -1) {};
		\node [style=none] (41) at (11, -1) {};
		\node [style=none] (42) at (8, -1) {};
		\node [style=none] (43) at (5, -1) {};
		\node [style=none] (44) at (-1, -1) {};
		\node [style=none] (45) at (18, 1) {};
		\node [style=none] (47) at (18, 6) {};
		\node [style=none] (48) at (18, 15) {};
		\node [style=none] (49) at (18, 18) {};
		\node [style=none] (53) at (18, 9) {};
		\node [style=none] (54) at (18, 12) {};
		\node [style=none] (57) at (2, -1) {};
		\node [style=none] (63) at (15, 18) {};
		\node [style=none] (64) at (-1, 2) {};
		\node [style=none] (65) at (-1, 5) {};
		\node [style=none] (66) at (12, 18) {};
		\node [style=none] (67) at (-1, 8) {};
		\node [style=none] (70) at (9, 18) {};
		\node [style=none] (71) at (6, 18) {};
		\node [style=none] (75) at (1, 18) {};
		\node [style=none] (80) at (-1, 16) {};
		\node [style=none] (84) at (-1, 11) {};
		\node [style=green dot] (85) at (17, 9) {};
		\node [style=none] (86) at (-1, 10) {};
		\node [style=none] (87) at (7, 18) {};
		\node [style=none] (88) at (10, 18) {};
		\node [style=none] (89) at (-1, 7) {};
		\node [style=none] (90) at (-1, 4) {};
		\node [style=none] (91) at (13, 18) {};
		\node [style=none] (92) at (18, 10) {};
		\node [style=none] (93) at (6.75, -1) {};
		\node [style=none] (94) at (10, -1) {};
		\node [style=none] (95) at (18, 7) {};
		\node [style=none] (96) at (18, 13) {};
		\node [style=none] (97) at (4, -1) {};
		\node [style=none] (98) at (3.75, -14) {};
	\end{pgfonlayer}
	\begin{pgfonlayer}{edgelayer}
		\draw [style=Thick arrow] (35.center) to (34.center);
		\draw [style=Thick arrow] (36.center) to (37.center);
		\draw (40.center) to (45.center);
		\draw [in=225, out=45] (41.center) to (47.center);
		\draw (53.center) to (42.center);
		\draw [style=dashes red] (54.center) to (43.center);
		\draw [style=dashes red] (57.center) to (48.center);
		\draw [style=dashes red] (44.center) to (49.center);
		\draw [style=dashes red] (64.center) to (63.center);
		\draw [style=dashes red] (66.center) to (65.center);
		\draw (67.center) to (70.center);
		\draw [in=45, out=-135] (71.center) to (84.center);
		\draw (80.center) to (75.center);
		\draw (86.center) to (87.center);
		\draw (89.center) to (88.center);
		\draw (90.center) to (91.center);
		\draw (95.center) to (94.center);
		\draw (93.center) to (92.center);
		\draw (97.center) to (96.center);
	\end{pgfonlayer}
\end{tikzpicture}}
    \caption[Spectral space of $\mathcal H^{\otimes 2}$ for unevenly spaced eigenvalues]{Spectral space of $\mathcal H^{\otimes 2}$ for unevenly spaced eigenvalues. Black solid lines indicate directions parallel to $\va{u}$ that intersect the grid at a single point only, leading to zero-variance states with $I=0$. Red dashed lines correspond to diagonals intersecting at least two points and thus to nontrivial states saturating the inequality~\eqref{eq: general bound collective global}. Adapted from Ref.~\cite{descamps_measuring_2025}. \copyright 2025 American Physical Society.}
    \label{fig:not evenly spaced grid}
\end{figure}
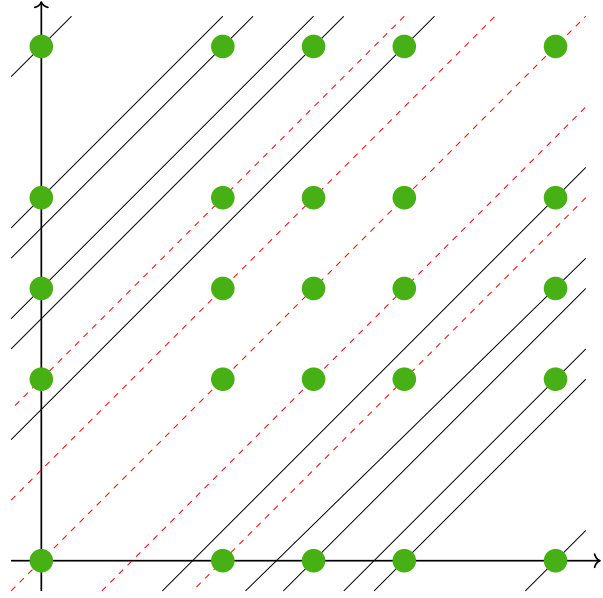

$\blacktriangleright$ Denoting by $h_{\max}$ and $h_{\min}$ the maximal and minimal eigenvalues of $\hat H$, the states saturating the simplified bound~\eqref{eq: general bound in finite dim} are given by
\begin{equation}\label{eq: saturate simplified eq}
    \ket{\psi} = \frac{1}{\sqrt{2}}\left[\ket{\psi_{\min}}^{\otimes n} + e^{i\varphi} \ket{\psi_{\max}}^{\otimes n}\right],
\end{equation}
where $\ket{\psi_{\min}}$ and $\ket{\psi_{\max}}$ are the eigenvectors associated with $h_{\min}$ and $h_{\max}$, respectively, and $\varphi \in \R$ is an arbitrary phase. Details for the derivation of these states are provided in Appendix~\ref{app: collective var derivation}, Result~\ref{res: popoviciu's}.

Up to the relative phase $e^{i\varphi}$, the states in Eq.~\eqref{eq: saturate simplified eq} correspond to Greenberger-Horne-Zeilinger-like states~\cite{greenberger_going_1989}, previously discussed in Ref.~\cite{pezze_quantum_2014}. There, it was shown that these states saturate Eq.~\eqref{eq: general bound in finite dim}. The present work provides a complete characterization of all states saturating the more general inequality~\eqref{eq: general bound collective global}. GHZ states, as well as their qudit generalizations,\footnote{For a $d$-dimensional Hilbert space with basis $\ket{1},\dots,\ket{d}$, GHZ-like states are commonly defined as $\ket{\text{GHZ}_d} = \sum_{j=1}^d \ket{j\cdots j}/\sqrt{d}$. They form a discrete subclass of the states in Eq.~\eqref{eq: first diagonal states} with uniform coefficients $C_\lambda = 1/\sqrt{d}$, which maximize $I$.} play a central role in quantum metrology~\cite{toth_quantum_2014}, quantum communication~\cite{durt_security_2004,cerf_security_2002}, and quantum computing~\cite{raussendorf_one-way_2001}. The quantifier introduced here could thus be used to assess the quality of experimentally generated states and to characterize imperfections in practical implementations.

It is worth emphasizing that the state in Eq.~\eqref{eq: saturate simplified eq} also saturates the original inequality~\eqref{eq: general bound collective global}. However, as shown above, the latter admits a significantly richer set of optimal states.

\paragraph{\publi Time-frequency case}
Time-frequency states saturating Eq.~\eqref{eq: general bound collective global} were analyzed in Sec.~\ref{subsec: TF optimal states}. Here, we reinterpret these results within the general geometric framework developed above. As discussed, the spectrum of the local operator plays a crucial role in determining the set of saturating states. In the present case, the spectrum of $\hat\omega$ is continuous and given by $\Lambda = \R$.

Since the analytical derivation has already been presented in Sec.~\ref{subsec: TF optimal states}, we focus here on the geometric interpretation. States saturating the bound must satisfy $\Delta^2 \hat P_j = 0$ for all $j \geq 2$, implying that the spectral amplitude $F(\omega_1,\dots,\omega_n)$ has support on a line directed by $\va{u} = (1,\dots,1)$ in $\R^n$. Unlike the finite-dimensional case, where the admissible support is restricted to discrete points, the full continuum of $\R^n$ is accessible here. This situation is illustrated in two dimensions in Fig.~\ref{fig: saturation continuous}.

As a result, any function $F$ whose support is concentrated along the collective direction $\va{u}$ defines a state saturating the bound. This leads to the general expression
\begin{equation}
    \ket{\psi} = \int \dd\Omega \, F(\Omega) \ket{\Omega + \omega_1^0,\dots,\Omega + \omega_n^0},
\end{equation}
where $F$ specifies the amplitude and phase distribution along the line, and the constants $\omega_i^0$ determine its global offset in $\R^n$.

\begin{figure}[ht]
    \centering
    \scalebox{0.8}{\begin{tikzpicture}
	\begin{pgfonlayer}{nodelayer}
		\node [style=none] (10) at (0, 17.5) {};
		\node [style=none] (11) at (0, -1) {};
		\node [style=none] (12) at (-1, 0) {};
		\node [style=none] (13) at (17.5, 0) {};
		\node [style=none] (68) at (4, 8) {};
		\node [style=none] (69) at (6, 10) {};
		\node [style=none] (70) at (4.5, 9.25) {$\va{u}$};
		\node [style=none] (71) at (6, 2) {};
		\node [style=none] (72) at (9, 5) {};
		\node [style=none] (73) at (12, 3) {};
		\node [style=none] (74) at (16, 7) {};
		\node [style=none] (75) at (9, 11) {};
		\node [style=none] (76) at (13, 15) {};
		\node [style=none] (77) at (1, 1) {};
		\node [style=none] (78) at (15, 15) {};
		\node [style=none] (79) at (1, 9) {};
		\node [style=none] (80) at (2, 10) {};
		\node [style=none] (81) at (2, 14) {};
		\node [style=none] (82) at (5, 17) {};
		\node [style=none] (83) at (11, 7) {};
		\node [style=none] (84) at (12, 8) {};
		\node [style=none] (85) at (13, 9) {};
		\node [style=none] (86) at (14, 10) {};
		\node [style=none] (87) at (6, 14) {};
		\node [style=none] (88) at (9, 17) {};
	\end{pgfonlayer}
	\begin{pgfonlayer}{edgelayer}
		\draw [style=Thick arrow] (11.center) to (10.center);
		\draw [style=Thick arrow] (12.center) to (13.center);
		\draw [style=Thick arrow] (68.center) to (69.center);
		\draw [style=red line] (71.center) to (72.center);
		\draw [style=red line] (73.center) to (74.center);
		\draw [style=red line] (75.center) to (76.center);
		\draw [style=Blue line] (78.center) to (77.center);
		\draw [style=red line] (79.center) to (80.center);
		\draw [style=red line] (81.center) to (82.center);
		\draw [style=red line] (83.center) to (84.center);
		\draw [style=red line] (85.center) to (86.center);
		\draw [style=dashes red] (72.center) to (83.center);
		\draw [style=dashes red] (84.center) to (85.center);
		\draw [style=red line] (87.center) to (88.center);
		\draw [style=dashes red] (80.center) to (87.center);
	\end{pgfonlayer}
\end{tikzpicture}}
    \caption[Spectral space for two time-frequency single-photon states]{Spectral space for two time-frequency single-photon states. Colored lines indicate possible supports of spectral amplitudes corresponding to states saturating Eq.~\eqref{eq: general bound collective global}. The blue line corresponds to the main diagonal, while red lines illustrate shifted collective directions. Dotted segments emphasize that the spectral support need not be connected. Adapted from Ref.~\cite{descamps_measuring_2025}. \copyright 2025 American Physical Society.}
    \label{fig: saturation continuous}
\end{figure}
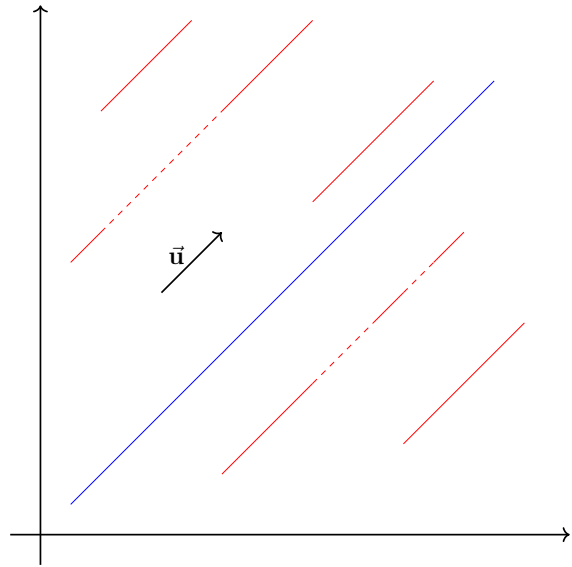

\subsection{Adding thickness}
\label{subsec: adding thickness}

\paragraph{\publi Thickness condition} In Sec.~\ref{subsec: TF optimal states}, we acknowledged that the time-frequency states saturating Eq.~\eqref{eq: general bound collective global} are unphysical, as they require infinite energy to be generated. More generally, states saturating Eq.~\eqref{eq: general bound collective global} are characterized by a vanishing thickness along all directions orthogonal to the collective variable $\hat H_\text{coll}$. While this condition can be met in finite-dimensional Hilbert spaces, it may be unattainable in practice due to experimental constraints.

In this section, we investigate how the presence of a non-zero thickness along these orthogonal directions modifies the upper bound on $I$, following an approach similar to that used in the time-frequency case. We model this constraint through the set of inequalities
\begin{equation}
    \forall j\geq 2,\,\Delta^2 \hat P_j \geq \zeta\, \Delta^2 \hat P_1,
\end{equation}
where the operators $\hat P_j$ were introduced in the proof of Eq.~\eqref{eq: general bound collective global}. The positive real parameter $\zeta$ controls how thin the states are allowed to be, and is related to the parameter $\eta$ introduced in Sec.~\ref{subsec: TF optimal states} via $\eta = 1 - \zeta$.

Although the operators $\hat P_j$ were initially introduced as mathematical tools for the derivation, their specific definition becomes relevant in the present context. Indeed, the condition $\Delta^2 \hat P_j \geq \zeta\, \Delta^2 \hat P_1$ is not invariant under a redefinition of the operators $\hat P_j$. The spectral representation provides an intuitive interpretation of this non-zero thickness condition. Figure~\ref{fig: non-zero thickness example} illustrates such states in spectral space for both a finite-dimensional system and a time-frequency system, in the case $n=2$. While the states are allowed to be predominantly elongated along the direction associated with the main collective variable, they must retain a finite extension along the orthogonal directions.

\paragraph{\publi Modified bound} Adapting the final step of the previous proof (see Eq.~\eqref{eq: last step proof inequ}), one obtains
\begin{equation}
    \max_j \Delta^2\hat H_j \geq \frac{1}{n}\Big(1 + \zeta (n-1)\Big)\Delta^2\hat P_1,
\end{equation}
which can be rearranged as
\begin{equation}
    I_{\ket{\psi}} \leq \frac{n^2}{(1-\zeta) + \zeta n}.
\end{equation}

A first insight into this bound can be gained by considering specific values of $\zeta$. For $\zeta = 0$, the constraint becomes trivial and one recovers Eq.~\eqref{eq: general bound collective global}. For $\zeta = 1$, the bound reduces to $I \leq n$, corresponding to a state $\ket{\psi}$ that is equally distributed along all directions. Finally, in the limit $\zeta \to \infty$, one finds $I \to 0$, which corresponds to a state that is infinitely thin along the direction of the collective operator $\hat H_\text{coll}$.

\begin{figure}[ht]
    \centering
    \footnotesize
    \scalebox{1.1}{\begin{tikzpicture}
	\begin{pgfonlayer}{nodelayer}
		\node [style=none] (0) at (-1, 0) {};
		\node [style=none] (1) at (0, -1) {};
		\node [style=none] (2) at (0, 9.5) {};
		\node [style=none] (3) at (9.5, 0) {};
		\node [style=none] (4) at (1.5, -0.25) {};
		\node [style=none] (5) at (1.5, 0) {};
		\node [style=none] (6) at (3.5, -0.25) {};
		\node [style=none] (7) at (3.5, 0) {};
		\node [style=none] (8) at (4.5, -0.25) {};
		\node [style=none] (9) at (4.5, 0) {};
		\node [style=none] (10) at (8, -0.25) {};
		\node [style=none] (11) at (8, 0) {};
		\node [style=none] (12) at (1.5, -0.25) {};
		\node [style=none] (13) at (0, 1.5) {};
		\node [style=none] (14) at (-0.25, 1.5) {};
		\node [style=none] (15) at (0, 3.5) {};
		\node [style=none] (16) at (-0.25, 3.5) {};
		\node [style=none] (17) at (0, 4.5) {};
		\node [style=none] (18) at (-0.25, 4.5) {};
		\node [style=none] (19) at (0, 8) {};
		\node [style=none] (20) at (-0.25, 8) {};
		\node [style=none] (21) at (1.25, 8) {};
		\node [style=none] (22) at (1.75, 8) {};
		\node [style=none] (23) at (1.5, 8.25) {};
		\node [style=none] (24) at (1.5, 7.75) {};
		\node [style=none] (25) at (1.25, 4.5) {};
		\node [style=none] (26) at (1.75, 4.5) {};
		\node [style=none] (27) at (1.5, 4.75) {};
		\node [style=none] (28) at (1.5, 4.25) {};
		\node [style=none] (29) at (1.25, 3.5) {};
		\node [style=none] (30) at (1.75, 3.5) {};
		\node [style=none] (31) at (1.5, 3.75) {};
		\node [style=none] (32) at (1.5, 3.25) {};
		\node [style=none] (37) at (3.25, 8) {};
		\node [style=none] (38) at (3.75, 8) {};
		\node [style=none] (39) at (3.5, 8.25) {};
		\node [style=none] (40) at (3.5, 7.75) {};
		\node [style=none] (49) at (3.25, 1.5) {};
		\node [style=none] (50) at (3.75, 1.5) {};
		\node [style=none] (51) at (3.5, 1.75) {};
		\node [style=none] (52) at (3.5, 1.25) {};
		\node [style=none] (53) at (4.25, 8) {};
		\node [style=none] (54) at (4.75, 8) {};
		\node [style=none] (55) at (4.5, 8.25) {};
		\node [style=none] (56) at (4.5, 7.75) {};
		\node [style=none] (57) at (4.25, 4.5) {};
		\node [style=none] (58) at (4.75, 4.5) {};
		\node [style=none] (59) at (4.5, 4.75) {};
		\node [style=none] (60) at (4.5, 4.25) {};
		\node [style=none] (61) at (4.25, 3.5) {};
		\node [style=none] (62) at (4.75, 3.5) {};
		\node [style=none] (63) at (4.5, 3.75) {};
		\node [style=none] (64) at (4.5, 3.25) {};
		\node [style=none] (65) at (4.25, 1.5) {};
		\node [style=none] (66) at (4.75, 1.5) {};
		\node [style=none] (67) at (4.5, 1.75) {};
		\node [style=none] (68) at (4.5, 1.25) {};
		\node [style=none] (73) at (7.75, 4.5) {};
		\node [style=none] (74) at (8.25, 4.5) {};
		\node [style=none] (75) at (8, 4.75) {};
		\node [style=none] (76) at (8, 4.25) {};
		\node [style=none] (77) at (7.75, 3.5) {};
		\node [style=none] (78) at (8.25, 3.5) {};
		\node [style=none] (79) at (8, 3.75) {};
		\node [style=none] (80) at (8, 3.25) {};
		\node [style=none] (81) at (7.75, 1.5) {};
		\node [style=none] (82) at (8.25, 1.5) {};
		\node [style=none] (83) at (8, 1.75) {};
		\node [style=none] (84) at (8, 1.25) {};
		\node [style=none] (85) at (0, 10.25) {$\lambda_2$};
		\node [style=none] (86) at (10.25, 0) {$\lambda_1$};
		\node [style=red dot] (87) at (1.5, 1.5) {};
		\node [style=red dot] (88) at (3.5, 3.5) {};
		\node [style=red dot] (89) at (3.5, 4.5) {};
		\node [style=red dot] (90) at (8, 8) {};
		\node [style=none] (94) at (5.25, 6.25) {};
		\node [style=none] (95) at (5.75, 5.75) {};
		\node [style=none] (96) at (6, 5.5) {};
		\node [style=none] (97) at (3.5, 6.5) {\scalebox{0.7}{$\Delta\hat P_2=\sqrt{\zeta}\Delta\hat P_1$}};
		\node [style=none] (98) at (1.5, 0.75) {};
		\node [style=none] (99) at (3.5, 2.75) {};
		\node [style=none] (100) at (8.75, 8) {};
		\node [style=none] (101) at (6, 4.25) {\scalebox{0.7}{$\Delta \hat  P_1$}};
		\node [style=none] (102) at (11.25, 0) {};
		\node [style=none] (103) at (12.25, -1) {};
		\node [style=none] (104) at (12.25, 9.5) {};
		\node [style=none] (105) at (21.75, 0) {};
		\node [style=none] (171) at (12.25, 10.25) {$\omega_2$};
		\node [style=none] (172) at (22.5, 0) {$\omega_1$};
		\node [style=none] (173) at (14, 0.5) {};
		\node [style=none] (174) at (20, 6.5) {};
		\node [style=none] (175) at (16.75, 3.75) {};
		\node [style=none] (176) at (17.25, 3.25) {};
		\node [style=none] (177) at (19, 3.25) {\scalebox{0.7}{$\Delta\hat P_2=\sqrt{\zeta}\Delta\hat P_1$}};
		\node [style=none] (178) at (18, 5.25) {};
		\node [style=none] (179) at (18.5, 4.75) {};
		\node [style=none] (180) at (18.75, 4.5) {};
		\node [style=none] (181) at (13.5, 1) {};
		\node [style=none] (182) at (15.5, 3) {};
		\node [style=none] (183) at (19.5, 7) {};
		\node [style=none] (184) at (15, 4.25) {\scalebox{0.7}{$\Delta \hat  P_1$}};
	\end{pgfonlayer}
	\begin{pgfonlayer}{edgelayer}
		\draw [style=Thick arrow] (0.center) to (3.center);
		\draw [style=Thick arrow] (1.center) to (2.center);
		\draw (5.center) to (12.center);
		\draw (7.center) to (6.center);
		\draw (9.center) to (8.center);
		\draw (11.center) to (10.center);
		\draw (20.center) to (19.center);
		\draw (18.center) to (17.center);
		\draw (16.center) to (15.center);
		\draw (14.center) to (13.center);
		\draw [style=Gray line] (23.center) to (24.center);
		\draw [style=Gray line] (21.center) to (22.center);
		\draw [style=Gray line] (27.center) to (28.center);
		\draw [style=Gray line] (25.center) to (26.center);
		\draw [style=Gray line] (31.center) to (32.center);
		\draw [style=Gray line] (29.center) to (30.center);
		\draw [style=Gray line] (39.center) to (40.center);
		\draw [style=Gray line] (37.center) to (38.center);
		\draw [style=Gray line] (51.center) to (52.center);
		\draw [style=Gray line] (49.center) to (50.center);
		\draw [style=Gray line] (55.center) to (56.center);
		\draw [style=Gray line] (53.center) to (54.center);
		\draw [style=Gray line] (59.center) to (60.center);
		\draw [style=Gray line] (57.center) to (58.center);
		\draw [style=Gray line] (63.center) to (64.center);
		\draw [style=Gray line] (61.center) to (62.center);
		\draw [style=Gray line] (67.center) to (68.center);
		\draw [style=Gray line] (65.center) to (66.center);
		\draw [style=Gray line] (75.center) to (76.center);
		\draw [style=Gray line] (73.center) to (74.center);
		\draw [style=Gray line] (79.center) to (80.center);
		\draw [style=Gray line] (77.center) to (78.center);
		\draw [style=Gray line] (83.center) to (84.center);
		\draw [style=Gray line] (81.center) to (82.center);
		\draw [style=Arrow] (95.center) to (96.center);
		\draw [style=Arrow] (95.center) to (94.center);
		\draw [style=Arrow] (99.center) to (100.center);
		\draw [style=Arrow] (99.center) to (98.center);
		\draw [style=Thick arrow] (102.center) to (105.center);
		\draw [style=Thick arrow] (103.center) to (104.center);
		\draw [style=Fill pink] (176.center)
			 to [in=-45, out=-135, looseness=0.25] (173.center)
			 to [in=-135, out=135, looseness=0.25] (175.center)
			 to [in=135, out=45, looseness=0.25] (174.center)
			 to [in=45, out=-45, looseness=0.25] cycle;
		\draw [style=Arrow] (179.center) to (180.center);
		\draw [style=Arrow] (179.center) to (178.center);
		\draw [style=Arrow] (182.center) to (183.center);
		\draw [style=Arrow] (182.center) to (181.center);
	\end{pgfonlayer}
\end{tikzpicture}}
    \caption[Examples of states with non-zero width along secondary variables in two dimensions]{Examples of states with non-zero width along secondary variables in two dimensions. Left: finite-dimensional Hilbert space. The crosses represent the spectral space, that is, all possible pairs $(\lambda_1,\lambda_2)$ of eigenvalues of the operators $\hat H_j$, while the filled red circle indicates the spectral support of a particular state. Right: time-frequency Hilbert space of two single-photon states. In both panels, the orthogonal arrows indicate the geometric extent of the state in spectral space, which is related to the variances of the corresponding operators. Adapted from Ref.~\cite{descamps_measuring_2025}. \copyright 2025 American Physical Society.}
    \label{fig: non-zero thickness example}
\end{figure}
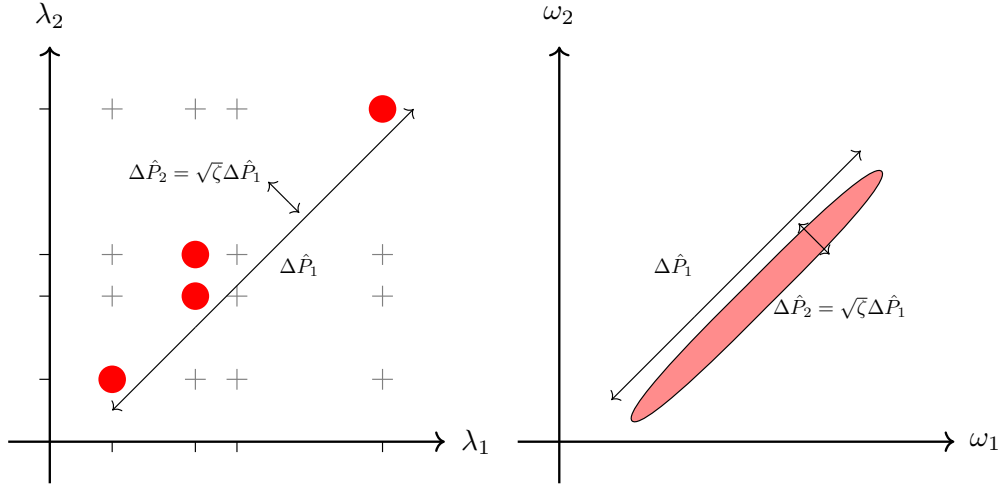

\subsection{Extension to mixed states}
\label{subsec: extension to mixed states}

\paragraph{\publi Extension guidelines} The bound of Eq.~\eqref{eq: general bound collective global} and the quantity $I$ can be extended to mixed states in several ways. Since the derivation of Eq.~\eqref{eq: general bound collective global} relies only on the bilinearity and positivity of the covariance, any construction based on such properties can be directly generalized. This applies in particular to the quantum variance, which naturally extends to a mixed state $\hat\rho$ as
\begin{equation}
    \Delta^2_{\hat \rho}\hat A=\Tr(\hat A^2\hat \rho)-\Tr(\hat A\hat \rho)^2,
\end{equation}
and to the quantum Fisher information $\mathcal Q_{\hat\rho}(\hat A)$, introduced and discussed in Sec.~\ref{subsec: quantum metrology}. Using the inequality $4\Delta^2_{\hat \rho}\hat A\geq \mathcal Q_{\hat \rho}(\hat A)$~\cite{toth_extremal_2013,yu_quantum_2013} (see also Appendix~\ref{app: formalism and framework}, Result~\ref{res: QFI variance inequality}), we obtain
\begin{equation}
     \frac{\mathcal Q_{\hat \rho}(\hat H_\text{coll})}{4\max\limits_j \Delta^2_{\hat \rho}(\hat H_j)}
     \leq
     \frac{\mathcal Q_{\hat \rho}(\hat H_\text{coll})}{\max\limits_j \mathcal Q_{\hat \rho}(\hat H_j)}
     \leq n^2,
\end{equation}
as well as
\begin{equation}
    \frac{\Delta^2_{\hat \rho}\hat H_\text{coll}}{\max\limits_j \Delta^2_{\hat \rho}\hat H_j}\leq n^2.
\end{equation}

In the case of non-zero thickness, assuming that $\mathcal Q(\hat P_j)\geq \zeta\, \mathcal Q(\hat P_1)$ for $j\geq 2$, or similarly $\Delta^2(\hat P_j)\geq \zeta\, \Delta^2(\hat P_1)$ for $j\geq 2$, allows one to modify the right-hand side in a similar way, namely $n^2\mapsto \frac{n^2}{(1-\zeta)+\zeta n}$.

The strategy we adopt to extend $I$ to mixed states follows closely the approach of Ref.~\cite{chen_wigner-yanase_2005}. That work begins by noting that for pure states of qubit systems, the variance of the collective spin observable $\hat H_\text{coll}=\hat Z_1+\cdots+\hat Z_n$ satisfies $\Delta^2_{\ket{\psi}}\hat H_\text{coll}\leq n^2$. The extension to mixed states is then achieved by identifying a convex generalization of the variance. One such extension is the quantum Fisher information, which is proportional to the convex roof of the variance~\cite{yu_quantum_2013} (see also Appendix~\ref{app: formalism and framework}, Result~\ref{res: convex roof} for a brief description of the convex roof construction). We adopt this construction in the following.

An alternative is provided by the Wigner-Yanase skew information~\cite{wigner_information_1963}, which is also convex and coincides with the variance for pure states. Although this choice is less powerful for entanglement detection, since it is always bounded from above by the quantum Fisher information, it has the advantage of being computationally simpler.

In the following, our goal is to construct quantities that are convex functions of the quantum state. While the quantum Fisher information itself is known to be convex~\cite{yu_quantum_2013} (see also Appendix~\ref{app: formalism and framework}, Result~\ref{res: convexity QFI}), the presence of a state-dependent denominator makes it non-trivial to preserve convexity for the previously proposed expressions. To address this issue, we propose three distinct constructions.

\paragraph{\publi Convex roof approach}
The first approach enforces convexity explicitly by considering the largest convex extension of $I$. This is achieved via the convex roof construction~\cite{toth_extremal_2013,yu_quantum_2013,uhlmann_roofs_2010}, defined as
\begin{align}\label{eq: convex roof def of F}
    I^\text{CR}_{\hat \rho}
    &=\inf_{\{p_j,\ket{\psi_j}\}} \sum_j p_j I_{\ket{\psi_j}}=\inf_{\{p_j,\ket{\psi_j}\}} \sum_j p_j
    \frac{\Delta^2_{\ket{\psi_j}}\hat H_\text{coll}}{\max_k\Delta^2_{\ket{\psi_j}}\hat H_k},
\end{align}
where the infimum is taken over all decompositions $\hat\rho=\sum_j p_j\ketbra{\psi_j}$ with $p_j>0$. Since a state $\ket{\psi_j}$ appearing in such a decomposition may satisfy
\begin{equation}
    \max_k\Delta^2_{\ket{\psi_j}}\hat H_k=0,
\end{equation}
we adopt the convention that $I=0$ for such states. One can verify that $I^\text{CR}_{\hat\rho}$
\begin{itemize}
    \item is upper bounded by $n^2$,
    \item reduces to the original definition for pure states,
    \item is convex in the quantum state, and
    \item is the largest function satisfying the two previous properties.
\end{itemize}
The first property follows directly from its validity on pure states and the fact that convex combinations cannot increase the value. The remaining properties are consequences of general results on convex roof constructions (see Appendix~\ref{app: collective var derivation}, Result~\ref{res: convex roof}).

While this construction is of fundamental theoretical importance, as it provides the optimal convex extension, it is difficult to interpret intuitively and is generally intractable from a computational perspective.

\paragraph{\publi Utilizing the support of $\hat\rho$}
The main obstacle to convexity lies in the denominator. The following definition preserves convexity by fixing the denominator through the support of the state:
\begin{equation}\label{eq: def F supp}
    I^\text{S}_{\hat \rho}
    =
    \frac{\mathcal Q_{\hat \rho}(\hat H_\text{coll})}
    {4\sup\limits_{\ket{\psi}\in\operatorname{Supp}(\hat\rho)} \max_j \Delta^2_{\ket{\psi}}\hat H_j}.
\end{equation}
The factor $4$ reflects the relation $\mathcal Q_{\ket{\psi}}(\hat H_\text{coll})=4\Delta^2_{\ket{\psi}}\hat H_\text{coll}$ for pure states. For finite-rank $\hat\rho$, the support $\operatorname{Supp}(\hat\rho)$ admits the equivalent characterizations (see Appendix~\ref{app: collective var derivation}, Result~\ref{res: def support})
\begin{itemize}
    \item $\ket{\psi}\in(\ker\hat\rho)^\perp$,
    \item $\ket{\psi}$ appears in a convex decomposition of $\hat\rho$ with non-zero weight.
\end{itemize}

One can verify that $I^\text{S}_{\hat\rho}$ is convex (see Appendix~\ref{app: collective var derivation}, Result~\ref{res: formula support conv}). For finite-rank states, the optimization is performed over a compact set, ensuring finiteness of the denominator and making this definition more tractable than Eq.~\eqref{eq: convex roof def of F}. For infinite-rank states, however, the supremum may diverge, in which case $I^\text{S}_{\hat\rho}=0$ and the construction loses operational meaning.

\paragraph{\publi Bounding the denominator}
A simpler, but less general, approach consists in bounding the denominator by a state-independent constant. If we restrict attention to states satisfying
\begin{equation}
    \max_j\Delta^2_{\hat \rho}(\hat H_j)\leq A,
\end{equation}
for some constant $A$, we can define
\begin{equation}
    I^\text{R}_{\hat\rho}
    =
    \frac{\mathcal Q_{\hat\rho}(\hat H_\text{coll})}{4A},
\end{equation}
which is convex by construction. This definition is computationally straightforward, but applies only to restricted classes of states and contains less detailed information about local fluctuations. In finite-dimensional Hilbert spaces, one may choose
\begin{equation}
    A=\frac{(h_{\max}-h_{\min})^2}{4},
\end{equation}
where $h_{\max}$ and $h_{\min}$ are the extremal eigenvalues of $\hat H$, yielding
\begin{equation}
    I^\text{R}_{\hat \rho}
    =
    \frac{\mathcal Q_{\hat \rho}(\hat H_\text{coll})}{(h_{\max}-h_{\min})^2},
\end{equation}
which satisfies $I^\text{R}_{\hat\rho}\leq n^2$.

\paragraph{\publi Discussion on the extensions}
We now briefly compare the advantages and limitations of the three extensions. They are ordered by decreasing computational complexity, with $I^\text{CR}$ being the most demanding and $I^\text{R}$ the simplest. They satisfy
\begin{equation}
    I^\text{R}\leq I^\text{S}\leq I^\text{CR}.
\end{equation}
The left inequality follows directly from bounding the denominator in $I^\text{S}$, while the right inequality reflects the fact that $I^\text{CR}$ is the largest convex extension coinciding with $I$ on pure states. Since violations of $I\leq n$ witness entanglement, larger extensions are more effective for entanglement detection, making $I^\text{CR}$ the most informative, albeit the least tractable.

To illustrate the behavior of these extensions under noise, we consider a simple toy model. Let $\ket{\psi}$ be an $n$-qubit state with $I_{\ket{\psi}}=n^2$. For $\epsilon\in[0,1]$, define
\begin{align}
    \hat \rho_\epsilon=(1-\epsilon)\ketbra{\psi}{\psi}+\epsilon\frac{\1}{2^n},
    &&
    \hat \sigma_\epsilon=(1-\epsilon)\ketbra{\psi}{\psi}+\epsilon\ketbra{\varphi}{\varphi},
\end{align}
where $\ket{\varphi}$ is any state such that $I_{\ket{\varphi}}=0$. The state $\hat\rho_\epsilon$ corresponds to depolarizing noise, while $\hat\sigma_\epsilon$ represents mixing with a minimally correlated state.

Details of the computation are given in Appendix~\ref{app: collective var derivation}, Result~\ref{res: mixed computation}. Although $I^\text{CR}_{\hat\rho_\epsilon}$ cannot be computed explicitly, we obtain
\begin{subequations}\label{eq: mixed state computation}
    \begin{equation}
        I_{\hat \rho_\epsilon}^\text{S} = I_{\hat \rho_\epsilon}^\text{R} = \frac{(1-\epsilon)^2}{(1-\epsilon)+\epsilon/2^{n-1}}\,n^2,
    \end{equation}
    \begin{equation}
        I_{\hat \sigma_\epsilon}^\text{CR} = I_{\hat \sigma_\epsilon}^\text{S} = I_{\hat \sigma_\epsilon}^\text{R} = (1-\epsilon)n^2.
    \end{equation}
\end{subequations}
These expressions interpolate smoothly between the maximal value $n^2$ at $\epsilon=0$ and zero as $\epsilon\to1$, confirming that noise degrades the detectable entanglement. The apparent equivalence of the extensions in this example is an artifact of working with qubit systems, where local variances are uniformly bounded. A more detailed comparison in higher-dimensional or continuous-variable systems is left for future work.

\subsection{$k$-entanglement}
\label{subsec: k sep}

\paragraph{\publi Original statement in a general setting}
\label{subsec: original statement in a general setting}
Following ideas developed in~\cite{guhne_entanglement_2009,hyllus_fisher_2012,chen_wigner-yanase_2005,pezze_quantum_2014}, we use the formula obtained previously to analyze the notion of $k$-entanglement. One can refine the usual separable-versus-entangled dichotomy by introducing a hierarchy of multipartite entanglement classes for states living in $\mathcal H^{\otimes n}$. The notion of $k$-entanglement is based on the following intuitive question: given $n$ parties, into how many groups can they be partitioned such that no quantum correlations are present between different groups?

Multipartite entanglement is a key resource in many protocols for sensing, computation, and communication. Understanding the degree of $k$-entanglement of a state is therefore essential, as it allows one to determine whether the available entanglement is sufficient for a given task. Moreover, this hierarchy can guide the design of protocols that operate with weakly entangled resources, which are often easier to prepare experimentally.

Formally, a pure state $\ket{\psi}$ is called $k$-separable, for an integer $1\leq k\leq n$, if it can be written as a tensor product of states, each acting on at most $k$ subsystems. That is, after a suitable permutation of the subsystems,
\begin{equation}
    \ket{\psi}=\ket{\phi_1}\otimes \cdots\otimes \ket{\phi_l},
\end{equation}
where each $\ket{\phi_j}\in\mathcal H^{\otimes r_j}$ with $r_j\leq k$. A state is said to be $k$-entangled if it is $k$-separable but not $(k-1)$-separable.

These definitions extend naturally to mixed states. A density operator $\hat\rho$ is $k$-separable if it can be written as a convex combination of pure $k$-separable states,
\begin{equation}
    \hat \rho = \sum_j p_j \ketbra{\psi_j},
\end{equation}
where each $\ket{\psi_j}$ is $k$-separable. As in the pure-state case, a mixed state is $k$-entangled if it is $k$-separable but not $(k-1)$-separable.

Following the proofs of~\cite{hyllus_fisher_2012,chen_wigner-yanase_2005,pezze_quantum_2014}, and using the fact that all $k$-partite pure states $\ket{\psi}$ satisfy $I_{\ket{\psi}}\leq k^2$, it follows that any $k$-separable pure state $\ket{\psi}$ must satisfy
\begin{equation}\label{eq: bound k entanglement}
    I_{\ket{\psi}}\leq \left\lfloor \frac{n}{k}\right\rfloor k^2
    +\left(n-\left\lfloor\frac{n}{k}\right\rfloor k\right)^2
    \leq nk.
\end{equation}
The left inequality coincides with bounds previously derived for the quantum Fisher information~\cite{hyllus_fisher_2012,pezze_quantum_2014} and for the Wigner-Yanase skew information~\cite{chen_wigner-yanase_2005}. The right inequality provides a simpler, although slightly weaker, upper bound that avoids integer parts. Equality between the two bounds holds whenever $n/k$ is an integer. 

As a consequence, one obtains a $k$-entanglement criterion: if inequality~\eqref{eq: bound k entanglement} is violated for some value of $k$, then the state must be at least $(k+1)$-entangled. As before, this statement extends to mixed states whenever the chosen extension of $I$ is convex.

\paragraph{\publi General statement}
The phenomenon described above admits a natural generalization. Consider again a Hilbert space $\mathcal H$ and two maps $I$ and $g$ defined as
\begin{equation}
    \begin{aligned}
        I&:\bigcup_{n=1}^{+\infty} \mathcal D(\mathcal H^{\otimes n}) \to \R,\\
        g&:\R\to \R,
    \end{aligned}
\end{equation}
where $\mathcal D(\cdot)$ denotes the set of density operators on the corresponding Hilbert space. Assume that the following conditions hold:
\begin{itemize}
    \item For all integers $n$, the map $I$ is convex on $\mathcal D(\mathcal H^{\otimes n})$.
    \item $I$ is sub-additive on pure product states, namely
    $I(\ketbra{\psi}\otimes\ketbra{\phi})\leq I(\ketbra{\psi})+I(\ketbra{\phi})$.
    \item The function $g$ is convex and satisfies $g(0)=0$.
    \item For all integers $n$, there exists a subset $\Gamma_n\subset\mathcal D(\mathcal H^{\otimes n})$ such that
    $I(\hat\rho)\leq g(n)$ for all $\hat\rho\in\Gamma_n$.
\end{itemize}
We define $k$-separability and $k$-entanglement over the set $\Gamma=\bigcup_n \Gamma_n$ by requiring that all states appearing in the corresponding decompositions belong to some $\Gamma_n$. Under these assumptions, the following general result holds. If $\hat\rho\in\mathcal D(\mathcal H^{\otimes n})$ is a $k$-separable state over $\Gamma$, then
\begin{equation}\label{eq: k sep general ineq}
    I(\hat \rho)\leq \left\lfloor \frac{n}{k}\right\rfloor g(k)
    + g\!\left(n-\left\lfloor \frac{n}{k}\right\rfloor k\right)
    \leq \frac{n}{k} g(k).
\end{equation}

A detailed proof is given in Appendix~\ref{app: collective var derivation}, Result~\ref{res: k sep}. The sets $\Gamma_n$ may coincide with the full state space $\mathcal D(\mathcal H^{\otimes n})$, or they may be strict subsets on which stronger bounds of the form $I(\hat\rho)\leq g(n)$ hold. This latter situation arises, for instance, for the states discussed in Sec.~\ref{subsec: adding thickness}, where an additional constraint on the thickness is imposed.

The rightmost inequality in Eq.~\eqref{eq: k sep general ineq} generalizes the simplified bound in Eq.~\eqref{eq: bound k entanglement}, yielding $I\leq \frac{n}{k}g(k)$. While this bound is generally less tight, it coincides with the optimal one whenever $k$ divides $n$.

In the remainder of this section, the quantifier $I$ is any extension to mixed states that satisfies both convexity and sub-additivity. Taking $\Gamma_n=\mathcal D(\mathcal H^{\otimes n})$ and $g:x\mapsto x^2$, the remaining assumptions are also fulfilled, and the general theorem reduces to inequality~\eqref{eq: bound k entanglement}. The flexibility of this formulation allows one to restrict attention to smaller classes of states $\Gamma_n$, for which stronger bounds may hold with different choices of $g$. In the following, we exploit this freedom in the context of non-zero thickness.

The detection of $k$-entanglement has been previously addressed in~\cite{hong_measure_2012}. Building on generalizations of concurrence~\cite{wootters_entanglement_1998}, the authors introduced a family of quantifiers $C_{k-\text{ME}}$ that depend explicitly on $k$ and are capable of detecting $k$-separable states. Determining the degree of entanglement of an unknown state thus requires the evaluation of several such quantities. By contrast, our approach relies on a single quantifier $I$, whose value provides partial but direct information about the entanglement structure. Moreover, the definition of $C_{k-\text{ME}}$ involves an optimization over all possible partitions of the $n$ subsystems, which grows rapidly with both $n$ and $k$, making the computation demanding even for pure states. From this perspective, the present criterion offers a computationally simpler alternative for witnessing $k$-entanglement.

\paragraph{\publi $k$-entanglement and non-zero thickness}
We now combine $k$-entanglement and the notion of non-zero thickness introduced in Sec.~\ref{subsec: adding thickness}. As discussed previously, both a low degree of multipartite entanglement (small $k$) and a large thickness in the orthogonal direction (large $\zeta$) tend to reduce the value of $I$. In realistic experimental settings, it may be difficult to prepare states that are simultaneously highly entangled and exhibit a very small thickness. Exploring the trade-off between these two features can therefore reveal experimentally accessible states that still achieve large values of $I$.

Consider a $k$-separable state $\ket{\psi}$ such that all states appearing in its decomposition satisfy the non-zero thickness condition $\Delta^2 \hat P_j \geq \zeta\, \Delta^2 \hat P_1$. One then obtains
\begin{equation}\label{eq: ineq for k sep and non-zero thickness}
    \begin{aligned}
        I &\leq \left\lfloor \frac{n}{k}\right\rfloor
        \frac{k^2}{(1-\zeta)+k\zeta}
        + \frac{\left(n-\left\lfloor\frac{n}{k}\right\rfloor k\right)^2}
        {(1-\zeta)+\left(n-\left\lfloor\frac{n}{k}\right\rfloor k\right)\zeta},\\
        &\leq \frac{kn}{(1-\zeta)+k\zeta}.
    \end{aligned}
\end{equation}
This situation fits naturally into the general framework described above, with $\Gamma_n$ defined as the set of states satisfying the thickness constraint and
$g:x\mapsto \frac{x^2}{(1-\zeta)+x\zeta}$, which is convex and satisfies $g(0)=0$.

Inequality~\eqref{eq: ineq for k sep and non-zero thickness} thus acts as a witness of at least $(k+1)$-entanglement under the assumption of non-zero thickness $\zeta$. This discussion is particularly relevant for time-frequency states. In finite-dimensional systems, all states are physical and, in principle, experimentally accessible. In contrast, in the time-frequency setting, perfectly correlated states that saturate the bound are non-normalizable and therefore unphysical. Only approximations can be realized experimentally, and the parameter $\zeta$ quantifies how close one can approach the ideal limit.

For a $k$-separable mixed state whose pure-state decomposition satisfies
$\Delta^2 \hat P_j \geq \zeta\, \Delta^2 \hat P_1$ for all $j\geq 2$, one obtains
\begin{equation}\label{eq: inequality with non-zero thickness}
    I\leq \frac{kn}{(1-\zeta)+k\zeta}.
\end{equation}
A violation of this inequality certifies $(k+1)$-entanglement. Since $I$ has a metrological interpretation as the ratio between the quantum Fisher information and a local variance, this bound quantifies the maximal metrological advantage achievable under simultaneous constraints on entanglement and thickness.

A particularly instructive comparison is between the situations $(k,\zeta=0)$ and $(k=n,\zeta)$. The former corresponds to a $k$-separable state without thickness constraints, while the latter describes a fully entangled state whose thickness is limited by $\zeta$. Requiring these two cases to yield the same upper bound on $I$ leads to
\begin{equation}
    kn=\frac{n^2}{(1-\zeta)+n\zeta}.
\end{equation}
Solving for $k$ or $\zeta$ gives
\begin{align}\label{eq: extreme tradeoff n zeta}
    k=\frac{n}{(1-\zeta)+n\zeta}, &&
    \zeta=\frac{n-k}{k(n-1)}.
\end{align}

\begin{figure}[ht]
    \centering
    \begin{tabular}{c c}
        \includegraphics[width=0.4\linewidth]{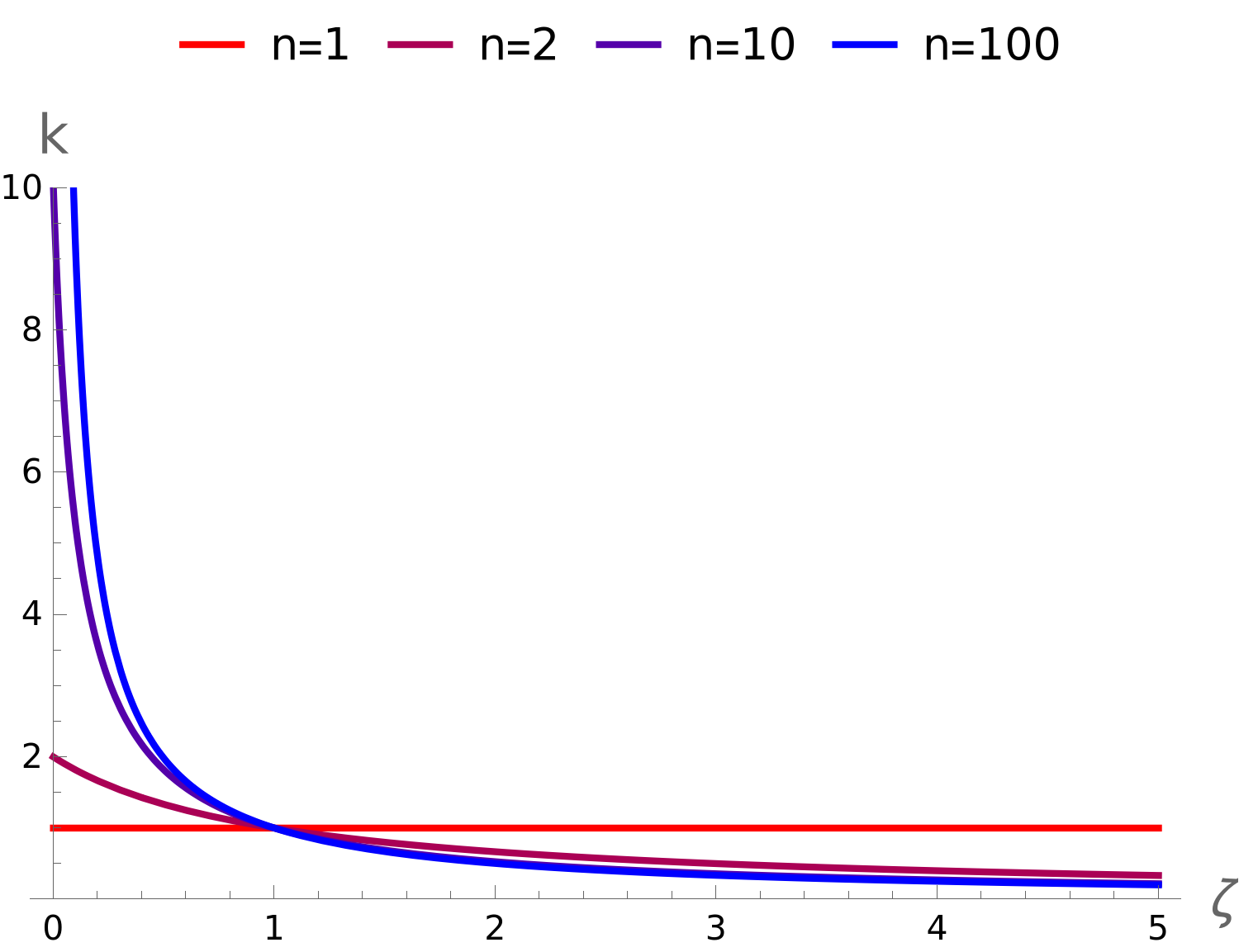}
        &
        \includegraphics[width=0.4\linewidth]{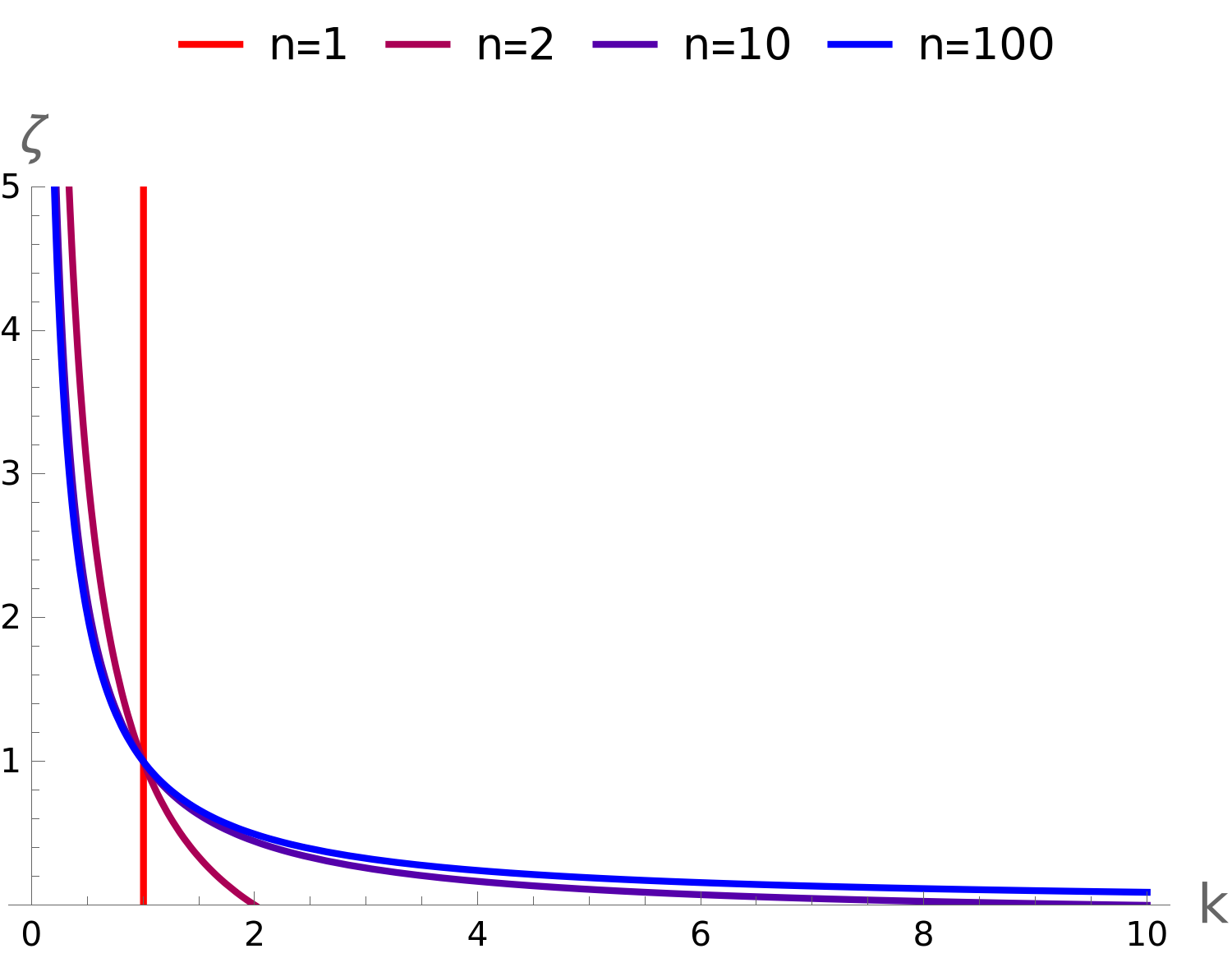}
    \end{tabular}
    \caption[Trade-off between $k$-entanglement and thickness $\zeta$]{Trade-off between $k$-entanglement and thickness $\zeta$ obtained from Eq.~\eqref{eq: extreme tradeoff n zeta}. The left panel shows $k$ as a function of $\zeta$, while the right panel shows $\zeta$ as a function of $k$, for different values of $n$. The two plots are related by inversion and can be viewed as reflections across the line $y=x$. All curves are monotonically decreasing, illustrating that larger thickness reduces the amount of multipartite entanglement required to achieve the same upper bound on $I$. Adapted from Ref.~\cite{descamps_measuring_2025}. \copyright 2025 American Physical Society.}
    \label{fig: plot varying n}
\end{figure}

The extreme cases are instructive. For $k=n$, one recovers $\zeta=0$, corresponding to identical situations. At the other extreme, $k=1$ yields $\zeta=1$, which corresponds to fully separable states exhibiting equal variance in all directions, $\Delta^2 \hat P_j=\Delta^2 \hat P_1$. In Fig. \ref{fig: plot varying n} we plot the relations of Eq. \eqref{eq: extreme tradeoff n zeta}.

More generally, for any target value $f\in[0,n^2]$, one may determine the pairs $(k,\zeta)$ satisfying
\begin{equation}
    \frac{kn}{(1-\zeta)+k\zeta}=f.
\end{equation}
Solving for $\zeta$ gives $\zeta=\frac{kn-f}{f(k-1)}$, which requires $kn\geq f$. Conversely, solving for $k$ yields $k=\frac{f(1-\zeta)}{n-f\zeta}$, with the constraint $k\leq n$ implying $f\leq \frac{n^2}{(1-\zeta)+n\zeta}$. Both conditions have a clear interpretation as upper bounds imposed by entanglement and thickness, respectively. We plot in Fig.~\ref{fig: plot varying f} both relations.

\begin{figure}[ht]
    \centering
    \begin{tabular}{c c}
        \includegraphics[width=0.4\linewidth]{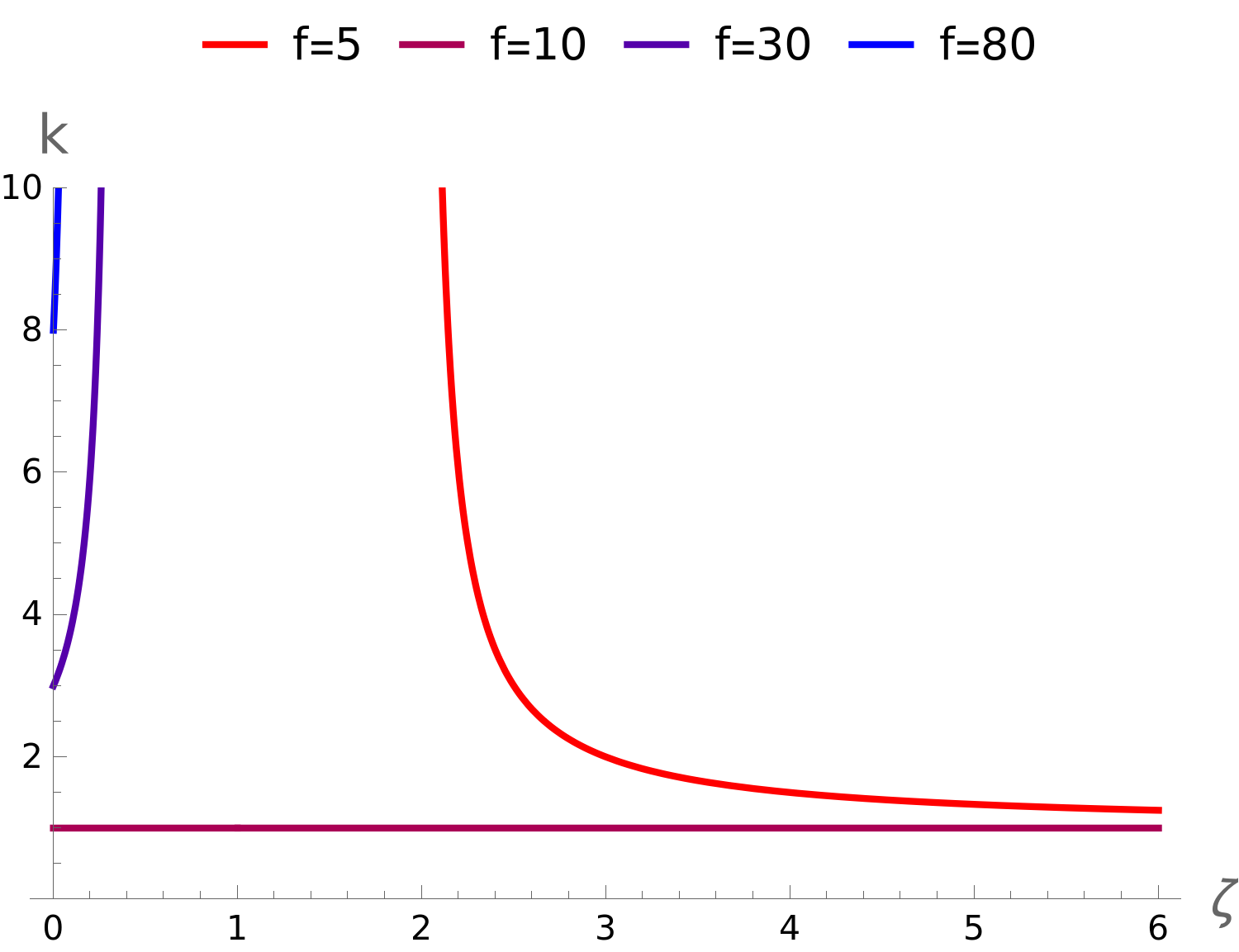}
        &
        \includegraphics[width=0.4\linewidth]{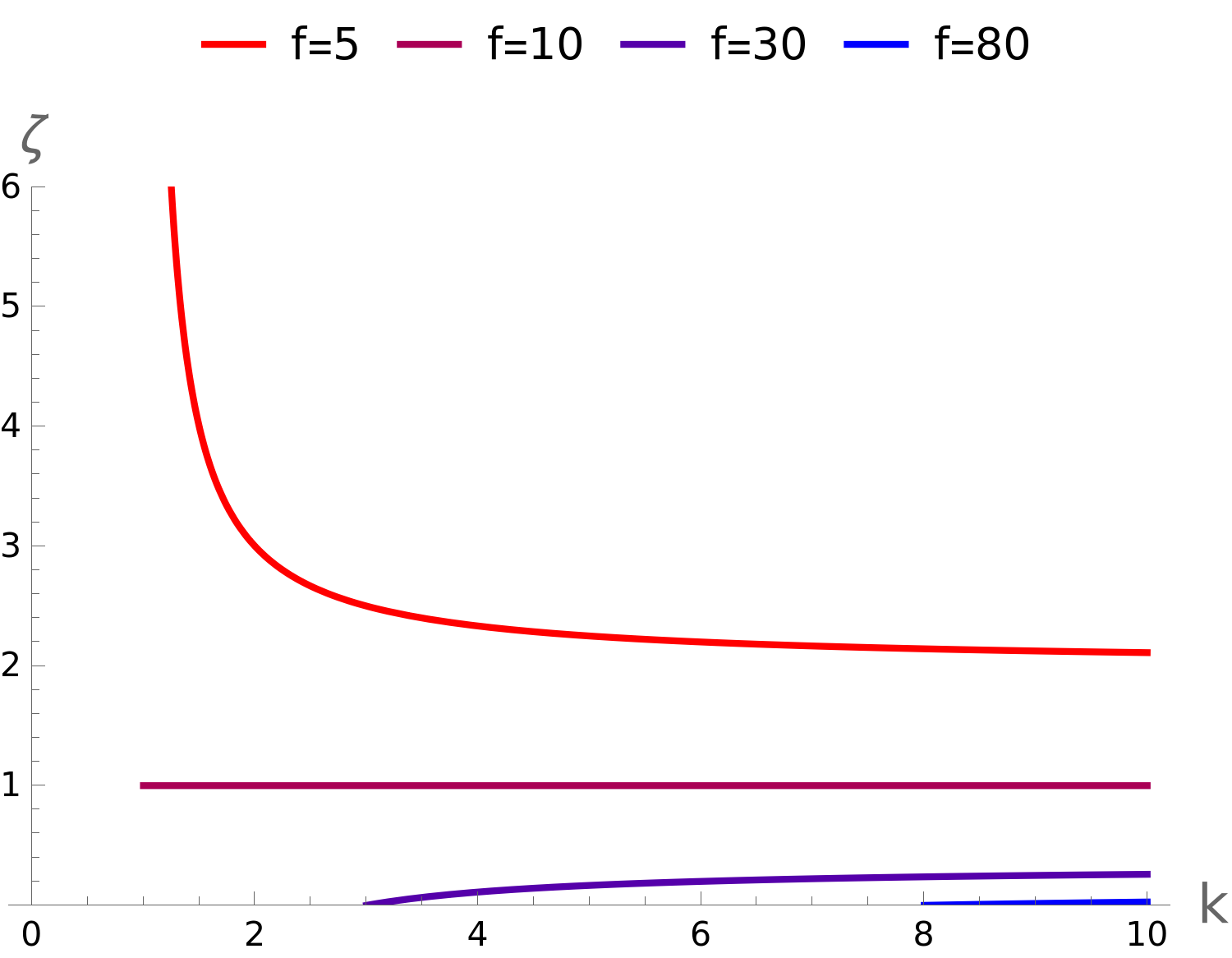}
    \end{tabular}
    \caption[Values of $k$ and $\zeta$ required to achieve a fixed value $I=f$ for $n=10$]{Values of $k$ and $\zeta$ required to achieve a fixed value $I=f$ for $n=10$. The left (right) panel shows the minimal value of $k$ (maximal value of $\zeta$) as a function of $\zeta$ ($k$). The two plots are related by inversion. The special case $f=n$ corresponds to either $k=1$ or $\zeta=1$, for which $I$ is independent of the remaining parameter. Larger target values of $f$ require simultaneously smaller thickness and larger multipartite entanglement. Adapted from Ref.~\cite{descamps_measuring_2025}. \copyright 2025 American Physical Society.}
    \label{fig: plot varying f}
\end{figure}

These results show that, in order to achieve a given value of the collective quantifier $I$, one may either increase the size $k$ of multipartite entanglement or reduce the thickness $\zeta$. In realistic scenarios, these two strategies may compete, and the relations derived above provide a quantitative guide for optimizing this trade-off.

\clearpage
\section{Encoding information in collective variables and GKP codes}
\label{sec: collective GKP codes}

\emph{Expanding on the notion of collective variables, this section demonstrates how one can robustly encode quantum information in collective degrees of freedom of time-frequency single photons with an adapted GKP code. This section is mainly based on \hyperlink{Article: prl GKP}{Gottesman-Kitaev-Preskill encoding in continuous modal variables of single photons}~\cite{descamps_gottesman-kitaev-preskill_2024} which was published during the PhD thesis.}

\subsection{Motivations}
\label{subsec: motivations GKP collective}

In the previous sections of this chapter, we have emphasized the importance of collective variables and entanglement for metrology and time estimation with time-frequency single-photon systems (Sec.~\ref{sec: TF entanglement and metrology}), and we have developed a general understanding of single-mode entanglement (Sec.~\ref{sec: Collective entanglement}). Beyond metrological applications, collective variables also provide a powerful resource for other quantum information tasks. In particular, they can be exploited for the robust encoding of quantum information, which is the focus of the present section.

Quantum error correction and stabilizer codes were introduced in Sec.~\ref{subsec: error correction}, where both discrete-variable and continuous-variable formulations were discussed. In particular, the Gottesman-Kitaev-Preskill (GKP) code~\cite{gottesman_encoding_2001} provides a continuous-variable analogue of discrete stabilizer codes~\cite{calderbank_quantum_1997,gottesman_class_1996}, enabling the correction of small displacement errors in phase space (see, e.g.,~\cite{grimsmo_quantum_2021}). We introduced GKP codes in the continuous-variable (CV) setting, where the two conjugate field quadratures play the role of position and momentum. Within this framework, physical noise processes such as photon loss or unwanted coupling to classical fields can be modeled as phase-space displacements~\cite{albert_performance_2018, joshi_quantum_2021}.

Despite their conceptual appeal, quadrature-based GKP codes pose significant experimental challenges. The preparation of highly non-classical grid states typically requires non-Gaussian resources, such as Schrödinger cat-like states, which are themselves difficult to generate deterministically~\cite{tzitrin_progress_2020}. While substantial progress has been made, with several theoretical proposals~\cite{walshe_continuous-variable_2020, bourassa_blueprint_2021, larsen_fault-tolerant_2021, zheng_gaussian_2023, eaton_non-gaussian_2019, dahan_creation_2023} and a first experimental realization~\cite{konno_logical_2024}, the implementation of GKP codes in propagating optical fields remains experimentally demanding. Alternative bosonic platforms, including superconducting circuits~\cite{campagne-ibarcq_quantum_2020} and trapped-ion motional modes~\cite{fluhmann_encoding_2019, fluhmann_sequential_2018}, offer complementary advantages but are less naturally suited for long-distance quantum communication or multi-user networks~\cite{appas_flexible_2021}.

The analogy between CV and time-frequency continuous variables of single photons, introduced in Sec.~\ref{subsec: TF analogy CV}, naturally permits the realization of GKP encoding based on collective time-frequency degrees of freedom of single photons. Beyond single-spatial-mode encoding, in this section, we explore GKP encoding in quantum optics, using the collective variables of single photons. Specifically, we consider encoding logical information in the time and frequency degrees of freedom of multiple photons, treated as collective continuous variables. This approach builds on the formalism of collective modes previously introduced and can be extended to other single-photon continuous degrees of freedom, such as transverse position and momentum~\cite{tasca_continuous-variable_2011}, propagation direction~\cite{solntsev_generation_2014}, or even collective modes of massive particles, such as the normal modes of trapped ions~\cite{steane_ion_1997, james_quantum_1998}.

A key conceptual difference with respect to quadrature-based GKP encoding lies in the physical origin of the protected phase space. In the time-frequency scheme, the conjugate variables are not associated with the amplitude and phase of a single bosonic mode, but instead emerge as collective degrees of freedom of several photons in distinguishable modes. As a consequence, error mechanisms, loss processes, and the scaling of protection with the number of physical constituents acquire a different interpretation. In particular, photon loss does not merely correspond to a displacement error but directly affects the structure of the collective variable, leading to distinct error-correction properties compared to CV encodings.

Using this collective-variable framework, we analyze how GKP codewords can be defined in a system of $n$ photons and investigate several key properties of the resulting encoded states, including the scaling of logical error rates with photon number, robustness against photon loss, and the impact of imperfect state preparation and measurement. Remarkably, many of the characteristic features of GKP codes are recovered, while their physical implementation relies on experimentally accessible resources.

Importantly, the generation and manipulation of time-frequency GKP-like states are already within reach of current experimental platforms. Entangled photon pairs exhibiting comb-like temporal or spectral correlations have been observed in a variety of contexts~\cite{appas_flexible_2021, fabre_generation_2020, maltese_generation_2020, kues_-chip_2017, imany_50-ghz-spaced_2018, olislager_frequency-bin_2010, yamazaki_massive-mode_2022, barros_free-space_2017, morrison_frequency-bin_2022, lukens_frequency-encoded_2017, sabattoli_silicon_2022, zhang_integrated_2018, clementi_programmable_2023, kaneda_direct_2019}. These systems naturally realize the type of collective continuous-variable structure required for time-frequency GKP encoding, making them promising candidates for near-term demonstrations of error-protected quantum information processing with propagating single photons.

\subsection{Time-frequency GKP codes}
\label{subsec: TF GKP}

\paragraph{\lit Single-spatial-mode code}
To understand how GKP encoding can be adapted to the time-frequency regime, we start with a simple example. We rely on the analogy between CV quadratures and time-frequency variables introduced in Sec.~\ref{subsec: TF analogy CV}, and closely follow the construction presented in Sec.~\ref{subsec: error correction}. A single qubit encoded in the time-frequency degree of freedom of a single photon is described by the two logical states
\begin{align}
    \ket{\overline 0}=\sum_{k\in\Z}\ket{2k\Omega_0}, && \ket{\overline 1}=\sum_{k\in\Z}\ket{(2k+1)\Omega_0}.
\end{align}
The spectrum of each logical state is therefore a comb of Dirac delta peaks separated by $2\Omega_0$. The associated code space is stabilized by the operators
\begin{align}
    \hat S_1=e^{-2i\Omega_0\hat t}, && \hat S_2=e^{-2i\pi\hat \omega/\Omega_0},
\end{align}
while the logical Pauli operators are implemented as
\begin{align}
    \overline X=e^{-i\Omega_0\hat t}, && \overline Z=e^{-i\pi\hat \omega/\Omega_0}.
\end{align}
This code is naturally suited to correct small time shifts $\delta t$ and frequency shifts $\delta \omega$, generated respectively by
\begin{align}
    \hat D_{\omega}(\delta t)= e^{-i\delta t \hat \omega}, && \hat D_{t}(\delta \omega)= e^{-i\delta \omega \hat t}.
\end{align}
Such errors can be detected via stabilizer measurements and corrected by appropriate displacements, provided that they are sufficiently small, \ie,
\begin{align}
    \abs{\delta t}<\frac{\pi}{2\Omega_0}, && \abs{\delta \omega}<\frac{\Omega_0}{2}.
\end{align}

It is important to emphasize that, although the mathematical structure of time-frequency GKP codes closely resembles that of quadrature-based GKP codes, the physical interpretation of errors and of their correction mechanisms is fundamentally different. In the quadrature formulation, errors are modeled as displacements in the quadrature phase space. In contrast, in the time-frequency formulation, errors correspond to shifts in the time-frequency phase space. As discussed in Sec.~\ref{subsec: TF Wigner function}, the quadrature and time-frequency phase spaces describe different physical aspects of the same underlying system. In particular, a time shift acts as a translation in the time-frequency phase space. However, when expressed in the quadrature representation, the same operation corresponds to a rotation in quadrature phase space. Time-frequency GKP codes are then intrinsically robust against rotations when viewed from the quadrature perspective. In other words, operations that would appear as rotational distortions in the quadrature phase space correspond to correctable shift errors in the time-frequency phase space.

\paragraph{\lit A two-mode GKP code}
Before introducing diagonal GKP codes, we show how a simple argument leads to a GKP-like construction that encodes a single qubit in the joint spectrum of two single photons. The idea is to use Dirac peaks arranged periodically along both frequency directions $\omega_1$ and $\omega_2$. Choosing a period $2\Omega_0$, the code words must be stabilized by the operators $\hat S_1=e^{-2i\Omega_0\hat t_1}$ and $\hat S_2=e^{-2i\Omega_0\hat t_2}$.

These two stabilizers alone are not sufficient to restrict the code space to a finite dimension. Additional stabilizers generated by frequency operators are required. One could choose $\hat\omega_1$ and $\hat\omega_2$, corresponding to $e^{-2i\pi\hat\omega_1/\Omega_0}$ and $e^{-2i\pi\hat \omega_2/\Omega_0}$, but this would lead to a four-dimensional code space, equivalent to two independent copies of the single-mode code. To obtain a two-dimensional code space, we instead use diagonal generators, namely $\hat \omega_1+\hat\omega_2$ and $\hat \omega_1-\hat\omega_2$.

Consider the stabilizer $\hat S_3=e^{-i\pi(\hat\omega_1+\hat\omega_2)/\Omega_0}$. Requiring that a generic state $\ket{\psi}=\int \dd\omega_1 \dd\omega_2\, F(\omega_1,\omega_2)\ket{\omega_1,\omega_2}$ be stabilized by $\hat S_3$ implies
\begin{equation}
    \int \dd\omega_1 \dd\omega_2\, e^{-i(\omega_1+\omega_2)\pi/\Omega_0}F(\omega_1,\omega_2)\ket{\omega_1,\omega_2}
    =
    \int \dd\omega_1 \dd\omega_2\, F(\omega_1,\omega_2)\ket{\omega_1,\omega_2}.
\end{equation}
Hence, for all $\omega_1$ and $\omega_2$, one must have $e^{-i(\omega_1+\omega_2)\pi/\Omega_0}F(\omega_1,\omega_2)=F(\omega_1,\omega_2)$. As in the single-mode case, this condition implies that $F(\omega_1,\omega_2)$ can be nonzero only when $\pi(\omega_1+\omega_2)/\Omega_0=0\,[2\pi]$, \ie, $\omega_2=2k\Omega_0-\omega_1$ for $k\in\Z$. This constraint corresponds to the red dotted lines in Fig.~\ref{fig: JSA 2D gkp code}. A similar argument for $\hat S_4=e^{-i\pi(\hat\omega_1-\hat\omega_2)/\Omega_0}$ shows that stabilized states must satisfy $\omega_2=2k\Omega_0+\omega_1$, corresponding to the blue dotted lines in Fig.~\ref{fig: JSA 2D gkp code}.

The joint spectral amplitude (JSA) of a state stabilized by both $\hat S_3$ and $\hat S_4$ is therefore supported only on the intersection points of these red and blue lines. Together with the $2\Omega_0$ periodicity imposed by $\hat S_1$ and $\hat S_2$, the code space stabilized by the four operators
\begin{align}
    \hat S_1=e^{-2i\Omega_0\hat t_1}, &&
    \hat S_2=e^{-2i\Omega_0\hat t_2}, &&
    \hat S_3=e^{-i\pi(\hat \omega_1+\hat \omega_2)/\Omega_0}, &&
    \hat S_4=e^{-i\pi(\hat\omega_1-\hat\omega_2)/\Omega_0},
\end{align}
is two-dimensional, with logical basis
\begin{align}
    \ket{\overline 0}=\sum_{k,l\in\Z} \ket{2\Omega_0k,2\Omega_0l}, &&
    \ket{\overline 1}=\sum_{k,l\in\Z}\ket{(2k+1)\Omega_0,(2l+1)\Omega_0}.
\end{align}

The logical operators can be chosen as
\begin{align}
    \hat Z= e^{-i\pi\hat \omega_1/\Omega_0}, && \hat X=e^{-i\Omega_0(\hat t_1+\hat t_2)}.
\end{align}
A straightforward check of the commutation relations shows that all stabilizers commute, that $\hat X$ and $\hat Z$ anticommute, and that both commute with the stabilizers.

For error correction, we consider errors of the form
\begin{align}
    \hat D_t(\alpha,\beta)=e^{-i(\alpha\hat t_1+\beta \hat t_2)}, && \hat D_\omega(\alpha,\beta)=e^{-i(\alpha\hat \omega_1+\beta\hat \omega_2)},
\end{align}
or products thereof. As in the single-mode case, one can determine correctability by analyzing how these errors commute with the stabilizers. A more intuitive and faster approach is graphical. In Fig.~\ref{fig: JSA 2D gkp code}, the central gray region corresponds to points closer to the central red node than to any other, in particular the surrounding green ones. A two-mode frequency shift can be corrected provided it lies within this region. From the equations of the boundaries, one finds that the error $\hat D_t(\alpha,\beta)$ is correctable if
\begin{align}
    \abs{\alpha+\beta}<\Omega_0, && \abs{\alpha-\beta}<\Omega_0.
\end{align}

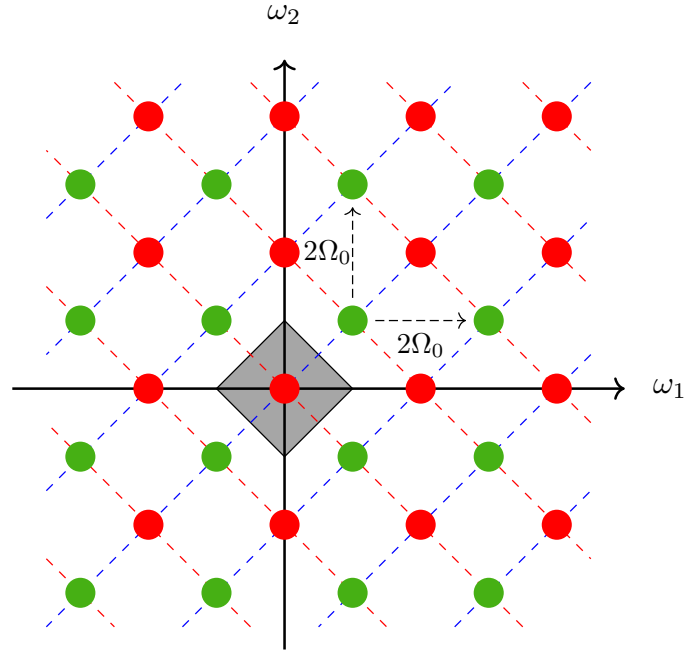
\begin{figure}[ht]
    \centering
    \footnotesize
    \scalebox{1.2}{\begin{tikzpicture}
	\begin{pgfonlayer}{nodelayer}
		\node [style=none] (0) at (6, 7.25) {};
		\node [style=none] (1) at (6, -5.75) {};
		\node [style=none] (2) at (0, 0) {};
		\node [style=none] (3) at (13.5, 0) {};
		\node [style=red dot] (297) at (6, 0) {};
		\node [style=red dot] (298) at (6, 3) {};
		\node [style=red dot] (299) at (9, 3) {};
		\node [style=red dot] (300) at (9, 0) {};
		\node [style=green dot] (301) at (7.5, 1.5) {};
		\node [style=green dot] (302) at (7.5, 4.5) {};
		\node [style=green dot] (303) at (10.5, 1.5) {};
		\node [style=green dot] (304) at (10.5, 4.5) {};
		\node [style=red dot] (307) at (3, 3) {};
		\node [style=red dot] (308) at (3, 0) {};
		\node [style=green dot] (309) at (1.5, 1.5) {};
		\node [style=green dot] (310) at (1.5, 4.5) {};
		\node [style=green dot] (311) at (4.5, 1.5) {};
		\node [style=green dot] (312) at (4.5, 4.5) {};
		\node [style=red dot] (316) at (3, 6) {};
		\node [style=red dot] (321) at (6, 6) {};
		\node [style=red dot] (324) at (9, 6) {};
		\node [style=red dot] (331) at (3, -3) {};
		\node [style=green dot] (333) at (1.5, -4.5) {};
		\node [style=green dot] (334) at (1.5, -1.5) {};
		\node [style=green dot] (335) at (4.5, -4.5) {};
		\node [style=green dot] (336) at (4.5, -1.5) {};
		\node [style=red dot] (338) at (6, -3) {};
		\node [style=red dot] (339) at (9, -3) {};
		\node [style=green dot] (341) at (7.5, -4.5) {};
		\node [style=green dot] (342) at (7.5, -1.5) {};
		\node [style=green dot] (343) at (10.5, -4.5) {};
		\node [style=green dot] (344) at (10.5, -1.5) {};
		\node [style=red dot] (345) at (12, 6) {};
		\node [style=red dot] (353) at (12, 0) {};
		\node [style=red dot] (354) at (12, 3) {};
		\node [style=red dot] (360) at (12, -3) {};
		\node [style=none] (389) at (6, 1.5) {};
		\node [style=none] (390) at (7.5, 0) {};
		\node [style=none] (391) at (6, -1.5) {};
		\node [style=none] (392) at (4.5, 0) {};
		\node [style=none] (393) at (14.5, 0) {$\omega_1$};
		\node [style=none] (394) at (6, 8.25) {$\omega_2$};
		\node [style=none] (396) at (6.95, 3) {\scalebox{0.9}{$2\Omega_0$}};
		\node [style=none] (397) at (0.75, 5.25) {};
		\node [style=none] (398) at (2.25, 6.75) {};
		\node [style=none] (399) at (5.25, 6.75) {};
		\node [style=none] (400) at (8.25, 6.75) {};
		\node [style=none] (401) at (11.25, 6.75) {};
		\node [style=none] (404) at (12.75, 5.25) {};
		\node [style=none] (405) at (12.75, 2.25) {};
		\node [style=none] (406) at (12.75, -0.75) {};
		\node [style=none] (407) at (12.75, -3.75) {};
		\node [style=none] (408) at (11.25, -5.25) {};
		\node [style=none] (409) at (8.25, -5.25) {};
		\node [style=none] (410) at (5.25, -5.25) {};
		\node [style=none] (411) at (2.25, -5.25) {};
		\node [style=none] (416) at (0.75, -3.75) {};
		\node [style=none] (417) at (0.75, -0.75) {};
		\node [style=none] (418) at (0.75, 2.25) {};
		\node [style=none] (422) at (3.75, 6.75) {};
		\node [style=none] (423) at (6.75, 6.75) {};
		\node [style=none] (424) at (9.75, 6.75) {};
		\node [style=none] (425) at (12.75, 6.75) {};
		\node [style=none] (426) at (12.75, 3.75) {};
		\node [style=none] (427) at (12.75, 0.75) {};
		\node [style=none] (428) at (12.75, -2.25) {};
		\node [style=none] (431) at (9.75, -5.25) {};
		\node [style=none] (432) at (6.75, -5.25) {};
		\node [style=none] (433) at (3.75, -5.25) {};
		\node [style=none] (434) at (0.75, -5.25) {};
		\node [style=none] (435) at (0.75, -2.25) {};
		\node [style=none] (436) at (0.75, 0.75) {};
		\node [style=none] (437) at (0.75, 3.75) {};
		\node [style=none] (440) at (8, 1.5) {};
		\node [style=none] (441) at (10, 1.5) {};
		\node [style=none] (442) at (7.5, 4) {};
		\node [style=none] (443) at (7.5, 2) {};
		\node [style=none] (444) at (9, 0.95) {\scalebox{0.9}{$2\Omega_0$}};
	\end{pgfonlayer}
	\begin{pgfonlayer}{edgelayer}
		\draw [style=Fill grey] (391.center)
			 to (392.center)
			 to (389.center)
			 to (390.center)
			 to cycle;
		\draw [style=Thick arrow] (2.center) to (3.center);
		\draw [style=Thick arrow] (1.center) to (0.center);
		\draw [style=dashes red] (401.center) to (404.center);
		\draw [style=dashes red] (400.center) to (405.center);
		\draw [style=dashes red] (406.center) to (399.center);
		\draw [style=dashes red] (398.center) to (407.center);
		\draw [style=dashes red] (408.center) to (397.center);
		\draw [style=dashes red, in=135, out=-45] (418.center) to (409.center);
		\draw [style=dashes red] (410.center) to (417.center);
		\draw [style=dashes red] (416.center) to (411.center);
		\draw [style=dashes blue] (437.center) to (422.center);
		\draw [style=dashes blue] (423.center) to (436.center);
		\draw [style=dashes blue] (435.center) to (424.center);
		\draw [style=dashes blue] (425.center) to (434.center);
		\draw [style=dashes blue] (433.center) to (426.center);
		\draw [style=dashes blue] (427.center) to (432.center);
		\draw [style=dashes blue] (431.center) to (428.center);
		\draw [style=dashed arrow] (443.center) to (442.center);
		\draw [style=dashed arrow] (440.center) to (441.center);
	\end{pgfonlayer}
\end{tikzpicture}}
    \caption[Joint spectral amplitude of a two-mode GKP code]{Joint spectral amplitude of a two-mode GKP code. Red points represent the logical state $\ket{\overline 0}$, while green points correspond to $\ket{\overline 1}$. The gray region indicates the domain in which a frequency shift error can be uniquely identified and corrected. The code words are $2\Omega_0$-periodic along both frequency axes, as indicated by the dotted arrows.}
    \label{fig: JSA 2D gkp code}
\end{figure}

To analyze the correction of time-shift errors $\hat D_\omega(\alpha,\beta)$, similar reasoning applies, now in the temporal domain. Rather than explicitly computing Fourier transforms, one can again rely on the stabilizers. The operators $\hat S_3$ and $\hat S_4$ impose a $\sqrt{2}\pi/\Omega_0$ periodicity along the two diagonal temporal directions, as shown by the dotted arrows in Fig.~\ref{fig: JTA 2D gkp code}. Moreover, $\hat S_1=e^{-2i\Omega_0\hat t_1}$ constrains the temporal spectrum to nonzero values only at $t_1=k\pi/\Omega_0$ for $k\in\Z$, corresponding to vertical dotted lines, while $\hat S_2=e^{-2i\Omega_0\hat t_2}$ similarly enforces $t_2=k\pi/\Omega_0$, yielding horizontal dotted lines. Together, these constraints leave only two degrees of freedom, represented by the red and green peaks. Consistency with the action of $\hat X$ and $\hat Z$ identifies these as the logical states $\ket{\overline +}$ (red) and $\ket{\overline -}$ (green). The region of correctability for time shifts, shown in gray, is again defined by the points closest to the central node. One finds that $\hat D_\omega(\alpha,\beta)$ can be corrected provided
\begin{align}
    \abs{\alpha}<\frac{\pi}{2\Omega_0}, && \abs{\beta}<\frac{\pi}{2\Omega_0}.
\end{align}
As a recap, a general error $\hat D_t(\alpha,\beta)\hat D_\omega(\gamma,\delta)$ is correctable if
\begin{align}
    \abs{\alpha+\beta}<\Omega_0 &&
    \abs{\alpha-\beta}<\Omega_0 &&
    \abs{\gamma}<\frac{\pi}{2\Omega_0} &&
    \abs{\delta}<\frac{\pi}{2\Omega_0}.
\end{align}

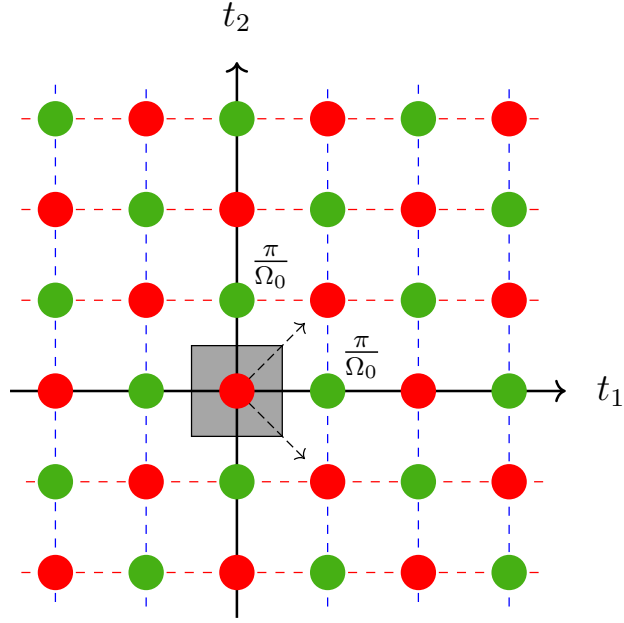
\begin{figure}[ht]
    \centering
    \scalebox{1.2}{\begin{tikzpicture}
	\begin{pgfonlayer}{nodelayer}
		\node [style=none] (0) at (6, 7.25) {};
		\node [style=none] (1) at (6, -5) {};
		\node [style=red dot] (297) at (6, 0) {};
		\node [style=red dot] (298) at (6, 4) {};
		\node [style=red dot] (299) at (10, 0) {};
		\node [style=red dot] (300) at (8, 2) {};
		\node [style=green dot] (302) at (12, -4) {};
		\node [style=red dot] (307) at (4, -2) {};
		\node [style=red dot] (308) at (4, 2) {};
		\node [style=green dot] (310) at (12, 0) {};
		\node [style=green dot] (312) at (12, 4) {};
		\node [style=red dot] (316) at (4, 6) {};
		\node [style=red dot] (321) at (10, 4) {};
		\node [style=red dot] (324) at (8, 6) {};
		\node [style=red dot] (331) at (2, 0) {};
		\node [style=red dot] (338) at (6, -4) {};
		\node [style=red dot] (339) at (8, -2) {};
		\node [style=red dot] (345) at (12, 6) {};
		\node [style=red dot] (353) at (12, -2) {};
		\node [style=red dot] (354) at (12, 2) {};
		\node [style=red dot] (360) at (10, -4) {};
		\node [style=none] (389) at (5, 1) {};
		\node [style=none] (390) at (7, 1) {};
		\node [style=none] (391) at (7, -1) {};
		\node [style=none] (392) at (5, -1) {};
		\node [style=none] (393) at (14.25, 0) {$t_1$};
		\node [style=none] (394) at (6, 8.25) {$t_2$};
		\node [style=none] (395) at (8.75, 0.75) {$\frac{\pi}{\Omega_0}$};
		\node [style=none] (396) at (6.75, 2.75) {$\frac{\pi}{\Omega_0}$};
		\node [style=none] (397) at (12.75, -4) {};
		\node [style=none] (398) at (12.75, -2) {};
		\node [style=none] (399) at (12.75, 2) {};
		\node [style=none] (400) at (1.25, 4) {};
		\node [style=none] (401) at (1.25, 6) {};
		\node [style=none] (404) at (12.75, 6) {};
		\node [style=none] (405) at (12.75, 4) {};
		\node [style=none] (406) at (1.25, 2) {};
		\node [style=none] (407) at (1.25, -2) {};
		\node [style=none] (408) at (1.25, -4) {};
		\node [style=none] (424) at (12, 6.75) {};
		\node [style=none] (425) at (10, 6.75) {};
		\node [style=none] (426) at (8, 6.75) {};
		\node [style=none] (427) at (4, -4.75) {};
		\node [style=none] (428) at (2, -4.75) {};
		\node [style=none] (431) at (2, 6.75) {};
		\node [style=none] (432) at (4, 6.75) {};
		\node [style=none] (433) at (8, -4.75) {};
		\node [style=none] (434) at (10, -4.75) {};
		\node [style=none] (435) at (12, -4.75) {};
		\node [style=red dot] (438) at (2, 4) {};
		\node [style=red dot] (439) at (2, -4) {};
		\node [style=green dot] (440) at (4, 0) {};
		\node [style=green dot] (441) at (4, 4) {};
		\node [style=green dot] (442) at (8, 0) {};
		\node [style=green dot] (443) at (6, 2) {};
		\node [style=green dot] (444) at (2, -2) {};
		\node [style=green dot] (445) at (2, 2) {};
		\node [style=green dot] (446) at (2, 6) {};
		\node [style=green dot] (447) at (8, 4) {};
		\node [style=green dot] (448) at (6, 6) {};
		\node [style=green dot] (450) at (4, -4) {};
		\node [style=green dot] (451) at (6, -2) {};
		\node [style=green dot] (452) at (10, 6) {};
		\node [style=green dot] (453) at (10, -2) {};
		\node [style=green dot] (454) at (10, 2) {};
		\node [style=green dot] (455) at (8, -4) {};
		\node [style=none] (456) at (13.25, 0) {};
		\node [style=none] (457) at (1, 0) {};
		\node [style=none] (458) at (6.25, 0.25) {};
		\node [style=none] (459) at (7.5, 1.5) {};
		\node [style=none] (460) at (6.25, -0.25) {};
		\node [style=none] (461) at (7.5, -1.5) {};
	\end{pgfonlayer}
	\begin{pgfonlayer}{edgelayer}
		\draw [style=Fill grey] (391.center)
			 to (392.center)
			 to (389.center)
			 to (390.center)
			 to cycle;
		\draw [style=Thick arrow] (1.center) to (0.center);
		\draw [style=dashes red] (401.center) to (404.center);
		\draw [style=dashes red] (400.center) to (405.center);
		\draw [style=dashes red] (406.center) to (399.center);
		\draw [style=dashes red] (398.center) to (407.center);
		\draw [style=dashes red] (408.center) to (397.center);
		\draw [style=dashes blue] (435.center) to (424.center);
		\draw [style=dashes blue] (425.center) to (434.center);
		\draw [style=dashes blue] (433.center) to (426.center);
		\draw [style=dashes blue] (427.center) to (432.center);
		\draw [style=dashes blue] (431.center) to (428.center);
		\draw [style=Thick arrow] (457.center) to (456.center);
		\draw [style=dashed arrow] (458.center) to (459.center);
		\draw [style=dashed arrow] (460.center) to (461.center);
	\end{pgfonlayer}
\end{tikzpicture}}
    \caption[Joint temporal amplitude of a two-mode GKP code]{Joint temporal amplitude of a two-mode GKP code. Red points represent the logical state $\ket{\overline +}$, while green points correspond to $\ket{\overline -}$. The gray region indicates the domain in which a time-shift error can be uniquely identified and corrected. The code words are $\sqrt{2}\pi/\Omega_0$-periodic along the diagonal temporal directions, as indicated by the dotted arrows.}
    \label{fig: JTA 2D gkp code}
\end{figure}

Although simple, this construction reveals the geometric structure underlying higher-dimensional generalizations of GKP codes. Systematizing these ideas naturally leads to lattice-based GKP codes, such as those developed in the quadrature regime in Refs.~\cite{conrad_gottesman-kitaev-preskill_2022,lin_closest_2023}. However, in the following, we will present a simpler construction based on collective variables.

\subsection{Collective GKP code}
\label{subsec: collective GKP code}

\paragraph{\publi Definition}
We fix a number $n$ of single photons and define a code that is robust against time and frequency displacements generated by
\begin{align}
    \hat D_\omega^j(\delta t)= e^{-i\delta t \hat \omega_j}, && \hat D_{t}^j(\delta \omega)= e^{-i\delta \omega \hat t_j}.
\end{align}
Guided by the use of collective variables, we propose the following form for the logical state $\ket{\overline k}$ ($k=0,1$),
\begin{equation}\label{eq: GKP multi general}
    \ket{\overline k}=\int \dd\Omega_1\cdots\dd\Omega_n\, F_k (\Omega_1)\prod_{j=2}^n G_j(\Omega_j)\ket{\omega_1,\dots,\omega_n},
\end{equation}
where $\Omega_j=\sum_{l=1}^n \alpha_{j,l} \omega_l$ are collective variables defined analogously to the $\hat P_j$ operators introduced in Sec.~\ref{subsec: entanglement measure}, via an orthogonal matrix $A=(\alpha_{j,l})$. As before, we assume $\alpha_{1,j}=1/\sqrt{n}$, such that
\begin{equation}
    \Omega_1=\frac{1}{\sqrt{n}}\sum_{j=1}^n \omega_j.
\end{equation}
Note the factor $\sqrt{n}$ in the definition of $\Omega_1$, which was absent in the previous section when manipulating collective variables. This is merely a matter of convention and has no practical consequences for the results. For simplicity, one may further require $\alpha_{j,l}=\pm1/\sqrt{n}$, which can simplify some computations. This choice is not possible for arbitrary $n$, as it requires $A$ to be a Hadamard matrix, known to exist for all powers of $2$ and conjectured to exist for all multiples of $4$~\cite{sylvester_thoughts_1908,hedayat_hadamard_1978}. By orthogonality of $A$, we have
\begin{equation}
    \omega_j=\sum_{l=1}^n \alpha_{l,j} \Omega_l.
\end{equation}

An ideal $n$-photon GKP state $\ket{\overline k}$ is obtained by choosing
\begin{equation}
    F_{k}(\Omega_1)=\sum_{s\in\Z}\delta(\Omega_1-(2s+k)\Omega_0),
\end{equation}
where $\Omega_0$ is an arbitrary frequency. The logical states $\ket{\overline k}$ are therefore non-physical, consisting of an infinite comb of delta peaks located at frequencies separated by $2\Omega_0$. Importantly, the logical information is encoded solely in the collective variable $\Omega_1$. The functions $G_j$ are arbitrary and play no essential role in the basic working principles of the code; their effect will be discussed later. Consequently, all information relevant for error diagnosis and correction is contained in $\Omega_1$, while the degrees of freedom associated with $\Omega_{j>1}$ are disregarded.

Situations of this kind are common in quantum optics, where different physical properties are associated with distinct collective variables. For instance, in the Hong-Ou-Mandel experiment~\cite{hong_measurement_1987} (see also Chap.~\ref{chap: HOM interferometry and metrology} for details and generalizations to many photons), the variable $\Omega_1=(\omega_1-\omega_2)/\sqrt{2}$ is directly measured, whereas the information contained in $\Omega_2=(\omega_1+\omega_2)/\sqrt{2}$ is discarded. More generally, by combining interferometric techniques~\cite{branning_simultaneous_2000}, it is possible to access different collective variables, probing not only frequency but also other continuous degrees of freedom such as transverse position and momentum~\cite{tasca_continuous-variable_2011, barros_free-space_2017}.

Time-frequency GKP states are intrinsically multimode states that rely on particle-mode non-separability. They are therefore fundamentally different from optical frequency combs, which are single-mode states obtained via spectral engineering of classical (coherent) fields or individual photons~\cite{fabre_generation_2020, yamazaki_linear_2023}. Owing to the encoding in collective variables of individual photons, time-frequency GKP states exhibit genuinely multiphotonic features, such as an error tolerance that scales with the photon number $n$. For this reason, they are also fundamentally distinct from CV GKP states, including their multi-dimensional generalizations~\cite{royer_encoding_2022}.

As an illustrative example, choosing $G_j(\Omega_j)=\delta(\Omega_j)$ for all $j=2,\dots,n$ in Eq.~\eqref{eq: GKP multi general} yields
\begin{equation}\label{eq: GKP multi thin}
    \ket{\overline k}=\sum_{s\in\Z}\ket{(2s+k)\frac{\Omega_0}{\sqrt{n}},\cdots,(2s+k)\frac{\Omega_0}{\sqrt{n}}}.
\end{equation}
This expression makes explicit that all photons share identical spectra, with peaks separated by $\Omega_0/\sqrt{n}$. This particular case corresponds to states with zero thickness in the orthogonal collective directions $\Omega_{j>1}$, \ie, $\Delta^2 \hat P_j=0$ for $j\geq 2$. More general states with non-zero thickness are obtained by choosing broader functions $G_j$. The impact of this thickness on error correction is discussed below.

\paragraph{\publi Logical gates}
In close analogy with CV GKP states, one can identify non-Hermitian operators that act as Pauli matrices on the logical subspace spanned by $\ket{\overline k}$~\cite{fabre_generation_2020}. A convenient approach uses displacements in the collective time-frequency variables,
$\hat D_T^1(\delta\omega)=e^{-i  \hat T_1 \delta\omega}$ and $\hat D_\Omega^1(\delta t)=e^{-i \hat \Omega_1 \delta t}$,
defined as in Sec.~\ref{subsec: Multimode TF operators}. The collective operators are
\begin{align}
    \hat \Omega_1= \sum_{j=1}^n\hat \omega_j/\sqrt{n},&& \hat T_1 = \sum_{j=1}^n \hat t_j/\sqrt{n},
\end{align}
which satisfy $[\hat \Omega_1,\hat T_1 ]=i\1$ on the single-photon subspace. Similar collective operators can be defined for $\Omega_{j>1}$. By choosing appropriate values of $\delta\omega$ and $\delta t$, we define Pauli-like operators on the time-frequency GKP code space as
\begin{align}
    \hat X=e^{-i \hat T_1 \Omega_0}, && \hat Z=e^{-i \hat \Omega_1 T_0},
\end{align}
with $T_0=\pi/\Omega_0$, and $\hat Y=i\hat X \hat Z$, such that $\ket{\overline 1}=\hat X\ket{\overline 0}$. Notably, the logical operators $\hat X$ and $\hat Z$ are not unique. One may alternatively define local operators
\begin{align}
    \label{eq: GKP logical operators local}
    \hat X_{j}=e^{-i\hat t_j \Omega_0\sqrt{n}}, && \hat Z_{j}=e^{-i\hat \omega_j T_0\sqrt{n}},
\end{align}
and verify that $\hat X_j\ket{\overline k}$ acts on the collective variable $\Omega_1$ in the same way as $\hat X$. This equivalence is illustrated in Fig.~\ref{fig: GKP different logical gates}; see also Appendix~\ref{app: collective var derivation}, Result~\ref{res: GKP global} and Result~\ref{res: GKP local}, for detailed derivations.

\begin{figure}[ht]
    \centering
    \includegraphics[width=\columnwidth]{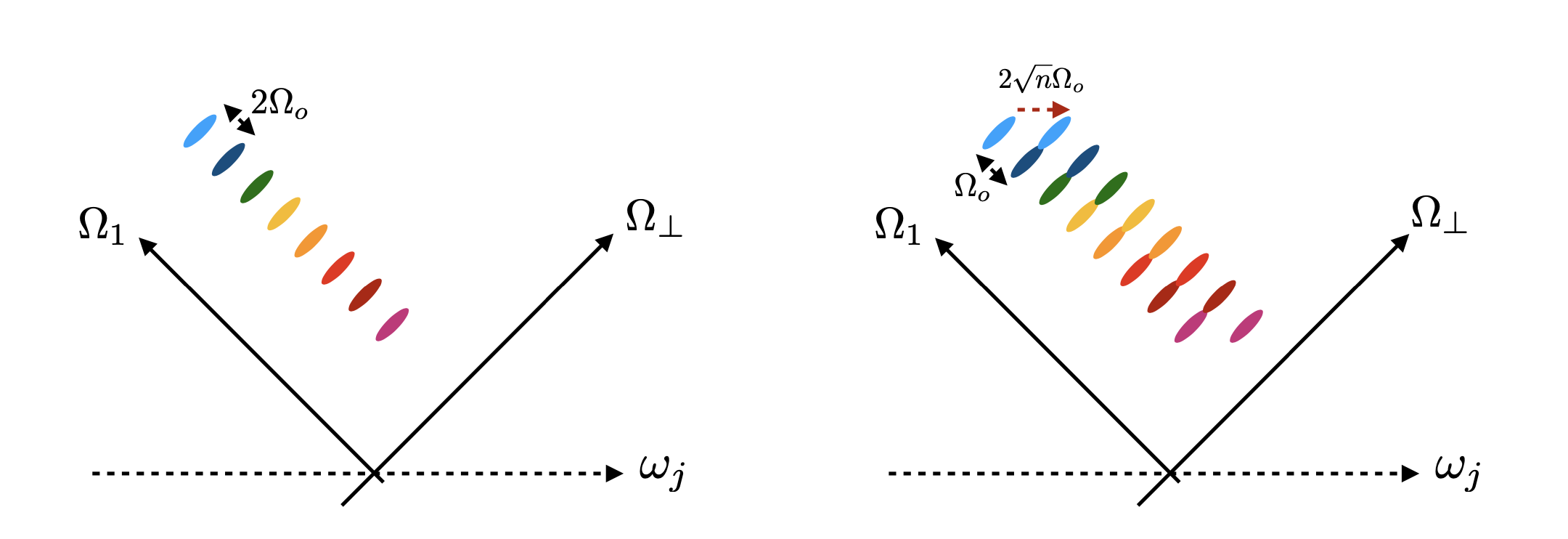}
    \caption[Ideal time-frequency GKP state exhibiting a comb structure in the collective variable and its displacement in a local frequency]{Left: Ideal time-frequency GKP state exhibiting a comb structure with peak spacing $2\Omega_0$ in the collective variable $\Omega_1$. $\Omega_{\perp}$ denotes any collective variable orthogonal to $\Omega_1$. Right: A displacement of $2\sqrt{n}\Omega_0$ in a local frequency variable $\omega_j$ induces an equivalent displacement in the $\Omega_1$ direction. The associated displacement in the orthogonal direction $\Omega_{\perp}$ is irrelevant, as the information it carries is not used for encoding. Adapted from Ref.~\cite{descamps_gottesman-kitaev-preskill_2024}. \copyright 2024 American Physical Society.}
  \label{fig: GKP different logical gates}
\end{figure}

\subsection{Error correction}
\label{subsec: error correction collective GKP}
\paragraph{\publi Shifts}
As explained above and detailed further in Appendix~\ref{app: collective var derivation}, see Result~\ref{res: GKP global}, collective translation operators $\hat D_T^1(\delta\omega)$ and $\hat D_\Omega^1(\delta t)$ act directly on the collective variable $\Omega_1$, in which the logical quantum information is encoded. Consequently, a measurement of the stabilizers
\begin{align}
    \hat S_t=e^{-2i\Omega_0 \hat T_1}, && \hat S_\omega=e^{-2i\pi \hat \Omega_1/\Omega_0},
\end{align}
allows one to correct \emph{collective} time and frequency errors corresponding to time-frequency displacements satisfying $\abs{\delta\omega}<\Omega_0/2$ and $\abs{\delta t}<\pi/(2\Omega_0)$. These bounds delimit a correctable region in time-frequency phase space with area
\begin{equation}
    4\abs{\delta\omega}\abs{\delta t}<\pi.
\end{equation}
Physically, such displacement errors may arise from imperfect interferometric alignment or from the action of nonlinear optical devices such as fibers.

Restricting the analysis to collective errors does not address our main objective, namely the construction of states that are robust against \emph{local} displacements in the time-frequency variables of individual photons. This corresponds to the experimentally relevant scenario in which photons propagate through distinct spatial or temporal modes and experience independent time and frequency shifts. Such time-frequency displacements correspond to rotations in quadrature phase space, a class of errors against which standard CV GKP codes are known to be inefficient~\cite{grimsmo_quantum_2021}, unless one resorts to rotation-symmetric encodings~\cite{grimsmo_quantum_2020}. By contrast, the time-frequency GKP states considered here are composed of single photons and are therefore invariant under both global and local rotations by construction.

We now analyze the effect of local noise on time-frequency GKP states. From Appendix~\ref{app: collective var derivation}, Result~\ref{res: GKP local}, one finds
\begin{equation}
    \hat D_\omega^l(\delta t)\ket{\overline k}
    =\int \dd\Omega_1\cdots\dd\Omega_n \,
    e^{-i\frac{\Omega_1\delta t}{\sqrt{n}}}
    F_k (\Omega_1)\prod_{j>1}^n \tilde G_j(\Omega_j)
    \ket{\omega_1,\dots,\omega_n},
\end{equation}
where $\tilde G_j(\Omega_j)= e^{-i\alpha_{j,l}\Omega_j \delta t} G_j(\Omega_j)$. Similarly,
\begin{equation}
    \hat D_t^l(\delta\omega)\ket{\overline k}
    =\int \dd\Omega_1\cdots\dd\Omega_n \,
    F_k \left (\Omega_1-\frac{\delta\omega}{\sqrt{n}}\right )
    \prod_{j>1}^n G_j(\Omega_j-\bar \Omega_j)
    \ket{\omega_1,\dots,\omega_n},
\end{equation}
where $\bar \Omega_j = \alpha_{j,l}\delta \omega$. These expressions lead to a central result: the logical states $\ket{\overline k}$ suppress local displacements in the variables $\omega_j$ with a protection that scales as $\sqrt{n}$ with the number of photons. For example, the following error patterns are correctable:
\begin{itemize}
    \item a single photon in mode $j$ displaced by $\abs{\delta t}<\sqrt{n}\pi/(2\Omega_0)$,
    \item up to approximately $\sqrt{n}$ photons each displaced by $\abs{\delta t}<\pi/(2\Omega_0)$,
    \item all $n$ photons equally displaced by $\abs{\delta t}<\pi/(2\Omega_0\sqrt{n})$.
\end{itemize}
More generally, focusing on \emph{local} errors, the logical states $\ket{\overline k}$ protect against them provided that~\cite{albert_performance_2018}
\begin{equation}
    \abs{\sum_{j=1}^n \delta\omega_j}<\frac{\sqrt{n}\Omega_0}{2}, \qquad \abs{\sum_{j=1}^n \delta t_j}<\frac{\sqrt{n}\pi}{2\Omega_0}.
\end{equation}
This can be interpreted as an effective rescaling of the code: the overall protection corresponds to that of a single-photon GKP state with lattice spacings $2\sqrt{n}\Omega_0$ in frequency and $2\pi\sqrt{n}/\Omega_0$ in time. Remarkably, unlike the classical Fourier relation where dilation in frequency entails contraction in time and vice versa, the effective phase-space dilation observed here arises geometrically from encoding information in collective variables while errors act locally.

\paragraph{\publi Photon loss}
The non-uniqueness of the logical operators $\hat X$ and $\hat Z$ can be exploited to detect the loss of a photon in an unknown mode $j'$ and to adapt the correction strategy using only time and frequency displacement measurements. Assuming a frequency-independent photon loss rate~\cite{gardiner_input_1985}, the loss of a photon leads to the state
\begin{equation}
    \ket{\overline k}_{-1}
    =\int \dd\omega \,\hat a_{j}(\omega)\ket{\overline k}
    =\hat {\cal E}_j\ket{\overline k}.
\end{equation}
We define the operators
\begin{equation}
    \hat S_j
    = e^{-i \eta_j\Omega_0\hat t_j}\hat Z_{j}^2 e^{i \eta_j\Omega_0\hat t_j}
    =e^{-i2(\hat \omega_j-\hat n_j\eta_j\Omega_0) T_0\sqrt{n}},
\end{equation}
where $\eta_j\in\mathbb{R}$ and $\hat Z_j$ is defined in Eq.~\eqref{eq: GKP logical operators local}. Restricting attention to the collective variable $\Omega_1$, the operator $\hat S_j\hat S_{j+1}$ stabilizes $\ket{\overline k}$ provided that
\begin{equation}
    (\eta_j+\eta_{j+1})\sqrt{n}\in\mathbb{Z},
\end{equation}
for all $j$. Moreover, it also stabilizes $\hat{\cal E}_{j'}\ket{\overline k}$ if $j'\neq j$ and $j'\neq j+1$. In contrast,
\begin{equation}
    \hat S_j \hat S_{j+1}\hat {\cal E}_j\ket{\overline k}
    =\hat {\cal E}_j\hat S_{j+1}\ket{\overline k}
    =e^{-2i\eta_j \pi\sqrt{n}}\hat {\cal E}_j\ket{\overline k},
\end{equation}
so that, by appropriately choosing the parameters $\eta_j$, one can detect both the occurrence of a photon loss and the mode from which it originated. Defining the global stabilizer $\hat S=\prod_j \hat S_j$, which stabilizes $\ket{\overline k}$, one finds
\begin{equation}
    \hat S \hat {\cal E}_j \ket{\overline k}
    =e^{-2i\eta_j\pi\sqrt{n}}\hat {\cal E}_j \ket{\overline k}.
\end{equation}
Thus, provided that $\sum_{j=1}^n \eta_j\sqrt{n}\in\Z$, a single stabilizer measurement suffices to identify both the loss event and the affected mode. This information allows one to adapt subsequent operations to an $(n-1)$-photon configuration. Since collective displacement effects scale with the number of photons, the stabilizer parameters $\Omega_0$ and $T_0$ must be rescaled accordingly. To illustrate this rescaling, consider the loss of a photon in the first mode. One finds
\begin{subequations}
    \begin{align}
        e^{-i\hat T_1\Omega_0'} \int \hat a_1(\omega)\,\dd\omega\ket{\overline k}
        &=\int \dd\Omega_1\cdots \dd\Omega_n\,
        F_k (\Omega_1)\prod_{j=2}^nG_j(\Omega_j)
        \vac_1\otimes\ket{\omega_2+\frac{\Omega_0'}{\sqrt{n}},\dots,\omega_n+\tfrac{\Omega_0'}{\sqrt{n}}},\\
        &=\int \dd\Omega_1\cdots\dd\Omega_n\,
        F_k \left(\Omega_1-\frac{\Omega_0'(n-1)}{n}\right)
        \prod_{j=2}^n G_j(\Omega_j')
        \vac_1\otimes\ket{\omega_2,\dots,\omega_n},\\
        &=\int \dd\Omega_1\cdots\dd\Omega_n\,
        F_{(k+1)\,\mathrm{mod}\,2} (\Omega_1)
        \prod_{j=2}^n G_j(\Omega_j')
        \vac_1\otimes\ket{\omega_2,\dots,\omega_n}.
    \end{align}
\end{subequations}
Setting $\Omega_0'=\Omega_0 n/(n-1)$ therefore requires redefining the logical operator as
\begin{equation}
    \hat X= e^{-i\hat T_1 \Omega_0 n/(n-1)},
\end{equation}
which ensures that $\hat X\ket{\overline k}_{-1}=\ket{\overline{(k+1)\,\mathrm{mod}\,2}}_{-1}$. An analogous argument applies to the logical operator $\hat Z$, yielding the same rescaling.

More complex operations, such as entangling gates, are also affected by photon loss. However, if the affected mode is known, these errors can be mitigated by adapting the operations to exclude that mode. Alternatively, the lost photon can be reinserted using entangling gates~\cite{fabre_time_2022,le_jeannic_dynamical_2022} together with temporal or frequency displacements~\cite{kurzyna_variable_2022}. For instance, if the photon lost from the first mode is reintroduced at frequency $\omega'$, one defines
\begin{equation}
    \ket{\overline k}_r
    =\hat a_1^\dagger(\omega')\ket{\overline k}_{-1}
    =\ket{\omega'}_1
    \int \dd\Omega_1\cdots\dd\Omega_n\,
    F_k (\Omega_1)\prod_{j=2}^n G_j(\Omega_j)
    \ket{\omega_2,\dots,\omega_n}.
\end{equation}
Applying a conditional displacement yields
\begin{equation}\label{eq: conditionnal displaced corrected GKP}
    e^{-i\hat t_1 \hat \omega_2}\ket{\overline k}_r
    =\int \dd\Omega_1\cdots\dd\Omega_n\,
    F_k (\Omega_1)\prod_{j=2}^n G_j(\Omega_j)
    \ket{\omega'+\omega_2, \omega_2,\dots,\omega_n}.
\end{equation}
Since $\omega_1=\sum_i \alpha_{i1}\Omega_i$ and $\omega_2=\sum_i \alpha_{i2}\Omega_i$, full recovery of the original $n$-photon time-frequency GKP state depends on the choice of the functions $G_j$. In the simple case of delta functions, as in Eq.~\eqref{eq: GKP multi thin}, it suffices to apply a frequency displacement of $-\omega'$ to mode $1$.

In this analysis we assumed a frequency-independent loss channel, for which highly frequency-correlated time-frequency GKP states are most easily correctable. In contrast, for a monochromatic loss channel such states would be irreversibly destroyed, necessitating instead encodings with strong temporal correlations. Thus, depending on the experimental context, one can tailor the time-frequency structure of the GKP code to optimize resilience against photon loss and enable recovery via photon addition.

Finally, it is important to emphasize that photon loss has qualitatively different effects in time-frequency GKP codes and in standard CV GKP codes. In the former, loss corresponds to the removal of an entire propagation mode and its associated information, whereas in the latter it manifests as attenuation of a quadrature mode, leading to a deformation of the GKP lattice itself.

\paragraph{\publi Comparison with other time-frequency codes}
We can compare our construction to other encodings based on GKP-like states within the continuous modes of single photons. As mentioned in the previous section, it is possible to define GKP qubits using a frequency comb spectral distribution in single photons. In this case, one photon corresponds to one qubit, and the spectral function of each photon defines a two-level-like system that exhibits local robustness against time-frequency displacements. This is a single-mode classical-like effect independent of the number of photons involved, and if the photon's peaks are separated by $2\Omega_0/\sqrt{n}$, the protection against displacement errors is limited to amplitudes $\delta_\omega \sim \Omega_0/(2\sqrt{n})$ \emph{per} photon. Of course, it's possible to enhance protection against local (single-photon) qubit flipping and dephasing by using entangled photons. However, the codes constructed in this way operate similarly to discrete ones~\cite{shor_scheme_1995,steane_error_1996,laflamme_perfect_1996,gottesman_stabilizer_1997} (and see Sec.~\ref{subsec: error correction}), and correction for qubit flips and dephasing requires using entangled states of at least five photons. In contrast, the encoding proposed here demonstrates enhanced protection starting from $n=2$.

\subsection{Approximated codes}
\label{subsec: approximated codes}

\paragraph{\publi Frequency width}
The states defined in Eq.~\eqref{eq: GKP multi thin} are not physical, as they are not normalizable. A physical version of these states, denoted $\ket{\tilde k}$ with $k \in \{0,1\}$, can be constructed by replacing the Dirac comb with a comb of Gaussian peaks,
\begin{equation}\label{eq: GKP spectrum with Delta}
    \tilde F_k(\Omega_1)=\sum_{s\in\Z} 
    e^{-\kappa^2 (2s+k)^2\Omega_0^2}
    e^{-\frac{(\Omega_1-(2s+k)\Omega_0)^2}{\Delta^2}} .
\end{equation}
Each peak of the time-frequency GKP code now has a Gaussian spectral profile of width $\Delta \ll \Omega_0$ in the variable $\Omega_1$, while the overall comb is modulated by a Gaussian envelope of width $\kappa^{-1} \gg \Omega_0$~\cite{gottesman_encoding_2001}. For simplicity, we will assume $\Delta=\kappa$ in the following.

A finite spectral width induces an intrinsic error probability due to the imperfect orthogonality of the logical states $\ket{\tilde 0}$ and $\ket{\tilde 1}$. This leads to an error probability ${\cal E}(\Delta/\Omega_0)$ associated with the error correction protocol. The finite spectral width can be modeled as independent displacements of individual photons with Gaussian-distributed amplitudes of width $\Delta$. Each photon $j$ is then described by the state
\begin{align}
    \ket{\widetilde{(2s+k)\frac{\Omega_0}{\sqrt{n}}}}_j
    = \int \dd\omega \,
    e^{-\Delta_j^2 (2s+k)^2\left(\frac{\Omega_0}{\sqrt{n}}\right)^2}
    e^{-\frac{\left(\omega-\frac{(2s+k)\Omega_0}{\sqrt{n}}\right)^2}{\Delta_j^2}}
    \ket{\omega}_j .
\end{align}
The local variables $\omega_j$ behave as independent random variables. Hence, in contrast with the CV GKP encoding, the spectral width does not depend on the average photon number. The probability of mistaking $\ket{\tilde 0}$ and $\ket{\tilde 1}$ is therefore given by
\begin{equation}
    {\cal E}(\Delta/\Omega_0)
    = \frac{\Delta}{\pi\Omega_0}
    e^{-\frac{\pi \Omega_0^2}{4\Delta^2}},
\end{equation}
as derived in Ref.~\cite{gottesman_encoding_2001}.

\paragraph{\publi Thickness in the orthogonal direction}
We now examine the effect of a finite spectral width in the collective variables $\Omega_{j>1}$. For simplicity, we focus on a single orthogonal variable $\Omega_\perp$ and neglect the others, considering a state with spectral amplitude of the form $F_k(\Omega_1)G(\Omega_\perp)$,
\begin{equation}
    \ket{\psi}
    = \int \dd\Omega_1 \cdots \dd\Omega_n \,
    F_k(\Omega_1)G(\Omega_\perp)
    \ket{\omega_1,\dots,\omega_n} .
\end{equation}
We take $F_k$ to be a Dirac comb as before and assume a Gaussian profile for $G$,
\begin{equation}
    F_k(\Omega_1)G(\Omega_\perp)
    = \sum_{s\in\Z}
    \delta(\Omega_1-(2s+k)\Omega_0)
    e^{-\frac{(\Omega_\perp-\omega_\perp^0)^2}{\sigma^2}} .
\end{equation}
The widths of $F_k$ and $G$ are independent and may arise from different physical constraints, such as energy conservation or phase-matching conditions in spontaneous parametric down-conversion, as discussed in Sec.~\ref{subsec: SPDC}. When the spectral function is separable in $\Omega_1$ and $\Omega_\perp$, and measurements are performed only on $\Omega_1$, the width $\sigma$ and the specific shape of $G$ are irrelevant.

In practice, however, state preparation imperfections may introduce correlations between $\Omega_1$ and $\Omega_\perp$. As a toy model, we assume that the state is separable in rotated variables
\begin{align}
        \Omega_1'=\cos\theta\,\Omega_1-\sin\theta\,\Omega_\perp, && \Omega_\perp'=\cos\theta\,\Omega_\perp+\sin\theta\,\Omega_1 ,
\end{align}
for an arbitrary rotation angle $\theta$. This model also captures imperfect measurements, where $\Omega_1'$ is measured instead of $\Omega_1$. The state then reads
\begin{equation}
    \ket{\psi}
    = \int \dd\Omega_1 \dd\Omega_\perp \,
    F_k(\cos\theta\,\Omega_1-\sin\theta\,\Omega_\perp)
    G(\cos\theta\,\Omega_\perp+\sin\theta\,\Omega_1)
    \ket{\omega_1,\dots,\omega_n} .
\end{equation}
Explicitly,
\begin{align}
    &F_k(\cos\theta\,\Omega_1-\sin\theta\,\Omega_\perp)
    G(\cos\theta\,\Omega_\perp+\sin\theta\,\Omega_1) \notag \\
    &\quad = \sum_{s\in\Z}
    \delta(\cos\theta\,\Omega_1-\sin\theta\,\Omega_\perp-(2s+k)\Omega_0)
    e^{-\frac{(\cos\theta\,\Omega_\perp+\sin\theta\,\Omega_1-\omega_\perp^0)^2}{\sigma^2}} .
\end{align}
For $\theta=-\pi/2$, the measured distribution reduces to a Gaussian of width $\sigma$, corresponding to the original $\Omega_\perp$ distribution. For clarity, we set $\omega_\perp^0=0$ in the following, noting that this shift can be reinstated straightforwardly. In this case, the imperfect state can be interpreted as an ideal time-frequency GKP state subject to a displacement $\sin\theta\,\Omega_\perp/\cos\theta$, distributed according to a Gaussian of width $\sigma$. This displacement acts as an effective error. Due to the rotation, both the effective peak spacing and peak width are modified, as illustrated in Fig.~\ref{fig: rotation GKP}.

\begin{figure}[ht]
    \centering
    \includegraphics[width=8cm]{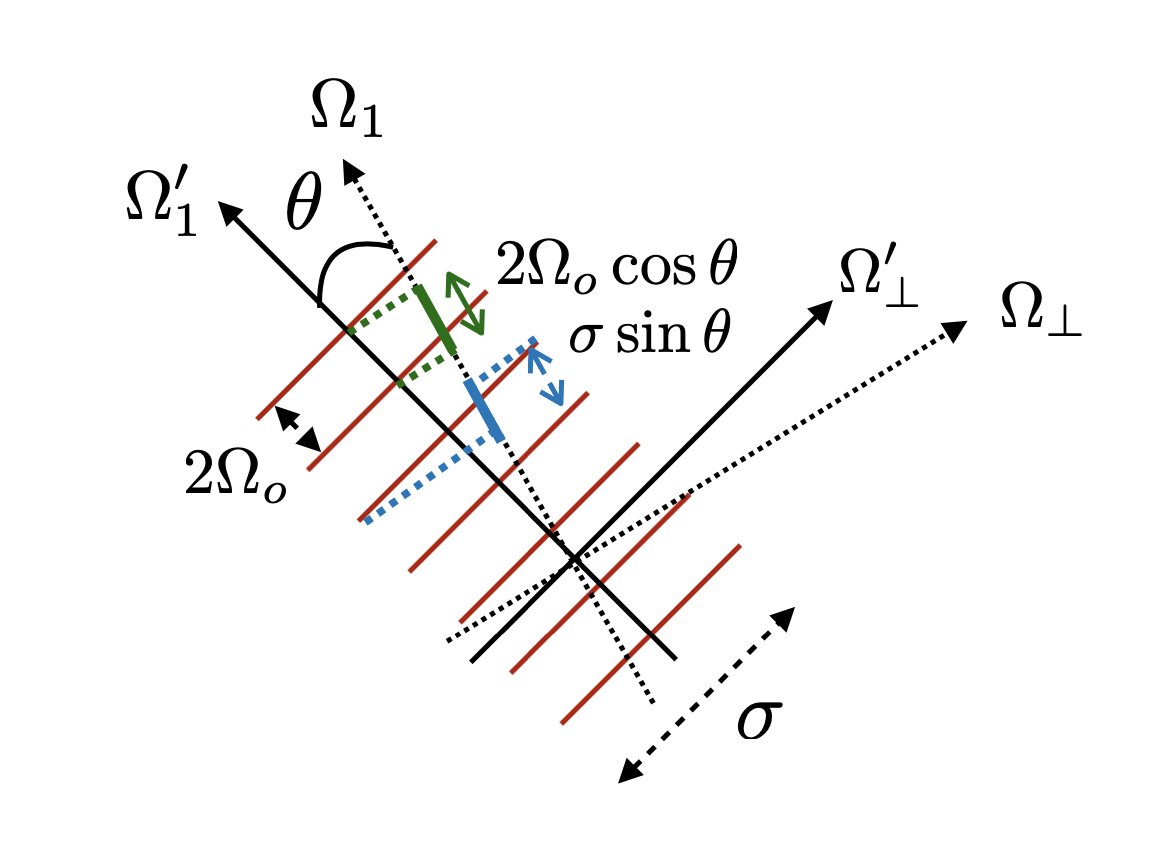}
    \caption[Effect of imperfect state preparation or measurement of a time-frequency GKP state in rotated variables]{Effect of imperfect state preparation or measurement in rotated variables $\Omega_1'$ and $\Omega_\perp'$. The time-frequency GKP peaks along $\Omega_1'$ are shown as red lines. Rotation broadens the peaks by an effective width $\sigma\sin\theta$ and rescales the peak spacing to $2\Omega_0\cos\theta$ in the $\Omega_1$ direction. Adapted from Ref.~\cite{descamps_gottesman-kitaev-preskill_2024}. \copyright 2024 American Physical Society.}
    \label{fig: rotation GKP}
\end{figure}

The effective width $\sigma\sin\theta$ must satisfy $\sigma\sin\theta \ll \Omega_0\cos\theta/4$, or equivalently $\tan\theta \ll \Omega_0/(4\sigma)$, to maintain a low error rate. Additionally, to ensure that the rescaled peak spacing remains within the code space, we require $\Omega_0(1-\cos\theta) \ll \sigma\sin\theta$, leading to
\begin{equation}
    \tan\theta \ll \frac{\sin\theta}{1-\cos\theta} .
\end{equation}

By Fourier duality, an analogous analysis applies in the time domain, with the substitution $\sigma \to \sigma^{-1}$. A low error rate is obtained provided
\begin{equation}
    \tan\theta \ll \sigma T_0/4
    = \pi\sigma/(4\Omega_0) .
\end{equation}
These conditions set bounds on the correctable imperfections in state preparation. In systems such as SPDC, one typically has $\sigma \ll \Omega_0$, implying that constraints arising from collective time variables are more restrictive than those from frequency measurements. Consequently, preparation-induced errors are usually negligible.

Finally, when the peaks have a finite width $\Delta$, as in Eq.~\eqref{eq: GKP spectrum with Delta}, the measured peak width is increased by an additive factor $\sigma\sin\theta$, while the peak spacing is rescaled to $2\Omega_0\cos\theta$. These effects can be tolerated provided the cumulative change in peak spacing $\Omega_0(1-\cos\theta)/(2\Omega_0\Delta)$ remains within the half-width $\Delta/2$. The associated error probabilities are ${\cal E}(\sigma\tan\theta/\Omega_0)$ in frequency and ${\cal E}(\tan\theta/(\sigma T_0))$ in time, leading to the overall condition
\begin{equation}
    \tan\theta \ll \min\{\pi\sigma/(2\Omega_0),\Omega_0/(2\sigma)\} .
\end{equation}

\subsection{Entanglement}
\label{subsec: GKP entangling gates}

\paragraph{\publi Entangling operation} We first note that controlled time-frequency interactions naturally implement conditional frequency displacements. In particular, the unitary operator
\begin{equation}
    e^{-i\hat\omega_1\hat t_2}\ket{\omega_1,\omega_2}=\ket{\omega_1,\omega_1+\omega_2}
\end{equation}
shifts the frequency of the second mode by an amount conditioned on the frequency of the first one. As a consequence, two $n$-photon time-frequency GKP states can be entangled by applying frequency CNOT gates $\hat C_{j,l}=e^{-i\hat \omega_j\otimes \hat t_l}$~\cite{fabre_generation_2020,fabre_time_2022} on suitable pairs of photons.

We define the multipartite entangling operation
\begin{equation}
    \hat {\cal D}= \bigotimes_{j=1}^n \hat C_{(j,1),(j,2)},
\end{equation}
where $(j,l)$ denotes the $j$-th spatial mode of the $l$-th time-frequency GKP qubit, with $l=1$ labeling the control qubit and $l=2$ the target qubit. In Appendix~\ref{app: collective var derivation}, Result~\ref{res: GKP entangling}, we show that
\begin{equation}
    \hat {\cal D} \ket{\bar k_1}_1\ket{\bar k_2}_2
    =\ket{\overline k_1}_1\ket{\overline{ (k_2+k_1)~{\rm mod} ~2}}_2.
\end{equation}
This equality holds, for the general state of Eq.~\eqref{eq: GKP multi general}, when the functions $F_k$ form a Dirac comb spectrum and the variables $\Omega_{j>1}$ are ignored. It becomes exact when considering the state of Eq.~\eqref{eq: GKP multi thin}, where the functions $G_j$ are taken as Dirac delta distributions. Figure~\ref{fig: Ent Gate GKP} schematically represents this entangling gate between two time-frequency GKP qubits. Frequency-controlled two-photon gates have been experimentally demonstrated in Refs.~\cite{lu_controlled-not_2019, le_jeannic_dynamical_2022}, and promising proposals also exist for cavity QED platforms, for example Ref.~\cite{alushi_waveguide_2023}.

\begin{figure}[ht]
    \centering
    \includegraphics[width=8cm]{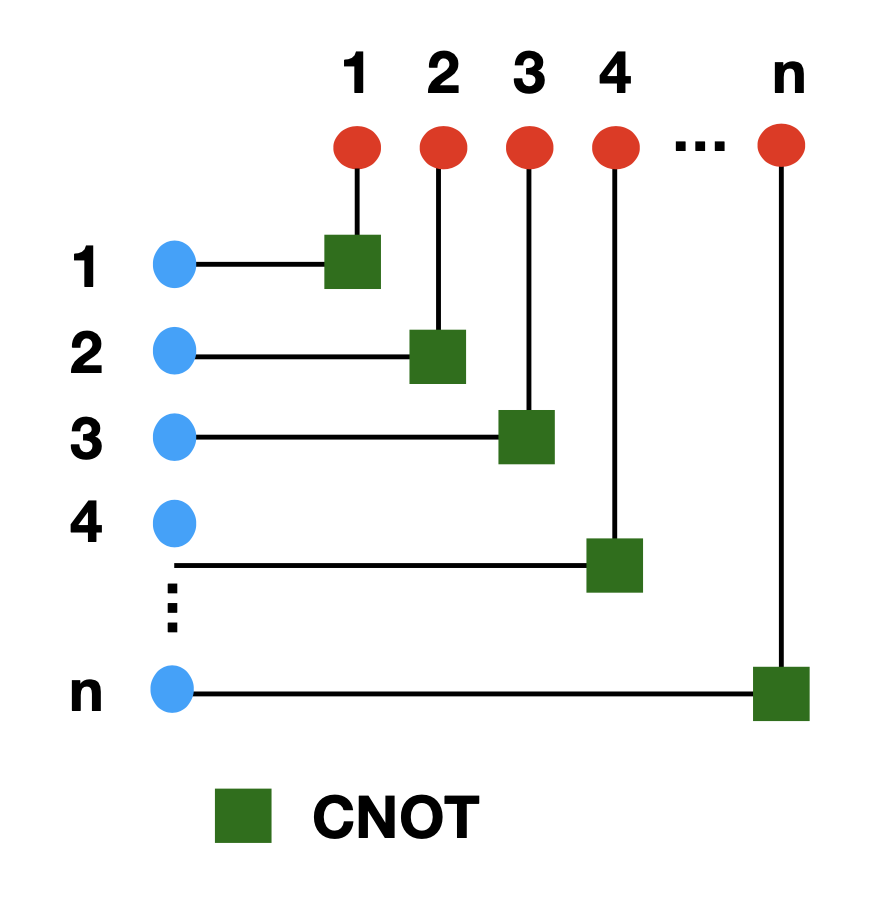}
    \caption[Schematic representation of a CNOT gate between two time-frequency GKP states]{Schematic representation of a CNOT gate $\hat {\cal D}_{1,2}$ between two time-frequency GKP states (blue and red), each comprising $n$ photons distributed over $n$ spatial modes. Photons, represented by blue and red circles, interact pairwise, with each photon pair interacting only once via a time-frequency CNOT gate $\hat C_{j,l}$ (green box). Different photon-pairing configurations across spatial modes are possible. The configuration shown here is an arbitrary choice that simplifies the notation and presentation. Adapted from Ref.~\cite{descamps_gottesman-kitaev-preskill_2024}. \copyright 2024 American Physical Society.}
    \label{fig: Ent Gate GKP}
\end{figure}

\subsection{Conclusion on collective variables}
\label{subsec:conclusion-collective-var}

Throughout this chapter, we have explored the central role of collective variables in a range of applications, including metrology and error correction. When applied to the time-frequency setting, this perspective reveals fundamental properties of photonic systems and sheds new light on the roles of modes, entanglement, photon-number distributions, and physical resources. These observations form the foundation for the analysis presented in Chap.~\ref{chap: SSR}, where an alternative formulation of the continuous-variable formalism in quantum optics, based on a photon-number superselection rule, is introduced and developed.

Beyond their application to time-frequency systems, we have shown that the proposed framework can be generalized to provide a unified description of specific entanglement phenomena. This generalization led to the introduction of a new entanglement quantifier, tailored to the detection and characterization of $k$-entanglement. Altogether, the ideas presented here offer a renewed perspective on the significance of collective variables in quantum optics and quantum information. Further investigations along this line appear both natural and promising.

\fi

\ifnum \theShowChapfour=1
\chapter{Hong-Ou-Mandel interferometry and metrology}
\setcurrentanchor{chap4}
\label{chap: HOM interferometry and metrology}
\emph{This chapter discusses linear interferometry in quantum optics and its connection to quantum metrology. We focus in particular on an extensive analysis of the Hong-Ou-Mandel (HOM) interferometer, highlighting its fundamental properties and presenting general results concerning its relevance for parameter estimation. Extending the ideas developed for the HOM interferometer naturally leads to more elaborate interferometric schemes that are directly suited for metrological applications.}

\localtableofcontents

\section{The Hong-Ou-Mandel interferometer}
\label{sec: HOM interferometer}
\emph{In this section, we review the basic elements of Hong-Ou-Mandel interferometry and emphasize its intimate connection with the modal distinguishability of the photons and time-delay estimation.}

\paragraph{\lit Introduction} 
The Hong-Ou-Mandel (HOM) interferometer is a two-photon interference device that reveals the bunching behavior of two independent bosonic particles, such as single photons~\cite{hong_measurement_1987}. Beyond its original role as a demonstration of bosonic exchange statistics, the HOM interferometer has become a versatile tool across several domains of quantum science~\cite{fabre_hongoumandel_2022}. The coincidence signal has been exploited as an entanglement witness~\cite{brecht_characterizing_2013, eckstein_broadband_2008, ray_verifying_2011}, as a probe of phase-space and spectral properties of quantum states~\cite{douce_direct_2013, tischler_measurement_2015, kurzyna_variable_2022} as well as spatial ones~\cite{dambrosio_tunable_2019}, and as a platform for simulating different quantum exchange statistics~\cite{francesconi_anyonic_2021, francesconi_engineering_2020}, among many other applications~\cite{dorfman_hong-ou-mandel_2021, ndagano_quantum_2022, lee_second-order_2012}. 

In quantum metrology, the HOM interferometer plays a particularly prominent role. It enables quantum parameter estimation in both time-unresolved measurement regimes~\cite{ fabre_parameter_2021, jordan_quantum_2022, lyons_attosecond-resolution_2018, chen_hong-ou-mandel_2019, chen_spectrally_2023, johnson_toward_2023} and time-resolved measurement schemes~\cite{scott_beyond_2020}. Owing to its operation in the low-intensity regime, it is well suited for probing fragile systems, including biological samples~\cite{taylor_quantum_2016}. Moreover, since the HOM effect relies on two-photon interference, it exhibits intrinsic robustness against background noise, group-velocity dispersion~\cite{steinberg_dispersion_1992}, and phase perturbations~\cite{scott_noise_2021}. Both theoretical and experimental studies have shown that HOM-based protocols can approach the quantum precision limit in time-delay estimation~\cite{chen_hong-ou-mandel_2019, johnson_toward_2023, fabre_parameter_2021, scott_beyond_2020}. 

Given its foundational and practical relevance, a detailed understanding of the physical principles underlying HOM interference is essential. In the following section, we therefore review the fundamental mechanisms governing two-photon interference at a beam splitter, establishing the theoretical framework necessary for the subsequent analysis.

\subsection{Interference of independent photons}
\label{subsec: independent HOM interference}

\paragraph{\lit Photons in identical modes}
The seminal work of Hong, Ou and Mandel in 1987~\cite{hong_measurement_1987} revealed a fundamental manifestation of quantum interference: when two photons, whose degrees of freedom are perfectly indistinguishable, impinge on a balanced beam splitter (BS), they always exit together through the same output port (see Fig.~\ref{fig: HOM interferometer}). This phenomenon, known as the Hong-Ou-Mandel (HOM) effect, is a direct consequence of the bosonic nature of photons and has profound implications for quantum optics and quantum information processing. A simple derivation of this effect can be obtained as follows. We consider two input spatial modes described by the annihilation operators $\hat a$ and $\hat b$, each initially containing one photon. The input state is
\begin{equation}
    \ket{\psi}_\text{in}=\hat a^\dagger \hat b^\dagger \vac.
\end{equation}

\begin{figure}[ht]
    \centering
    \footnotesize
    \scalebox{1.2}{\begin{tikzpicture}
	\begin{pgfonlayer}{nodelayer}
		\node [style=none] (0) at (-1.5, 0) {};
		\node [style=none] (1) at (0, 1.5) {};
		\node [style=none] (2) at (1.5, 0) {};
		\node [style=none] (3) at (0, -1.5) {};
		\node [style=none] (5) at (1, 1) {};
		\node [style=none] (6) at (1, -1) {};
		\node [style=none] (8) at (3, 3) {};
		\node [style=none] (9) at (3, -3) {};
		\node [style=none] (15) at (3, 4) {};
		\node [style=none] (16) at (4, 3) {};
		\node [style=none] (18) at (4, -3) {};
		\node [style=none] (19) at (3, -4) {};
		\node [style=none] (20) at (4, 4) {};
		\node [style=none] (21) at (5.75, 4) {};
		\node [style=none] (22) at (6.25, 2.25) {};
		\node [style=none] (23) at (6.5, 1.25) {};
		\node [style=none] (24) at (4, -4) {};
		\node [style=none] (25) at (5.75, -3.75) {};
		\node [style=none] (26) at (7.5, -2.5) {};
		\node [style=none] (27) at (6.5, -1.25) {};
		\node [style=none] (29) at (0, -2.5) {BS};
		\node [style=Photon] (54) at (-5.25, 3) {~~~};
		\node [style=Photon] (55) at (-5, -3) {~~~};
		\node [style=none] (64) at (10.5, 0) {$P_c$};
		\node [style=none] (65) at (4.5, 1.25) {};
		\node [style=none] (66) at (8.5, 1.25) {};
		\node [style=none] (67) at (4.5, -1.25) {};
		\node [style=none] (68) at (8.5, -1.25) {};
		\node [style=none] (69) at (6.5, 0.5) {Coincidence};
		\node [style=none] (70) at (6.5, -0.5) {detection};
		\node [style=none] (71) at (8.75, 0) {};
		\node [style=none] (72) at (10, 0) {};
		\node [style=none] (80) at (-1, -1) {};
		\node [style=none] (81) at (-3, -3) {};
		\node [style=none] (89) at (-4, -3) {};
		\node [style=none] (90) at (-1, 1) {};
		\node [style=none] (91) at (-3, 3) {};
		\node [style=none] (93) at (-4, 3) {};
		\node [style=none] (94) at (-3, 2) {$\hat a$};
		\node [style=none] (95) at (-3, -2) {$\hat b$};
		\node [style=none] (97) at (1.5, 2.5) {$\hat a$};
		\node [style=none] (99) at (1.5, -2.5) {$\hat b$};
	\end{pgfonlayer}
	\begin{pgfonlayer}{edgelayer}
		\draw [style=Fill grey] (2.center)
			 to (3.center)
			 to (0.center)
			 to (1.center)
			 to cycle;
		\draw (0.center) to (2.center);
		\draw [style=Fill red] (16.center)
			 to (15.center)
			 to [bend left=90, looseness=1.75] cycle;
		\draw [style=Fill red] (19.center)
			 to (18.center)
			 to [bend left=90, looseness=1.75] cycle;
		\draw [in=165, out=45, looseness=1.75] (20.center) to (21.center);
		\draw [in=60, out=-15, looseness=1.25] (21.center) to (22.center);
		\draw [in=90, out=-105] (22.center) to (23.center);
		\draw [in=165, out=-45] (24.center) to (25.center);
		\draw [in=-60, out=-15, looseness=1.50] (25.center) to (26.center);
		\draw [in=-90, out=120, looseness=1.25] (26.center) to (27.center);
		\draw (65.center) to (66.center);
		\draw (66.center) to (68.center);
		\draw (68.center) to (67.center);
		\draw (67.center) to (65.center);
		\draw [style=Thick arrow] (71.center) to (72.center);
		\draw [style=dashed arrow] (5.center) to (8.center);
		\draw [style=dashed arrow] (6.center) to (9.center);
		\draw [style=dashed arrow] (89.center)
			 to (81.center)
			 to [in=-120, out=0, looseness=0.75] (80.center);
		\draw [style=dashed arrow] (93.center)
			 to (91.center)
			 to [in=120, out=0, looseness=0.75] (90.center);
	\end{pgfonlayer}
\end{tikzpicture}}
    \caption[Schematic representation of the Hong-Ou-Mandel interferometer]{Schematic representation of the Hong-Ou-Mandel interferometer. Two photons, depicted as yellow circles, enter a balanced beam splitter (gray box) from two distinct input ports. The outputs are monitored by single-photon detectors (pink). When the photons are in indistinguishable states, they always leave together through the same output port, resulting in the complete suppression of coincidence events.}
    \label{fig: HOM interferometer}
\end{figure}
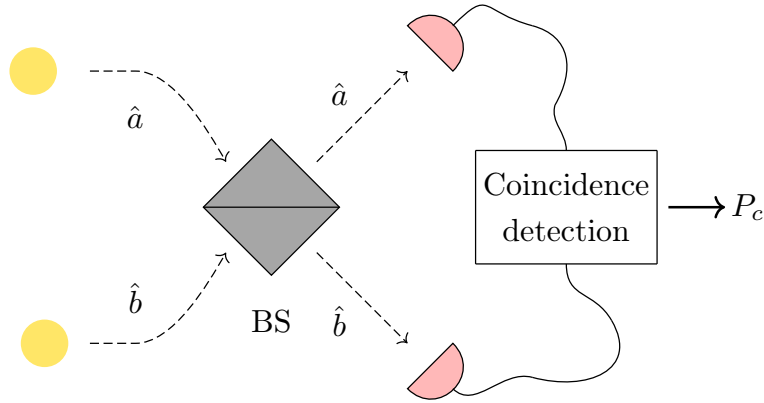

The action of a balanced BS on the input modes is described by the transformations
\begin{align}
    \hat a &\mapsto \frac{\hat a+\hat b}{\sqrt{2}}, &
    \hat b &\mapsto \frac{\hat a-\hat b}{\sqrt{2}},
\end{align}
which, as discussed in Sec.~\ref{subsec: mode transformation}, can be compactly written in matrix form as
\begin{equation}
    U=\frac{1}{\sqrt{2}}\begin{pmatrix}
    1 & 1\\ 1 & -1
    \end{pmatrix}.
\end{equation}
This description implicitly assumes that, apart from their spatial degree of freedom, all other degrees of freedom of the two photons are identical. Applying the BS transformation to the input state yields~\cite{fearn_theory_1989}
\begin{equation}
    \ket{\psi}_\text{out}
    =\frac{1}{2}\left(\hat a^\dagger+\hat b^\dagger\right)\left(\hat a^\dagger-\hat b^\dagger\right)\vac
    =\frac{1}{2}\hat a^{\dagger 2}\vac-\frac{1}{2}\hat b^{\dagger 2}\vac.
\end{equation}
The cross terms proportional to $\hat a^\dagger \hat b^\dagger$ cancel exactly due to destructive quantum interference. The output state is therefore a coherent superposition of both photons exiting through the same port, either in the spatial mode described by $\hat a$ or $\hat b$. This is the hallmark of the HOM effect.

Although photons do not interact directly during propagation, this example illustrates how quantum interference can lead to correlations that are fundamentally different from the behavior of non-interacting classical particles. The destructive interference responsible for the cancellation of coincidence events is illustrated schematically in Fig.~\ref{fig: HOM cancelation}. In this context, one often says that the photons bunch together. Equivalently, since the HOM effect is experimentally observed by monitoring coincidence events between the two output detectors, it can be stated that the coincidence probability $P_c$ vanishes.

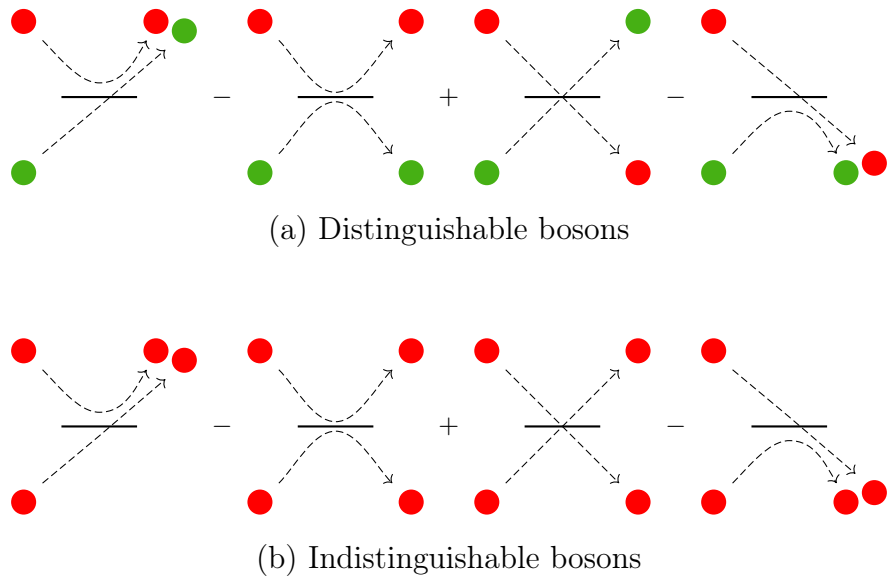
\begin{figure}[ht]
    \renewcommand{\arraystretch}{2}
    \centering
    \footnotesize
    \begin{tabular}{c}
        \scalebox{1}{\begin{tikzpicture}
	\begin{pgfonlayer}{nodelayer}
		\node [style=red dot] (0) at (-5, 2) {};
		\node [style=green dot] (1) at (-5, -2) {};
		\node [style=none] (2) at (-4, 0) {};
		\node [style=none] (3) at (-2, 0) {};
		\node [style=red dot] (4) at (-1, 2) {};
		\node [style=green dot] (5) at (-1, -2) {};
		\node [style=none] (6) at (-4.5, 1.5) {};
		\node [style=none] (7) at (-3, 0.125) {};
		\node [style=none] (8) at (-1.5, 1.5) {};
		\node [style=none] (9) at (-4.5, -1.5) {};
		\node [style=none] (10) at (-3, -0.125) {};
		\node [style=none] (11) at (-1.5, -1.5) {};
		\node [style=none] (12) at (0, 0) {$+$};
		\node [style=red dot] (13) at (1, 2) {};
		\node [style=green dot] (14) at (1, -2) {};
		\node [style=none] (15) at (2, 0) {};
		\node [style=none] (16) at (4, 0) {};
		\node [style=red dot] (17) at (5, -2) {};
		\node [style=green dot] (18) at (5, 2) {};
		\node [style=none] (19) at (1.5, 1.5) {};
		\node [style=none] (20) at (3, 0) {};
		\node [style=none] (21) at (4.5, -1.5) {};
		\node [style=none] (22) at (1.5, -1.5) {};
		\node [style=none] (23) at (3, 0) {};
		\node [style=none] (24) at (4.5, 1.5) {};
		\node [style=none] (25) at (6, 0) {$-$};
		\node [style=red dot] (26) at (-11.25, 2) {};
		\node [style=green dot] (27) at (-11.25, -2) {};
		\node [style=none] (28) at (-10.25, 0) {};
		\node [style=none] (29) at (-8.25, 0) {};
		\node [style=red dot] (30) at (-7.75, 2) {};
		\node [style=green dot] (31) at (-7, 1.75) {};
		\node [style=none] (32) at (-10.75, 1.5) {};
		\node [style=none] (33) at (-9.25, 0.375) {};
		\node [style=none] (34) at (-8, 1.5) {};
		\node [style=none] (35) at (-10.75, -1.5) {};
		\node [style=none] (36) at (-9.25, -0.25) {};
		\node [style=none] (37) at (-7.5, 1.25) {};
		\node [style=none] (38) at (-6, 0) {$-$};
		\node [style=green dot] (39) at (7, -2) {};
		\node [style=red dot] (40) at (7, 2) {};
		\node [style=none] (41) at (8, 0) {};
		\node [style=none] (42) at (10, 0) {};
		\node [style=green dot] (43) at (10.5, -2) {};
		\node [style=red dot] (44) at (11.25, -1.75) {};
		\node [style=none] (45) at (7.5, -1.5) {};
		\node [style=none] (46) at (9, -0.375) {};
		\node [style=none] (47) at (10.25, -1.5) {};
		\node [style=none] (48) at (7.5, 1.5) {};
		\node [style=none] (49) at (9, 0.25) {};
		\node [style=none] (50) at (10.75, -1.25) {};
	\end{pgfonlayer}
	\begin{pgfonlayer}{edgelayer}
		\draw [style=Thick] (2.center) to (3.center);
		\draw [style=dashed arrow] (9.center)
			 to [in=-180, out=45, looseness=0.75] (10.center)
			 to [in=135, out=0, looseness=0.75] (11.center);
		\draw [style=dashed arrow] (6.center)
			 to [in=180, out=-45, looseness=0.75] (7.center)
			 to [in=225, out=0, looseness=0.75] (8.center);
		\draw [style=Thick] (15.center) to (16.center);
		\draw [style=dashed arrow] (22.center)
			 to [in=-135, out=45, looseness=0.75] (23.center)
			 to [in=-135, out=45, looseness=0.50] (24.center);
		\draw [style=dashed arrow] (19.center)
			 to [in=135, out=-45, looseness=0.75] (20.center)
			 to [in=135, out=-45, looseness=0.50] (21.center);
		\draw [style=Thick] (28.center) to (29.center);
		\draw [style=dashed arrow] (35.center)
			 to (36.center)
			 to (37.center);
		\draw [style=dashed arrow] (32.center)
			 to [in=-180, out=-45, looseness=0.75] (33.center)
			 to [in=-120, out=0] (34.center);
		\draw [style=Thick] (41.center) to (42.center);
		\draw [style=dashed arrow] (48.center)
			 to (49.center)
			 to (50.center);
		\draw [style=dashed arrow] (45.center)
			 to [in=-180, out=45, looseness=0.75] (46.center)
			 to [in=120, out=0] (47.center);
	\end{pgfonlayer}
\end{tikzpicture}}\\
        \normalsize (a) Distinguishable bosons\\
        ~\\
        \footnotesize\scalebox{1}{\begin{tikzpicture}
	\begin{pgfonlayer}{nodelayer}
		\node [style=red dot] (0) at (-5, 2) {};
		\node [style=red dot] (1) at (-5, -2) {};
		\node [style=none] (2) at (-4, 0) {};
		\node [style=none] (3) at (-2, 0) {};
		\node [style=red dot] (4) at (-1, 2) {};
		\node [style=red dot] (5) at (-1, -2) {};
		\node [style=none] (6) at (-4.5, 1.5) {};
		\node [style=none] (7) at (-3, 0.125) {};
		\node [style=none] (8) at (-1.5, 1.5) {};
		\node [style=none] (9) at (-4.5, -1.5) {};
		\node [style=none] (10) at (-3, -0.125) {};
		\node [style=none] (11) at (-1.5, -1.5) {};
		\node [style=none] (12) at (0, 0) {$+$};
		\node [style=red dot] (13) at (1, 2) {};
		\node [style=red dot] (14) at (1, -2) {};
		\node [style=none] (15) at (2, 0) {};
		\node [style=none] (16) at (4, 0) {};
		\node [style=red dot] (17) at (5, -2) {};
		\node [style=red dot] (18) at (5, 2) {};
		\node [style=none] (19) at (1.5, 1.5) {};
		\node [style=none] (20) at (3, 0) {};
		\node [style=none] (21) at (4.5, -1.5) {};
		\node [style=none] (22) at (1.5, -1.5) {};
		\node [style=none] (23) at (3, 0) {};
		\node [style=none] (24) at (4.5, 1.5) {};
		\node [style=none] (25) at (6, 0) {$-$};
		\node [style=red dot] (26) at (-11.25, 2) {};
		\node [style=red dot] (27) at (-11.25, -2) {};
		\node [style=none] (28) at (-10.25, 0) {};
		\node [style=none] (29) at (-8.25, 0) {};
		\node [style=red dot] (30) at (-7.75, 2) {};
		\node [style=red dot] (31) at (-7, 1.75) {};
		\node [style=none] (32) at (-10.75, 1.5) {};
		\node [style=none] (33) at (-9.25, 0.375) {};
		\node [style=none] (34) at (-8, 1.5) {};
		\node [style=none] (35) at (-10.75, -1.5) {};
		\node [style=none] (36) at (-9.25, -0.25) {};
		\node [style=none] (37) at (-7.5, 1.25) {};
		\node [style=none] (38) at (-6, 0) {$-$};
		\node [style=red dot] (39) at (7, -2) {};
		\node [style=red dot] (40) at (7, 2) {};
		\node [style=none] (41) at (8, 0) {};
		\node [style=none] (42) at (10, 0) {};
		\node [style=red dot] (43) at (10.5, -2) {};
		\node [style=red dot] (44) at (11.25, -1.75) {};
		\node [style=none] (45) at (7.5, -1.5) {};
		\node [style=none] (46) at (9, -0.375) {};
		\node [style=none] (47) at (10.25, -1.5) {};
		\node [style=none] (48) at (7.5, 1.5) {};
		\node [style=none] (49) at (9, 0.25) {};
		\node [style=none] (50) at (10.75, -1.25) {};
	\end{pgfonlayer}
	\begin{pgfonlayer}{edgelayer}
		\draw [style=Thick] (2.center) to (3.center);
		\draw [style=dashed arrow] (9.center)
			 to [in=-180, out=45, looseness=0.75] (10.center)
			 to [in=135, out=0, looseness=0.75] (11.center);
		\draw [style=dashed arrow] (6.center)
			 to [in=180, out=-45, looseness=0.75] (7.center)
			 to [in=225, out=0, looseness=0.75] (8.center);
		\draw [style=Thick] (15.center) to (16.center);
		\draw [style=dashed arrow] (22.center)
			 to [in=-135, out=45, looseness=0.75] (23.center)
			 to [in=-135, out=45, looseness=0.50] (24.center);
		\draw [style=dashed arrow] (19.center)
			 to [in=135, out=-45, looseness=0.75] (20.center)
			 to [in=135, out=-45, looseness=0.50] (21.center);
		\draw [style=Thick] (28.center) to (29.center);
		\draw [style=dashed arrow] (35.center)
			 to (36.center)
			 to (37.center);
		\draw [style=dashed arrow] (32.center)
			 to [in=-180, out=-45, looseness=0.75] (33.center)
			 to [in=-120, out=0] (34.center);
		\draw [style=Thick] (41.center) to (42.center);
		\draw [style=dashed arrow] (48.center)
			 to (49.center)
			 to (50.center);
		\draw [style=dashed arrow] (45.center)
			 to [in=-180, out=45, looseness=0.75] (46.center)
			 to [in=120, out=0] (47.center);
	\end{pgfonlayer}
\end{tikzpicture}}\\
        \normalsize (b) Indistinguishable bosons
    \end{tabular}
    \caption[Interference mechanisms in a HOM interferometer]{Interference mechanisms in a HOM interferometer. (a) For distinguishable classical-like particles, the two processes leading to coincidence events do not interfere, yielding a coincidence probability $P_c=1/2$. (b) For photons in indistinguishable states, the two anti-bunching amplitudes interfere destructively, resulting in the complete suppression of coincidence events and thus $P_c=0$.}
    \label{fig: HOM cancelation}
\end{figure}

\paragraph{\lit Photons in arbitrary modes}
In practice, preparing two photons in exactly the same quantum state is experimentally demanding, as it requires precise control over all non-spatial degrees of freedom, such as frequency, temporal profile, polarization, and spatial mode structure. It is therefore natural to ask how the HOM effect is modified when the incoming photons are not in identical states. To address this question, we consider two photons prepared in arbitrary internal states $\ket{\psi}$ and $\ket{\varphi}$. The corresponding input state of the interferometer can be written as (see Sec.~\ref{subsec: FockSpace})
\begin{equation}
    \ket{\psi}_\text{in}=\hat a^\dagger_\psi\hat b^\dagger_\varphi\vac,
\end{equation}
where the operators $\hat a^\dagger$ and $\hat b^\dagger$ label the spatial modes, while the subscripts denote the internal states of the photons. After the BS, the output state becomes
\begin{equation}
    \ket{\psi}_\text{out}
    =\frac{1}{2}\left(
    \hat a_\psi^\dagger\hat a_\varphi^\dagger
    -\hat b_\psi^\dagger\hat b_\varphi^\dagger
    -\hat a_\psi^\dagger\hat b_\varphi^\dagger
    +\hat a_\varphi^\dagger\hat b_\psi^\dagger
    \right)\vac.
\end{equation}
This state can be naturally decomposed into a bunching component and an anti-bunching one,
\begin{align}
    \ket{\psi}_\text{bunch}
    &=\frac{1}{2}\left(\hat a_\psi^\dagger\hat a_\varphi^\dagger-\hat b_\psi^\dagger\hat b_\varphi^\dagger\right)\vac,\\
    \ket{\psi}_\text{anti-bunch}
    &=\frac{1}{2}\left(\hat a_\varphi^\dagger\hat b_\psi^\dagger-\hat a_\psi^\dagger\hat b_\varphi^\dagger\right)\vac.
\end{align}
The coincidence probability is given by the squared norm of the anti-bunching component. A straightforward calculation (see Appendix~\ref{app: hom interferometry}, Result~\ref{res: HOM derivation}) yields~\cite{ou_observation_1988,grice_spectral_1997}
\begin{equation}
    P_c=\frac{1}{2}\left(1-\abs{\braket{\psi}{\varphi}}^2\right).
\end{equation}

Several important conclusions follow from this expression. First, the derivation relies crucially on the bosonic commutation relations of the field operators, confirming that the HOM effect is intrinsically quantum in nature. Second, the coincidence probability is directly governed by the distinguishability of the photon states. For identical states, $\braket{\psi}{\varphi}=1$, one recovers the ideal HOM effect with $P_c=0$. For orthogonal states, $\braket{\psi}{\varphi}=0$, the photons behave as distinguishable particles and $P_c=1/2$, as expected classically. Intermediate overlaps correspond to partial distinguishability and lead to coincidence probabilities between these two extremes.

This quantitative link between HOM interference and the distinguishability of the photon degrees of freedom has been extensively confirmed experimentally and constitutes a central ingredient in the use of HOM interferometry for characterizing the properties of photon sources.

\subsection{Clarification on (in)distinguishability}
\label{subsec: distinguishability clarification}

\paragraph{\lit Two notions of distinguishability}
In the literature and in the quantum optics community, two different notions of distinguishability coexist.
\begin{itemize}
    \item From an experimental and application-oriented perspective, one is typically concerned with producing photons whose relevant properties exhibit a high degree of \emph{indistinguishability}, as this constitutes a key resource for quantum interference and quantum information processing. In this context, distinguishability is assessed only with respect to a restricted and explicitly chosen set of degrees of freedom, and is therefore operationally defined relative to a given experimental configuration. It is common, in this setting, to conflate modal indistinguishability with the indistinguishability of the photons themselves. For example, one may state that two red photons emitted at different times are indistinguishable, whereas a red and a green photon are distinguishable. In this comparison, only the spectral degree of freedom is taken into account, while the temporal degree of freedom, which would in practice allow one to tell the two red photons apart, is deliberately ignored. In this sense, only those degrees of freedom that are deemed relevant for a given task or application are retained to quantify the ``distinguishability of the photons''. 
    
    \item From a more formal perspective, the construction of bosonic Fock spaces requires that physical multiphoton states belong to the symmetric subspace of the tensor-product Hilbert space (see Sec.~\ref{subsec: FockSpace}). In this description, one may temporarily introduce particle labels to construct the tensor-product space, but these labels have no physical meaning once the symmetrization postulate is imposed. Consequently, photons cannot be individually tracked or assigned a permanent identity.

    To illustrate this point, consider two orthogonal modes $a$ and $b$. The state containing one photon in each mode is written in second quantization as
    \begin{equation}
        \hat a^\dagger \hat b^\dagger \ket{\mathrm{vac}}.
    \end{equation}
    In the corresponding first-quantized description, the same state is
    \begin{equation}
        \ket{\psi}=\frac{1}{\sqrt{2}}\left(\ket{a}\otimes\ket{b}+\ket{b}\otimes\ket{a}\right).
    \end{equation}
    If one introduces temporary particle labels $1$ and $2$, this can be written as
    \begin{equation}
        \ket{\psi}=\frac{1}{\sqrt{2}}\left(\ket{1{:}a}\otimes\ket{2{:}b}+\ket{1{:}b}\otimes\ket{2{:}a}\right).
    \end{equation}
    The symmetrized form of the state shows that no physical measurement can determine which photon occupies mode $a$ and which occupies mode $b$. The particle labels are therefore not physical observables, even though the occupied modes are perfectly distinguishable.

    For this reason, it is useful to distinguish between \emph{particle indistinguishability}, which follows from the bosonic exchange symmetry, and \emph{modal distinguishability}, which refers to the distinguishability of the single-particle modes occupied by the photons. In the above example, the photons occupy perfectly distinguishable modes, yet the quantum state does not allow one to assign a definite identity to either photon. The distinguishability therefore originates from the orthogonality of the modes rather than from any intrinsic property of the photons themselves.
\end{itemize}

These two viewpoints reveal that photon degrees of freedom are naturally separated into \emph{internal} and \emph{external} ones. This distinction is not fundamental but depends on the physical process under consideration. In the context of multiphoton interference, external degrees of freedom are those that are actively manipulated and mixed by the interferometric transformation. Internal degrees of freedom, by contrast, are not affected by the transformation and act as spectator variables. Their role is nevertheless crucial, as the overlap of the corresponding modal states determines the visibility of the interference effects.

For example, in linear optical interferometry, spatial modes are typically treated as external degrees of freedom since beam splitters and phase shifters act directly on them. Degrees of freedom such as frequency, time, or polarization are then regarded as internal and their overlap controls the degree of photon indistinguishability. In other physical settings, however, the roles may be reversed. One may equally well consider interferometers acting on frequency bins, temporal modes, polarization modes, orbital angular momentum modes, or any other set of optical modes. In such cases, the degree of freedom on which the transformation acts naturally becomes the external one, while the remaining degrees of freedom play the role of internal variables.

Throughout this chapter, and unless stated otherwise, we will adopt the conventional setting of linear optical interferometry and treat the spatial degree of freedom as external. All other degrees of freedom, such as frequency, time, or polarization, will be regarded as internal. This choice is made for convenience and because it corresponds to the physical situations considered in the following sections; none of the formal developments fundamentally relies on this particular assignment.

As we show explicitly below, the second notion of indistinguishability is the fundamental one, as it follows directly from the bosonic symmetrization postulate and is required for a correct description of multiphoton interference phenomena such as the Hong--Ou--Mandel effect. The first notion is operationally useful because it identifies which degrees of freedom must be controlled in a given experimental setting in order to observe or exploit interference.

However, it should be emphasized that this operational notion is inherently context dependent. Whether two photons are declared distinguishable or indistinguishable depends on which degrees of freedom are retained in the description and which are ignored. Consequently, the resulting notion of distinguishability is not an intrinsic property of the photons themselves, nor even of the full quantum state, but rather of a reduced description adapted to a particular experimental task.

For this reason, speaking of \emph{distinguishable photons} constitutes, strictly speaking, an abuse of language. What is actually being assessed is the distinguishability of the modal quantum states occupied by the photons within a chosen set of degrees of freedom. Throughout this manuscript, we therefore strive to use precise terminology and emphasize that (in)distinguishability refers to the relevant modal degrees of freedom, while the photons themselves remain fundamentally indistinguishable bosonic particles.

\paragraph{\lit HOM and the fundamental indistinguishability of photons}
To emphasize the crucial role played by the bosonic nature of photons, we rederive the HOM effect from a first-quantized perspective. In Sec.~\ref{subsec: independent HOM interference}, the derivation relied on creation and annihilation operators, with bosonic symmetry implicitly encoded in their commutation relations. While this formalism is powerful and convenient, it can obscure the fact that the HOM effect is a direct consequence of exchange symmetry. To make this point explicit, we now derive the effect using the two-photon wavefunction in the formal Fock space, rather than second-quantized operators.

To highlight the importance of symmetrization, we first consider what happens if it is (incorrectly) omitted. We take the initial state to be
\begin{equation}
    \ket{\psi}_\text{in}=\ket{1:a}\ket{2:b},
\end{equation}
where the notation explicitly labels the two photons and indicates that the first occupies mode $a$ while the second occupies mode $b$. The two modes are assumed to differ only by their spatial degree of freedom. In this formalism, the action of a balanced BS is described by
\begin{align}
    \ket{j:a}\mapsto\frac{1}{\sqrt{2}}\big(\ket{j:a}+\ket{j:b}\big), &&
    \ket{j:b}\mapsto\frac{1}{\sqrt{2}}\big(\ket{j:a}-\ket{j:b}\big),
\end{align}
for $j=1,2$. See Sec.~\ref{subsec: mode transformation} for an explanation of the connection between the descriptions of mode transformation in the first- and second-quantization pictures. Applying this transformation to the initial state yields
\begin{equation}
    \ket{\psi}_\text{out}=\frac{1}{2}\big(\ket{1:a}\ket{2:a}-\ket{1:a}\ket{2:b}+\ket{1:b}\ket{2:a}-\ket{1:b}\ket{2:b}\big).
\end{equation}
Within this non-symmetrized description, the two contributions leading to coincidence events, $\ket{1:a}\ket{2:b}$ and $\ket{1:b}\ket{2:a}$, correspond to distinct configurations and therefore do not interfere. One then finds a coincidence probability $P_c=1/2$, in clear disagreement with the HOM effect.

Let us now treat the problem correctly by enforcing bosonic symmetry. The properly symmetrized initial state is
\begin{equation}
    \ket{\psi}_\text{in}=\frac{1}{\sqrt{2}}\big(\ket{1:a}\ket{2:b}+\ket{1:b}\ket{2:a}\big).
\end{equation}
Applying the BS transformation leads to
\begin{subequations}
    \begin{align}
        \ket{\psi}_\text{out}
        &=\frac{1}{2\sqrt{2}}\Big[\big(\ket{1:a}+\ket{1:b}\big)\big(\ket{2:a}-\ket{2:b}\big)
        +\big(\ket{1:a}-\ket{1:b}\big)\big(\ket{2:a}+\ket{2:b}\big)\Big],\\
        &=\frac{1}{2\sqrt{2}}\Big[\ket{1:a}\ket{2:a}-\cancel{\ket{1:a}\ket{2:b}}
        +\cancel{\ket{1:b}\ket{2:a}}-\ket{1:b}\ket{2:b}\notag\\
        &\qquad+\ket{1:a}\ket{2:a}+\cancel{\ket{1:a}\ket{2:b}}
        -\cancel{\ket{1:b}\ket{2:a}}-\ket{1:b}\ket{2:b}\Big],\\
        &=\frac{1}{\sqrt{2}}\big(\ket{1:a}\ket{2:a}-\ket{1:b}\ket{2:b}\big).
    \end{align}
\end{subequations}
In this case, the amplitudes leading to coincidence events interfere destructively and cancel exactly, resulting in $P_c=0$, which is the hallmark of the HOM effect.

This derivation makes explicit that the HOM effect is a direct consequence of the bosonic exchange symmetry of photons and of the required symmetrization of their joint wavefunction. It serves as a reminder that, although the operator formalism is extremely effective, a clear physical interpretation of multiphoton interference phenomena ultimately rests on the fundamental indistinguishability of photons and on the modal structure of their quantum states. These considerations will play a central role in Chap.~\ref{chap: SSR}.

\subsection{Time dependence and metrology}
\label{subsec: time dependence and metrology}
So far, we have described the HOM effect and related it to modal indistinguishability of single-photon states, without any explicit reference to time. In practice, however, HOM interference is most commonly observed by introducing a controllable time delay $\tau$ between the two photons before they impinge on the BS. The coincidence probability then becomes a function of $\tau$. Its dependence on this parameter provides direct access to the temporal properties of the photons and enables precise time-delay estimation, giving rise to the characteristic HOM interferogram known as the HOM dip. We discuss these aspects in more detail below.

\paragraph{\lit Time-frequency description}
Temporal properties of single photons are most conveniently described within the time-frequency formalism. It is therefore natural to express the two-photon input state in the spectral domain in order to make the time dependence of the coincidence probability explicit. We assume that the two photons occupy identical spectral modes described by the normalized spectral amplitude $F$, such that the initial state reads
\begin{equation}
    \ket{\psi}_\text{in}=\int \dd\omega_1 \dd\omega_2\, F(\omega_1)F(\omega_2)\ket{\omega_1,\omega_2}.
\end{equation}
We now introduce a tunable time delay $\tau$ on the first photon, for instance by means of an optical delay line. The state incident on the BS then becomes
\begin{equation}
    \ket{\psi(\tau)}_\text{in}=e^{-i\hat\omega_1\tau}\ket{\psi}_\text{in}
    =\int \dd\omega_1 \dd\omega_2\, F(\omega_1)F(\omega_2)e^{-i\omega_1\tau}\ket{\omega_1,\omega_2}.
\end{equation}
The coincidence probability can be obtained by explicitly reproducing the previous derivation. Alternatively, one may note that after the delay the two photons occupy spectral modes $F(\omega)e^{-i\omega\tau}$ and $F(\omega)$, respectively. The coincidence probability is then directly given by the overlap between these two modes~\cite{legero_quantum_2004,mosley_heralded_2008},
\begin{align}
    P_c(\tau)=\frac{1}{2}\left(1-\abs{\int \dd\omega\, \abs{F(\omega)}^2 e^{-i\omega\tau}}^2\right).
\end{align}
Equivalently, introducing the temporal wave packet $\tilde F(t)$, one finds
\begin{equation}
    P_c(\tau)=\frac{1}{2}\left(1-\abs{\int \dd t\, \tilde F(t)^*\tilde F(t-\tau)}^2\right).
\end{equation}
These expressions make it clear that HOM interference does not merely require the photons to arrive simultaneously at the BS ($\tau=0$), but also that they occupy the same temporal and spectral mode. More precisely, the depth of the HOM dip is governed by the overlap of the internal quantum states of the photons, and reduced temporal mode overlap leads to a reduction of quantum interference.

For a Gaussian spectrum of central frequency $\omega_0$ and width $\sigma$,
\begin{equation}
    F(\omega)=\frac{1}{(2\pi\sigma^2)^{1/4}}e^{-\frac{(\omega-\omega_0)^2}{4\sigma^2}},
\end{equation}
the coincidence probability takes the simple form
\begin{equation}
    P_c(\tau)=\frac{1}{2}\left(1-e^{-\sigma^2\tau^2}\right).
\end{equation}
This function is plotted in Fig.~\ref{fig: HOM interferogram}. Notably, only the temporal width $1/\sigma$ of the wave packet affects the HOM interference pattern, while the central frequency $\omega_0$ does not influence the coincidence probability.

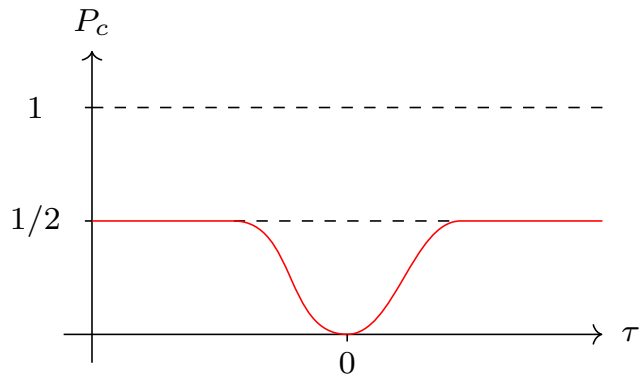
\begin{figure}[ht]
    \centering
    \scriptsize
    \scalebox{1.5}{\begin{tikzpicture}
	\begin{pgfonlayer}{nodelayer}
		\node [style=none] (0) at (-4.5, -2.5) {};
		\node [style=none] (1) at (-4.5, 3) {};
		\node [style=none] (2) at (-5, -2) {};
		\node [style=none] (3) at (4.5, -2) {};
		\node [style=none] (4) at (-4.625, 2) {};
		\node [style=none] (5) at (-4.5, 2) {};
		\node [style=none] (6) at (-5.5, 2) {$1$};
		\node [style=none] (7) at (-4.625, 0) {};
		\node [style=none] (8) at (-4.5, 0) {};
		\node [style=none] (9) at (-5.5, 0) {$1/2$};
		\node [style=none] (10) at (0, -2.5) {$0$};
		\node [style=none] (11) at (5, -2) {$\tau$};
		\node [style=none] (12) at (4.5, 2) {};
		\node [style=none] (13) at (-4.5, 3.5) {$P_c$};
		\node [style=none] (20) at (0, -2) {};
		\node [style=none] (21) at (4.5, 0) {};
		\node [style=none] (22) at (2, 0) {};
		\node [style=none] (23) at (-4.5, 0) {};
		\node [style=none] (24) at (-2, 0) {};
		\node [style=none] (25) at (0, -2) {};
		\node [style=none] (26) at (0, -2.125) {};
	\end{pgfonlayer}
	\begin{pgfonlayer}{edgelayer}
		\draw [style=dashes] (24.center) to (22.center);
		\draw [style=Arrow] (2.center) to (3.center);
		\draw [style=Arrow] (0.center) to (1.center);
		\draw (4.center) to (5.center);
		\draw (7.center) to (8.center);
		\draw [style=dashes] (5.center) to (12.center);
		\draw [style=red line] (23.center)
			 to (24.center)
			 to [in=-180, out=0] (20.center)
			 to [in=-180, out=0, looseness=0.75] (22.center)
			 to (21.center);
		\draw (25.center) to (26.center);
	\end{pgfonlayer}
\end{tikzpicture}}
    \caption[Coincidence probability as a function of the time delay between two photons occupying identical modes]{Coincidence probability $P_c$ as a function of the time delay $\tau$ between two photons occupying identical Gaussian temporal modes. The HOM dip is centered at $\tau=0$, where the temporal mode overlap is maximal and $P_c$ reaches its minimum value of zero. As $\tau$ increases, the overlap between the temporal modes decreases and $P_c$ approaches the classical value of $1/2$. The width of the dip is set by the inverse spectral width $1/\sigma$ of the photons.}
    \label{fig: HOM interferogram}
\end{figure}

\paragraph{\lit Time-delay metrology}
Since the coincidence probability depends explicitly on the time delay $\tau$, the HOM interferometer can be used as a sensitive tool for time-delay estimation. A simple heuristic approach consists in measuring $P_c$ experimentally and inverting its known functional dependence on $\tau$ to infer an estimate of the delay~\cite{giovannetti_advances_2011}. While a full metrological treatment is deferred to later sections, useful insight into the achievable precision can already be gained from elementary considerations.

Each experimental run results in either a coincidence or a non-coincidence event, corresponding to a Bernoulli random variable with parameter $P_c$ and variance $P_c(1-P_c)$. After $N$ repetitions, the statistical uncertainty on the estimated coincidence probability is therefore $\sqrt{P_c(1-P_c)/N}$. Propagating this uncertainty to the estimation of $\tau$ yields
\begin{equation}\label{eq: precision uncertainty propag}
    \delta\tau=\frac{1}{\sqrt{N}}\frac{\sqrt{P_c(1-P_c)}}{\abs{\partial P_c/\partial \tau}}.
\end{equation}
Applying this expression to the Gaussian case discussed above, one finds
\begin{equation}
    \delta\tau=\frac{1}{\sqrt{N}}\frac{\sqrt{e^{2\sigma^2\tau^2}-1}}{2\sigma^2\abs{\tau}}.
\end{equation}
An analytical analysis shows that the precision is optimized at $\tau=0$, leading to
\begin{equation}
    \delta\tau_{\min}=\frac{1}{\sigma\sqrt{2N}}.
\end{equation}
In this simple scenario, the achievable precision is directly set by the temporal width $1/\sigma$ of the photon wave packets and by the number of experimental runs $N$. Extremely high precision can therefore be achieved using temporally short photons, as already demonstrated in the seminal experiment by Hong, Ou and Mandel, where sub-picosecond resolution was reported. Later works have shown that even higher precision can be obtained by tailoring the spectral structure of the photons or by exploiting entangled states. These developments will be discussed in detail in the following sections of this chapter. We note that the optimal precision is reached at a point where the above error-propagation expression becomes singular. The theoretical and practical implications of this feature are discussed in Sec.~\ref{subsec: visibility impact on precision}.

\paragraph{\lit Visibility}
Achieving the optimal precision requires a high-contrast HOM interferogram, which in turn relies on a high degree of overlap between the internal quantum states of the photons. In practice, imperfections in state preparation, residual mode mismatch and experimental noise reduce the visibility of the HOM interference pattern~\cite{mandel_coherence_1991}, thereby degrading the achievable estimation precision~\cite{lyons_attosecond-resolution_2018}. Several definitions of the visibility exist in the literature. Here we adopt a definition that is particularly suited for analyzing its impact on metrological performance (see Sec.~\ref{subsec: general visi model})\footnote{The absolute value in the definition of the visibility is not necessary in the present case, since for independent photons the coincidence probability is bounded between $0$ and $1/2$. As discussed later, entangled states may lead to coincidence probabilities exceeding $1/2$, in which case the absolute value becomes relevant.}
\begin{equation}
    V=2\max_{\tau}\abs{P_c(\infty)-P_c(\tau)}=\max_\tau \abs{1-2P_c(\tau)},
\end{equation}
where $P_c(\infty)$ denotes the coincidence probability at large delays, corresponding to vanishing temporal mode overlap, for which $P_c(\infty)=1/2$. This definition quantifies the maximum deviation of the coincidence probability from its classical value and thus directly measures the strength of two-photon quantum interference. A visibility $V=1$ corresponds to perfect temporal mode overlap and an ideal HOM effect, while $V=0$ indicates the absence of interference. High visibility is therefore a crucial requirement for exploiting HOM interferometry in metrological applications. The effect of reduced visibility on the HOM interferogram is illustrated in Fig.~\ref{fig: HOM visibility}.

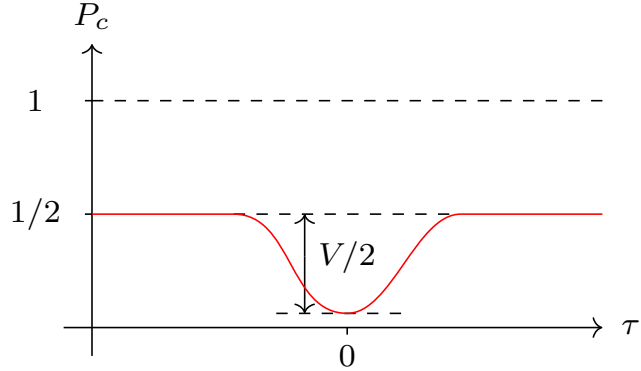
\begin{figure}[ht]
    \centering
    \scriptsize
    \scalebox{1.5}{\begin{tikzpicture}
	\begin{pgfonlayer}{nodelayer}
		\node [style=none] (0) at (0, -0.5) {};
		\node [style=none] (1) at (0, 5) {};
		\node [style=none] (2) at (-0.5, 0) {};
		\node [style=none] (3) at (9, 0) {};
		\node [style=none] (4) at (-0.125, 4) {};
		\node [style=none] (6) at (0, 4) {};
		\node [style=none] (7) at (-1, 4) {$1$};
		\node [style=none] (8) at (-0.125, 2) {};
		\node [style=none] (9) at (0, 2) {};
		\node [style=none] (10) at (-1, 2) {$1/2$};
		\node [style=none] (11) at (4.5, -0.5) {$0$};
		\node [style=none] (12) at (9.5, 0) {$\tau$};
		\node [style=none] (17) at (9, 4) {};
		\node [style=none] (29) at (0, 5.5) {$P_c$};
		\node [style=none] (45) at (3.75, 2) {};
		\node [style=none] (46) at (3.75, 0.25) {};
		\node [style=none] (47) at (3.75, 1.25) {};
		\node [style=none] (48) at (4.5, 1.25) {$V/2$};
		\node [style=none] (49) at (3.25, 0.25) {};
		\node [style=none] (50) at (5.5, 0.25) {};
		\node [style=none] (14) at (4.5, 0.25) {};
		\node [style=none] (16) at (9, 2) {};
		\node [style=none] (20) at (6.5, 2) {};
		\node [style=none] (38) at (0, 2) {};
		\node [style=none] (51) at (2.5, 2) {};
		\node [style=none] (52) at (4.5, 0) {};
		\node [style=none] (53) at (4.5, -0.125) {};
	\end{pgfonlayer}
	\begin{pgfonlayer}{edgelayer}
		\draw [style=dashes] (51.center) to (20.center);
		\draw [style=Arrow] (2.center) to (3.center);
		\draw [style=Arrow] (0.center) to (1.center);
		\draw (4.center) to (6.center);
		\draw (8.center) to (9.center);
		\draw [style=dashes] (6.center) to (17.center);
		\draw [style=Arrow] (47.center) to (46.center);
		\draw [style=Arrow] (47.center) to (45.center);
		\draw [style=dashes] (49.center) to (50.center);
		\draw [style=red line] (38.center)
			 to (51.center)
			 to [in=-180, out=0] (14.center)
			 to [in=-180, out=0, looseness=0.75] (20.center)
			 to (16.center);
		\draw (52.center) to (53.center);
	\end{pgfonlayer}
\end{tikzpicture}}
    \caption[Coincidence probability as a function of the time delay for modes with reduced visibility]{Coincidence probability $P_c$ as a function of the time delay $\tau$ for Gaussian temporal modes with reduced visibility $V<1$. The minimum of the HOM dip no longer reaches zero, reflecting imperfect overlap between the temporal modes of the two photons.}
    \label{fig: HOM visibility}
\end{figure}

\subsection{Experimental considerations}
\label{subsec: HOM experimental considerations}

Throughout this chapter, and as already done above, we mostly consider idealized assumptions on the experimental setup. These assumptions allow us to clearly identify the fundamental physical mechanisms underlying HOM interferometry and to derive analytical bounds on estimation performance. Deviations from these ideal conditions are briefly discussed below, together with comments on how they can be accounted for or mitigated in realistic experiments.

\begin{itemize}
    \item {\bf No loss and perfect detection.} We assume that all photons entering the interferometer are detected with unit efficiency and that no photons are lost during propagation. In practice, propagation losses and imperfect detection reduce the visibility of the HOM interference pattern and thus degrade the estimation precision. We also assume detectors free of dark counts, \ie, detectors that only register the photons of interest. While these effects are unavoidable in realistic implementations, they can be strongly reduced using high-quality optical components and state-of-the-art detectors. Moreover, losses can be explicitly included in the theoretical model, leading to modified expressions for the coincidence probability and the achievable precision, as discussed for instance in~\cite{chen_hong-ou-mandel_2019,scott_beyond_2020}.

    \item {\bf Access to number-resolved detectors.} Measuring the coincidence probability requires distinguishing events where both photons exit the interferometer through the same output port from events where they exit through different ports. This can be directly achieved using number-resolved detectors, which can discriminate between one- and two-photon detection events. Although such detectors are available, they may be costly and exhibit non-ideal performance. When only bucket detectors (\ie, non-number-resolving detectors) are used, dark counts can mimic genuine bunching events and thus reduce the estimation precision. Nevertheless, coincidence events are largely insensitive to dark counts. In practice, the coincidence probability is often inferred by first estimating the source generation rate, a procedure that can be robust against dark counts and does not require number-resolved detection.

    \item {\bf Perfect beam splitter implementation.} We assume a perfectly balanced beam splitter implementing the ideal unitary transformation introduced above. Deviations from perfect balance or additional phase shifts can alter the HOM interference pattern and affect the achievable precision. However, beam splitters with very high balance and stability are commercially available, and residual imperfections can typically be compensated through calibration.

    \item {\bf Perfect state preparation.} We assume that the input photons are prepared in the intended quantum states with high fidelity. In practice, imperfections in state preparation lead to a reduced overlap of the internal (spectral, temporal, polarization) quantum states of the photons, resulting in partial modal distinguishability and a reduced HOM visibility. Significant progress in single-photon sources and quantum state engineering has nevertheless enabled the generation of high-quality states suitable for HOM-based metrology.

    \item {\bf Negligible multiphoton contributions.} We assume that the photon sources emit single photons with negligible multiphoton components. Sources such as spontaneous parametric down-conversion can occasionally produce multiple photon pairs, leading to spurious coincidence events and degraded performance. Experimentally, one can operate in a low-gain regime where such contributions are negligible or apply post-selection strategies to suppress their impact.

    \item {\bf No time information required.} We assume that the detection process is not time resolved and that detectors only register the presence or absence of photons. In this idealized abstraction, this corresponds to assuming that a single photon pair is generated over the entire duration of the experiment. In practice, a large number of pairs are generated, and time-resolved detection is usually required to associate detection events with the same photon pair. Importantly, the timing jitter of the detectors does not fundamentally limit the estimation precision. HOM interferometry can achieve sub-picosecond precision even with detectors exhibiting jitters of tens of picoseconds, provided that the photons themselves have sufficiently short temporal wave packets. The ultimate precision is determined by the temporal structure of the photon quantum state, not by the detector timing resolution, which is only needed to identify coincident events. Including detection times in the analysis can further enhance the estimation precision, as shown for example in~\cite{scott_beyond_2020}. However, as will be shown below, in a wide range of situations non-time-resolved HOM measurements already achieve the optimal precision, as quantified by the QFI, within our idealized framework.
\end{itemize}

\clearpage
\section{Time-frequency HOM}
\label{sec: TF HOM}
\emph{In this section, we develop in detail the analysis of the HOM effect and its application to quantum metrology in the case of time-frequency input states. In particular, we go beyond the assumption of independent input states, provide general expressions based on symmetry considerations, and derive metrological bounds for parameter estimation beyond simple time-delay measurements. Importantly, our results first assume ideal visibility, and we then quantitatively analyse the impact of imperfect visibility. This section is mainly based on \hyperlink{Article: pra HOM}{Time-frequency metrology with two single-photon states: Phase-space picture and the Hong-Ou-Mandel interferometer}~\cite{descamps_time-frequency_2023} and \hyperlink{Article: prl visibility}{Approaching maximal precision of Hong-Ou-Mandel interferometry with nonperfect visibility}~\cite{meskine_approaching_2024}, which were published during the PhD thesis.}

\subsection{Motivations, setup and assumptions}
\label{subsec: TF HOM setup}
In the previous section, we assumed that the two photons impinging on the beam splitter are prepared in independent states, \ie, that the two-photon state can be written as a product of single-photon states. This assumption is widespread in the literature and is often motivated by the use of independent photon sources. However, for time-frequency states generated by SPDC, as discussed in Sec.~\ref{subsec: SPDC}, the two-photon state generally exhibits correlations and cannot be factorized into a product of single-photon states.

Beyond this practical consideration, correlations in the internal (time-frequency) quantum state of the two photons play a central role in HOM interferometry. In particular, for time-delay estimation, it has been shown that entanglement between the two incoming photons can be exploited as a resource to enhance the estimation precision in HOM-based metrology~\cite{chen_hong-ou-mandel_2019,lyons_attosecond-resolution_2018,lang_correlations_2013}. While the analysis of specific correlated or entangled states is relatively straightforward, such case-by-case studies do not provide a comprehensive understanding of how correlations influence HOM interference, nor do they allow one to derive general metrological bounds.

The goal of this section is therefore to provide a unified and general analysis of HOM interferometry and of the associated precision limits, without restricting ourselves to independent input states. We consider the most general pure two-photon single-pair state in the frequency domain, described by the joint spectral amplitude (JSA) $F$,
\begin{equation}
    \ket{\psi}=\int \dd\omega_1 \dd\omega_2\, F(\omega_1,\omega_2)\ket{\omega_1,\omega_2}.
\end{equation}

The analysis is structured in two main steps. First, we study the HOM interferometric effect itself and derive a general expression for the coincidence probability for an arbitrary two-photon state (see Fig.~\ref{fig: general HOM} (a)). Second, in order to address parameter estimation beyond time-delay measurements, we assume that an unknown parameter $\theta$ is encoded in the two-photon state through a general unitary evolution $\hat V(\theta)=e^{-i\theta\hat H}$, where $\hat H$ is an arbitrary Hamiltonian acting on the time-frequency degrees of freedom. Time-delay estimation corresponds to the specific choice $\hat H=\hat \omega_1$, but more general evolutions can be considered, such as those discussed in Sec.~\ref{sec: TF evolutions} (see Fig.~\ref{fig: general HOM} (b)).

\begin{figure}[ht]
    \centering
    \footnotesize
    \renewcommand{\arraystretch}{2}
    \begin{tabular}{c}
        \scalebox{1}{\begin{tikzpicture}
	\begin{pgfonlayer}{nodelayer}
		\node [style=none] (0) at (-1.5, 0) {};
		\node [style=none] (1) at (0, 1.5) {};
		\node [style=none] (2) at (1.5, 0) {};
		\node [style=none] (3) at (0, -1.5) {};
		\node [style=none] (4) at (-1, 1) {};
		\node [style=none] (5) at (1, 1) {};
		\node [style=none] (6) at (1, -1) {};
		\node [style=none] (8) at (3, 3) {};
		\node [style=none] (9) at (3, -3) {};
		\node [style=none] (15) at (3, 4) {};
		\node [style=none] (16) at (4, 3) {};
		\node [style=none] (18) at (4, -3) {};
		\node [style=none] (19) at (3, -4) {};
		\node [style=none] (20) at (4, 4) {};
		\node [style=none] (21) at (5.75, 4) {};
		\node [style=none] (22) at (5.75, 2.25) {};
		\node [style=none] (23) at (6, 1.25) {};
		\node [style=none] (24) at (4, -4) {};
		\node [style=none] (25) at (5.75, -3.75) {};
		\node [style=none] (26) at (7, -2.5) {};
		\node [style=none] (27) at (6, -1.25) {};
		\node [style=none] (29) at (0, -2.5) {BS};
		\node [style=none] (31) at (2.5, 1.5) {1};
		\node [style=none] (33) at (2.5, -1.5) {2};
		\node [style=none] (35) at (-3, 2) {};
		\node [style=Photon] (54) at (-4, 1.75) {~~};
		\node [style=Photon] (55) at (-4, -1.75) {~~};
		\node [style=none] (57) at (-2, 1.25) {1};
		\node [style=none] (59) at (-2, -1.25) {2};
		\node [style=none] (64) at (10, 0) {$P_c$};
		\node [style=none] (65) at (4, 1.25) {};
		\node [style=none] (66) at (8, 1.25) {};
		\node [style=none] (67) at (4, -1.25) {};
		\node [style=none] (68) at (8, -1.25) {};
		\node [style=none] (69) at (6, 0.5) {Coincidence};
		\node [style=none] (70) at (6, -0.5) {detection};
		\node [style=none] (71) at (8.25, 0) {};
		\node [style=none] (72) at (9.5, 0) {};
		\node [style=none] (80) at (-1, -1) {};
		\node [style=none] (81) at (-3, -2) {};
		\node [style=none] (112) at (-4.25, 0.5) {};
		\node [style=none] (113) at (-3.75, 0.5) {};
		\node [style=none] (114) at (-4, 0.75) {};
		\node [style=none] (115) at (-4, 0) {};
		\node [style=none] (116) at (-4.25, -0.5) {};
		\node [style=none] (117) at (-3.75, -0.5) {};
		\node [style=none] (118) at (-4, -0.75) {};
		\node [style=none] (119) at (-5.5, 0) {$\ket{\psi}$};
	\end{pgfonlayer}
	\begin{pgfonlayer}{edgelayer}
		\draw [style=Fill grey] (2.center)
			 to (3.center)
			 to (0.center)
			 to (1.center)
			 to cycle;
		\draw (0.center) to (2.center);
		\draw [style=Fill red] (16.center)
			 to (15.center)
			 to [bend left=90, looseness=1.75] cycle;
		\draw [style=Fill red] (19.center)
			 to (18.center)
			 to [bend left=90, looseness=1.75] cycle;
		\draw [in=165, out=45, looseness=1.75] (20.center) to (21.center);
		\draw [in=60, out=-15, looseness=1.25] (21.center) to (22.center);
		\draw [in=90, out=-105] (22.center) to (23.center);
		\draw [in=165, out=-45] (24.center) to (25.center);
		\draw [in=-60, out=-15, looseness=1.50] (25.center) to (26.center);
		\draw [in=-90, out=120, looseness=1.25] (26.center) to (27.center);
		\draw (65.center) to (66.center);
		\draw (66.center) to (68.center);
		\draw (68.center) to (67.center);
		\draw (67.center) to (65.center);
		\draw [style=Thick arrow] (71.center) to (72.center);
		\draw [style=dashed arrow] (5.center) to (8.center);
		\draw [style=dashed arrow] (6.center) to (9.center);
		\draw [style=dashed arrow, in=120, out=0, looseness=0.75] (35.center) to (4.center);
		\draw [style=dashed arrow, in=-120, out=0, looseness=0.75] (81.center) to (80.center);
		\draw [in=180, out=105] (112.center) to (114.center);
		\draw [in=75, out=0] (114.center) to (113.center);
		\draw [in=90, out=-120, looseness=0.75] (115.center) to (116.center);
		\draw [in=-180, out=-105] (116.center) to (118.center);
		\draw [in=-75, out=0] (118.center) to (117.center);
		\draw [in=-60, out=90, looseness=0.75] (117.center) to (115.center);
		\draw [in=-90, out=120, looseness=0.75] (115.center) to (112.center);
		\draw [in=60, out=-90, looseness=0.75] (113.center) to (115.center);
	\end{pgfonlayer}
\end{tikzpicture}}\\
        \normalsize (a) \\
        \footnotesize \scalebox{1}{\begin{tikzpicture}
	\begin{pgfonlayer}{nodelayer}
		\node [style=none] (0) at (-1.5, 0) {};
		\node [style=none] (1) at (0, 1.5) {};
		\node [style=none] (2) at (1.5, 0) {};
		\node [style=none] (3) at (0, -1.5) {};
		\node [style=none] (4) at (-1, 1) {};
		\node [style=none] (5) at (1, 1) {};
		\node [style=none] (6) at (1, -1) {};
		\node [style=none] (8) at (3, 3) {};
		\node [style=none] (9) at (3, -3) {};
		\node [style=none] (15) at (3, 4) {};
		\node [style=none] (16) at (4, 3) {};
		\node [style=none] (18) at (4, -3) {};
		\node [style=none] (19) at (3, -4) {};
		\node [style=none] (20) at (4, 4) {};
		\node [style=none] (21) at (5.75, 4) {};
		\node [style=none] (22) at (5.75, 2.25) {};
		\node [style=none] (23) at (6, 1.25) {};
		\node [style=none] (24) at (4, -4) {};
		\node [style=none] (25) at (5.75, -3.75) {};
		\node [style=none] (26) at (7, -2.5) {};
		\node [style=none] (27) at (6, -1.25) {};
		\node [style=none] (29) at (0, -2.5) {BS};
		\node [style=none] (31) at (2.5, 1.5) {1};
		\node [style=none] (33) at (2.5, -1.5) {2};
		\node [style=none] (35) at (-3, 3) {};
		\node [style=Photon] (54) at (-7.5, 1.5) {~~~};
		\node [style=Photon] (55) at (-7.5, -1.5) {~~~};
		\node [style=none] (57) at (-2, 1.5) {1};
		\node [style=none] (59) at (-2, -1.5) {2};
		\node [style=none] (64) at (10, 0) {$P_c$};
		\node [style=none] (65) at (4, 1.25) {};
		\node [style=none] (66) at (8, 1.25) {};
		\node [style=none] (67) at (4, -1.25) {};
		\node [style=none] (68) at (8, -1.25) {};
		\node [style=none] (69) at (6, 0.5) {Coincidence};
		\node [style=none] (70) at (6, -0.5) {detection};
		\node [style=none] (71) at (8.25, 0) {};
		\node [style=none] (72) at (9.5, 0) {};
		\node [style=none] (77) at (-4, -1) {$e^{-i\theta\hat H}$};
		\node [style=none] (78) at (-7, 2) {};
		\node [style=none] (79) at (-5, 3) {};
		\node [style=none] (80) at (-1, -1) {};
		\node [style=none] (81) at (-3, -3) {};
		\node [style=none] (88) at (-7, -2) {};
		\node [style=none] (89) at (-5, -3) {};
		\node [style=none] (95) at (-4.75, 3.5) {};
		\node [style=none] (96) at (-3.25, 3.5) {};
		\node [style=none] (97) at (-3.25, -3.5) {};
		\node [style=none] (98) at (-4.75, -3.5) {};
		\node [style=none] (101) at (-5, 3.25) {};
		\node [style=none] (102) at (-3, 3.25) {};
		\node [style=none] (103) at (-3, -3.25) {};
		\node [style=none] (104) at (-5, -3.25) {};
		\node [style=none] (105) at (-4, 1) {$\hat V(\theta)$};
		\node [style=none] (106) at (-4, 0) {$=$};
		\node [style=none] (107) at (-7.75, 0.5) {};
		\node [style=none] (108) at (-7.25, 0.5) {};
		\node [style=none] (109) at (-7.5, 0.75) {};
		\node [style=none] (110) at (-7.5, 0) {};
		\node [style=none] (111) at (-7.75, -0.5) {};
		\node [style=none] (112) at (-7.25, -0.5) {};
		\node [style=none] (113) at (-7.5, -0.75) {};
		\node [style=none] (114) at (-8.5, 0) {$\ket{\psi}$};
	\end{pgfonlayer}
	\begin{pgfonlayer}{edgelayer}
		\draw [style=Fill grey] (2.center)
			 to (3.center)
			 to (0.center)
			 to (1.center)
			 to cycle;
		\draw (0.center) to (2.center);
		\draw [style=Fill red] (16.center)
			 to (15.center)
			 to [bend left=90, looseness=1.75] cycle;
		\draw [style=Fill red] (19.center)
			 to (18.center)
			 to [bend left=90, looseness=1.75] cycle;
		\draw [in=165, out=45, looseness=1.75] (20.center) to (21.center);
		\draw [in=60, out=-15, looseness=1.25] (21.center) to (22.center);
		\draw [in=90, out=-105] (22.center) to (23.center);
		\draw [in=165, out=-45] (24.center) to (25.center);
		\draw [in=-60, out=-15, looseness=1.50] (25.center) to (26.center);
		\draw [in=-90, out=120, looseness=1.25] (26.center) to (27.center);
		\draw (65.center) to (66.center);
		\draw (66.center) to (68.center);
		\draw (68.center) to (67.center);
		\draw (67.center) to (65.center);
		\draw [style=Thick arrow] (71.center) to (72.center);
		\draw [style=dashed arrow] (5.center) to (8.center);
		\draw [style=dashed arrow] (6.center) to (9.center);
		\draw [style=dashed arrow, in=120, out=0, looseness=0.75] (35.center) to (4.center);
		\draw [style=dashed arrow, in=-180, out=45, looseness=0.75] (78.center) to (79.center);
		\draw [style=dashed arrow, in=180, out=-45, looseness=0.75] (88.center) to (89.center);
		\draw [style=dashed arrow, in=-120, out=0, looseness=0.75] (81.center) to (80.center);
		\draw (97.center)
			 to (98.center)
			 to [bend right=315] (104.center)
			 to (101.center)
			 to [bend left=45] (95.center)
			 to (96.center)
			 to [bend left=45] (102.center)
			 to (103.center)
			 to [bend left=45] cycle;
		\draw [in=180, out=105] (107.center) to (109.center);
		\draw [in=75, out=0] (109.center) to (108.center);
		\draw [in=90, out=-120, looseness=0.75] (110.center) to (111.center);
		\draw [in=-180, out=-105] (111.center) to (113.center);
		\draw [in=-75, out=0] (113.center) to (112.center);
		\draw [in=-60, out=90, looseness=0.75] (112.center) to (110.center);
		\draw [in=-90, out=120, looseness=0.75] (110.center) to (107.center);
		\draw [in=60, out=-90, looseness=0.75] (108.center) to (110.center);
	\end{pgfonlayer}
\end{tikzpicture}}\\
        \normalsize (b)
    \end{tabular}
    \caption[General HOM interferometer and its extension for parameter estimation]{(a) General HOM interferometer fed by two photons prepared in an arbitrary two-photon state $\ket{\psi}$, fully characterized by the joint spectral amplitude $F$. (b) Extension of the HOM interferometer in which an unknown parameter $\theta$ is encoded in the two-photon state via a unitary evolution $\hat V(\theta)=e^{-i\theta\hat H}$ acting on the time-frequency degrees of freedom.}
    \label{fig: general HOM}
\end{figure}
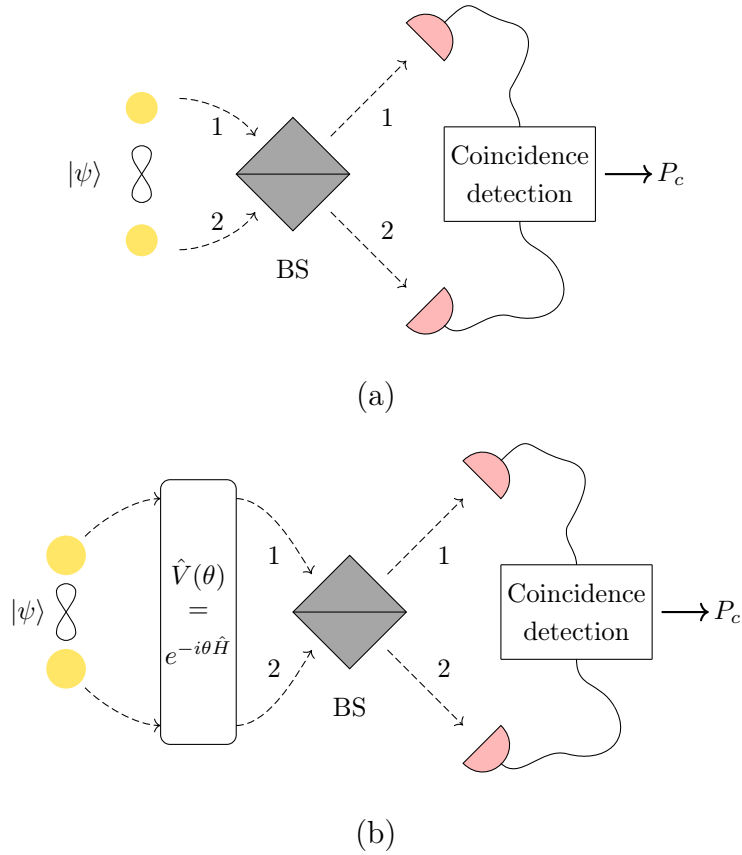

\paragraph{\newc Comment on time detection}

As discussed in the previous section, we disregard explicit time-of-detection information. In other words, only the global information of whether the two photons arrived at the same port or not is considered. Clearly, incorporating the individual arrival times of the photons into the analysis could, in principle, increase the estimation precision. Although we do not perform the full general analysis here, several important remarks can be made.

First, for the additional information carried by time-resolved detection to be operationally useful, the detector time resolution, or equivalently its timing jitter, must be at least of the same order of magnitude as the intrinsic temporal sensitivity of the HOM interferometer. Otherwise, the gain would be negligible. Since attosecond-scale precision has been demonstrated using HOM interferometry~\cite{lyons_attosecond-resolution_2018}, exploiting such an advantage via direct time detection would require substantial further improvements in single-photon detector technology. More fundamentally, time detection is sensitive only to the temporal intensity profile of the photon wavepacket, whereas HOM interferometry probes the full complex structure of the two-photon state. Consequently, there exist situations in which HOM-based estimation can {\it a priori} achieve a higher precision than time-resolved detection, even assuming the latter is performed with infinite temporal resolution.

To make this statement explicit, consider a simplified scenario in which a single photon is used to estimate a time delay and is measured with an infinitely precise time detector. Although this setup is slightly simpler than the two-photon HOM configuration supplemented with time-resolved detection, it already captures the essential mechanism responsible for the difference in performance. Using the temporal representation of the photon state,
\begin{equation}
    \ket{\psi}=\int \dd t\, \tilde F(t)\ket{t},
\end{equation}
we can compute both the quantum Fisher information and the Fisher information associated with ideal time detection. As shown in Appendix~\ref{app: hom interferometry}, Result~\ref{res: time detection precision}, they are given by
\begin{align}
    \mathcal Q&= 4\left[\int \dd t\, \abs{\partial_t \tilde F(t)}^2 - \left(i\int \dd t\, \tilde F(t)^* \partial_t \tilde F(t)\right)^2\right], &
    \mathcal F&= \int \dd t\, \frac{1}{\abs*{\tilde F(t)}^2}\left(\partial_t \abs{\tilde F(t)}^2\right)^2.
\end{align}

In this result, we further show that $\mathcal F=\mathcal Q$ if and only if the phase of $\tilde F$ is an affine function of time, that is, $t\mapsto at+b$. This condition admits a clear interpretation. Ideal time detection accesses only the probability density $\abs{\tilde F(t)}^2$ and is therefore completely insensitive to the phase of $\tilde F(t)$. If the phase exhibits a non-affine temporal dependence, a temporal delay $\tau$ modifies it in a non-trivial manner. In such a situation, part of the information about $\tau$ is encoded in the local phase structure of the wavepacket. Although analyzing the phase would provide additional information about $\tau$, this contribution is irretrievably lost in a time-of-arrival measurement, resulting in $\mathcal F<\mathcal Q$.

In contrast, HOM interferometry is sensitive to both the modulus and the phase of $\tilde F(t)$ through two-photon interference. It can therefore exploit the full information content of the wavepacket and attain the quantum limit even when the phase of $\tilde F$ is not affine. This demonstrates that HOM interferometry can outperform ideal time detection, not because of technical imperfections in the latter, but because it probes a strictly larger set of state properties relevant for time-delay estimation.

\subsection{Coincidence probability}
\label{subsec: TF HOM coincidence probability}

\paragraph{\lit Expression}
Starting from a general two-photon state in the time-frequency domain,
\begin{equation}
    \ket{\psi} = \int \dd\omega_1 \dd\omega_2\, F(\omega_1,\omega_2)\ket{\omega_1,\omega_2} 
    = \int \dd\omega_1 \dd\omega_2\, F(\omega_1,\omega_2)\hat a_1^\dagger(\omega_1)\hat a_2^\dagger(\omega_2)\vac,
\end{equation}
the derivation of the coincidence probability follows straightforwardly, in a manner analogous to the previous derivation. Assuming the beam splitter does not mix frequency modes, the beam splitter transformation reads
\begin{align}
    \hat a_1^\dagger(\omega) &\mapsto \frac{1}{\sqrt{2}}\big(\hat a_1^\dagger(\omega) + \hat a_2^\dagger(\omega)\big), &
    \hat a_2^\dagger(\omega) &\mapsto \frac{1}{\sqrt{2}}\big(\hat a_1^\dagger(\omega) - \hat a_2^\dagger(\omega)\big).
\end{align}
The state at the output of the beam splitter is then
\begin{align}
    \ket{\psi}_\text{out} &= \frac{1}{2}\int \dd\omega_1 \dd\omega_2\, F(\omega_1,\omega_2)
    \Big(\hat a_1^\dagger(\omega_1)\hat a_1^\dagger(\omega_2) - \hat a_2^\dagger(\omega_1)\hat a_2^\dagger(\omega_2) \notag \\
    &\qquad - \hat a_1^\dagger(\omega_1)\hat a_2^\dagger(\omega_2) + \hat a_1^\dagger(\omega_2)\hat a_2^\dagger(\omega_1)\Big)\vac.
\end{align}
Post-selecting on coincidences yields
\begin{subequations}
\begin{align}
    \ket{\psi}_c &= \frac{1}{2}\int \dd\omega_1 \dd\omega_2\, F(\omega_1,\omega_2)\big(\hat a_1^\dagger(\omega_1)\hat a_2^\dagger(\omega_2) - \hat a_1^\dagger(\omega_2)\hat a_2^\dagger(\omega_1)\big)\vac, \\
    &= \frac{1}{2}\int \dd\omega_1 \dd\omega_2\, \big(F(\omega_1,\omega_2) - F(\omega_2,\omega_1)\big) \hat a_1^\dagger(\omega_1)\hat a_2^\dagger(\omega_2)\vac, \\
    &= \frac{1}{2}\int \dd\omega_1 \dd\omega_2\, \big(F(\omega_1,\omega_2) - F(\omega_2,\omega_1)\big) \ket{\omega_1,\omega_2}.
\end{align}
\end{subequations}
The corresponding coincidence probability is then
\begin{subequations}
\begin{align}
    P_c &=  \prescript{}{c}{\langle} \psi {| \psi \rangle}_{c}, \\
    &= \frac{1}{4}\int \dd\omega_1 \dd\omega_2\, \abs{F(\omega_1,\omega_2) - F(\omega_2,\omega_1)}^2, \\
    &= \frac{1}{2}\left(1 - \int \dd\omega_1 \dd\omega_2\, F(\omega_1,\omega_2) F^*(\omega_2,\omega_1)\right).
\end{align}
\end{subequations}
In particular, for a factorizable two-photon state, $F(\omega_1,\omega_2) = f(\omega_1) f(\omega_2)$, one recovers the previous expression of the coincidence probability for independent photons. 

\paragraph{\publi Symmetry}
To provide a more general interpretation of the coincidence probability in the HOM interferometer, highlighting its relation to symmetry, we introduce the operator $\hat S$ defined by its action on the vacuum and creation operators:
\begin{align}
    \hat S \vac = \vac, &&
    \hat S \hat a_1^\dagger(\omega) = \hat a_2^\dagger(\omega) \hat S, &&
    \hat S \hat a_2^\dagger(\omega) = \hat a_1^\dagger(\omega) \hat S.
\end{align}
As shown in Appendix~\ref{app: hom interferometry}, Result~\ref{res: def and ppt of S}, $\hat S$ is well defined, Hermitian, unitary, and satisfies $\hat S^2 = \1$. Its physical interpretation is simple: it corresponds to exchanging the two spatial input modes of the HOM interferometer.  

Basic linear algebra ensures that any state can be decomposed into symmetric and anti-symmetric parts:
\begin{equation}
    \ket{\psi} = \ket{\psi_s} + \ket{\psi_a} = \frac{\ket{\psi} + \hat S\ket{\psi}}{2} + \frac{\ket{\psi} - \hat S\ket{\psi}}{2},
\end{equation}
with $\hat S \ket{\psi_s} = \ket{\psi_s}$ and $\hat S \ket{\psi_a} = - \ket{\psi_a}$. The action of $\hat S$ on a Hermitian operator $\hat H$ can be analogously defined as $\hat S \hat H \hat S$, which corresponds to swapping the roles of the two spatial modes in $\hat H$.  For any state $\ket{\psi}$, the Cauchy-Schwarz inequality shows that the expectation value $\bra{\psi} \hat S \ket{\psi}$ satisfies
\begin{equation}
    -1 \leq \bra{\psi} \hat S \ket{\psi} \leq 1.
\end{equation}
It equals $1$ for purely symmetric states and $-1$ for purely anti-symmetric states, providing a continuous measure of the symmetry of the state. Importantly, noting that
\begin{equation}
    \hat S \ket{\psi} = \int \dd\omega_1 \dd\omega_2\, F(\omega_2,\omega_1)\ket{\omega_1,\omega_2},
\end{equation}
the coincidence probability can be compactly written as
\begin{equation}\label{eq: PC with S}
    P_c = \frac{1}{2} \left( 1 - \bra{\psi} \hat S \ket{\psi} \right).
\end{equation}

\paragraph{\publi Interpretation}
The form of the coincidence probability in Eq.~\eqref{eq: PC with S}, expressed as the expectation value of the symmetry operator $\hat S$, highlights several fundamental aspects of the HOM effect:

\begin{itemize}
    \item First, it provides a universal expression for the coincidence probability that is valid for \emph{any} two-photon state, without assuming the presence or absence of correlations between the photons. This general expression can be directly applied to a wide range of scenarios, avoiding the need to reproduce the full derivation in each case. Moreover, for evolved states of the form $e^{-i\hat H\theta}\ket{\psi}$, the coincidence probability can be immediately obtained by substituting the evolved state, yielding
    \begin{equation}\label{eq: pc with s and evol}
        P_c(\theta)=\frac{1}{2}\left(1-\bra{\psi}e^{i\hat H\theta}\hat S e^{-i\hat H\theta}\ket{\psi}\right),
    \end{equation}
    which is particularly convenient for deriving metrological bounds for parameter estimation, as will be discussed in the next section.

    \item Second, this formulation reveals that the HOM effect is fundamentally linked to the \emph{symmetry properties} of the two-photon state. This perspective goes beyond the common interpretation that the HOM effect is tied to the notion of photon distinguishability. Indeed, there is no contradiction: distinguishability is only defined for independent incoming photons. If the two-photon state factorizes as $\ket{\psi}=\ket{\phi_1}\otimes\ket{\phi_2}$, one finds
    \begin{equation}
        \bra{\psi}\hat S\ket{\psi}=\bra{\phi_1}\otimes\bra{\phi_2}\hat S\ket{\phi_1}\otimes\ket{\phi_2}=\bra{\phi_1}\ket{\phi_2}\bra{\phi_2}\ket{\phi_1}=\abs{\braket{\phi_1}{\phi_2}}^2.
    \end{equation}
    The concept of symmetry, although not expressed in full generality, was already present in the seminal work by Hong, Ou, and Mandel~\cite{hong_measurement_1987}. Because of its practical importance, for instance in characterizing single-photon sources, the HOM effect is often interpreted in terms of modal overlap between independent photons, which has become the standard heuristic and has overshadowed the more fundamental symmetry-based interpretation. However, the latter is more general and provides deeper insight into the underlying physics of the HOM effect, as well as a powerful tool for analyzing its metrological applications.

    \item While we derived the coincidence probability within the framework of time-frequency systems, as we show in Sec.~\ref{sec: generalized HOM multiphoton}, this expression is universal and holds for photons in any degree of freedom. Consequently, the symmetry-based interpretation of the HOM effect applies to all implementations of the HOM interferometer, including polarization or transverse spatial modes.
\end{itemize}

As discussed above, the average symmetry of an incoming two-photon state can range from $-1$ to $1$. Hence, Eq.~\eqref{eq: PC with S} shows that for arbitrary input states, the coincidence probability can take any value in $[0,1]$. This exceeds the range allowed for independent photons, where the coincidence probability is bounded between $0$ and $1/2$. In particular, the HOM effect can lead to coincidence probabilities larger than $1/2$, signaling correlations in the input state and thus serving as an entanglement witness~\cite{brecht_characterizing_2013,eckstein_broadband_2008,ray_verifying_2011}.  

The extreme case corresponds to anti-symmetric states, which reach a coincidence probability of $1$. In this case, the usual HOM dip is replaced by a HOM peak, as illustrated in Fig.~\ref{fig: PC HOM with both sym}. An example of an anti-symmetric state is given by a joint spectral amplitude of the form
\begin{equation}
    F(\omega_1,\omega_2)=f_+(\omega_+)f_-(\omega_-),
\end{equation}
where $f_-$ is any odd function and $\omega_\pm=(\omega_1\pm \omega_2)/\sqrt{2}$. Such states can be generated in SPDC with appropriate phase-matching conditions~\cite{boucher_toolbox_2015}.

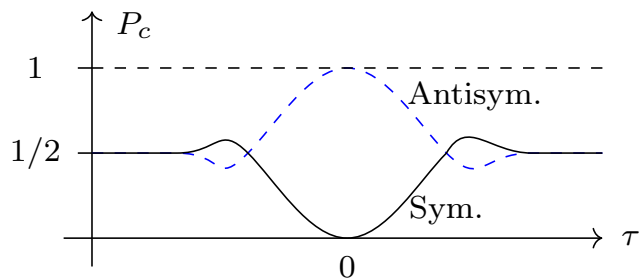
\begin{figure}[ht]
    \centering
    \scriptsize
    \scalebox{1.5}{\begin{tikzpicture}
	\begin{pgfonlayer}{nodelayer}
		\node [style=none] (0) at (0, -0.5) {};
		\node [style=none] (1) at (0, 4) {};
		\node [style=none] (2) at (-0.5, 0) {};
		\node [style=none] (3) at (9, 0) {};
		\node [style=none] (4) at (-0.25, 3) {};
		\node [style=none] (5) at (4.5, 0) {};
		\node [style=none] (6) at (0, 3) {};
		\node [style=none] (7) at (-1, 3) {$1$};
		\node [style=none] (8) at (-0.25, 1.5) {};
		\node [style=none] (9) at (0, 1.5) {};
		\node [style=none] (10) at (-1, 1.5) {$1/2$};
		\node [style=none] (11) at (4.5, -0.5) {$0$};
		\node [style=none] (12) at (9.5, 0) {$\tau$};
		\node [style=none] (13) at (2.75, 1.5) {};
		\node [style=none] (14) at (4.5, 3) {};
		\node [style=none] (15) at (6.25, 1.5) {};
		\node [style=none] (16) at (9, 1.5) {};
		\node [style=none] (17) at (9, 3) {};
		\node [style=none] (19) at (1.5, 1.5) {};
		\node [style=none] (20) at (7.75, 1.5) {};
		\node [style=none] (29) at (0.75, 3.75) {$P_c$};
		\node [style=none] (30) at (2.75, 1.5) {};
		\node [style=none] (31) at (4.5, 0) {};
		\node [style=none] (32) at (6.25, 1.5) {};
		\node [style=none] (33) at (1.5, 1.5) {};
		\node [style=none] (34) at (7.75, 1.5) {};
		\node [style=none] (35) at (6.25, 0.5) {Sym.};
		\node [style=none] (36) at (6.75, 2.5) {Antisym.};
	\end{pgfonlayer}
	\begin{pgfonlayer}{edgelayer}
		\draw [style=Arrow] (2.center) to (3.center);
		\draw [style=Arrow] (0.center) to (1.center);
		\draw (4.center) to (6.center);
		\draw (8.center) to (9.center);
		\draw [style=dashes blue] (9.center)
			 to (19.center)
			 to [in=-135, out=0, looseness=1.75] (13.center)
			 to [in=180, out=45, looseness=0.75] (14.center)
			 to [in=135, out=0, looseness=0.75] (15.center)
			 to [in=180, out=-45, looseness=1.50] (20.center)
			 to (16.center);
		\draw [style=dashes] (6.center) to (17.center);
		\draw (9.center)
			 to (33.center)
			 to [in=135, out=0, looseness=1.50] (30.center)
			 to [in=180, out=-45, looseness=0.75] (31.center)
			 to [in=-135, out=0, looseness=0.75] (32.center)
			 to [in=180, out=60, looseness=1.25] (34.center)
			 to (16.center);
	\end{pgfonlayer}
\end{tikzpicture}}
    \caption[Coincidence probability as a function of the relative delay]{Coincidence probability $P_c$ as a function of the relative delay. The solid black curve shows a typical HOM dip for a symmetric state, while the dashed blue curve illustrates a HOM peak for an anti-symmetric state.}
    \label{fig: PC HOM with both sym}
\end{figure}

\subsection{HOM and chrono-cyclic Wigner function}
\label{subsec: TF HOM Wigner function}

Let us consider an initial two-photon state whose joint spectral amplitude (JSA) is factorizable in the variables $\omega_\pm$, namely
\begin{equation}
    F(\omega_1,\omega_2)=f_+(\omega_+)f_-(\omega_-).
\end{equation}
As shown in Sec.~\ref{subsec: Wigner function for typical TF states}, this structure directly translates into a factorization of the corresponding chrono-cyclic Wigner function,
\begin{equation}
    W(\varphi_1,\varphi_2,\tau_1,\tau_2)=W_+(\varphi_+,\tau_+)W_-(\varphi_-,\tau_-).
\end{equation}
The functions $W_\pm$ describe, respectively, the diagonal and anti-diagonal components of the state in the rotated time-frequency coordinates.

A remarkable property of HOM interference is that it provides direct access to the Wigner function associated with the anti-diagonal component, $W_-(\varphi_-,\tau_-)$, as first observed in~\cite{douce_direct_2013}. This component encodes the symmetry properties of the two-photon state under exchange and therefore fully determines the HOM interference pattern. Consider the interferometric scheme depicted in Fig.~\ref{fig: HOM setup for Wigner}, where a time delay $\tau$ is applied in the first arm and a frequency shift $\varphi$ is applied in the second arm before the photons interfere on a balanced beam splitter. In this configuration, the coincidence probability at the output ports reads
\begin{equation}
    P_c(\tau,\varphi)=\frac{1}{2}\left(1-\pi W_-\left(\frac{\varphi}{\sqrt{2}},\frac{\tau}{\sqrt{2}}\right)\right).
\end{equation}
A detailed derivation is provided in Appendix~\ref{app: hom interferometry}, Result~\ref{res: PC with Wigner}. This expression shows that the HOM interferometer acts as a direct probe of the anti-diagonal Wigner function. By scanning the time delay and frequency shift, one effectively samples $W_-(\varphi_-,\tau_-)$ point by point in phase space. In particular, when only a time delay is varied, the resulting HOM interferogram corresponds to a cut of the Wigner function along the $\tau_-$ axis. This provides a clear geometric interpretation of the HOM dip: it reflects the structure of the anti-diagonal component of the two-photon state in the time-frequency phase space. Such a phase-space perspective is especially useful for assessing the symmetry properties of the state and for analyzing its metrological potential. Additionally, it shows that the HOM interferometer is sufficient to extract all the information contained in the anti-diagonal component of the state as shown in~\cite{douce_direct_2013}, where the Wigner function is reconstructed from HOM interferograms. 

\begin{figure}[ht]
    \centering
    \scriptsize
    \scalebox{1.2}{\begin{tikzpicture}
	\begin{pgfonlayer}{nodelayer}
		\node [style=none] (0) at (-1.5, 0) {};
		\node [style=none] (1) at (0, 1.5) {};
		\node [style=none] (2) at (1.5, 0) {};
		\node [style=none] (3) at (0, -1.5) {};
		\node [style=none] (4) at (-1, 1) {};
		\node [style=none] (5) at (1, 1) {};
		\node [style=none] (6) at (1, -1) {};
		\node [style=none] (8) at (3, 3) {};
		\node [style=none] (9) at (3, -3) {};
		\node [style=none] (15) at (3, 4) {};
		\node [style=none] (16) at (4, 3) {};
		\node [style=none] (18) at (4, -3) {};
		\node [style=none] (19) at (3, -4) {};
		\node [style=none] (20) at (4, 4) {};
		\node [style=none] (21) at (5.75, 4) {};
		\node [style=none] (22) at (5.75, 2.25) {};
		\node [style=none] (23) at (6, 1.25) {};
		\node [style=none] (24) at (4, -4) {};
		\node [style=none] (25) at (5.75, -3.75) {};
		\node [style=none] (26) at (7, -2.5) {};
		\node [style=none] (27) at (6, -1.25) {};
		\node [style=none] (29) at (0, -2.5) {BS};
		\node [style=none] (31) at (2.5, 1.5) {1};
		\node [style=none] (33) at (2.5, -1.5) {2};
		\node [style=none] (35) at (-2.75, 3) {};
		\node [style=Photon] (54) at (-7.5, 1.5) {~~~};
		\node [style=Photon] (55) at (-7.5, -1.5) {~~~};
		\node [style=none] (57) at (-2, 1.5) {1};
		\node [style=none] (59) at (-2, -1.5) {2};
		\node [style=none] (64) at (10, 0) {$P_c$};
		\node [style=none] (65) at (4, 1.25) {};
		\node [style=none] (66) at (8, 1.25) {};
		\node [style=none] (67) at (4, -1.25) {};
		\node [style=none] (68) at (8, -1.25) {};
		\node [style=none] (69) at (6, 0.5) {Coincidence};
		\node [style=none] (70) at (6, -0.5) {detection};
		\node [style=none] (71) at (8.25, 0) {};
		\node [style=none] (72) at (9.5, 0) {};
		\node [style=none] (77) at (-4, -3) {$e^{-i\varphi\hat t_2}$};
		\node [style=none] (78) at (-7, 2) {};
		\node [style=none] (79) at (-5.25, 3) {};
		\node [style=none] (80) at (-1, -1) {};
		\node [style=none] (81) at (-2.75, -3) {};
		\node [style=none] (88) at (-7, -2) {};
		\node [style=none] (89) at (-5.25, -3) {};
		\node [style=none] (95) at (-4.75, -2) {};
		\node [style=none] (96) at (-3.25, -2) {};
		\node [style=none] (97) at (-3.25, -4) {};
		\node [style=none] (98) at (-4.75, -4) {};
		\node [style=none] (101) at (-5, -2.25) {};
		\node [style=none] (102) at (-3, -2.25) {};
		\node [style=none] (103) at (-3, -3.75) {};
		\node [style=none] (104) at (-5, -3.75) {};
		\node [style=none] (107) at (-7.75, 0.5) {};
		\node [style=none] (108) at (-7.25, 0.5) {};
		\node [style=none] (109) at (-7.5, 0.75) {};
		\node [style=none] (110) at (-7.5, 0) {};
		\node [style=none] (111) at (-7.75, -0.5) {};
		\node [style=none] (112) at (-7.25, -0.5) {};
		\node [style=none] (113) at (-7.5, -0.75) {};
		\node [style=none] (114) at (-8.5, 0) {$\ket{\psi}$};
		\node [style=none] (115) at (-4, 3) {$e^{-i\hat\omega_1 \tau}$};
		\node [style=none] (118) at (-4.75, 4) {};
		\node [style=none] (119) at (-3.25, 4) {};
		\node [style=none] (120) at (-3.25, 2) {};
		\node [style=none] (121) at (-4.75, 2) {};
		\node [style=none] (122) at (-5, 3.75) {};
		\node [style=none] (123) at (-3, 3.75) {};
		\node [style=none] (124) at (-3, 2.25) {};
		\node [style=none] (125) at (-5, 2.25) {};
		\node [style=none] (126) at (-4, 4.75) {Time shift $\tau$};
		\node [style=none] (127) at (-4, -4.75) {Frequency shift $\varphi$};
	\end{pgfonlayer}
	\begin{pgfonlayer}{edgelayer}
		\draw [style=Fill grey] (2.center)
			 to (3.center)
			 to (0.center)
			 to (1.center)
			 to cycle;
		\draw (0.center) to (2.center);
		\draw [style=Fill red] (16.center)
			 to (15.center)
			 to [bend left=90, looseness=1.75] cycle;
		\draw [style=Fill red] (19.center)
			 to (18.center)
			 to [bend left=90, looseness=1.75] cycle;
		\draw [in=165, out=45, looseness=1.75] (20.center) to (21.center);
		\draw [in=60, out=-15, looseness=1.25] (21.center) to (22.center);
		\draw [in=90, out=-105] (22.center) to (23.center);
		\draw [in=165, out=-45] (24.center) to (25.center);
		\draw [in=-60, out=-15, looseness=1.50] (25.center) to (26.center);
		\draw [in=-90, out=120, looseness=1.25] (26.center) to (27.center);
		\draw (65.center) to (66.center);
		\draw (66.center) to (68.center);
		\draw (68.center) to (67.center);
		\draw (67.center) to (65.center);
		\draw [style=Thick arrow] (71.center) to (72.center);
		\draw [style=dashed arrow] (5.center) to (8.center);
		\draw [style=dashed arrow] (6.center) to (9.center);
		\draw [style=dashed arrow, in=120, out=0, looseness=0.75] (35.center) to (4.center);
		\draw [style=dashed arrow, in=-180, out=45, looseness=0.75] (78.center) to (79.center);
		\draw [style=dashed arrow, in=180, out=-45, looseness=0.75] (88.center) to (89.center);
		\draw [style=dashed arrow, in=-120, out=0, looseness=0.75] (81.center) to (80.center);
		\draw (97.center)
			 to (98.center)
			 to [bend right=315] (104.center)
			 to (101.center)
			 to [bend left=45] (95.center)
			 to (96.center)
			 to [bend left=45] (102.center)
			 to (103.center)
			 to [bend left=45] cycle;
		\draw [in=180, out=105] (107.center) to (109.center);
		\draw [in=75, out=0] (109.center) to (108.center);
		\draw [in=90, out=-120, looseness=0.75] (110.center) to (111.center);
		\draw [in=-180, out=-105] (111.center) to (113.center);
		\draw [in=-75, out=0] (113.center) to (112.center);
		\draw [in=-60, out=90, looseness=0.75] (112.center) to (110.center);
		\draw [in=-90, out=120, looseness=0.75] (110.center) to (107.center);
		\draw [in=60, out=-90, looseness=0.75] (108.center) to (110.center);
		\draw (120.center)
			 to (121.center)
			 to [bend right=315] (125.center)
			 to (122.center)
			 to [bend left=45] (118.center)
			 to (119.center)
			 to [bend left=45] (123.center)
			 to (124.center)
			 to [bend left=45] cycle;
	\end{pgfonlayer}
\end{tikzpicture}}
    \caption[HOM interferometer used to probe the anti-diagonal chrono-cyclic Wigner function]{HOM interferometer used to probe the anti-diagonal chrono-cyclic Wigner function. A time delay $\tau$ is applied in the first arm, while a frequency shift $\varphi$ is applied in the second arm before interference on a balanced beam splitter. The measured coincidence probability gives direct access to $W_-(\varphi_-,\tau_-)$ evaluated at $\left(\varphi/\sqrt{2},\tau/\sqrt{2}\right)$, enabling phase-space sampling of the anti-diagonal component of the two-photon state.}
    \label{fig: HOM setup for Wigner}
\end{figure}
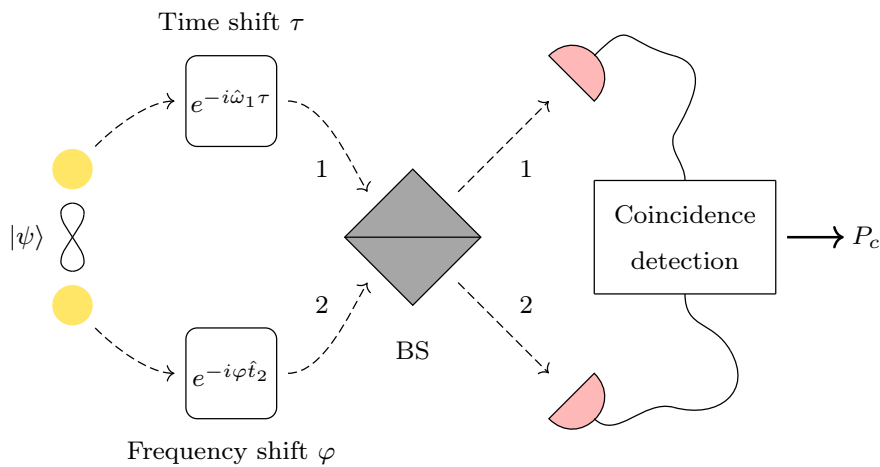

\subsection{General metrological analysis}
\label{subsec: TF HOM analysis}

\paragraph{\lit Framework and reminders}
The expression for the coincidence probability given in Eq.~\eqref{eq: PC with S}, together with its extension to parameter-dependent states in Eq.~\eqref{eq: pc with s and evol}, provides a general framework for analysing the metrological performance of HOM-based parameter estimation schemes. As already discussed in Sec.~\ref{subsec: TF HOM setup} and illustrated in Fig.~\ref{fig: general HOM} (b), we consider a generic unitary evolution of the form
\begin{equation}
    \hat V(\theta)=e^{-i\theta\hat H},
\end{equation}
where $\hat H$ is an arbitrary Hamiltonian acting on the time-frequency degrees of freedom of the two photons. The aim is to estimate the unknown parameter $\theta$ by measuring the coincidence probability at the output of the HOM interferometer for an arbitrary input state $\ket{\psi}$. We recall that, as introduced in Sec.~\ref{sec: metrology}, quantum metrology relies on two central quantities:
\begin{itemize}
    \item The (classical) Fisher information $\mathcal F$ associated with a specific measurement, which quantifies the precision achievable by that measurement for estimating the parameter $\theta$. This is the relevant quantity for assessing the efficiency of a HOM-based estimation strategy.
    \item The quantum Fisher information $\mathcal Q$ associated with the parameter-dependent quantum state $\ket{\psi_\theta}=e^{-i\hat H\theta}\ket{\psi}$, which characterises the maximum achievable precision over all possible measurements. It therefore provides a fundamental benchmark for evaluating the optimality of a HOM-based protocol. For pure states undergoing unitary evolution, the QFI takes the simple form
    \begin{equation}
        \mathcal Q=4\Delta^2 \hat H.
    \end{equation}
\end{itemize}

The optimality of a HOM-based estimation scheme is assessed by comparing the Fisher information $\mathcal F$ associated with coincidence measurements to the quantum Fisher information $\mathcal Q$ of the evolved state. When $\mathcal F=\mathcal Q$, the HOM measurement is optimal and achieves the best precision allowed by quantum mechanics for the given probe state and evolution.

\paragraph{\publi Precision bound}
Experimentally, HOM interferometry relies solely on coincidence detection. From a probabilistic point of view, the measurement therefore has only two possible outcomes: a coincidence event with probability $P_c$ and an anti-coincidence event with probability $P_a=1-P_c$. Applying the general expression of the Fisher information given in Eq.~\eqref{eq: Fisher information expression}, one obtains
\begin{equation}
    \mathcal F
    =\frac{1}{P_c}\left(\frac{\partial P_c}{\partial \theta}\right)^2
    +\frac{1}{P_a}\left(\frac{\partial P_a}{\partial \theta}\right)^2
    =\frac{1}{P_c(1-P_c)}\left(\frac{\partial P_c}{\partial \theta}\right)^2.
\end{equation}

Interestingly, this expression applied to the Cramér-Rao bound coincides exactly with the precision bound derived in Eq.~\eqref{eq: precision uncertainty propag} for time-delay estimation using a simple error-propagation argument. This agreement is not accidental: the error-propagation formula is a particular case of the more general Fisher-information formalism, applicable to arbitrary measurements and parameter-estimation problems.

In principle, one could insert the full expression of $P_c(\theta)$ into the formula above. However, this leads to lengthy expressions that are better suited for numerical evaluation than for analytical insight. Instead, following methods similar to those used to demonstrate the saturation of the Cramér-Rao bound for pure states under unitary evolution (see Appendix~\ref{app: formalism and framework}, Result~\ref{res: saturation QCRB pure states}), one can derive a compact and physically transparent expression under some assumptions.

Assuming first that the input state is perfectly symmetric or anti-symmetric under spatial mode exchange, $\hat S\ket{\psi}=\pm\ket{\psi}$, the coincidence probability can be expanded to second order in $\theta$ as (see Appendix~\ref{app: hom interferometry}, Result~\ref{res: expansion pc})
\begin{equation}
    P_c(\theta)=\frac{1}{2}\left(1\mp 1 \pm \frac{\theta^2}{2}\Delta^2(\hat H-\hat S\hat H\hat S)+O(\theta^3)\right).
\end{equation}
This directly leads to the following expression for the Fisher information in the limit $\theta\to 0$
\begin{equation}\label{eq: limit FI}
    \lim_{\theta\rightarrow 0} \mathcal F
    =\Delta^2(\hat H-\hat S\hat H\hat S).
\end{equation}
The detailed derivation is provided in Appendix~\ref{app: hom interferometry}, Result~\ref{res: Fisher info with S}. This result constitutes a general precision bound for HOM-based estimation strategies, valid for arbitrary generators $\hat H$ and for any input state satisfying the symmetry condition. It is important to stress that the Fisher information is obtained here as a limit at $\theta=0$. Technically, this arises from lifting an indeterminate form: the precision is evaluated at a point of the interferogram where both the coincidence probability and its first derivative vanish. The practical consequences of operating in this regime are discussed in Sec.~\ref{subsec: visibility impact on precision}.

\paragraph{\publi Symmetry and optimality}
Equation~\eqref{eq: limit FI} highlights the central role played by symmetry in HOM interferometry. First, it requires the input two-photon state to be either symmetric or anti-symmetric under spatial mode exchange. Second, it shows that the symmetry properties of the evolution generator $\hat H$ crucially determine the achievable precision.

If $\hat H$ is symmetric, \ie, if $\hat S\hat H\hat S=\hat H$, then the Fisher information vanishes and the HOM interferometer provides no sensitivity to the parameter $\theta$. Physically, such an evolution preserves the exchange symmetry of the two-photon quantum state and therefore does not modify the interference behavior at the beam splitter. Conversely, if $\hat H$ is anti-symmetric, \ie, if $\hat S\hat H\hat S=-\hat H$, the Fisher information reaches its maximum value $4\Delta^2\hat H$, for all perfectly symmetric or anti-symmetric states, which coincides with the quantum Fisher information. In this case, the HOM-based estimation strategy is optimal.

More generally, any Hamiltonian can be decomposed into a symmetric component and an anti-symmetric component,
\begin{equation}
    \hat H=\hat H_s+\hat H_a,
\end{equation}
where $\hat H_s=(\hat H+\hat S\hat H\hat S)/2$ and $\hat H_a=(\hat H-\hat S\hat H\hat S)/2$. The precision achievable with a HOM-based strategy is then governed solely by the variance of the anti-symmetric component,
\begin{equation}
    \lim_{\theta\rightarrow 0} \mathcal F
    =4\Delta^2 \hat H_a
    \leq 4\Delta^2 \hat H
    =\mathcal Q.
\end{equation}

HOM-based parameter-estimation protocols can therefore be interpreted intuitively as follows. The HOM effect probes the exchange symmetry of the two-photon quantum state. Maximum sensitivity is achieved by preparing an initially symmetric state and applying an evolution that perturbs this symmetry as strongly as possible, a role played precisely by the anti-symmetric component of the generator $\hat H$. 

It is interesting to note the similarity between the role played by the symmetry operator $\hat S$ in the present framework and the notion developed in~\cite{frerot_symmetry_2024}. In that work, closely related symmetry conditions imposed on both the probe state and the parameter-encoding evolution were shown to guarantee optimality in parameter estimation. This parallel highlights that symmetry considerations can constitute a fundamental resource that determines whether optimal precision can be achieved.

\subsection{Explicit examples}
\label{subsec: TF HOM examples}
To provide concrete illustrations of the general results derived above, we now consider specific evolutions that are central to time-frequency metrology: time-delay estimation, dispersion and rotations in the time-frequency phase space. For each example, we give the value of the Fisher information, obtained for parameter estimation at the origin, and we compare the performance of HOM-based strategies to the quantum Fisher information, while commenting on the different cases.

%To provide concrete expression we explicitly derive the form of the different variance for two different bi-photon states $\ket{\psi_G}$ and $\ket{\psi_C}$, where we assume that the JSA of each states can be factored in the $\omega_\pm$ variables as
%\begin{align}
%    F_G(\omega_1,\omega_2)=f_G(\omega_+)g_G(\omega_-), && F_C(\omega_1,\omega_2)=f_C(\omega_+)g_C(\omega_-),
%\end{align}
%with
%\begin{subequations}
%    \begin{align}
%        f_G(\omega_+)&=\frac{1}{(2\pi\sigma_+^2)^{1/4}}e^{-\frac{(\omega_+-\omega_p)^2}{4\sigma_+^2}}, & g_G(\omega_-)&=\frac{1}{(2\pi\sigma_-^2)^{1/4}}e^{-\frac{\omega_-^2}{4\sigma_-^2}}, \\
%        f_C(\omega_+)&=g_G(\omega_+), & g_C(\omega_-)&=\frac{1}{\sqrt{2}}\big(g_c(\omega_+-\Delta/2)-g_c(\omega_++\Delta/2)\big),
%    \end{align}    
%\end{subequations}
%where $\Delta$ is the distance between the two Gaussian peaks of $g_C$. We assume that the two peaks are well separated: $\Delta\gg \sigma_-$. Consequently, to simplify the computation, we consider that $g_C$ is approximately normalized to one. We can note, as already noticed before, that the spatial mode exchange $\omega_1\leftrightarrow\omega_2$ corresponds to mapping $(\omega_+,\omega_-)$ to $(\omega_+,-\omega_-)$. As such $\ket{\psi_G}$ is symmetric under spatial mode exchange, while $\ket{\psi_C}$ is anti-symmetric, and thus both states satisfies the condition necessary to apply the precision bound given in Eq.~\eqref{eq: limit FI}.

\paragraph{\publi Time-delay estimation}

Time-delay evolutions are generated by frequency operators. As discussed in previous chapters, several physically relevant generators can be considered. The most direct ones are $\hat \omega_1$ and $\hat \omega_2$, corresponding to the application of a time delay on one of the two spatial modes. It is also possible to implement correlated time delays acting on both photons, for instance through the collective generators $\hat \omega_\pm = \hat \omega_1 \pm \hat \omega_2$.\footnote{Notice the choice of normalization, which differs from that used in Sec.~\ref{subsec: Multimode TF operators}. The normalization depends on the experimental convention and does not lead to any fundamental difference in the attainable precision.}

Applying the general expressions for the Fisher information $\mathcal F$ and the quantum Fisher information $\mathcal Q$, we obtain the precision values reported in Table~\ref{tab: time delay precision}. Several remarks can be made.

\begin{itemize}

    \item The estimation precision for $\hat \omega_1$ and $\hat \omega_2$ has the same structure, and both lead to the same Fisher information $\mathcal F$. This confirms the usual interferometric intuition that, for symmetric configurations, the specific arm on which the delay is applied is irrelevant.

    \item For the local generators $\hat \omega_1$ and $\hat \omega_2$, one generally has $\mathcal F \leq \mathcal Q$, so the HOM scheme is not guaranteed to be optimal. However, for the class of states $\ket{\psi}$ that are factorizable in the $\omega_\pm$ variables,
    \begin{equation}
        \ket{\psi} = \int \dd \omega_1 \dd \omega_2 \, F_+(\omega_+) F_-(\omega_-) \ket{\omega_1,\omega_2},
    \end{equation}
    which, as discussed in Sec.~\ref{subsec: SPDC}, is a common structure for SPDC-generated states, the variance of $\hat \omega_j$ can be written as
    \begin{equation}
        4 \Delta^2 \hat \omega_j = \Delta^2 \hat \omega_+ + \Delta^2 \hat \omega_-.
    \end{equation}
    The estimation becomes optimal if and only if $\Delta^2 \hat \omega_+ = 0$, which requires a perfectly localized distribution $F_+(\omega_+) = \delta(\omega_+ - \omega_+^0)$ for some central frequency $\omega_+^0$. 

    When $\Delta^2 \hat \omega_+$ is non-zero, the non-optimality of the HOM scheme admits a clear interpretation. The HOM interferometer is sensitive only to relative delays between the two spatial modes. A finite variance $\Delta^2 \hat \omega_+$ corresponds to fluctuations associated with a global time reference, which do contribute to the quantum Fisher information but do not affect the exchange symmetry probed by the beam splitter. In this sense, the interferometric scheme does not exploit the full metrological content of the probe state.

    \item For the collective generator $\hat \omega_+$, the Fisher information vanishes, $\mathcal F = 0$. Indeed, $\hat \omega_+$ is symmetric under spatial mode exchange and implements the same time delay on both photons. Such a global delay leaves the symmetry properties of the joint internal state unchanged and therefore does not modify the coincidence probability at the beam splitter.

    \item In contrast, for $\hat \omega_-$ one finds $\mathcal F = \mathcal Q$. The generator $\hat \omega_-$ is anti-symmetric under exchange of the spatial modes and thus induces the strongest possible perturbation of the exchange symmetry of the internal state. Since the HOM interferometer precisely probes this symmetry, the corresponding estimation is optimal. Physically, $\hat \omega_-$ implements a relative time delay between the two modes, which is exactly the parameter to which HOM interference is maximally sensitive.

\end{itemize}

\begin{table}
    \setlength{\tabcolsep}{10pt}
    \renewcommand{\arraystretch}{1.3}
    \centering
    \begin{tabular}{|c|c|c|}
        \hline
        Generator & $\mathcal Q$ & $\mathcal F$\\
        \hline \hline
        $\hat \omega_1$ & $4\Delta^2 \hat \omega_1$ & $\Delta^2 \hat \omega_-$ \\
        \hline
        $\hat \omega_2$ & $4\Delta^2 \hat \omega_2$ & $\Delta^2 \hat \omega_-$ \\
        \hline
        $\hat \omega_+$ & $4\Delta^2 \hat \omega_+$ & $0$ \\
        \hline
        $\hat \omega_-$ & $4\Delta^2 \hat \omega_-$ & $4\Delta^2 \hat \omega_-$\\
        \hline
    \end{tabular}
    \caption[Quantum Fisher information and Fisher information for time-delay evolutions]{Quantum Fisher information $\mathcal Q$ and Fisher information $\mathcal F$, evaluated at the origin of the parameter, for different generators of time-delay evolutions. The HOM scheme is only sensitive to anti-symmetric generators that modify the exchange symmetry of the internal time-frequency state. Adapted from Ref.~\cite{descamps_time-frequency_2023}. \copyright 2023 American Physical Society.}
    \label{tab: time delay precision}
\end{table}

\paragraph{\newc Dispersion}

As introduced in Sec.~\ref{subsec: TF dispersion and time lens}, dispersion is implemented through quadratic frequency operators. With two spatial modes, several collective variants can be considered: $\hat \omega_1^2$, $\hat \omega_2^2$, $\hat \omega_\pm^2$, and $\hat \omega_1^2 \pm \hat \omega_2^2$. The corresponding precision values are given in Table~\ref{tab: dispersion precision}. The structure closely parallels the case of time delays.

\begin{itemize}

    \item The roles of the two spatial modes are again symmetric, and $\hat \omega_1^2$ and $\hat \omega_2^2$ lead to identical Fisher information values.

    \item For the local generators $\hat \omega_1^2$ and $\hat \omega_2^2$, the estimation is not necessarily optimal, since $\mathcal F \leq \mathcal Q$ in general. Nevertheless, they provide non-trivial sensitivity and may be relevant from an experimental perspective.

    \item The collective generators $\hat \omega_+^2$ and $\hat \omega_-^2$ are both symmetric under exchange of the spatial modes. Although $\hat \omega_-$ is anti-symmetric, the square removes this property. Consequently, these operations do not modify the exchange symmetry of the internal state and lead to $\mathcal F = 0$. The same holds for the correlated shear $\hat \omega_1^2 + \hat \omega_2^2$, which corresponds to a form of global dispersion.

    \item Among the quadratic generators considered, only $\hat \omega_1^2 - \hat \omega_2^2$ is anti-symmetric and yields $\mathcal F = \mathcal Q$. This operator implements a relative dispersion between the two modes. Since the HOM interferometer is sensitive only to transformations that alter the exchange symmetry of the internal state, relative dispersion is the only quadratic transformation that can be estimated optimally within this scheme.

\end{itemize}

\begin{table}
    \setlength{\tabcolsep}{10pt}
    \renewcommand{\arraystretch}{1.3}
    \centering
    \begin{tabular}{|c|c|c|}
        \hline
        Generator & $\mathcal Q$ & $\mathcal F$\\
        \hline \hline
        $\hat \omega_1^2$ & $4\Delta^2 \hat \omega_1^2$ & $\Delta^2 (\hat \omega_1^2 - \hat \omega_2^2) = \Delta^2 (\hat \omega_+ \hat \omega_-)$ \\
        \hline
        $\hat \omega_2^2$ & $4\Delta^2 \hat \omega_2^2$ & $\Delta^2 (\hat \omega_1^2 - \hat \omega_2^2) = \Delta^2 (\hat \omega_+ \hat \omega_-)$ \\
        \hline
        $\hat \omega_+^2$ & $4\Delta^2 \hat \omega_+^2$ & $0$ \\
        \hline
        $\hat \omega_-^2$ & $4\Delta^2 \hat \omega_-^2$ & $0$\\
        \hline
        $\hat \omega_1^2 + \hat \omega_2^2$ & $4\Delta^2 (\hat \omega_1^2 + \hat \omega_2^2)$ & $0$\\
        \hline
        $\hat \omega_1^2 - \hat \omega_2^2$ & $4\Delta^2 (\hat \omega_1^2 - \hat \omega_2^2)$ & $4\Delta^2 (\hat \omega_1^2 - \hat \omega_2^2)$\\
        \hline
    \end{tabular}
    \caption[Quantum Fisher information and Fisher information for different dispersion evolutions]{Quantum Fisher information $\mathcal Q$ and Fisher information $\mathcal F$, evaluated at the origin of the parameter, for different quadratic generators describing dispersion. Only the anti-symmetric combination $\hat \omega_1^2 - \hat \omega_2^2$, corresponding to relative dispersion, is estimated optimally with HOM interference.}
    \label{tab: dispersion precision}
\end{table}

\paragraph{\publi Rotations}

Finally, we consider rotations in time-frequency phase space. As discussed in Sec.~\ref{subsec: Multimode TF operators}, typical generators include $\hat R_1$, $\hat R_2$, $\hat R_\pm$, and $\hat R_1 \pm \hat R_2$. Owing to the quadratic structure of these operators, the qualitative behavior closely resembles that of dispersion. The precision values are summarized in Table~\ref{tab: rotation precision}.

As in the previous cases, only the anti-correlated rotation $\hat R_1 - \hat R_2$ is anti-symmetric under spatial mode exchange and therefore achieves $\mathcal F = \mathcal Q$. Local rotations $\hat R_1$ and $\hat R_2$ generally yield non-zero but suboptimal Fisher information. Collective rotations such as $\hat R_1 + \hat R_2$ and $\hat R_\pm$ are symmetric operations that leave the exchange symmetry of the internal quantum state unchanged, and consequently provide no sensitivity within the HOM scheme.

\begin{table}
    \setlength{\tabcolsep}{10pt}
    \renewcommand{\arraystretch}{1.3}
    \centering
    \begin{tabular}{|c|c|c|}
        \hline
        Generator & $\mathcal Q$ & $\mathcal F$\\
        \hline \hline
        $\hat R_1$ & $4\Delta^2 \hat R_1$ & $\Delta^2 (\hat R_1 - \hat R_2)$ \\ 
        \hline
        $\hat R_2$ & $4\Delta^2 \hat R_2$ & $\Delta^2 (\hat R_1 - \hat R_2)$ \\ 
        \hline
        $\hat R_+$ & $4\Delta^2 \hat R_+$ & $0$ \\
        \hline
        $\hat R_-$ & $4\Delta^2 \hat R_-$ & $0$\\
        \hline
        $\hat R_1 + \hat R_2$ & $4\Delta^2 (\hat R_1 + \hat R_2)$ & $0$\\
        \hline
        $\hat R_1 - \hat R_2$ & $4\Delta^2 (\hat R_1 - \hat R_2)$ & $4\Delta^2 (\hat R_1 - \hat R_2)$\\
        \hline
    \end{tabular}
    \caption[Quantum Fisher information and Fisher information for different time-frequency rotations]{Quantum Fisher information $\mathcal Q$ and Fisher information $\mathcal F$, evaluated at the origin of the parameter, for different generators of rotations in time-frequency phase space. As for dispersion, only the anti-symmetric combination corresponding to a relative rotation is estimated optimally using HOM interference. Adapted from Ref.~\cite{descamps_time-frequency_2023}. \copyright 2023 American Physical Society.}
    \label{tab: rotation precision}
\end{table}

\subsection{Visibility impact on precision}
\label{subsec: visibility impact on precision}

\paragraph{\publi General problem} 
In the previous section, we analyzed the precision of a HOM-based estimation scheme. A crucial step in the derivation of the expression of the FI was the assumption that the initial two-photon state is either symmetric or anti-symmetric under exchange of the spatial modes. This symmetry condition is in fact equivalent to requiring unit visibility of the interferogram. Recall that, in Sec.~\ref{subsec: time dependence and metrology}, we define the visibility as
\begin{equation}
    V=\max_\theta\abs{2P_c(\theta)-1},
\end{equation}
where $P_c(\theta)$ denotes the parameter-dependent coincidence probability,
\begin{equation}
    P_c(\theta)=\frac{1}{2}\left(1-\bra{\psi}e^{i\hat H\theta}\hat S e^{-i\hat H\theta}\ket{\psi}\right).
\end{equation}
The visibility is equal to one if and only if there exists a value of the parameter $\theta$ such that
\begin{equation}
    \bra{\psi}e^{i\hat H\theta}\hat S e^{-i\hat H\theta}\ket{\psi}=\pm 1.
\end{equation}
Choosing the origin of the parameter at this value of $\theta$, this condition is equivalent to
\begin{equation}
    \hat S\ket{\psi}=\pm\ket{\psi},
\end{equation}
which is precisely the symmetry requirement used to derive the precision bound given in Eq.~\eqref{eq: limit FI}. Therefore, unit visibility is not merely a convenient feature of the interferogram, but a direct signature that the two-photon state lies entirely in a definite symmetry sector with respect to spatial mode exchange.

In practice, concrete experiments never achieve perfect visibility. Any mismatch in temporal, spectral, spatial, or polarization degrees of freedom prevents the two-photon state from being purely symmetric or anti-symmetric, thereby lowering the visibility. This has a direct consequence for the computation of the FI at the origin. In the ideal case, the expression was obtained by evaluating the limit of a quotient of two vanishing quantities,
\begin{equation}
    \mathcal F=\lim_{\theta\rightarrow 0} 
    \frac{1}{P_c(\theta)(1-P_c(\theta))}
    \left(\frac{\partial P_c}{\partial \theta}\right)^2,
\end{equation}
with $P_c(1-P_c)\to 0$ and $\partial_\theta P_c\to 0$. The non-trivial finite value arises from the precise balance between numerator and denominator ensured by perfect symmetry. For non-unit visibility, however, $P_c(0)$ does not reach $0$ or $1$, so that $P_c(0)(1-P_c(0))\neq 0$ while $\partial_\theta P_c|_{\theta=0}=0$. As a consequence,
\begin{equation}
    \lim_{\theta\to 0}\mathcal F=0.
\end{equation}
Hence, the origin is no longer an optimal operating point. In Fig.~\ref{fig: FI realistic} we illustrate the typical shape of the FI as a function of the parameter $\theta$ for an imperfect HOM interferometer. The optimal estimation point shifts away from $\theta=0$, and the achievable precision is reduced compared to the ideal case.

\begin{figure}[ht]
    \centering
    \scriptsize
    \scalebox{1.8}{\begin{tikzpicture}
	\begin{pgfonlayer}{nodelayer}
		\node [style=none] (0) at (0, -0.5) {};
		\node [style=none] (1) at (0, 5) {};
		\node [style=none] (2) at (-0.5, 0) {};
		\node [style=none] (3) at (7.5, 0) {};
		\node [style=none] (6) at (0, 4) {};
		\node [style=none] (7) at (-1, 4) {$\mathcal F_{\text{ideal}}$};
		\node [style=none] (11) at (3.75, -0.5) {$0$};
		\node [style=none] (12) at (8, 0) {$\theta$};
		\node [style=none] (13) at (3.5, 3.25) {};
		\node [style=none] (14) at (3.75, 0) {};
		\node [style=none] (15) at (4, 3.25) {};
		\node [style=none] (19) at (2, 0) {};
		\node [style=none] (20) at (5.5, 0) {};
		\node [style=none] (29) at (0.75, 4.75) {$\mathcal F$};
		\node [style=none] (30) at (7.5, 4) {};
		\node [style=none] (31) at (3.75, 4) {};
		\node [style=none] (32) at (-1, 3.25) {$\mathcal F_{\text{max}}$};
		\node [style=none] (33) at (0, 3.25) {};
		\node [style=none] (34) at (7.5, 3.25) {};
		\node [style=none] (35) at (0, 0) {};
		\node [style=none] (36) at (7, 2.25) {};
		\node [style=none] (37) at (7.75, 2.25) {};
		\node [style=none] (38) at (7, 1.5) {};
		\node [style=none] (39) at (7.75, 1.5) {};
		\node [style=none] (40) at (8.5, 2.25) {\scalebox{0.8}{$V=1$}};
		\node [style=none] (41) at (8.5, 1.5) {\scalebox{0.8}{$V<1$}};
	\end{pgfonlayer}
	\begin{pgfonlayer}{edgelayer}
		\draw [style=Arrow, in=180, out=0] (2.center) to (3.center);
		\draw [style=Arrow] (0.center) to (1.center);
		\draw (19.center)
			 to [in=180, out=0, looseness=0.25] (13.center)
			 to [in=-180, out=0, looseness=0.25] (14.center)
			 to [in=-180, out=0, looseness=0.25] (15.center)
			 to [in=180, out=0, looseness=0.25] (20.center);
		\draw [style=dashes] (6.center) to (30.center);
		\draw [style=dashes blue] (35.center)
			 to (19.center)
			 to [in=180, out=0, looseness=0.25] (31.center)
			 to [in=-180, out=0, looseness=0.25] (20.center)
			 to (3.center);
		\draw [style=dashes] (33.center) to (34.center);
		\draw [style=dashes blue] (36.center) to (37.center);
		\draw (38.center) to (39.center);
	\end{pgfonlayer}
\end{tikzpicture}}
    \caption[Typical behavior of the Fisher information as a function of the estimation parameter for a HOM interferometer]{Typical behavior of the Fisher information $\mathcal F$ as a function of the parameter $\theta$ for a HOM interferometer. Dashed blue line: ideal visibility $V=1$, for which $\mathcal F$ reaches its maximal value $\mathcal F_\text{ideal}$ at $\theta=0$. Solid black line: reduced visibility $V<1$, corresponding to residual modal distinguishability in the two-photon state or imperfect symmetry preparation. The central peak splits into two symmetric maxima located at non-zero values of $\theta$, while $\mathcal F(0)=0$. The maximal value $\mathcal F_\text{max}$ is typically smaller than $\mathcal F_\text{ideal}$.}
    \label{fig: FI realistic}
\end{figure}
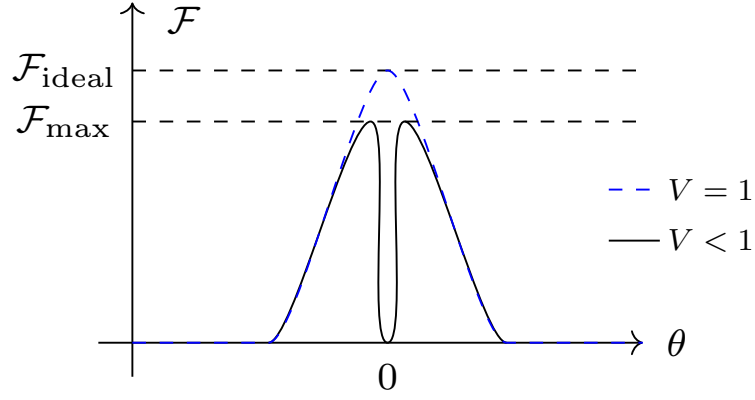

\paragraph{\publi Illustrative example: time-frequency cat-like state}

To make the link between visibility and estimation precision explicit, we consider a concrete example: a time-frequency cat-like state with a JSA $F$ factorizable in the diagonal variables,
\begin{equation}
    F(\omega_1,\omega_2)=F_+(\omega_+)F_-(\omega_-),
\end{equation}
where $F_+$ is arbitrary and does not affect the HOM interference, while $F_-$ is chosen as
\begin{equation}
    F_-(\omega)=\frac{1}{\mathcal N}
    \left[
    \exp\left(-\frac{(\omega-\Delta/2)^2}{4\sigma^2}\right)
    +
    \exp\left(-\frac{(\omega+\Delta/2)^2}{4\sigma^2}\right)
    \right],
\end{equation}
with normalization factor $\mathcal N$, peak separation $\Delta$, and width $\sigma$. This state is symmetric under $\omega_-\to -\omega_-$ and therefore yields perfect visibility $V=1$ for time-delay estimation.

To deliberately reduce the visibility, we apply a frequency shift on the second mode generated by $e^{-i\hat t_2\varphi}$. This operation modifies the relative-frequency structure of the two-photon state and breaks the parity symmetry of $F_-$. Considering an unknown delay $\tau$ applied in the first arm, the coincidence probability can be expressed in terms of the Wigner function as shown in Sec.~\ref{subsec: TF HOM Wigner function},
\begin{equation}
    P_c(\tau)=\frac{1}{2}
    \left(
    1-\pi W_-\left(\frac{\varphi}{\sqrt{2}},\frac{\tau}{\sqrt{2}}\right)
    \right),
\end{equation}
where $W_-$ is the Wigner function associated with $F_-$. Using the result derived in Sec.~\ref{subsec: Wigner function for typical TF states}, we obtain
\begin{align}
    P_c(\tau)=\frac{1}{2}\Bigg( 1-\frac{e^{-\sigma^2\tau^2}}{\mathcal N^2}
    \Bigg[ e^{-\frac{(\varphi/\sqrt{2}-\Delta/2)^2}{2\sigma^2}} +e^{-\frac{(\varphi/\sqrt{2}+\Delta/2)^2}{2\sigma^2}}  +2e^{-\frac{\varphi^2}{4\sigma^2}} \cos\left(\frac{\tau\Delta}{\sqrt{2}}\right) \Bigg] \Bigg).
\end{align}

In the regime of well-separated peaks $\Delta\gg\sigma$ and small displacement $\varphi\ll\Delta$, this simplifies to
\begin{equation}
    P_c(\tau)\approx 
    \frac{1}{2}
    \left(
    1-
    e^{-\sigma^2\tau^2}
    e^{-\frac{\varphi^2}{4\sigma^2}}
    \cos\left(\frac{\tau\Delta}{\sqrt{2}}\right)
    \right).
\end{equation}
The visibility is then
\begin{equation}
    V=\max_{\tau}\abs{2P_c-1}
    =
    e^{-\frac{\varphi^2}{4\sigma^2}},
\end{equation}
showing that the frequency shift $\varphi$ continuously tunes the visibility. The Fisher information can be computed in closed form,
\begin{equation}
    \mathcal F=
    \frac{
    V^2
    \left(
    2\sigma^2\tau\cos(\Delta\tau/\sqrt{2})
    +
    \frac{\Delta}{\sqrt{2}}\sin(\Delta\tau/\sqrt{2})
    \right)^2
    }{
    e^{2\sigma^2\tau^2}
    -
    V^2\cos^2(\Delta\tau/\sqrt{2})
    }.
\end{equation}

As a consistency check, in the case of perfect visibility $V=1$ and in the limit $\tau\to 0$, one recovers the ideal HOM precision,
\begin{equation}
    \mathcal F_\text{ideal}
    =
    \Delta^2(\hat \omega_1-\hat \omega_2)
    =
    \frac{1}{2}\Delta^2+2\sigma^2
    =
    \lim_{\tau\to 0}\mathcal F(V=1).
\end{equation}

In the regime of large $\Delta$, one obtains the approximate bound
\begin{equation}
    \mathcal F
    \lesssim
    \frac{V^2\Delta^2}{2}
    e^{-2\sigma^2\tau^2}
    \sin^2\left(\frac{\Delta\tau}{\sqrt{2}}\right),
\end{equation}
which explicitly shows that the achievable precision is reduced by a factor $V^2$ compared to the ideal case. Moreover, the optimal working point is no longer $\tau=0$, but approximately
\begin{equation}
    \tau_\text{opt}\approx \pm\frac{\pi\sqrt{2}}{2\Delta}.
\end{equation}

In Fig.~\ref{fig: grid FI cat state} we display the Fisher information as a function of the delay $\tau$ for different values of the visibility $V$ and peak separation $\Delta$. The reduction of the maximal precision and the shift of the optimal operating point are clearly visible.

\begin{figure}[ht]
    \centering
    \includegraphics[width=\linewidth]{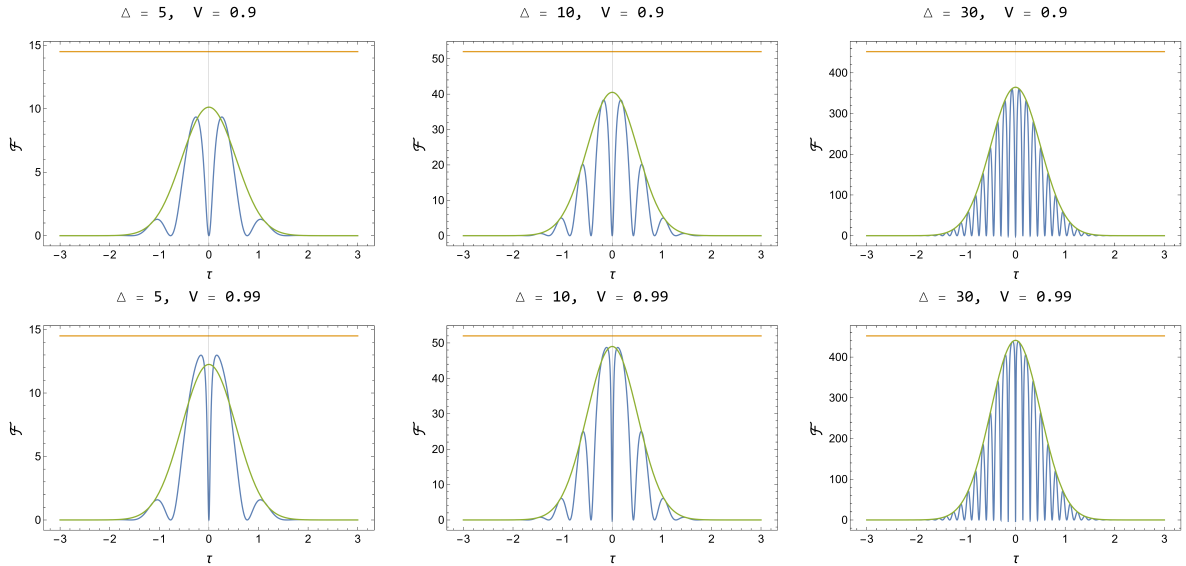}
    \caption[Fisher information as a function of the delay for different values of the visibility and peak separation]{Fisher information $\mathcal F$ as a function of the delay $\tau$ for different values of the visibility $V$ and peak separation $\Delta$, with $\sigma=1$. Blue curves: exact expression. Orange line: ideal limit $\mathcal F_\text{ideal}=\Delta^2(\hat \omega_1-\hat \omega_2)$. Green line: approximate Gaussian envelope $V^2\Delta^2 e^{-2\sigma^2\tau^2}/2$. The reduction of the maximal value and the displacement of the optimal estimation point away from $\tau=0$ become more pronounced as $V$ decreases.}
    \label{fig: grid FI cat state}
\end{figure}

\subsection{General visibility model}
\label{subsec: general visi model}

\paragraph{\publi General model}
While the imperfect symmetry preparation of the time-frequency cat-like state was initially justified by the application of a frequency shift, it is noteworthy that, under reasonable assumptions, the coincidence probability can be expressed in a form where the visibility appears with a simple dependence. This observation motivates the following general parametrization:
\begin{equation}\label{eq: general visibility model}
    P_c(\tau)=\frac{1}{2}\left(1-Vf(\tau)\right),
\end{equation}
where $f:\R\to[-1,1]$ is any function such that $\frac{1}{2}\left(1-f(\tau)\right)$ represents the ideal reference interferogram with perfect visibility $V=1$. It is important to emphasize that, although this general model was derived for the cat-like state using a frequency shift assumption, other physical mechanisms can also lead to such a visibility parametrization. Notably, imperfect polarization alignment of the two photons or any other imperfect preparation of a secondary degree of freedom will automatically reduce the visibility, producing a dependence of the form above.

Within this parametrization, the Fisher information simplifies to
\begin{equation}
    \mathcal F(\tau)=\frac{V^2 f'(\tau)^2}{1-V^2 f(\tau)^2}.
\end{equation}
Since $V\leq 1$ by definition, we have the simple inequality
\begin{equation}
    \mathcal F(\tau)\leq V^2\frac{ f'(\tau)^2}{1- f(\tau)^2}=V^2\mathcal F_{V=1}(\tau).
\end{equation}
Maximizing over the optimal estimation point $\tau$ leads to the general scaling bound
\begin{equation}\label{eq: scaling FI visi}
    \mathcal F_\text{max}(V)\leq V^2 \mathcal F_\text{ideal},
\end{equation}
where $\mathcal F_\text{ideal}$ denotes the maximal precision achievable with perfect visibility. This scaling law shows that under non-ideal visibility conditions, the HOM interferometer can never reach the QCR bound, even in the absence of losses.

Interestingly, as observed above, for the time-frequency cat-like state in the limit of large peak separation, this scaling law is saturated. This is particularly relevant, as similar analyses for other spectra $F_-$ show different behaviors. For example, a Gaussian spectrum would yield a precision loss much worse than the quadratic scaling predicted by Eq.~\eqref{eq: scaling FI visi}. This illustrates that the impact of visibility is highly dependent on the specific JSA of the states and highlights a new key property of cat states: their exceptional resilience against visibility degradation.

It is important to note that the quadratic scaling of visibility loss is highly dependent on the chosen parametrization of Eq.~\eqref{eq: general visibility model}. In particular, we assume that the realistic situation ($V<1$) can be compared with an ideal reference obtained by rescaling the interferogram. While this is physically meaningful when a specific mechanism is identified as the source of visibility reduction, for probe states with inherently imperfect symmetry and no natural ideal reference, extrapolating precision from hypothetical improved visibility can lead to misleading conclusions.

\paragraph{\publi Analysis of the robustness of the cat state}
To understand the origin of the robustness of the time-frequency cat-like state, it is instructive to examine the derivation of the general quadratic scaling. This scaling arises from replacing the denominator $1-V^2 f(\tau)^2$ with its upper bound $1-f(\tau)^2$. This is a conservative bound, which is only saturated when $f(\tau)\approx 0$. In the context of HOM estimation, the symmetry condition allows us to derive a precision bound specifically at $\tau=0$, where $f(\tau)$ reaches an extremum, $f(0)=\pm1$. Optimal scaling is achieved if the precision for perfect visibility is the same at $\tau=0$ and at points where $f(\tau)\approx 0$. This naturally raises the question: which functions $f(\tau)$ yield a constant Fisher information? In Appendix~\ref{app: hom interferometry}, Result~\ref{res: constant FI}, we show that the only functions $f$ leading to constant Fisher information $\mathcal F$ are cosines~\cite{cafaro_decrease_2018}:
\begin{equation}
    f(\tau)=\cos(\sqrt{\mathcal F}\,\tau+\phi).
\end{equation}
This is precisely the form of the interferogram obtained for the time-frequency cat-like state in the regime of large peak separation. Therefore, the robustness of the cat-like state is not accidental: it follows directly from the structure of its JSA, which shapes the Wigner function to produce a cosine interferogram. Consequently, the Fisher information remains constant near $\tau=0$, leading to a saturation of the visibility scaling.

\subsection{Experimental verification of the visibility model}
\label{subsec: Visi experimental verification}
\paragraph{\publi Engineered two-photon states and quantum source}

The validity of the model is assessed through an experiment designed to engineer two-photon states characterized by different anti-diagonal spectral functions $F_-(\omega_-)$, chosen to exhibit distinct scaling behaviors of the HOM visibility $V$. The experiment was performed by our collaborators Othmane Meskine, Sara Ducci and Florent Baboux using a quantum source based on an AlGaAs Bragg reflector waveguide, generating polarization-entangled photon pairs via type-II spontaneous parametric down-conversion at telecom wavelengths and operating at room temperature~\cite{appas_nonlinear_2022}. A schematic of the experimental setup is shown in Fig.~\ref{fig: experimental HOM}. A continuous-wave laser at $\lambda_\text{pump}=772.42\text{ nm}$ is coupled into the waveguide using a microscope objective. The output light is collected by a second objective, and the residual pump is removed with a long-pass filter. The photon pairs are then coupled into a single-mode fiber and directed to a programmable filter (PF, Finisar 4000s), which enables full control of the joint spectral amplitude (JSA).

\begin{figure}[ht]
    \centering
    \includegraphics[width=0.7\linewidth]{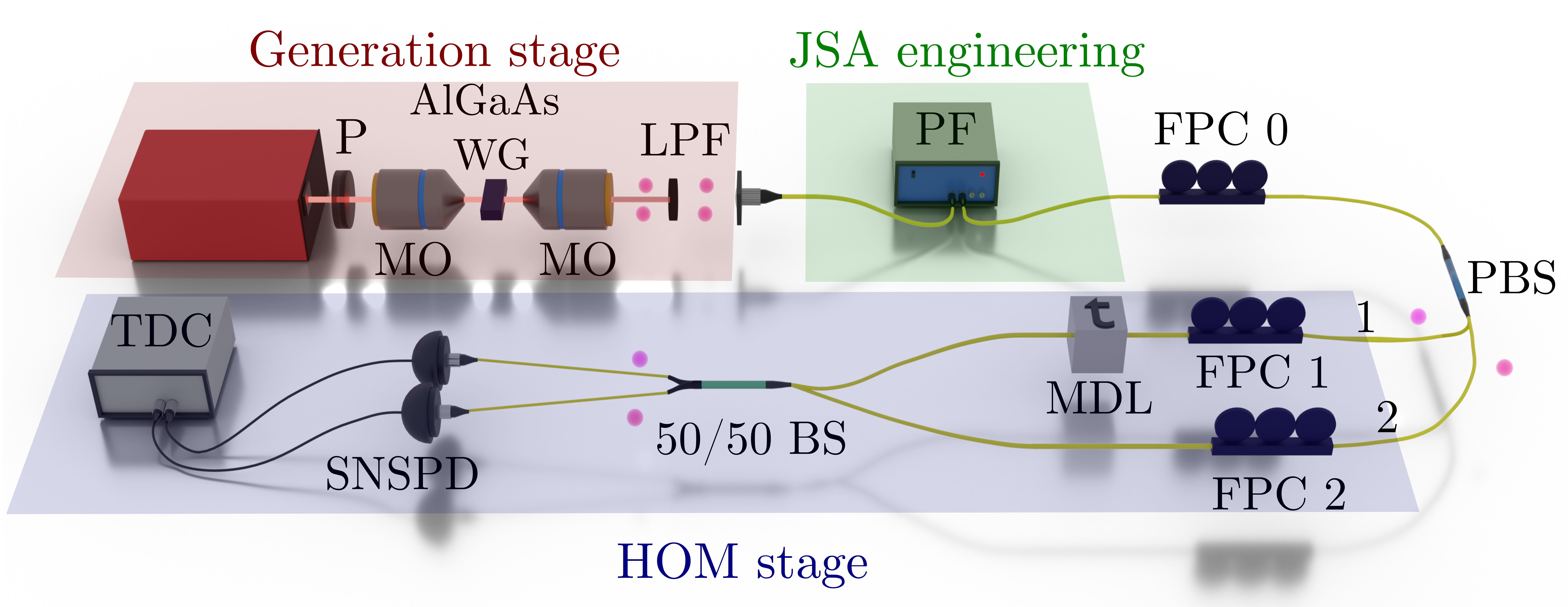}
    \caption[Experimental setup used to investigate the metrological performance of the HOM interferometer]{Experimental setup used to investigate the metrological performance of the HOM interferometer. The scheme highlights three stages: generation of polarization-entangled photon pairs in an AlGaAs Bragg reflector waveguide, spectral engineering of the JSA using a programmable filter, and interferometric characterization in a HOM configuration. P: Polarizer. MO: Microscope Objective. LPF: Long-Pass Filter. PF: Programmable Filter. FPC: Fibered Polarization Controller. PBS: Polarizing Beam Splitter. MDL: Motorized Delay Line. BS: Beam Splitter. SNSPD: Superconducting Nanowire Single-Photon Detector. TDC: Time-to-Digital Converter. Adapted from Ref.~\cite{meskine_approaching_2024}. \copyright 2024 American Physical Society.}
    \label{fig: experimental HOM}
\end{figure}

In the absence of spectral filtering, the generated state is well described by
\begin{equation}
    F_-(\omega_-)\propto\operatorname{sinc}(a\omega_-^2+b\omega_-+c),
\end{equation}
where the coefficients $a$, $b$ and $c$ depend on material properties such as birefringence and chromatic dispersion~\cite{maltese_generation_2019}. In addition to this intrinsic ``Sinc'' state, three filtering configurations were implemented:
\begin{itemize}
    \item[(i)] a $15\text{ nm}$-wide rectangular filter centered at the wavelength $\lambda_{\text{deg}}=1544.8\text{ nm}$,
    \item[(ii)] a Gaussian filter with identical width and center wavelength,
    \item[(iii)] a combination of two $5\text{ nm}$-wide rectangular filters centered at $\lambda_1=1560\text{ nm}$ and $\lambda_2=1530\text{ nm}$, corresponding to energy-matched channels and enabling the generation of a time-frequency cat-like state.
\end{itemize}
The associated functions $F_-(\omega_-)$ and the resulting coincidence probabilities $P_c(\tau)$ are summarized in Table~\ref{tab: different states HOM visi}. 

\begin{table}[th]
    \renewcommand{\arraystretch}{2}
    \begin{center}
        \begin{tabular}{|c|c|c|}
            \hline
            State & $F_-(\omega_-)$ &$P_c(\tau)$ \\ 
            \hline
            Sinc & $\operatorname{sinc}(a\omega_-^2+b\omega_-+c)$ &$\displaystyle\frac{1}{2}\left(1-\int \dd\omega_-\,F_-(\omega_-)F_-^*(-\omega_-)e^{-i\omega_- \tau} \right)$ \\ 
            \hline
            Gauss &$\exp(-\omega_-^2/4\sigma_{\omega_-}^2)$ &$\frac{1}{2}\Big(1-\exp(-\tau^2\sigma_{\omega_-}^2/2)\Big)$\\
            \hline
            Rect & $\Pi\!\left(\frac{\omega_-}{\Delta\omega_-}\right)$& $\frac{1}{2}\Big(1-\operatorname{sinc}\!\left(\frac{\Delta\omega_-\tau}{2}\right)\Big)$\\ 
            \hline
            SC & $\Pi\!\left(\frac{\omega_--\omega'}{\Delta\omega'}\right)+\Pi\!\left(\frac{\omega_-+\omega'}{\Delta\omega'}\right)$ & $\frac{1}{2}\Big(1-\operatorname{sinc}\!\left(\frac{\Delta\omega'\tau}{2}\right)\cos(\omega'\tau)\Big)$\\
            \hline
        \end{tabular}
    \end{center}
    \caption[Spectral functions and corresponding coincidence probabilities]{Spectral functions $F_-(\omega_-)$ and corresponding coincidence probabilities $P_c(\tau)$ for the four engineered states. Here $\omega_-=\omega_1-\omega_2$, and $\sigma_{\omega_-}$ denotes the rms spectral width of the Gaussian state, $\Delta\omega_-$ and $\Delta\omega'$ are the rectangular bandwidths, and $\omega'$ is the frequency separation between the two spectral components of the time-frequency cat-like state. The different variables are defined in the left column of Fig.~\ref{fig: HOM visi exp graph}. Adapted from Ref.~\cite{meskine_approaching_2024}. \copyright 2024 American Physical Society.}
    \label{tab: different states HOM visi}
\end{table}

\paragraph{\publi Interferometric characterization and Fisher information}

After spectral engineering, the photon pairs are separated by a polarizing beam splitter. The $H$ (resp. $V$) polarized photon is injected into arm 1 (resp. 2) of a HOM interferometer. Independent fiber polarization controllers allow fine adjustment of the polarization distinguishability, and thus of the HOM visibility $V$. The temporal delay $\tau$ between the photons is tuned using a motorized delay line. The two paths are recombined on a 50/50 beam splitter, and the outputs are detected by superconducting nanowire single-photon detectors. Coincidence events are recorded with a time-to-digital converter.

For each of the four states, a series of measurements was performed while systematically varying $V$ in order to investigate the scaling of the ratio $\mathcal F_\text{max}(V)/\mathcal F_\text{ideal}$. The coincidence data, shown as red points in Fig.~\ref{fig: HOM visi exp graph}, are fitted using the theoretical expression of $P_c(\tau)$ based on the engineered wavefunctions. The Fisher information $\mathcal F(V,\tau)$, displayed as blue lines, is then calculated from its definition. Error bars are estimated assuming Poissonian counting statistics. As expected, decreasing visibility reduces $\mathcal F(V,\tau)$, and for finite $V$ the FI vanishes at $\tau=0$. 

Remarkably high visibilities exceeding $99\%$ are obtained for the Gaussian, rectangular, and time-frequency cat-like states. For the full unfiltered Sinc state, a maximum visibility of $94.9\%$ is achieved, limited by residual modal birefringence of the AlGaAs source. Owing to its broad spectral bandwidth of approximately $100\text{ nm}$, this state produces a very narrow HOM dip and consequently a large FI, reaching $2100\text{ ps}^{-2}$, which is about two orders of magnitude higher than for the Gaussian and rectangular states.

\begin{figure}[ht]
    \centering
    \includegraphics[width=0.7\linewidth]{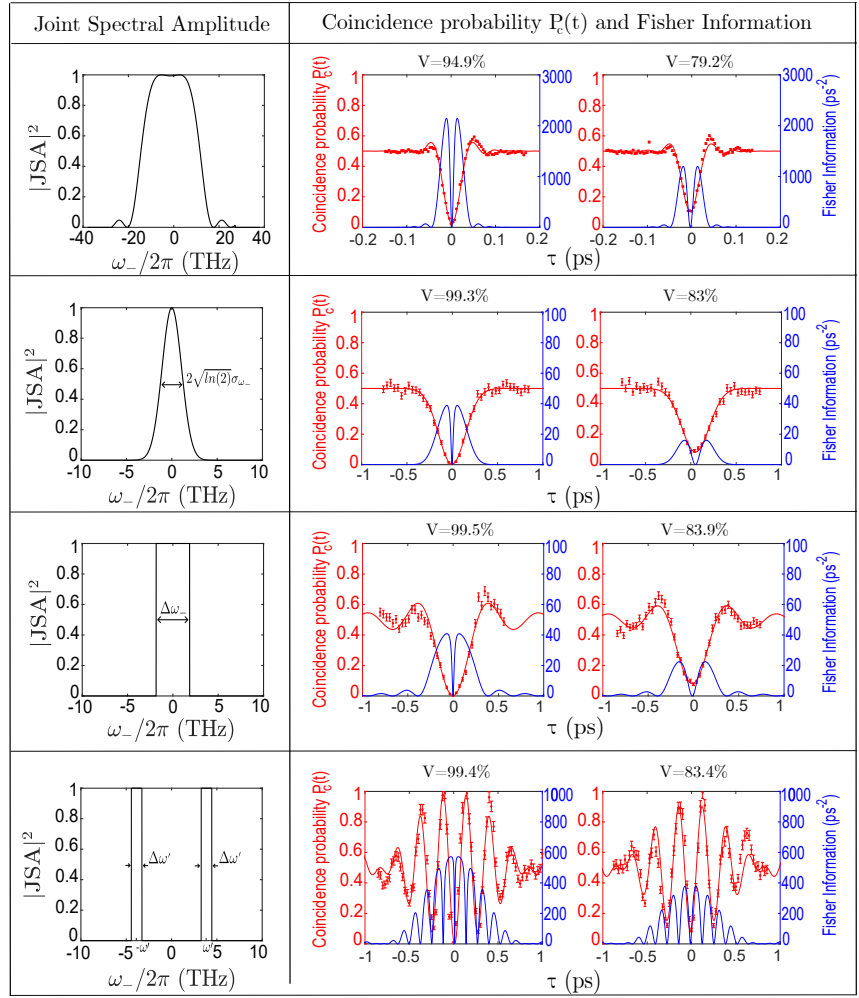}
    \caption[Various joint spectral amplitudes and the corresponding HOM coincidence probability and Fisher information for different values of the visibility $V$]{Left column: Joint spectral amplitude of the four states analyzed in this work. Middle and right columns: corresponding HOM coincidence probability $P_c(\tau)$ and Fisher information $\mathcal F(V,\tau)$ for different values of the visibility $V$. Experimental data are shown as points and theoretical curves as solid lines. Adapted from Ref.~\cite{meskine_approaching_2024}. \copyright 2024 American Physical Society.}
    \label{fig: HOM visi exp graph}
\end{figure}

\paragraph{\publi Scaling of the metrological performance}

Although spectral filtering reduces the detected photon flux and therefore affects the absolute metrological performance for a fixed integration time, the present proof-of-principle experiment focuses on validating the predicted scaling of $\mathcal F_\text{max}(V)/\mathcal F_\text{ideal}$ with visibility. Figure~\ref{fig: HOM Visi scalings} shows the evolution of this ratio for the four engineered states, comparing experimental data points with theoretical predictions derived from the functions listed in Table~\ref{tab: different states HOM visi}. The experimental error bars are obtained from the uncertainty on the visibility extracted from the HOM fits presented in Fig.~\ref{fig: HOM visi exp graph}. A clear hierarchy in the scaling behavior is observed. The time-frequency cat-like state exhibits the most favorable robustness against imperfect visibility, whereas the Gaussian state shows the strongest degradation. For instance, at $V\simeq 99.4\%$ (resp. $83\%$), the ratio $\mathcal F_\text{max}(V)/\mathcal F_\text{ideal}$ decreases to $0.97$ (resp. $0.64$) for the time-frequency cat-like state, compared to $0.85$ (resp. $0.35$) for the Gaussian state. These observations are in quantitative agreement with the theoretical analysis of Sec.~\ref{subsec: visibility impact on precision}, which predicts a quadratic scaling of the precision with visibility for the cat-like state, while other spectral shapes can exhibit a more severe dependence. 

Overall, the results demonstrate that tailoring the spectral structure of the two-photon state provides a powerful lever to enhance robustness against visibility degradation, and identify time-frequency cat-like states as particularly promising candidates for practical HOM-based quantum metrology.

\begin{figure}[ht]
    \centering
    \includegraphics[width=0.8\linewidth]{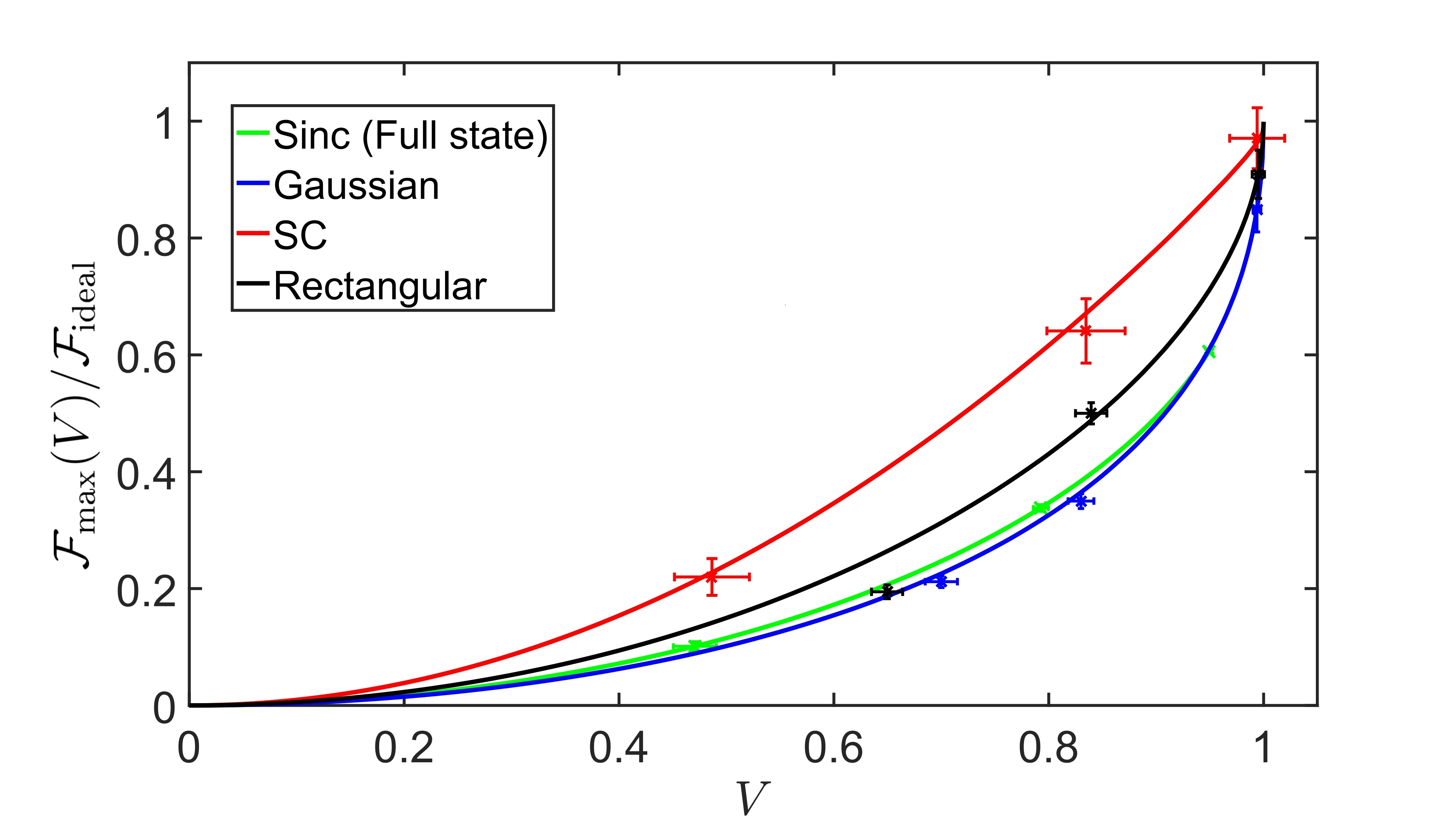}
    \caption[Scaling of the ratio $\mathcal F_\text{max}(V)/\mathcal F_\text{ideal}$ as a function of the HOM visibility]{Scaling of the ratio $\mathcal F_\text{max}(V)/\mathcal F_\text{ideal}$ as a function of the HOM visibility $V$ for the four biphoton states. Points correspond to experimental data and solid lines to theoretical predictions. For the time-frequency cat-like state, a maximal value $\mathcal F_\text{max}(V)/\mathcal F_\text{ideal}=0.97$ is obtained at $V=99.4\%$. Note that the absolute values of $\mathcal F_\text{ideal}$ differ for each state. Adapted from Ref.~\cite{meskine_approaching_2024}. \copyright 2024 American Physical Society.}
    \label{fig: HOM Visi scalings}
\end{figure}

\clearpage
\section{Generalized HOM with multiphoton input states}
\label{sec: generalized HOM multiphoton}

\emph{The aim of this section is to generalize the results obtained in the previous section to situations beyond the single-photon regime. We show how a balanced beam splitter naturally links spatial symmetry and parity properties, enabling the derivation of formulas predicting the interferometric outcome as well as the associated precision. We also discuss related topics, such as the effect of loss and a general analysis of the Mach-Zehnder interferometer. This section presents part of the results described in \hyperlink{Article: prl extended HOM}{Role of Symmetry in Generalized Hong-Ou-Mandel Interference and Quantum Metrology}~\cite{descamps_role_2026}, which was published during the PhD thesis.}

\subsection{Multi-photon interferometry in the literature}
\label{subsec: multiphoton interferometry in the literature}

In the previous section, we explicitly demonstrated the crucial role played by spatial symmetry in the HOM interference effect. In particular, we obtained the formula
\begin{equation}
    P_c=\frac{1}{2}\left(1-\bra{\psi}\hat S \ket{\psi}\right),
\end{equation}
which directly links the symmetry of the incoming state under exchange of the spatial modes to the coincidence probability at the output of a balanced beam splitter. The purpose of this section is to develop a general understanding of interference effects obtained with arbitrary two-mode optical states, beyond the single-photon regime. We consider general continuous-variable states with arbitrary photon-number distributions and investigate how their symmetry properties determine the interferometric outcome. A natural question emerging from the HOM analysis is whether a general relation exists between the average symmetry $\expval{\hat S}$ of an arbitrary input state and specific photon-number statistics at the output of a balanced beam splitter. Rather than reconstructing the path that led us to the conjectures of the result we will present below by examining low photon-number cases, we provide a concise overview of multiphoton interferometry in the literature, where systematic and rigorous frameworks have been developed.

\paragraph{\lit The Mach-Zehnder interferometer}

While the Hong-Ou-Mandel effect concerns interference at a single beam splitter, multiphoton interference has been most extensively studied in the context of the Mach-Zehnder interferometer (MZI)~\cite{ernst_dorn_zeitschrift_1891}. In its standard configuration, the MZI consists of two balanced beam splitters separated by a phase shift $\varphi$ applied in one arm. Historically, it was introduced to interfere a coherent beam with itself after spatial splitting.

Denoting by $\hat a$ and $\hat b$ the annihilation operators associated with the two spatial modes, we consider the situation where a coherent state $\ket{\alpha}$ is injected in one input port while the second port receives the vacuum. The setup is shown in Fig.~\ref{fig: Simple MZI}. A straightforward calculation yields the output intensities after the second beam splitter,
\begin{align}
    I_1=\expval{\hat a^\dagger\hat a}=\abs{\alpha}^2\cos^2(\varphi/2), 
    && 
    I_2=\expval{\hat b^\dagger\hat b}=\abs{\alpha}^2\sin^2(\varphi/2).
\end{align}
The phase $\varphi$ can therefore be estimated with a precision scaling as $1/\sqrt{\abs{\alpha}^2}$, which corresponds to the shot-noise limit. The natural question that arose was whether nonclassical states of light could surpass this scaling. This line of research led to the identification of quantum resources such as squeezed states~\cite{caves_quantum-mechanical_1981,pezze_mach-zehnder_2008} and NOON states, capable of achieving enhanced phase sensitivity. The MZI thus provides a paradigmatic framework for studying multiphoton interference and quantum-enhanced metrology.

\begin{figure}[ht]
    \centering
    \scriptsize
    \scalebox{1}{\begin{tikzpicture}
	\begin{pgfonlayer}{nodelayer}
		\node [style=none] (0) at (-10, 0) {};
		\node [style=none] (1) at (-8.5, 1.5) {};
		\node [style=none] (2) at (-7, 0) {};
		\node [style=none] (3) at (-8.5, -1.5) {};
		\node [style=none] (4) at (-9.5, 1) {};
		\node [style=none] (5) at (-7.5, 1) {};
		\node [style=none] (7) at (-5, 3) {};
		\node [style=none] (8) at (-5, -3) {};
		\node [style=none] (21) at (-8.5, -2.5) {BS};
		\node [style=none] (24) at (-12, 3) {};
		\node [style=none] (27) at (-10.75, 1.5) {$\hat a$};
		\node [style=none] (41) at (-9.5, -1) {};
		\node [style=none] (42) at (-12, -3) {};
		\node [style=none] (63) at (-1, 0) {};
		\node [style=none] (64) at (0.5, 1.5) {};
		\node [style=none] (65) at (2, 0) {};
		\node [style=none] (66) at (0.5, -1.5) {};
		\node [style=none] (67) at (-0.5, 1) {};
		\node [style=none] (68) at (1.5, 1) {};
		\node [style=none] (69) at (1.5, -1) {};
		\node [style=none] (70) at (3.5, 3) {};
		\node [style=none] (71) at (3.5, -3) {};
		\node [style=none] (72) at (0.5, -2.5) {BS};
		\node [style=none] (75) at (-3, 3) {};
		\node [style=none] (78) at (-0.5, -1) {};
		\node [style=none] (79) at (-3, -3) {};
		\node [style=none] (80) at (-4.75, 3.25) {};
		\node [style=none] (81) at (-4.5, 3.5) {};
		\node [style=none] (82) at (-3.5, 3.5) {};
		\node [style=none] (83) at (-3.25, 3.25) {};
		\node [style=none] (84) at (-3.25, 2.75) {};
		\node [style=none] (85) at (-3.5, 2.5) {};
		\node [style=none] (86) at (-4.5, 2.5) {};
		\node [style=none] (87) at (-4.75, 2.75) {};
		\node [style=none] (130) at (-5, -3) {};
		\node [style=none] (141) at (3.5, 4) {};
		\node [style=none] (142) at (4.5, 3) {};
		\node [style=none] (143) at (4.5, -3) {};
		\node [style=none] (144) at (3.5, -4) {};
		\node [style=none] (145) at (4.5, 4) {};
		\node [style=none] (146) at (6.25, 4) {};
		\node [style=none] (147) at (6.25, 2.25) {};
		\node [style=none] (148) at (6.5, 1.25) {};
		\node [style=none] (149) at (4.5, -4) {};
		\node [style=none] (150) at (6.25, -3.75) {};
		\node [style=none] (151) at (7.5, -2.5) {};
		\node [style=none] (152) at (6.5, -1.25) {};
		\node [style=none] (153) at (10.5, 0) {$\varphi$};
		\node [style=none] (154) at (4.5, 1.25) {};
		\node [style=none] (155) at (8.5, 1.25) {};
		\node [style=none] (156) at (4.5, -1.25) {};
		\node [style=none] (157) at (8.5, -1.25) {};
		\node [style=none] (158) at (6.5, 0.5) {Intensity};
		\node [style=none] (159) at (6.5, -0.5) {Measurement};
		\node [style=none] (160) at (8.75, 0) {};
		\node [style=none] (161) at (10, 0) {};
		\node [style=none] (162) at (-4, 3) {$\varphi$};
		\node [style=none] (167) at (-10.75, -1.5) {$\hat b$};
		\node [style=none] (171) at (-1.75, 1.5) {$\hat a$};
		\node [style=none] (172) at (3, 1.5) {$\hat a$};
		\node [style=none] (173) at (-6.25, 1.5) {$\hat a$};
		\node [style=none] (174) at (3, -1.5) {$\hat b$};
		\node [style=none] (175) at (-1.75, -1.5) {$\hat b$};
		\node [style=none] (176) at (-6.25, -1.5) {$\hat b$};
		\node [style=none] (177) at (-7.5, -1) {};
		\node [style=none] (178) at (-10, 0) {};
		\node [style=none] (179) at (-7, 0) {};
		\node [style=none] (180) at (-13, 3) {$\ket{\alpha}$};
		\node [style=none] (181) at (-13, -3) {$\ket{\text{vac}}$};
	\end{pgfonlayer}
	\begin{pgfonlayer}{edgelayer}
		\draw [style=Fill grey] (2.center)
			 to (3.center)
			 to (0.center)
			 to (1.center)
			 to cycle;
		\draw [style=dashed arrow, in=-180, out=60, looseness=0.75] (5.center) to (7.center);
		\draw [style=dashed arrow] (177.center)
			 to [in=-180, out=-60, looseness=0.75] (130.center)
			 to (79.center)
			 to [in=-120, out=0, looseness=0.75] (78.center);
		\draw [style=dashed arrow, in=120, out=0, looseness=0.75] (24.center) to (4.center);
		\draw [style=dashed arrow, in=-120, out=0, looseness=0.75] (42.center) to (41.center);
		\draw [style=Fill grey] (64.center)
			 to (65.center)
			 to (66.center)
			 to (63.center)
			 to cycle;
		\draw (63.center) to (65.center);
		\draw [style=dashed arrow] (68.center) to (70.center);
		\draw [style=dashed arrow] (69.center) to (71.center);
		\draw [style=dashed arrow, in=120, out=0, looseness=0.75] (75.center) to (67.center);
		\draw (80.center)
			 to [bend left=45, looseness=1.25] (81.center)
			 to (82.center)
			 to [bend left=45, looseness=1.25] (83.center)
			 to (84.center)
			 to [bend left=45, looseness=1.25] (85.center)
			 to (86.center)
			 to [bend right=315, looseness=1.25] (87.center)
			 to cycle;
		\draw [style=Fill red] (142.center)
			 to (141.center)
			 to [bend left=90, looseness=1.75] cycle;
		\draw [style=Fill red] (144.center)
			 to (143.center)
			 to [bend left=90, looseness=1.75] cycle;
		\draw [in=165, out=45, looseness=1.75] (145.center) to (146.center);
		\draw [in=60, out=-15, looseness=1.25] (146.center) to (147.center);
		\draw [in=90, out=-105] (147.center) to (148.center);
		\draw [in=165, out=-45] (149.center) to (150.center);
		\draw [in=-60, out=-15, looseness=1.50] (150.center) to (151.center);
		\draw [in=-90, out=120, looseness=1.25] (151.center) to (152.center);
		\draw (154.center) to (155.center);
		\draw (155.center) to (157.center);
		\draw (157.center) to (156.center);
		\draw (156.center) to (154.center);
		\draw [style=Thick arrow] (160.center) to (161.center);
		\draw (178.center) to (179.center);
	\end{pgfonlayer}
\end{tikzpicture}}
    \caption[Mach-Zehnder interferometer]{Mach-Zehnder interferometer composed of two balanced beam splitters separated by a phase shift $\varphi$ in one arm. A coherent state $\ket{\alpha}$ enters one input port while the other receives the vacuum. The output intensities $I_1$ and $I_2$ depend sinusoidally on $\varphi$, enabling phase estimation at the shot-noise limit for classical coherent illumination.}
    \label{fig: Simple MZI}
\end{figure}
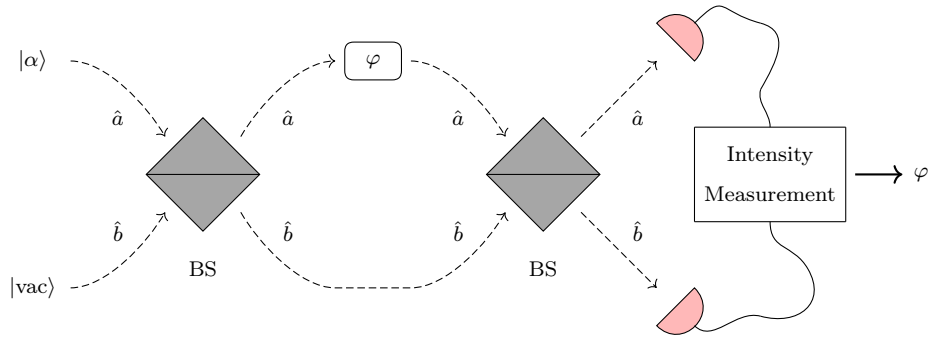

\paragraph{\lit Parity measurement}

In quantum metrology, two distinct notions of precision must be distinguished: the interferometer precision, quantified by the Fisher information associated with a specific measurement, and the ultimate precision set by the quantum Fisher information. Achieving sub-shot-noise scaling requires both a suitable probe state with large QFI and a measurement capable of extracting this enhanced sensitivity.

For squeezed states, intensity measurements at the output of the MZI are sufficient to reach the optimal scaling. In contrast, for NOON states\footnote{Here the NOON state is considered after the first beam splitter. One may equivalently view the first beam splitter as part of the state-preparation stage or assume that the appropriate entangled state is directly injected. This viewpoint highlights the close connection between the MZI and generalized HOM interference, where a single balanced beam splitter already generates nontrivial multiphoton interference.} of the form
\begin{equation}
    \ket{\psi}=\frac{1}{\sqrt{2}}\left(\ket{N,0}+\ket{0,N}\right),
\end{equation}
simple intensity measurements are not sufficient to extract the Heisenberg-limited sensitivity. It has been shown that full photon-number-resolved detection provides an optimal measurement strategy for NOON and related states~\cite{ben-aryeh_phase_2012}. Remarkably, a simpler observable, photon-number parity, also achieves optimal sensitivity for a broad class of states~\cite{gerry_heisenberg-limit_2000,pezze_mach-zehnder_2008,anisimov_quantum_2010}. The parity operators are defined as
\begin{align}
    \hat \Pi_a=\exp(i\pi \hat a^\dagger \hat a), 
    &&
    \hat \Pi_b=\exp(i\pi \hat b^\dagger \hat b),
\end{align}
and measure whether the detected photon number in a given output mode is even or odd. Parity detection has been shown to be optimal for many relevant quantum states, including dual Fock states, Yurke states~\cite{yurke_su2_1986}, and Yuen states~\cite{yuen_generation_1986}. In each case, one can explicitly compute the associated Fisher information and demonstrate that the MZI combined with parity measurement saturates the QFI.

\paragraph{\lit Path-symmetry}

Path symmetry provides the structural condition under which conventional photon counting at the output of an MZI can saturate the quantum Cramér-Rao bound. A pure two-mode state is said to be \emph{path-symmetric} if it is invariant, up to a global phase, under exchange of the two arms combined with complex conjugation of the amplitudes.\footnote{Path symmetry, as defined here, corresponds to an unphysical symmetry acting on the amplitudes: exchange of the two paths combined with complex conjugation. This operation does not coincide with the physical swap operator $\hat S$ introduced in the previous section. In particular, a physical exchange of spatial modes may introduce additional phase factors depending on the chosen reference frame, whereas path symmetry is defined independently of such conventions.} In the photon-number basis, this condition reads
\begin{equation}
    \langle n_1,n_2|\psi\rangle  = e^{-2 i \chi_0} \langle n_2,n_1|\psi\rangle^{*}.
\end{equation}

As shown by Hofmann~\cite{hofmann_all_2009}, this symmetry is a necessary and sufficient condition for standard photon counting at the MZI outputs to attain the QCRB. Since the symmetry is preserved under relative phase shifts, a path-symmetric state achieves its maximal phase sensitivity independently of the interferometer bias phase. A further important result~\cite{seshadreesan_phase_2013} establishes that parity measurement at one output port saturates the QCRB for the entire class of path-symmetric states. All pure optical states proposed so far for sub-shot-noise interferometry, including NOON states, twin-Fock states, and coherent states combined with squeezed vacuum, satisfy this symmetry. Parity detection therefore constitutes a universal optimal measurement for local phase estimation with these resources. For such states, the attainable precision is entirely determined by the variance of the phase generator, so that surpassing the shot-noise limit reduces to engineering large photon-number fluctuations between the two paths.

\paragraph{\lit Generalized HOM interference}

Although the MZI provides a canonical setting for multiphoton interference, a single balanced beam splitter already exhibits remarkably rich interference phenomena. Early work by Ou~\cite{ou_quantum_1996} analyzed the output statistics of a general two-mode input state
\begin{equation}
    \ket{\psi}=\sum_{n_a,n_b=0}^\infty c_{n_a,n_b}\ket{n_a,n_b},
\end{equation}
and showed that interference effects generalizing the HOM mechanism arise even when only one input contains a single photon while the other carries an arbitrary state. In particular, destructive interference can suppress specific output photon-number configurations.

More recently, the notion of \emph{extended} or generalized HOM interference has been developed~\cite{alsing_extending_2022,alsing_hong-ou-mandel_2024,alsing_examination_2025}. These works demonstrate that the photon-number parity of a nonclassical input state plays a central role in determining the output statistics at a balanced beam splitter, independently of the second input. If one input has odd parity, for instance a Fock state $\ket{n}$ with odd $n$, the joint photon-number distribution at the output exhibits a central nodal line along the main diagonal, corresponding to complete suppression of equal photon-number events in both output arms. The familiar two-photon HOM dip therefore appears as a special case of a more general multiphoton interference principle.

For arbitrary Fock inputs $\ket{n,m}$, the extended HOM effect produces destructive interference patterns that generalize the original two-photon mechanism to higher photon numbers. Besides exact zeros forming central nodal lines, one also observes pseudonodal curves where probabilities are strongly suppressed though not strictly vanishing. Remarkably, even the interference of a single photon with a classical state such as a coherent or thermal field can imprint clear nonclassical signatures in the output distribution, highlighting the fundamental role of photon-number discreteness and parity in multiphoton interference.

\subsection{Generalized HOM and symmetry}
\label{subsec: generalized HOM and symmetry}

\paragraph{\publi Framework}
In this section, we study generalized HOM interference, considering an incoming state on the beam splitter that is as general as possible, encompassing arbitrary superpositions of photon-number states. We note that in the previous Sec.~\ref{sec: TF HOM}, only frequency was treated as an internal degree of freedom. However, examining the formulas developed there, one can see that the physical nature of the frequency plays no explicit role, indicating that any other internal degree of freedom could be used instead. Therefore, we denote by $\lambda$ an abstract degree of freedom representing any desired internal property, such as frequency, polarization, spatial mode, etc. We thus manipulate the annihilation operators $\hat a_1(\lambda)$ and $\hat a_2(\lambda)$ associated with the two spatial modes and the internal degree of freedom $\lambda$. The incoming state on the beam splitter can then be expressed as\footnote{In the following equation, we treat the variables $\lambda$ as continuous, integrating over them. The same formulas hold if $\lambda$ is discrete, in which case integrals are replaced by sums, or if $\lambda$ encompasses both discrete and continuous degrees of freedom.}
\begin{align}
    \ket{\psi}&=\sum_{n,m=0}^\infty \int \dd\lambda_1\dots \dd\lambda_{n+m}\, F_{n,m}(\lambda_1,\dots,\lambda_{n+m})\hat a_1(\lambda_1)^\dagger\cdots\hat a_1(\lambda_n)^\dagger \notag\\
    &\qquad\times\hat a_2(\lambda_{n+1})^\dagger\cdots\hat a_2(\lambda_{n+m})^\dagger\vac,
\end{align}
where the functions $F_{n,m}$ can be taken symmetric under exchange of the first $n$ variables and under exchange of the last $m$ variables, due to the bosonic nature of photons.

Importantly, the spatial symmetry operator $\hat S$, already defined in Sec.~\ref{subsec: TF HOM coincidence probability}, is well defined for any state of this form. We can thus compute its expectation value $\bra{\psi}\hat S\ket{\psi}$. We will show that this expectation value is directly linked to the photon-number statistics at the output of the beam splitter, providing a powerful tool to predict interferometric outcomes for arbitrary input states. The experimental setup we now analyze is illustrated in Fig.~\ref{fig: multi photon HOM}: a general state impinges on a balanced beam splitter with matrix $\frac{1}{\sqrt{2}}\left(\begin{smallmatrix} 1 & 1\\ 1 & -1\end{smallmatrix}\right)$, and photon-number-resolving detectors measure the probabilities of detecting $m_1$ and $m_2$ photons in the first and second arms, respectively.

\begin{figure}[ht]
    \centering
    \scriptsize
    \scalebox{1.2}{\begin{tikzpicture}
	\begin{pgfonlayer}{nodelayer}
		\node [style=none] (0) at (-1.5, 0) {};
		\node [style=none] (1) at (0, 1.5) {};
		\node [style=none] (2) at (1.5, 0) {};
		\node [style=none] (3) at (0, -1.5) {};
		\node [style=none] (5) at (1, 1) {};
		\node [style=none] (6) at (1, -1) {};
		\node [style=none] (8) at (3, 3) {};
		\node [style=none] (9) at (3, -3) {};
		\node [style=none] (15) at (3, 4) {};
		\node [style=none] (16) at (4, 3) {};
		\node [style=none] (18) at (4, -3) {};
		\node [style=none] (19) at (3, -4) {};
		\node [style=none] (20) at (4, 4) {};
		\node [style=none] (21) at (5, 2.75) {};
		\node [style=none] (24) at (4, -4) {};
		\node [style=none] (25) at (5, -2.5) {};
		\node [style=none] (29) at (0, -2.5) {BS};
		\node [style=Photon] (55) at (-4.75, 3.75) {~};
		\node [style=none] (80) at (-1, -1) {};
		\node [style=none] (81) at (-3, -3) {};
		\node [style=none] (89) at (-4, -3) {};
		\node [style=none] (90) at (-1, 1) {};
		\node [style=none] (91) at (-3, 3) {};
		\node [style=none] (93) at (-4, 3) {};
		\node [style=Photon] (94) at (-5.75, -3.25) {~};
		\node [style=Photon] (95) at (-4.75, -2.5) {~};
		\node [style=Photon] (96) at (-7, 2.25) {~};
		\node [style=Photon] (97) at (-5.25, 2.5) {~};
		\node [style=Photon] (98) at (-8, 3.25) {~};
		\node [style=none] (101) at (-6, 0) {$\ket{\psi}$};
		\node [style=none] (102) at (0, 2.5) {$\hat U$};
		\node [style=none] (103) at (5, 2) {$m_1$};
		\node [style=none] (104) at (5, -2) {$m_2$};
	\end{pgfonlayer}
	\begin{pgfonlayer}{edgelayer}
		\draw [style=Fill grey] (2.center)
			 to (3.center)
			 to (0.center)
			 to (1.center)
			 to cycle;
		\draw (0.center) to (2.center);
		\draw [style=Fill red] (16.center)
			 to (15.center)
			 to [bend left=90, looseness=1.75] cycle;
		\draw [style=Fill red] (19.center)
			 to (18.center)
			 to [bend left=90, looseness=1.75] cycle;
		\draw [in=90, out=45, looseness=2.00] (20.center) to (21.center);
		\draw [in=-90, out=-45, looseness=1.75] (24.center) to (25.center);
		\draw [style=dashed arrow] (5.center) to (8.center);
		\draw [style=dashed arrow] (6.center) to (9.center);
		\draw [style=dashed arrow] (89.center)
			 to (81.center)
			 to [in=-120, out=0, looseness=0.75] (80.center);
		\draw [style=dashed arrow] (93.center)
			 to (91.center)
			 to [in=120, out=0, looseness=0.75] (90.center);
	\end{pgfonlayer}
\end{tikzpicture}}
    \caption[Generalized HOM interference with an arbitrary two-mode input state]{Generalized HOM interference with an arbitrary two-mode input state. The state $\ket{\psi}$ impinges on a balanced beam splitter, and the output photon-number distribution is measured with photon-number-resolving detectors. The multiple yellow dots indicate that the input state may be a superposition of different photon-number states, and the internal degree of freedom $\lambda$ is not explicitly represented for simplicity.}
    \label{fig: multi photon HOM}
\end{figure}
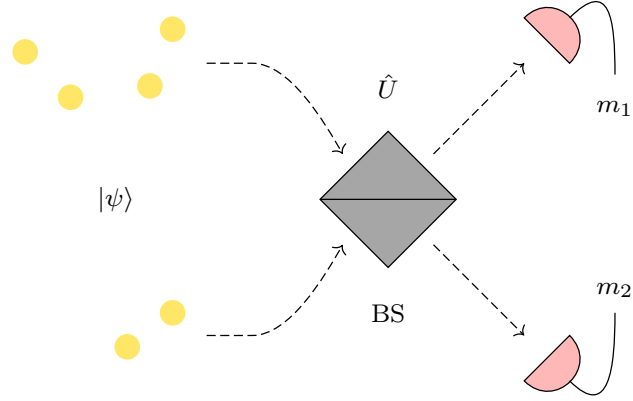

\paragraph{\publi Symmetry and parity}
We now show that Eq.~\eqref{eq: PC with S} admits a generalization in this setting. The core insight is the following: since both the beam splitter $\hat U$ and the symmetry operator $\hat S$ are passive linear transformations that leave the frequency degree of freedom unchanged, they can be described by $2\times 2$ matrices (see Sec.~\ref{subsec: mode transformation}):
\begin{align}
    \hat S: \begin{pmatrix} 0 & 1 \\ 1 & 0 \end{pmatrix}, && \hat U: \frac{1}{\sqrt{2}} \begin{pmatrix} 1 & 1 \\ 1 & -1 \end{pmatrix},
\end{align}
which are, respectively, the Pauli $X$ and the Hadamard matrix $H$.\footnote{To emphasize the difference between matrices and operations on the optical Hilbert space, we omit the hats on the Pauli and Hadamard matrices.} Importantly, as we have shown in Sec.~\ref{subsec: mode transformation}, the association between the matrices and the passive linear transformations is compatible with composition. This implies that the usual relation $H X H = Z$ shows that $\hat U^\dagger \hat S \hat U$ is represented by
\begin{equation}
    \begin{pmatrix}1 & 0 \\ 0 & -1\end{pmatrix},
\end{equation}
which corresponds to the parity operator $\hat \Pi_2$ on the second mode. In other words,
\begin{equation}\label{eq: relation sym parity}
    \hat U \hat S \hat U^\dagger = \hat \Pi_2 = \exp\!\Big(i\pi \int \dd\lambda\, \hat a_2^\dagger(\lambda) \hat a_2(\lambda)\Big).
\end{equation}
For states with a fixed number of photons, $\hat \Pi_2$ returns $+1$ for an even number of photons in mode 2, and $-1$ otherwise and is insensitive to the photon number in mode 1. It is important to emphasize the significance of this relation: symmetry with respect to spatial mode exchange is directly transformed into parity properties at the output of the beam splitter. This principle underlies multiphoton interference phenomena and provides a powerful tool to analyze generalized HOM interference for arbitrary input states.

The connection between parity and the symmetry operator has been partially observed in~\cite{gao_super-resolution_2010}, leading to direct Wigner function measurement~\cite{douce_direct_2013}. However, to our knowledge, no analysis including internal degrees of freedom has been performed, nor has $\hat S$ been explicitly identified as a symmetry operator. Equation~\eqref{eq: relation sym parity} allows a generalization of Eq.~\eqref{eq: PC with S}: if a state $\ket{\psi}$ (not necessarily a single-photon state) impinges on a beam splitter, the probability $\mathbb P(n_2 \equiv 0 \,[2])$ of measuring an even number of photons in the second spatial mode is
\begin{align}\label{eq: Peven HOM}
    \mathbb P(n_2 \equiv 0 \,[2]) &= \frac{1}{2} \left(1 + \bra{\psi} \hat U^\dagger \hat \Pi_2 \hat U \ket{\psi} \right)  = \frac{1}{2} \left(1 + \bra{\psi} \hat S \ket{\psi} \right).
\end{align}
This demonstrates that parity is a key observable at the output of a beam splitter when photon-number resolution is available~\cite{birrittella_parity_2021, alsing_extending_2022, alsing_hong-ou-mandel_2024, chiruvelli_parity_2011}. Applying this formula to a product state recovers the possibility of performing a continuous-variable swap test with a beam splitter and parity measurement~\cite{volkoff_ancilla-free_2022, garcia-escartin_swap_2013}.

\paragraph{\publi Discussion on the beam splitter implementation}
In this section, we considered a beam splitter described by the Hadamard matrix
\begin{equation}
    \frac{1}{\sqrt{2}}\begin{pmatrix} 1 & 1\\ 1 & -1 \end{pmatrix},
\end{equation}
which introduces an asymmetry between the two spatial modes. This explains why formulas involving the parity operator $\hat \Pi_2$ show an explicit asymmetry between the spatial modes. We now discuss alternative choices and their consequences. A general parametrization of a balanced beam splitter is
\begin{equation}
    \frac{e^{i\tau}}{\sqrt{2}}\begin{pmatrix} e^{i\theta} & e^{-i\phi}\\ e^{i\phi} & -e^{-i\theta} \end{pmatrix},
\end{equation}
for $\theta,\phi,\tau \in [-\pi,\pi]$, ensuring that all matrix elements have equal modulus. Common parameter choices yield matrices frequently used in practice:
\begin{align}
    \frac{1}{\sqrt{2}} \begin{pmatrix} 1 & 1\\ 1 & -1 \end{pmatrix}, &&
    \frac{1}{\sqrt{2}} \begin{pmatrix} 1 & 1\\ -1 & 1 \end{pmatrix}, &&
    \frac{1}{\sqrt{2}} \begin{pmatrix} 1 & i\\ i & 1 \end{pmatrix}.
\end{align}

We now ask: if we compute $\mathbb P(n_2 \equiv 0\,[2])$ using a different beam splitter, which symmetry $\hat{\mathfrak S}$ of the state are we effectively measuring? Suppose the beam splitter acts via the general matrix above. Then the relevant symmetry operator is
\begin{equation}
    \hat{\mathfrak S} = \hat{U}^\dagger \hat{\Pi}_2 \hat{U},
\end{equation}
corresponding to the matrix product
\begin{equation}
    \left[\frac{e^{i\tau}}{\sqrt{2}}\begin{pmatrix} e^{i\theta} & e^{-i\phi}\\ e^{i\phi} & -e^{-i\theta} \end{pmatrix}\right]^\dagger 
    \begin{pmatrix} 1 & 0\\ 0 & -1 \end{pmatrix}
    \frac{e^{i\tau}}{\sqrt{2}}\begin{pmatrix} e^{i\theta} & e^{-i\phi}\\ e^{i\phi} & -e^{-i\theta} \end{pmatrix}
    = \begin{pmatrix} 0 & e^{-i(\theta+\phi)}\\ e^{i(\theta+\phi)} & 0 \end{pmatrix}.
\end{equation}
This implies
\begin{align}
    \hat{\mathfrak S} \hat{a}_1^\dagger(\lambda) \hat{\mathfrak S}^\dagger = e^{i(\theta+\phi)} \hat{a}_2^\dagger(\lambda), &&
    \hat{\mathfrak S} \hat{a}_2^\dagger(\lambda) \hat{\mathfrak S}^\dagger = e^{-i(\theta+\phi)} \hat{a}_1^\dagger(\lambda).
\end{align}
Computing $\mathbb P(n_2 \equiv 0 \,[2])$ then probes the symmetry of $\ket{\psi}$ with respect to $\hat{\mathfrak S}$. By redefining the phases of the creation operators as
\begin{align}
    \hat{b}_1^\dagger(\lambda) = \hat{a}_1^\dagger(\lambda), &&
    \hat{b}_2^\dagger(\lambda) = e^{i(\theta+\phi)} \hat{a}_2^\dagger(\lambda),
\end{align}
the action of $\hat{\mathfrak S}$ on $\hat{b}_i^\dagger(\lambda)$ reduces to
\begin{align}
    \hat{b}_1^\dagger(\lambda) \mapsto \hat{b}_2^\dagger(\lambda), &&
    \hat{b}_2^\dagger(\lambda) \mapsto \hat{b}_1^\dagger(\lambda),
\end{align}
which corresponds to the standard symmetry defined by $\hat S$. The difference lies solely in the phase conventions of the spatial modes.

If $\ket{\psi}$ contains the same number of photons in each mode (or a coherent superposition of such states), the action of $\hat{\mathfrak S}$ coincides with that of $\hat{S}$. The phase factors $e^{\pm i(\theta+\phi)}$ then cancel. This is particularly relevant for the standard HOM experiment, where two single photons enter different arms. Hence, the specific implementation of a balanced beam splitter does not affect the coincidence probability, explaining why this point is often omitted. In more general settings, however, the beam splitter implementation can have a significant effect.

\paragraph{\newc Generalized HOM and oddness}
The link between symmetry and parity provides an intuitive understanding of phenomena previously derived by more complex calculations~\cite{alsing_extending_2022, alsing_hong-ou-mandel_2024, alsing_examination_2025}. Consider an input state
\begin{equation}
    \ket{\psi_\text{in}} = \ket{\psi_g} \otimes \ket{\psi_\text{odd}},
\end{equation}
where $\ket{\psi_g} = \sum_{n=0}^\infty c_n \ket{n}$ is an arbitrary single-mode state and $\ket{\psi_\text{odd}} = \sum_{n=0}^\infty d_n \ket{2n+1}$ is a single-mode state containing only odd photon-number components. For this input, the probability of measuring an even number of photons in the second output mode is completely suppressed:
\begin{equation}
    \mathbb P(m_1 = m_2) = 0.
\end{equation}
While previous analyses derive this result via binomial expansions and term manipulations, it follows immediately from the symmetry-parity relation: since $\ket{\psi_\text{odd}}$ contains an odd number of photons, the initial state has odd parity,
\begin{equation}
    \bra{\psi_\text{in}}\hat \Pi_2\ket{\psi_\text{in}} = -1.
\end{equation}
Applying the beam splitter yields an output state $\ket{\psi_\text{out}} = \hat U \ket{\psi_\text{in}}$ that is anti-symmetric:
\begin{equation}
    \bra{\psi_\text{out}}\hat S \ket{\psi_\text{out}} = \bra{\psi_\text{in}} \hat U^\dagger \hat S \hat U \ket{\psi_\text{in}} = \bra{\psi_\text{in}} \hat \Pi_2 \ket{\psi_\text{in}} = -1.
\end{equation}
Writing
\begin{equation}
    \ket{\psi_\text{out}} = \sum_{m_1,m_2} c_{m_1,m_2} \ket{m_1,m_2},
\end{equation}
the anti-symmetry condition implies $c_{m_1,m_2} = -c_{m_2,m_1}$, so $c_{m,m} = 0$ for all $m$, exactly enforcing the vanishing probability of measuring the same number of photons in both modes. This example illustrates how the symmetry-parity relation provides a clear and intuitive explanation for complex interference phenomena.

It is important to note that this result holds only when internal degrees of freedom are not considered. When internal degrees of freedom are included, the anti-symmetry condition on $\ket{\psi_\text{out}}$ requires $c_{m,m}$ to be an anti-symmetric function of the internal degree of freedom, which can admit non-zero solutions.

\subsection{Generalized HOM and metrology}
\label{subsec: generalized HOM and metrology}

\paragraph{\publi Precision bound}

Given the similarity between the formula obtained in the previous section for arbitrary input states and the one obtained for the single-photon HOM interference,
\begin{align}
    \mathbb P(n_2 \equiv 0 \,[2]) = \frac{1}{2} \left(1 + \bra{\psi} \hat S \ket{\psi} \right), 
    && 
    P_c = \frac{1}{2} \left(1 - \bra{\psi} \hat S \ket{\psi} \right),
\end{align}
one naturally sees that the same metrological ideas can be used to derive precision bounds in a wide range of scenarios. As in Sec.~\ref{sec: TF HOM}, we consider a single parameter $\theta$ encoded into the state $\ket{\psi}$ via a unitary transformation $\hat V(\theta) = e^{-i\theta \hat H}$, where $\hat H$ is the generator of the transformation. Photon-number-resolved detection is performed at the output such that the parity of the photon number in the second mode can be reconstructed. See Fig.~\ref{fig: multi photon HOM evolution} for a schematic representation of the protocol.

Because the expression of $\mathbb P(n_2 \equiv 0 \,[2])$ is structurally identical to the one for $P_c$, adapting Results~\ref{res: expansion pc} and \ref{res: Fisher info with S} and copying the proof \textit{verbatim} allows one to deduce the following bound. Assuming that the initial state $\ket{\psi}$ has perfect symmetry, namely $\hat S\ket{\psi}=\pm \ket{\psi}$, the Fisher information associated with the measurement of $\mathbb P(n_2 \equiv 0 \,[2])$, for estimation around $\theta=0$, is given by
\begin{equation}
    \label{eq: Peven HOM Fisher info}
    \mathcal F = \Delta^2(\hat H-\hat S\hat H\hat S).
\end{equation}

As the formula coincides with the one obtained in the single-photon case, similar comments apply regarding the role of the symmetry properties of the generator $\hat H$ and the importance of the symmetry condition for the state. Explicit examples of generators can be analyzed in exactly the same manner. However, since the present framework applies to arbitrary photon-number distributions and to arbitrary internal degrees of freedom, several additional remarks are in order:
\begin{itemize}
    \item Most results in the literature~\cite{anisimov_quantum_2010, ben-aryeh_phase_2012, birrittella_multiphoton_2012, chiruvelli_parity_2011}, such as those discussed in Sec.~\ref{subsec: multiphoton interferometry in the literature}, can be reanalyzed from this perspective. In many cases, the symmetry of the probe state can be assessed with minimal computation, which automatically guarantees the optimality of the protocol within the considered measurement scheme.
    
    \item While our result may appear reminiscent of path-symmetry arguments, we emphasize once again that we are not considering the same notion of symmetry. Path-symmetry corresponds to an exchange of spatial modes supplemented by an additional phase modulation, and is therefore associated with a different physical transformation. In contrast, our symmetry operator $\hat S$ acts directly on the internal degrees of freedom and is defined independently of any specific phase convention. Moreover, the range of application of our result is significantly broader, as we are not restricted to relative time-delay estimation and we allow for arbitrary internal mode structures.
    
    \item The Fisher information $\mathcal F$ must always be compared with the quantum Fisher information $\mathcal Q=4\Delta^2\hat H$. As we are now dealing with states with arbitrary photon-number distributions, as discussed in Sec.~\ref{subsec: multimode case}, the variance $\Delta^2\hat H$ depends both on the modal structure and on the photon-number statistics of the state. In particular, fluctuations in photon number \textit{a priori} allow for an increase of the achievable precision compared to the single-photon regime.
    
    \item Since the result holds for arbitrary photon number, a promising platform for experimentally verifying Eq.~\eqref{eq: Peven HOM Fisher info} and demonstrating metrological protocols beyond the single-photon regime is based on spontaneous parametric down-conversion sources~\cite{boucher_toolbox_2015, meskine_approaching_2024} operated in a medium-power pulsed-pump regime. In such a regime, a limited number of photon pairs are generated simultaneously, enabling photon-number resolution with currently available photon-number-resolving detectors. This makes it possible to probe properties of multimode, multiphoton states and to explore photon-number-enhanced metrology within the generalized HOM framework.
\end{itemize}

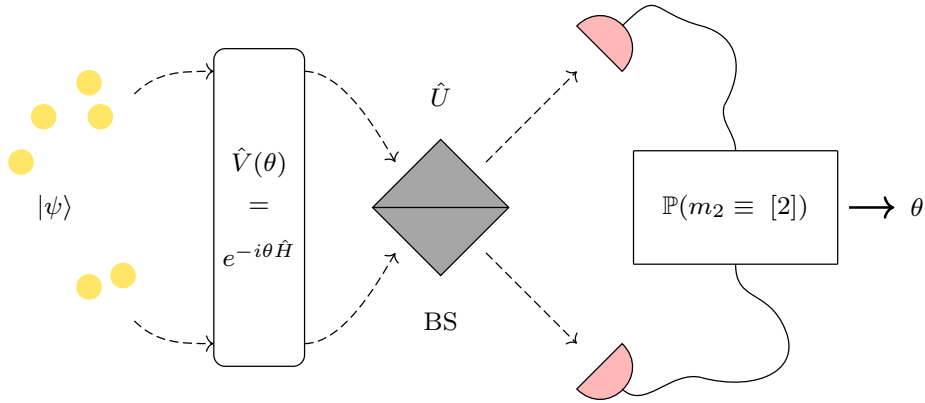
\begin{figure}[ht]
    \centering
    \scriptsize
    \scalebox{1.2}{\begin{tikzpicture}
	\begin{pgfonlayer}{nodelayer}
		\node [style=none] (0) at (-1.5, 0) {};
		\node [style=none] (1) at (0, 1.5) {};
		\node [style=none] (2) at (1.5, 0) {};
		\node [style=none] (3) at (0, -1.5) {};
		\node [style=none] (5) at (1, 1) {};
		\node [style=none] (6) at (1, -1) {};
		\node [style=none] (8) at (3, 3) {};
		\node [style=none] (9) at (3, -3) {};
		\node [style=none] (15) at (3, 4) {};
		\node [style=none] (16) at (4, 3) {};
		\node [style=none] (18) at (4, -3) {};
		\node [style=none] (19) at (3, -4) {};
		\node [style=none] (20) at (4, 4) {};
		\node [style=none] (21) at (5.75, 4) {};
		\node [style=none] (22) at (6.25, 2.25) {};
		\node [style=none] (23) at (6.5, 1.25) {};
		\node [style=none] (24) at (4, -4) {};
		\node [style=none] (25) at (5.75, -3.75) {};
		\node [style=none] (26) at (7.5, -2.5) {};
		\node [style=none] (27) at (6.5, -1.25) {};
		\node [style=none] (29) at (0, -2.5) {BS};
		\node [style=Photon] (55) at (-7.75, 2.75) {~};
		\node [style=none] (64) at (10.5, 0) {$\theta$};
		\node [style=none] (65) at (4.25, 1.25) {};
		\node [style=none] (66) at (8.75, 1.25) {};
		\node [style=none] (67) at (4.25, -1.25) {};
		\node [style=none] (68) at (8.75, -1.25) {};
		\node [style=none] (70) at (6.5, 0) {$\mathbb P(m_2\equiv\;[2])$};
		\node [style=none] (71) at (9, 0) {};
		\node [style=none] (72) at (10, 0) {};
		\node [style=none] (80) at (-1, -1) {};
		\node [style=none] (81) at (-3, -3) {};
		\node [style=none] (90) at (-1, 1) {};
		\node [style=none] (91) at (-3, 3) {};
		\node [style=Photon] (94) at (-7.75, -1.75) {~};
		\node [style=Photon] (95) at (-7, -1.5) {~};
		\node [style=Photon] (96) at (-9.25, 1) {~};
		\node [style=Photon] (97) at (-7.5, 2) {~};
		\node [style=Photon] (98) at (-8.75, 2) {~};
		\node [style=none] (101) at (-8.5, 0) {$\ket{\psi}$};
		\node [style=none] (102) at (0, 2.5) {$\hat U$};
		\node [style=none] (103) at (-3, 3) {};
		\node [style=none] (106) at (-4, -1) {$e^{-i\theta\hat H}$};
		\node [style=none] (107) at (-6.75, 2.5) {};
		\node [style=none] (108) at (-5, 3) {};
		\node [style=none] (109) at (-3, -3) {};
		\node [style=none] (110) at (-6.75, -2.5) {};
		\node [style=none] (111) at (-5, -3) {};
		\node [style=none] (112) at (-4.75, 3.5) {};
		\node [style=none] (113) at (-3.25, 3.5) {};
		\node [style=none] (114) at (-3.25, -3.5) {};
		\node [style=none] (115) at (-4.75, -3.5) {};
		\node [style=none] (116) at (-5, 3.25) {};
		\node [style=none] (117) at (-3, 3.25) {};
		\node [style=none] (118) at (-3, -3.25) {};
		\node [style=none] (119) at (-5, -3.25) {};
		\node [style=none] (120) at (-4, 1) {$\hat V(\theta)$};
		\node [style=none] (121) at (-4, 0) {$=$};
	\end{pgfonlayer}
	\begin{pgfonlayer}{edgelayer}
		\draw [style=Fill grey] (2.center)
			 to (3.center)
			 to (0.center)
			 to (1.center)
			 to cycle;
		\draw (0.center) to (2.center);
		\draw [style=Fill red] (16.center)
			 to (15.center)
			 to [bend left=90, looseness=1.75] cycle;
		\draw [style=Fill red] (19.center)
			 to (18.center)
			 to [bend left=90, looseness=1.75] cycle;
		\draw [in=165, out=45, looseness=1.75] (20.center) to (21.center);
		\draw [in=60, out=-15, looseness=1.25] (21.center) to (22.center);
		\draw [in=90, out=-105] (22.center) to (23.center);
		\draw [in=165, out=-45] (24.center) to (25.center);
		\draw [in=-60, out=-15, looseness=1.50] (25.center) to (26.center);
		\draw [in=-90, out=120, looseness=1.25] (26.center) to (27.center);
		\draw (65.center) to (66.center);
		\draw (66.center) to (68.center);
		\draw (68.center) to (67.center);
		\draw (67.center) to (65.center);
		\draw [style=Thick arrow] (71.center) to (72.center);
		\draw [style=dashed arrow] (5.center) to (8.center);
		\draw [style=dashed arrow] (6.center) to (9.center);
		\draw [style=dashed arrow, in=-120, out=0, looseness=0.75] (81.center) to (80.center);
		\draw [style=dashed arrow, in=120, out=0, looseness=0.75] (91.center) to (90.center);
		\draw [style=dashed arrow, in=-180, out=45] (107.center) to (108.center);
		\draw [style=dashed arrow, in=180, out=-45] (110.center) to (111.center);
		\draw (114.center)
			 to (115.center)
			 to [bend right=315] (119.center)
			 to (116.center)
			 to [bend left=45] (112.center)
			 to (113.center)
			 to [bend left=45] (117.center)
			 to (118.center)
			 to [bend left=45] cycle;
	\end{pgfonlayer}
\end{tikzpicture}}
    \caption[Generalized HOM interference with parameter encoding]{Generalized HOM interference with parameter encoding. A two-mode state $\ket{\psi}$ with arbitrary photon-number distribution undergoes a unitary transformation $\hat V(\theta) = e^{-i\theta \hat H}$ that encodes the parameter $\theta$. The state then impinges on a balanced beam splitter. Photon-number-resolving detectors measure the photon statistics at the output, allowing one to reconstruct the probability $\mathbb P(n_2 \equiv 0 \,[2])$ of detecting an even number of photons in the second mode. This probability provides the metrological signal used to estimate $\theta$.}
    \label{fig: multi photon HOM evolution}
\end{figure}

\paragraph{\newc Multi-pair model}

Understanding how the precision $\mathcal Q$ scales with the number of photons is a non-trivial problem, as one must consider explicit and physically relevant states. Furthermore, when internal degrees of freedom are taken into account, the explicit computation of the variance $\Delta^2\hat H$ can rapidly become intractable. In order to nevertheless gain insight into how an increase in photon number can enhance the measurement precision, we introduce the following toy model.

\begin{itemize}
    \item We assume that successive single-photon pairs are generated during the duration of the experiment. The probability that $n$ pairs are generated is given by a generic distribution $\{A_n\}$,
    \begin{equation}
        \P[n\text{ pairs}]=A_n.
    \end{equation}
    A relevant case for SPDC generation is a Poissonian distribution,
    \begin{equation}
        A_n = e^{-\mu}\frac{\mu^n}{n!},
    \end{equation}
    where $\mu$ is the mean number of pairs generated during the experiment.
    
    \item Different pairs do not interfere with each other at the beam splitter. This can result, for instance, from sufficiently distinct arrival times or from other distinguishability-inducing effects that we leave unspecified.
    
    \item Each pair is perfectly symmetric\footnote{Interestingly, the symmetric assumption is required, as the anti-symmetric one cannot guarantee the ability to perform efficient parameter estimation from the collective data.} and yields the same individual precision contribution $\Delta^2(\hat H-\hat S\hat H\hat S)$.
\end{itemize}

Under these assumptions, the precision obtained with our interferometric protocol is
\begin{equation}
    \mathcal F
    =
    \Delta^2(\hat H-\hat S\hat H\hat S)
    \sum_{n=0}^\infty n A_n
    =
    \Delta^2(\hat H-\hat S\hat H\hat S)
    \,\mathbb E(\text{nbr. pairs}),
\end{equation}
see Appendix~\ref{app: hom interferometry}, Result~\ref{res: multi-pair model} for details. This expression shows that the precision increases linearly with the mean number of generated pairs. Such a linear scaling is consistent with the shot-noise limit expected for independent resources.

Importantly, although the scaling is intuitive, the protocol mixes the information associated with each individual pair into a single parity-based observable. It is therefore not \textit{a priori} obvious that this limited information is sufficient to retain the full shot-noise scaling. The model is relevant for regimes in which the average time between two generated pairs is larger than the coherence time of the source, so that simultaneous multi-pair interference can be neglected, but smaller than the temporal resolution of the detectors, so that different pairs cannot be resolved experimentally.

\paragraph{\publi Photon losses}

A crucial experimental requirement for the viability of the interferometric and metrological results presented above is the ability to perform parity measurements. Parity is particularly sensitive to photon losses, since the loss of a single photon changes the parity of the detected photon number. This sensitivity can significantly degrade the performance of protocols relying on parity measurements, especially when losses are non-negligible and when large photon numbers are involved.

Although we do not provide a comprehensive quantitative model of the impact of losses on the precision, several qualitative and semi-quantitative insights can be obtained:

\begin{itemize}
    \item Losses may occur at any stage of the experiment, either before or after the evolution $\hat V(\theta)$ and the interferometer $\hat U$. As shown in Appendix~\ref{app: hom interferometry}, Result~\ref{res: losses commute}, losses commute with mode transformations. Consequently, the same output statistics are obtained whether the loss occurs before or after a mode transformation. Since the beam splitter and many parameter-encoding operations such as phase shifts or time delays are mode transformations, losses can equivalently be pushed to the state-preparation stage or to the detection stage. This viewpoint simplifies the analysis of their effect.
    
    \item When modeling losses at the detection stage, the ideal probabilities $P_j$ and the lossy probabilities $Q_k$ are related by a Bernoulli convolution. If each photon is detected independently with probability $\eta$, the probability of detecting $k$ photons is
    \begin{equation}
        Q_k=\sum_{j=k}^\infty \binom{j}{k}\eta^k(1-\eta)^{j-k}P_j,
    \end{equation}
    where $P_j$ denotes the ideal probability of measuring $j$ photons. This relation can be inverted to reconstruct $P_j$ from $Q_k$, allowing a direct comparison between ideal and lossy Fisher information values. See Appendix~\ref{app: hom interferometry}, Result~\ref{res: losses probabilities} for the derivation of both the forward and inverse relations.
    
    \item In experiments with a fixed total photon number $N$, one may post-select on events in which all photons are detected. The corresponding probability of success scales as $\eta^N$, which is not scalable, but this strategy can nevertheless be useful in proof-of-principle experiments with small $N$, such as those performed with SPDC sources in the low-gain regime.
    
    \item For metrological estimation in this lossy regime, a trade-off arises between two competing effects. Increasing the photon number enhances the precision, as shown by the multi-pair model, but simultaneously increases the sensitivity to losses. For a given detection efficiency $\eta$, there therefore exists an optimal photon number that maximizes the achievable precision. A quantitative analysis of this trade-off requires specifying both the probe state and the generator $\hat H$, and is left for future work.
\end{itemize}

\subsection{Single-photon MZI}
\label{subsec: single photon MZI}

\paragraph{\newc Setting}
To further demonstrate the versatility of the introduced framework, we analyze the Mach-Zehnder interferometer probed by a time-frequency single-photon state, where coincidences are recorded at the output ports. The setup is depicted in Fig.~\ref{fig: MZI single photon}. The input state is a pure two-mode single-photon state $\ket{\psi}$, which can be written in the frequency domain as
\begin{equation}
    \ket{\psi}=\int \dd \omega_1 \dd \omega_2\, F(\omega_1,\omega_2)\,\hat a_1^\dagger(\omega_1)\hat a_2^\dagger(\omega_2)\vac,
\end{equation}
where $F(\omega_1,\omega_2)$ denotes the joint spectral amplitude and satisfies the normalization condition
\begin{equation}
    \int \dd\omega_1 \dd\omega_2\, \abs{F(\omega_1,\omega_2)}^2=1.
\end{equation}
We consider a general unitary encoding of the parameter $\theta$ given by
\begin{equation}
    \hat V(\theta)=e^{-i\hat H\theta},
\end{equation}
where $\hat H$ is an arbitrary Hermitian generator acting on the two-mode Hilbert space.

\begin{figure}[ht]
    \centering
    \footnotesize
    \scalebox{1.2}{\begin{tikzpicture}
	\begin{pgfonlayer}{nodelayer}
		\node [style=none] (0) at (-10, -0.25) {};
		\node [style=none] (1) at (-8.5, 1.25) {};
		\node [style=none] (2) at (-7, -0.25) {};
		\node [style=none] (3) at (-8.5, -1.75) {};
		\node [style=none] (4) at (-9.5, 0.75) {};
		\node [style=none] (5) at (-7.5, 0.75) {};
		\node [style=none] (6) at (-7.5, -1.25) {};
		\node [style=none] (7) at (-4.5, 2.75) {};
		\node [style=none] (8) at (-4.5, -3.25) {};
		\node [style=none] (21) at (-8.5, -2.75) {BS};
		\node [style=none] (24) at (-12, 2.75) {};
		\node [style=Photon] (25) at (-12.75, 2.75) {~};
		\node [style=Photon] (26) at (-12.75, -3.25) {~};
		\node [style=none] (27) at (-10.75, 1.25) {1};
		\node [style=none] (41) at (-9.5, -1.25) {};
		\node [style=none] (42) at (-12, -3.25) {};
		\node [style=none] (63) at (0, -0.25) {};
		\node [style=none] (64) at (1.5, 1.25) {};
		\node [style=none] (65) at (3, -0.25) {};
		\node [style=none] (66) at (1.5, -1.75) {};
		\node [style=none] (67) at (0.5, 0.75) {};
		\node [style=none] (68) at (2.5, 0.75) {};
		\node [style=none] (69) at (2.5, -1.25) {};
		\node [style=none] (70) at (4.5, 2.75) {};
		\node [style=none] (71) at (4.5, -3.25) {};
		\node [style=none] (72) at (1.5, -2.75) {BS};
		\node [style=none] (75) at (-2, 2.75) {};
		\node [style=none] (78) at (0.5, -1.25) {};
		\node [style=none] (79) at (-2, -3.25) {};
		\node [style=none] (130) at (-4.5, -3.25) {};
		\node [style=none] (141) at (4.5, 3.75) {};
		\node [style=none] (142) at (5.5, 2.75) {};
		\node [style=none] (143) at (5.5, -3.25) {};
		\node [style=none] (144) at (4.5, -4.25) {};
		\node [style=none] (145) at (5.5, 3.75) {};
		\node [style=none] (146) at (7.25, 3.75) {};
		\node [style=none] (147) at (7.25, 2) {};
		\node [style=none] (148) at (7.5, 1) {};
		\node [style=none] (149) at (5.5, -4.25) {};
		\node [style=none] (150) at (7.25, -4) {};
		\node [style=none] (151) at (8.5, -2.75) {};
		\node [style=none] (152) at (7.5, -1.5) {};
		\node [style=none] (153) at (11.5, -0.25) {$P_c$};
		\node [style=none] (154) at (5.5, 1) {};
		\node [style=none] (155) at (9.5, 1) {};
		\node [style=none] (156) at (5.5, -1.5) {};
		\node [style=none] (157) at (9.5, -1.5) {};
		\node [style=none] (158) at (7.5, 0.25) {Coincidence};
		\node [style=none] (159) at (7.5, -0.75) {detection};
		\node [style=none] (160) at (9.75, -0.25) {};
		\node [style=none] (161) at (11, -0.25) {};
		\node [style=none] (164) at (4, 1.25) {1};
		\node [style=none] (165) at (-0.75, 1.25) {1};
		\node [style=none] (166) at (-5.875, 1.25) {1};
		\node [style=none] (167) at (-10.75, -1.75) {2};
		\node [style=none] (168) at (-5.875, -1.75) {2};
		\node [style=none] (169) at (-0.75, -1.75) {2};
		\node [style=none] (170) at (4, -1.75) {2};
		\node [style=none] (193) at (-13, -0.25) {$\ket{\psi}$};
		\node [style=none] (194) at (-13.75, 4.25) {};
		\node [style=none] (195) at (-7.25, -5.25) {};
		\node [style=none] (196) at (-13.25, -5.25) {};
		\node [style=none] (197) at (-13.75, -4.75) {};
		\node [style=none] (198) at (-4.5, 4.75) {};
		\node [style=none] (199) at (-5, 4.25) {};
		\node [style=none] (200) at (-5, -4.75) {};
		\node [style=none] (201) at (-4.5, -5.25) {};
		\node [style=none] (202) at (-10.25, -6) {Preparation stage};
		\node [style=none] (203) at (3.5, -6) {HOM stage};
		\node [style=none] (204) at (-13.25, 4.75) {};
		\node [style=none] (205) at (-7.25, 4.75) {};
		\node [style=none] (206) at (-6.75, 4.25) {};
		\node [style=none] (207) at (-6.75, -4.75) {};
		\node [style=none] (208) at (11.5, 4.75) {};
		\node [style=none] (209) at (12, 4.25) {};
		\node [style=none] (210) at (12, -4.75) {};
		\node [style=none] (211) at (11.5, -5.25) {};
		\node [style=none] (213) at (-3.25, -1.25) {$e^{-i\theta\hat H}$};
		\node [style=none] (217) at (-4, 3.25) {};
		\node [style=none] (218) at (-2.5, 3.25) {};
		\node [style=none] (219) at (-2.5, -3.75) {};
		\node [style=none] (220) at (-4, -3.75) {};
		\node [style=none] (221) at (-4.25, 3) {};
		\node [style=none] (222) at (-2.25, 3) {};
		\node [style=none] (223) at (-2.25, -3.5) {};
		\node [style=none] (224) at (-4.25, -3.5) {};
		\node [style=none] (225) at (-3.25, 0.75) {$\hat V(\theta)$};
		\node [style=none] (226) at (-3.25, -0.25) {$=$};
		\node [style=none] (227) at (-5.75, -0.25) {$\ket{\psi_\text{in}}$};
	\end{pgfonlayer}
	\begin{pgfonlayer}{edgelayer}
		\draw [style=Fill grey] (2.center)
			 to (3.center)
			 to (0.center)
			 to (1.center)
			 to cycle;
		\draw (0.center) to (2.center);
		\draw [style=dashed arrow, in=-180, out=60, looseness=0.75] (5.center) to (7.center);
		\draw [style=dashed arrow, in=-180, out=-60, looseness=0.75] (6.center) to (8.center);
		\draw [style=dashed arrow, in=120, out=0, looseness=0.75] (24.center) to (4.center);
		\draw [style=dashed arrow, in=-120, out=0, looseness=0.75] (42.center) to (41.center);
		\draw [style=Fill grey] (64.center)
			 to (65.center)
			 to (66.center)
			 to (63.center)
			 to cycle;
		\draw (63.center) to (65.center);
		\draw [style=dashed arrow] (68.center) to (70.center);
		\draw [style=dashed arrow] (69.center) to (71.center);
		\draw [style=dashed arrow, in=120, out=0, looseness=0.75] (75.center) to (67.center);
		\draw [style=dashed arrow, in=-120, out=0, looseness=0.75] (79.center) to (78.center);
		\draw [style=Fill red] (142.center)
			 to (141.center)
			 to [bend left=90, looseness=1.75] cycle;
		\draw [style=Fill red] (144.center)
			 to (143.center)
			 to [bend left=90, looseness=1.75] cycle;
		\draw [in=165, out=45, looseness=1.75] (145.center) to (146.center);
		\draw [in=60, out=-15, looseness=1.25] (146.center) to (147.center);
		\draw [in=90, out=-105] (147.center) to (148.center);
		\draw [in=165, out=-45] (149.center) to (150.center);
		\draw [in=-60, out=-15, looseness=1.50] (150.center) to (151.center);
		\draw [in=-90, out=120, looseness=1.25] (151.center) to (152.center);
		\draw (154.center) to (155.center);
		\draw (155.center) to (157.center);
		\draw (157.center) to (156.center);
		\draw (156.center) to (154.center);
		\draw [style=Thick arrow] (160.center) to (161.center);
		\draw [style=dashes red] (206.center)
			 to (207.center)
			 to [bend left=45, looseness=1.25] (195.center)
			 to (196.center)
			 to [bend right=315, looseness=1.25] (197.center)
			 to (194.center)
			 to [bend left=45, looseness=1.25] (204.center)
			 to (205.center)
			 to [bend left=45, looseness=1.25] cycle;
		\draw [style=dashes blue] (210.center)
			 to [bend left=45, looseness=1.25] (211.center)
			 to (201.center)
			 to [bend left=45, looseness=1.25] (200.center)
			 to (199.center)
			 to [bend left=45, looseness=1.25] (198.center)
			 to (208.center)
			 to [bend left=45, looseness=1.25] (209.center)
			 to cycle;
		\draw (219.center)
			 to (220.center)
			 to [bend right=315] (224.center)
			 to (221.center)
			 to [bend left=45] (217.center)
			 to (218.center)
			 to [bend left=45] (222.center)
			 to (223.center)
			 to [bend left=45] cycle;
	\end{pgfonlayer}
\end{tikzpicture}}
    \caption[Single-photon Mach-Zehnder interferometer]{Single-photon Mach-Zehnder interferometer. A two-mode single-photon state $\ket{\psi}$ with arbitrary time-frequency structure is injected into the interferometer. A general unitary transformation $\hat V(\theta)=e^{-i\hat H\theta}$ encodes the parameter $\theta$ to be estimated. Coincidence measurements are performed at the output ports. The interferometer is conceptually divided into two parts. In red, the state-preparation stage, which includes the source and the first beam splitter. In blue, the HOM stage, which comprises the parameter-dependent evolution, the second beam splitter, and the coincidence measurement. This separation allows one to reinterpret the MZI as a generalized HOM interferometer fed with an effective two-photon input state.}
    \label{fig: MZI single photon}
\end{figure}
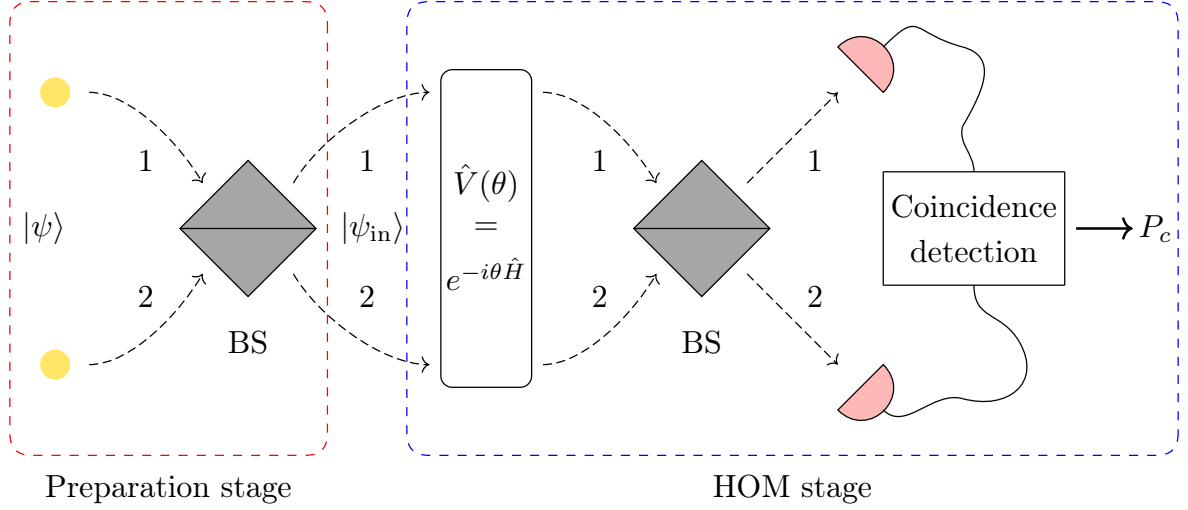

A direct derivation of the coincidence probability, the Fisher information, and the quantum Fisher information is in principle straightforward. However, after the first beam splitter, the photons are redistributed over the two arms in all possible ways, which leads to lengthy intermediate expressions. The general framework developed in the previous sections provides a significantly more transparent route. By incorporating the first beam splitter into the state-preparation stage, the remaining part of the interferometer can be viewed as a generalized HOM interferometer, fed with an effective input state $\ket{\psi_\text{in}}$ that contains two photons arbitrarily distributed in the two arms. One can therefore directly apply the results of Sec.~\ref{subsec: generalized HOM and symmetry} and Sec.~\ref{subsec: generalized HOM and metrology} to obtain compact expressions for the coincidence probability, the Fisher information, and the quantum Fisher information.

\paragraph{\newc General result}
Beyond simplifying the calculations, this perspective reveals several non-trivial features.

\begin{itemize}
    \item The probe state contains exactly one photon per spatial input mode and therefore has odd parity with respect to $\hat \Pi_2$. As a consequence, the effective state entering the HOM stage, $\ket{\psi_\text{in}}$, is anti-symmetric under spatial mode exchange. It follows that, for zero time delay, the coincidence probability reaches its maximal value $P_c=1$ irrespective of the properties of $\ket{\psi}$. This behavior contrasts with the standard HOM effect, where only a symmetric input state leads to vanishing coincidences.

    \item The anti-symmetry of $\ket{\psi_\text{in}}$ implies that, for estimation around $\theta=0$, the Fisher information is given by
    \begin{equation}
        \mathcal F=\Delta^2_{\ket{\psi_\text{in}}}(\hat H-\hat S\hat H\hat S),
    \end{equation}
    where $\hat S$ denotes the spatial mode-swap operator. Importantly, the variance must be evaluated on the state entering the HOM stage, $\ket{\psi_\text{in}}$, and not on the initial state $\ket{\psi}$.

    \item The role of the symmetry properties of $\hat H$ is identical to the general case discussed previously. For a local time delay generated by $\hat H=\hat\omega_1$, the estimation is in general not optimal. In contrast, for a collective relative delay generated by $\hat H=\hat\omega_1-\hat\omega_2$, the symmetry guarantees optimality.

    \item The precision can be expressed directly in terms of the initial probe state by using
    \begin{equation}
        \Delta^2_{\ket{\psi_\text{in}}}(\hat H-\hat S\hat H\hat S)
        =
        \Delta^2_{\ket{\psi}}(\hat U\hat H\hat U^\dagger-\hat \Pi_2\hat U\hat H\hat U^\dagger\hat \Pi_2),
    \end{equation}
    where $\hat U$ denotes the first beam-splitter operator and $\hat \Pi_2$ is the parity operator on the second mode. This relation allows one to work entirely with the original state $\ket{\psi}$ if desired.
\end{itemize}

\paragraph{\newc Explicit computations}

The detailed derivations are presented in Appendix~\ref{app: hom interferometry}, Result~\ref{res: MZI single photon}. We summarize and discuss here the main expressions. A useful starting point is to decompose the JSA $F$ into its symmetric and anti-symmetric components under exchange of the frequency variables,
\begin{align}
    F^s(\omega_1,\omega_2) = \frac{F(\omega_1,\omega_2) + F(\omega_2,\omega_1)}{2}, 
    &&
    F^a(\omega_1,\omega_2) = \frac{F(\omega_1,\omega_2) - F(\omega_2,\omega_1)}{2}.
\end{align}
The normalization of $F$ implies
\begin{equation}
    \int \dd\omega_1 \dd\omega_2\,  \left(\abs{F^s(\omega_1,\omega_2)}^2 + \abs{F^a(\omega_1,\omega_2)}^2 \right) = 1.
\end{equation}

After the first beam splitter, the effective input state of the HOM stage can be written as
\begin{equation}
    \ket{\psi_\text{in}}=\ket{\psi^b}-\ket{\psi^a},
\end{equation}
where the bunching and anti-bunching contributions read
\begin{align}
    \ket{\psi^b} =& \frac{1}{2} \int \dd\omega_1 \dd\omega_2\, F^s(\omega_1,\omega_2)\left( \hat a_1^\dagger(\omega_1)\hat a_1^\dagger(\omega_2) - \hat a_2^\dagger(\omega_1)\hat a_2^\dagger(\omega_2)\right)\vac,\\\ket{\psi^a} =& \int \dd\omega_1 \dd\omega_2\, F^a(\omega_1,\omega_2)\hat a_1^\dagger(\omega_1)\hat a_2^\dagger(\omega_2)\vac.
\end{align}
In agreement with the usual HOM intuition, the symmetric part $F^s$ is associated with photon bunching, whereas the anti-symmetric part $F^a$ is associated with anti-bunching. It is important to stress that both $\ket{\psi^b}$ and $\ket{\psi^a}$ are anti-symmetric under spatial mode exchange, even though $F^s$ is a symmetric function of its arguments.\footnote{The generalized HOM effect probes the symmetry of the full two-photon state under spatial mode exchange. Only in the standard HOM configuration, where the two photons occupy different spatial modes, does this symmetry coincide directly with the symmetry of the JSA. For the state $\ket{\psi^b}$, the two photons occupy the same spatial mode, and the description involves two spectral amplitudes ($F^s$ and $-F^s$) associated with the two modes. The swap operator $\hat S$ therefore compares the joint spectral distributions in the two arms rather than the symmetry of a single JSA.} This is consistent with the fact that $\ket{\psi_\text{in}}$ is anti-symmetric, as discussed above.

Using Eq.~\eqref{eq: pc with s and evol} for the state $\ket{\psi_\text{in}}$ and a local delay $V(\tau)=e^{-i\tau\hat\omega_1}$, one obtains
\begin{equation}
    P_c = \int \dd\omega_1 \dd\omega_2\, \abs{F^s(\omega_1,\omega_2)}^2 \cos^2\left(\tfrac{(\omega_1+\omega_2)\tau}{2}\right)+ \int \dd\omega_1 \dd\omega_2\, \abs{F^a(\omega_1,\omega_2)}^2 \cos^2\left(\tfrac{(\omega_1-\omega_2)\tau}{2}\right).
\end{equation}
The symmetric contribution depends on the sum frequency $\omega_1+\omega_2$, whereas the anti-symmetric contribution depends on the difference frequency $\omega_1-\omega_2$. In contrast to the standard HOM effect, which always probes the JSA along the $\omega_1-\omega_2$ direction, the single-photon MZI can probe either the $\omega_1-\omega_2$ or the $\omega_1+\omega_2$ direction, depending on the symmetry component of the JSA.

For a local delay generated by $\hat\omega_1$, the Fisher information around $\tau=0$ reads
\begin{equation}
    \mathcal F= \expval{(\hat \omega_1+\hat\omega_2)^2}_s  + \expval{(\hat \omega_1-\hat \omega_2)^2}_a,
\end{equation}
while the quantum Fisher information is
\begin{equation}
    \mathcal Q = 4\Delta^2_{\ket{\psi_\text{in}}}(\hat \omega_1) = 2\expval{(\hat \omega_1+\hat \omega_2)^2}_s  + 4\expval{\hat \omega_1^2}_a  - \left(\expval{\hat \omega_1+\hat \omega_2}_s + 2\expval{\hat \omega_1}_a \right)^2,
\end{equation}
where $\expval{\cdot}_s$ and $\expval{\cdot}_a$ denote expectation values evaluated on the effective single-photon states
\begin{align}
    \ket{\psi_s}= \int \dd\omega_1 \dd\omega_2\,  F^s(\omega_1,\omega_2) \ket{\omega_1,\omega_2},  && \ket{\psi_a}= \int \dd\omega_1 \dd\omega_2\,  F^a(\omega_1,\omega_2) \ket{\omega_1,\omega_2}.
\end{align}
In general, the measurement is not optimal, in the sense that $\mathcal F<\mathcal Q$. To proceed further, one may consider specific models for the JSA $F$. A common and analytically convenient assumption would be to assume, as we have already done, that $F$ factorizes in the variables $\omega_\pm$, which allows one to evaluate the above expressions explicitly, facilitating the comparison between $\mathcal F$ and $\mathcal Q$.

\clearpage
\section{Multimode generalization}
\label{sec: multimode generalization}

\emph{In this section, we generalize the results presented in the previous section to the multimode setting. We provide an explicit $n$-mode interferometer that constitutes a direct extension of the HOM interferometer. This extension remains centered around the notion of symmetry and leads to general metrological bounds for parameter estimation. We also discuss how time information can be incorporated into the formalism and present the extension to mixed states. This section mainly contains the remaining results of \hyperlink{Article: prl extended HOM}{Role of Symmetry in Generalized Hong-Ou-Mandel Interference and Quantum Metrology}~\cite{descamps_role_2026}, published during the PhD thesis.}

\subsection{Extension to $n$ modes}
\label{subsec: n-mode extension}

\paragraph{\publi Motivation and guidelines}

While the previous section introduced a powerful perspective to analyse the interferometric effect associated with a balanced beam splitter and its consequences for metrological parameter estimation, it remained restricted to two spatial modes. In order to provide a more general framework, we now extend these results to the case of $n$ spatial modes. To reach such a generalization, the main ingredients of the two-mode analysis must be adapted as follows:

\begin{itemize}
    \item In the previous section, we demonstrated how spatial symmetry, embodied in the symmetry operator $\hat S$, plays a central role in the analysis of the HOM effect and its metrological implications. In the regime of $n$ modes, many permutations of the $n$ modes are possible. The first step is therefore to identify the relevant symmetry operation that generalizes the two-mode swap operator $\hat S$.
    
    \item The usual HOM effect relies on coincidence detection, which we understood as a parity measurement. In the $n$-mode regime, one must therefore identify the appropriate generalization of parity measurement. We expect that this will be obtained as a suitable post-processing of the photon-number-resolved detection statistics.
    
    \item The HOM effect and its generalization rely on the use of a balanced beam splitter. In the $n$-mode regime, one must therefore identify the appropriate generalization of a balanced beam splitter, which we expect to be an interferometer that symmetrically mixes all the modes.
\end{itemize}

A useful insight is obtained by considering the elementary but fundamental identity for Pauli matrices
\begin{equation}
    Z = H X H,
\end{equation}
whose natural extension at the level of passive linear transformations links symmetry and parity properties through a beam splitter. Viewing $X$ as a permutation matrix, the above formula expresses the diagonalization of the permutation matrix $X$ by the matrix $H$. Recalling that permutation matrices are diagonalized by discrete Fourier transform matrices, this observation naturally leads us to consider
\begin{align}
    P = \scalebox{0.9}{$\begin{pmatrix} 
        0 & 1 & & \\
        & \ddots & \ddots & \\
        & & \ddots & 1 \\
        1 & & & 0 
    \end{pmatrix}$}, 
    &&
    U=\frac{1}{\sqrt{n}}\scalebox{0.9}{$\begin{pmatrix}
        1 & 1 & \cdots & 1 \\
        1 & \omega & \cdots & \omega^{n-1} \\
        \vdots & \vdots & \ddots & \vdots \\
        1 & \omega^{n-1} & \cdots & \omega^{(n-1)(n-1)}
    \end{pmatrix}$},
    &&
    D=\scalebox{0.9}{$\begin{pmatrix}
        1 & & & \\
        & \omega & & \\
        & & \ddots & \\
        & & & \omega^{n-1}
    \end{pmatrix}$},
\end{align}
where $P$ consists of $n$ ones and zeros elsewhere, with the ones placed along the first upper diagonal ($n-1$ entries) and a single $1$ in the lower-left corner. The matrix $D$ is diagonal, and $\omega = e^{2\pi i/n}$ denotes the primitive $n$-th root of unity. One can verify that these matrices satisfy
\begin{equation}
    P = U D U^\dagger.
\end{equation}
See Appendix~\ref{app: hom interferometry}, Result~\ref{res: permutation diagonalisation} for a verification. This matrix product relation provides answers to the three questions raised above:

\begin{itemize}
    \item Associating the matrix $P$ with the corresponding passive linear transformation $\hat P$, defined through
    \begin{equation}
        \hat P \, \hat a_j^\dagger(\lambda) \, \hat P^\dagger = \hat a_{j-1}^\dagger(\lambda),
    \end{equation}
    where, on the right-hand side, $j-1$ is taken modulo $n$, we identify the relevant symmetry operation as the cyclic permutation of the modes. This is the natural generalization of the two-mode swap operator $\hat S$.
    
    \item The matrix $D$ is diagonal with diagonal entries given by the $n$-th roots of unity. The associated passive linear transformation acts as\footnote{For ease of notation, we label spatial modes from $0$ to $n-1$, as made explicit below.}
    \begin{equation}
        \hat D \, \hat a_j^\dagger(\lambda) \, \hat D^\dagger = \omega^j \hat a_j^\dagger(\lambda).
    \end{equation}
    This suggests that the relevant generalization of parity measurement is obtained by post-processing the photon-number-resolved statistics in such a way that the complex amplitudes $\omega^j$ interfere. The appropriate combination will emerge naturally below.
    
    \item The matrix $U$ is the discrete Fourier transform matrix, which symmetrically mixes all the modes. The corresponding passive linear transformation $\hat U$ therefore constitutes the natural generalization of a balanced beam splitter. Such interferometers have been extensively studied, notably in the context of boson sampling~\cite{tichy_zero-transmission_2010,lim_generalized_2005,tichy_many-particle_2012} and in metrological applications~\cite{motes_linear_2015,olson_linear_2017}. Interestingly, such an interferometer can be implemented in practice with a reasonable $O(n^2)$ number of optical elements, as shown in~\cite{reck_experimental_1994}.
\end{itemize}

\paragraph{\publi Definition of the interferometer}

We now define the extension of the HOM interferometer to $n$ orthogonal spatial modes. Consider $n$ spatial modes with annihilation operators
\begin{equation}
    \hat a_j(\lambda),
\end{equation}
where the spatial index $j$ ranges from $0$ to $n-1$, and $\lambda$ denotes any internal degree of freedom. We consider as input a general state $\ket{\psi}$ over the $n$ spatial modes, with arbitrary photon-number statistics and arbitrary internal degree-of-freedom distributions. The state is injected into a DFT interferometer $\hat U$, which transforms the creation operators as
\begin{equation}
    \hat U \, \hat a_j^\dagger(\lambda) \, \hat U^\dagger = \frac{1}{\sqrt{n}} \sum_{k=0}^{n-1} \omega^{jk} \hat a_k^\dagger(\lambda).
\end{equation}
At the output, photon-number-resolved detection is performed and the number $m_j$ of photons detected in the $j$-th mode is recorded. A schematic representation of the setup is shown in Fig.~\ref{fig: n-mode HOM}.

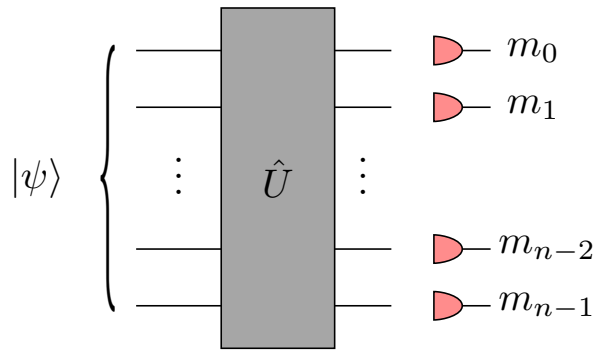
\begin{figure}[ht]
    \centering
    \footnotesize
    \scalebox{1.5}{\begin{tikzpicture}
	\begin{pgfonlayer}{nodelayer}
		\node [style=none] (0) at (1.25, 0) {$\hat U$};
		\node [style=none] (1) at (0.25, -3) {};
		\node [style=none] (2) at (2.25, -3) {};
		\node [style=none] (3) at (0.25, 3) {};
		\node [style=none] (4) at (2.25, 3) {};
		\node [style=none] (5) at (0.25, 2.25) {};
		\node [style=none] (6) at (0.25, 1.25) {};
		\node [style=none] (7) at (0.25, -2.25) {};
		\node [style=none] (8) at (0.25, -1.25) {};
		\node [style=none] (9) at (-1.25, 2.25) {};
		\node [style=none] (10) at (-1.25, 1.25) {};
		\node [style=none] (11) at (-1.25, -1.25) {};
		\node [style=none] (12) at (-1.25, -2.25) {};
		\node [style=none] (13) at (-0.5, 0.25) {$\vdots$};
		\node [style=none] (14) at (3.25, 2.25) {};
		\node [style=none] (15) at (3.25, 1.25) {};
		\node [style=none] (16) at (3.25, -2.25) {};
		\node [style=none] (17) at (3.25, -1.25) {};
		\node [style=none] (18) at (2.25, 2.25) {};
		\node [style=none] (19) at (2.25, 1.25) {};
		\node [style=none] (20) at (2.25, -1.25) {};
		\node [style=none] (21) at (2.25, -2.25) {};
		\node [style=none] (22) at (2.75, 0.25) {$\vdots$};
		\node [style=none] (23) at (4, 2.5) {};
		\node [style=none] (24) at (4, 2) {};
		\node [style=none] (25) at (4.5, 2.25) {};
		\node [style=none] (26) at (5, 2.25) {};
		\node [style=none] (27) at (5.75, 2.25) {$m_0$};
		\node [style=none] (28) at (4, 1.5) {};
		\node [style=none] (29) at (4, 1) {};
		\node [style=none] (30) at (4.5, 1.25) {};
		\node [style=none] (31) at (5, 1.25) {};
		\node [style=none] (32) at (5.75, 1.25) {$m_1$};
		\node [style=none] (37) at (6, -1.25) {$m_{n-2}$};
		\node [style=none] (42) at (6, -2.25) {$m_{n-1}$};
		\node [style=none] (43) at (-3, 0) {$\ket{\psi}$};
		\node [style=none] (44) at (-1.75, 0) {\scalebox{1}[6.7]{$\{$}};
		\node [style=none] (59) at (4, -1) {};
		\node [style=none] (60) at (4, -1.5) {};
		\node [style=none] (61) at (4.5, -1.25) {};
		\node [style=none] (62) at (5, -1.25) {};
		\node [style=none] (63) at (4, -2) {};
		\node [style=none] (64) at (4, -2.5) {};
		\node [style=none] (65) at (4.5, -2.25) {};
		\node [style=none] (66) at (5, -2.25) {};
	\end{pgfonlayer}
	\begin{pgfonlayer}{edgelayer}
		\draw [style=Fill grey] (2.center)
			 to (1.center)
			 to (3.center)
			 to (4.center)
			 to cycle;
		\draw (9.center) to (5.center);
		\draw (6.center) to (10.center);
		\draw (11.center) to (8.center);
		\draw (7.center) to (12.center);
		\draw (18.center) to (14.center);
		\draw (15.center) to (19.center);
		\draw (20.center) to (17.center);
		\draw (16.center) to (21.center);
		\draw [style=Fill pink] (24.center)
			 to (23.center)
			 to [in=90, out=0, looseness=0.75] (25.center)
			 to [in=0, out=-90, looseness=0.75] cycle;
		\draw (25.center) to (26.center);
		\draw [style=Fill pink] (29.center)
			 to (28.center)
			 to [in=90, out=0, looseness=0.75] (30.center)
			 to [in=0, out=-90, looseness=0.75] cycle;
		\draw (30.center) to (31.center);
		\draw [style=Fill pink] (60.center)
			 to (59.center)
			 to [in=90, out=0, looseness=0.75] (61.center)
			 to [in=0, out=-90, looseness=0.75] cycle;
		\draw (61.center) to (62.center);
		\draw [style=Fill pink] (64.center)
			 to (63.center)
			 to [in=90, out=0, looseness=0.75] (65.center)
			 to [in=0, out=-90, looseness=0.75] cycle;
		\draw (65.center) to (66.center);
	\end{pgfonlayer}
\end{tikzpicture}}
    \caption[Multimode extension of the HOM interferometer]{$n$-mode extension of the HOM interferometer. An arbitrary input state $\ket{\psi}$, with arbitrary photon-number statistics and internal degree-of-freedom distribution, is injected into a DFT interferometer $\hat U$ that symmetrically mixes all modes. Photon-number-resolved detection is performed at the output, and the numbers $m_j$ of detected photons in each mode are recorded. This setup acts as a symmetry analyzer with respect to cyclic permutations of the spatial modes. Adapted from Ref.~\cite{descamps_role_2026}. \copyright 2026 American Physical Society.}
    \label{fig: n-mode HOM}
\end{figure}

\paragraph{\publi Interferometric results}

Restricting first to states with fixed total photon number and using the geometric series identity
\begin{equation}
    \frac{1}{n} \sum_{k=0}^{n-1} \omega^{kj} = \delta_{j,0},
\end{equation}
one finds (see Appendix~\ref{app: hom interferometry}, Result~\ref{res: D and photon number}) that for a general state
\begin{equation}
    \frac{1}{n} \sum_{l=0}^{n-1} \expval{\hat D^l}
    = \mathbb{P}\left( \sum_{k=0}^{n-1} k m_k \equiv 0\,[n] \right).
\end{equation}
This relation links the expectation values of powers of $\hat D$ to the probability of detecting photons in configurations satisfying a specific modular constraint. Inserting the DFT transformation and using $\hat P = \hat U \hat D \hat U^\dagger$, we obtain
\begin{equation}\label{eq: probability Fourier}
    \frac{1}{n} \sum_{l=0}^{n-1} \expval{\hat P^l}
    = \mathbb{P}\left( \sum_{k=0}^{n-1} k m_k \equiv 0\,[n] \right),
\end{equation}
which links the symmetry of the input state with respect to cyclic permutations to the probability of specific output detection patterns. This is the direct generalization of the two-mode HOM formula of Eq.~\eqref{eq: Peven HOM}, relating swap symmetry to even-photon detection. The operator
\begin{equation}
    \hat \Pi = \frac{1}{n} \sum_{l=0}^{n-1} \hat P^l
\end{equation}
is the projector onto the eigenspace of $\hat P$ associated with eigenvalue $+1$. The above probability therefore evaluates the squared norm of the component of $\ket{\psi}$ in this eigenspace. Other values of $\sum_{k=0}^{n-1} k m_k$ probe the remaining eigenspaces; see Appendix~\ref{app: hom interferometry}, Result~\ref{res: proba fourier other}. The interferometer thus acts as a symmetry analyzer with respect to cyclic permutations.

\paragraph{\publi Metrology}

The interferometric result can be exploited for parameter estimation. Consider a parameter-dependent evolution $\hat V(\theta) = e^{-i \hat H \theta}$ applied before $\hat U$; see Fig.~\ref{fig: n-mode HOM evol}. Under the assumption that either $\hat \Pi \ket{\psi} = \ket{\psi}$ or $\hat \Pi \ket{\psi} = 0$, one finds
\begin{equation}
    \lim_{\theta \rightarrow 0} \mathcal F
    = 4 \Delta^2 \big( i [\hat H, \hat \Pi] \big).
\end{equation}
This expression quantifies how strongly the generator $\hat H$ drives the state out of the symmetry sector selected by $\hat \Pi$. Assuming a symmetric probe state, $\hat P \ket{\psi} = \ket{\psi}$, one obtains
\begin{equation}\label{eq: fourier fisher sym}
    \mathcal F
    = \frac{4}{n^2} \Delta^2 \Big( n \hat H - \sum_{l=0}^{n-1} \hat P^l \hat H \hat P^{-l} \Big),
\end{equation}
while for an anti-symmetric probe state,\footnote{This is possible only if $n$ is even.} $\hat P \ket{\psi} = -\ket{\psi}$,
\begin{equation}\label{eq: fourier fisher anti-sym}
    \mathcal F
    = \frac{4}{n^2} \Delta^2 \Big( \sum_{l=0}^{n-1} (-1)^l \hat P^l \hat H \hat P^{-l} \Big).
\end{equation}

Comparing with the QFI $\mathcal Q = 4 \Delta^2(\hat H)$ allows one to assess optimality. If $\hat P \hat H \hat P^\dagger = -\hat H$, then $\mathcal F = \mathcal Q$. If $\hat P \hat H \hat P^\dagger = \hat H$, then $\mathcal F = 0$. More generally, the symmetry properties of $\hat H$ with respect to cyclic permutations determine the amount of extractable information. The corresponding derivations are provided in Appendix~\ref{app: hom interferometry}, Result~\ref{res: n-mode HOM Fisher info}.

\begin{figure}[ht]
    \centering
    \footnotesize
    \scalebox{1.5}{\begin{tikzpicture}
	\begin{pgfonlayer}{nodelayer}
		\node [style=none] (67) at (1.75, 0) {$\hat U$};
		\node [style=none] (68) at (0.75, -3) {};
		\node [style=none] (69) at (2.75, -3) {};
		\node [style=none] (70) at (0.75, 3) {};
		\node [style=none] (71) at (2.75, 3) {};
		\node [style=none] (72) at (0.75, 2.25) {};
		\node [style=none] (73) at (0.75, 1.25) {};
		\node [style=none] (74) at (0.75, -2.25) {};
		\node [style=none] (75) at (0.75, -1.25) {};
		\node [style=none] (76) at (-0.75, 2.25) {};
		\node [style=none] (77) at (-0.75, 1.25) {};
		\node [style=none] (78) at (-0.75, -1.25) {};
		\node [style=none] (79) at (-0.75, -2.25) {};
		\node [style=none] (80) at (0, 0.25) {$\vdots$};
		\node [style=none] (81) at (3.75, 2.25) {};
		\node [style=none] (82) at (3.75, 1.25) {};
		\node [style=none] (83) at (3.75, -2.25) {};
		\node [style=none] (84) at (3.75, -1.25) {};
		\node [style=none] (85) at (2.75, 2.25) {};
		\node [style=none] (86) at (2.75, 1.25) {};
		\node [style=none] (87) at (2.75, -1.25) {};
		\node [style=none] (88) at (2.75, -2.25) {};
		\node [style=none] (89) at (3.25, 0.25) {$\vdots$};
		\node [style=none] (90) at (4.5, 2.5) {};
		\node [style=none] (91) at (4.5, 2) {};
		\node [style=none] (92) at (5, 2.25) {};
		\node [style=none] (93) at (5.5, 2.25) {};
		\node [style=none] (94) at (6.25, 2.25) {$m_0$};
		\node [style=none] (95) at (4.5, 1.5) {};
		\node [style=none] (96) at (4.5, 1) {};
		\node [style=none] (97) at (5, 1.25) {};
		\node [style=none] (98) at (5.5, 1.25) {};
		\node [style=none] (99) at (6.25, 1.25) {$m_1$};
		\node [style=none] (100) at (6.5, -1.25) {$m_{n-2}$};
		\node [style=none] (101) at (6.5, -2.25) {$m_{n-1}$};
		\node [style=none] (102) at (-5.25, 0) {$\ket{\psi}$};
		\node [style=none] (103) at (-4, 0) {\scalebox{1}[6.7]{$\{$}};
		\node [style=none] (104) at (-1.75, -0.75) {$e^{-i\theta\hat H}$};
		\node [style=none] (105) at (-2.5, -3) {};
		\node [style=none] (106) at (-1, -3) {};
		\node [style=none] (107) at (-2.5, 3) {};
		\node [style=none] (108) at (-1, 3) {};
		\node [style=none] (109) at (-2.75, 2.25) {};
		\node [style=none] (110) at (-2.75, 1.25) {};
		\node [style=none] (111) at (-2.75, -2.25) {};
		\node [style=none] (112) at (-2.75, -1.25) {};
		\node [style=none] (113) at (-3.75, 2.25) {};
		\node [style=none] (114) at (-3.75, 1.25) {};
		\node [style=none] (115) at (-3.75, -1.25) {};
		\node [style=none] (116) at (-3.75, -2.25) {};
		\node [style=none] (117) at (-3.25, 0.25) {$\vdots$};
		\node [style=none] (118) at (4.5, -1) {};
		\node [style=none] (119) at (4.5, -1.5) {};
		\node [style=none] (120) at (5, -1.25) {};
		\node [style=none] (121) at (5.5, -1.25) {};
		\node [style=none] (122) at (4.5, -2) {};
		\node [style=none] (123) at (4.5, -2.5) {};
		\node [style=none] (124) at (5, -2.25) {};
		\node [style=none] (125) at (5.5, -2.25) {};
		\node [style=none] (126) at (-2.75, 2.75) {};
		\node [style=none] (127) at (-0.75, 2.75) {};
		\node [style=none] (128) at (-0.75, -2.75) {};
		\node [style=none] (129) at (-2.75, -2.75) {};
		\node [style=none] (130) at (-1.75, 0) {$=$};
		\node [style=none] (131) at (-1.75, 1) {$\hat V(\theta)$};
	\end{pgfonlayer}
	\begin{pgfonlayer}{edgelayer}
		\draw [style=Fill grey] (69.center)
			 to (68.center)
			 to (70.center)
			 to (71.center)
			 to cycle;
		\draw (76.center) to (72.center);
		\draw (73.center) to (77.center);
		\draw (78.center) to (75.center);
		\draw (74.center) to (79.center);
		\draw (85.center) to (81.center);
		\draw (82.center) to (86.center);
		\draw (87.center) to (84.center);
		\draw (83.center) to (88.center);
		\draw [style=Fill pink] (91.center)
			 to (90.center)
			 to [in=90, out=0, looseness=0.75] (92.center)
			 to [in=0, out=-90, looseness=0.75] cycle;
		\draw (92.center) to (93.center);
		\draw [style=Fill pink] (96.center)
			 to (95.center)
			 to [in=90, out=0, looseness=0.75] (97.center)
			 to [in=0, out=-90, looseness=0.75] cycle;
		\draw (97.center) to (98.center);
		\draw (127.center)
			 to [bend right=45, looseness=1.25] (108.center)
			 to (107.center)
			 to [bend right=45, looseness=1.25] (126.center)
			 to (129.center)
			 to [bend right=45] (105.center)
			 to (106.center)
			 to [bend right=45] (128.center)
			 to cycle;
		\draw (113.center) to (109.center);
		\draw (110.center) to (114.center);
		\draw (115.center) to (112.center);
		\draw (111.center) to (116.center);
		\draw [style=Fill pink] (119.center)
			 to (118.center)
			 to [in=90, out=0, looseness=0.75] (120.center)
			 to [in=0, out=-90, looseness=0.75] cycle;
		\draw (120.center) to (121.center);
		\draw [style=Fill pink] (123.center)
			 to (122.center)
			 to [in=90, out=0, looseness=0.75] (124.center)
			 to [in=0, out=-90, looseness=0.75] cycle;
		\draw (124.center) to (125.center);
	\end{pgfonlayer}
\end{tikzpicture}}
    \caption[Multimode interferometer with parameter-dependent evolution]{$n$-mode interferometer with parameter-dependent evolution. The input state $\ket{\psi}$ undergoes $\hat V(\theta) = e^{-i \hat H \theta}$ followed by the DFT interferometer $\hat U$. Photon-number-resolved detection yields outcomes $m_j$, from which the quantity $\sum_{k=0}^{n-1} k m_k$ modulo $n$ is computed. Estimating the associated probabilities enables parameter estimation governed by the symmetry of $\hat H$. Adapted from Ref.~\cite{descamps_role_2026}. \copyright 2026 American Physical Society.}
    \label{fig: n-mode HOM evol}
\end{figure}
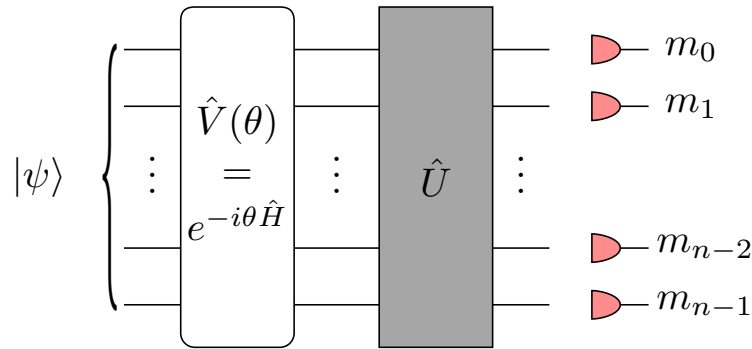

\paragraph{\publi Application to measuring collective delays}
To illustrate the versatility of the framework developed above, we now discuss how it provides an explicit measurement scheme capable of saturating the QFI analyzed in Sec.~\ref{sec: TF entanglement and metrology}. Recall that in that section we investigated the metrological properties of $n$ time-frequency single photons, each propagating in a distinct spatial mode. In particular, we analyzed the QFI associated with the collective time-delay generator
\begin{equation}
    \hat \Omega=\sum_{l=0}^{n-1} \hat \omega_l,
\end{equation}
where $\hat \omega_l$ denotes the local time-evolution generator acting on mode $l$. The discussion in Sec.~\ref{sec: TF entanglement and metrology} remained at a theoretical level. We derived general bounds on the achievable precision and identified the states that attain the ultimate precision limit. This analysis provided insight into the role of frequency correlations and entanglement in quantum metrology and quantum optics. However, the explicit construction of a measurement setup capable of optimally probing the evolution generated by $\hat \Omega$ was not addressed.

Let us now connect those results to the interferometric framework developed in the present section. The $n$ single-photon time-frequency states considered previously, and in particular the optimal state introduced in Sec.~\ref{subsec: TF optimal states}, can be chosen to be either symmetric or anti-symmetric under mode permutations. Therefore, our general interferometric result applies: a DFT interferometer followed by photon-number-resolving detection constitutes a natural candidate measurement scheme to probe the corresponding evolution.

However, an important subtlety arises. The generator $\hat \Omega$ is symmetric under the action of the permutation operator $\hat P$. As a consequence, the measurement is insensitive to the parameter encoded through $\hat \Omega$. In this case, the classical Fisher information extracted from the interferometric measurement vanishes, $\mathcal F=0$, despite the fact that the QFI may be non-zero.

To overcome this limitation and achieve an optimal measurement, one can instead consider the modified generator
\begin{equation}
    \hat \Omega' = \sum_{l=0}^{n-1} (-1)^l \hat \omega_l.
\end{equation}
For even $n$, the operator $\hat \Omega'$ is anti-symmetric under the action of $\hat P$. In this situation, the interferometric measurement becomes sensitive to the parameter and saturates the quantum Cramér-Rao bound, yielding $\mathcal F = \mathcal Q$. Importantly, as discussed in Sec.~\ref{sec: TF entanglement and metrology}, the specific choice of $\hat \Omega$ is not essential for the general metrological analysis. All results derived there can be straightforwardly adapted to the generator $\hat \Omega'$ without altering the underlying physical conclusions.

This example illustrates concretely how the general framework developed in this section enables the design of explicit measurement schemes tailored to a broad class of collective evolutions. In particular, it applies to evolutions that are not generated by a single local mode operator, but instead arise from structured multi-mode generators encoding global time-delay patterns.

\subsection{Incorporating time information}
\label{subsec: with time information}

\paragraph{\newc Motivations}

So far, we have completely ignored the time information, effectively assuming that the detectors run for an infinitely long time and that only the total number of detected photons is recorded. In the standard HOM case, we have argued that time information is only required to separate photon pairs during post-processing of the experimental data, and not to provide directly exploitable information for enhancing the time-delay estimation itself. In particular, optimal precision can be achieved without time-resolved detection. 

The situation becomes more intricate in the multimode and multiphoton regime. Now that we have developed tools capable of analyzing scenarios in which more than one photon pair is measured simultaneously, it is natural to investigate settings involving more complex probe states. Beyond idealized situations in which the probe state is assumed to be fully controlled, a particularly relevant experimental platform for multiphoton interference is SPDC operated closer to the squeezing regime, where multiple photon pairs are generated with non-negligible probability. In this regime, the probabilistic nature of pair generation leads to events in which different pairs temporally overlap and can no longer be unambiguously separated and treated independently. 

This naturally motivates a formal analysis of how time information can be used to partition the experimental data into smaller, more manageable subsets. Moreover, in realistic scenarios where photon loss cannot be neglected, the use of time-resolved detection enables the restriction of the analysis to sufficiently small temporal windows during which only a limited number of photons are produced. This can mitigate the impact of loss on the data processing and, potentially, on the attainable estimation precision.

To formalize this idea, we fix two times $T_1 < T_2$ and assume throughout that only detection events occurring within the time window $[T_1,T_2]$ are recorded. By partitioning the real line into equal-size time windows, analyzing the data independently within each window, and subsequently combining the results, one can reconstruct the total information contained in the full data set while keeping the analysis of each window tractable. 

Since time information now plays a fundamental role, we assume that the internal parameter $\lambda$ introduced previously corresponds to the photon arrival time, and we denote it by $t$ instead of $\lambda$.\footnote{Other internal degrees of freedom, such as polarization, could be incorporated in a similar manner to the treatment developed previously for arbitrary internal variables. For simplicity, we restrict the discussion here to the time of arrival as the only internal degree of freedom.} Furthermore, we assume that the detectors, although not necessarily perfect, possess sufficient temporal resolution to reliably assign detection events to the corresponding time windows.

\paragraph{\newc Windowed symmetry}

The key ingredient that allows us to generalize the previous results is the introduction of a windowed symmetry operator $\hat P_w$. This operator acts on the time-resolved creation operators $\hat a_j^\dagger(t)$ conditionally on the arrival time:
\begin{equation}
    \hat P_w \hat a_j^\dagger(t) \hat P_w^\dagger = 
    \begin{cases}
        \hat a_{j-1}^\dagger(t) & \text{if } t \in [T_1,T_2],\\
        \hat a_j^\dagger(t) & \text{otherwise}.
    \end{cases}
\end{equation}
Hence, $\hat P_w$ acts non-trivially only on creation operators associated with photons arriving inside the time window $[T_1,T_2]$, while leaving all other modes unaffected.

The previous analysis can be adapted to this modified symmetry. Consider a general probe state $\ket{\psi}$ injected into a DFT interferometer, and suppose that photon numbers $m_k$ are detected in the output modes during the time window $[T_1,T_2]$. One then obtains
\begin{equation}
    \frac{1}{n} \sum_{l=0}^{n-1} \expval{\hat P_w^l}
    = \mathbb{P}_{[T_1,T_2]}\left( \sum_{k=0}^{n-1} k m_k \equiv 0\,[n]\right).
\end{equation}
The proof directly follows from the derivation of Eq.~\eqref{eq: probability Fourier} presented in Appendix~\ref{app: hom interferometry}, Result~\ref{res: D and photon number}. The essential observation is that the DFT relates $\hat P_w$ to a windowed diagonal operator $\hat D_w$ defined through
\begin{equation}
    \hat D_w \hat a_j^\dagger(t) \hat D_w^\dagger = 
    \begin{cases}
        \omega^j \hat a_j^\dagger(t) & \text{if } t \in [T_1,T_2],\\
        \hat a_j^\dagger(t) & \text{otherwise},
    \end{cases}
\end{equation}
and that the expectation values $\expval{\hat D_w^l}$ are sensitive only to the number of photons arriving within the time window $[T_1,T_2]$.

By adapting the metrological derivations established in the time-insensitive case, the Fisher information associated with a parameter encoded via the unitary evolution $e^{-i\hat H \theta}$ can be expressed as
\begin{equation}
    \mathcal F = \frac{4}{n^2} \Delta^2\Big( n \hat H - \sum_{l=0}^{n-1} \hat P_w^l \hat H \hat P_w^{-l} \Big),
\end{equation}
for a symmetric probe state satisfying $\hat P_w\ket{\psi}=\ket{\psi}$, or
\begin{equation}
    \mathcal F = \frac{4}{n^2} \Delta^2\Big( \sum_{l=0}^{n-1} (-1)^l \hat P_w^l \hat H \hat P_w^{-l} \Big),
\end{equation}
for an anti-symmetric probe state obeying $\hat P_w\ket{\psi}=-\ket{\psi}$. Here, the symmetry is defined with respect to the windowed permutation operator $\hat P_w$, which acts exclusively on photons arriving within $[T_1,T_2]$. 

The measurement therefore probes the symmetry properties of the generator $\hat H$ relative to the windowed permutation $\hat P_w$. This provides a direct way to analyze how the information about the parameter is distributed in time and how it can be accessed through suitably chosen time-resolved measurements.

\paragraph{\newc Discussion}

The introduction of the windowed symmetry operator offers both conceptual clarity and technical flexibility. By restricting the action of the permutation to photons arriving within a prescribed temporal interval, the formalism effectively decomposes a global estimation task into temporally localized subproblems. This decomposition can substantially simplify the analysis in regimes where multi-pair emission and photon loss would otherwise render a complete treatment exceedingly involved. Moreover, it establishes a direct link between symmetry properties and experimentally accessible, time-resolved detection records, thereby giving an operational interpretation to the temporal structure of the probe state.

At the same time, this windowed approach raises several non-trivial issues that warrant further investigation. In particular, determining whether an incoming state is symmetric or anti-symmetric with respect to $\hat P_w$ generally requires detailed knowledge of the full temporal structure of the multiphoton wavefunction. A simplified description based on successive independent pairs is typically insufficient, especially when photon pairs are generated with a temporal distribution that partially overlaps the chosen window. In such cases, correlations may extend across different time windows, and the effective symmetry within a given interval cannot be inferred without a continuous-time description of the state.

Furthermore, even if one computes the Fisher information associated with each individual window, it is not immediately clear how these window-dependent contributions should be combined. The additivity of the corresponding precisions is not guaranteed, since temporal correlations and shared quantum resources may preclude a simple summation rule from reproducing the true global Fisher information. Consequently, while the windowed decomposition renders the analysis more manageable, it is likely to lead to suboptimal estimation strategies compared to a fully time-resolved treatment of the complete state. A detailed assessment of these aspects, in the context of realistic experimental implementations of multiphoton interference, is left for future work.

\subsection{Symmetry discussions}
\label{subsec: symmetry discussions}

In this section we provide several complementary discussions related to symmetry considerations and their implications for metrology.

\paragraph{\publi Symmetry generation}

A crucial assumption for the metrological result to hold is that the initial state must satisfy a strong symmetry condition,
\begin{equation}
    \hat P\ket{\psi}=\pm\ket{\psi}.
\end{equation}
While such a condition can be met in simple situations, for instance with identical independent single photons or entangled biphotons generated by SPDC sources~\cite{boucher_toolbox_2015}, the experimental complexity of our protocol is justified only when more structured probe states are considered, in particular those that provide a quantum advantage in terms of the QFI. The generation of highly entangled multimode states remains an active area of research~\cite{bensemhoun_multipartite_2025}, and we do not enter into these technical aspects here. 

Instead, we emphasize that there exists a natural way to impose the required symmetry condition, which generalizes the observation made in the single-photon Mach-Zehnder interferometer discussed in Sec.~\ref{subsec: single photon MZI}. There, we noticed that the state after the first beam splitter is always anti-symmetric, independently of the JSA of the incoming single-photon state. This illustrates the fact that a suitable interferometer can enforce symmetry constraints irrespective of the detailed internal structure of the state.

Since the DFT interferometer links symmetry properties and photon-number statistics, one can generate the desired symmetry by appropriately engineering the input photon-number distribution. Denoting by $m_k'$ the photon-number distribution of the incoming state, the condition
\begin{equation}
    \sum_{k=0}^{n-1} k m_k' \equiv 0 \,[n],
\end{equation}
guarantees that the DFT interferometer transforms the input into a state that is symmetric under the cyclic shift generated by $\hat P$. In contrast, for even $n$, the condition
\begin{equation}
    \sum_{k=0}^{n-1} k m_k' \equiv \frac{n}{2} \,[n],
\end{equation}
leads to the generation of a state with odd symmetry with respect to the same cyclic shift. Engineering the symmetry in this manner, prior to a parameter-dependent evolution followed by our measurement scheme, leads to an architecture realizing a generalized Mach-Zehnder interferometer, as illustrated in Fig.~\ref{fig: generalized MZI}.

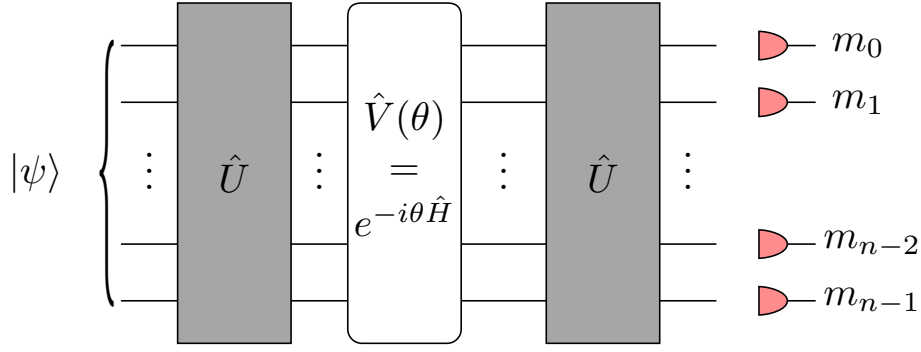
\begin{figure}[ht]
    \centering
    \footnotesize
    \scalebox{1.5}{\begin{tikzpicture}
	\begin{pgfonlayer}{nodelayer}
		\node [style=none] (67) at (1.75, 0) {$\hat U$};
		\node [style=none] (68) at (0.75, -3) {};
		\node [style=none] (69) at (2.75, -3) {};
		\node [style=none] (70) at (0.75, 3) {};
		\node [style=none] (71) at (2.75, 3) {};
		\node [style=none] (72) at (0.75, 2.25) {};
		\node [style=none] (73) at (0.75, 1.25) {};
		\node [style=none] (74) at (0.75, -2.25) {};
		\node [style=none] (75) at (0.75, -1.25) {};
		\node [style=none] (76) at (-0.75, 2.25) {};
		\node [style=none] (77) at (-0.75, 1.25) {};
		\node [style=none] (78) at (-0.75, -1.25) {};
		\node [style=none] (79) at (-0.75, -2.25) {};
		\node [style=none] (80) at (0, 0.25) {$\vdots$};
		\node [style=none] (81) at (3.75, 2.25) {};
		\node [style=none] (82) at (3.75, 1.25) {};
		\node [style=none] (83) at (3.75, -2.25) {};
		\node [style=none] (84) at (3.75, -1.25) {};
		\node [style=none] (85) at (2.75, 2.25) {};
		\node [style=none] (86) at (2.75, 1.25) {};
		\node [style=none] (87) at (2.75, -1.25) {};
		\node [style=none] (88) at (2.75, -2.25) {};
		\node [style=none] (89) at (3.25, 0.25) {$\vdots$};
		\node [style=none] (90) at (4.5, 2.5) {};
		\node [style=none] (91) at (4.5, 2) {};
		\node [style=none] (92) at (5, 2.25) {};
		\node [style=none] (93) at (5.5, 2.25) {};
		\node [style=none] (94) at (6.25, 2.25) {$m_0$};
		\node [style=none] (95) at (4.5, 1.5) {};
		\node [style=none] (96) at (4.5, 1) {};
		\node [style=none] (97) at (5, 1.25) {};
		\node [style=none] (98) at (5.5, 1.25) {};
		\node [style=none] (99) at (6.25, 1.25) {$m_1$};
		\node [style=none] (100) at (6.5, -1.25) {$m_{n-2}$};
		\node [style=none] (101) at (6.5, -2.25) {$m_{n-1}$};
		\node [style=none] (104) at (-1.75, -0.75) {$e^{-i\theta\hat H}$};
		\node [style=none] (105) at (-2.5, -3) {};
		\node [style=none] (106) at (-1, -3) {};
		\node [style=none] (107) at (-2.5, 3) {};
		\node [style=none] (108) at (-1, 3) {};
		\node [style=none] (109) at (-2.75, 2.25) {};
		\node [style=none] (110) at (-2.75, 1.25) {};
		\node [style=none] (111) at (-2.75, -2.25) {};
		\node [style=none] (112) at (-2.75, -1.25) {};
		\node [style=none] (113) at (-3.75, 2.25) {};
		\node [style=none] (114) at (-3.75, 1.25) {};
		\node [style=none] (115) at (-3.75, -1.25) {};
		\node [style=none] (116) at (-3.75, -2.25) {};
		\node [style=none] (117) at (-3.25, 0.25) {$\vdots$};
		\node [style=none] (118) at (4.5, -1) {};
		\node [style=none] (119) at (4.5, -1.5) {};
		\node [style=none] (120) at (5, -1.25) {};
		\node [style=none] (121) at (5.5, -1.25) {};
		\node [style=none] (122) at (4.5, -2) {};
		\node [style=none] (123) at (4.5, -2.5) {};
		\node [style=none] (124) at (5, -2.25) {};
		\node [style=none] (125) at (5.5, -2.25) {};
		\node [style=none] (126) at (-2.75, 2.75) {};
		\node [style=none] (127) at (-0.75, 2.75) {};
		\node [style=none] (128) at (-0.75, -2.75) {};
		\node [style=none] (129) at (-2.75, -2.75) {};
		\node [style=none] (130) at (-1.75, 0) {$=$};
		\node [style=none] (131) at (-1.75, 1) {$\hat V(\theta)$};
		\node [style=none] (132) at (-4.75, 0) {$\hat U$};
		\node [style=none] (133) at (-5.75, -3) {};
		\node [style=none] (134) at (-3.75, -3) {};
		\node [style=none] (135) at (-5.75, 3) {};
		\node [style=none] (136) at (-3.75, 3) {};
		\node [style=none] (137) at (-8.25, 0) {$\ket{\psi}$};
		\node [style=none] (138) at (-7, 0) {\scalebox{1}[6.7]{$\{$}};
		\node [style=none] (139) at (-5.75, 2.25) {};
		\node [style=none] (140) at (-5.75, 1.25) {};
		\node [style=none] (141) at (-5.75, -2.25) {};
		\node [style=none] (142) at (-5.75, -1.25) {};
		\node [style=none] (143) at (-6.75, 2.25) {};
		\node [style=none] (144) at (-6.75, 1.25) {};
		\node [style=none] (145) at (-6.75, -1.25) {};
		\node [style=none] (146) at (-6.75, -2.25) {};
		\node [style=none] (147) at (-6.25, 0.25) {$\vdots$};
	\end{pgfonlayer}
	\begin{pgfonlayer}{edgelayer}
		\draw [style=Fill grey] (69.center)
			 to (68.center)
			 to (70.center)
			 to (71.center)
			 to cycle;
		\draw (76.center) to (72.center);
		\draw (73.center) to (77.center);
		\draw (78.center) to (75.center);
		\draw (74.center) to (79.center);
		\draw (85.center) to (81.center);
		\draw (82.center) to (86.center);
		\draw (87.center) to (84.center);
		\draw (83.center) to (88.center);
		\draw [style=Fill pink] (91.center)
			 to (90.center)
			 to [in=90, out=0, looseness=0.75] (92.center)
			 to [in=0, out=-90, looseness=0.75] cycle;
		\draw (92.center) to (93.center);
		\draw [style=Fill pink] (96.center)
			 to (95.center)
			 to [in=90, out=0, looseness=0.75] (97.center)
			 to [in=0, out=-90, looseness=0.75] cycle;
		\draw (97.center) to (98.center);
		\draw (127.center)
			 to [bend right=45, looseness=1.25] (108.center)
			 to (107.center)
			 to [bend right=45, looseness=1.25] (126.center)
			 to (129.center)
			 to [bend right=45] (105.center)
			 to (106.center)
			 to [bend right=45] (128.center)
			 to cycle;
		\draw (113.center) to (109.center);
		\draw (110.center) to (114.center);
		\draw (115.center) to (112.center);
		\draw (111.center) to (116.center);
		\draw [style=Fill pink] (119.center)
			 to (118.center)
			 to [in=90, out=0, looseness=0.75] (120.center)
			 to [in=0, out=-90, looseness=0.75] cycle;
		\draw (120.center) to (121.center);
		\draw [style=Fill pink] (123.center)
			 to (122.center)
			 to [in=90, out=0, looseness=0.75] (124.center)
			 to [in=0, out=-90, looseness=0.75] cycle;
		\draw (124.center) to (125.center);
		\draw [style=Fill grey] (134.center)
			 to (133.center)
			 to (135.center)
			 to (136.center)
			 to cycle;
		\draw (143.center) to (139.center);
		\draw (140.center) to (144.center);
		\draw (145.center) to (142.center);
		\draw (141.center) to (146.center);
	\end{pgfonlayer}
\end{tikzpicture}}
    \caption[Generalized multimode Mach-Zehnder interferometer]{An initial state $\ket{\psi}$ with photon-number distribution $\{m_k'\}$ is injected into a DFT interferometer $\hat U$, which enforces a well-defined symmetry according to the modular condition on $\sum_{k=0}^{n-1} k m_k'$. The resulting symmetric or anti-symmetric state then undergoes a parameter-dependent evolution $e^{-i\hat H \theta}$, followed by a second DFT interferometer and photon-number-resolved detection. This setup generalizes the single-photon Mach-Zehnder interferometer, where the first beam splitter generates the symmetry required for optimal estimation. Adapted from Ref.~\cite{descamps_role_2026}. \copyright 2026 American Physical Society.}
    \label{fig: generalized MZI}
\end{figure}

It is worth stressing that, although generating complex photon-number distributions can be experimentally demanding, already the single-photon regime provides a sufficient condition for producing the required symmetry. In that case $m_k'=1$ for all $k$, and one readily checks that
\begin{equation}
    \sum_{k=0}^{n-1} k \equiv 
    \begin{cases}
        0 \,[n] & \text{if } n \text{ is odd},\\
        \frac{n}{2} \,[n] & \text{if } n \text{ is even},
    \end{cases}
\end{equation}
which results, after the DFT interferometer, in either a symmetric or an anti-symmetric state, respectively.

It is however important to note that passing through the DFT interferometer does not only modify the symmetry property of the probe state. It also reshapes the photon-number statistics and the distribution over internal degrees of freedom in a non-trivial way, which in turn affects the QFI. Therefore, while the DFT interferometer can be used to generate the required symmetry, the resulting state is not necessarily optimal (or even advantageous) for parameter estimation. In particular, there is no general guarantee that the DFT interferometer maps an optimal probe state to another optimal one. The systematic design of optimal states where the DFT interferometer is used to impose the symmetry remains an open problem.

\paragraph{\publi Imperfect symmetry}
Although the DFT interferometer provides, in principle, a way to engineer the required symmetry, the generation of perfectly symmetric states is, strictly speaking, unattainable in realistic experiments. Imperfections such as fabrication errors, mode mismatch, loss, or deviations from an ideal DFT transformation inevitably degrade the symmetry of the state. This naturally raises the question of the robustness of the metrological advantage against such imperfections. Since the measurement strategy consists of dichotomic outcomes, in close analogy with the standard HOM scenario, one can introduce a notion of visibility quantifying the quality of the interference. This enables us to adapt the analysis developed in Sec.~\ref{subsec: visibility impact on precision}, \ref{subsec: general visi model}, and \ref{subsec: Visi experimental verification} to the present, more general setting. Under reasonable assumptions on how the visibility enters the measurement statistics, one expects the precision to scale quadratically with the visibility, so that a high but imperfect visibility still preserves a significant fraction of the metrological gain. Nevertheless, a complete quantitative analysis of the interplay between imperfect symmetry, visibility, and QFI remains to be carried out, and is left for future work.

\paragraph{\publi Permutations beyond cycles}
Throughout this section, we have focused on the operator $\hat P$ associated with a cyclic permutation of the modes. While this choice is natural and leads to a particularly transparent structure, it is legitimate to ask whether other permutations could serve as a starting point. Let $\sigma \in \mathfrak S_n$ be a generic permutation of the set $\{0,\dots,n-1\}$, and define the passive linear transformation operator $\hat{P}_\sigma$ through
\begin{equation}
    \hat{P}_\sigma \, \hat{a}_j^\dagger(\lambda) \, \hat{P}_\sigma^\dagger 
    = \hat{a}_{\sigma(j)}^\dagger(\lambda).
\end{equation}
We denote by $P_\sigma$ the associated matrix representation, which is a permutation matrix. A standard result in the theory of permutations states that any $\sigma$ can be decomposed into a product of disjoint cycles,
\begin{equation}
    \sigma = c_1 \circ c_2 \circ \cdots \circ c_k,
\end{equation}
where each $c_j$ is a cycle, \ie, a permutation of the form $\alpha_1 \mapsto \alpha_2 \mapsto \cdots \mapsto \alpha_\ell \mapsto \alpha_1$. Up to a relabeling of the spatial modes, the matrix $P_\sigma$ can therefore be brought into a block-diagonal form,
\begin{equation}
    P_\sigma = 
    \begin{pmatrix}
        \boxed{P_1} & 0 & \cdots & 0 \\
        0 & \boxed{P_2} & \cdots & 0 \\
        \vdots & \vdots & \ddots & \vdots \\
        0 & 0 & \cdots & \boxed{P_k}
    \end{pmatrix},
\end{equation}
where each block $P_j$ corresponds to a single cycle and has the same cyclic structure as the matrix $P$ considered before,
\begin{equation}
    P_j = 
    \begin{pmatrix} 
        0 & 1 & & \\
          & \ddots & \ddots & \\
          &        & \ddots & 1 \\
        1 &        &        & 0 
    \end{pmatrix}.
\end{equation}
Each block $P_j$ is diagonalized by a DFT matrix of the corresponding size. Consequently, $P_\sigma$ itself is diagonalized by the direct sum $U_1 \oplus \cdots \oplus U_k$ of smaller DFT matrices. In operator language, the interferometer adapted to the transformation $\hat{P}_\sigma$ is therefore composed of $k$ independent DFT interferometers acting on disjoint subsets of modes, as illustrated in Fig.~\ref{fig: many-fourier}. Such an interferometer does not mix different mode groups.

From a metrological perspective, this raises a subtle question. If the initial state $\ket{\psi}$ contains entanglement across different mode groups, an interferometer that acts independently on each group may fail to exploit these correlations. It is therefore not clear whether restricting ourselves to block-diagonal DFT architectures preserves all the potentially useful quantum resources for estimation. Clarifying this point may lead to alternative, possibly simpler, interferometric architectures tailored to specific permutation symmetries.

\begin{figure}[ht]
    \centering
    \footnotesize
    \scalebox{1.5}{\begin{tikzpicture}
	\begin{pgfonlayer}{nodelayer}
		\node [style=none] (0) at (0.75, 2) {$\hat U_1$};
		\node [style=none] (1) at (-0.25, 1) {};
		\node [style=none] (2) at (1.75, 1) {};
		\node [style=none] (3) at (-0.25, 3) {};
		\node [style=none] (4) at (1.75, 3) {};
		\node [style=none] (5) at (-0.25, 2.5) {};
		\node [style=none] (6) at (-0.25, 1.5) {};
		\node [style=none] (7) at (-0.25, -2) {};
		\node [style=none] (8) at (-0.25, -1) {};
		\node [style=none] (9) at (-1.75, 2.5) {};
		\node [style=none] (10) at (-1.75, 1.5) {};
		\node [style=none] (11) at (-1.75, -1) {};
		\node [style=none] (12) at (-1.75, -2) {};
		\node [style=none] (13) at (-1, 0.5) {$\vdots$};
		\node [style=none] (14) at (2.75, 2.5) {};
		\node [style=none] (15) at (2.75, 1.5) {};
		\node [style=none] (16) at (2.75, -2) {};
		\node [style=none] (17) at (2.75, -1) {};
		\node [style=none] (18) at (1.75, 2.5) {};
		\node [style=none] (19) at (1.75, 1.5) {};
		\node [style=none] (20) at (1.75, -1) {};
		\node [style=none] (21) at (1.75, -2) {};
		\node [style=none] (22) at (2.25, 0.5) {$\vdots$};
		\node [style=none] (23) at (3.5, 2.75) {};
		\node [style=none] (24) at (3.5, 2.25) {};
		\node [style=none] (25) at (4, 2.5) {};
		\node [style=none] (26) at (4.5, 2.5) {};
		\node [style=none] (27) at (5.25, 2.5) {$m_0$};
		\node [style=none] (28) at (3.5, 1.75) {};
		\node [style=none] (29) at (3.5, 1.25) {};
		\node [style=none] (30) at (4, 1.5) {};
		\node [style=none] (31) at (4.5, 1.5) {};
		\node [style=none] (32) at (5.25, 1.5) {$m_l$};
		\node [style=none] (33) at (5.5, -1) {$m_j$};
		\node [style=none] (34) at (5.5, -2) {$m_{n-1}$};
		\node [style=none] (35) at (-6.25, 0.25) {$\ket{\psi}$};
		\node [style=none] (36) at (-5, 0.25) {\scalebox{1}[6.7]{$\{$}};
		\node [style=none] (37) at (-2.75, 0.25) {$e^{-i\kappa\hat H}$};
		\node [style=none] (38) at (-3.5, -2.75) {};
		\node [style=none] (39) at (-2, -2.75) {};
		\node [style=none] (40) at (-3.5, 3.25) {};
		\node [style=none] (41) at (-2, 3.25) {};
		\node [style=none] (42) at (-3.75, 2.5) {};
		\node [style=none] (43) at (-3.75, 1.5) {};
		\node [style=none] (44) at (-3.75, -2) {};
		\node [style=none] (45) at (-3.75, -1) {};
		\node [style=none] (46) at (-4.75, 2.5) {};
		\node [style=none] (47) at (-4.75, 1.5) {};
		\node [style=none] (48) at (-4.75, -1) {};
		\node [style=none] (49) at (-4.75, -2) {};
		\node [style=none] (50) at (-4.25, 0.5) {$\vdots$};
		\node [style=none] (51) at (3.5, -0.75) {};
		\node [style=none] (52) at (3.5, -1.25) {};
		\node [style=none] (53) at (4, -1) {};
		\node [style=none] (54) at (4.5, -1) {};
		\node [style=none] (55) at (3.5, -1.75) {};
		\node [style=none] (56) at (3.5, -2.25) {};
		\node [style=none] (57) at (4, -2) {};
		\node [style=none] (58) at (4.5, -2) {};
		\node [style=none] (59) at (-3.75, 3) {};
		\node [style=none] (60) at (-1.75, 3) {};
		\node [style=none] (61) at (-1.75, -2.5) {};
		\node [style=none] (62) at (-3.75, -2.5) {};
		\node [style=none] (63) at (-4.5, -0.25) {};
		\node [style=none] (66) at (-4.25, 2.125) {\scalebox{0.6}{$\vdots$}};
		\node [style=none] (67) at (2.25, 2.125) {\scalebox{0.6}{$\vdots$}};
		\node [style=none] (68) at (-1, 2.125) {\scalebox{0.6}{$\vdots$}};
		\node [style=none] (69) at (0.75, -1.5) {$\hat U_k$};
		\node [style=none] (70) at (-0.25, -2.5) {};
		\node [style=none] (71) at (-0.25, -0.5) {};
		\node [style=none] (72) at (1.75, -0.5) {};
		\node [style=none] (73) at (-4.25, -1.375) {\scalebox{0.6}{$\vdots$}};
		\node [style=none] (74) at (2.25, -1.375) {\scalebox{0.6}{$\vdots$}};
		\node [style=none] (75) at (-1, -1.375) {\scalebox{0.6}{$\vdots$}};
		\node [style=none] (76) at (1.75, -2.5) {};
	\end{pgfonlayer}
	\begin{pgfonlayer}{edgelayer}
		\draw [style=Fill grey] (2.center)
			 to (1.center)
			 to (3.center)
			 to (4.center)
			 to cycle;
		\draw (9.center) to (5.center);
		\draw (6.center) to (10.center);
		\draw (11.center) to (8.center);
		\draw (7.center) to (12.center);
		\draw (18.center) to (14.center);
		\draw (15.center) to (19.center);
		\draw (20.center) to (17.center);
		\draw (16.center) to (21.center);
		\draw [style=Fill pink] (24.center)
			 to (23.center)
			 to [in=90, out=0, looseness=0.75] (25.center)
			 to [in=0, out=-90, looseness=0.75] cycle;
		\draw (25.center) to (26.center);
		\draw [style=Fill pink] (29.center)
			 to (28.center)
			 to [in=90, out=0, looseness=0.75] (30.center)
			 to [in=0, out=-90, looseness=0.75] cycle;
		\draw (30.center) to (31.center);
		\draw (60.center)
			 to [bend right=45, looseness=1.25] (41.center)
			 to (40.center)
			 to [bend right=45, looseness=1.25] (59.center)
			 to (62.center)
			 to [bend right=45] (38.center)
			 to (39.center)
			 to [bend right=45] (61.center)
			 to cycle;
		\draw (46.center) to (42.center);
		\draw (43.center) to (47.center);
		\draw (48.center) to (45.center);
		\draw (44.center) to (49.center);
		\draw [style=Fill pink] (52.center)
			 to (51.center)
			 to [in=90, out=0, looseness=0.75] (53.center)
			 to [in=0, out=-90, looseness=0.75] cycle;
		\draw (53.center) to (54.center);
		\draw [style=Fill pink] (56.center)
			 to (55.center)
			 to [in=90, out=0, looseness=0.75] (57.center)
			 to [in=0, out=-90, looseness=0.75] cycle;
		\draw (57.center) to (58.center);
		\draw [style=Fill grey] (76.center)
			 to [in=0, out=180] (70.center)
			 to (71.center)
			 to (72.center)
			 to cycle;
	\end{pgfonlayer}
\end{tikzpicture}}
    \caption[Interferometer adapted to a general permutation symmetry]{Interferometer adapted to a general permutation symmetry $\hat{P}_\sigma$. After decomposing $\sigma$ into disjoint cycles, the corresponding transformation is implemented by independent DFT interferometers acting on each cycle subspace. The overall device is therefore block-diagonal in the mode basis and does not mix different groups of modes. Adapted from Ref.~\cite{descamps_role_2026}. \copyright 2026 American Physical Society.}
    \label{fig: many-fourier}
\end{figure}
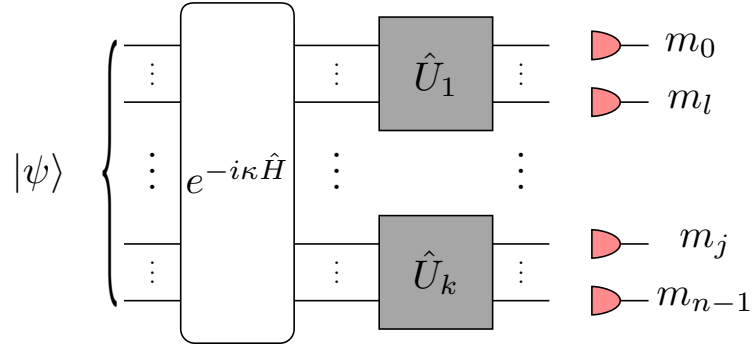

\subsection{Extension to mixed states}
\label{subsec: mixed states interferometry}
Throughout this whole chapter, for simplicity, we have only considered the pure-state case. However, as we now demonstrate, most of the results, from the interferometric ones to the metrological ones, can be adapted to mixed states. We thus consider a mixed state $\hat \rho$ with a pure-state decomposition
\begin{equation}
    \hat\rho=\sum_j p_j \ketbra{\psi_j},
\end{equation}
with $p_j>0$ and $\sum_j p_j=1$. This decomposition can be taken, for instance, as the spectral decomposition of the density matrix, in which case $\braket{\psi_j}{\psi_k}=\delta_{j,k}$, but the results presented below hold for any decomposition, not necessarily with orthogonal pure states.

\paragraph{\publi Symmetry of mixed states}

The natural extension of the symmetry measure $\bra{\psi}\hat P\ket{\psi}$ to mixed states is given by the expectation value
\begin{equation}
    \Tr(\hat \rho\hat P).
\end{equation}
Using the linearity of the trace, we have
\begin{equation}
    \Tr(\hat \rho\hat P)=\sum_j p_j \bra{\psi_j}\hat P\ket{\psi_j},
\end{equation}
so that $\Tr(\hat \rho\hat P)$ can be interpreted as the average symmetry of the pure components of $\hat \rho$, weighted by the probabilities $p_j$.

An important situation arises when $\Tr(\hat \rho \hat P)$ takes an extremal value, \ie, when $\abs{\Tr(\hat \rho\hat P)}=1$. In Appendix~\ref{app: hom interferometry}, Result~\ref{res: extremal symmetry}, we show that this condition implies a strong structural constraint: all pure states appearing in any decomposition of $\hat \rho$ must share the same symmetry, namely
\begin{equation}
    \hat P\ket{\psi_j}=e^{i\phi}\ket{\psi_j},
\end{equation}
with a phase $e^{i\phi}$ independent of $j$. In other words, the support of $\hat \rho$ is entirely contained in a single eigenspace of $\hat P$. As an immediate consequence, if $\Tr(\hat \rho\hat S)=1$ (resp. $-1$), then all pure states appearing in any decomposition of $\hat \rho$ are necessarily symmetric (resp. anti-symmetric).

\paragraph{\publi Interferometric probabilities for mixed states}

If the state $\hat \rho$ is injected into the DFT interferometer, the photon-number distribution in the output ports is a convex combination of the distributions obtained for each pure component. Exploiting linearity, we obtain
\begin{subequations}
    \label{eq: fourier proba mixed states}
    \begin{align}
        \frac{1}{n}\sum_{l=0}^{n-1}\Tr(\hat P^l\hat \rho)
        &=\sum_j p_j \frac{1}{n}\sum_{l=0}^{n-1}\expval{\hat P^l}_{\ket{\psi_j}}, \\
        &=\sum_j p_j\mathbb P_{\ket{\psi_j}}\left(\sum_{k=0}^{n-1} k m_k\equiv0\,[n]\right), \\
        &=\mathbb P_{\hat \rho}\left(\sum_{k=0}^{n-1} k m_k\equiv 0\,[n]\right),
    \end{align}
\end{subequations}

where the notation $\mathbb P_{\hat \sigma}[\cdots]$ denotes the probability of observing a given photon-number distribution at the output of the DFT interferometer when the state $\hat \sigma$ is injected. Equation~\eqref{eq: fourier proba mixed states} is directly analogous to Eq.~\eqref{eq: probability Fourier} for pure states, and shows that the connection between output photon-number statistics and symmetry properties of the input state extends straightforwardly to mixed states.

In the particular case $n=2$, the cyclic permutation operator reduces to the swap operator $\hat S$, and the previous expression becomes
\begin{equation}
    \mathbb P_{\hat \rho}(n_1\equiv 0 \,[2])=\frac{1}{2}\left(1+\Tr(\hat \rho\hat S)\right),
\end{equation}
which is the natural extension of Eq.~\eqref{eq: Peven HOM} to mixed states. Hence, the probability of detecting an even photon number in one output port of the HOM interferometer directly measures the symmetry of the mixed input state.

\paragraph{\publi Metrology}

We finally consider the metrological task of estimating a parameter $\theta$ encoded through the unitary evolution $e^{-i\hat H\theta}$, applied before the DFT interferometer, when the initial probe is a mixed state $\hat \rho$. For the measurement defined by the projectors associated with the symmetry sectors, the corresponding Fisher information reads
\begin{equation}
    \lim_{\theta\to 0}\mathcal F=4\sum_j p_j \Delta^2_{\ket{\psi_j}}\!\left(i[\hat H,\hat \Pi]\right), \qquad \Tr(\hat\rho\hat\Pi)\in\{0,1\},
\end{equation}
showing that the precision achieved by the protocol is the weighted average of the precisions associated with each pure component.

A more physically transparent expression can be obtained by imposing additional symmetry constraints on $\hat \rho$, in particular in the extremal cases $\Tr(\hat \rho\hat P)=\pm 1$. As discussed above, this implies that $\hat \rho$ is entirely supported in a single symmetry sector. In this situation, the Fisher information can be written as
\begin{align}
    \mathcal F&= \mathcal Q\left(\hat H-\frac{1}{n}\sum_{l=0}^{n-1} \hat P^l\hat H\hat P^{-l}\right), && \text{if } \Tr(\hat \rho\hat P)=1, \\
    \mathcal F&= \mathcal Q\left(\frac{1}{n}\sum_{l=0}^{n-1}(-1)^l \hat P^l\hat H\hat P^{-l}\right), && \text{if } \Tr(\hat \rho\hat P)=-1,
\end{align}
where $\mathcal Q(\hat H')$ denotes the QFI associated with the generator $\hat H'$ and the state $\hat \rho$. These expressions are direct generalizations of the pure-state formulas, with the variance replaced by the QFI.

The interpretation remains analogous to the pure-state case: the measurement implemented by the interferometer is sensitive to the symmetry properties of $\hat H$ with respect to $\hat P$, and the attainable precision depends on how strongly $\hat H$ drives the state out of the symmetry sector selected by $\hat \Pi$. Optimality is again achieved when $\hat H$ is fully anti-symmetric under cyclic permutation, in which case $\mathcal F=\mathcal Q$. The detailed derivation of these results is provided in Appendix~\ref{app: hom interferometry}, Result~\ref{res: metrology mixed state extension}.

\subsection{Conclusion on HOM interferometry and metrology}

Throughout this chapter, we have presented a detailed analysis of the HOM interferometer and its generalizations. Importantly, we have shown that adopting a symmetry-based perspective on the HOM interferometric effect not only clarifies its physical origin and interpretation but also provides a fertile framework for understanding its metrological applications. We have derived general precision bounds for a wide range of scenarios and investigated the impact of imperfect visibility, or equivalently imperfect symmetry of the probe state, on the achievable estimation precision.

Guided by this symmetry viewpoint, we have demonstrated that HOM effects and their metrological applications naturally extend to the multiphoton and multimode regimes. Beyond offering additional interferometric configurations for state analysis and enhanced sensing, this perspective further elucidates the connections between HOM interferometry and other areas of quantum interferometry, such as boson sampling.

A significant next step for developing the concepts presented in this chapter is the realization of explicit experimental demonstrations of the newly proposed effects. As an initial goal, one could envision demonstrating multiphoton interference effects, ultimately aiming to implement a comprehensive estimation procedure that leverages the enhanced precision of multiphoton probes. Although we have discussed several strategies to mitigate experimental noise and imperfections, a number of questions remain open: which multiphoton states can be practically prepared, what are their metrological capabilities, how should events with different photon numbers be combined when using probabilistic sources, and how can losses be effectively managed? Addressing these questions will require close collaboration between theory and experiment and will likely provide further insights into the fundamental physics of multiphoton interference and its applications in quantum metrology.

\fi

\ifnum \theShowChapfive=1
\chapter{Superselection rules}
\setcurrentanchor{chap5}
\label{chap: SSR}
\emph{In this chapter, we first describe the theoretical formalism and tools that are useful for the analysis of optical superselection rules (SSR). Beyond merely providing an equivalent picture of the usual continuous-variable formalism, we then demonstrate how the SSR framework offers an alternative point of view providing a deeper understanding of fundamental questions in quantum optics such as computational resources, universality, and metrology.}

\localtableofcontents

\section{Formalism of superselection rules}
\label{sec: SSR formalism}
\emph{In this section, we review and present the major mathematical tools useful for manipulating superselection-rules-compliant (SSRC) systems. Sticking to a simple but paradigmatic bimodal situation, and using the Schwinger representation, we demonstrate a strong link with the formalism commonly used in solid-state physics. Adapting these mathematical tools, we discuss the notion of mode transformation, rotations, spin coherent states, Möbius transforms, Majorana polynomials, spherical Wigner functions and spin squeezing. Most of the content presented in this section is well known in the literature, but we provide a self-contained presentation with a focus on the applications that will be developed in the next sections.}

\subsection{Formalism and Schwinger representation}
\label{subsec: SSRC and Schwinger}

\paragraph{\publi Motivations}

In the first chapter, Sec.~\ref{subsec: Phase Reference and SSR}, we argued that quantum optics, and especially the usual continuous-variable formalism, requires the implicit assumption of the existence of a phase reference. This assumption is often justified by the fact that, in practice, one can always engineer a suitable phase reference to perform the desired operations. Experimentally, this is reflected by the fact that either a single laser beam is used at the origin of all manipulated light fields, or phase-locking techniques are implemented to ensure the existence of a stable common relative phase reference. While this usually does not constitute a practical problem, since experiments are designed so as to guarantee the existence of a phase reference, the latter is not explicitly included in the theoretical formalism.

In this whole chapter we argue that the role of the phase reference goes beyond a mere technicality and that it is in fact a fundamental ingredient of the theory. In particular, consistency with the SSR requires that the phase reference be explicitly described within the theoretical framework, rather than implicitly assumed.

\paragraph{\publi Paradigmatic model}

To describe a single-mode quantum optical state while explicitly accounting for the phase reference, we consider a bimodal system with two orthogonal modes $a$ and $b$. The first mode $a$ corresponds to the physical mode of interest, while the second mode $b$ serves as the phase reference. We denote by $\hat a$ and $\hat b$ the annihilation operators associated with these modes, satisfying the canonical commutation relations $[\hat a,\hat a^\dagger]=1$ and $[\hat b,\hat b^\dagger]=1$, with all other commutators vanishing.

As the optical superselection rules impose conservation of the total photon number $N$, a general pure state in a fixed-$N$ sector can be expressed as\footnote{Throughout this section, we use indices on the ket notation to distinguish between representations. Here $\ket{k_1,k_2}_p$ denotes a state with $k_1$ photons in mode $a$ and $k_2$ photons in mode $b$, the index `$p$' standing for `photon'.}
\begin{equation}
    \ket{\psi}=\sum_{k=0}^N c_k\frac{(\hat a^\dagger)^k}{\sqrt{k!}}\frac{(\hat b^\dagger)^{N-k}}{\sqrt{(N-k)!}}\vac
    =\sum_{k=0}^N c_k \ket{k,N-k}_p,
\end{equation}
where $\ket{k,N-k}_p$ denotes the state with $k$ photons in mode $a$ and $N-k$ photons in mode $b$.

As will be described in more detail later in this chapter, in the limit where $N$ is large and where the average photon number in mode $b$ is much larger than that in mode $a$, the state $\ket{\psi}$ can be interpreted as an effective single-mode state with a well-defined phase reference. In this regime, it becomes equivalent to the single-mode state
\begin{equation}
    \ket{\psi}=\sum_{k=0}^\infty c_k \frac{(\hat a^\dagger)^k}{\sqrt{k!}}\vac.
\end{equation}

\paragraph{\lit Relation to spin systems}

The Schwinger representation~\cite{schwinger_angular_2015} provides a powerful tool that permits the use of angular momentum physics to analyse two optical modes. In this section, we recall the fundamental formulas and properties of this representation, as well as the relevant properties of spin systems. The Schwinger representation reveals that operators satisfying the canonical angular momentum commutation relations can be defined as\footnote{Note that there are multiple conventions for the definition of these operators. In particular, the sign of $\hat J_y$ and $\hat J_z$ could be reversed. The choice we make here is the most convenient for the following considerations.}
\begin{align}\label{eq: Schwinger representation}
    \hat J_x=\frac{1}{2}\left(\hat a^\dagger\hat b+\hat a\hat b^\dagger\right), &&
    \hat J_y=\frac{i}{2}\left(\hat a^\dagger\hat b-\hat a\hat b^\dagger\right), &&
    \hat J_z=\frac{1}{2}\left(\hat b^\dagger\hat b-\hat a^\dagger\hat a\right).
\end{align}
A direct computation, or the general argument provided in Appendix~\ref{app: transferring op to Fock}, shows that these operators indeed satisfy
\begin{align}
     [\hat J_x,\hat J_y]=i\hat J_z, &&
     [\hat J_y,\hat J_z]=i\hat J_x, &&
     [\hat J_z,\hat J_x]=i\hat J_y.
\end{align}
Additionally, one verifies that the total squared angular momentum operator admits the compact expression
\begin{equation}
    \hat J^2=\hat J_x^2+\hat J_y^2+\hat J_z^2
    =\frac{\hat N}{2}\left(\frac{\hat N}{2}+1\right),
\end{equation}
where $\hat N=\hat n_a+\hat n_b$ is the total photon number operator. For compactness, we group the three operators into a vector
\begin{equation}
    \vec{\hat J}=(\hat J_x,\hat J_y,\hat J_z),
\end{equation}
so that $\hat J^2=\norm*{\vec{\hat J}}^2$.

We now recall some fundamental properties of the angular momentum operators. It is common to analyse spin systems in the basis obtained by co-diagonalizing the commuting operators $\hat J^2$ and $\hat J_z$, with associated quantum numbers $j\in\frac{1}{2}\N$ and $m\in\{-j,-j+1,\dots,j\}$. One thus considers the states $\ket{j,m}_s$ such that
\begin{align}
    \hat J^2\ket{j,m}_s=j(j+1)\ket{j,m}_s, &&
    \hat J_z\ket{j,m}_s=m\ket{j,m}_s,
\end{align}
where the `$s$' index indicates that we are using the spin basis. Given the expression of $\hat J^2$, we obtain the relation $j=N/2$, where $N$ is the total photon number. Moreover, since $\hat J_z=(\hat n_b-\hat n_a)/2$, we have $m=(n_b-n_a)/2$, where $n_a$ and $n_b$ denote the photon numbers in each mode. We thus obtain the simple correspondence
\begin{align}
    \ket{j,m}_s=\ket{j-m,j+m}_p, &&
    \ket{n_a,n_b}_p=\ket{\tfrac{n_a+n_b}{2},\tfrac{n_b-n_a}{2}}_s.
\end{align}

We define the spin ladder operators as
\begin{align}
    \hat J_+=\hat J_x+i\hat J_y=\hat a \hat b^\dagger, &&
    \hat J_-=\hat J_x-i\hat J_y=\hat J_+^\dagger=\hat a^\dagger\hat b,
\end{align}
with inverse relations
\begin{align}
    \hat J_x=\frac{\hat J_++\hat J_-}{2}, &&
    \hat J_y=\frac{\hat J_+-\hat J_-}{2i}.
\end{align}
From spin theory, these ladder operators act as
\begin{equation}
    \hat J_\pm\ket{j,m}_s=\sqrt{j(j+1)-m(m\pm1)}\ket{j,m\pm 1}_s,
\end{equation}
which can also be verified directly in the photon number representation.

\paragraph{\lit Spins and rotations}

Spin systems and the corresponding angular momentum operators are naturally associated with rotations. Indeed, the angular momentum operators are generators of $SU_2(\C)$, which forms a double cover of the rotation group $SO(3)$. This link is made explicit by the relation
\begin{equation}\label{eq: rotation operator J}
    e^{-i\theta \vec{\hat J}\cdot\vec n}\vec{\hat J}e^{i\theta \vec{\hat J}\cdot\vec n}
    =\vec n(\vec n\cdot\vec{\hat J})
    +\cos(\theta)\left[\vec{\hat J}-\vec n(\vec n\cdot\vec{\hat J})\right]
    -\sin(\theta)\vec n\times \vec{\hat J},
\end{equation}
where $\vec a\cdot\vec b$ denotes the usual dot product, $\vec a\times\vec b$ the vector cross product, $\vec n\in\R^3$ is a unit vector and $\theta\in\R$. Taking the scalar product with a unit vector $\vec m$, we see that the angular momentum operator $\vec m\cdot\vec{\hat J}$ is transformed into $\vec m'\cdot\vec{\hat J}$ with\footnote{It is important to notice that $e^{-i\theta\vec n\cdot\vec{\hat J}}$ does not act in the same way on $\vec{\hat J}$ and on $\vec m$. From Eq.~\eqref{eq: rotation operator J}, we see that
\begin{equation}
    e^{-i\theta \vec n\cdot \vec{\hat J}}\vec{\hat J}e^{i\theta \vec n\cdot \vec{\hat J}}=R\vec{\hat J},
\end{equation}
where $R$ is a rotation matrix. Taking the scalar product, we obtain
\begin{equation}
    e^{-i\theta \vec n\cdot \vec{\hat J}}(\vec m\cdot \vec{\hat J})e^{i\theta \vec n\cdot \vec{\hat J}}
    =\vec m\cdot (R\vec{\hat J})
    =(R^T\vec m)\cdot \vec{\hat J}.
\end{equation}
Thus, the vector $\vec m$ is transformed by the transposed rotation $R^T$, corresponding to a rotation with opposite angle.}
\begin{equation}\label{eq: rotation vector}
    \vec m'=\vec n(\vec n\cdot\vec m)
    +\cos(\theta)\left[\vec m-\vec n(\vec n\cdot\vec m)\right]
    +\sin(\theta)\vec n\times \vec m,
\end{equation}
which is Rodrigues' rotation formula~\cite{rodrigues_lois_1840}. It describes the rotation of the vector $\vec m$ around the axis defined by $\vec n$ in the positive direction by an angle $\theta$. For completeness, especially regarding the convention with the sign in front of the generators in the exponential, we provide the verification of both formulas in Appendix~\ref{app: SSRC}, Result~\ref{res: angular momentum and rotation}.

Provided that $\vec n$ and $\vec m$ have unit norm, $\vec m'$ is also a unit vector, meaning that the rotation of an angular momentum operator in any direction remains an angular momentum operator. Rodrigues' formula admits the following geometric interpretation:
\begin{itemize}[label=$\bullet$]
    \item The term $\vec n(\vec n\cdot\vec m)$ is the projection of $\vec m$ onto the direction defined by $\vec n$. For a rotation around this axis, this component must remain invariant, which is indeed the case.
    \item The term $\vec m-\vec n(\vec n\cdot\vec m)$ is the component of $\vec m$ orthogonal to $\vec n$. Under a rotation around $\vec n$ by an angle $\theta$, this component is modulated by $\cos(\theta)$.
    \item The vector $\vec n\times\vec m$ is orthogonal to both $\vec n$ and $\vec m-\vec n(\vec n\cdot\vec m)$. For a rotation around $\vec n$, this contribution evolves with $\sin(\theta)$. Using the right-hand rule, one verifies that it implements a rotation in the positive direction around $\vec n$.
\end{itemize}

\paragraph{\lit Spin and mode transformation}

Acting on the creation operators $\hat a^\dagger$ and $\hat b^\dagger$ with the unitary operator $e^{-i\theta\vec n\cdot\vec{\hat J}}$, for $\theta\in\R$, yields
\begin{subequations}\label{eq: rotation creation op}
    \begin{align}
        e^{-i\theta\vec n\cdot \vec{\hat J}}\hat a^\dagger e^{i\theta \vec n\cdot \vec{\hat J}}
        &=\hat a'^\dagger
        =\left[\cos\left(\tfrac{\theta}{2}\right)+i n_z \sin\left(\tfrac{\theta}{2}\right)\right]\hat a^\dagger
        -i\sin\left(\tfrac{\theta}{2}\right)(n_x -i n_y)\hat b^\dagger,\\
        e^{-i\theta\vec n\cdot \vec{\hat J}}\hat b^\dagger e^{i\theta \vec n\cdot \vec{\hat J}}
        &=\hat b'^\dagger
        =\left[\cos\left(\tfrac{\theta}{2}\right)-i n_z \sin\left(\tfrac{\theta}{2}\right)\right]\hat b^\dagger
        -i\sin\left(\tfrac{\theta}{2}\right)(n_x +i n_y)\hat a^\dagger,
    \end{align}
\end{subequations}
where we have denoted $\vec n=(n_x,n_y,n_z)$ and assumed that $\vec n$ has unit norm.

From their definition as conjugation of $\hat a^\dagger$ and $\hat b^\dagger$ by a unitary transformation, one verifies that the operators $\hat a'^\dagger$ and $\hat b'^\dagger$ satisfy the canonical commutation relations and thus define legitimate creation operators. This transformation therefore implements a mode transformation. One can verify that, up to a global phase acting identically on both modes, any two-mode linear optical transformation can be expressed in this way. This establishes the correspondence between mode transformations and rotations in the spin picture.

Furthermore, in a multimode setting, any passive linear optical transformation can be decomposed into a sequence of two-mode rotations of the above form. The derivation of Eqs.~\eqref{eq: rotation creation op} is provided in Appendix~\ref{app: SSRC}, Result~\ref{res: roation creation op}.

\subsection{Stereographic projection and Riemann sphere}
\label{subsec: stereographic projection}

\paragraph{\lit The stereographic projection}

As we have seen in the previous section, the sphere and its associated rotations are fundamental to the study of spin and two-mode optical systems. A convenient way to describe points on the sphere is to use the stereographic projection, which maps points from the sphere to the complex plane. This construction provides a natural parametrization in terms of complex numbers of objects that we will consider later, such as the spin coherent states (Sec.~\ref{subsec: spin coherent states}) and Majorana polynomials (Sec.~\ref{subsec: majorana polynomial}).

The south-based stereographic projection of a point on the sphere is defined as the intersection between the equatorial plane and the straight line linking the south pole to the point on the sphere.\footnote{There are multiple conventions for defining the stereographic projection. The reference point can be chosen as the north pole instead of the south pole, and the projection plane can be shifted. All these constructions rely on the same geometric principle but lead to different analytical expressions. The present convention is the most convenient for our later developments. In the following, the term stereographic projection will implicitly refer to the south-based projection onto the equatorial plane.} 
Using the complex coordinate $z$ to parametrize the plane, this construction allows one to describe points on the sphere by complex numbers. The corresponding geometry is depicted in Fig.~\ref{fig: sterographic proj}.

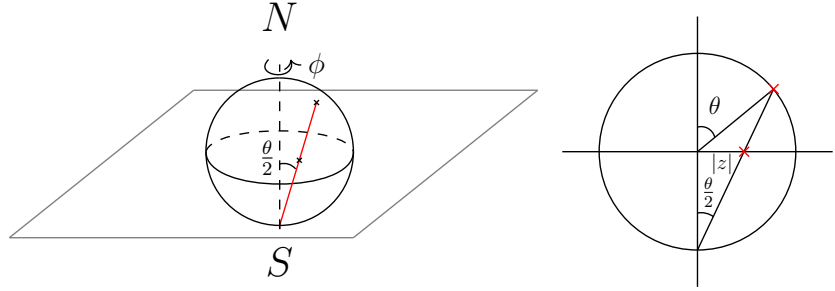
\begin{figure}[ht]
    \centering
    \scalebox{1.3}{\begin{tikzpicture}
	\begin{pgfonlayer}{nodelayer}
		\node [style=none] (0) at (-7.75, -1) {};
		\node [style=none] (1) at (-4, 2) {};
		\node [style=none] (2) at (-0.75, -1) {};
		\node [style=none] (3) at (3, 2) {};
		\node [style=none] (4) at (-3.75, 0.75) {};
		\node [style=none] (5) at (-0.75, 0.75) {};
		\node [style=none] (6) at (-2.25, 2.25) {};
		\node [style=none] (7) at (-2.25, -0.75) {};
		\node [style=none] (8) at (-1.5, 1.75) {};
		\node [style=none] (9) at (-1.45, 1.7) {};
		\node [style=none] (10) at (-1.55, 1.8) {};
		\node [style=none] (11) at (-1.55, 1.7) {};
		\node [style=none] (12) at (-1.45, 1.8) {};
		\node [style=none] (13) at (-1.8, 0.525) {};
		\node [style=none] (14) at (-1.9, 0.625) {};
		\node [style=none] (15) at (-1.9, 0.525) {};
		\node [style=none] (16) at (-1.8, 0.625) {};
		\node [style=none] (17) at (-2.25, -0.675) {};
		\node [style=none] (18) at (-2.25, -0.825) {};
		\node [style=none] (20) at (-2.25, 2.525) {};
		\node [style=none] (21) at (-2.25, 3.5) {$N$};
		\node [style=none] (22) at (-2.25, -1.5) {$S$};
		\node [style=none] (23) at (6.25, 3.5) {};
		\node [style=none] (24) at (6.25, -2) {};
		\node [style=none] (25) at (3.5, 0.75) {};
		\node [style=none] (26) at (9, 0.75) {};
		\node [style=none] (27) at (6.25, 2.75) {};
		\node [style=none] (28) at (8.25, 0.75) {};
		\node [style=none] (29) at (4.25, 0.75) {};
		\node [style=none] (30) at (6.25, -1.25) {};
		\node [style=none] (31) at (7.8, 2.025) {};
		\node [style=none] (32) at (6.25, 0.75) {};
		\node [style=none] (33) at (6.25, 1.25) {};
		\node [style=none] (34) at (6.6, 1.05) {};
		\node [style=none] (35) at (6.25, -0.5) {};
		\node [style=none] (36) at (6.575, -0.575) {};
		\node [style=none] (37) at (6.6, 1.7) {\scalebox{0.6}{$\theta$}};
		\node [style=none] (38) at (6.475, -0.075) {\scalebox{0.6}{$\tfrac{\theta}{2}$}};
		\node [style=none] (40) at (7.9, 1.925) {};
		\node [style=none] (41) at (7.7, 2.125) {};
		\node [style=none] (42) at (7.7, 1.925) {};
		\node [style=none] (43) at (7.9, 2.125) {};
		\node [style=none] (44) at (7.3, 0.65) {};
		\node [style=none] (45) at (7.1, 0.85) {};
		\node [style=none] (46) at (7.1, 0.65) {};
		\node [style=none] (47) at (7.3, 0.85) {};
		\node [style=none] (48) at (-2.25, 0.5) {};
		\node [style=none] (49) at (-1.925, 0.4) {};
		\node [style=none] (51) at (-2.5, 0.625) {\scalebox{0.7}{$\tfrac{\theta}{2}$}};
		\node [style=none] (52) at (-2.475, 2.625) {};
		\node [style=none] (53) at (-2.025, 2.625) {};
		\node [style=none] (55) at (-1.5, 2.5) {\scalebox{0.7}{$\phi$}};
		\node [style=none] (57) at (6.25, 3.5) {};
		\node [style=none] (58) at (6.75, 0.5) {\scalebox{0.5}{$\abs{z}$}};
	\end{pgfonlayer}
	\begin{pgfonlayer}{edgelayer}
		\draw [style=Gray line] (1.center) to (3.center);
		\draw [style=Gray line] (3.center) to (2.center);
		\draw [style=Gray line] (0.center) to (2.center);
		\draw [style=Gray line] (0.center) to (1.center);
		\draw [bend left=315] (6.center) to (4.center);
		\draw [bend left=45] (6.center) to (5.center);
		\draw [bend left=45] (5.center) to (7.center);
		\draw [bend left=45] (7.center) to (4.center);
		\draw [bend right=90, looseness=0.75] (4.center) to (5.center);
		\draw [style=dashes, bend left=75, looseness=0.50] (4.center) to (5.center);
		\draw [style=red line] (7.center) to (8.center);
		\draw (10.center) to (9.center);
		\draw (11.center) to (12.center);
		\draw (14.center) to (13.center);
		\draw (15.center) to (16.center);
		\draw (17.center) to (18.center);
		\draw (23.center) to (24.center);
		\draw (25.center) to (26.center);
		\draw [bend left=45] (29.center) to (27.center);
		\draw [bend right=45] (28.center) to (27.center);
		\draw [bend right=45] (29.center) to (30.center);
		\draw [bend right=45] (30.center) to (28.center);
		\draw (30.center) to (31.center);
		\draw (32.center) to (31.center);
		\draw [bend left=45] (33.center) to (34.center);
		\draw [bend left, looseness=0.75] (35.center) to (36.center);
		\draw [style=red line] (41.center) to (40.center);
		\draw [style=red line] (42.center) to (43.center);
		\draw [style=red line] (45.center) to (44.center);
		\draw [style=red line] (46.center) to (47.center);
		\draw [style=dashes] (18.center) to (20.center);
		\draw [bend left] (48.center) to (49.center);
		\draw [style=Arrow, in=-75, out=-135, looseness=2.75] (52.center) to (53.center);
	\end{pgfonlayer}
\end{tikzpicture}}
    \caption[Stereographic projection of the unit sphere onto the equatorial plane based on the south pole]{Stereographic projection of the unit sphere onto the equatorial plane based on the south pole. Left: three-dimensional view. The red line connects the south pole to the point on the sphere and intersects the plane at its image. Right: two-dimensional cross-section in a vertical plane containing the south pole and the point to be projected, illustrating the geometric construction.}
    \label{fig: sterographic proj}
\end{figure}

This geometric definition leads to simple analytical relations. Parametrizing the sphere in terms of the usual spherical coordinates $(\theta,\phi)$, the stereographic projection reads
\begin{equation}
    z = e^{i\phi}\tan\!\left(\tfrac{\theta}{2}\right).
\end{equation}
A derivation of this relation is provided in Appendix~\ref{app: SSRC}, Result~\ref{res: spherical stereographic}. In Cartesian coordinates, the stereographic projection and its inverse can be written as
\begin{align}
    \Phi : (x,y,z) \mapsto \frac{x+iy}{1+z}, 
    && 
    \Psi : z \mapsto \frac{1}{1+\abs{z}^2}\left(2\Re z, 2\Im z, 1-\abs{z}^2\right),
\end{align}
where $\Phi$ denotes the projection from the sphere to the plane and $\Psi$ the inverse map from the plane to the sphere. A derivation in arbitrary dimension is given in Appendix~\ref{app: SSRC}, Result~\ref{res: cartesian stereographic}, which can be straightforwardly specialized to the two-dimensional case considered here using complex coordinates.

In either parametrization, the south pole has no image in the complex plane, as expected from the geometric construction. The stereographic projection therefore defines a non-global chart of the sphere, with a single singularity at the south pole. By adjoining a single point at infinity to the complex plane, one obtains a global parametrization of the sphere, known as the Riemann sphere. In this extended picture, the south pole corresponds to the point at infinity.

\paragraph{\lit Simple change of coordinates}

In Tab.~\ref{tab: sphere vs plane transforms}, we list several simple transformations of the sphere and their counterparts in the complex plane under the stereographic projection. In particular, one immediately deduces that antipodal points $z$ and $w$ satisfy the relation $w = -1/z^*$. The derivation of the results presented in this table is shown in Appendix~\ref{app: SSRC}, Result~\ref{res: reflections and stereographic}.

\begin{table}[ht]
    \setlength{\tabcolsep}{6pt}
    \renewcommand{\arraystretch}{1.8}
    \centering
    \begin{tabular}{|c|c|}
        \hline
        Transformation of the sphere & Transformation of the complex plane \\
        \hline\hline
        $(x,y,z)\leftrightarrow (-x,y,z)$ & $z\leftrightarrow -z^*$ \\
        \hline
        $(x,y,z)\leftrightarrow (x,-y,z)$ & $z\leftrightarrow z^*$ \\
        \hline
        $(x,y,z)\leftrightarrow (x,y,-z)$ & $z\leftrightarrow \frac{1}{z^*}$ \\
        \hline\hline
        $(x,y,z)\leftrightarrow (-x,-y,z)$ & $z\leftrightarrow -z$ \\
        \hline
        $(x,y,z)\leftrightarrow (x,-y,-z)$ & $z\leftrightarrow \frac{1}{z}$ \\
        \hline
        $(x,y,z)\leftrightarrow (-x,y,-z)$ & $z\leftrightarrow -\frac{1}{z}$ \\
        \hline\hline
        $(x,y,z)\leftrightarrow (-x,-y,-z)$ & $z\leftrightarrow -\frac{1}{z^*}$ \\
        \hline        
    \end{tabular}
    \caption[Examples of reflections of the sphere and the corresponding transformations of the plane under the stereographic projection]{Examples of simple reflections of the sphere and the corresponding transformations in the complex plane under the south-based stereographic projection.}
    \label{tab: sphere vs plane transforms}
\end{table}

One may also consider the stereographic projection based on the north pole. Repeating the same derivation, one obtains in arbitrary dimension
\begin{align}
    \Phi_N : (x,y,z) \mapsto \frac{x+iy}{1-z}, 
    && 
    \Psi_N : z \mapsto \frac{1}{1+\abs{z}^2}\left(2\Re z, 2\Im z, \abs{z}^2 - 1\right).
\end{align}
Comparing these expressions with those obtained for the south-based projection shows that the two constructions differ by the inversion $(x,y,z)\leftrightarrow (x,y,-z)$ in three-dimensional space. In the complex plane, this corresponds to the transformation $z\mapsto 1/z^*$, as already indicated in Tab.~\ref{tab: sphere vs plane transforms}.

\paragraph{\lit Möbius transformations}

As emphasized above, rotations play a central role in the geometry of angular momentum operators. Once the stereographic projection has been introduced, it is natural to ask how rotations of the sphere are expressed in terms of complex coordinates. The appropriate framework is provided by Möbius transformations.\footnote{In a more general setting, Möbius transformations are maps of $\R^n\cup\{\infty\}$ obtained by finite compositions of inversions with respect to lines and circles. In two dimensions, they can be written as
\begin{equation}
    M : z \mapsto \frac{az+b}{cz+d},
\end{equation}
with $ad-bc\neq 0$. Here we focus on the subclass corresponding specifically to rotations of the sphere.} As shown in Appendix~\ref{app: SSRC}, Result~\ref{res: möbius and rotation}, rotations of the sphere are represented, in the complex plane, by maps of the form
\begin{equation}
    M : z \mapsto \frac{az-c^*}{cz+a^*},
\end{equation}
where $a,c\in\C$ satisfy $\abs{a}^2+\abs{c}^2=1$. Since the Riemann sphere includes the point at infinity, this formula is understood with the following conventions:
\begin{itemize}
    \item If $c\neq 0$, then $M(-a^*/c)=\infty$ and $M(\infty)=a/c$.
    \item If $c=0$, in which case $a\neq 0$, then $M(\infty)=\infty$.
\end{itemize}

We denote by $\mathcal M_R$ the set of Möbius transformations associated with rotations, where the index $R$ stands for rotation. For simplicity, in the following, the term Möbius transformation will implicitly refer to this subclass. The transformation $M : z \mapsto \frac{az-c^*}{cz+a^*}$ is naturally associated with the unitary matrix
\begin{equation}
    \begin{pmatrix}
        a & -c^* \\
        c & a^*
    \end{pmatrix}.
\end{equation}
In Appendix~\ref{app: SSRC}, Result~\ref{res: möbius and SU2}, we show that this correspondence defines a group homomorphism from the group of $2\times 2$ unitary matrices to the group $\mathcal M_R$, with kernel $\{\pm \1\}$. Consequently, the composition of two Möbius transformations corresponds to the matrix product of the associated unitary matrices. The fact that the kernel is nontrivial reveals that there is a degeneracy in the representation of Möbius transformations as unitary matrices, as the two matrices
\begin{align} 
    \begin{pmatrix} 
        a & -c^*\\ 
        c & a^* 
    \end{pmatrix}, && \begin{pmatrix} 
        -a & c^*\\
         -c & -a^* 
        \end{pmatrix}
    \end{align}
define the same Möbius transformation.

The relation between three-dimensional rotations and Möbius transformations can be made explicit as follows. Let $\vec n=(n_x,n_y,n_z)$ be a unit vector and $\theta$ an angle. The rotation of the sphere around the axis $\vec n$ by an angle $\theta$ is represented by the Möbius transformation $z\mapsto \frac{az-c^*}{cz+a^*}$ with coefficients
\begin{align}\label{eq: Möbius coefficients from rotations}
    a = \cos\!\left(\frac{\theta}{2}\right) + i n_z \sin\!\left(\frac{\theta}{2}\right),
    &&
    c = -i(n_x - i n_y)\sin\!\left(\frac{\theta}{2}\right).
\end{align}
It is important to stress that the explicit expressions of $a$ and $c$ depend on the chosen convention for the stereographic projection. For instance, adopting the north-based projection would lead to different coefficients. As a consequence, our expressions may differ from those found in the literature, such as in~\cite{vourdas_analytic_2006,john_olsen_geometry_2010,pennestri_dual_2016}. The derivation is provided in Appendix~\ref{app: SSRC}, Result~\ref{res: möbius and rotation}. It relies on the explicit computation of the composition
\begin{equation}
    z \mapsto \Psi(z)=\vec m \mapsto \vec m' = R_{\vec n}(\theta)\vec m \mapsto z'=\Phi(\vec m'),
\end{equation}
where $R_{\vec n}(\theta)$ denotes the rotation of the sphere about the axis $\vec n$ with angle $\theta$, and $\Phi$ and $\Psi$ are the stereographic projection and its inverse. Although conceptually straightforward, this computation is however an algebraic test of faith, and it is not {\it a priori} evident that the resulting expression can be reduced to such a compact form.

\paragraph{\lit Example of Möbius rotations}

For practical use, we list a few explicit examples obtained from the general formula. For a rotation around the $z$ axis, one has $a=e^{i\theta/2}$ and $c=0$, hence
\begin{equation}
    f^\theta_z(w)=\frac{e^{i\theta/2} w}{e^{-i\theta/2}}=e^{i\theta} w,
\end{equation}
with the special case $f_z^{\pi/2}(w)=iw$.

For a rotation around the $x$ axis, one finds $a=\cos(\frac{\theta}{2})$ and $c=-i\sin(\frac{\theta}{2})$, leading to
\begin{equation}
    f_x^\theta(w)=\frac{\cos(\frac{\theta}{2}) w - i \sin(\frac{\theta}{2})}{-i \sin(\frac{\theta}{2}) w + \cos(\frac{\theta}{2})},
\end{equation}
with the special case $f_x^{\pi/2}(w)=\dfrac{w - i}{-i w + 1}$.

Finally, for a rotation around the $y$ axis, one has $a=\cos(\frac{\theta}{2})$ and $c=-\sin(\frac{\theta}{2})$, so that
\begin{equation}
    f_y^\theta(w)=\frac{\cos(\frac{\theta}{2}) w + \sin(\frac{\theta}{2})}{-\sin(\frac{\theta}{2}) w + \cos(\frac{\theta}{2})},
\end{equation}
with the special case $f_y^{\pi/2}(w)=\dfrac{w + 1}{- w + 1}$.

\subsection{Spin coherent states}
\label{subsec: spin coherent states}

\paragraph{\lit Definitions}

Analogous to the coherent states of continuous-variable quantum optics, spin coherent states aim at capturing the notion of most classical states for a finite-dimensional spin system~\cite{radcliffe_properties_1971,arecchi_atomic_1972}. In our superselection-rules-compliant formalism, the total photon number $N$ is fixed. The Hilbert space therefore coincides with the spin-$j$ representation with $j=N/2$, and spin coherent states naturally arise as the most classical pure states of this system. Recall that, throughout this section, we restrict ourselves to two orthogonal modes, described by the creation operators $\hat a^\dagger$ and $\hat b^\dagger$. The following definitions are equivalent characterizations of spin coherent states.

\begin{enumerate}
    \item In the first-quantization picture, optical states with fixed total photon number are constructed as symmetrized states of $N$ identical photons, as done in the Fock space construction presented in Sec.~\ref{subsec: FockSpace}. In this framework, spin coherent states correspond precisely to particle-separable states, that is, states in which all individual photons occupy the same single-photon state.

    More concretely, for a normalized single-photon state $\ket{\varphi}=\alpha \ket{0}+\beta\ket{1}$, the associated spin coherent state is
    \begin{equation}
        \ket{\psi}=\ket{\varphi}\otimes\cdots\otimes\ket{\varphi}.
    \end{equation}
    \label{def: spin particle sep}

    \item A spin coherent state is any state that can, up to a global phase, be written in the form\footnote{By convention, the value $z=\infty$ corresponds to the state $\ket{N,0}_p$.}
    \begin{subequations}
        \begin{align}
            \ket{\psi}
            &=\frac{1}{(1+\abs{z}^2)^{N/2}}e^{z\hat a^\dagger \hat b}\ket{0,N}_p \\
            &=\frac{1}{(1+\abs{z}^2)^{N/2}}\sum_{k=0}^N z^k\sqrt{\binom{N}{k}}\ket{k,N-k}_p,
        \end{align}
    \end{subequations}
    for some complex number $z$. We then call $z$ the complex amplitude of the spin coherent state.
    \label{def: spin complex}

    \item A state $\ket{\psi}$ is a spin coherent state if and only if there exists a unique mode
    \begin{equation}
        \hat c^\dagger=\alpha \hat a^\dagger +\beta \hat b^\dagger,
    \end{equation}
    such that $\ket{\psi}$ is an $N$-photon Fock state in this mode, namely
    \begin{equation}
        \ket{\psi}=\frac{(\hat c^\dagger)^N}{\sqrt{N!}}\vac
        =\sum_{k=0}^N \sqrt{\binom{N}{k}}\alpha^k \beta^{N-k}\ket{k,N-k}_p.
    \end{equation}
    \label{def: spin monomode}

    \item The state $\ket{\psi}$ is a spin coherent state if and only if there exist spherical coordinates $(\theta,\phi)$, necessarily unique, such that, up to a global phase,
    \begin{equation}
        \ket{\psi}=\hat R(\theta,\phi)\ket{0,N}_p,
    \end{equation}
    where
    \begin{equation}
        \hat R(\theta,\phi)=\exp\big(-i\theta\big(-\sin(\phi)\hat J_x+\cos(\phi)\hat J_y\big)\big)
    \end{equation}
    is the rotation operator that maps the $z$ axis onto the direction of spherical coordinates $(\theta,\phi)$.
    \label{def: spin rotation}

    \item Finally, $\ket{\psi}$ is a spin coherent state if and only if there exists a unique unit vector $\vec m$ such that $\ket{\psi}$ is an eigenvector of $\vec m\cdot\vec{\hat J}$ with maximal eigenvalue $N/2$. Up to a global phase, $\ket{\psi}$ is the only such eigenvector.
    \label{def: spin operator} 

\end{enumerate}

The equivalence between these definitions is established in Appendix~\ref{app: SSRC}, Result~\ref{res: def spin coherent states}. Each definition provides a complementary perspective. Definition~\ref{def: spin particle sep} emphasizes the absence of particle entanglement, while Definition~\ref{def: spin monomode} highlights that these states are simple $N$-photon Fock states in a single mode\footnote{In the SSRC framework, the total photon number is fixed. A Fock state containing all photons in a single mode therefore represents the simplest possible photon distribution.}, which justifies their interpretation as the most classical states of the system. Definitions~\ref{def: spin rotation} and~\ref{def: spin operator} relate spin coherent states to the geometry of the Bloch sphere. Finally, Definition~\ref{def: spin complex} provides a convenient parametrization in terms of a single complex number $z$, which corresponds to the stereographic projection of the point $\vec m$ on the sphere.

For a spin coherent state, the different parametrizations are related by
\begin{align}
    &\begin{array}{rl}
        z\!\!\!\!&=\Phi(\vec n),\\
        &=e^{i\phi}\tan(\theta/2),
    \end{array}
    &&
    \begin{cases}
        \alpha=\dfrac{z}{\sqrt{1+\abs{z}^2}},\\[0.4em]
        \beta=\dfrac{1}{\sqrt{1+\abs{z}^2}},
    \end{cases} \\
    &
    \begin{cases}
        \alpha=e^{i\phi}\sin(\theta/2),\\[0.4em]
        \beta=\cos(\theta/2),
    \end{cases}
    &&
    \begin{array}{rl}
        \vec n\!\!\!\!&=\Psi(z),\\
        &=(\sin\theta \cos\phi,\sin\theta \sin\phi,\cos\theta).
    \end{array}
\end{align}

We use the notations
\begin{equation}
    \ket{z}_{sc}=\ket{\hat c}_{sc}=\ket{\theta,\phi}_{sc}=\ket{\vec m}_{sc},
\end{equation}
to denote a spin coherent state in its various parametrizations. There is no ambiguity in this notation, since the nature of the parameter (complex number, mode operator, pair of angles, or three-dimensional unit vector) uniquely determines the intended representation.

Definition~\ref{def: spin operator} implies the following geometrical property. For any unit vector $\vec n$
\begin{equation}\label{eq: rotation coherent state with propto}
    e^{-i\theta \vec{\hat J}\cdot \vec n}\ket{\vec m}_{sc}\propto\ket{R_{\vec n}(\theta)\vec m}_{sc},
\end{equation}
that is, the unitary generated by $\vec{\hat J}\cdot\vec n$ rotates the coherent state around the axis $\vec n$ by an angle $\theta$, up to a global phase. A detailed proof is provided in Appendix~\ref{app: SSRC}, Result~\ref{res: spin coherent rotation}.

\paragraph{\lit Mathematical properties}

The scalar product between two spin coherent states in the complex parametrization reads
\begin{equation}
    \prescript{}{sc}{\langle} z' {| z \rangle}_{sc}=\frac{(1+z'^* z)^N}{(1+\abs{z'}^2)^{N/2}(1+\abs{z}^2)^{N/2}},
\end{equation}
for any complex numbers $z$ and $z'$. For $z=z'$, we recover $\prescript{}{sc}{\langle} z {| z \rangle}_{sc}=1$, as required by normalization. Moreover, two spin coherent states are orthogonal if and only if $z'=-1/z^*$, which corresponds to antipodal points on the sphere. This contrasts with coherent states of continuous-variable quantum optics, which are never exactly orthogonal. Using the associated unit vectors $\vec n$ and $\vec n'$, the squared modulus of the overlap takes the simple form
\begin{equation}
    \abs\big{\prescript{}{sc}{\langle} \vec n' {| \vec n \rangle}_{sc}}^2
    =\left(\frac{1+\vec n\cdot\vec n'}{2}\right)^N.
\end{equation}
The overlap thus depends only on the angle between the two directions. The explicit dependence on $N$ shows that, for large $N$, two coherent states become nearly orthogonal as soon as the angle between their associated vectors is not too small. This increasing distinguishability for large $N$ is consistent with their interpretation as the most classical states of the system. The derivation is given in Appendix~\ref{app: SSRC}, Result~\ref{res: scalar product spin coherent}.

Spin coherent states satisfy the completeness relation~\cite{vourdas_analytic_2006}
\begin{equation}\label{eq: closure relation spin coherent states}
    \frac{N+1}{\pi}\int_{\C} \frac{\dd^2 z}{(1+\abs{z}^2)^2}{| z \rangle}_{sc}\prescript{}{sc}{\langle z |}=\1, 
\end{equation}
which is the analogue of the closure relation for continuous-variable coherent states. The proof is provided in Appendix~\ref{app: SSRC}, Result~\ref{res: closure relation spin coherent}.

\paragraph{\lit Spin coherent states and Möbius transformations}

The different equivalent definitions of spin coherent states naturally connect them to rotations. Since rotations act as Möbius transformations in the complex plane, it is instructive to describe explicitly how spin coherent states transform under such maps. Up to a global phase, Eq.~\eqref{eq: rotation coherent state with propto} and the link between Möbius transformations and rotations directly give
\begin{equation}
    e^{-i\theta\vec n\cdot\vec{\hat J}}\ket{z}_{sc}\propto\ket{f(z)}_{sc},
\end{equation}
where $f$ is the Möbius transformation associated with the rotation $R_{\vec n}(\theta)$. A more detailed computation yields the explicit formula
\begin{equation}\label{eq: rotation coherent state with z}
    e^{-i\theta\vec n\cdot\vec{\hat J}}\ket{z}_{sc}
    =(cz+a^*)^N\left(\frac{1+\abs{f(z)}^2}{1+\abs{z}^2}\right)^{N/2}\ket{f(z)}_{sc}
    =\left(\frac{cz+a^*}{\abs{cz+a^*}}\right)^N\ket{f(z)}_{sc},
\end{equation}
where the second equality makes the global phase explicit. The derivation is given in Appendix~\ref{app: SSRC}, Result~\ref{res: rotation coherent state}.

Beyond its computational relevance, this formula has an additional conceptual advantage. Indeed, as shown in Result~\ref{res: rotation coherent state}, the derivation does not rely on the explicit correspondence between spatial rotations and Möbius transformations given in Eq.~\eqref{eq: Möbius coefficients from rotations}. In other words, the construction of rotation coherent states can be carried out independently of the detailed coefficient identification appearing in Eq.~\eqref{eq: Möbius coefficients from rotations}. This observation has an important consequence. Since the argument developed in Result~\ref{res: rotation coherent state} does not make use of the explicit relation between rotations and Möbius maps, the result can in fact be turned around: starting from the properties of rotation coherent states alone, one can recover the explicit link between rotations and Möbius transformations. In this way, Eq.~\eqref{eq: Möbius coefficients from rotations} can be rederived with significantly fewer algebraic manipulations and with substantially reduced computational complexity. 

\paragraph{\lit SSRC Cat-like states}

Analogues of continuous-variable cat states can be constructed as superpositions of spin coherent states. We define
\begin{equation}
    \ket{z_1,z_2}_{cat}^\pm
    =\frac{1}{\mathcal N^\pm}\left(\ket{z_1}_{sc}\pm\ket{z_2}_{sc}\right),
\end{equation}
with normalization factor
\begin{equation}
    \mathcal N_\pm
    =\sqrt{2\pm\frac{2\Re\left[(1+z_1^* z_2)^N\right]}{(1+\abs{z_1}^2)^{N/2}(1+\abs{z_2}^2)^{N/2}}}.
\end{equation}

We consider two particularly relevant cases.

\begin{itemize}

    \item If $z_1=-z_2=z$, we obtain a cat state symmetric around $z=0$
    \begin{equation}
        \ket{z,-z}_{cat}^\pm
        =\frac{1}{\sqrt{2\pm\frac{2(1-\abs{z}^2)^N}{(1+\abs{z}^2)^N}}}
        \left(\ket{z}_{sc}\pm\ket{-z}_{sc}\right).
    \end{equation}
    The quantity $\frac{1-\abs{z}^2}{1+\abs{z}^2}$ coincides with the $z$ coordinate of the corresponding point on the sphere. When $\abs{z}=1$, the two coherent states correspond to antipodal points on the equator and are orthogonal. In this situation, $\mathcal N_\pm=\sqrt{2}$. Any cat state of two coherent states can be brought to this form by an appropriate rotation, so that, without loss of generality, $z$ may be chosen real.

    \item If $z_1=-\frac{1}{z_2^*}=z$, the two coherent states correspond to antipodal points on the sphere. In this case,
    \begin{equation}
        \ket{z,-\tfrac{1}{z^*}}_{cat}^\pm
        =\frac{1}{\sqrt{2}}
        \left(\ket{z}_{sc}\pm\ket{-\tfrac{1}{z^*}}_{sc}\right),
    \end{equation}
    since the two components are exactly orthogonal.
\end{itemize}

\subsection{Majorana polynomial}
\label{subsec: majorana polynomial}

In quantum physics, alternative mathematical objects associated with a quantum state often provide complementary points of view. In continuous-variable systems, this is for instance the case of the Wigner function introduced in Sec.~\ref{subsec: continuous variables}, which encodes all the information of a quantum state in a phase-space representation. Other objects, such as the Bargmann function~\cite{bargmann_hilbert_1961}, offer an analytic representation of continuous-variable quantum states. In the context of spin systems, the Majorana polynomial~\cite{majorana_atomi_1932,serrano-ensastiga_majorana_2020} is a complex polynomial associated with any spin $j$ quantum state, which contains its full information content. As we will see below, its structure as a polynomial provides a powerful tool to analyse and interpret the properties of spin $j$ systems within our SSRC framework.

\paragraph{\lit Definition}
Adapting the original definition to our SSRC setting, for any state $\ket{\psi}$ with $N$ total photons we define the associated Majorana polynomial evaluated at $z$ using the scalar product with the spin coherent state $\ket{z^*}_{sc}$. Explicitly,
\begin{equation}
    P_{\ket{\psi}}(z) = (1+\abs{z}^2)^{N/2}\prescript{}{sc}{\bra{z^*}\ket{\psi}}, 
\end{equation}
where the normalisation factor ensures that $P_{\ket{\psi}}(z)$ is a polynomial function of $z$ of degree at most $N$. Expanding the spin coherent state in the photon number basis, we obtain
\begin{equation}
    P_{\ket{\psi}}(z) = \sum_{k=0}^N c_k z^k \sqrt{\binom{N}{k}}.
\end{equation}
It is crucial to take the scalar product with $\ket{z^*}_{sc}$ rather than with $\ket{z}_{sc}$. Indeed, $\ket{z^*}_{sc}$ depends on $z^*$, and upon taking the Hermitian conjugate to form the corresponding bra, we obtain an expression that depends analytically on $z$. This guarantees that $P_{\ket{\psi}}(z)$ is a polynomial in the complex variable $z$. Since the binomial factor $\sqrt{\binom{N}{k}}$ is non-zero for all $0 \leq k \leq N$, the coefficients $c_k$ can be uniquely recovered from the polynomial. Therefore, $\ket{\psi}$ and $P_{\ket{\psi}}$ are in one-to-one correspondence.

It is instructive to compare this definition with that of the Bargmann function for a continuous-variable state
\begin{equation}
    \ket{\psi} = \sum_{n=0}^\infty c_n \ket{n},
\end{equation}
for which one defines
\begin{equation}
    F_B(z) = e^{\abs{z}^2/2}\bra{z^*}\ket{\psi} 
    = \sum_{n=0}^\infty \frac{z^n c_n}{\sqrt{n!}},
\end{equation}
where $\bra{z^*}$ denotes the coherent state of amplitude $z^*$. Both constructions associate to a quantum state an analytic function of $z$ from which the expansion coefficients can be recovered. 

\paragraph{\lit Properties}
We have seen that the Majorana polynomial contains the full information about the quantum state, since the expansion coefficients of the state can be reconstructed from the polynomial. A more explicit reconstruction formula reads
\begin{equation}
    \label{eq: reconstructing state from Majo}
    \ket{\psi} = P_{\ket{\psi}}(\hat a^\dagger {\hat b^\dagger}^{-1})\ket{0,N}_p,
\end{equation}
where ${\hat b^\dagger}^{-1}$ denotes the left inverse of $\hat b^\dagger$, satisfying ${\hat b^\dagger}^{-1}\hat b^\dagger = \1$, while $\hat b^\dagger {\hat b^\dagger}^{-1} \neq \1$. It is formally defined on mode $b$ by
\begin{equation}
    {\hat b^\dagger}^{-1} :
    \begin{cases}
        \ket{n} \mapsto \dfrac{1}{\sqrt{n}} \ket{n-1}, \\
        \ket{0} \mapsto 0.
    \end{cases}
\end{equation}
The verification of Eq.~\eqref{eq: reconstructing state from Majo} is provided in Appendix~\ref{app: SSRC}, Result~\ref{res: reconstructing state from Majo}.

The scalar product of two states $\ket{\psi_1}$ and $\ket{\psi_2}$ can also be expressed directly in terms of their respective Majorana polynomials $P_{\ket{\psi_1}}$ and $P_{\ket{\psi_2}}$ as
\begin{equation}
    \bra{\psi_1}\ket{\psi_2}
    = \frac{N+1}{\pi}
    \int_{\C} \frac{\dd^2 z}{(1+\abs{z}^2)^{N+2}}
    P_{\ket{\psi_1}}(z)^*
    P_{\ket{\psi_2}}(z).
\end{equation}
The derivation of this formula is given in Appendix~\ref{app: SSRC}, Result~\ref{res: scalar product from majo}. This expression highlights that the Hilbert space structure can be entirely reformulated in terms of polynomials equipped with a suitable measure on the complex plane.

We now describe how the Majorana polynomial transforms under a rotation generated by $e^{-i\theta\vec n\cdot\vec{\hat J}}$. One finds
\begin{equation}
    \label{eq: rotation Majo poly}
    P_{e^{-i\theta\vec n\cdot\vec{\hat J}}\ket{\psi}}(z)
    = (a^* - c^* z)^N 
    P_{\ket{\psi}}\!\left(\frac{a z + c}{-c^* z + a^*}\right),
\end{equation}
as shown in Appendix~\ref{app: SSRC}, Result~\ref{res: Majo and rotations}. By definition, the left-hand side is a polynomial in $z$, hence the right-hand side must also be polynomial. Although the composition
\begin{equation}
P_{\ket{\psi}}\!\left(\frac{a z + c}{-c^* z + a^*}\right)
\end{equation}
is not itself polynomial because of the denominator, multiplication by $(a^* - c^* z)^N$ cancels the denominator exactly. Since $P_{\ket{\psi}}$ has degree at most $N$, the final expression is indeed a polynomial of degree at most $N$.

As illustrative examples, we give the transformation rules for rotations around the principal axes.

\begin{itemize}
    \item Rotation around $z$. Recalling that $f_z^\theta(w) = w e^{i\theta}$, one obtains
    \begin{equation}
        P_{e^{-i\theta \hat J_z}\ket{\psi}}(z) = e^{-i N \theta/2} P_{\ket{\psi}}(z e^{i\theta}),
    \end{equation}
    and in particular, for a $90^\circ$ rotation with $\theta = \pi/2$,
    \begin{equation}
        P_{e^{-i\frac{\pi}{2} \hat J_z}\ket{\psi}}(z) = e^{-i N \pi/4} P_{\ket{\psi}}(i z).
    \end{equation}

    \item Rotation around $x$. Using 
    \begin{equation}
        f_x^\theta(w) = \frac{\cos(\frac{\theta}{2}) w - i \sin(\frac{\theta}{2})}
     {-i \sin(\frac{\theta}{2}) w + \cos(\frac{\theta}{2})},
    \end{equation}
    we find
    \begin{equation}
        P_{e^{-i\theta \hat J_x}\ket{\psi}}(z) = \left(\cos(\tfrac{\theta}{2}) - i \sin(\tfrac{\theta}{2}) z\right)^N P_{\ket{\psi}}\!\left( \frac{\cos(\tfrac{\theta}{2}) z - i \sin(\tfrac{\theta}{2})}{-i \sin(\tfrac{\theta}{2}) z + \cos(\tfrac{\theta}{2})}
        \right),
    \end{equation}
    and for $\theta = \pi/2$,
    \begin{equation}
        P_{e^{-i\frac{\pi}{2} \hat J_x}\ket{\psi}}(z) = \left(\frac{1 - i z}{\sqrt{2}}\right)^N P_{\ket{\psi}}\!\left(\frac{z - i}{-i z + 1}\right).
    \end{equation}

    \item Rotation around $y$. Using 
    \begin{equation}
    f_y^\theta(w) = \frac{\cos(\frac{\theta}{2}) w + \sin(\frac{\theta}{2})}{-\sin(\frac{\theta}{2}) w + \cos(\frac{\theta}{2})},
    \end{equation}
    we obtain
    \begin{equation}
        P_{e^{-i\theta \hat J_y}\ket{\psi}}(z) = \left(\cos(\tfrac{\theta}{2}) + \sin(\tfrac{\theta}{2}) z\right)^N P_{\ket{\psi}}\!\left( \frac{\cos(\tfrac{\theta}{2}) z - \sin(\tfrac{\theta}{2})}{\sin(\tfrac{\theta}{2}) z + \cos(\tfrac{\theta}{2})}
        \right),
    \end{equation}
    and for $\theta = \pi/2$,
    \begin{equation}
        P_{e^{-i\frac{\pi}{2} \hat J_y}\ket{\psi}}(z) = \left(\frac{1 + z}{\sqrt{2}}\right)^N P_{\ket{\psi}}\!\left(\frac{z - 1}{z + 1}\right).
    \end{equation}
\end{itemize}

\paragraph{\lit Majorana stars}
So far, we have used the Majorana polynomial as an alternative representation of quantum states, suitable for computations and transformations. We now exploit its analytic structure. By the fundamental theorem of algebra, any complex polynomial $P$ of degree $d$ admits $d$ complex roots, counted with multiplicity, and can be factorized as
\begin{equation}
    P(z) = c_d \prod_{k=1}^d (z - \alpha_k),
\end{equation}
where $c_d$ is the leading coefficient and the $\alpha_k$ are the roots. Applying this to $P_{\ket{\psi}}$, we obtain a set of $d$ complex numbers $\alpha_1,\dots,\alpha_d$. Through stereographic projection, each $\alpha_k$ is associated with a point on the Bloch sphere, reminiscent of stars distributed on a celestial sphere, hence the name Majorana stars.

The degree of $P_{\ket{\psi}}$ is not necessarily equal to $N$. If some coefficients in the photon number basis vanish, the degree may be $d < N$. In that case, we interpret $z = \infty$ as a root with multiplicity $N-d$, so that the total number of roots, including those at infinity, is always $N$.

Combining the factorisation of $P_{\ket{\psi}}$ with Eq.~\eqref{eq: reconstructing state from Majo}, one obtains\footnote{This formula remains valid when the degree $d<N$, in which case some roots are at infinity, corresponding to $\theta_k = 0$.}
\begin{equation}\label{eq: state from Majo stars}
    \ket{\psi} = \Lambda \prod_{k=1}^N \left( e^{i\phi_k}\sin(\theta_k/2)\hat a^\dagger + \cos(\theta_k/2)\hat b^\dagger \right)\vac  = \Lambda \prod_{k=1}^N \hat c_k^\dagger \vac,
\end{equation}
where 
\begin{equation}
    \hat c_k^\dagger = e^{i\phi_k}\sin(\theta_k/2)\hat a^\dagger + \cos(\theta_k/2)\hat b^\dagger,
\end{equation}
$\Lambda$ is a normalization factor, and
\begin{equation}
    \alpha_k = -\frac{1}{\Phi(\theta_k,\phi_k)} = - e^{-i\phi_k} \cot(\theta_k/2).
\end{equation}
Thus, any state can be written as a product of $N$ creation operators associated with its Majorana stars, acting on the vacuum. The derivation is given in Appendix~\ref{app: SSRC}, Result~\ref{res: Majo stars and state decomposition}.

Each star $\alpha_k$ is associated with a creation operator $\hat c_k^\dagger$ whose complex parameter is $\gamma_k = -1/\alpha_k$. This point is almost antipodal to $\alpha_k$. Indeed, antipodal points satisfy $z' = -1/z^*$, and the absence of complex conjugation in $\gamma_k = -1/\alpha_k$ introduces a discrepancy. This originates from the extra conjugation in the definition of the Majorana polynomial, required to obtain a polynomial in $z$. As discussed in Tab.~\ref{tab: sphere vs plane transforms}, complex conjugation corresponds to a reflection of the $y$ coordinate. Therefore, Majorana stars are not exactly located at the antipodal point of the associated creation operator, but rather at the point obtained by reflecting the $y$ coordinate of that antipodal point. This subtle distinction is important when interpreting the geometric representation.

Finally, the factorized form also clarifies how Majorana stars transform under rotations. Since the operators $\hat c_k^\dagger$ transform under $e^{-i\theta\vec n\cdot\vec{\hat J}}$ according to the associated Möbius transformation,
\begin{equation}
    \gamma_k \mapsto f(\gamma_k),
\end{equation}
and since $\alpha_k = -1/\gamma_k$, it follows that
\begin{equation}
    \alpha_k\mapsto -\frac{1}{f(-1/\alpha_k)}=f(\alpha_k^*)^*=\frac{a^*\alpha_k-c}{c^*\alpha_k+a}.
\end{equation}
This transformation rule can equivalently be recovered from Eq.~\eqref{eq: rotation Majo poly}, which describes how the polynomial transforms under rotations, by analyzing how the zeros of the polynomial are mapped. The derivation based on the factorized form is more rigorous, as it properly accounts for multiplicities and possible roots at infinity.

\subsection{Spherical Wigner function}
\label{subsec: spherical Wigner function}
The Wigner functions we introduced for both CV systems and time-frequency single photons provide powerful geometrical representations of quantum states on the plane. For spin systems, an analogous Wigner function can be introduced, which encodes the quantum state as a real function on the sphere. It turns out that the expression of the spherical Wigner function is much more complicated than its CV counterpart, and its construction is more subtle. To provide intuition about the construction, in this section we give a pedagogical introduction to the spherical Wigner function. This presentation is based on various references \cite{davis_wigner_2021,davis_wigner_2023,dowling_wigner_1994,tilma_wigner_2016}.

\paragraph{\lit Aim and Structural Requirements}
For an SSRC quantum system, the Hilbert space $\mathcal{H}_N$ of total photon number $N$ has finite dimension $N+1$ and carries an irreducible unitary representation of $SU_2(\C)$. As seen above, the 2-sphere appears naturally in this context and can be identified with the classical phase space of the system. Explicitly, parametrizing
\begin{equation}
    S^2 = \{ \Omega = (\theta,\phi) \},
\end{equation}
it is natural to seek a phase-space representation of quantum spin states as real functions on $S^2$, in close analogy with the standard Wigner function $W(q,p)$ for continuous variables. Our goal is to construct a mapping
\begin{equation}
    \hat \rho \,\longmapsto\, W_{\hat \rho}(\Omega),
\end{equation}
for all $\Omega \in S^2$, which assigns to each density operator $\hat \rho$ on $\mathcal{H}_N$ a real function on the sphere. This mapping should preserve the essential structural features of the quantum theory while reflecting the geometry and symmetry of the classical phase space. The construction is guided by the following requirements, commonly referred to as the Stratonovich-Weyl conditions~\cite{stratonovich_distributions_1957}

\begin{enumerate}
    \item \textbf{Linearity:}
    \begin{equation}
        W_{a\hat \rho_1 + b\hat \rho_2}(\Omega) = a\, W_{\hat \rho_1}(\Omega) + b\, W_{\hat \rho_2}(\Omega),
    \end{equation}
    for all density operators $\hat \rho_1,\hat \rho_2$ and complex numbers $a,b$.

    \item \textbf{Reality:}
    \begin{equation}
        W_{\hat \rho}(\Omega) \in \mathbb{R},
    \end{equation}
    for all $\Omega \in S^2$.

    \item \textbf{Normalization:}
    \begin{equation}
        \int_{S^2} W_{\hat \rho}(\Omega)\, \dd\Omega = \operatorname{Tr}(\hat \rho),
    \end{equation}
    where $\dd\Omega = \sin\theta\, \dd\theta\, \dd\phi$ denotes the invariant measure on the sphere.

    \item \textbf{Traciality:}\footnote{The prefactor is imposed by the following argument: by rotational covariance, the Wigner function of the identity $\1$ must be a constant function on the sphere, and by normalization, it must integrate to $\operatorname{Tr}(\1) = N+1$. Since the area of the sphere is $4\pi$, we must have $W_{\1}(\Omega) = \frac{N+1}{4\pi}$. Now comparing the Normalization condition and traciality for any state $\rho=\rho\1$ fixes the prefactor in the traciality condition.}
    \begin{equation}
        \Tr(\hat \rho_1\hat \rho_2) = \frac{4\pi}{N+1}\int_{S^2} W_{\hat \rho_1}(\Omega)\, W_{\hat \rho_2}(\Omega)\, \dd\Omega.
    \end{equation}

    \item \textbf{Rotational covariance:} For every rotation $R \in SU_2(\C)$,
    \begin{equation}
        \rho \longmapsto U(R)\rho U^\dagger(R) \quad \Longrightarrow \quad W_\rho(\Omega) \longmapsto W_\rho(R^{-1}\Omega).
    \end{equation}
\end{enumerate}

Among these conditions, rotational covariance plays a central role. Since both the quantum system and its classical counterpart share the symmetry group $SU_2(\C)$, the phase-space representation must intertwine the adjoint action of $SU_2(\C)$ on operators with its natural action on the sphere. As will become clear in the following sections, this symmetry requirement essentially determines the structure of the spherical Wigner function.

\paragraph{\lit Spherical harmonics and the action of $SU_2(\C)$}

To construct a function on $S^2$ compatible with rotational covariance, we first recall the harmonic decomposition of functions on the sphere~\cite{hobson_theory_nodate,muller_spherical_2006}. The space $L^2(S^2,\dd\Omega)$ carries a natural unitary representation of $SU_2(\C)$ defined by
\begin{equation}
    \big(U(R)f\big)(\Omega) = f(R^{-1}\Omega), \qquad R\in SU_2(\C).
\end{equation}
The irreducible subspaces of this representation are spanned by the spherical harmonics $Y_{\ell m}(\Omega)$, which are eigenfunctions of the Laplace operator on the sphere
\begin{equation}
    \Delta_{S^2} Y_{\ell m} = -\ell(\ell+1) Y_{\ell m},
\end{equation}
for $\ell \in \N$ and $m=-\ell,\dots,\ell$. Equivalently, recalling that the orbital angular momentum operator satisfies
\begin{equation}
    \hat L^2 = -\Delta_{S^2},
\end{equation}
the spherical harmonics are simultaneous eigenfunctions of $\hat L^2$ and $\hat L_z$
\begin{align}
    \hat L^2 Y_{\ell m} = \ell(\ell+1) Y_{\ell m}, && \hat L_z Y_{\ell m} = m\, Y_{\ell m},
\end{align}
where $\hat L_x$, $\hat L_y$, and $\hat L_z$ are the usual 3D angular momentum operators
\begin{equation}
    \begin{pmatrix}
        \hat L_x\\\hat L_y\\\hat L_z
    \end{pmatrix}= -i \begin{pmatrix} 
        \hat x\\ \hat y \\ \hat z
    \end{pmatrix} \times \begin{pmatrix}
        \partial_x\\ \partial_y \\ \partial_z
    \end{pmatrix},
\end{equation}
with $\hat x$, $\hat y$, and $\hat z$ denoting position operators in $\R^3$. For fixed $\ell$, the set $\{Y_{\ell m}\}_{m=-\ell}^{\ell}$ spans a $(2\ell+1)$-dimensional irreducible representation of $SU_2(\C)$. Under a rotation $R \in SU_2(\C)$, they transform according to
\begin{equation}
    Y_{\ell m}(\Omega) \longmapsto \sum_{m'=-\ell}^{\ell} D^{(\ell)}_{m'm}(R)\, Y_{\ell m'}(\Omega),
\end{equation}
where $D^{(\ell)}(R)$ denotes the Wigner $D$-matrix of weight $\ell$.\footnote{The Wigner $D$-matrix is defined as the matrix elements of the unitary representation of $SU_2(\C)$ in the basis of eigenstates of $\hat J^2$ and $\hat J_z$. More explicitly, for a rotation $R$ with axis $\vec n$ and angle $\theta$, implemented by $e^{-i\theta \vec n\cdot\vec{\hat J}}$, we have
\begin{equation}
    D^{(j)}_{m'm}(R) = \prescript{}{s}{\bra{j, m'}} e^{-i\theta \vec n\cdot\vec{\hat J}} \ket{j, m}_s.
\end{equation}
This transformation rule shows that the spherical harmonics transform under rotation in the same way as any angular momentum objects.}
Moreover, the spherical harmonics form an orthonormal basis of $L^2(S^2,\dd\Omega)$
\begin{equation}
    \int_{S^2} Y_{\ell m}^*(\Omega)\, Y_{\ell' m'}(\Omega) \,\dd\Omega = \delta_{\ell\ell'} \delta_{mm'},
\end{equation}
and satisfy the conjugation relation
\begin{equation}\label{eq: spherical harmonics and conjugation}
    Y_{\ell m}^*(\Omega) = (-1)^m Y_{\ell,-m}(\Omega).
\end{equation}
Hence, any square-integrable function on $S^2$ admits the harmonic expansion
\begin{equation}
    f(\Omega) = \sum_{\ell=0}^{\infty} \sum_{m=-\ell}^{\ell} f_{\ell m}\, Y_{\ell m}(\Omega).
\end{equation}

The key structural point is that the decomposition of functions on the sphere into spherical harmonics precisely corresponds to the decomposition of $L^2(S^2)$ into irreducible representations of $SU_2(\C)$. 

\paragraph{\lit Irreducible tensor operators and operator space}
We now turn to properties of quantum operators. The space of linear operators on $\mathcal{H}_N$, denoted $\mathcal{O}(\mathcal{H}_N)$, has dimension $(N+1)^2$ and carries a natural representation of $SU_2(\C)$ given by the adjoint action
\begin{equation}
    \hat A \longmapsto U(R)\, \hat A \, U^\dagger(R),
\end{equation}
for $R\in SU_2(\C)$. This representation decomposes into irreducible components. Viewing the operator space as the tensor product
\begin{equation}
    \mathcal{O}(\mathcal{H}_N) \simeq \mathcal{H}_N \otimes \mathcal{H}_N^{*},
\end{equation}
using the equivalence $\mathcal{H}_N^{*} \simeq \mathcal{H}_N$ for $SU_2(\C)$ representations, and employing the tensor product decomposition of spins, we obtain the Clebsch-Gordan decomposition
\begin{equation}
    \mathcal{O}(\mathcal{H}_N)\simeq \mathcal H_N \otimes \mathcal{H}_N = \bigoplus_{k=0}^{N} \mathcal H_{2k},
\end{equation}
which implies the dimension identity
\begin{equation}
    (N+1)^2 = \sum_{k=0}^{N} (2k+1).
\end{equation}
Thus, the operator space decomposes into irreducible subspaces labelled by $k=0,\dots,N$, each transforming as an angular momentum-$k$ representation. The corresponding operator basis is provided by the \emph{irreducible tensor operators} $\hat T_{kq}$, defined for $k=0,\dots,N$ and $q=-k,\dots,k$, and characterized by the covariance property
\begin{equation}
    U(R)\, \hat T_{kq}\, U^\dagger(R) = \sum_{q'=-k}^{k} D^{(k)}_{q'q}(R)\, \hat T_{kq'}.
\end{equation}
Analogously to the spherical harmonics, the set $\{\hat T_{kq}\}_{q=-k}^{k}$ spans a $(2k+1)$-dimensional irreducible representation of $SU_2(\C)$ inside operator space. A convenient explicit realization is obtained by coupling two spin-$j$ representations using Clebsch-Gordan coefficients. In the standard angular momentum basis $\{\ket{j,m}_s\}_{m=-j}^{j}$ of $\mathcal{H}_N$, they are defined by
\begin{equation}
    \hat T_{kq} = \sqrt{\frac{2k+1}{2j+1}} \sum_{m,m'=-j}^{j} \braket{j,m';\, k,q }{ j,m}\, \ket{j,m}_s\prescript{}{s}{\bra{j,m'}}.
\end{equation}
Here, $\braket{j,m';\, k,q }{ j,m}$ denotes the Clebsch-Gordan coefficient coupling a spin-$j$ and a spin-$k$ representation into total spin $j$.\footnote{Formally, from the tensor decomposition of angular momentum
\begin{equation}
    j \otimes k = \bigoplus_{J=\abs{j-k}}^{j+k} J,
\end{equation}
two bases can be used to span the total space: the uncoupled basis $\ket{j,m}_s \otimes \ket{k,q}_s$ ($m=-j,\dots,j$, $q=-k,\dots,k$), and the coupled basis $\ket{J,M}_s$ ($J=\abs{j-k},\dots,j+k$, $M=-J,\dots,J$). The Clebsch-Gordan coefficients are the change-of-basis coefficients
\begin{equation}
    \ket{J,M}_s = \sum_{m=-j}^{j} \sum_{q=-k}^{k} \braket{j,m;\, k,q }{J,M}\, \ket{j,m}_s \otimes \ket{k,q}_s.
\end{equation}
}
By construction, the operators $\hat T_{kq}$ satisfy the orthonormality relations
\begin{equation}
    \Tr(\hat T_{kq}^\dagger \hat T_{k'q'}) = \delta_{kk'}\delta_{qq'},
\end{equation}
as well as the Hermiticity condition
\begin{equation}\label{eq: tensor operators and conjugation}
    \hat T_{kq}^\dagger = (-1)^q \hat T_{k,-q}.
\end{equation}
Any operator $\hat A \in \mathcal{O}(\mathcal{H}_N)$ therefore admits the irreducible expansion
\begin{align}
    \hat A = \sum_{k=0}^{N} \sum_{q=-k}^{k} A_{kq}\, \hat T_{kq}, && A_{kq} = \Tr(\hat A\, \hat T_{kq}^\dagger).
\end{align}
This decomposition is the quantum analogue of the harmonic expansion of functions on the sphere and allows us to establish a symmetry-preserving correspondence between quantum operators and classical spherical harmonics.

\paragraph{\lit Construction of the spherical Wigner function}
We are now in a position to construct a rotationally covariant correspondence between operators on $\mathcal{H}_N$ and functions on the sphere $S^2$. Comparing the harmonic expansion of spherical functions,
\begin{equation}
    f(\Omega) = \sum_{\ell=0}^{\infty} \sum_{m=-\ell}^{\ell} f_{\ell m}\, Y_{\ell m}(\Omega),
\end{equation}
with the irreducible expansion of operators,
\begin{equation}
    \hat A = \sum_{k=0}^{N} \sum_{q=-k}^{k} A_{kq}\, \hat T_{kq},
\end{equation}
we observe that both decompositions are organized according to the same irreducible representations of $SU_2(\C)$. Since $\mathcal{O}(\mathcal{H}_N)$ contains angular momenta only up to $k=N$, a natural identification is obtained by truncating the harmonic expansion at $\ell=N$ and matching equal-rank components. We therefore define the \emph{spherical Wigner function} associated with an operator $\hat A$ as\footnote{The normalization factor $\sqrt{\frac{N+1}{4\pi}}$ ensures that both the Normalization and Traciality conditions are satisfied.}
\begin{equation}
    W_{\hat A}(\Omega) = \sqrt{\frac{N+1}{4\pi}}\sum_{k=0}^{N} \sum_{q=-k}^{k} A_{kq}\, Y_{kq}(\Omega),
\end{equation}
where the coefficients $A_{kq}$ are those of the tensor decomposition:
\begin{equation}
    A_{kq} = \Tr\!\left(\hat A\, \hat T_{kq}^\dagger\right).
\end{equation}
Equivalently, introducing the operator-valued kernel
\begin{equation}
    \hat \Delta(\Omega) = \sqrt{\frac{N+1}{4\pi}}\sum_{k=0}^{N} \sum_{q=-k}^{k} \hat T_{kq}^\dagger\, Y_{kq}(\Omega),
\end{equation}
the Wigner function can be written compactly as
\begin{equation}
    W_{\hat A}(\Omega) = \Tr(\hat A\, \hat \Delta(\Omega)).
\end{equation}
By construction, this map satisfies rotational covariance:
\begin{equation}
    W_{U(R)\hat A U^\dagger(R)}(\Omega) = W_{\hat A}(R^{-1}\Omega),
\end{equation}
which follows directly from the identical transformation laws of $Y_{kq}$ and $\hat T_{kq}$. Moreover, Eqs.~\eqref{eq: spherical harmonics and conjugation} and~\eqref{eq: tensor operators and conjugation} show that the kernel is Hermitian, which guarantees the reality of the Wigner function. Normalization and traciality also follow from the orthonormality of spherical harmonics and tensor operators. Thus, the spherical Wigner function provides a symmetry-preserving correspondence between operators on $\mathcal{H}_N$ and truncated harmonic functions on the sphere.

\paragraph{\lit Spin-$\frac12$ case ($N=1$)}
We first consider the simplest non-trivial case, corresponding to $j=\frac12$ and $N=1$. In this situation, the operator space has dimension $(N+1)^2=4$ and decomposes as
\begin{equation}
    \tfrac12 \otimes \tfrac12 = 0 \oplus 1.
\end{equation}
Consequently, only $k=0$ and $k=1$ contribute to the Wigner expansion. For $k=0$, we have
\begin{equation}
    \hat T_{00} = \frac{1}{\sqrt{2}} \1.
\end{equation}
For $k=1$, the three operators $\hat T_{1q}$ form a vector operator. Using either Clebsch-Gordan coefficients or direct evaluation, one finds that they are proportional to the Pauli matrices:
\begin{align}
    \hat T_{10} = \frac{1}{\sqrt{2}}\, \sigma_z, && 
    \hat T_{1,\pm 1} = \mp \sigma_\pm,
\end{align}
where $\sigma_\pm = \frac12(\sigma_x \pm i \sigma_y)$. Any density operator on $\mathcal H_1$ can then be expressed in the familiar Bloch form
\begin{equation}
    \hat\rho = \frac12 \big( \1 + \vec r \cdot \vec\sigma \big),
\end{equation}
where $\vec r$ is the Bloch vector satisfying $\abs{\vec r} \leq 1$. Using the definition of the spherical Wigner function,
\begin{equation}
    W_{\hat\rho}(\Omega) = \sqrt{\frac{2}{4\pi}} \sum_{k=0}^{1} \sum_{q=-k}^{k} A_{kq} Y_{kq}(\Omega),
\end{equation}
a straightforward calculation yields
\begin{equation}
    W_{\hat\rho}(\Omega) = \frac{1}{4\pi} \big( 1 + \sqrt{3}\, \vec r \cdot \vec n(\Omega) \big),
\end{equation}
where $\vec n(\Omega)$ is the unit vector on the sphere defined by the angles $\Omega=(\theta,\phi)$:
\begin{equation}
    \vec n(\Omega) = (\sin\theta\cos\phi, \, \sin\theta\sin\phi, \, \cos\theta).
\end{equation}
Thus, for a qubit, the Wigner function is simply an affine function on the sphere. Negativity occurs whenever $\abs{\vec r} > \frac{1}{\sqrt{3}}$ in some direction, highlighting the non-classical nature of pure states.

\paragraph{\lit Wigner function of spin coherent states}
As introduced in Sec.~\ref{subsec: spin coherent states}, spin coherent states are defined as the orbit of the highest-weight state $\ket{0,N}_p$ under the action of $SU_2(\C)$
\begin{equation}
    \ket{\Omega}_{sc} = \hat R(\theta,\phi)\ket{0,N}_p,
\end{equation}
where $\hat R(\theta,\phi)$ rotates the north pole into the direction $\Omega$. By covariance of the kernel, the Wigner function of a coherent state satisfies
\begin{equation}
    W_{\hat\rho_\Omega}(\Omega') = W_{\ket{0,N}_p\prescript{}{p}{\bra{0,N}}}\big(R(\theta,\phi)^{-1}\Omega'\big).
\end{equation}
Hence it is sufficient to compute the north-pole case $\ket{0,N}_p$. The exact Wigner function is given by the finite sum over irreducible tensor ranks
\begin{equation}
    W_{\hat\rho_\Omega}(\Omega')  = \sum_{k=0}^{N} \frac{2k+1}{4\pi} \, \braket{j,j; k,0 }{ j,j} \, P_k\big (\cos\gamma\big),
\end{equation}
where $P_k$ are Legendre polynomials and $\gamma$ is the angle between $\Omega$ and $\Omega'$. The Clebsch-Gordan coefficient $\braket{j,j; k,0}{ j,j}$ selects the contribution of each tensor rank. In the large-$N$ limit, this sum becomes sharply peaked around $\Omega = \Omega'$, and can be approximated by
\begin{equation}
    W_{\hat\rho_\Omega}(\Omega') \approx \frac{N+1}{4\pi} \left( \frac{1+\vec n(\Omega)\cdot \vec n(\Omega')} {2} \right)^N,
\end{equation}
which is positive everywhere and increasingly localized as $N$ grows, reproducing the classical phase-space picture. This illustrates how spin coherent states converge to classical points on the Bloch sphere in the large-spin limit.

\subsection{Spin squeezing}
\label{subsec: spin squeezing}

In the continuous-variable setting, the notion of squeezed states refers to states for which the uncertainty of one quadrature is reduced below the vacuum level, at the expense of an increased uncertainty in the conjugate quadrature. This concept is rooted in the Heisenberg uncertainty relation
\begin{equation}
    \Delta^2\hat x \, \Delta^2\hat p \geq \frac{1}{4}.
\end{equation}
For the vacuum state, one has $\Delta^2 \hat x = \Delta^2 \hat p = 1/2$, so that the inequality is saturated. A state is said to be squeezed if the variance of one quadrature is reduced below the vacuum value $1/2$. The uncertainty relation then implies that the variance of the conjugate quadrature must increase accordingly. Squeezed states constitute a fundamental resource for quantum metrology and quantum information processing, as they allow one to enhance the precision of measurements beyond the standard quantum limit.

In the present context, spin-squeezed states are defined in an analogous way by replacing the quadrature operators with angular momentum operators~\cite{kitagawa_squeezed_1993}. Since the commutation relations of angular momentum components differ from those of canonical quadratures, the definition must be adapted accordingly. We therefore begin by recalling the general uncertainty relation.

\paragraph{\lit Heisenberg-Robertson inequality}
For two arbitrary observables $\hat A$ and $\hat B$, the uncertainty or Heisenberg-Robertson inequality reads
\begin{equation}
    \Delta^2 A \, \Delta^2 B \geq \frac{1}{4}\abs{\ev{[\hat A,\hat B]}}^2.
\end{equation}
This is a standard result \cite{robertson_uncertainty_1929}, which for completeness we rederive in Appendix~\ref{app: SSRC}, Result~\ref{res: heisenberg robertson}. Applying this relation to angular momentum operators and using $[\hat J_x,\hat J_y] = i \hat J_z$, we obtain
\begin{equation}
    \Delta^2 \hat J_x \, \Delta^2 \hat J_y \geq \frac{1}{4}\abs{\ev{\hat J_z}}^2.
\end{equation}

As an illustration, let us verify that the spin coherent state $\ket{\vec e_z}_{sc} = \ket{0}_{sc} = \ket{0,N}_p$ satisfies this inequality. One readily finds
\begin{align}
    \hat J_z\ket{0,N}_p &= \frac{N}{2}\ket{0,N}_p, &
    \hat J_x\ket{0,N}_p &= \frac{\sqrt{N}}{2}\ket{1,N-1}_p, &
    \hat J_y\ket{0,N}_p &= \frac{i\sqrt{N}}{2}\ket{1,N-1}_p.
\end{align}
Taking the scalar product of these expressions with $\ket{0,N}_p$ yields
\begin{align}
    \ev{J_z} = \frac{N}{2}, && \ev{J_x} = 0,  && \ev{J_y} = 0.
\end{align}
Taking instead the scalar product of each expression with itself gives
\begin{align}
    \ev{J_x^2} = \frac{N}{4}, && \ev{J_y^2} = \frac{N}{4}.
\end{align}
Hence
\begin{align}
    \Delta^2 J_x = \frac{N}{4},  && \Delta^2 J_y = \frac{N}{4}.
\end{align}
The Heisenberg inequality then reads
\begin{equation}
    \underbrace{\Delta^2 J_x}_{N/4} \underbrace{\Delta^2 J_y}_{N/4} \geq \frac{1}{4} \underbrace{\abs{\ev{J_z}}^2}_{N^2/4},
\end{equation}
which is saturated in this case. By applying an arbitrary rotation to the angular momentum operators, the Heisenberg inequality can be written in its most general form as
\begin{equation}
    \Delta^2 (\vec u \cdot \vec{\hat J}) \, \Delta^2 (\vec v \cdot \vec{\hat J}) \geq \frac{1}{4} \abs{\ev{\vec w \cdot \vec{\hat J}}}^2,
\end{equation}
where $\vec u$, $\vec v$ and $\vec w$ are three orthonormal unit vectors.

\subsubsection{\lit General spin-squeezing}
As discussed above, the central idea of squeezing is to reduce the uncertainty of one observable below a reference level, at the price of increasing the uncertainty of a conjugate observable. The Heisenberg inequality provides the appropriate framework to formalize this notion for collective spin systems. We say that a quantum state $\ket{\psi}$ is spin squeezed if
\begin{equation}
\label{eq: general squeezing criterion}
    \Delta^2 (\vec n_\perp \cdot \vec{\hat J}) < \frac{1}{2} \abs{\ev{\vec n \cdot \vec{\hat J}}},
\end{equation}
where $\vec n \propto \ev*{\vec{\hat J}}$ defines the mean spin direction, and $\vec n_\perp$ is any unit vector orthogonal to $\vec n$. Importantly, this means that for states with zero mean spin, the notion of squeezing is ill-defined. This should always be kept in mind when applying the squeezing criterion, as it can lead to erroneous conclusions if not properly taken into account.

It is crucial to require that $\vec n$ be aligned with the mean spin direction. If one were to test squeezing along arbitrary pairs of orthogonal directions, coherent states would incorrectly be identified as spin squeezed. Indeed, for a coherent state polarized along $\vec n$ and for an arbitrary unit vector $\vec m$, one finds
\begin{align}
    \expval{\vec m \cdot \vec{\hat J}}_{\ket{\vec n}_{sc}}
    &= \frac{N}{2} (\vec n \cdot \vec m), &
    \Delta^2 (\vec m \cdot \vec{\hat J})_{\ket{\vec n}_{sc}}
    &= \frac{N}{4} \left( 1 - (\vec n \cdot \vec m)^2 \right).
\end{align}
A detailed derivation of these expressions is given in Appendix~\ref{app: SSRC}, Result~\ref{res: variance of arbitrary spin component}. With these general expressions, we can analyse the consequence of not defining the notion of spin squeezing with respect to the mean spin orientation. Up to a global rotation, we consider a coherent state $\ket{\theta,\phi}_{sc}$ characterized by spherical angles $(\theta,\phi)$ and evaluate $\expval{\hat J_z}$ and $\Delta^2 \hat J_x$. One obtains
\begin{align}
    \expval{\hat J_z}_{\ket{\theta,\phi}_{sc}}
    &= \frac{N}{2}\cos(\theta), &
    \Delta^2_{\ket{\theta,\phi}_{sc}} \hat J_x
    &= \frac{N}{4}\left(1 - \sin^2(\theta)\cos^2(\phi)\right).
\end{align}
While the expectation value is independent of $\phi$, the variance depends on this angle and is minimized for $\phi = 0$. In that case,
\begin{equation}
    \Delta^2_{\ket{\theta,0}_{sc}} \hat J_x
    =
    \frac{N}{4}\cos^2(\theta).
\end{equation}
The squeezing criterion of Eq.~\eqref{eq: general squeezing criterion} then becomes
\begin{equation}
    \frac{N}{4}\cos^2(\theta)
    <
    \frac{N}{4}\abs{\cos(\theta)}.
\end{equation}
For $\cos(\theta) \neq 0$, this inequality reduces to $\abs{\cos(\theta)} < 1$, which is satisfied for all $\theta \neq 0,\pi$. This shows that coherent states are `squeezed' if we consider squeezing along any direction except the mean spin direction.

A commonly accepted way~\cite{wineland_spin_1992,carrera_testing_2025,ma_quantum_2011} to deal with this problem of needing to align the squeezing direction with the mean spin, which poses problems when the mean spin direction vanishes, is to replace the right-hand side of the squeezing criterion by a fixed reference value, such as $N/4$, which corresponds to the variance of a coherent state. This leads to the alternative squeezing criterion
\begin{equation*}
    \Delta^2 (\vec n \cdot \vec{\hat J}) < \frac{N}{4},
\end{equation*}
known as the Wineland criterion~\cite{wineland_spin_1992}.

\paragraph{\lit Example of spin-squeezed states}
There exists a vast zoology of spin-squeezed states with evocative names such as One-Axis Twisting, Two-Axis Counter-Twisting, Planar Spin-Squeezed States, Twist-and-Turn states, and others. In this paragraph we provide an overview of some of the most common examples of spin-squeezed states, without entering into the technical details of their construction or a full characterization of their properties. For comprehensive treatments we refer the reader to the literature~\cite{ma_quantum_2011,kitagawa_squeezed_1993}. A general strategy to construct spin-squeezed states consists in starting from a spin coherent state and applying a non-local unitary transformation in order to generate particle entanglement. By finely tuning the interaction and its orientation, one can engineer squeezing along a desired direction, thus obtaining a wide variety of spin-squeezed states.

A paradigmatic example is the One-Axis Twisting (OAT) state, obtained by applying the unitary evolution generated by $\hat J_z^2$ to a coherent state polarized along the $x$-direction
\begin{equation}
    \ket{\psi}_{\text{OAT}} = e^{-i\mu \hat J_z^2}\ket{\vec e_x}_{sc},
\end{equation}
where the parameter $\mu$ must be optimized in order to achieve maximal squeezing. The quadratic evolution admits an interpretation analogous to a shear in phase space: it implements a rotation around the $z$-axis by an angle proportional to the $z$-component of the spin. This justifies the terminology ``twisting''.\footnote{The geometrical interpretation of this transformation is however less robust than in the case of planar shears. While it provides an intuitive picture at short times, it does not faithfully describe the full evolution of the Wigner function on the Bloch sphere. In particular, as can be verified in the Dicke basis, the evolution generated by $e^{-i\mu \hat J_z^2}$ is periodic in $\mu$, whereas a continuous classical twist of the sphere would not be periodic. Care must therefore be taken when using the geometrical picture, as it may lead to incorrect conclusions if extrapolated beyond its regime of validity.} This twisting deforms the spin distribution by elongating it along one direction and compressing it along the orthogonal one. The analytical study of OAT states was carried out in~\cite{kitagawa_squeezed_1993}. %Adapting their notation to our setting, the optimally squeezed variance can be written as
%\begin{equation}
%    \Delta^2 J_{\min} = \frac{N}{16}\Big[4 + (N-1)\big(A - \sqrt{A^2 + B^2}\big)\Big],
%\end{equation}
%with
%\begin{align}
%    A = 1 - \cos^{N-2}(2\mu), && B = 4\sin(\mu)\cos^{N-2}(\mu),
%\end{align}
%while the mean spin components read
%\begin{align}
%    \expval{\hat J_x} = \frac{N}{2}\cos^{N-1}(\mu),  && \expval{\hat J_y} = 0, && \expval{\hat J_z} = 0.
%\end{align}
%From these expressions one can directly compute the squeezing parameters introduced previously. The optimal value of $\mu$ can then be obtained numerically. An asymptotic analysis in the large-$N$ limit shows that
%\begin{equation}
%    \mu_{\text{opt}} \sim N^{-2/3},
%\end{equation}
%and
%\begin{equation}
%    \Delta^2 J_{\min} \sim N^{1/3}.
%\end{equation}
%In this regime, all squeezing parameters exhibit the same scaling,
%\begin{equation}
%    \xi_H \sim \xi_{KU} \sim \xi_W \sim N^{-2/3},
%\end{equation}
%demonstrating that OAT states are genuinely spin squeezed in the large-$N$ limit.

In the same work, Kitagawa and Ueda introduced a second important class of spin-squeezed states, the Two-Axis Counter-Twisting (TACT) states. These are generated by the Hamiltonian
\begin{equation}
    \hat H_{\text{TACT}} = i\frac{\chi}{2}(\hat J_+^2 - \hat J_-^2)
    = \chi\Big[(\vec n_{\frac{\pi}{2},-\frac{\pi}{4}}\cdot \vec{\hat J})^2
    - (\vec n_{\frac{\pi}{2},\frac{\pi}{4}}\cdot \vec{\hat J})^2\Big],
\end{equation}
acting on a spin coherent state polarized along the $z$-axis. This Hamiltonian can be interpreted as a simultaneous twisting around two orthogonal axes in the $xy$-plane, one acting clockwise and the other counter-clockwise, which explains the name ``counter-twisting''. The corresponding state reads
\begin{equation}
    \ket{\psi_{\text{TACT}}(\mu)} 
    = e^{\frac{\mu}{2}(\hat J_+^2 - \hat J_-^2)}\ket{\vec e_z}_{sc}.
\end{equation}

In contrast to the OAT case, no simple closed analytical expressions are known for the expectation values and variances for arbitrary $N$. However, analytical approximations and numerical simulations show that one spin component in the $xy$-plane becomes squeezed while the orthogonal one is anti-squeezed, the roles being interchanged depending on the sign of $\mu$. %Asymptotically, the minimal variance reaches
%\begin{equation}
%    \Delta^2 J_{\min} \sim \frac{1}{2},
%\end{equation}
%so that the squeezing parameters scale as
%\begin{equation}
%    \xi_H \sim \xi_{KU} \sim \xi_W \sim \frac{1}{N},
%\end{equation}
%which is a better scaling than in the OAT case. This improved behavior stems from the fact that the two-axis counter-twisting mechanism redistributes quantum fluctuations more efficiently by acting simultaneously along two orthogonal directions, whereas one-axis twisting deforms the distribution along a single axis only.

To conclude this section, we introduce a final class of squeezed states that is less commonly discussed in the spin squeezing literature, but will play an important role in the latter part of this chapter (see Sec.~\ref{subsec: angular momentum universality and bosonic resources}). These states are most conveniently described through their Majorana representation. For even $N$, we define
\begin{equation}
    \ket{z_1,z_2}_\text{sq} = \frac{1}{\mathcal N} (\hat c^\dagger)^{N/2}(\hat d^\dagger)^{N/2}\vac,
\end{equation}
where
\begin{align}
    \hat c^\dagger &= \frac{1}{\sqrt{1+\abs{z_1}^2}} (z_1\hat a^\dagger + \hat b^\dagger), &
    \hat d^\dagger &= \frac{1}{\sqrt{1+\abs{z_2}^2}} (z_2\hat a^\dagger + \hat b^\dagger),
\end{align}
are creation operators associated with the complex numbers $z_1$ and $z_2$, and $\mathcal N$ ensures normalization. Notice that contrary to the operators $\hat a$ and $\hat b$, the operators $\hat c$ and $\hat d$ are not necessarily associated with orthogonal modes as
\begin{equation}
    [\hat c,\hat d^\dagger]=\frac{1 + z_1^* z_2}{\sqrt{(1+\abs{z_1}^2)(1+\abs{z_2}^2)}}.
\end{equation}
The corresponding Majorana stars are directly located at $-1/z_1$ and $-1/z_2$. Hence, $\ket{z_1,z_2}_\text{sq}$ possesses only two distinct Majorana stars, each with degeneracy $N/2$. Since the spin properties are invariant under global rotations, it is convenient to choose the parametrization\footnote{A suitable global rotation allows one to choose $z_1$ and $z_2$ with opposite signs and both in the northern hemisphere, ensuring $\abs{z_j}\leq 1$ and justifying the parametrization below. The limit $r\to\infty$ corresponds to $z_1$ and $z_2$ approaching the equator.}
\begin{align}
    z_1 = e^{i\varphi/2}\sqrt{\tanh(r)}, && z_2 = -z_1.
\end{align}
With this choice one obtains
\begin{subequations}
    \begin{align}
        \hat c^\dagger &= e^{-r/2}\left(e^{i\varphi/2}\sqrt{\sinh(r)}\hat a^\dagger + \sqrt{\cosh(r)}\hat b^\dagger\right), \\
        \hat d^\dagger &= e^{-r/2}\left(-e^{i\varphi/2}\sqrt{\sinh(r)}\hat a^\dagger + \sqrt{\cosh(r)}\hat b^\dagger\right).
    \end{align}
\end{subequations}
Their product takes the simple form
\begin{equation}
    \hat c^\dagger \hat d^\dagger  = e^{-r}\left( - e^{i\varphi}\hat a^{\dagger 2}\sinh(r)  + \hat b^{\dagger 2}\cosh(r)\right),
\end{equation}
showing explicitly that $\ket{z_1,z_2}_\text{sq}$ contains only even photon numbers in each mode. Its expansion reads
\begin{align}\label{eq: squeezed state expansion}
    \ket{z_1,z_2}_\text{sq} &= \frac{e^{-rN/2}}{\mathcal N}\sum_{k=0}^{N/2} \binom{N/2}{k} (-e^{i\varphi})^k \sinh^k(r)\cosh^{N/2-k}(r)\notag\\
    &\quad \times\sqrt{(2k)!}\sqrt{(N-2k)!}\ket{2k,N-2k}_p.
\end{align}
The normalization constant is
\begin{equation}
    \mathcal N^2  = e^{-Nr}\cosh^N(r)(N/2)!^2 \sum_{k=0}^{N/2}  \binom{2k}{k} \binom{N-2k}{N/2-k}  \tanh^{2k}(r),
\end{equation}
which does not admit a simple closed form. Although $\ket{z_1,z_2}_\text{sq}$ contains only even photon numbers, similarly to TACT states, the two families are distinct, as can be verified explicitly for small values of $N$. Nevertheless, analytical considerations and numerical evidence indicate that $\ket{z_1,z_2}_\text{sq}$ are also spin-squeezed states. In this case, the squeezed and anti-squeezed directions lie in the $xy$-plane, with orientations determined by the angle $\varphi$.

\clearpage
\section{Continuous-variable systems as a limit of super\-selection-rules-compliant ones}
\label{sec: CV as limit of SSRC}

\emph{In this short section we show how continuous-variable systems can be obtained as a limit of superselection-rules-compliant ones. This construction provides a rigorous justification for the common practice of treating continuous-variable systems as idealized limits of finite-dimensional systems, and grounds even more firmly the importance of the SSRC framework to understand the properties of optical systems.}

\subsection{Formal limit}
\label{subsec: formal limit}

\paragraph{\publi Description of the limit} 
As discussed at the end of the first chapter (see Sec.~\ref{subsec: Phase Reference and SSR}) as well as at the beginning of this chapter (see Sec.~\ref{subsec: SSRC and Schwinger}), two conceptually opposite models can be used to describe quantum optical systems. The standard continuous-variable (CV) description relies on the existence of a phase reference and therefore does not make the particle-number superselection rule (SSR) explicit. In contrast, the superselection-rules-compliant (SSRC) framework incorporates the phase reference as a physical degree of freedom and restricts states to sectors of fixed total photon number.

In the standard CV description of quantum optical systems, single-mode states are written as coherent superpositions of Fock states,
\begin{equation}
    \ket{\psi} = \sum_{n=0}^{\infty} c_n \ket{n},
\end{equation}
implicitly assuming the existence of a phase reference and therefore not making the particle-number superselection rule explicit. In contrast, within the SSRC framework the phase reference is modeled as an additional mode and global states are restricted to a sector of fixed total photon number. A general pure SSRC state with total photon number $N$ can be written as\footnote{From now on, notation of the form $\ket{n}_q$ will denote a Fock state with $n$ photons in a mode labeled by $q$.}
\begin{equation}
    \ket{\Psi} = \sum_{n=0}^{N} c_n \, \ket{n}_A \ket{N-n}_B,
\end{equation}
where modes $A$ and $B$ are orthogonal and the second mode plays the role of an internalized phase reference. This representation is exact and involves no approximation. The CV formalism is recovered as a particular limit of this construction. When the photon-number distribution is strongly imbalanced, such that the average occupation of mode $A$ satisfies
\begin{equation}
    \sum_{n=0}^{N} n \abs{c_n}^2 \ll N,
\end{equation}
most photons populate the reference mode $B$. In this regime one can formally consider $N \to \infty$ while keeping the energy in mode $A$ finite. Introducing a cutoff $n_{\max}$ such that $n_{\max} \ll N$ and $\sum_{n=0}^{n_{\max}}  \abs{c_n}^2 \simeq 1$, the state can be approximated as
\begin{equation}\label{eq: SSRC to CV}
    \sum_{n=0}^{N} c_n \ket{n}_A \ket{N-n}_B  \,\longrightarrow\, \sum_{n=0}^{n_{\max}} c_n \ket{n}_A,
\end{equation}
where the reference mode becomes effectively classical and can be omitted at the level of the reduced description of mode $A$. Taking subsequently $n_{\max} \to \infty$ yields the usual CV expression. Continuous-variable systems therefore appear not as fundamentally distinct objects, but as a large-$N$ limit of finite-dimensional SSRC states with a strongly populated phase reference mode. This establishes the CV formalism as an emergent description within the more general and physically consistent SSRC framework.

It is important to emphasize that the mapping of Eq.~\eqref{eq: SSRC to CV} is only a formal limit stating that both descriptions are equivalent in the regime of large $N$ and strongly imbalanced photon-number distributions. While the left-hand side SSRC state generally exhibits entanglement between the system and the reference, the right-hand side CV state is a pure state of the system alone. Hence, the mapping formally removes the explicit entanglement between the system and its reference: although it is not present in the CV description, it is physically encoded in the global state and is replaced at the effective level by the notion of coherence with respect to a phase reference.

\paragraph{\publi Mathematical subtleties}
While the mapping of Eq.~\eqref{eq: SSRC to CV} intuitively provides the physical condition under which an SSRC state can be approximated by a CV one, it does not mathematically specify how the transition from one description to the other can be made rigorous. In particular, the limit $N \to \infty$ is not well defined as it stands, since we have not specified how the coefficients $c_n$ should behave as $N$ tends to infinity. Without delving into all mathematical details, several notions of convergence can be considered.

\begin{itemize}
    \item First, one may assume that the sequence $\{c_n\}$ is independent of $N$, and that $c_n$ is non-zero only up to some fixed index $n_{\max}$. In that case the value of $N$ can be increased arbitrarily in the expression
    \begin{equation}
        \ket{\psi_N}=\sum_{n=0}^{n_{\max}} c_n \ket{n}_A \ket{N-n}_B,
    \end{equation}
    and the limiting CV state naturally reads
    \begin{equation}
        \ket{\psi}=\sum_{n=0}^{n_{\max}} c_n \ket{n}_A.
    \end{equation}
    This notion of limit is simple, as it involves only the manipulation of a finite and constant number of terms. However, it cannot be used to recover CV states with an infinite Fock-state support.

    \item In the most general case, one needs to consider sequences $\{c_n^{(N)}\}$ that explicitly depend on $N$. In this situation, the limit $N \to \infty$ is implemented by associating to the sequence of SSRC states
    \begin{equation}
        \ket{\psi_N}=\sum_{n=0}^{N} c_n^{(N)} \ket{n}_A \ket{N-n}_B,
    \end{equation}
    the CV state
    \begin{equation}
        \ket{\psi}=\sum_{n=0}^{\infty} \lim_{N\to\infty}c_n^{(N)} \ket{n}_A.
    \end{equation}
    Formally, we are thus considering the convergence of a sequence of sequences. Since these are infinite-dimensional objects, it is necessary to specify which notion of convergence is used. The most common ones in this context are the following:
    \begin{itemize}
        \item Pointwise convergence, which requires that for each fixed $n$, the sequence $\{c_n^{(N)}\}_N$ converges to a limit as $N \to \infty$. This notion of convergence is relatively weak, as it allows for different rates of convergence for different values of $n$ and does not guarantee uniform control over the entire sequence. For ease of mathematical manipulation, this is the notion we will mostly use.

        \item $L^2$ convergence, or convergence in the Hilbert space of square-summable sequences, which requires that
        \begin{equation}
            \sum_{n=0}^{\infty}\abs{c_n^{(N)}-c_n}^2\to 0,
        \end{equation}
        where for $n>N$ we set $c_n^{(N)}=0$. This notion of convergence is stronger than pointwise convergence, as it requires that the entire sequence $\{c_n^{(N)}\}$ converges to $\{c_n\}$ in the norm of the Hilbert space. It ensures that the limit state $\ket{\psi}$ is well defined and belongs to the right Hilbert space.
    \end{itemize}
\end{itemize}

\paragraph{\publi Convergence of the operators}
As discussed above, the SSRC and the CV regimes are endowed with different sets of fundamental operators. For SSRC systems, the angular momentum operators obtained via the Schwinger transformation yield the generators of rotations,
\begin{align}
    \hat J_x = \frac{1}{2}(\hat a^\dagger \hat b + \hat b^\dagger \hat a), && 
    \hat J_y = \frac{i}{2}(\hat a^\dagger \hat b - \hat b^\dagger \hat a), && 
    \hat J_z = \frac{1}{2}(\hat b^\dagger \hat b-\hat a^\dagger \hat a),
\end{align}
whereas for CV systems the quadrature operators are defined as
\begin{align}
    \hat x = \frac{1}{\sqrt{2}}(\hat a + \hat a^\dagger), && 
    \hat p = \frac{i}{\sqrt{2}}(\hat a^\dagger - \hat a).
\end{align}
Fixing $k\leq N$, we observe that in the limit of large $N$
\begin{equation}
    \frac{\hat a\hat b^\dagger}{\sqrt{N}} \ket{k,N-k} =\sqrt{\frac{N-k+1}{N}}\sqrt{k}\ket{k-1,N-k+1} \,\longrightarrow\, \sqrt{k}\ket{k-1},
\end{equation}
where in the last expression the reference mode is omitted in accordance with the mapping of Eq.~\eqref{eq: SSRC to CV}. This coincides with the action of $\hat a$ on the CV Fock state $\ket{k}$. Hence, at the operator level the SSRC-CV correspondence suggests
\begin{equation}
    \frac{1}{\sqrt{N}}\hat a\hat b^\dagger \,\longrightarrow\, \hat a,
\end{equation}
and similarly for the Hermitian conjugate. Applying this relation to the angular momentum operators $\hat J_x$ and $\hat J_y$ yields
\begin{align}
    \frac{1}{\sqrt{N}}\hat J_x &\,\longrightarrow\, \frac{1}{2}(\hat a + \hat a^\dagger)=\frac{1}{\sqrt{2}}\hat x, \\
    \frac{1}{\sqrt{N}}\hat J_y  &\,\longrightarrow\, \frac{i}{2}(\hat a^\dagger - \hat a) = \frac{1}{\sqrt{2}}\hat p.
\end{align}
Additionally, we observe that
\begin{subequations}
    \begin{align}
        \frac{\hat b^\dagger\hat b}{N}\ket{k,N-k} &=\frac{N-k}{N}\ket{k,N-k} \,\longrightarrow\, \ket{k},\\
        \frac{\hat a^\dagger\hat a}{N}\ket{k,N-k} &=\frac{k}{N}\ket{k,N-k} \,\longrightarrow\, 0,
    \end{align}
\end{subequations}
so that
\begin{equation}
    \frac{1}{N}\hat J_z  = \frac{1}{2N}(\hat b^\dagger\hat b-\hat a^\dagger \hat a) \,\longrightarrow\, \frac{1}{2}\1.
\end{equation}
These computations reveal two important aspects.
\begin{itemize}
    \item There is a natural correspondence between the angular momentum operators of the Schwinger representation and the generators of the Heisenberg-Weyl algebra. The operators $\hat J_x$ and $\hat J_y$, properly rescaled by $\sqrt{N}$, are associated with the quadrature operators, while $\hat J_z$ becomes proportional to the identity in the large-$N$ limit. Applying the angular momentum uncertainty relation,
    \begin{equation}
        \Delta^2 \hat J_x \Delta^2 \hat J_y 
        \geq \frac{1}{4}\abs{\ev{\hat J_z}}^2,
    \end{equation}
    and using $\ev{\hat J_z}\simeq N/2$ in the strongly imbalanced regime, we obtain after rescaling by $1/N$
    \begin{equation}
        \Delta^2 \hat x \Delta^2 \hat p \geq \frac{1}{4},
    \end{equation}
    thus recovering the Heisenberg uncertainty relation for quadrature operators.

    \item The normalization factor $\sqrt{N}$ appears naturally in the correspondence. As will be confirmed in further derivations and computations, this suggests that this factor provides the natural rescaling that bridges the large-$N$ limit of SSRC systems and CV ones.
\end{itemize}

The above limit of operators has been computed by considering their direct action on basis states. If one instead looks at the action of unitary operators, obtained by taking the complex exponential, we have
\begin{equation}
    e^{-i\hat J_z \theta}\ket{k,N-k}= e^{-i\theta(N-2k)/2}\ket{k,N-k} \,\longrightarrow\, e^{-i\theta N/2}e^{i\theta k}\ket{k}.
\end{equation} 
As the phase $e^{-i\theta N/2}$ is independent of $k$, it can be ignored and we see the operator correspondence
\begin{equation}
    e^{-i\hat J_z \theta} \,\longrightarrow\, e^{i\theta \hat a^\dagger \hat a},
\end{equation}
showing that rotations around the $z$-axis in the SSRC picture correspond to phase shifts in the CV picture.

\subsection{Examples of limits}
\label{subsec: examples of SSRC CV limits}

\paragraph{\publi From spin coherent states to CV coherent states}

Beyond their analogy as the most classical states of their respective Hilbert spaces, spin coherent states and quadrature coherent states exhibit a much closer relation. As we now show, CV coherent states arise as a natural large-$N$ limit of a suitable family of SSRC spin coherent states.

Consider an initial Fock state $\ket{N}$ impinging on a weakly reflecting beam splitter. The photon number distribution in the reflected mode follows a binomial distribution, which, in the limit of large $N$, tends to a Poissonian distribution. See Fig.~\ref{fig: N and weak BS}. More explicitly, setting the reflectivity to
\begin{equation}
    r=\alpha/\sqrt{N},
\end{equation}
the probability to detect $n$ photons in the reflected mode reads
\begin{equation}
    P(n) = \binom{N}{n} \abs{r}^{2n}(1-\abs{r}^2)^{N-n} \,\longrightarrow\,  e^{-\abs{\alpha}^2}\frac{\abs{\alpha}^{2n}}{n!},
\end{equation}
in the limit $N\to\infty$. The expression on the right-hand side is precisely the Poissonian distribution associated with a coherent state of amplitude $\alpha$,
\begin{equation}
    \ket{\alpha} = e^{-\abs{\alpha}^2/2}\sum_{n=0}^{\infty} \frac{\alpha^n}{\sqrt{n!}}\ket{n}.
\end{equation}
This simple scaling, in which the reflectivity vanishes as $1/\sqrt{N}$ while the input photon number diverges, therefore provides a natural mechanism through which CV coherent states emerge from SSRC spin coherent states.

\begin{figure}[ht]
    \centering
    \scalebox{1.6}{\begin{tikzpicture}
	\begin{pgfonlayer}{nodelayer}
		\node [style=none] (4) at (-1.75, 0.75) {};
		\node [style=none] (5) at (-1.75, -0.75) {};
		\node [style=none] (6) at (-4.75, 0.75) {};
		\node [style=none] (7) at (-4.75, -0.75) {};
		\node [style=none] (0) at (-1, 1) {};
		\node [style=none] (1) at (-1, -1) {};
		\node [style=none] (2) at (1, 1) {};
		\node [style=none] (3) at (1, -1) {};
		\node [style=none] (16) at (-5.5, 0) {\scalebox{0.6}{$\ket{N}$}};
		\node [style=none] (17) at (0, 1.75) {\scalebox{0.6}{$r=\alpha/\sqrt{N}$}};
		\node [style=none] (18) at (0, -3.5) {\scalebox{0.6}{$P(n)\sim e^{-\abs{\alpha}^2}\frac{\abs{\alpha}^{2n}}{n!}$}};
		\node [style=none] (19) at (-1.75, 1.125) {};
		\node [style=none] (20) at (-1.25, 0) {};
		\node [style=none] (21) at (-1.75, -1.125) {};
		\node [style=none] (22) at (4.25, 0.625) {};
		\node [style=none] (23) at (4.25, -0.625) {};
		\node [style=none] (24) at (1.25, 0.625) {};
		\node [style=none] (25) at (1.25, -0.625) {};
		\node [style=none] (26) at (4.25, 1) {};
		\node [style=none] (27) at (4.75, 0) {};
		\node [style=none] (28) at (4.25, -1) {};
		\node [style=none] (29) at (0.075, -2.75) {};
		\node [style=none] (30) at (-0.075, -2.75) {};
		\node [style=none] (31) at (0.075, -1.25) {};
		\node [style=none] (32) at (-0.075, -1.25) {};
		\node [style=none] (33) at (0.325, -2.75) {};
		\node [style=none] (34) at (0, -3) {};
		\node [style=none] (35) at (-0.35, -2.75) {};
	\end{pgfonlayer}
	\begin{pgfonlayer}{edgelayer}
		\draw [style=Fill red] (19.center)
			 to (20.center)
			 to (21.center)
			 to (5.center)
			 to (7.center)
			 to (6.center)
			 to (4.center)
			 to cycle;
		\draw [style=Fill grey] (0.center)
			 to (2.center)
			 to (3.center)
			 to (1.center)
			 to cycle;
		\draw (0.center) to (3.center);
		\draw [style=Fill red] (26.center)
			 to (27.center)
			 to (28.center)
			 to (23.center)
			 to (25.center)
			 to (24.center)
			 to (22.center)
			 to cycle;
		\draw [style=Fill red] (33.center)
			 to (34.center)
			 to (35.center)
			 to (30.center)
			 to (32.center)
			 to (31.center)
			 to (29.center)
			 to cycle;
	\end{pgfonlayer}
\end{tikzpicture}}
    \caption[Photon number distribution in the reflected mode of a weakly reflecting beam splitter fed with a Fock state]{Photon number distribution in the reflected mode of a weakly reflecting beam splitter fed with a Fock state $\ket{N}$. For a reflectivity $r=\alpha/\sqrt{N}$, the binomial distribution converges, as $N$ increases, towards a Poissonian distribution of mean $\abs{\alpha}^2$, characteristic of a coherent state of amplitude $\alpha$.}
    \label{fig: N and weak BS}
\end{figure}
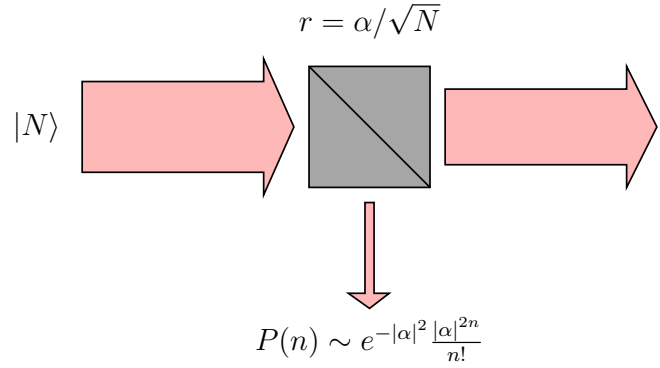

This intuitive argument can be made fully explicit within the SSRC framework, which clarifies the precise correspondence between spin coherent states and CV coherent states. In the SSRC picture, the input state $\ket{N}$ is represented as $\ket{N}_B$, where mode $B$ plays the role of phase reference. Fixing $\alpha\in\C$, the beam splitter unitary $\hat U$ is defined through
\begin{align}
    \hat U\hat a^\dagger\hat U^\dagger &= \sqrt{1-\frac{\abs{\alpha}^2}{N}}\hat a^\dagger -\frac{\alpha^*}{\sqrt{N}}\hat b^\dagger, & \hat U\hat b^\dagger\hat U^\dagger = \frac{\alpha}{\sqrt{N}}\hat a^\dagger +\sqrt{1-\frac{\abs{\alpha}^2}{N}}\hat b^\dagger.
\end{align}
This transformation corresponds to a change of mode basis, that is, a rotation in the SSRC phase space, by an angle $\theta$ around the axis $\vec n=(n_x,n_y,0)$ where
\begin{align}
    \cos(\theta/2)=\sqrt{1-\frac{\abs{\alpha}^2}{N}}, && n_x = -\frac{\Im(\alpha)}{\abs{\alpha}}, && n_y = \frac{\Re(\alpha)}{\abs{\alpha}}.
\end{align}

Applying $\hat U$ to $\ket{N}_B$ yields
\begin{equation}
    \hat U\ket{N}_B = \sum_{k=0}^N  \binom{N}{k} \frac{\sqrt{k!}\sqrt{(N-k)!}}{\sqrt{N!}} \frac{\alpha^k}{\sqrt{N}^{\,k}} \left(1-\frac{\abs{\alpha}^2}{N}\right)^{\frac{N-k}{2}} \ket{k}_A\ket{N-k}_B.
\end{equation}
Taking the limit $N\to\infty$ at fixed $\alpha$, each term of the sum admits a well-defined asymptotic expansion. One obtains
\begin{equation}
    \hat U\ket{N}_B \simeq e^{-\abs{\alpha}^2/2} \sum_{k=0}^N \frac{\alpha^k}{\sqrt{k!}} \ket{k}_A\ket{N-k}_B,
\end{equation}
and in the large-$N$ limit this state is the SSRC counterpart of the CV coherent state of amplitude $\alpha$,
\begin{equation}
    \ket{\alpha} = e^{-\abs{\alpha}^2/2} \sum_{k=0}^{\infty} \frac{\alpha^k}{\sqrt{k!}} \ket{k}_A.
\end{equation}
The detailed derivation is given in Appendix~\ref{app: SSRC}, Result~\ref{res: SSRC to CV coherent states}. This construction shows explicitly how CV coherent states arise as limits of SSRC spin coherent states. The beam splitter transformation, which implements an infinitesimal rotation in SSRC phase space, generates the required superposition of Fock states in mode $A$ while mode $B$ remains macroscopically populated and thus acts as an effective internal phase reference.

Beyond this formal correspondence, the construction also illustrates how coherence between two beams can emerge without invoking an external phase reference. One may assume, for instance, that the initial beam is not in a Fock state but in an incoherent Poissonian mixture,
\begin{equation}
    \rho = e^{-\abs{\alpha}^2} \sum_{n=0}^{\infty} \frac{\abs{\alpha}^{2n}}{n!} \ket{n}\bra{n}.
\end{equation}
A similar calculation then shows that coherence between the two output modes is nevertheless present, even though the input state lacks phase coherence. The output is well approximated by a coherent state in mode $A$ together with its strongly populated reference mode $B$, emphasizing the relational character of phase in the SSRC framework.

\paragraph{\publi SSRC rotation and CV displacement}

In the previous paragraph, the transformation $\hat U$ maps $\ket{N}_B$, which is the SSRC analogue of the CV vacuum $\ket{0}_A$, to a state that becomes, in the large-$N$ limit, the SSRC version of a coherent state of amplitude $\alpha$. As discussed in Sec.~\ref{subsec: continuous variables}, coherent states can equivalently be defined as displaced vacuum states. The displacement operator
\begin{equation}
    \hat D(\alpha) = e^{\alpha\hat a^\dagger - \alpha^*\hat a}
\end{equation}
acts as
\begin{equation}
    \hat D(\alpha)\ket{0}_A = \ket{\alpha}_A.
\end{equation}
The previous construction therefore indicates that, when acting on $\ket{N}_B$ and in the limit of large $N$, the $\alpha$-dependent beam splitter $\hat U$ reproduces the action of the displacement operator $\hat D(\alpha)$ on the vacuum.

This correspondence can in fact be extended beyond the vacuum. As detailed in Appendix~\ref{app: SSRC}, Result~\ref{res: SSRC rotation to CV displacement}, one may compare the action of both transformations on arbitrary Fock states under the identification
\begin{equation}
    \ket{k}_A\ket{N-k}_B \,\longleftrightarrow\, \ket{k}_A.
\end{equation}
On the CV side, one has
\begin{align}
    \hat D(\alpha)\ket{k}_A = e^{-\abs{\alpha}^2/2} \sum_{l=0}^k \sum_{j=0}^\infty \frac{(-\alpha^*)^{k-l}\alpha^j}{(k-l)!j!} \frac{\sqrt{k!}\sqrt{(l+j)!}}{l!} \ket{l+j}_A,
\end{align}
while in the SSRC framework,
\begin{align}
    \hat U\ket{k}_A\ket{N-k}_B &= \sum_{l=0}^k \sum_{j=0}^{N-k} \binom{k}{l} \binom{N-k}{j} \frac{\sqrt{(j+l)!}\sqrt{(N-j-l)!}}{\sqrt{k!}\sqrt{(N-k)!}} \left(1-\frac{\abs{\alpha}^2}{N}\right)^{\frac{N-j+l-k}{2}} \notag\\
    &\qquad\times \frac{\alpha^j(-\alpha^*)^{k-l}}{N^{(j+k-l)/2}} \ket{j+l}_A\ket{N-j-l}_B.
\end{align}
One verifies that, for each fixed $l$ and $j$, the corresponding terms coincide in the large-$N$ limit. This establishes the precise correspondence between the SSRC rotation generated by $\hat U$ and the CV displacement operator $\hat D(\alpha)$.

\paragraph{\publi Spin squeezing and quadrature squeezing}

A similar relation can be established between spin-squeezed states and CV quadrature-squeezed states. In Sec.~\ref{subsec: spin squeezing}, we introduced different classes of spin-squeezed states. In particular, the states
\begin{equation}
    \ket{z_1,z_2}_{\text{sq}} = \frac{1}{\mathcal N} (\hat c^\dagger)^{N/2} (\hat d^\dagger)^{N/2} \vac,
\end{equation}
where $\hat c^\dagger$ and $\hat d^\dagger$ are the creation operators associated with the points $z_1$ and $z_2$ on the Bloch sphere, are naturally related, in the large-$N$ limit, to CV squeezed vacuum states, provided one chooses
\begin{align}
    z_1 = e^{i\varphi/2}\sqrt{\tanh(r)}, && z_2 = -z_1.
\end{align}
In the large-$N$ limit, the expansion of Eq.~\eqref{eq: squeezed state expansion} becomes
\begin{align}
    \ket{z_1,z_2}_\text{sq} &= \frac{e^{-rN/2}}{\mathcal N} \sum_{k=0}^{N/2} \binom{N/2}{k} (-e^{i\varphi})^k \sinh^k(r) \cosh^{N/2-k}(r) \notag\\
    &\quad\times \sqrt{(2k)!}\sqrt{(N-2k)!} \ket{2k}_A\ket{N-2k}_B.
\end{align}
Using Stirling's formula $n!\sim \sqrt{2\pi n}(n/e)^n$, one verifies that for each fixed $k$ the $k$-dependent terms converge to those of a CV squeezed vacuum state. Although the normalization factor $\mathcal N$ has a non-trivial expression, normalization of the state ensures convergence of the $N$-dependent terms towards the CV normalization. One obtains
\begin{equation}
    \ket{z_1,z_2}_\text{sq} \simeq \frac{1}{\sqrt{\cosh r}} \sum_{k=0}^{N/2} \frac{\sqrt{(2k)!}}{2^k k!} \left(-e^{i\varphi}\tanh r\right)^k \ket{2k}_A\ket{N-2k}_B,
\end{equation}
which is the SSRC analogue of the CV squeezed vacuum state
\begin{equation}
    \ket{\text{sq}(r,\varphi)} = \frac{1}{\sqrt{\cosh r}} \sum_{k=0}^{\infty} \frac{\sqrt{(2k)!}}{2^k k!} \left(-e^{i\varphi}\tanh r\right)^k \ket{2k}_A.
\end{equation}
This establishes, at the level of explicit expansions, the convergence of SSRC spin-squeezed states towards quadrature-squeezed states in the continuous-variable limit.

\subsection{The Majorana and stellar representations}
\label{subsec: Majorana and stellar representation}

\paragraph{\publi The Bargmann representation of CV states}
As introduced in Sec.~\ref{subsec: majorana polynomial}, continuous-variable states
\begin{equation}
    \ket{\psi}=\sum_{n=0}^\infty c_n \ket{n}_A,
\end{equation}
admit a useful analytic representation via the Bargmann or stellar function~\cite{vourdas_analytic_2006,motamedi_stellar_2026}, defined in close analogy with the Majorana polynomial as
\begin{equation}
    F_B(z)=e^{\abs{z}^2/2}\braket{z^*}{\psi}=\sum_{n=0}^\infty c_n \frac{z^n}{\sqrt{n!}}.
\end{equation}
In contrast to the Majorana function, which is a polynomial of finite degree, the Bargmann function is in general an entire analytic function. It is therefore a slightly more involved mathematical object, although its analytic nature allows it to be analysed by powerful tools from complex analysis.

In direct analogy with the Majorana stars, defined as the roots of the Majorana polynomial, one defines the Bargmann stars as the zeros of the Bargmann function. However, contrary to the finite-dimensional case, the Bargmann function may possess infinitely many zeros, and hence infinitely many Bargmann stars. The distribution of these zeros in the complex plane encodes relevant structural properties of the state.

A particularly important subclass is that of finite stellar rank states, defined as those states for which the Bargmann function admits only a finite number of zeros. For these states, a remarkable factorized form holds~\cite{chabaud_stellar_2020}
\begin{equation}
    \ket{\psi}=\frac{1}{\mathcal N}\prod_{k=1}^{r^*}\hat D(z_k^*)\hat a^\dagger\hat D(z_k^*)^\dagger\ket{G},
\end{equation}
where $\mathcal N$ is a normalization constant, $\hat D(z_k^*)$ is the CV displacement operator, $\{z_k\}$ are the roots of the Bargmann function, $\ket{G}$ is a Gaussian state, and $r^*$ is the stellar rank of the state $\ket{\psi}$. This expression is the analogue of the factorized form of SSRC states given in Eq.~\eqref{eq: state from Majo stars}, which can be rewritten as
\begin{equation}
    \ket{\psi}=\frac{1}{\mathcal N}\prod_{k=1}^N \hat R(\theta_k,\phi_k)\hat b^\dagger\hat R(\theta_k,\phi_k)^\dagger\vac.
\end{equation}
Within this analogy, both similarities and differences can be identified:
\begin{itemize}
    \item The overall structure is closely related: in both cases, the state is written as a product of creation operators, one for each root of the corresponding function, each obtained from a reference operator by a rotation or a translation. As discussed previously, the SSRC rotation and the CV displacement are intimately connected, which explains the structural similarity between the two forms.
    \item The reference state on which this factorization acts plays a different role in the two settings. In the SSRC picture, the operators act on the vacuum state. In the CV case, the operators act on a Gaussian state $\ket{G}$, which can be non-trivial by containing squeezings or displacements.
    \item The number of factors also differs. In the SSRC case, there are always $N$ terms, independently of the state. In the CV case, the number of factors is given by the stellar rank $r^*$, which depends on the state. For example, a coherent state, which has stellar rank zero, contains no such factors, whereas a Fock state $\ket{n}$, which has stellar rank $n$, contains $n$ factors.
\end{itemize}

Although the analogy is striking, the differences highlight that additional structure is hidden in the correspondence. In particular, only some CV states possess a finite stellar rank, whereas in the SSRC description the number of Majorana stars is always equal to $N$ and thus diverges when considering the limit $N\to\infty$. This already indicates that part of the information carried by the Majorana representation is not explicitly visible in the CV description. This is consistent with the fact that the CV picture obscures the relation with the phase reference, so that certain properties which are explicit in the SSRC picture become implicit in the CV one.

\paragraph{\publi Convergence}

The formal correspondence between the Majorana polynomial and the Bargmann function in the large-$N$ limit can be made more explicit by evaluating the Majorana polynomial at $z/\sqrt{N}$. One obtains
\begin{equation}
    P_{\ket{\psi}}\left(\frac{z}{\sqrt{N}}\right) =\sum_{k=0}^N c_k\frac{z^k}{\sqrt{k!}} \sqrt{\frac{N(N-1)\cdots(N-k+1)}{N^k}} \,\simeq\, \sum_{k=0}^N c_k \frac{z^k}{\sqrt{k!}},
\end{equation}
where the last approximation holds for fixed $k$ in the limit $N\to\infty$, since
\begin{equation}
    \sqrt{\frac{N(N-1)\cdots(N-k+1)}{N^k}} \longrightarrow 1.
\end{equation}
This suggests the formal correspondence
\begin{equation}
    P_{\ket{\psi}}\left(\frac{z}{\sqrt{N}}\right)\,\longrightarrow\, F_B(z),
\end{equation}
in the regime where $N$ is large and $k\ll N$. Several remarks are in order:
\begin{itemize}
    \item The appearance of the $1/\sqrt{N}$ scaling can be compared with the $1/\sqrt{N}$ factor that arises in the correspondence between spin coherent states and CV coherent states: a CV coherent state of amplitude $\alpha$ corresponds to an SSRC spin coherent state with parameter $\alpha/\sqrt{N}$. Since the Majorana polynomial and the Bargmann function are constructed from overlaps with spin coherent states and CV coherent states respectively, the presence of the same scaling factor in both correspondences is fully consistent.
    \item For the SSRC and CV correspondence on coherent states to remain valid, the magnitude of the complex parameter must satisfy $\abs{z}\lesssim \sqrt{N}$. This sets a fundamental scale on the amplitudes that can be faithfully represented when passing from the SSRC to the CV picture. In this sense, the Bargmann function $F_B$ can be viewed as the large-$N$ limit of the Majorana polynomial, but effectively restricted to the region $\abs{z}\lesssim \sqrt{N}$. In realistic experimental situations, $N$ is macroscopic and far larger than the excitation numbers that are experimentally accessible, so that this restriction is not directly perceived. Information contained in the Majorana polynomial at larger values of $\abs{z}$ is therefore effectively ignored in the CV description. As such, physical CV systems can be seen as states that explore only a limited region of the Bloch sphere located around the north pole.
    \item The roots of the Majorana polynomial can accordingly be separated into two classes, depending on their magnitude relative to $\sqrt{N}$. In the limit $N\to\infty$, roots whose magnitude scales faster than $\sqrt{N}$ are pushed to infinity in the rescaled variable and disappear from the Bargmann function, while those whose magnitude remains below this scale survive in the CV description. This explains how a finite number of roots may remain visible in the CV picture, despite the fact that the SSRC representation always contains exactly $N$ roots. The roots that are lost in the CV limit are precisely those whose effect is tied to the phase reference structure, which is not explicitly retained in the CV framework.
    \item A more formal analysis of the transition from the SSRC to the CV picture can be carried out and provides further insight into the properties of physical states in the CV regime~\cite{moulonguet_non-gaussianity_2026}. However, such an analysis lies beyond the already extensive scope of this thesis.
\end{itemize}

\subsection{Phase spaces and geometric picture}
\label{subsec: phase spaces and geometry}

As introduced in Sec.~\ref{subsec: continuous variables}, CV systems can be described in terms of a phase space through the Wigner function. This representation provides a geometric and visual interpretation of states and transformations. In the SSRC regime, in Sec.~\ref{subsec: spherical Wigner function}, we have also shown how a Wigner function can be constructed. While the CV phase space is the 2D plane, the natural phase space of SSRC systems is the sphere. Since CV systems can be interpreted as limits of SSRC systems exhibiting a large photon number imbalance between the two modes, it is natural to ask how these two phase spaces are related.

From an intuitive point of view, SSRC states that admit a well-defined CV limit are those that explore only a limited region of the Bloch sphere located around the north pole. In this regime, the spherical phase space can be approximated by its tangent plane at the north pole, which coincides with the CV phase space. This geometric picture is consistent with the fact that the SSRC-CV correspondence involves a scaling of the complex parameter by $1/\sqrt{N}$, which effectively restricts the relevant region of phase space to a neighborhood of size $\sqrt{N}$ around the north pole.

More precisely, it has been shown in~\cite{amiet_contracting_2000} that the kernels used to construct the Wigner functions in both regimes are related through a suitable contraction procedure. As a consequence, in the large-$N$ limit, the spherical Wigner function of an SSRC state converges to the planar Wigner function of its CV counterpart. This result provides a rigorous mathematical underpinning for the geometric picture described above.

This connection also helps identify the relation between CV and SSRC transformations. While angular momentum operators generate rotations on the spherical phase space, taking the CV limit leads to the following correspondences:
\begin{itemize}
    \item Rotation around the $z$ axis, generated by $\hat J_z$, corresponds to a rotation in the planar phase space. This confirms the correspondence between $\hat J_z$ and $\hat n_a=\hat a^\dagger\hat a$ observed in Sec.~\ref{subsec: formal limit}. Notice that the change of sign observed in the correspondence between $\hat J_z$ and $\hat n_a$ is consistent. Indeed, fixing the convention for the orientation of the planar phase space such that the coherent state $\ket{\alpha}$ is centered at the coordinates $\alpha$, we see that while $e^{-i\hat J_z\theta}$ yields a rotation in the positive direction in the SSRC picture, one must consider $e^{i\hat n_a\theta}$ to obtain a rotation in the positive direction in the plane.
    
    \item Rotation around an axis in the $xy$ equatorial plane, generated by $n_x\hat J_x+n_y\hat J_y$, corresponds to a translation in the planar phase space. This confirms the correspondence between $\hat J_x$ and $\hat x$, and between $\hat J_y$ and $\hat p$, observed in Sec.~\ref{subsec: formal limit}. Once again, comparing the expression of the CV limit of $n_x\hat J_x+n_y\hat J_y$ with the CV translation operator $\hat D(\alpha)= e^{\alpha\hat a^\dagger - \alpha^*\hat a}$, we observe a full consistency between the correspondence of operators and their action on the respective phase spaces.
\end{itemize}

In Fig.~\ref{fig: phase space limits}, we illustrate this geometric picture by representing the spherical phase space and the planar geometry emerging from the vicinity of the north pole.

\begin{figure}[ht]
    \centering
    \scalebox{1}{\begin{tikzpicture}
	\begin{pgfonlayer}{nodelayer}
		\node [style=none] (0) at (-5, 0) {};
		\node [style=none] (1) at (5, 0) {};
		\node [style=none] (4) at (0, 0) {};
		\node [style=none] (5) at (0, 5.75) {};
		\node [style=none] (6) at (5.75, 0) {};
		\node [style=none] (7) at (2.75, 1.75) {};
		\node [style=none] (8) at (0, 6) {$z$};
		\node [style=none] (9) at (3.75, 2.25) {$y$};
		\node [style=none] (10) at (6.25, 0) {$x$};
		\node [style=none] (11) at (0, -5) {};
		\node [style=none] (12) at (0, -5.25) {};
		\node [style=none] (17) at (-2.425, 4.5) {};
		\node [style=none] (18) at (2.475, 4.5) {};
		\node [style=none] (19) at (10.75, -0.75) {};
		\node [style=none] (20) at (19.25, -0.75) {};
		\node [style=none] (21) at (13.25, 2.25) {};
		\node [style=none] (22) at (21.75, 2) {};
		\node [style=none] (23) at (16.25, 1.5) {};
		\node [style=none] (24) at (16.25, -1.75) {};
		\node [style=none] (25) at (16.25, 5.25) {};
		\node [style=none] (26) at (-0.325, 6.85) {};
		\node [style=none] (27) at (0, 6.5) {};
		\node [style=none] (28) at (7.275, 0.025) {};
		\node [style=none] (29) at (6.925, -0.25) {};
		\node [style=none] (30) at (2.85, 2.125) {};
		\node [style=none] (31) at (3.25, 1.75) {};
		\node [style=none] (32) at (11.5, -4.75) {};
		\node [style=none] (33) at (16.2, 6.1) {};
		\node [style=none] (34) at (16.5, 5.75) {};
		\node [style=none] (35) at (9.5, -1.25) {};
		\node [style=none] (36) at (10.75, 0.5) {};
		\node [style=none] (37) at (12.5, -1.25) {};
		\node [style=none] (38) at (11, 1) {$y$};
		\node [style=none] (39) at (13, -1.25) {$x$};
		\node [style=none] (40) at (19, 2) {};
		\node [style=none] (41) at (18.25, 0.5) {};
		\node [style=none] (42) at (14.75, 2.25) {};
		\node [style=none] (43) at (17.5, 2.25) {};
	\end{pgfonlayer}
	\begin{pgfonlayer}{edgelayer}
		\draw [bend left=90, looseness=1.75] (0.center) to (1.center);
		\draw [bend right=90, looseness=1.75] (0.center) to (1.center);
		\draw [bend right=90, looseness=0.50] (0.center) to (1.center);
		\draw [style=dashes, bend left=90, looseness=0.50] (0.center) to (1.center);
		\draw [style=Arrow] (4.center) to (5.center);
		\draw [style=Arrow] (4.center) to (6.center);
		\draw [style=Arrow] (4.center) to (7.center);
		\draw (11.center) to (12.center);
		\draw [style=dashes] (18.center)
			 to [bend left=90, looseness=0.20] (17.center)
			 to [bend left=90, looseness=0.20] cycle;
		\draw [bend left=15] (19.center) to (20.center);
		\draw [in=60, out=-150, looseness=0.50] (21.center) to (19.center);
		\draw [bend left=15] (21.center) to (22.center);
		\draw [in=60, out=180, looseness=0.50] (22.center) to (20.center);
		\draw [style=Arrow] (23.center) to (25.center);
		\draw [style=dashes] (24.center) to (23.center);
		\draw [style=arrow, in=-165, out=-165, looseness=3.00] (26.center) to (27.center);
		\draw [style=arrow, in=105, out=105, looseness=3.00] (28.center) to (29.center);
		\draw [style=arrow, in=60, out=60, looseness=2.25] (30.center) to (31.center);
		\draw [style=arrow, in=-165, out=-165, looseness=3.00] (33.center) to (34.center);
		\draw [style=Arrow] (35.center) to (36.center);
		\draw [style=Arrow] (35.center) to (37.center);
		\draw [style=arrow, in=75, out=-135] (40.center) to (41.center);
		\draw [style=arrow, bend left=15] (42.center) to (43.center);
	\end{pgfonlayer}
\end{tikzpicture}}
    \caption[Geometric picture of the transition from the SSRC spherical phase space to the CV planar phase space]{Geometric picture of the transition from the SSRC spherical phase space to the CV planar phase space. Left: the spherical phase space associated with SSRC systems. Right: the planar phase space emerging in the CV limit, obtained by approximating the sphere by its tangent plane at the north pole. In this limit, rotations around the $z$ axis remain rotations in phase space, while rotations around axes in the equatorial plane become translations in the planar phase space. The red arrows depict the direction of rotation/translation of the respective transforms.}
    \label{fig: phase space limits}
\end{figure}
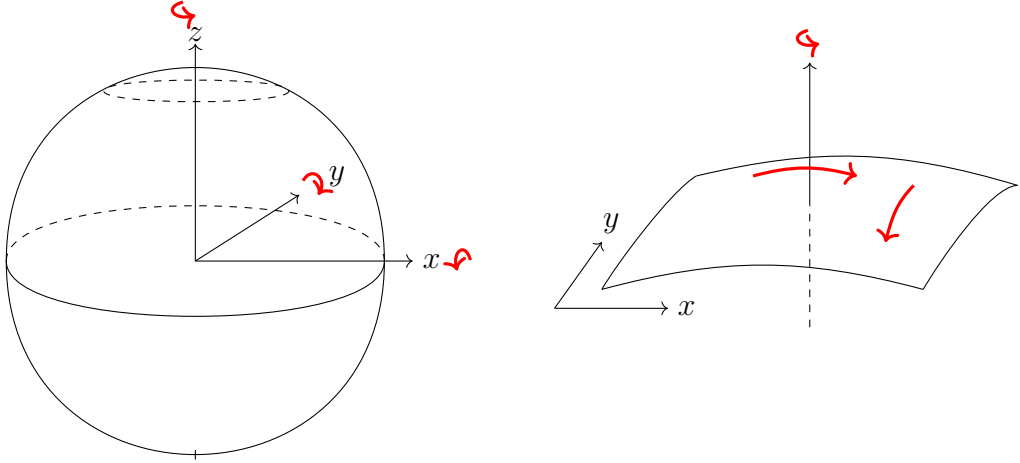

\clearpage
\section{Superselection rules and quantum computational resources in bosonic
systems}
\label{sec: SSR and resources in bosonic systems}
\emph{In this section we investigate the relation between superselection rules and the general problem of quantum computation in bosonic systems. Adopting the superselection-rules-compliant (SSRC) framework, we clarify how physically constrained bosonic states and operations give rise to computational resources, and how notions such as Gaussianity, non-classicality, and universality depend on the underlying representation. We show that discrete- and continuous-variable encodings can be understood within a unified structure, and that continuous-variable systems emerge as particular limits of fixed-particle-number ones. This perspective provides a physically grounded characterization of the resources required for universal quantum computation with bosonic platforms. This section is based on the results of \hyperlink{Article: prl SSRC}{Superselection Rules and Bosonic Quantum Computational Resources}~\cite{descamps_superselection_2024}, and \hyperlink{Article: optica SSRC}{Unified framework for bosonic quantum information encoding, resources and universality from superselection rules}~\cite{descamps_unified_2026} published during the PhD thesis.}

\subsection{Computational resources, non-classicality and physical constraints}
\label{subsec: resources, non-classicality and physical constraints}

\paragraph{\lit Quantum advantage and non-classicality}
Quantum theory gives rise to phenomena that defy classical intuition not only in physics but also in computer science~\cite{nielsen_quantum_2010}. In particular, specific quantum algorithms have been shown to outperform their classical counterparts~\cite{aaronson_computational_2013, shor_polynomial-time_1997}, leading to the notion of \emph{quantum advantage}, namely the impossibility of efficiently simulating a given protocol using classical computational resources. While this notion provides a clear computational criterion for non-classicality, the latter is less sharply defined from a physical perspective. In quantum optics, non-classicality has been associated with entanglement and non-locality~\cite{bell_einstein_1964, clauser_proposed_1969}, as well as with statistical and phase-space properties of the electromagnetic field~\cite{kimble_photon_1977,short_observation_1983,slusher_observation_1985, hertz_relating_2020}. Numerous criteria have been proposed~\cite{de_bievre_measuring_2019,innocenti_nonclassicality_2022,kuriakose_few-photon_2022,lachman_faithful_2019}, often based on properties of quasi-probability distributions~\cite{chabaud_stellar_2020,kenfack_negativity_2004,richter_nonclassicality_2002,de_bievre_complete_2021,fabre_majorana_2023} or metrological performance~\cite{hillery_total_1989,yadin_operational_2018,kwon_nonclassicality_2019,bouchard_quantum_2017}, sometimes leading to seemingly paradoxical classifications~\cite{goldberg_how_2021}. A central question is therefore whether a unified notion of non-classicality can simultaneously capture physical features and computational power.

\paragraph{\lit Universality in abstract quantum computation}
In the standard qubit model, quantum computation is formulated in a finite-dimensional Hilbert space, where universal gate sets have been identified~\cite{lloyd_almost_1995}. Combining single- and two-qubit gates, any unitary operation can be approximated efficiently, as guaranteed by the Solovay-Kitaev theorem~\cite{kitaev_quantum_1997,nielsen_quantum_2010,dawson_solovay-kitaev_2005,gottesman_heisenberg_1998}. A paradigmatic universal set is given by Clifford gates together with a non-Clifford resource such as the $\hat T$ gate. Clifford circuits acting on computational basis states can be efficiently classically simulated, as formalized by the Gottesman-Knill theorem~\cite{gottesman_theory_1998, aaronson_improved_2004}. The addition of a non-Clifford element, either directly through the $\hat T$ gate or indirectly via magic states~\cite{bravyi_universal_2005, zhou_methodology_2000}, promotes the model to universality and potentially to quantum advantage. Identifying the resources that elevate classically simulable models to universal quantum computation is thus a central theme in quantum information theory.

\paragraph{\lit Bosonic encodings and Gaussian paradigms}
When quantum information is encoded in bosonic systems, additional structure and constraints enter the picture. In CV architectures, universality can be achieved by combining quadrature-Gaussian operations with at least one quadrature non-Gaussian element~\cite{lloyd_quantum_1999}. Purely Gaussian states and operations, associated with non-negative Wigner functions~\cite{wigner_quantum_1932}, can be efficiently classically simulated~\cite{mari_positive_2012}, implying that non-Gaussianity is a necessary condition for quantum computational advantage in this setting. However, non-Gaussianity alone is not sufficient~\cite{calcluth_efficient_2022,calcluth_sufficient_2024,chabaud_classical_2021}. More refined analyses relate computational power to quantities such as stellar rank~\cite{chabaud_stellar_2020,chabaud_resources_2023} and non-Gaussian entanglement~\cite{walschaers_non-gaussian_2021}. These results highlight the subtle interplay between physical properties and computational resources in bosonic systems.

\paragraph{\publi Physical constraints and the need for a unified framework}
Unlike abstract qubit models, physical bosonic systems are subject to constraints such as symmetrization and conservation laws. These restrictions shape both the accessible state space and the set of implementable operations, leading to a nontrivial relation between modes, particles, and information encoding. While different bosonic encodings, ranging from fixed-particle-number schemes to continuous-variable ones, have been developed, it remains unclear how to classify general bosonic states in terms of their computational power in a representation-independent manner. This motivates the introduction of a systematic framework in which bosonic states can be interpreted through their induced action on a digital quantum computer, thereby allowing one to identify which physical resources promote a classically simulable architecture to universality.

\subsection{Bosonic encodings and quantum computation}
\label{subsec: bosonic encodings and quantum computation}
\paragraph{\lit Fixed-particle-number multimode encoding}
A first and paradigmatic strategy to encode quantum information in bosonic systems is the single-photon, fixed-particle-number regime. In a bosonic quantum computer (BQC), $N$ qubits are encoded in $2N$ orthogonal modes occupied by $N$ photons, with at most one photon per mode~\cite{knill_scheme_2001}. For instance, in dual-rail encoding~\cite{chuang_simple_1995}, the logical states of the $i$-th qubit are defined as
\begin{align}
    \ket{0}_i = \hat a^\dagger_{i,0}\vac, && \ket{1}_i = \hat a^\dagger_{i,1}\vac,
\end{align}
where the indices label orthogonal modes ({\it e.g.}, polarization or path). Although photons are bosons and cannot be individually identified with qubits, the restriction to one photon per pair of modes makes the system behave effectively as an $N$-qubit register spanning a $2^N$-dimensional Hilbert space.

In this encoding, one-qubit gates correspond to mode manipulations. Linear optical elements implement passive, photon-number-preserving operations, while scalable two-qubit entangling gates require nonlinear photon-photon interactions, \ie, quadrature non-Gaussian (QNG) operations. Hence, achieving universality demands physical resources beyond passive Gaussian transformations. The structure of computational resources is therefore intimately tied to the multimode architecture and to the fixed total particle number.

\paragraph{\lit Single-mode continuous-variable encoding}
A conceptually distinct approach consists in encoding qubits within orthogonal states of a \emph{single} bosonic mode. In continuous-variable (CV) encodings such as cat codes~\cite{mirrahimi_dynamically_2014, leghtas_deterministic_2013} or the Gottesman-Kitaev-Preskill (GKP) code~\cite{gottesman_encoding_2001}, logical states are defined as
\begin{align}
    \ket{\overline{0}}_i = \sum_{n=0}^\infty c_n^{(0)} \ket{n}_i, && \ket{\overline{1}}_i = \sum_{n=0}^\infty c_n^{(1)} \ket{n}_i,
\end{align}
with orthogonality ensured within an infinite-dimensional single-mode Hilbert space. Here, the continuous nature of quadrature variables and the unbounded Fock basis play a central role. Universality is formulated in terms of operations generated by polynomials of quadrature operators~\cite{lloyd_quantum_1999}, and the distinction between quadrature-Gaussian (QG) and QNG operations becomes central for assessing classical simulability.

Unlike the fixed-particle-number case, the CV encoding explicitly exploits the infinite dimensionality of the mode. Ideal code states often require infinite energy and must be approximated in practice, introducing intrinsic errors. Moreover, the identification of the minimal physical resources required for universality, such as the exact combination of QG and QNG elements, depends subtly on the encoding and is not yet fully clarified.

\paragraph{\publi Apparent incompatibility and need for unification}
These two encoding paradigms rely on seemingly different physical features: in the single-photon case, logical states correspond to the \emph{same} field excitation distributed over different modes, yielding a finite-dimensional effective Hilbert space; in the CV case, logical states correspond to different superpositions within a \emph{single} mode, and infinite dimensionality appears as an essential ingredient. Consequently, notions such as coherence, entanglement, Gaussianity, and non-Gaussianity do not play identical roles across the two frameworks.

This apparent mismatch reflects a deeper tension between abstract computational resources and their physical implementation. While both encodings can in principle support universal quantum computation, the identification of necessary and sufficient bosonic resources, independently of the chosen encoding, remains unclear. This motivates the search for a unifying structure capable of encompassing fixed-particle-number and continuous-variable approaches within a single, physically consistent framework.

\subsection{Angular Momentum Universality and Bosonic Resources}
\label{subsec: angular momentum universality and bosonic resources}

\paragraph{\publi SSRC representation and angular momentum structure.}
Superselection-rules-compliant (SSRC) systems have been extensively introduced in the previous sections, albeit restricted to the two-mode setting. We now generalize this construction to an arbitrary number $K$ of orthogonal modes, each characterized by a creation operator $\hat a_k^\dagger$. A general pure SSRC state with fixed total photon number $N$ can be written as
\begin{equation}
    \ket{\Psi} = \sum_{\{n_k\}:\,\sum_{k=1}^{K} n_k = N}  
    c_{n_1,\dots,n_K}  
    \ket{n_1}_1 \cdots \ket{n_K}_K,
\end{equation}
with normalized complex coefficients, where $\ket{n_k}_k$ denotes a Fock state in mode $k$. This representation explicitly enforces the global $U(1)$ symmetry associated with particle-number conservation and incorporates all phase references as internal quantum degrees of freedom.

For any pair of modes $(k,l)$, one can define angular momentum operators through the Schwinger construction,
\begin{align}
    \hat J_x^{(k,l)} = \frac{1}{2}\left(\hat a_k^\dagger \hat a_l + \hat a_l^\dagger \hat a_k \right), &&
    \hat J_y^{(k,l)} = \frac{i}{2}\left(\hat a_k^\dagger \hat a_l - \hat a_l^\dagger \hat a_k \right), &&
    \hat J_z^{(k,l)} = \frac{1}{2}\left(\hat a_l^\dagger \hat a_l - \hat a_k^\dagger \hat a_k \right),
\end{align}
which satisfy the angular momentum commutation relations. In this way, multimode SSRC bosonic systems can be viewed as symmetric spin systems, extending the two-mode spin correspondence to a fully multimode setting and providing a natural algebraic structure for the analysis of bosonic computational resources.

\paragraph{\publi Linear and quadratic generators: universality and Gaussianity.}
Results established for symmetric spin systems~\cite{gutman_universal_2024} imply that the unitaries generated by linear and quadratic functions of the angular momentum operators form a universal gate set within each fixed-$N$ symmetric subspace. Concretely, the operations
\begin{align}
    \exp(-i\chi \hat J_{\vec n}^{(k,l)}),
    && 
    \exp(-i\chi \left(\hat J_{\vec n}^{(k,l)}\right)^2),
\end{align}
where $\hat J_{\vec n}^{(k,l)}$ denotes the angular momentum component along a unit vector $\vec n$ built from two arbitrary orthogonal modes $k$ and $l$, allow one to approximate any unitary transformation in the SSRC Hilbert space with arbitrary precision, using a number of gates that scales polynomially with $N$.

This structure naturally leads to an SSRC-adapted notion of Gaussianity. The linear generators $\hat J_{\vec n}^{(k,l)}$ implement passive, photon-number-preserving transformations, physically corresponding to beam splitters and phase shifters, and act as mode basis changes. In first-quantization language, they operate independently on each particle and do not generate particle-particle interactions. We therefore refer to the associated unitaries as \emph{superselection-rules-compliant Gaussian} (SG) operations.  

In contrast, the quadratic generators $\left(\hat J_{\vec n}^{(k,l)}\right)^2$ correspond to effective particle interactions (spin-squeezing or Kerr-like terms). These operations modify intrinsic state properties rather than merely redefining the mode basis, and they are responsible for entanglement generation within the symmetric subspace. We denote them as \emph{superselection-rules-compliant non-Gaussian} (SNG) operations. Within the SSRC framework, the distinction between SG and SNG operations is therefore grounded in a clear physical criterion: single-particle mode transformations versus genuine interaction terms.

\paragraph{\publi Representation dependence of Gaussianity and unification.}
This perspective sheds new light on the usual Gaussian/non-Gaussian distinction of the continuous-variable (CV) framework. In the quadrature representation, Gaussianity is defined in terms of the preservation of Gaussian Wigner functions and is tied to transformations generated by operators at most quadratic in $\hat x$ and $\hat p$. However, when re-expressed in the SSRC picture, some operations that are Gaussian in the CV sense correspond simply to mode basis changes (SG), while others may effectively encode interaction-like processes once particle-number conservation is enforced. Conversely, operations that appear non-Gaussian in the CV description must reduce to interaction terms that are naturally classified as SNG in the SSRC algebra.

The SSRC/CV squeezing transition discussed in Sec.~\ref{subsec: examples of SSRC CV limits} explicitly illustrates this ambiguity. In the CV framework, squeezed vacuum states are Gaussian states generated by QG operations. In contrast, their SSRC counterparts can be represented as states with two Majorana stars,
\begin{equation}
    \ket{z_1,z_2}_{\text{sq}} = \frac{1}{\mathcal N} (\hat c^\dagger)^{N/2} (\hat d^\dagger)^{N/2} \vac ,
\end{equation}
where $\mathcal N$ is a normalization constant and $\hat c^\dagger$, $\hat d^\dagger$ denote the creation operators associated with the two points $z_1$ and $z_2$ on the Bloch sphere. Importantly, such a state cannot be produced from a spin coherent state using SG operations alone. Consequently, starting from an initial classical state $\ket{N}_B$, its preparation requires the application of SNG operations, which we denote by $\hat U_{\cal S}$. However, the situation changes when considering the CV limit. In this limit, both $\ket{N}_B$ and $\ket{z_1,z_2}_{\text{sq}}$ become QG states. As a result, the operation $\hat U_{\cal S}$ that connects them, which is SNG in the SSRC description, effectively becomes a QG operation in the CV description. 

This observation highlights an important conceptual point: the operational distinction between SG and SNG transformations, which is well defined in the finite spin representation, is not preserved under the CV limit. In other words, operations that are classified as non-Gaussian within the SSRC framework can correspond to Gaussian operations in the CV regime. This example therefore explicitly illustrates how the physical consistency of the SG/SNG classification can be lost when passing to the CV limit.

Gaussianity thus becomes representation-dependent: it is not an intrinsic property of the physical process alone, but of the chosen description (CV versus SSRC). The SSRC framework resolves this ambiguity by tying the classification directly to the underlying particle structure and symmetry constraints.  In this way, both fixed-particle-number multimode encodings and continuous-variable schemes can be embedded into the same algebraic structure. The CV description appears as a particular limit of the SSRC formalism, while the spin-universality result ensures that linear (SG) and quadratic (SNG) angular momentum generators suffice to connect arbitrary states within a fixed-$N$ sector. This establishes a unified, physically consistent characterization of universality in bosonic systems, bridging discrete- and continuous-variable paradigms through the common language of symmetric spin dynamics.

\subsection{Extraction protocol and computational interpretation}
\label{subsec: extraction protocol}

\paragraph{\publi Motivation and operational strategy.}
We now establish a precise operational bridge between SSRC states and the standard model of quantum computation. The central idea is to define a physically consistent criterion of classicality and non-classicality by mapping SSRC states to states of an $N$-qubit Bosonic Quantum Computer (BQC). This mapping allows us to import notions such as separability, entanglement, universality, and classical simulability into the SSRC framework.

The tool enabling this connection is an extraction protocol, inspired by~\cite{killoran_extracting_2014}, that converts states of $N$ indistinguishable bosons into states of $N$ distinguishable modes. Crucially, this protocol transforms \emph{particle separability} into \emph{modal separability} without introducing additional physical resources. As a consequence, computational resources in the BQC can be directly associated with intrinsic properties of SSRC states.

\paragraph{\publi Mode expansion and qubit encoding.}
Consider a single bosonic mode ${\bf A}$ containing $N$ photons. We expand this mode in a $2N$-dimensional orthogonal basis,
\begin{equation}
    \hat a_{\vec q} = \sum_{k=1}^{N}\left( q_k \hat b_k(0) + \overline q_k \hat b_k(1)\right),
\end{equation}
where $k=1,\dots,N$ labels an external degree of freedom (for instance, propagation direction), and $p\in\{0,1\}$ denotes an internal degree of freedom (such as polarization or a spatial sub-mode). The coefficients satisfy $\sum_{k} \abs{q_k}^2 + \abs{\overline q_k}^2 = 1$. A single-mode Fock state then reads
\begin{equation}
    \ket{N}_{\vec q} = \frac{(\hat a_{\vec q}^\dagger)^N}{\sqrt{N!}}\vac.
\end{equation}

To complete the extraction, we project onto the subspace containing exactly one excitation in each external mode,
\begin{equation}
    \mathcal P_{\mathcal S_N} = \prod_{k=1}^{N} \mathcal P_k,
\end{equation}
where 
\begin{equation}
    \mathcal{P}_k  = \left(\hat{b}_k^{\dagger}(0)\vac\bra{\text{vac}} \hat{b}_k(0) + \hat{b}_k^{\dagger}(1)\vac\bra{\text{vac}} \hat{b}_k(1)\right)\otimes\1_{\ell\ne k},
\end{equation}
acts on the pair of modes $\hat b_k(0),\hat b_k(1)$ and as the identity on all other pairs, enforcing single occupancy of mode $k$. One obtains
\begin{equation}\label{eq: extraction projection on fock state}
    \mathcal P_{\mathcal S_N} (\hat a_{\vec q}^\dagger)^N \vac \propto \prod_{k=1}^{N} \left( \cos\theta_k \hat b_k^\dagger(0) + e^{i\varphi_k} \sin\theta_k \hat b_k^\dagger(1) \right)\vac.
\end{equation}

\paragraph{\publi Particle separability versus modal separability.}
A single-mode state $\ket{N}_{\vec{q}}$, which describes a separable particle state, is therefore converted into a mode-separable state. Moreover, the state in Eq.~\eqref{eq: extraction projection on fock state} is a separable state of a BQC, where each external mode $k$, $(k=1,2,\cdots,N)$ labels a qubit prepared in the state 
\begin{equation}
    \left(\cos{\theta_k}\hat b_k^{\dagger}(0)+e^{i\varphi_k}\sin{\theta_k}\hat b_k^{\dagger}(1)\right)\vac.
\end{equation}
The $2N$ modes $\{k\}\times \{0,1\}$ are orthogonal, satisfying 
\begin{align}
    [\hat{b}_k(p),\hat{b}_{k'}(p')] = 0, && [\hat{b}_{k}(p),\hat{b}^{\dagger}_{k'}(p')] = \delta_{kk'}\delta_{p p'}.
\end{align}
With this encoding, the computational basis $\{\ket{x}\}$ is given by
\begin{equation}
    \ket{x} = \prod_{k=1}^{N}\hat{b}^\dagger_k(x_k)\vac,
\end{equation}
where $x\in \{0,\cdots, 2^{N}-1 \}$ and $x_1x_2\cdots x_k \cdots x_{N}$ is the binary representation of $x$. Notice that the above projection transforms particle separability into modal separability~\cite{morris_entanglement_2020} without creating additional resources.

In the conversion protocol described above we have considered a single-mode Fock state $\ket{N}_{\vec{q}}$. As we will see in the following, the protocol can also be executed for arbitrary states of the form
\begin{equation}
    \ket{n}_{\vec{q}}\ket{N-n}_{\vec{w}},
\end{equation}
where $\vec{w}$ is an arbitrary mode orthogonal to $\vec{q}$ that serves as a phase reference. Finally, the described mapping can also be applied to superpositions
\begin{equation}
    \ket{\Psi}=\sum_{n=0}^{N} c_n\ket{n}_{\vec{q}}\ket{N-n}_{\vec{w}}.
\end{equation}
In this case, the coefficients $c_n$ are directly related to the non-classical properties of states in the SSRC representation and of the BQC, such as entanglement, as will be discussed in the following.

\paragraph{\publi Effect of SSRC Gaussian (SG) operations.}
We now analyze how SSRC operations are mapped into the BQC picture. First consider SG operations, generated by linear angular momentum operators. These correspond to passive mode transformations
\begin{equation}
    e^{-2 i \eta \hat J_x} (\hat a_{\vec q_1}^\dagger)^N \vac 
    = (\cos\eta \, \hat a_{\vec q_1}^\dagger - i\sin\eta \, \hat a_{\vec w}^\dagger)^N \vac,
\end{equation}
where $\hat J_x$ denotes the $x$ angular momentum operator coupling the two modes associated with $\vec q_1$ and $\vec w$. Such rotations merely redefine the mode $\vec q_1 \to \vec q$ and preserve the intrinsic single-mode structure of the state. No superposition over different occupation sectors is created. 

Under the extraction protocol, these transformations correspond to local single-qubit rotations in the BQC. Since local operations cannot generate entanglement, the BQC remains within the separable subspace. Consequently, Fock states acted upon solely by SG operations are efficiently classically simulable. This identifies the \emph{classical sector} of the SSRC representation: although these states may appear non-Gaussian in the quadrature sense, their manipulation by SG operations does not promote the associated BQC beyond separable states.

\paragraph{\publi Effect of SSRC non-Gaussian (SNG) operations.}
We now move to non-Gaussian operations, for instance
\begin{equation}
    e^{-4i \eta  \hat J_{z}^2 }=e^{-i\eta( \hat n_{\vec{q}_1}-\hat n_{\vec{w}})^2 },
\end{equation}
where $\hat n_{\vec{q}_1}$ and $\hat n_{\vec{w}}$ are the number operators of the two modes. This operation can be applied to a rotated state $\ket{N}_{\vec{q}}$, where  $\hat a_{\vec{q}}^{\dagger}= u_1\hat a_{\vec{q}_1}^{\dagger}+u_w\hat a_{\vec{w}}^{\dagger}$ with $\abs{u_1}^2+\abs{u_{w}}^2=1$. We obtain 
\begin{align}\label{eq: quadratic evolution on spin N}
    e^{-4i\eta  \hat J_{z}^2 }\ket{N}_{\vec{q}} 
    &\propto e^{-4i\eta  \hat J_{z}^2 }\left (u_1\hat a_{\vec{q}_1}^{\dagger}+u_w\hat a_{\vec{w}}^{\dagger}\right )^N\vac\notag\\
    &= \sum_{n=0}^N\binom{N}{n}\left ( u_1e^{-i\eta n}\hat a_{\vec{q}_1}^{\dagger}\right )^n 
    \left (u_w e^{-i\eta(N-3n)} \hat a_{\vec{w}}^{\dagger}\right )^{N-n}\vac.
\end{align} 

The state of Eq.~\eqref{eq: quadratic evolution on spin N} cannot in general be transformed by rotations into a single-mode state. It is a non-Gaussian particle-entangled state that is transferred to a mode-entangled state of the BQC after the extraction protocol. Notice that this state is analogous to an OAT spin-squeezed state (see Sec.~\ref{subsec: spin squeezing}).

To better analyze how the unitary $e^{-4i\eta  \hat J_{z}^2 }$ can create mode entanglement and highlight the importance of a phase reference, we study the example $N=2$ (see~\cite{descamps_superselection_2024} for the general case which follows the same principles as for $N=2$). We use, for instance, $\hat a_{\vec{q}}=\left ( \hat a_{\vec{q}_1} +\hat a_{\vec{w}} \right )/\sqrt{2}$ (leading to a $\hat J_x$ eigenstate), which gives 
\begin{equation}
    e^{-4i\eta  \hat J_{z}^2 }\ket{2}_{\vec{q}}
    =\frac{1}{2}\left ( e^{-4i\eta}\ket{2}_{\vec{q}_1}\ket{0}_{\vec{w}}
    +e^{-4i\eta}\ket{0}_{\vec{q}_1}\ket{2}_{\vec{w}}
    +\sqrt{2} \ket{1}_{\vec{q}_1}\ket{1}_{\vec{w}} \right ).
\end{equation}
This state cannot be written as a single-mode state unless $\eta = n\pi/4$, $n \in \N$. Physically, the quadratic interaction in Eq.~\eqref{eq: quadratic evolution on spin N} corresponds to a cross-Kerr interaction~\cite{noauthor_optical_2010, kok_linear_2007, scala_deterministic_2024, spagnolo_non-linear_2023}.

The non-classical properties of this final state and, more generally, states such as the one in Eq.~\eqref{eq: quadratic evolution on spin N}, can be extracted and converted into quantum computational resources of the BQC, which provides an alternative way to assess them. For instance, setting, in the previous example,
\begin{align}
    \hat a_{\vec{q}_1}&=\frac{1}{2}\left (  \hat b_1(0)+ \hat b_2(0)+  \hat b_1(1)+ \hat b_2(1) \right ), &
    \hat a_{\vec{w}}&=\frac{1}{2}\left (  \hat b_1(0)+ \hat b_2(0)-\hat b_1(1)- \hat b_2(1) \right ),
\end{align}
we find 
\begin{equation}
    {\cal P}_{{\cal S}_2}e^{-4i\eta  J_{z}^2 }\ket{2}_{\vec{q}} 
    \propto 
    \left ( \hat b_1^{\dagger}(0)\hat b_2^{\dagger}(0)(e^{-4i\eta }+1)
    + \hat b_1^{\dagger}(1)\hat b_2^{\dagger}(1)(e^{-4i\eta }-1) \right )\vac.
\end{equation}

If $\eta = n\pi/4$, for $n\in\N$, the state is separable. Otherwise, it is a particle- and mode-entangled state of the BQC as well. Moreover, if $\eta =\pi/8$, $e^{-4i\eta  J_{z}^2 }$ is mapped into a conditional operation that creates maximally entangled states in the BQC. Combined with rotations, this gate enables the efficient exploration of the full Hilbert space of the BQC by moving it outside of the subspace of separable states, and completes the set of universal gates in the BQC~\cite{shor_scheme_1995,kok_linear_2007,howard_qudit_2012,brylinski_universal_2001}.

\paragraph{\publi Non-classicality as computational promotion and unification.}
The extraction protocol allows us to formulate a physically grounded and operational criterion of non-classicality within the SSRC framework. Single-mode Fock states and their manipulation by SSRC Gaussian (SG) operations are mapped to separable states of the BQC and to local qubit operations. The associated dynamics therefore remains confined to a classically simulable sector, despite the fact that such states may appear non-Gaussian in the usual quadrature representation.

In contrast, SSRC non-Gaussian (SNG) operations, generated by quadratic angular momentum terms and corresponding physically to particle interactions, produce mode-entangled states which, after extraction, become entangled qubit states. When combined with SG operations, these SNG transformations complete a universal gate set in the BQC.

Non-classicality can therefore be defined operationally as the ability of a bosonic state or operation to promote the extracted BQC from a separable, efficiently simulable regime to a universal one. In this sense, non-classical states constitute genuine computational resources. A universal bosonic quantum computer requires, as elementary resources, Fock states distributed across different modes together with at least one non-classical (interaction-generated) operation capable of inducing entanglement in the BQC.

Beyond providing a classification of resources, this construction unifies previously distinct hardware descriptions. The SSRC representation (and its continuous-variable limit), the BQC model, and symmetric spin systems can all be embedded into the same formal structure. Different bosonic encodings, including continuous-variable codes, can be systematically represented in the SSRC picture and translated, via extraction, into corresponding multi-qubit states. The total photon number then acquires a direct informational meaning: it determines the maximal number of physical qubits available in the associated BQC, establishing a transparent link between bosonic resources and qubit computational capacity.

This unified viewpoint clarifies the role of Gaussian and non-Gaussian resources in quantum advantage. Gaussian transformations correspond to mode basis changes and remain computationally local in the extracted picture, whereas non-Gaussian interaction terms generate the entanglement necessary for universality. The SSRC framework thus provides essential conceptual tools for analyzing bosonic quantum states in terms of their computational power, while explicitly accounting for phase references and symmetry constraints, and bridging discrete-variable, continuous-variable, and spin-based approaches within a single consistent formalism.

Importantly, the extraction protocol introduced in this section establishes a link between the SSRC formalism and a specific model of computation. The bosonic quantum computer corresponds to the particular case of single-photon encoding, in which at most one photon is present in each orthogonal mode. Within this framework, the protocol allows us to establish a direct correspondence between particle entanglement and computational resources. Indeed, for single-photon encoding, particle entanglement is equivalent to mode entanglement, which provides a clear operational interpretation of the resources within the SSRC formalism. However, this perspective remains model dependent. The interpretation of complexity and universality obtained in this setting relies on the assumption of single-photon encoding and therefore on the specific structure of the associated Hilbert space. As a consequence, although the correspondence between particle entanglement and computational resources is well defined in this context, it does not immediately generalize to other bosonic encodings in which several photons may occupy the same mode. The next section aims precisely at lifting this constraint by developing an encoding-independent understanding of this problem.

\subsection{Encoding-independent conditions for universality}
\label{subsec: encoding independent universality}

\paragraph{\publi Motivation.}
Having established an operational connection between SSRC states and qubit quantum computation through the extraction protocol, we now turn to a more general question: what are the fundamental physical resources required for universal quantum computation in bosonic systems? Specifically, we ask whether these resources depend on the particular encoding of information (discrete-variable, continuous-variable, cat codes, GKP codes, etc.), or whether they can be formulated in an encoding-independent manner.

The SSRC framework provides a natural language to address this question, because it separates intrinsic particle properties from mode-dependent descriptions and allows all bosonic encodings to be treated within the same finite-dimensional Hilbert space associated with a fixed total photon number $N$.

\paragraph{\publi Measurement basis and computational resources.}
We begin with a remark on measurement-related resources in bosonic quantum information. Measurements themselves constitute quantum resources, as they are formally described by positive operator-valued measures (POVMs). For instance, homodyne and heterodyne detections are often classified as QG operations, while photon-number-resolving measurements are typically regarded as QNG. However, this distinction relies on an approximation valid only in the weak-coupling limit between light and matter. In general, photodetection involves interactions that are not strictly Gaussian, and treating them as such neglects important physical subtleties, as illustrated in previous examples.

This point has been emphasized in earlier works~\cite{tyc_operational_2004, molmer_optical_1997}, where homodyne detection can be interpreted as a non-Gaussian measurement associated with POVM elements of the form $\hat{\Pi}_n=\ketbra{n}$. This observation is particularly relevant in protocols such as Boson Sampling~\cite{aaronson_computational_2013}, where mode-resolved detection reveals computational resources or entanglement~\cite{killoran_extracting_2014, morris_entanglement_2020} that may not be apparent from the state description alone.

From the perspective of quantum computation, measurements primarily define the computational basis. Accordingly, in this work we assume that detection is performed in the chosen computational basis. Under this assumption, measurements do not introduce additional physical or computational resources beyond those already present in the encoded states, but rather reflect the resources inherent to the encoding itself.

Consequently, our focus is on the SG and SNG resources required to implement universal gate sets independently of the chosen encoding, rather than on the specific states used to realize that encoding. In this sense, the choice of an encoding, and hence a corresponding class of states, plays a role analogous to selecting a classicality criterion or defining a pointer basis~\cite{zurek_pointer_1981}.

\paragraph{\publi General bosonic encodings.}
Within the SSRC representation, a general bosonic qubit encoding can be constructed from two orthogonal modes $(a_k,b_k)$ associated with the $k$-th logical qubit. The logical basis states can be written as
\begin{align}
    \ket{\overline 0}_{k}=\hat U^{(a_k,b_k)}_0\ket{N}_{b_k}, && \ket{\overline 1}_k=\hat U^{(a_k,b_k)}_1\ket{N}_{a_k},
\end{align}
where for $j=0,1$, $\hat U^{(a_k,b_k)}_{j}=e^{-i {\cal H}^{(k)}_{j}(\hat J^{(a_k,b_k)}_{\vec n})}$ are unitary operators generated by polynomials in angular momentum operators coupling only the two orthogonal modes $(a_k,b_k)$. 

This description encompasses a wide range of bosonic encodings. For example, the choice $\hat U_0=\hat U_1=\1$ corresponds to a Fock-state encoding, while coherent-state encodings and cat codes arise in the continuous-variable limit of the same construction. More elaborate encodings, such as GKP states, can be expressed through suitable nonlinear generators acting on the SSRC Hilbert space. The SSRC framework thus provides a unified description of both discrete-variable and continuous-variable encodings.

\paragraph{\publi Encoding-independent universality conditions.}
We can now state the central result: the physical resources required for universality in bosonic quantum computation are independent of the chosen encoding. For any bosonic qubit encoding based on orthogonal modes, two conditions must be satisfied, extending and generalizing the observations made via the extraction protocol of Sec.~\ref{subsec: extraction protocol}.

\begin{itemize}
    \item First, as shown in Appendix~\ref{app: SSRC}, Result~\ref{res: SNG required for local gates}, the set of available local gates must include SNG operations. Linear mode transformations alone are insufficient to achieve universality. Equivalently, this implies that the physical implementation of both the Clifford Hadamard gate $\hat H$ and the non-Clifford phase gate $\hat T$ cannot be realized solely with SG operations. Since these two gates can generate arbitrary single-qubit rotations, if both could be implemented by SG operations alone, any single-qubit rotation could be realized with SG operations, which is not the case.
    \item Second, as shown in Appendix~\ref{app: SSRC}, Result~\ref{res: SNG required for cnot gates}, entangling operations between qubits must also be realized through SNG interactions. Unlike abstract qubit models, where entangling gates can be postulated independently of the underlying physics, bosonic architectures require interaction-generated operations to produce conditional dynamics between photons.
\end{itemize}
An important exception arises in the case of single-photon encodings ($N=1$), where linear mode manipulations coincide with transformations of the encoding states. In this situation, SG operations alone can implement arbitrary single-qubit gates.

\paragraph{\publi Relation to continuous-variable quantum computing.}
In this perspective, CV quantum computing is not a fundamentally distinct paradigm but rather a limit of the SSRC framework in which the total photon number is large compared to the number of occupied excitations. In this limit, certain SG operations correspond to the familiar Gaussian operations in the quadrature formalism (such as displacements), while some SNG interactions may appear Gaussian in the quadrature description, such as squeezing as we saw earlier. Consequently, the Gaussian/non-Gaussian distinction becomes representation-dependent and may obscure the underlying physical resources. The SSRC perspective resolves this ambiguity by identifying the fundamental resource as interaction-generated operations, independent of the encoding or representation used to describe the bosonic field.

In the context of CV quantum computing with optical systems, it is often claimed that beam splitter operations (or more generally, passive linear optics such as the SG gate) generate entanglement. While such operations can produce mode entanglement, this corresponds merely to a change of basis in the mode space and does not generate qubit entanglement. Intuitively, this can be understood by noting that these operations are local in the first-quantization picture; they do not allow for conditional interactions between photons in different modes.

\subsection{Implications for Quantum Advantage and Resource Theories}
\label{subsec: quantum advantage and resources}

The framework developed in this section provides a physically grounded and unifying perspective on the resources necessary for quantum advantage in bosonic systems. By formulating bosonic quantum states in the SSRC representation and linking them to a bosonic quantum computer (BQC) via the extraction protocol, we have identified that the fundamental resources are not representation-dependent notions such as quadrature Gaussianity or Wigner negativity, but rather the interaction structure encoded in SNG operations. Linear, mode-preserving SG operations are insufficient for universality; they only transform modes without generating particle interactions or increasing the informational complexity of a state. In contrast, SNG operations correspond to genuine particle interactions, induce particle entanglement, and are necessary for creating both local and non-local correlations that enable a BQC to achieve universality.

This distinction clarifies several apparent paradoxes in continuous-variable (CV) quantum information. Common criteria, such as Wigner negativity or the stellar rank, while often associated with non-classicality and computational advantage, can depend sensitively on the choice of representation or basis. For instance, Fock states are non-Gaussian in the quadrature representation but are classical-like in the SSRC sense, whereas coherent states appear Gaussian yet correspond to rotated Fock states relative to a phase reference. By grounding the analysis in SSRC, we show that the intrinsic physical properties of bosonic states, particle separability and interaction-induced entanglement, are the true determinants of computational resources, rather than the superficial classification of states in a particular phase-space representation.

Our results also demonstrate that necessary and sufficient resources for universality are largely independent of the encoding, with the exception of single-photon encodings. Both discrete-variable (DV) and CV encodings can be embedded into the same SSRC framework, revealing that CV systems are not a fundamentally distinct paradigm, but rather a limiting case of multimode SSRC states with highly populated reference modes. This unification provides a clear operational meaning to notions of classicality and non-classicality: excluding the single-photon encoding, states that fail to generate entangling SNG operations remain efficiently classically simulable, whereas states that enable SNG-induced entanglement promote the system to universality and potential quantum advantage.

The SSRC formalism also provides a resolution to longstanding conceptual ambiguities. By working in a finite-dimensional, particle-number-conserving Hilbert space, we circumvent the mathematical and physical limitations of conventional CV approaches, such as unbounded Hilbert spaces or reliance on unphysical, infinitely squeezed states. Physical operations and states are described consistently, and universal gates can be implemented and analyzed without appealing to abstract, representation-dependent constructs. Consequently, our framework recasts previously studied CV-based measures, such as Wigner negativity, non-Gaussianity, and stellar rank, in a physically meaningful context, where their role as indicators of computational advantage is fully captured by the presence of SNG-induced interactions.

Finally, the SSRC perspective provides a versatile foundation for analyzing quantum protocols across bosonic platforms, including optical, atomic, and spin-based systems. By explicitly accounting for phase references and the particle-number structure, it bridges DV and CV regimes, unifies resource identification, and clarifies the origin of quantum advantage in bosonic quantum computation. In doing so, it lays the groundwork for systematic studies of error correction, multimode encodings, and generalized resource theories that are encoding-independent, physically consistent, and directly applicable to experimental architectures.

\clearpage
\section{Superselection rules and quantum metrology}
\label{sec: SSRC and metrology}
\emph{In this section, we discuss the connection between SSRC and metrological estimation problems. In particular, we show how adopting the SSRC point of view clarifies the notion of resource, explicitly identifying the roles of mode and particle entanglement, and allowing a unified description of seemingly incompatible bosonic regimes within a single framework. This section is based on the results of \hyperlink{Article: PRA SSRC metro}{Quantum-enhanced precision from a superselection rules perspective}~\cite{saharyan_resources_2026}, published during the PhD thesis.}

\subsection{Bosonic resources for quantum metrology}
\label{subsec: bosonic resources for quantum metrology}

As introduced in Sec.~\ref{sec: metrology}, quantum metrology investigates how quantum states and dynamics can be exploited to enhance the precision with which physical parameters are estimated. In classical estimation strategies employing independent or classically correlated probes, the achievable precision is fundamentally limited by the shot-noise scaling, which decreases as the inverse square root of the available resources such as energy or probe number. Quantum protocols can overcome this limitation and reach quadratic improvements in precision for a fixed resource budget~\cite{giovannetti_quantum-enhanced_2004, giovannetti_quantum_2006}. Such quantum-enhanced metrological strategies have already been demonstrated in a wide variety of physical platforms, ranging from gravitational-wave detectors and atomic sensors to biological measurements~\cite{ligo_scientific_collaboration_and_virgo_collaboration_observation_2016, toussaint_quantum_2004, jones_magnetic_2009, taylor_quantum_2016, crespi_measuring_2012}.

\paragraph{\publi Bosonic platforms for quantum-enhanced sensing}

A particularly broad class of systems used in quantum metrology is provided by bosonic degrees of freedom. These include single photons and photonic fields~\cite{olson_linear_2017, giovannetti_quantum-enhanced_2001}, intense optical modes~\cite{pezze_mach-zehnder_2008, dowling_quantum_2008, polino_photonic_2020, duivenvoorden_single-mode_2017, fadel_quantum_2025, holland_interferometric_1993, kwon_quantum_2022, pinel_ultimate_2012}, atomic ensembles~\cite{pezze_quantum_2018, sinatra_spin-squeezed_2022}, and vibrational modes of trapped ions or massive oscillators~\cite{zhang_noon_2018, qvarfort_gravimetry_2018}. In these platforms, quantum states can be engineered that enable estimation precisions beyond the shot-noise limit, a fact that has been both theoretically predicted and experimentally verified~\cite{giovannetti_advances_2011}.

Despite this common objective, the physical mechanisms responsible for quantum-enhanced precision can appear very different depending on the regime considered. In systems where the total particle number is well defined and bosons occupy several orthogonal modes, quantum advantage is typically associated with correlations between particles and modes, often described in terms of particle or mode entanglement~\cite{benatti_sub-shot-noise_2010, dalton_new_2014, benatti_sub-shot_2011, morris_entanglement_2020, pezze_quantum_2018}. In contrast, in the CV regime, where states may not possess a definite particle number, even single-mode states such as quadrature-squeezed states can provide quantum-enhanced precision~\cite{dowling_quantum_2008, polino_photonic_2020, duivenvoorden_single-mode_2017, fadel_quantum_2025, kwon_quantum_2022, pinel_ultimate_2012, safranek_estimation_2018}. In such situations, the relevant resources are often described in terms of field quadratures, squeezing, or phase-space structure rather than particle correlations.

\paragraph{\publi Different representations of bosonic resources}

These different perspectives arise largely from the representation used to describe the bosonic field. In interferometric settings with a fixed particle number, the Schwinger representation provides a natural framework in which pairs of bosonic modes are mapped onto angular-momentum operators~\cite{schwinger_angular_2015} (see also Sec.~\ref{subsec: SSRC and Schwinger}). For two modes $k$ and $l$, one defines
\begin{align}
    \hat J_x^{(k,l)} = \frac{1}{2}\left(\hat a_k^\dagger \hat a_l + \hat a_l^\dagger \hat a_k \right), &&
    \hat J_y^{(k,l)} = \frac{i}{2}\left(\hat a_k^\dagger \hat a_l - \hat a_l^\dagger \hat a_k \right), &&
    \hat J_z^{(k,l)} = \frac{1}{2}\left(\hat a_l^\dagger \hat a_l - \hat a_k^\dagger \hat a_k \right),
\end{align}
where $\hat a_k$ and $\hat a_k^{\dagger}$ denote the annihilation and creation operators of mode $k$. In this picture, phase shifts correspond to unitary evolutions generated by $\hat J_z^{(k,l)}$, while beam splitters implement rotations of the form
\begin{equation}
    \hat U = e^{- i \theta \hat J_y^{(k,l)}},
\end{equation}
with $\theta$ determined by the optical transmissivity. In this context, the achievable precision, quantified by the QFI or equivalently by the variance of the evolution generator through the Cramér-Rao bound (see Sec.~\ref{subsec: quantum metrology pure states}), depends on the total particle number $N$.

By contrast, in the CV description commonly used in quantum optics, a single-mode field is represented as a superposition of Fock states,
\begin{equation}
    \ket{\psi}_1 = \sum_{n=0}^{\infty} c_n \ket{n}_1 ,
\end{equation}
with $\sum_n \abs{c_n}^2 = 1$. Within this framework, metrological resources are typically expressed in terms of quadrature fluctuations, squeezing, or phase-space nonclassicality. The relevant evolution generators are then the quadrature operators $\hat x_1$, $\hat p_1$, or the number or phase operators $\hat n_1=\hat a_1^\dagger \hat a_1$. In this regime, where the particle nature of light is not explicitly emphasized, the connection between metrological enhancement and particle-number correlations or multimode structures is therefore less direct.

Furthermore, the natural quantifier of resource is the mean photon number
\begin{equation}
    \overline n = \bra{\psi}_1 \hat a_1^\dagger \hat a_1 \ket{\psi}_1=\sum_{n=0}^\infty \abs{c_n}^2 n,
\end{equation}
which, although the natural analogue of $N$, plays a very different operational role. Indeed, in the SSRC regime $N$ is fixed, and the precision scaling can only be improved by introducing suitably engineered correlations between individual particles. By contrast, in the CV regime $\overline n$ is not fixed, and the precision scaling can be improved simply by increasing the mean photon number of the state, apparently without the need to introduce correlations between photons.

\paragraph{\publi Motivation for a unified description}

The coexistence of these descriptions raises a conceptual challenge for bosonic quantum metrology. While both particle-based and continuous-variable approaches successfully predict quantum-enhanced precision, they attribute this enhancement to seemingly different resources. As a consequence, identifying the fundamental origin of metrological advantage, and comparing protocols across distinct physical platforms, can become difficult. A unified description capable of consistently capturing both regimes is therefore highly desirable. Such a framework should allow one to identify, within a single formalism, how particle statistics, mode structure, and field fluctuations contribute to precision scaling across different bosonic systems.

\paragraph{\publi Superselection-rules-compliant representation}

Following the framework developed throughout this chapter, we adopt a superselection-rules-compliant (SSRC) description of the bosonic field, in which the phase reference is explicitly treated as a physical degree of freedom rather than implicitly assumed. This approach restores particle-number conservation and automatically reunifies fixed-particle-number states and continuous-variable states within a single compact framework.

In the simplest case, a pure state of $N$ bosons distributed over two modes can be written as
\begin{equation}
    \ket{\Psi} = \sum_{n=0}^{N} c_n \ket{n}_1 \ket{N-n}_2 ,
\end{equation}
where the two orthogonal modes act as mutual phase references. In regimes where one mode contains a macroscopic occupation, this representation effectively reduces to the usual single-mode continuous-variable description, while remaining fully consistent with the photon-number superselection rules~\cite{sanders_photon-number_2003, bartlett_reference_2007} (see also Sec.~\ref{sec: CV as limit of SSRC}).

The SSRC framework therefore provides a natural bridge between the discrete and continuous-variable descriptions of bosonic systems. In the following sections, we use this representation to identify the fundamental resources responsible for quantum-enhanced metrology and to clarify the respective roles played by particle correlations, mode structure, and reference frames.

\subsection{Two-mode analysis}
\label{subsec: two mode analysis}

We now illustrate how this unified description enables a consistent analysis of the different limits of quantum metrology. As an example, consider a two-mode rotation, $e^{-i\hat J_{\vec{n}}^{(1,2)}\theta}$, corresponding to a combination of a phase shift and a beam splitter. In metrological applications, one may seek to estimate the parameter $\theta$, for an arbitrary rotation axis. Here we focus, for instance, on the case of rotations around the $y$ axis.

\paragraph{\publi Continuous-variable limit of the $y$ rotations}

In the CV limit, as we have argued in Sec.~\ref{subsec: Majorana and stellar representation}, the dynamics on the Bloch sphere are restricted to the neighborhood of the north pole, which, as discussed in Sec.~\ref{subsec: phase spaces and geometry}, can be approximated by its tangent plane~\cite{pezze_quantum_2018}, \ie, $\theta \ll 1$. Furthermore, we have seen that the correct scaling factor for going from the SSRC representation to its CV limit is $\sqrt{N}$. As such, we can introduce the corresponding CV parameter $q$ as
\begin{equation}
    \theta = q/\sqrt{N}.
\end{equation}

The corresponding precision in estimating $q$, or equivalently $\theta$, given by the Cramér-Rao bound (see Sec.~\ref{subsec: quantum metrology pure states}) then reads
\begin{equation}\label{eq: precision rotation}
    \delta \theta = \delta\left(\frac{q}{\sqrt{N}}\right) \geq \frac{1}{\sqrt{4 \Delta^2 \hat J_y^{(1,2)}}}.
\end{equation}
Remembering the SSRC-CV correspondence for the operators presented in Sec.~\ref{subsec: formal limit}
\begin{equation}
    \frac{1}{\sqrt{N}}\hat J_y\longrightarrow \frac{1}{\sqrt{2}}\hat p, 
\end{equation}
in the CV limit this inequality reduces to
\begin{equation}
    \delta q \ge 1/\sqrt{2\Delta^2\hat p},
\end{equation}
which is consistent with the fact that rotations around $\vec{x}$ and $\vec{y}$ correspond, in this limit, to translations in quadrature phase-space (see Sec.~\ref{subsec: formal limit}). Consequently, Eq.~\eqref{eq: precision rotation}, which can be generalized to arbitrary rotation axes $\vec n$, holds for all states and regimes.

Importantly, making the phase reference explicit reveals that particle entanglement, often hidden in the CV representation, is required to surpass the shot-noise limit. Indeed, within the SSRC framework, the only particle-separable states are spin-coherent states (see Sec.~\ref{subsec: spin coherent states}), which, in the CV limit, reduce to coherent states (see Sec.~\ref{subsec: examples of SSRC CV limits}), for which one always finds $\Delta^2\hat J_{\vec n}^{(1,2)} \propto N$ (or $\overline n$ in the CV limit).

Interestingly, while the estimation of $\theta$ can \textit{a priori} reach the Heisenberg limit with a scaling as $1/N$, the constraint to small rotation angles in the CV limit yields a quantum Fisher information that scales at most linearly with the mean photon number of mode~$1$, $\overline n$. This is consistent with the usual CV observation where the precision for quadrature evolution scales at most linearly with the mean photon number.\footnote{This follows from the simple chain of inequalities
\begin{equation}
    \Delta^2 \hat x \leq \expval{\hat x^2} \leq \expval{\hat x^2+\hat p^2} = 2\overline n + 1.
\end{equation}
}
Thus one obtains
\begin{equation}
    \Delta^2\hat x \leq 2\overline n +1 .
\end{equation}

\paragraph{\publi Continuous-variable limit of the $z$ rotation}
The previous result contrasts with rotations around the $z$ axis, or phase evolution in the CV limit. While states remain restricted to a small region around the north pole, which constrains the accessible rotation angles for $x$ and $y$ rotations, rotations around the $z$ axis can occur with arbitrary angles. More explicitly, observe that for a CV state $\ket{\psi}=\sum_{n=0}^\infty c_n \ket{n}_1$, we have
\begin{equation}
    e^{-i\hat n_1\theta}\ket{\psi} = \sum_{n=0}^\infty c_n e^{-i n \theta} \ket{n}_1 ,
\end{equation}
while for the corresponding SSR state $\ket{\psi}=\sum c_n \ket{n}_1 \ket{N-n}_2$
\begin{equation}
    e^{i\hat J_z^{(1,2)}\theta}\ket{\psi} = \sum_{n=0}^N c_n e^{-i (n-N/2) \theta} \ket{n}_1 \ket{N-n}_2=e^{iN\theta/2}\sum_{n=0}^N c_n e^{-i n \theta} \ket{n}_1 \ket{N-n}_2.
\end{equation}
Thus up to the unimportant global phase $e^{iN\theta/2}$, the dynamics generated by $\hat J_z^{(1,2)}$ and $\hat n_1$ are identical. We thus expect that both operators yield the same precision scaling in the CV limit, which can be verified by explicitly computing the variances
\begin{equation}
    \Delta^2 \hat J_z^{(1,2)} = \Delta^2\left(\frac{1}{2}(\hat n_2-\hat n_1)\right)=\Delta^2\left(\frac{1}{2}(N-2\hat n_1 )\right)=\Delta^2 \hat n_1,
\end{equation}
and in the CV limit, the scaling is not restricted to a linear scaling. The SSRC representation therefore provides a unified framework for quantum metrology, explicitly treating the two modes as mutual phase references. For balanced photon numbers, the total photon number $N$ is the relevant resource. In contrast, in the CV limit, \ie, when there is a strong photon-number imbalance between the modes, the dynamics effectively reduce to a single-mode evolution, with $\overline n$ as the relevant resource.

\paragraph{\publi Example of correspondence: NOON and cat states}

This ``replacement'' of $N$ by $\overline n$ in the CV limit has important physical implications for how modes and states function as resources in metrology. To illustrate this point, we consider two paradigmatic examples for quantum metrology: the NOON state and the Schrödinger cat state. These states are often compared in the literature~\cite{volkoff_measurement-_2014, huang_achieving_2018, maleki_quantum_2021, fischer_photonic_2015} and both lead to a quadratic enhancement in precision with respect to the shot-noise limit. While NOON states are two-mode states,
\begin{equation}
    \ket{{\cal N}} = \frac{1}{\sqrt{2}}\left(\ket{N}_1\ket{0}_2 + \ket{0}_1\ket{N}_2\right),
\end{equation}
Schrödinger cat-like states in the CV representation are superpositions of single-mode coherent states,
\begin{equation}
    \ket{{\cal C}} = \frac{1}{{\cal K}}\left(\ket{0}_1 + \ket{\alpha}_1\right),
\end{equation}
where ${\cal K}$ is the normalization factor. Usually, the QFI is computed for both states using an evolution generated by $\hat H \propto \hat n_1$, leading respectively to
\begin{align}
    \mathcal Q_{\cal N} \propto N^2, &&
    \mathcal Q_{\cal C} \propto \frac{2\abs{\alpha}^2}{1+e^{-\abs{\alpha}^2/2}} + \frac{\abs{\alpha}^4\left(1+2e^{-\abs{\alpha}^2/2}\right)}{\left(1+e^{-\abs{\alpha}^2/2}\right)^2},
\end{align}
\ie, both states permit reaching the Heisenberg scaling when $\abs{\alpha}^2 \simeq 2\overline n \gg 1$. However, within the SSRC formalism, the states $\ket{{\cal C}}$ appear as the large-$N$ limit of the states studied in~\cite{sanders_connection_2014}. The SSRC/CV coherent-state correspondence demonstrated in Sec.~\ref{subsec: examples of SSRC CV limits} shows that, up to a normalization factor,
\begin{equation}
    \ket{N}_2 + \ket{N}_{\theta,\phi} \to \ket{0}_1 + \ket{\alpha}_1
\end{equation}
in the CV limit. Notably, the left-hand side is similar to a NOON state, but with the $N$ photons distributed in non-orthogonal modes. This observation reveals a deep connection between NOON states and cat states. While the resources and evolutions in both cases appear very different, the SSRC perspective shows that the enhanced precision in both scenarios arises from the same underlying physical mechanism: the presence of a superposition of two states with different total photon numbers in the mode used for estimation. In the NOON case the mode overlap is ideal, leading to a scaling determined by $N$. In the CV case the mode overlap is not ideal, which effectively replaces $N$ by $\overline n$ in the precision scaling.

\subsection{Multimode analysis}
\label{subsec: multimode analysis SSRC metro}
We now extend our analysis to the multimode regime and show how the SSRC representation clarifies the respective roles of mode correlations and particle statistics in precision enhancement. As an illustrative example, we first consider a three-mode state, which corresponds to a two-mode configuration in the CV limit, before generalizing to an arbitrary number of modes. The most general pure state with a fixed total particle number $N$ can be written as
\begin{equation}
    \ket{\psi} = \sum_{\substack{n_1,n_2=0 \\ n_1+n_2 \leq N}} c_{n_1,n_2} \ket{n_1}_1 \ket{n_2}_2 \ket{N-n_1-n_2}_3 ,
\end{equation}
where modes $1$, $2$, and $3$ are modes on which the probe state is defined. In the CV limit, modes $1$ and $2$ constitute the physical system while mode $3$ plays the role of a phase reference. The state evolves under a unitary transformation $e^{-i\hat H\theta}$ where the parameter $\theta$ is to be estimated. To illustrate our method, we restrict our analysis to rotations around the $z$ axis and therefore consider the generator
\begin{equation}
    \hat H = \alpha_{1,2}\hat J_{z}^{(1,2)} + \alpha_{1,3}\hat J_{z}^{(1,3)} + \alpha_{2,3}\hat J_{z}^{(2,3)},
\end{equation}
where the $\alpha_{j,k}$ are arbitrary constant coefficients. Writing this operator in terms of number operators yields
\begin{equation}
    \hat H = \beta_1\hat n_1 + \beta_2\hat n_2 + \beta_3\hat n_3 ,
\end{equation}
where the coefficients $\beta_j$ satisfy the constraint $\beta_1+\beta_2+\beta_3=0$ due to the expression of the angular momentum operator $\hat J_z^{(j,k)}$ in terms of number operators. 

This construction and the reasoning below mirror the ideas on collective variables and operators presented in Chap.~\ref{chap: collective entanglement}, especially Sec.~\ref{sec: Collective entanglement}. Here we adapt them to the case of SSRC systems where the total photon number is fixed and where the phase reference plays an explicit role. Our goal is therefore to identify the interplay between the choice of a combination of modes and the choice of probe state such that the quantum Fisher information $\mathcal Q=4\Delta^2 \hat H$ is maximized. Importantly, to allow for a fair comparison, a normalization condition must be imposed on the coefficients $\beta_j$, as otherwise the precision could be arbitrarily increased by simply increasing their magnitude. In the following, we therefore impose $\sum_j \beta_j^2=1$.

\paragraph{\publi  Geometric optimization of collective observables.}
Previous studies have analyzed the dependence of the QFI on the modal structure in the CV limit~\cite{gessner_estimation_2023} and identified optimal two-mode interferometric transformations for symmetric states~\cite{hyllus_not_2010}. The key observation is that, since the coefficients $\beta_j$ satisfy a normalization condition corresponding to points on a sphere, the operator $\hat H$ can be interpreted as a rotation of the operator $\hat n_1$ in mode space. Applying the same rotation to the remaining number operators yields two additional observables, denoted $\hat H^{(2)}$ and $\hat H^{(3)}$. Their variances satisfy the identity\footnote{This identity can be verified explicitly by providing a parametrization of the corresponding rotation. It also follows directly from the invariance of the trace of the covariance matrix under orthogonal transformations.}
\begin{equation}
    \Delta^2 H + \Delta^2 H^{(2)} + \Delta^2 H^{(3)} = \Delta^2 \hat n_1 + \Delta^2 \hat n_2 + \Delta^2 \hat n_3 .
\end{equation}
Maximizing the metrological sensitivity associated with $\hat H$ by optimizing the parameters $\beta_j$ therefore amounts to minimizing the complementary quantity $\Delta^2 H^{(2)} + \Delta^2 H^{(3)}$. The optimal situation corresponds to
\begin{equation}
    \Delta^2 H^{(2)} + \Delta^2 H^{(3)} = 0,
\end{equation}
which identifies conserved observables associated with the underlying physical symmetries~\cite{frerot_symmetry_2024}.  As an illustration, consider the two-mode maximally correlated state
\begin{equation}
    \ket{\mathcal C_2} = \sum_{\substack{n=0 \\ 2n\leq N}}^N c_n \ket{n}_1 \ket{n}_2 \ket{N-2n}_3 .
\end{equation}
As can be verified explicitly, and as will follow from the more general result presented in the next paragraph, the optimal parameters for these states are
\begin{align}
    \beta_1=\beta_2=-\sqrt{\frac{1}{6}}, && \beta_3=\sqrt{\frac{2}{3}} .
\end{align}
This corresponds to the optimal evolution generator
\begin{equation}
    \hat H=\sqrt{\frac{2}{3}}\Big(\hat J_z^{(1,3)}+\hat J_z^{(2,3)}\Big).
\end{equation}
In this case, the variance is
\begin{equation}
    \Delta^2\hat H=6\Delta^2\hat n,
\end{equation}
where $\Delta^2\hat n=\Delta^2\hat n_1=\Delta^2\hat n_2$ denotes the photon-number variance, which is identical in the first two modes.

\paragraph{\newc Generalisation to an arbitrary number of modes}
The above observation can be generalized to an arbitrary number of modes for SSRC states of the form
\begin{equation}
    \ket{\psi} = \sum_{\substack{n_1,n_2,\dots,n_k=0 \\ n_1+n_2+\dots+n_k \leq N}} c_{n_1,n_2,\dots,n_k} \ket{n_1}_1 \ket{n_2}_2 \dots \ket{n_k}_k \ket{N-n_1-n_2-\dots-n_k}_{k+1} .
\end{equation}
Considering the most general evolution generator for $z$ rotations,
\begin{equation}
    \hat H=\sum_{1\leq j<j'\leq k+1} \alpha_{j,j'}\hat J_z^{(j,j')},
\end{equation}
we seek to optimize the evolution direction by tuning both the initial state and the coefficients $\alpha_{j,j'}$ in order to maximize the QFI and determine the best scaling in terms of the available resources: namely the photon-number distribution and the number of modes. Following the reasoning above, we can rewrite $\hat H$ in terms of number operators
\begin{equation}
    \hat H = \sum_{j=1}^{k+1} \beta_j \hat n_j,
\end{equation}
where the coefficients $\beta_j$ must satisfy $\sum_j \beta_j=0$. As before, a normalization condition must be imposed on these coefficients, since otherwise the precision could be arbitrarily increased by simply scaling their magnitude. We thus fix $\sum_j \beta_j^2=1$. 

As we explicitly demonstrate in Appendix~\ref{app: SSRC}, Result~\ref{res: optimization SSRC measurement direction}, the variance of $\hat H$ can be bounded as
\begin{equation}
    \Delta^2 \hat H \leq k(k+1) \Delta^2 n,
\end{equation}
where $\Delta^2 n=\max_{j=1,\dots,k}\Delta^2\hat n_j$ is the maximal photon-number variance among the first $k$ modes, \ie excluding the phase reference. Furthermore, we show that the optimal strategy consists in choosing the coefficients $\beta_j$ such that $\hat H$ corresponds to a rotation of the form
\begin{equation}
    \hat H = \frac{2}{\sqrt{k(k+1)}}\sum_{\alpha=1}^k \hat J_z^{(\alpha,k+1)}.
\end{equation}
For such evolutions, the optimal probe states are maximally correlated states such as
\begin{equation}\label{eq: SSRC correlated states}
    \ket{\psi} = \sum_{\substack{n=0 \\ kn\leq N}}^N c_n \ket{n}_1 \ket{n}_2 \dots \ket{n}_k \ket{N-kn}_{k+1},
\end{equation}
which saturate the QFI with $\mathcal Q = 4 k(k+1) \Delta^2 \hat n$.

\paragraph{\newc Comments}

Several remarks help clarify the role of mode and particle correlations in quantum metrology, as well as the role played by the phase reference.

\begin{itemize}
    \item First, notice that this framework closely mirrors the considerations of Sec.~\ref{sec: Collective entanglement}, in which we studied the optimization of the variance of collective operators of the form
    \begin{equation}
        \hat H_\text{coll}=\sum_{j=1}^k \beta_j \hat H_j .
    \end{equation}
    In Sec.~\ref{subsec: remark def coll op} and Appendix~\ref{app: collective var derivation}, Result~\ref{res: cauchy-schwarz general}, we showed that optimality of Eq.~\eqref{eq: general bound collective global} required $\abs{\beta_j}=\text{cst.}$, so that the evolution acts with equal weight on each subsystem. In that case the optimal precision scales as
    \begin{equation}
        \Delta^2 \hat H_\text{coll} \leq k^2 \max \Delta^2 \hat H_j,
    \end{equation}
    for the normalization choice $\abs{\beta_j}=1$.

    In the present context $\hat H_j=\hat n_j$. However, while we ask a similar question, we are now constrained by superselection rules. Namely, we have the two constraints
    \begin{align}
        \sum_{j=1}^{k+1}\beta_j=0, && \sum_{j=1}^{k+1}\hat n_j=N,
    \end{align}
    where the first equality follows from the structure of the generator $\hat H$ expressed in terms of angular momentum operators, while the second is a direct consequence of the superselection rule fixing the total particle number. These constraints modify the final result, yielding a tighter bound, and the structure of the optimal states.

    \item While we \textit{a priori} allow evolution generators that couple any pair of modes, the optimal evolution involves only rotations that couple each probe mode to the reference mode. Since the evolution generator $\hat J_z^{(j,k+1)}$ mimics the operator $\hat n_j$ in the CV limit,
    \begin{equation}
        -\frac{2}{k+1}\left(\sum_{j=1}^k \hat J_z^{(j,k+1)}-\frac{kN}{2}\1\right) \to \sum_{j=1}^k \hat n_j,
    \end{equation}
    we recover the unconstrained result of collective variables, which requires that the evolution be performed with equal weight on each mode. Furthermore, the optimal SSRC states, which are perfectly correlated across all non-reference modes, reduce in the CV limit to states exhibiting perfect correlations between all considered modes,
    \begin{equation}
        \ket{\psi} = \sum_{\substack{n=0 \\ kn\leq N}}^N c_n \ket{n}_1 \ket{n}_2 \cdots \ket{n}_k \ket{N-kn}_{k+1}
        \to
        \ket{\psi} = \sum_{n=0}^\infty c_n \ket{n}_1 \ket{n}_2 \cdots \ket{n}_k .
    \end{equation}

    \item Although the scalings in the SSRC and CV cases appear qualitatively consistent, as both exhibit quadratic behavior, a more careful examination of the normalization conventions is required. In the CV case, the optimal scaling
    \begin{equation}
        \Delta^2\hat H_\text{coll} \leq k^2 \max \Delta^2 \hat n_j
    \end{equation}
    is obtained with the normalization
    \begin{equation}
        \sum_{j=1}^k \beta_j^2=k ,
    \end{equation}
    while in the SSRC case we obtain
    \begin{equation}
        \Delta^2 \hat H \leq k(k+1) \max \Delta^2 \hat n_j
    \end{equation}
    for
    \begin{equation}
        \sum_{j=1}^{k+1} \beta_j^2=1 .
    \end{equation}

    Rescaling the SSRC result by a factor $k$ in order to match the CV normalization yields
    \begin{equation}
        \Delta^2 \hat H \leq k^2(k+1) \max \Delta^2 \hat n_j ,
    \end{equation}
    which seems to suggest a larger-than-quadratic scaling. However, there is no mathematical inconsistency. The SSRC result can be viewed as a constrained version of the collective-variable scenario. The apparent super-quadratic scaling arises because the reference quantity $\Delta^2 n$ is defined as the maximum variance among only the first $k$ modes, while in the collective-variable framework all modes should be considered. For states of the form Eq.~\eqref{eq: SSRC correlated states}, one finds
    \begin{equation}
        \Delta^2 \hat n_{k+1}=k^2 \Delta^2 n ,
    \end{equation}
    so that when the reference mode is also included the overall scaling remains consistent with the collective-variable bound and is actually only linear. This makes intuitive sense as, from the collective-variable point of view, the optimal scaling requires equal weight on each local evolution generator, while in the SSRC case, the reference mode operator $\hat n_{k+1}$ is highly favored.

    \item These observations allow us to clarify the role played by the phase reference. In the SSRC description the phase reference acts as an explicit resource that contributes to the estimation precision. To illustrate the difference, one may compare the action of $\hat n_1+\hat n_2$ on a CV state with that of $\hat J_z^{(1,3)}+\hat J_z^{(2,3)}$ on the corresponding SSRC state. In the first case,
    \begin{equation}
        e^{-i(\hat n_1+\hat n_2)\theta}\ket{\psi} = \sum_{n_1,n_2=0}^{\infty} e^{-i(n_1+n_2)\theta} c_{n_1,n_2} \ket{n_1}_1 \ket{n_2}_2 ,
    \end{equation}
    while for SSRC systems
    \begin{equation}
        e^{i(\hat J_z^{(1,3)}+\hat J_z^{(2,3)})\theta}\ket{\psi} = e^{iN\theta} \sum_{\substack{n_1,n_2=0 \\ n_1+n_2\leq N}}^{N} e^{-i\frac{3}{2}(n_1+n_2)\theta} c_{n_1,n_2} \ket{n_1}_1 \ket{n_2}_2 \ket{N-n_1-n_2}_{3}.
    \end{equation}

    This shows that although for two modes the generators $\hat n$ and $\hat J_z$ are effectively equivalent, this correspondence does not extend straightforwardly to larger multimode configurations. When several modes interact coherently with the same reference mode, the evolution can be effectively amplified by the presence of the reference, which modifies the scaling of the variance. In this sense the reference mode is not only a passive resource, but also an active one that can influence the dynamics and thus the achievable precision.

    A possible way to remove this amplification and recover a structure closer to the CV scenario is to associate a different phase reference to each probe mode. In that case, the different $\hat J_z$ operations couple distinct pairs of modes without overlap, thereby preventing the collective amplification mechanism described above.

    Alternatively, one may note that the evolution generated by $\hat J_z^{(j,k+1)}$ explicitly couples the considered probe mode to the phase reference by construction. While this description is reasonable when the phase reference is physically accessible, it is important to recall that in most realistic situations the phase reference is effectively inaccessible. Indeed, a shared reference beam may at first appear to be a suitable candidate for a phase reference. However, if, as typically done, this beam is described by a coherent state, its phase is defined only relative to another reference. In other words, specifying the coherence of the beam implicitly assumes the existence of an additional phase reference with respect to which this coherence is defined. As discussed by~\cite{molmer_optical_1997}, the atomic ensemble forming the laser source provides a natural phase reference. In this picture, the laser field and the system of interest must be entangled, and the phase of the emitted light is defined relative to the internal degrees of freedom of the source. Consequently, the relevant phase reference is not directly accessible to the experimenter. Within this perspective, a typical CV protocol can be described as follows. Only operations acting on the modes of interest are allowed, while operations involving the reference mode are forbidden. This corresponds to a constrained version of the SSRC scenario, in which the evolution is generated by operators of the form $\hat n_j$ ($j<k+1$) rather than $\hat J_z^{(j,k+1)}$. Since these generators do not couple the probe modes to an explicit reference mode, the collective amplification mechanism discussed above does not arise. As a result, the scaling of the variance is consistent with the standard CV case.
\end{itemize}

\subsection{Conclusion on superselection rules}

Throughout this chapter we have explored superselection-rules-compliant systems. We have shown that, beyond their appealing mathematical properties, which have been widely studied by the spin community, they also provide an alternative point of view on optical systems. By incorporating the phase reference as an explicit physical object within the formalism, they offer a coherent and physically transparent framework that naturally accounts for the problem of optical coherence. Importantly, this is not merely an equivalent reformulation of the traditional continuous-variable formalism in which the phase reference is made explicit. Rather, SSRC systems provide an alternative conceptual perspective on quantum optics that offers important and useful insight into the properties and roles played by optical modes and statistical distributions.

More broadly, this perspective reveals that coherence and phase references are not only practical concerns for experimental implementations. Instead, treating them explicitly at the theoretical level provides significant physical and conceptual insight into the structure of optical quantum systems.

The inclusion of an explicit phase reference forces a reformulation of quantum optics in a regime where the first-quantized nature of light, in terms of identical particles, becomes particularly transparent. In the CV framework, especially in the single-mode regime, the particle nature of light is often hidden by the field-based description. In contrast, SSRC representations demonstrate that particle entanglement is a fundamental feature of optical systems and that it can play the role of a genuine resource.

We have shown how the SSRC framework allows one to obtain new insight into informational and computational considerations by providing a clearer picture of the role played by particle entanglement and by mode transformations. We have also clarified the role of Gaussian resources and shown how, in the SSRC picture, they are more constrained due to the explicit representation of the phase reference. This analysis reveals that although CV systems can be obtained as limits of SSRC ones, part of the information about the correlations and structural properties of the system is typically lost when this limit is taken. In particular, the relation of the system to the phase reference is usually overlooked and effectively ignored in the limiting description.

Finally, we applied the SSRC point of view to quantum metrology and showed how it allows one to obtain a unified description of the different regimes of quantum metrology. Within this framework it becomes possible to clarify the respective roles played by particle correlations, mode structure, and phase references in precision enhancement. In particular, we showed how the SSRC framework clarifies the connection between NOON states and cat states. These states are often compared in the literature, but their relationship is not always clearly articulated. We also showed how the SSRC framework allows one to identify optimal evolutions and probe states for multimode metrology, while simultaneously clarifying the role played by the phase reference in this context.

It is also worth noting that further fundamental analyses of the role played by phase references can be carried out within this framework and that they reveal additional and interesting perspectives on several problems. In~\cite{descamps_heisenberg-weyl_2025}, we studied the different phase spaces that are typically used in quantum information and quantum optics and showed how they naturally emerge when studying SSRC systems. This provides a natural connection between these phase-space representations and reveals interesting geometrical insights. In Ref.~\cite{moulonguet_non-gaussianity_2026}, we exploit the connection between Majorana polynomials and Bargmann functions in order to obtain new insight into the structure of physical states and the notion of Gaussianity. 

However, in order to keep the scope of this thesis bounded, we have chosen not to include these works in the present document.

\fi

\ifnum \theShowFront = 1
\chapter*{Conclusion and perspectives}
\setcurrentanchor{conclusion}
\addcontentsline{toc}{chapter}{Conclusion and perspectives}
\markboth{Conclusion and perspectives}{Conclusion and perspectives}

Throughout this manuscript, we have studied the notions of modes, states, and symmetries in quantum optics across a variety of contexts and applications. The analyses presented throughout this work shed new light on the interplay between modal structure, particle statistics, and symmetry transformations. In particular, we revealed the ubiquitous role played by particle entanglement, a property that is often overlooked in the quantum optics literature.

Quantum metrology served as a guiding framework for much of our development. Beyond being a fruitful research field with promising practical applications, it provides a powerful lens through which the properties of quantum systems and their manipulation can be understood. In this regard, we introduced new metrological constructs and protocols that could potentially enhance parameter estimation tasks. More importantly, metrological considerations provide direct access to the structural properties of quantum states and operations. From this perspective, quantum metrology has proven instrumental in revealing the intricate interplay between modes, states, and symmetries in quantum optical systems.

In Chap.~\ref{chap: Time-Frequency Systems}, we developed a formal description of time-frequency single-photon systems, which constitute a promising platform for quantum technologies. Although descriptions of quantum light in terms of time and frequency variables are common, particularly for single photons, the conceptual contribution of our work lies in the introduction of an operator formalism for time and frequency detection. This formalism explicitly reveals the correspondence with the usual quadrature representation, thereby opening both conceptual and practical perspectives for exploiting time-frequency systems in quantum information processing and quantum metrology. Importantly, the formulation we present is not entirely new and has previously been explored in~\cite{fabre_quantum_2020,fabre_time_2022}. In this chapter, however, we extended this description and emphasized its connection with quantum metrology.

In Chap.~\ref{chap: collective entanglement}, we developed a general framework for collective-variable observables and their relation to the structure of multipartite entanglement. This framework makes it possible to identify optimal probe states and scaling laws for precision, while also extending the analysis to mixed states and partially entangled systems. Guided by the metrological perspective, we obtained a deeper understanding of multipartite entanglement and of the role played by frequency as a quantum degree of freedom. Our analysis unifies several existing approaches used to study precision scaling. Finally, we demonstrated that collective variables and their associated entanglement can serve as valuable resources for other areas of quantum information by constructing a multipartite GKP-inspired code.

In Chap.~\ref{chap: HOM interferometry and metrology}, we investigated the role of symmetry in interferometric phenomena and quantum metrology. We showed how the symmetry properties of both the input state and the interferometer determine the interference visibility and the attainable metrological precision. We further extended our analysis to multiphoton and multimode interferometric settings, revealing the crucial role played by indistinguishability and symmetry in determining achievable precision. Beyond introducing a broad class of interferometric configurations and metrological protocols, our results shed new light on a wide range of existing phenomena, bridging Hong-Ou-Mandel interferometry, its various generalizations, Mach-Zehnder interference, and boson sampling experiments.

In Chap.~\ref{chap: SSR}, we studied the role of superselection rules in quantum optics and quantum metrology. We showed how superselection-rules-compliant (SSRC) descriptions provide a unified framework for discrete-variable and continuous-variable encodings, while clarifying the respective roles of Gaussian and non-Gaussian resources. Although a complete understanding of bosonic quantum computing remains elusive, we demonstrated how the SSRC framework offers new conceptual insights. Furthermore, continuing our metrological analysis, we showed that SSRC considerations provide a unified description of quantum metrology, clarifying the roles played by particle correlations, modal structure, and phase references in precision enhancement.

\medbreak
With this overview in mind, we can now discuss the role played by the three central conceptual elements of this thesis.
\begin{itemize}

    \item \emph{Modes}. By defining the accessible degrees of freedom, modes provide the canvas on which quantum information is encoded and manipulated. For some applications, such as interferometry and metrology, particular modes, typically associated with external degrees of freedom, can be naturally identified. However, the SSRC analysis revealed that the notion of a mode is fundamentally relative. Mode transformations, such as those implemented by beam splitters, do not create resources but rather reveal them in an exploitable form. From the SSRC perspective, describing the phase reference as an accessible mode further clarifies this relativity. In the continuous-variable limit, large photon-number imbalances effectively restrict the accessible region of the Bloch sphere to its northern hemisphere. By contrast, the SSRC description allows access to the entire sphere, thereby restoring the fundamental spherical symmetry of the modal structure.

    \item \emph{States}. Quantum states are the carriers of quantum information and resources. The analyses presented in the different chapters revealed that the internal structure of states, and in particular their entanglement structure, plays a crucial role in determining the performance of quantum protocols. In particular, we showed that particle entanglement constitutes a fundamental resource in quantum optics and can play a decisive role in quantum metrology as well as in quantum information processing.

    \item \emph{Symmetries}. Symmetries play a fundamental role in determining both the properties and the dynamics of quantum systems. In this thesis, symmetries appear at two different levels: the intrinsic symmetries of the physical systems under consideration and the symmetries of the interferometric transformations applied to them. Both types of symmetry strongly constrain the structure of quantum states and the performance of quantum protocols. The analysis of HOM interferometry and its extensions demonstrated how symmetry properties determine interference visibility and metrological precision, while the SSRC framework revealed how symmetry constraints shape the structure of states and the limits of achievable precision.

\end{itemize}

Taken together, these observations highlight that the interplay between modes, states, and symmetries is a fundamental aspect of quantum optics and quantum metrology. Understanding this interplay is essential both for the development of new quantum technologies and for gaining deeper insight into the nature of quantum systems.

\medbreak

We now discuss the limitations of the present work and outline several perspectives for future research. As mentioned in the conclusion of Chap.~\ref{chap: HOM interferometry and metrology}, many theoretical and experimental challenges must still be addressed before an explicit demonstration of the extended HOM effect proposed in this thesis can be realized. Practical issues such as noise, losses, probe-state generation, and precise experimental control represent important obstacles. On the theoretical side, several promising directions remain to be explored, including extensions to multiparameter estimation and generalizations to interferometers beyond those based on discrete Fourier transforms. Furthermore, although connections with the boson sampling problem were highlighted, strengthening these connections and investigating the implications of our results for questions of computational complexity may provide valuable insights.

While Chap.~\ref{chap: collective entanglement} provided conceptual insights into the role of modal entanglement, it also proposed concrete experimental protocols for quantum metrology and quantum error correction. Although we showed that the interferometer introduced in Chap.~\ref{chap: HOM interferometry and metrology} could in principle implement the measurement of the collective observables considered in this framework, such an implementation remains challenging with current experimental capabilities. Developing simplified and experimentally feasible versions of these protocols therefore constitutes an important direction for future work. In addition, the experimental realization of the proposed multipartite GKP code has already been discussed with experimental collaborators. Although explicit generation and control have not yet been achieved, we hope that future experiments will soon demonstrate the potential of this code for quantum error correction.

Finally, another promising direction concerns the further development of the superselection-rules-compliant framework and its applications. Despite the extensive discussion presented in this thesis, several conceptual challenges remain. In particular, while we showed how continuous-variable systems can emerge as a limit of SSRC systems, we did not establish a fully rigorous mathematical correspondence. Because the SSRC framework more faithfully captures physical constraints, a deeper understanding of this limit could provide new insights into the nature of physical states in the continuous-variable regime. Indeed, although the continuous-variable formalism has proven extremely successful in describing experimental systems, its infinite-dimensional Hilbert space structure introduces significant mathematical challenges. These difficulties make it hard to obtain rigorous results concerning the structure of physical states and the role of Gaussian resources. This situation has motivated several recent attempts to address or ``patch'' the infinities inherent to continuous-variable descriptions, such as the approach in~\cite{arzani_effective_2025}. While our ongoing work~\cite{descamps_heisenberg-weyl_2025} and~\cite{moulonguet_non-gaussianity_2026} already moves in this direction, much further research will likely be required to fully understand the continuous-variable limit and its implications for the structure of physical states and quantum resources.

\medbreak

In conclusion, this thesis highlights how a unified perspective based on modes, states, and symmetries can clarify the structure of quantum optical systems and their potential for quantum technologies. By combining conceptual analysis with metrological considerations, we have shown how seemingly different phenomena, from time-frequency encoding to collective entanglement, interferometric symmetries, and superselection rules, can be understood within a common framework. We hope that the ideas developed in this work will contribute to a deeper theoretical understanding of quantum optics while also inspiring new experimental and technological developments in quantum information science.

\fi

\appendix

\ifnum \theShowChaptwo=1
\chapter{Transferring operations to Fock space}
\setcurrentanchor{App A}
\label{app: transferring op to Fock}
\emph{In this short appendix we consider the problem of understanding the link between operations in a Hilbert space and those performed inside the associated Fock space.}
\localtableofcontents
\section{Setting and problem}
\paragraph{Definition}
We consider a Hilbert space $\mathcal H$ and its associated Fock space $\mathcal F$ as introduced in Sec.~\ref{subsec: FockSpace}. By definition $\mathcal F$ is the direct sum of all symmetric tensor products of $\mathcal H$; as such, the single-photon subspace is naturally isomorphic to $\mathcal H$. Introducing a basis of $\mathcal H$, $\{k\}$, we can make this connection explicit
\begin{align}
    \Lambda: \mathcal H&\rightarrow\mathcal F,\\
    \ket{\psi}=\sum_k \psi_k\ket{k}&\mapsto \sum_k \psi_k\hat a_k^\dagger\vac,
\end{align}
where we have introduced the creation operators $\{\hat a_k^\dagger\}$ associated with the basis $\{k\}$, satisfying the commutation relation $[\hat a_k,\hat a_l^\dagger]=\delta_{k,l}$. Additionally, there is a natural way to transport operators acting on $\mathcal H$ to operators acting on $\mathcal F$. For simplicity, we still denote this map $\Lambda$
\begin{align}
    \Lambda: \mathcal L(\mathcal H)&\rightarrow\mathcal L(\mathcal F),\\
    \hat U=\sum_{k,l} u_{k,l}\ket{k}\bra{l}&\mapsto \sum_{k,l} u_{k,l}\hat a_k^\dagger\hat a_l,
\end{align}

\paragraph{Compatibility}
It is then easy to verify that the map $\Lambda$ is compatible with the action of the operators on the state. More concretely, for any operator $\hat U\in\mathcal{L(H)}$ and state $\ket{\psi}\in\mathcal H$ we have
\begin{align}
    \Lambda(\hat U)\Lambda(\ket{\psi})=\Lambda(\hat U\ket{\psi}).
\end{align}
This mathematically encodes the fact that the Hilbert space $\mathcal H$ and the single-photon subspace of $\mathcal F$ are operationally equivalent.

\begin{derivation}
    Indeed, on the one hand
    \begin{subequations}
        \begin{align}
            \Lambda(\hat U\ket{\psi})=&\Lambda\left(\sum_{k,l}\sum_m u_{k,l}\psi_m \ket{k}\braket{l}{m}\right),\\
            &=\Lambda\left(\sum_{k,m} u_{k,m}\psi_m \ket{k}\right),\\
            &=\sum_{k,m} u_{k,m}\psi_m \hat a_k^\dagger\vac.
        \end{align}
    \end{subequations}
    And on the other, we indeed have
    \begin{subequations}
        \begin{align}
            \Lambda(\hat U)\Lambda(\ket{\psi})=&\left(\sum_{k,l} u_{k,l}\hat a_k^\dagger\hat a_l\right)\left(\sum_m \psi_m \hat a_m^\dagger\vac\right),\\
            =&\sum_{k,l,m} u_{k,l}\psi_m \hat a_k^\dagger\hat a_l\hat a_m^\dagger\vac,\\
            =&\sum_{k,l,m} u_{k,l}\psi_m \hat a_k^\dagger[\hat a_l,\hat a_m^\dagger]\vac,\\
            =&\sum_{k,m} u_{k,m}\psi_m \hat a_k^\dagger\vac.
        \end{align}
    \end{subequations}    
    \end{derivation}

\paragraph{Result}
Now that we have defined with $\Lambda$ operators that act on the whole Fock space $\mathcal F$, one can wonder what properties they satisfy. One could for example wonder if $\Lambda$ is a multiplicative morphism. Sadly, this is not the case in general. However, the following formula does hold
\begin{equation}
     \Lambda([\hat U,\hat V])=[\Lambda(\hat U),\Lambda(\hat V)],
\end{equation}
meaning that the action of the operators on the whole Fock space is compatible with the commutator.

\begin{derivation}
    We take $\hat U=\sum_{k,l} u_{k,l}\ketbra{k}{l}$ and $\hat V=\sum_{k,l}v_{k,l}\ketbra{k}{l}$ to be two operators of $\mathcal L(\mathcal H)$. We have
    \begin{align}
        \Lambda(\hat U)=\sum_{k,l} u_{k,l}\hat a_k^\dagger \hat a_l, && \Lambda(\hat V)=\sum_{k,l}v_{k,l}\hat a_k^\dagger \hat a_l,\\
        \Lambda(\hat U\hat V)=\sum_{k,l,m} u_{k,m}v_{m,l}\hat a_k^\dagger \hat a_l, && \Lambda(\hat V\hat U)=\sum_{k,l,m} v_{k,m}u_{m,l}\hat a_k^\dagger \hat a_l.
    \end{align}
    So that
    \begin{subequations}
        \begin{align}
            \Lambda(\hat U)\Lambda(\hat V)&=\sum_{j,k,l,m} u_{j,k}v_{l,m}\hat a_j^\dagger \hat a_k\hat a_l^\dagger \hat a_m,\\
            &=\sum_{j,k,l,m} u_{j,k}v_{l,m}\hat a_j^\dagger (\hat a_l^\dagger\hat a_k+\delta_{l,k})\hat a_m,\\
            &=\sum_{j,k,l,m} u_{j,k}v_{l,m}\hat a_l^\dagger\hat a_j^\dagger \hat a_m\hat a_k+\sum_{j,k,m} u_{j,k}v_{k,m}\hat a_j^\dagger \hat a_m,\\
            &=\sum_{j,k,l,m} u_{j,k}v_{l,m}\hat a_l^\dagger(\hat a_m\hat a_j^\dagger -\delta_{m,j})\hat a_k+\sum_{j,k,m} u_{j,k}v_{k,m}\hat a_j^\dagger \hat a_m,\\
            &=\sum_{j,k,l,m} u_{j,k}v_{l,m}\hat a_l^\dagger\hat a_m\hat a_j^\dagger \hat a_k-\sum_{j,k,l} u_{j,k}v_{l,j}\hat a_l^\dagger\hat a_k+\sum_{j,k,m} u_{j,k}v_{k,m}\hat a_j^\dagger \hat a_m,\\
            &=\Lambda(\hat V)\Lambda(\hat U)-\sum_{j,k,l} v_{j,l}u_{l,k}\hat a_j^\dagger\hat a_k+\sum_{j,k,m} u_{j,m}v_{m,k}\hat a_j^\dagger \hat a_k,\\
            &=\Lambda(\hat V)\Lambda(\hat U)-\Lambda(\hat V\hat U)+\Lambda(\hat U\hat V),\\
            &=\Lambda(\hat V)\Lambda(\hat U)+\Lambda([\hat U,\hat V]).
        \end{align}
    \end{subequations}
    Thus, we indeed get
    \begin{equation}
        [\Lambda(\hat U),\Lambda(\hat V)]=\Lambda([\hat U,\hat V]).
    \end{equation}
\end{derivation}

\section{Consequences}
\paragraph{Time-frequency commutation relation}
At least two consequences can be drawn from this result. First, if we assume that it can be generalized to the case of an infinite-dimensional Hilbert space, we recover the commutation relation of the operators $\hat \omega$ and $\hat t$ mentioned in Sec.~\ref{subsec: TF operators} and proven in Appendix~\ref{app: TF basic computations}. Indeed, here the mapping is for $\mathcal H$, the CV Hilbert space, with
\begin{equation}
    \Lambda\left(\int\dd x\, f(x)\ket{x}\right)=\int\dd\omega\, f(\omega)\hat a^\dagger(\omega)\vac
\end{equation}
We have $\hat \omega=\Lambda(\hat x)$, $\hat t=\Lambda(\hat p)$ and $\Lambda(\1)=\hat N$, so that
\begin{equation}
    [\hat \omega,\hat t]=[\Lambda(\hat x),\Lambda(\hat p)]=\Lambda([\hat x,\hat p])=\Lambda(i\1)=i\hat N
\end{equation}
With this we recover very quickly the commutation relation of the time and frequency operators in the general situation. 

\paragraph{Schwinger relations}
A second example is obtained in the case of two modes. If we consider a qubit and the three Pauli matrices $\hat X$, $\hat Y$ and $\hat Z$, we can associate a two-mode Fock space with ladder operators $\hat a_1$ and $\hat a_2$ and three operators
\begin{subequations}
    \begin{align}
        \hat J_z =\Lambda(\hat Z/2)=  \frac{1}{2}(\hat a_1^\dagger \hat a_1-\hat a_2^\dagger \hat a_2),\\
        \hat J_x=\Lambda(\hat X/2)=\frac{1}{2}(\hat a_2^\dagger \hat a_1+\hat a_1^\dagger \hat a_2),\\
        \hat J_y=\Lambda(\hat Y/2)=\frac{1}{2}(i\hat a_2^\dagger\hat a_1-i\hat a_1^\dagger \hat a_2),
    \end{align}
\end{subequations}
These are the operators introduced by Schwinger to represent angular momentum operators in terms of two bosonic modes~\cite{schwinger_angular_2015}. While verifying that these three operators indeed satisfy the angular momentum commutation relations is not hard, it is extremely tedious and all three commutation relations have to be verified, as the computation differs slightly each time.

With the general result from the previous section, this computation is not necessary anymore. From the commutation relations of the Pauli matrices,
\begin{align}
    [\hat X,\hat Y]=2i \hat Z, && [\hat Y,\hat Z]=2i \hat X, && [\hat Z,\hat X]=2i \hat Y,
\end{align}
we directly recover the angular momentum commutation relations
\begin{equation}
    [\hat J_x,\hat J_y]=i \hat J_z, \quad [\hat J_y,\hat J_z]=i \hat J_x, \quad [\hat J_z,\hat J_x]=i \hat J_y.
\end{equation}

\section{Mathematical digression on representation theory}
Since the formula proved in the previous section involves the commutator, and the commutator is the basic building block of Lie algebras, one can wonder if the previous statement can be rephrased in a Lie-theoretic way. It turns out to be the case. Let's first reintroduce the basic definition. A (complex) Lie algebra is a linear vector space $\mathfrak{g}$ which is endowed with a bilinear operation
\begin{align}
    \mathfrak g\times\mathfrak g:&\to \mathfrak g,\notag\\
    (x,y)&\mapsto [x,y],
\end{align}
which verifies for $x,y,z\in\mathfrak g$ the following two identities,
\begin{align}
    [x,x]=0, && [x,[y,z]]+[y,[z,x]]+[z,[x,y]]=0,
\end{align}
the last identity being called the Jacobi identity. The first identity and the linearity show that the Lie bracket is anti-symmetric $[x,z]=-[z,x]$. Given two Lie algebras $\mathfrak g$ and $\mathfrak h$, one can define the notion of Lie algebra morphism. It is a linear map $f:\mathfrak g\to\mathfrak h$ which is compatible with the Lie bracket. For $x,y\in \mathfrak g$,
\begin{equation}
    f([x,y])=[f(x),f(y)].
\end{equation}
The last definition we give is that of a representation. Given a vector space $V$, a representation of $\mathfrak g$ is a morphism $\rho: \mathfrak g\to\mathcal L(V)$, where $\mathcal L(V)$ is the Lie algebra of the linear operators acting on $V$, for which the Lie bracket is given by the usual commutator: $[f,g]=f\circ g-g\circ f$. When studying representation theory one can search for tools that allow the construction of new representations. A method to do so is the tensor product. Starting from two representations $\rho:\mathfrak g\to \mathcal L(V)$ and $\sigma:\mathfrak g\to \mathcal L(W)$, one can define a new representation on the space $V\otimes W$. It is given by the morphism $\rho\otimes\sigma$ defined, for $x\in\mathfrak g$, by
\begin{equation}
    (\rho\otimes\sigma)(x)=\rho(x)\otimes\1_W+\1_V\otimes\sigma(x).
\end{equation}
\begin{derivation}
    To verify that this is indeed a Lie algebra morphism, we mainly have to verify the compatibility with the commutator. If we take $x, y\in\mathfrak g$, we have
    \begin{subequations}
        \begin{align}
            [(\rho\otimes\sigma)(x),(\rho\otimes\sigma)(y)]&=(\rho\otimes\sigma)(x)\circ (\rho\otimes\sigma)(y)-(\rho\otimes\sigma)(y)\circ(\rho\otimes\sigma)(x),\\
            &=\Big(\rho(x)\otimes\1_W+\1_V\otimes\sigma(x)\Big)\circ\Big(\rho(y)\otimes\1_W+\1_V\otimes\sigma(y)\Big)\notag\\
            &\qquad-\Big(\rho(y)\otimes\1_W+\1_V\otimes\sigma(y)\Big)\notag\\
            &\qquad\circ\Big(\rho(x)\otimes\1_W+\1_V\otimes\sigma(x)\Big),\\
            &=(\rho(x)\circ\rho(y))\otimes\1_W+\rho(x)\otimes\sigma(y)+\rho(y)\otimes\sigma(x)\notag\\
            &\qquad+\1_V\otimes(\sigma(x)\circ\sigma(y))-(\rho(y)\circ\rho(x))\otimes\1_W\notag\\
            &\qquad-\rho(y)\otimes\sigma(x)-\rho(x)\otimes\sigma(y)-\1_V\otimes(\sigma(y)\circ\sigma(x))\\
            &=(\rho(x)\circ\rho(y))\otimes\1_W-(\rho(y)\circ\rho(x))\otimes\1_W\notag\\
            &\qquad+\1_V\otimes(\sigma(x)\circ\sigma(y))-\1_V\otimes(\sigma(y)\circ\sigma(x)),\\
            &=[\rho(x),\rho(y)]\otimes\1_W+\1_V\otimes[\sigma(x),\sigma(y)],\\
            &=\rho([x,y])\otimes\1_W+\1_V\otimes[\sigma(x),\sigma(y)],\\
            &=(\rho\otimes\sigma)([x,y]).
        \end{align}
    \end{subequations}
    So $\rho\otimes\sigma$ is indeed a Lie algebra morphism and thus a representation on $V\otimes W$.
\end{derivation}

If one has only one representation $(\rho,V)$ at hand, one can consider the tensor product with itself,
\begin{align}
    (\rho\otimes\rho): \mathfrak g&\to \mathcal L(V\otimes V),\notag\\
    x&\mapsto \rho(x)\otimes\1 +\1\otimes \rho(x).
\end{align}
Finally, if one considers tensor powers of $V$, one can iterate the construction and get a representation on $V^{\otimes n}$ defined by
\begin{align}
    \rho^{\otimes n}: \mathfrak g&\to \mathcal L(V^{\otimes n}),\notag\\
    x&\mapsto \rho(x)\otimes\1\otimes\cdots\otimes \1+\1\otimes\rho(x)\otimes\cdots\otimes\1+\cdots+\1\otimes\cdots\otimes\1\otimes\rho(x).
\end{align}
This construction is very similar to the formula obtained for the action of $\Lambda(\hat U)$ on the subspace of $n$ photons in $\mathcal F$. Let us make this link even clearer. But before doing that, we have to consider a sub-representation of $V^{\otimes n}$. Indeed, in the Fock space $\mathcal F$ the photons are indistinguishable. This means that whenever the quantum state contains multiple photons, it has to be invariant under the exchange of photons (no minus sign since we are dealing with bosons). We denote by $\mathfrak S_n$ the permutation group on the $n$ elements $1,\dots,n$. We can define a linear group action of $\mathfrak S_n$ on $V^{\otimes n}$ by its action on pure tensors and then extending it by linearity. Taking $v_1,\dots,v_n\in V$, and $\sigma\in \mathfrak S_n$ we define
\begin{equation}
    \sigma\cdot(v_1\otimes\cdots\otimes v_n)=v_{\sigma^{-1}(1)}\otimes\cdots\otimes v_{\sigma^{-1}(n)}.
\end{equation}

We can verify that the action of $\mathfrak S_n$ is compatible with the representation $\rho$ of the Lie algebra $\mathfrak g$. This means that for any $v\in V^{\otimes n}$, $x\in\mathfrak g$ and $\sigma\in\mathfrak S_n$, we have
\begin{equation}
    \sigma\cdot \rho^{\otimes n}(x)[v]=\rho^{\otimes n}(x)[\sigma\cdot v].
\end{equation}

\begin{derivation}
    To prove it, it is sufficient to verify this identity on the pure tensors by taking $v=v_1\otimes\cdots\otimes v_n$. Furthermore, since the symmetric group $\mathfrak S_n$ is generated by transpositions, we can assume that $\sigma$ is a transposition and by symmetry of the indices, we can take $\sigma=(1~2)$ (the permutation that exchanges the first two indices). So we compute
    \begin{subequations}
        \begin{align}
            \sigma\cdot(\rho^{\otimes n}(x)[v])&=\sigma\cdot(\rho^{\otimes n}(x)[v_1\otimes\cdots\otimes v_n]),\\
            &=\sigma\cdot(\rho(x)[v_1]\otimes\cdots\otimes v_n+v_1\otimes\rho(x)[v_2]\otimes\cdots\otimes v_n+\cdots\notag\\
            &\qquad+v_1\otimes\cdots\otimes v_{n-1} \otimes \rho(x)[v_n]),\\
            &=v_2\otimes\rho(x)[v_1]\otimes\cdots\otimes v_n+\rho(x)[v_2]\otimes v_1\otimes\cdots\otimes v_n +\cdots\notag\\
            &\qquad+v_2\otimes v_1\otimes\cdots\otimes v_{n-1} \otimes \rho(x)[v_n],\\
            &=\rho^{\otimes n}(x)[v_2\otimes v_1\otimes\cdots\otimes v_n],\\
            &=\rho^{\otimes n}(x)[\sigma\cdot(v_1\otimes\cdots\otimes v_n)],\\
            &=\rho^{\otimes n}(x)[\sigma\cdot v].
        \end{align}
    \end{subequations}
\end{derivation}

We now define the symmetric subspace $S_nV$ by
\begin{equation}
    S_nV=\{v\in V^{\otimes n}|\forall \sigma\in\mathfrak S_n,~\sigma\cdot v= v\}.
\end{equation}
The compatibility condition between the action of $\mathfrak S_n$ and $\mathfrak g$ shows that $S_nV$ is stable under the action of $\mathfrak g$. Indeed, for $v\in S_nV$, and $x\in\mathfrak g$ we have $\rho^{\otimes n}(x)[v]\in S_nV$ since for $\sigma\in\mathfrak S_n$ we have
\begin{equation}
    \sigma\cdot(\rho^{\otimes n}(x)[v])=\rho^{\otimes n}(x)[\sigma\cdot v]=\rho^{\otimes n}(x)[v].
\end{equation}
This means that the space $S_nV$ is a sub-representation of $V^{\otimes n}$. Now, we can make explicit the link with $\mathcal F$. We consider the Lie group to be $\mathcal {L(H)}$ and the vector space to be $V=\mathcal H$. The representation $\rho$ of $\mathcal {L(H)}$ on $\mathcal H$ is simply the defining representation, where the action is simply given by operator multiplication (so $\rho=\1$). One can then extend the representation to $S_n \mathcal H$ by the previous construction. Denoting by $\mathcal F_n$ the subspace of $n$ photons, we can extend the isomorphism $\Lambda$ to $\mathcal F_n$
\begin{align}
    \Lambda:S_n \mathcal H&\to \mathcal F_n\notag\\
     \sum_{j_1,\dots,j_n}\phi_{j_1\cdots j_n}\ket{j_1}\otimes\cdots\otimes\ket{j_n}&\mapsto\frac{1}{\sqrt{n!}}\sum_{j_1,\dots,j_n}\phi_{j_1\cdots j_n}\hat a_{j_1}^\dagger \cdots \hat a_{j_n}^\dagger \vac
\end{align}
Furthermore, the previous computations show that for $\hat U\in \mathcal L(\mathcal H)$, the morphism $\Lambda$ is compatible with the action of $\hat U$. So that for $\ket{\psi}\in S_n\mathcal H$,
\begin{equation}
    \Lambda\big(\rho^{\otimes n}(\hat U)\ket{\psi}\big)=\Lambda(\hat U)\Lambda(\ket{\psi})
\end{equation}
Finally, by considering the direct sum of all the $S_n \mathcal H$, we get an isomorphism 
\begin{equation}
    \Lambda: \bigoplus_{n\geq 0} S_n\mathcal H  	\tilde{\rightarrow} \mathcal F
\end{equation}
where $S_1\mathcal H=\mathcal H$ and $S_0\mathcal H=\C$. This isomorphism gives a formal, mathematically rigorous presentation of the Fock space $\mathcal F$. And the construction of the representation over the tensor power of $\mathcal H$ explains why the commutator is compatible with the action of the operators on $\mathcal F$. Notably, all the arguments of this more mathematical presentation do not use finite-dimensional assumptions and thus provide a proper justification of the $\hat \omega$ and $\hat t$ commutation relation.

\fi

\ifthenelse{
    \value{ShowChapone}=1 \OR
    \value{ShowChaptwo}=1 \OR
    \value{ShowChapthree}=1 \OR
    \value{ShowChapfour}=1  \OR
    \value{ShowChapfive}=1
}{
    \chapter{Mathematical derivations}

    \emph{This appendix collects the different mathematical derivations supporting equations and statements presented in the main text.}

    \localtableofcontents
}{}

\ifnum \theShowChapone=1

\section{Formalism and framework}
\label{app: formalism and framework}
\setcurrentanchor{app formalism}
\emph{This section of the appendix collects all the results of Chap.~\ref{chap: Formalism and framework} and provides the associated proofs.}

\vspace*{1em}
\par\noindent
\textbf{\large Results in this section}\par
\vspace{-0.8em}
\noindent\rule{\textwidth}{0.8pt}\par

\resultentry{res: BCH}{Baker-Campbell-Hausdorff}
\resultentry{res: unbaised estimator may not exist}{An unbiased estimator may not exist}
\resultentry{res: bias and variance of a repeated measurement}{Bias and variance of a repeated measurement}
\resultentry{res: Fi coin flip}{Fisher information for coin flipping}
\resultentry{res: additivity of Fi}{Additivity of the Fisher information}
\resultentry{res: convexity of Fi}{Convexity of the Fisher information}
\resultentry{res: CRB}{The Cramér-Rao bound}
\resultentry{res: QCRB pure states}{Quantum Cramér-Rao bound for pure states}
\resultentry{res: saturation QCRB pure states}{Saturation of the QCRB for pure states}
\resultentry{res: qubit variance computations}{Qubit variance computations}
\resultentry{res: QCRB mixed states}{QCRB for mixed states}
\resultentry{res: SLD and QFI mixed states}{Expression of the SLD and QFI}
\resultentry{res: saturation QCRB mixed states}{Saturation of QCRB for mixed states}
\resultentry{res: additivity QFI}{Additivity of the QFI}
\resultentry{res: convexity QFI}{Convexity of the QFI}
\resultentry{res: QFI variance inequality}{Inequality between QFI and quantum variance}
\resultentry{res: convex roof QFI}{Quantum Fisher information and convex roof}
\resultentry{res: equivalence of mode transformation definitions}{Equivalence of mode transformation definitions}

\vspace{-0.2em}
\par\noindent\rule{\textwidth}{0.8pt}\par

\begin{result}[Baker-Campbell-Hausdorff]~\\
    \label{res: BCH}
    A particular case of the Baker-Campbell-Hausdorff formula states that for two operators $\hat A$ and $\hat B$ such that their commutator $[\hat A, \hat B]$ commutes with both $\hat A$ and $\hat B$, the following identity holds
    \begin{equation}\label{eq: BCH1}
        e^{\hat A} e^{\hat B} = e^{\hat A+\hat B} e^{\frac{1}{2}[\hat A, \hat B]},
    \end{equation}
    from which we deduce
    \begin{equation}\label{eq: BCH2}
        e^{\hat A} e^{\hat B} = e^{\hat B} e^{\hat A} e^{[\hat A, \hat B]}.
    \end{equation}
\end{result}

\begin{derivation}
    We first show that for $\hat A$, $\hat B$ commuting with $[\hat A,\hat B]$ and a real parameter $t$, we have
    \begin{equation}
        e^{t\hat A} \hat Be^{-t\hat A}=\hat B+t[\hat A,\hat B].
    \end{equation}
    To do so, we simply introduce the functions $f(t)=e^{t\hat A} \hat Be^{-t\hat A}$ and $g(t)=\hat B+t[\hat A,\hat B]$ and compute the derivatives
    \begin{subequations}
        \begin{align}
            f'(t)&=e^{t\hat A}\hat A\hat Be^{-t\hat A}-e^{t\hat A}\hat B\hat Ae^{-t\hat A}=e^{t\hat A}[\hat A,\hat B]e^{-t\hat A}=[\hat A,\hat B],\\
            g'(t)&=[\hat A,\hat B].
        \end{align}
    \end{subequations}
    Thus, $f$ and $g$ have the same derivatives; since they take the same value at $t=0$ we have $f=g$. We then show the first result \eqref{eq: BCH1} by setting $f(t)=e^{t\hat A}e^{t\hat B}$ and $g(t)=e^{t(\hat A+\hat B)}e^{\frac{t^2}{2}[\hat A,\hat B]}$. We once again compute the respective derivatives:
    \begin{subequations}
        \begin{align}
            f'(t)&=\hat Ae^{t\hat A}e^{t\hat B}+e^{t\hat A}\hat Be^{t\hat B},\\
            &=\left(\hat A+e^{t\hat A}\hat Be^{-t\hat A}\right)e^{t\hat A}e^{t\hat B},\\
            &=(\hat A+\hat B+t[\hat A,\hat B])f(t),\\
            g'(t)&=(\hat A+\hat B)e^{t(\hat A+\hat B)}e^{\frac{t^2}{2}[\hat A,\hat B]}+e^{t(\hat A+\hat B)}t[\hat A,\hat B]e^{\frac{t^2}{2}[\hat A,\hat B]},\\
            &=(\hat A+\hat B+t[\hat A,\hat B])g(t).
        \end{align}
    \end{subequations}
    This means that $f$ and $g$ satisfy the same first-order linear differential equation. Since $f(0)=g(0)=1$, by the Cauchy-Lipschitz theorem, we indeed have $f=g$. Setting $t=1$ gives us the desired result. Finally the second equality \eqref{eq: BCH2} is obtained as
    \begin{equation}
        e^{\hat A}e^{\hat B}=e^{\hat A+\hat B}e^{[\hat A,\hat B]/2}=e^{\hat B+\hat A}e^{[\hat A,\hat B]/2}=e^{\hat B}e^{\hat A}e^{[\hat A,\hat B]/2}e^{-[\hat B,\hat A]/2}=e^{\hat B}e^{\hat A} e^{[\hat A,\hat B]}.
    \end{equation}
\end{derivation}

\begin{result}[An unbiased estimator may not exist]~\\
    \label{res: unbaised estimator may not exist}
    No estimator that is unbiased for all values of $\theta$ exists for determining $\theta$ in a coin flip scenario where the probability of getting heads is $\theta^2$.
\end{result}

\begin{derivation}
    As an estimator is a map from $\{H,T\}$ to $\R$, we can parametrize all possibilities by defining $a=\overline{\theta}(H)$ and $b=\overline{\theta}(T)$. With this, we see that the expected value is
    \begin{equation}
        \mathbb E(\overline{\theta})=\overline{\theta}(H)P_\theta(H)+\overline{\theta}(T)P_\theta(T)=a\theta^2+b(1-\theta^2).
    \end{equation}
    Requiring $\mathbb E(\overline{\theta})=\theta$ for all parameter values gives the equations
    \begin{equation}
        \left\{\begin{array}{cc}
            b=0 & \text{for $\theta=0$}, \\
            a=1 & \text{for $\theta=1$}, \\
            \frac{a}{4}+\frac{3b}{4}=\frac{1}{2} & \text{for $\theta=\frac{1}{2}$}, \\
        \end{array}\right.
    \end{equation}
    which obviously have no solutions.
\end{derivation}

\begin{result}[Bias and variance of a repeated measurement]~\\
    \label{res: bias and variance of a repeated measurement}
    For $n$ repetitions of a measurement process associated with the set of outcomes $X$ with probability distribution $P_\theta$, described by the collective outcomes $\tilde{X}=X^n=\{(x_1,\dots,x_n)|x_j\in X\}$ and probability distribution
    \begin{equation}
        \tilde{P}_\theta(\tilde{x})=P_\theta(x_1)\times\cdots\times P_\theta(x_n),
    \end{equation}
    for $\tilde{x}=(x_1,\dots,x_n)\in\tilde X$, the estimator $\overline{\tilde{\theta}}$ given by 
    \begin{equation}
            \overline{\tilde{\theta}}(\tilde x)=\frac{1}{n}\sum_{j=1}^n \overline{\theta}(x_j).
    \end{equation}
    is unbiased if  $\overline{\theta}$ is unbiased and satisfies
    \begin{equation}
            \mathbb V(\overline{\tilde{\theta}})=\frac{\mathbb V( \overline{\theta})}{n}.
    \end{equation}
\end{result}

\begin{derivation}
    By linearity of the expectation value we get
    \begin{subequations}
        \allowdisplaybreaks
        \begin{align}
            \mathbb E(\overline{\tilde{\theta}})&=\sum_{\tilde x\in \tilde X} \tilde P_\theta(\tilde x)\overline{\tilde{\theta}}(\tilde x),\\
            &=\frac{1}{n}\sum_{x_1,\dots,x_n\in X}P_\theta(x_1)\cdots P_\theta(x_n)[\overline{\theta}(x_1)+\cdots +\overline{\theta}(x_n)],\\
            &=\frac{1}{n}\left[\sum_{x_1,\dots,x_n\in X}P_\theta(x_1)\cdots P_\theta(x_n)\overline{\theta}(x_1)+\cdots+\sum_{x_1,\dots,x_n\in X}P_\theta(x_1)\cdots P_\theta(x_n)\overline{\theta}(x_n)\right],\\
            &=\frac{1}{n}\left[\sum_{x_1\in X} P_\theta(x_1)\overline{\theta}(x_1)+\cdots+\sum_{x_n\in X} P_\theta(x_n)\overline{\theta}(x_n)\right],\\
            &=\frac{1}{n}[\mathbb E(\overline{\theta})+\cdots+\mathbb E(\overline{\theta})],\\
            &=\mathbb E(\overline{\theta}) = \theta.
        \end{align}
    \end{subequations}
    
    So that $\overline{\tilde{\theta}}$ is indeed unbiased. We also compute
    \begin{subequations}
        \begin{align}
            \mathbb V(\overline{\tilde{\theta}})&=\mathbb E[\overline{\tilde{\theta}}^2]-\mathbb E(\overline{\tilde{\theta}})^2,\\
            &=\frac{1}{n^2}\sum_{x_1,\dots,x_n\in X}P_\theta(x_1)\cdots P_\theta(x_n)[\overline{\theta}(x_1)+\cdots +\overline{\theta}(x_n)]^2-\mathbb E(\overline{\theta})^2,\\
            &=\frac{1}{n^2}\sum_{j=1}^n \sum_{x_j\in X} P_\theta(x_j)\overline{\theta}(x_j)^2+\frac{1}{n^2}\sum_{j\neq k}\sum_{x_j,x_k\in X} P_\theta(x_j)P_\theta(x_k)\overline{\theta}(x_j)\overline{\theta}(x_k)-\mathbb E(\overline{\theta})^2,\\
            &=\frac{1}{n}\mathbb E(\overline{\theta}^2)+\frac{n(n-1)}{n^2}\mathbb E(\overline{\theta})^2-\mathbb E(\overline{\theta})^2,\\
            &=\frac{1}{n}\Big[\mathbb E(\overline{\theta}^2)-\mathbb E(\overline{\theta})^2\Big],\\
            &=\frac{\mathbb V( \overline{\theta})}{n}.
        \end{align}
    \end{subequations}
    {\it We note that these two computations could have been done trivially using properties of the expectation value and variance by noting that $\overline{\tilde\theta}$ is the sum of $n$ independent and identically distributed random variables.}
\end{derivation}

\begin{result}[Fisher information for coin flipping]~\\
    \label{res: Fi coin flip}
    In the case of a coin flip, where the probability distribution is $P_\theta(H)=\theta$ and $P_\theta(T)=1-\theta$, the estimator $\overline{\theta}$ defined by $\overline{\theta}(H)=1$ and $\overline{\theta}(T)=0$ saturates the Cramér-Rao bound.
\end{result}

\begin{derivation}
    First, we have computed in the main text that the associated variance is $\mathbb V(\overline{\theta})=\theta(1-\theta)$. We now compute the Fisher information. By definition, we have
    \begin{align}
        \mathcal F_\theta &=\sum_{x\in\{H,T\}} \frac{1}{P_\theta(x)}\left(\frac{\partial P_\theta(x)}{\partial \theta}\right)^2,\\
        &=\frac{1}{P_\theta(H)}\left(\frac{\partial P_\theta(H)}{\partial \theta}\right)^2+\frac{1}{P_\theta(T)}\left(\frac{\partial P_\theta(T)}{\partial \theta}\right)^2,\\
        &=\frac{1}{\theta}\left(1\right)^2+\frac{1}{1-\theta}\left( -1\right)^2,\\
        &=\frac{1-\theta+\theta}{\theta(1-\theta)},\\
        &=\frac{1}{\theta(1-\theta)}.
    \end{align}
    We then indeed have
    \begin{equation}
        \mathbb V(\overline{\theta})=\theta(1-\theta)=\frac{1}{\mathcal F_\theta },
    \end{equation}
    so that the Cramér-Rao bound is saturated.
\end{derivation}

\begin{result}[Additivity of the Fisher information]~\\
    \label{res: additivity of Fi}
    The Fisher information is additive in two independent experiments. That is, if $X=X_1\times X_2$, and $P_\theta(x_1,x_2)=P^{(1)}_\theta(x_1)P_\theta^{(2)}(x_2)$ where $P_\theta$, $P^{(1)}_\theta$ and $P^{(2)}_\theta$ are the probabilities to get the respective outcomes given that the true value of the parameter is $\theta$, then the associated Fisher information $\mathcal F_\theta$, $\mathcal F^{(1)}_\theta$ and $\mathcal F^{(2)}_\theta$ satisfy
    \begin{equation}
        \mathcal F_\theta=\mathcal F^{(1)}_\theta+\mathcal F^{(2)}_\theta.
    \end{equation}
\end{result}

\begin{derivation}
    We compute
    \allowdisplaybreaks
    \begin{align}
        \mathcal F_\theta &=\sum_{x\in X}\frac{1}{P_\theta(x)}\left(\frac{\partial P_\theta(x)}{\partial \theta}\right)^2,\\
        &=\sum_{x_1\in X_1,\, x_2\in X_2}\frac{1}{P^{(1)}_\theta(x_1)P^{(2)}_\theta(x_2)}\left(\frac{\partial }{\partial \theta}\Big[P^{(1)}_\theta(x_1)P^{(2)}_\theta(x_2)\Big]\right)^2,\\
        &=\sum_{x_1\in X_1,\, x_2\in X_2}\frac{1}{P^{(1)}_\theta(x_1)P^{(2)}_\theta(x_2)}\left(\tfrac{\partial P^{(1)}_\theta(x_1)}{\partial \theta}P^{(2)}_\theta(x_2)+P^{(1)}_\theta(x_1)\tfrac{\partial P^{(2)}_\theta(x_2)}{\partial \theta}\right)^2,\\
        &=\sum_{x_1\in X_1,\, x_2\in X_2}\frac{1}{P^{(1)}_\theta(x_1)P^{(2)}_\theta(x_2)}\left(\tfrac{\partial P^{(1)}_\theta(x_1)}{\partial \theta}P^{(2)}_\theta(x_2)\right)^2\\
        &\qquad+2\sum_{x_1\in X_1,\, x_2\in X_2}\frac{1}{P^{(1)}_\theta(x_1)P^{(2)}_\theta(x_2)}\left(P^{(1)}_\theta(x_1)P^{(2)}_\theta(x_2)\tfrac{\partial P^{(1)}_\theta(x_1)}{\partial \theta}\tfrac{\partial P^{(2)}_\theta(x_2)}{\partial \theta}\right)\\
        &\qquad+\sum_{x_1\in X_1,\, x_2\in X_2}\frac{1}{P^{(1)}_\theta(x_1)P^{(2)}_\theta(x_2)}\left(\tfrac{\partial P^{(2)}_\theta(x_2)}{\partial \theta}P^{(1)}_\theta(x_1)\right)^2,\\
        &=\sum_{x_1\in X_1,\, x_2\in X_2}\frac{P^{(2)}_\theta(x_2)}{P^{(1)}_\theta(x_1)}\left(\tfrac{\partial P^{(1)}_\theta(x_1)}{\partial \theta}\right)^2+2\sum_{x_1\in X_1,\, x_2\in X_2}\tfrac{\partial P^{(1)}_\theta(x_1)}{\partial \theta}\tfrac{\partial P^{(2)}_\theta(x_2)}{\partial \theta}\\
        &\qquad+\sum_{x_1\in X_1,\, x_2\in X_2}\frac{P^{(1)}_\theta(x_1)}{P^{(2)}_\theta(x_2)}\left(\tfrac{\partial P^{(2)}_\theta(x_2)}{\partial \theta}\right)^2,\\
        &=\sum_{x_1\in X_1}\frac{1}{P^{(1)}_\theta(x_1)}\left(\tfrac{\partial P^{(1)}_\theta(x_1)}{\partial \theta}\right)^2\sum_{x_2\in X_2}P^{(2)}_\theta(x_2)+2\sum_{x_1\in X_1}\tfrac{\partial P^{(1)}_\theta(x_1)}{\partial \theta}\sum_{x_2\in X_2}\tfrac{\partial P^{(2)}_\theta(x_2)}{\partial \theta}\\
        &\qquad+\sum_{x_1\in X_1}P^{(1)}_\theta(x_1)\sum_{x_2\in X_2}\frac{1}{P^{(2)}_\theta(x_2)}\left(\tfrac{\partial P^{(2)}_\theta(x_2)}{\partial \theta}\right)^2,\\
        &=\mathcal F^{(1)}_\theta\times 1+2\frac{\partial }{\partial \theta}\sum_{x_1\in X_1}P^{(1)}_\theta(x_1)\frac{\partial}{\partial \theta}\sum_{x_2\in X_2} P^{(2)}_\theta(x_2)+1\times \mathcal F^{(2)}_\theta,\\
        &=\mathcal F^{(1)}_\theta+\mathcal F^{(2)}_\theta+2\frac{\partial }{\partial \theta}1\frac{\partial}{\partial \theta}1,\\
        &=\mathcal F^{(1)}_\theta+\mathcal F^{(2)}_\theta.
    \end{align}
\end{derivation}

\begin{result}[Convexity of the Fisher information]~\\
    \label{res: convexity of Fi}
    The Fisher information is convex. That is, considering a probability distribution $\{p_i\}$ such that 
    \begin{equation}
        P_\theta(x)=\sum_j p_j P_\theta^{(j)}(x),
    \end{equation}
    where $P_\theta^{(j)}(x)$ are some parameter-dependent probability distributions on $X$ (while the $p_j$ do not depend on $\theta$), we have
    \begin{equation}
        \mathcal F_\theta\leq \sum_j p_j \mathcal F_\theta^{(j)}.
    \end{equation}
\end{result}

\begin{derivation}
    We introduce the inner product on $\R^n$ given by $(\vec x,\vec y)=\sum_j p_j x_j y_j$, and use the Cauchy-Schwarz inequality to bound the term $\left(\frac{\partial P_\theta(x)}{\partial \theta}\right)^2$.
    \begin{subequations}
        \allowdisplaybreaks
        \begin{align}
            \left(\frac{\partial P_\theta(x)}{\partial \theta}\right)^2&=\left(\sum_j p_j\frac{\partial P_\theta^{(j)}(x)}{\partial \theta}\right)^2,\\
            &=\left(\sum_j p_j\,\sqrt{P_\theta^{(j)}(x)}\cdot\frac{1}{\sqrt{P_\theta^{(j)}(x)}}\frac{\partial P_\theta^{(j)}(x)}{\partial \theta}\right)^2,\\
            &\leq\left(\sum_j p_j P_\theta^{(j)}(x)\right)\left(\sum_j p_j \frac{1}{P_\theta^{(j)}(x)}\left(\frac{\partial P_\theta^{(j)}(x)}{\partial \theta}\right)^2\right),\\
            &=P_\theta(x) \sum_j p_j \frac{1}{P_\theta^{(j)}(x)}\left(\frac{\partial P_\theta^{(j)}(x)}{\partial \theta}\right)^2.
        \end{align}
    \end{subequations}
    Dividing by $P_\theta(x)$ and summing over $x$ gives
    \begin{equation}
        \mathcal F_\theta\leq \sum_j p_j \mathcal F_\theta^{(j)}.
    \end{equation}
\end{derivation}

\begin{result}[The Cramér-Rao bound]~\\
    \label{res: CRB}
    Given a probability distribution $P_\theta(x)$ over a set of outcomes $X$ depending on a parameter $\theta\in \R$, with associated Fisher information and any estimator $\overline{\theta}$, we have
    \begin{equation}
        \mathbb V(\overline{\theta})\geq \frac{\left(\frac{\partial}{\partial \theta}\mathbb E(\overline{\theta})\right)^2}{\mathcal F_\theta }.
    \end{equation}
\end{result}

\begin{derivation}
    We first define the score function $S:X\to \R$ as
    \begin{equation}
        S(x)=\frac{1}{P_\theta(x)}\frac{\partial P_\theta(x)}{\partial \theta}.
    \end{equation}
    We first notice that
    \begin{equation}
        \mathbb E(S)=\sum_{x\in X} P_\theta(x) S(x)=\sum_{x\in X}\frac{\partial}{\partial \theta} P_\theta(x)=\frac{\partial}{\partial\theta}\sum_{x\in X} P_\theta(x)=\frac{\partial}{\partial\theta} 1=0.
    \end{equation}
    Then we compute
    \begin{equation}
        \mathbb E(S^2)=\sum_{x\in X} P_\theta(x) S(x)^2=\sum_{x\in X}\frac{1}{P_\theta(x)}\left(\frac{\partial P_\theta(x)}{\partial \theta}\right)^2=\mathcal F_\theta .
    \end{equation}
    So that we have $\mathbb V(S)=\mathbb E(S^2)-\mathbb E(S)^2=\mathbb E(S^2)=\mathcal F_\theta $. We then compute the covariance $\Cov (S,\overline{\theta})$
    \begin{equation}
        \mathbb E(S\overline{\theta})=\sum_{x\in X}P_\theta(x)S(x)\overline{\theta}(x)=\sum_{x\in X}\frac{\partial}{\partial \theta}P_\theta(x)\overline{\theta}(x)=\frac{\partial}{\partial \theta}\sum_{x\in X}P_\theta(x)\overline{\theta}(x)=\frac{\partial}{\partial \theta}\mathbb E(\overline{\theta}),
    \end{equation}
    by moving the derivative in front of the sum since $\overline{\theta}(x)$ does not depend on $\theta$. Finally, we get the expression of the covariance
    \begin{equation}
        \Cov (S,\overline{\theta})=\mathbb E(S\overline{\theta})-\underbrace{\mathbb E(S)}_{=0}\mathbb E (\overline{\theta})=\frac{\partial}{\partial \theta}\mathbb E(\overline{\theta}).
    \end{equation}
    For two random variables $A,B:X\to \R$, the Cauchy-Schwarz inequality states that
    \begin{equation}
        \abs{\Cov (A,B)}^2\leq \mathbb V(A)\mathbb V(B).
    \end{equation}
    Applying this to $S$ and $\overline \theta$ gives
    \begin{equation}
        \abs{\Cov (S,\overline{\theta})}^2\leq \mathbb V(S)\mathbb V(\overline{\theta}),
    \end{equation}
    which then becomes
    \begin{equation*}
        \left(\frac{\partial}{\partial\theta}\mathbb E(\overline{\theta})\right)^2\leq \mathcal F_\theta  \mathbb V(\overline{\theta}).
    \end{equation*}
    Rearranging the terms gives the Cramér-Rao bound.
\end{derivation}

\begin{result}[Quantum Cramér-Rao bound for pure states]~\\
    \label{res: QCRB pure states}
    For a pure state $\ket{\psi_0}$ and a Hermitian operator $\hat H$, if the probability distribution $P_\theta(x)$ has the form
    \begin{equation}
        P_\theta(x)=\bra{\psi_0}e^{i\theta \hat H}\hat E_xe^{-i\theta \hat H}\ket{\psi_0}.
    \end{equation}
    for some POVM $\{\hat E_x\}_{x\in X}$, then the associated Fisher information satisfies
    \begin{equation}
        \mathcal F_\theta\leq 4\Delta^2 \hat H.
    \end{equation}
\end{result}

\begin{derivation}
    We begin by computing the derivative of $P_\theta(x)$ with respect to $\theta$. A direct differentiation yields
    \begin{align}
        \partial_\theta P_\theta(x)
        &= \partial_\theta \bra{\psi_0} e^{i \hat H \theta}\hat E_x e^{-i \hat H \theta}\ket{\psi_0}, \notag \\
        &= -i \bra{\psi_\theta}\hat E_x \hat H\ket{\psi_\theta}
        + i \bra{\psi_\theta}\hat H \hat E_x\ket{\psi_\theta},\notag\\
        &= 2\Im(\bra{\psi_\theta}\hat E_x \hat H\ket{\psi_\theta}),
    \end{align}
    since $\bra{\psi_\theta}\hat E_x \hat H\ket{\psi_\theta}^*= \bra{\psi_\theta}\hat H \hat E_x\ket{\psi_\theta}$. Let $g\in\mathbb{R}$ be an arbitrary real constant. Since $g\,\bra{\psi_\theta}\hat E_x\ket{\psi_\theta} \in \mathbb{R}$, it does not contribute to the imaginary part, and we may equivalently write
    \begin{equation}
        \partial_\theta P_\theta(x)
        = 2\,\Im\!\left(\bra{\psi_\theta}\hat E_x (\hat H - g)\ket{\psi_\theta}\right).
    \end{equation}
    Using the elementary inequality $\abs{\Im(z)} \leq \abs{z}$ for any $z\in\mathbb{C}$, we obtain
    \begin{equation}
        \left(\partial_\theta P_\theta(x)\right)^2
        \leq 4\,\abs{\bra{\psi_\theta}\hat E_x (\hat H - g)\ket{\psi_\theta}}^2.
    \end{equation}
    We now define the vectors
    \begin{align}
        \ket{\Psi_\theta} = \sqrt{\hat E_x}\ket{\psi_\theta}, &&
        \ket{\Phi_\theta} = \sqrt{\hat E_x}(\hat H - g)\ket{\psi_\theta}.
    \end{align}
    With these definitions, we can rewrite $\bra{\psi_\theta}\hat E_x (\hat H - g)\ket{\psi_\theta} = \braket{\Psi_\theta}{\Phi_\theta}$. Applying the Cauchy-Schwarz inequality, we obtain
    \begin{align}
        \left(\partial_\theta P_\theta(x)\right)^2
        &\leq 4\,\braket{\Psi_\theta}\braket{\Phi_\theta}, \notag \\
        &= 4\,\bra{\psi_\theta}\hat E_x\ket{\psi_\theta}
        \bra{\psi_\theta}(\hat H - g)\hat E_x(\hat H - g)\ket{\psi_\theta}.
    \end{align}
    Recognizing $P_\theta(x)=\bra{\psi_\theta}\hat E_x\ket{\psi_\theta}$, we arrive at
    \begin{equation}
        \left(\partial_\theta P_\theta(x)\right)^2
        \leq 4\,P_\theta(x)\,
        \bra{\psi_\theta}(\hat H - g)\hat E_x(\hat H - g)\ket{\psi_\theta}.
        \label{eq:key-inequality}
    \end{equation}
    The classical Fisher information is defined as
    \begin{equation}
        \mathcal F_\theta = \sum_{x\in X} \frac{1}{P_\theta(x)}
        \left(\partial_\theta P_\theta(x)\right)^2.
    \end{equation}
    Using inequality~\eqref{eq:key-inequality}, we find
    \begin{align}
        \mathcal F_\theta&\leq 4 \sum_{x\in X}
        \bra{\psi_\theta}(\hat H - g)\hat E_x(\hat H - g)\ket{\psi_\theta}, \notag \\
        &= 4\,\bra{\psi_\theta}(\hat H - g)
        \left(\sum_{x\in X}\hat E_x\right)
        (\hat H - g)\ket{\psi_\theta}.
    \end{align}
    Since the POVM satisfies the normalization condition $\sum_{x\in X}\hat E_x = \1$, this simplifies to
    \begin{equation}
        \mathcal F_\theta \leq 4\,\bra{\psi_\theta}(\hat H - g)^2\ket{\psi_\theta}.
    \end{equation}
    The right-hand side is minimized by choosing $g = \bra{\psi_\theta}\hat H\ket{\psi_\theta}$, which yields
    \begin{equation}
        \mathcal F_\theta \leq 4\left(\bra{\psi_\theta}\hat H^2\ket{\psi_\theta}
            - \bra{\psi_\theta}\hat H\ket{\psi_\theta}^2
        \right)
        = 4\,\Delta^2 \hat H.
    \end{equation}
\end{derivation}

\begin{result}[Saturation of the QCRB for pure states]~\\
    \label{res: saturation QCRB pure states}
    For a pure state $\ket{\psi_0}$ and a Hermitian operator $\hat H$, and a fixed value of the parameter $\theta_0$, there exists a POVM $\{\hat E_x\}_{x\in X}$ such that the Fisher information $\mathcal F_\theta$ of the probability distribution
    \begin{equation}
        P_\theta(x)=\bra{\psi_0}e^{i\theta \hat H}\hat E_xe^{-i\theta \hat H}\ket{\psi_0}.
    \end{equation}
    satisfies
    \begin{equation}
        \mathcal F_{\theta_0}= 4\Delta^2 \hat H.
    \end{equation}
\end{result}

\begin{derivation}
    For simplicity, we consider the case $\theta_0=0$. The general case is treated similarly, or is obtained by performing the evolution $e^{-i\hat H\theta_0}$ on the probe state $\ket{\psi_0}$ and the POVM $\{\hat E_x\}_{x\in X}$, which does not change the variance of $\hat H$ or the Fisher information. We consider the two-outcome POVM
    \begin{align}
        \hat E_1 = \ketbra{\psi_0}, &&
        \hat E_2 = \1 - \ketbra{\psi_0}.
    \end{align}
    The corresponding probabilities are
    \begin{align}
        P_\theta(1) = \abs{\braket{\psi_0}{\psi_\theta}}^2
        = \abs{\bra{\psi_0}e^{-i\hat H\theta}\ket{\psi_0}}^2, &&
        P_\theta(2) = 1 - P_\theta(1).
    \end{align}
    By definition, the Fisher information reads
    \begin{align}\label{eq:fisher-two-outcomes}
        \mathcal F_\theta &= \frac{1}{P_\theta(1)}\left(\partial_\theta P_\theta(1)\right)^2
        + \frac{1}{P_\theta(2)}\left(\partial_\theta P_\theta(2)\right)^2,\notag\\
        &= \frac{1}{P_\theta(1)\big(1-P_\theta(1)\big)}
        \left(\partial_\theta P_\theta(1)\right)^2,
    \end{align}
    since $P_\theta(2)=1-P_\theta(1)$ and $\partial_\theta P_\theta(2)=-\partial_\theta P_\theta(1)$. We now expand $P_\theta(1)$ for small $\theta$. Using the Taylor expansion $e^{-i\hat H\theta}
        = \1 - i\hat H\theta - \frac{1}{2}\hat H^2\theta^2 + o(\theta^2)$, we find
    \begin{align}
        P_\theta(1) &= \abs{1 - i\theta\bra{\psi_0}\hat H\ket{\psi_0} - \frac{\theta^2}{2}\bra{\psi_0}\hat H^2\ket{\psi_0} + o(\theta^2)}^2, \notag \\ &= 1 - \theta^2 \left(\bra{\psi_0}\hat H^2\ket{\psi_0} - \bra{\psi_0}\hat H\ket{\psi_0}^2 \right) + o(\theta^2), \notag \\ &= 1 - \theta^2 \Delta^2 \hat H + o(\theta^2).
    \end{align}
    Differentiating, we obtain
    \begin{equation}
        \partial_\theta P_\theta(1)
        = -2\theta\,\Delta^2 \hat H + o(\theta).
    \end{equation}
    Substituting these expansions into Eq.~\eqref{eq:fisher-two-outcomes}, we find
    \begin{align}
        \mathcal F_\theta
        &= \frac{\left(-2\theta\,\Delta^2 \hat H + o(\theta)\right)^2}
        {\left[1-\theta^2\Delta^2 \hat H+o(\theta^2)\right]
        \left[\theta^2\Delta^2 \hat H+o(\theta^2)\right]}, \notag \\
        &= \frac{4\theta^2(\Delta^2 \hat H)^2 + o(\theta^2)}
        {\theta^2\Delta^2 \hat H + o(\theta^2)}, \notag \\
        &= 4\,\Delta^2 \hat H + o(1).
    \end{align}
    Taking the limit $\theta \to 0$, we conclude that
    \begin{equation}
        \lim_{\theta\to 0} \mathcal F_\theta = 4\Delta^2 \hat H.
    \end{equation}

    This demonstrates that the two-outcome POVM $\{\hat E_1,\hat E_2\}$ saturates the quantum Cramér-Rao bound locally around $\theta=0$.
\end{derivation}

\begin{result}[Qubit variance computations]~\\
    \label{res: qubit variance computations}
    \begin{enumerate}
        \item[(i)] The variance of $\hat H=\hat Z$ for the single-qubit probe state $\ket{+}=(\ket{0}+\ket{1})/\sqrt{2}$ is
        \begin{equation}
            \Delta^2\hat Z=1.
        \end{equation}
        \item[(ii)] The variance of $\hat H=\sum_j \hat Z_j$ for the $n$-qubit probe state $\ket{\phi_1}=\ket{+}^{\otimes n}$ is
        \begin{equation}
            \Delta^2\hat H=n.
        \end{equation}
        \item[(iii)] The variance of $\hat H=\sum_j \hat Z_j$ for the $n$-qubit GHZ probe state $\ket{\phi_2}=(\ket{0\cdots0}+\ket{1\cdots1})/\sqrt{2}$ is
        \begin{equation}
            \Delta^2\hat H=n^2.
        \end{equation}
    \end{enumerate}
\end{result}
    
\begin{derivation}
    \textbf{(i) Single-qubit probe.}
    Let $\ket{+} = (\ket{0}+\ket{1})/\sqrt{2}$ and $\hat H=\hat Z$. Since $\hat Z^2=\1$, one finds
    \begin{align}
        \bra{+}\hat Z^2\ket{+}=1,
        &&
        \bra{+}\hat Z\ket{+}=0,
    \end{align}
    and therefore
    \begin{equation}
        \Delta^2 \hat Z
        = \bra{+}\hat Z^2\ket{+}
        - \bra{+}\hat Z\ket{+}^2
        = 1.
    \end{equation}
    \medbreak
    \textbf{(ii) Additivity of the variance for product states.}
    Let $\ket{\psi}=\ket{\psi_1}\otimes\ket{\psi_2}$ and
    \begin{equation}
        \hat H=\hat H_1\otimes\1+\1\otimes\hat H_2.
    \end{equation}
    A direct computation yields
    \begin{subequations}
        \begin{align}
            \bra{\psi}\hat H\ket{\psi}
            &= \bra{\psi_1}\hat H_1\ket{\psi_1}
            + \bra{\psi_2}\hat H_2\ket{\psi_2}, \\
            \bra{\psi}\hat H^2\ket{\psi}
            &= \bra{\psi_1}\hat H_1^2\ket{\psi_1}
            + \bra{\psi_2}\hat H_2^2\ket{\psi_2}
            + 2\bra{\psi_1}\hat H_1\ket{\psi_1}
            \bra{\psi_2}\hat H_2\ket{\psi_2}.
        \end{align}    
    \end{subequations}
    Subtracting the square of the mean, one obtains
    \begin{equation}
        \Delta^2 \hat H
        = \Delta^2 \hat H_1 + \Delta^2 \hat H_2.
    \end{equation}
    By iteration, for a product state of $n$ qubits and a collective generator
    $\hat H=\sum_{j=1}^n \hat Z_j$, one finds $\Delta^2 \hat H = n$ for the state $\ket{\phi_1}$.

    \medbreak
    \textbf{(iii) GHZ state and quadratic enhancement.}
    Since $\hat Z_j\ket{0\cdots0}=\ket{0\cdots0}$ and
    $\hat Z_j\ket{1\cdots1}=-\ket{1\cdots1}$, $\ket{\phi_2}$ is always orthogonal to $\hat Z_j\ket{\phi_2}$ leading to
    \begin{equation}
        \bra{\phi_2}\hat H\ket{\phi_2}=0.
    \end{equation}
    Moreover, for any indices $j$ and $k$,
    \begin{align}
        \hat Z_j\hat Z_k\ket{0\cdots0}=\ket{0\cdots0}, && \hat Z_j\hat Z_k\ket{1\cdots1}=\ket{1\cdots1},
    \end{align}
    which implies that $\hat Z_j\hat Z_k$ acts trivially on $\ket{\phi_2}$. Hence,
    \begin{equation}
        \bra{\phi_2}\hat H^2\ket{\phi_2}
        = \sum_{j,k=1}^n \bra{\phi_2}\hat Z_j\hat Z_k\ket{\phi_2}
        = n^2.
    \end{equation}
    Hence,
    \begin{equation}
        \Delta^2 \hat H = n^2.
    \end{equation}
\end{derivation}

\begin{result}[QCRB for mixed states]~\\
    \label{res: QCRB mixed states}
    For a mixed state $\hat \rho_\theta=e^{-i\hat H\theta}\hat \rho_0 e^{i\hat H\theta}$ undergoing unitary parameter encoding, and any POVM elements $\{\hat E_x\}_{x\in X}$, the associated Fisher information $\mathcal F_\theta$ satisfies the quantum Cramér-Rao bound
    \begin{equation}
        \mathcal F_\theta \leq \Tr(\hat \rho_\theta \hat L_\theta^2),
    \end{equation}
    where $\hat L_\theta$ is the symmetric logarithmic derivative (SLD) defined implicitly by $\partial_\theta \hat \rho_\theta = \frac{1}{2}(\hat L_\theta \hat \rho_\theta + \hat \rho_\theta \hat L_\theta).$    
\end{result}

\begin{derivation}
    The outcome probabilities are
    \begin{equation}
        P_\theta(x) = \Tr(\hat E_x \hat \rho_\theta).
    \end{equation}
    Differentiating $P_\theta(x)$ and using cyclicity of the trace yields
    \begin{subequations}
        \begin{align}
            \partial_\theta P_\theta(x)
            &= \Tr(\hat E_x \partial_\theta \hat \rho_\theta),\\
            &= \frac{1}{2}\Tr(\hat E_x \hat L_\theta \hat \rho_\theta)
            + \frac{1}{2}\Tr(\hat E_x \hat \rho_\theta \hat L_\theta),\\
            &= \Re\!\left\{\Tr(\hat \rho_\theta \hat L_\theta \hat E_x)\right\}.
        \end{align}     
    \end{subequations}
    Using the inequality $\abs{\Re(z)}^2 \le \abs{z}^2$ and writing
    \begin{equation}
        \Tr(\hat \rho_\theta \hat L_\theta \hat E_x)
        = \Tr\!\left(
            \sqrt{\hat \rho_\theta}\hat L_\theta \sqrt{\hat E_x}
            \sqrt{\hat E_x}\sqrt{\hat \rho_\theta}
        \right),
    \end{equation}
    we apply the Cauchy-Schwarz inequality in the operator Hilbert space
    \begin{equation}
        \abs{\Tr(\hat A^\dagger \hat B)}^2
        \le \Tr(\hat A^\dagger \hat A)\Tr(\hat B^\dagger \hat B),
    \end{equation}
    with
    \begin{align}
        \hat A = \sqrt{\hat E_x}\hat L_\theta\sqrt{\hat \rho_\theta}, && \hat B = \sqrt{\hat E_x}\sqrt{\hat \rho_\theta}.
    \end{align}
    This gives
    \begin{equation}
        \bigl(\partial_\theta P_\theta(x)\bigr)^2
        \le P_\theta(x)\,
        \Tr(\hat \rho_\theta \hat L_\theta \hat E_x \hat L_\theta).
    \end{equation}
    Substituting into the definition of the classical Fisher information
    \begin{subequations}
        \begin{align}
            \mathcal F_\theta
            &= \sum_x \frac{1}{P_\theta(x)}
            \bigl(\partial_\theta P_\theta(x)\bigr)^2 ,\\
            &\le \sum_x \Tr(\hat \rho_\theta \hat L_\theta \hat E_x \hat L_\theta) ,\\
            &= \Tr(\hat \rho_\theta \hat L_\theta \1 \hat L_\theta) ,\\
            &= \Tr(\hat \rho_\theta \hat L_\theta^2),
        \end{align}    
    \end{subequations}
    
    which proves the QCRB.   
\end{derivation}

\begin{result}[Expression of the SLD and QFI]~\\
    \label{res: SLD and QFI mixed states}
    For a mixed state $\hat \rho_\theta=e^{-i\hat H\theta}\hat \rho_0 e^{i\hat H\theta}$ undergoing unitary parameter encoding, the SLD $\hat L_0$ and QFI $Q(\hat \rho_0)=\Tr(\hat \rho_0 \hat L_0^2)$ admit the explicit forms
    \begin{equation}
        \hat L_0
        = 2i\sum_{j,k\,:\,p_j+p_k>0} \frac{p_j-p_k}{p_j+p_k} \bra{\psi_j}\hat H \ket{\psi_k}
        \ketbra{\psi_j}{\psi_k},
    \end{equation}
    \begin{equation}
        \mathcal Q (\hat \rho_0)
        = 2\sum_{j,k\,:\,p_j+p_k>0} \frac{(p_j-p_k)^2}{p_j+p_k} \abs{\bra{\psi_j}\hat H\ket{\psi_k}}^2,
    \end{equation}
    where $\hat \rho_0=\sum_j p_j \ketbra{\psi_j}{\psi_j}$ is the spectral decomposition of the initial state. Moreover, the SLD transforms as $\hat L_\theta = e^{-i\hat H\theta}\hat L_0 e^{i\hat H\theta}$, implying that the QFI is independent of $\theta$, \ie, $Q(\hat \rho_\theta)=Q(\hat \rho_0)$.    
\end{result}

\begin{derivation}
    We denote by $\Lambda_{j,k}$ the coefficient of $\hat L_0$ in the basis $\{\ket{\psi_j}\}$
    \begin{equation}
        \hat L_0=\sum_{j,k}\Lambda_{j,k} \ketbra{\psi_j}{\psi_k}.
    \end{equation}
    Now, we consider the SLD equation $\partial_\theta \hat \rho_\theta\big|_{\theta=0}=\frac{1}{2}(\hat L_0\hat \rho_0+\hat \rho_0\hat L_0)$. Since $\hat \rho_\theta=e^{-i\theta \hat H}\hat \rho_0 e^{i\theta \hat H}$, we have 
    \begin{align}
        \partial_\theta \hat \rho_\theta\Big|_{\theta=0}&=-i\hat H\hat \rho_0+i\hat \rho_0 \hat H,\notag\\
        &=-i\sum_k p_k \hat H\ketbra{\psi_k}+i\sum_k p_k \ketbra{ \psi_k} \hat H.
    \end{align}
    On the other hand, we have
    \begin{subequations}
        \begin{align}
            \frac{1}{2}(\hat \rho_0\hat L_0+\hat L_0\hat \rho_0)&=\frac{1}{2}\left(\sum_{j,k,l} \Lambda_{k,l}p_j\ketbra{\psi_j}\ketbra{\psi_k}{\psi_l}+\sum_{j,k,l} \Lambda_{j,k}p_l\ketbra{\psi_j}{\psi_k}\ketbra{\psi_l}\right),\\
            &=\frac{1}{2}\left(\sum_{j,k,l} \Lambda_{k,l}p_j\ketbra{\psi_j}{\psi_l}\delta_{j,k}+\sum_{j,k,l} \Lambda_{j,k}p_l\ketbra{\psi_j}{\psi_l}\delta_{k,l}\right),\\
            &=\frac{1}{2}\left(\sum_{j,l} \Lambda_{j,l}p_j\ketbra{\psi_j}{\psi_l}+\sum_{j,l} \Lambda_{j,l}p_l\ketbra{\psi_j}{\psi_l}\right),\\
            &=\frac{1}{2}\sum_{j,l} \Lambda_{j,l}(p_j+p_l)\ketbra{\psi_j}{\psi_l}.
        \end{align}    
    \end{subequations}
    As such we have the equation
    \begin{subequations}
        \begin{align}
            \frac{1}{2}\Lambda_{j,l}(p_j+p_l)&=\frac{1}{2}\bra{\psi_j}\hat \rho_0\hat L_0+\hat L_0\hat \rho_0\ket{\psi_l},\\
            &=\bra{\psi_j}\partial_\theta \hat \rho_\theta\big|_{\theta=0}\ket{\psi_l},\\
            &=-i\sum_k p_k\bra{\psi_j}\hat H\ket{\psi_k}\braket{\psi_k}{\psi_l}+i\sum_k p_k\braket{\psi_j}{\psi_k}\bra{\psi_k}\hat H\ket{\psi_l},\\
            &=-i p_l \bra{\psi_j}\hat H\ket{\psi_l}+i p_j\bra{\psi_j}\hat H\ket{\psi_l},\\
            &=i(p_j-p_l)\bra{\psi_j}\hat H\ket{\psi_l}.
        \end{align}    
    \end{subequations}
    If $p_j+p_l>0$, this equation is solved as
    \begin{equation}
        \Lambda_{j,l}=2i\frac{p_j-p_l}{p_j+p_l}\bra{\psi_j}\hat H\ket{\psi_l}.
    \end{equation}
    Otherwise, if $p_j+p_l=0$, any value for $\Lambda_{j,l}$ will satisfy the equation. As such we do not have a unique solution. If we fix $\Lambda_{j,l}=0$ in these cases, we get
    \begin{equation}
        \hat L_0=2i\sum_{j,k}\frac{p_j-p_k}{p_j+p_k}\bra{\psi_j}\hat H\ket{\psi_k}\ketbra{\psi_j}{\psi_k},
    \end{equation}
    where the sum runs only over pairs of indices $(j,k)$ such that $p_j+p_k>0$. For this expression of $\hat L_0$ we verify that it is indeed Hermitian
    \begin{align}
        \hat L_0^\dagger&=-2i\sum_{j,k}\frac{p_j-p_k}{p_j+p_k}\bra{\psi_j}\hat H\ket{\psi_k}^*(\ketbra{\psi_j}{\psi_k})^\dagger,\notag\\
        &=-2i\sum_{j,k}\frac{p_j-p_k}{p_j+p_k}\bra{\psi_k}\hat H\ket{\psi_j}\ketbra{\psi_k}{\psi_j}=\hat L_0,
    \end{align}
    by exchange of the indices $j\leftrightarrow k$. To obtain the QFI we simply need to compute $Q(\hat H,\hat \rho_0)=\Tr(\hat L_0^2\hat \rho_0)$. To simplify this, we first stick to the general expression of $\hat L_0=\sum_{j,k} \Lambda_{j,k}\ketbra{\psi_j}{\psi_k}$.
    \begin{subequations}
        \begin{align}
            Q(\hat H,\rho_0)&=\Tr(\hat L_0^2\hat \rho_0),\\
            &=\sum_{a,b,c,d,e}\Lambda_{a,b}\Lambda_{c,d}p_e \Tr\Big(\ketbra{\psi_a}{\psi_b}\ketbra{\psi_c}{\psi_d}\ketbra{\psi_e}{\psi_e}\Big),\\
            &=\sum_{a,b,c,d,e}\Lambda_{a,b}\Lambda_{c,d}p_e  \delta_{b,c}\delta_{d,e}\delta_{e,a},\\
            &=\sum_{a,b}\Lambda_{a,b}\Lambda_{b,a}p_a,\\
            &=\frac{1}{2}\sum_{a,b}(p_a+p_b)\Lambda_{a,b}\Lambda_{b,a},
        \end{align}    
    \end{subequations}
    where we use the trick that $S=\frac{1}{2}(S+S)$ and where we exchanged the indices $a\leftrightarrow b$ in the second sum. Now, we can make explicit the expression of the coefficients, noting that since $\hat L_0$ is Hermitian $\Lambda_{b,a}=\Lambda_{a,b}^*$, and now running the sums only over indices such that $p_a+p_b>0$.
    \begin{subequations}
        \begin{align*}
            Q(\hat H,\hat \rho_0)&=\frac{1}{2}\sum_{a,b}(p_a+p_b)\abs{\Lambda_{a,b}}^2,\\
            &=2\sum_{a,b}(p_a+p_b)\frac{(p_a-p_b)^2}{(p_a+p_b)^2}\abs{\bra{\psi_a}H\ket{\psi_b}}^2,\\
            &=2\sum_{a,b}\frac{(p_a-p_b)^2}{p_a+p_b}\abs{\bra{\psi_a}H\ket{\psi_b}}^2.
        \end{align*}
    \end{subequations}

    We have seen that $\hat L_0$ is not unique, but that the only freedom comes from the choice of the coefficients $\Lambda_{j,k}$ when the indices $j$ and $k$ are both such that $p_j=p_k=0$. Looking at the expression
    \begin{equation*}
        Q(\hat H,\hat \rho_0)=\frac{1}{2}\sum_{a,b}(p_a+p_b)\Lambda_{a,b}\Lambda_{b,a},
    \end{equation*}
    we see that such coefficients will not contribute. As such the quantum Fisher information is independent of the choice of the SLD.

    The expression of $\hat L_\theta$ for $\theta$ arbitrary is obtained by noting that the state $\rho_\theta$ satisfies the same equation as $\rho_0$, so we simply need to modify the eigenbasis as $\ket{\psi_j}\mapsto e^{-i\theta \hat H}\ket{\psi_j}$. With this change, we get
    \begin{equation*}
        L_\theta=2i\sum_{j,k}\frac{p_j-p_k}{p_j+p_k}\bra{\psi_j}e^{i\theta \hat H}\hat He^{-i\theta \hat H}\ket{\psi_k}e^{-i\theta \hat H}\ketbra{\psi_j}{\psi_k}e^{i\theta\hat H}=e^{-i\theta \hat H}L_0e^{i\theta \hat H}.
    \end{equation*}
    Now we compute
    \begin{align*}
        Q(\hat H,\hat \rho_\theta)&=\Tr(\hat L_\theta^2\hat \rho_\theta)=\Tr(e^{-i\theta \hat H}\hat L_0e^{i\theta \hat H}e^{-i\theta \hat H}\hat L_0e^{i\theta \hat H}e^{-i\theta \hat H}\hat \rho_0e^{i\theta \hat H}),\notag\\
        &=\Tr(\hat L_0^2\hat \rho_0)=Q(\hat H,\hat \rho_0).
    \end{align*}
\end{derivation}

\begin{result}[Saturation of QCRB for mixed states]~\\
    \label{res: saturation QCRB mixed states}
    For a mixed state $\hat \rho_{\theta_0}=e^{-i\hat H\theta_0}\hat\rho e^{i\hat H\theta_0}$, the POVM $\{\hat \Pi_j\}$ has associated Fisher information $\mathcal F_{\theta_0}=Q(\hat \rho_{\theta_0})$ at parameter value $\theta_0$, where
    \begin{equation}
        \hat L_{\theta_0} = \sum_j \lambda_j \hat\Pi_j,
    \end{equation}
    is the spectral decomposition of the SLD at $\theta_0$ with orthogonal projectors $\hat\Pi_j$.
\end{result}

\begin{derivation}
    Let the SLD at $\theta_0$ admit the spectral decomposition
    \begin{equation}
        \hat L_{\theta_0} = \sum_j \lambda_j \hat\Pi_j,
    \end{equation}
    where $\hat\Pi_j$ are orthogonal projectors. Choosing the POVM $\{\hat\Pi_j\}$, the outcome probabilities read
    \begin{equation}
        P_\theta(j) = \Tr(\hat \rho_\theta \hat\Pi_j).
    \end{equation}
    Using the SLD defining equation and orthogonality of the projectors, one finds
    \begin{subequations}
        \allowdisplaybreaks
        \begin{align}
            \partial_\theta P_\theta(j)\Big|_{\theta={\theta_0}}&=\Tr(\partial_\theta\hat \rho_\theta\Big|_{\theta={\theta_0}} \hat \Pi_j),\\
            &=\frac{1}{2}\Tr(\hat L_{\theta_0}\hat \rho_{\theta_0}\hat \Pi_j)+\frac{1}{2}\Tr(\hat \rho_{\theta_0}\hat L_{\theta_0}\hat \Pi_j),\\
            &=\frac{1}{2}\Tr(\lambda_j\hat \rho_{\theta_0}\hat \Pi_j)+\frac{1}{2}\Tr(\hat \rho_{\theta_0}\lambda_j\hat \Pi_j),\\
            &=\lambda_j\Tr(\hat \rho_{\theta_0}\hat \Pi_j),
        \end{align}
    \end{subequations}
    since by orthogonality of the projectors, we have 
    \begin{equation}
        \hat L_{\theta_0}\hat \Pi_j=\sum_k \lambda_k \hat \Pi_k\hat \Pi_j=\sum_k \lambda_k\hat \Pi_k\delta_{j,k}=\lambda_j\hat \Pi_j.
    \end{equation}
    The associated classical Fisher information is therefore
    \begin{align}
        \mathcal F_{\theta_0}
        &= \sum_j
        \frac{1}{\Tr(\hat \rho_{\theta_0}\hat\Pi_j)}
        \lambda_j^2
        \Tr(\hat \rho_{\theta_0}\hat\Pi_j)^2 \notag\\
        &= \sum_j \lambda_j^2 \Tr(\hat \rho_{\theta_0}\hat\Pi_j) \notag\\
        &= \Tr(\hat \rho_{\theta_0}\hat L_{\theta_0}^2) \notag\\
        &= Q(\hat \rho_{\theta_0}),
    \end{align}
    since $\hat L_{\theta_0}^2=\sum_{j,k}\lambda_j\lambda_k\hat \Pi_j\hat \Pi_k=\sum_{j,k}\lambda_j\lambda_k\hat \Pi_j\delta_{j,k}=\sum_j \lambda_j^2 \hat \Pi_j$.
\end{derivation}

\begin{result}[Additivity of the QFI]~\\
    \label{res: additivity QFI}
    For two quantum states $\hat \rho_1$ and $\hat \rho_2$, and two generators $\hat H_1$ and $\hat H_2$, the QFI of the joint state $\hat \rho = \hat \rho_1 \otimes \hat \rho_2$ with respect to the generator $\hat H = \hat H_1 \otimes \1 + \1 \otimes \hat H_2$ satisfies
    \begin{equation}
        Q(\hat \rho, \hat H) = Q(\hat \rho_1, \hat H_1) + Q(\hat \rho_2, \hat H_2).
    \end{equation}    
\end{result}

\begin{derivation}
    We start by showing that if $\hat L_1$ and $\hat L_2$ are SLDs of $\hat \rho_1$ and $\hat \rho_2$ with respect to $\hat H_1$ and $\hat H_2$, respectively, then $\hat L=\hat L_1 \otimes \1 + \1 \otimes \hat L_2$ provides an SLD of $\hat \rho = \hat \rho_1 \otimes \hat \rho_2$ with respect to $\hat H = \hat H_1 \otimes \1 + \1 \otimes \hat H_2$. Indeed, by the derivative of a product
    \begin{equation}
        \partial_\theta\hat\rho=\partial_\theta\hat\rho_1 \otimes \hat\rho_2 + \hat\rho_1 \otimes \partial_\theta\hat\rho_2.
    \end{equation}
    Inserting the equation of the SLD, we get
    \begin{subequations}
        \begin{align}
            \partial_\theta\hat \rho&= \frac{1}{2}(\hat L_1\hat \rho_1+\hat \rho_1\hat L_1)\otimes \hat \rho_2 + \frac{1}{2}\hat \rho_1 \otimes (\hat L_2\hat \rho_2+\hat \rho_2\hat L_2),\\
            &= \frac{1}{2}(\hat L_1 \otimes \1 + \1 \otimes \hat L_2)(\hat \rho_1 \otimes \hat \rho_2)+\frac{1}{2}(\hat \rho_1 \otimes \hat \rho_2)(\hat L_1 \otimes \1 + \1 \otimes \hat L_2),\\
            &= \frac{1}{2}(\hat L \hat \rho + \hat \rho \hat L).
        \end{align}
    \end{subequations}
    So $\hat L$ indeed satisfies the SLD equation. Inserting this in the definition of the QFI, we get
    \begin{subequations}
        \allowdisplaybreaks
        \begin{align}
            \mathcal Q(\hat H,\hat \rho)&=\Tr((\hat L_1\otimes\1+\1\otimes\hat L_2)^2\hat \rho),\\
            &=\Tr((\hat L_1^2\otimes\1 + \1\otimes\hat L_2^2 + \hat L_1\otimes \hat L_2 + \hat L_1\otimes \hat L_2)\hat \rho),\\
            &=\Tr(\hat L_1^2\hat \rho_1)\Tr(\hat \rho_2)+\Tr(\hat \rho_1)\Tr(\hat L_2^2\hat \rho_2)+2\Tr(\hat L_1\hat \rho_1)\Tr(\hat L_2\hat \rho_2),\\
            &=\Tr(\hat L_1^2\hat \rho_1)+\Tr(\hat L_2^2\hat \rho_2),\\
            &=\mathcal Q(\hat H_1,\hat \rho_1)+\mathcal Q(\hat H_2,\hat \rho_2),
        \end{align}
    \end{subequations}
    as we have shown previously that $\Tr(\hat L_j\hat \rho_j)=0$ for $j=1,2$.
\end{derivation}

\begin{result}[Convexity of the QFI]~\\
    \label{res: convexity QFI}
    For a set of quantum states $\{\hat \rho_j\}$, probabilities $\{p_j\}$ and Hamiltonian $\hat H$, the QFI is convex, \ie,
    \begin{equation}
        Q\left(\sum_j p_j \hat \rho_j, \hat H\right) \leq \sum_j p_j Q(\hat \rho_j, \hat H).
    \end{equation}    
\end{result}

\begin{derivation}
    We denote $\hat \rho=\sum_j p_j \hat \rho_j$ and we provide two derivations.

    {\bf (i) Using the convexity of the FI}
    
    Let $\{\hat E_x\}$ be the optimal POVM associated with $\hat \rho$ such that $Q(\hat \rho,\hat H)=\mathcal F_\theta(\{\hat E_x\},\hat \rho)$. Using the convexity of the classical Fisher information, we have
    \begin{equation}
        Q(\hat \rho,\hat H)
        = \mathcal F_\theta(\{\hat E_x\},\hat \rho)
        \leq \sum_j p_j \mathcal F_\theta(\{\hat E_x\},\hat \rho_j),
    \end{equation}
    where $\mathcal F_\theta(\{\hat E_x\},\hat \rho_j)$ is the Fisher information associated with the POVM $\{\hat E_x\}$ and the initial state $\hat \rho_j$ evolved with the Hamiltonian $\hat H$ and the parameter $\theta$. Since the QFI is the maximum of the Fisher information over all POVMs, we have $\mathcal F_\theta(\{\hat E_x\},\hat \rho_j) \leq Q(\hat \rho_j,\hat H)$, which concludes the proof.

    \medbreak

    {\bf (ii) Algebraic proof}
    
    We denote by $\hat L$ and $\hat L_j$ the SLDs of $\hat \rho$ and $\hat \rho_j$, respectively. By linearity of the derivative we have $\partial \hat \rho = \sum_j p_j \partial \hat \rho_j$. Inserting the SLD equations, we get
    \begin{equation}
        \sum_j p_j (\hat L\hat \rho_j+\hat \rho_j \hat L)=\hat L\hat \rho+\hat \rho \hat L=2\partial_\theta\hat\rho=\sum_j p_j (\hat L_j\hat \rho_j+\hat \rho_j \hat L_j).
    \end{equation}
    This formula allows us to compute the FI associated with $\hat \rho$
    \begin{subequations}
        \allowdisplaybreaks
        \begin{align}
            \Tr(\hat L^2\hat \rho)&=\sum_j p_j \Tr(\hat L^2\hat \rho_j),\\
            &=\frac{1}{2}\sum_j p_j \Tr\left(\hat L(\hat L_j\hat \rho_j+\hat \rho_j \hat L_j)\right),\\
            &=\frac{1}{2}\Tr(\hat L\sum_j p_j(\hat L_j\hat \rho_j+\hat \rho_j \hat L_j)),\\
            &=\frac{1}{2}\sum_j p_j \Tr(\hat L\hat L_j\hat \rho_j)+\frac{1}{2}\sum_j p_j \Tr(\hat L\hat \rho_j \hat L_j),\\
            &=\Re(\sum_j p_j \Tr(\hat L\hat L_j\hat \rho_j)),
        \end{align}
    \end{subequations}
    by noting that the sums are complex conjugates of each other. We thus get
    \begin{equation}
        \Tr(\hat L^2\hat \rho)=\Re(\sum_j p_j \Tr(\hat L\hat L_j\hat \rho_j))\leq \abs{\sum_j p_j \Tr(\hat L\hat L_j\hat \rho_j)}.
    \end{equation}
    We introduce an inner product on $\mathcal L(\mathcal H)^n$ via the formula $(\vec{\hat A}, \vec{\hat B}) = \sum_j p_j \Tr(\hat A_j^\dagger \hat B_j)$ where $\vec{\hat A}$ and $\vec{\hat B}$ are vectors of operators. Using the Cauchy-Schwarz inequality, we get
    \begin{equation}
        \abs{\sum_j p_j\Tr(\hat A_j^\dagger \hat B_j)}\leq \sqrt{\sum_j p_j \Tr(\hat A_j^\dagger \hat A_j)}\sqrt{\sum_j p_j \Tr(\hat B_j^\dagger \hat B_j)},
    \end{equation}
    which we apply to $\hat A_j=\sqrt{\hat \rho_j}\hat L$ and $\hat B_j=\sqrt{\hat \rho_j}\hat L_j$. We get
    \begin{equation}
        \Tr(\hat L^2\hat \rho)\leq \sqrt{\sum_j p_j \Tr(\hat L^2\hat \rho_j)}\sqrt{\sum_j p_j \Tr(\hat L_j^2\hat \rho_j)}=\sqrt{\Tr(\hat L^2\hat \rho)}\sqrt{\sum_j p_j \Tr(\hat L_j^2\hat \rho_j)}.
    \end{equation}
    Squaring and simplifying by $\Tr(\hat L^2\hat \rho)$, we get 
    \begin{equation}
        \Tr(\hat L^2\hat \rho)\leq \sum_j p_j \Tr(\hat L_j^2\hat \rho_j),
    \end{equation}
    which concludes the proof.
\end{derivation}

\begin{result}[Inequality between QFI and quantum variance]~\\
    \label{res: QFI variance inequality}
    For a quantum state $\hat \rho$ and a Hermitian operator $\hat H$, the QFI and the quantum variance satisfy the inequality
    \begin{equation}
        Q(\hat \rho, \hat H) \leq 4\,\Delta^2_{\hat\rho}\hat H.
    \end{equation}
\end{result}

\begin{derivation}
    We denote by $\hat \rho=\sum_j p_j \ketbra{\psi_j}{\psi_j}$ the spectral decomposition of $\hat \rho$. By convexity of the QFI, we first have
    \begin{equation}
        Q(\hat \rho, \hat H)
        \leq \sum_j p_j Q(\ketbra{\psi_j}{\psi_j}, \hat H)= 4\sum_j p_j \Delta^2_{\ket{\psi_j}} \hat H.
    \end{equation}
    We thus need to show that the quantum variance is concave, \ie,
    \begin{equation}
        \sum_j p_j \Delta^2_{\ket{\psi_j}} \hat H \leq \Delta^2_{\hat \rho} \hat H,
    \end{equation}
    which would conclude the proof. To see why, we expand both sides of the concavity inequality.
    \begin{equation}
        \Delta^2_{\hat\rho}\hat H=\Tr(\hat \rho \hat H^2)-\Tr(\hat \rho \hat H)^2=\sum_j p_j \bra{\psi_j}\hat H^2\ket{\psi_j}-\left(\sum_j p_j \bra{\psi_j}\hat H\ket{\psi_j}\right)^2.
    \end{equation}
    While,
    \begin{equation}
        \sum_j p_j \Delta^2_{\ket{\psi_j}}\hat H=\sum_j p_j \bra{\psi_j}\hat H^2\ket{\psi_j}-\sum_j p_j \bra{\psi_j}\hat H\ket{\psi_j}^2.
    \end{equation}
    Introducing the real vectors 
    \begin{align}
        \vec u = \begin{pmatrix}
            \sqrt{p_1} \bra{\psi_1}\hat H\ket{\psi_1} \\
            \sqrt{p_2} \bra{\psi_2}\hat H\ket{\psi_2} \\
            \vdots
        \end{pmatrix}, && \vec v = \begin{pmatrix}
            \sqrt{p_1} \\
            \sqrt{p_2} \\
            \vdots
        \end{pmatrix},
    \end{align}
    we see, by applying the Cauchy-Schwarz inequality, $\langle \vec u, \vec v \rangle^2 \leq \langle \vec u, \vec u \rangle \langle \vec v, \vec v \rangle$, that
    \begin{equation}
        \left(\sum_j p_j \bra{\psi_j}\hat H\ket{\psi_j}\right)^2 \leq \sum_j p_j \bra{\psi_j}\hat H\ket{\psi_j}^2,
    \end{equation}
    which concludes the proof.
\end{derivation}

\begin{result}[Quantum Fisher information and convex roof]
    \label{res: convex roof QFI}
    For any observable $\hat H$ and quantum state $\hat \rho$, the quantum Fisher information $\mathcal Q(\hat H,\hat \rho)$ satisfies 
    \begin{equation}
        \mathcal Q(\hat H, \hat \rho) = 4\inf_{\{p_j, \ket{\phi_j}\}} \sum_j p_j \Delta^2_{\ket{\phi_j}} \hat H,
    \end{equation}
    where the infimum is taken over all pure-state convex decompositions $\hat \rho = \sum_j p_j \ketbra{\phi_j}{\phi_j}$ of $\hat \rho$.
\end{result}

\begin{derivation}
    $\blacktriangleright$ {\bf Objective.} Due to the convexity of the quantum Fisher information, we have
    \begin{equation}
        \mathcal Q(\hat H, \hat \rho) \leq  4\inf_{\{p_j, \ket{\phi_j}\}} \sum_j p_j \Delta^2_{\ket{\phi_j}} \hat H.
    \end{equation}
    We thus need to prove the converse inequality, which amounts to constructing an explicit ensemble $\{p_j, \ket{\phi_j}\}$ such that $\mathcal Q(\hat H, \hat \rho) = 4\sum_j p_j \Delta^2_{\ket{\phi_j}} \hat H$. 

    As announced in the main part of the thesis, we reproduce the proof of~\cite{yu_quantum_2013}. However, although efficient, the original argument introduces several objects whose role is not immediately transparent. In the present derivation, we aim at providing an alternative and more structurally motivated perspective. In particular, instead of working in the eigenbasis of $\hat \rho$, we formulate the construction in a basis-independent way, which makes the underlying algebraic properties more explicit.

    \medskip

    \noindent
    $\blacktriangleright$ {\bf Properties of the anti-commutator.} We start with a preliminary analysis. Recall that the quantum Fisher information is defined in terms of the SLD $\hat L$ which satisfies
    \begin{equation}
        \partial \hat \rho = \frac{1}{2}(\hat L\hat \rho + \hat \rho \hat L)=\frac{1}{2}\{\hat L, \hat \rho\},
    \end{equation}
    where $\{\cdot, \cdot\}$ denotes the anti-commutator. This naturally motivates the study of the linear map associated with anti-commutation by $\hat \rho$. For any operator $\hat O$, we define
    \begin{equation}
        \mathcal A(\hat O) = \{\hat O, \hat \rho\}.
    \end{equation}
    Expressing $\hat O$ in the eigenbasis of $\hat \rho=\sum_j \lambda_j \ketbra{\psi_j}{\psi_j}$, we obtain
    \begin{equation}
        \mathcal A(\hat O) = \sum_{j,k} (\lambda_j+\lambda_k) \bra{\psi_j}\hat O\ket{\psi_k} \ketbra{\psi_j}{\psi_k}.
    \end{equation}
    From this expression, we see that $\mathcal A$ is injective if and only if $\lambda_j+\lambda_k\neq 0$ for all $(j,k)$, which is equivalent to $\hat \rho$ having no vanishing eigenvalues. When $\hat \rho$ has zero as an eigenvalue, $\mathcal A$ is no longer invertible. Nevertheless, we can define a pseudo-inverse by
    \begin{equation}
        \mathcal A^{-1}(\hat O) = \sum_{j,k} \frac{1}{\lambda_j+\lambda_k} \bra{\psi_j}\hat O\ket{\psi_k} \ketbra{\psi_j}{\psi_k},
    \end{equation}
    where the sum is restricted to pairs $(j,k)$ such that $\lambda_j+\lambda_k\neq 0$. In general,
    \begin{align}
        \mathcal A^{-1}(\mathcal A(\hat O))=\mathcal A(\mathcal A^{-1}(\hat O))\neq\hat O,
    \end{align}
    since the components corresponding to $\lambda_j+\lambda_k=0$ are lost under the action of $\mathcal A$. However, the following identities hold:
    \begin{align}\label{eq: cancelation for A}
        \hat \rho\mathcal A(\mathcal A^{-1}(\hat O))=\hat \rho \hat O, && \mathcal A(\mathcal A^{-1}(\hat O))\hat \rho = \hat O \hat \rho.
    \end{align}
    These relations follow from the fact that multiplication by $\hat \rho$ removes precisely the matrix elements that cannot be recovered by $\mathcal A^{-1}$. Similarly, the same holds by replacing $\hat\rho$ by its square root $\sqrt{\hat \rho}$.

    We endow the space of operators with the scalar product
    \begin{equation}
        \langle \hat O_1,\hat O_2\rangle = \Tr(\hat O_1\hat O_2^\dagger).
    \end{equation}
    In the orthonormal basis $\{\ketbra{\psi_j}{\psi_k}\}$, the map $\mathcal A$ is diagonal with eigenvalues $\lambda_j+\lambda_k\geq 0$, hence it is positive semi-definite. As a consequence, it is Hermitian with respect to this scalar product
    \begin{equation}
        \Tr(\hat O_1 \mathcal A(\hat O_2)^\dagger) = \Tr(\mathcal A(\hat O_1) \hat O_2^\dagger),
    \end{equation}
    which can also be verified explicitly. Similarly, $\mathcal A^{-1}$ is also Hermitian and thus satisfies the same trace property.

    Finally, $\mathcal A$ commutes with Hermitian conjugation,
    \begin{equation}
        \mathcal A(\hat O^\dagger)=\mathcal A(\hat O)^\dagger,
    \end{equation}
    and with multiplication by $\hat \rho$,
    \begin{align}
        \mathcal A(\hat \rho \hat O)=\hat \rho \mathcal A(\hat O), && \mathcal A(\hat O\hat \rho)=\mathcal A(\hat O)\hat \rho,
    \end{align}
    as well as with powers of $\hat \rho$. These properties also extend to $\mathcal A^{-1}$. In particular,
    \begin{equation}
        \mathcal A^{-1}([\hat \rho,\hat O]) = [\hat \rho, \mathcal A^{-1}(\hat O)].
    \end{equation}

    \medskip

    \noindent
    $\blacktriangleright$ {\bf Alternative expression for the QFI.} 
    Starting from the definition of the QFI, we first exploit the interplay between the commutator and the map $\mathcal A$ in order to obtain an alternative expression. Remembering that, for unitary parametrization, $\partial \hat\rho=i[\hat \rho,\hat H]$, the symmetric logarithmic derivative reads
    \begin{equation}
        \hat L = 2\mathcal A^{-1}(\partial \hat \rho) = 2i\mathcal A^{-1}([\hat \rho,\hat H])=2i[\hat \rho, \mathcal A^{-1}(\hat H)].
    \end{equation}
    As such, the quantum Fisher information admits the expression
    \begin{equation}
        \mathcal Q(\hat H, \hat \rho) = -4\Tr(\hat \rho [\hat \rho, \mathcal A^{-1}(\hat H)]^2).
    \end{equation}

    We now observe the following identity relating the squares of commutators and anti-commutators,
    \begin{equation}
        [\hat A,\hat B]^2+\{\hat A,\hat B\}^2=2(\hat A\hat B\hat A\hat B+\hat B\hat A\hat B\hat A).
    \end{equation}
    Using this identity, the quantum Fisher information can be rewritten as
    \begin{subequations}
        \begin{align}
            \mathcal Q(\hat H, \hat \rho) 
            &= -4\Tr(\hat \rho [\hat \rho, \mathcal A^{-1}(\hat H)]^2),\\
            &= 4\Tr(\hat \rho \{\hat \rho,\mathcal A^{-1}(\hat H)\}^2)-16\Tr(\hat \rho^2 \mathcal A^{-1}(\hat H)\hat \rho \mathcal A^{-1}(\hat H)),\\
            &=4\Tr(\hat \rho \mathcal A(\mathcal A^{-1}(\hat H))^2)-16\Tr(\hat \rho \sqrt{\hat \rho} \mathcal A^{-1}(\hat H)\hat \rho \mathcal A^{-1}(\hat H)\sqrt{\hat \rho}),\\
            &=4\Tr(\hat \rho \hat H^2)-4\Tr(\hat \rho\hat Y_H^2),
        \end{align}
    \end{subequations}
    where we have used the inversion formula of Eq.~\eqref{eq: cancelation for A} to cancel the $\mathcal A$ in the first term. We have also introduced the Hermitian operator\footnote{This is the same operator as the one introduced in~\cite{yu_quantum_2013}. However, the definition we provide is not given directly in terms of its matrix elements, but rather emerges naturally from the algebraic rewriting of the QFI.}
    \begin{equation}
        \hat Y_H = 2\sqrt{\hat \rho} \mathcal A^{-1}(\hat H)\sqrt{\hat \rho}.
    \end{equation}

    The above rewriting of the quantum Fisher information is extremely useful. Indeed, it allows us to isolate a first term linear in $\hat\rho$, which exactly matches the term appearing in the definition of the quantum variance. When comparing the value of the QFI with the weighted average of the variance over a pure-state ensemble, we obtain
    \begin{equation}
        4\sum_j p_j \bra{\phi_j}\hat H^2\ket{\phi_j}=4\Tr(\hat \rho \hat H^2).
    \end{equation}
    This contribution is therefore invariant over all pure-state decompositions of $\hat \rho$. Consequently, we can focus our analysis on the second term and seek an ensemble such that
    \begin{equation}
        \sum_j p_j \bra{\phi_j}\hat H\ket{\phi_j}^2 = \Tr(\hat \rho\hat Y_H^2).
    \end{equation}

    \medskip

    \noindent
    $\blacktriangleright$ {\bf Ensemble construction.} 
    Having introduced the operator $\hat Y_H$, it is natural to construct the ensemble using the eigenbasis of $\hat Y_H$. We thus write the spectral decomposition
    \begin{equation}
        \hat Y_H = \sum_j \alpha_j \ketbra{y_j}{y_j}.
    \end{equation}
    Although there is no reason for the vectors $\ket{y_j}$ themselves to form a convex decomposition of $\hat \rho$, they can nevertheless be used to define it. First define the scalars
    \begin{equation}
        u_j = \bra{y_j}\hat \rho\ket{y_j}.
    \end{equation}
    If $u_j$ is nonzero, then we define the states
    \begin{align}
        \ket{U_j} = \frac{1}{\sqrt{u_j}} \sqrt{\hat \rho} \ket{y_j},
    \end{align}
    and otherwise we set $\ket{U_j}$ arbitrarily. We now verify that $\{u_j, \ket{U_j}\}$ indeed forms a convex decomposition of $\hat \rho$. We have
    \begin{subequations}
        \begin{align}
            \sum_j u_j \ketbra{U_j}{U_j}&=\sum_{u_j\neq 0} \sqrt{\hat \rho}\ketbra{y_j}{y_j}\sqrt{\hat \rho},\\
            &=\sum_j \sqrt{\hat \rho}\ketbra{y_j}{y_j}\sqrt{\hat \rho},\\
            &=\sqrt{\hat \rho} \left(\sum_j \ketbra{y_j}{y_j}\right) \sqrt{\hat \rho}=\hat \rho.
        \end{align}
    \end{subequations}
    Indeed, $u_j=0$ if and only if $\sqrt{\hat \rho}\ket{y_j}=0$, so the terms corresponding to $u_j=0$ do not contribute to the sum and can be added so that the sum is taken over all indices $j$. The remaining terms sum up to $\hat \rho$
    since $\{\ket{y_j}\}$ forms an orthonormal basis and therefore satisfies the closure relation $\sum_j \ketbra{y_j}{y_j}=\1$. Finally, we verify that this ensemble achieves the desired value of the variance. Observe that for indices $j$ such that $u_j\neq 0$, we have
    \begin{subequations}
        \allowdisplaybreaks
        \begin{align}
            \bra{U_j}\hat H\ket{U_j}
            &=\frac{1}{u_j}\bra{y_j}\sqrt{\hat \rho}\hat H\sqrt{\hat \rho}\ket{y_j},\\
            &=\frac{1}{u_j}\Tr(\sqrt{\hat \rho} \hat H \sqrt{\hat \rho}\ketbra{y_j}{y_j}),\\
            &=\frac{1}{u_j}\Tr(\sqrt{\hat \rho} \hat H \sqrt{\hat \rho}\mathcal A^{-1}(\mathcal A(\ketbra{y_j}{y_j}))),\label{eq: deriv conv roof step 1}\\
            &=\frac{1}{u_j}\Tr(\mathcal A^{-1}(\sqrt{\hat \rho} \hat H \sqrt{\hat \rho})\mathcal A(\ketbra{y_j}{y_j})),\label{eq: deriv conv roof step 2}\\
            &=\frac{1}{u_j}\Tr(\sqrt{\hat \rho} \mathcal A^{-1}(\hat H) \sqrt{\hat \rho}\mathcal A(\ketbra{y_j}{y_j})),\label{eq: deriv conv roof step 3}\\
            &=\frac{1}{2u_j}\Tr(\hat Y_H\{\hat \rho,\ketbra{y_j}{y_j}\}),\label{eq: deriv conv roof step 4}\\
            &=\frac{1}{2u_j}\Tr(\hat Y_H\hat \rho\ketbra{y_j}{y_j}+\hat Y_H\ketbra{y_j}{y_j}\hat \rho),\\
            &=\frac{1}{u_j}\alpha_j \Tr(\hat \rho\ketbra{y_j}{y_j})=\alpha_j,\label{eq: deriv conv roof step 5}
        \end{align}
    \end{subequations}
    where at step~\eqref{eq: deriv conv roof step 1} we have used the pseudo-inversion property of $\mathcal A$, guaranteed by the presence of $\sqrt{\hat\rho}$, which allows us to insert $\mathcal A^{-1}\circ \mathcal A$. At step~\eqref{eq: deriv conv roof step 2} we have used the Hermiticity of $\mathcal A^{-1}$. At step~\eqref{eq: deriv conv roof step 3} we have used the commutation of $\mathcal A^{-1}$ with multiplication by $\sqrt{\hat\rho}$. At step~\eqref{eq: deriv conv roof step 4} we have used the definition of $\hat Y_H$ together with that of $\mathcal A$. Finally, at step~\eqref{eq: deriv conv roof step 5} we have used the fact that $\ket{y_j}$ is an eigenvector of $\hat Y_H$. We therefore obtain
    \begin{subequations}
        \begin{align}
            \sum_j u_j \bra{U_j}\hat H\ket{U_j}^2&=\sum_{u_j\neq 0} u_j \bra{U_j}\hat H\ket{U_j}^2,\\
            &=\sum_{u_j\neq 0} u_j \alpha_j^2,\\
            &=\sum_j u_j \alpha_j^2,\\
            &=\Tr\!\left(\sum_j \alpha_j^2\ketbra{y_j}{y_j}\hat \rho\right),\\ 
            &=\Tr(\hat Y_H^2 \hat \rho),
        \end{align}
    \end{subequations}
    where once again, the terms for which $u_j=0$ do not contribute and can thus be added and removed freely. This concludes the proof, since the ensemble $\{u_j, \ket{U_j}\}$ satisfies
    \begin{equation}
        4\sum_j u_j \Delta^2_{\ket{U_j}} \hat H 
        = 4\Tr(\hat \rho\hat H^2)-4\Tr(\hat \rho\hat Y_H^2)
        =\mathcal Q(\hat H, \hat \rho).
    \end{equation}
\end{derivation}

\begin{result}[Equivalence of mode transformation definitions]~\\
   \label{res: equivalence of mode transformation definitions}
    Let $\mathcal H$ be a Hilbert space representing the space of modes, or equivalently the single-particle Hilbert space, and let $\mathcal F$ denote the associated Fock space. Given an operator $M$\footnote{We omit the hat notation in order to distinguish operators acting on $\mathcal H$ from those acting on $\mathcal F$.} acting on $\mathcal H$, we define its extension to $\mathcal F$ by
    \begin{equation}
        \hat M\mathcal S(\psi_1\otimes\cdots\otimes \psi_n)=\mathcal S\Big((M\psi_1)\otimes\cdots\otimes (M\psi_n)\Big),
    \end{equation}
    where $\mathcal S$ denotes the symmetrization operator. By linearity, this defines $\hat M$ on every fixed-photon-number sector, and therefore on the whole Fock space. We have the relation
    \begin{equation}
        \hat M \hat a^\dagger_{\psi} = \hat a^\dagger_{M\psi} \hat M.
    \end{equation}
    Consequently, if $M$ is invertible and $\{\hat a_1,\dots,\hat a_n\}$ is an operator basis, we have the transformation rule
    \begin{equation}
        \hat M \hat a_j^\dagger \hat M^{-1}= \sum_{k=1}^n M_{k,j} \hat a_k^\dagger ,
    \end{equation}
    where $M_{k,j}$ denotes the coordinates of $M$ in the corresponding basis of $\mathcal H$.
\end{result}

\begin{derivation}
    With all the definitions in place, the verification is very simple. Taking $\psi$ and $\psi_1,\dots,\psi_n\in\mathcal H$, we can compute the action of both $\hat M \hat a^\dagger_{\psi}$ and $\hat a^\dagger_{M\psi} \hat M$ on the symmetrized state $\mathcal S(\psi_1\otimes\cdots\otimes \psi_n)$ as follows. First recall that 
    \begin{equation*}
        \hat a^\dagger_{\psi} \mathcal S(\psi_1\otimes\cdots\otimes \psi_n) = \sqrt{n+1}\,\mathcal S(\psi\otimes\psi_1\otimes\cdots\otimes \psi_n).
    \end{equation*}
    As such, on the one hand, we have
    \begin{subequations}
        \begin{align}
            \hat M \hat a^\dagger_{\psi} \mathcal S(\psi_1\otimes\cdots\otimes \psi_n) 
            &= \sqrt{n+1}\,\hat M \mathcal S(\psi\otimes\psi_1\otimes\cdots\otimes \psi_n),\\
            &= \sqrt{n+1}\,\mathcal S((M\psi)\otimes(M\psi_1)\otimes\cdots\otimes (M\psi_n)).
        \end{align}
    \end{subequations}
    And on the other
    \begin{subequations}
        \begin{align}
            \hat a^\dagger_{M\psi} \hat M \mathcal S(\psi_1\otimes\cdots\otimes \psi_n) 
            &= \hat a^\dagger_{M\psi} \mathcal S((M\psi_1)\otimes\cdots\otimes (M\psi_n)),\\
            &= \sqrt{n+1}\,\mathcal S((M\psi)\otimes(M\psi_1)\otimes\cdots\otimes (M\psi_n)).
        \end{align}
    \end{subequations}
    As the states of the form $\mathcal S(\psi_1\otimes\cdots\otimes \psi_n)$ generate the whole Fock space $\mathcal F$, this ensures that we indeed have the relation
    \begin{equation}
        \hat M \hat a^\dagger_{\psi} = \hat a^\dagger_{M\psi} \hat M.
    \end{equation}
\end{derivation}

\fi

\ifnum \theShowChaptwo=1
\ifthenelse{
    \value{ShowChapone}=1
}{
\clearpage
}{}
\section{Time-frequency systems}
\label{app: TF basic computations}
\setcurrentanchor{app tf systems}
\emph{This section of the appendix collects all the results of Chap.~\ref{chap: Time-Frequency Systems} and provides the associated proofs.}

\vspace*{1em}
\par\noindent
\textbf{\large Results in this section}\par
\vspace{-0.8em}
\noindent\rule{\textwidth}{0.8pt}\par

\resultentry{res: Time creation op commutation}{Commutation relation of time creation/annihilation operators}
\resultentry{res: Single photon TF commutation}{Single-photon time-frequency commutation relation}
\resultentry{res: General TF commutator}{General time-frequency commutator}
\resultentry{res: alternate wigner}{Alternate expression of the chrono-cyclic Wigner function}
\resultentry{res: Wigner GKP}{Wigner function of GKP states}
\resultentry{res: Wigner local separable}{Wigner function of two-mode separable states}
\resultentry{res: Wigner diagonal separable}{Wigner function of two-mode diagonally separable states}
\resultentry{res: TF coherent state}{Time-frequency coherent state}
\resultentry{res: TF coherent states scalar product}{Scalar product of time-frequency coherent states}
\resultentry{res: Translation creation operators}{Translation of creation operators}
\resultentry{res: Wigner translation}{Wigner function and translation}
\resultentry{res: Quadratic gate creation op}{Quadratic transformation of creation operators}
\resultentry{res: Wigner quadratic transformation}{Quadratic transformation of the Wigner function}
\resultentry{res: Rotation of TF operators}{Rotation of time-frequency operators}
\resultentry{res: rotation wigner}{Wigner function and rotation}
\resultentry{res: Partial Fourier}{Partial Fourier transform and shears}
\resultentry{res: variance coherent states}{Variance of coherent states}

\vspace{-0.2em}
\par\noindent\rule{\textwidth}{0.8pt}\par

\begin{result}[Commutation relation of time creation/annihilation operators]~\\
    \label{res: Time creation op commutation}
    Defining the time annihilation operator as
    \begin{equation}
        \hat b_\alpha(t) = \frac{1}{\sqrt{2\pi}} \int \dd \omega\, \hat a_\alpha(\omega) e^{-i\omega t},
    \end{equation} 
    where $\hat a_\alpha(\omega)$ is the frequency annihilation operator on spatial mode $\alpha$ satisfying the commutation relation $[\hat a_\alpha(\omega), \hat a_\beta^\dagger(\omega')]=\delta_{\alpha,\beta}\delta(\omega-\omega')$, we have 
    \begin{equation}
        [\hat b_\alpha(t), \hat b_\beta^\dagger(t')]
        = \delta_{\alpha,\beta} \delta(t-t').
    \end{equation}    
\end{result}

\begin{derivation}
    We simply expand the commutator using the definition of the operators in the temporal domain
	\begin{subequations}
        \begin{align}
            [\hat{b}_\alpha(t),\hat{b}_\beta^\dagger(t')]&=\frac{1}{2\pi}\int\limits_{\R^2}\dd\omega \dd\omega'\, [\hat{a}_\alpha(\omega),\hat{a}_\beta^\dagger(\omega')] e^{i(\omega' t'-\omega t)},\\
            &=\frac{1}{2\pi}\int\limits_{\R^2}\dd\omega \dd\omega'\, \delta(\omega-\omega')\delta_{\alpha,\beta} e^{i(\omega' t'-\omega t)},\\
            &=\frac{\delta_{\alpha,\beta}}{2\pi}\int\limits_{\R}\dd\omega\, e^{i\omega(t'-t)},\\
            &=\frac{\delta_{\alpha,\beta}}{2\pi} 2\pi \delta(t-t'),\\
            &=\delta(t-t')\delta_{\alpha,\beta},
        \end{align}
	\end{subequations}
    where we have used the Fourier transform relation $\int \dd \omega\, e^{i\omega t} = 2\pi \delta(t)$.
\end{derivation}

\begin{result}[Single-photon time-frequency commutation relation]~\\
    \label{res: Single photon TF commutation}
    The action of $\hat \omega$ and $\hat t$ on the spectrum $F$ of a single-photon state is given by
    \begin{align}
        \hat \omega : F(\omega)\mapsto \omega F(\omega), && \hat t : F(\omega)\mapsto -i F'(\omega).
    \end{align}
    Thus the commutator of these two operators satisfies $[\hat \omega, \hat t] = i\1$, on the single-photon subspace.
\end{result}

\begin{derivation}
    We recall that a single-mode single-photon state can be written as
    \begin{equation}
        \ket{\psi}=\int \dd\omega\, F(\omega)\hat{a}^\dagger(\omega)\vac,
    \end{equation}
    where the wavefunction $F(\omega)$ is supposed to be sufficiently regular and to vanish at infinity. The action of $\hat \omega$ is given by
    \begin{subequations}
        \begin{align}
            \hat{\omega}\ket{\psi}&=\int \dd\omega \dd\omega'\, \omega F(\omega')\hat{a}^\dagger(\omega)\hat{a}(\omega)\hat{a}^\dagger(\omega')\vac,\\
            &=\int \dd\omega \dd\omega'\, \omega F(\omega')\delta(\omega-\omega')\hat{a}^\dagger(\omega)\vac,\\
            &=\int \dd\omega\, \omega F(\omega) \hat{a}^\dagger(\omega) \vac.
        \end{align}
    \end{subequations}
    Thus 
    \begin{equation}
        \hat{\omega} : F(\omega)\mapsto \omega F(\omega).
    \end{equation}
    We do the same for the operator $\hat{t}$
    \begin{subequations}
        \allowdisplaybreaks
        \begin{align}
            \hat{t}\ket{\psi}&=\int \dd t \dd \omega\, t F(\omega) \hat{b}^\dagger(t) \hat{b}(t)\hat{a}^\dagger(\omega)\vac,\\
            &=\frac{1}{2\pi}\int \dd t \dd \omega \dd\omega_1 \dd\omega_2\, t F(\omega) e^{i(\omega_1-\omega_2)t}\hat{a}^\dagger(\omega_1) \underbrace{\hat{a}(\omega_2)\hat{a}^\dagger(\omega)\vac}_{\delta(\omega_2-\omega)\vac},\\
            &=\frac{1}{2\pi}\int \dd t  \dd \omega \dd\omega_1\, t F(\omega)e^{i(\omega_1-\omega)t}\hat{a}^\dagger(\omega_1)\vac,\\
            &=\frac{i}{2\pi}\int \dd t \dd \omega_1 \, e^{i\omega_1 t}\underbrace{\int \dd \omega\, (-it)e^{-i\omega t}F(\omega)}_{\left[e^{-i\omega t}F(\omega)\right]^{+\infty}_{-\infty}-\int \dd \omega\, e^{-i\omega t}F'(\omega)} \hat{a}^\dagger(\omega_1)\vac,\\
            \intertext{$F'(\omega)$ being the derivative of $F(\omega)$. The fully integrated term coming from the integration by part vanishes under the regularity assumption on $F(\omega)$.}
            &=\frac{-i}{2\pi}\int \dd t \dd \omega \dd \omega_1\, e^{i(\omega_1-\omega)t}F'(\omega) \hat{a}^\dagger(\omega_1)\vac,\\
            &=-i\int \dd \omega \dd\omega_1\,\delta(\omega-\omega_1)F'(\omega)\hat{a}^\dagger(\omega_1)\vac,\\
            &=-i\int \dd \omega\, F'(\omega) \hat{a}^\dagger(\omega)\vac,
        \end{align}
    \end{subequations}
    Thus 
    \begin{equation}
        \hat{t} : F(\omega)\mapsto -iF'(\omega).
    \end{equation}
    The final result then directly follows from the transformation rule of $F$
    \begin{subequations}
    \begin{align}
        [\hat{\omega},\hat{t}]\ket{\psi}&=\hat{\omega}\hat{t}\int \dd \omega\, F(\omega)\ket{\omega}-\hat{t}\hat{\omega}\int \dd \omega\, F(\omega)\ket{\omega},\\
        &=-i\hat{\omega}\int \dd \omega\, F'(\omega)\ket{\omega}-\hat{t}\int \dd \omega\, \omega F(\omega)\ket{\omega},\\
        &=-i\int \dd \omega\, \omega F'(\omega)\ket{\omega}+i\int \dd \omega\, \underbrace{(\omega F(\omega))'}_{F(\omega)+ \omega F'(\omega)}\ket{\omega},\\
        &=i\int \dd \omega\, F(\omega)\ket{\omega}= i\ket{\psi}.
    \end{align}
    \end{subequations}
    This proves that on the single-photon space, the commutator $ [\hat{\omega},\hat{t}]$ indeed acts as $i\1$.
\end{derivation}

\begin{result}[General time-frequency commutator]~\\
    \label{res: General TF commutator}
    The action of $\hat \omega$ and $\hat t$ on the generalized JSA of a time-frequency single-mode state $\ket{\psi_k}$ with $k$ photons in total is
    \begin{align}
        \hat \omega : F(\omega_1,\dots,\omega_k)\mapsto (\omega_1+\cdots+\omega_k) F(\omega_1,\dots,\omega_k), \\
        \hat t : F(\omega_1,\ldots,\omega_k)\mapsto -i (\partial_1+\cdots+\partial_k) F(\omega_1,\ldots,\omega_k),
    \end{align}
    where 
    \begin{equation}
        \ket{\psi_k} = \int \dd \omega_1 \cdots \dd\omega_k\, F(\omega_1,\ldots,\omega_k) \hat a^\dagger(\omega_1)\cdots \hat a^\dagger(\omega_k)\vac.
    \end{equation}
    It follows that the commutator satisfies $[\hat \omega, \hat t] = i \hat N $ on the general time-frequency space.
\end{result}

\begin{derivation}
    First we observe the following commutation relation
    \begin{align}
        [\hat \omega,\hat a^\dagger(\omega)]=\int \dd \omega'\, \omega' \hat a^\dagger(\omega')[\hat a(\omega'),\hat a^\dagger(\omega)]=\int \dd \omega'\, \omega' \hat a^\dagger(\omega')\delta(\omega'-\omega)=\omega \hat a^\dagger(\omega),
    \end{align}
    which implies that by repeated swaps, we have the formula
    \begin{equation}
        \hat \omega \hat a^\dagger(\omega_1)\cdots\hat a^\dagger(\omega_k)=(\omega_1+\cdots+\omega_k)\hat a^\dagger(\omega_1)\cdots\hat a^\dagger(\omega_k)+\hat a^\dagger(\omega_1)\cdots\hat a^\dagger(\omega_k)\hat \omega.
    \end{equation}
    Noticing that $\hat \omega\vac=0$, we get
    \begin{equation}
        \hat \omega\ket{\psi_k}=\int \dd \omega_1 \cdots \dd\omega_k\, (\omega_1+\cdots+\omega_k) F(\omega_1,\ldots,\omega_k) \hat a^\dagger(\omega_1)\cdots \hat a^\dagger(\omega_k)\vac,
    \end{equation}
    which proves the action of $\hat \omega$. We do the same for $\hat t$. First, the commutator is
    \begin{align}
        [\hat t,\hat a^\dagger(\omega_j)]&=\frac{1}{2\pi}\int \dd t\dd\omega\dd\omega'\, te^{i(\omega-\omega')t}\hat a^\dagger(\omega)[\hat a(\omega'),\hat a^\dagger(\omega_j)],\notag\\
        &=\frac{1}{2\pi}\int \dd t\dd\omega\, te^{i(\omega-\omega_j)t}\hat a^\dagger(\omega).
    \end{align}
    Thus by repeated swaps, we get
    \begin{align}
        \hat t \hat a^\dagger(\omega_1)&\cdots\hat a^\dagger(\omega_k)=\hat a^\dagger(\omega_1)\cdots\hat a^\dagger(\omega_k)\hat t\notag\\
        &+\frac{1}{2\pi}\sum_{j=1}^k\int \dd t\dd\omega\, te^{i(\omega-\omega_j)t}\hat a^\dagger(\omega_1)\cdots\hat a^\dagger(\omega_{j-1})\hat a^\dagger(\omega)\hat a^\dagger(\omega_{j+1})\cdots\hat a^\dagger(\omega_k).
    \end{align}
    Injecting this in the expression of $\hat t\ket{\psi_k}$
    \begin{align}
        \hat t\ket{\psi_k}&=\frac{1}{2\pi}\sum_{j=1}^k \int \dd t\dd\omega\dd\omega_1\cdots\dd\omega_k\, F(\omega_1,\ldots,\omega_k) te^{i(\omega-\omega_j)t}\notag\\
        &\qquad\times\hat a^\dagger(\omega_1)\cdots\hat a^\dagger(\omega_{j-1})\hat a^\dagger(\omega)\hat a^\dagger(\omega_{j+1})\cdots\hat a^\dagger(\omega_k)\vac
    \end{align}
    We perform an integration by parts with respect to the variable $\omega_j$, while noticing that the fully integrated term is zero
    \begin{subequations}
        \allowdisplaybreaks
        \begin{align}
            \hat t\ket{\psi_k}&=-\frac{i}{2\pi}\sum_{j=1}^k \int \dd t\dd\omega\dd\omega_1\cdots\dd\omega_k\, e^{i(\omega-\omega_j)t}\partial_j F(\omega_1,\ldots,\omega_k)\notag\\
            &\qquad\times \hat a^\dagger(\omega_1)\cdots\hat a^\dagger(\omega_{j-1})\hat a^\dagger(\omega)\hat a^\dagger(\omega_{j+1})\cdots\hat a^\dagger(\omega_k)\vac,\\
            &=-i\sum_{j=1}^k \int \dd\omega\dd\omega_1\cdots\dd\omega_k\, \delta(\omega-\omega_j)\partial_j F(\omega_1,\ldots,\omega_k)\notag\\
            &\qquad\times \hat a^\dagger(\omega_1)\cdots\hat a^\dagger(\omega_{j-1})\hat a^\dagger(\omega)\hat a^\dagger(\omega_{j+1})\cdots\hat a^\dagger(\omega_k)\vac,\\
            &=-i\sum_{j=1}^k \int \dd\omega_1\cdots\dd\omega_k\, \partial_j F(\omega_1,\ldots,\omega_k)\hat a^\dagger(\omega_1)\cdots\hat a^\dagger(\omega_k)\vac.
        \end{align}
    \end{subequations}
    We thus indeed get the transformation rule for $F$. We can now compute the commutator
    \begin{align}
        \hat\omega\hat t: F(\omega_1,\ldots,\omega_k)&\mapsto -i\sum_{j,l=1}^k\omega_j\partial_l F(\omega_1,\ldots,\omega_k),\\
        \hat t\hat\omega: F(\omega_1,\ldots,\omega_k)&\mapsto -i\sum_{j,l=1}^k\partial_j(\omega_l F(\omega_1,\ldots,\omega_k))\notag\\
        &\qquad= -i\sum_{j,l=1}^k(\delta_{j,l} F(\omega_1,\ldots,\omega_k)+\omega_l\partial_j F(\omega_1,\ldots,\omega_k)).
    \end{align}
    Thus
    \begin{equation}
        [\hat \omega,\hat t]: F(\omega_1,\ldots,\omega_k) \mapsto i k F(\omega_1,\ldots,\omega_k),
    \end{equation}
    meaning that $[\hat \omega,\hat t] \ket{\psi_k}= i k \ket{\psi_k}$. Since this relation holds for each sector of fixed number of photons, we indeed get
    \begin{equation}
        [\hat \omega,\hat t] = i \hat N .
    \end{equation}
\end{derivation}

\begin{result}[Alternate expression of the chrono-cyclic Wigner function]
    \label{res: alternate wigner}
    The chrono-cyclic Wigner function defined for a single-photon state in a single spatial mode $\hat\rho$ as
    \begin{equation}
    W(\varphi,\tau) = \frac{1}{\pi} \int \dd \omega\, e^{2i\omega \tau}\bra{\varphi-\omega}\hat \rho\ket{\varphi+\omega},
    \end{equation}
    can also be expressed as
    \begin{equation}
        W(\varphi,\tau) = \frac{1}{\pi} \int \dd t\, e^{-2i\varphi t}\bra{\tau-t}\hat \rho\ket{\tau+t}.
    \end{equation}
\end{result}

\begin{derivation}
    We use the closure relation for single-photon states $\int \dd t\, \ketbra{t}=\1$, which by change of variable also gives
    \begin{align}
        \int \dd t_1\, \ketbra{\tau-t_1}=\1,&& \int \dd t_2\, \ketbra{\tau+t_2}=\1.
    \end{align}
    We can thus compute, noting that the scalar product between a time and a frequency eigenstate is
    \begin{equation}
        \braket{\omega}{t}=\frac{1}{\sqrt{2\pi}}e^{i\omega t},
    \end{equation}
    \begin{subequations}
        \allowdisplaybreaks
        \begin{align}
            W(\varphi,\tau) &= \frac{1}{\pi} \int \dd \omega\, e^{2i\omega \tau}\bra{\varphi-\omega}\hat \rho\ket{\varphi+\omega},\\
            &=\frac{1}{\pi} \int \dd \omega\dd t_1\dd t_2\, e^{2i\omega \tau}\bra{\varphi-\omega}\ket{\tau-t_1}\bra{\tau-t_1}\hat \rho\ket{\tau+t_2}\bra{\tau+t_2}\ket{\varphi+\omega},\\
            &=\frac{1}{2\pi^2}\int \dd\omega\dd t_1\dd t_2\, e^{2i\omega \tau}e^{i(\varphi-\omega)(\tau-t_1)}e^{-i(\tau+t_2)(\varphi+\omega)}\bra{\tau-t_1}\hat \rho\ket{\tau+t_2},\\
            &=\frac{1}{2\pi^2}\int \dd\omega\dd t_1\dd t_2\,\operatorname{exp}(\cancel{2i\omega\tau}+\cancel{i\varphi\tau}-i\varphi t_1-\cancel{i\omega\tau}+i\omega t_1-\cancel{i\tau\varphi}-\cancel{i\tau\omega}\notag\\
            &\qquad-i t_2\varphi-i t_2\omega)\bra{\tau-t_1}\hat \rho\ket{\tau+t_2},\\
            &=\frac{1}{2\pi^2}\int \dd\omega\dd t_1\dd t_2\, e^{i\omega(t_1-t_2)}e^{-i\varphi(t_1+t_2)}\bra{\tau-t_1}\hat \rho\ket{\tau+t_2},\\
            &=\frac{1}{\pi}\int \dd t_1\dd t_2\,\delta(t_1-t_2)e^{-i\varphi(t_1+t_2)}\bra{\tau-t_1}\hat \rho\ket{\tau+t_2},\\
            &=\frac{1}{\pi}\int \dd t\, e^{-2i\varphi t}\bra{\tau-t}\hat \rho\ket{\tau+t}.
        \end{align}
    \end{subequations}
\end{derivation}

\begin{result}[Wigner function of GKP states]~\\
    \label{res: Wigner GKP}
    The Wigner function of the GKP state
    \begin{equation}
        \ket{\psi_\text{GKP}} = \sum_{k=-\infty}^\infty \ket{\omega_0+k\Delta},
    \end{equation}
    is
    \begin{align}
        W_\text{GKP}(\varphi,\tau)&=\frac{1}{2\Delta}\sum_{u,v\in\Z}\delta(\varphi-\omega_0-v\Delta)\delta\left(\tau-\frac{\pi}{\Delta}u\right)\notag\\
        &\qquad+\frac{1}{2\Delta}\sum_{u,v\in\Z}(-1)^u\delta\left(\varphi-\omega_0-\frac{2v+1}{2}\Delta\right)\delta\left(\tau-\frac{\pi}{\Delta}u\right).
    \end{align}    
\end{result}

\begin{derivation}
    $\blacktriangleright$ {\bf Poisson summation formula.} To derive the expression of the Wigner function of the GKP states we first need an intermediary result, which is an alternative way to phrase the Poisson summation formula
    \begin{equation}\label{eq: poisson}
        \frac{1}{2\pi}\sum_{k\in\Z} e^{ikx}=\sum_{k\in\Z}\delta(x-2k\pi).
    \end{equation}
    The first way to interpret this formula is to say that the Fourier transform of a Dirac comb is a re-scaled Dirac comb. Indeed
    \begin{subequations}
        \allowdisplaybreaks
        \begin{align}
            \frac{1}{\sqrt{2\pi}}\int\dd x\, e^{ixy}\sum_{k\in\Z}\delta(x-k\Delta)  &= \frac{1}{\sqrt{2\pi}}\sum_{k\in \Z}e^{i\Delta ky},\\
            &=\sqrt{2 \pi} \sum_{k\in\Z} \delta(\Delta y-2k\pi),\\
            &=\frac{\sqrt{2\pi}}{\Delta}\sum_{k\in\Z}\delta\left(y-\frac{2k\pi}{\Delta}\right),
        \end{align}
    \end{subequations}
    which is indeed a Dirac comb in re-scaled variables. We see that if $\Delta=\sqrt{2\pi}$ then the Dirac comb is exactly its own Fourier transform.

    An informal proof of the formula of Eq.~\eqref{eq: poisson} is quite easy. Indeed if $x$ is a multiple of $2\pi$ then the sum only adds an infinite number of $1$, which should ``sum'' to $\infty$. Otherwise, the sum will add a ``balanced'' number of positive and negative terms, leading to a zero contribution. This exactly corresponds to the intuitive values taken by the Dirac comb: infinity at integer multiples of $2\pi$ and zero everywhere else. However this approach is really sketchy, and does not provide an explanation on the normalization factor $\frac{1}{2\pi}$. Obviously the left-hand side does not converge in the usual sense. Thus the equality in Eq.~\eqref{eq: poisson} has to be understood in a broader sense: in terms of distributions.

    Wanting to show this equality in terms of distributions means that whenever we take a test function $f$ which is sufficiently well behaved at infinity (for example with compact support, or in Schwartz space), we have the equality
    \begin{equation}
        \int\dd x\, f(x)\left(\frac{1}{2\pi}\sum_{k\in\Z} e^{ikx}\right) =\int \dd x\, f(x)\left(\sum_{k\in\Z} \delta(x-2k\pi)\right) ,
    \end{equation}
    which is equivalent to
    \begin{equation}
        \frac{1}{\sqrt{2\pi}}\sum_{k\in\Z}\hat f(k)=\sum_{k\in\Z} f(2k\pi),
    \end{equation}
    where $\hat f (y)=\frac{1}{\sqrt{2\pi}}\int \dd x\, f(x)e^{ixy} $ is the Fourier transform of $f$. We introduce the function
    \begin{equation}
        F(x)=\sum_{k\in\Z}f(x+2k\pi),
    \end{equation}
    for which it is easy to see that it is $2\pi$-periodic. We can thus consider its expansion in terms of its (discrete) Fourier coefficients:
    \begin{equation}
        F(x)=\frac{1}{\sqrt{2\pi}}\sum_{n\in\Z}\hat F(n)e^{inx},
    \end{equation}
    where the Fourier inversion formula for periodic functions tells us that
    \begin{equation}
        \hat F(n)=\frac{1}{\sqrt{2\pi}}\int\limits_0^{2\pi}\dd x\, F(x)e^{-inx} .
    \end{equation}
    We then compute $\hat F(n)$:
    \begin{subequations}
        \allowdisplaybreaks
        \begin{align}
            \hat F(n)&=\frac{1}{\sqrt{2\pi}}\int_0^{2\pi}\dd x \, F(x)e^{-inx} ,\\
            &=\frac{1}{\sqrt{2\pi}}\sum_{k\in\Z}\int_0^{2\pi} \dd x\, f(x+2k\pi)e^{-inx} ,\\
            &=\frac{1}{\sqrt{2\pi}}\sum_{k\in\Z}\int_{2k\pi}^{2(k+1)\pi}\dd x \, f(x)e^{-in(x-2k\pi)} ,\\
            &=\frac{1}{\sqrt{2\pi}}\sum_{k\in\Z}\int_{2k\pi}^{2(k+1)\pi}\dd x\, f(x)e^{-inx} ,\\
            &=\frac{1}{\sqrt{2\pi}}\int_{-\infty}^{\infty} \dd x\, f(x)e^{-inx} ,\\
            &=\hat f(-n),
        \end{align}
    \end{subequations}
    using Chasles' relation to combine all integrals into one. Evaluating both expressions of $F$ at zero gives the result:
    \begin{subequations}
        \begin{align}
            F(0)&=\sum_{k\in\Z}f(2k\pi),\\
            F(0)&=\frac{1}{\sqrt{2\pi}}\sum_{n\in\Z} \hat F(n)=\frac{1}{\sqrt{2\pi}}\sum_{n\in\Z} \hat f(-n),\\
            &=\frac{1}{\sqrt{2\pi}}\sum_{n\in\Z} \hat f(n),
        \end{align}
    \end{subequations}
    which concludes the proof.

    \medskip

    \noindent
    $\blacktriangleright$ {\bf Computation of the Wigner function.} After this identity, we can now compute the Wigner function of the Dirac comb state. By going back to the definition we compute:
    \begin{subequations}
        \begin{align}
            W(\varphi,\tau)&=\frac{1}{\pi}\int\dd\omega\, e^{2i\omega\tau}\bra{\varphi-\omega}\ket{\psi_\text{GKP}}\bra{\psi_\text{GKP}}\ket{\varphi+\omega},\\
            &=\frac{1}{\pi}\int \dd\omega\, e^{2i\omega\tau}\sum_{k\in\Z}\delta(\varphi-\omega-\omega_0-k\Delta)\sum_{l\in\Z}\delta(\varphi+\omega-\omega_0-l\Delta),\\
            &=\frac{1}{\pi}\sum_{k,l\in\Z}\int \dd\omega\, e^{2i\omega\tau}\delta(\varphi-\omega-\omega_0-k\Delta)\delta(\varphi+\omega-\omega_0-l\Delta),\\
            &=\frac{1}{\pi}\sum_{k,l\in\Z}e^{2i(l\Delta+\omega_0-\varphi)\tau}\delta\Big(\varphi-(l\Delta+\omega_0-\varphi)-\omega_0-k\Delta\Big),\\
            &=\frac{1}{2\pi}\sum_{k,l\in\Z}e^{i(l-k)\Delta\tau}\delta\left(\varphi-\omega_0-\frac{(k+l)\Delta}{2}\right).
        \end{align}
    \end{subequations}
    Since the terms inside the sum are expressed in terms of the diagonal variables, $u=l-k$ and $v=k+l$, one wants to perform a change of variables so that we can decouple both sums. However since we are dealing with sums over integers, this change of variables is more subtle. Indeed, $\sum\limits_{k,l\in\Z} f(l-k,k+l)\neq \sum\limits_{u,v\in\Z}f(u,v)$. Indeed a quick analysis reveals that $k+l$ and $l-k$ have the same parity. Thus the term $f(0,3)$ does not appear in the left sum. In fact this is the only constraint, meaning that the correct way to perform the change of variable is $\sum\limits_{k,l\in\Z} f(l-k,k+l)= \sum\limits_{u,v\in2\Z}f(u,v)+\sum\limits_{u,v\in2\Z+1}f(u,v)=\sum\limits_{u,v\in\Z}f(2u,2v)+\sum\limits_{u,v\in\Z}f(2u+1,2v+1)$. With this, we can continue the computation while using Eq.~\eqref{eq: poisson}:
    \begin{subequations}
        \begin{align}
            W(\varphi,\tau)&=\frac{1}{2\pi}\sum_{u,v\in\Z}e^{i2u\Delta\tau}\delta\left(\varphi-\omega_0-v\Delta\right)\notag\\
            &\qquad+\frac{1}{2\pi}\sum_{u,v\in\Z}e^{i(2u+1)\Delta\tau}\delta\left(\varphi-\omega_0-\frac{2v+1}{2}\Delta\right),\\
            &=\sum_{u,v\in\Z}\delta(2\Delta\tau-2\pi u)\delta(\varphi-\omega_0-v\Delta) \notag\\
            &\qquad +e^{i\Delta\tau}\sum_{u,v\in\Z}\delta(2\Delta\tau-2\pi u) \delta\left(\varphi-\omega_0-\frac{2v+1}{2}\Delta\right),\\
            &=\frac{1}{2\Delta}\sum_{u,v\in\Z}\delta\left(\tau-\frac{\pi}{\Delta} u\right)\delta(\varphi-\omega_0-v\Delta)\notag\\
            &\qquad+\frac{1}{2\Delta} \sum_{u,v\in\Z}(-1)^u\delta\left(\tau-\frac{\pi}{\Delta} u\right) \delta\left(\varphi-\omega_0-\frac{2v+1}{2}\Delta\right).
        \end{align}
    \end{subequations}
\end{derivation}

\begin{result}[Wigner function of two-mode separable states]~\\
    \label{res: Wigner local separable}
    If a two-mode single-photon state $\ket{\psi}$ has its JSA factorizable as
    \begin{equation}
        F(\omega_s,\omega_i)=F_s(\omega_s)F_i(\omega_i),
    \end{equation}
    then its Wigner function can be factored as
    \begin{equation}
        W(\varphi_s,\varphi_i,\tau_s,\tau_i)=W_s(\varphi_s,\tau_s)W_i(\varphi_i,\tau_i),
    \end{equation}
    where
    \begin{equation}
        W_j(\varphi_j,\tau_j)=\frac{1}{\pi}\int\dd \omega_j\, e^{2i\omega_j \tau_j} F_j(\varphi_j-\omega_j)F_j^\ast(\varphi_j+\omega_j),
    \end{equation}
    is the Wigner function of the single-mode state with spectrum $F_j$ for $j=s,i$.    
\end{result}

\begin{derivation}
    We simply need to perform the computation. We start with the definition
    \begin{subequations}
        \allowdisplaybreaks
        \begin{align}
            W(\varphi_s,\varphi_i,\tau_s,\tau_i)&=\frac{1}{\pi^2}\int e^{2i(\omega_s\tau_s+\omega_i\tau_i)}\bra{\varphi_s-\omega_s,\varphi_i-\omega_i}\ket{\psi}\bra{\psi}\ket{\varphi_s+\omega_s,\varphi_i+\omega_i}\, \dd \omega_s \dd \omega_i,\\
            &=\frac{1}{\pi^2}\int e^{2i(\omega_s\tau_s+\omega_i\tau_i)} F_s(\varphi_s-\omega_s)F_i(\varphi_i-\omega_i) \notag\\
            &\qquad\times F_s^\ast(\varphi_s+\omega_s)F_i^\ast(\varphi_i+\omega_i)\,\dd \omega_s \dd \omega_i,\\
            &=\left(\frac{1}{\pi}\int e^{2i\omega_s\tau_s} F_s(\varphi_s-\omega_s)F_s^\ast(\varphi_s+\omega_s)\, \dd \omega_s\right)\notag\\
            &\qquad\times \left(\frac{1}{\pi}\int e^{2i\omega_i\tau_i} F_i(\varphi_i-\omega_i)F_i^\ast(\varphi_i+\omega_i)\, \dd \omega_i\right),\\
            &=W_s(\varphi_s,\tau_s)W_i(\varphi_i,\tau_i).
        \end{align}
    \end{subequations}
\end{derivation}

\begin{result}[Wigner function of two-mode diagonally separable states]~\\
    \label{res: Wigner diagonal separable}
    If a two-mode single-photon state $\ket{\psi}$ has its JSA factorizable as
    \begin{equation}
        F(\omega_s,\omega_i)=F_+(\omega_+)F_-(\omega_-),
    \end{equation}
    where $\omega_\pm=\frac{\omega_s\pm\omega_i}{\sqrt{2}}$, then its Wigner function can be factored as
    \begin{equation}
        W(\varphi_s,\varphi_i,\tau_s,\tau_i)=W_+(\varphi_+,\tau_+)W_-(\varphi_-,\tau_-),
    \end{equation}
    where we have introduced the variables $\varphi_\pm=\frac{\varphi_s\pm\varphi_i}{\sqrt{2}}$ and $\tau_\pm=\frac{\tau_s\pm\tau_i}{\sqrt{2}}$, and the local Wigner functions are defined as
    \begin{equation}
        W_\pm(\varphi_\pm,\tau_\pm)=\frac{1}{\pi}\int e^{2i\omega_\pm \tau_\pm} F_\pm(\varphi_\pm-\omega_\pm)F_\pm^\ast(\varphi_\pm+\omega_\pm)\,\dd \omega_\pm.
    \end{equation}    
\end{result}

\begin{derivation}
    We simply need to perform the computation. We start with the definition and perform a change of variables
    \begin{subequations}
        \begin{align}
            W(\varphi_s,\varphi_i,\tau_s,\tau_i)&=\frac{1}{\pi^2}\int e^{2i(\omega_s\tau_s+\omega_i\tau_i)}\bra{\varphi_s-\omega_s,\varphi_i-\omega_i}\ket{\psi}\bra{\psi}\ket{\varphi_s+\omega_s,\varphi_i+\omega_i}\,\dd \omega_s \dd \omega_i,\\
            &=\frac{1}{\pi^2}\int e^{2i(\omega_s\tau_s+\omega_i\tau_i)} F_+\left(\tfrac{\varphi_s+\varphi_i}{\sqrt{2}}-\tfrac{\omega_s+\omega_i}{\sqrt{2}}\right) F_+^*\left(\tfrac{\varphi_s+\varphi_i}{\sqrt{2}}+\tfrac{\omega_s+\omega_i}{\sqrt{2}}\right)\notag\\
            &\qquad\times F_-\left(\tfrac{\varphi_s-\varphi_i}{\sqrt{2}}-\tfrac{\omega_s-\omega_i}{\sqrt{2}}\right) F_-^*\left(\tfrac{\varphi_s-\varphi_i}{\sqrt{2}}+\tfrac{\omega_s-\omega_i}{\sqrt{2}}\right)\, \dd \omega_s \dd \omega_i,\\
            &=\frac{1}{\pi^2}\int e^{2i(\omega_+\tau_++\omega_-\tau_-)} F_+(\varphi_+-\omega_+) F_+^*(\varphi_++\omega_+)\notag\\
            &\qquad\times F_-(\varphi_--\omega_-) F_-^*(\varphi_-+\omega_-)\,\dd \omega_+ \dd \omega_-,\\
            &=\left(\frac{1}{\pi}\int e^{2i\omega_+\tau_+} F_+(\varphi_+-\omega_+) F_+^*(\varphi_++\omega_+)\,\dd \omega_+\right)\notag\\
            &\qquad\times \left(\frac{1}{\pi}\int e^{2i\omega_-\tau_-} F_-(\varphi_--\omega_-) F_-^*(\varphi_-+\omega_-)\,\dd \omega_-\right),\\
            &=W_+(\varphi_+,\tau_+)W_-(\varphi_-,\tau_-),
        \end{align}
    \end{subequations}
    by noticing that
    \begin{equation}
        \omega_s\tau_s+\omega_i\tau_i=\frac{(\omega_++\omega_-)(\tau_++\tau_-)}{2}+\frac{(\omega_+-\omega_-)(\tau_+-\tau_-)}{2}=\omega_+\tau_++\omega_-\tau_-.
    \end{equation}
\end{derivation}

\begin{result}[Time-frequency coherent state]~\\
    \label{res: TF coherent state}
    For $F:\R\to\C$ a normalized function, $I>0$ and the creation operator
    \begin{equation}
        \hat a_F^\dagger=\int\dd\omega \,F(\omega)\hat a^\dagger(\omega),
    \end{equation}
    the state 
    \begin{equation}
        \ket{\psi}=e^{-I/2}e^{\sqrt{I}\hat a_F^\dagger}\vac,
    \end{equation}
    satisfies the eigenvalue equation
    \begin{equation}
        \hat a(\omega)\ket{\psi}=\sqrt{I} F(\omega)\ket{\psi}.
    \end{equation}    
\end{result}

\begin{derivation}
    We first compute the commutator
    \begin{subequations}
        \begin{align}
            \left[\hat a(\omega),\hat a_F^\dagger\right]&=\int \dd \omega'\, F(\omega') \left[\hat a(\omega),\hat a(\omega')^\dagger\right],\\
            &=\int \dd \omega'\, F(\omega')\delta(\omega-\omega')\1,\\
            &=F(\omega)\1.
        \end{align}
    \end{subequations}
    From this we can deduce the commutation relation between $\hat a(\omega)$ and $ (\hat a_F^\dagger)^n$. Indeed
    \begin{subequations}
        \begin{align}
            \hat a(\omega)(\hat a_F^\dagger)^n&=(\hat a_F^\dagger \hat a(\omega)+F(\omega)\1)(\hat a_F^\dagger)^{n-1},\\
            &=\hat a_F^\dagger \hat a(\omega)(\hat a_F^\dagger)^{n-1}+ F(\omega)(\hat a_F^\dagger)^{n-1},\\
            &=(\hat a_F^\dagger)^2  \hat a(\omega)(\hat a_F^\dagger)^{n-2}+ 2F(\omega)(\hat a_F^\dagger)^{n-1},\\
            &=\cdots= (\hat a_F^\dagger)^n\hat a(\omega)+nF(\omega)(\hat a_F^\dagger)^{n-1}.
        \end{align}
    \end{subequations}
    So that $\left[\hat a(\omega),(\hat a_F^\dagger)^n\right]=n F(\omega)(\hat a_F^\dagger)^{n-1}$. This allows us to compute
    \begin{subequations}
        \allowdisplaybreaks
        \begin{align}
            \hat a(\omega) \ket{\psi}&=e^{-I/2}\hat a(\omega)e^{\sqrt{I}\hat a_F^\dagger}\vac,\\
            &=e^{-I/2}\hat a(\omega)\sum_{k=0}^\infty \frac{\sqrt{I}^k}{k!}(\hat a_F^\dagger)^k\vac,\\
            &=e^{-I/2}\sum_{k=0}^\infty \frac{\sqrt{I}^k}{k!}(\hat a_F^\dagger)^k\underbrace{\hat a(\omega)\vac}_{=0}+e^{-I/2}\sum_{k=0}^\infty \frac{\sqrt{I}^k}{k!}kF(\omega)(\hat a_F^\dagger)^{k-1}\vac,\\
            &=e^{-I/2}\sum_{k=1}^\infty \frac{\sqrt{I}^k}{(k-1)!}F(\omega)(\hat a_F^\dagger)^{k-1}\vac,\\
            &=e^{-I/2}\sqrt{I} F(\omega)\sum_{k=0}^\infty \frac{\sqrt{I}^k}{k!}(\hat a_F^\dagger)^{k}\vac,\\
            &=e^{-I/2}\sqrt{I} F(\omega)e^{\sqrt{I}\hat a_F^\dagger}\vac,\\
            &=\sqrt{I} F(\omega)\ket{\psi}.
        \end{align}
    \end{subequations}
\end{derivation}

\begin{result}[Scalar product of time-frequency coherent states]~\\
    \label{res: TF coherent states scalar product}
    The scalar product between two time-frequency coherent states $\ket{\alpha(\omega)}$ and $\ket{\beta(\omega)}$ is given by
    \begin{equation}
        \bra{\alpha(\omega)}\ket{\beta(\omega)}=\exp\left(-\frac{1}{2}\int \dd\omega\,\left[\abs{\alpha(\omega)}^2+\abs{\beta(\omega)}^2-2\alpha^\ast(\omega)\beta(\omega)\right]\right).
    \end{equation}
    Squaring this formula leads to
    \begin{equation}
        \abs{\bra{\alpha(\omega)}\ket{\beta(\omega)}}^2=\exp\left(-\int \dd\omega\, \abs{\alpha(\omega)-\beta(\omega)}^2\right).
    \end{equation}    
\end{result}

\begin{derivation}
    Using the explicit representation of time-frequency coherent states, we introduce the single-mode operators $\hat a_F^\dagger$ and $\hat a_G^\dagger$ associated with the normalized spectra
    \begin{align}
        F(\omega) =\frac{1}{\sqrt{I_a}}\alpha(\omega), && G(\omega)=\frac{1}{\sqrt{I_b}}\beta(\omega),
    \end{align}
    with $I_a=\int \abs{\alpha(\omega)}^2\, \dd\omega$ and $I_b=\int \abs{\beta(\omega)}^2\,\dd\omega$, so that $\ket{\alpha(\omega)}=e^{-I_a/2}e^{\sqrt{I_a}\hat a_F^\dagger}\vac$ and $\ket{\beta(\omega)}=e^{-I_b/2}e^{\sqrt{I_b}\hat a_G^\dagger}\vac$. We verify that the creation operators are correctly normalized as well as computing the commutator $[\hat a_F,\hat a_G^\dagger]$
    \begin{subequations}
        \begin{align}
            [\hat a_F,\hat a_F^\dagger]&=\frac{1}{I_a}\int \dd\omega_1 \dd\omega_2\, \alpha^*(\omega_1)\alpha(\omega_2)[\hat a(\omega_1),\hat a(\omega_2)^\dagger],\\
            &=\frac{1}{I_a}\int \dd\omega_1 \dd\omega_2\, \alpha^\ast (\omega_1)\alpha(\omega_2)\delta(\omega_1-\omega_2)\1,\\
            &=\frac{1}{I_a}\int \dd \omega\,\abs{\alpha(\omega)}^2\1,\\
            &=\1.
        \end{align}
    \end{subequations}
    And the same for $[\hat a_G,\hat a_G^\dagger]=\1$. Then
    \begin{subequations}
        \begin{align}
            [\hat a_F,\hat a_G^\dagger]&=\frac{1}{\sqrt{I_aI_b}}\int \dd\omega_1 \dd\omega_2\, \alpha^\ast(\omega_1) \beta(\omega_2) [\hat a(\omega_1),\hat a(\omega_2)^\dagger],\\
            &=\frac{1}{\sqrt{I_aI_b}}\int \dd\omega_1 \dd\omega_2\, \alpha^\ast (\omega_1)\beta(\omega_2)\delta(\omega_1-\omega_2)\1,\\
            &=\frac{1}{\sqrt{I_aI_b}}\int \dd \omega\,\alpha^\ast(\omega)\beta(\omega)\1.
        \end{align}
    \end{subequations}
    Thanks to the commutation relation between $\hat a_F$ and $\hat a_G^\dagger$, we can use the Baker-Campbell-Hausdorff (see Appendix~\ref{app: formalism and framework}, Result~\ref{res: BCH}) formula to get
    \begin{equation}
        e^{\sqrt{I_a}\hat a_F}e^{\sqrt{I_b} \hat a_G^\dagger}=e^{\sqrt{I_b} \hat a_G^\dagger}e^{\sqrt{I_a}\hat a_F}e^{\sqrt{I_aI_b}[\hat a_F,\hat a_G^\dagger]}.
    \end{equation}
    Now we can compute
    \begin{subequations}
        \begin{align}
            \bra{\alpha(\omega)}\ket{\beta(\omega)}&=e^{-I_a/2-I_b/2}\bra{\text{vac}}e^{\sqrt{I_a}\hat a_F}e^{\sqrt{I_b}\hat a_G^\dagger}\vac,\\
            &=e^{-I_a/2-I_b/2}e^{\int \dd\omega\,\alpha^\ast(\omega)\beta(\omega)}\bra{\text{vac}}e^{\sqrt{I_b}\hat a_G^\dagger}e^{\sqrt{I_a}\hat a_F}\vac,\\
            &=e^{-I_a/2-I_b/2}e^{\int \dd\omega\,\alpha^\ast(\omega)\beta(\omega)}\braket{\text{vac}},\\
            &=\exp(-\frac{1}{2}\int \dd\omega\, \Big[\abs{\alpha(\omega)}^2+\abs{\beta(\omega)}^2-2\alpha^\ast(\omega)\beta(\omega)\Big]).
        \end{align}
    \end{subequations}
\end{derivation}

\begin{result}[Translation of creation operators]~\\
    \label{res: Translation creation operators}
    For any real parameters $\tau$ and $\varphi$, we have the relations
    \begin{align}
        e^{-i\hat \omega \tau}\hat a^\dagger(\omega)e^{i\hat \omega \tau}&=e^{-i\omega \tau}\hat a^\dagger(\omega), &
        e^{-i\hat \omega \tau}\hat b^\dagger(t)e^{i\hat \omega \tau}&=\hat b^\dagger(t-\tau),\\
        e^{-i\varphi \hat t}\hat a^\dagger(\omega)e^{i\varphi \hat t}&=\hat a^\dagger(\omega+\varphi), &
        e^{-i\varphi \hat t}\hat b^\dagger(t)e^{i\varphi\hat t} &=e^{-i\varphi t}\hat b^\dagger(t).
    \end{align}
\end{result}

\begin{derivation}
    $\blacktriangleright$ {\bf Time evolution of $\hat a^\dagger(\omega)$.} The first equation is obtained using the commutator
    \begin{equation}
        [\hat \omega,\hat a^\dagger(\omega)]=\int \dd \omega'\, \omega'[\hat a^\dagger(\omega')\hat a(\omega'),\hat a^\dagger(\omega)]=\int \dd \omega'\, \omega' \delta(\omega-\omega')\hat a^\dagger(\omega)=\omega \hat a^\dagger(\omega),
    \end{equation}
    which implies $\hat \omega\hat a^\dagger(\omega)= \hat a^\dagger(\omega)(\hat \omega+\omega)$ or $\hat \omega^k\hat a^\dagger(\omega)= \hat a^\dagger(\omega)(\hat \omega+\omega)^k$. We thus obtain
    \begin{subequations}
        \begin{align}
            e^{-i\hat \omega \tau}\hat a^\dagger(\omega)e^{i\hat \omega \tau}&=\sum_{k=0}^\infty \frac{(-i\tau\hat \omega)^k}{k!}\hat a^\dagger(\omega)e^{i\hat \omega \tau},\\
            &=\hat a^\dagger(\omega)\sum_{k=0}^\infty \frac{(-i\tau)^k}{k!}(\hat \omega+\omega)^k e^{i\hat \omega \tau},\\
            &=\hat a^\dagger(\omega)e^{-i\omega \tau}\sum_{k=0}^\infty \frac{(-i\tau\hat \omega)^k}{k!} e^{i\hat \omega \tau},\\
            &=e^{-i\omega \tau}\hat a^\dagger(\omega).
        \end{align}
    \end{subequations}
    \medskip

    \noindent
    $\blacktriangleright$ {\bf Time evolution of $\hat b^\dagger(t)$.} The transformation of $\hat b^\dagger(t)$ under an evolution generated by $\hat \omega$ is then obtained by expressing $\hat b^\dagger(t)$ in terms of frequency operators
    \begin{subequations}
        \begin{align}
            e^{-i\hat \omega \tau}\hat b^\dagger(t)e^{i\hat \omega \tau}&=\frac{1}{\sqrt{2\pi}}\int \dd\omega\, e^{i\omega t} e^{-i\hat \omega \tau}\hat a^\dagger(\omega)e^{i\hat \omega \tau},\\
            &=\frac{1}{\sqrt{2\pi}}\int \dd\omega\,  e^{i\omega (t-\tau)}\hat a^\dagger(\omega),\\
            &=\hat b^\dagger(t-\tau).
        \end{align}
    \end{subequations}
    \medskip

    \noindent
    $\blacktriangleright$ {\bf Frequency evolution of $\hat b^\dagger(t)$.} We proceed similarly with the time operator. We first compute the commutator
    \begin{equation}
        [\hat t,\hat b^\dagger(t)]=\int \dd t'\, t'[\hat b^\dagger(t')\hat b(t'),\hat b^\dagger(t)]=\int \dd t'\, t' \delta(t-t')\hat b^\dagger(t)= t \hat b^\dagger(t),
    \end{equation}
    leading to $\hat t^k\hat b^\dagger(t)= \hat b^\dagger(t)(\hat t+t)^k$. Thus
    \begin{subequations}
        \begin{align}
            e^{-i\varphi \hat t}\hat b^\dagger(t)e^{i\varphi \hat t}&=\sum_{k=0}^\infty \frac{(-i\varphi\hat t)^k}{k!}\hat b^\dagger(t)e^{i\varphi \hat t},\\
            &=\hat b^\dagger(t)\sum_{k=0}^\infty \frac{(-i\varphi)^k}{k!}(\hat t+t)^k e^{i\varphi \hat t},\\
            &=\hat b^\dagger(t)e^{-i\varphi t}\sum_{k=0}^\infty \frac{(-i\varphi\hat t)^k}{k!} e^{i\varphi \hat t},\\
            &=e^{-i\varphi t}\hat b^\dagger(t).
        \end{align}
    \end{subequations}
    \medskip

    \noindent
    $\blacktriangleright$ {\bf Frequency evolution of $\hat a^\dagger(\omega)$.} Finally we express $\hat a^\dagger(\omega)$ in terms of time operators to get
    \begin{subequations}
        \begin{align}
            e^{-i\varphi \hat t}\hat a^\dagger(\omega)e^{i\varphi \hat t}&=\frac{1}{\sqrt{2\pi}}\int \dd t\, e^{-i\omega t} e^{-i\varphi \hat t}\hat b^\dagger(t)e^{i\varphi \hat t},\\
            &=\frac{1}{\sqrt{2\pi}}\int \dd t\,  e^{-i(\omega+\varphi) t}\hat b^\dagger(t),\\
            &=\hat a^\dagger(\omega+\varphi).
        \end{align}
    \end{subequations}
\end{derivation}

\begin{result}[Wigner function and translation]~\\
    \label{res: Wigner translation}
    Given an initial single-mode single-photon time-frequency state $\hat \rho$, the action of the evolution $e^{-i\hat \omega\tau_0}$ and $e^{-i\varphi_0\hat t}$ on the Wigner function $W_{\hat \rho}$ is given by
    \begin{align}
        W_{e^{-i\hat \omega \tau_0}\hat \rho e^{i\hat \omega \tau_0}}(\varphi,\tau)&=W_{\hat \rho}(\varphi,\tau+\tau_0),\\
        W_{e^{-i\varphi_0 \hat t}\hat \rho e^{i\varphi_0 \hat t}}(\varphi,\tau)&=W_{\hat \rho}(\varphi-\varphi_0,\tau).
    \end{align}
\end{result}

\begin{derivation}
    This is a simple computation using $e^{-i\hat \omega \tau}\ket{\omega}= e^{-i\omega \tau}\ket{\omega}$
    \begin{subequations}
        \begin{align}
            W_{e^{-i\hat \omega \tau_0}\hat \rho e^{i\hat \omega \tau_0}}(\varphi,\tau)&=\frac{1}{\pi}\int e^{2i\omega \tau}\bra{\varphi-\omega}e^{-i\hat \omega \tau_0}\hat \rho e^{i\hat \omega \tau_0}\ket{\varphi+\omega}\,\dd \omega,\\
            &=\frac{1}{\pi}\int e^{2i\omega \tau-i(\varphi-\omega)\tau_0+i(\varphi+\omega)\tau_0}\bra{\varphi-\omega}\hat \rho \ket{\varphi+\omega}\,\dd \omega,\\
            &=\frac{1}{\pi}\int e^{2i\omega (\tau+\tau_0)}\bra{\varphi-\omega}\hat \rho \ket{\varphi+\omega}\,\dd \omega,\\
            &=W_{\hat \rho}(\varphi,\tau+\tau_0).
        \end{align}
    \end{subequations}
    And, using $e^{-i\varphi\hat t}\ket{\omega}= \ket{\omega+\varphi}$
    \begin{subequations}
        \begin{align}
            W_{e^{-i\varphi_0 \hat t}\hat \rho e^{i\varphi_0 \hat t}}(\varphi,\tau)&=\frac{1}{\pi}\int e^{2i\omega \tau}\bra{\varphi-\omega}e^{-i\varphi_0 \hat t}\hat \rho e^{i\varphi_0 \hat t}\ket{\varphi+\omega}\,\dd \omega,\\
            &=\frac{1}{\pi}\int e^{2i\omega \tau}\bra{\varphi-\omega-\varphi_0}\hat \rho \ket{\varphi+\omega-\varphi_0}\,\dd \omega,\\
            &=\frac{1}{\pi}\int e^{2i\omega \tau}\bra{(\varphi-\varphi_0)-\omega}\hat \rho \ket{(\varphi-\varphi_0)+\omega}\,\dd \omega,\\
            &=W_{\hat \rho}(\varphi-\varphi_0,\tau).
        \end{align}
    \end{subequations}
\end{derivation}

\begin{result}[Quadratic transformation of creation operators]~\\
    \label{res: Quadratic gate creation op}
    Under the quadratic evolution generated by $\hat \omega^2$ and $\hat t^2$, the time and frequency creation operators transform as
    \begin{align}
        e^{-i\alpha \hat \omega^2}\hat a^\dagger(\omega)e^{i\alpha \hat \omega^2}&=e^{-i\alpha \omega^2}\hat a^\dagger(\omega)e^{-2i\alpha\hat \omega\omega}, &
        e^{-i\beta \hat t^2}\hat b^\dagger(t)e^{i\beta \hat t^2} &=e^{-i\beta t^2}\hat b^\dagger(t)e^{-2i\beta \hat t t}.
    \end{align}
    As such single-photon states transform as
    \begin{align}
        e^{-i\alpha \hat \omega^2}\ket{\omega}&=e^{-i\alpha \omega^2}\ket{\omega}, &
        e^{-i\beta \hat t^2}\ket{t}&=e^{-i\beta t^2}\ket{t}.
    \end{align}
\end{result}

\begin{derivation}
    $\blacktriangleright$ {\bf Time evolution.} We use once again the commutation relation $\hat\omega^k\hat a^\dagger(\omega)=\hat a^\dagger(\omega)(\hat\omega+\omega)^k$, which we verified in the proof of Result~\ref{res: Translation creation operators}. The computation yields
    \begin{subequations}
        \begin{align}
            e^{-i\alpha \hat \omega^2}\hat a^\dagger(\omega)e^{i\alpha \hat \omega^2}&=\sum_{k=0}^\infty \frac{(-i\alpha \hat \omega^2)^k}{k!}\hat a^\dagger(\omega)e^{i\alpha \hat \omega^2},\\
            &=\hat a^\dagger(\omega)\sum_{k=0}^\infty \frac{(-i\alpha)^k}{k!}(\hat \omega+\omega)^{2k} e^{i\alpha \hat \omega^2},\\
            &=\hat a^\dagger(\omega)e^{-i\alpha(\hat\omega+\omega)^2} e^{i\alpha \hat \omega^2},\\
            &=e^{-i\alpha \omega^2}\hat a^\dagger(\omega)e^{-2i\alpha\hat \omega\omega}.
        \end{align}
    \end{subequations}
    The action on a single-photon state is then given by
    \begin{align}
        e^{-i\alpha \hat \omega^2}\ket{\omega}=e^{-i\alpha \hat \omega^2}\hat a^\dagger(\omega)e^{i\alpha \hat \omega^2}e^{-i\alpha \hat \omega^2}\vac=e^{-i\alpha \omega^2}\hat a^\dagger(\omega)e^{-2i\alpha\hat \omega\omega}\vac=e^{-i\alpha \omega^2}\ket{\omega},
    \end{align}
    since the vacuum is invariant under translation.
    \medskip

    \noindent
    $\blacktriangleright$ {\bf Frequency evolution.} The reasoning for the quadratic time gate is similar. Following the formula $\hat t^k\hat b^\dagger(t)=\hat b^\dagger(t)(\hat t+t)^k$, we have
    \begin{subequations}
        \allowdisplaybreaks
        \begin{align}
            e^{-i\beta \hat t^2}\hat b^\dagger(t)e^{i\beta \hat t^2}&=\sum_{k=0}^\infty \frac{(-i\beta \hat t^2)^k}{k!}\hat b^\dagger(t)e^{i\beta \hat t^2},\\
            &=\hat b^\dagger(t)\sum_{k=0}^\infty \frac{(-i\beta)^k}{k!}(\hat t+t)^{2k} e^{i\beta \hat t^2},\\
            &=\hat b^\dagger(t)e^{-i\beta(\hat t+t)^2} e^{i\beta \hat t^2},\\
            &=e^{-i\beta t^2}\hat b^\dagger(t)e^{-2i\beta \hat t t}.
        \end{align}
    \end{subequations}
    Leading to
    \begin{equation}
        e^{-i\beta \hat t^2}\ket{t}=e^{-i\beta t^2}\ket{t}.
    \end{equation}
\end{derivation}

\begin{result}[Quadratic transformation of the Wigner function]~\\
    \label{res: Wigner quadratic transformation}
    For a single-photon state $\hat\rho$, its Wigner function $W_{\hat \rho}$ is transformed as 
    \begin{align}
        W_{e^{-i\alpha \hat \omega^2}\hat \rho e^{i\alpha \hat \omega^2}}(\varphi,\tau)&=W_{\hat \rho}(\varphi,\tau+2\alpha \varphi),\\
        W_{e^{-i\beta \hat t^2}\hat \rho e^{i\beta \hat t^2}}(\varphi,\tau)&=W_{\hat \rho}(\varphi-2\beta\tau,\tau),
    \end{align}
    under the quadratic time or frequency evolutions.
\end{result}

\begin{derivation}
    $\blacktriangleright$ {\bf Frequency shear.} We start with the quadratic frequency gate, and compute
    \begin{subequations}
        \begin{align}
            W_{e^{-i\alpha \hat \omega^2}\hat \rho e^{i\alpha \hat \omega^2}}(\varphi,\tau)&=\frac{1}{\pi}\int e^{2i\omega \tau}\bra{\varphi-\omega}e^{-i\alpha \hat \omega^2}\hat \rho e^{i\alpha \hat \omega^2}\ket{\varphi+\omega}\,\dd \omega,\\
            &=\frac{1}{\pi}\int e^{2i\omega \tau}e^{-i\alpha (\varphi-\omega)^2}e^{i\alpha (\varphi+\omega)^2}\bra{\varphi-\omega}\hat \rho \ket{\varphi+\omega}\,\dd \omega,\\
            &=\frac{1}{\pi}\int e^{2i\omega (\tau+2\alpha \varphi)}\bra{\varphi-\omega}\hat \rho \ket{\varphi+\omega}\,\dd \omega,\\
            &=W_{\hat \rho}(\varphi,\tau+2\alpha \varphi).
        \end{align}
    \end{subequations}
    \medskip

    \noindent
    $\blacktriangleright$ {\bf Time shear.} For the quadratic time gate, we use the alternate form of the Wigner function
    \begin{subequations}
        \allowdisplaybreaks
        \begin{align}
            W_{e^{-i\beta \hat t^2}\hat \rho e^{i\beta \hat t^2}}(\varphi,\tau)&=\frac{1}{\pi}\int e^{-2i t \varphi}\bra{\tau -t}e^{-i\beta \hat t^2}\hat \rho e^{i\beta \hat t^2}\ket{\tau +t}\,\dd t,\\
            &=\frac{1}{\pi}\int e^{-2i t \varphi}e^{-i\beta (\tau -t)^2}e^{i\beta (\tau +t)^2}\bra{\tau -t}\hat \rho \ket{\tau +t}\,\dd t,\\
            &=\frac{1}{\pi}\int e^{-2i t (\varphi -2\beta \tau)}\bra{\tau -t}\hat \rho \ket{\tau +t}\,\dd t,\\
            &=W_{\hat \rho}(\varphi-2\beta\tau,\tau).
        \end{align}
    \end{subequations}
\end{derivation}

\begin{result}[Rotation of time-frequency operators]~\\
    \label{res: Rotation of TF operators}
    Defining $\hat R=(\hat \omega^2+\hat t^2)/2$, the time and frequency operators transform as
    \begin{align}
        e^{-i\theta \hat R}\hat \omega e^{i\theta \hat R}=\cos\theta\hat \omega  -\sin\theta\hat t , && 
        e^{-i\theta \hat R}\hat t e^{i\theta \hat R}=\sin\theta\hat \omega  +\cos\theta\hat t .
    \end{align}
\end{result}

\begin{derivation}
    We first compute the commutation relations between $\hat{R}$ and $\hat{\omega}$ or $\hat{t}$
    \begin{subequations}
        \begin{align}
            [\hat{\omega},\hat{R}]&=\frac{1}{2}\Big[\cancel{\hat{\omega}^3}+\hat{\omega}\hat{t}^2-\cancel{\hat{\omega}^3}-\hat{t}^2\hat{\omega}\Big],\\
            &=\frac{1}{2}\Big[\hat{t}\hat{\omega}\hat{t}+i\hat{t}-\hat{t}^2\hat{\omega}\Big],\\
            &=\frac{1}{2}\Big[\cancel{\hat{t}^2\hat{\omega}}+2i\hat{t}-\cancel{\hat{t}^2\hat{\omega}}\Big],\\
            &=i\hat{t}.
        \end{align}
    \end{subequations}
    And
    \begin{subequations}
        \begin{align}
            [\hat{t},\hat{R}]&=\frac{1}{2}\Big[\hat{t}\hat{\omega}^2+\cancel{\hat{t}^3}-\hat{\omega}^2\hat{t}-\cancel{\hat{t}^3}\Big],\\
            &=\frac{1}{2}\Big[\hat{\omega}\hat{t}\hat{\omega}-i\hat{\omega}-\hat{\omega}^2\hat{t}\Big],\\
            &=\frac{1}{2}\Big[\cancel{\hat{\omega}^2\hat{t}}-2i\hat{\omega}-\cancel{\hat{\omega}^2\hat{t}}\Big],\\
            &=-i\hat{\omega}.
        \end{align}
    \end{subequations}
    
    Now we introduce the parameter-dependent operators
    \begin{subequations}
        \begin{align}
            \hat{\omega}(\theta)&=e^{-i\theta\hat{R}}\hat{\omega}e^{i\theta\hat{R}}=\cos\theta\hat{\omega}-\sin\theta\hat{t},\\
            \hat{t}(\theta)&=e^{-i\theta\hat{R}}\hat{t}e^{i\theta\hat{R}}=\cos\theta\hat{t}+\sin\theta\hat{\omega}.
        \end{align}
    \end{subequations}
    And we want to show that they equal the following operators
    \begin{align}
        \hat{A}(\theta)=\cos\theta\hat{\omega}-\sin\theta\hat{t}, &&
        \hat{B}(\theta)=\cos\theta\hat{t}+\sin\theta\hat{\omega}.
    \end{align}
    To do this we will use differential equations and the Cauchy-Lipschitz theorem. We can observe that
	\begin{equation}
	    \frac{\dd\hat{\omega}(\theta)}{\dd\theta}=-i\hat{R}\hat{\omega}(\theta)+i\hat{\omega}(\theta)\hat{R}=i[\hat{\omega}(\theta),\hat{R}].
	\end{equation}
	On the other hand we can compute the following commutator
	\begin{equation}
	    [\hat{A}(\theta),\hat{R}]=i\cos\theta\hat{t}+i\sin\theta\hat{\omega}.
	\end{equation}
	As well as the derivative of $\hat{A}(\theta)$
	\begin{equation}
	    \frac{\dd\hat{A}(\theta)}{\dd\theta}=-\sin\theta\hat{\omega}-\cos\theta\hat{t}=i[\hat{A}(\theta),\hat{R}].
	\end{equation}
	We thus observe that $\hat{A}(\theta)$ and $\hat{\omega}(\theta)$ satisfy the same first-order linear differential equation; since they take the same value for $\theta =0$ both functions are equal. We do the same thing for $\hat{t}(\theta)$:
	\begin{subequations}
        \begin{align}
            \frac{d\hat{t}(\theta)}{d\theta}&=-i\hat{R}\hat{t}(\theta)+i\hat{t}(\theta)\hat{R}=i[\hat{t}(\theta),\hat{R}],\\
            [\hat{B}(\theta),\hat{R}]&=-i\cos\theta\hat{\omega}+i\sin\theta\hat{t},\\
            \frac{d\hat{B}(\theta)}{d\theta}&=-\sin\theta\hat{t}+\cos\theta\hat{\omega}=i[\hat{B}(\theta),\hat{R}].
        \end{align}
	\end{subequations}
    This concludes the proof.
\end{derivation}

\begin{result}[Wigner function and rotation]~\\
    \label{res: rotation wigner}
    The Wigner function of a single-photon state $\hat \rho$ evolves under rotation as
    \begin{equation}
        W_{e^{-i\theta \hat R}\hat \rho e^{i\theta \hat R}}(\varphi,\tau)=W_{\hat \rho}(\varphi\cos\theta-\tau\sin\theta, \tau\cos\theta+\varphi\sin\theta).
    \end{equation}
\end{result}

\begin{derivation}
    We consider the two functions 
    \begin{align}
        W_\theta(\varphi,\tau)&=W_{e^{-i\theta \hat R}\hat \rho e^{i\theta \hat R}}(\varphi,\tau), \\
        \tilde{W}_\theta(\varphi,\tau)&=W_{\hat{\rho}}(\varphi\cos\theta-\tau\sin\theta,\tau\cos\theta+\varphi\sin\theta),
    \end{align}
    and using once again the Cauchy-Lipschitz theorem (or more precisely the Cauchy-Kovalevskaya theorem as the differential equation we will write involves derivatives with respect to $\varphi$ and $\tau$) we will show that they are equal as they satisfy the same first-order partial differential equation with the same initial condition at $\theta=0$.

    \medskip

    \noindent
    
    $\blacktriangleright$ {\bf Differential action.} To study $W_\theta(\varphi,\tau)$, we first need to understand the differential action of the Wigner function. We observe that under the parametrization
    \begin{equation}
        \hat{\rho}(\theta)=e^{-i\theta\hat{R}}\hat{\rho}e^{i\theta\hat{R}},
    \end{equation}
    the infinitesimal action is given by the expansion, for small $\theta$
    \begin{equation}
        \hat{\rho}(\theta)=(1-i\theta\hat{R})\hat{\rho}(1+i\theta\hat{R})+O(\theta^2)=\hat{\rho}+i\theta[\hat{\rho},\hat{R}]+O(\theta^2).
    \end{equation}
    Or equivalently, by linearity of the Wigner function
    \begin{equation}\label{evolutionWigner}
        \pdv{\theta}W_\theta=iW_{[\hat{\rho}(\theta),\hat{R}]}.
    \end{equation}
    We thus need to compute $W_{[\hat{\rho}(\theta),\hat{R}]}$. Splitting $\hat R$ in two parts, we compute
    \begin{subequations}
        \begin{align}
            W_{[\hat{\rho}(\theta),\hat\omega^2]}(\varphi,\tau)&=\frac{1}{\pi}\int \dd\omega\, e^{2i\omega\tau}\bra{\varphi-\omega}(\hat{\rho}(\theta)\hat{\omega}^2-\hat{\omega}^2\hat{\rho}(\theta))\ket{\varphi+\omega},\\
            &=\frac{1}{\pi}\int \dd\omega\, e^{2i\omega\tau}\Big[(\varphi+\omega)^2-(\varphi-\omega)^2\Big]\bra{\varphi-\omega}\hat{\rho}(\theta)\ket{\varphi+\omega},\\
            &=-2i\varphi\pdv{\tau}W_\theta(\varphi,\tau).
        \end{align}
    \end{subequations}
    Then
    \begin{subequations}\label{actiont2}
        \begin{align}
            W_{[\hat{\rho}(\theta),\hat t^2]}(\varphi,\tau)&=\frac{1}{\pi}\int \dd t\, e^{-2it\varphi}\bra{\tau-t}(\hat{\rho}(\theta)\hat{t}^2-\hat{t}^2\hat{\rho}(\theta))\ket{\tau+t},\\
            &=\frac{1}{\pi}\int \dd t\, e^{-2it\varphi}\Big[(\tau+t)^2-(\tau-t)^2\Big]\bra{\tau-t}\hat{\rho}(\theta)\ket{\tau+t},\\
            &=2i\tau\pdv{\varphi}W_\theta(\varphi,\tau).
        \end{align}
    \end{subequations}
    We thus get
    \begin{equation}
        \pdv{\theta}W_\theta(\varphi,\tau)=iW_{[\hat{\rho}(\theta),\hat{R}]}(\varphi,\tau)=\varphi\pdv{\tau}W_\theta(\varphi,\tau)-\tau\pdv{\varphi}W_\theta(\varphi,\tau).
    \end{equation}

    \medskip

    \noindent
    
    $\blacktriangleright$ {\bf Differential equation and conclusion.} For $\tilde{W}_\theta(\varphi,\tau)$, we differentiate using the chain rule. We introduce the rotated variables $\varphi^{(\theta)}=\cos\theta\varphi-\sin\theta\tau$ and $\tau^{(\theta)}=\cos\theta\tau+\sin\theta\varphi$, which verify
    \begin{align}
        \pdv{\varphi^{(\theta)}}{\theta}=-\tau^{(\theta)}, && \pdv{\tau^{(\theta)}}{\theta}=\varphi^{(\theta)}.
    \end{align}
    Writing $\partial_1 W_{\hat\rho}$ and $\partial_2 W_{\hat\rho}$ for derivatives with respect to its first and second arguments, evaluated at $(\varphi^{(\theta)},\tau^{(\theta)})$, we obtain
    \begin{align}
        \pdv{\tilde W_\theta}{\theta}&=-\tau^{(\theta)}\partial_1 W_{\hat\rho}+\varphi^{(\theta)}\partial_2 W_{\hat\rho},\\
        \pdv{\tilde W_\theta}{\varphi}&=\cos\theta\partial_1 W_{\hat\rho}+\sin\theta\partial_2 W_{\hat\rho},\\
        \pdv{\tilde W_\theta}{\tau}&=-\sin\theta\partial_1 W_{\hat\rho}+\cos\theta\partial_2 W_{\hat\rho}.
    \end{align}
    Inverting the last two relations gives
    \begin{align}
        \partial_1 W_{\hat\rho}&=\cos\theta\pdv{\tilde W_\theta}{\varphi}-\sin\theta\pdv{\tilde W_\theta}{\tau},\\
        \partial_2 W_{\hat\rho}&=\sin\theta\pdv{\tilde W_\theta}{\varphi}+\cos\theta\pdv{\tilde W_\theta}{\tau}.
    \end{align}
    Substituting into the derivative with respect to $\theta$, and using
    \begin{align}
        \tau^{(\theta)}\cos\theta-\varphi^{(\theta)}\sin\theta&=\tau,\\
        \varphi^{(\theta)}\cos\theta+\tau^{(\theta)}\sin\theta&=\varphi,
    \end{align}
    we obtain
    \begin{equation}
        \pdv{\theta}\tilde W_\theta(\varphi,\tau)=\varphi\pdv{\tau}\tilde W_\theta(\varphi,\tau)-\tau\pdv{\varphi}\tilde W_\theta(\varphi,\tau).
    \end{equation}
    Both functions $W_\theta$ and $\tilde W_\theta$ satisfy the same linear first-order partial differential equation, with the same initial condition. Thus $W_\theta=\tilde W_\theta$, which proves the claim.
\end{derivation}

\begin{result}[Partial Fourier transform and shears]~\\
    \label{res: Partial Fourier}
    We have the relation
    \begin{equation}
        e^{-i a\frac{\hat t^2}{2}}e^{-ib\frac{\hat \omega^2}{2}}e^{-ia\frac{\hat t^2}{2}}=e^{-i\theta\hat R},
    \end{equation}
    provided that the coefficients satisfy
    \begin{align}
        a=\frac{1-\cos(\theta)}{\sin(\theta)}, && b=\sin(\theta).
    \end{align}
\end{result}

\begin{derivation}
    In full generality we can consider the following transformation
    \begin{equation}
        e^{-i a\frac{\hat t^2}{2}}e^{-ib\frac{\hat \omega^2}{2}}e^{-ic\frac{\hat t^2}{2}},
    \end{equation}
    We use the Wigner function point of view: indeed, the operators $\hat \omega^2$, $\hat t^2$ and $\hat R$ act on the phase-space coordinates as linear transformations.
    \begin{subequations}
        \begin{align}
            e^{-i\kappa\hat\omega^2}:~~&\begin{pmatrix}
                \varphi\\
                \tau 
            \end{pmatrix}\mapsto \begin{pmatrix}
            \varphi\\
            \tau+2\kappa\varphi
            \end{pmatrix}=\begin{pmatrix}
                1 & 0\\
                2\kappa & 1
            \end{pmatrix}\begin{pmatrix}
                \varphi\\
                \tau
            \end{pmatrix},\\
            e^{-i\kappa\hat t^2}:~~&\begin{pmatrix}
                \varphi\\
                \tau 
            \end{pmatrix}\mapsto \begin{pmatrix}
            \varphi-2\tau\kappa\\
            \tau
            \end{pmatrix}=\begin{pmatrix}
                1 & -2\kappa\\
                0 & 1
            \end{pmatrix}\begin{pmatrix}
                \varphi\\
                \tau
            \end{pmatrix},\\
            e^{-i\kappa\hat R}:~~&\begin{pmatrix}
                \varphi\\
                \tau 
            \end{pmatrix}\mapsto \begin{pmatrix}
            \cos(\kappa)\varphi-\sin(\kappa)\tau\\
            \cos(\kappa)\tau+\sin(\kappa)\varphi
            \end{pmatrix}=\begin{pmatrix}
                \cos(\kappa) & -\sin(\kappa)\\
                \sin(\kappa) & \cos(\kappa)
            \end{pmatrix}\begin{pmatrix}
                \varphi\\
                \tau
            \end{pmatrix}.
        \end{align}
    \end{subequations}
    So we simply need to solve the following matrix equation:
    \begin{equation}
        \begin{pmatrix}
            1 & -a\\
            0 & 1
        \end{pmatrix}\begin{pmatrix}
            1 & 0\\
            b & 1
        \end{pmatrix}\begin{pmatrix}
            1 & -c\\
            0 & 1
        \end{pmatrix}=\begin{pmatrix}
            \cos(\theta) & -\sin(\theta)\\
            \sin(\theta) & \cos(\theta)
        \end{pmatrix}.
    \end{equation}
    And manually, or with the help of a numerical solver, we get:
    \begin{align}
        a=\frac{1-\cos(\theta)}{\sin(\theta)}, && b=\sin(\theta), && c=a=\frac{1-\cos(\theta)}{\sin(\theta)}.
    \end{align}
\end{derivation}

\begin{result}[Variance of coherent states]~\\
    \label{res: variance coherent states}
    The coherent state $\ket{\alpha(\omega)}$ with intensity $I=\int \abs{\alpha(\omega)}^2 \,\dd \omega$ and spectrum $F(\omega)=\alpha(\omega)/\sqrt{I}$ has frequency variances given by
    \begin{equation}
        \Delta^2\hat\omega = I\int \omega^2 \abs{F(\omega)}^2 \,\dd \omega.
    \end{equation}
\end{result}

\begin{derivation}
    We use the defining equation of the coherent state $\hat a(\omega)\ket{\alpha(\omega)}=\alpha(\omega)\ket{\alpha(\omega)}$ to compute the expectation value
    \begin{subequations}
        \begin{align}
            \bra{\alpha(\omega)}\hat{\omega}\ket{\alpha(\omega)}&= \int \dd\omega\, \omega\bra{\alpha(\omega)}\hat{a}^\dagger(\omega)\hat{a}(\omega)\ket{\alpha(\omega)},\\
            &=\int \dd\omega\, \omega \alpha^\ast(\omega)\braket{\alpha(\omega)}\alpha(\omega),\\
            &=\int \dd\omega\, \omega \abs{\alpha(\omega)}^2,\\
            &= I \int \dd\omega\, \omega \abs{F(\omega)}^2.
        \end{align}
    \end{subequations}
    Then the expectation value of $\hat\omega^2$ is
    \begin{subequations}
        \begin{align}
            \bra{\alpha(\omega)}\hat{\omega}^2\ket{\alpha(\omega)}&=\int \dd\omega_1 \dd\omega_2\, \omega_1 \omega_2 \bra{\alpha(\omega)}\hat{a}^\dagger(\omega_1)\hat{a}(\omega_1)\hat{a}^\dagger(\omega_2)\hat{a}(\omega_2)\ket{\alpha(\omega)},\\
            &=\int \dd\omega_1 \dd\omega_2\, \omega_1 \omega_2 \alpha^\ast(\omega_1)\alpha(\omega_2) \bra{\alpha(\omega)}\underbrace{\hat{a}(\omega_1)\hat{a}^\dagger(\omega_2)}_{\hat{a}^\dagger(\omega_2)\hat{a}(\omega_1)+\delta(\omega_2-\omega_1)}\ket{\alpha(\omega)},\\
            &=\int \dd\omega_1 \dd\omega_2\, \omega_1 \omega_2 \alpha^\ast(\omega_1)\alpha(\omega_2) [\alpha^\ast(\omega_2)\alpha(\omega_1)+\delta(\omega_2-\omega_1)],\\
            &=\int \dd\omega_1 \dd\omega_2\,\omega_1 \omega_2 \abs{\alpha(\omega_1)}^2\abs{\alpha(\omega_2)}^2+\int \dd\omega \omega^2\abs{\alpha(\omega)}^2,\\
            &=\Big(\int \dd\omega\,  \omega \abs{\alpha(\omega)}^2\Big)^2+\int \dd\omega\,  \omega^2\abs{\alpha(\omega)}^2,\\
            &= I^2\Big(\int \dd\omega\,  \omega \abs{F(\omega)}^2\Big)^2+I\int \dd\omega \, \omega^2\abs{F(\omega)}^2.
        \end{align}
    \end{subequations}
    So the variance is simply given by 
    \begin{equation}
        \Delta^2\hat\omega = \bra{\alpha(\omega)}\hat{\omega}^2\ket{\alpha(\omega)}-\bra{\alpha(\omega)}\hat{\omega}\ket{\alpha(\omega)}^2=I\int \dd\omega\, \omega^2 \abs{F(\omega)}^2.
    \end{equation}
\end{derivation}

\fi

\ifnum \theShowChapthree=1
\ifthenelse{
    \value{ShowChapone}=1 \OR
    \value{ShowChaptwo}=1 
}{
\clearpage
}{}
\section{Collective-variable entanglement}
\setcurrentanchor{app coll var}
\emph{This section of the appendix collects all the results of Chap.~\ref{chap: collective entanglement} and provides the associated proofs.}

\vspace*{1em}
\par\noindent
\textbf{\large Results in this section}\par
\vspace{-0.8em}
\noindent\rule{\textwidth}{0.8pt}\par

\resultentry{res: single mode variance}{Single-mode state frequency variance}
\resultentry{res: physical approx}{Variance of states with nonzero width}
\resultentry{res: ideal correlated wigner}{Wigner function of an ideal frequency-correlated state}
\resultentry{res: popoviciu's}{General bound on the variance}
\resultentry{res: cauchy-schwarz general}{Saturation condition}
\resultentry{res: convex roof}{Convex roof construction}
\resultentry{res: def support}{Equivalent definition of the support}
\resultentry{res: formula support conv}{Convexity of $I^{\text{S}}$}
\resultentry{res: mixed computation}{Example of computation of extensions of $I$}
\resultentry{res: k sep}{General $k$-entanglement inequality}
\resultentry{res: GKP global}{Collective GKP displacement}
\resultentry{res: GKP local}{Local GKP displacement}
\resultentry{res: GKP entangling}{GKP and entangling gates}

\vspace{-0.2em}
\par\noindent\rule{\textwidth}{0.8pt}\par

\label{app: collective var derivation}
\begin{result}[Single-mode state frequency variance]~\\
    \label{res: single mode variance}
    Let $\ket{\psi}$ be a time-frequency single-mode state of spectrum $F$, meaning that defining
    \begin{equation}
        \hat a_F^\dagger =\int \dd \omega\, F(\omega) \hat a^\dagger(\omega),
    \end{equation}
    we can write
    \begin{equation}
        \ket{\psi}=\sum_{m=0}^\infty c_m \frac{(\hat a_F^\dagger)^m}{\sqrt{m!}}\vac.
    \end{equation}
    The variance of $\hat \omega$ can be written 
    \begin{equation}
        \Delta^2 \hat \omega = \overline n \Delta^2\omega +{\overline \omega}^2\Delta^2 n,
    \end{equation}
    where $\overline n$ denotes the average photon number, and $\Delta^2 n$ its variance
    \begin{align}
        \overline n&=\sum_{n=0}^\infty n\abs{c_n}^2, & \Delta^2 n&=\sum_{n=0}^\infty (n-\overline n)^2\abs{c_n}^2,
    \end{align}
    while $\overline \omega$ denotes the average frequency and $\Delta^2 \omega$ the variance of the frequency distribution $\abs{F(\omega)}^2$
    \begin{align}
        \overline \omega&=\int \dd \omega\, \abs{F(\omega)}^2 \omega, & \Delta^2 \omega&=\int \dd \omega\, \abs{F(\omega)}^2 (\omega-\overline\omega)^2.
    \end{align}
\end{result}

\begin{derivation}
    Expanding the expression of $\hat a_F^\dagger$, we can write
    \begin{equation}
        \ket{\psi}=\sum_{m=0}^\infty c_m \frac{1}{\sqrt{m!}}\int \dd \omega_1\cdots\dd\omega_m\, F(\omega_1)\cdots F(\omega_m)\hat a^\dagger(\omega_1)\cdots \hat a^\dagger(\omega_m)\vac.
    \end{equation}
    Remembering the action of the operator $\hat \omega$ on such a description (see Result~\ref{res: General TF commutator}) we deduce
    \begin{align}
        \hat\omega\ket{\psi}&=\sum_{m=0}^\infty c_m \frac{1}{\sqrt{m!}}\int \dd \omega_1\cdots\dd\omega_m\, (\omega_1+\cdots+\omega_m) F(\omega_1)\cdots F(\omega_m)\notag\\
        &\qquad\times\hat a^\dagger(\omega_1)\cdots \hat a^\dagger(\omega_m)\vac.
    \end{align}
    Now recall the general formula for computing the scalar product for symmetric spectral distributions
    \begin{equation}
        \braket{\psi_f}{\psi_g}=k!\int \dd\omega_1\cdots \dd\omega_k\, f^\ast(\omega_1,\cdots,\omega_k)g(\omega_1,\cdots,\omega_k),
    \end{equation}
    whenever $f$ and $g$ are symmetric functions of their variables and are the spectral distributions of states with the same number of photons $k$. Using this we get
    \begin{subequations}
        \begin{align}
            \bra{\psi}\hat\omega\ket{\psi}&=\sum_{m=0}^\infty \abs{c_m}^2 \int \dd \omega_1\cdots\dd\omega_m\, (\omega_1+\cdots+\omega_m) \abs{F(\omega_1)}^2\cdots \abs{F(\omega_m)}^2,\\
            &=\sum_{m=0}^\infty \abs{c_m}^2 m \overline \omega=\overline n \overline \omega.
        \end{align}
    \end{subequations}
    Similarly, we get
    \begin{subequations}
        \begin{align}
            \bra{\psi}\hat\omega^2\ket{\psi}&=\sum_{m=0}^\infty \abs{c_m}^2 \int \dd \omega_1\cdots\dd\omega_m\, (\omega_1+\cdots+\omega_m)^2 \abs{F(\omega_1)}^2\cdots \abs{F(\omega_m)}^2,\\
            &=\sum_{m=0}^\infty \abs{c_m}^2 \Big[m \overline{\omega^2} + m(m-1) \overline \omega^2\Big]=\overline n \overline{\omega^2} +\overline{n^2} \overline\omega^2-\overline n\, \overline\omega^2,
        \end{align}
    \end{subequations}
    leading to
    \begin{equation}
        \Delta^2 \hat \omega = \overline n\overline{\omega^2}+\overline{n^2}\overline\omega^2-\overline n\overline\omega^2-\overline\omega^2\overline n^2=\overline n \Delta^2\omega +{\overline \omega}^2\Delta^2 n.
    \end{equation}
\end{derivation}

\begin{result}[Variance of states with nonzero width]~\\
    \label{res: physical approx}
    A physical approximation of the state 
    \begin{equation}
        \ket{\psi_\text{ideal}}=\int \dd\Omega\, f(\Omega)\ket{\Omega+\omega_1^0,\dots,\Omega+\omega_n^0},
    \end{equation}
    of the form
    \begin{equation}
        \ket{\psi}=\int\dd\Omega_1\cdots\dd\Omega_n\,f(\Omega_1)g(\Omega_2)\cdots g(\Omega_n)\ket{\omega_1,\dots,\omega_n},
    \end{equation}
    where $\Omega_j$ for $j\geq 1$ are collective orthogonal directions, where $\Omega_1=\omega_1+\cdots+\omega_n$ and $g$ is a spectral distribution satisfying
    \begin{equation}
        (\Delta^2\Omega)_g=(1-\eta)(\Delta^2\Omega)_f,
    \end{equation}
    satisfies
    \begin{equation}
        \Delta^2\hat\Omega=\frac{n^2}{n(1-\eta)+\eta}\Delta^2\omega.
    \end{equation}
\end{result}

\begin{derivation}
    For such a state, which is separable in the collective variables, we have
    \begin{align}
        \Delta^2\Omega_1=\Delta^2, && \Delta^2\hat\Omega_j=\sigma^2\quad(j\geq2), && \Cov(\Omega_j,\Omega_k)=0\quad(j\neq k),
    \end{align}
    where the assumption means $\sigma^2=(1-\eta)\Delta^2$. We can then compute the local variances. To do so, note that we have assumed that the $\Omega_j$ variables are orthogonal. As they are normalized to $n$, we can write
    \begin{equation}
        \omega_j=\frac{1}{n}\Big[(\omega_j\mid \Omega_1) \Omega_1+\cdots+(\omega_j\mid \Omega_n) \Omega_n\Big].
    \end{equation}
    Where $(\cdot,\cdot)$ denotes the usual scalar product on $\R^n$ of the associated direction vectors. So
    \begin{subequations}
        \begin{align}
            \Delta^2\omega_j&=\frac{1}{n^2}\Delta^2\big[(\omega_j\mid \Omega_1) \Omega_1+\cdots+(\omega_j\mid \Omega_n) \Omega_n\big],\\
            &=\frac{1}{n^2} (\omega_j\mid \Omega_1)^2\underbrace{\Delta^2 \Omega_1}_{\Delta^2}+\frac{1}{n^2}(\omega_j\mid \Omega_2)^2\underbrace{\Delta^2 \Omega_2}_{\sigma^2}+\cdots+\frac{1}{n^2}(\omega_j\mid \Omega_n)^2 \underbrace{\Delta^2 \Omega_n}_{\sigma^2},\\
            &=\frac{1}{n^2}\underbrace{(\omega_j\mid \Omega_1)^2}_{1}(\Delta^2-\sigma^2)+\frac{\sigma^2}{n^2}\underbrace{\sum_k (\omega_j\mid \Omega_k)^2}_{n},\\
            &=\frac{1}{n^2}(\Delta^2-\sigma^2)+\frac{1}{n}\sigma^2,\\
            &=\frac{1}{n^2}(\Delta^2-(1-\eta)\Delta^2)+\frac{1}{n}(1-\eta)\Delta^2,\\
            &=\frac{\Delta^2}{n^2}(\eta + n(1-\eta)).
        \end{align}
    \end{subequations}
    Inverting the formula indeed yields
    \begin{equation}
        \Delta^2\hat\Omega=\Delta^2=\frac{n^2}{n(1-\eta)+\eta}\Delta^2\omega,
    \end{equation}
    since all local variances are equal.
\end{derivation}

\begin{result}[Wigner function of an ideal frequency-correlated state]~\\
    \label{res: ideal correlated wigner}
    The Wigner function of the nonphysical state
    \begin{equation}
        \ket{\psi}=\int \dd\Omega\, f(\Omega)\ket{\Omega+\omega_1^0,\dots,\Omega+\omega_n^0},
    \end{equation}
    can be expressed as
    \begin{align}
        W(\varphi_1+\omega_1^0,&\dots,\varphi_n+\omega_n^0,\tau_1,\dots,\tau_n)\notag\\
        &=\frac{1}{(2\pi)^{n-1}}W_1(\varphi_1,\tau_1+\tau_2 +\cdots + \tau_n)\delta(\varphi_2-\varphi_1)\cdots\delta(\varphi_n-\varphi_1),
    \end{align}
    where
    \begin{equation}
    W_1(\varphi,t) = \frac{1}{\pi}\int \dd\Omega\, e^{2i\Omega t}f(\varphi-\Omega)f^\ast(\varphi+\Omega),
    \end{equation}
    is the Wigner function associated with the single-spatial-mode state $\int \dd\omega\, f(\omega)\ket{\omega}$.
\end{result}

\begin{derivation}
    \begin{subequations}
        \allowdisplaybreaks
        \begin{align}
            W(\varphi_1,&\dots,\varphi_n,\tau_1,\dots,\tau_n)\notag\\
            &=\frac{1}{\pi^n}\int \dd\omega_1\cdots\dd\omega_n\, e^{2i(\omega_1\tau_1+\cdots+\omega_n\tau_n)}\bra{\varphi_1-\omega_1,\dots,\varphi_n-\omega_n}\ket{\psi}\notag\\
            &\qquad\times\bra{\psi}\ket{\varphi_1+\omega_1,\dots,\varphi_n+\omega_n},\\
            &=\frac{1}{\pi^n}\int \dd\omega_1\cdots\dd\omega_n\, \dd\Omega \dd\Omega' e^{2i(\omega_1\tau_1+\cdots+\omega_n\tau_n)}f(\Omega)f^\ast(\Omega')\notag\\
            &\qquad\times\braket{\varphi_1-\omega_1}{\Omega+\omega_1^0}\cdots\braket{\varphi_n-\omega_n}{\Omega+\omega_n^0}\notag\\
            &\qquad\times\braket{\Omega'+\omega_1^0}{\varphi_1+\omega_1}\cdots\braket{\Omega'+\omega_n^0}{\varphi_n+\omega_n},\\
            &=\frac{1}{\pi^n}\int \dd\omega_1\cdots\dd\omega_n\, \dd\Omega \dd\Omega' e^{2i(\omega_1\tau_1+\cdots+\omega_n\tau_n)}f(\Omega)f^\ast(\Omega')\notag\\
            &\qquad\times\delta(\varphi_1-\omega_1-\Omega-\omega_1^0)\cdots\delta(\varphi_n-\omega_n-\Omega-\omega_n^0)\notag\\
            &\qquad\times\delta(\Omega'+\omega_1^0-\varphi_1-\omega_1)\cdots\delta(\Omega'+\omega_n^0-\varphi_n-\omega_n),\\
            &=\frac{1}{\pi^n}\int \dd\Omega \dd\Omega'\,\exp[2i(\Omega'+\omega_1^0-\varphi_1)\tau_1+\cdots+2i(\Omega'+\omega_n^0-\varphi_n)\tau_n]\notag\\
            &\qquad\times f(\Omega)f^\ast(\Omega')\delta(2\varphi_1-2\omega_1^0-\Omega-\Omega')\cdots\delta(2\varphi_n-2\omega_n^0-\Omega-\Omega'),\\
            &=\frac{1}{\pi^n}\int \dd\Omega\, \exp[2i(\varphi_1-\omega_1^0-\Omega)\tau_1+\cdots+2i(2\varphi_1-2\omega_1^0-\Omega-\varphi_n+\omega_n^0)\tau_n]\notag\\
            &\qquad\times f(\Omega)f^\ast(2\varphi_1-2\omega^0_1-\Omega)\delta(2\varphi_2-2\omega_2^0-2\varphi_1+2\omega_1^0)\notag\\
            &\qquad\cdots\delta(2\varphi_n-2\omega_n^0-2\varphi_1+2\omega_1^0),\\
            &=\frac{1}{\pi^n 2^{n-1}}\int \dd\Omega\, e^{2i(\varphi_1-\omega_1^0-\Omega)\tau_1+\cdots+2i(2\varphi_1-2\omega_1^0-\Omega-\varphi_n+\omega_n^0)\tau_n} f(\Omega)f^\ast(2\varphi_1-2\omega^0_1-\Omega)\notag\\
            &\qquad\times\delta(\varphi_2-\omega_2^0-(\varphi_1-\omega_1^0))\cdots\delta(\varphi_n-\omega_n^0-(\varphi_1-\omega_1^0))\\
            &=\frac{1}{\pi^n2^{n-1}}\int \dd\Omega\, e^{2i\Omega(\tau_1+\cdots+\tau_n)}f(\varphi_1-\omega_1^0-\Omega)f^\ast(\varphi_1-\omega^0_1+\Omega)\notag\\
            &\qquad\times\delta(\varphi_2-\omega_2^0-(\varphi_1-\omega_1^0))\cdots\delta(\varphi_n-\omega_n^0-(\varphi_1-\omega_1^0)),
        \end{align}
    \end{subequations}
    by using the Dirac delta function transformation rule with constant factors $\delta(2x)=\delta(x)/2$, and in the last step we performed the change of variable $\Omega\to \varphi_1-\omega_1^0-\Omega$.
    This Wigner function can be rewritten as
    \begin{align}
        W(\varphi_1+\omega_1^0,&\dots,\varphi_n+\omega_n^0,\tau_1,\dots,\tau_n)\notag\\
        &=\frac{1}{(2\pi)^{n-1}}W_1(\varphi_1,\tau_1+\tau_2 +\cdots + \tau_n)\delta(\varphi_2-\varphi_1)\cdots\delta(\varphi_n-\varphi_1),
    \end{align}
    where
    \begin{equation}
    W_1(\varphi,t) = \frac{1}{\pi}\int \dd\Omega\, e^{2i\Omega t}f(\varphi-\Omega)f^\ast(\varphi+\Omega),
    \end{equation}
    is the Wigner function associated with the spatial single-mode state $\int \dd\omega\, f(\omega)\ket{\omega}$.
\end{derivation}

\begin{result}[General bound on the variance]~\\
    \label{res: popoviciu's}
    For a Hilbert space $\mathcal H$ and a Hamiltonian $\hat H$ with largest and smallest eigenvalues $h_{\max}$, $h_{\min}$, we have
    \begin{equation}
        \Delta^2 \hat H\leq \frac{(h_{\max}-h_{\min})^2}{4},
    \end{equation}
    with equality for states $\ket{\psi}=\frac{1}{\sqrt{2}}\left[\ket{\psi_{\min}}+\ket{\psi_{\max}}\right]$ where $\ket{\psi_{\min}}$ (resp. $\ket{\psi_{\max}}$) is any eigenstate with eigenvalue $h_{\min}$ (resp. $h_{\max}$).
\end{result}

\begin{derivation}
    $\blacktriangleright$ {\bf Probability reformulation.} This statement is a direct consequence of a classical probability result known as Popoviciu's inequality~\cite{bhatia_better_2000}: For any real random variable $X$ that is bounded (almost surely), we set $m=\inf X$, $M=\sup X$. (More precisely, we set $m=\sup\{x\in\R|\mathbb P(X\geq x)=1\}$ and $M=\inf\{x\in\R|\mathbb P(X\leq x)=1\}$, which is rigorously not the same thing). Then we have
    \begin{equation}
        \mathbb V(X)\leq \frac{(M-m)^2}{4}.
    \end{equation}
    Furthermore, we have equality exactly when $\mathbb P(X=m)=\mathbb P(X=M)=1/2$.

    \medskip

    \noindent

    $\blacktriangleright$ {\bf Proof of Popoviciu's inequality.} Since we have $\mathbb P(m\leq X\leq M)=1$, the random variable $(X-m)(M-X)$ is non-negative almost surely. It thus has a non-negative expectation value
    \begin{equation}
        \mathbb E\big[(X-m)(M-X)\big]\geq 0.
    \end{equation}
    Expanding the expectation value and subtracting $\mathbb E[X]^2$ on both sides gives
    \begin{subequations}
        \begin{align}
            &\mathbb E\big[(X-m)(M-X)\big]\geq 0,\\
            \Rightarrow& \mathbb E\big[MX-X^2-mM+mX\big]\geq 0,\\
            \Rightarrow& M\mathbb E(X)-\mathbb E(X^2)-mM+m\mathbb E(X)\geq 0,\\
            \Rightarrow& \mathbb E(X^2)-\mathbb E(X)^2\leq M\mathbb E(X)-mM+m\mathbb E(X)-\mathbb E(X)^2,\\
            \Rightarrow& \mathbb V(X)\leq (\mathbb E(X)-m)(M-\mathbb E(X)).
        \end{align}
    \end{subequations}
    Finally we can use the classic inequality $ab\leq \frac{(a+b)^2}{4}$ for $a=\mathbb E(X)-m$ and $b=M-\mathbb E(X)$ to get
    \begin{equation}
        \mathbb V(X)\leq \frac{(\mathbb E(X)-m+M-\mathbb E(X))^2}{4}=\frac{(M-m)^2}{4}.
    \end{equation}
    To understand the case of equality, we simply have to look at the two steps where we put an inequality. First, we said that the random variable $(X-m)(M-X)$ is non-negative; thus, its expectation value has to be non-negative. However, it is a well-known result that for a non-negative random variable $Y$ if $\mathbb E(Y)=0$ then $\mathbb P(Y=0)=1$. In our case, this means that
    \begin{equation}
        \mathbb P[(X-m)(M-X)=0]=1.
    \end{equation}
    As a product is zero if and only if one of the factors is zero, we know that almost surely either $X=m$ or $X=M$. Denoting $p=\mathbb P[X=m]$ we get that $\mathbb P[X=M]=1-p$.\\
    Now, we can consider the second inequality we used: $ab\leq (a+b)^2/4$. It is easy to see that we have equality only when $a=b$, which imposes
    \begin{equation}
        \mathbb E(X)-m=M-\mathbb E(X)\Rightarrow \mathbb E(X)=\frac{m+M}{2}.
    \end{equation}
    On the other hand, we can compute the expectation value since we know the exact distribution of $X$
    \begin{equation}
        \mathbb E(X)=m\mathbb P[X=m]+M\mathbb P[X=M]=pm+(1-p)M.
    \end{equation}
    Requiring both expressions to be the same yields $p=1/2$.
\end{derivation}

\begin{result}[Saturation condition]~\\
    \label{res: cauchy-schwarz general}
    For the collective operator
    \begin{equation}
        \hat H_\text{coll}=\sum_{j=1}^n c_j \hat H_j,
    \end{equation}
    with coefficients $c_j\in\R$ satisfying $\sum_{j=1}^n c_j^2=n$, the inequality
    \begin{equation}
        \Delta^2\hat H_\text{coll}\leq n^2\max_j\Delta^2 \hat H_j,
    \end{equation}
    can be saturated if and only if we have $c_j=\pm1$ for all $j$.
\end{result}

\begin{derivation}
    The idea is to look at the derivation of this inequality via the use of the Cauchy-Schwarz inequality which is done as follows
    \begin{subequations}
        \allowdisplaybreaks
        \begin{align}
            \Delta^2\hat H_\text{coll}&=\sum_{j,k=1}^n c_jc_k\Cov(\hat H_j,\hat H_k),\\
            &\leq\sum_{j,k=1}^n \abs{c_jc_k}\sqrt{\Delta^2\hat H_j\Delta^2 \hat H_k},\\
            &\leq\left(\sum_{j=1}^n\abs{c_j}\right)^2\max_k\Delta^2 \hat H_k,\\
            &\leq \left(\sum_{j=1}^n\abs{c_j}^2\right)\left(\sum_{j=1}^n1\right)\max_k\Delta^2 \hat H_k,\\
            &=n^2\max_j\Delta^2 \hat H_j,
        \end{align}
    \end{subequations}
    where the first inequality follows from the application of the Cauchy-Schwarz inequality to the quantum covariance. The second inequality is obtained by bounding $\Delta^2 \hat H_i$ by the maximal local variance, while the third inequality is again derived using the Cauchy-Schwarz inequality, this time applied to the standard inner product on $\R^n$. The second use of the Cauchy-Schwarz inequality is independent of the states considered. Therefore, if this inequality is strict, no state can saturate the inequality in Eq.~\eqref{eq: general bound collective global}. By analyzing the condition for equality in this step, we find that the inequality can only be saturated if $(\abs{c_1},\dots,\abs{c_n})\propto (1,\dots,1)$, which, given the normalization condition, implies that $c_i=\pm1$.  Sec.~\ref{subsec: equality case} then shows that in this situation, optimal states (if unphysical states are allowed) indeed exist.
\end{derivation}

\begin{result}[Convex roof construction]~\\
    \label{res: convex roof}
    Let's consider a convex set $\mathcal C$ and define its extreme points $\delta\mathcal{C}$ as
    \begin{equation}
        \delta\mathcal C=\left\{x\in\mathcal C\,\middle|\, \forall y,z\in\mathcal C,\,\forall t\in(0,1),\, x=ty+(1-t)z\Rightarrow x=y=z\right\}.
    \end{equation}
    Intuitively, this corresponds to the points that cannot be written as a non-trivial convex combination. In the context of quantum mechanics, $\mathcal C$ is the set of all mixed states, while $\delta\mathcal C$ is the set of the pure states.  One can also note that {\it a priori} $\delta\mathcal C\neq \partial \mathcal C$ where $\partial \mathcal C$ is the topological boundary of $\mathcal C$. We further assume that $\delta\mathcal C$ generates the set $\mathcal C$
    \begin{equation}
        \forall x\in\mathcal C,\, \exists x_1,\dots,x_n\in\delta\mathcal C\!:\, \exists p_1,\dots,p_n\geq0\!:\, \sum_{j=1}^n p_j=1\text{ and } x=\sum_{j=1}^n p_j x_j,
    \end{equation}
    meaning that each point of $\mathcal C$ can be written as a convex combination of extreme points. If $f: \delta\mathcal C\to \R$ is an arbitrary real function defined on the extreme points of $\mathcal C$, one can define 
    \begin{align}
        F: \mathcal C &\to \R\\
        x&\mapsto \inf_{\{p_j|x_j\}}  \sum_j p_jf(x_j),
    \end{align}
    where the infimum is taken over all convex decompositions of $x$: $x=\sum p_j x_j$ with $p_j\geq 0$. One can note that the hypothesis that $\mathcal C$ is generated by $\delta\mathcal C$ is not strictly necessary. With the convention that $\inf \emptyset=+\infty$, we still get a valid formula for $F$ that may take the value $+\infty$. Such a construction satisfies the following properties
    \begin{enumerate}
        \item For $x\in\delta\mathcal C$, $F(x)=f(x)$. On the extreme points, $F$ reduces to $f$.
        \item $F$ is convex.
        \item $F$ is the largest convex function that is equal to $f$ on $\delta\mathcal C$. This means that if $G$ is a convex function such that $\forall x\in\delta\mathcal C,\, G(x)=f(x)$ then $\forall x\in\mathcal C,\, G(x)\leq F(x)$ 
    \end{enumerate}
\end{result}

\begin{derivation}
    \begin{enumerate}
        \item Due to the definition of $\delta\mathcal C$, for any $x\in\delta\mathcal C$, there exists only one convex decomposition: the trivial one $x=x$. As such it follows that $F(x)=f(x)$.
        
        \item $F$ is convex since if we take $x,y\in\mathcal C$ and $t\in[0,1]$ then for any convex decomposition of $x=\sum_j p_j x_j$ and $y=\sum_k q_k y_k$
        \begin{align}
            t \sum_j p_j f(x_j) + (1-t)\sum_k q_k f(y_k)&=  \sum_j tp_j f(x_j) + \sum_k (1-t)q_k f(y_k)\notag\\
            &\geq F( tx+(1-t)y),
        \end{align}
        since $tx+(1-t)y=\sum_j tp_j x_j+\sum_k (1-t)q_k y_k$ is a valid decomposition. By taking the infimum over all decompositions of $x$ and $y$, we indeed have
        \begin{equation}
            tF(x)+(1-t)F(y)\geq F(tx+(1-t)y),
        \end{equation}
        Thus $F$ is indeed convex.

        \item Finally if $G$ is a convex function that equals $f$ on $\delta\mathcal C$, for any decomposition of $x\in\mathcal C$, $x=\sum_j p_j x_j$, with $x_j\in\delta\mathcal C$, we have
        \begin{equation}
            G(x)\leq \sum_j p_j G(x_j)= \sum_j p_j f(x_j).
        \end{equation}
        The inequality comes from the convexity of $G$, while the equality comes from the fact that $G$ equals $f$ on $\delta\mathcal C$. By optimizing all possible such decompositions, we indeed get 
        \begin{equation}
            G(x) \leq \inf_{\{p_j|x_j\}} \sum_j p_j f(x_j)=F(x).
        \end{equation}     
    \end{enumerate}
\end{derivation}

\begin{result}[Equivalent definition of the support]~\\
    \label{res: def support}
    For a finite-rank density matrix $\hat\rho$, the support of $\hat\rho$ is defined by the two equivalent definitions: $\ket{\psi}\in\operatorname{Supp}(\hat\rho)$ if and only if
    \begin{itemize}
    \item $\ket{\psi}\in(\ker\hat\rho)^\perp$,
    \item if and only if $\ket{\psi}$ appears in a convex decomposition of $\hat\rho$ with non-zero weight.
\end{itemize}
    
\end{result}

\begin{derivation}
    \underline{$\Rightarrow$:} If $\ket{\psi}\in(\ker \hat\rho)^\perp$ then by definition for all states $\ket{\sigma}\in\ker\hat\rho$, we have $\bra{\psi}\ket{\sigma}=0$. Let us now consider a diagonalization of $\hat \rho$
    \begin{equation}
        \hat \rho=\sum_j \lambda_j \ketbra{\psi_j},
    \end{equation}
    where $\lambda_j>0$ and the sum is finite since the rank of $\hat\rho$ is finite. We can then define $\lambda=\min_j \lambda_j$. Since there is only a finite number of non-zero eigenvalues, we know that $\lambda>0$. Then we claim that $\hat\rho-\lambda\ketbra{\psi}$ is a positive operator. Indeed, for any state $\ket{\phi}$ we verify that $\bra{\phi}(\hat\rho-\lambda\ketbra{\psi})\ket{\phi}\geq0$. To show this, we decompose $\ket{\phi}=\ket{\sigma}+\ket{\tau}$ where $\ket{\sigma}\in\ker\hat\rho$ and $\ket{\tau}\in(\ker\hat\rho)^\perp$. We get
    \begin{subequations}
        \begin{align}
            \bra{\phi}(\hat\rho-\lambda\ketbra{\psi})\ket{\phi}&=\bra{\sigma}\hat\rho\ket{\sigma}-\lambda\bra{\sigma}\ket{\psi}\bra{\psi}\ket{\sigma}+\bra{\sigma}\hat\rho\ket{\tau}-\lambda\bra{\sigma}\ket{\psi}\bra{\psi}\ket{\tau}\notag\\
            &\qquad+\bra{\tau}\hat\rho\ket{\sigma}-\lambda\bra{\tau}\ket{\psi}\bra{\psi}\ket{\sigma}+\bra{\tau}\hat\rho\ket{\tau}-\lambda\bra{\tau}\ket{\psi}\bra{\psi}\ket{\tau},\\
            &=\bra{\tau}\hat\rho\ket{\tau}-\lambda\bra{\tau}\ket{\psi}\bra{\psi}\ket{\tau},
        \end{align}
    \end{subequations}
    since $\ket{\sigma}\in\ker\hat\rho$ thus $\hat\rho\ket{\sigma}=0$ and $\bra{\psi}\ket{\sigma}=0$. Since $\ket{\tau}\in(\ker\hat\rho)^\perp$ we can decompose it over the states $\ket{\psi_j}$: $\ket{\tau}=\sum \tau_j\ket{\psi_j}$. Thus
    \begin{equation}
        \bra{\tau}\hat\rho\ket{\tau}=\sum_j \abs{\tau_j}^2\lambda_j\geq \lambda\sum_j \abs{\tau_j}^2=\lambda\bra{\tau}\ket{\tau}.
    \end{equation}
    Finally, by Cauchy-Schwarz inequality, we have $\bra{\tau}\ket{\psi}\bra{\psi}\ket{\tau}\leq \braket{\tau}$. In the end, we indeed have
    \begin{equation}
        \bra{\phi}(\hat\rho-\lambda\ketbra{\psi})\ket{\phi}\geq0.
    \end{equation}
    To finish the proof, we simply have to say that since $\hat\rho-\lambda\ketbra{\psi}$ is a positive operator, we can diagonalize it and get $\hat\rho-\lambda\ketbra{\psi}=\sum_j\alpha_j\ketbra{\psi_j'}$ and we obtain
    \begin{equation}
        \hat\rho =\lambda\ketbra{\psi}+\sum_j\alpha_j\ketbra{\psi_j'}.
    \end{equation}

    \underline{$\Leftarrow$:} We now assume that we can write $\hat\rho$ as a convex decomposition containing $\ket{\psi}$
    \begin{equation}
        \hat\rho=\alpha\ketbra{\psi}+\sum_j\alpha_j\ketbra{\psi_j}.
    \end{equation}
    If we consider $\ket{\phi}\in\ker\hat\rho$ we have 
    \begin{equation}
        \bra{\phi}\hat\rho\ket{\phi}=\alpha\abs{\bra{\psi}\ket{\phi}}^2+\sum\alpha_j \abs{\bra{\psi_j}\ket{\phi}}^2=0.
    \end{equation}
    Since we are only summing non-negative terms, this means that all the terms in the sum are zero, specifically we have $\bra{\psi}\ket{\phi}=0$. Since this is true for all states $\ket{\phi}$ in $\ker\hat\rho$, this means that $\ket{\psi}\in(\ker\hat\rho)^\perp$.
\end{derivation}

\begin{result}[Convexity of $I^{\text{S}}$]~\\
    \label{res: formula support conv}
    Defining $I^{\text{S}}$ by 
    \begin{equation}
        I^\text{S}_{\hat \rho}=\frac{\mathcal Q_{\hat \rho}(\hat H_\text{coll})}{4\sup\limits_{\ket{\psi}\in\operatorname{Supp}(\hat\rho)} \max_i \Delta^2_{\ket{\psi}}\hat H_i},
    \end{equation}
    $I^\text{S}$ is convex in the quantum state $\hat\rho$.
\end{result}

\begin{derivation}
    Let us consider two mixed states $\hat \rho_1$ and $\hat\rho_2$ and a convex combination $\hat\rho=t\hat\rho_1+(1-t)\hat\rho_2$ with $0<t<1$ (the endpoint cases are equalities). Since the quantum Fisher information is convex, we have
    \begin{equation}
        \mathcal Q_{\hat\rho}(\hat H_\text{coll})\leq t\mathcal Q_{\hat\rho_1}(\hat H_\text{coll})+(1-t)\mathcal Q_{\hat\rho_2}(\hat H_\text{coll}).
    \end{equation}
    Moreover, due to the definition of the support in terms of convex decomposition, it is clear that 
    \begin{equation}
        \operatorname{Supp}(\hat\rho_j)\subset \operatorname{Supp}(\hat\rho),\qquad\text{(for $j=1,2$)}.
    \end{equation}
    Indeed, any decomposition of $\hat\rho_1$ and $\hat\rho_2$ induces a decomposition of $\hat \rho$. So if $\ket{\psi}$ is part of a decomposition of $\hat\rho_1$ or $\hat\rho_2$ then it is immediately part of a decomposition of $\hat\rho$. Taking the supremum, it follows that 
    \begin{equation}
        \sup\limits_{\ket{\psi}\in\operatorname{Supp}(\hat\rho)} \max_k \Delta^2_{\ket{\psi}} H_k\geq \sup\limits_{\ket{\psi}\in\operatorname{Supp}(\hat\rho_j)} \max_k \Delta^2_{\ket{\psi}} H_k,
    \end{equation}
    for $j=1,2$. With all of this, we can finally write
    \begin{subequations}
        \begin{align}
            I^\text{S}_{\hat\rho}&=\frac{\mathcal Q_{\hat \rho}(\hat H_\text{coll})}{4\sup\limits_{\ket{\psi}\in\operatorname{Supp}(\hat\rho)} \max_j \Delta^2_{\ket{\psi}}\hat H_j},\\
            &\leq \frac{t\mathcal Q_{\hat \rho_1}(\hat H_\text{coll})+(1-t)\mathcal Q_{\hat \rho_2}(\hat H_\text{coll})}{4\sup\limits_{\ket{\psi}\in\operatorname{Supp}(\hat\rho)} \max_j \Delta^2_{\ket{\psi}}\hat H_j},\\
            &\leq t\frac{\mathcal Q_{\hat \rho_1}(\hat H_\text{coll})}{4\sup\limits_{\ket{\psi}\in\operatorname{Supp}(\hat\rho_1)} \max_j \Delta^2_{\ket{\psi}}\hat H_j}+(1-t)\frac{\mathcal Q_{\hat \rho_2}(\hat H_\text{coll})}{4\sup\limits_{\ket{\psi}\in\operatorname{Supp}(\hat\rho_2)} \max_j \Delta^2_{\ket{\psi}}\hat H_j},\\
            &=tI^\text{S}_{\hat\rho_1}+(1-t)I^\text{S}_{\hat\rho_2}.
        \end{align}
    \end{subequations}    
\end{derivation}

\begin{result}[Example of computation of extensions of $I$]~\\
    \label{res: mixed computation}
    For qubits, with $\mathcal{H} = \mathbb{C}^2$ and the collective Hamiltonian $\hat{H}_\text{coll} = \hat{Z}_1 + \cdots + \hat{Z}_n$ we define the following states
    \begin{align}
        \ket{\psi}&=\frac{1}{\sqrt{2}}(\ket{0\cdots0}+\ket{1\cdots1}),\notag\\
        \ket{\varphi}=(\1^{\otimes \frac{n}{2}}\otimes\hat X^{\otimes \frac{n}{2}})\ket{\psi}&=\frac{1}{\sqrt{2}}(\ket{0\cdots01\cdots1}+\ket{1\cdots10\cdots0}),
    \end{align}
    where $\ket{\varphi}$ is defined only if $n$ is even. Defining for $\epsilon\in[0,1]$
    \begin{align}
        \hat\rho_\epsilon=(1-\epsilon)\ketbra{\psi}+\epsilon\frac{\1}{2^n}, && \hat\sigma_\epsilon=(1-\epsilon)\ketbra{\psi}+\epsilon\ketbra{\varphi},
    \end{align}
    we have
    \begin{align}
        I_{\hat\rho_\epsilon}^\text{S}=I_{\hat\rho_\epsilon}^\text{R}=\frac{(1-\epsilon)^2n^2}{(1-\epsilon)+\epsilon/2^{n-1}}, && I_{\hat\sigma_\epsilon}^\text{CR}=I_{\hat\sigma_\epsilon}^\text{S}=I_{\hat\sigma_\epsilon}^\text{R}=(1-\epsilon)n^2.
    \end{align}
\end{result}

\begin{derivation}
     To perform the computations, we introduce a basis for the Hilbert space $(\mathbb{C}^2)^{\otimes n}$. For a binary string $s = s_1\cdots s_n$ with $s_j \in {0,1}$, we define
    \begin{equation}
        \ket{\psi_s^\pm}=\frac{1}{\sqrt{2}}(\ket{s}\pm\ket{\overline{s}}),
    \end{equation}
    where $\overline{s}$ is the string obtained by flipping each bit of $s$: $\overline{s}_k = 1 - s_k$. Thus, we can rewrite the states as $\ket{\psi} = \ket{\psi_0^+}$ and $\ket{\varphi} = \ket{\psi_c^+}$, where $0$ represents the string of all zeros, and $c = 0\cdots01\cdots1$ is the string consisting of $n/2$ zeros followed by $n/2$ ones. Since $\ket{\psi_s^\pm} = \pm \ket{\psi_{\overline{s}}^\pm}$, we restrict our attention to strings that begin with a $0$. We define the set $\mathcal{S}$ as follows: $\mathcal S=\{0s_2\cdots s_n|s_k=0,1\}$. For $s,s'\in\mathcal S$ and $j,k=\pm1$ we can verify the following orthogonality relation
    \begin{equation}
        \braket{\psi_s^j}{\psi_{s'}^k}=\delta_{s,s'}\delta_{j,k}.
    \end{equation}
    Since $\abs{\mathcal S\times \{1,-1\}}=2^{n}$, the states $\ket{\psi_s^\pm}$ then form a basis of $(\C^2)^{\otimes n}$. The quantum Fisher information appears in the definitions of $I^\text{S}$ and $I^\text{R}$, so we first compute $\mathcal Q_{\hat\rho_\epsilon}(\hat H_\text{coll})$ and $\mathcal Q_{\hat\sigma_\epsilon}(\hat H_\text{coll})$. We recall the expression of the quantum Fisher information for mixed states
    \begin{equation}
        \mathcal Q_{\hat \rho}(\hat A)=2\sum_{\lambda_j+\lambda_k>0}\frac{(\lambda_j-\lambda_k)^2}{\lambda_j+\lambda_k}\abs{\bra{j}\hat A \ket{k}}^2,
    \end{equation}
    where the density matrix has been diagonalized as $\hat \rho=\sum_j \lambda_j\ketbra{j}$, and the sum runs over all pairs of indices $(j,k)$ satisfying $\lambda_j+\lambda_k>0$. Additionally, we observe that
    \begin{equation}
        \bra{\psi_s^j}\hat Z_l\ket{\psi_{s'}^k}=(-1)^{s'_l}\braket{\psi_s^j}{\psi_{s'}^{-k}}=(-1)^{s_l}\delta_{s,s'}\delta_{j,-k},
    \end{equation}
    which leads to $\bra{\psi_s^j}\hat H_\text{coll}\ket{\psi_{s'}^k}=\delta_{s,s'}\delta_{j,-k}\sum_{l=1}^n (-1)^{s_l}$. With this, we can now compute $\mathcal Q_{\hat \rho_\epsilon}(\hat H_\text{coll})$ given the eigendecomposition of $\hat \rho_\epsilon$. Using the closure relation $\1=\sum_{s\in\mathcal S,j=\pm} \ketbra{\psi_s^j}$, we have
    \begin{equation}
        \hat \rho_\epsilon=\left[(1-\epsilon)+\frac{\epsilon}{2^n}\right]\ketbra{\psi_0^+}+\frac{\epsilon}{2^n}\ketbra{\psi_0^-}+\frac{\epsilon}{2^n}\sum_{s\in \mathcal S\setminus\{0\}, j=\pm}\ketbra{\psi_s^j}.
    \end{equation}
    Since $\hat \rho_\epsilon$ has only two distinct eigenvalues, and only pairs of eigenvectors with different eigenvalues contribute to the quantum Fisher information, we can proceed with the calculation as follows
    \begin{subequations}
        \begin{align}
            \mathcal Q_{\hat \rho_\epsilon}(\hat H_\text{coll})&=4\sum_{(s,j)\in\mathcal S\times\{\pm\}\setminus\{(0\cdots0,+)\}}\frac{[(1-\epsilon)+\epsilon/2^n-\epsilon/2^n]^2}{(1-\epsilon)+\epsilon/2^n+\epsilon/2^n}\abs{\bra{\psi_0^+}\hat H_\text{coll}\ket{\psi_s^j}}^2,\\
            &=4\sum_{(s,j)\in\mathcal S\times\{\pm\}\setminus\{(0\cdots0,+)\}}\frac{(1-\epsilon)^2}{(1-\epsilon)+\epsilon/2^{n-1}}\left(\delta_{0,s}\delta_{+,-j}\sum_{k=1}^n(-1)^{s_k}\right)^2,\\
            &=4\frac{(1-\epsilon)^2}{(1-\epsilon)+\epsilon/2^{n-1}}n^2.
        \end{align}
    \end{subequations}
    We follow the same procedure to compute $\mathcal Q_{\hat \sigma_\epsilon}(\hat H_\text{coll})$. First, we diagonalize $\hat\sigma_\epsilon$, explicitly identifying the eigenvectors corresponding to the eigenvalue 0.
    \begin{equation}
        \hat \sigma_\epsilon=(1-\epsilon)\ketbra{\psi_0^+}+\epsilon\ketbra{\psi_c^+}+ 0\sum_{(s,j)\in\mathcal S\times\{\pm\}\setminus\{(0,+),(c,+)\}}\ketbra{\psi_s^j}.
    \end{equation}
    In this case, there are three distinct eigenvalues. We then have
    \begin{subequations}
        \begin{align}
            \mathcal Q_{\hat \sigma_\epsilon}(\hat H_\text{coll})&=4\frac{(1-2\epsilon)^2}{1}\abs{\underbrace {\bra{\psi_0^+}\hat H_\text{coll}\ket{\psi_c^+}}_{0}}^2\notag\\
            &\qquad+4(1-\epsilon)\sum_{(s,j)\in\mathcal S\times\{\pm\}\setminus\{(0,+),(c,+)\}} \abs{\underbrace{\bra{\psi_0^+}\hat H_\text{coll}\ket{\psi_s^j}}_{n\delta_{s,0}\delta_{j,-}}}^2\notag\\
            &\qquad+4\epsilon\sum_{(s,j)\in\mathcal S\times\{\pm\}\setminus\{(0,+),(c,+)\}}\abs{\underbrace{\bra{\psi_c^+}\hat H_\text{coll}\ket{\psi_s^j}}_{0}}^2,\\
            &=4(1-\epsilon)n^2.
        \end{align}
    \end{subequations}
    Since we are working with qubits, the local variance of any pure state $\Delta^2_{\ket{\phi}}\hat Z_k\leq 1$, which allows us to set $A=1$ in the definition of $I^\text{R}$. This leads to the following expressions
    \begin{align}
        I_{\hat\rho_\epsilon}^\text{R}=\frac{(1-\epsilon)^2}{(1-\epsilon)+\epsilon/2^{n-1}}n^2, && I_{\hat\sigma_\epsilon}^\text{R}=(1-\epsilon)n^2.
    \end{align}
    For similar reasons, we have $\sup\limits_{\ket{\psi}\in\operatorname{Supp}(\hat\tau)} \max_j \Delta^2_{\ket{\psi}}\hat Z_j=1$ for $\hat \tau=\hat\rho_\epsilon,\hat \sigma_\epsilon$, as $\ket{\psi_0^+}$ (or $\ket{\psi_c^+}$ for $\hat\tau=\hat\sigma_1$) is a state in the support with maximal variance, so
    \begin{align}
        I_{\hat\rho_\epsilon}^\text{S}=\frac{(1-\epsilon)^2}{(1-\epsilon)+\epsilon/2^{n-1}}n^2, && I_{\hat\sigma_\epsilon}^\text{S}=(1-\epsilon)n^2.
    \end{align}

    Finally, we demonstrate that $I^\text{CR}_{\hat\sigma_\epsilon}=(1-\epsilon)n^2$. By definition, we have
    \begin{equation}
        I^\text{CR}_{\hat\sigma_\epsilon}=\inf_{\{p_j,\ket{\psi_j}\}} \sum_j p_j \frac{\Delta^2_{\ket{\psi_j}}(\hat H_\text{coll})}{\max_k\Delta^2_{\ket{\psi_j}}(\hat Z_k)}
    \end{equation}
    where the optimization is done over all decompositions of the form $\hat \sigma_\epsilon=\sum_i p_i\ketbra{\psi_i}$. As shown in Appendix~\ref{app: collective var derivation}, Result~\ref{res: def support}, the states appearing in the decomposition must belong to $(\ker\hat \sigma_\epsilon)^\perp\subseteq\operatorname{Vect}(\ket{\psi_0^+},\ket{\psi_c^+})$. Since we have
    \begin{align}
        \hat Z_k\ket{\psi_0^+}=\ket{\psi_0^-}, && \hat Z_k\ket{\psi_c^+}=\left\{\begin{array}{cc} \ket{\psi_c^-}, & \text{if $k\leq n/2$},\\ -\ket{\psi_c^-}, & \text{if $k>n/2$},  \end{array}\right.
    \end{align}
    we can compute the variance of $\hat Z_k$ on any state $\ket{\phi}=\alpha\ket{\psi_0^+}+\beta\ket{\psi_c^+}$. We have
    \begin{subequations}
        \begin{align}
            \bra{\phi}\hat Z_k\ket{\phi}&=\abs{\alpha}^2\bra{\psi_0^+}\hat Z_k\ket{\psi_0^+}+\abs{\beta}^2\bra{\psi_c^+}\hat Z_k\ket{\psi_c^+}+\alpha^*\beta\bra{\psi_0^+}\hat Z_k\ket{\psi_c^+}\notag\\
            &\qquad+\alpha\beta^*\bra{\psi_c^+}\hat Z_k\ket{\psi_0^+},\\
            &=\abs{\alpha}^2\bra{\psi_0^+}\ket{\psi_0^-}\pm\abs{\beta}^2\bra{\psi_c^+}\ket{\psi_c^-}\pm\alpha^*\beta\bra{\psi_0^+}\ket{\psi_c^-}+\alpha\beta^*\bra{\psi_c^+}\ket{\psi_0^-},\\
            &=0.
        \end{align} 
    \end{subequations}
    And
    \begin{subequations}
        \begin{align}
            \bra{\phi}\hat Z_k^2\ket{\phi}&=\abs{\alpha}^2\bra{\psi_0^+}\hat Z_k^2\ket{\psi_0^+}+\abs{\beta}^2\bra{\psi_c^+}\hat Z_k^2\ket{\psi_c^+}+\alpha^*\beta\bra{\psi_0^+}\hat Z_k^2\ket{\psi_c^+}\notag\\
            &\qquad+\alpha\beta^*\bra{\psi_c^+}\hat Z_k^2\ket{\psi_0^+},\\
            &=\abs{\alpha}^2\bra{\psi_0^+}\ket{\psi_0^+}+\abs{\beta}^2\bra{\psi_c^+}\ket{\psi_c^+}+\alpha^*\beta\bra{\psi_0^+}\ket{\psi_c^+}+\alpha\beta^*\bra{\psi_c^+}\ket{\psi_0^+},\\
            &=\abs{\alpha}^2+\abs{\beta}^2=1.
        \end{align} 
    \end{subequations}
    So $\Delta^2_{\ket{\phi}}\hat Z_k=\bra{\phi}\hat Z_k^2\ket{\phi}-\bra{\phi}\hat Z_k\ket{\phi}^2=1$, which means that all pure states in the support of $\hat \sigma_\epsilon$ have the same local variance $\Delta^2\hat Z_k$. From this, it follows that 
    \begin{equation}
        I^\text{CR}_{\hat\sigma_\epsilon}=\inf_{\{p_j,\ket{\psi_j}\}} \sum_j p_j \frac{\Delta^2_{\ket{\psi_j}}(\hat H_\text{coll})}{1}=\inf_{\{p_j,\ket{\psi_j}\}} \sum_j p_j \Delta^2_{\ket{\psi_j}}(\hat H_\text{coll})=\frac{1}{4}\mathcal Q_{\hat\sigma_\epsilon}(\hat H_\text{coll})
    \end{equation}
    where in the final step we use the fact that the quantum Fisher information is four times the convex roof of the variance~\cite{yu_quantum_2013} (see also Appendix~\ref{app: formalism and framework}, Result~\ref{res: convex roof QFI}). Using the previously established results allows us to conclude the proof.
\end{derivation}

\begin{result}[General $k$-entanglement inequality]~\\
    \label{res: k sep}
    For a Hilbert space  $\mathcal H$ and two maps $I$ and $g$ defined on 
    \begin{align}
            I&:\bigcup_{n=1}^{+\infty} \mathcal D(\mathcal H^{\otimes n}) \to \R, &
            g&:\R\to \R,
    \end{align}
    where $\mathcal D(\cdot)$ denotes the set of mixed states of the corresponding Hilbert space, we assume that
    \begin{itemize}
        \item For every integer $n$, $I$ is convex on $\mathcal D(\mathcal H^{\otimes n})$.
        \item $I$ is sub-additive on pure product states: $I(\ketbra{\psi}\otimes\ketbra{\phi})\leq I(\ketbra{\psi})+I(\ketbra{\phi})$.
        \item The function $g$ is convex and $g(0)=0$.
        \item For every integer $n$, there exists a subset $\Gamma_n\subset\mathcal D(\mathcal H^{\otimes n})$ on which $I(\hat \rho)\leq g(n)$.
    \end{itemize}
    If $\hat \rho\in \mathcal D(\mathcal H^{\otimes n})$ is $k$-separable over $\Gamma$ then
    \begin{equation}
        I(\hat \rho)\leq \left\lfloor \frac{n}{k}\right\rfloor g(k)+g\left(n-\left\lfloor \frac{n}{k}\right\rfloor k\right)\leq \frac{n}{k}g(k).
    \end{equation}
\end{result}

\begin{derivation}
    First, the convexity hypothesis on $I$ allows us to consider only pure states, as the extension to mixed states is immediate. If a pure state $\ket{\psi}$ can be decomposed as
    \begin{equation}
        \ket{\psi}=\ket{\phi_1}\otimes \cdots\otimes \ket{\phi_l}
    \end{equation}
    where $\ket{\phi_j}$ is a pure state in $\mathcal H^{\otimes k_j}$ with $\ketbra{\phi_j}\in\Gamma_{k_j}$, the sub-additivity of $I$ shows that
    \begin{equation}
        I(\ketbra{\psi})\leq I(\ketbra{\phi_1})+\cdots+I(\ketbra{\phi_l})\leq g(k_1)+\cdots +g(k_l)
    \end{equation}
    We then need to show that for any finite family $k_1,\dots,k_r$ of positive real numbers, such that $k_j\leq k$ for all $j$ and $k_1+\cdots+k_r=n$, we have 
    \begin{equation}
        g(k_1)+\cdots+g(k_r)\leq \left\lfloor \frac{n}{k}\right\rfloor g(k)+g\left(n-\left\lfloor \frac{n}{k}\right\rfloor k\right)\leq \frac{n}{k}g(k)
    \end{equation}

    For the first inequality, we reason by induction on $r\in\N^*$. For $r=1$, if $k_1\leq k$ is such that $k_1=n$, then $k\geq n$, and the inequality is simply verified, provided $g(0)=0$. For $r>1$, we consider the set
    \begin{equation}
        C^n_r=\Big\{(x_1,\dots,x_r)\Big|x_1+\cdots +x_r=n\text{ and } 0\leq x_i\leq k\Big\}
    \end{equation}
    which is a compact convex set, and the function $g_r$ defined on $C^n_r$ by $g_r(x_1,\dots,x_r)=g(x_1)+\cdots+g(x_r)$. Since $g$ is convex, it follows that $g_r$ is also convex. By the Bauer principle (a convex function on a compact convex set attains a maximum on one of its extreme points), we know that $g_r$ attains a maximum on an extreme point of $C_r^n$. Extreme points of $C^n_r$ are given by points $(k_1,\dots,k_r)$ where at least one $k_j$ is equal to $0$ or $k$.
    \begin{itemize}
        \item If $k_j=0$ then the maximum of $g_r$ on $C^n_r$ is the same as the one of $g_{r-1}$ on $C^n_{r-1}$, where we remove the $j$-th coordinate to lower the dimension.
        \item If $k_j=k$ then the maximum of $g_r$ on $C^n_r$ is $\max\limits_{C^{n-k}_{r-1}} g_{r-1}+g(k)$.
    \end{itemize}
    In both cases, we conclude by induction.

    For the second inequality, the case $n-\lfloor n/k\rfloor k=0$ gives equality since $g(0)=0$. Otherwise, we notice that it is satisfied if and only if
    \begin{equation}
        \frac{g(k)}{k}\geq \frac{g\left(n-\left\lfloor \frac{n}{k}\right\rfloor k\right)}{n-\left\lfloor \frac{n}{k}\right\rfloor k}
    \end{equation}
    Recall a useful property of convex functions, usually called the three-secant inequality: if $x<y<z$ then
    \begin{equation}
        \frac{g(y)-g(x)}{y-x}\leq \frac{g(z)-g(x)}{z-x}\leq \frac{g(z)-g(y)}{z-y}
    \end{equation}
    Using the left inequality for $x=0$, (with $g(0)=0$), $y=n-\left\lfloor \frac{n}{k}\right\rfloor k$ and $z=k$, directly yields the result.
\end{derivation}

\begin{result}[Collective GKP displacement]~\\
    \label{res: GKP global}
    Defining the logical states for $k=0,1$
    \begin{equation}
        \ket{\overline k}=\sum_{s\in\Z}\int\dd\Omega_1\cdots\dd\Omega_n\, \delta(\Omega_1-(2s+k)\Omega_0)\prod_{j=2}^n G_j(\Omega_j)\ket{\omega_1,\cdots,\omega_n},
    \end{equation}
    one can observe that
    \begin{subequations}
        \begin{align}
            e^{-i\varphi\hat T_1}\ket{\overline k}&=\sum_{s\in\Z}\int\dd\Omega_1\cdots\dd\Omega_n\, \delta(\Omega_1-\varphi-(2s+k)\Omega_0)\prod_{j=2}^n G_j(\Omega_j)\ket{\omega_1,\cdots,\omega_n},\\
            e^{-i\tau\hat \Omega_1}\ket{\overline k}&=\sum_{s\in\Z}\int\dd\Omega_1\cdots\dd\Omega_n\, e^{-i\tau(2s+k)\Omega_0}\delta(\Omega_1-(2s+k)\Omega_0)\prod_{j=2}^n G_j(\Omega_j)\ket{\omega_1,\cdots,\omega_n},
        \end{align}
    \end{subequations}
    which indeed reveals that collective transformations generated by $\hat \Omega_1$ and $\hat T_1$ act as displacements. Setting $\varphi=\Omega_0$ and $\tau=\pi/\Omega_0$ yields the logical transformations $\hat X$ and $\hat Z$. Similarly, setting $\varphi=2\Omega_0$ and $\tau=2\pi/\Omega_0$ provides stabilizers.
\end{result}

\begin{derivation}
    We simply need to perform each computation one by one.
    \begin{subequations}
        \allowdisplaybreaks
        \begin{align}
            e^{-i\varphi\hat T_1}&\ket{\overline k}\notag\\
            &=e^{-i\varphi\hat T_1}\sum_{s\in\Z}\int\dd\Omega_1\cdots\dd\Omega_n \,\delta(\Omega_1-(2s+k)\Omega_0)\prod_{j=2}^n G_j(\Omega_j)\ket{\omega_1,\cdots,\omega_n},\\
            &=\sum_{s\in\Z}\int\dd\Omega_1\cdots\dd\Omega_n\, \delta(\Omega_1-(2s+k)\Omega_0)\prod_{j=2}^n G_j(\Omega_j)\ket{\omega_1+\tfrac{\varphi}{\sqrt{n}},\cdots,\omega_n+\tfrac{\varphi}{\sqrt{n}}},\\
            &=\sum_{s\in\Z}\int\dd\Omega_1\cdots\dd\Omega_n\, \delta(\Omega_1-\varphi-(2s+k)\Omega_0)\prod_{j=2}^n G_j(\Omega_j)\ket{\omega_1,\cdots,\omega_n},
        \end{align}
    \end{subequations}
    by performing the change of variable $\Omega_1\to \Omega_1-\varphi$, which transforms the local frequencies as $\omega_j\to\omega_j-\tfrac{\varphi}{\sqrt{n}}$. Then

    \begin{subequations}
        \begin{align}
            e^{-i\tau\hat \Omega_1}&\ket{\overline k}\notag\\
            &=e^{-i\tau\hat \Omega_1}\sum_{s\in\Z}\int\dd\Omega_1\cdots\dd\Omega_n\, \delta(\Omega_1-(2s+k)\Omega_0)\prod_{j=2}^n G_j(\Omega_j)\ket{\omega_1,\cdots,\omega_n},\\
            &=\sum_{s\in\Z}\int\dd\Omega_1\cdots\dd\Omega_n\,e^{-i\tau\Omega_1}\delta(\Omega_1-(2s+k)\Omega_0)\prod_{j=2}^n G_j(\Omega_j)\ket{\omega_1,\cdots,\omega_n},\\
            &=\sum_{s\in\Z}\int\dd\Omega_1\cdots\dd\Omega_n\, e^{-i\tau(2s+k)\Omega_0}\delta(\Omega_1-(2s+k)\Omega_0)\prod_{j=2}^n G_j(\Omega_j)\ket{\omega_1,\cdots,\omega_n}.
        \end{align}
    \end{subequations}
\end{derivation}

\begin{result}[Local GKP displacement]~\\
    \label{res: GKP local}
    Defining the logical states for $k=0,1$
    \begin{equation}
        \ket{\overline k}=\sum_{s\in\Z}\int\dd\Omega_1\cdots\dd\Omega_n\, \delta(\Omega_1-(2s+k)\Omega_0)\prod_{j=2}^n G_j(\Omega_j)\ket{\omega_1,\cdots,\omega_n},
    \end{equation}
    one can observe that
    \begin{subequations}
        \begin{align}
            e^{-i\varphi\hat t_l}&\ket{\overline k}=\sum_{s\in\Z}\int\dd\Omega_1\cdots\dd\Omega_n\, \delta(\Omega_1-\tfrac{\varphi}{\sqrt{n}}-(2s+k)\Omega_0)\prod_{j=2}^n \tilde G_j(\Omega_j)\ket{\omega_1,\cdots,\omega_n},\\
            e^{-i\tau\hat \omega_l}&\ket{\overline k}=\sum_{s\in\Z}\int\dd\Omega_1\cdots\dd\Omega_n\, e^{-i\tau(2s+k)\Omega_0/\sqrt{n}}\delta(\Omega_1-(2s+k)\Omega_0)\prod_{j=2}^n \tilde G_j(\Omega_j)\ket{\omega_1,\cdots,\omega_n},
        \end{align}
    \end{subequations}
    where on this first line the modified function is defined by $\tilde G_j(\Omega_j)=G_j(\Omega_j-\alpha_{j,l}\varphi)$ while on the second line $\tilde G_j(\Omega_j)=e^{-i\alpha_{j,l}\tau\Omega_j}G_j(\Omega_j)$. Disregarding the information encoded in the collective variables for $j>1$, this shows that local transformations generated by $\hat \omega_l$ and $\hat t_l$ also act as displacements on the logical states, but with an amplitude reduced by $1/\sqrt{n}$. Setting $\varphi=\sqrt{n}\Omega_0$ and $\tau=\sqrt{n}\pi/\Omega_0$ yields again logical transformations $\hat X$ and $\hat Z$. Similarly, setting $\varphi=2\sqrt{n}\Omega_0$ and $\tau=2\sqrt{n}\pi/\Omega_0$ provides stabilizers.
\end{result}

\begin{derivation}
    We simply need to perform each computation one by one. We then have for $l=1,\dots,n$
    \begin{subequations}
        \begin{align}
            e^{-i\varphi\hat t_l}&\ket{\overline k}\notag\\
            &=\sum_{s\in\Z}\int\dd\Omega_1\cdots\dd\Omega_n\, \delta(\Omega_1-(2s+k)\Omega_0)\prod_{j=2}^n G_j(\Omega_j)\ket{\omega_1,\cdots,\omega_l+\varphi,\cdots,\omega_n},\\
            &=\sum_{s\in\Z}\int\dd\Omega_1\cdots\dd\Omega_n\, \delta(\Omega_1-\tfrac{\varphi}{\sqrt{n}}-(2s+k)\Omega_0)\notag\\
            &\qquad\times\prod_{j=2}^n G_j(\Omega_j-\alpha_{j,l}\varphi)\ket{\omega_1,\cdots,\omega_l,\cdots,\omega_n},\\
            &=\sum_{s\in\Z}\int\dd\Omega_1\cdots\dd\Omega_n\, \delta(\Omega_1-\tfrac{\varphi}{\sqrt{n}}-(2s+k)\Omega_0)\prod_{j=2}^n \tilde G_j(\Omega_j)\ket{\omega_1,\cdots,\omega_n},
        \end{align}
    \end{subequations}
    by performing the change of variable $\omega_l\to \omega_l-\varphi$. This transforms $\Omega_1\to \Omega_1-\frac{\varphi}{\sqrt{n}}$ and $\Omega_j\to \tilde\Omega_j=\Omega_j-\alpha_{j,l}\varphi$. We thus define $\tilde G_j(\Omega_j)=G_j(\Omega_j-\alpha_{j,l}\varphi)$. Finally
    \begin{subequations}
        \begin{align}
            e^{-i\tau\hat \omega_l}&\ket{\overline k}\notag\\
            &=\sum_{s\in\Z}\int\dd\Omega_1\cdots\dd\Omega_n\, e^{-i\tau \omega_l}\delta(\Omega_1-(2s+k)\Omega_0)\prod_{j=2}^n G_j(\Omega_j)\ket{\omega_1,\cdots,\omega_n},\\
            &=\sum_{s\in\Z}\int\dd\Omega_1\cdots\dd\Omega_n\, e^{-i\tau\Omega_1/\sqrt{n}}\delta(\Omega_1-(2s+k)\Omega_0)\notag\\
            &\qquad\times\prod_{j=2}^n e^{-i\alpha_{j,l}\tau\Omega_j}G_j(\Omega_j)\ket{\omega_1,\cdots,\omega_n},\\
            &=\sum_{s\in\Z}\int\dd\Omega_1\cdots\dd\Omega_n\, e^{-i\tau(2s+k)\Omega_0/\sqrt{n}}\delta(\Omega_1-(2s+k)\Omega_0)\prod_{j=2}^n \tilde G_j(\Omega_j)\ket{\omega_1,\cdots,\omega_n},
        \end{align}
    \end{subequations}
    by defining in this case $\tilde G_j(\Omega_j)=e^{-i\alpha_{j,l}\tau\Omega_j}G_j(\Omega_j)$.
\end{derivation}

\begin{result}[GKP and entangling gates]~\\
    \label{res: GKP entangling}
    The gate
    \begin{equation}
        \hat{\mathcal D}=\prod_{j=1}^n e^{-i\hat\omega_{j,1}\hat t_{j,2}},
    \end{equation}
    acts on the time-frequency GKP state with encoding
    \begin{equation}
        \ket{\overline k}=\sum_{s\in\Z}\int\dd\Omega_1\cdots\dd\Omega_n\, \delta(\Omega_1-(2s+k)\Omega_0)\prod_{j=2}^n G_j(\Omega_j)\ket{\omega_1,\cdots,\omega_n},
    \end{equation}
    as
    \begin{align}
        \hat{\mathcal D}\ket{\overline{k_1},\overline{k_2}}&=\int\dd\Omega_1\cdots\dd\Omega_n\dd\Omega_1'\cdots\dd\Omega_n'\, \sum_{s_1\in\Z}\delta(\Omega_1-(2s_1+k_1)\Omega_0)\prod_{j=2}^n G_j(\Omega_j)\ket{\omega_1,\dots,\omega_n}\notag\\
        &\qquad\otimes \sum_{s_2\in\Z}\delta(\Omega_1'-(2s_2+k_1+k_2)\Omega_0)\prod_{j=2}^n G_j(\Omega_j'-\Omega_j)\ket{\omega_1',\dots,\omega_n'},
    \end{align}
    which implements an entangling gate provided that the information in the collective variables for $j>1$ is disregarded.

    For $G_j(\Omega_j)=\delta(\Omega_j)$, this simplifies to the transformation of the encoding 
    \begin{equation}
        \ket{\overline k}=\sum_{s\in\Z}\ket{\tfrac{2s+k}{\sqrt{n}}\Omega_0,\dots,\tfrac{2s+k}{\sqrt{n}}\Omega_0},
    \end{equation}
    to
    \begin{equation}
        \hat{\mathcal D}\ket{\overline{k_1},\overline{k_2}}=\ket{\overline{k_1},\overline{k_1+k_2~\rm mod~2}},
    \end{equation}
    which is the perfect logical CNOT gate transformation.
\end{result}

\begin{derivation}
    We start with the general case, and compute
    \begin{subequations}
        \begin{align}
            \hat{\mathcal D}&\ket{\overline{k_1},\overline{k_2}}\notag\\
            &=\int\dd\Omega_1\cdots\dd\Omega_n\dd\Omega_1'\cdots\dd\Omega_n'\, \sum_{s_1\in\Z}\delta(\Omega_1-(2s_1+k_1)\Omega_0)\prod_{j=2}^n G_j(\Omega_j)\ket{\omega_1,\dots,\omega_n}\notag\\
            &\qquad\otimes \sum_{s_2\in\Z}\delta(\Omega_1'-(2s_2+k_2)\Omega_0)\prod_{j=2}^n G_j(\Omega_j')\ket{\omega_1'+\omega_1,\dots,\omega_n'+\omega_n},\\
            &=\int\dd\Omega_1\cdots\dd\Omega_n\dd\Omega_1'\cdots\dd\Omega_n'\, \sum_{s_1\in\Z}\delta(\Omega_1-(2s_1+k_1)\Omega_0)\prod_{j=2}^n G_j(\Omega_j)\ket{\omega_1,\dots,\omega_n}\notag\\
            &\qquad\otimes \sum_{s_2\in\Z}\delta(\Omega_1'-\Omega_1-(2s_2+k_2)\Omega_0)\prod_{j=2}^n G_j(\Omega_j'-\Omega_j)\ket{\omega_1',\dots,\omega_n'},\\
            &=\int\dd\Omega_1\cdots\dd\Omega_n\dd\Omega_1'\cdots\dd\Omega_n'\, \sum_{s_1\in\Z}\delta(\Omega_1-(2s_1+k_1)\Omega_0)\prod_{j=2}^n G_j(\Omega_j)\ket{\omega_1,\dots,\omega_n}\notag\\
            &\qquad\otimes \sum_{s_2\in\Z}\delta(\Omega_1'-(2s_1+2s_2+k_2+k_1)\Omega_0)\prod_{j=2}^n G_j(\Omega_j'-\Omega_j)\ket{\omega_1',\dots,\omega_n'},\\
            &=\int\dd\Omega_1\cdots\dd\Omega_n\dd\Omega_1'\cdots\dd\Omega_n'\, \sum_{s_1\in\Z}\delta(\Omega_1-(2s_1+k_1)\Omega_0)\prod_{j=2}^n G_j(\Omega_j)\ket{\omega_1,\dots,\omega_n}\notag\\
            &\qquad\otimes \sum_{s_2\in\Z}\delta(\Omega_1'-(2s_2+k_2+k_1)\Omega_0)\prod_{j=2}^n G_j(\Omega_j'-\Omega_j)\ket{\omega_1',\dots,\omega_n'},
        \end{align}
    \end{subequations}
    where on the second line we performed the change of variable $\omega_j'\to\omega_j'-\omega_j$, on the third we use the Dirac delta to replace $\Omega_1\to (2s_1+k_1)\Omega_0$ in the argument of the second delta, and on the last line we re-index the sum over $s_2$.

    Applying this relation to the case $G_j(\Omega_j)=\delta(\Omega_j)$ for $j\geq 2$, we obtain
    \begin{subequations}
        \begin{align}
            \hat{\mathcal D}&\ket{\overline{k_1},\overline{k_2} }\notag\\
            &=\int\dd\Omega_1\cdots\dd\Omega_n\dd\Omega_1'\cdots\dd\Omega_n'\, \sum_{s_1\in\Z}\delta(\Omega_1-(2s_1+k_1)\Omega_0)\prod_{j=2}^n \delta(\Omega_j)\ket{\omega_1,\dots,\omega_n}\notag\\
            &\qquad\otimes \sum_{s_2\in\Z}\delta(\Omega_1'-(2s_2+k_2+k_1)\Omega_0)\prod_{j=2}^n \delta(\Omega_j'-\Omega_j)\ket{\omega_1',\dots,\omega_n'},\\
            &=\int\dd\Omega_1\cdots\dd\Omega_n\dd\Omega_1'\cdots\dd\Omega_n'\, \sum_{s_1\in\Z}\delta(\Omega_1-(2s_1+k_1)\Omega_0)\prod_{j=2}^n \delta(\Omega_j)\ket{\omega_1,\dots,\omega_n}\notag\\
            &\qquad\otimes \sum_{s_2\in\Z}\delta(\Omega_1'-(2s_2+k_2+k_1)\Omega_0)\prod_{j=2}^n \delta(\Omega_j')\ket{\omega_1',\dots,\omega_n'},\\
            &=\ket{\overline{k_1},\overline{k_1+k_2~\rm mod~2}},
        \end{align}
    \end{subequations}  
    as the Dirac delta $\delta(\Omega_j)$ now imposes $\Omega_j=0$ for all $j\geq 2$, which allows us to decouple the two codes in the orthogonal variables.
\end{derivation}

\fi

\ifnum \theShowChapfour=1
\ifthenelse{
    \value{ShowChapone}=1 \OR
    \value{ShowChaptwo}=1 \OR
    \value{ShowChapthree}=1
}{
\clearpage
}{}
\section{HOM interferometry and metrology}
\setcurrentanchor{app HOM}
\emph{This section of the appendix collects all the results of Chap.~\ref{chap: HOM interferometry and metrology} and provides the associated proofs.}

\vspace*{1em}
\par\noindent
\textbf{\large Results in this section}\par
\vspace{-0.8em}
\noindent\rule{\textwidth}{0.8pt}\par

\resultentry{res: HOM derivation}{HOM derivation}
\resultentry{res: time detection precision}{Optimality of direct time detection}
\resultentry{res: def and ppt of S}{Definition and properties of $\hat S$}
\resultentry{res: PC with Wigner}{Coincidence probability and Wigner function}
\resultentry{res: expansion pc}{Second-order expansion of the coincidence probability}
\resultentry{res: Fisher info with S}{Limit of the FI at $\theta=0$ for HOM parameter estimation}
\resultentry{res: constant FI}{Probabilities with constant Fisher information}
\resultentry{res: multi-pair model}{Multi-pair model estimation precision}
\resultentry{res: losses commute}{Losses commutation with mode transformations}
\resultentry{res: losses probabilities}{Losses and probabilities}
\resultentry{res: MZI single photon}{Single-photon MZI computation}
\resultentry{res: permutation diagonalisation}{Permutation matrix diagonalisation}
\resultentry{res: D and photon number}{Expection value of $\hat D$ and photon number}
\resultentry{res: proba fourier other}{Probabilities and eigenspaces of $\hat P$}
\resultentry{res: n-mode HOM Fisher info}{$n$-mode HOM Fisher information}
\resultentry{res: extremal symmetry}{Extremal values of $\Tr(\hat \rho \hat P)$}
\resultentry{res: metrology mixed state extension}{Mixed-state extension of the metrological bounds}

\vspace{-0.2em}
\par\noindent\rule{\textwidth}{0.8pt}\par

\label{app: hom interferometry}

\begin{result}[HOM derivation]~\\
    \label{res: HOM derivation}
    For an input state 
    \begin{equation}
        \ket{\psi}_\text{in}=\hat a_\psi^\dagger\hat b_\varphi^\dagger\vac,
    \end{equation}
    entering a balanced BS given by the matrix $\frac{1}{\sqrt{2}}\begin{pmatrix} 1&1\\1&-1 \end{pmatrix}$, the coincidence probability is given by
    \begin{equation}
        P_c=\frac{1}{2}\left(1-\abs{\bra{\psi}\ket{\varphi}}^2\right),
    \end{equation}
    where $\ket{\psi}$ and $\ket{\varphi}$ are the quantum states associated with each incoming photon, and the letters $a$ and $b$ denote the two input/output modes of the BS.
\end{result}

\begin{derivation}
    One first expands the output state
    \begin{subequations}
        \allowdisplaybreaks
        \begin{align}
            \ket{\psi}_\text{out}&=\frac{1}{2}\left(\hat a_\psi^\dagger+\hat b_\psi^\dagger\right)\left(\hat a_\varphi^\dagger-\hat b_\varphi^\dagger\right)\vac,\\
            &=\frac{1}{2}\left(\hat a_\psi^\dagger\hat a_\varphi^\dagger-\hat a_\psi^\dagger\hat b_\varphi^\dagger+\hat b_\psi^\dagger\hat a_\varphi^\dagger-\hat b_\psi^\dagger\hat b_\varphi^\dagger\right)\vac,\\
            &=\frac{1}{2}\left(\hat a_\psi^\dagger\hat a_\varphi^\dagger-\hat b_\psi^\dagger\hat b_\varphi^\dagger\right)\vac+\frac{1}{2}\left(-\hat a_\psi^\dagger\hat b_\varphi^\dagger+\hat b_\psi^\dagger\hat a_\varphi^\dagger\right)\vac,\\
            &=\ket{\psi}_\text{bunch}+\ket{\psi}_\text{coinc}.
        \end{align}    
    \end{subequations}
    The coincidence probability is then given by
    \begin{subequations}
        \begin{align}
            P_c&= \prescript{}{\text{coinc}}{\langle} \psi {| \psi \rangle}_{\text{coinc}},\\
            &=\frac{1}{4}\bra{\text{vac}}\left(-\hat a_\psi\hat b_\varphi+\hat b_\psi\hat a_\varphi\right)\left(-\hat a_\psi^\dagger\hat b_\varphi^\dagger+\hat b_\psi^\dagger\hat a_\varphi^\dagger\right)\vac,\\
            &=\frac{1}{4}\bra{\text{vac}}\left(\hat a_\psi\hat b_\varphi\hat a_\psi^\dagger\hat b_\varphi^\dagger+\hat b_\psi\hat a_\varphi\hat b_\psi^\dagger\hat a_\varphi^\dagger-\hat a_\psi\hat b_\varphi\hat b_\psi^\dagger\hat a_\varphi^\dagger-\hat b_\psi\hat a_\varphi\hat a_\psi^\dagger\hat b_\varphi^\dagger\right)\vac,\\
            &=\frac{1}{4}\bra{\text{vac}}\left(\hat a_\psi\hat a_\psi^\dagger\hat b_\varphi\hat b_\varphi^\dagger+\hat a_\varphi\hat a_\varphi^\dagger\hat b_\psi\hat b_\psi^\dagger-\hat a_\psi\hat a_\varphi^\dagger\hat b_\varphi\hat b_\psi^\dagger-\hat a_\varphi\hat a_\psi^\dagger\hat b_\psi\hat b_\varphi^\dagger\right)\vac,\\
            &=\frac{1}{4}\bra{\text{vac}}\Big[\Big(\hat a_\psi^\dagger\hat a_\psi+[\hat a_\psi,\hat a_\psi^\dagger]\Big)\Big(\hat b_\varphi^\dagger\hat b_\varphi+[\hat b_\varphi,\hat b_\varphi^\dagger]\Big)+\Big(\hat a_\varphi^\dagger\hat a_\varphi+[\hat a_\varphi,\hat a_\varphi^\dagger]\Big)\notag\\
            &\qquad\times\Big(\hat b_\psi^\dagger\hat b_\psi+[\hat b_\psi,\hat b_\psi^\dagger]\Big)-\Big(\hat a_\varphi^\dagger\hat a_\psi+[\hat a_\psi,\hat a_\varphi^\dagger]\Big)\Big(\hat b_\psi^\dagger\hat b_\varphi+[\hat b_\varphi,\hat b_\psi^\dagger]\Big)\notag\\
            &\qquad-\Big(\hat a_\psi^\dagger\hat a_\varphi+[\hat a_\varphi,\hat a_\psi^\dagger]\Big)\Big(\hat b_\varphi^\dagger\hat b_\psi+[\hat b_\psi,\hat b_\varphi^\dagger]\Big)\Big]\vac,\\
            &=\frac{1}{4}\bra{\text{vac}}\Big[[\hat a_\psi,\hat a_\psi^\dagger][\hat b_\varphi,\hat b_\varphi^\dagger]+[\hat a_\varphi,\hat a_\varphi^\dagger][\hat b_\psi,\hat b_\psi^\dagger]-[\hat a_\psi,\hat a_\varphi^\dagger][\hat b_\varphi,\hat b_\psi^\dagger]-[\hat a_\varphi,\hat a_\psi^\dagger][\hat b_\psi,\hat b_\varphi^\dagger]\Big]\vac,\\
            &=\frac{1}{4}\Big[1+1-\abs{\bra{\psi}\ket{\varphi}}^2-\abs{\bra{\varphi}\ket{\psi}}^2\Big],\\
            &=\frac{1}{2}\left(1-\abs{\bra{\psi}\ket{\varphi}}^2\right),
        \end{align}
    \end{subequations}
    where we used the following commutation relation: $a$ and $b$ operators commute, while for any state $\ket{\xi}$, $[\hat a_\xi,\hat a_\xi^\dagger]=1$ and $[\hat a_\psi,\hat a_\varphi^\dagger]=\bra{\psi}\ket{\varphi}$, and similarly for $b$ operators.
\end{derivation}

\begin{result}[Optimality of direct time detection]~\\
    \label{res: time detection precision}
    For time-delay estimation with a single photon with temporal distribution
    \begin{equation}
        \ket{\psi}=\int\dd t\, \tilde F(t)\ket{t},
    \end{equation}
    time-delay estimation generated by $\hat \omega$ and measured with perfect time-sensitive detectors leads to the Fisher information and quantum Fisher information
    \begin{align}
        \mathcal Q&= 4\left[\int \dd t\, \abs{\partial_t \tilde F(t)}^2 - \left(i\int \dd t\, \tilde F(t)^* \partial_t \tilde F(t)\right)^2\right],&
        \mathcal F&= \int \dd t\, \frac{1}{\abs*{\tilde F(t)}^2}\left(\partial_t \abs{\tilde F(t)}^2\right)^2.
    \end{align}
    Additionally, writing the temporal distribution as
    \begin{equation}
        \tilde F(t)=r(t)e^{i\varphi(t)},
    \end{equation}
    in terms of its amplitude and phase, we have the equality $\mathcal Q=\mathcal F$, if and only if $\varphi$ is affine
    \begin{equation}
        \varphi(t)=a t+b,
    \end{equation}
    for some real numbers $a$ and $b$.
\end{result}

\begin{derivation}
    $\blacktriangleright$ {\bf Expression of the FI and QFI.} Recalling that the operator $\hat \omega$ acts on the temporal distribution (see Eq.~\eqref{eq: tf op actions})
    \begin{equation}
        \hat \omega\ket{\psi}=i\int\dd t\, \partial_t \tilde F(t)\ket{t},
    \end{equation}
    the quantum Fisher information is given by  
    \begin{equation}
        \mathcal Q=4\Delta^2\hat \omega= 4\left[\int \dd t\, \abs{\partial_t \tilde F(t)}^2 - \left(i\int \dd t\, \tilde F(t)^* \partial_t \tilde F(t)\right)^2\right].
    \end{equation}
    On the other hand, the evolution $e^{-i\hat \omega\tau}$ transforms the state as (see Eq.~\eqref{eq: translation TF basis states})
    \begin{equation}
        e^{-i\hat \omega\tau}\ket{\psi}=\int\dd t\, \tilde F(t+\tau)\ket{t}.
    \end{equation}
    For time detection with perfect sensitivity, the POVM elements are given by $\hat \Pi_t=\ketbra{t}$, leading to the probability distribution
    \begin{equation}
        P_\tau(t)=\abs{\tilde F(t+\tau)}^2.
    \end{equation}
    As such the Fisher information can be written as
    \begin{equation}
        \mathcal F=\int \dd t\, \frac{1}{P_\tau(t)}\left(\frac{\partial P_\tau(t)}{\partial \tau}\right)^2=\int \dd t\, \frac{1}{\abs*{\tilde F(t)}^2}\left(\partial_t \abs{\tilde F(t)}^2\right)^2.
    \end{equation}

    \medskip

    \noindent

    $\blacktriangleright$ {\bf Optimality condition.} We now write $\tilde F(t)=r(t)e^{i\varphi(t)}$, in terms of its amplitude and phase and analyse the optimality condition. Under this ansatz, the Fisher information can be simplified to
    \begin{equation}
        \mathcal F=\int\dd t \frac{1}{r(t)^2}\left(\frac{\partial}{\partial t}r(t)^2\right)^2 =4\int\dd t\, r'(t)^2.
    \end{equation}
    Similarly the variance can be computed as
    \begin{subequations}
        \allowdisplaybreaks
        \begin{align}
            \Delta^2\hat \omega&=\int \dd t\, \abs{\partial_t \tilde F(t)}^2 - \left(i\int \dd t\, \tilde F(t)^* \partial_t \tilde F(t)\right)^2,\\
            &=\int \dd t\, \abs{r'(t)e^{i\varphi(t)}+i\varphi'(t)r(t)e^{i\varphi(t)}}^2\notag\\
            &\qquad-\left(i\int \dd t\, r(t)e^{-i\varphi(t)}\left[r'(t)e^{i\varphi(t)}+i\varphi'(t)r(t)e^{i\varphi(t)}\right]\right)^2,\\
            &=\int \dd t\,\abs{r'(t)+i\varphi'(t)r(t)}^2-\left(i\int \dd t\, r(t)\left[r'(t)+i\varphi'(t)r(t)\right]\right)^2,\\
            &=\int\dd t\, r'(t)^2+\int\dd t\, \varphi'(t)^2r(t)^2-\left(i\underbrace{\int\dd t\, r(t)r'(t)}_{=0} -\int \dd t\,\varphi'(t)r(t)^2\right)^2,\\
            &=\int\dd t\, r'(t)^2+\int\dd t\, \varphi'(t)^2r(t)^2-\left(\int\dd t\, \varphi'(t)r(t)^2\right)^2,
        \end{align}
    \end{subequations}
    where $\int\dd t\, r(t)r'(t)=[r(t)^2/2]_{-\infty}^\infty=0$, as $r$ is assumed to vanish at infinity. To provide a compact formula we introduce an inner product on real bounded functions as 
    \begin{equation}
        (g,h)=\int\dd t\, g(t)h(t)r(t)^2,
    \end{equation}
    which allows us to rewrite
    \begin{equation}
        \mathcal Q=4\Delta^2\hat\omega=\mathcal F+4(\varphi',\varphi')-4(\varphi',1)^2,
    \end{equation}
    where $1$ denotes the constant function equal to $1$ everywhere. Since $r$ is normalized, we have $(1,1)=1$ and the Cauchy-Schwarz inequality implies that 
    \begin{equation}
        (\varphi',1)^2\leq (\varphi',\varphi')(1,1)=(\varphi',\varphi'),
    \end{equation}
    which allows us to recover, as it should be, the inequality $\mathcal Q\geq \mathcal F$. Additionally, the equality in the Cauchy-Schwarz inequality is achieved if and only if $\varphi'$ is proportional to $1$, which means that $\varphi$ is affine, as claimed.
\end{derivation}

\begin{result}[Definition and properties of $\hat S$]~\\
    \label{res: def and ppt of S}
    The following properties 
    \begin{align}
        \hat S\vac =\vac, && \hat S\hat a_1^\dagger(\omega)=\hat a_2^\dagger(\omega)\hat S, && \hat S\hat a_2^\dagger(\omega)=\hat a_1^\dagger(\omega)\hat S,
    \end{align}
    uniquely define the symmetry operator $S$, which is then automatically Hermitian and unitary and satisfies $\hat S^2=1$.
\end{result}

\begin{derivation}
    $\blacktriangleright$ {\bf Uniqueness.} On the one hand, assuming that $\hat S$ satisfies the above properties, we see that its action on a general two-spatial-mode time-frequency state with a fixed number of photons
    \begin{align}
        \ket{\psi}&=\int\dd\omega_1\cdots\dd\omega_n \dd \theta_1\cdots\dd\theta_k\, F(\omega_1,\dots,\omega_n,\theta_1,\dots,\theta_k)\hat a_1^\dagger(\omega_1)\cdots \hat a_1^\dagger(\omega_n)\notag\\
        &\qquad\times\hat a_2^\dagger(\theta_1)\cdots \hat a_2^\dagger(\theta_k)\vac,
    \end{align}
    is given by
    \begin{align}
        \hat S\ket{\psi}&=\int\dd\omega_1\cdots\dd\omega_n \dd \theta_1\cdots\dd\theta_k\, F(\omega_1,\dots,\omega_n,\theta_1,\dots,\theta_k)\hat a_2^\dagger(\omega_1)\cdots \hat a_2^\dagger(\omega_n)\notag\\
        &\qquad\hat a_1^\dagger(\theta_1)\cdots \hat a_1^\dagger(\theta_k)\vac,
    \end{align}
    which is obtained simply by using the properties of $\hat S$ one by one to shift $\hat S$ from left to right. As such there is only one way such an operator can act on any state, and thus it is uniquely defined by the above properties.

    \medskip

    \noindent

    $\blacktriangleright$ {\bf Existence.} On the other hand, we can define $\hat S$ by defining its action on any state with fixed photon number as above, and then extend it by linearity to the whole Hilbert space. By construction, it leaves the vacuum invariant. Additionally, the action on the monomode states necessarily means that
    \begin{align}
        \hat S \hat a_1^\dagger(\omega)\hat S=\hat a_2^\dagger(\omega), && \hat S \hat a_2^\dagger(\omega)\hat S=\hat a_1^\dagger(\omega),
    \end{align}
    which shows that $\hat S$ indeed satisfies the required properties, as it is clear in this case that $\hat S^2=\1$. 

    \medskip

    \noindent

    $\blacktriangleright$ {\bf Properties.} By its action on the creation operators, it is clear that $\hat S^2=\1$. The adjoint of $\hat S$ is defined as the operator satisfying 
    \begin{equation}
        \bra{\psi}\hat S^\dagger\ket{\phi}=\bra{\phi}\hat S\ket{\psi}^*,
    \end{equation}
    for all states $\ket{\psi}$ and $\ket{\phi}$. By looking at this equation for states with fixed photon number, it is clear that $\hat S$ satisfies the property of the adjoint and thus $\hat S$ is Hermitian. Finally, the properties $\hat S^2=\1$ and $\hat S$ Hermitian imply that $\hat S$ is also unitary, as $\hat S^{-1}=\hat S^\dagger=\hat S$.
\end{derivation}

\begin{result}[Coincidence probability and Wigner function]~\\
    \label{res: PC with Wigner}
    For a state whose JSA can be factored as
    \begin{equation}
        F(\omega_1,\omega_2)=F_+(\omega_+)F_-(\omega_-),
    \end{equation}
    with $\omega_\pm=\frac{\omega_1\pm\omega_2}{\sqrt{2}}$, under the evolution $e^{-i(\hat\omega_1\tau+\varphi\hat t_2)}$, the coincidence probability at the output of the HOM interferometer is given by
    \begin{equation}
        P_c(\tau,\varphi)=\frac{1}{2}\left(1-\pi W_-\left(\frac{\varphi}{\sqrt{2}},\frac{\tau}{\sqrt{2}}\right)\right),
    \end{equation}
    where 
    \begin{equation}
        W_-(\varphi,\tau)=\frac{1}{\pi}\int\dd\omega\, e^{2i\omega\tau}F_-(\varphi-\omega)F_-^*(\varphi+\omega),
    \end{equation}
    is the Wigner function associated with the JSA in the anti-symmetric variable $\omega_-$.
\end{result}

\begin{derivation}
    We simply use the general formula for the coincidence probability
    \begin{equation}
        P_c=\frac{1}{2}\left(1-\bra{\psi}\hat S\ket{\psi}\right),
    \end{equation}
    applied to the evolved state
    \begin{equation}
        e^{-i(\hat\omega_1\tau+\varphi\hat t_2)}\ket{\psi}=\int\dd\omega_1\dd\omega_2\, e^{-i\omega_1\tau}F(\omega_1,\omega_2-\varphi)\ket{\omega_1,\omega_2}.
    \end{equation}
    We thus need to compute the overlap 
    \begin{subequations}
        \begin{align}
            \bra{\psi}&e^{i(\hat\omega_1\tau+\varphi\hat t_2)}\hat S e^{-i(\hat\omega_1\tau+\varphi\hat t_2)}\ket{\psi}\notag\\
            &=\int \dd\omega_1\dd\omega_2\, e^{i\omega_1\tau}F^*(\omega_1,\omega_2-\varphi)e^{-i\omega_2\tau}F(\omega_2,\omega_1-\varphi),\\
            &=\int\dd\omega_+\dd\omega_-\, e^{i\sqrt{2}\omega_-\tau}F_+^*(\omega_+-\tfrac{\varphi}{\sqrt{2}})F_-^*(\omega_-+\tfrac{\varphi}{\sqrt{2}})F_+(\omega_+-\tfrac{\varphi}{\sqrt{2}})F_-(-\omega_-+\tfrac{\varphi}{\sqrt{2}}),\\
            &=\int\dd\omega_+\, \abs{F_+(\omega_+-\tfrac{\varphi}{\sqrt{2}})}^2\int\dd\omega_-\, e^{i\sqrt{2}\omega_-\tau}F_-^*(\omega_-+\tfrac{\varphi}{\sqrt{2}})F_-(-\omega_-+\tfrac{\varphi}{\sqrt{2}}),\\
            &=\pi W_-\left(\frac{\varphi}{\sqrt{2}},\frac{\tau}{\sqrt{2}}\right),
        \end{align}
    \end{subequations}
    which when plugged back into the expression of $P_c$ yields the desired result.
\end{derivation}

\begin{result}[Second-order expansion of the coincidence probability]~\\
    \label{res: expansion pc}
    For an initial state $\ket{\psi}$ following the symmetry condition
    \begin{equation}
        \hat S\ket{\psi}=\pm\ket{\psi},
    \end{equation}
    the coincidence probability at the output of the HOM interferometer, where the evolution $e^{-i\theta\hat H}$ was first applied to the state, is given by the second-order expansion
    \begin{equation}
        P_c(\theta)=\frac{1}{2}\left(1\mp 1 \pm \frac{\theta^2}{2}\Delta^2(\hat H-\hat S\hat H\hat S)+O(\theta^3)\right).
    \end{equation}
\end{result}

\begin{derivation}
    The coincidence probability is given by Eq.~\eqref{eq: pc with s and evol}
    \begin{subequations}
        \begin{align}
            P_c(\theta)&=\frac{1}{2}\left(1-\bra{\psi}e^{i\theta\hat H}\hat Se^{-i\theta\hat H}\ket{\psi}\right),\\
            &=\frac{1}{2}\left(1\mp\bra{\psi}e^{i\theta\hat H}\hat Se^{-i\theta\hat H}\hat S\ket{\psi}\right),\\
            &=\frac{1}{2}\left(1\mp\bra{\psi}e^{i\theta\hat H} e^{-i\theta\hat S\hat H\hat S}\ket{\psi}\right),
        \end{align}    
    \end{subequations}
    where we used the symmetry condition to replace $\pm\ket{\psi}$ by $\hat S\ket{\psi}$, and then moved $\hat S$ inside the exponential using $\hat S^{-1}=\hat S$. We can then expand the expectation value as
    \begin{subequations}
        \begin{align}
            \expval{e^{i\theta\hat H} e^{-i\theta\hat S\hat H\hat S}}&=\left\langle\left(1+i\theta\hat H+\frac{(i\theta)^2}{2}\hat H^2+O(\theta^3)\right)\right.\notag\\
            &\qquad\times\left.\left(1-i\theta\hat S\hat H\hat S+\frac{(-i\theta)^2}{2}(\hat S\hat H\hat S)^2+O(\theta^3)\right)\right\rangle,\\
            &=1+i\theta\expval{\hat H-\hat S\hat H\hat S}+\frac{(i\theta)^2}{2}\expval{\hat H^2+(\hat S\hat H\hat S)^2-2\hat H\hat S\hat H\hat S}+O(\theta^3).
        \end{align}
    \end{subequations}
    The symmetry condition imposes that $\expval{\hat S\hat H\hat S}=\expval{\hat H}$ such that the first-order term vanishes. Additionally, we also have $\expval{\hat H\hat S\hat H\hat S}=\expval{\hat S\hat H\hat S\hat H}$, so that
    \begin{equation}
        \expval{\hat H^2+(\hat S\hat H\hat S)^2-2\hat H\hat S\hat H\hat S}=\expval{(\hat H-\hat S\hat H\hat S)^2}=\Delta^2(\hat H-\hat S\hat H\hat S),
    \end{equation}
    since once again $\expval{\hat H-\hat S\hat H\hat S}=0$. This finally yields the second-order expansion of expectation value
    \begin{equation}
        \expval{e^{i\theta\hat H} e^{-i\theta\hat S\hat H\hat S}}=1-\frac{\theta^2}{2}\Delta^2(\hat H-\hat S\hat H\hat S)+O(\theta^3).
    \end{equation}
    Plugging this back into the expression of $P_c(\theta)$ yields the desired expansion
    \begin{equation}
        P_c(\theta)=\frac{1}{2}\left(1\mp 1 \pm \frac{\theta^2}{2}\Delta^2(\hat H-\hat S\hat H\hat S)+O(\theta^3)\right).
    \end{equation}
\end{derivation}

\begin{result}[Limit of the Fisher information at $\theta=0$ for HOM parameter estimation]~\\
    \label{res: Fisher info with S}
    For an initial state $\ket{\psi}$ following the symmetry condition $\hat S\ket{\psi}=\pm\ket{\psi}$, the Fisher information for the estimation of $\theta$ at $\theta=0$ is given by
    \begin{equation}
    \lim_{\theta\rightarrow 0} \mathcal F = \Delta^2(\hat H-\hat S\hat H\hat S).
    \end{equation}
\end{result}

\begin{derivation}
    To do the derivation, we use the expansion obtained in Appendix~\ref{app: hom interferometry}, Result~\ref{res: expansion pc} for the coincidence probability
    \begin{equation}
        P_c(\theta)=\frac{1}{2}\left(1\mp 1 \pm \frac{\theta^2}{2}\Delta^2(\hat H-\hat S\hat H\hat S)+O(\theta^3)\right).
    \end{equation}
    We thus see that depending on the input state symmetry we either get
    \begin{align}
        P_c=\frac{\theta^2}{4}\Delta^2(\hat H-\hat S\hat H\hat S)+O(\theta^3), && \text{or,} && P_c=1-\frac{\theta^2}{4}\Delta^2(\hat H-\hat S\hat H\hat S)+O(\theta^3).
    \end{align}
    We then observe the following general result: if $P_c(\theta)=\alpha \theta^2+O(\theta^3)$, then $\lim\limits_{\theta\to 0}\mathcal F=4\alpha$. This can be shown by using the definition of the Fisher information
    \begin{subequations}
        \begin{align}
            \mathcal F&=\frac{(\partial_\theta P_c)^2}{P_c(1-P_c)},\\
            &=\frac{(2\alpha\theta+O(\theta^2))^2}{(\alpha\theta^2+O(\theta^3))(1-\alpha\theta^2+O(\theta^3))},\\
            &=\frac{4\alpha^2\theta^2+O(\theta^3)}{\alpha\theta^2+O(\theta^3)},\\
            &=4\alpha+O(\theta),
        \end{align}
    \end{subequations}
    leading to the limit $\lim_{\theta\to 0}\mathcal F=4\alpha$. Applying this result to $P_c$ directly, or by exchanging the role of $P_c$ and $1-P_c$ depending on the symmetry of the input state, we obtain
    \begin{equation}
        \lim_{\theta\rightarrow 0} \mathcal F = \Delta^2(\hat H-\hat S\hat H\hat S).
    \end{equation}
\end{derivation}

\begin{result}[Probabilities with constant Fisher information]~\\
    \label{res: constant FI}
    The two-outcome metrological scenarios leading to a constant value of the Fisher information are those where the outcome probabilities are parametrized as
    \begin{align}
        P_1(\theta)=\frac{1}{2}\left(1+ \sin(\alpha\theta+\beta)\right), && P_2(\theta)=1-P_1(\theta),
    \end{align}
    for which the Fisher information equals
    \begin{equation}
        \mathcal F(\theta)=\alpha^2.
    \end{equation}
\end{result}

\begin{derivation}
    With the ansatz 
    \begin{equation}
        P_1(\theta)=\frac{1}{2}\left(1+ f(\theta)\right),
    \end{equation}
    the Fisher information is expressed as
    \begin{equation}
        \mathcal F(\theta)=\frac{f'(\theta)^2}{1-f(\theta)^2}.
    \end{equation}
    Under the assumption of constant Fisher information $\mathcal F(\theta)=\alpha^2$, the expression of $f$ is obtained by solving the differential equation
    \begin{equation}
        f'(\theta)^2=\alpha^2(1-f(\theta)^2).
    \end{equation}
    Differentiating yields,
    \begin{equation}
        2f'(\theta)f''(\theta)=-2\alpha^2 f(\theta)f'(\theta).
    \end{equation}
    $f'=0$ yields constant probabilities, which is a trivial case. Otherwise, we can divide by $2f'(\theta)$ to obtain
    \begin{equation}
        f''(\theta)=-\alpha^2 f(\theta),
    \end{equation}
    which is solved by cosine solutions. Plugging into the initial form yields
    \begin{equation}
        P_1(\theta)=\frac{1}{2}\left(1+ \sin(\alpha\theta+\beta)\right).
    \end{equation}
\end{derivation}

\begin{result}[Multi-pair model estimation precision]~\\
    \label{res: multi-pair model}
    Consider an experiment in which $n$ independent and non-interfering single-photon pairs are generated with probability distribution $\{A_n\}_{n\ge 0}$, where
    \begin{equation}
        \mathbb{P}[n\ \text{pairs}] = A_n,
    \end{equation}
    and denote by $\mathbb{E}(n) = \sum_{n=0}^\infty n A_n$ the mean number of generated pairs. Assume that each photon pair:
    \begin{itemize}
        \item evolves independently of the others,
        \item is perfectly symmetric,
        \item yields the same individual contribution 
        $\Delta^2(\hat H - \hat S \hat H \hat S)$ to the Fisher information.
    \end{itemize}
    Then, the total precision (Fisher information) achieved by the interferometric protocol is
    \begin{equation}
        \lim_{\theta\to0}\mathcal F = \Delta^2(\hat H - \hat S \hat H \hat S) \mathbb{E}(n).
    \end{equation}
\end{result}

\begin{derivation}
    $\blacktriangleright$ {\bf Expression of $P_n(\theta)$.} We first compute the probability $P_n(\theta)$ to detect a total of $n$ photons in the second arm. Let $\{A_k\}_{k\ge 0}$ be the probability distribution of the number of generated pairs. Conditioned on the event ``$k$ pairs are generated'', the pairs are
    independent and each pair contributes $0,1,$ or $2$ photons to the second arm
    with probabilities $R_0(\theta),R_1(\theta),R_2(\theta)$. We compute the total probability $P_n(\theta)$ to detect $n$ photons by
    conditioning on the number of generated pairs:
    \begin{equation}
        P_n(\theta) = \sum_{k=0}^\infty A_k\, \mathbb P[n \text{ photons}\mid k \text{ pairs}].
    \end{equation}
    Fix $k$. To obtain a total of $n$ detected photons, suppose that
    \begin{itemize}
        \item $l$ pairs contribute $2$ photons each,
        \item $m$ pairs contribute $1$ photon each,
        \item the remaining $k-l-m$ pairs contribute $0$ photons.
    \end{itemize}
    These integers must satisfy
    \begin{align}
        2l+m=n, && l,m\ge 0, && l+m\le k.
    \end{align}
    For a fixed $l$, we necessarily have $m=n-2l$, hence the constraints become
    \begin{align}
        0\le l\le \lfloor n/2\rfloor, && k\ge n-l.
    \end{align}
    For such $k$ and $l$, the probability of the corresponding configuration is
    obtained as follows
    \begin{itemize}
        \item choose $l$ pairs among the $k$ that will contribute $2$ photons:
        $\binom{k}{l}$,
        \item choose among the remaining $k-l$ pairs those $n-2l$ that will
        contribute $1$ photon:
        $\binom{k-l}{n-2l}$,
        \item multiply by the corresponding probabilities
        $R_2^l R_1^{n-2l} R_0^{k+l-n}$.
    \end{itemize}
    Therefore,
    \begin{equation}
        \mathbb P[n \text{ photons}\mid k \text{ pairs}] = \sum_{l=0}^{\lfloor n/2\rfloor} \delta_{k\ge n-l}\, \binom{k}{l} \binom{k-l}{\,n-2l\,} R_2^l R_1^{\,n-2l} R_0^{\,k+l-n},
    \end{equation}
    and hence
    \begin{equation}
        P_n(\theta) = \sum_{l=0}^{\lfloor n/2\rfloor} \sum_{k=n-l}^{\infty} A_k \binom{k}{l} \binom{k-l}{\,n-2l\,} R_2(\theta)^l R_1(\theta)^{\,n-2l} R_0(\theta)^{\,k+l-n}.
    \end{equation}
    \medskip

    \noindent
    $\blacktriangleright$ {\bf Single-pair probabilities.} Since the pairs are symmetric, the single-pair probabilities have the expansions
    \begin{align}
        R_1(\theta) &= \frac{\theta^2}{4} \Delta^2(\hat H-\hat S\hat H\hat S) +o(\theta^2), \\ R_0(\theta)=R_2(\theta) &= \frac12 - \frac{\theta^2}{8} \Delta^2(\hat H-\hat S\hat H\hat S) +o(\theta^2),
    \end{align}
     which are taken from Appendix~\ref{app: hom interferometry}, Result~\ref{res: expansion pc}.
    \medskip

    \noindent
    $\blacktriangleright$ {\bf Precision.} For $n=2m$, the lowest-order contribution comes from $l=m$, giving
    \begin{equation}
        P_{2m}(\theta) = \sum_{n=m}^\infty \binom{n}{m}\frac{A_n}{2^n} + \mathcal O(\theta^2),
    \end{equation}
    yielding the expansion for the probability to measure an even number of photons
    \begin{subequations}
        \allowdisplaybreaks
        \begin{align}
            P_\text{even}(\theta) &= \sum_{m=0}^\infty \sum_{n=m}^\infty \binom{n}{m}\frac{A_n}{2^n} + \mathcal O(\theta^2),\\
            &= \sum_{n=0}^\infty \frac{A_n}{2^n} \sum_{m=0}^n \binom{n}{m} + \mathcal O(\theta^2),\\
            &= \sum_{n=0}^\infty A_n + \mathcal O(\theta^2),\\
            &= 1 + \mathcal O(\theta^2).
        \end{align}
    \end{subequations}
    For $n=2m+1$, the leading term is linear in $R_1$:
    \begin{align}
        P_{2m+1}(\theta) &= \sum_{n=m+1}^\infty \binom{n}{m} \binom{n-m}{1} A_n R_1 R_0^{\,n-1} +o(\theta^2), \\
        &= \frac{\theta^2}{4} \Delta^2(\hat H-\hat S\hat H\hat S) \sum_{n=m+1}^\infty n\binom{n-1}{m} \frac{A_n}{2^{n-1}} +o(\theta^2).
    \end{align}
    Hence
    \begin{subequations}
        \begin{align}
            P_\text{odd}(\theta)&= \sum_{m=0}^\infty P_{2m+1}(\theta),\\
            &= \frac{\theta^2}{4} \Delta^2(\hat H-\hat S\hat H\hat S) \sum_{m=0}^\infty \sum_{n=m+1}^\infty n\binom{n-1}{m} \frac{A_n}{2^{n-1}} +o(\theta^2),\\
            &= \frac{\theta^2}{4} \Delta^2(\hat H-\hat S\hat H\hat S) \sum_{n=1}^\infty n\frac{A_n}{2^{n-1}} \sum_{m=0}^{n-1} \binom{n-1}{m} +o(\theta^2),\\
            &= \frac{\theta^2}{4} \Delta^2(\hat H-\hat S\hat H\hat S) \sum_{n=1}^\infty n A_n +o(\theta^2),\\
            &= \frac{\theta^2}{4} \Delta^2(\hat H-\hat S\hat H\hat S) \mathbb E(n) +o(\theta^2).
        \end{align}
    \end{subequations}
    A computation of the Fisher information similar to the one in Appendix~\ref{app: hom interferometry}, Result~\ref{res: Fisher info with S} yields the result
    \begin{equation}
        \lim_{\theta\to0}\mathcal F = \Delta^2(\hat H - \hat S \hat H \hat S) \mathbb{E}(n).
    \end{equation}
\end{derivation}

\begin{result}[Losses commutation with mode transformations]~\\
    \label{res: losses commute}
    Assuming a loss modeled by a lossy beam splitter which transforms the creation operators as\footnote{Note that we put no internal degree of freedom as an argument of the creation operators. The index $j$ then encompasses both internal and external degrees of freedom, allowing us to extend the result beyond passive linear transformations acting only on external degrees of freedom.}
    \begin{equation}
        \hat a_j^\dagger \to \sqrt{\eta} \hat a_j^\dagger + \sqrt{1-\eta} \hat b_j^\dagger
    \end{equation}
    where $\hat b_j^\dagger$ denotes the creation operator of an environmental mode initially in the vacuum state, losses commute with any mode transformation.
\end{result}

\begin{derivation}
    Consider the mode transformation $\hat M$ defined by
    \begin{equation}
        \hat a_j^\dagger \mapsto \sum_{k=0}^{n-1} M_{k,j} \hat a_k^\dagger.
    \end{equation}
    Depending on the order of composition, the operator $\hat a_j^\dagger$ transforms as
    \begin{equation}
        \hat a_j^\dagger \mapsto \sqrt{\eta} \sum_{k=0}^{n-1} M_{k,j} \hat a_k^\dagger + \sqrt{1-\eta} \hat b_j^\dagger,
    \end{equation}
    if losses are considered before $\hat M$, and as
    \begin{equation}
        \hat a_j^\dagger \mapsto \sqrt{\eta} \sum_{k=0}^{n-1} M_{k,j} \hat a_k^\dagger + \sqrt{1-\eta} \sum_{k=0}^{n-1} M_{k,j} \hat b_k^\dagger,
    \end{equation}
    if losses are considered after $\hat M$. In these two expressions, the only difference lies in the environmental modes: $\hat b_j^\dagger$ versus $\hat c_j^\dagger=\sum_{k=0}^{n-1} M_{k,j}\hat b_k^\dagger$. However, since these modes are unobserved, the distinction is irrelevant: the particular mode in which photons are lost does not affect the measurable output statistics. Assuming that the environmental modes are initially in the vacuum state, we can compute the output probabilities for a general input state $\ket{\psi}$ by expanding the transformed creation operators. Because of the equivalence between the environmental modes in the two scenarios, the output statistics are identical. Hence, losses commute with linear optical operations without affecting the measurement outcomes.
\end{derivation}

\begin{result}[Losses and probabilities]~\\
    \label{res: losses probabilities}
    Denoting by $P_j$ the ideal probability to detect $j$ photons, assuming an independent rate $\eta$ to keep each photon, the lossy probability $Q_k$ to detect $k$ photons is given by
    \begin{equation}
        Q_k = \sum_{j=k}^\infty \binom{j}{k} \eta^k (1-\eta)^{j-k} P_j.
    \end{equation}
    For $\eta>0$, this formula can be inverted to
    \begin{equation}
        P_k = \sum_{j=k}^\infty \binom{j}{k} \frac{(\eta-1)^{j-k}}{\eta^j} Q_j.
    \end{equation}
\end{result}

\begin{derivation}
    The expression for $Q_k$ can be obtained by the following combinatorial argument: if $k$ photons are detected after losses, they could have originated from any initial number $j\geq k$ of photons, of which $k$ were transmitted and $j-k$ lost. The binomial coefficient counts the number of possible transmitted subsets, while the factors $\eta^k$ and $(1-\eta)^{j-k}$ represent the transmission and loss probabilities, respectively. To invert the relation we use generating functions. Define
    \begin{align}
        F(x) = \sum_{k=0}^\infty P_k x^k, && G(x) = \sum_{k=0}^\infty Q_k x^k.
    \end{align}
    From the relation above, we obtain
    \begin{subequations}
        \begin{align}
            G(x) &= \sum_{k=0}^\infty Q_k x^k,\\
            &= \sum_{k=0}^\infty \sum_{j=k}^\infty \binom{j}{k} \eta^k (1-\eta)^{j-k} P_j x^k,\\
            &= \sum_{j=0}^\infty \sum_{k=0}^j \binom{j}{k} \eta^k (1-\eta)^{j-k} P_j x^k,\\
            &= \sum_{j=0}^\infty \big[1-\eta + \eta x\big]^j P_j,\\
            &= F\big(1-\eta + \eta x\big).
        \end{align}
    \end{subequations}
    Thus, the inverse map $x \mapsto \frac{x-(1-\eta)}{\eta}$ gives
    \begin{subequations}
        \begin{align}
            F(x) &= G\left(\frac{x-(1-\eta)}{\eta}\right),\\
            &= \sum_{j=0}^\infty \left(\frac{x-(1-\eta)}{\eta}\right)^j Q_j,\\
            &= \sum_{j=0}^\infty \frac{1}{\eta^j} \sum_{k=0}^j \binom{j}{k} (\eta-1)^{j-k} x^k Q_j,\\
            &= \sum_{k=0}^\infty \sum_{j=k}^\infty \binom{j}{k} \frac{(\eta-1)^{j-k}}{\eta^j} Q_j x^k.
        \end{align}
    \end{subequations}
    By identifying the coefficients we recover
    \begin{equation}
        P_k = \sum_{j=k}^\infty \binom{j}{k} \frac{(\eta-1)^{j-k}}{\eta^j} Q_j.
    \end{equation}
\end{derivation}

\begin{result}[Single-photon MZI computation]~\\
    \label{res: MZI single photon}
    For an initial single-photon state $\ket{\psi}$,
    \begin{equation}
        \ket{\psi}=\int \dd\omega_1\dd\omega_2\, F(\omega_1,\omega_2)\hat a_1^\dagger(\omega_1)\hat a_2^\dagger(\omega_2)\vac,
    \end{equation}
    and a time-delay evolution of the form $\hat V=e^{-i\tau\hat \omega_1}$ placed between the two beam splitters, the coincidence probability at the output of the MZI is given by
    \begin{equation}
        P_c=\int \dd\omega_1 \dd\omega_2\, \abs{F^s(\omega_1,\omega_2)}^2 \cos^2\left(\tfrac{(\omega_1+\omega_2)\tau}{2}\right)+\int \dd\omega_1 \dd\omega_2\, \abs{F^a(\omega_1,\omega_2)}^2 \cos^2\left(\tfrac{(\omega_1-\omega_2)\tau}{2}\right),
    \end{equation}
    where
    \begin{align}
        F^s=\frac{F(\omega_1,\omega_2)+F(\omega_2,\omega_1)}{2}, && F^a=\frac{F(\omega_1,\omega_2)-F(\omega_2,\omega_1)}{2},
    \end{align}
    are the symmetric and anti-symmetric parts of the JSA, respectively. Additionally, for every JSA $F$, the MZI provides a parameter estimation protocol, whose sensitivity around $\tau=0$ is given by
    \begin{equation}
        \lim_{\tau\to 0} \mathcal F = \expval{(\hat \omega_1+\hat \omega_2)^2}_s+\expval{(\hat \omega_1-\hat \omega_2)^2}_a,
    \end{equation}
    and the quantum Fisher information is given by
    \begin{equation}
        \mathcal Q=4\Delta^2_{\ket{\psi_\text{in}}}(\hat \omega_1) = 2\expval{(\hat \omega_1+\hat \omega_2)^2}_s + 4\expval{\hat \omega_1^2}_a - \left(\expval{\hat \omega_1+\hat \omega_2}_s + 2\expval{\hat \omega_1}_a\right)^2,
    \end{equation}
    where the expectation values $\expval{\cdot}_s$ and $\expval{\cdot}_a$ are the ones computed on the symmetric and anti-symmetric single-photon states
    \begin{align}
        \ket{\psi_s}=\int \dd\omega_1 \dd\omega_2\, F^s(\omega_1,\omega_2) \ket{\omega_1,\omega_2}, &&
        \ket{\psi_a}=\int \dd\omega_1 \dd\omega_2\, F^a(\omega_1,\omega_2) \ket{\omega_1,\omega_2}.
    \end{align}
\end{result}

\begin{derivation}
    $\blacktriangleright$ {\bf Computation of the coincidence probability.} Applying the beam splitter relation to the initial probe state yields the state
    \begin{equation}
        \ket{\psi_\text{in}}=\ket{\psi^b}-\ket{\psi^a},
    \end{equation}
    where 
    \begin{align}
        \ket{\psi^b} =& \frac{1}{2} \int \dd\omega_1 \dd\omega_2\, F^s(\omega_1,\omega_2) \left( \hat a_1^\dagger(\omega_1) \hat a_1^\dagger(\omega_2) - \hat a_2^\dagger(\omega_1) \hat a_2^\dagger(\omega_2) \right) \vac,\\
        \ket{\psi^a} =& \int \dd\omega_1 \dd\omega_2\, F^a(\omega_1,\omega_2) \hat a_1^\dagger(\omega_1) \hat a_2^\dagger(\omega_2) \vac.
    \end{align}
    The coincidence probability is then given by 
    \begin{equation}
        P_c=\frac{1}{2}\left(1-\bra{\psi_\text{in}}\hat V^\dagger\hat S \hat V\ket{\psi_\text{in}}\right).
    \end{equation}
    As we have mentioned in the main body of this thesis or as it can explicitly be verified, the state $\ket{\psi_\text{in}}$ is anti-symmetric. As such the coincidence probability can be simplified to
    \begin{subequations}
        \begin{align}
            P_c&=\frac{1}{2}\left(1+\bra{\psi_\text{in}}e^{i\hat \omega_1\tau}\hat Se^{-i\hat\omega_1\tau}\hat S\ket{\psi_\text{in}}\right),\\
            &=\frac{1}{2}\left(1+\bra{\psi_\text{in}}e^{i\hat \omega_1\tau} e^{-i\hat S\hat \omega_1 \hat S \tau}\ket{\psi_\text{in}}\right),\\
            &=\frac{1}{2}\left(1+\bra{\psi_\text{in}}e^{i (\hat\omega_1-\hat\omega_2 )\tau}\ket{\psi_\text{in}}\right).
        \end{align}    
    \end{subequations}
    Notice that the states $\ket{\psi^b}$ and $\ket{\psi^a}$ contain different distributions of photons. As such they are orthogonal. Furthermore, the operator $e^{i(\hat\omega_1-\hat\omega_2)\tau}$ does not change the photon distribution. We can thus compute the expectation value by summing the bunched and anti-bunched components independently. For $\ket{\psi^b}$, we have
    \begin{align}
        e^{i(\hat\omega_1-\hat\omega_2)\tau}\ket{\psi^b}&=\frac{1}{2} \int \dd\omega_1 \dd\omega_2\, F^s(\omega_1,\omega_2) \left( e^{i(\omega_1+\omega_2)\tau}\hat a_1^\dagger(\omega_1) \hat a_1^\dagger(\omega_2)\right.\notag\\
        &\qquad\left. - e^{-i(\omega_1+\omega_2)\tau}\hat a_2^\dagger(\omega_1) \hat a_2^\dagger(\omega_2) \right) \vac.
    \end{align}
    The scalar product with $\ket{\psi^b}$ thus yields
    \begin{subequations}
        \begin{align}
            \bra{\psi^b}e^{i(\hat\omega_1-\hat\omega_2)\tau}\ket{\psi^b}&=\frac{1}{2} \int \dd\omega_1 \dd\omega_2\, \abs{F^{s}(\omega_1,\omega_2)}^2  \left( e^{i(\omega_1+\omega_2)\tau} + e^{-i(\omega_1+\omega_2)\tau} \right),\\
            &=\int \dd\omega_1 \dd\omega_2 \,\abs{F^s(\omega_1,\omega_2)}^2 \cos\left((\omega_1+\omega_2)\tau\right).
        \end{align}
    \end{subequations}
    Using the trigonometric formula $\cos(2x)=2\cos^2(x)-1$, we can rewrite this as
    \begin{align}
        \bra{\psi^b}e^{i(\hat\omega_1-\hat\omega_2)\tau}\ket{\psi^b}&=2\int \dd\omega_1 \dd\omega_2\, \abs{F^s(\omega_1,\omega_2)}^2 \cos^2\left(\tfrac{(\omega_1+\omega_2)\tau}{2}\right)\notag\\
        &\qquad-\int \dd\omega_1 \dd\omega_2\, \abs{F^s(\omega_1,\omega_2)}^2.
    \end{align}
    Similarly for the anti-symmetric part
    \begin{subequations}
        \begin{align}
            \bra{\psi^a}e^{i(\hat\omega_1-\hat\omega_2)\tau}\ket{\psi^a}&=\int \dd\omega_1 \dd\omega_2 \,\abs{F^a(\omega_1,\omega_2)}^2  e^{i(\omega_1-\omega_2)\tau},\\
            &=\frac{1}{2}\int \dd\omega_1 \dd\omega_2\, \abs{F^a(\omega_1,\omega_2)}^2\left(e^{i(\omega_1-\omega_2)\tau}+e^{-i(\omega_1-\omega_2)\tau}\right),\\
            &=\int \dd\omega_1 \dd\omega_2\, \abs{F^a(\omega_1,\omega_2)}^2 \cos\left((\omega_1-\omega_2)\tau\right),\\
            &=2\int \dd\omega_1 \dd\omega_2\, \abs{F^a(\omega_1,\omega_2)}^2 \cos^2\left(\tfrac{(\omega_1-\omega_2)\tau}{2}\right)\notag\\
            &\qquad-\int \dd\omega_1 \dd\omega_2 \,\abs{F^a(\omega_1,\omega_2)}^2.
        \end{align}
    \end{subequations}
    Summing both contributions while using the normalization relation
    \begin{equation}
        \int\dd\omega_1\dd\omega_2\, \abs{F^s(\omega_1,\omega_2)}^2 + \int\dd\omega_1\dd\omega_2\, \abs{F^a(\omega_1,\omega_2)}^2 = 1,
    \end{equation}
    we get
    \begin{equation}
        P_c=\int \dd\omega_1 \dd\omega_2\, \abs{F^s(\omega_1,\omega_2)}^2 \cos^2\left(\tfrac{(\omega_1+\omega_2)\tau}{2}\right)+\int \dd\omega_1 \dd\omega_2\, \abs{F^a(\omega_1,\omega_2)}^2 \cos^2\left(\tfrac{(\omega_1-\omega_2)\tau}{2}\right).
    \end{equation}
    \medskip

    \noindent
    $\blacktriangleright$ {\bf Computation of the FI.} As the state $\ket{\psi_\text{in}}$ is anti-symmetric, the precision for the estimation of $\tau$ at the origin is given by
    \begin{equation}
        \lim_{\tau\to 0} \mathcal F = \Delta^2_{\ket{\psi_\text{in}}}(\hat \omega_1 - \hat \omega_2).
    \end{equation}
    To compute the required expectation values notice that once again, as the operators $\hat\omega_1$ and $\hat\omega_2$ do not change the photon number distribution, the bunched and anti-bunched parts can be treated independently. We thus have, for $k=1,2$,
    \begin{equation}
        \bra{\psi_\text{in}}(\hat \omega_1 - \hat \omega_2)^k\ket{\psi_\text{in}} = \bra{\psi^b}(\hat \omega_1 - \hat \omega_2)^k\ket{\psi^b} + \bra{\psi^a}(\hat \omega_1 - \hat \omega_2)^k\ket{\psi^a}.
    \end{equation}
    For the bunched part
    \begin{subequations}
        \begin{align}
            (\hat \omega_1-\hat \omega_2)^k\ket{\psi^b}&=\frac{1}{2} \int \dd\omega_1 \dd\omega_2\, F^s(\omega_1,\omega_2) \left( (\hat \omega_1+\hat \omega_2)^k\hat a_1^\dagger(\omega_1) \hat a_1^\dagger(\omega_2) \right.\notag\\
            &\qquad\left.- (\hat \omega_1-\hat \omega_2)^k\hat a_2^\dagger(\omega_1) \hat a_2^\dagger(\omega_2) \right) \vac,\\
            &=\frac{1}{2} \int \dd\omega_1 \dd\omega_2\, F^s(\omega_1,\omega_2) \left( (\omega_1+\omega_2)^k\hat a_1^\dagger(\omega_1) \hat a_1^\dagger(\omega_2)\right.\notag\\
            &\qquad\left. - (-1)^k(\omega_1+\omega_2)^k\hat a_2^\dagger(\omega_1) \hat a_2^\dagger(\omega_2) \right) \vac,\\
            &=\frac{1}{2} \int \dd\omega_1 \dd\omega_2\, (\omega_1+\omega_2)^k F^s(\omega_1,\omega_2) \left( \hat a_1^\dagger(\omega_1) \hat a_1^\dagger(\omega_2) \right.\notag\\
            &\qquad\left.- (-1)^k\hat a_2^\dagger(\omega_1) \hat a_2^\dagger(\omega_2) \right) \vac.
        \end{align}
    \end{subequations}
    Leading to
    \begin{equation}
        \bra{\psi^b}(\hat \omega_1 - \hat \omega_2)^k\ket{\psi^b} = \frac{1}{2}\int \dd\omega_1 \dd\omega_2\, \abs{F^s(\omega_1,\omega_2)}^2 (\omega_1+\omega_2)^k(1+(-1)^k),
    \end{equation}
    and we have
    \begin{align}
        \bra{\psi^b}(\hat \omega_1 - \hat \omega_2)\ket{\psi^b} &= 0, & \bra{\psi^b}(\hat \omega_1 - \hat \omega_2)^2\ket{\psi^b} &= \expval{(\hat \omega_1+\hat \omega_2)^2}_s.
    \end{align}
    Similarly
    \begin{equation}
        (\hat\omega_1-\hat \omega_2)^k\ket{\psi^a}=\int \dd\omega_1 \dd\omega_2\, F^a(\omega_1,\omega_2) (\omega_1-\omega_2)^k \ket{\omega_1,\omega_2},
    \end{equation}
    leading to
    \begin{equation}
        \bra{\psi^a}(\hat \omega_1 - \hat \omega_2)^k\ket{\psi^a} = \int \dd\omega_1 \dd\omega_2\, \abs{F^a(\omega_1,\omega_2)}^2 (\omega_1-\omega_2)^k,
    \end{equation}
    and thus
    \begin{align}
        \bra{\psi^a}(\hat \omega_1 - \hat \omega_2)\ket{\psi^a} &= 0, & \bra{\psi^a}(\hat \omega_1 - \hat \omega_2)^2\ket{\psi^a} &= \expval{(\hat \omega_1-\hat \omega_2)^2}_a.
    \end{align}
    We thus get the final expression
    \begin{equation}
        \lim_{\tau\to 0} \mathcal F = \expval{(\hat \omega_1+\hat \omega_2)^2}_s+\expval{(\hat \omega_1-\hat \omega_2)^2}_a.
    \end{equation}
    \medskip

    \noindent
    $\blacktriangleright$ {\bf Computation of the QFI.} The quantum Fisher information is by definition 
    \begin{equation}
        \mathcal Q = 4\Delta^2_{\ket{\psi_\text{in}}}(\hat \omega_1).
    \end{equation}
    We can compute it similarly, expanding the bunched and anti-bunched parts independently. For the bunched part, we have
    \begin{equation}
        \hat\omega_1^k\ket{\psi^b}=\frac{1}{2} \int \dd\omega_1 \dd\omega_2\, F^s(\omega_1,\omega_2) (\omega_1+\omega_2)^k\hat a_1^\dagger(\omega_1) \hat a_1^\dagger(\omega_2) \vac,
    \end{equation}
    and thus
    \begin{equation}
        \bra{\psi^b}\hat\omega_1^k\ket{\psi^b} = \frac{1}{2}\int \dd\omega_1 \dd\omega_2\, \abs{F^s(\omega_1,\omega_2)}^2 (\omega_1+\omega_2)^k,
    \end{equation}
    leading to
    \begin{align}
        \bra{\psi^b}\hat\omega_1\ket{\psi^b} &= \frac{1}{2}\expval{\hat \omega_1+\hat \omega_2}_s,&
        \bra{\psi^b}\hat\omega_1^2\ket{\psi^b} &= \frac{1}{2}\expval{(\hat \omega_1+\hat \omega_2)^2}_s.
    \end{align}
    For the anti-bunched part
    \begin{equation}
        \hat\omega_1^k\ket{\psi^a}=\int \dd\omega_1 \dd\omega_2\, F^a(\omega_1,\omega_2) \omega_1^k \ket{\omega_1,\omega_2},
    \end{equation}
    leading to
    \begin{equation}
        \bra{\psi^a}\hat\omega_1^k\ket{\psi^a} = \int \dd\omega_1 \dd\omega_2\, \abs{F^a(\omega_1,\omega_2)}^2 \omega_1^k,
    \end{equation}
    and
    \begin{align}
        \bra{\psi^a}\hat\omega_1\ket{\psi^a} &= \expval{\hat \omega_1}_a, & \bra{\psi^a}\hat\omega_1^2\ket{\psi^a} &= \expval{\hat \omega_1^2}_a.
    \end{align}
    Finally, we have
    \begin{equation}
        \mathcal Q=4\Delta^2_{\ket{\psi_\text{in}}}(\hat \omega_1) = 2\expval{(\hat \omega_1+\hat \omega_2)^2}_s + 4\expval{\hat \omega_1^2}_a - \left(\expval{\hat \omega_1+\hat \omega_2}_s + 2\expval{\hat \omega_1}_a\right)^2.
    \end{equation}
\end{derivation}

\begin{result}[Permutation matrix diagonalisation]~\\
    \label{res: permutation diagonalisation}
    The three matrices
    \begin{align}
        P = \scalebox{0.9}{$\begin{pmatrix} 
            0 & 1 & & \\
            & \ddots & \ddots & \\
            & & \ddots & 1 \\
            1 & & & 0 
        \end{pmatrix}$}, && U=\frac{1}{\sqrt{n}}\scalebox{0.9}{$\begin{pmatrix}
            1 & 1 & \cdots & 1 \\
            1 & \omega & \cdots & \omega^{n-1} \\
            \vdots & \vdots & \ddots & \vdots \\
            1 & \omega^{n-1} & \cdots & \omega^{(n-1)(n-1)}
        \end{pmatrix}$}, && D=\scalebox{0.9}{$\begin{pmatrix}
            1 & & & \\
            & \omega & & \\
            & & \ddots & \\
            & & & \omega^{n-1}
        \end{pmatrix}$},
    \end{align}
    where $\omega = e^{2\pi i/n}$ is the primitive $n$-th root of unity, follow the matrix product relations
    \begin{align}
        P=UDU^\dagger, && P=U^\dagger D^\dagger U.
    \end{align}
\end{result}

\begin{derivation}
    Interpreting $P$ as a discrete translation operator with periodic boundary conditions, it is natural to seek its eigenvectors in the form of (discrete) plane waves. Define $e_j = (1, \omega^j, \omega^{2j}, \dots, \omega^{(n-1)j})^T$. A straightforward computation shows that
    \begin{equation}
        P e_j = P 
        \begin{pmatrix} 
            1 \\ \omega^j \\ \vdots \\ \omega^{(n-1)j} 
        \end{pmatrix} 
        = \begin{pmatrix} 
            \omega^j \\ \vdots \\ \omega^{(n-1)j} \\ 1 
        \end{pmatrix} 
        = \omega^j 
        \begin{pmatrix} 
            1 \\ \vdots \\ \omega^{(n-2)j} \\ \omega^{(n-1)j} 
        \end{pmatrix} 
        = \omega^j e_j.
    \end{equation}
    Thus, the vectors $(e_0, \dots, e_{n-1})$ form an eigenbasis of $P$. Normalizing by $1/\sqrt{n}$ and collecting the vectors in a single matrix, we recover $U = \frac{1}{\sqrt{n}} \begin{pmatrix} e_0 & e_1 & \cdots & e_{n-1} \end{pmatrix}$. We then obtain the diagonalization
    \begin{equation}
        P = U D U^{-1},
    \end{equation}
    where $D = \operatorname{diag}(1, \omega, \dots, \omega^{n-1})$. It is well known (or can be readily verified) that $U$ is unitary. We thus obtain
    \begin{equation}
        P = U D U^\dagger.
    \end{equation}
    Taking the complex conjugate (not the Hermitian conjugate) and noting that $U^* = U^\dagger$ and $D^* = D^\dagger$, we also find
    \begin{equation}
        P = U^\dagger D^\dagger U.
    \end{equation}
\end{derivation}

\begin{result}[Expectation value of $\hat D$ and photon number]~\\
    \label{res: D and photon number}
    For any general state over $n$ spatial modes $\ket{\psi}$ with arbitrary photon number distribution, we have the relation
    \begin{equation}
        \frac{1}{n}\sum_{l=0}^{n-1}\expval{\hat D^l}= \mathbb{P}\left(\sum_{k=0}^{n-1}km_k \equiv 0\,[n] \right),
    \end{equation}
    where $m_k$ is the number of photons in spatial mode $k$. Inserting the state through a DFT interferometer, the photon number distribution at the output and the input state symmetry are linked by
    \begin{equation}
        \mathbb{P}\left(\sum_{k=0}^{n-1}km_k \equiv 0\,[n] \right)= \frac{1}{n}\sum_{l=0}^{n-1} \expval{\hat P^l}_{\ket{\psi}}.
    \end{equation}
\end{result}

\begin{derivation}
    $\blacktriangleright$ {\bf Expectation values of $\hat D$.} Temporarily setting aside the dependence of the operator on $\lambda$, we observe that for integers $m_0,\dots,m_{n-1}$,
    \begin{subequations}
        \begin{align}
            \hat D (\hat a_0^\dagger)^{m_0}\cdots (\hat a_{n-1}^\dagger)^{m_{n-1}}\hat D^\dagger &= (\hat D\hat a_0^\dagger\hat D^\dagger)^{m_0}\cdots (\hat D\hat a_{n-1}^\dagger\hat D^\dagger)^{m_{n-1}}, \\
            &= (\omega^0\hat a_0^\dagger)^{m_0}\cdots (\omega^{n-1}\hat a_{n-1}^\dagger)^{m_{n-1}}, \\
            &= \omega^{\sum_{k=0}^{n-1}km_k}(\hat a_0^\dagger)^{m_0}\cdots (\hat a_{n-1}^\dagger)^{m_{n-1}}.
        \end{align}
    \end{subequations}
    This implies that if 
    \begin{equation}
        \ket{\psi} = \frac{1}{\sqrt{m_0!\cdots m_{n-1}!}}(\hat a_0^\dagger)^{m_0}\cdots (\hat a_{n-1}^\dagger)^{m_{n-1}}\vac,
    \end{equation}
    is a Fock state with a definite number of photons in each spatial mode, then
    \begin{subequations}
        \begin{align}
            \hat D\ket{\psi} &= \frac{1}{\sqrt{m_0!\cdots m_{n-1}!}}\hat D(\hat a_0^\dagger)^{m_0}\cdots (\hat a_{n-1}^\dagger)^{m_{n-1}}\hat D^\dagger \hat D\vac, \\
            &= \frac{\omega^{\sum_{k=0}^{n-1}km_k}}{\sqrt{m_0!\cdots m_{n-1}!}}(\hat a_0^\dagger)^{m_0}\cdots (\hat a_{n-1}^\dagger)^{m_{n-1}}\vac, \\
            &= \omega^{\sum_{k=0}^{n-1}km_k}\ket{\psi},
        \end{align}
    \end{subequations}
    so that $\expval*{\hat D} = \bra{\psi}\hat D\ket{\psi} = \omega^{\sum_{k=0}^{n-1}km_k}$.

    The extension to states including internal degrees of freedom is similar, though more cumbersome to write explicitly. A general state $\ket{\psi}$ with $m_j$ photons in mode $j$ and an arbitrary wavefunction is written as
    \begin{equation}\label{eq: equation n mode tf state with fixed nbr of photons}
        \ket{\psi} = \int \dd\{\lambda_{j,k}\}\, F(\{\lambda_{j,k}\})\prod_{j,k} \hat a_j(\lambda_{j,k})^\dagger\vac,
    \end{equation}
    where the integration is over all variables $\lambda_{j,k}$ with $j=0,\dots,n-1$ and $k=1,\dots, m_j$. The function $F$ depends on all these variables. Since the action of $\hat D$ is independent of the $\lambda_{j,k}$, only the number of creation operators in each spatial mode matters. Therefore, we again have $\hat D\ket{\psi} = \omega^{\sum_{k=0}^{n-1}km_k}\ket{\psi}$, and consequently $\expval*{\hat D} = \omega^{\sum_{k=0}^{n-1}km_k}$. Since $\hat D^l\ket{\psi} = \omega^{l\sum_{k=0}^{n-1}km_k}\ket{\psi}$, it follows that $\expval*{\hat D^l} = \omega^{l\sum_{k=0}^{n-1}km_k}$. Summing over $l=0,\dots,n-1$ and using the geometric series formula, we obtain
    \begin{equation}
        \frac{1}{n}\sum_{l=0}^{n-1}\expval*{\hat D^l} = \frac{1}{n}\sum_{l=0}^{n-1} \left(\omega^{\sum_{k=0}^{n-1}km_k}\right)^l = \left\{\begin{array}{cc}
            1 & \text{if } \sum_{k=0}^{n-1}km_k \equiv 0\, [n], \\
            0 & \text{otherwise}.
        \end{array}\right.
    \end{equation}

    The most general pure state over $n$ time-frequency modes can be written as
    \begin{equation}
        \ket{\psi} = \sum_{m_0,\dots,m_{n-1}=0}^\infty \alpha_{\vec{m}}\ket{\psi_{\vec{m}}},
    \end{equation}
    where $\ket{\psi_{\vec{m}}}$ is some state with $m_j$ photons in mode $j$, of the form given in Eq.~\eqref{eq: equation n mode tf state with fixed nbr of photons}, and $\alpha_{\vec{m}} \in \C$ with $\sum_{\vec{m}}\abs{\alpha_{\vec{m}}}^2 = 1$. As states with different photon numbers are orthogonal, $\braket{\psi_{\vec m}}{\psi_{\vec m'}} = \delta_{\vec m, \vec m'}$, and the action of $\hat D$ preserves photon number, we compute
    \begin{subequations}
        \allowdisplaybreaks
        \begin{align}
            \frac{1}{n}\sum_{l=0}^{n-1}\expval{\hat D^l}_{\ket{\psi}} &= \frac{1}{n}\sum_{l=0}^{n-1} \sum_{\vec m,\vec m'} \alpha_{\vec m}^*\alpha_{\vec m'}\bra{\psi_{\vec m}}\hat D^l\ket{\psi_{\vec m'}}, \\
            &= \frac{1}{n}\sum_{l=0}^{n-1} \sum_{\vec m} \abs{\alpha_{\vec m}}^2 \omega^{l\sum_{k=0}^{n-1}km_k}, \\
            &= \sum_{\vec m} \abs{\alpha_{\vec m}}^2 \delta_{\sum_{k=0}^{n-1}km_k \equiv 0\,[n]}, \\
            &= \sum_{\substack{\vec m \\ \sum_{k=0}^{n-1}km_k \equiv 0\,[n]}} \abs{\alpha_{\vec m}}^2,\\
            &= \mathbb{P}\left(\sum_{k=0}^{n-1}km_k \equiv 0\,[n] \right).
        \end{align}
    \end{subequations}
    \medskip

    \noindent
    $\blacktriangleright$ {\bf Interferometric formula.} We now insert the interferometer $\hat U$ between the state and number-resolving photon detection. Based on the result above and using the identity $\hat P = \hat U^\dagger \hat D^\dagger \hat U$, the probability that the measured photon numbers satisfy $\sum_{k=0}^{n-1} k m_k \equiv 0\,[n]$ is
    \begin{subequations}
        \begin{align}
            \mathbb{P}\left(\sum_{k=0}^{n-1}km_k \equiv 0\,[n] \right) &= \frac{1}{n}\sum_{l=0}^{n-1} \expval{\hat D^l}_{\hat U\ket{\psi}} = \frac{1}{n}\sum_{l=0}^{n-1} \bra{\psi} \hat U^\dagger \hat D^l \hat U \ket{\psi} \\
            &= \frac{1}{n}\sum_{l=0}^{n-1} \bra{\psi} \hat U^\dagger (\hat D^\dagger)^{-l} \hat U \ket{\psi} = \frac{1}{n}\sum_{l=0}^{n-1} \bra{\psi} \hat P^{-l} \ket{\psi} \\
            &= \frac{1}{n}\sum_{l=0}^{n-1} \expval{\hat P^l}_{\ket{\psi}},
        \end{align}
    \end{subequations}
    where we used $\hat P^n = \1$ and the change of index $l \to n - l$.
\end{derivation}

\begin{result}[Probabilities and eigenspaces of $\hat P$]~\\
    \label{res: proba fourier other}
    Inputting a general state $\ket{\psi}$ into an $n$-mode DFT interferometer and measuring the number of photons $m_k$ at the output of each mode, we have
    \begin{equation}
        \mathbb{P}\left(\sum_{k=0}^{n-1} k m_k \equiv -j \,[n]\right) 
        = \expval{\hat \Pi_j}_{\ket{\psi}},
    \end{equation}
    where 
    \begin{equation}
        \hat \Pi_j = \frac{1}{n} \sum_{l=0}^{n-1} (\omega^{-j} \hat P)^l,
    \end{equation}
    is the projector onto the eigenspace of $\hat P$ with eigenvalue $\omega^j$.
\end{result}

\begin{derivation}
    As $\hat P$ is conjugated to the diagonal operator $\hat D$, its eigenvalues are precisely $1, \omega, \dots, \omega^{n-1}$. If $\ket{\psi}$ is an eigenstate of $\hat P$ with eigenvalue $\omega^\ell$, we have
    \begin{equation}
        \hat \Pi_j \ket{\psi} = \frac{1}{n} \sum_{l=0}^{n-1} (\omega^{-j} \hat P)^l \ket{\psi} 
        = \frac{1}{n} \sum_{l=0}^{n-1} (\omega^{-j} \omega^\ell)^l \ket{\psi}
        = \left\{
        \begin{array}{cl}
            \ket{\psi}, & \text{if } j = \ell, \\
            0, & \text{otherwise}.
        \end{array}
        \right.
    \end{equation}

    Now consider an initial state $\ket{\psi}$ for which $\ket{\varphi} = \hat U \ket{\psi}$ has exactly $m_k$ photons in output mode $k$. Then,
    \begin{subequations}
        \begin{align}
            \expval{\hat \Pi_j}_{\ket{\psi}} 
            &= \frac{1}{n} \sum_{l=0}^{n-1} \omega^{-j l} \bra{\psi} \hat P^l \ket{\psi} 
            = \frac{1}{n} \sum_{l=0}^{n-1} \omega^{-j l} \bra{\psi} \hat U^\dagger \hat U \hat P^l \hat U^\dagger \hat U \ket{\psi}, \\
            &= \frac{1}{n} \sum_{l=0}^{n-1} \omega^{-j l} \bra{\varphi} (\hat D^\dagger)^l \ket{\varphi} 
            = \frac{1}{n} \sum_{l=0}^{n-1} \omega^{-j l} \left( \omega^{-\sum_{k=0}^{n-1} k m_k} \right)^l, \\
            &= \frac{1}{n} \sum_{l=0}^{n-1} \left( \omega^{-j - \sum_{k=0}^{n-1} k m_k} \right)^l, \\
            &= \left\{
            \begin{array}{cl}
                1, & \text{if } \sum_{k=0}^{n-1} k m_k \equiv -j \,[n], \\
                0, & \text{otherwise}.
            \end{array}
            \right.
        \end{align}
    \end{subequations}
    Following the reasoning presented in Result~\ref{res: D and photon number}, this implies that
    \begin{equation}
        \mathbb{P}\left(\sum_{k=0}^{n-1} k m_k \equiv -j \,[n]\right) 
        = \expval{\hat \Pi_j}_{\ket{\psi}},
    \end{equation}
    demonstrating that values of $\sum_{k=0}^{n-1} k m_k$ modulo $n$ probe the squared norm of the component of $\ket{\psi}$ in the other eigenspaces of $\hat P$.
\end{derivation}

\begin{result}[$n$-mode HOM Fisher information]~\\
    \label{res: n-mode HOM Fisher info}
    Consider an initial probe state $\ket{\psi}$ fed into a generalized $n$-mode HOM interferometer, with evolution $\hat V(\theta)=e^{-i\hat H\theta}$ followed by a DFT interferometer and a measurement of the photon numbers $m_k$ selecting the outcomes where $\sum_{k=0}^{n-1} k m_k \equiv 0\,[n]$. If $\hat \Pi\ket{\psi}=\ket{\psi}$ then the estimation around zero is given by
    \begin{equation}
        \lim_{\theta\to 0} \mathcal F = 4\Delta^2(i[\hat H,\hat \Pi])=\frac{4}{n^2} \Delta^2\left(n \hat{H} - \sum_{l=0}^{n-1} \hat{P}^l \hat{H} \hat{P}^{-l} \right).
    \end{equation}
    If $\hat \Pi\ket{\psi}=0$, the precision is given by 
    \begin{equation}
        \lim_{\theta\to 0} \mathcal F = 4\Delta^2(i[\hat H,\hat \Pi]),
    \end{equation}
    which if $\hat P\ket{\psi}=-\ket{\psi}$ can be rewritten as
    \begin{equation}
        \lim_{\theta\to 0} \mathcal F =\frac{4}{n^2} \Delta^2\left(\sum_{l=0}^{n-1} (-1)^l \hat{P}^l \hat{H} \hat{P}^{-l}\right).
    \end{equation}
\end{result}

\begin{derivation}
    $\blacktriangleright$ {\bf General expansion result.} The derivation is based on Taylor expanding the probabilities and the result, already used in Result~\ref{res: Fisher info with S} for a binary measurement with probabilities $p(\theta)$ and $q(\theta)$, where $p(0) = 1$. More generally, we assume that the probabilities can be expanded near $\theta = 0$ as
    \begin{equation}
        p(\theta) = 1 - \theta^2 A + o(\theta^2), \quad q(\theta) = \theta^2 A + o(\theta^2)
    \end{equation}
    for some arbitrary non-negative constant $A$. In this case, for $\theta \neq 0$ and assuming differentiability of the Taylor expansion, the FI is given by
    \begin{subequations}
        \begin{align}
            \mathcal{F} &= \frac{1}{p(\theta)}\left(\frac{\dd p(\theta)}{\dd\theta}\right)^2 + \frac{1}{q(\theta)}\left(\frac{\dd q(\theta)}{\dd\theta}\right)^2 \\
            &= \frac{(p'(\theta))^2}{p(\theta)(1 - p(\theta))} \\
            &= \frac{(2\theta A + o(\theta))^2}{(\theta^2 A + o(\theta^2))(1 - \theta^2 A + o(\theta^2))} \\
            &= \frac{4\theta^2 A^2 + o(\theta^2)}{\theta^2 A + o(\theta^2)} \\
            &= 4A + o(1)
        \end{align}
    \end{subequations}
    so that
    \begin{equation}
        \lim_{\theta \to 0} \mathcal{F}(\theta) = 4A.
    \end{equation}
    The precision is thus directly proportional to the parameter $A$ defined in the expansion above. 
    \medskip

    \noindent
    $\blacktriangleright$ {\bf Expansion of the probability.} We begin by considering a state arriving at the DFT interferometer of the form $e^{-i\theta \hat{H}} \ket{\psi}$ and expand the expression of the measurement probability
    \begin{subequations}\label{eq: expansion second order pure}
        \begin{align}
            \mathbb{P}\left[\sum_{k=0}^{n-1} k m_k \equiv 0\,[n] \,\middle|\, \theta\right] 
            &= \frac{1}{n}\sum_{l=0}^{n-1} \bra{\psi} e^{i\theta \hat{H}} \hat{P}^l e^{-i\theta \hat{H}} \ket{\psi},\\
            &= \bra{\psi} e^{i\theta \hat{H}} \hat{\Pi} e^{-i\theta \hat{H}} \ket{\psi} ,\\
            &= \expval{ \left(1 + i\theta \hat{H} - \tfrac{\theta^2}{2} \hat{H}^2 \right) \hat{\Pi} \left(1 - i\theta \hat{H} - \tfrac{\theta^2}{2} \hat{H}^2 \right) } + o(\theta^2), \\
            &= \expval{ \hat{\Pi} + i\theta ( \hat{H}\hat{\Pi} -  \hat{\Pi}\hat{H}) - \tfrac{\theta^2}{2} \left( \hat{H}^2 \hat{\Pi} + \hat{\Pi} \hat{H}^2 - 2\hat{H} \hat{\Pi} \hat{H} \right)}\notag\\
            &\qquad + o(\theta^2), \\
            &= \expval{\hat{\Pi}} + i\theta \left( \expval{ \hat{H}\hat{\Pi}} - \expval{ \hat{\Pi}\hat{H}} \right) - \tfrac{\theta^2}{2} \left( \expval{\hat{H}^2 \hat{\Pi}} + \expval{\hat{\Pi} \hat{H}^2}\right.\notag\\
            &\qquad \left.- 2\expval{\hat{H} \hat{\Pi} \hat{H}} \right) + o(\theta^2).
        \end{align}
    \end{subequations}
    We want this equation to take the form $p(\theta)$ or $q(\theta)$, which requires that $\expval{\hat{\Pi}} = 1$ or $0$. As the computations differ slightly, we treat each case separately.
    \medskip

    \noindent
    $\blacktriangleright$ {\bf Symmetric case.} Let's first assume that $\expval{\hat{\Pi}} = 1$. Since $\hat{\Pi}$ is a projector, this implies that $\hat{\Pi} \ket{\psi} = \ket{\psi}$. As such, we can simplify the previous expression using the fact that each occurrence of $\hat{\Pi}$ to the left or right of an expectation value can be removed. We obtain
    \begin{equation}
        \mathbb{P}\left[\sum_{k=0}^{n-1} k m_k \equiv 0\,[n] \,\middle|\, \theta\right] = 1 - \theta^2 \left( \expval{\hat{H}^2} - \expval{\hat{H} \hat{\Pi} \hat{H}} \right) + o(\theta^2).
    \end{equation}
    We observe that the term $\expval*{\hat{H}^2} - \expval*{\hat{H} \hat{\Pi} \hat{H}}$ can be expressed as the variance of two different operators
    \begin{equation}\label{eq: second order for sym}
        \expval*{\hat{H}^2} - \expval*{\hat{H} \hat{\Pi} \hat{H}} = \Delta^2(i[\hat{H}, \hat{\Pi}]) = \frac{1}{n^2} \Delta^2\left(n \hat{H} - \sum_{l=0}^{n-1} \hat{P}^l \hat{H} \hat{P}^{-l} \right).
    \end{equation}
    Let us first verify the left-hand identity. The commutator of two Hermitian operators is anti-Hermitian, hence, we use $i[\hat{H}, \hat{\Pi}]$ to obtain a Hermitian operator. Since $\hat{\Pi} \ket{\psi} = \ket{\psi}$, we find:
    \begin{equation}
        \expval{i[\hat{H}, \hat{\Pi}]} = i\expval{\hat{H} \hat{\Pi}} - i\expval{\hat{\Pi} \hat{H}} = i\expval{\hat{H}} - i\expval{\hat{H}} = 0.
    \end{equation}
    Then,
    \begin{subequations}
        \begin{align}
            \expval{(i[\hat{H}, \hat{\Pi}])^2} &= -\expval{(\hat{H} \hat{\Pi} - \hat{\Pi} \hat{H})^2}, \\
            &= -\expval{ \hat{H} \hat{\Pi} \hat{H} \hat{\Pi} - \hat{H} \hat{\Pi}^2 \hat{H} - \hat{\Pi} \hat{H}^2 \hat{\Pi} + \hat{\Pi} \hat{H} \hat{\Pi} \hat{H} }, \\
            &= -\expval{\hat{H} \hat{\Pi} \hat{H}} + \expval{\hat{H} \hat{\Pi} \hat{H}} + \expval{\hat{H}^2} - \expval{\hat{H} \hat{\Pi} \hat{H}}, \\
            &= \expval{\hat{H}^2} - \expval{\hat{H} \hat{\Pi} \hat{H}}.
        \end{align}
    \end{subequations}
    To verify the second expression, we use the identity $\hat{P} \ket{\psi} = \ket{\psi}$, which implies that we can remove $\hat{P}$ when it appears on the left or right of an expectation value
    \begin{equation}
        \expval{n\hat{H} - \sum_{l=0}^{n-1} \hat{P}^l \hat{H} \hat{P}^{-l}} = n \expval{\hat{H}} - \sum_{l=0}^{n-1} \expval{\hat{H}} = 0.
    \end{equation}
    Then,
    \begin{subequations}
        \begin{align}
            \expval{\left(n\hat{H} - \sum_{l=0}^{n-1} \hat{P}^l \hat{H} \hat{P}^{-l} \right)^2} &= n^2 \expval{\hat{H}^2} - 2n \sum_{l=0}^{n-1} \expval{\hat{H} \hat{P}^{-l} \hat{H}} + \sum_{k,l=0}^{n-1} \expval{\hat{H} \hat{P}^{l-k} \hat{H}}, \\
            &= n^2 \expval{\hat{H}^2} - 2n \sum_{l=0}^{n-1} \expval{\hat{H} \hat{P}^{-l} \hat{H}} + n \sum_{l=0}^{n-1} \expval{\hat{H} \hat{P}^{l} \hat{H}} ,\\
            &= n^2 \left( \expval{\hat{H}^2} - \expval{\hat{H} \hat{\Pi} \hat{H}} \right).
        \end{align}
    \end{subequations}
    \medskip

    \noindent
    $\blacktriangleright$ {\bf Non-symmetric case.} Now assume $\expval{\hat{\Pi}} = 0$. Since $\hat{\Pi}$ is a projector, this implies $\hat{\Pi} \ket{\psi} = 0$. Thus, any expectation value involving $\hat{\Pi}$ at either end vanishes, leading to
    \begin{equation}
        \mathbb{P}\left[\sum_{k=0}^{n-1} k m_k \equiv 0\,[n] \,\middle|\, \theta\right] = \theta^2 \expval*{\hat{H} \hat{\Pi} \hat{H}} + o(\theta^2).
    \end{equation}
    As before, the quadratic term can be expressed as a variance
    \begin{equation}\label{eq: second order non sym}
        \expval*{\hat{H} \hat{\Pi} \hat{H}} = \Delta^2(i[\hat{H}, \hat{\Pi}]).
    \end{equation}
    Indeed, 
    \begin{equation}
        \expval{i[\hat{H}, \hat{\Pi}]} = i\expval{\hat{H} \hat{\Pi}} - i\expval{\hat{\Pi} \hat{H}} = 0,
    \end{equation}
    and
    \begin{align}
        \expval{(i[\hat{H}, \hat{\Pi}])^2} &= -\expval{(\hat{H} \hat{\Pi} - \hat{\Pi} \hat{H})^2}= \expval{\hat{H} \hat{\Pi} \hat{H}}.
    \end{align}

    To derive a second formula similar to the case $\hat{\Pi} \ket{\psi} = \ket{\psi}$, we assume $\hat P\ket{\psi}=-\ket{\psi}$ (hence $n$ is even) and show that
    \begin{equation}
        \frac{1}{n^2} \Delta^2\left(\sum_{l=0}^{n-1} (-1)^l \hat{P}^l \hat{H} \hat{P}^{-l}\right) = \expval{\hat{H} \hat{\Pi} \hat{H}}.
    \end{equation}
    Using the anti-symmetry condition:
    \begin{equation}
        \expval{\sum_{l=0}^{n-1} (-1)^l\hat{P}^l \hat{H} \hat{P}^{-l}} = \sum_{l=0}^{n-1} (-1)^l (-1)^l (-1)^l \expval{\hat{H}} = \expval{\hat{H}} \sum_{l=0}^{n-1} (-1)^l = 0,
    \end{equation}
    and
    \begin{subequations}
        \begin{align}
            \expval{\left(\sum_{l=0}^{n-1} (-1)^l \hat{P}^l \hat{H} \hat{P}^{-l}\right)^2} &= \sum_{k,l=0}^{n-1} (-1)^{k+l} \expval{ \hat{P}^k \hat{H} \hat{P}^{l-k} \hat{H} \hat{P}^{-l} }, \\
            &= \sum_{k,l=0}^{n-1} (-1)^{k+l+k+l} \expval{ \hat{H} \hat{P}^{k-l} \hat{H} }, \\
            &= n \sum_{l=0}^{n-1} \expval{\hat{H} \hat{P}^l \hat{H}} = n^2 \expval{\hat{H} \hat{\Pi} \hat{H}}.
        \end{align}
    \end{subequations}
\end{derivation}

\begin{result}[Extremal values of $\Tr(\hat \rho \hat P)$]~\\
    \label{res: extremal symmetry}
    For a mixed state $\hat \rho$ decomposed as a pure state superposition
    \begin{equation}
        \hat \rho = \sum_j p_j \ket{\psi_j}\bra{\psi_j},
    \end{equation}
    with only nonzero $p_j$, we have $\Tr(\hat \rho\hat P)=e^{i\phi}$ if and only if for all $j$
    \begin{equation}
        \hat P\ket{\psi_j} = e^{i\phi}\ket{\psi_j} .
    \end{equation}
\end{result}

\begin{derivation}
    The reverse result is obvious, so let us assume $\Tr(\hat \rho\hat P)=e^{i\phi}$. It is then natural to expect that all pure-state components of $\hat\rho$ satisfy $\bra{\psi_j}\hat P\ket{\psi_j}=e^{i\phi}$. By a clever use of the Cauchy-Schwarz inequality, we get
    \begin{equation}
        1=\abs{\sum_j p_j \bra{\psi_j}\hat P\ket{\psi_j}}^2\leq \sum_k p_k\sum_j p_j \abs{\bra{\psi_j}\hat P\ket{\psi_j}}^2\leq1,
    \end{equation}
    where we have used $\sum_j p_j=1$ and $\abs{\bra{\psi_j}\hat P\ket{\psi_j}}\leq 1$ since the eigenvalues of $\hat P$ are roots of unity, as $\hat P^n=\1$. The condition for equality in the Cauchy-Schwarz inequality implies that the following vectors are proportional
    \begin{equation}
        \begin{pmatrix}
            \sqrt{p_1}\\
            \vdots\\
            \sqrt{p_n}
        \end{pmatrix}\propto \begin{pmatrix}
            \sqrt{p_1}\bra{\psi_1}\hat P\ket{\psi_1}\\
            \vdots\\
            \sqrt{p_n}\bra{\psi_n}\hat P\ket{\psi_n}
        \end{pmatrix}.
    \end{equation}
    As all $p_j$ are assumed to be nonzero, this implies that all $\bra{\psi_j}\hat P\ket{\psi_j}$ are equal, necessarily to $e^{i\phi}$.
\end{derivation}

\begin{result}[Mixed-state extension of the metrological bounds]~\\
    \label{res: metrology mixed state extension}
    Consider an initial mixed state $\hat\rho$, with a pure-state decomposition
    \begin{equation}
        \hat \rho = \sum_j p_j \ket{\psi_j}\bra{\psi_j},
    \end{equation}
    evolved through $e^{-i\hat H\theta}$, fed into a DFT interferometer and measured with number-resolving photon detection, where the probability
    \begin{equation}
        \mathbb{P}\left[\sum_{k=0}^{n-1} k m_k \equiv 0\,[n] \,\middle|\, \theta\right],
    \end{equation}
    is experimentally measured. Assuming $\Tr(\hat \rho\hat\Pi)=1$ or $0$, the precision of estimating $\theta$ around $0$ is given by
    \begin{equation}
        \lim_{\theta\to0}\mathcal F=4\sum_j p_j \Delta^2_{\ket{\psi_j}}(i[\hat H,\hat \Pi]).
    \end{equation}
    Assuming perfect symmetry, $\Tr(\hat\rho\hat P)=1$, the precision can be rewritten as
    \begin{equation}
        \mathcal F= \mathcal Q\left(\hat H-\frac{1}{n}\sum_{l=0}^{n-1} \hat P^l\hat H\hat P^{-l}\right),
    \end{equation}
    where $\mathcal Q(\hat H')$ denotes the QFI in the situation where the operator $\hat H'$ is used for the evolution instead of $\hat H$. Assuming perfect anti-symmetry, $\Tr(\hat\rho\hat P)=-1$, the precision can be rewritten as
    \begin{equation}
        \mathcal F= \mathcal Q\left(\frac{1}{n}\sum_{l=0}^{n-1}(-1)^l \hat P^l\hat H\hat P^{-l}\right).
    \end{equation}
\end{result}

\begin{derivation}
    Since Eq.~\eqref{eq: fourier proba mixed states} is linear in the quantum state, we can use the pure-state expansions of Eq.~\eqref{eq: expansion second order pure} to develop the probability to second-order in $\theta$:
    \begin{align}
        \mathbb{P}\left[\sum_{k=0}^{n-1} k m_k \equiv 0\,[n] \,\middle|\, \theta\right] 
            &= \sum_j p_j \Bigg[\expval{\hat{\Pi}}_{\ket{\psi_j}} + i\theta \left( \expval{\hat{H}\hat{\Pi} }_{\ket{\psi_j}} - \expval{ \hat{\Pi}\hat{H}}_{\ket{\psi_j}} \right)\notag\\
            &\quad- \tfrac{\theta^2}{2} \left( \expval{\hat{H}^2 \hat{\Pi}}_{\ket{\psi_j}} + \expval{\hat{\Pi} \hat{H}^2}_{\ket{\psi_j}} - 2\expval{\hat{H} \hat{\Pi} \hat{H}}_{\ket{\psi_j}} \right) + o(\theta^2)\Bigg].
    \end{align}
    Following the same reasoning as in Result~\ref{res: n-mode HOM Fisher info}, we require the constant term to be either $0$ or $1$, which means that $\Tr(\hat \rho\hat \Pi)=1$ or $\Tr(\hat \rho\hat \Pi)=0$. In both cases, using Eq.~\eqref{eq: second order for sym} and Eq.~\eqref{eq: second order non sym}, we obtain
    \begin{equation}
        \lim_{\theta\to0}\mathcal F=4\sum_j p_j \Delta^2_{\ket{\psi_j}}(i[\hat H,\hat \Pi]).
    \end{equation}
    We can provide a more physically concrete formula if we make further assumptions on the symmetry of $\hat \rho$. If all pure components of $\hat\rho$ are symmetric, which is equivalent (as argued in Result~\ref{res: extremal symmetry}) to $\Tr(\hat \rho\hat P)=1$, we get
    \begin{equation}
        \mathcal F=\frac{4}{n^2}\sum_j p_j \Delta^2_{\ket{\psi_j}}\left(n\hat H-\sum_{l=0}^{n-1} \hat P^l\hat H\hat P^{-l}\right).
    \end{equation}
    Conversely, if all pure components of $\hat\rho$ are anti-symmetric, which is equivalent to $\Tr(\hat \rho\hat P)=-1$, we have
    \begin{equation}
        \mathcal F=\frac{4}{n^2}\sum_j p_j \Delta^2_{\ket{\psi_j}}\left(\sum_{l=0}^{n-1} (-1)^l\hat P^l\hat H\hat P^{-l}\right).
    \end{equation}
    Knowing that the QFI is convex, we obtain
    \begin{align}\label{eq: ineq precision mixed state}
        \mathcal F\geq \mathcal Q\left(\hat H-\frac{1}{n}\sum_{l=0}^{n-1} \hat P^l\hat H\hat P^{-l}\right) &&\text{or} && \mathcal F\geq \mathcal Q\left(\frac{1}{n}\sum_{l=0}^{n-1}(-1)^l \hat P^l\hat H\hat P^{-l}\right),
    \end{align}
    depending on whether $\Tr(\hat \rho\hat P)$ equals $1$ or $-1$, where $\mathcal Q(\hat H')$ denotes the QFI in the situation where the operator $\hat H'$ is used for the evolution instead of $\hat H$. As we now argue, we have equality in both equations above. Define $\hat H'$ as $\hat H-\frac{1}{n}\sum_{l=0}^{n-1} \hat P^l\hat H\hat P^{-l}$ or $\frac{1}{n}\sum_{l=0}^{n-1}(-1)^l \hat P^l\hat H\hat P^{-l}$, depending on whether we are considering the first or the second case. We can consider the alternative metrological situation where $\hat H'$ is used for the evolution instead of $\hat H$. Applying Eq.~\eqref{eq: ineq precision mixed state} with $\hat H'$ instead of $\hat H$ bounds the FI as $\mathcal F(\hat H')\geq \mathcal Q(\hat H')$. Since the QFI gives the optimal precision obtainable, we necessarily have the reverse inequality $\mathcal F(\hat H')\leq \mathcal Q(\hat H')$. With a straightforward computation, we can verify that in both cases, $\mathcal F(\hat H')=\mathcal F(\hat H)$. Consequently, the precision of the protocol when the original generator $\hat H$ is used is
    \begin{align}\label{eq: eq precision mixed state}
        \mathcal F= \mathcal Q\left(\hat H-\frac{1}{n}\sum_{l=0}^{n-1} \hat P^l\hat H\hat P^{-l}\right) &&\text{or} && \mathcal F= \mathcal Q\left(\frac{1}{n}\sum_{l=0}^{n-1}(-1)^l \hat P^l\hat H\hat P^{-l}\right),
    \end{align}
    depending on whether $\Tr(\hat \rho\hat P)$ equals $1$ or $-1$. These formulas can be analyzed in the same spirit as those valid for pure states. In particular, if $\hat H$ is anti-symmetric under cyclic permutation ($\hat P\hat H\hat P^\dagger=-\hat H$), the measurement device is optimal with $\mathcal F=\mathcal Q$.
\end{derivation}
\fi

\ifnum \theShowChapfive=1
\ifthenelse{
    \value{ShowChapone}=1 \OR
    \value{ShowChaptwo}=1 \OR
    \value{ShowChapthree}=1 \OR
    \value{ShowChapfour}=1
}{
\clearpage
}{}
\section{Superselection rules}
\setcurrentanchor{app SSR}
\emph{This section of the appendix collects all the results of Chap.~\ref{chap: SSR} and provides the associated proofs.}

\vspace*{1em}
\par\noindent
\textbf{\large Results in this section}\par
\vspace{-0.8em}
\noindent\rule{\textwidth}{0.8pt}\par

\resultentry{res: angular momentum and rotation}{Angular momentum operators and rotations}
\resultentry{res: roation creation op}{Rotation of creation operators}
\resultentry{res: spherical stereographic}{Stereographic projection in spherical coordinates}
\resultentry{res: cartesian stereographic}{Stereographic projection in Cartesian coordinates in dimension $n$}
\resultentry{res: reflections and stereographic}{Reflections and stereographic projection}
\resultentry{res: möbius and SU2}{Möbius transformation and $SU_2(\C)$}
\resultentry{res: möbius and rotation}{Möbius transformations and rotations of the sphere}
\resultentry{res: def spin coherent states}{Equivalent definitions of spin coherent states}
\resultentry{res: spin coherent rotation}{Spin coherent states and rotations}
\resultentry{res: scalar product spin coherent}{Scalar product of spin coherent states}
\resultentry{res: closure relation spin coherent}{Closure relation of spin coherent states}
\resultentry{res: rotation coherent state}{Rotation of spin coherent states}
\resultentry{res: reconstructing state from Majo}{Reconstructing a state from its Majorana polynomial}
\resultentry{res: scalar product from majo}{Scalar product expressed with the Majorana Polynomial}
\resultentry{res: Majo and rotations}{Rotation of Majorana polynomials}
\resultentry{res: Majo stars and state decomposition}{Majorana stars and state decomposition}
\resultentry{res: heisenberg robertson}{Heisenberg-Robertson uncertainty relation}
\resultentry{res: variance of arbitrary spin component}{Expectation value and variance of an arbitrary spin component in a spin coherent state}
\resultentry{res: SSRC to CV coherent states}{SSRC to CV coherent states}
\resultentry{res: SSRC rotation to CV displacement}{SSRC rotation to CV displacement}
\resultentry{res: SNG required for local gates}{Necessity of non-Gaussian operations for logical single-qubit gates}
\resultentry{res: SNG required for cnot gates}{Gaussian SSRC operations cannot be used to implement entangling two-qubit gates}
\resultentry{res: optimization SSRC measurement direction}{Optimization of SSRC measurement direction}

\vspace{-0.2em}
\par\noindent\rule{\textwidth}{0.8pt}\par

\label{app: SSRC}

\begin{result}[Angular momentum operators and rotations]~\\
    \label{res: angular momentum and rotation}
    For $\hat J_x$, $\hat J_y$, and $\hat J_z$ which satisfy the canonical angular momentum commutation relations, $\vec n$ of unit norm and $\theta\in\R$, we have
    \begin{equation}
        e^{-i\theta \vec{\hat J}\cdot\vec{n}}\vec{\hat J}e^{i\theta \vec{\hat J}\cdot\vec{n}}=\vec{n}(\vec{n}\cdot\vec{\hat J})+\cos(\theta)\left[\vec{\hat J}-\vec{n}(\vec{n}\cdot\vec{\hat J})\right]-\sin(\theta)\vec{n}\times \vec{\hat J}.
    \end{equation}
    Taking the scalar product with a unit vector $\vec m$, the angular momentum operator $\vec m\cdot\vec{\hat J}$ are transformed into another angular momentum operator $\vec m'\cdot\vec{\hat J}$ with
    \begin{equation}
        \vec m'=\vec{n}(\vec{n}\cdot\vec m)+\cos(\theta)\left[\vec m-\vec{n}(\vec{n}\cdot\vec m)\right]+\sin(\theta)\vec{n}\times \vec m.
    \end{equation}
\end{result}

\begin{derivation}
    $\blacktriangleright$ {\bf Transformation of $\vec{\hat J}$.} Deriving this formula from scratch is quite cumbersome as it requires solving a first-order vector differential equation, by diagonalizing a 3 by 3 matrix and would provide an expression not as compact and intuitive as the one expressed with the dot and cross products. However, given the formula, it is quite simple to verify that it indeed holds. We introduce two functions that correspond to both sides of the equality and we show that they are equal by virtue of the Cauchy-Lipschitz theorem. We define
    \begin{align}
        \vec{\varphi}(\theta)=e^{-i\theta \vec{\hat J}\cdot\vec{n}}\vec{\hat J}e^{i\theta \vec{\hat J}\cdot\vec{n}},&& \vec{\psi}(\theta)=\vec{n}(\vec{n}\cdot\vec{\hat J})+\cos(\theta)\left[\vec{\hat J}-\vec{n}(\vec{n}\cdot\vec{\hat J})\right]-\sin(\theta)\vec{n}\times \vec{\hat J}.
    \end{align}
    We first differentiate $\vec{\varphi}$ to get
    \begin{equation}
            \vec{\varphi}'(\theta)=-ie^{-i\theta \vec{\hat J}\cdot\vec{n}}(\vec{n}\cdot\vec{\hat J})\vec{\hat J}e^{i\theta \vec{\hat J}\cdot\vec{n}}+ie^{-i\theta \vec{\hat J}\cdot\vec{n}}\vec{\hat J}(\vec{n}\cdot\vec{\hat J})e^{i\theta \vec{\hat J}\cdot\vec{n}}=-ie^{-i\theta \vec{\hat J}\cdot\vec{n}}[(\vec{n}\cdot\vec{\hat J}),\vec{\hat J}]e^{i\theta \vec{\hat J}\cdot\vec{n}}.
    \end{equation}
    Using the compact expression of the commutator in terms of the Levi-Civita symbol, $[\hat J_a,\hat J_b]=i\epsilon_{abc}\hat J_c$, we can provide a compact expression (using Einstein summation convention)
    \begin{equation}
        [(\vec{n}\cdot\vec{\hat J}),\vec{\hat J}]_a=[n_b\hat J_b,\hat J_a]=i n_b\epsilon_{bac}\hat J_c=-i\epsilon_{abc}n_b\hat J_c=-i(\vec{n}\times \vec{\hat J})_a,
    \end{equation}
    where the components of a cross product can be written in the compact way $(\vec{n}\times \vec{m})_a=\epsilon_{abc}n_bm_c$. Thus
    \begin{equation}
        \vec{\varphi}'(\theta)=-e^{-i\theta \vec{\hat J}\cdot\vec{n}}(\vec{n}\times \vec{\hat J})e^{i\theta \vec{\hat J}\cdot\vec{n}}=-\vec{n}\times (e^{-i\theta \vec{\hat J}\cdot\vec{n}}\vec{\hat J}e^{i\theta \vec{\hat J}\cdot\vec{n}})=-\vec{n}\times \vec{\varphi}(\theta).
    \end{equation}
    Which is a first-order linear differential equation. We thus want to show that $\vec{\psi}$ satisfies the same equation. On the one hand we have
    \begin{equation}
        \vec{\psi}'(\theta)=-\sin(\theta)\left[\vec{\hat J}-\vec{n}(\vec{n}\cdot\vec{\hat J})\right]-\cos(\theta)\vec n\times \vec{\hat J}.
    \end{equation}
    And on the other we verify that
    \begin{subequations}
        \begin{align}
            -\vec n\times\vec{\psi(\theta)}&=-\cos(\theta)\vec n\times \vec{\hat J}+\cos(\theta)(\vec n\times \vec n)(\vec n\cdot\vec{\hat J})+\sin(\theta)\vec n\times(\vec n\times \vec{\hat J}),\\
            &=\sin(\theta)\left[\vec n(\vec n\cdot\vec{\hat J})-\vec{\hat J}(\vec n\cdot \vec n)\right]-\cos(\theta)\vec n\times \vec{\hat J}=\vec{\psi}'(\theta),
        \end{align}
    \end{subequations}
    where we use the cross product properties $\vec n\times \vec n=0$ and $\vec a\times(\vec b\times\vec c)=\vec b(\vec a\cdot \vec c)-\vec c(\vec a\cdot \vec b)$ and the normalisation $\vec n\cdot \vec n=1$. As we can easily check that $\vec{\varphi}(0)=\vec{\psi}(0)=\vec{\hat J}$, the Cauchy-Lipschitz theorem guarantees that
    \begin{equation}
        \vec{\varphi}=\vec{\psi},
    \end{equation}
    which concludes the proof.
    \medskip

    \noindent
    $\blacktriangleright$ {\bf Transformation of $\vec m$.} The transformation rule for $\vec m'$ can be obtained by taking the scalar product of both sides of the transformation rule for $\vec{\hat J}$
    \begin{equation}
        e^{-i\theta \vec{\hat J}\cdot\vec{n}}(\vec m\cdot\vec{\hat J})e^{i\theta \vec{\hat J}\cdot\vec{n}}=\vec m\cdot\left[\vec{n}(\vec{n}\cdot\vec{\hat J})+\cos(\theta)\left[\vec{\hat J}-\vec{n}(\vec{n}\cdot\vec{\hat J})\right]-\sin(\theta)\vec{n}\times \vec{\hat J}\right].
    \end{equation}
    Using the bilinearity of the dot product we get
    \begin{align}
        e^{-i\theta \vec{\hat J}\cdot\vec{n}}(\vec m\cdot\vec{\hat J})e^{i\theta \vec{\hat J}\cdot\vec{n}}&=\left[(\vec m\cdot \vec n)(\vec n\cdot \vec{\hat J})+\cos(\theta)\left[\vec m\cdot \vec{\hat J}-(\vec m\cdot \vec n)(\vec n\cdot \vec{\hat J})\right]\right.\notag\\
        &\qquad-\left.\sin(\theta)\vec m\cdot(\vec n\times \vec{\hat J})\right].
    \end{align}
    Using the property of the triple product, $\vec a\cdot (\vec b\times \vec c)=(\vec a\times \vec b)\cdot \vec c$, we have
    \begin{equation}
        e^{-i\theta \vec{\hat J}\cdot\vec{n}}(\vec m\cdot\vec{\hat J})e^{i\theta \vec{\hat J}\cdot\vec{n}}=\left[\vec n(\vec n\cdot \vec m)+\cos(\theta)\left[\vec m-\vec n(\vec n\cdot \vec m)\right]-\sin(\theta)(\vec m\times \vec n)\right]\cdot \vec{\hat J}.
    \end{equation}
    Thus, by definition
    \begin{equation}
        \vec m'=\vec n(\vec n\cdot \vec m)+\cos(\theta)\left[\vec m-\vec n(\vec n\cdot \vec m)\right]-\sin(\theta)(\vec m\times \vec n).
    \end{equation}
    Finally, by using the anti-commutativity of the cross product, we get the desired result.
\end{derivation}

\begin{result}[Rotation of creation operators]~\\
    \label{res: roation creation op}
    Defining the angular momentum operator $\vec{\hat J}$ via the Schwinger relation on two modes $\hat a^\dagger$ and $\hat b^\dagger$ shown in Eq.~\eqref{eq: Schwinger representation}, for $\vec n=(n_x,n_y,n_z)$ of unit norm and $\theta\in\R$, we have
    \begin{subequations}
        \begin{align}
            e^{-i\theta\vec n\cdot \vec {\hat J}}\hat a^\dagger e^{i\theta \vec n\cdot \vec{\hat J}}&=\left[\cos(\tfrac{\theta}{2})+i n_z \sin(\tfrac{\theta}{2})\right]\hat a^\dagger -i\sin(\tfrac{\theta}{2})(n_x -i n_y)\hat b^\dagger,\\
            e^{-i\theta\vec n\cdot \vec {\hat J}}\hat b^\dagger e^{i\theta \vec n\cdot \vec{\hat J}}&=\left[\cos(\tfrac{\theta}{2})-i n_z \sin(\tfrac{\theta}{2})\right]\hat b^\dagger-i\sin(\tfrac{\theta}{2})(n_x +i n_y)\hat a^\dagger.
        \end{align}
    \end{subequations}
\end{result}

\begin{derivation}
    We introduce the two functions of $\theta$
    \begin{align}
        \varphi(\theta)=e^{-i\theta \vec n\cdot \vec{\hat J}}\hat a^\dagger e^{i\theta \vec n\cdot \vec{\hat J}},&& \psi(\theta)=e^{-i\theta \vec n\cdot \vec{\hat J}}\hat b^\dagger e^{i\theta \vec n\cdot \vec{\hat J}}.
    \end{align}
    The idea is to find a differential equation involving $\varphi$ and $\psi$. As the derivatives will depend on the commutators $[\vec n\cdot\vec{\hat J},\hat a^\dagger]$ and $[\vec n\cdot\vec{\hat J},\hat b^\dagger]$, we compute them first. We have
    \begin{subequations}
        \allowdisplaybreaks
        \begin{align}
            [\hat J_x,\hat a^\dagger]&=\frac{1}{2}[\hat a^\dagger \hat b +\hat a \hat b^\dagger,\hat a^\dagger]=\frac{1}{2}[\hat a,\hat a^\dagger]\hat b^\dagger=\frac{1}{2}\hat b^\dagger,\\
            [\hat J_y,\hat a^\dagger]&=\frac{i}{2}[\hat a^\dagger \hat b -\hat a \hat b^\dagger,\hat a^\dagger]=-\frac{i}{2}[\hat a,\hat a^\dagger]\hat b^\dagger=-\frac{i}{2}\hat b^\dagger,\\
            [\hat J_z,\hat a^\dagger]&=\frac{1}{2}[\hat b^\dagger \hat b -\hat a^\dagger \hat a,\hat a^\dagger]=-\frac{1}{2}[\hat a^\dagger \hat a,\hat a^\dagger]=-\frac{1}{2}\hat a^\dagger,\\
            [\hat J_x,\hat b^\dagger]&=\frac{1}{2}[\hat a^\dagger \hat b +\hat a \hat b^\dagger,\hat b^\dagger]=\frac{1}{2}\hat a^\dagger,\\
            [\hat J_y,\hat b^\dagger]&=\frac{i}{2}[\hat a^\dagger \hat b -\hat a \hat b^\dagger,\hat b^\dagger]=\frac{i}{2}\hat a^\dagger,\\
            [\hat J_z,\hat b^\dagger]&=\frac{1}{2}[\hat b^\dagger \hat b -\hat a^\dagger \hat a,\hat b^\dagger]=\frac{1}{2}\hat b^\dagger.
        \end{align}
    \end{subequations}
    We thus have
    \begin{equation}
        \varphi'(\theta)=-ie^{-i\theta\vec n\cdot\vec{\hat J}}[\vec n\cdot\vec{\hat J},\hat a^\dagger]e^{i\theta\vec n\cdot\vec{\hat J}}=-\frac{i}{2}\Big(n_x\psi(\theta)-in_y\psi(\theta)-n_z\varphi(\theta)\Big).
    \end{equation}
    And,
    \begin{equation}
        \psi'(\theta)=-ie^{-i\theta\vec n\cdot\vec{\hat J}}[\vec n\cdot\vec{\hat J},\hat b^\dagger]e^{i\theta\vec n\cdot\vec{\hat J}}=-\frac{i}{2}\Big(n_x\varphi(\theta)+in_y\varphi(\theta)+n_z\psi(\theta)\Big).
    \end{equation}
    We thus get a $2\times 2$ linear system of differential equations, which can be written in matrix form as
    \begin{equation}
        \begin{pmatrix}
            \psi'\\
            \varphi'
        \end{pmatrix}=-\frac{i}{2}
        \begin{pmatrix}
            n_z & n_x +i n_y\\
            n_x -i n_y & -n_z
        \end{pmatrix}
        \begin{pmatrix}
            \psi\\
            \varphi
        \end{pmatrix}.
    \end{equation}
    Introducing the Pauli matrices, we can write this compactly as
    \begin{equation}
        \begin{pmatrix}
            \psi'\\
            \varphi'
        \end{pmatrix}=-\frac{i}{2}\left(n_x \sigma_x - n_y \sigma_y + n_z \sigma_z\right)\begin{pmatrix}
            \psi\\
            \varphi
        \end{pmatrix}=-\frac{i}{2}(\vec n\cdot \vec \sigma^T)\begin{pmatrix}
            \psi\\
            \varphi
        \end{pmatrix},
    \end{equation}
    by noting that $\sigma_y$ is the only Pauli matrix whose transpose is its opposite. This equation is solved by considering the matrix exponential
    \begin{equation}
        \begin{pmatrix}
            \psi(\theta)\\
            \varphi(\theta)
        \end{pmatrix}=\exp\left(-\frac{i\theta}{2}(\vec n\cdot \vec \sigma^T)\right)\begin{pmatrix}
            \psi(0)\\
            \varphi(0)
        \end{pmatrix}=\exp\left(-\frac{i\theta}{2}(\vec n\cdot \vec \sigma^T)\right)\begin{pmatrix}
            \hat b^\dagger\\
            \hat a^\dagger
        \end{pmatrix}.
    \end{equation}
    It remains to compute the matrix exponential. By remembering that for any unit vector $\vec n$, we have $(\vec n\cdot \vec \sigma)^2=\1$, we also have $(\vec n\cdot \vec \sigma^T)^2=\1$. Thus, using the Taylor expansion of the exponential function, we get
    \begin{subequations}
        \begin{align}
            \exp\left(-\frac{i\theta}{2}(\vec n\cdot \vec \sigma^T)\right)&= \sum_{k=0}^\infty \frac{1}{k!}\left(\frac{-i\theta}{2}\right)^k (\vec n\cdot \vec \sigma^T)^k,\\
            &=\sum_{m=0}^\infty \frac{1}{(2m)!}\left(\frac{-i\theta}{2}\right)^{2m}\1 +\sum_{m=0}^\infty \frac{1}{(2m+1)!}\left(\frac{-i\theta}{2}\right)^{2m+1}(\vec n\cdot \vec \sigma^T),\\
            &=\cos(\tfrac{\theta}{2})\1 - i\sin(\tfrac{\theta}{2})(\vec n\cdot \vec \sigma^T),\\
            &=\begin{pmatrix}
                \cos(\tfrac{\theta}{2})-i n_z \sin(\tfrac{\theta}{2}) & -i\sin(\tfrac{\theta}{2})(n_x +i n_y)\\
                -i\sin(\tfrac{\theta}{2})(n_x -i n_y) & \cos(\tfrac{\theta}{2})+i n_z \sin(\tfrac{\theta}{2})
            \end{pmatrix}.
        \end{align}
        Expanding the matrix product, we thus indeed get the advertised result.
    \end{subequations}
\end{derivation}

\begin{result}[Stereographic projection in spherical coordinates]~\\
    \label{res: spherical stereographic}
    The south-based stereographic projection of the sphere in spherical coordinates is given by
    \begin{equation}
    z=e^{i\phi}\tan(\tfrac{\theta}{2}),
    \end{equation}
    where $\theta$ is the polar angle and $\phi$ the azimuthal angle and $z$ the complex coordinate of the stereographic projection.
\end{result}

\begin{derivation}
    Based on these two panels of Fig.~\ref{fig: stereo appendix}, already present in the main content of this manuscript, the relation $z=e^{i\phi}\tan(\tfrac{\theta}{2})$ becomes clear really quickly. Indeed, the coordinate $\phi$ naturally corresponds to the rotation angle around the vertical axis, which is naturally associated with the argument of the complex number $z$. For the modulus, it is clear that it can be obtained by looking at the right picture which is a cut of the sphere by the vertical plane at angle $\phi$. As the circle has radius 1, $\abs{z}$ is obtained via the tangent of the small lower angle. A simple geometric observation reveals that this angle is equal to $\theta/2$. Thus
    \begin{equation}
        \abs{z}=\tan(\frac{\theta}{2}).
    \end{equation}
\end{derivation}
\begin{figure}[ht]
    \centering
    \scalebox{1.3}{\begin{tikzpicture}
	\begin{pgfonlayer}{nodelayer}
		\node [style=none] (0) at (-7.75, -1) {};
		\node [style=none] (1) at (-4, 2) {};
		\node [style=none] (2) at (-0.75, -1) {};
		\node [style=none] (3) at (3, 2) {};
		\node [style=none] (4) at (-3.75, 0.75) {};
		\node [style=none] (5) at (-0.75, 0.75) {};
		\node [style=none] (6) at (-2.25, 2.25) {};
		\node [style=none] (7) at (-2.25, -0.75) {};
		\node [style=none] (8) at (-1.5, 1.75) {};
		\node [style=none] (9) at (-1.45, 1.7) {};
		\node [style=none] (10) at (-1.55, 1.8) {};
		\node [style=none] (11) at (-1.55, 1.7) {};
		\node [style=none] (12) at (-1.45, 1.8) {};
		\node [style=none] (13) at (-1.8, 0.525) {};
		\node [style=none] (14) at (-1.9, 0.625) {};
		\node [style=none] (15) at (-1.9, 0.525) {};
		\node [style=none] (16) at (-1.8, 0.625) {};
		\node [style=none] (17) at (-2.25, -0.675) {};
		\node [style=none] (18) at (-2.25, -0.825) {};
		\node [style=none] (20) at (-2.25, 2.525) {};
		\node [style=none] (21) at (-2.25, 3.5) {$N$};
		\node [style=none] (22) at (-2.25, -1.5) {$S$};
		\node [style=none] (23) at (6.25, 3.5) {};
		\node [style=none] (24) at (6.25, -2) {};
		\node [style=none] (25) at (3.5, 0.75) {};
		\node [style=none] (26) at (9, 0.75) {};
		\node [style=none] (27) at (6.25, 2.75) {};
		\node [style=none] (28) at (8.25, 0.75) {};
		\node [style=none] (29) at (4.25, 0.75) {};
		\node [style=none] (30) at (6.25, -1.25) {};
		\node [style=none] (31) at (7.8, 2.025) {};
		\node [style=none] (32) at (6.25, 0.75) {};
		\node [style=none] (33) at (6.25, 1.25) {};
		\node [style=none] (34) at (6.6, 1.05) {};
		\node [style=none] (35) at (6.25, -0.5) {};
		\node [style=none] (36) at (6.575, -0.575) {};
		\node [style=none] (37) at (6.6, 1.7) {\scalebox{0.6}{$\theta$}};
		\node [style=none] (38) at (6.475, -0.075) {\scalebox{0.6}{$\tfrac{\theta}{2}$}};
		\node [style=none] (40) at (7.9, 1.925) {};
		\node [style=none] (41) at (7.7, 2.125) {};
		\node [style=none] (42) at (7.7, 1.925) {};
		\node [style=none] (43) at (7.9, 2.125) {};
		\node [style=none] (44) at (7.3, 0.65) {};
		\node [style=none] (45) at (7.1, 0.85) {};
		\node [style=none] (46) at (7.1, 0.65) {};
		\node [style=none] (47) at (7.3, 0.85) {};
		\node [style=none] (48) at (-2.25, 0.5) {};
		\node [style=none] (49) at (-1.925, 0.4) {};
		\node [style=none] (51) at (-2.5, 0.625) {\scalebox{0.7}{$\tfrac{\theta}{2}$}};
		\node [style=none] (52) at (-2.475, 2.625) {};
		\node [style=none] (53) at (-2.025, 2.625) {};
		\node [style=none] (55) at (-1.5, 2.5) {\scalebox{0.7}{$\phi$}};
		\node [style=none] (57) at (6.25, 3.5) {};
		\node [style=none] (58) at (6.75, 0.5) {\scalebox{0.5}{$\abs{z}$}};
	\end{pgfonlayer}
	\begin{pgfonlayer}{edgelayer}
		\draw [style=Gray line] (1.center) to (3.center);
		\draw [style=Gray line] (3.center) to (2.center);
		\draw [style=Gray line] (0.center) to (2.center);
		\draw [style=Gray line] (0.center) to (1.center);
		\draw [bend left=315] (6.center) to (4.center);
		\draw [bend left=45] (6.center) to (5.center);
		\draw [bend left=45] (5.center) to (7.center);
		\draw [bend left=45] (7.center) to (4.center);
		\draw [bend right=90, looseness=0.75] (4.center) to (5.center);
		\draw [style=dashes, bend left=75, looseness=0.50] (4.center) to (5.center);
		\draw [style=red line] (7.center) to (8.center);
		\draw (10.center) to (9.center);
		\draw (11.center) to (12.center);
		\draw (14.center) to (13.center);
		\draw (15.center) to (16.center);
		\draw (17.center) to (18.center);
		\draw (23.center) to (24.center);
		\draw (25.center) to (26.center);
		\draw [bend left=45] (29.center) to (27.center);
		\draw [bend right=45] (28.center) to (27.center);
		\draw [bend right=45] (29.center) to (30.center);
		\draw [bend right=45] (30.center) to (28.center);
		\draw (30.center) to (31.center);
		\draw (32.center) to (31.center);
		\draw [bend left=45] (33.center) to (34.center);
		\draw [bend left, looseness=0.75] (35.center) to (36.center);
		\draw [style=red line] (41.center) to (40.center);
		\draw [style=red line] (42.center) to (43.center);
		\draw [style=red line] (45.center) to (44.center);
		\draw [style=red line] (46.center) to (47.center);
		\draw [style=dashes] (18.center) to (20.center);
		\draw [bend left] (48.center) to (49.center);
		\draw [style=Arrow, in=-75, out=-135, looseness=2.75] (52.center) to (53.center);
	\end{pgfonlayer}
\end{tikzpicture}}
    \caption[Stereographic projection of the unit sphere onto the equatorial plane based on the south pole]{Left: Stereographic projection of the sphere on the plane based on the south pole. The red line links the point to be projected with the south pole. Right: 2D cut of the projection in the vertical plane.}
    \label{fig: stereo appendix}
\end{figure}
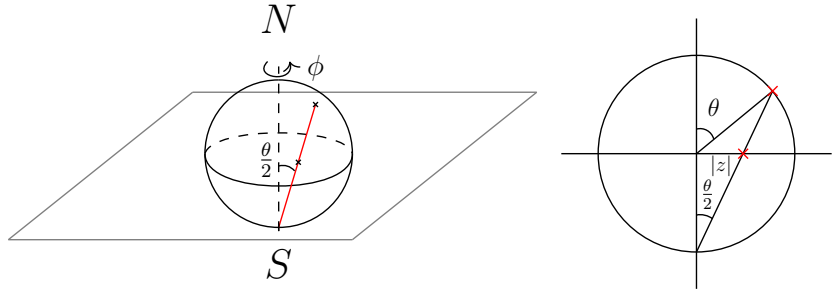

\begin{result}[Stereographic projection in Cartesian coordinates in dimension $n$]~\\
    \label{res: cartesian stereographic}
    In dimension $n$, we denote by $\Phi$ the stereographic projection map and by $\Psi$ its inverse
    \begin{align}
        \Phi: \mathcal S^n\setminus \{(\vec 0,-1)\}\to \R^n, && \Psi:\R^n \to \mathcal S^n\setminus \{(\vec 0,-1)\},
    \end{align}
    where $\mathcal S^n=\{(\vec x,y)\in\R^{n+1}|\norm{\vec x}^2+y^2=1\}\subset \R^{n+1}$ is the $n$-dimensional sphere. These two maps admit the following explicit expressions.
    \begin{align}
        \Phi:\begin{cases}
            \mathcal S^n\setminus\{(\vec 0,-1)\}\to \R^n,\\
            (\vec x,y)\mapsto \dfrac{\vec x}{y+1},
        \end{cases}&&\Psi:\begin{cases}
            \R^n\to \mathcal S^n\setminus\{(\vec 0,-1)\},\\
            \vec \alpha\mapsto \left(\dfrac{2\vec \alpha}{1+\norm{\vec \alpha}^2},\dfrac{1-\norm{\vec \alpha}^2}{1+\norm{\vec \alpha}^2}\right).
        \end{cases}
    \end{align}
\end{result}

\begin{derivation}
    The image of $\Phi$ is obtained by considering the $\vec x$-intercept of the line linking $(\vec x,y)$ with the south pole $(\vec 0,-1)$. The parametric equation of this line is
    \begin{equation}
        \ell(t)=(t\vec x,(y+1)t-1),
    \end{equation}
    as we indeed have $\ell(0)=(\vec 0,-1)$ and $\ell(1)=(\vec x,y)$. Searching for the value of $t_0$ for which the last component of $\ell(t_0)$ is zero yields by definition $\ell(t_0)=(\Phi(\vec x,y),0)$. A quick computation gives
    \begin{align}
        (y+1)t_0-1=0 && \Rightarrow && t_0=\frac{1}{y+1} && \Rightarrow && \Phi(\vec x,y)=t_0\vec x=\frac{\vec x}{y+1}.
    \end{align}
    Thus we have
    \begin{equation}
        \Phi:\begin{cases}
            \mathcal S^n\setminus\{(\vec 0,-1)\}\to \R^n,\\
            (\vec x,y)\mapsto \dfrac{\vec x}{y+1}.
        \end{cases}
    \end{equation}
    To construct the inverse stereographic projection, we consider the map $\Psi$ inverse of $\Phi$. It is obtained by solving for $\vec \alpha\in\R^n$ the equation
    \begin{equation}
        \Phi(\vec x,y)=\frac{\vec x}{y+1}=\vec \alpha,
    \end{equation}
    with unknowns $(\vec x,y)\in\mathcal S^n\setminus\{(\vec 0,-1)\}$. Taking the squared norm we get the equation
    \begin{equation}
        \norm{\vec \alpha}^2(y+1)^2=\norm{\vec x}^2.
    \end{equation}
    Since $(\vec x, y)$ is a point on the sphere we must have $\norm{\vec x}^2+y^2=1$. Thus we can get an equation only on $y$.
    \begin{equation}
        \norm{\vec \alpha}^2(y+1)^2+y^2=1.
    \end{equation}
    Expanding and grouping the terms we get a second-order polynomial equation in $y$
    \begin{equation}
        (\norm{\vec \alpha}^2+1)y^2+2\norm{\vec \alpha}^2 y +(\norm{\vec \alpha}^2-1)=0.
    \end{equation}
    Its discriminant is 
    \begin{equation}
        \Delta=4\norm{\alpha}^4-4(\norm{\alpha}^2+1)(\norm{\alpha}^2-1)=4\norm{\alpha}^4-4(\norm{\alpha}^4-1)=4.
    \end{equation}
    We thus have two solutions
    \begin{equation}
        y_{\pm}=\frac{-2\norm{\vec \alpha}^2\pm 2}{2(1+\norm{\vec \alpha}^2)}=\begin{cases}
            -1,\\
            \dfrac{1-\norm{\vec \alpha}^2}{1+\norm{\vec \alpha}^2}.
        \end{cases}
    \end{equation}
    The solution $y=-1$ corresponds to the south pole of the sphere that we need to ignore. We thus keep as only solution $y=\frac{1-\norm{\vec \alpha}^2}{1+\norm{\vec \alpha}^2}$. Finally injecting into $\vec x=(1+y)\vec \alpha$, we get
    \begin{equation}
        \Psi:\begin{cases}
            \R^n\to \mathcal S^n\setminus\{(\vec 0,-1)\},\\
            \vec \alpha\mapsto \left(\dfrac{2\vec \alpha}{1+\norm{\vec \alpha}^2},\dfrac{1-\norm{\vec \alpha}^2}{1+\norm{\vec \alpha}^2}\right),
        \end{cases}
    \end{equation}
    where a quick sanity check allows us to verify that the image has indeed unit norm.
\end{derivation}

\begin{result}[Reflections and stereographic projection]~\\
    \label{res: reflections and stereographic}
    Denoting by $(\alpha, \beta,\gamma)$ the Cartesian coordinates of a point on the sphere and by $z$ the complex coordinate of its stereographic projection, we have the following correspondence between simple transformations of the sphere and the complex plane.
    \begin{subequations}
        \begin{align}
            (\alpha,\beta,\gamma)\leftrightarrow (-\alpha,\beta,\gamma) &&\text{corresponds to} && z\leftrightarrow -z^*,\\
            (\alpha,\beta,\gamma)\leftrightarrow (\alpha,-\beta,\gamma) &&\text{corresponds to} && z\leftrightarrow z^*,\\
            (\alpha,\beta,\gamma)\leftrightarrow (\alpha,\beta,-\gamma) &&\text{corresponds to} && z\leftrightarrow \frac{1}{z^*}.
        \end{align}
    \end{subequations}
    Composing these, we recover the relations between the transformations in Tab.~\ref{tab: sphere vs plane transforms}.

\end{result}

\begin{derivation}
    For $z=\Phi(\alpha,\beta,\gamma)=\frac{\alpha+i\beta}{1+\gamma}$, it is simple to observe that complex conjugation corresponds to changing $\beta\leftrightarrow -\beta$. This gives the second line of the table. Additionally, 
    \begin{equation}
        \frac{-\alpha+i\beta}{1+\gamma}=-\left(\frac{\alpha+i\beta}{1+\gamma}\right)^*=-z^*.
    \end{equation}
    So we get the first line of the table. For the third we observe that
    \begin{equation}
        \frac{\alpha+i\beta}{1-\gamma}=\frac{\alpha+i\beta}{1-\gamma}\frac{\alpha-i\beta}{\alpha-i\beta}\frac{1+\gamma}{1+\gamma}=\frac{\alpha^2+\beta^2}{1-\gamma^2}\frac{1+\gamma}{\alpha-i\beta}=\frac{1+\gamma}{\alpha-i\beta}=\frac{1}{z^*},
    \end{equation}
    as $\alpha^2+\beta^2=1-\gamma^2$ for a point on the sphere.
\end{derivation}

\begin{result}[Möbius transformation and $SU_2(\C)$]~\\
    \label{res: möbius and SU2}
    The mapping
    \begin{equation}
        \begin{cases}
            SU_2(\C)&\to \mathcal M_R,\\
            U=\begin{pmatrix} a & -c^*\\ c & a^* \end{pmatrix}&\mapsto \left(f_U: z\mapsto \dfrac{az-c^*}{cz+a^*}\right),
        \end{cases} 
    \end{equation}
    is a group morphism with kernel $\{\pm \1\}$.
\end{result}

\begin{derivation}
    $\blacktriangleright$ {\bf Composition stability.} We first verify that $\mathcal M_R$ is stable under composition. To do so, we consider $f,g\in\mathcal M_R$, parametrized as
    \begin{align}
        f(z)=\frac{az-c^*}{cz+a^*}, && \text{with } \abs{a}^2+\abs{c}^2=1,\\
        g(z)=\frac{bz-d^*}{dz+b^*}, && \text{with } \abs{b}^2+\abs{d}^2=1.
    \end{align}
    And we verify that 
    \begin{equation}
        (f\circ g)(z)=h(z)=\frac{(ab-c^*d)z-(a^*d+bc)^*}{(a^*d+cb)z+(ab-c^*d)^*}.
    \end{equation}
    First note that $h$ is indeed an element of $\mathcal M_R$, since
    \begin{subequations}
        \begin{align}
            \abs{ab-c^*d}^2+\abs{a^*d+bc}^2&=(ab-c^*d)(a^*b^*-cd^*)+ (a^*d+bc)(ad^*+b^* c^*),\\
            &= \abs{a}^2\abs{b}^2-abcd^* -a^*b^* c^* d+\abs{c}^2\abs{d}^2+\abs{a}^2\abs{d}^2\notag\\
            &\qquad +abcd^* +a^*b^* c^* d+\abs{b}^2\abs{c}^2,\\
            &=(\abs{a}^2+\abs{c}^2)(\abs{b}^2+\abs{d}^2),\\
            &=1.
        \end{align}
    \end{subequations}
    \begin{itemize}[label=$\bullet$]
        \item First case: $z\neq \infty$, $z\neq -b^*/d$ and $g(z)\neq -a^*/c$ (in case $c$ or $d$ are not zero). In this case, we can simply compute the composition of the functions
        \begin{subequations}
            \begin{align}
                f(g(z))&=f\left(\frac{bz-d^*}{dz+b^*}\right),\\
                &=\frac{a\frac{bz-d^*}{dz+b^*}-c^*}{c\frac{bz-d^*}{dz+b^*}+a^*},\\
                &=\frac{(ab-c^*d)z+( -ad^*-b^* c^*)}{(cb+da^*)z+( -cd^*+a^*b^*)}.
            \end{align}
        \end{subequations}
        Comparing with the expression of the matrix product we indeed have
        \begin{equation}
            f(g(z))=h(z),
        \end{equation}
        as it is easy to verify that the condition $g(z)\neq -a^*/c$ implies that the denominator of $h(z)$ is not zero.
        \item Second case: $z=\infty$ and $d=0$ (and thus $b\neq 0$). In this case we have $g(\infty)=\infty$ and thus 
        \begin{align}
            f(\infty)=\begin{cases}
                a/c, \text{ if }c\neq 0,\\
                \infty,\text{ if } c=0.
            \end{cases} && h(\infty)=\begin{cases}
                (ab)/(cb), \text{ if } cb\neq 0,\\
                \infty,\text{ if } cb=0.
            \end{cases}
        \end{align}
        Since $b\neq 0$ these two equations are equivalent and thus we indeed have
        \begin{equation}
            f(g(\infty))=h(\infty).
        \end{equation}
        \item Third case: $z=\infty$ and $d\neq 0$. In this case we have $g(\infty)=b/d$. We thus have
        \begin{align}
            f(b/d)=\begin{cases}
                \dfrac{ab/d-c^*}{cb/d+a^*}, \text{ if } cb/d+a^*\neq 0,\\
                \infty,\text{ if } cb/d+a^*=0.
            \end{cases}\\ h(\infty)=\begin{cases}
                \dfrac{(ab-c^*d)}{(cb+da^*)}, \text{ if } (cb+da^*)\neq 0,\\
                \infty,\text{ if } (cb+da^*)=0.
            \end{cases}
        \end{align}
        Once again by comparing we indeed get $f(g(\infty))=h(\infty)$.
        \item Fourth case: $z=-b^*/d$ and $d\neq 0$. In this case, $g(z)=\infty$, and thus
        \begin{align}
            &f(\infty)=\begin{cases}
                a/c, \text{ if }c\neq 0,\\
                \infty,\text{ if } c=0.
            \end{cases}\notag\\
            &h(-b^*/d)=\begin{cases}
                \dfrac{(ab-c^*d)(-b^*/d)+(-ad^*-b^* c^*)}{(cb+da^*)(-b^*/d)+(-cd^*+a^*b^*)}, \\
                \qquad\text{ if } (cb+da^*)(-b^*/d)+(-cd^*+a^*b^*)\neq 0,\\
                \infty,\text{ if } (cb+da^*)(-b^*/d)+(-cd^*+a^*b^*)=0.
            \end{cases}
        \end{align}
        As we observe that we can simplify
        \begin{equation}
            (cb+da^*)\frac{-b^*}{d}+(-cd^*+a^*b^*)=-c\frac{\abs{b}^2}{d}-a^*b^*-cd^*+a^*b^*=-\frac{c}{d}(\abs{b}^2+\abs{d}^2)=-\frac{c}{d},
        \end{equation}
        and
        \begin{equation}
            (ab-c^*d)\frac{-b^*}{d}+(-ad^*-b^* c^*)= -\frac{ab b^*}{d}+c^* b^*-ad^*-b^* c^*=-\frac{a}{d}(\abs{b}^2+\abs{d}^2)=-\frac{a}{d},
        \end{equation}
        we see that the expression of $h(-b^*/d)$ can be simplified and we indeed have $f(g(-b^*/d))=h(-b^*/d)$.
    \end{itemize}
    We have thus checked all cases and can thus conclude in full generality that $f\circ g=h$.
    \medskip

    \noindent
    $\blacktriangleright$ {\bf Möbius transformations form a group.} Based on the expression of the product, it is easy to verify the axioms of a group. Indeed for two matrices $U,V\in SU_2(\C)$, parametrized by $(a,c)$ and $(b,d)$ respectively, we have
    \begin{equation}
        UV=\begin{pmatrix}
            ab - c^* d & -a d^* - c^* b^* \\ 
            cb + a^* d & -c d^* + a^* b^*
        \end{pmatrix},
    \end{equation}
    which shows, when compared with the expression of the composition of Möbius transformations, that the map $U\mapsto f_U$ is a group morphism. Additionally, the identity element is obviously the map $z\mapsto z$ obtained for parameter choice $a=1$ and $c=0$. The associativity stems from the associativity of function composition and the inverse of a map $z\mapsto\frac{az-c^*}{cz+a^*}$ is given by
    \begin{equation}
        z\mapsto\frac{a^*z+c^*}{-cz+a}.
    \end{equation}
    \medskip

    \noindent
    $\blacktriangleright$ {\bf Study of the Kernel.} We now study the kernel, and thus consider a matrix such that 
    \begin{equation}
        z\mapsto \frac{az-c^*}{cz+a^*},
    \end{equation}
    is the identity.
    \begin{itemize}
        \item First case: if $c=0$ then necessarily, $d=0$ (otherwise $f(\infty)=\infty$ but $g(\infty)=b/d$). We thus have the equality for all $z\in\C$
        \begin{equation}
            \frac{a}{a^*}z=\frac{b}{b^*}z,
        \end{equation}
        and thus $a/a^*=b/b^*$, which can be rewritten as $a/b=(a/b)^*$, meaning the quotient is real. Since we have $\abs{a}=\abs{b}=1$, the quotient is also of modulus 1, and thus $a=\pm b$ and we indeed have $(a,c)=\pm (b,d)$.
        \item Second case: if $a=0$ then $b=0$ too (otherwise $f(0)=-c^*/a^*=\infty$ but $g(0)=-d^*/b^*$ is finite). We thus have the equality for all $z\in\C$
        \begin{equation}
            -\frac{c^*}{cz}=-\frac{d^*}{dz},
        \end{equation}
        which can be rewritten as $c/c^*=d/d^*$. Following the same reasoning as before we get $c=\pm d$ and thus $(a,c)=\pm(b,d)$.
        \item Third case: if $c\neq 0$ and $a\neq 0$ then $d\neq 0$ and $b\neq 0$ too. The two rational fractions should then have the same pole, so
        \begin{equation}
            -\frac{a^*}{c}=-\frac{b^*}{d},
        \end{equation}
        Rearranging we get
        \begin{equation}
            \lambda=\frac{a^*}{b^*}=\frac{c}{d}
        \end{equation}
        We thus have both $a=\lambda^*b$ and $c=\lambda d$. Injecting into the equality of the two maps we get
        \begin{equation}
            \frac{bz - d^*}{dz + b^*}=\frac{\lambda^* b z - (\lambda d)^*}{\lambda d  z + (\lambda^* b)^*}=\frac{\lambda^*}{\lambda}\frac{bz-d^*}{dz+b^*}.
        \end{equation}
        As the equality should hold for all $z$, we have $\lambda^*/\lambda=1$, forcing $\lambda$ to be real. Finally, the normalisation condition reads
        \begin{equation}
            \abs{a}^2+\abs{c}^2=\abs{\lambda}^2(\abs{b}^2+\abs{d}^2)=\abs{\lambda}^2=1,
        \end{equation}
        forcing $\lambda=\pm 1$. We thus have $(a,c)=\pm(b,d)$.
    \end{itemize}
\end{derivation}

\begin{result}[Möbius transformations and rotations of the sphere]~\\
    \label{res: möbius and rotation}
    The Möbius transformation 
    \begin{equation}
        z\mapsto \frac{az-c^*}{cz+a^*}
    \end{equation}
    corresponds, via the stereographic projection, to the rotation of the sphere with rotation axis $\vec n$ and angle $\theta$ given by the following relations
    \begin{align}
        a=\cos(\frac{\theta}{2})+i n_z \sin(\frac{\theta}{2}), && 
        c=-i(n_x - i n_y)\sin(\frac{\theta}{2}).
    \end{align}
\end{result}

\begin{derivation}
    The derivation is based on simply computing the compositions
    \begin{equation}
        z\mapsto \Psi(z)=\vec m\mapsto \vec m'=R_{\vec n}(\theta)\vec m\mapsto z'=\Phi(\vec m'),
    \end{equation}
    where $R_{\vec n}(\theta)$ is the rotation matrix of angle $\theta$ around the axis $\vec n$. We then verify that $z'=\frac{az-c^*}{cz+a^*}$ with the advertised expressions of $a$ and $c$. We start from $z\in\C$. Following the expression of the stereographic projection we get
    \begin{equation}
        \vec m=\Psi(z)=\frac{1}{1+\abs{z}^2}\begin{pmatrix} 2\Re z \\ 2\Im z \\ 1-\abs{z}^2 \end{pmatrix}=\frac{1}{1+\abs{z}^2}\begin{pmatrix} z+z^* \\ -i(z-z^*) \\ 1-\abs{z}^2 \end{pmatrix}.
    \end{equation}
    We can then obtain the vector $\vec m'$ by applying the rotation matrix $R_{\vec n}(\theta)$ to $\vec m$. Using the expression of the rotation matrix, obtained by the result of Eq.~\eqref{eq: rotation vector}, we get
    \begin{align}
        \vec m'&=\frac{1}{1+\abs{z}^2}\scalebox{0.82}{$\begin{pmatrix}
        \cos\theta+n_x^2(1-\cos\theta) & n_x n_y(1-\cos\theta) - n_z \sin\theta & n_x n_z(1-\cos\theta) + n_y \sin\theta\\
        n_y n_x(1-\cos\theta) + n_z \sin\theta & \cos\theta+n_y^2(1-\cos\theta) & n_y n_z(1-\cos\theta) - n_x \sin\theta\\
        n_z n_x(1-\cos\theta) - n_y \sin\theta & n_z n_y(1-\cos\theta) + n_x \sin\theta & \cos\theta+n_z^2(1-\cos\theta)
        \end{pmatrix}$}\notag\\
        &\qquad\times\begin{pmatrix}
            z+z^* \\ -i(z-z^*) \\ 1-\abs{z}^2
        \end{pmatrix}.
    \end{align}
    We can express each component of $\vec m'$ by carefully collecting the terms as factors of either $z$, $z^*$ or $1-\abs{z}^2$. We get
    \begin{subequations}
        \begin{align}
            m_x'&=\frac{z}{1+\abs{z}^2}\Big[ 
            \cos\theta+n_x(n_x-in_y)(1-\cos\theta)+i n_z\sin\theta
            \Big]\notag\\
            &\qquad+\frac{z^*}{1+\abs{z}^2}\Big[
            \cos\theta+n_x(n_x+in_y)(1-\cos\theta)-i n_z\sin\theta
            \Big]\notag\\
            &\qquad+\frac{1-\abs{z}^2}{1+\abs{z}^2}\Big[
            n_x n_z(1-\cos\theta)+ n_y \sin\theta
            \Big],\\
            m_y'&=\frac{z}{1+\abs{z}^2}\Big[ 
            -i\cos\theta+n_y(n_x-in_y)(1-\cos\theta)+ n_z\sin\theta
            \Big]\notag\\
            &\qquad+\frac{z^*}{1+\abs{z}^2}\Big[
            i\cos\theta+n_y(n_x+in_y)(1-\cos\theta)+ n_z\sin\theta
            \Big]\notag\\
            &\qquad+\frac{1-\abs{z}^2}{1+\abs{z}^2}\Big[
            n_y n_z(1-\cos\theta)- n_x \sin\theta
            \Big],\\
            m_z'&=\frac{z}{1+\abs{z}^2}\Big[ 
            n_z(n_x - in_y)(1-\cos\theta)-i(n_x-i n_y) \sin\theta
            \Big]\notag\\
            &\qquad+\frac{z^*}{1+\abs{z}^2}\Big[
            n_z(n_x + in_y)(1-\cos\theta)+i(n_x+i n_y) \sin\theta
            \Big]\notag\\
            &\qquad+\frac{1-\abs{z}^2}{1+\abs{z}^2}\Big[
            \cos\theta+n_z^2(1-\cos\theta)
            \Big].
        \end{align}
    \end{subequations}
    We can thus have an expression for $m_x'+im_y'$
    \begin{align}
        m_x'+im_y'&=\frac{z}{1+\abs{z}^2}\Big[
        2\cos\theta+(n_x^2+n_y^2)(1-\cos\theta)+2i n_z \sin\theta\Big]\notag\\
        &\qquad+\frac{z^*}{1+\abs{z}^2}
        (n_x + in_y)^2(1-\cos\theta)\notag\\
        &\qquad+\frac{1-\abs{z}^2}{1+\abs{z}^2}(n_x+in_y)\Big[
         n_z(1-\cos\theta)-i\sin\theta\Big].
    \end{align}
    If we now compute the stereographic projection of $\vec m'$, $z'=\Phi(\vec m')=\frac{m_x'+im_y'}{1+m_z'}$, we will get a complex expression of the form
    \begin{equation}
        z'=\frac{A z+Bz^*+(1-\abs{z}^2)C}{Dz+D^* z^* +E+F\abs{z}^2},
    \end{equation}
    where the coefficients $A$, $B$, $C$, $D$ and $E$ are complex numbers depending on $\vec n$ and $\theta$. Directly reading on the different previous expressions gives
    \begin{subequations}
        \begin{align}
            A&=2\cos\theta+(n_x^2+n_y^2)(1-\cos\theta)+2i n_z \sin\theta\notag\\
            &\qquad=2\cos\theta+(1-n_z^2)(1-\cos\theta)+2i n_z \sin\theta\notag\\
            &\qquad=1+\cos\theta-n_z^2(1-\cos\theta)+2i n_z \sin\theta\\
            B&= (n_x + in_y)^2(1-\cos\theta),\\
            C&=(n_x + in_y)(n_z(1-\cos\theta)-i\sin\theta),\\
            D&=(n_x - in_y)(n_z(1-\cos\theta)-i\sin\theta),\\
            E&=1+\cos\theta+n_z^2(1-\cos\theta),\\
            F&=1-\cos\theta-n_z^2(1-\cos\theta)=(1-\cos\theta)(1-n_z^2).
        \end{align}
    \end{subequations}
    Since the function we obtained at the end should take the form of a Möbius transformation, we expect that an identical factor of $\alpha+\beta z^*$ should be factorizable from both the numerator and the denominator so that the expression takes the desired form. Utilizing the trigonometric identities for doubling angles
    \begin{align}
        1+\cos\theta=2\cos^2(\theta/2), && 1-\cos\theta=2\sin^2(\theta/2), && \sin\theta=2\sin(\theta/2)\cos(\theta/2).
    \end{align}
    We can thus provide alternative expressions for all coefficients
    \begin{subequations}
        \begin{align}
            A&=2\cos^2(\theta/2)-2 n_z^2 \sin^2(\theta/2)+4i n_z \sin(\theta/2)\cos(\theta/2),\notag\\
            &\qquad=2(\cos(\theta/2)+in_z\sin(\theta/2))^2\\
            B&= 2(n_x + in_y)^2\sin^2(\theta/2),\\
            C&=2(n_x + in_y)\sin(\theta/2)(n_z \sin(\theta/2)-i\cos(\theta/2)),\\
            D&=2(n_x - in_y)\sin(\theta/2)(n_z \sin(\theta/2)-i\cos(\theta/2)),\\
            E&= 2\cos^2(\theta/2)+2n_z^2\sin^2(\theta/2),\\
            F&=2\sin^2(\theta/2)(1-n_z^2).
        \end{align}
    \end{subequations}
    Introducing the two quantities
    \begin{align}
        a=\cos(\theta/2)+i n_z \sin(\theta/2), && c=-i(n_x - in_y)\sin(\theta/2),
    \end{align}
    we observe that every coefficient admits a simple expression
    \begin{align}
        A=2a^2,&& B=-2{c^*}^2 , && C=-2c^*a  ,&& D= 2ca , && E=2\abs{a}^2 , && F= 2\abs{c}^2.
    \end{align}
    Thus, we have
    \begin{equation}
        z'=\frac{2a^2 z - 2{c^*}^2 z^* -(1-\abs{z}^2)2c^* a}{2ca z + 2c^* a^* z^* + 2\abs{a}^2 + 2\abs{c}^2 \abs{z}^2}=\frac{(az-c^*)(c^*z^*+a)}{(cz+a^*)(c^*z^*+a)}=\frac{(az-c^*)}{(cz+a^*)}.
    \end{equation}
    Which is indeed a Möbius transformation. To be completely rigorous, we have to check the two special cases in the definition of a Möbius map. Indeed, our computation is valid for any value of $z$ except for $z=\infty$ and, when taking the stereographic projection again, it is possible only if $m_z'\neq -1$. Geometrically, we need to deal with the south pole either when it is the starting point or the ending point of the transformation.
    \begin{itemize}
        \item First, if $z=\infty$, the stereographic projection associate to the south pole of the sphere: $\vec m=(0,0,-1)$. The point after rotation is then
        \begin{subequations}
            \begin{align}
                \vec m'&=R_{\vec n}(\theta)\begin{pmatrix} 0 \\ 0 \\ -1 \end{pmatrix}=\begin{pmatrix} 
                    -n_xn_z(1-\cos\theta)-n_y\sin\theta\\
                    -n_yn_z(1-\cos\theta)+n_x\sin\theta\\
                    -\cos\theta-n_z^2(1-\cos\theta)
                \end{pmatrix},\\
                &=\begin{pmatrix} 
                    -2n_xn_z\sin^2(\theta/2)-2n_y\sin(\theta/2)\cos(\theta/2)\\
                    -2n_yn_z\sin^2(\theta/2)+2n_x\sin(\theta/2)\cos(\theta/2)\\
                    2\sin^2(\theta/2)(1-n_z^2)-1
                \end{pmatrix}.
            \end{align}
        \end{subequations}
        This point remains at infinity if and only if $2\sin^2(\theta/2)(1-n_z^2)-1=-1$ which requires either $\theta\equiv 0\,[2\pi]$ (complete rotation of the sphere) or $n_z=\pm 1$ (rotation around the $Z$ axis in either direction). In both cases the south pole is left invariant by the rotation, and the Möbius map should send $\infty \mapsto\infty$ which is the case when $c=-i(n_x-in_y)\sin(\theta/2)=0$, this condition being equivalent to the previous ones.
        
        On the contrary, if the south pole has been moved, we can project it back to a finite value
        \begin{subequations}
            \allowdisplaybreaks
            \begin{align}
                z'&=\frac{m_x'+im_y'}{1+m_z'},\\
                &=\frac{-2n_x n_z \sin^2(\theta/2)-2 n_y \sin(\theta/2)\cos(\theta/2)}{2\sin^2(\theta/2)(1-n_z^2)}\notag\\
                &\qquad +i\frac{-2n_y n_z \sin^2(\theta/2)+2 n_x \sin(\theta/2)\cos(\theta/2)}{2\sin^2(\theta/2)(1-n_z^2)},\\
                &=\frac{-n_x n_z \sin(\theta/2)- n_y \cos(\theta/2)+i(-n_y n_z \sin(\theta/2)+ n_x \cos(\theta/2))}{\sin(\theta/2)(n_x^2+n_y^2)},\\
                &=\frac{-n_z(n_x+in_y)\sin(\theta/2)+i(n_x+in_y)\cos(\theta/2)}{\sin(\theta/2)(n_x+in_y)(n_x-in_y)},\\
                &=\frac{- n_z \sin(\theta/2)+i\cos(\theta/2)}{\sin(\theta/2)(n_x - in_y)},\\
                &=\frac{\cos(\theta/2)+in_z\sin(\theta/2)}{-i(n_x-in_y)\sin(\theta/2)},\\
                &=\frac{a}{c}.
            \end{align}
        \end{subequations}
        which is indeed the prescribed value for the image of $\infty$ in this case.

        \item If after rotation $\vec m'$ is at the south pole, we cannot stereographically project it to a finite point. This happens when $m_z'=-1$, or when the denominator of 
        \begin{equation}
            \frac{A z+Bz^*+(1-\abs{z}^2)C}{Dz+D^* z^* +E+F\abs{z}^2}
        \end{equation}
        vanishes. Based on the previous computation this denominator can simply be expressed as
        \begin{equation}
            2(cz+a^*)(c^*z^*+a)=2\abs{cz+a^*}^2.
        \end{equation}
        This happens exactly when $z=-a^*/c$, in which case, the Möbius transformation maps $z$ to $\infty$. So this particular case is also coherent.
    \end{itemize}
\end{derivation}

\begin{result}[Equivalent definitions of spin coherent states]~\\
    \label{res: def spin coherent states}
    A spin coherent state in the two-mode SSRC framework admits the following equivalent definitions.
    \begin{enumerate}
        \item In the first-quantization picture, optical states with fixed total photon number are constructed as symmetrized states of $N$ identical photons, as done in the Fock space construction presented in Sec.~\ref{subsec: FockSpace}. In this framework, spin coherent states correspond precisely to particle-separable states, that is, states in which all individual photons occupy the same single-photon state.

        More concretely, for a normalized single-photon state $\ket{\varphi}=\alpha \ket{0}+\beta\ket{1}$, the associated spin coherent state is
        \begin{equation}
            \ket{\psi}=\ket{\varphi}\otimes\cdots\otimes\ket{\varphi}.
        \end{equation}\label{def app: spin particle sep}

        \item A spin coherent state is any state that can, up to a global phase, be put into the form\footnote{The value $z=\infty$ corresponds by convention to the state $\ket{N,0}_p$.}
        \begin{subequations}
            \begin{align}
                \ket{\psi}&=\frac{1}{(1+\abs{z}^2)^{N/2}}e^{z\hat a^\dagger \hat b}\ket{0,N}_p=\frac{1}{(1+\abs{z}^2)^{N/2}}\sum_{k=0}^N z^k\sqrt{\binom{N}{k}}\ket{k,N-k}_p,
            \end{align}
        \end{subequations}
        for some complex number $z$. In this case we call $z$ the complex amplitude of the coherent state.\label{def app: spin complex}

        \item A state $\ket{\psi}$ is a spin coherent state if and only if there exists a unique mode $\hat c^\dagger =\alpha \hat a^\dagger +\beta \hat b^\dagger$ such that $\ket{\psi}$ is an $N$-photon Fock state in this mode
        \begin{equation}
            \ket{\psi}=\frac{(\hat c^\dagger)^N}{\sqrt{N!}}\vac=\sum_{k=0}^N \sqrt{\binom{N}{k}}\alpha^k \beta^{N-k}\ket{k,N-k}_p.
        \end{equation}\label{def app: spin monomode}

        \item The state $\ket{\psi}$ is a coherent state if and only if there exist spherical coordinates $(\theta,\phi)$, necessarily unique, such that, up to a global phase
        \begin{equation}
            \ket{\psi}=\hat R(\theta,\phi)\ket{0,N}_p,
        \end{equation}
        where $\hat R(\theta,\phi)=\exp(-i\theta\left( -\sin(\phi)\hat J_x+\cos(\phi)\hat J_y\right))$ is a rotation operator whose effect is to rotate the $z$ axis to the axis at coordinates $(\theta,\phi)$.\label{def app: spin rotation}

        \item Finally, we say that $\ket{\psi}$ is a spin coherent state if and only if there exists a unit vector $\vec m$, necessarily unique, such that $\ket{\psi}$ is an eigenvector of $\vec m\cdot\vec{\hat J}$ with eigenvalue $N/2$. Additionally, up to a global phase, $\ket{\psi}$ is the only such eigenvector.\label{def app: spin operator}
    \end{enumerate}
\end{result}

\begin{derivation}
\begin{itemize}
        \item \ref{def app: spin particle sep} $\Leftrightarrow$ \ref{def app: spin monomode}. It is clear that $N$ photons put into the mode $\hat c^\dagger =\alpha \hat a^\dagger +\beta \hat b^\dagger$ correspond in the Fock space picture to the state
        \begin{equation}
            \ket{\psi}=\ket{\varphi}\otimes\cdots\otimes\ket{\varphi},
        \end{equation}
        for $\ket{\varphi}=\alpha \ket{0}+\beta \ket{1}$ which is clearly particle separable. Conversely, an $N$-partite separable state is necessarily of the form
        \begin{equation}
            \ket{\psi}=\ket{\varphi_1}\otimes \cdots \otimes \ket{\varphi_N},
        \end{equation}
        The symmetrization of the Fock space automatically guarantees that all $\ket{\varphi_i}$ are equal, and thus $\ket{\psi}=\ket{\varphi}\otimes\cdots\otimes\ket{\varphi}$ for some single-photon state $\ket{\varphi}$. This single-photon state can be written as $\ket{\varphi}=\alpha \ket{0}+\beta \ket{1}$, and thus $\ket{\psi}$ is an $N$-photon Fock state in the mode $\hat c^\dagger =\alpha \hat a^\dagger +\beta \hat b^\dagger$.

        \item \ref{def app: spin complex} $\Leftrightarrow$ \ref{def app: spin monomode}. The special case $z=\infty$ is clear. By comparing the expanded form of both definitions we see that a coherent state of amplitude $z$ is an $N$-photon Fock state in mode $\hat c^\dagger=(\alpha\hat a^\dagger +\beta\hat b^\dagger)$ if and only if, for all integers $k$, we have the relation
        \begin{equation}
            \frac{z^k}{(1+\abs{z}^2)^{N/2}}=\alpha^k \beta^{N-k}.
        \end{equation}
        For $k=0$, we get
        \begin{equation}
            \beta=\frac{1}{\sqrt{1+\abs{z}^2}}.
        \end{equation}
        Then for $k=1$, we get
        \begin{equation}
            \frac{z}{(1+\abs{z}^2)^{N/2}}=\alpha \beta^{N-1}= \frac{\alpha}{(1+\abs{z}^2)^{(N-1)/2}},
        \end{equation}
        Thus $\alpha=\frac{z}{\sqrt{1+\abs{z}^2}}$. Reciprocally, it is clear that based on these expressions for $\alpha$ and $\beta$, the relation $\frac{z^k}{(1+\abs{z}^2)^{N/2}}=\alpha^k \beta^{N-k}$ is satisfied for all $k$.

        \item \ref{def app: spin monomode} $\Leftrightarrow$ \ref{def app: spin rotation}. We apply Eq.~\eqref{eq: rotation creation op} for the vector $\vec n=(-\sin(\phi),\cos(\phi),0)$ and angle $\theta$. We have
        \begin{subequations}
            \begin{align}
                \hat R(\theta,\phi)\hat b^\dagger \hat R(\theta,\phi)^\dagger&=\left[\cos(\tfrac{\theta}{2})-i n_z \sin(\tfrac{\theta}{2})\right]\hat b^\dagger-i\sin(\tfrac{\theta}{2})(n_x +i n_y)\hat a^\dagger,\\
                &=\cos(\tfrac{\theta}{2})\hat b^\dagger-i\sin(\tfrac{\theta}{2})(-\sin\phi +i\cos\phi)\hat a^\dagger,\\
                &=\cos(\tfrac{\theta}{2})\hat b^\dagger +e^{i\phi}\sin(\tfrac{\theta}{2})\hat a^\dagger.
            \end{align}
        \end{subequations}
        It is clear that $\hat R(\theta,\phi)\ket{0,N}_p$ is an $N$-photon Fock state in the mode $\hat c^\dagger=\alpha \hat a^\dagger +\beta \hat b^\dagger$ with $\alpha=e^{i\phi}\sin(\theta/2)$ and $\beta=\cos(\theta/2)$. Reciprocally, given such a mode, we can always find $(\theta,\phi)$ such that those relations are satisfied.

        \item \ref{def app: spin rotation} $\Leftrightarrow$ \ref{def app: spin operator}. First we observe that we have $\hat J_z\ket{0,N}_p=\frac{\hat b\hat b^\dagger-\hat a\hat a^\dagger}{2}\ket{0,N}_p=\frac{N}{2}\ket{0,N}_p$. As $\hat R(\theta,\phi)=e^{-i\theta\vec{\hat J}\cdot \vec n}$ for $\vec n=(-\sin(\phi),\cos(\phi),0)$, we see that the coherent state $\hat R(\theta,\phi)\ket{0,N}_p$ is an eigenvector with eigenvalue $N/2$ of
        \begin{equation}
            \hat R(\theta,\phi)\hat J_z\hat R(\theta,\phi)^\dagger=\vec{\hat J}\cdot\vec m,
        \end{equation}
        where $\vec m$ can be computed using Eq.~\eqref{eq: rotation vector}. We have
        \begin{equation}
            \vec m=(\sin(\theta)\cos(\phi),\sin(\theta)\sin(\phi),\cos(\theta)),
        \end{equation}
        which is exactly the point on the sphere with coordinates $(\theta,\phi)$ associated with the complex number $z$. Conversely, if a state $\ket{\psi}$ is an eigenvector with eigenvalue $N/2$ of $\vec{\hat J}\cdot \vec m$, with $\vec m$ with spherical coordinates $(\theta,\phi)$, then by performing a rotation $\hat R(-\theta,\phi)$, we would get the relation
        \begin{equation}
            \hat J_z \hat R(-\theta,\phi)\ket{\psi}=\frac{N}{2}\hat R(-\theta,\phi)\ket{\psi},
        \end{equation}
        This forces $\hat R(-\theta,\phi)\ket{\psi}=\ket{0,N}_p$, thus $\ket{\psi}=\hat R(\theta,\phi)\ket{0,N}_p$.
        
        To check that a coherent state is associated with a unique direction $\vec m$, we need to show that for $\vec n\neq \vec m$, $\vec{\hat J}\cdot \vec n$, $\vec{\hat J}\cdot\vec m$ cannot share the same eigenvector associated with the value $N/2$. As unitary transformations do not change the spectrum, by performing a suitable rotation, we can assume $\vec n=\vec e_z$ and $\vec m=\cos\alpha \vec e_z+\sin\alpha \vec e_x$. As
        \begin{equation}
            \left(\cos\alpha\hat J_z+\sin\alpha\hat J_x\right)\ket{0,N}_p=\frac{N\cos\alpha}{2}\ket{0,N}_p+\frac{\sin\alpha\sqrt{N}}{2}\ket{1,N-1}_p,
        \end{equation}
        it is clear that $\hat J_z$ and $\cos\alpha\hat J_z+\sin\alpha\hat J_x$ share an eigenvector with eigenvalue $N/2$ if and only if $\alpha=0$.
    \end{itemize}
\end{derivation}

\begin{result}[Spin coherent states and rotations]~\\
    \label{res: spin coherent rotation}
    The action of a rotation on a spin coherent state is to rotate the associated point on the sphere. More concretely, if $\ket{\vec m}_{sc}$ is the coherent state associated with the point $\vec m$ on the sphere, then for any rotation of angle $\theta$ around an axis $\vec n$, we have
    \begin{equation}
        e^{-i\theta \vec{\hat J}\cdot \vec n}\ket{\vec m}_{sc}\propto\ket{R_{\vec n}(\theta)\vec m}_{sc}.
    \end{equation}

\end{result}

\begin{derivation}
    By the point \ref{def app: spin operator} of Result~\ref{res: def spin coherent states}, the spin coherent state in the direction $m$ is characterized by the equation
    \begin{equation}
        (\vec{\hat J}\cdot \vec m) \ket{\vec m}_{sc}=\frac{N}{2}\ket{\vec m}_{sc}.
    \end{equation}
    By applying the unitary transformation $e^{-i\theta \vec{\hat J}\cdot \vec n}$ on both sides, we get
    \begin{equation}
        e^{-i\theta \vec{\hat J}\cdot \vec n}(\vec{\hat J}\cdot \vec m) \ket{\vec m}_{sc}=\frac{N}{2}e^{-i\theta \vec{\hat J}\cdot \vec n}\ket{\vec m}_{sc}.
    \end{equation}
    Using Eq.~\eqref{eq: rotation vector}, we see that this can be written as
    \begin{equation}
        (\vec{\hat J}\cdot R_{\vec n}(\theta)\vec m)e^{-i\theta \vec{\hat J}\cdot \vec n}\ket{\vec m}_{sc}=\frac{N}{2}e^{-i\theta \vec{\hat J}\cdot \vec n}\ket{\vec m}_{sc}.
    \end{equation}
    Thus, by definition, $e^{-i\theta \vec{\hat J}\cdot \vec n}\ket{\vec m}_{sc}$ is a spin coherent state associated with the vector $R_{\vec n}(\theta)\vec m$. By uniqueness of the association between spin coherent states and unit vectors, we get the desired result, noting that we get only a proportionality relation, as two eigenvectors of the same nondegenerate eigenvalue may differ by a global phase.
\end{derivation}

\begin{result}[Scalar product of spin coherent states]~\\
    \label{res: scalar product spin coherent}
    The scalar product between two spin coherent states $\ket{z}_{sc}$ and $\ket{z'}_{sc}$ is given by the following expression
    \begin{equation}
        \prescript{}{sc}{\langle} z' {| z \rangle}_{sc}=\frac{(1+z'^* z)^N}{(1+\abs{z'}^2)^{N/2}(1+\abs{z}^2)^{N/2}}.
    \end{equation}
    Additionally, we can write
    \begin{equation}
        \abs\big{\prescript{}{sc}{\langle} \vec n' {| \vec n \rangle}_{sc}}^2=\left(\frac{1+\vec n\cdot\vec n'}{2}\right)^N,
    \end{equation}
    if we use the parametrization of the spin coherent states in terms of unit 3D vectors
\end{result}

\begin{derivation}
    $\blacktriangleright$ {\bf Complex parametrization.} Using the expansion of spin coherent states in the Fock basis, we have directly
    \begin{align}
        \prescript{}{sc}{\langle} z' {| z \rangle}_{sc}&=\frac{1}{(1+\abs{z'}^2)^{N/2}(1+\abs{z}^2)^{N/2}}\sum_{k=0}^N (z'^*)^k z^k \binom{N}{k},\\
        &=\frac{1}{(1+\abs{z'}^2)^{N/2}(1+\abs{z}^2)^{N/2}}(1+z'^* z)^N.
    \end{align}
    \medskip

    \noindent
    $\blacktriangleright$ {\bf Vector parametrization.} For the second relation, we observe that, considering $z=\Phi(\vec n)$ and $z'=\Phi(\vec n')$, we have the following identity
\begin{equation}
    \frac{\abs{1+z'^* z}^2}{(1+\abs{z'}^2)(1+\abs{z}^2)}
    =
    \frac{1+\vec n(z')\cdot\vec n(z)}{2}.
\end{equation}
    Indeed, using the explicit expression
    \begin{equation}
        \vec n=\frac{1}{1+\abs{z}^2}\left(2\Re z,\,2\Im z,\,1-\abs{z}^2\right),
    \end{equation}
    and similarly for $z'$, we compute
    \begin{equation}
        \vec n\cdot\vec n'=\frac{4\Re z'\Re z +4\Im z'\Im z +(\abs{z'}^2-1)(\abs{z}^2-1)}{(1+\abs{z'}^2)(1+\abs{z}^2)}.
    \end{equation}
    Since
    \begin{equation}
        4\Re z'\Re z+4\Im z'\Im z = 2(z'^* z+z' z^*),
    \end{equation}
    we obtain
    \begin{equation}
        \vec n\cdot\vec n'=\frac{2(z'^* z+z' z^*)+(\abs{z'}^2-1)(\abs{z}^2-1)}{(1+\abs{z'}^2)(1+\abs{z}^2)}.
    \end{equation}
    Adding $1$ gives
    \begin{equation}
        1+\vec n\cdot\vec n' = \frac{ (1+\abs{z'}^2)(1+\abs{z}^2) +2(z'^* z+z' z^*) +(\abs{z'}^2-1)(\abs{z}^2-1) } {(1+\abs{z'}^2)(1+\abs{z}^2)}.
    \end{equation}
    A straightforward expansion of the numerator yields
    \begin{equation}
        (1+\abs{z'}^2)(1+\abs{z}^2) +(\abs{z'}^2-1)(\abs{z}^2-1) = 2\bigl(1+\abs{z'}^2\abs{z}^2\bigr),
    \end{equation}
    so that
    \begin{equation}
        1+\vec n\cdot\vec n' = \frac{ 2\bigl(1+\abs{z'}^2\abs{z}^2 +z'^* z+z' z^*\bigr) } {(1+\abs{z'}^2)(1+\abs{z}^2)}.
    \end{equation}
    Recognizing
    \begin{equation}
        1+\abs{z'}^2\abs{z}^2+z'^* z+z' z^* = \abs{1+z'^* z}^2,
    \end{equation}
    we conclude that
    \begin{equation}
        \frac{1+\vec n(z')\cdot\vec n(z)}{2} = \frac{\abs{1+z'^* z}^2} {(1+\abs{z'}^2)(1+\abs{z}^2)},
    \end{equation}
    which proves the identity, and directly yields the second expression for the scalar product of spin coherent states.
    \medskip

    \noindent
    $\blacktriangleright$ {\bf Alternate point of view for the vector parametrization.} Alternatively, the scalar product in terms of Bloch vectors can easily be obtained, with the point of view that spin coherent states are, in the first-quantization picture, states of the form
    \begin{equation}
        \ket{\psi}=\ket{\varphi}\otimes\cdots\otimes\ket{\varphi}.
    \end{equation}
    Knowing that the scalar product of two pure qubit states has the following expression in terms of the Bloch vector
    \begin{equation}
        \abs{\braket{\varphi'}{\varphi}}^2=\frac{1+\vec n\cdot\vec n'}{2},
    \end{equation}
    the result for spin coherent states directly follows.
\end{derivation}

\begin{result}[Closure relation of spin coherent states]~\\
    \label{res: closure relation spin coherent}
    Spin coherent states satisfy the following closure relation
    \begin{equation}
        \frac{N+1}{\pi}\int_{\C} \frac{\dd^2 z}{(1+\abs{z}^2)^2}{| z \rangle}_{sc}\prescript{}{sc}{\langle z |}=\1.
    \end{equation}
\end{result}

\begin{derivation}
    This is simply an integral computation. First we expand the expression of the spin coherent state
    \begin{align}
        \frac{N+1}{\pi}\int_{\C}& \frac{\dd^2 z}{(1+\abs{z}^2)^2}{| z \rangle}_{sc}\prescript{}{sc}{\langle z |}\notag\\
        &=\frac{N+1}{\pi}\sum_{k,l=0}^N \sqrt{\binom{N}{k}\binom{N}{l}}\int_\C \frac{\dd ^2z}{(1+\abs{z}^2)^{N+2}}z^k {z^*}^l\ket{k,N-k}_{p}\prescript{}{p}{\bra{l,N-l}}.
    \end{align}
    We thus need to compute for two integers $k$ and $l$ the integral $\int_\C \frac{\dd ^2z}{(1+\abs{z}^2)^{N+2}}z^k {z^*}^l$. By going to polar coordinates $(r,\phi)$ with $z=re^{i\phi}$ and $\dd^2 z=r\dd r\dd\phi$, we have
    \begin{subequations}
        \begin{align}
            \int_\C \frac{\dd ^2z}{(1+\abs{z}^2)^{N+2}}z^k {z^*}^l&=\int \frac{r\dd r\dd \phi}{(1+r^2)^{N+2}}r^{k+l}e^{i\phi(k-l)},\\
            &=\int_0^\infty \frac{r^{k+l+1}\dd r}{(1+r^2)^{N+2}}\int_0^{2\pi}\dd\phi e^{i\phi(k-l)}.
        \end{align}
    \end{subequations}
    The second integral equals $2\pi\delta_{k,l}$; we thus need to compute the integral in $r$, in the case $k=l$.  By performing the change of variable $u=r^2$ we see that it equals
    \begin{equation}
        \int_0^\infty \frac{r^{2k+1}\dd r}{(1+r^2)^{N+2}}=\frac{1}{2}\int_0^\infty \frac{u^k\dd u}{(1+u)^{N+2}}
    \end{equation}
    This integral can be computed in 3 ways. 
    \begin{itemize}
        \item First, we clearly see that it is related to the beta function defined as
        \begin{equation}
            B(x,y)=\int_0^\infty \frac{t^{x-1}}{(1+t)^{x+y}}\,\dd t,
        \end{equation}
        if $x=k+1$ and $y=N+1-k$. Using the relation between the beta and gamma functions $B(x,y)=\frac{\Gamma(x)\Gamma(y)}{\Gamma(x+y)}$ and the property of the gamma function $\Gamma(n+1)=n!$ for $n\in\N$, we get
        \begin{subequations}
            \begin{align}
                \int_0^\infty \frac{u^k\dd u}{(1+u)^{N+2}}&=B(k+1,N+1-k)=\frac{\Gamma(k+1)\Gamma(N+1-k)}{\Gamma(N+2)},\\
                &=\frac{k!(N-k)!}{(N+1)!},\\
                &=\frac{1}{N+1}\binom{N}{k}^{-1}.
            \end{align}
        \end{subequations}

        \item In the second approach, we essentially rederive in this particular case the properties of the beta and gamma functions that we used. Denoting our integral $I_{N,k}=\int_0^\infty \frac{u^k\dd u}{(1+u)^{N+2}}$, we perform an integration by parts to observe a recurrence relation (for $1\leq k\leq N$)
        \begin{equation}
            I_{N,k}=\frac{-1}{N+1}\left[\frac{u^{k}}{(1+u)^{N+1}}\right]_0^\infty+\frac{k}{N+1}\int_0^\infty \frac{u^{k-1}}{(1+u)^{N+1}}\,\dd u=\frac{k}{N+1}I_{N-1,k-1}.
        \end{equation}
        By iterating, we get
        \begin{equation}
            I_{N,k}=\frac{k}{N+1}\cdots\frac{1}{N+2-k}I_{N-k,0}=\frac{k!(N+1-k)!}{(N+1)!}I_{N-k,0}.
        \end{equation}
        When the second parameter is zero, the integral is easy to compute
        \begin{equation}
            I_{N-k,0}=\int_0^\infty \frac{\dd u}{(1+u)^{N-k+2}}=\left[\frac{-(1+u)^{-N-1+k}}{N+1-k}\right]_0^\infty=\frac{1}{N+1-k}.
        \end{equation}
        Thus 
        \begin{equation}
            \int_0^\infty \frac{u^k\dd u}{(1+u)^{N+2}}=\frac{k!(N+1-k)!}{(N+1)!}\frac{1}{N+1-k}=\frac{k!(N-k)!}{(N+1)!}=\frac{1}{N+1}\binom{N}{k}^{-1}.
        \end{equation}

        \item The third proof we provide is way more technical. It stems from the realization that the function we wish to integrate is a rational function, which we know can always be computed. We thus try to apply the general method: polynomial Euclidean division followed by partial fraction decomposition. A clever use of the binomial formula allows us to perform this in practice. Since $u^k=(1+u-1)^k=\sum_{l=0}^k \binom{k}{l}(1+u)^l(-1)^{k-l}$ we have
        \begin{subequations}
            \allowdisplaybreaks
            \begin{align}
                \int_0^\infty\frac{u^k\dd u}{(1+u)^{N+2}}&=\sum_{l=0}^k \binom{k}{l}(-1)^{k-l}\int_0^\infty\frac{\dd u}{(1+u)^{N+2-l}},\\
                &=\sum_{l=0}^k \binom{k}{l}(-1)^{k-l}\left[\frac{-(1+u)^{-N-1+l}}{N+1-l}\right]_0^\infty,\\
                &=\sum_{l=0}^k \binom{k}{l}\frac{(-1)^{k-l}}{N+1-l}.
            \end{align}
        \end{subequations}
        The tricky part is to deal with this sum and show that it admits a simple expression in terms of binomial coefficients. The use of partial fraction decomposition provides a very clever answer. Observe that the following function is a rational fraction with simple poles
        \begin{equation}
            F(x)=\frac{k!}{x(x-1)(x-2)\cdots (x-k)}.
        \end{equation}
        As such, it admits a partial fraction decomposition of the form $F(x)=\sum_{l=0}^k\frac{ A_l}{x-l}$ where the coefficients $A_l$ can be computed as
        \begin{subequations}
            \begin{align}
                A_l&=\lim_{x\to l}(x-l)F(x),\\
                &=\frac{k!}{l(l-1)\cdots 1\times (-1)\cdots (l-k)},\\
                &=\frac{(-1)^{k-l}k!}{l!(k-l)!},\\
                &=(-1)^{k-l}\binom{k}{l}.
            \end{align}
        \end{subequations}
        As such our sum admits the simple expression
        \begin{subequations}
            \begin{align}
                \sum_{l=0}^k \binom{k}{l}\frac{(-1)^{k-l}}{N+1-l}&=F(N+1),\\
                &=\frac{k!}{(N+1)\cdots (N+1-k)},\\
                &=\frac{k!(N-k)!}{(N+1)!},\\
                &=\frac{1}{N+1}\binom{N}{k}^{-1}.
            \end{align}
        \end{subequations}
    \end{itemize}
    Based on the expression obtained the original integral can simply be evaluated to
    \begin{equation}
        \int_\C \frac{\dd ^2z}{(1+\abs{z}^2)^{N+2}}(-z^*)^k (-z)^l=2\pi \delta_{k,l}\frac{1}{2}\frac{1}{N+1}\binom{N}{k}^{-1}=\frac{\pi}{N+1}\binom{N}{k}^{-1}\delta_{k,l}.
    \end{equation}
    And we thus indeed have the closure relation
    \begin{equation}
        \frac{N+1}{\pi}\int_{\C} \frac{\dd^2 z}{(1+\abs{z}^2)^2}\ket{z}_{sc}\prescript{}{sc}{\bra{z}}=\sum_{k=0}^N \ket{k,N-k}_{p}\prescript{}{p}{\bra{k,N-k}}=\1.
    \end{equation}
\end{derivation}

\begin{result}[Rotation of spin coherent states]~\\
    \label{res: rotation coherent state}
    Denoting by $f$ the Möbius transformation associated with the rotation $R_{\vec n}(\theta)$, where the explicit expression of $f$ is given by Result~\ref{res: möbius and rotation}, we have
    \begin{equation}
        e^{-i\theta\vec n\cdot\vec{\hat J}}\ket{z}_{sc}=(cz+a^*)^N\left(\frac{1+\abs{f(z)}^2}{1+\abs{z}^2}\right)^{N/2}\ket{f(z)}_{sc}=\left(\frac{cz+a^*}{\abs{cz+a^*}}\right)^N\ket{f(z)}_{sc}.
    \end{equation}
\end{result}

\begin{derivation}
    We start by recalling that 
    \begin{equation}
        \ket{z}_{sc}=\frac{1}{(1+\abs{z}^2)^{N/2}}\frac{\left(z\hat a^\dagger +\hat b^\dagger\right)^N}{\sqrt{N!}}\vac.
    \end{equation}
    Using the transformation rule of Eq.~\eqref{eq: rotation creation op}, which can be rewritten in terms of the Möbius transformation parameters
    \begin{align}
        e^{-i\theta\vec n\cdot\vec{\hat J}}\hat a^\dagger e^{i\theta\vec n\cdot\vec{\hat J}}=a \hat a^\dagger + c \hat b^\dagger, && e^{-i\theta\vec n\cdot\vec{\hat J}}\hat b^\dagger e^{i\theta\vec n\cdot\vec{\hat J}}=-c^* \hat a^\dagger + a^* \hat b^\dagger,
    \end{align}
    we see that
    \begin{subequations}
        \begin{align}
            e^{-i\theta\vec n\cdot\vec{\hat J}}\ket{z}_{sc}&=\frac{1}{(1+\abs{z}^2)^{N/2}}\frac{\left(z(a \hat a^\dagger + c \hat b^\dagger) +(-c^* \hat a^\dagger + a^* \hat b^\dagger)\right)^N}{\sqrt{N!}}\vac,\\
            &=\frac{1}{(1+\abs{z}^2)^{N/2}}\frac{\left((za-c^*)\hat a^\dagger +( cz+ a^* )\hat b^\dagger\right)^N}{\sqrt{N!}}\vac,\\
            &=\frac{(cz+ a^*)^N}{(1+\abs{z}^2)^{N/2}}\frac{\left(\frac{az - c^*}{cz+ a^*}\hat a^\dagger + \hat b^\dagger\right)^N}{\sqrt{N!}}\vac,\\
            &=(cz+a^*)^N\left(\frac{1+\abs{f(z)}^2}{1+\abs{z}^2}\right)^{N/2}\frac{1}{(1+\abs{f(z)}^2)^{N/2}}\frac{(f(z)\hat a^\dagger+\hat b^\dagger)^N}{\sqrt{N!}}\vac,\\
            &=(cz+a^*)^N\left(\frac{1+\abs{f(z)}^2}{1+\abs{z}^2}\right)^{N/2}\ket{f(z)}_{sc}.
        \end{align}
    \end{subequations}
    It remains to simplify the proportionality coefficient. We first observe that
    \begin{subequations}
        \allowdisplaybreaks
        \begin{align}
            1+\abs{f(z)}^2&=1+\frac{\abs{az-c^*}^2}{\abs{cz+a^*}^2},\\
            &=\frac{(cz+a^*)(c^*z^*+a)+(az-c^*)(a^*z^*-c)}{\abs{cz+a^*}^2},\\
            &=\frac{\abs{c}^2\abs{z}^2+caz+a^*c^*z^*+\abs{a}^2+\abs{a}^2\abs{z}^2-caz-c^*a^*z^*+\abs{c}^2}{\abs{cz+a^*}^2},\\
            &=\frac{(\abs{a}^2+\abs{c}^2)(1+\abs{z}^2)}{\abs{cz+a^*}^2},\\
            &=\frac{1+\abs{z}^2}{\abs{cz+a^*}^2}.
        \end{align}
    \end{subequations}
    Thus we indeed have
    \begin{equation}
        (cz+a^*)^N\left(\frac{1+\abs{f(z)}^2}{1+\abs{z}^2}\right)^{N/2}=\left(\frac{cz+a^*}{\abs{cz+a^*}}\right)^N.
    \end{equation}
    And, as announced
    \begin{equation}
        e^{-i\theta\vec n\cdot\vec{\hat J}}\ket{z}_{sc}=\left(\frac{cz+a^*}{\abs{cz+a^*}}\right)^N\ket{f(z)}_{sc}.
    \end{equation}
\end{derivation}

\begin{result}[Reconstructing a state from its Majorana polynomial]~\\
    \label{res: reconstructing state from Majo}
    A two-mode quantum state $\ket{\psi}$ with $N$ photons in total can be expressed explicitly in terms of its Majorana polynomial as
    \begin{equation}
        \ket{\psi}=P_{\ket{\psi}}(\hat a^\dagger {\hat b^\dagger}^{-1})\ket{0,N}_p,
    \end{equation}
    where ${\hat b^\dagger}^{-1}$ is the left inverse (\ie, ${\hat b^\dagger}^{-1}\hat b^\dagger=\1$, but $\hat b^\dagger{\hat b^\dagger}^{-1}\neq\1$) of $\hat b^\dagger$ formally defined on the mode $b$ by
    \begin{equation}
        {\hat b^\dagger}^{-1}:\begin{cases}
            \ket{n}\mapsto \frac{1}{\sqrt{n}}\ket{n-1}\quad(n\geq1),\\
            \ket{0}\mapsto 0.
        \end{cases}
    \end{equation}
\end{result}

\begin{derivation}
    It is a simple verification.
    \begin{subequations}
        \begin{align}
            P_{\ket{\psi}}(\hat a^\dagger {\hat b^\dagger}^{-1})\ket{0,N}_p&=\sum_{k=0}^N c_k \sqrt{\binom{N}{k}} (\hat a^\dagger {\hat b^\dagger}^{-1})^k\ket{0,N}_p,\\
            &=\sum_{k=0}^N c_k \sqrt{\binom{N}{k}}  \frac{\sqrt{k!}}{\sqrt{N(N-1)\cdots(N-k+1)}} \ket{k,N-k}_p,\\
            &=\sum_{k=0}^N c_k \ket{k,N-k}_p=\ket{\psi}.
        \end{align}
    \end{subequations}
\end{derivation}

\begin{result}[Scalar product expressed with the Majorana Polynomial]~\\
    \label{res: scalar product from majo}
    The scalar product of two states $\ket{\psi_1}$ and $\ket{\psi_2}$ can be expressed in terms of their respective Majorana polynomials $P_{\ket{\psi_1}}$ and $P_{\ket{\psi_2}}$ as
    \begin{equation}
        \bra{\psi_1}\ket{\psi_2}=\frac{N+1}{\pi}\int_{\C} \frac{d^2 z}{(1+\abs{z}^2)^{N+2}} P_{\ket{\psi_1}}(z)^* P_{\ket{\psi_2}}(z).
    \end{equation}
\end{result}

\begin{derivation}
    This is a simple consequence of the resolution of the identity in terms of coherent states of Eq.~\eqref{eq: closure relation spin coherent states}. We have
    \begin{subequations}
        \begin{align}
            \bra{\psi_1}\ket{\psi_2}&=\frac{N+1}{\pi}\int_{\C} \frac{d^2 z}{(1+\abs{z}^2)^{2}} \bra{\psi_1}\ket{z}_{sc}\prescript{}{sc}{\braket{z}{\psi_2}},\\
            &=\frac{N+1}{\pi}\int_{\C} \frac{d^2 z}{(1+\abs{z}^2)^{N+2}} (1+\abs{z}^2)^{N/2}\prescript{}{sc}{\braket{z^*}{\psi_1}}^* (1+\abs{z}^2)^{N/2}\prescript{}{sc}{\braket{z^*}{\psi_2}},\\
            &=\frac{N+1}{\pi}\int_{\C} \frac{d^2 z}{(1+\abs{z}^2)^{N+2}} P_{\ket{\psi_1}}(z)^* P_{\ket{\psi_2}}(z).
        \end{align}
    \end{subequations}
\end{derivation}

\begin{result}[Rotation of Majorana polynomials]~\\
    \label{res: Majo and rotations}
    Finally, we show how the Majorana polynomial transforms under the action of a rotation $e^{-i\theta\vec n\cdot\vec{\hat J}}$. We have
    \begin{equation}
        P_{e^{-i\theta\vec n\cdot\vec{\hat J}}\ket{\psi}}(z)=(a^*-c^*z)^N P_{\ket{\psi}}(f^{-1}(z^*)^*)=(a^*-c^*z)^N P_{\ket{\psi}}\left(\frac{az+c}{-c^*z+a^*}\right).
    \end{equation}
\end{result}

\begin{derivation}
    We essentially only need to use the definition of the Majorana polynomial and use the transformation rule of the spin coherent states under rotation. As we will use the functions $f$, $f^{-1}$ and its composition with the conjugation, we first recall
    \begin{align}
        f:z\mapsto\frac{a z-c^*}{cz+a^*}, && f^{-1}:z\mapsto \frac{a^* z + c^*}{-c z + a}, && f^{-1}(z^*)^*=\frac{a z + c}{-c^* z + a^*}.
    \end{align}
    We then consider the following
    \begin{equation}
        e^{i\theta\vec n\cdot\vec{\hat J}}\ket{z^*}_{sc}=(-cz^*+a)^N\left(\frac{1+\abs{f^{-1}(z^*)}^2}{1+\abs{z}^2}\right)^{N/2} \ket{f^{-1}(z^*)}_{sc},
    \end{equation}
    which we obtain from Eq.~\eqref{eq: rotation coherent state with z} by noting that replacing $\theta$ by $-\theta$ corresponds to considering the inverse transformation $f^{-1}$. We can now compute
    \begin{subequations}
        \begin{align}
            P_{e^{-i\theta\vec n\cdot\vec{\hat J}}\ket{\psi}}(z)&=(1+\abs{z}^2)^{N/2}\prescript{}{sc}{\bra{z^*}}e^{-i\theta\vec n\cdot\vec{\hat J}}\ket{\psi},\\
            &=(1+\abs{z}^2)^{N/2}\left(e^{i\theta\vec n\cdot\vec{\hat J}}\ket{z^*}_{sc}\right)^\dagger \ket{\psi},\\
            &=(1+\abs{z}^2)^{N/2}(-cz^*+a)^{*N}\left(\frac{1+\abs{f^{-1}(z^*)}^2}{1+\abs{z}^2}\right)^{N/2} \prescript{}{sc}{\bra{f^{-1}(z^*)}\ket{\psi}},\\
            &= (a^*-c^* z)^N P_{\ket{\psi}}(f^{-1}(z^*)^*),\\
            &= (a^*-c^* z)^N P_{\ket{\psi}}\left(\frac{az+c}{-c^*z+a^*}\right).
        \end{align}
    \end{subequations}
\end{derivation}

\begin{result}[Majorana stars and state decomposition]~\\
    \label{res: Majo stars and state decomposition}
    A pure two-mode state with $N$ single photons can always be written as
    \begin{equation}
        \ket{\psi}=\Lambda\prod_{k=1}^N \left(e^{i\phi_k}\sin(\theta_k/2)\hat a^\dagger + \cos(\theta_k/2) \hat b^\dagger\right)\vac=\Lambda\prod_{k=1}^N \hat c_k^\dagger\vac,
    \end{equation}
    with $\hat c_k^\dagger=e^{i\phi_k}\sin(\theta_k/2)\hat a^\dagger + \cos(\theta_k/2) \hat b^\dagger$, where the relation to the roots $\alpha_k$ of the Majorana polynomial is
    \begin{equation}
        \alpha_k=-e^{-i\phi_k}\operatorname{cotan}(\theta_k/2).
    \end{equation}
\end{result}

\begin{derivation}
    We introduce the Majorana polynomial $P_{\ket{\psi}}$, which we factor in terms of its roots
    \begin{equation}
        P_{\ket{\psi}}(z)=\gamma_d \prod_{k=1}^d (z-\alpha_k),
    \end{equation}
    with $d\leq N$ the degree of the polynomial, $\gamma_d=c_d\sqrt{\binom{N}{d}}$ the leading coefficient, and $\alpha_k$ the (finite) roots counted with multiplicity. Using Result~\ref{res: reconstructing state from Majo}, we obtain the relation
    \begin{equation}
        \ket{\psi}=P_{\ket{\psi}}(\hat a^\dagger{\hat b^\dagger}^{-1})\ket{0,N}_p=\gamma_d \prod_{k=1}^d (\hat a^\dagger{\hat b^\dagger}^{-1}-\alpha_k)\ket{0,N}_p,
    \end{equation}
    where ${\hat b^\dagger}^{-1}$ is the left inverse of $\hat b^\dagger$ satisfying ${\hat b^\dagger}^{-1}\hat b^\dagger=\1$. Using $\ket{0,N}_p=\frac{1}{\sqrt{N!}}{\hat b^\dagger}^N\vac$, we have
    \begin{subequations}
        \begin{align}
            \ket{\psi}&=\frac{\gamma_d}{\sqrt{N!}} \prod_{k=1}^d (\hat a^\dagger{\hat b^\dagger}^{-1}-\alpha_k){\hat b^\dagger}^N\vac,\\
            &=\frac{\gamma_d}{\sqrt{N!}} \prod_{k=1}^d (\hat a^\dagger -\alpha_k \hat b^\dagger){\hat b^\dagger}^{N-d}\vac.
        \end{align}
    \end{subequations}
    For $k\leq d$, we parametrize
    \begin{equation}
        \alpha_k=-e^{-i\phi_k}\operatorname{cotan}(\theta_k/2),
    \end{equation}
    which allows us to rewrite
    \begin{equation}
        \hat a^\dagger -\alpha_k \hat b^\dagger=\frac{e^{-i\phi_k}}{\sin(\theta_k/2)}\left(e^{i\phi_k}\sin(\theta_k/2)\hat a^\dagger + \cos(\theta_k/2) \hat b^\dagger\right).
    \end{equation}
    Additionally for $k>d$ we define $\theta_k=0$ and $\phi_k=0$, so that we can rewrite
    \begin{equation}
        \hat b^\dagger=\left(e^{i\phi_k}\sin(\theta_k/2)\hat a^\dagger + \cos(\theta_k/2) \hat b^\dagger\right).
    \end{equation}
    We thus get
    \begin{equation}
        \ket{\psi}=\Lambda\prod_{k=1}^N \left(e^{i\phi_k}\sin(\theta_k/2)\hat a^\dagger + \cos(\theta_k/2) \hat b^\dagger\right)\vac,
    \end{equation}
    where we have taken
    \begin{equation}
        \Lambda=\frac{\gamma_d}{\sqrt{N!}}\prod_{k=1}^d \frac{e^{-i\phi_k}}{\sin(\theta_k/2)}.
    \end{equation}
\end{derivation}

\begin{result}[Heisenberg-Robertson uncertainty relation]~\\
    \label{res: heisenberg robertson}
    For two general observables $\hat A$ and $\hat B$, we have
    \begin{equation}
        \Delta^2 A \Delta^2 B\geq \frac{1}{4}\abs{\ev{[\hat A,\hat B]}}^2,
    \end{equation}
    where $\Delta^2 A=\ev{\hat A^2}-\ev{\hat A}^2$ is the variance of the observable $\hat A$. 
\end{result}

\begin{derivation}
    $\blacktriangleright$ {\bf First proof.} A compact proof can be made with a clever use of the Cauchy-Schwarz inequality. Notice that the mapping $(\hat A,\hat B)\mapsto \ev{\hat A^\dagger \hat B}$ is a complex inner product of the space of operators. Applying the Cauchy-Schwarz inequality to the operators $\delta \hat A=\hat A-\ev{\hat A}$ and $\delta \hat B=\hat B-\ev{\hat B}$, we get
    \begin{equation}
        \Delta^2\hat A\Delta^2\hat B=\ev{(\delta \hat A)^\dagger \delta \hat A}\ev{(\delta \hat B)^\dagger \delta \hat B}\geq \abs{\ev{(\delta \hat A)^\dagger \delta \hat B}}^2.
    \end{equation} 
    We finally use the inequality on complex numbers $\abs{\Im z}\leq \abs{z}$. As 
    \begin{equation}
        \Im(\ev{\delta \hat A\delta \hat B})=\frac{1}{2i}\ev{[\delta \hat A,\delta \hat B]}=\frac{1}{2i}\ev{[\hat A,\hat B]}.
    \end{equation}
    We thus get the desired result.
    \medskip

    \noindent
    $\blacktriangleright$ {\bf Second proof.} An alternate proof can be provided that avoids the use of the Cauchy-Schwarz inequality. In fact it is based on the proof idea of the Cauchy-Schwarz inequality.  We introduce the operators $\delta \hat A=\hat A-\ev{\hat A}$  and $\delta\hat B=\hat B-\ev{\hat B}$. We notice that for all $\lambda\in\R$, the operator $(\delta \hat A+i\lambda \delta \hat B)^\dagger(\delta \hat A+i\lambda \delta \hat B)$ is positive semi-definite. Thus for any state $\ket{\psi}$, its expectation value is positive. By expanding, we get
    \begin{equation}
        \ev{(\delta \hat A+i\lambda \delta \hat B)^\dagger(\delta \hat A+i\lambda \delta \hat B)}=\Delta^2 A +\lambda^2 \Delta^2 B +i\lambda \ev{[\hat A,\hat B]}\geq 0.
    \end{equation}
    This is a second-order polynomial in $\lambda$ which must be positive for all $\lambda$. This is only possible if its discriminant is negative
    \begin{equation}
        -\ev{[\hat A,\hat B]}^2 -4 \Delta^2 A \Delta^2 B\leq 0.
    \end{equation}
    Notice that $[\hat A,\hat B]$ is anti-Hermitian and thus the corresponding expectation value is purely imaginary, thus $\ev{[\hat A,\hat B]}^2=-\abs{\ev{[\hat A,\hat B]}}^2$. We thus get the desired result. 
\end{derivation}

\begin{result}[Expectation value and variance of an arbitrary spin component of a spin coherent state]~\\
    \label{res: variance of arbitrary spin component}
    For a spin coherent state $\ket{\vec n}_{sc}$ in an arbitrary direction $\vec n$, the expectation value and the variance of the angular momentum operator $\vec m\cdot\vec{\hat J}$ are
    \begin{align}
        \expval{\vec m\cdot\vec{\hat J}}_{\ket{\vec n}_{sc}}=\frac{N}{2}(\vec n\cdot \vec m), && \Delta^2 (\vec m\cdot\vec{\hat J})_{\ket{\vec n}_{sc}}=\frac{N}{4}(1-(\vec n\cdot \vec m)^2).
    \end{align}
\end{result}

\begin{derivation}
    As this computation is invariant under rotation, it is sufficient to consider the case $\vec n=\vec e_z$ (\ie, $\ket{\vec n}_{sc}=\ket{0,N}_p$) and $\vec m=(\sin(\theta),0,\cos(\theta))$. Considering
    \begin{equation}
        \left(\cos(\theta)\hat J_z+\sin(\theta)\hat J_x\right)\ket{0,N}_p=\frac{N}{2}\cos(\theta)\ket{0,N}_p+\sin(\theta)\frac{\sqrt{N}}{2}\ket{1,N-1}_p,
    \end{equation}
    we obtain the expectation value by taking the scalar product with $\ket{0,N}$
    \begin{equation}
        \expval{\cos(\theta)\hat J_z+\sin(\theta)\hat J_x}=\frac{N}{2}\cos(\theta),
    \end{equation}
    and the quadratic expectation value is obtained by considering the scalar product with itself
    \begin{equation}
        \expval{(\cos(\theta)\hat J_z+\sin(\theta)\hat J_x)^2}=\frac{N^2}{4}\cos^2(\theta)+\frac{N}{4}\sin^2(\theta),
    \end{equation}
    Leading to the variance
    \begin{equation}
        \Delta^2 (\cos(\theta)\hat J_z+\sin(\theta)\hat J_x)=\frac{N}{4}\sin^2(\theta).
    \end{equation}
    Going back to the general case simply requires replacing $\cos(\theta)=\vec n\cdot \vec m$ and $\sin^2(\theta)=1-(\vec n\cdot \vec m)^2$.
\end{derivation}

\begin{result}[SSRC to CV coherent states]~\\
    \label{res: SSRC to CV coherent states}
    Inputting the state $\ket{0}_A\ket{N}_B$ into the beam splitter transformation $\hat U$ defined by
    \begin{align}
        \hat U\hat a^\dagger\hat U^\dagger=\sqrt{1-\frac{\abs{\alpha}^2}{N}}\hat a^\dagger-\frac{\alpha^*}{\sqrt{N}}\hat b^\dagger, && \hat U\hat b^\dagger\hat U^\dagger =\frac{\alpha}{\sqrt{N}}\hat a^\dagger+\sqrt{1-\frac{\abs{\alpha}^2}{N}}\hat b^\dagger,
    \end{align}
    approximately yields
    \begin{equation}
        \hat U\ket{0}_A\ket{N}_B\simeq e^{-\abs{\alpha}^2/2}\sum_{k=0}^N \frac{\alpha^k}{\sqrt{k!}}\ket{k}_A\ket{N-k}_B.
    \end{equation}
\end{result}

\begin{derivation}
    We simply compute
    \begin{subequations}
        \allowdisplaybreaks
        \begin{align}
            \hat U\ket{0}_A\ket{N}_B&=\frac{(\hat U\hat b^\dagger\hat U^\dagger)^N}{\sqrt{N!}}\vac,\\
            &=\frac{1}{\sqrt{N!}}\left(\frac{\alpha}{\sqrt{N}}\hat a^\dagger+\sqrt{1-\frac{\abs{\alpha}^2}{N}}\hat b^\dagger\right)^N\vac,\\
            &=\frac{1}{\sqrt{N!}}\sum_{k=0}^N \binom{N}{k}\frac{\alpha^k}{\sqrt{N}^k}\sqrt{1-\frac{\abs{\alpha}^2}{N}}^{N-k} (a^\dagger)^k(\hat b^\dagger)^{N-k}\vac,\\
            &=\sum_{k=0}^N \binom{N}{k}\frac{\sqrt{k!}\sqrt{(N-k)!}}{\sqrt{N!}}\frac{\alpha^k}{\sqrt{N}^k}\sqrt{1-\frac{\abs{\alpha}^2}{N}}^{N-k} \ket{k}_A\ket{N-k}_B,\\
            &= \sum_{k=0}^N \sqrt{\frac{N(N-1)\cdots(N-k+1)}{k!}}\frac{\alpha^k}{\sqrt{N}^k}\sqrt{1-\frac{\abs{\alpha}^2}{N}}^{N}\notag\\
            &\qquad\times\sqrt{1-\frac{\abs{\alpha}^2}{N}}^{-k}\ket{k}_A\ket{N-k}_B,\\
            &\simeq \sum_{k=0}^N \frac{\sqrt{N}^k}{\sqrt{k!}}\frac{\alpha^k}{\sqrt{N}^k} e^{-\abs{\alpha}^2/2}\ket{k}_A\ket{N-k}_B,\\
            &=e^{-\abs{\alpha}^2/2}\sum_{k=0}^N \frac{\alpha^k}{\sqrt{k!}}\ket{k}_A\ket{N-k}_B.
        \end{align}    
    \end{subequations}
    where we have used the pointwise approximations
    \begin{align}
        N(N-1)\cdots(N-k+1)\simeq N^k, &&\sqrt{1-\frac{\abs{\alpha}^2}{N}}^N\simeq e^{-\abs{\alpha}^2/2}, &&\sqrt{1-\frac{\abs{\alpha}^2}{N}}^k\simeq 1.
    \end{align}
\end{derivation}

\begin{result}[SSRC rotation to CV displacement]~\\
    \label{res: SSRC rotation to CV displacement}
    Considering the mode rotation $\hat U$ defined by
    \begin{align}
        \hat U\hat a^\dagger\hat U^\dagger=\sqrt{1-\frac{\abs{\alpha}^2}{N}}\hat a^\dagger-\frac{\alpha^*}{\sqrt{N}}\hat b^\dagger, && \hat U\hat b^\dagger\hat U^\dagger =\frac{\alpha}{\sqrt{N}}\hat a^\dagger+\sqrt{1-\frac{\abs{\alpha}^2}{N}}\hat b^\dagger,
    \end{align}
    for any fixed integer $0\leq k\leq N$, the state
    \begin{equation}
        \hat U\ket{k}_A\ket{N-k}_B, 
    \end{equation}
    is equivalent in the large-$N$ limit to the state
    \begin{align}
        \hat D(\alpha)\ket{k}_A,
    \end{align}
    where $\hat D(\alpha)=e^{\alpha \hat a^\dagger - \alpha^* \hat a}$ is the displacement operator of amplitude $\alpha$.
\end{result}

\begin{derivation}
    To do the derivation, we have on the one hand to compute the expression of a displaced $k$-photon Fock state in the CV picture. To do so, we use the Baker-Campbell-Hausdorff formula (since $\hat a$ and $\hat a^\dagger$ commute with their commutator $[\hat a,\hat a^\dagger]=1$; see Result~\ref{res: BCH}) to rewrite the displacement operator as
    \begin{equation}
        \hat D(\alpha)=e^{\alpha\hat a^\dagger-\alpha^*\hat a}=e^{-\abs{\alpha}^2/2}e^{\alpha\hat a^\dagger}e^{-\alpha^*\hat a}.
    \end{equation}
    Additionally, we recall the action of the ladder operator on the Fock states
    \begin{align}
        \hat a^\dagger\ket{l}_A=\sqrt{l+1}\ket{l+1}_A, && \hat a\ket{l}_A=\sqrt{l}\ket{l-1}_A.
    \end{align}
    Iterating these formulas, we get the action of powers of the ladder operators onto the Fock states
    \begin{align}
        (\hat a^\dagger)^j\ket{l}_A=\frac{\sqrt{(j+l)!}}{\sqrt{l!}}\ket{l+j}_A, && (\hat a)^j\ket{l}_A=\frac{\sqrt{l!}}{\sqrt{(l-j)!}}\ket{l-j}_A,
    \end{align}
    where the second formula is valid for $j\leq l$ and yields zero otherwise. With this, we can compute
    \begin{subequations}
        \begin{align}
            \hat D(\alpha)\ket{k}_A&=e^{-\abs{\alpha}^2/2}e^{\alpha\hat a^\dagger}e^{-\alpha^*\hat a}\ket{k}_A,\\
            &=e^{-\abs{\alpha}^2/2}e^{\alpha\hat a^\dagger}\sum_{l=0}^\infty \frac{(-\alpha^*\hat a)^l}{l!}\ket{k}_A,\\
            &=e^{-\abs{\alpha}^2/2}e^{\alpha\hat a^\dagger}\sum_{l=0}^k \frac{(-\alpha^*)^l}{l!}\frac{\sqrt{k!}}{\sqrt{(k-l)!}}\ket{k-l}_A,\\
            &=e^{-\abs{\alpha}^2/2}e^{\alpha\hat a^\dagger}\sum_{l=0}^k \frac{(-\alpha^*)^{k-l}}{(k-l)!}\frac{\sqrt{k!}}{\sqrt{l!}}\ket{l}_A,\\
            &=e^{-\abs{\alpha}^2/2}e^{\alpha\hat a^\dagger}\sum_{l=0}^k \frac{(-\alpha^*)^{k-l}}{(k-l)!}\frac{\sqrt{k!}}{\sqrt{l!}}\ket{l}_A,\\
            &=e^{-\abs{\alpha}^2/2}\sum_{l=0}^k \frac{(-\alpha^*)^{k-l}}{(k-l)!}\frac{\sqrt{k!}}{\sqrt{l!}}\sum_{j=0}^\infty\frac{\alpha^j}{j!}(\hat a^\dagger)^j\ket{l}_A,\\
            &=e^{-\abs{\alpha}^2/2}\sum_{l=0}^k \frac{(-\alpha^*)^{k-l}}{(k-l)!}\frac{\sqrt{k!}}{\sqrt{l!}}\sum_{j=0}^\infty\frac{\alpha^j}{j!}\frac{\sqrt{(l+j)!}}{\sqrt{l!}}\ket{l+j}_A,\\
            &=e^{-\abs{\alpha}^2/2}\sum_{l=0}^k \sum_{j=0}^\infty\frac{(-\alpha^*)^{k-l}\alpha^j}{(k-l)!j!}\frac{\sqrt{k!}\sqrt{(l+j)!}}{l!}\ket{l+j}_A.
        \end{align}
    \end{subequations}
    On the other hand, we can compute the effect of the mode basis change on the state $\ket{k,N-k}$ which is the SSRC equivalent of a $k$-photon state.
    \begin{subequations}
        \allowdisplaybreaks
        \begin{align}
            \hat U\ket{k}_A\ket{N-k}_B&=\frac{1}{\sqrt{(N-k)!}}\frac{1}{\sqrt{k!}}(\hat U\hat a^\dagger\hat U^\dagger)^k(\hat U\hat b^\dagger\hat U^\dagger)^{N-k}\vac,\\
            &=\frac{1}{\sqrt{(N-k)!}}\frac{1}{\sqrt{k!}}\left(\sqrt{1-\frac{\abs{\alpha}^2}{N}}\hat a^\dagger -\frac{\alpha^*}{\sqrt{N}}\hat b^\dagger\right)^k\notag\\
            &\qquad\times\left(\frac{\alpha}{\sqrt{N}}\hat a^\dagger +\sqrt{1-\frac{\abs{\alpha}^2}{N}}\hat b^\dagger\right)^{N-k}\vac,\\
            &=\frac{1}{\sqrt{(N-k)!}}\frac{1}{\sqrt{k!}}\sum_{l=0}^k\binom{k}{l}\sqrt{1-\frac{\abs{\alpha}^2}{N}}^l\frac{(-\alpha^*)^{k-l}}{\sqrt{N}^{k-l}}(\hat a^\dagger)^l (\hat b^\dagger)^{k-l}\notag\\
            &\qquad\times\sum_{j=0}^{N-k}\binom{N-k}{j}\frac{\alpha^j}{\sqrt{N}^j}\sqrt{1-\frac{\abs{\alpha}^2}{N}}^{N-k-j}(\hat a^\dagger)^j (\hat b^\dagger)^{N-k-j}\vac,\\
            &=\frac{1}{\sqrt{(N-k)!}}\frac{1}{\sqrt{k!}}\sum_{l=0}^k\sum_{j=0}^{N-k}\binom{k}{l}\binom{N-k}{j}\left(1-\frac{\abs{\alpha}^2}{N}\right)^{\frac{N-j+l-k}{2}}\notag\\
            &\qquad\times\frac{\alpha^j(-\alpha^*)^{k-l}}{N^{\frac{j+k-l}{2}}}(\hat a^\dagger)^{l+j} (\hat b^\dagger)^{N-j-l}\vac,\\
            &=\sum_{l=0}^k\sum_{j=0}^{N-k}\binom{k}{l}\binom{N-k}{j}\frac{\sqrt{(j+l)!}}{\sqrt{(N-k)!}}\frac{\sqrt{(N-j-l)!}}{\sqrt{k!}}\notag,\\
            &\qquad\times\left(1-\frac{\abs{\alpha}^2}{N}\right)^{\frac{N-j+l-k}{2}}\frac{\alpha^j(-\alpha^*)^{k-l}}{N^{\frac{j+k-l}{2}}}\ket{j+l}_A\ket{N-j-l}_B.
        \end{align}
    \end{subequations}
    Comparing the equations for $\hat D(\alpha)\ket{k}$ and $\hat U\ket{k}_A\ket{N-k}_B$, we see that for fixed $k$, $l$, $j$ and for $N\to\infty$
    \begin{itemize}
        \item The two sums in $j$ and $l$ match if we take the upper limit $N-k\to\infty$.
        \item The powers of $\alpha$ and $\alpha^*$ are the same.
        \item The limit of $(1-\abs{\alpha}^2/N)^{(N-j+l-k)/2}$ corresponds to the normalisation factor $e^{-\abs{\alpha}^2/2}$.
    \end{itemize}
    It remains to check the combinatorial coefficients
    \begin{subequations}
        \begin{align}
            \binom{k}{l}\binom{N-k}{j}&\frac{\sqrt{(j+l)!}}{\sqrt{(N-k)!}}\frac{\sqrt{(N-j-l)!}}{\sqrt{k!}N^{\frac{j+k-l}{2}}}\notag\\
            &=\frac{k!}{l!(k-l)!}\frac{(N-k)!}{j!(N-k-j)!}\frac{\sqrt{(j+l)!}}{\sqrt{(N-k)!}}\frac{\sqrt{(N-j-l)!}}{\sqrt{k!}N^{\frac{j+k-l}{2}}},\\
            &=\frac{\sqrt{k!}\sqrt{(j+l)!}}{(k-l)!j!l!}\frac{\sqrt{(N-k)!}\sqrt{(N-j-l)!}}{(N-k-j)!N^{\frac{j+k-l}{2}}},
        \end{align}
    \end{subequations}
    where we have put in the first fraction all the $N$-independent terms and in the second fraction, all the $N$-dependent ones. Comparing this with the expansion of $\hat D(\alpha)\ket{k}$, we see that all the $N$-independent terms match. It only remains to show that the $N$-dependent ones converge to $1$.
    \begin{subequations}
        \begin{align}
            &\frac{\sqrt{(N-k)!}\sqrt{(N-j-l)!}}{(N-k-j)!N^{\frac{j+k-l}{2}}}\notag\\
            &\qquad=\sqrt{\frac{(N-k)!}{N!}}\sqrt{\frac{(N-j-l)!}{N!}}\frac{N!}{(N-k-j)!}\frac{1}{N^{\frac{j+k-l}{2}}},\\
            &\qquad=\frac{N(N-1)\cdots(N-k-j+1)}{\sqrt{N(N-1)\cdots(N-k+1)}\sqrt{N(N-1)\cdots(N-j-l+1)}}\frac{1}{N^{\frac{j+k-l}{2}}},\\
            &\qquad\sim\frac{N^{k+j}}{\sqrt{N^k}\sqrt{N^{j+l}}}\frac{1}{N^{\frac{j+k-l}{2}}}=1.        
        \end{align}
    \end{subequations}
    This concludes the proof.
\end{derivation}

\begin{result}[Necessity of non-Gaussian operations for logical single-qubit gates]~\\
    \label{res: SNG required for local gates}
    Let $\mathcal H_N$ be the fixed total photon-number subspace of two bosonic modes, with $N\geq1$, and let $\mathcal C\subseteq\mathcal H_N$ be an arbitrary two-dimensional logical code. For $N>1$, Gaussian SSRC operations alone cannot implement a universal set of logical single-qubit gates on $\mathcal C$. In particular, they cannot implement all rotations $\mathcal R_\alpha(\theta)=e^{-i\theta\sigma_\alpha/2}$ for $\alpha\in\{x,y,z\}$ and arbitrary $\theta$. This holds for every encoding, including encodings prepared by non-Gaussian operations.
    
    For $N=1$, arbitrary logical single-qubit gates can be implemented by passive mode transformations.
\end{result}

\begin{derivation}
    $\blacktriangleright$ \textbf{The group preserving the code.} As we have seen in Sec.~\ref{sec: SSR formalism}, the Schwinger representation demonstrate that, on $\mathcal H_N$, passive two-mode transformations act, up to a global phase, through the irreducible spin-$N/2$ representation
    \begin{equation}
        D_N:SU_2(\C)\longrightarrow U(\mathcal H_N),\qquad \dim\mathcal H_N=N+1.
    \end{equation}
    Define the subgroup of rotations preserving the code by
    \begin{equation}
        K=\left\{U\in SU_2(\C)\,\middle|\,D_N(U)\mathcal C=\mathcal C\right\}.
    \end{equation}
    If $\hat\Pi_{\mathcal C}$ is the orthogonal projector onto $\mathcal C$, this condition is equivalent to $[D_N(U),\hat\Pi_{\mathcal C}]=0$. It is therefore closed, so $K$ is a compact Lie subgroup of $SU_2(\C)$. Restriction to the code, followed by identification of global phases, gives a smooth group homomorphism
    \begin{equation}
        r:K\longrightarrow PU(\mathcal C),\qquad U\longmapsto\left[D_N(U)\big|_{\mathcal C}\right],
    \end{equation}
    where $PU(\mathcal C)=U(\mathcal C)/U(1)$ is the projective unitary group of the logical qubit and has real dimension three.

    \medskip
    \noindent
    $\blacktriangleright$ \textbf{Universality forces invariance under all rotations.} If every logical single-qubit rotation were Gaussian, $r$ would be surjective. The same conclusion holds if Gaussian logical gates form an approximately universal set: their products belong to $r(K)$, and $r(K)$ is compact, hence closed, so it cannot be a proper dense subgroup of $PU(\mathcal C)$. Surjectivity implies
    \begin{equation}
        3=\dim PU(\mathcal C)\leq\dim K\leq\dim SU_2(\C)=3.
    \end{equation}
    Thus $K$ has the same Lie algebra as $SU_2(\C)$. Since $SU_2(\C)$ is connected, this implies $K=SU_2(\C)$. Consequently, $\mathcal C$ is invariant under the entire representation $D_N$. Irreducibility then forces
    \begin{equation}
        \mathcal C=\mathcal H_N,\qquad 2=N+1.
    \end{equation}
    Hence Gaussian single-qubit universality is possible only for $N=1$. Conversely, for $N=1$, $\mathcal H_1$ is two-dimensional and the defining representation of $SU_2(\C)$ implements every logical single-qubit unitary up to phase.

    This conclusion does not exclude individual Gaussian logical gates or continuous rotations about one logical axis. For example, for $\ket{\overline0}=\ket{0,N}_p$ and $\ket{\overline1}=\ket{N,0}_p$, the convention $\hat J_z=(\hat n_B-\hat n_A)/2$ gives
    \begin{align}
        e^{-i\theta\hat J_z/N}\ket{\overline0}&=e^{-i\theta/2}\ket{\overline0},\\
        e^{-i\theta\hat J_z/N}\ket{\overline1}&=e^{i\theta/2}\ket{\overline1}.
    \end{align}
    Thus every logical $z$ rotation is Gaussian for this encoding, while a universal set of logical single-qubit gates requires non-Gaussian operations when $N>1$.
\end{derivation}

\begin{result}[Gaussian SSRC operations cannot be used to implement entangling two-qubit gates]~\\
    \label{res: SNG required for cnot gates}
    Consider two arbitrary two-dimensional logical codes $\mathcal C_A$ and $\mathcal C_B$, each encoded in two bosonic modes with fixed total photon number $N\geq1$. A passive linear transformation $\hat R$ acting on the four modes and preserving $\mathcal C_A\otimes\mathcal C_B$ induces local logical unitaries, possibly composed with an exchange of the two logical subsystems. It therefore cannot implement a deterministic entangling two-qubit gate, such as CNOT.
\end{result}

\begin{derivation}
    $\blacktriangleright$ \textbf{Fixed photon number in each partition.} Write the annihilation operators of the two mode pairs as $\vec{\hat a}=(\hat a_1,\hat a_2)^T$ and $\vec{\hat b}=(\hat b_1,\hat b_2)^T$, and define $\hat N_A=\hat a_1^\dagger\hat a_1+\hat a_2^\dagger\hat a_2$ and $\hat N_B=\hat b_1^\dagger\hat b_1+\hat b_2^\dagger\hat b_2$. Both operators have eigenvalue $N$ on $\mathcal C_A\otimes\mathcal C_B$. Let $U\in U(4)$ describe the passive transformation, with the convention
    \begin{equation}
        \hat R^\dagger\vec{\hat d}\hat R=U\vec{\hat d},\qquad \vec{\hat d}=\begin{pmatrix}\hat a_1\\\hat a_2\\\hat b_1\\\hat b_2\end{pmatrix}.
    \end{equation}
    For the one-particle projector $Q_A=\operatorname{diag}(I_2,0_2)$, write
    \begin{equation}
        U^\dagger Q_AU=\begin{pmatrix}A&C\\C^\dagger&B\end{pmatrix}.
    \end{equation}
    The transformed number observable is then
    \begin{equation}
        \hat R^\dagger\hat N_A\hat R=\vec{\hat a}^\dagger A\vec{\hat a}+\vec{\hat b}^\dagger B\vec{\hat b}+\hat T+\hat T^\dagger,\qquad \hat T=\vec{\hat a}^\dagger C\vec{\hat b}.
    \end{equation}
    Preservation of the code implies, for every $\ket{\psi}\in\mathcal C_A$ and $\ket{\phi}\in\mathcal C_B$,
    \begin{equation}
        \hat R^\dagger\hat N_A\hat R\bigl(\ket{\psi}\otimes\ket{\phi}\bigr)=N\ket{\psi}\otimes\ket{\phi}.
    \end{equation}
    The two diagonal terms preserve the photon-number sector $(N,N)$, whereas $\hat T$ and $\hat T^\dagger$ map it into $(N+1,N-1)$ and $(N-1,N+1)$, respectively. Since these sectors are mutually orthogonal, their contributions cannot cancel. In particular,
    \begin{equation}
        \hat T\bigl(\ket{\psi}\otimes\ket{\phi}\bigr)=0
    \end{equation}
    for every code product state.

    \medskip
    \noindent
    $\blacktriangleright$ \textbf{Vanishing of the off-diagonal block.} Let $\hat\rho_A=\hat\Pi_{\mathcal C_A}/2$ and $\hat\rho_B=\hat\Pi_{\mathcal C_B}/2$ be the maximally mixed states on the local codes. Define the positive semidefinite one-particle density matrices
    \begin{equation}
        (\Gamma_A)_{pj}=\Tr(\hat\rho_A\hat a_j^\dagger\hat a_p),\qquad (\Gamma_B)_{pj}=\Tr(\hat\rho_B\hat b_j^\dagger\hat b_p).
    \end{equation}
    In fact, $\Gamma_B$ is positive definite. If a nonzero vector $v$ belonged to its kernel, the normalized mode $\hat b_v=\sum_p v_p^*\hat b_p/\norm{v}$ would have zero mean occupation in $\hat\rho_B$. Positivity would then imply $\hat b_v\ket{\phi}=0$ for every $\ket{\phi}\in\mathcal C_B$. With two modes and fixed total photon number $N$, all photons would have to occupy the single orthogonal mode. This space is one-dimensional, contradicting $\dim\mathcal C_B=2$.

    Expanding $\hat T^\dagger\hat T$ and using $\hat a_p\hat a_k^\dagger=\delta_{pk}+\hat a_k^\dagger\hat a_p$ gives
    \begin{align}
        0&=\Tr\bigl[(\hat\rho_A\otimes\hat\rho_B)\hat T^\dagger\hat T\bigr]\\
        &=\sum_{p,j,k,l=1}^2 C_{pj}^*C_{kl}\bigl[\delta_{pk}+(\Gamma_A)_{pk}\bigr](\Gamma_B)_{lj}\\
        &=\operatorname{tr}\bigl[(I_2+\Gamma_A)C\Gamma_BC^\dagger\bigr]\\
        &\geq\operatorname{tr}(C\Gamma_BC^\dagger)\geq0,
    \end{align}
    where $\operatorname{tr}$ denotes the trace over the two-dimensional matrices. The first inequality follows because both $\Gamma_A$ and $C\Gamma_BC^\dagger$ are positive semidefinite. Since $\Gamma_B$ is positive definite, equality forces $C=0$.

    \medskip
    \noindent
    $\blacktriangleright$ \textbf{Classification of the remaining blocks.} The matrix $U^\dagger Q_AU$ is a rank-two orthogonal projector. With $C=0$, its diagonal blocks $A$ and $B$ are themselves orthogonal projectors, with
    \begin{equation}
        (\operatorname{rank}A,\operatorname{rank}B)\in\{(2,0),(1,1),(0,2)\}.
    \end{equation}
    Suppose first that both ranks are one. Then $\vec{\hat a}^\dagger A\vec{\hat a}$ and $\vec{\hat b}^\dagger B\vec{\hat b}$ are the occupation operators $\hat n_u$ and $\hat n_v$ of one selected mode in each partition. Every code product state must be an eigenstate of $\hat n_u+\hat n_v$ with eigenvalue $N$. Additivity of the variance on product states gives
    \begin{equation}
        0=\Delta^2_{\ket{\psi}\otimes\ket{\phi}}(\hat n_u+\hat n_v)=\Delta^2_{\ket{\psi}}\hat n_u+\Delta^2_{\ket{\phi}}\hat n_v.
    \end{equation}
    Thus every vector in each code is an eigenvector of the corresponding occupation operator. Fixing $\ket{\phi}$ also fixes the eigenvalue of $\hat n_u$ to $N-\expval{\hat n_v}_{\ket{\phi}}$, independently of $\ket{\psi}$. Hence $\mathcal C_A$ lies in a single eigenspace of $\hat n_u$. Within two modes of fixed total photon number $N$, such an eigenspace is one-dimensional. This contradicts $\dim\mathcal C_A=2$ and excludes the rank pair $(1,1)$.

    The remaining cases are
    \begin{equation}
        U^\dagger Q_AU=Q_A\qquad\text{or}\qquad U^\dagger Q_AU=Q_B,\qquad Q_B=I_4-Q_A.
    \end{equation}
    In the first case, $U$ is block diagonal and preserves the two mode partitions. In the second, $U$ is block anti-diagonal and exchanges them. After second quantization, the corresponding operation therefore has, up to a global phase, one of the forms
    \begin{equation}
        \hat R=\hat R_A\otimes\hat R_B\qquad\text{or}\qquad\hat R=(\hat R_A\otimes\hat R_B)\hat W,
    \end{equation}
    where $\hat R_A$ and $\hat R_B$ are passive two-mode transformations and $\hat W$ exchanges the two mode pairs. Both forms map every product state to a product state. Restricting to the preserved logical code consequently yields local logical unitaries, possibly composed with a logical swap, and cannot generate logical entanglement.
\end{derivation}

\begin{result}[Optimization of SSRC measurement direction]~\\
    \label{res: optimization SSRC measurement direction}
    Fix an integer $k$. Consider the Hamiltonian
    \begin{equation}
        \hat H=\sum_{j=1}^{k+1} \beta_j \hat n_j,
    \end{equation}
    where the coefficients $\beta_j$ are arbitrary real coefficients satisfying
    \begin{align}
        \sum_{j=1}^{k+1} \beta_j=0, && \sum_{j=1}^{k+1} \beta_j^2=1,
    \end{align}
    with $\hat n_j$ the number operator of mode $j$. For any SSRC state over $k+1$ modes
    \begin{equation}
        \ket{\psi}=\sum_{n_1+\cdots + n_{k+1}=N} c_{n_1,\ldots,n_{k+1}}\ket{n_1}\cdots\ket{n_{k+1}},
    \end{equation}
    the variance of $\hat H$ satisfies 
    \begin{equation}
        \Delta^2\hat H \leq k(k+1)\Delta^2 n,
    \end{equation}
    where $\Delta^2 n=\max_{j=1,\dots,k}\Delta^2\hat n_j$ is the maximal variance excluding the reference mode.\footnote{Note that the reference mode has arbitrarily been chosen as the last one. The same statement holds by fixing another mode as the reference.} Additionally, saturation requires
    \begin{align}
        \beta_1=\cdots=\beta_k=\pm\frac{1}{\sqrt{k(k+1)}}, && \beta_{k+1}=\mp\sqrt{\frac{k}{k+1}},
    \end{align}
    meaning 
    \begin{equation}
        \hat H=\pm\frac{2}{\sqrt{k(k+1)}}\sum_{j=1}^k \hat J_z^{(j,k+1)},
    \end{equation}
    where $\hat J_z^{(j,k+1)}=\frac{1}{2}(\hat n_j-\hat n_{k+1})$ is the $z$ component of the angular momentum operator between mode $j$ and the reference mode $k+1$. Furthermore, the optimal states are the perfectly correlated states on the first $k$ modes such as
    \begin{equation}
        \ket{\psi}=\sum_{n=0}^{\lfloor N/k\rfloor} c_n \ket{n}\cdots\ket{n}\ket{N-kn}.
    \end{equation}
\end{result}

\begin{derivation}
    $\blacktriangleright$ {\bf Inequality.} We start by showing the inequality. We observe that for SSRC states we have the equality
    \begin{equation}
        \sum_{j=1}^{k+1} \hat n_j=N,
    \end{equation}
    which means that we can rewrite 
    \begin{equation}
        \hat H=\sum_{j=1}^{k} \beta_j \hat n_j + \beta_{k+1}\left(N-\sum_{j=1}^k \hat n_j\right)=\sum_{j=1}^k (\beta_j-\beta_{k+1})\hat n_j + \beta_{k+1}N.
    \end{equation}
    As $N\beta_{k+1}$ is a constant, the variance can be rewritten as
    \begin{equation}
        \Delta^2\hat H = \Delta^2\left(\sum_{j=1}^k (\beta_j-\beta_{k+1})\hat n_j\right).
    \end{equation}
    Developing in terms of the covariance, and using the Cauchy-Schwarz inequality for the covariance
    \begin{equation}
        \abs{\operatorname{Cov}(\hat n_j,\hat n_{j'})}\leq \sqrt{\Delta^2\hat n_j \Delta^2\hat n_{j'}}\leq \Delta^2 n,
    \end{equation}
    by setting $\Delta^2 n=\max_{j=1,\dots,k}\Delta^2\hat n_j$, we get
    \begin{subequations}\label{eq: chain inequ collective SSRC}
        \begin{align}
            \Delta^2\hat H &=\sum_{j,j'=1}^k (\beta_j-\beta_{k+1})(\beta_{j'}-\beta_{k+1}) \operatorname{Cov}(\hat n_j,\hat n_{j'}),\\
            &\leq \sum_{j,j'=1}^k \abs{\beta_j-\beta_{k+1}}\abs{\beta_{j'}-\beta_{k+1}} \Delta^2 n,\\
            &= \left(\sum_{j=1}^k \abs{\beta_j-\beta_{k+1}}\right)^2 \Delta^2 n,\\
            &\leq k\sum_{j=1}^k (\beta_j-\beta_{k+1})^2\Delta^2 n,\\
            &=k\left[1+(k+1)\beta_{k+1}^2\right]\Delta^2 n,
        \end{align}
    \end{subequations}
    where the second inequality is obtained by using the Cauchy-Schwarz inequality $(\sum_{j=1}^k x_j\times 1)^2\leq k\sum_{j=1}^kx_j^2$ and the last line is obtained by using the constraints on the $\beta_j$ coefficients. It remains to find the largest possible value for $\beta_{k+1}$ under the constraints
    \begin{align}
        \sum_{j=1}^{k+1} \beta_j=0, && \sum_{j=1}^{k+1} \beta_j^2=1.
    \end{align}
    To solve this properly, we use Lagrange multipliers, and seek to optimize
    \begin{equation}
        \mathcal L(\beta_1,\ldots,\beta_{k+1},\lambda,\mu) = \beta_{k+1} - \lambda \sum_{j=1}^{k+1} \beta_j - \mu\left(\sum_{j=1}^{k+1} \beta_j^2-1\right).
    \end{equation}
    Derivatives with respect to $\beta_j$ for $j=1,\dots,k$ give
    \begin{equation}
        \frac{\partial \mathcal L}{\partial \beta_j} = -\lambda - 2\mu \beta_j =0,
    \end{equation}
    Meaning that $\beta_1=\cdots=\beta_k=-\lambda/(2\mu)$. The derivative with respect to $\beta_{k+1}$ gives
    \begin{equation}
        \frac{\partial \mathcal L}{\partial \beta_{k+1}} = 1 - \lambda - 2\mu \beta_{k+1} =0,
    \end{equation}
    yielding  $\beta_{k+1}=\frac{1-\lambda}{2\mu}$. Finally, the derivatives with respect to $\lambda$ and $\mu$ recover the constraints on the coefficients $\beta_j$. We thus have
    \begin{equation}
        \sum_{j=1}^{k+1} \beta_j=-\frac{k\lambda}{2\mu}+\frac{1-\lambda}{2\mu}=\frac{1-(k+1)\lambda}{2\mu}=0,
    \end{equation}
    giving $\lambda=\frac{1}{k+1}$, and
    \begin{equation}
        \sum_{j=1}^{k+1} \beta_j^2 = k\frac{\lambda^2}{4\mu^2}+\frac{(1-\lambda)^2}{4\mu^2}=\frac{k}{(k+1)^2}\frac{1}{4\mu^2}+\frac{k^2}{(k+1)^2}\frac{1}{4\mu^2}=\frac{k}{k+1}\frac{1}{4\mu^2}=1,
    \end{equation}
    yielding $\mu=\pm\frac{\sqrt{k}}{2\sqrt{k+1}}$. We thus have
    \begin{align}
        \beta_1=\cdots=\beta_k&=\pm\frac{1}{\sqrt{k(k+1)}}, & \beta_{k+1}&=\mp\sqrt{\frac{k}{k+1}}.
    \end{align}
    Inserting this in the expression for $\Delta^2\hat H$ indeed gives
    \begin{equation}
        \Delta^2\hat H \leq k(k+1)\Delta^2 n.
    \end{equation}

    \medskip

    \noindent
    $\blacktriangleright$ {\bf Saturation of the inequality.} For the case of equality, notice that the chain of inequalities in Eqs.~\eqref{eq: chain inequ collective SSRC} requires for all $j$ and $j'$ less than $(k+1)$
    \begin{equation}
        \operatorname{Cov}(\hat n_j,\hat n_{j'})=\Delta^2 n,
    \end{equation}
    which by Cauchy-Schwarz implies that the variables $\hat n_j$ must be perfectly correlated. As such optimal states must take the form
    \begin{equation}
        \ket{\psi}=\sum_{n=0}^{\lfloor (N-n_1-\cdots-n_k)/k\rfloor} c_n \ket{n+n_1}\cdots\ket{n+n_k}\ket{N-kn-(n_1+\cdots+n_k)},
    \end{equation}
    for non-negative integer offsets $n_1,\ldots,n_k$ with $n_1+\cdots+n_k\leq N$. The particular case $n_1=\cdots=n_k=0$ gives the perfectly correlated states shown in the statement of the result. Furthermore, clearly optimality requires $\beta_{k+1}=\mp\sqrt{k/(k+1)}$, which means that, due to the above derivation we indeed also have
    \begin{equation}
        \beta_j-\beta_{k+1}=\pm\sqrt{\frac{k+1}{k}},
    \end{equation}
    yielding the expression for $\hat H$
    \begin{equation}
        \hat H=\pm\frac{2}{\sqrt{k(k+1)}}\sum_{j=1}^k \hat J_z^{(j,k+1)}.
    \end{equation}
\end{derivation}

\fi

\pagestyle{backmatterstyle}
\small
\bibliographystyle{mystyle}
\bibliography{refs}

\end{document}